%% file: lheccdr.tex
\pdfoutput=1
\documentclass{LHeC}
\usepackage{cite}
\usepackage[]{lineno}
\usepackage{float}

\usepackage{machinedefs}

\newcommand{\as}{$\alpha_s\,$}
 
\newcommand{\bs}{\overline{s}}
\newcommand{\bc}{\overline{c}}
\newcommand{\bu}{\overline{u}}
\newcommand{\bd}{\overline{d}}

\newcommand{\bU}{\overline{U}}
\newcommand{\bD}{\overline{D}}

\newcommand{\pdff}{$\partial F_{2} / \partial \ln Q^{2}\,$ }
\def\gsim{\mathrel{\rlap{\lower4pt\hbox{\hskip1pt$\sim$}}}}
\def\MS{\hbox{$\overline{\rm MS}$}}
\def\QMS{Q$_0$\MS}
\newcommand{\xpom}{x_{\mathbb P}}

\newcommand{\sqrtsgN}{\sqrt{s_{_{\gamma N}}}}
\providecommand{\jpsi}{J/\psi}
\providecommand{\ups}{\Upsilon}
\providecommand{\dNdeta}{dN_{ch}/d\eta|_{\eta=0}}

\graphicspath{{introduction/figures/}{physics/figures/} {machine/figures/} {detector/figures/}}

\begin{document}
%
%
\noindent
CERN-OPEN-2012-015\\
LHeC-Note-2012-002 GEN \\
Geneva, June 13, 2012 \\

\begin{figure}[h]
\vspace{-2.5cm}
\hspace{12.cm}
\includegraphics[clip=,width=.15\textwidth]{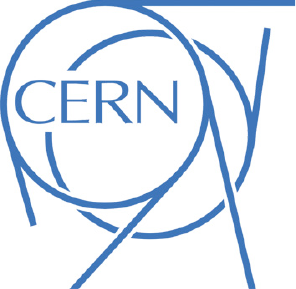}
\end{figure}
\begin{figure}[h]
\vspace{-1.3cm}
\hspace{3.6cm}
\includegraphics[clip=,width=0.45\textwidth]{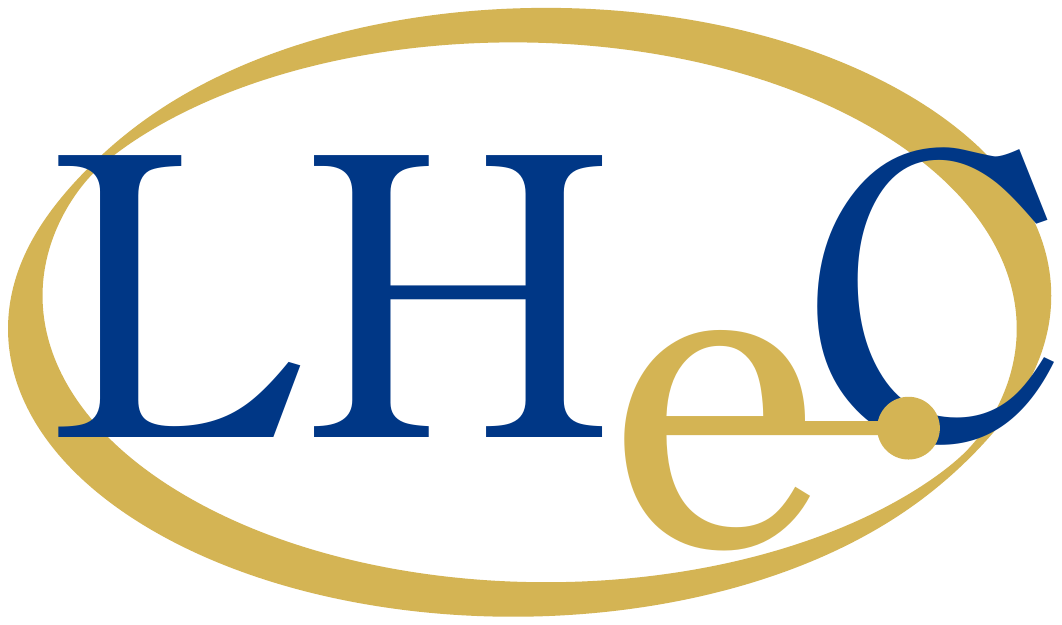}
\end{figure}
\begin{center}
\vspace{-.5cm}
\begin{LARGE}
\bf{A Large Hadron Electron Collider  
at CERN} \\
\end{LARGE}
\vspace{0.5cm}
\begin{LARGE}
 Report on the Physics and Design \\
Concepts for Machine and Detector \\
 \end{LARGE}
 \vspace{1cm}
\begin{Large}
\bf{LHeC Study Group} \\ 
\end{Large}
%
\vspace{0.1cm}
\begin{figure}[hb]
\vspace{0.1cm}
\hspace{2.cm}
\includegraphics[clip=,width=0.7\textwidth]{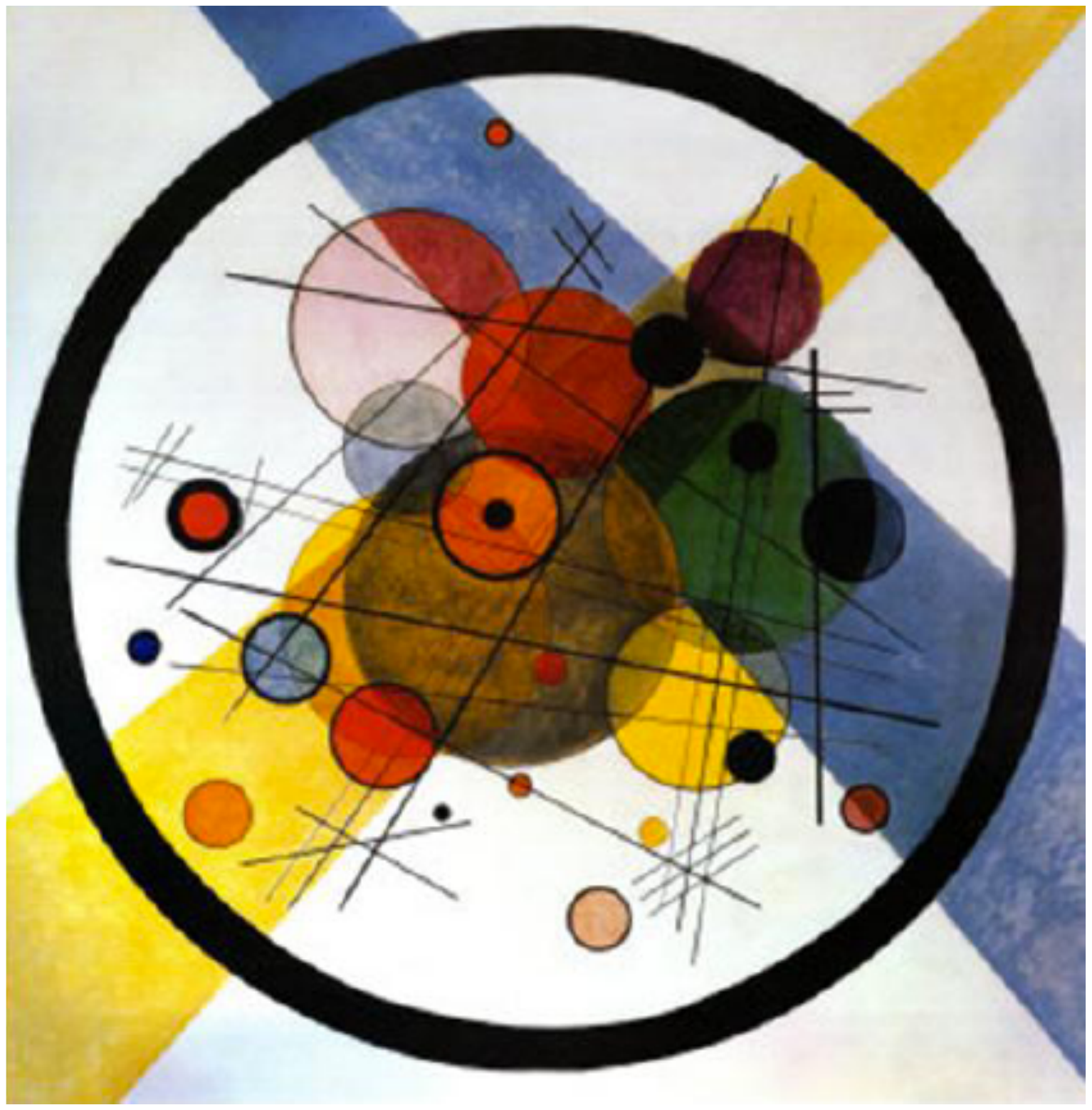}
\end{figure}
%
\vspace{0.3cm}
\end{center}
\newpage
\noindent
The Kandinsky painting, ``Circles in a circle" (1923), is taken from a talk on gluon saturation and $5D$ black hole duality as presented at the first CERN-ECFA-NuPECC Workshop on the
LHeC held at  Divonne near to CERN in September 2008~\cite{lhecvera}. We thank  
the Philadelphia Museum of Art, USA, for the 
permission to reproduce it here.

\noindent
Kreise im Kreis, 923 (R.702), Vassily Kandinsky

\noindent
Oil on canvas, 38 7/8 x 37 5/8 inches (98.7 x 95.6 cm) 

\noindent
Philadelphia Museum
of Art: The Louise and Walter Arensberg Collection, 1950

\noindent
\copyright ADAGP, Paris and DACS, London 2012.

\newpage
\vspace{5cm}
\begin{abstract}
\noindent The physics programme and the design are described
of a new  collider for particle and nuclear  physics,
the Large Hadron Electron Collider (LHeC),
in which a newly built electron beam of $60$\,GeV, 
to possibly $140$\,GeV, energy collides with the intense 
hadron beams of the LHC.  Compared to the first 
$ep$ collider, HERA,
the kinematic range covered  is extended 
by a factor of twenty in the
negative four-momentum squared, 
$Q^2$, and in the inverse Bjorken $x$,
while with the design luminosity of
$10^{33}$\,cm$^{-2}$s$^{-1}$ the LHeC
is projected to exceed the
integrated HERA luminosity by two orders of magnitude. 
The physics programme is devoted to an exploration
of the energy frontier, complementing the LHC and
its discovery potential for physics beyond the Standard Model
with high precision deep inelastic scattering
measurements.  These are designed to
investigate a variety of fundamental questions in strong
and electroweak interactions. The LHeC thus 
continues the path of deep inelastic
scattering (DIS) into unknown areas of physics
and kinematics. The physics programme also includes 
electron-deuteron and electron-ion
scattering in a $(Q^2,~1/x)$
range extended by four orders of magnitude
as compared to previous lepton-nucleus DIS experiments
for novel investigations of neutron's and nuclear structure, 
the initial conditions  of  Quark-Gluon Plasma formation
and further quantum chromodynamic phenomena.
The LHeC may be realised either as a ring-ring or as a linac-ring
collider.  Optics and beam dynamics
studies are presented for both versions, along with technical design
considerations on the interaction region, magnets including
new dipole prototypes,
cryogenics, RF, and further components.
A design study is also presented of a detector suitable
to perform high precision DIS measurements in a wide
range of acceptance using state-of-the art detector
technology, which is modular and of limited size enabling
its fast installation. The detector includes
tagging devices for electron, photon, proton
and neutron detection near to the beam pipe.
Civil engineering and installation studies are
presented for the accelerator and the detector.
The LHeC can be built within a decade and thus
be operated while the LHC runs in its high-luminosity phase.
It so represents a major opportunity for progress in particle physics
exploiting   the investment made in the LHC.

%
\begin{figure}[h]
\vspace{0.1cm}
\hspace{4.2cm}
\includegraphics[clip=,width=0.5\textwidth]{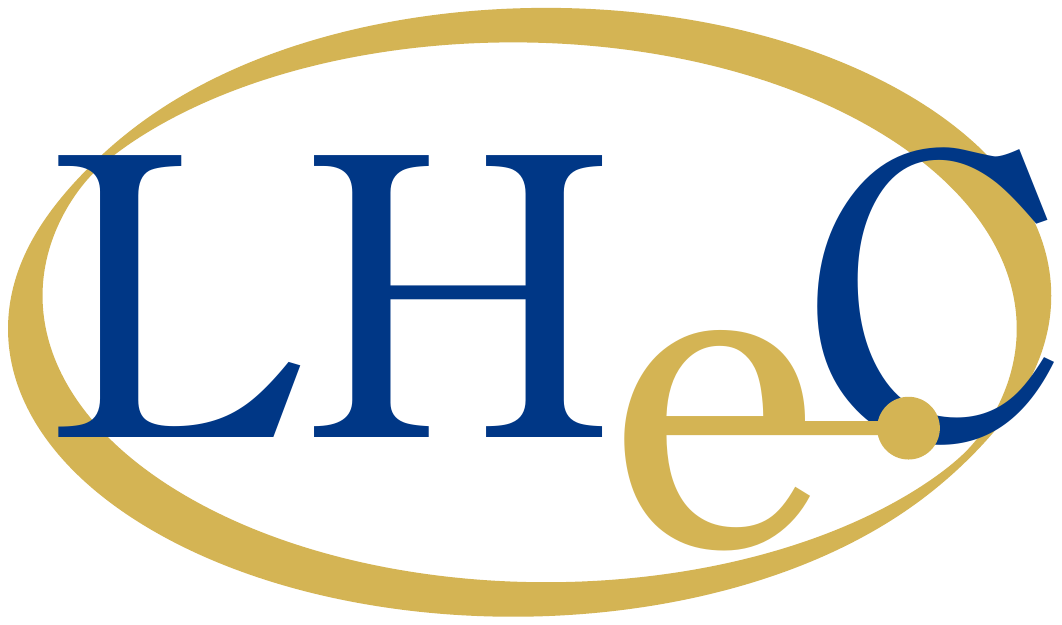}
\end{figure}
\end{abstract}
%
\pagestyle{plain}
\newpage
\section*{LHeC Study Group}
\input{summary/lhec_authors}

\newpage
\chapter*{Foreword}
\input{introduction/sbfw}
\chapter*{Preface}
\input{introduction/preface}
%
\newpage
\tableofcontents
\newpage
\input{introduction/introduction}

\part{Introduction}
%
%
%
\chapter{Lepton-Hadron Scattering}
\input{introduction/pqcd}

%
\chapter{Design Considerations}
\input{introduction/design}
\newpage
\part{Physics}
\chapter{Precision QCD and Electroweak Physics}
\label{chapter:qce}
\input{physics/qce}

%
\chapter{Physics at High Parton Densities}
\input{physics/CDR_lowx}
%
\chapter{New Physics at High Energy}
\input{physics/bsm}

%
\part{Accelerator}
\input{machine/ring}

%
\input{machine/linac}

%
\input{machine/systemdesign}

%
\input{machine/civilengineering}
%
\input{machine/planning}

\newpage
\part{Detector}

\chapter{Detector Requirements}

\input{detector/preamble}
\input{detector/detr}

\chapter{Central Detector}
\label{LHEC:Detector:Maindet}

\input{detector/det}
%
\chapter{Forward and Backward Detectors}
\label{LHEC:Detector:ForwardBackward}
\input{detector/fwdbkwd}

\chapter{Detector Assembly and Integration}
\label{LHEC:Detector:Assembly+Integration}
\input{detector/detector_assembly_integration}

%
\newpage
\part{Conclusion}
\chapter{Executive Summary}
\input{introduction/exsummary}

\chapter{Committees, Convenors and Referees}
%
\input{summary/appendix2}

\chapter*{Acknowledgement}
\input{summary/acknowledgements}
\newpage
\footnotesize
\bibliographystyle{atlasstylem}
\footnotesize
\bibliography{bibliography}
%
%

%
\end{document}

%% file: summary/lhec_authors.tex
J.L.Abelleira Fernandez$^{16,23}$, 
C.Adolphsen$^{57}$, 
A.N.Akay$^{03}$, 
H.Aksakal$^{39}$, 
J.L.Albacete$^{52}$, 
S.Alekhin$^{17,54}$, 
P.Allport$^{24}$, 
V.Andreev$^{34}$, 
R.B.Appleby$^{14,30}$,
E.Arikan$^{39}$
% ...

\bigskip{\it\noindent
$^{01}$ Ankara University, Turkey \\
$^{02}$ SANAEM Ankara, Turkey \\
$^{03}$ TOBB University of Economics and Technology, Ankara, Turkey\\
$^{04}$ LAPP, Annecy, France\\
$^{05}$ University of Antwerp, Belgium\\
$^{06}$ University of Birmingham, UK\\
$^{07}$ INFN Bologna, Italy\\
$^{08}$ IIHE, Universit\'e Libre de Bruxelles, Belgium, supported by the FNRS \\
$^{09}$ University of Buenos Aires, Argentina \\
$^{10}$ INFN Gruppo Collegato di Cosenza and Universita della Calabria, Italy \\
$^{11}$ Massachusetts Institute of Technology, Cambridge, USA\\
$^{12}$ Charkow National University, Ukraine\\
$^{13}$ University of Copenhagen, Denmark \\
$^{14}$ Cockcroft Institute, Daresbury, UK\\
$^{15}$ Rutherford Appleton Laboratory, Didcot, UK \\
$^{16}$ CERN, Geneva, Switzerland\\
$^{17}$ DESY, Hamburg and Zeuthen, Germany\\
$^{18}$ University of Hamburg, Germany\\
$^{19}$ Hampton University, USA \\
$^{20}$ University of California, Irvine, USA \\
$^{21}$ University of Jyv\"askyl\"a, Finland\\
$^{22}$ Kastamonu University, Turkey \\
$^{23}$ EPFL, Lausanne, Switzerland\\
$^{24}$ University of Liverpool, UK\\
$^{25}$ University of California, Los Angeles, USA\\
$^{26}$ Lund University, Sweden\\
$^{27}$ University of Wisconsin-Madison, USA\\
$^{28}$ Universidad Aut\'onoma de Madrid, Spain\\
$^{29}$ University of Mainz, Germany\\
$^{30}$ The University of Manchester, UK \\
$^{31}$ INFN Milano, Italy\\
$^{32}$ University of Milano, Italy \\
$^{33}$ University of Montr\'eal, Canada\\
$^{34}$ LPI Moscow, Russia\\
$^{35}$ Tata Institute, Mumbai, India\\
$^{36}$ Jefferson Lab, Newport News, VA 23606, USA \\
$^{37}$ Brookhaven National Laboratory, New York, USA\\
$^{38}$ Columbia University, New York, USA\\
$^{39}$ Nigde University, Turkey\\
$^{40}$ Budker Institute of Nuclear Physics SB RAS, Novosibirsk, 630090 Russia\\
$^{41}$ Orissa University, India\\
$^{42}$ LAL, Orsay, France\\
$^{43}$ Laboratoire de Physique Th\'eorique, Universit\'e Paris XI, Orsay, France \\
$^{44}$ University of Oxford, UK\\
$^{45}$ CPHT, \'Ecole Polytechnique, CNRS, 91128 Palaiseau, France \\
$^{46}$ UPMC University of Paris 06, Facult\'e de Physique, Paris, France \\
$^{47}$ LPNHE University of Paris 06 and 07, CNRS/IN2P3, 75252 Paris, France \\
$^{48}$ Pennsylvania State University, USA\\
$^{49}$ University of Pisa, Italy\\
$^{50}$ University of Pittsburgh, USA\\
$^{51}$ Charles University, Praha, Czech Republic \\
$^{52}$ IPhT Saclay, France\\
$^{53}$ University of Santiago de Compostela, Spain\\
$^{54}$ Serpukhov Institute, Russia\\
$^{55}$ University of Siegen, Germany \\
$^{56}$ University of Southampton, UK \\
$^{57}$ SLAC National Accelerator Laboratory, Stanford, USA\\
$^{58}$ Tokyo Institute of Technology, Japan\\
$^{59}$ University of Torino and INFN Torino, Italy\\
$^{60}$ NTNU, Trondheim, Norway\\
$^{61}$ KEK, Tsukuba, Japan\\
$^{62}$ Uludag University, Turkey\\
$^{63}$ Uppsala University, Sweden  \\
$^{64}$ TRIUMF, Vancouver, Canada \\
$^{65}$ Paul Scherrer Institute, Villigen, Switzerland\\
$^{66}$ National Center for Nuclear Research (NCBJ), Warsaw, Poland \\
$^{67}$ University of Warsaw, Poland\\
$^{68}$ ETH Zurich, Switzerland\\
$^{69}$ University of Zurich, Switzerland
}

\bigskip{\it\noindent
$^a$ supported by European Research Council grant HotLHC ERC-2011-StG-279579 and\\
MiCinn of Spain grants FPA2008-01177, FPA2009-06867-E and Consolider-Ingenio 2010 CPAN CSD2007-00042,
Xunta de Galicia grant PGIDIT10PXIB206017PR, and FEDER.\\
$^b$ supported by the U.S. Department of Energy,contract DE--AC02--76SF00515.\\
$^c$ supported by the Academy of Finland, project no. 141555.\\
$^d$ supported by the Sloan Foundation,
DOE OJI grant No. DE - SC0002145 and \\
Polish NCN grant DEC-2011/01/B/ST2/03915.\\
$^e$ supported by the Turkish Atomic Energy Authority (TAEK).\\
$^f$ supported by the P2IO consortium.\\
}

%% file: introduction/sbfw.tex
The traditions of CERN on deep inelastic lepton-hadron scattering
date back to the discovery of weak neutral currents by the Gargamelle
collaboration and, subsequently, the exploration of the valence and sea-quark
contents of the nucleon, tests of Quantum Chromodynamics and
electroweak phenomena and the observation of unexpected effects
in the behaviour of quarks in protons and nuclei, made in a series
of neutrino and muon scattering experiments. Following HERA,
the first electron-proton collider built at DESY, with the LHeC
there is an opportunity for energy frontier deep inelastic scattering
to return to CERN in order to enrich the physics which has been
made accessible by the Large Hadron Collider. Using a novel
high energy electron beam scattered off LHC protons and also ions,
the LHeC would represent the cleanest high resolution microscope
in the world, based on new principles which deserve to be developed.
The design report, available herewith, covering concepts of the accelerator
and detector, together with an evaluation of the physics potential,
had been initiated by the CERN Science Policy Committee and been
worked out by an international study group, supported by CERN,
the European Committee for Future Accelerators, ECFA, and
the Nuclear Physics European Collaboration Committee, NuPECC.

The report describes a challenging new opportunity for European and
global particle physics. Looking forward to the further development of
the LHeC project, CERN with international partners is now evaluating
ways of cooperation towards technical designs of the highest energy
electron linac, with power recovery, and of a new detector which
would enable ultra-precise, large acceptance deep inelastic scattering
measurements. By the time the LHC will provide its first luminous
results at the design beam energy, in around 2015, a possible upgrade
of the LHC as is proposed here may advance. For now, CERN has to thank
the scientists and engineers involved, the members of the Scientific
Advisory Committee, of ECFA and NuPECC and especially the many expert
referees which in the final phase of this study helped in scrutinising
the LHeC design.
\\
\vspace{0.4cm}
\\
Sergio Bertolucci  (Director of Research and Computing of CERN)

\newpage

%% file: introduction/preface.tex
%
%
Preparations for new, big machines take time. The
idea of an electron-proton ($ep$) collider in the LEP-LHC
tunnel was  discussed  as early as 1984~\cite{Altarelli:1984rn},
 at the first LHC workshop at Lausanne. This was the
time when the first ever built $ep$ collider, HERA,
was approved by the German government.
HERA was a machine of about $30$\,GeV electron beam energy and
nearly $1$\,TeV proton beam energy, a combination of
a warm dipole  electron ring with a superconducting
dipole proton ring, in a $6$\,km circumference tunnel.
The machine started operation $8$ years after its approval.
It reached luminosities of  $10^{31}$\,cm$^{-2}$s$^{-1}$
in its first phase of operation which were increased
by about a factor of $4$ in the subsequent, upgraded
configuration. HERA never attempted to collide
electrons with deuterons nor with ions.

The realisation of HERA at DESY had followed a number of attempts to
realise $ep$ interactions in collider mode, mainly driven
by the unforgettable Bjoern Wiik: since the late 1960s,
he and his colleagues had considered such machines and proposed
to probe the proton's structure more deeply with an $ep$ collider
at DORIS\,\cite{Febel:1973ej}, later at
PETRA (PROPER)\,\cite{proper} and subsequently
at the SPS at CERN
(CHEEP)\,\cite{Ellis:1978ms}.  Further $ep$ collider studies were made
for  PEP\,\cite{pepep}, TRISTAN\,\cite{tristanep} and also
the Tevatron (CHEER)\,\cite{Blackmore:1981ch}.

In 1990, at a workshop at Aachen, the combination of
LEP with the LHC was discussed, with
studies~\cite{Verdiern,Barteln,Ruckl:1990eh}
on the luminosity, interaction region, a detector
and the physics as seen with the knowledge of that time,
 before HERA. Following a request of
the CERN Science Policy Committee (SPC), a brief
study of the ring-ring $ep$ collider in the LEP tunnel
was performed~\cite{ekeil} leading to an estimated luminosity
of about $10^{32}$\,cm$^{-2}$s$^{-1}$.

At the end of the eighties it had been anticipated that there
was a possible end to the  increase of the energy of $ep$
colliders in the ring-ring configuration,
because of the synchrotron radiation losses of an
electron ring accelerator. The classic SLAC
fixed target $ep$ experiment had already  used a $2$ mile
linac. For $ep$ linac-ring collider configurations,  two
design sketches considering
electron beam energies up to a few hundred GeV were published,
in 1988~\cite{pgw} and in 1990~\cite{Tigner:1991wt}.
As part of the TESLA linear collider proposal,
an option (THERA) was studied~\cite{Katz:2001kz}
to collide electrons of a few hundred GeV energy
with protons and  ions from HERA.
Later, in 2003, the possibility was
evaluated to combine LHC protons with CLIC
electrons~\cite{Schulte:2004uj}. It was yet realised, that the
bunch structures of the LHC and CLIC were not compliant
with the need for high luminosities.

In September 2007,  the SPC again asked
whether one could realise an $ep$ collider at CERN.
Some of us had written a paper~\cite{Dainton:2006wd}
in the year before, that had shown in detail, for the first time,
that a luminosity of $10^{33}$\,cm$^{-2}$s$^{-1}$
was achievable. This appeared possible in a ring-ring
configuration  based on the ``ultimate'' LHC beam, with
$1.7 \cdot 10^{11}$ protons in bunches $25$\,ns apart.
Thanks to the small beam-beam tune-shift, it was found to be feasible
to simultaneously operate $pp$ in the LHC and
$ep$ in the new machine, which in 2005 was termed the
Large Hadron Electron Collider
(LHeC)~\cite{Klein:2005wj}. Thus it appeared possible to realise
an $ep$ collider that was complementary to the LHC,
just as HERA was to the Tevatron.  The  integrated
luminosity was projected to be O(100)\,fb$^{-1}$, 
a factor of a hundred more
than HERA had collected over its lifetime of 15 years.

It was clear that with a centre-of-mass energy of about
$\sqrt{s} \simeq 1.5$\,TeV an exciting programme of
deep inelastic scattering (DIS) measurements  at the energy-frontier
was in reach.  This would comprise searches and analyses for physics
beyond the Standard Model, novel measurements in QCD and
electroweak physics to unprecedented precision,
as well as DIS physics at such low Bjorken $x$,
that all the known laws of parton and gluon interactions would
have to be modified to account for non-linear parton interaction effects.
It had also been realised that the kinematic
region, in terms of negative four-momentum-transfer squared, $Q^2$,
and $1/x$, accessed in lepton-nucleus interactions could be
extended by $4$ orders of magnitude using the ion beams
of the LHC. A salient theme of the LHeC therefore is the
precise mapping of the gluon field, over six orders of magnitude in
Bjorken $x$, in protons, neutrons and nuclei, with unprecedented
sensitivity.

In the autumn of 2007, (r)ECFA and CERN invited us to work out  the
LHeC concept to a degree, which would allow one to understand its physics
programme, evaluate the accelerator options and their technical
realisation.  The detector design should be affordable
and capable of realising a high precision, large acceptance
experimental programme of deep inelastic scattering at
the energy frontier. The electron beam energy range was set
to be between about $50-150$\,GeV. The wall plug power consumed for the
electron beam was limited to $100$\,MW.

For the installation of the LHC it had been decided to remove
LEP from the tunnel and to  re-use the injector chain.
To realise an $ep$ collider based
on the LHC, a new electron accelerator has to be built.
The following report details two solutions for
the chosen default electron beam energy of $E_e=60$\,GeV.
One option is to  build and install a new ring, with
modern magnet technology, on top of the LHC,
using a new $10$\,GeV injector. Alternatively, one can build
a ``linac'', actually two $10$\,GeV superconducting linacs in a
racetrack configuration.  By employing energy recovery
techniques, this configuration could provide the equivalent of about $1$\,GW available power
and reach $10^{33}$\,cm$^{-2}$s$^{-1}$ luminosity.
The LHeC linac would be of about the same length as the
one used for the discovery of quarks at
SLAC~\cite{Breidenbach:1969kd,Bloom:1969kc},
but capable of probing parton interactions with a $Q^2$ exceeding that of the 1969 machine by a factor of nearly $10^5$.

It was agreed early on to devote a few years to the report,
also because none  of the people involved could work
anything near to full time for this endeavour.  Three
workshops were held in 2008-2010, that annually assembled
about a hundred experts on theory, experiment and accelerator
to develop the LHeC design concepts.  The project was presented
annually to ECFA and in 2008 to ICFA, see~\cite{lhecweb}. In view of the unique
electron-ion scattering programme of the LHeC, the design effort
became also supported by NuPECC, and the LHeC is now
part of the NuPECC roadmap for European nuclear physics
as released in 2010~\cite{nupecc}. Following an intermediate
report to the Science Policy Committee of CERN, in July 2010,
the SPC considered the LHeC  ``an option for a future project
at CERN''.

In August 2011, a first complete draft of this conceptual design report was
handed to more than twenty experts on various aspects of the physics and
technology of the LHeC, which CERN had invited to referee the project and
scrutinise its motivation and its design. The  report has been
completed following often close interactions with the referees and
due consideration of their observations.

The LHeC by its nature is an upgrade of the LHC. It substantially enriches
the physics harvest related to the gigantic investment in
the LHC.   Whatever the outcome of the searches at the LHC
for physics beyond the Standard Model turns out to be, an $ep$ collider
operating at the energy frontier is guaranteed to deepen
the understanding of TeV scale  physics
 and thus will support the development of the
theory of elementary particles and their interactions.

The LHeC needs the
LHC proton and ion beams to be operational and so the
design is made for synchronous $pp$ and $ep$ operation,
as well as $AA$ and $eA$, including deuterons. Should
the LHC eventually be upgraded to even higher
beam energy, beyond $7$\,TeV per beam~\cite{myersHE},
or a new proton collider be built,
it would open an even higher energy reach for $ep$ also.
There certainly is a future for deep inelastic scattering at the energy frontier.
It is herewith envisaged to begin with the LS3 shutdown of the LHC,
in the early twenties, likely leading
into further decades. As Frank Wilczek put
it, ``one of the joys of our subject is the continuing of our
culture that bridges continents and generations''~\cite{frankw09}.

Our science is driven by curiosity, by theoretical
expectations, sometimes too great, but also by experiment
and technology, and the authors of this study therefore hope that the LHeC
may be given the chance to contribute to the common
efforts of our community for a deeper understanding of nature.
\\
\vspace{0.1cm}
\\
Max Klein (Chair of the LHeC Steering Committee)

%% file: introduction/introduction.tex
%
%
\noindent
The present document is a detailed presentation of the
physics, the accelerator options and a detector
design comprising the LHeC project. It has been
developed under the auspices and with support
of CERN, ECFA and NuPECC, between 2008 and now.
The paper is organised as follows:

$\bf{Part~I}$, the introduction, summarises
cornerstones of deep inelastic scattering and
the main considerations for the design of the LHeC 
are summarised. The emphasis is
on adding a $60$\,GeV energy electron beam to
the existing proton and ion beams of the LHC,
in a manner which foresees the simultaneous
$ep$ and $pp$ operation for the realisation
of a luminous DIS programme while minimising the
interference with the LHC.

$\bf{Part~II}$  presents selected subjects, 
with related simulation studies and theoretical 
considerations, in order to sketch the physics programme
of the LHeC. These subjects are grouped into three
main, though related areas: 
high precision QCD and electroweak physics,
the physics of high parton densities
at low Bjorken x, in protons and in nuclei, and finally 
the potential for searches for phenomena beyond
the standard model and its relation to the LHC.
It  has rarely been possible, fortunately,
to accurately predict nor to
fully simulate the physics of a new
machine at much enlarged energies. Equally, the subjects
here presented are not supposed to cover the complete
field as it is known today. However, for a new laboratory
of particle physics as the LHeC represents, 
a broad view must be taken to what it most likely comprises.

$\bf{Part~III}$ is devoted to the accelerator design, 
with studies presenting the ring-ring and linac-ring 
concepts, optics etc. and in a third section the
various technical systems which often are
common to both accelerator options. The emphasis here is 
on an  understanding of the main challenges and
characteristics of both options and not on
discussing their relative merits. The accelerator
part is concluded with separate sections on the
civil engineering and a tentative time schedule
for the realisation of the LHeC within about the next
ten years.

$\bf{Part~IV}$ presents the design
considerations for a detector with its challenging
central part and further systems to tag forward
nucleons and backward scattered electrons and 
photons, including a study for a high precision
measurement of the lepton beam polarisation. 
The salient feature of the detector baseline design
is its silicon tracker surrounded by an electromagnetic
liquid argon calorimeter inside a superconducting coil
which uses a tile hadron calorimeter for the flux return.
The detector part concludes with a first study
of the installation of the apparatus, with premounting
on the surface, lowering and integration underground.

$\bf{Part~V}$ contains a summary of the main 
results and considerations of this report with
the intention of providing a brief overview
on the LHeC design and possible prospect.
This design study has been organised jointly by a steering 
group and convenors for the various physics,
accelerator and detector parts.
It was accompanied by a scientific advisory 
committee. A first draft was handed to more than 20 referees,
which were nominated by the CERN directorate for a detailed
evaluation of the design and a corresponding update.
The composition of these groups is listed below the summary 
of the paper.
Various members of the
advisory committee have made direct scientific
contributions to the LHeC design as presented here.
They therefore also appear among the authors of
this study which are representing a group of
nearly $200$ physicists and engineers from $70$
institutes.

While this report is being published, the
first luminous results from the LHC have become
available, and HERA publishes its final papers. 
The interest in a TeV energy scale DIS collider
of high luminosity has grown. The LHeC
development will continue with a view
to come to a technical design within a few years.

%% file: introduction/pqcd.tex
%
%
\section{Development and contributions}
It is almost exactly 100 years since the birth of the scattering
experiment as a means of revealing the structure of matter.
Geiger and Marsden's experiment \cite{geiger}
and its interpretation by Rutherford \cite{rutherf}
set the scene for a century of ever-deeper and more precise
resolution of the constituents of the atom, the nucleus and 
the nucleon. Lepton-hadron scattering
has played a crucial role in this exploration
over the past 55 years. The finite radius of the proton 
of about $1$\,fm
was first established 
through elastic electron-proton scattering 
experiments \cite{Hofstadter:1955ae}. Later, 
through deep inelastic 
electron proton scattering at 
Stanford \cite{Bloom:1969kc,Breidenbach:1969kd},
proton structure 
was understood in terms of quarks, still the
smallest known constituents of matter. 
With the discovery of Bjorken scaling of the proton 
structure function $F_2(x,Q^2)$,
its quark model interpretation,
and the subsequent discovery of scaling violation in support
of asymptotic freedom \cite{Gross:1973id,Politzer:1973fx}, 
deep inelastic scattering (DIS) became a 
field of fundamental theoretical importance~\cite{Feynman:1973xc}
to the understanding of the strong interaction.
Precise measurements of the parton momentum distributions
of the nucleon became a major testing ground
for the selection and development of 
Quantum Chromodynamics (QCD)~\cite{Fritzsch:1973pi}
as the appropriate theory of the strong interaction.
Prior to these developments, the theory of strong interactions was
of merely phenomenological nature, built around S matrix
theory and general amplitude features and various
concepts such as Regge, bootstrap or further models~\cite{froissart}. 

Quantum Chromodynamics is a Yang-Mills gauge theory, in which the interaction 
between confined quarks proceeds via coloured gluon exchange. 
With improved resolution, as provided
by increased $Q^2$, quarks can be resolved
as quarks radiating gluons, whilst gluons may split into 
quark-antiquark pairs or, due to the non-abelian 
nature of the underlying gauge field theory, into 
pairs of gluons~\cite{gribovlipatov,Dokshitzer:1977sg,Altarelli:1977zs}.
The development of QCD calculations beyond leading 
order~\cite{Moch:2004pa,Vogt:2004mw} is
one of the most remarkable recent achievements of 
particle physics theory supported by experiment. It
leads to a consistent description of all perturbatively
accessible hadron observables in DIS (and beyond), 
as has recently been  established over the kinematic
range accessible to HERA \cite{:2009wt}. This includes the unexpected
observation of deep inelastic diffractive scattering at HERA,
where in a significant fraction of violent DIS interactions
the proton remains intact, mediated by an exchange of vacuum
quantum numbers which often is termed ``Pomeron exchange''.

Despite previous successes, many fundamental areas of QCD
have not been verified experimentally, 
with instantons~\cite{Schrempp:2005vc} as only one example.
Even the classic areas related to quarks and gluons have not been
exploited as required due to limited precision, range and variation
of initial conditions.
Meanwhile the theory underlying  DIS
experiences further fundamental  developments.  Four-dimensional
conformal field theory is seen to be related to  
superstring theory in the anti-de Sitter
space in ten dimensions, which  relates
the $N=4$ supersymmetric pomeron to the graviton
in this space~\cite{Lipatov:2011ab}. The evolution of partons
is expected to obey different laws than explored hitherto
at HERA, as at small $x$ their interactions have to be
damped for the occurrence of non-linear interactions
and possibly the restoration of unitarity, see \cite{Ioffe:2010zz}
for a review. 

\begin{figure}[t]
\centerline{\includegraphics[clip=,width=0.7\textwidth]{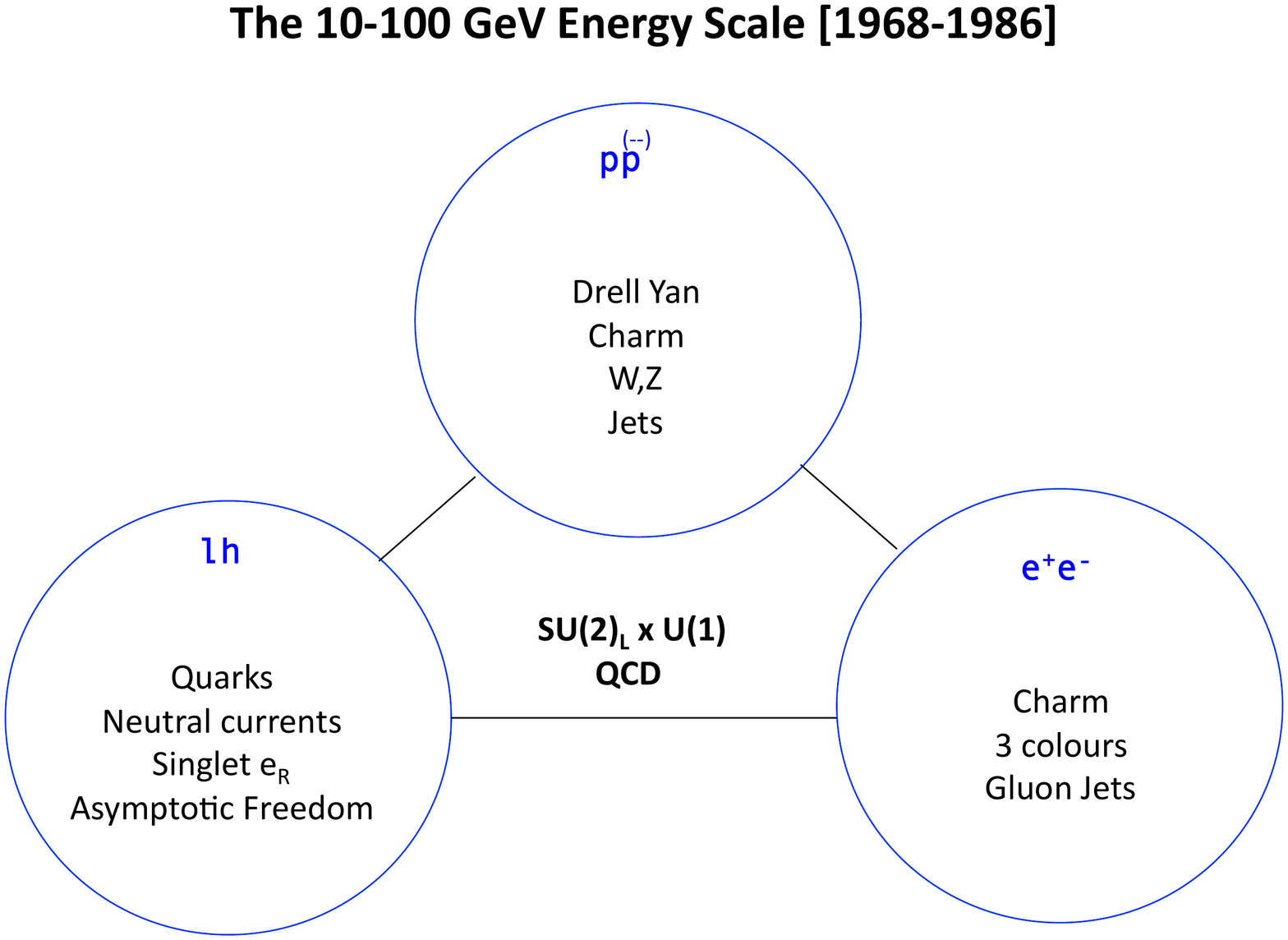}}
\caption{Key results of 
the exploration of the $10-100$\,GeV energy scale 
in hadron-hadron (top), deep inelastic lepton-hadron (lh) (bottom left) and
$e^+e^-$ scattering (bottom right). These and further
important results selected the SU(2)$_L$\,x\,U(1) 
and QCD as the appropriate theories for the electroweak and
the strong interaction, 
 respectively, of leptons and quarks 
transmitted by the photon, the $W^{\pm},~Z$ bosons
and gluons.
}
   \label{fig:scale10}
\end{figure}
\begin{figure}[t]
\centerline{\includegraphics[clip=,width=0.7\textwidth]{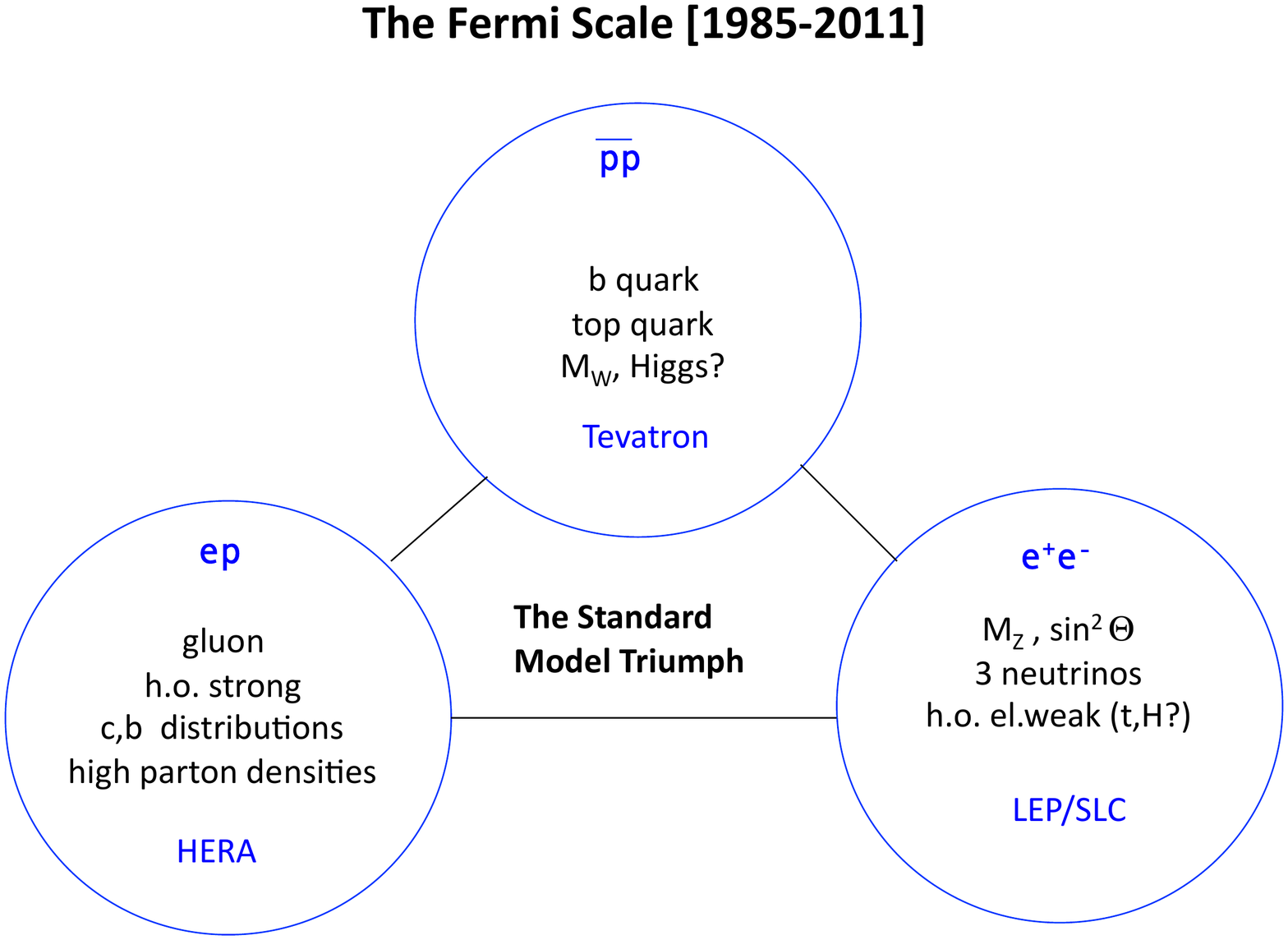}}
\caption{Key results of 
the exploration of the Fermi energy scale 
in $p \overline{p}$ (top), deep inelastic (bottom left) and
$e^+e^-$ scattering (bottom right) with the energy
frontier colliders, the Tevatron, HERA and the SLC/LEP, respectively.
These and further important results established the Standard
Model of particle physics with six types of quarks and leptons
in three families, and the development of higher order
calculations used for the prediction of the top quark
and the Higgs mass, based mainly on $e^+e^-$ scattering 
results, and for the understanding of the partonic contents
of the proton to NNLO pQCD, based mainly on the results
from HERA and previous DIS fixed target experiments. 
Despite intense searches the Higgs particle was not found at
LEP and no clear signal was established at the Tevatron.
}
   \label{fig:scalefermi}
\end{figure}
\begin{figure}[t]
\centerline{\includegraphics[clip=,width=0.7\textwidth]{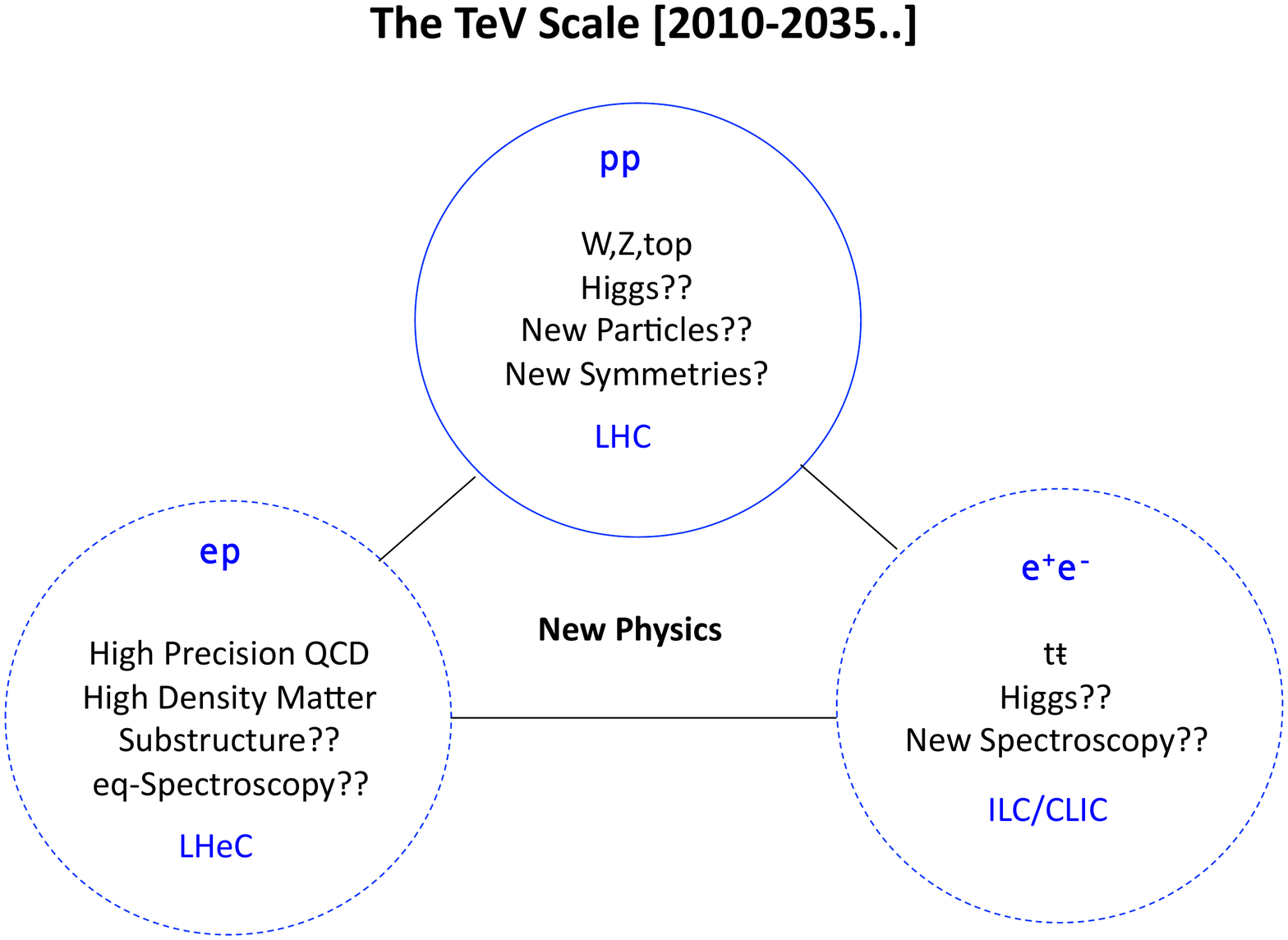}}
\caption{The exploration of the TeV energy scale has begun with the
LHC. The present document describes one of its complements, a new TeV scale
$ep$ and $eA$ collider, while intense work is continuing on the development
of concepts for new $e^+e^-$ and possibly $\mu^+ \mu^-$ colliders.
While each of the new machines has exciting standard model programmes
to pursue with higher precision and range, physics beyond the SM has been
elusive at the moment this report is released and the first few fb$^{-1}$ of
about half energy  LHC data have been analysed. A tantalising hint had
been observed in the $7$\,TeV data taken in 2011 at the LHC
for the Higgs particle to exist with a mass near to $125$\,GeV. No signs,
however, have been reported for SUSY or other new particles.}
   \label{fig:scaleTev}
\end{figure}

Particle physics in the past could profit very much from the
complementarity of hadron-hadron, DIS and $e^+e^-$ experiments.
Key observations were made in all three areas, and the overlap in physics
coverage was used to achieve confidence in new and precision results.
This  is sketched in Figure\,\ref{fig:scale10} for the experiments of
the seventies and eighties, which resulted in the birth of
the Standard Model. Fig.\,\ref{fig:scalefermi} illustrates this
for the experiments of the nineties until now, 
when the Tevatron, HERA and the SLC/LEP machines
determined the progress in the exploration of particle
physics at the energy frontier accessed with colliders.  
The present report deals with the reasons and possibility
to extend deep inelastic scattering experimentation
into an unexplored range of energy for which the
LHC at CERN provides a unique opportunity
for the next decades ahead. Simultaneous LHC and 
LHeC operation  would put the $ep$ part of the TeV scale
triangle, as shown in  Figure\,\ref{fig:scaleTev}, on a
firm ground.

\section{Open questions}
For a project of the dimension of 
the LHeC one needs to understand which fundamental
properties of nature it promises to deal with and which possibly specific
questions it is expected to answer. 

The Standard Model of particle physics contains a remarkable, but
unexplained, symmetry between quarks and leptons \cite{alvaro76}, 
with three generations, in each of
which two quarks and two leptons are embedded. It was
pointed out long ago \cite{salam76} that it appears somewhat
artificial that the basic building blocks of matter share the
electromagnetic and the weak interactions but differ in their sensitivity
to the strong interaction. 
Many theories which unify the 
quark and lepton sectors, such as models based on the E6
gauge group~\cite{Hewett:1988xc},
$R$-parity violating (RPV) supersymmetry and left-right symmetric
extensions of the Standard Model~\cite{Pati:1974yy}, predict
new resonant states with both lepton and baryon numbers, 
usually referred to as leptoquarks (LQ). 
In the technicolour theory, leptoquarks are bound states of
technifermions~\cite{Susskind:1978ms,Farhi:1980xs}. 
Although some of the specific
theories have not been supported by experiment, 
the search for leptoquarks has been a prime motivation for 
high energy scattering, especially DIS experiments. The 
present limits for leptoquark
states  from the LHC leave the possibility
of new LQ states at around $1$\,TeV mass open while the absence
of large missing energy may be seen as being compliant 
with RPV SUSY states in which there is no lightest, stable 
supersymmetric particle. 
An LHeC, in combination with the existing LHC programme, 
can extend this search into a previously
unexplored mass region, with the prospect of
deciphering the leptoquark
quantum numbers.

No analytic proof yet exists that QCD should exhibit the   
property of colour confinement, though it is reasonable to assume
that it is a consequence of gluon
dynamics, as reflected for example in
popular hadronisation models~\cite{Andersson:1983ia}
and Monte Carlo simulations on the lattice. Studying the
behaviour of gluons under new extreme conditions and
contrasting the conditions under which the proton stays intact with
those in which it is destroyed may help to shed
light on the precise mechanism at work.

The search for
the Higgs boson, which explains the masses of the
electroweak bosons,
and for the origin of electroweak symmetry breaking 
is currently the central focus 
of particle physics and is expected to be
principally resolved within
the near future by the ATLAS and CMS experiments.
If there indeed exists a Higgs particle at masses 
around $125$\,GeV, the determination of its properties
becomes an important issue. The LHeC, due to its
clean initial state and the absence of pile-up at high luminosity,
both in contrast to the LHC,
has an interesting potential to accurately determine
the Higgs particle coupling to $b \overline{b}$
and possibly further final states,
and to also investigate the HWW vertex, which provides
direct insight into the nature of electroweak symmetry
breaking and the CP properties of the Higgs field.

The question of hadronic mass deserves similar exploration.
The mass of baryons is almost entirely due
to strong interaction field energy, generated through 
quark and gluon vacuum condensates via
the self-interaction of gluons in a manner which is not yet well
understood. It may be accessible through a more detailed
exploration of QCD dynamics.

The salient theme of physics with the LHeC is the 
mapping of the gluon field. This is achieved with
precision measurements of the evolution of structure
functions over an unprecedented range of $\ln Q^2$.
It relates inclusive $ep$ DIS with jets and heavy flavour,
it also concerns  the unexplored role of the gluon in nuclei
and in deeply virtual Compton scattering. The 
gluon field is central to QCD but not directly measurable.
It may exhibit spots of maximum density (hot spots)
and it may also disappear (cold spots) as it does towards
low $Q^2$ and $x$, and possibly at the scaling point 
near $x \simeq 0.2$~\cite{Glazov:2010bw}. 
Knowing the gluon means understanding
the origin of baryonic matter, the production of the 
Higgs boson and of other new particles and, not least important,
understanding Quantum Chromodynamics.

The study of deep inelastic $ep$ scattering is important for the
investigation of the nature of the Pomeron and Odderon, which are Regge singularities 
of  the $t$-channel partial waves $f_j(t)$ in the complex plane of the
angular momentum $j$. The Pomeron is responsible for a growth of
total cross sections with energy. The Odderon describes the behaviour of
the difference of the cross sections for particle-particle and particle-antiparticle
scattering which obey the Pomeranchuck theorem. In perturbative QCD, the Pomeron 
and Odderon are the simplest colourless reggeons (families of glueballs) constructed 
from two and three reggeized
gluons, respectively. Their wave functions satisfy the generalised BFKL equation. 
In the next-to-leading approximation the solution of the BFKL equation contains
an infinite number of Pomerons and to verify this prediction of QCD one needs to increase
the energy of colliding particles. In the N=4 supersymmetric generalisation of
QCD, in the t'Hooft limit of large $N_c$, the BFKL Pomeron is equivalent to the
reggeized graviton living in the 10-dimensional anti-de-Sitter space. Therefore,
the Pomeron interaction  describing the screening corrections to the BFKL
predictions, at least in this model, should be based on a general covariant effective
theory being a generalisation of the Einstein-Hilbert action for  general relativity.  
Thus, the investigation of high energy $ep$ scattering could be
interesting for the construction of a non-perturbative approach to QCD based on
an effective string model in high dimensional spaces. 

The strong coupling constant $\alpha_s$
decreases as energy scales increase, in contrast to the
energy dependence of the weak coupling and the fine structure
constant. 
It appears possible in SUSY theories that the three constants
approach a common value at energies of order $10^{15}$\,GeV.
The distinctions we make between the electromagnetic,
weak and strong interactions may merely be a consequence of
the low energy scale at which we live.
The possible grand unification
of the known interactions has been one of the major goals
of modern particle physics theory and experiment. 
Progress in this area requires that
$\alpha_s$, by far the most poorly constrained of the
fundamental couplings,  is determined
much more accurately than is currently the case.
The LHeC promises a factor of ten 
reduction in the uncertainty on $\alpha_s$ based on a
major renewal and extension of the experimental
and the theoretical basis of the physics of deep inelastic scattering.

After quarks were discovered, a distinction was soon made between
valence and sea quarks~\cite{Kuti:1971ph}. However, it was not until
the high energy colliding beam configuration of HERA became available
that the rich partonic structure of the proton
was fully realised. 
Despite the resulting fast development
of the knowledge of the parton distribution functions (PDFs) in
the proton, there are still many outstanding important questions
regarding the quark contents of the nucleon which the LHeC would 
address. These regard for example:
i) the unresolved question of whether sea quarks and anti-quarks 
have the same momentum distributions;
ii)  the clarification of the role of heavy quarks in QCD, including the
search for their intrinsic states~\cite{Brodsky:1980pb}, the precision
measurement of the $b$ quark density or, owing  to the huge reach in
$Q^2$ of the LHeC,  the first exploration of top production in DIS and the transition
of top from a heavy to a light quark, for $Q^2 \gg m_t^2$;
iii) the partonic structure of the neutron, which is
to be resolved over many orders of magnitude
in $1/x$, and  the assumption of isospin 
symmetry, which relates  the neutron down-quark
distribution to the proton up-quark distribution. Modern fits of
PDFs use quite a number of symmetry assumptions and
exploit parameterisations which are to be questioned and
overcome by a new basis for the PDF determinations which
the LHeC uniquely provides as it constrains all quark distributions,
$u_v$, $d_v$, $u$, $\overline{u}$, $d$, $\overline{d}$, $s$, $\overline{s}$,
$c$, $b$ and likely $t$ and $\overline{t}$ over an unprecedented
range of $x$ and $Q^2$. The LHeC will put the whole of
PDF related physics on new, much firmer ground, which also
becomes crucial for searches for physics beyond the standard model,
as these move to higher and higher masses at the LHC.
It is also necessary for high precision tests of the 
electroweak theory, including the ultimate measurement of the
mass of the $W$ boson~\cite{Krasny:2010vd} 
as a test for the validity of the SM, especially
the relation to the masses of the top quark and the Higgs boson.

The structure of the neutron at low $x \leq 0.01$ in the
DIS region is experimentally unknown.
With no data on the scattering of leptons from heavy ions
with colliding beam kinematics, the knowledge of the modifications
to nucleon parton densities when they are bound inside nuclei,
rather than free, is also restricted to high $x$ values. This is reflected
in a lack of detailed understanding of shadowing phenomena, particularly
for the gluon density, and a corresponding lack of knowledge of the
initial state of heavy ion collisions at LHC energies.
The mechanism of shadowing at low $x$ can be tested for the first time
via  Gribov's fundamental relation to diffraction
and also via measurements with different light nuclei. 
 Antishadowing at larger 
$x$~\cite{Kovarik:2010uv} may possibly be non-universal and flavour specific.
Nuclear corrections at large $x$ may be dealt with in $eD$ scattering
at the LHeC by tagging the spectator nucleon and reconstructing its
momentum well enough to account for the disturbing effects 
of Fermi motion. This promises to overcome the uncertainty
from nuclear corrections which has been an obstacle for 
decades in the understanding of nucleon structure and
represents a formidable experimental task, see 
e.g.~\cite{Kovarik:2010uv} for a recent study.
Parton distributions in nuclei, for $x \lesssim 0.01$,
presently are based on HERA's proton data convoluted with theoretical
expectations. With the LHeC they will be determined down to below
$10^{-5}$ and largely flavour separated. It is unknown what will
be found from an experimental point of view, and it is critical
for the understanding of the quark gluon plasma.

There are various fundamental properties predicted in QCD
which have never been resolved or even tested so far and which
will become accessible with the LHeC. While
ordinary quark distributions correspond to an
incoherent sum of squared amplitudes, a new approach has been
developed, which uses quark amplitudes and Generalised
Parton Distributions (GPDs)
to understand proton structure in a new,
three-dimensional way~\cite{Mueller:1998fv,Belitsky:2005qn}.
The understanding of GPDs is limited by the relative paucity of 
experimental data on exclusive DIS channels.
The emission of partons is assumed in PDF fits
to be governed by the linear DGLAP evolution equations, 
an approximation to a full solution to QCD, in which 
parton cascades are ordered in transverse momentum. There
are good reasons to believe that the DGLAP approximation is
insufficient to describe the $Q^2$ evolution of low $x$ partons, 
possibly even within the $x$
range to which the LHC rapidity plateau corresponds
at lower masses produced in Drell-Yan scattering. 
Inclusive DIS and jet data in an extended low $x$ 
kinematic regime are required to resolve this situation.   

The rapid rise of the proton gluon density as $x$ decreases 
cannot continue indefinitely. At $x$ values within the reach of
LHeC $ep$ and $eA$ scattering, a transition takes place from the
currently known DIS
regime in which the proton behaves as a dilute system to
a new low $x$ domain in which parton densities 
are expected to saturate 
and the proton
approaches a 'black disk' limit \cite{Gribov:1968gs}. 
This latter region represents a
fundamentally new regime of strong interaction dynamics, for which a
rich phenomenology has been developed, but where the detailed mechanisms 
and the full consequences are not yet known. 
Experimental data at sufficiently low $x$ with scales which are large
enough to allow a partonic interpretation are 
required in order to test the models and fully understand 
the behaviour of partons at high densities. The so well known
DGLAP evolution should fail and non-linear evolution
equations will determine the parton distributions, for
which various untested predictions exist.

The high precision and range of the LHeC DIS measurements 
provide many further opportunities for explorations of fundamental interest.
With the $ep$ initial state any new phenomenon singly produced can be
investigated with particular sensitivity, for example
if excited leptons exist. Variations of beam charge and polarisation
allow the resolution of quantum numbers of  new, so-called contact interactions
of scales up to about $50$\,TeV,
and to novel precision measurements of the scale dependence of 
the weak mixing angle around the $Z$ pole.

Despite its huge success in describing existing high energy data, 
the Standard Model is known to
be incomplete, not only due to the absence of an 
experimentally established
mechanism for electroweak symmetry breaking.
As the exploitation of the TeV energy regime and the high luminosities
of the LHC era develop further, a full understanding 
requires challenging the existing theory through new
precision measurements, which are as broad in scope as possible, with initial
states involving leptons as well as quarks and gluons.
The LHeC will not just answer some of the currently
outstanding questions but represents the opportunity
to build a new laboratory for particle physics which
owing to its specific configuration, its enlarged
DIS energy range and unprecedented precision will
accompany the LHC, and any possibly built pure lepton
machines, in exploring the next layer of the
high energy frontier of physics.  

%% file: introduction/design.tex
%
%
The following sections describe briefly which general considerations
have determined the LHeC design as presented in this report.
Major changes to the underlying assumptions would naturally
require appropriate modifications to the design.
\section{Deep Inelastic Scattering and Particle Physics}
Deep inelastic scattering (DIS) experiments with charged leptons
may be classified as low energy, medium and high energy
experiments. The pioneering low energy 
DIS experiment, which discovered quarks,
 was performed at SLAC. Classic medium energy
experiments were the BCDMS and the NMC experiments
at CERN, while HERA, the first $ep$ collider ever built,
pushed the  DIS energy reach to the Fermi scale.
This allowed the field of deep inelastic scattering
to develop as part of the energy
frontier of particle physics,  complementary to
the Tevatron and LEP.
In all three areas, the field of DIS is
considering upgrade projects with the $12$\,GeV upgrade 
at Jlab, the medium energy colliders at Jlab and/or BNL,
possibly further fixed target neutrino experiments
and the LHeC. 

The LHeC provides the only realistic possibility
for an energy frontier $ep$ programme of experimentation
in the coming 
decades. Thanks to the unprecedented high energy and intense
LHC proton beams, there is a unique opportunity to 
complement the TeV scale $pp$ machine with
a TeV energy $ep$ collider, in addition to a pure lepton collider
in this energy range. It took about $30$ years for HERA, LEP and
the Tevatron to be built, operated and analysed. The exploration
of the tera energy scale is subject to similar time horizons.
\section{Synchronous $\bf pp$ and $\bf ep$ operation}
%
The LHeC  by its nature is an upgrade to
the LHC, which determines its site and also in a way
its dimensions. A first main design consideration builds
on the assumption that the LHC still runs in $pp$ mode
when an electron beam becomes operational. This
has several implications:
\begin{itemize}
\item The construction of the LHeC has to be
completed in about the next $10$ years, and it may operate for 
a similar time period.
\item The design has to be adapted for synchronous $pp$ and $ep$
     (and $AA$ and $eA$) operation, e.g. the magnets in the
        interaction region (IR)
  must steer three beams, while civil engineering and
        detector modularity requirements have to be compliant with the
       LHC operation and upgrade programme.
\item The synchronous operation of $pp$ and $ep$ allows 
        the collection of a high integrated luminosity, with the goal
        of a total of order $100$\,fb$^{-1}$, and makes the most
       efficient use of both the proton beams and the electron beam
       installation too.
\end{itemize}
It cannot realistically be assumed today that $ep$ physics would 
commence only after the $pp$ program has finished because several 
key LHC 
components have a limited lifetime, which is currently estimated to be 
about 20 years. Planning for an $ep$ run after the $pp$ program finishes 
therefore implies a significant risk of additional cost for the project 
due to a substantial consolidation effort in the LHC.
The LHeC aims to accompany the proton and the ion
physics programme of the LHC in its high luminosity phase,
now assumed to begin in 2024.
\section{Choice of electron beam energy}
\label{sect:elener}
The centre of mass energy squared of an $ep$ collider is
given by the electron beam energy, $E_e$, and the proton
beam energy, $E_p$, as  $s=4 E_e E_p$.
It determines the maximum negative 
four-momentum transfer squared, $Q^2$,
between the electron and the proton, 
since $Q^2 = s x y$ with $0 < x,y \leq 1$. Here
$x$ is the fraction of four momentum of the proton carried by the
struck parton while $y$ is the inelasticity of the scattering process, which
in the laboratory frame is the relative energy transfer.

HERA operated with a proton beam energy of $E_p = 0.92$\,TeV
and an electron (and positron) beam energy of $E_e = 27.5$\,GeV.
With Sokolov-Ternov build-up times of about half an hour, the electron
beam became polarised and mean polarisations of up to $40$\,\% were
achieved.  HERA did not accelerate any hadron beam other than protons.
The LHeC has to surpass these parameters significantly for a
unique and exciting programme to be pursued.

\begin{figure}[htbp]
\centerline{\includegraphics[clip=,width=0.62\textwidth]{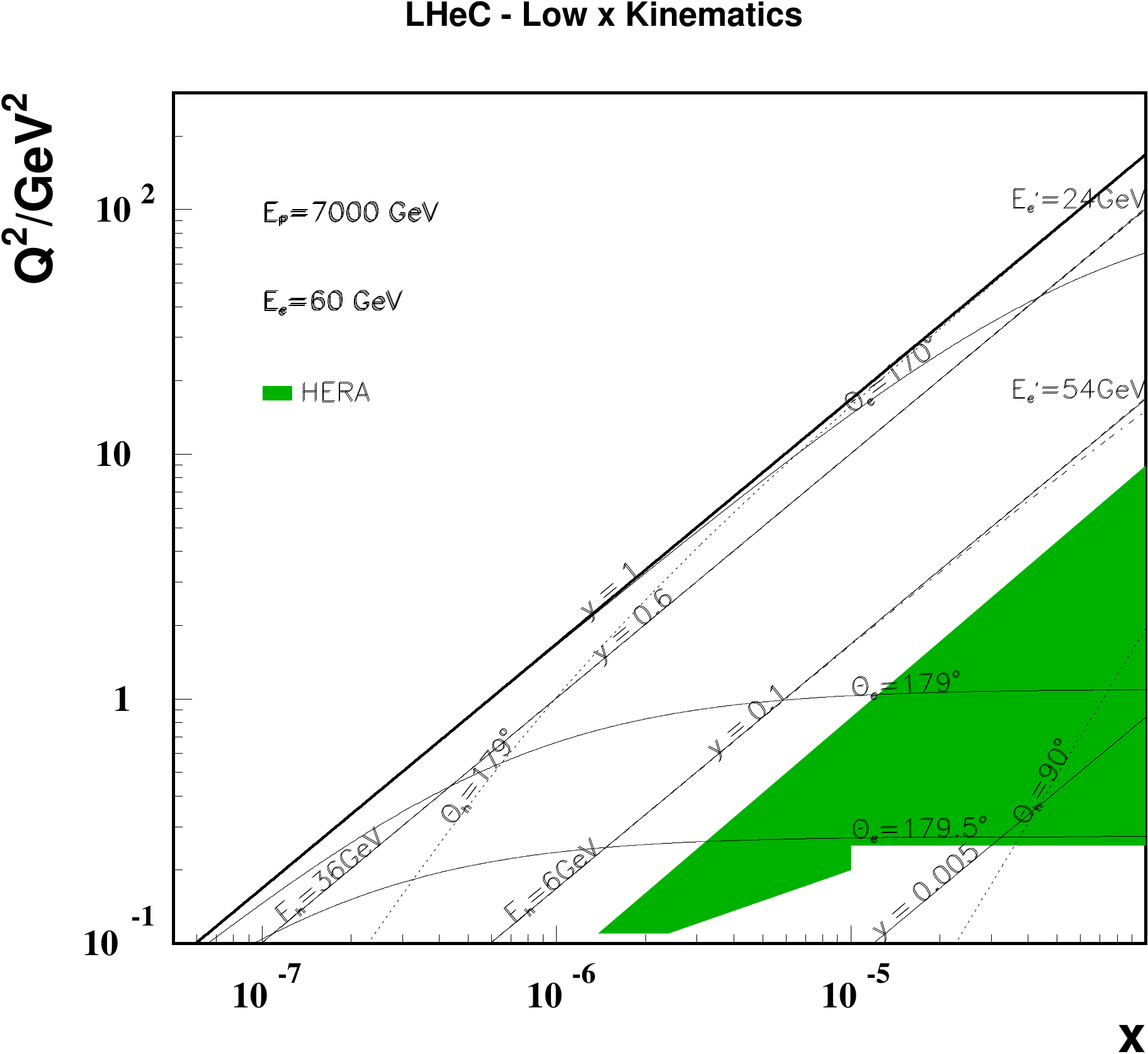}}
\centerline{\includegraphics[clip=,width=0.62\textwidth]{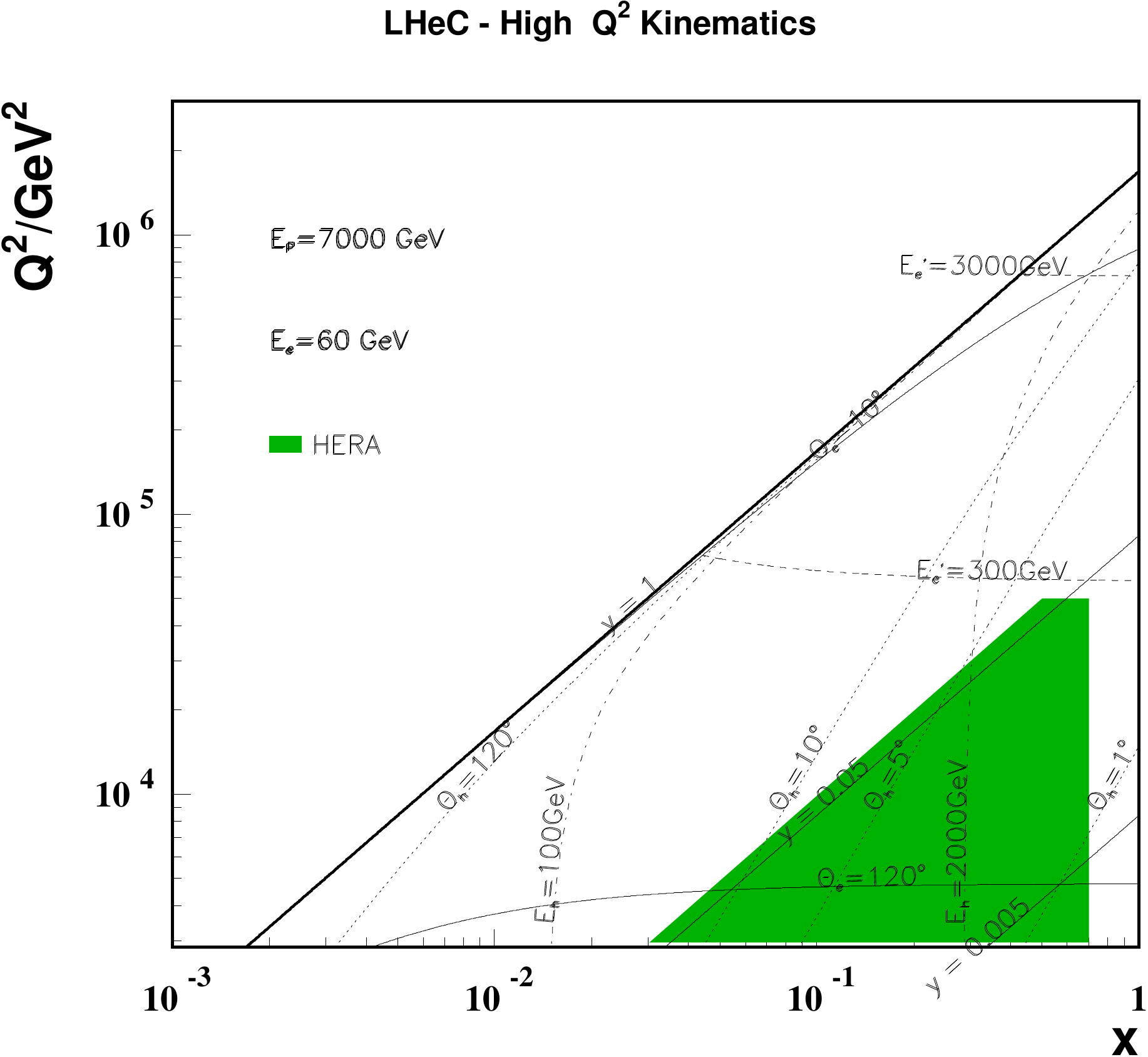}}
\caption{Kinematics of $ep$ scattering at the LHeC
 at low $x$ (top) and high $Q^2$ (bottom). 
Solid (dotted) curves correspond to constant polar
 angles $\theta_e$ ($\theta_h$) of the scattered electron (hadronic final
 state). The polar angle is defined with respect to the
 proton beam direction.   Dashed (dashed-dotted) curves
 correspond to constant energies $E_e'$ ($E_h$) of the scattered
 electron (hadronic final state).  The shaded (green) area illustrates
 the region of kinematic coverage in neutral current scattering at HERA. 
 The energy and angle isochrone lines are discussed in the detector
 design chapter in detail in Sec. $\ref{sec:detectorAcceptance}$.
}
   \label{fig:kinHL}
\end{figure}

The LHeC can use a proton beam with energy up to $7$\,TeV. For this design
study, the electron  beam energy  is set to $60$\,GeV. This implies
that the gain in $s$, or $Q^2$ at fixed $(x,~y)$, as compared to HERA 
will be a factor of $16.6$, or about $4$ in $\sqrt{s}$. 
The real gain in the range of
$Q^2$ and $x$ will be even larger as, with the superior LHeC
luminosity, even the highest $Q^2$ and values of 
$x$ very close to $1$ become
accessible.  The kinematic range of the LHeC as compared
to HERA at low $x$ and at high $Q^2$ is illustrated in Fig.\,\ref{fig:kinHL}.

The choice of a default $E_e=60$\,GeV for this design report is
dictated by physics and by practical considerations:
\begin{itemize}
\item{New physics has been assumed to appear at the TeV energy scale.
At the time of completion of this report, the LHC has excluded much of the
sub-TeV physics beyond the Standard Model  but leaves the possibility
open of resonant lepton-parton states with masses of larger than about 
$700$\,GeV, for which the LHeC would be a particularly suitable machine
with a range of up to $M \lesssim \sqrt{s}$.}
\item{High precision QCD and electroweak physics require a maximum
range in $\ln Q^2$ and highest $Q^2$, respectively. The unification
of electromagnetic and weak forces takes place at $Q^2 \simeq M_Z^2$
which is much exceeded with the LHeC energies. 
Part of the electroweak physics programme requires lepton beam polarisation,
which in a ring configuration is difficult to achieve for higher
energies than $60$\,GeV as is demonstrated in this report.} 
\item{
The discovery of gluon saturation requires to measure at typical
values of small $x \simeq 10^{-5}$ with 
 $Q^2 \gg M_p^2$, where $M_p$ is the mass of the proton.
The choice of energies ensures this discovery at the LHeC
in the DIS region, both in $ep$ and in $eA$, if this
phenomenon indeed exists.}
\item{Energy losses by synchrotron radiation,
$\propto E_e^4$, both in the ring and the return arcs for the linac,
can be kept at reasonable levels, in terms of the power, $P$,
needed to achieve high luminosity, and the radius of the
racetrack return arcs can be chosen such that the LHeC tunnel
in the linac configuration is only $1/3$ of the LHC.} 
\end{itemize}
Thus it appears that $60$\,GeV is an appropriate and affordable choice.
and yet it is well possible that $60$\,GeV  may not be the
final value of the electron beam energy, especially if the 
LHC would find non-SM physics just above the 
default energy range considered here.
The design therefore also considers a dedicated high energy
beam of $140$\,GeV as an option, which as yet has not been
worked out in comparable detail~\footnote{Such a large $E_e$ would
also fit better to a future  HE LHC of $E_p \simeq 16$\,TeV
or, looking even further into the future, to a proton collider of  $E_p \simeq 40$\,TeV
in a new $~70$\,km tunnel with stronger dipoles, as this would keep 
the $ep$ beam energy asymmetry at a tolerable level.}.

\section{Detector constraints}
It is easily recognised, in Fig.\,\ref{fig:kinHL}, 
that the asymmetry of the electron and proton
beam energies poses severe constraints on the detector design:
i) the ``whole'' low $Q^2$ and low $x$ physics programme requires to measure
the electron, of energy $E_e' \lesssim E_e$,
scattered in the backward direction between about $170^{\circ}$
and $179^{\circ}$, and ii) the forward scattered final state, of
energy comparable to $E_p$, needs to be reconstructed down
to very small angles in order to cover the high $x$ region in a
range of not too extreme $Q^2$. 

The current detector design  considers an
option to have split data taking phases, like HERA I 
and II, with different interaction
region configurations, 
 a high acceptance phase,
covering $1^{\circ} - 179^{\circ}$, at reduced luminosity
and a high luminosity phase, of acceptance limited
to $8^{\circ} - 172^{\circ}$.
In the course of the study, however, an optics was found for
the high acceptance configuration with only a factor of two reduction in
luminosity. It is likely, therefore, that the TDR will lead to a unification
of these configurations and correspondingly weakened demands on the
modularity of the inner detector region.

The joint
 $ep$ and $pp$ operation implies that at least one
of the four IPs, currently occupied by experiments, will have to be
made available for an LHeC detector~\footnote{The
four other principal possibilities are excluded because 
IP3 and IP7 have no cavern, IP6 houses the beam extraction (dump
area) and IP4 is filled with RF equipment.}  
It was decided to use for this report IP2 as  site
 and to limit the study of bypasses, in the ring option,
to IP1 and IP5. In the linac configuration, the racetrack tunnel is inside the
LHC ring and tangential to IP2. Access to the linac seems then
possible with shafts placed at CERN territory only, the Prevessin site.
IP8, which houses LHCb, is close to the airport which makes the
civil engineering and access impractical. It therefore has to be
tentatively recognised that the LHeC is an option for housing a new, fifth
experiment at the LHC, which would require to conclude the
ALICE experiment in due time.
 
There has often been a discussion about the
need for two detectors and ambitious detector push-pull concepts are
discussed for the Linear Collider. For the LHeC this would 
imply a major overhead of cost and delay in construction time.
The detector envisaged here will be challenging but also 
based on known technology. Truly independent reconstruction,
simulation and analysis software teams using 
one common facility may lead to sufficient confidence  when it
comes to crucial and the most precise results.

\section{Two electron beam options}
It was shown a few years ago~\cite{Dainton:2006wd} that an electron beam
in the LHC tunnel would allow to achieve an outstanding luminosity
of about $10^{33}$\,cm$^{-2}$s$^{-1}$ in $ep$
interactions for both electrons and positrons. It is obvious, however,
that while such a ring may be built without any major technical
obstacle, installing it on top of the
LHC magnet ring would be a non-trivial engineering and logistics task. 
Mostly for this
reason it was decided to  consider besides this ``ring-ring (RR)'' 
option also  a ``linac-ring (LR)'' configuration, with a
linear electron accelerator tangential to the LHC. For the comparison
of RR and LR options, $E_e$ was kept the same at $60$\,GeV.
The ring could extend to somewhat higher energies, but only
a Linac would allow $100$\,GeV  to be significantly exceeded. 
The potential for higher energy is not the only, and possibly not
the dominant reason for considering a linac-ring collider.
Other important  benefits include the potential for higher electron 
current than assumed in the LHeC baseline design  
and thus higher luminosity, and a construction phase that can overlap 
with LHC running prior to the LS3 shutdown. 

This report presents all major components and considerations
for both the RR and the LR configuration. A choice between the
two configurations is envisaged to be taken
soon after the appearance of the CDR. It is important to consider
that the RR configuration delivers high electron and also positron luminosity, 
with difficulties for high polarisation, while the LR configuration has
a high potential for polarised electrons, but difficulties
to deliver an intense positron beam, yet offering also a photon beam option.
The electrical power required for a ring-ring collider at constant beam
current increases with the fourth power of energy, while 
for a linac-ring collider the increase is roughly linear as long as
the synchrotron radiation loss in the return arcs remains a small fraction
of the total.
For higher electron energies in the ring the polarisation 
greatly decreases, whereas for the linac solution the polarisation
is independent of beam energy. 
A choice of one over the other option has to be based on
physics but also technical, cost and further considerations,
which is why considerable effort had been spent to develop both
options to the required level of detail. No attempt is made in the present report to
favour one over the other configuration. In the period of this design
study both options came into a very fruitful interaction and
occasional competition which stimulated both designs.
\section{Luminosity and power}
\begin{figure}[htbp]
\centerline{\includegraphics[clip=,width=0.62\textwidth]{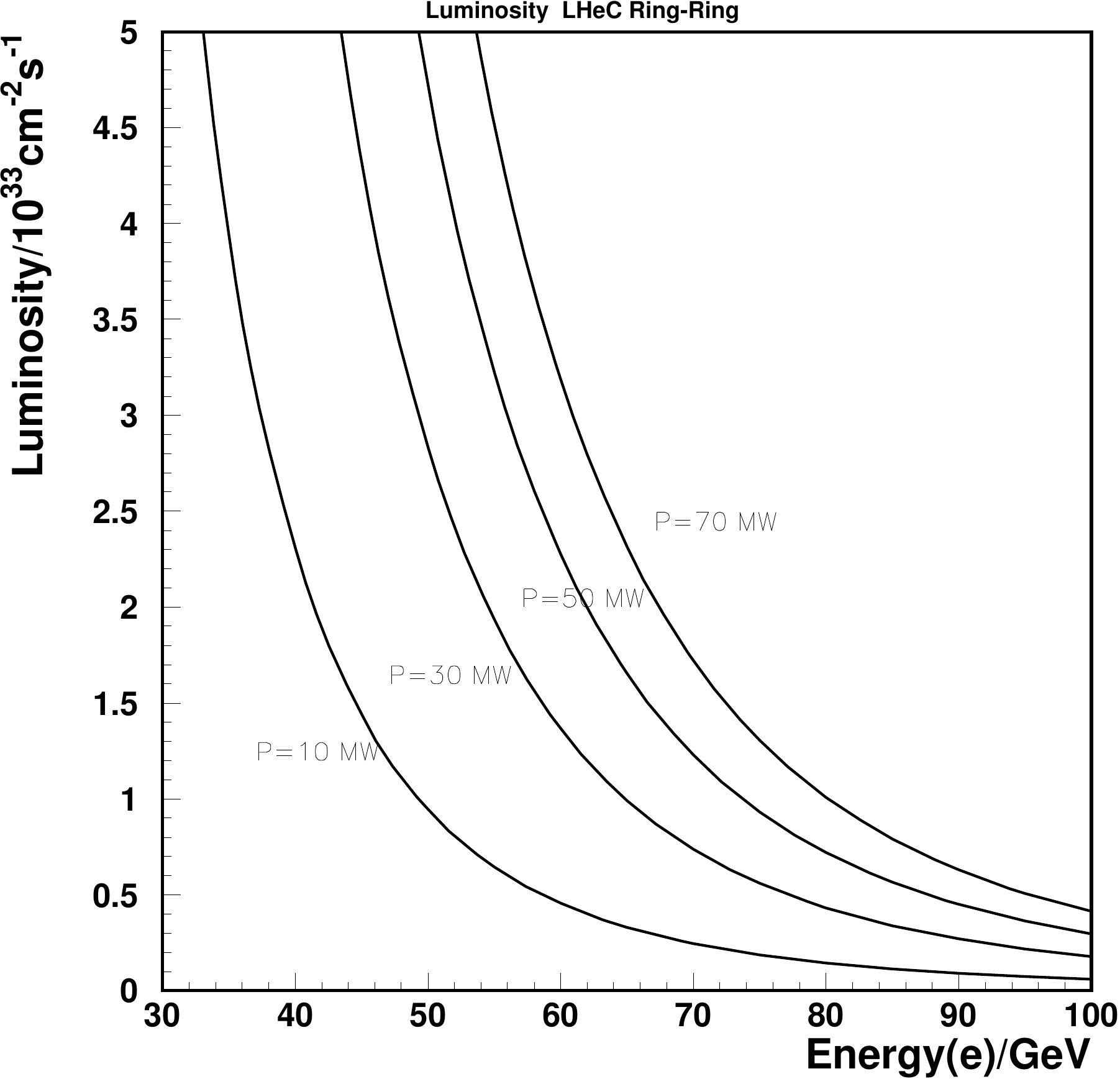}}
\centerline{\includegraphics[clip=,width=0.62\textwidth]{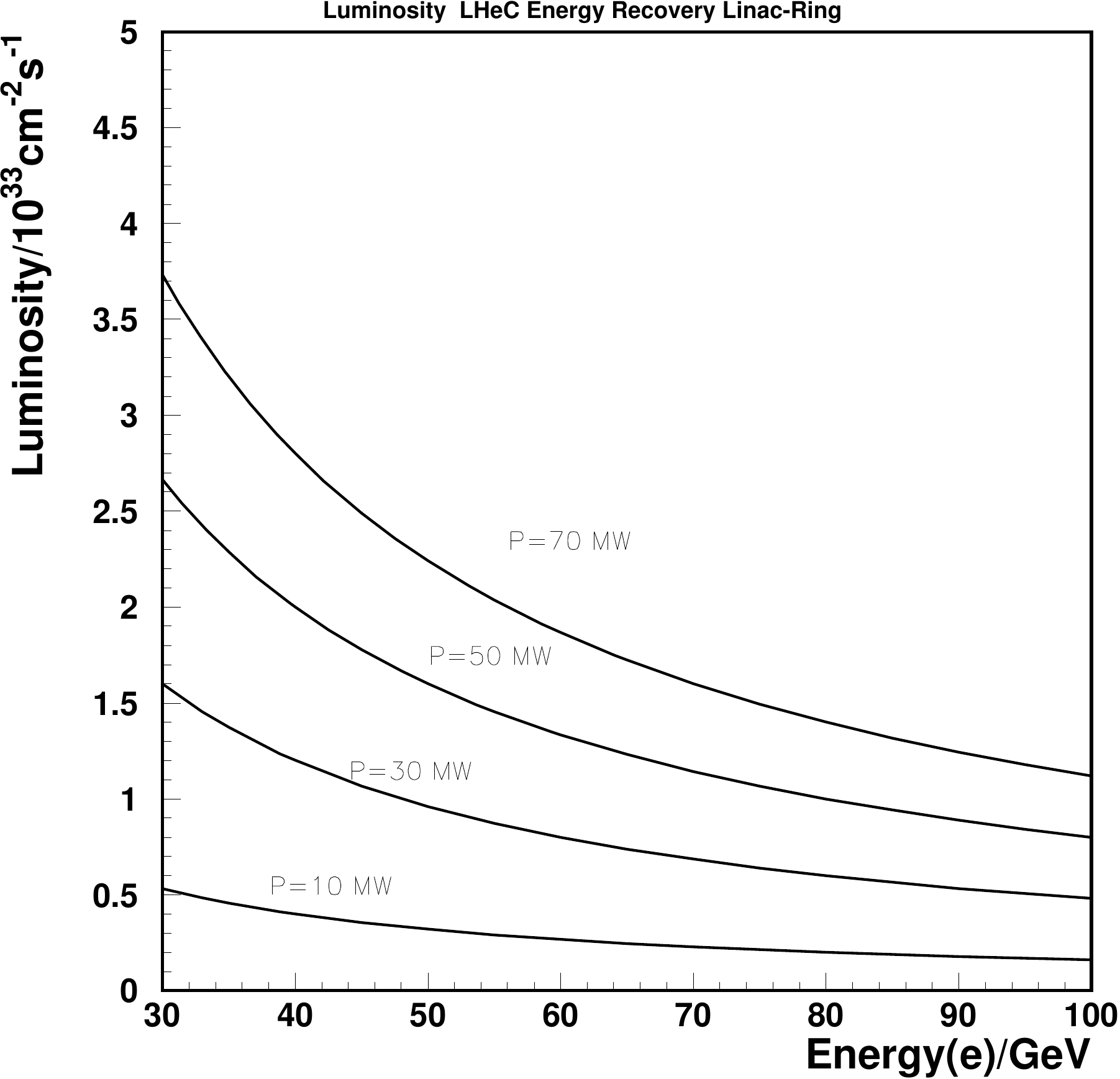}}
\caption{Estimated luminosity, in units of $10^{33}$\,cm$^{-2}$\,s$^{-1}$,
for the RR configuration (top) and the LR energy recovery
configuration (bottom), displayed as a function of the electron beam
energy with the beam power as a parameter, see text.
}
   \label{fig:lumiPE}
\end{figure}

The relation of the luminosity, power and energy differs for the
RR and LR configurations.
As for HERA,
the luminosity for matched beams is determined by the
number of protons per bunch ($N_p$), the normalised
proton beam emittance
($\epsilon_p$), the $x,y$ coordinates of the proton beam beta
function values at the interaction point 
 ($\beta_{x,y}$) and the electron beam current ($I_e$) as
\begin{equation}
\label{eq:LRR}
L=\frac{N_p \cdot \gamma }{4 \pi  e \epsilon_p}
  \cdot \frac{I_e}{ \sqrt {\beta _{px} \beta _{py} } },
\end{equation}
with $\gamma = E_p/M_p$.
The design luminosity assumes the so-called ultimate proton
beam parameters for $E_p = 7$\,TeV with $1.7 ~10^{11}$ protons
per bunch and $\epsilon_p = 3.8$\,$\mu$m. It is interesting to note that
already the first year of operating the LHC has indicated that
smaller emittance values are in reach and the bunch intensities
have exceeded $10^{11}$, for $50$\,ns spacing. Eq.\,\ref{eq:LRR} then
corresponds to 
\begin{equation}
\label{eq:LRRs}
L= 8.2 \cdot 10^{32} {\rm cm^{-2}s^{-1}} \cdot \frac{N_p}{1.7 \cdot 10^{11}}
  \cdot \frac{1 {\rm m}}{ \sqrt {\beta _{px} \beta _{py} } } \cdot \frac{I_e}{50 {\rm mA}},
\end{equation}
where the electron beam current is given by
\begin{equation}
\label{eq:IeR}
I_e = 0.35 {\rm mA} \cdot P {\rm [MW]} \cdot (\frac{100 {\rm GeV}}{E_e})^4.
\end{equation}
With $\beta_{x(y)} = 1.8 (0.5)$\,m, see the
optics section, one obtains a
typical value of $10^{33}$\,cm$^{-2}$s$^{-1}$ 
luminosity for $E_e = 60$\,GeV with $30$\,MW
of synchrotron-radiation power $P$. 
The dependence of $L(E,P)$ is shown in Fig.\,\ref{fig:lumiPE}
(top) for the RR configuration. While with the matching requirement 
for each $E_e$ a separate evaluation would have to be done of the
$\beta$ functions, it is evident that the RR option has 
a great potential to indeed achieve very high luminosities,
even exceeding $10^{33}$\,cm$^{-2}$s$^{-1}$ 
if $E_e$ was slightly lower or if $P$ was somewhat increased. 

For this design report, the wall-plug power limit for the LHeC was set
to $100$\,MW. With a $10$ years running period at such
a high luminosity and $N_p$ probably enlarged and the emittance
smaller than assumed here,
an integrated luminosity for
the LHeC of $O(100)$\,fb$^{-1}$ can be considered to be a realistic perspective in 
simultaneous operation with the LHC.  That is necessary
for exploiting the high $Q^2$, high mass and large $x$ boundaries.
It implies that, unlike at HERA,
the whole low $Q^2,~x$ physics program, with
the exception of rare processes such as DVCS and subject to trigger
acceptance considerations, may be pursued in a rather
short period of time.

A linear electron beam colliding with a storage ring proton beam
was considered quite some time ago~\cite{pgw}. Its luminosity,
for head-on collisions,
can be obtained from the following relation~\cite{Tigner:1991wt}
\begin{equation}
\label{eq:LLR}
L=\frac{N_p \cdot \gamma }{4 \pi  e \epsilon_p}
  \cdot \frac{I_e}{\beta ^* },
\end{equation}
which is Eq.\,\ref{eq:LRR} if one sets $\beta_x =  \beta_y$.
The luminosity scales as 
\begin{equation}
\label{eq:LLRs}
L= 8 \cdot 10^{31} {\rm cm^{-2}s^{-1}} \cdot \frac{N_p }{1.7 \cdot 10^{11} }
  \cdot \frac{0.2 {\rm m}}{ \beta^* } \cdot \frac{I_e}{1 {\rm mA}},
\end{equation}
where the electron beam current is given by
\begin{equation}
\label{eq:IeL}
I_e = {\rm mA} \cdot \frac{P {\rm [MW]} }{(1-\eta) E_e {\rm [GeV]}}.
\end{equation}
Here $\eta$ denotes the efficiency of the energy recovery process,
defined in terms of beam power 
at the collision point with and without recovery.
A pulsed linac without recovery is correspondingly lower, by about 
an order of magnitude, in luminosity compared to the RR configuration,
even for a demanding $\beta^*$ value of $0.1$\,m, which is 
introduced in the LR section. With energy
 recovery, however, and an assumed efficiency
of $90$\,\% 
luminosities of similar value to the RR case are obtained, see
Fig.\,\ref{fig:lumiPE}. The energy recovery linac (ERL) operates
the cavities in CW mode at moderate gradients of typically $20$\,MV/m.

The recovery of energy requires two linacs which can be of opposite orientation
as was originally considered~\cite{Tigner:1965wf}. 
In the racetrack geometry chosen here each linac, of $1$\,km length,
is passed three times
and is used for acceleration, by $10$\,GeV, and equally for
deceleration for recovering  power.
This introduces synchrotron
radiation losses as a parameter of concern also for the LR configuration.
A short linac passage is required to compensate for  these losses.
With the design proposed here, the arcs have a bending radius 
of $764$\,m, which leads to a LR accelerator of about $9$\,km length,
a bit larger than the SPS. The length is matched to $1/3$
of the LHC circumference.
A straight, high energy, pulsed linac is also considered, which at
$E_e=140$\,GeV reaches a luminosity of about  $5 \cdot 10^{31}$,
the design value of the HERA upgrade phase. The possibility of
having stages of ERL returns, which provide much higher luminosities
also in this case, is briefly demonstrated in this report too. 

%

%% file: physics/qce.tex
This chapter elucidates the physics prospects 
which are related to high precision measurements with the LHeC
to test and develop QCD and the electroweak theory.
Section\,\ref{sec:partalf} presents
inclusive deep inelastic scattering and consists of three parts: 
NC and CC cross sections and structure functions, 
the simulation of NC and CC data sets including estimates for
the expected systematic uncertainties, 
and the simulation of LHeC precision measurements of
the longitudinal structure function $F_L$.
The LHeC is the first DIS experiment which is able to completely
unfold the quark contents of the nucleon.  
Section\,\ref{sec:dirpart} 
introduces assumptions for the QCD fit, used for
illustrating the expected gain in precision at the LHeC
as compared to HERA, BCDMS 
and $W$, $Z$ electroweak data expected from the LHC.
Results are then presented first for the
determination of the valence quark and the strange
quark distributions, which are also compared with
the current information as contained in modern 
PDF determinations. A dedicated part is written for
top quark physics at the LHeC as at very high $Q^2$,
$t$ and $\overline{t}$ production in DIS become 
a new subject of research.
Sections\,\ref{sec:gluon} and \ref{sec:alphas} discuss in detail
the expected precision measurements of the gluon distribution
and of the strong coupling constant, respectively.
Section\,\ref{sec:deuterons} motivates the measurements 
with electron-deuteron scattering which extend current
experimental knowledge on the structure of the
neutron (and the deuteron) by nearly four orders
of magnitude in $Q^2$ and $1/x$.
Section\,\ref{sec:qce_hfl} introduces the measurements
of the charm and beauty densities. Owing to the much extended
range, higher energy (cross section) and dedicated Silicon tracking,
high precision measurements of the $c$ and
$b$ densities  will be provided
for the development of the QCD theory of heavy quarks
and for the description of new phenomena which may be expected
to be related especially to the $b$ density, such as the production
of the Higgs particle in MSSM SUSY.
Section\,\ref{sec:qce_jets}
illustrates the precision QCD tests that can be performed at the LHeC
with jets in the final state.
With the increased energy, new measurements of the 
total photoproduction cross section can be performed, as
discussed in Section\,\ref{sec:gammaptot}.
The Chapter is concluded with the electroweak physics
Section\,\ref{sec:eweak} which focuses on the precision
measurements of the light quark weak NC couplings and on the
scale dependence of the electroweak mixing angle,
as can be determined from polarisation asymmetries
in NC and the NC/CC cross section ratio.
\section{Inclusive deep inelastic scattering}
\label{sec:partalf}
\subsection{Cross sections and structure functions}
\label{sec:disformalism}
\input{physics/disform}

\subsection{Cross section simulation and uncertainties}
\label{sec:simNC}
\input{physics/simnccc}
\subsection{Longitudinal structure function $\bf{F_L}$}
\label{sec:flong}
\input{physics/flong}
\section{Determination of parton distributions}
\label{sec:dirpart}
\input{physics/dirpart}

\section{Gluon distribution}
\label{sec:gluon}
\input{physics/pgluon}
\section{Prospects to measure the strong coupling constant}
\label{sec:alphas}
\input{physics/alphajb}
\section{Electron-deuteron scattering}
\label{sec:deuterons}
\input{physics/deuterons}

\section{Charm and beauty production}
\label{sec:qce_hfl}
\input{physics/qce_hfl}
\section{High $p_t$ jets }
\label{sec:qce_jets}
\input{physics/qce_jets}

\section{Total photoproduction cross section}
\label{sec:gammaptot}
%
\input{physics/tex/gpcs.tex}
\section{Electroweak physics}
\label{sec:eweak}
\input{physics/electroweak}
\subsection{Determination of the weak mixing angle}
\input{physics/sinteta}

%

%% file: physics/disform.tex
%
%
The scattering amplitude for electron-proton scattering is a product
of lepton and hadron currents times the propagator characteristic
of the exchanged particle, a photon or $Z_0$ in neutral current scattering,
a $W^{\pm}$ in charged current scattering. The inclusive scattering
cross section therefore is given by the product of two tensors,
\begin{equation}
\frac{d^2\sigma}{dxdQ^2} = \frac{2 \pi \alpha^2}{Q^4 x} 
\sum_j{\eta_j L_j^{\mu \nu} W_j^{\mu \nu}},
\label{siglw}
\end{equation} 
where $j$ denotes the summation over $\gamma$, $Z_0$ exchange
and their interference for NC, and $j=W^+$ or $W^-$ for CC. 
The leptonic tensor $L_j^{\mu \nu}$ is related to the coupling 
of the electron with the
exchanged boson and contains the electromagnetic or the weak couplings,
such as the vector and axial-vector electron-$Z_0$ couplings, $v_e$ and $a_e$,
in the NC case.  This leptonic part of the cross
section can be calculated exactly in the standard electroweak $U_1 \times SU_2$
theory.  The hadronic tensor, however, describing the interaction of
the exchanged boson with the proton, can only be reduced to a sum of
structure functions, $F_i(x,Q^2)$, and cannot be fully calculated.
Conservation laws reduce the number of basic structure functions
in unpolarised $ep$ scattering to $i=1-3$.
In perturbative QCD the structure functions are related to 
parton distributions $f$ via coefficient functions $C$
\begin{equation} \label{factoreq}
 [F_{1,3},F_2] = \sum_i{\int_0^1{[1,z]\frac{dz}{z} C_{1,2,3}(\frac{x}{z},\frac{Q^2}{\mu_r^2},
 \frac{\mu_f^2}{\mu_r^2},\alpha_s(\mu_r^2)) \cdot f_{i}(z,\mu_f^2,\mu_r^2)}},
\end{equation} 
where $i$ sums the quark $q$, anti-quark $\overline{q}$ and gluon $g$ contributions
and $f_i(x)$ is the probability distribution
of the parton of type $i$ to carry a fraction $x$
of the proton's longitudinal momentum. 
The coefficient functions are  exactly calculable
but depend on the factorisation and renormalisation scales $\mu_f$ and 
$\mu_r$. The parton distributions are not calculable and have to be
determined by experiment. Their $Q^2$ dependence obeys evolution
equations.  A general factorisation
theorem, however, has proven the parton distributions 
to be universal, i.e. to be independent of 
the type of hard scattering process. This makes 
deep inelastic lepton-nucleon scattering a
most fundamental process: the parton distributions in the proton
are measured best with a lepton probe and may be used
to predict hard scattering cross sections at, for example, the LHC.
The parton distributions  are derived 
from measurements of the structure functions in NC and CC scattering,
as is discussed below.

\subsection{Neutral current}
The neutral  current deep inelastic $ep$ scattering cross section, at tree level,
is given by a sum of generalised structure functions
according to 
\begin{eqnarray} \label{ncsi}     
\frac{d^2\sigma_{NC}}{dxdQ^2} =  \frac{2\pi \alpha^2 Y_+}{Q^4 x} \cdot \sigma_{r,NC} \\  
 \sigma_{r,NC}  =     {\bf F_2} + \frac{Y_-}{Y_+} {\bf xF_3} -\frac{y^2}{Y_+} {\bf F_L},
\end{eqnarray}                                                                  
where the electromagnetic coupling constant $\alpha$, the photon              
propagator and a helicity factor are absorbed 
in the definition of a reduced cross section $\sigma_r$, and $Y_{\pm}=1 \pm (1-y)^2$.                                                                              
The functions ${\bf F_2}$ and  ${\bf xF_3}$ 
depend on the lepton beam charge and                  
polarisation ($P$) and on the electroweak parameters as~\cite{Klein:1983vs} 
\begin{eqnarray} \label{strf}                                                   
 {\bf F_2^ \pm} &=& F_2 + \kappa_Z(-v_e \mp P a_e) \cdot F_2^{\gamma Z} +                      
  \kappa_Z^2 (v_e^2 + a_e^2 \pm 2 P v_e a_e) \cdot F_2^Z \nonumber \\                     
 {\bf xF_3^ \pm} &=&  \kappa_Z( \pm a_e + P v_e) \cdot xF_3^{\gamma Z} +                       
  \kappa_Z^2( \mp 2 v_e a_e - P (v_e^2+a_e^2)) \cdot xF_3^Z.                                   
\end{eqnarray} 
In the on-mass shell $\overline{MS}$ scheme the
propagator function $\kappa_Z$ is given by                     
the weak boson masses  ($M_Z,~M_W$)
\begin{equation}
  \kappa_Z(Q^2) =  \frac{Q^2}{Q^2+M_Z^2} \cdot \frac{1}{4\sin^2 \Theta \cos^2 \Theta}
\end{equation}
with the weak mixing angle     $\sin^2  \Theta=1 -M^2_W /M^2_Z$.
In the hadronic tensor  decomposition~\cite{Derman:1973sp} the structure
functions are well defined                          
quantities. In the Quark Parton Model (QPM)  
the longitudinal structure function is zero\,\cite{Callan:1969uq} and the 
two other functions are given by the sums and differences of
quark ($q$) and anti-quark ($\overline{q}$) distributions as
\begin{eqnarray} \label{ncfu}                                                   
  (F_2, F_2^{\gamma Z}, F_2^Z) &=&x \sum (e_q^2, 2e_qv_q, v_q^2+a_q^2)(q+\bar{q})            
                                 \nonumber \\                                   
  (xF_3^{\gamma Z}, xF_3^Z) &=& 2x \sum (e_qa_q, v_qa_q) (q-\bar{q}),                         
\end{eqnarray} 
where the sum extends over all up and down type quarks
and $e_q =e_u,e_d$ denotes the electric charge of up- or
down-type quarks. The vector and axial-vector weak couplings of the fermions
($f=e,u,d$) to the $Z_0$ boson in the standard electroweak model
are given by
\begin{equation} \label{va}
 v_f= i_f - e_f 2 \sin^2 \Theta ~~~~~~~~ a_f=i_f
 \end{equation}
 where   $e_f = -1,2/3,-1/3$ 
 and $i_f=I(f)_{3,L}=-1/2,1/2,-1/2$ denotes the
 left-handed weak isospin charges. Thus the vector coupling of the
 electron, for example, is very small, $v_e=-1/2 + 2 \sin^2 \Theta \simeq 0$,
 since the weak mixing angle is roughly equal to 1/4.

At low $Q^2$ and low $y$ the reduced NC cross section, Eq.\,\ref{ncsi}, 
to a very good approximation is given by $\sigma_r = F_2(x,Q^2)$. 
 At $y > 0.5$, $F_L$ makes a sizeable
contribution to $\sigma_{r,NC}$. In the DGLAP
approximation of perturbative QCD, to
lowest order, the longitudinal structure function is given 
by~\cite{Altarelli:1978tq}
\begin{equation}
        F_L(x) = \frac{\alpha_s}{4 \pi} x^2
        \int_x^1 \frac{dz}{z^3} \cdot \left[ \frac{16}{3}F_2(z) + 8
        \sum e_q^2 \left(1-\frac{x}{z} \right) zg(z)
        \right],
\label{altmar}
\end{equation}
which at low $x$ is dominated by the gluon contribution. 
A measurement of $F_L$ requires a variation of the beam energy.

Two further structure functions can be accessed with 
cross section asymmetry measurements,
in which the charge and/or the polarisation of the lepton beam are varied.
A charge asymmetry measurement,  with polarisation values $P_{\pm}$
of the $e^{\pm}$ beam, 
determines the following structure function combination 
\begin{equation} \label{casy}
 \sigma_{r,NC}^+(P_+) - \sigma_{r,NC}^-(P_-) =
  - \kappa_Z  a_e (P_+ + P_-) \cdot  F_2^{\gamma Z}   
 + \frac{Y_-}{Y_+} \kappa_Z a_e \cdot [2 xF_3^{\gamma Z}  
  + (P_+ - P_-)  \kappa_Z  a_e xF_3^Z]
\end{equation}
neglecting terms $\propto v_e$ which can be easily obtained from
Eq.\,\ref{strf}. If data are taken with opposite polarisation and
charge, the asymmetry represents a measurement of
the difference of quark and anti-quark distributions in NC,
see Eq.\,\ref{ncfu}. In contrast to what is often stated,
the charge asymmetry is a parity conserving quantity $\propto a_e a_q$.
Assuming symmetry between sea and antiquarks,
it is a direct measure of the valence quarks, 
$xF_3^{\gamma Z}\simeq (2u_v+d_v)/3$ in $ep$. 
This function was measured 
for the first time in $\mu^{\pm}$ Carbon
scattering by the BCDMS 
Collaboration~\cite{Argento:1983dj}
at large $x > 0.2$ and for $Q^2$ of about 50\,GeV$^2$.
With the LHeC, for the first time, high precision measurements
of $xF_3$ in NC become possible as is demonstrated in
Sect.\,\ref{sec:valquarks}. These will access the
valence quarks at low $x \lesssim 0.001 $ for the first time in direct measurements.

A genuine polarisation asymmetry measurement, keeping the beam charge fixed,
according to  eqs.\,\ref{ncsi} and \ref{strf} determines a similar
combination of $F_2^{\gamma Z}$ and $xF_3^{\gamma Z}$
\begin{equation} \label{pasy}
\frac{ \sigma_{r,NC}^{\pm}(P_L) - \sigma_{r,NC}^{\pm}(P_R)}{P_L -P_R} 
 = \kappa_Z    [ \mp a_e F_2^{\gamma Z} +  \frac{Y_-}{Y_+} v_e xF_3^{\gamma Z}]
 \simeq  \mp \kappa_Z  a_e F_2^{\gamma Z} 
\end{equation}
neglecting again the term $\propto v_e$.
The product $a_e F_2^{\gamma Z}$ is proportional to 
combinations of $a_e v_q$ and thus provides
a direct measure of parity violation at very small distances.

\begin{figure}
\centerline{\includegraphics[clip=,angle=0.,width=0.9\textwidth]{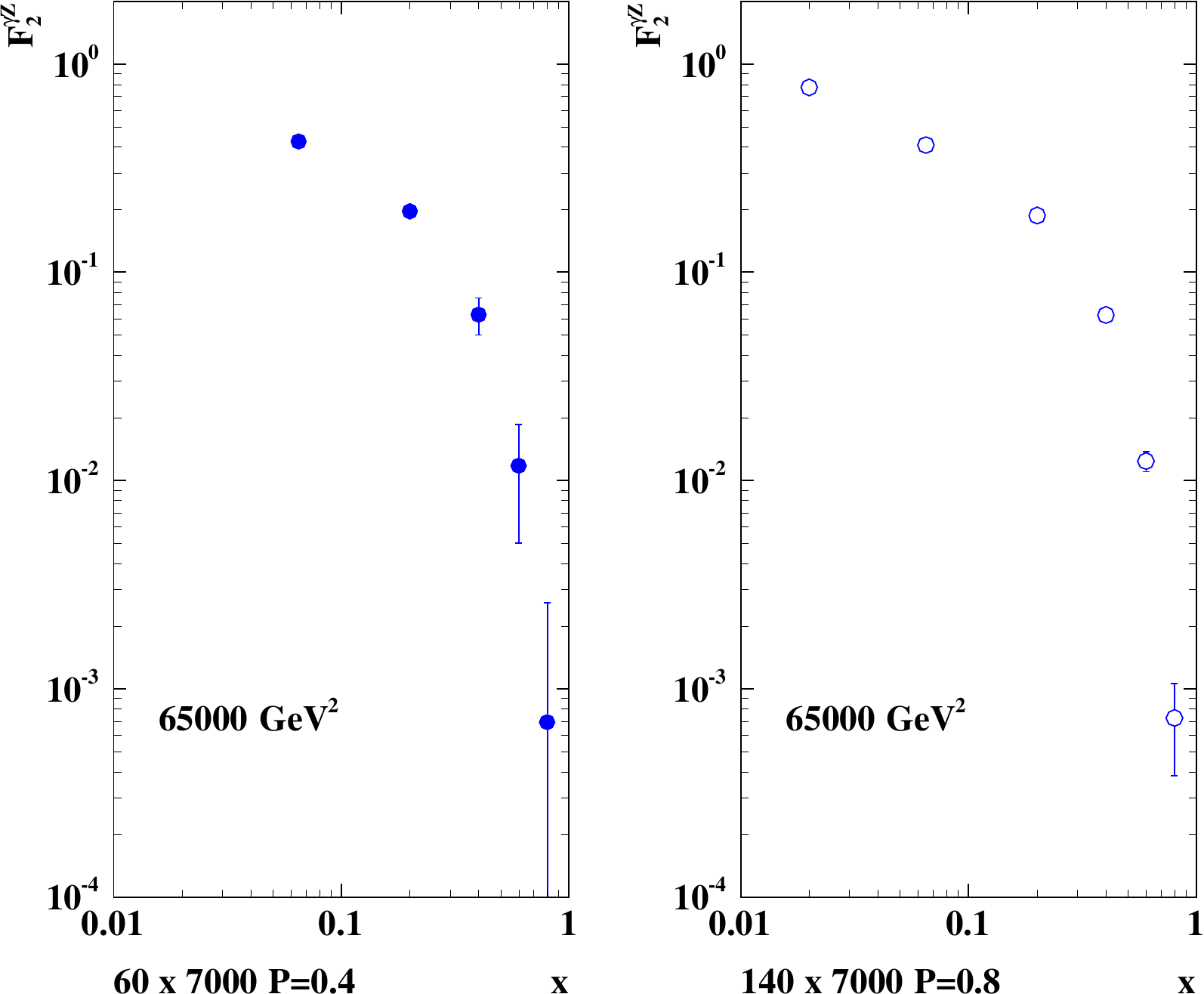}}
\vspace{0.3cm}
\caption{Simulation of the measurement of the $\gamma Z$ interference
structure function $F_2^{\gamma Z}$, shown as a function of $x$
for a typical high $Q^2$ value, for
two LHeC configurations ($E_e=60$\,GeV and $P= \pm 0.4$, left) and
($E_e=140$\,GeV and $P = \pm 0.9$, right). The proton beam energy is $7$\,TeV
and the luminosity assumed is $10$\,fb$^{-1}$ per polarisation state.
 This function 
is a measure of parity violation and provides additional information
on the quark distributions as it is proportional to $e_q v_q$ to be
compared with $e_q^2$ in the lowest order function $F_2$.
 Shown are statistical uncertainties
only. The systematic uncertainty can be expected to be small
as in the asymmetry many effects cancel and because at the LHeC
such asymmetries are large, and the polarisation possibly controlled
at the per mille level, as is discussed in the technical part of the CDR.
}
\label{fig:g2}
\end{figure}

The structure function $F_2^{\gamma Z}$ accesses a new
combination of quark distributions and is measurable
for the first time, and with high precision, at the LHeC, see
Fig.\,\ref{fig:g2}, in which the result is shown of its possible
measurement. The remarkable precision on $F_2^{\gamma Z}$ illustrates
the huge potential in precision and range which the LHeC brings.
For the study of electroweak effects it is clearly desirable to
have the maximum beam energy and polarisation available, as 
the comparison of the two results for different beam conditions
but the same luminosity in  Fig.\,\ref{fig:g2} shows.
%

The polarisation asymmetry also permits a high precision measurement of the weak
mixing angle $\sin^2{\Theta}$ at different $Q^2$ values,
complementing the precision measurements at $M_Z^2$ made at LEP and the SLC, and
extending to lower and much higher scales, see Sect.\,\ref{sec:sint}.

\subsection{Charged current}
The inclusive polarised charged current $e^{\pm} p$  scattering  cross section
can be written as
\begin{equation}
 \label{ccsi}
 \frac{{\rm d}^2 \sigma_{CC}^{\pm}}{{\rm d}x{\rm d}Q^2} =    \frac{1\pm P}{2} \cdot
 \frac{G_F^2}{2 \pi x} \cdot \left[ \frac {M_W^2} {M_W^2+Q^2}  \right]^2 Y_+ \cdot \sigma_{r,CC}. 
\end{equation}
The reduced charged current cross section, in analogy with the NC case given in Eq.~\ref{ncsi}, is
a sum of structure function terms 
\begin{eqnarray}
 \label{ccred}
 \sigma_{r,CC}^{\pm}=
 W_2^\pm   \mp \frac{Y_-}{Y_+} xW_3^\pm - \frac{y^2}{Y_+} W_L^\pm\,.
\end{eqnarray}
In the on-mass shell scheme, the Fermi constant 
$G_F$ is defined, see for example~\cite{Arbuzov:1995id},
 using the weak boson masses as
\begin{equation}
\label{equG}
  G_F = \frac{ \pi \alpha}{ \sqrt{2} M_W^2 \sin^2 \Theta (1- \Delta r)}
\end{equation}
with $\sin^2 \Theta = 1 - M_W^2/M_Z^2$ as above.  The higher order
correction term $\Delta r$ can be approximated~\cite{Nakamura:2010zzi} as $\Delta r = 1 - \alpha / \alpha(M_Z) -
0.0094 (m_t/173 {\rm GeV})^2 / \tan^2 \Theta $, and thus introduces a
dependence of the DIS cross section on the mass of the top quark.  The
choice of $G$ above allows the CC cross section, Eq.\,\ref{ccsi}, to
be rewritten as
\begin{equation}
\label{sicc}
 \frac{{\rm d}^2 \sigma_{CC}^{\pm}}{{\rm d}x{\rm d}Q^2} =    \frac{1\pm P}{2} \cdot 
 \frac{2 \pi \alpha^2 Y_+ }{Q^4 x} \cdot \kappa_W^2 \cdot \sigma_{r,CC}, 
\end{equation}
with 
\begin{equation}
\label{kappaw}
\kappa_W (Q^2) =  \frac{Q^2}{Q^2 + M_W^2} \cdot \frac{1}{4 \sin^2 \Theta},
\end{equation}
which is convenient for the consideration of NC/CC cross section ratios.

In the QPM (where $W_L^\pm = 0$),
the structure functions represent beam charge dependent
sums and differences of quark and anti-quark distributions and are given by
\begin{eqnarray}
 \label{ccstf}
    W_2^{+}  =  x (\bU+D)\hspace{0.05cm}\mbox{,}\hspace{0.1cm}
  xW_3^{+}  =  x (D-\bU)\hspace{0.05cm}\mbox{,}\hspace{0.1cm} 
    W_2^{-}  =  x (U+\bD)\hspace{0.05cm}\mbox{,}\hspace{0.1cm}
 xW_3^{-}  =  x (U-\bD)\,.
\end{eqnarray}
Using these equations one finds
\begin{eqnarray}
\label{ccupdo}
 \sigma_{r,CC}^+ \sim x\bU+ (1-y)^2xD, ~~\\
 \sigma_{r,CC}^- \sim xU +(1-y)^2 x\bD . ~~
\end{eqnarray}
Combined with Equation~\ref{strf}, this approximately reduces to
\begin{eqnarray} \label{f23ud}
\sigma^{\pm}_{r,NC} \simeq [c_u (U+\bU) +c_d(D+\bD)] + \kappa_Z [d_u(U-\bU) + d_d (D-\bD)] \nonumber \\
 c_{u,d} = e^2_{u,d} + \kappa_Z (-v_e \mp P a_e) e_{u,d}v_{u,d} 
\hspace{0.2cm}  d_{u,d} = \pm a_e a_{u,d} e_{u,d},
\end{eqnarray}
showing that the NC and CC cross section measurements at the LHeC
determine the complete set of quark-type distributions  $U$, $D$, $\bU$ and $\bD$,
i.e. the sum of up-type, of down-type and of their anti-quark-type distributions. 
Below the $b$ quark mass threshold,
these are related to the individual quark distributions as follows
\begin{equation}  \label{ud}
  U  = u + c    ~~~~~~~~
 \bU = \bu + \bc ~~~~~~~~
  D  = d + s    ~~~~~~~~
 \bD = \bd + \bs\,. 
\end{equation}
Assuming symmetry between sea quarks and anti-quarks, 
the valence quark distributions result from 
\begin{equation} \label{valq}
u_v = U -\bU ~~~~~~~~~~~~~ d_v = D -\bD.
\end{equation}

%% file: physics/simnccc.tex
%
%
The LHeC greatly extends the kinematic range compared to HERA. The
increase in negative momentum transfer squared $Q^2$ is from
a maximum of about $0.03$ at HERA to $1$\,TeV$^2$ at the LHeC, and 
in $x$, e.g. for $Q^2 = 3$\,GeV$^2$,
from about $4 \cdot 10^{-5}$ to $2 \cdot 10^{-6}$.
The projected increase of integrated luminosity
by a factor of $100$ allows to also extend the 
kinematic range at large $x$, 
in charged currents, from practically
about $0.4$ to $0.8$. Due to the enlarged electron 
beam energy $E_e$ the range of high inelasticity
$y \simeq 1 - E_e'/E_e$ 
should extend closer to $1$. A reduced
noise in the calorimeters may allow to reach lower
values of $y$ than at HERA, also because the hadronic
$y$ is determined as the sum over $E-p_z$ divided by
twice the (LHeC enhanced) electron beam energy.
Very recently it has been observed by H1 that the
reconstruction of the hadronic final state with jets
rather than the sum of all hadronic energy depositions
allows better control 
of the low $y$ region, i.e. scattering close
to the beam pipe. At the LHeC these jets are extremely
energetic and it would be expected, subject to detailed
simulation studies at a later stage of the project,
that kinematic reconstruction for
values of $y$ down to 0.001 or even below
could be trusted.

While the extensions 
of kinematic coverage and improvements of statistical
precision are impressive, an estimate of the impact
of LHeC NC and CC cross section measurements on
derived quantities such as structure functions and 
parton distributions also requires an estimate of  the
expected systematic measurement precision, as may be 
achieved with the detector described in  
Chapter\,\ref{LHEC:MainDetector}.
In the following the assumptions
and simulation results are presented for the
NC and the CC cross sections, which are subsequently used
in QCD fits and other analyses throughout this report.
%
%

The systematic uncertainties of the DIS cross sections
have a number of sources, which at HERA have broadly
been classified as uncorrelated and correlated across bin 
boundaries.  For the NC case,
the uncorrelated sources, apart from data and Monte
Carlo statistics, are a global efficiency uncertainty,
due for example to tracking or electron identification
errors, photoproduction background, calorimeter noise
and radiative corrections.
The correlated uncertainties result from
imperfect energy scale and angle calibrations.
In the classic kinematic reconstruction methods used
here, and described in Sect.\,\ref{LHEC:Detector:Requirements}
the scattered electron energy $E_e'$ and
polar angle $\theta_e$ are used, complemented by the energy of the
hadronic final state $E_h$~\footnote{
Briefly, $Q^2$ is best determined with the electron
kinematics and $x$ is calculated from $y=Q^2/sx$. At large
$y$ the inelasticity is essentially
measured with the electron energy $y_e \simeq 1 - E_e'/E_e$.
At low $y$ the relation $y_h = E_h \sin^2(\theta_h/2)/E_e$
is used, with the hadronic final state energy $E_h$ and
angle $\theta_h$, resulting in $\delta y_h /y_h \simeq \delta E_h/E_h$
to good approximation. There have been various refined methods
proposed to determine the DIS kinematics, such as the
double angle method or the so-called sigma method. For the
estimate of the cross section uncertainty behaviour as
functions of $Q^2$ and $x$, however, the simplest
method using $Q^2_e, y_e$ at large $y$ and 
$Q^2_e, y_h$ at low $y$ is transparent and accurate
to better than a factor of two. In much of
the phase space, moreover, it is rather the uncorrelated
efficiency or further specific errors than the kinematic correlations, 
which dominate the cross section measurement precision.}.
The correlated errors are due to scale uncertainties
of the electron energy $E_e'$ and of the hadronic
final state energy $E_h$. There are also systematic 
errors due to an uncertainty of the measurement
of the electron polar angle $\theta_e$. The assumptions used
in the simulation of pseudodata are summarised in
Table\,\ref{tab:sys}.
\begin{table}[h]
  \centering
  \begin{tabular}{|l|c|}
    \hline
source of uncertainty & error on the source or cross section \\ \hline
scattered electron energy scale $\Delta E_e' /E_e'$ & 0.1 \% \\
scattered electron polar angle  & 0.1\,mrad \\
hadronic energy scale $\Delta E_h /E_h$ & 0.5\,\% \\
calorimeter noise (only $y < 0.01$) & 1-3\,\% \\ 
radiative corrections & 0.5\% \\
photoproduction background (only $y > 0.5$) & 1\,\% \\
global efficiency error & 0.7\,\%  \\
 \hline
  \end{tabular}
\caption{
Assumptions used in the simulation of the NC cross sections
on the size of uncertainties from various sources. 
These assumptions correspond to typical best values
achieved in the H1 experiment. Note that in the cross section
measurement, the energy scale and angular uncertainties
are relative to the Monte Carlo and not to be confused with
resolution effects which determine the purity and stability
of binned cross sections. The total cross section error
due to these uncertainties, e.g. for $Q^2 = 100$\,GeV$^2$,
is about $1.2$, $0.7$ and $2.0$\,\% for $y=0.84,~0.1,~0.004$.
}
\label{tab:sys}
\end{table}

In the absence of a detailed detector simulation at this stage,
the systematic NC cross uncertainties due to $E_e'$, $\theta_e$
and $E_h$ are calculated, following~\cite{Blumlein:1992we},
from the derivatives of the NC cross section in the chosen bins
taking into account the Jacobians where needed. The results
have been compared, for the HERA kinematics, with the H1
MC simulation of systematic errors~\cite{mksys09} and found
to be in very good agreement for all three sources.
The resulting error depends much on the kinematics.
At low $Q^2$, for example, the systematic cross section
error due to the uncertainty of $\theta_e$ rises because
of $\delta Q^2 /Q^2 = \delta E_e' /E_e' \oplus
 \tan{(\theta_e/2)} \cdot \delta \theta_e $ while at high
$Q^2$ it is negligible. Low $Q^2$ is the backward
region, of large electron scattering angles with respect
to the proton beam direction.

\begin{figure}
\centerline{\includegraphics[clip=,angle=90.,width=.9\textwidth]{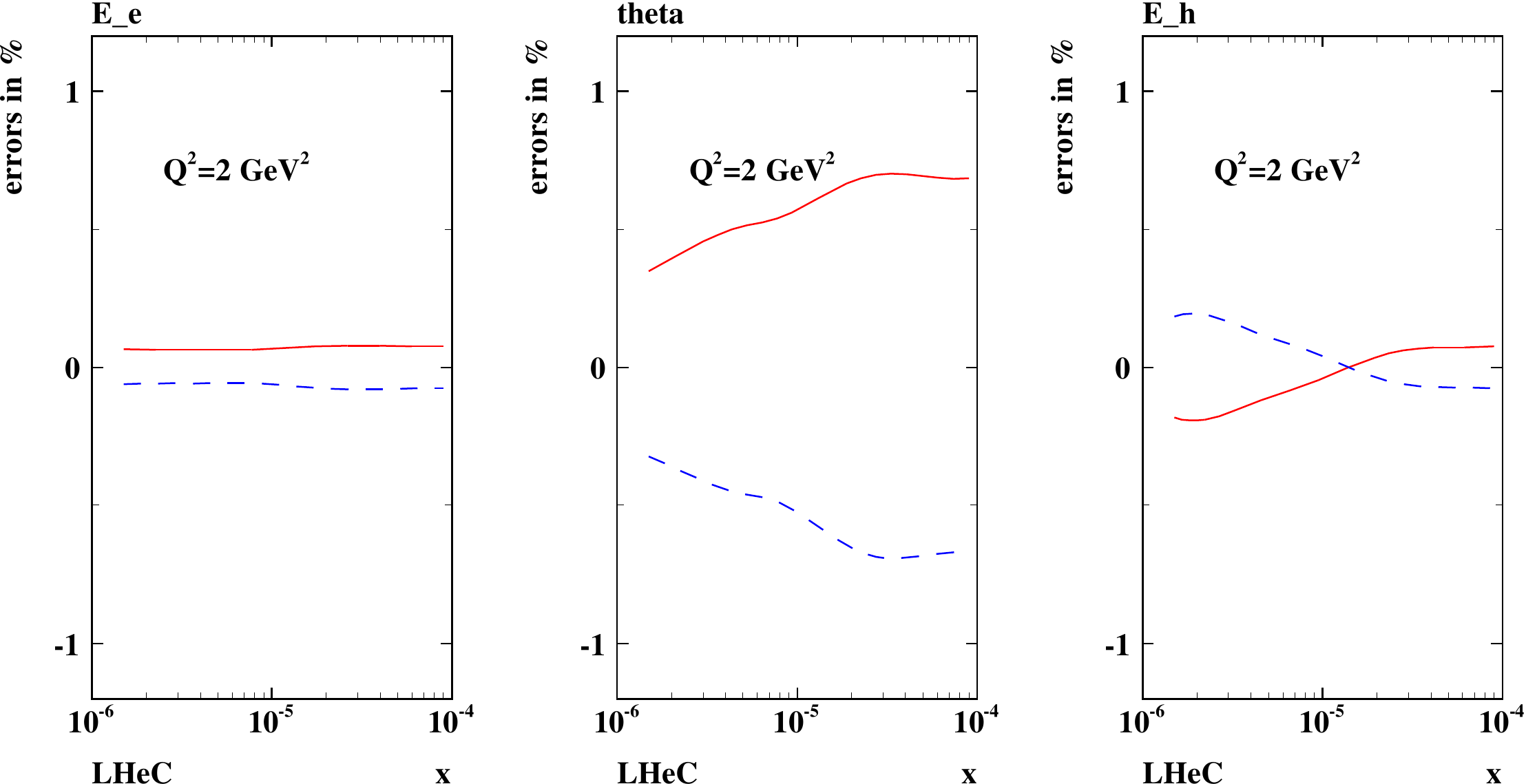}}
\caption{Neutral current cross section errors,
calculated for $60 \times 7000$\,GeV$^2$, resulting from scale
uncertainties of the scattered electron energy $\delta E_e' / E_e' = 0.1$\,\%, 
of its polar
angle $\delta \theta_e =0.1$\,mrad and 
the hadronic final state energy $\delta E_h /E_h =0.5$\,\%, at
low $Q^2 = 2$\,GeV$^2$ and correspondingly low $x$.  
}
\label{fig:sysq2}
\end{figure}

\begin{figure}
\centerline{\includegraphics[clip=,angle=90.,width=0.9\textwidth]{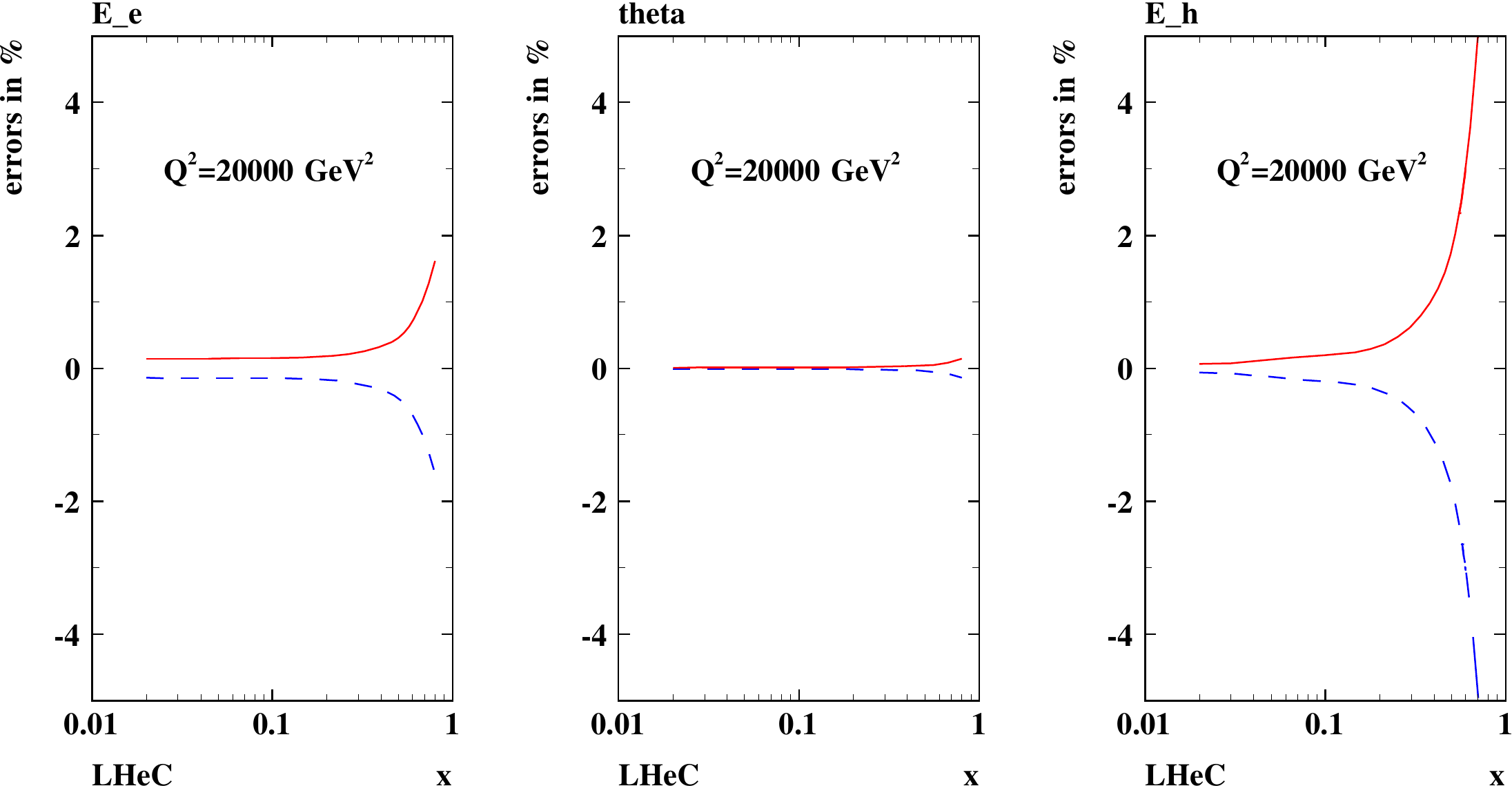}}        
\caption{Neutral current cross section errors, 
calculated for $60 \times 7000$\,GeV$^2$ unpolarised $e^-p$ scattering,
resulting from scale
uncertainties of the scattered electron energy $\delta E_e' / E_e' = 0.1$\,\%, 
of its polar
angle $\delta \theta_e =0.1$\,mrad and
the hadronic final state energy $\delta E_h /E_h =0.5$\,\%, at
large $Q^2 = 20000$\,GeV$^2$ and correspondingly large $x$. Note that the
characteristic behaviour of the relative uncertainty at large $x$, i.e.
to diverge
$\propto 1/(1-x)$, is independent of $Q^2$, i.e. persistently observed
at $Q^2 = 200000$\,GeV$^2$ for example too.
}
\label{fig:sysq20000}
\end{figure}

\begin{figure}
\centerline{\includegraphics[clip=,angle=0.,width=0.95\textwidth]{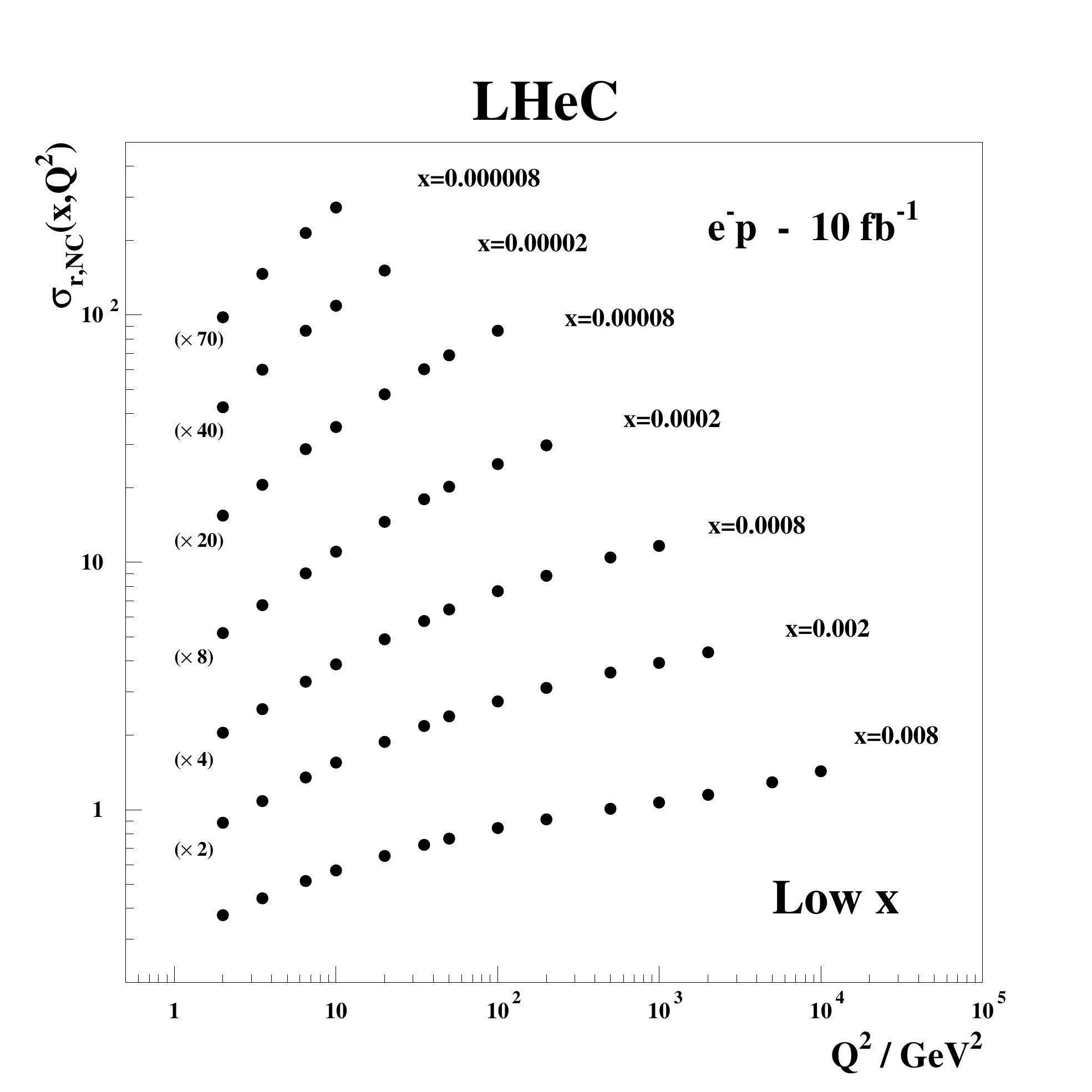}}
\caption{Simulated neutral current, inclusive reduced cross section measurement,
for an integrated luminosity of $10$\,fb$^{-1}$, in unpolarised
$e^-p$ scattering at $E_e=60$ and $E_p = 7000$\,GeV. The
DIS cross section is measurable at unprecedented precision
and range. The uncertainty is about or below $1$\,\% and thus
not visible on this plot. 
Departures from the strong rise of
the reduced cross section, $\sigma_r \simeq F_2$, at very low $x$
and $Q^2$ are expected to appear due to non-linear gluon-gluon
interaction effects in the so-called saturation region. 
}
\label{fig:sigpmlx}
\end{figure}

\begin{figure}
\centerline{\includegraphics[clip=,angle=0.,width=0.95\textwidth]{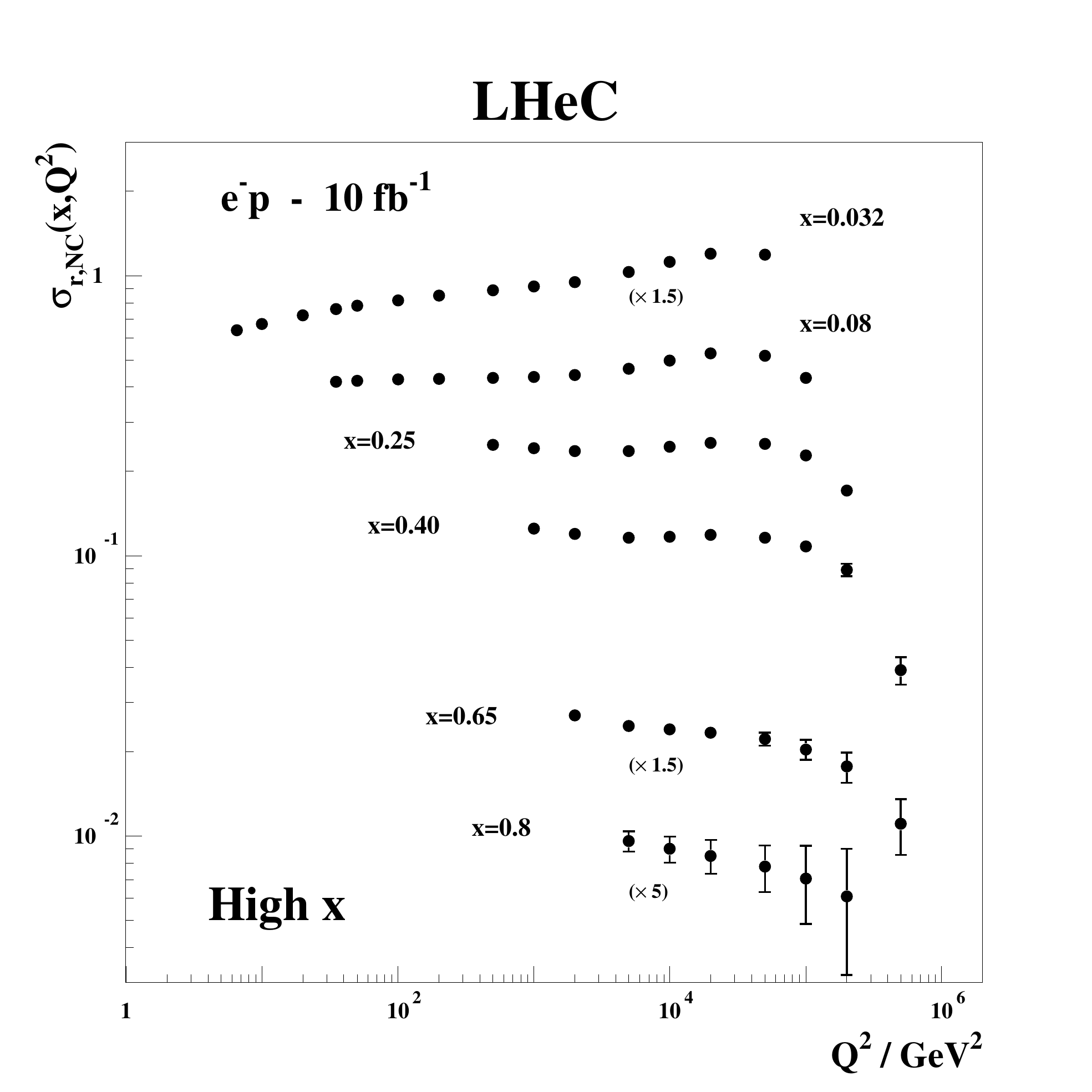}}
\caption{Simulated neutral current, inclusive reduced cross section measurement,
for an integrated luminosity of $10$\,fb$^{-1}$, in unpolarised
$e^-p$ scattering at $E_e=60$ and $E_p = 7000$\,GeV. The
DIS cross section is measurable at unprecedented precision
and range. Plotted is the total uncertainty which, where visible
at high $x$ and $Q^2$, is dominated by the statistical error.
Similar data sets are expected with different beam polarisations
and charges, and in CC scattering, for $Q^2 \geq 100$\,GeV$^2$.
The strong variations of $\sigma_r$ with $Q^2$, as at $x=0.25$,
are due to the effects of $Z$ exchange as is discussed and
illustrated subsequently.
}
\label{fig:sigpmhx}
\end{figure}

 A particular challenge is the
measurement at large $x$ because the
cross section varies as $(1-x)^c$, with $c \simeq 3$,
and thus the relative error is amplified $\propto 1/(1-x)$ as
$x$ approaches $1$. At high $x$ the hadronic final
state is scattered into the forward 
detector region where  the
energy calibration becomes challenging. 
The calculated correlated
NC cross section errors are illustrated in Figs.\,\ref{fig:sysq2}
and \ref{fig:sysq20000} for $Q^2 =2$ and $20000$\,GeV$^2$,
respectively. In the detector chapter these calculations
have been taken to define approximate requirements
on the scale calibrations in the different detector regions.
An example for the resulting cross section measurement is displayed
in Fig.\,\ref{fig:sigpmlx} for low $x$ and in Fig.\,\ref{fig:sigpmhx} for large $x$ .

\begin{figure}
\centerline{\includegraphics[clip=,angle=0.,width=1.0\textwidth]{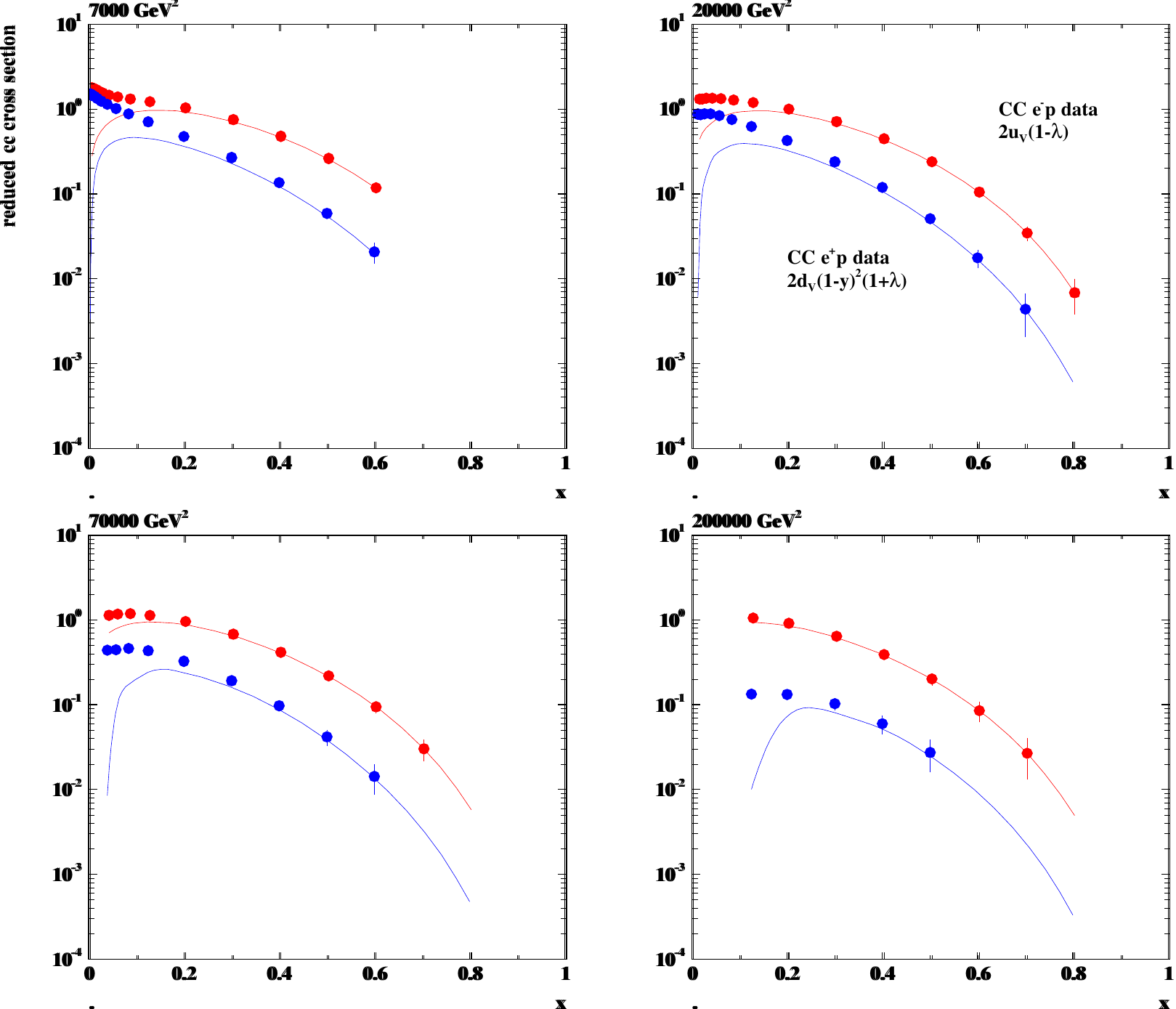}}
\vspace{0.01cm}
\caption{
Reduced charged current cross sections with statistical uncertainties
corresponding to $1$\,fb$^{-1}$ electron (top data points, red) 
and positron (lower data points, blue) proton scattering at the LHeC, 
The curves are determined by the dominant valence quark
distributions, $u_v$ for $e^-p$ and $d_v$ for $e^+p$. 
In the simulation the lepton polarisation is taken to be zero. 
The valence-quark approximation of the reduced cross section 
is seen to hold  at $x \geq 0.3$. A precise determination 
of the $u/d$ ratio up to large $x$ appears to be feasible 
at very high $Q^2$.
}
\label{fig:ccpm}
\end{figure}

For the CC case, a similar simulation was done, albeit with less
numeric effort. An illustration of the high precision and
large range of the inclusive CC cross section measurements
is presented in Fig.\,\ref{fig:ccpm}.
The systematic cross section error, based on the
H1 experience, was set to $2$\,\% and for larger $x > 0.3$ 
a term was added to allow the error
to rise linearly to $10$\,\% at $x=0.9$.
For both NC and CC cross sections the statistical error is given 
by the number of events but limited to $0.1$\,\% from below.
With these error assumptions a number of data sets was 
simulated, both for NC and CC, which is summarised in 
Table\,\ref{tab:datasets}. The energies of these sets
had been chosen prior to the final baseline energy choice.
For the simulation of the $F_L$ measurement, described below,
a separate set of beam energies is considered.
%
%
%
\begin{table}[h]
  \centering
  \begin{tabular}{|c|r|c|c|c|c|c|}
    \hline
Set & $E_e$/GeV & $E_N$/TeV & N & $L^+$/fb$^{-1}$  & $L^-$/fb$^{-1}$ & Pol \\ \hline
A   &  20       &     7     & 7 &   1            &    1          & 0   \\
B   &  50       &     7     & 7 &   50           &   50          & 0.4   \\
C   &  50       &     7     & 7 &   1            &    1          & 0.4   \\
D   &  100      &     7     & 7 &   5            &   10          & 0.9   \\
E   &  150      &     7     & 7 &   3            &    6          & 0.9   \\
F   &  50       &     3.5   & 7 &   1            &    1          & 0   \\
G   &  50       &     2.7   & 7 &   0.1          &  0.1          & 0.4   \\
H   &  50       &     1     & 7 &   -            &    1          & 0   \\ 
 \hline
  \end{tabular}
\caption{Conditions for simulated NC and CC data sets for LHeC physics studies. Here,
A defines a low electron beam energy option which is of interest to reach lowest $Q^2$ because
$Q^2_{min}$ decreases $\propto E_e^{-2}$; B is the standard set, with a total luminosity
split between different polarisation and charge states. C is a lower luminosity
version which was considered in case there was a need for a dedicated low/large 
angle acceptance configuration, which according to more recent findings
could be avoided since the luminosity in the restricted acceptance configuration
is estimated, from the $\beta$ functions obtained in the optics design,
to be half of the luminosity in the full acceptance 
configuration; D is an intermediate energy linac-ring version, while E is the
highest energy version considered, with the luminosities as given. It is likely
that the assumptions for D and E on the positron luminosity are
a bit optimistic. However, even with twenty times lower positron than electron luminosity
one would have $0.5$\,fb$^{-1}$, i.e. the total HERA luminosity equivalent available
in option D for example.  F is the deuteron
and G the lead option; finally H was simulated for a low proton beam energy
configuration as is of interest to maximise the acceptance at large $x$.
}
\label{tab:datasets}
\end{table}

%% file: physics/flong.tex
%
%
The inclusive, deep inelastic electron-proton scattering 
cross section at low $Q^2$,
\begin{equation}
 \frac{d^2\sigma}{dxdQ^2} =    \frac{2\pi \alpha^2 Y_+}{Q^4 x}  
   [ F_2(x,Q^2) - f(y) \cdot F_L(x,Q^2)],
       \label{sig}
  \end{equation}  
is defined by two proton structure functions, $F_2$ and $F_L$ with
$y=Q^2/sx$, $Y_+ = 1+ (1-y)^2$ and $f(y)=y^2/Y_+$. 
The two functions reflect the transverse and the
longitudinal polarisation state of the virtual photon probing the
proton structure, i.e. $F_T=F_2 -F_L$ and $F_L$, respectively.
The positivity of the transverse and longitudinal cross sections
requires $0 \leq F_L \leq F_2$. Since for most of the kinematic
range the $y$ dependent factor $f(y)$ is very small, there follows
that $F_L$ causes in most of the kinematic range
only a small correction to the reduced cross
section, which is governed by $F_2$, apart from the 
region of maximum $y$. At small $x$, the inelasticity is
given as $y \simeq 1 - E_e'/E_e$. Therefore,
in order to extract $F_L$, DIS 
has to be measured extremely precisely
at small scattered lepton energies.
Quite how small depends on
how large $E_e$ is, with related experimental difficulties being 
how to trigger on these events and how to control the
background from particle production at low energies.
A variation of the beam energies is required to separate
the two functions measured at the same $x$ and $Q^2$
by variation of $y=Q^2/sx$.

A first measurement of $F_L$ at low $x$ at HERA
has recently been performed by the ZEUS 
Collaboration~\cite{Chekanov:2009na}
and by the H1 Collaboration~\cite{Collaboration:2010ry}.
For the study of the gluon distribution at lowest $x$,
the H1 data are crucial as only H1 has measured $F_L$
below $Q^2$ of about $10$\,GeV$^2$ owing to their backward
detector constellation upgraded in the nineties.
The $F_L$ measurement at HERA was performed towards the
end of the accelerator operation and could only extend over
a period of  three months with about $10$\,pb$^{-1}$
of integrated luminosity spent at two reduced proton beam 
energies, $450$ and $565$\,GeV, besides the nominal $920$\,GeV.
The H1 result is consistent with pQCD predictions.
The ratio $R=F_L/(F_2-F_L)$ has been found to be
independent of $x$ and $Q^2$ to a precision of $20$\,\%, i.e.
 $R = 0.26 \pm 0.05$~\cite{Collaboration:2010ry}. This
interesting relation deserves a more precise investigation and 
may break when the region of saturation is entered
at lower $x$ than HERA could access.

The LHeC will extend this initial measurement by 
using higher luminosities and dedicated detector conditions
into a much enlarged kinematic range.  
Since the LHeC will run synchronously with the LHC,  the simulation presented
here has been made with reduced electron beam energies
keeping the proton beam energy untouched.
The following set of energies and integrated luminosities have been used:
(60, 1), (30, 0.3), (20, 0.1) and (10, 0.05) (GeV, fb$^{-1}$).
Note that the $F_L$ measurement also requires
data with the opposite beam charge in order to be able
to reliably subtract the non DIS background which at high $y$
is substantial. This has not been simulated here.

In the low $x$ studies below, a similar simulation was
used for which the luminosity assumptions were similar but 
a set of reduced proton beam energies was considered.
The advantage of lowering $E_p$ is that the maximum $y$
for all beam energy configurations can be high, e.g.
$0.95$ for $E_e=60$\,GeV. When $E_e$ is lowered instead,
a lower $y_{max}$ is achieved, as below a few GeV
of energy the background is too high for a reliable measurement
to be performed. The results of both $F_L$ simulations,
with reduced $E_e$ or $E_p$, come out to be very similar.

\begin{figure}[htbp]
\includegraphics[clip=,angle=90.,width=1.0\textwidth]{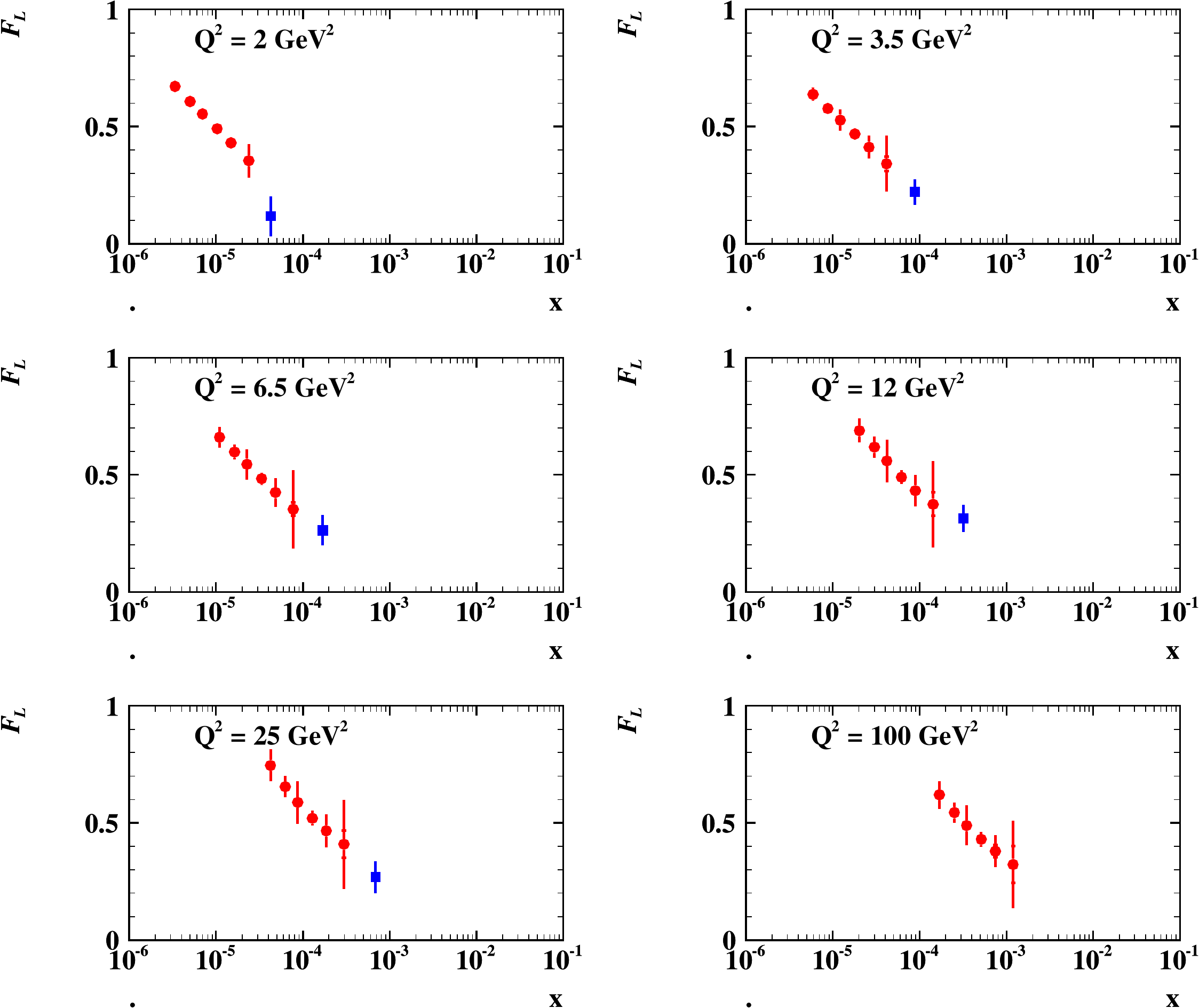}
\vspace{0cm}
\caption{Simulated measurement of the longitudinal structure 
function $F_L(x,Q^2)$ at the LHeC (red closed circles)
from a series of runs with
reduced electron beam energy, see text. The inner error
bars denote the statistical uncertainty, the outer error
bars are the total errors with the additional
uncorrelated and correlated systematic uncertainties added
in quadrature. The blue squares denote the recently published
result of the H1 Collaboration, plotting only
the $x$ averaged results as the more accurate ones,
see \cite{Collaboration:2010ry}. The LHeC extends the
measurement towards low $x$ and high $Q^2$ (not fully
illustrated here) with much improved precision.
}
   \label{fig:flong}
\end{figure}

The result of the simulation study is shown in 
Fig.\,\ref{fig:flong}. The technique applied is the conventional
separation of $F_2$ and $F_L$ by fitting a straight line
to the various reduced cross section data points at fixed $Q^2$
and $x$ with $f(y)$ as the parameter and separating the 
uncorrelated from the correlated systematic uncertainties which
partially cancel in such an analysis. The expected precision
on $F_L$ is typically $4$\,\% at $Q^2$ of $3.5$\,GeV$^2$
or $7$\,\% at $Q^2$ of $25$\,GeV$^2$ at a number of points
in $x$, usually with similar contributions from the
calculated correlated and the assumed uncorrelated
systematic uncertainties, with statistics being less important
until $Q^2 \geq 100$\,GeV$^2$.
The LHeC thus will provide the first precision measurement
of $F_L(x,Q^2)$ ever. The covered kinematic region
is of particular importance for testing QCD at extremely low
Bjorken $x$. When analysed jointly with the $F_2$ behaviour,
it will become possible to solve the question of whether
the gluon is negative or valence like at low $Q^2$. If
a saturation of the rise of the gluon density towards low $x$
occurs, it will not be missed with such a precision measurement. 

A related measurement of prime interest is the determination
of $F_L$ in diffraction, as is discussed in Section\,\ref{sec:diffpdfs}.

%
%
 
%

%% file: physics/dirpart.tex
Despite a series of deep inelastic scattering experiments with 
neutrinos, electrons and muons using stationary targets
and with HERA, and despite the addition of some
Drell Yan data, the knowledge of the quark distributions
in the proton is still limited. It often relies on pQCD
analyses using various assumptions on the Bjorken $x$ dependence
of the PDFs and their symmetries. The LHeC has the
potential to put the PDF knowledge on a qualitatively and
quantitatively new and superior basis. This is due to the
kinematic range, huge luminosity, availability of polarised
electron and positron beams, both proton and deuteron beams,
and to the anticipated very high
precision of the cross section measurements, as
discussed above. 

The LHeC has the potential to provide crucial constraints on
the parton distributions and determine them
completely, and to certain extent
independently of the conventional QCD fitting techniques.
For example, the valence quarks can be measured up to high  $x$,
and all heavy quark distributions, $s,~c,~b$ and $t$,
 can be  determined from dedicated
$c$ and $b$ tagging analyses with unprecedented precision.
Therefore, the QCD fits, which will necessarily evolve with and be 
based on real LHeC data, will be set-up with a
massively improved and better constrained input data base. 
Their eventual effect is thus not easy to simulate now,
and yet it may be illustrated based on the currently used 
procedures.

The striking potential of the determination of the quark
and gluon distributions will be discussed and illustrated
below.  For the various PDFs,
the current knowledge is illustrated with a series
of plots based on the world's best PDF determinations
available today. Simulations of essentially direct
quark distribution measurements, as for the charm quark,  will be shown.
Moreover, a consistent set of standard QCD fits has been
performed using the simulated LHeC and further data which is first
described in what follows. This is used to illustrate the
effect the inclusive NC and CC data from the LHeC are expected to have on
the PDF uncertainties.

Currently extensive work is being
performed to test and further constrain PDFs with Drell-Yan scattering
data from the LHC. This naturally focuses first 
on the $Z$ and $W^{\pm}$ production
and decay and will be extended to lower
and higher mass di-lepton production.
While such tests are undoubtedly of interest, they 
require an extremely high level of precision as at scales
$Q^2 \sim M_{W,Z}^2$ any effect due to PDF
differences at smaller scales is 
washed out by the overriding effect of quark-antiquark pair production 
from gluon emission, below the valence quark region.
Some of the present QCD fit results also use a set of simulated
$W^+ - W^-$ asymmetry data of ultimate precision
in order to be able to estimate the effect the Drell-Yan data
will have besides the LHeC in the determination of the
PDFs.  A brief study has also been made which 
illustrates the effect of the $W,~Z$ data, of ATLAS,
and of the LHeC on the determination of PDFs when
a maximum number of constraints, 
otherwise default to HERA analyses,
is released.

\subsection{QCD fit ansatz}
\label{sec:qcdfita}
NLO QCD fits are performed in order to study the effect
of the (simulated) LHeC data on the PDF knowledge.
Fits are done using the combined HERA data,
which is published and so available to date (HERA I), adding BCDMS proton
data as the most accurate fixed target structure function dataset of
importance at high $x$, simulated precision $W^+-W^-$ asymmetry
LHC data,  using the LHeC data alone and in combination.
In the fits, for the central values of the LHeC data,  the
Standard Model expectation is used,
smeared within the uncorrelated, Gaussian distributed uncertainties 
and taking into account the correlated  uncertainties as well.

The procedure used here is adopted from the HERA QCD fit
analysis~\cite{:2009wt}.
The QCD fit analysis to extract the proton's PDFs is 
performed imposing a $Q^2_{min}=3.5$~GeV$^2$  to restrict it
to the region where perturbative QCD can be assumed to be valid.
The fits are extended to lowest $x$ for systematic uncertainty
studies, even when at such low $x$ values non-linear
effects are expected to appear.

The fit procedure consists first in parameterising PDFs at a 
starting scale  $Q^2_0=1.9~ \rm GeV^2$, chosen to be below the charm mass threshold.
The parameterised PDFs are the valence distributions
$xu_v$ and  $xd_v$,  the gluon distribution $xg$, and the 
$x\bar{U}$ and $x\bar{D}$ distributions, where $x\bar{U} = x\bar{u}$, 
$x\bar{D} = x\bar{d} +x\bar{s}$. This ansatz is natural to the
extent that the NC and CC inclusive cross sections determine
the sums of up and down quark distributions, 
and their antiquark distributions, as the four independent
sets of PDFs, which may be transformed to the ones chosen
if one assumes $u_v = U -\overline{U}$ and $d_v = D - \overline{D}$,
i.e. the equality of anti- and sea quark distributions of given flavour.

The following standard functional form is used to parameterise them 
\begin{equation}
 xf(x) = A x^{B} (1-x)^{C} (1 + D x + E x^2),
\label{equ:pdf}
\end{equation}
where the normalisation parameters ($A_{uv}, A_{dv}, A_g$)  
are constrained by  quark counting and momentum  sum rules. 

The parameters  $B_{\bar{U}}$ and $B_{\bar{D}}$ are set equal,
 $B_{\bar{U}}=B_{\bar{D}}$, such that 
there is a single $B$ parameter for the sea distributions,
an assumption the validity of which will be settled with the LHeC.
The strange quark distribution  at the starting scale  
is  assumed to be a constant fraction of $\bar{D}$, 
$x\bar{s}= f_s  x\bar{D}$,
chosen to be $f_s=0.31$.
In addition, to ensure that $x\bar{u} \to x\bar{d}$ 
as $x \to 0$,  
$A_{\bar{U}}=A_{\bar{D}} (1-f_s)$.
The $D$ and $E$ are introduced one by one until no further
 improvement in $\chi^2$ is found.
The best fit  resulted in a total of $10$ free parameters~\cite{:2009wt},
while fits with a tested set of $14$ parameters lead
to very similar results.
As discussed above this will change considerably when
the LHeC data become available and
more flexible parameterisations and methods  can be tested.
This has been studied to some extent in the simulation
for $\alpha_s$ presented below.

The PDFs are then evolved using DGLAP evolution 
equations~\cite{qcdnum} at NLO in the $\overline{MS}$ scheme with the
renormalisation and factorisation scales set to $Q^2$
using standard sets of parameters as for $\alpha_s(M_Z)$.
These, as well as the exact treatment of the heavy quark
thresholds, have no significant influence on the
estimates of the PDF uncertainties to which the
subsequent analysis is only directed.
The experimental uncertainties on the PDFs 
are determined using the  $\Delta\chi^2=1$ criterion.
\subsection{Valence quarks}
\label{sec:valquarks}
\begin{figure}[htbp]
\centerline{\includegraphics[clip=,width=0.8\textwidth]{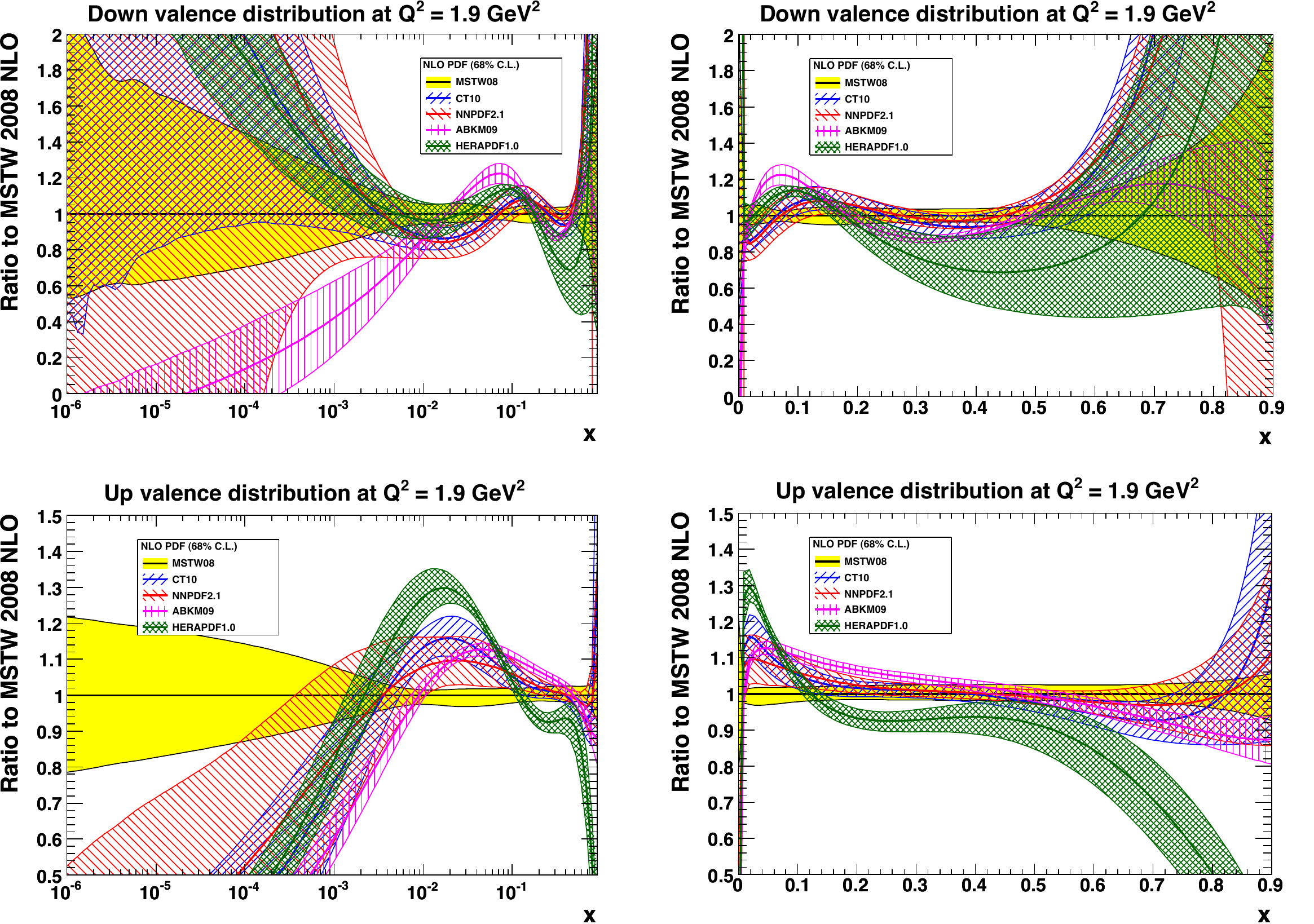}}
\caption{Ratios (to MSTW08) and uncertainty bands of 
valence quark distributions, at
$Q^2 = 1.9$\,GeV$^2$, for most of the available recent PDF determinations.
Top: up valence quark;
down: down valence quark; left: logarithmic $x$, right: linear~$x$.
}
   \label{fig:graval}
\end{figure}

\begin{figure}[htbp]
\centerline{\includegraphics[clip=,width=0.8\textwidth]{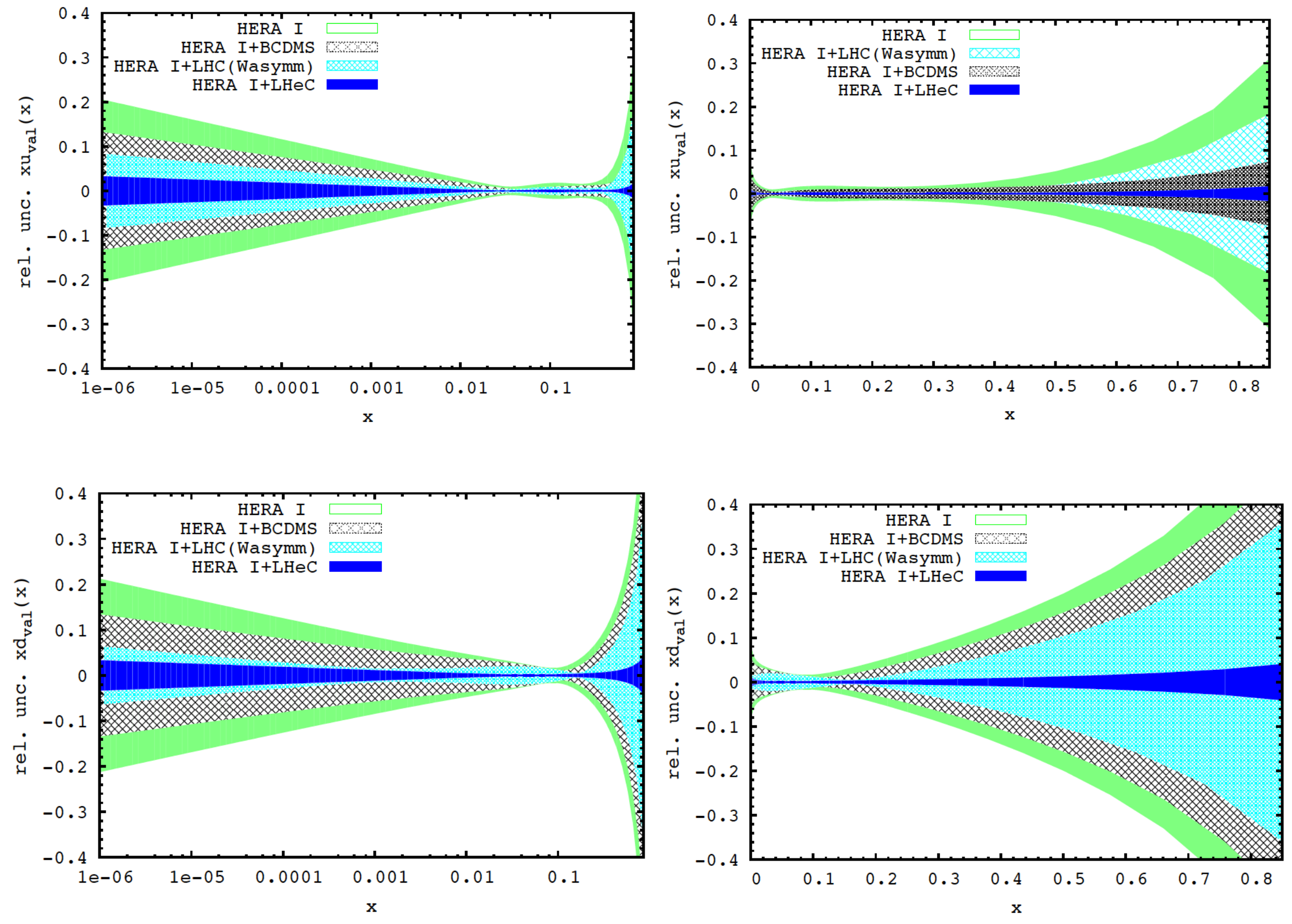}}
\caption{Uncertainty of valence quark distributions, at
$Q^2 = 1.9$\,GeV$^2$, as resulting from an NLO QCD fit to
 HERA (I) alone (green, outer), HERA and BCDMS (crossed),
HERA and LHC (light blue, crossed)
and the LHeC added (blue, dark). Top: up valence quark;
down: down valence quark; left: logarithmic $x$, right: linear~$x$.
}
   \label{fig:voival}
\end{figure}

\begin{figure}[htbp]
\centerline{\includegraphics[clip=,width=0.9\textwidth]{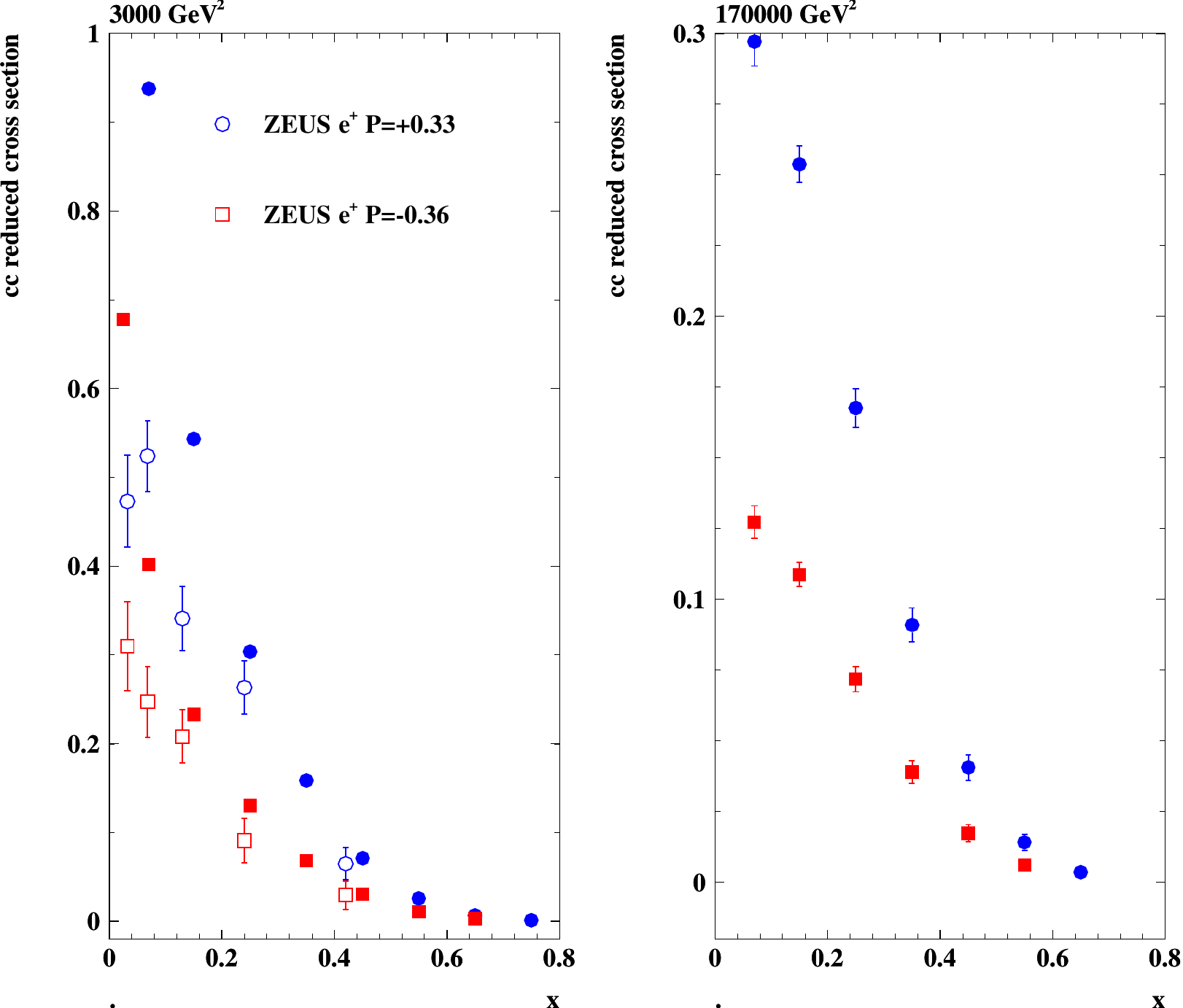}}
\caption{Reduced charged current $e^+p$ scattering cross section
versus Bjorken $x$ for different polarisations $\pm P$ and values of $Q^2$.
Closed points: LHeC simulations
for $10$\,fb$^{-1}$; open points: ZEUS measurements based on
the full HERA statistics of about $0.15$\,fb$^{-1}$ per polarisation
state. Note that the reduced CC cross section at fixed $x$ and $Q^2$ 
contains an explicit dependence on the beam energy via
the ratio of inelasticity dependent factors  $Y_-/Y_+$, which is at
the origin of the simulated and measured cross section differences
apparent at lower $x$. 
}
   \label{fig:supCC}
\end{figure}

The knowledge of the valence quark distributions, both
 at large and at low Bjorken $x$,
as derived in the current world data QCD fit analyses is amazingly 
limited, as is illustrated in Fig.\,\ref{fig:graval} from a comparison
of the leading determinations of PDF sets. 
This has to do, at high $x$, with the limited luminosity,
challenging systematics rising $\propto 1/(1-x)$ and
nuclear correction uncertainties, and, at low $x$, with the smallness
of the valence quark distributions as compared to the sea quarks.
The impressive
improvement expected 
from the LHeC is demonstrated in Fig.\,\ref{fig:voival}.
As can be seen, the uncertainty of the down valence quark
distribution at, for example, $x=0.7$ is reduced from
a level of   $50 - 100$\,\% 
to about $5$\,\%. The up valence quark
distribution is better known than $d_v$,
because it enters with a four-fold weight
in $F_2$, due to the electric quark charge ratio squared, and yet a big
improvement is also visible. These huge improvements
at large $x$ are a consequence of the high precision 
measurements of the NC and the
CC inclusive cross sections, which at high $x$ tend to
$4 u_v + d_v$ and $u_v$ ($d_v$) for electron (positron) scattering,
respectively. At HERA the luminosity and range had not been high enough
to allow a similar measurement as will be possible for the
first time with the LHeC. This is illustrated in Fig.\,\ref{fig:supCC}
which compares recent results of the ZEUS Collaboration, 
on the CC cross section with the LHeC simulation.

\begin{figure}[htbp]
\centerline{\includegraphics[clip=,angle=0.,width=0.8\textwidth]{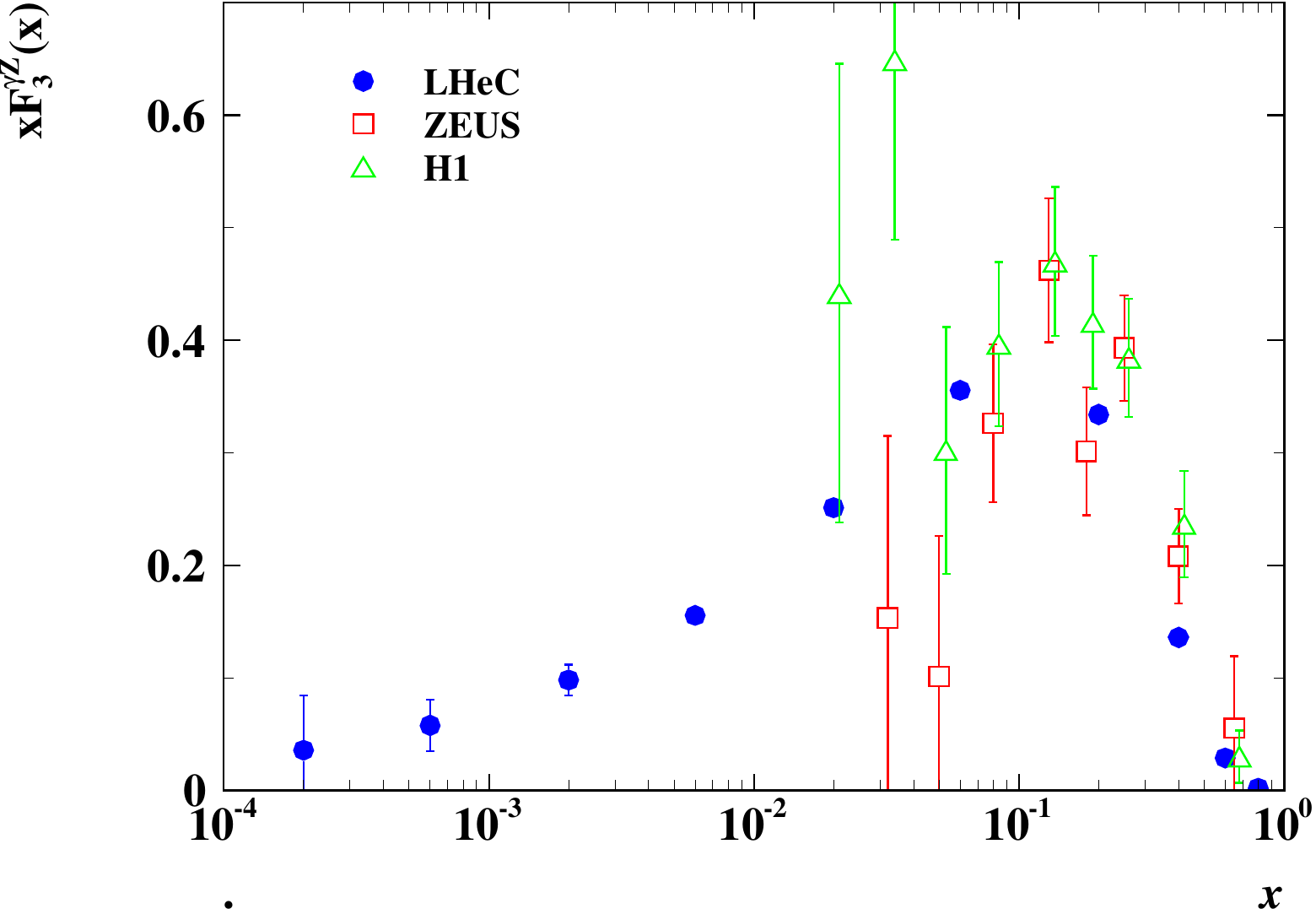}}
\vspace{0.3cm}
\centerline{\includegraphics[clip=,angle=0.,width=0.8\textwidth]{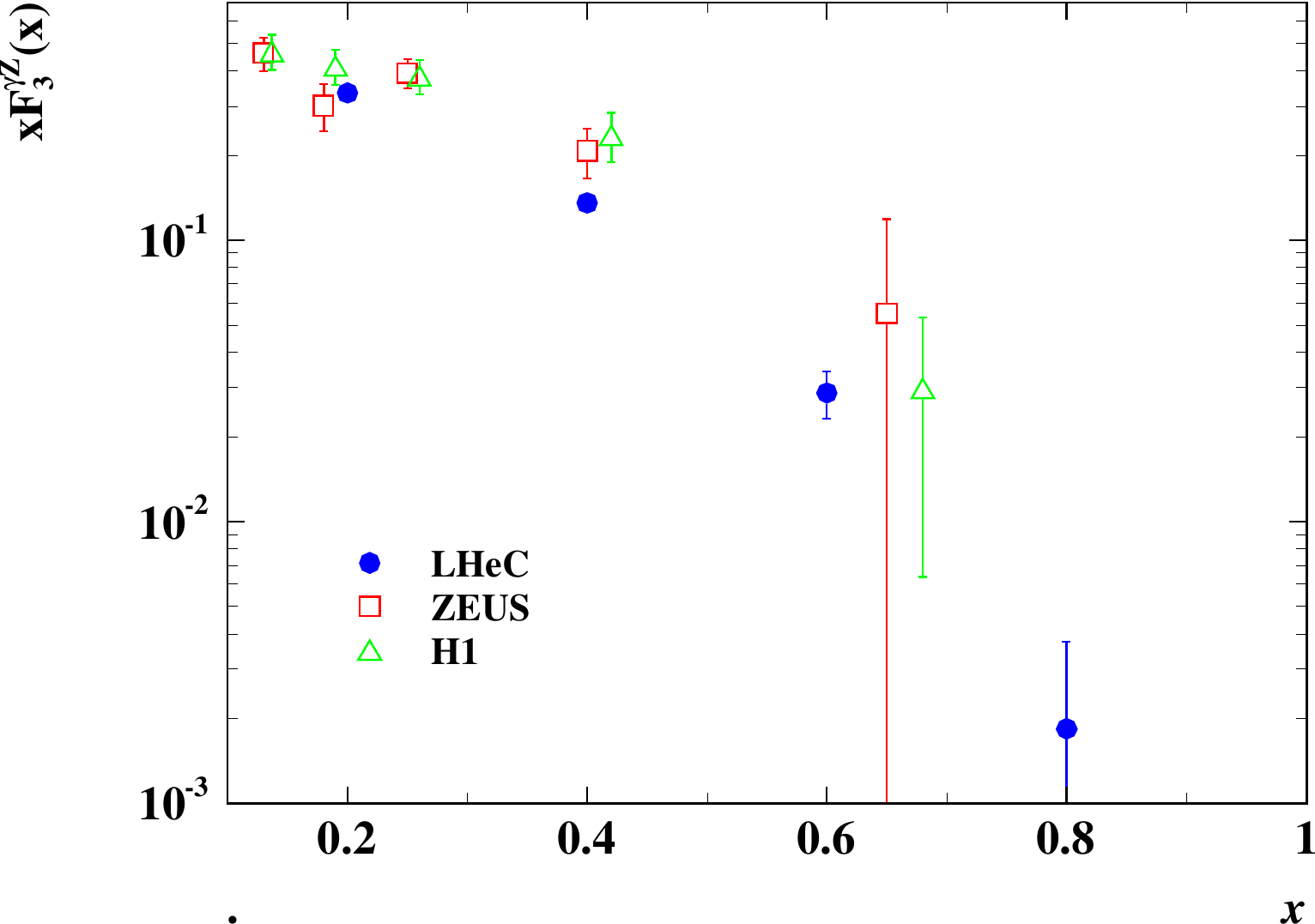}} 
\caption{Simulation of the LHeC measurement of the interference structure function
$xF_3^{\gamma Z}$ from unpolarised $e^{\pm}p$ scattering with $10$\,fb$^{-1}$
luminosity per beam  (blue, closed points) compared 
with the HERA II data as obtained by
H1 (preliminary, green triangles) and by ZEUS (red squares) 
 with about $0.15$\,fb$^{-1}$ luminosity per beam charge.
The H1 $x$ values are enlarged by $10$\,\% of their given values for clarity.
It should be noted that any significant deviation of sea from anti-quarks,
see Eq.\,\ref{eq:xfdel},  
would cause $xF_3^{\gamma Z}$ at low $x$ to not tend to zero. The top
plot shows an average of $xF_3^{\gamma Z}$ over $Q^2$ projected
to a chosen $Q^2$ value of $1500$\,GeV$^2$ exploiting the
fact that the valence quarks are approximately independent of $Q^2$. 
The lower plot is a zoom into the high $x$ region. 
}
   \label{fig:xf3}
\end{figure}

Access to valence quarks at low $x$ can be obtained from 
the $e^{\pm}p$ cross section difference as introduced above:
\begin{equation} \label{xf3}
 \sigma_{r,NC}^- - \sigma_{r,NC}^+ = 
2 \frac{Y_-}{Y_+} (-a_e \cdot \kappa_Z  xF_3^{\gamma Z} +  2v_e a_e \cdot \kappa_Z^2 xF_3^Z).
\end{equation}
Since the electron vector coupling, $v_e$,  is small and $\kappa_Z$ not much exceeding $1$, to a very good
approximation the cross section difference is equal to $-2\kappa_Z Y_- a_e xF_3^{\gamma Z}/Y_+$.
In leading order pQCD this ``interference structure function'' 
can be written as
\begin{equation}
 xF_3^{\gamma Z} =  2x [e_u a_u (U-\overline{U}) + e_d a_d (D - \overline{D})], 
\end{equation}
with $U=u+c$ and $D=d+s$ for four flavours. The $xF_3^{\gamma Z}$ structure
function thus provides information  about the light-quark
axial vector couplings ($a_u,\,a_d$) and the sign of the
electric quark charges ($e_u,\,e_d$).
Equivalently one can write
\begin{equation}
 xF_3^{\gamma Z} =  2x [e_u a_u (u_v +\Delta_u) + e_d a_d (d_v + \Delta_d)]. 
\label{eq:xfdel}
\end{equation}
In the naive parton model
as in conventional perturbative QCD, it is assumed that the differences
$\Delta_u =( u_{sea} -\overline{u} + c -\overline{c})$
 and $\Delta_d = (d_{sea} - \overline{d} + s - \overline{s})$
are zero~\footnote{However, in non-perturbative QCD there may occur differences,
for example
between the strange and anti-strange quark distributions, for which there are
some hints in DIS neutrino nucleon di-muon data and corresponding QCD fit analyses, 
see below.}.
Inserting the SM charge and axial coupling values one finds 
\begin{equation} \label{xg3}
xF_3^{\gamma Z} = \frac{x}{3}  (2 u_v + d_v + \Delta)
\end{equation}
with $\Delta = 2 \Delta_u + \Delta_d$. 
Neglecting $\Delta$ leads to a sum rule~\cite{Rizvi:2000qf}, which in leading order is 
\begin{equation}
\int_0^1{xF_3^{\gamma Z} \frac{dx}{x}} = \frac{1}{3} \int_0^1{(2u_v + d_v)dx} = \frac{5}{3}.
\end{equation}
The $xF_3^{\gamma  Z}$ structure function  thus is determined by the
valence quark distributions and predicted to be only very weakly dependent on $Q^2$.
Fig.\,\ref{fig:xf3} shows a simulation of $xF_3^{\gamma Z}$ and its comparison
with the most accurate measurement from HERA so far. With such a high precision,
interesting tests are possible of the relation of $xF_3^{\gamma Z}$
to $xW_3$, which should only differ by the weak couplings involved in NC and CC.
%
\subsection{Probing $q \ne \bar{q}$ and $u^p\ne d^n$}
For evolution at high $Q^2$, the transition $g \to q\bar{q}$ populates
the $q$ and $\bar{q}$ PDFs equally.  Of course, in the non-perturbative region 
there is no reason to have $q=\bar{q}$. Until recently, the lack of appropriate 
data has meant that this equality is assumed to be true for $s,c,..$ quarks, 
and that $ u=u_v+u_{\rm sea}, \;\;\;\bar{u}=u_{\rm sea}$,
and similarly for $d$.  Recent PDF analyses have attempted to determine 
$s$ and $\bar{s}$ separately, using dimuon production data, subject to the constraint
\begin{equation}
 \int^1_0(s(x,Q^2)-\bar{s}(x,Q^2))dx=0 
\end{equation}
which follows since protons have no valence strange quarks.  However 
the information obtained for $s-\bar{s}$ is very limited.
In this whole area the LHeC can dramatically transform the present knowledge. 
For the first time, it will be possible to explore
 $\bar{u}\ne u_{\rm sea},\;\bar{d}\ne d_{\rm sea},\; \bar{s}\ne s,\; \bar{c}\ne c...$ 
with high precision.

Moreover, by measuring the DIS processes $eN\to e\gamma X$, the LHeC has 
the unique opportunity to perform a precision measurement of the photon 
parton distributions of the proton and the neutron. Hence to quantify the 
amount of the corresponding isospin violations $u^p \ne d^n$ and $u^n \ne d^p$.

%
\subsection{Strange quarks}
\label{sec:strangeq}
The strange quark distribution in the proton is
one of the least well known PDFs. In flavour $SU(3)$,
the three light quark distributions are expected to be equal.
The larger mass of the strange quark, as compared to
up and down quarks, has been used to motivate 
its suppression. The strange-quark density is important
for many processes, as is the case for a precision measurement
of the $W$ boson mass~\cite{Krasny:2010vd}, 
for the formation of strange 
matter~\cite{Farhi:1984qu} and 
for neutrino interactions at ultra-high 
energies~\cite{Gandhi:1998ri}.

The strange quark distribution is accessible in charged current neutrino 
scattering through the subprocesses $W^+s \to c$ and $W^-\bar{s} \to \bar c$. 
This measurement has been made by the NuTeV~\cite{PhysRevLett.99.192001} and 
CCFR~\cite{Goncharov:2001qe} experiments,  
in the range of $x \sim 0.1$ and $Q^2 \sim 10$~GeV$^2$.
However, the interpretation of these data is sensitive to uncertainties 
from charm fragmentation and nuclear corrections. The analyses of 
MSTW and ABKM~\cite{MSTW2008,ABKM09} and of
the NNPDF group~\cite{Ball:2008by,Ball:2011uy} 
suggest strangeness suppression, 
with $\bar{s}/\bar{d} \lesssim 0.5$, whereas the analysis of 
CTEQ~\cite{Lai:2010vv} is consistent 
with $\bar{s}/\bar{d} \simeq 1$. 
Kaon multiplicity data analysed by HERMES~\cite{Airapetian:2008qf}
point to a striking $x$ dependence of the
strange quark density and a rather large value of $x(s+\bar{s})$
at $x \simeq 0.04$ and $Q^2 \simeq1.3$\,GeV$^2$.
A first NNLO QCD analysis jointly of the HERA DIS and the ATLAS
inclusive $W^{\pm}$ and $Z$ boson data, performed by the
ATLAS Collaboration~\cite{Aad:2012sb}, has most recently determined the
ratio of strange-to-anti-down quarks to
be $1.00^{+0.25}_{-0.28}$ at
$Q^2=1.9$~GeV$^2$ and $x=0.023$, in line with $SU(3)$.
 Some information on the 
strange density can be expected also from the $W s \to c$ production
at the LHC.
At low $x$ so far the light quark PDFs are solely fixed by the
accurate measurement of $F_2$, which determines a combination
of $4 u + d +s$. A significant enhancement of $s$ with this 
constraint diminishes the up and down quark distributions
and leads to an enhancement of the light sea by $8$\,\%,
as has been noted by ATLAS in \cite{Aad:2012sb}.

\begin{figure}[htbp]
\centerline{\includegraphics[clip=,width=1.0\textwidth]{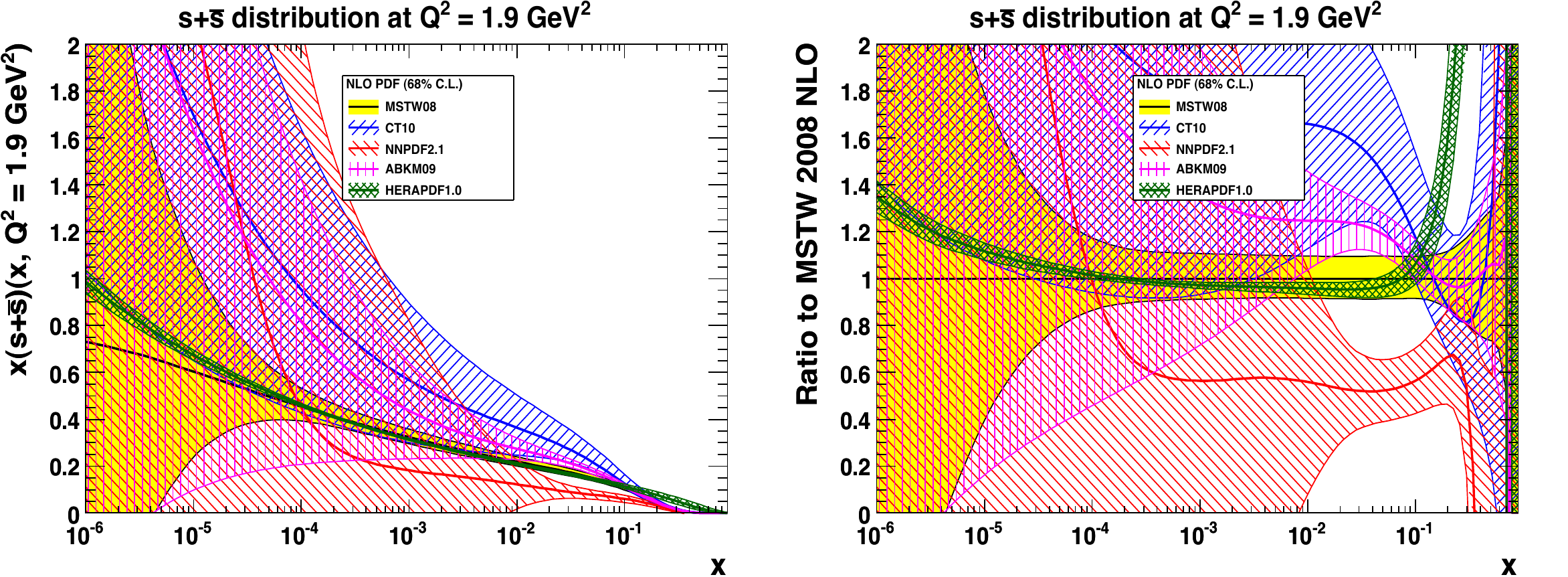}}
\caption{
Sum of the strange and anti-strange quark distribution
as embedded in the NLO QCD fit sets as noted in the legend.
Left: $s + \overline{s}$ versus Bjorken $x$ at $Q^2 = 1.9$\,GeV$^2$;
right: ratio of  $s + \overline{s}$ of various PDF determinations to
MSTW08.  In the HERAPDF1.0 analysis (green) the strange quark
distribution is assumed to be a fixed fraction of the down quark distribution
which is conventionally assumed to have the same low $x$ behaviour
as the up quark distribution, which results in a small uncertainty 
of $s + \overline{s}$.
}
   \label{fig:grasplus}
\end{figure}

\begin{figure}[htbp]
\centerline{\includegraphics[clip=,width=0.9\textwidth]{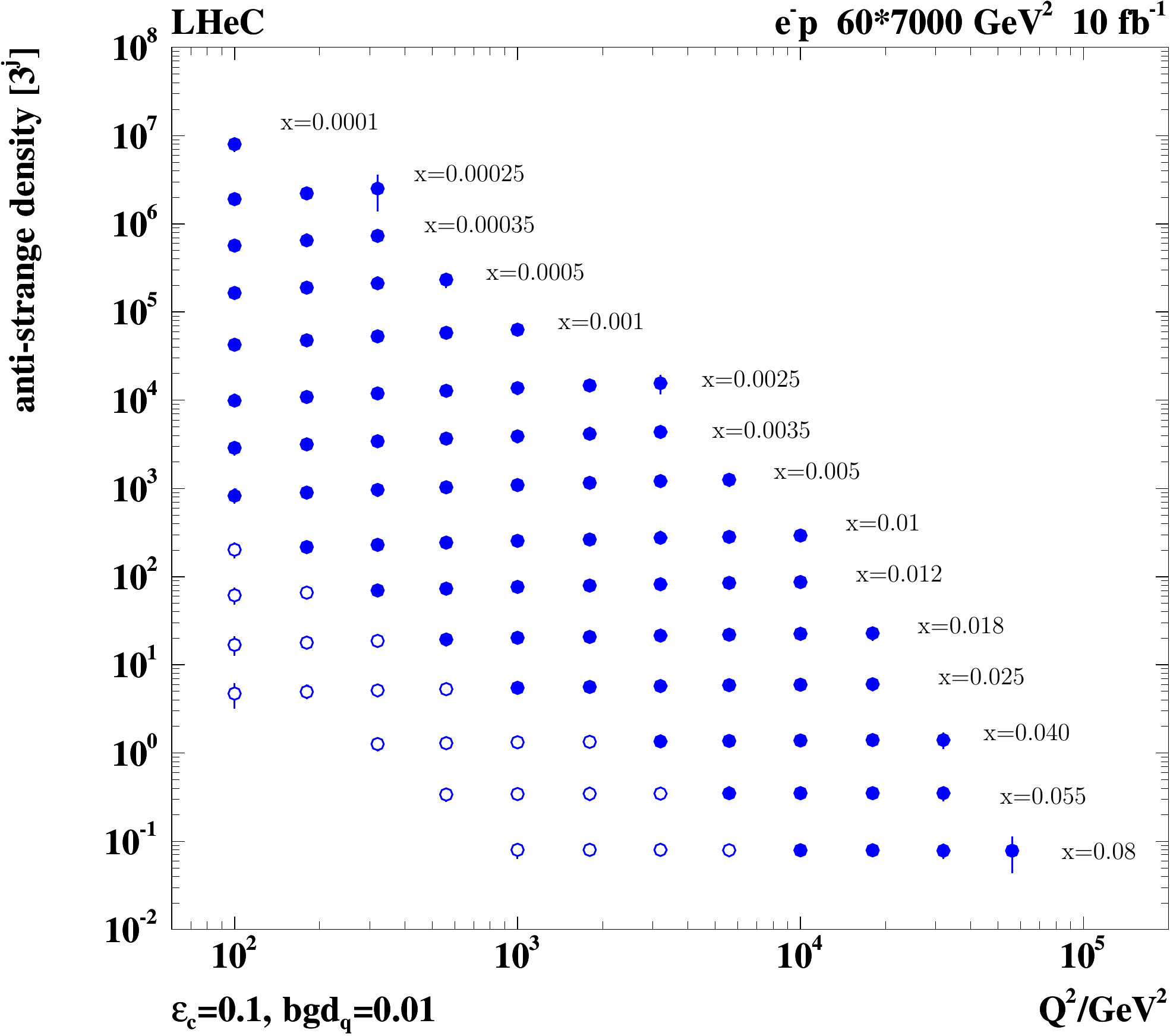}}
\caption{
Simulated measurement of the anti-strange quark density in CC $e^-p$
scattering with charm tagging
at the LHeC, for a luminosity of $10$\,fb$^{-1}$.
 Closed (open) points: tagging acceptance down
to $10$ ($1^{\circ}$).
The charm quark tagging efficiency is assumed to be 
$\epsilon_c = 10\%$ and the efficiency to keep light quark background
$\mbox{bgd}_{\mbox{q}}=1\%$.}
   \label{fig:strange}
\end{figure}

The existing information on the sum of the strange and anti-strange
quark distributions, prior to the ATLAS observation which is 
limited to $x \simeq 0.02$,
is plotted in Fig.\,\ref{fig:grasplus}. Clearly 
there is no real understanding of the strange quark distribution
in the proton available. This will change with the LHeC. Here
$s$ and $\overline{s}$ may be very well measured as
a function of $x$ and $Q^2$ from the
$W^+ s \rightarrow c$ and $W^- \overline{s} \rightarrow \overline{c}$
processes, i.e. with charmed quark tagging in CC DIS using
electron and positron beams, respectively. The precision for $s$
which may be
obtained is illustrated in Fig.\,\ref{fig:strange}. 
The systematic uncertainty, assumed to be $5$\,\%, is
included but not visible in this graph, and can serve only as
a rough estimate of such a determination. Based on the
high cross section, high luminosity, small beam spot (of
about $30 \times 10$ $\mu$m$^2$) and a modern Silicon
vertex detector, however, it is clear that
accurate measurements of the strange quark density
may be obtained for the first time. The simulation
of $xs$ leads to the same picture, subject to a possibly
reduced positron-proton luminosity in the linac-ring option.
Yet, over a wide kinematic range possible differences between 
$s$ and $\overline{s}$ may be established. 

\subsection{Releasing PDF constraints}
\begin{figure}[htbp]
\centerline{\includegraphics[clip=,width=0.8\textwidth]{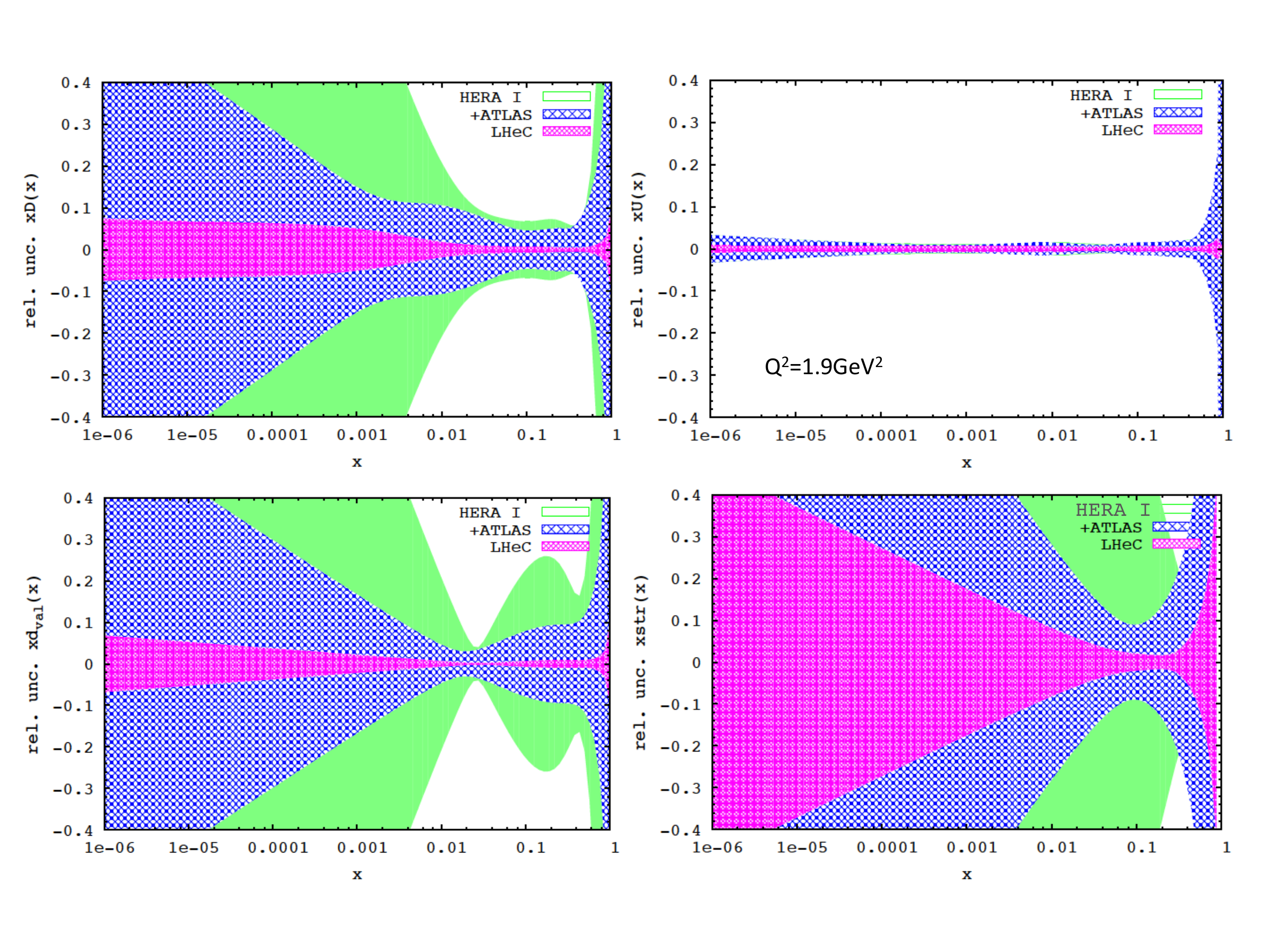}}
\caption{
Uncertainties of parton distributions in least constraint QCD fits to
the HERA data (green), the HERA and ATLAS $W,~Z$ data (blue)
and the simulated $ep$ data from the LHeC. Top:  $xD$ and $xU$;
Down: $xd_v$ and $xs$, at the initial scale $Q^2-0=1.9$\,GeV$^2$.
}
   \label{fig:freeuds}
\end{figure}

Based on the HERAPDF analyses, a QCD fit ansatz has been exploited
as is described in Sect.\,\ref{sec:qcdfita}, see Eq.\,\ref{equ:pdf}.
The results shown above
use a $10$ parameter fit, as also used in~\cite{:2009wt},
with five PDFs, $xg,~xu_v,~xd_v,~x\bar{U}, x\bar{D}$.
The following parameter changes have been made for the subsequent study:
an extra parameter $D_g$ is added for more freedom of $xg$ at 
larger $x$; the constraints $B_{u_v}=B_{d_v}$ and
$B_{\bar{U}}=B_{\bar{D}}$ 
are removed, and the relation $A_{\bar{U}}=A_{\bar{D}}(1-f_s)$
is given up, such that the up and
down valence and sea quark distributions become totally 
uncorrelated in the
analysis; free parameters $B_s$ and $C_s$ are introduced to
study the effect of the strange quark density on the 
inclusive NC and CC cross sections, complementary to the above where
the charm tagging result is used to access the strange quark density
directly from semi-inclusive data. 
Results are obtained using only the HERA data, adding the
recent $W,~Z$ cross section measurements from 
ATLAS~\cite{Aad:2012sb}, assuming only a
$1.4$\,\% normalisation error,
and considering the LHeC data.
One observes that the relaxation of the up-down quark parameter
relations leads for the HERA data to essentially no constraint
for $x < 0.01$ to the down-quark distributions as
shown for the total down-quark density, $xD$, and the
down valence quark  in Fig.\,\ref{fig:freeuds}.
The total up quark distribution is in any case rather well constrained
already by HERA at low $x$ because it dominates the $F_2$.
There is no significant sensitivity of the inclusive HERA data
on the strange density. ATLAS released a rather accurate measurement
of the $W$ and $Z$ rapidity dependent cross sections. These improve the
determination of the down quark densities and they also
can be seen to have a sensitivity to the strange density 
between $x$ of $0.01$ and $0.2$ which ATLAS has employed for
obtaining a constraint on the $s/\bar{d}$ ratio recently~\cite{Aad:2012sb}.
Fig.\,\ref{fig:freeuds} shows that the LHeC inclusive NC and CC
data lead to very precise determinations of all these PDFs.
It is worth noting that the $\sim 2$\,\% accuracy obtained for $xU$
can not be met fully by the $xD$ uncertainty, which, however,
will be about as precise if deuteron measurements become
available. The determination of the strange distribution from 
the inclusive fits is accurate to a few \% at $x \sim 0.1$ and can
complement the $xs$ determination from charm data presented above.
Clearly, such analyses are to certain extent only illustrative, yet
showing the unique potential of the DIS inclusive LHeC data 
for unfolding the nucleon quark contents. 
%
%
%
%

\subsection{Top quarks}
\begin{figure}[t]
\centerline{\includegraphics[clip=,width=.7\textwidth]{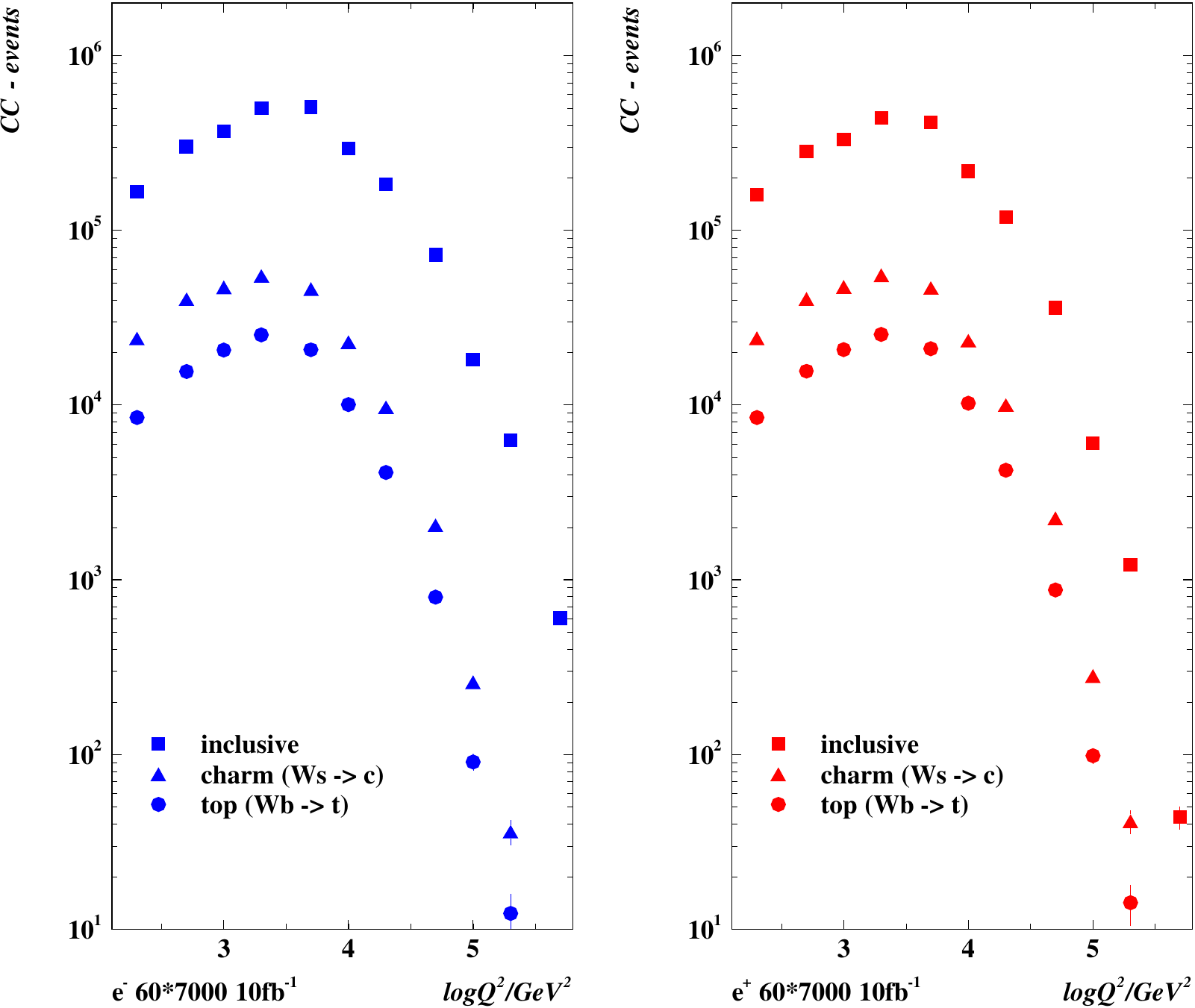}}
\caption{
Charged current event rates for unpolarised $e^-p$ (left) and $e^+p$ (right)
scattering in which  $\overline{c}$ and $c$ 
or $\overline{t}$ and $t$ are produced, respectively. 
Squares: inclusive CC rate vs. $Q^2$; triangles: charm production from $Ws$ 
fusion;
closed circles: top production from $Wb$ fusion, estimated in a massless
heavy flavour treatment. The rates are calculated for the default beam
energies for $10$\,fb$^{-1}$ of integrated luminosity. The errors
are only statistical.
}
   \label{fig:cctop}
\end{figure}

The top is the heaviest of the quarks. It decays before hadrons are
formed. It has not been explored in DIS
yet because the cross sections at HERA are too small~\cite{Baur:1987ai}.
This is different at the LHeC where top in charged currents is produced
with a cross section of order $5$\,pb as can easily be estimated
from the LO calculation of $Wb$ scattering. The energy dependence of
top production cross sections in $ep$ scattering is calculated and
shown in Fig.\,\ref{fig:hfl_proctot} below.
At the LHeC therefore, for the first time, top
quarks can be studied in deep inelastic scattering. Positron (electron) proton charged 
current scattering provides a clear distinction between top (anti-top) quark
production in $W b$ to $t$ fusion. The rates of this process
are very high, as is illustrated as a function of $Q^2$ in Fig.\,\ref{fig:cctop}.
Besides the rates and the charge tag it is notable that the absence of
pile-up and underlying event effects, characteristic for LHC measurements,
provide comfortable conditions for top quark physics at the LHeC.

Due to its large mass, the top quark may very well play a role in the
mechanism of electroweak symmetry breaking (EWSB)  both in
the Standard Model as well as BSM physics.  In  the Standard Model, a
precise measurement of single top production in DIS (see
for example \cite{Fritzsch:1999rd})
is sensitive to the $b$ quark content of the proton. 
In a BSM EWSB scenario, the top quark
couples to the new physics sector and gives rise to anomalous
production modes.  The LHeC is expected to provide competitive
sensitivity  to  flavour changing neutral currents  (FCNC)
especially anomalous $tu \gamma$ and $tuZ$ couplings.

\begin{figure}[htbp]
\centerline{\includegraphics[clip=,width=0.5\textwidth]{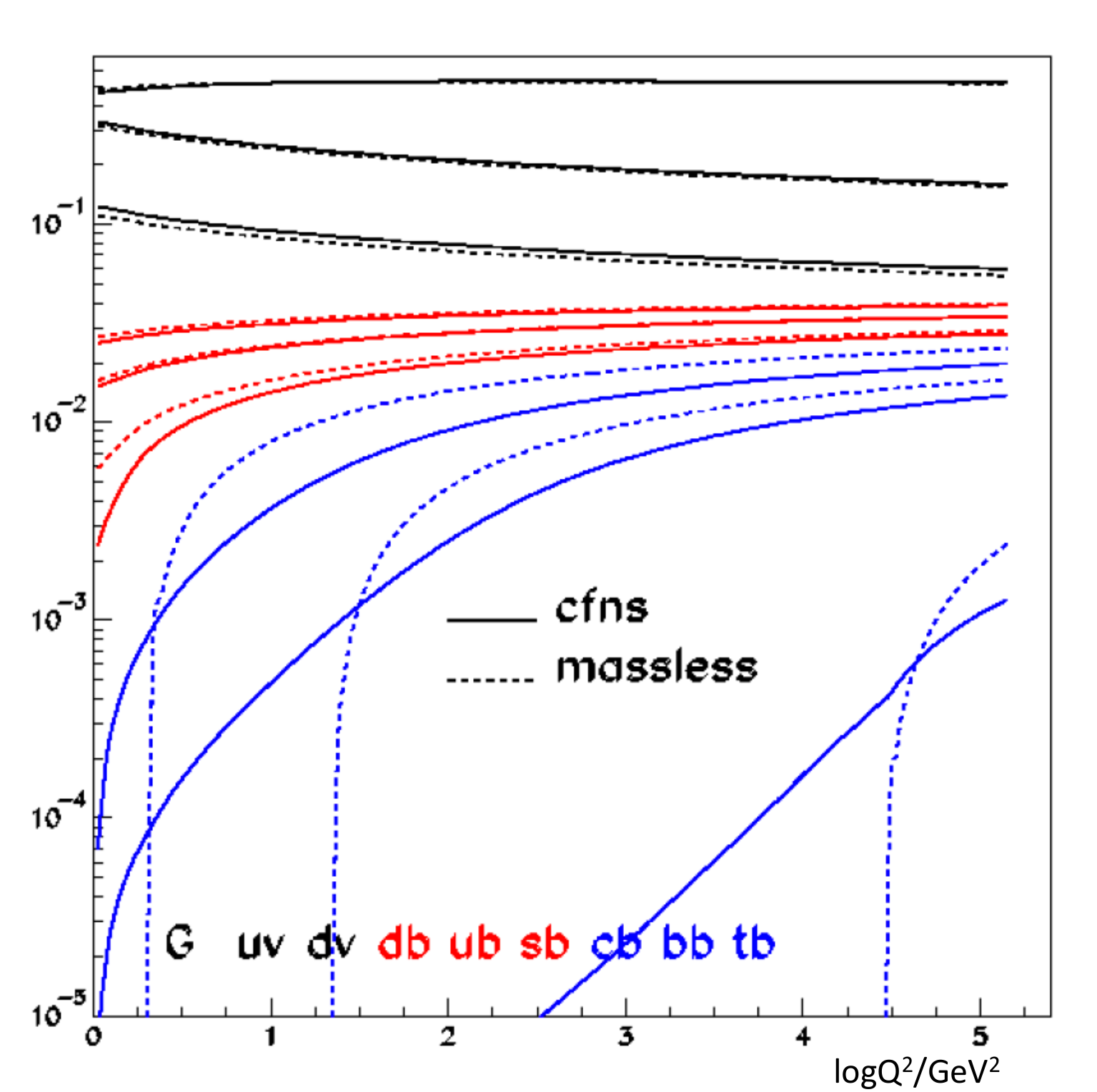}}
\caption{
Parton momentum fractions as a function of $Q^2$ in 
a novel six-flavour variable number scheme (CFNS),
solid curves, and in the massless scheme, dashed curves.
The scheme of~\cite{cpdis11} suggests that there is a very
early onset of top with measurable rates already at $Q^2$
values of only about one tenth of $m_t^2 \simeq 3~10^4$\,GeV$^2$.
}
   \label{fig:cptop}
\end{figure}

In the SM, top is produced dominantly in gluon-boson fusion at
$x \lesssim 0.1$. In CC this leads to a top-beauty final state while in 
NC this gives rise to pair produced top-antitop quarks, with
a cross section of order $10$ times lower than in CC~\cite{Baur:1987ai},
still sizeable at the LHeC.
The electron beam charge distinguishes  top and anti-top quark production
in CC.  Thus a unique SM top physics program can be performed at
the LHeC. This includes the consideration of a quark density for the top,
which at very high scales may be considered ``light''.
Recently a six-flavour variable number scheme has been proposed~\cite{cpdis11},
limited to leading order. The onset of top production in this model
 is illustrated in Fig.\,\ref{fig:cptop}. Naturally this is indicative
only of accurate higher order QCD calculations, in which 
heavy quarks are generated in the final state.
Due to the very high $Q^2$ and statistics,
the LHeC opens top quark PDF physics as a new field of research.

Top, including anomalous couplings, has been considered for the
CDR initially~\cite{brandt08}, based on some ANOTOP and PYTHIA
studies at generation level. With a full detector simulation
and in the light of the first top results provided by
the LHC experiments~\cite{Chatrchyan:2011vp},
the CC and NC top physics at the LHeC deserves
a more detailed study than was presented here.
This will include an analysis about the possible precision
measurement of the top (and anti) top quark mass, which at the
LHC may be determined with a precision of $1$\,GeV
and possibly better in $ep$. Independently of whether the SM Higgs particle is found,
or it remains elusive,
a high precision measurement of $m_t$ is of prime importance.

%% file: physics/pgluon.tex
%
%
There are many fundamental reasons to understand the gluon distribution
and the gluon interactions deeper than hitherto.
Half of the proton's momentum is carried by gluons.
The gluon self-interaction is responsible for the creation of baryonic
mass. The Higgs particle, should it exist, is predominantly
produced by gluon-gluon interactions. The rise of the gluon density
towards low Bjorken $x$ must be tamed
for unitarity reasons: there
is a new phase of hadronic matter to be discovered, in which
gluons interact non-linearly while $\alpha_s$ is smaller than $1$.
 
The LHeC, with precision and range of the most appropriate
process (DIS) to explore $xg(x,Q^2)$, will pin down the
gluon distribution much more accurately than could be
done before. This primarily comes from the extension of
range and precision in the measurement of \pdff
which at  small $x$ is a measure of $xg$. The
inclusive NC and CC measurements together provide a
fully constrained data base for the determination of the
quark distributions, which strongly constrains $xg$.
The addition of precision measurements of $F_L$, discussed
above and used in the small $x$ chapter of this document,
will unravel the saturating behaviour of $xg$.
High precision measurements of boson-gluon fusion
to heavy quark pairs will provide a complementary basis
for understanding the gluon and its parton interactions.

The peculiarity of the gluon density
is that it is defined and observable only in the context of 
a theory. Moreover, a crude data base and correspondingly rough
fit ansatz can screen local deviations from an otherwise
preferred smooth behaviour.
It has yet not been settled whether there are gluonic ``hot'' spots
in the proton or not. An example for possible surprises is
provided by the analysis~\cite{Glazov:2010bw}, in which Chebyshev
polynomials have been used to parameterise the parton distributions
in contrast to more conventional forms as in Eq.\,\ref{equ:pdf}.
Inspection of the gluon distribution obtained there reveals that
it seems to be vanishing at $x \simeq 0.2$, i.e. at the
point, in which scaling holds for $F_2(x,Q^2)$,
which one might term a ``cool'' spot in the proton. Much more
is still to be learned about the gluon, even when one is 
disregarding the yet to be explored role of the gluon in the theory 
of generalised and of unintegrated parton distributions.

\begin{figure}[htbp]
\begin{center}
\includegraphics[clip=,width=0.49\textwidth]{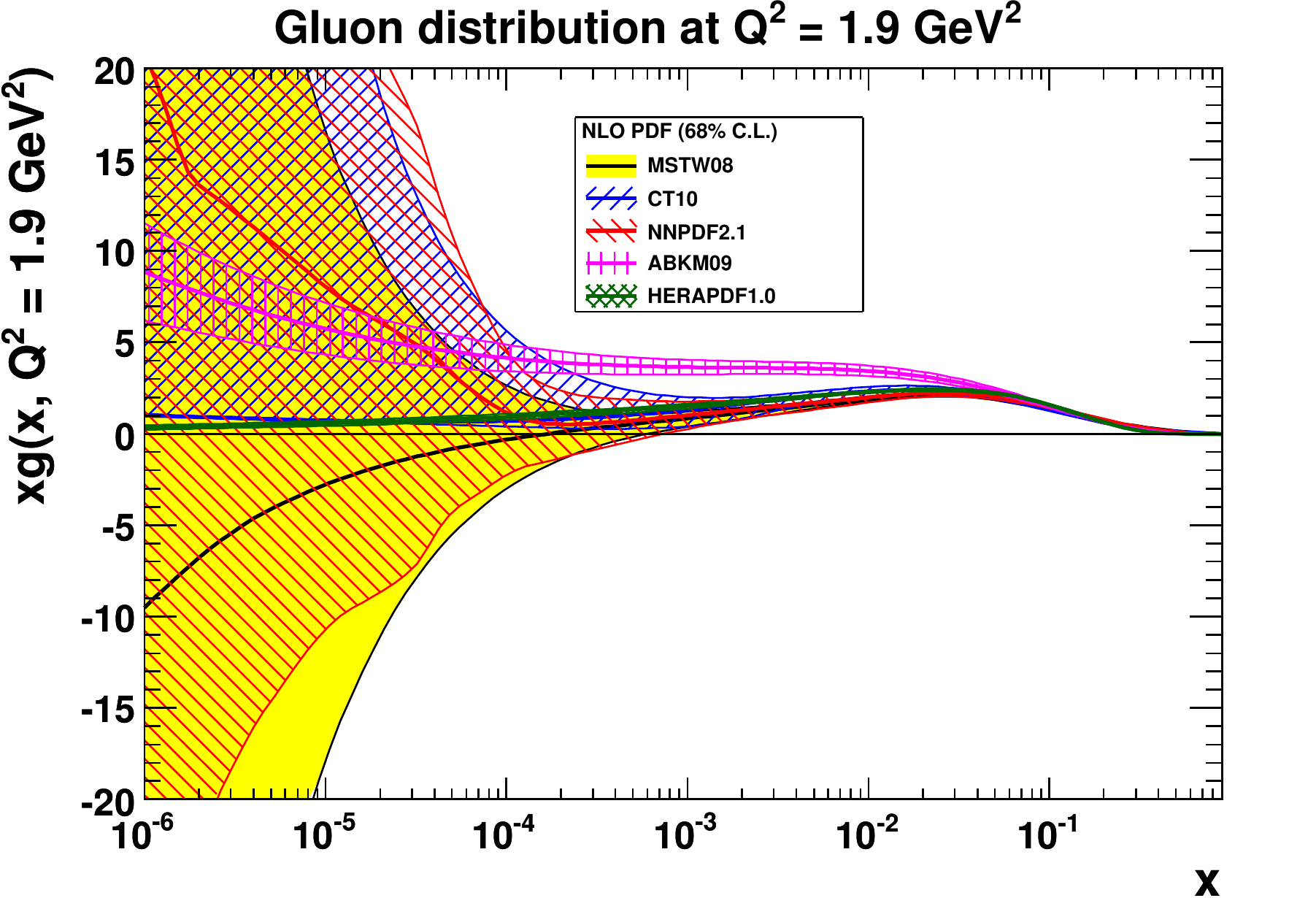}
\includegraphics[clip=,width=0.49\textwidth]{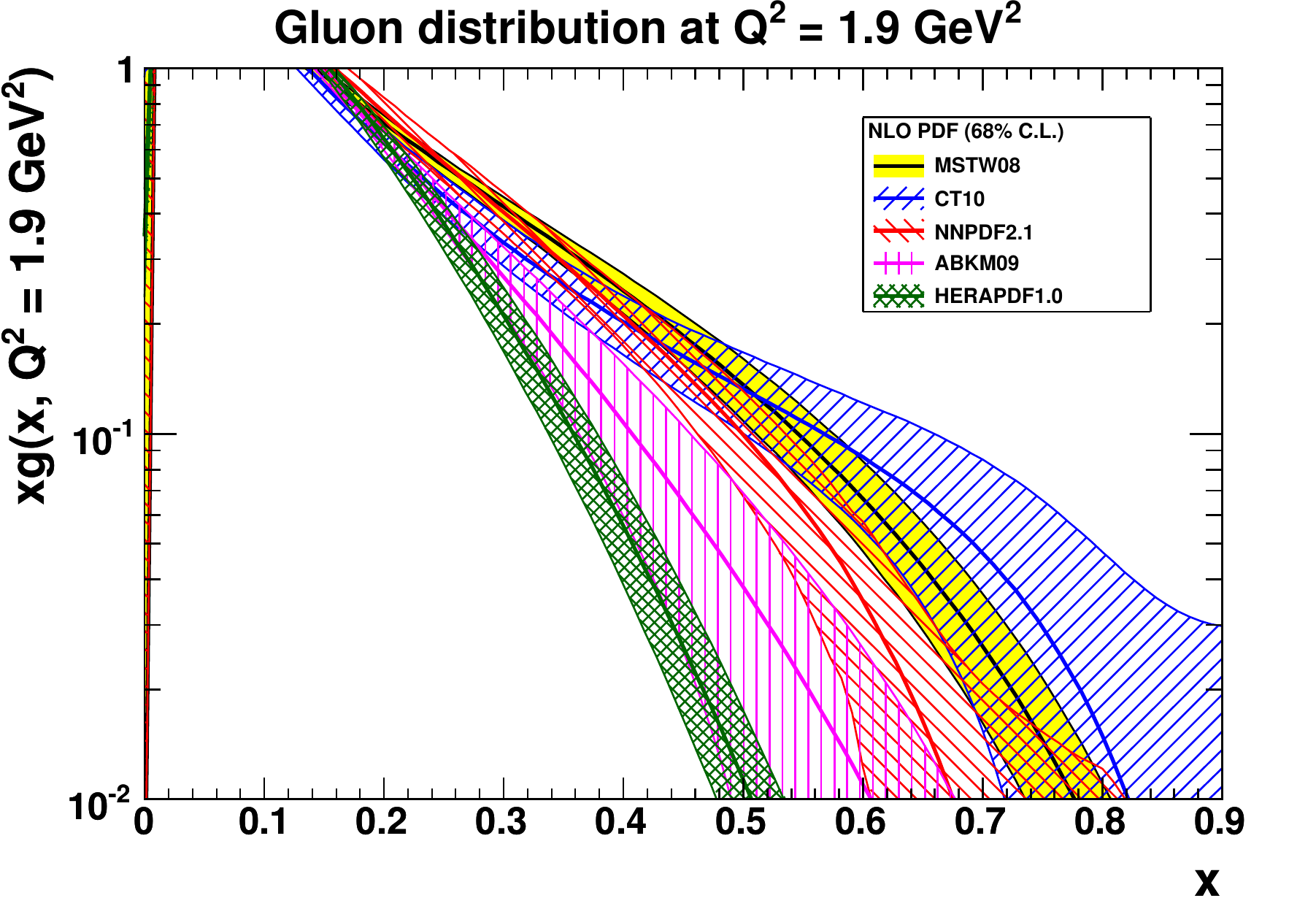}
\end{center}
\caption{Gluon distribution and uncertainty bands, at
$Q^2 = 1.9$\,GeV$^2$, for most of the available recent
PDF determinations.
Left: logarithmic $x$, right: linear~$x$.
}
   \label{fig:graglu}
\end{figure}

\begin{figure}[htbp]
\begin{center}
\includegraphics[clip=,width=0.49\textwidth]{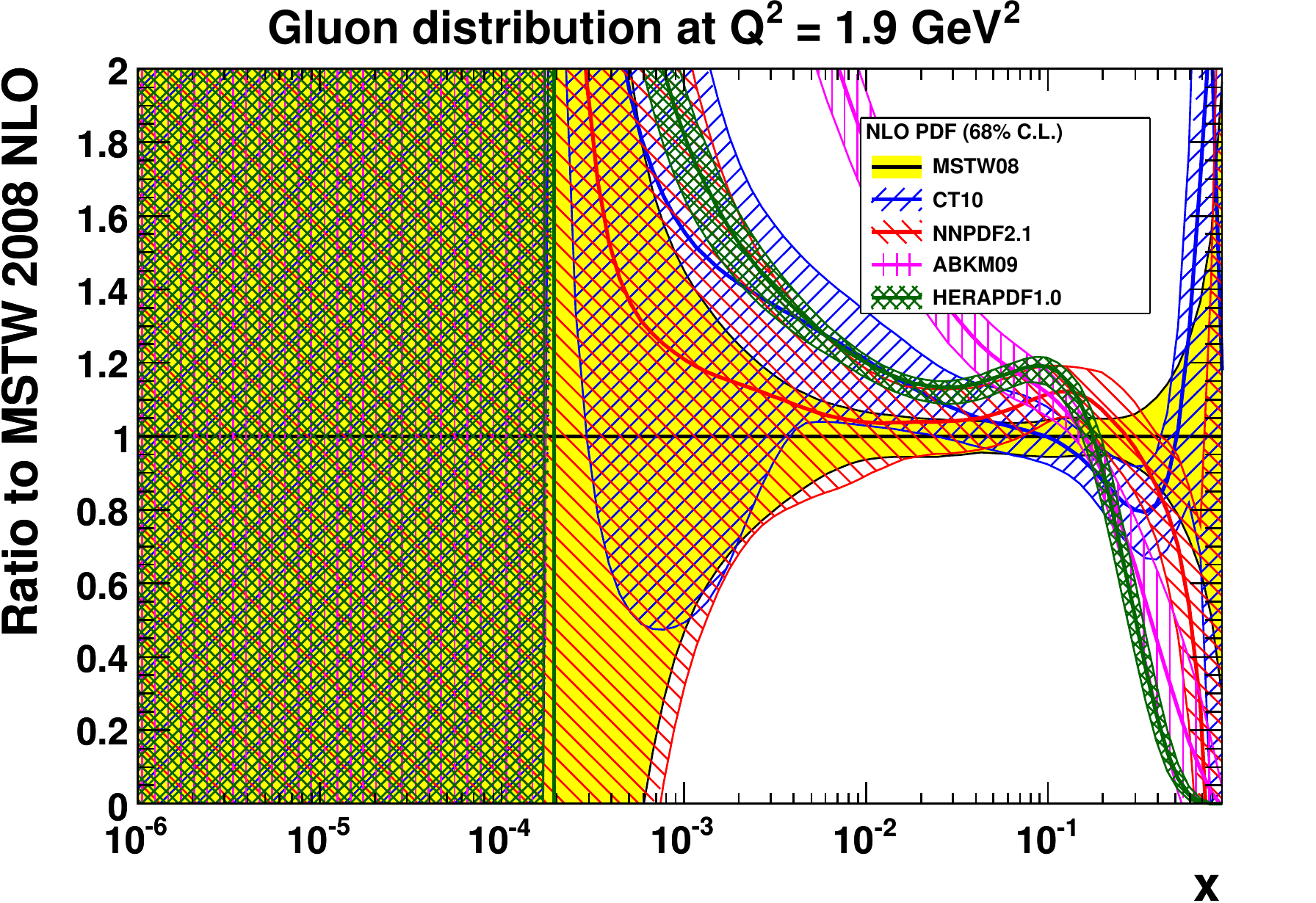}
\includegraphics[clip=,width=0.49\textwidth]{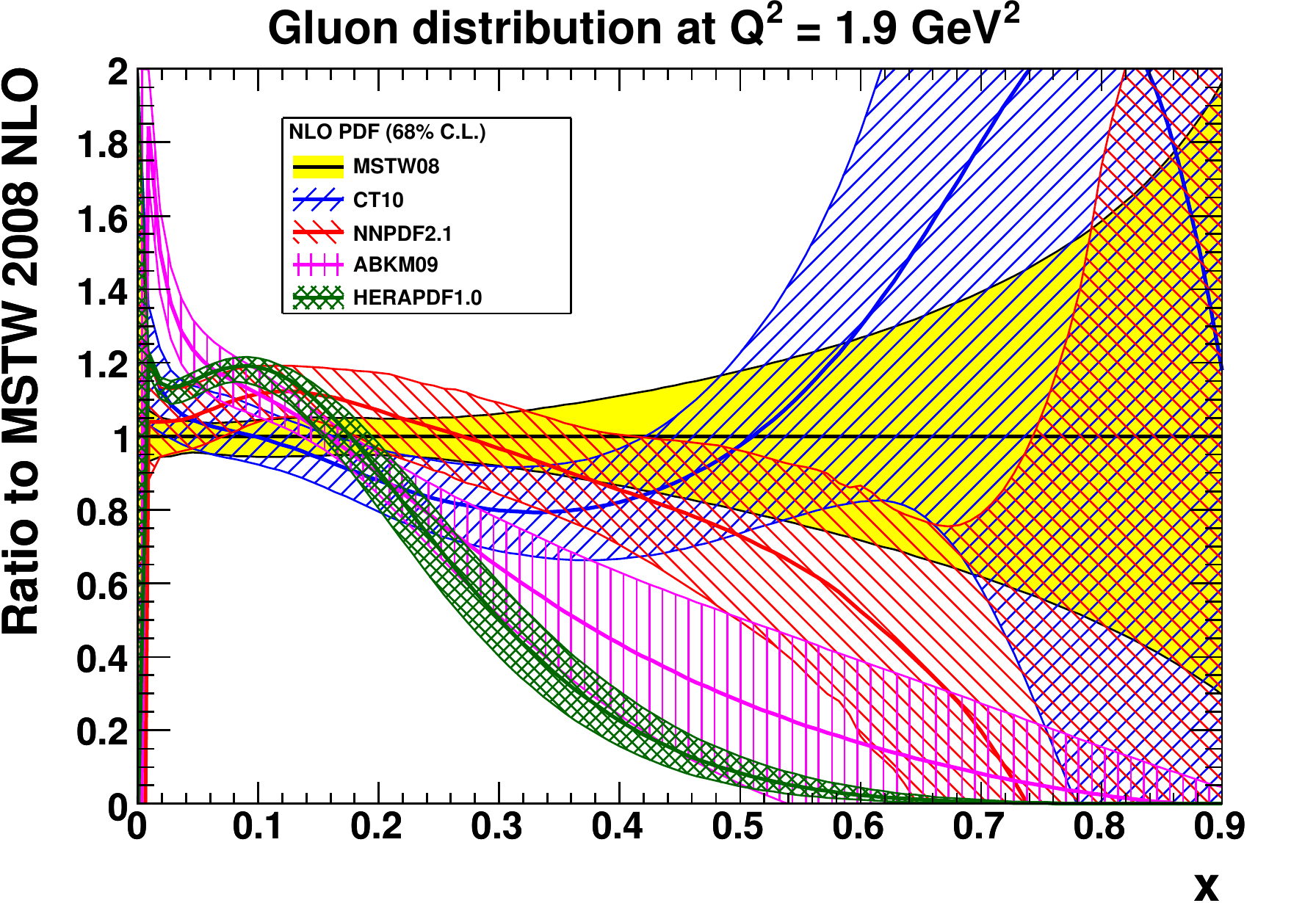}
\end{center}
\caption{Ratios to MSTW08 of
gluon distribution and uncertainty bands, at
$Q^2 = 1.9$\,GeV$^2$, for most of the available recent
PDF determinations.
Left: logarithmic $x$, right: linear~$x$.
}
   \label{fig:graratglu}
\end{figure}

The current knowledge of the gluon distribution in the proton 
is astonishingly limited as becomes clear from Fig.\,\ref{fig:graglu}
showing the world determinations, and their uncertainties,
of $xg(x,Q^2)$ at a typical initial, low scale, and from
 Fig.\,\ref{fig:graratglu} expressing this information
with ratios to one of the PDF sets. At low $x$ and $Q^2$
most but not all of the PDF sets predict $xg$ to be of
valence like type with very large uncertainties for
$x$ below a few times $10^{-4}$. At large $x$ inclusive 
DIS  has difficulties to pin down $xg$ because 
the  evolution of valence quarks
as non-singlet quantities in QCD is not directly coupled to 
the gluon and very weak. Yet, even the information from
jets, used in some of the PDF sets, does not lead to
a clear understanding of $xg$ at large $x$ as is illustrated too.
In fact, there is a tendency to obtain a smaller $xg$ at
large $x$ from HERA (I) data alone, see Fig.\,\ref{fig:graglu},
as compared to the other determinations, albeit with
large uncertainties.

\begin{figure}[htbp]
\centerline{\includegraphics[clip=,width=1.\textwidth]{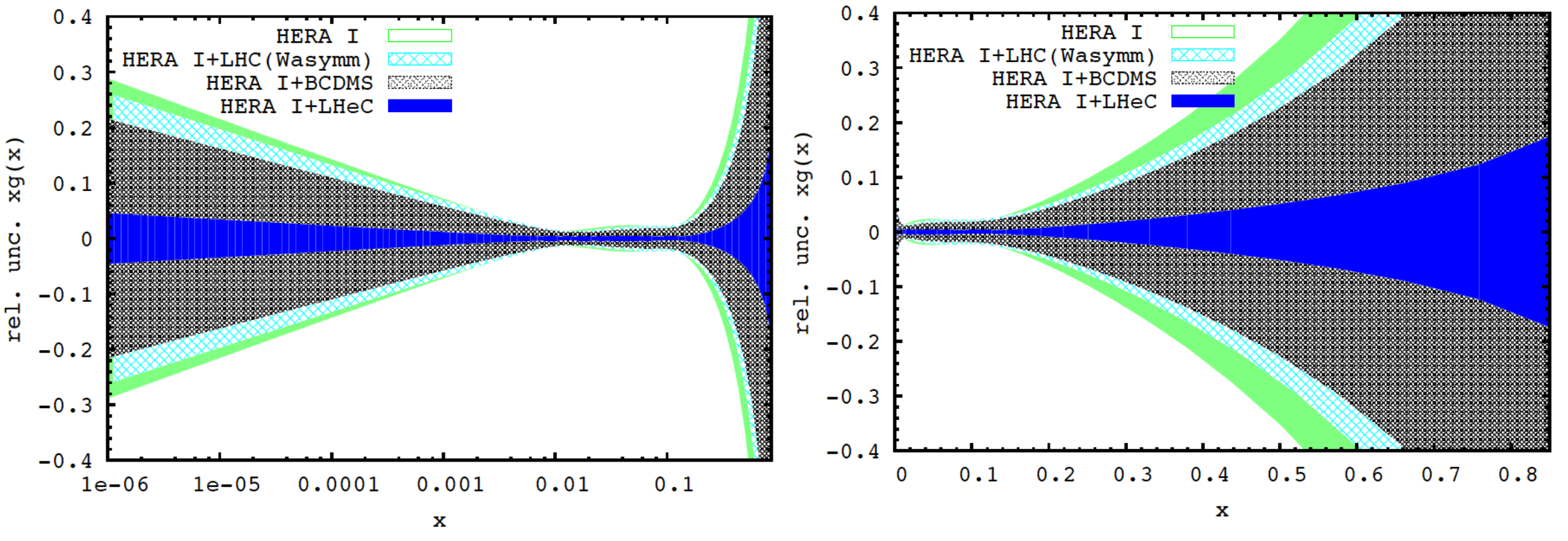}}
\caption{Relative uncertainty of the gluon distribution at
$Q^2 = 1.9$\,GeV$^2$, as resulting from an NLO QCD fit to
 HERA (I) alone (green, outer), HERA and BCDMS (crossed),
HERA and LHC (light blue, crossed)
and the LHeC added (blue, dark).
Left: logarithmic $x$, right: linear~$x$. 
}
   \label{fig:voiglu}
\end{figure}

The determination of $xg$ is predicted to be radically improved
with the LHeC precision data which extend  up to lowest
$x$ near to $10^{-6}$ and large $x \geq 0.7$. The result of the QCD fit
analysis for $xg$ as described above in Sect.\,\ref{sec:qcdfita}
is shown in Fig.\,\ref{fig:voiglu}. One observes a dramatic
improvement at low $x$, as must be expected from 
the extension of the kinematic range, but also at high $x$,
as is attributed to the high $x$ precision measurements of
the NC and CC cross sections. At $x=0.6$, for example,
the predicted experimental uncertainty of $xg$  is $5$\,\%, which is
about ten times more accurate than the results of 
MSTW08 or of the HERA fit indicate. 

It is worth noting that
the uncertainties considered here are restricted to those
related to the genuine cross section measurement errors.
There are further uncertainties, as discussed e.g. in~\cite{:2009wt},
related to the difficulty of parameterising the PDFs and choosing
the optimum solution in such a fit analysis. These will be
also considerably reduced with the LHeC extended data base.
Moreover,
 this analysis is not making
use of the plethora of extra information on $xg$, which the LHeC 
will provide with $F_L$, $F_2^{c,b}$ and jet cross section
measurements. The understanding of the gluon and its
 interactions is a primary task of the LHeC and
undoubtedly a new horizon in strong interaction physics
will be opened.

%% file: physics/alphajb.tex
%
%
The precise knowledge of $\alpha_s(M_Z^2)$ is
of instrumental importance for the correct prediction of the electroweak gauge
boson production cross sections and the Higgs boson cross section
at Tevatron and the LHC \cite{Alekhin:2010dd}. Independently of such applications,
the accurate determination of the coupling constants of the known fundamental forces is 
of importance in the search for their possible unification within a more fundamental 
theory. Among the coupling constants of the forces in the Standard Model, 
the strong coupling $\alpha_s$ exhibits the largest uncertainty, which is currently  
of the size of $\sim 1\%$. Any future improvement of this precision, along with the 
consolidation of the real central value, is one of the central issues of 
contemporary elementary particle physics. It demands deep experimental and 
theoretical efforts to obtain the required precision and especially to handle all 
essential systematic effects. 

Experimentation at the LHeC will allow to measure the strong coupling constant 
$\alpha_s(M_Z^2)$ at much higher precision than hitherto,
both from the scaling violations of the 
deep inelastic structure functions, as will be demonstrated below,
and using $ep$ multiple jet cross sections.
For the final inclusion of jet data  in global PDF analyses, both from $ep$
and from hadron colliders, their description
at NNLO is required. At the LHeC, similar to HERA, the measurement of the $ep$ jet cross
sections will form important data samples \footnote{
 These are presented below but have not been used in this document for a
 determination of the strong coupling constant. One knows of course that
 the use of jet data in DIS helps to resolve the $\alpha_s$-$xg$ correlation, especially
 at large $x$, and consequently leads to a significant reduction of the
 uncertainty on the coupling constant. This, however, tends to also change the
 central value. The LHeC as will be shown below determines $\alpha_s$ 
 to per mille precision already in inclusive scattering. Comparison with
 precise values from jets can be expected to shed light on the as yet unresolved
 question as to whether there is a theoretical or systematic effect which
 leads to different values in inclusive DIS and jets or not.}
 for the measurement of $\alpha_s(M_Z^2)$.

Subsequently, a brief account will be given on the status and the
complexity of determining $\alpha_s$ in DIS, followed by a presentation of the
study of the $\alpha_s$ measurement uncertainty with the inclusive NC and CC data
from the LHeC.
\subsection{Status of the DIS measurements of $\alpha_s$}
During the last 35 years the strong coupling constant has been measured with 
increasing precision in lepton-nucleon scattering in various experiments at CERN,
FERMILAB and DESY. The precision, which has been reached currently, requires
the description of the deep-inelastic scattering structure functions at 
$O(\alpha_s^3)$ \cite{Moch:2004pa,Vermaseren:2005qc,Bierenbaum:2009mv}.
{\small
\restylefloat{table}
\begin{table}[H]
{\renewcommand{\arraystretch}{1}{
\begin{center}
\begin{tabular}{|l|l|l|}
\hline
\multicolumn{1}{|c|}{ } &
\multicolumn{1}{c|}{$\alpha_s({M_Z^2})$} &
\multicolumn{1}{c|}{  } \\
\hline
BBG      & $0.1134 {\tiny{\begin{array}{c} +0.0019 \\
           -0.0021 \end{array}}}$
         & {\rm valence~analysis, NNLO}  \cite{Blumlein:2006be}           
\\
GRS      & $0.112 $ & {\rm valence~analysis, NNLO}  \cite{Gluck:2006yz}           
\\
ABKM           & $0.1135 \pm 0.0014$ & {\rm HQ:~FFNS~$N_f=3$} \cite{Alekhin:2009ni}             
\\
ABKM           & $0.1129 \pm 0.0014$ & {\rm HQ:~BSMN-approach} 
\cite{Alekhin:2009ni}             
\\
JR       & $0.1124 \pm 0.0020$ & {\rm
dynamical~approach} \cite{JimenezDelgado:2008hf}   
\\
JR       & $0.1158 \pm 0.0035$ & {\rm
standard~fit}  \cite{JimenezDelgado:2008hf}    
\\
MSTW & $0.1171\pm 0.0014$ &  \cite{Martin:2009bu}     \\
ABM            & $0.1147\pm 0.0012$ &   FFNS, incl. combined H1/ZEUS data   
\cite{Alekhin:2010iu}
\\
\hline
BBG & {{$
0.1141 {\tiny{\begin{array}{c} +0.0020 \\
-0.0022 \end{array}}}$}}
& {\rm valence~analysis, N$^3$LO}  \cite{Blumlein:2006be}            \\
\hline
{world average} & {$
0.1184 \pm 0.0007$  } & \cite{Bethke:2009jm}
\\
\hline
\end{tabular}
\end{center}
\renewcommand{\arraystretch}{1}   
\caption{
\label{tab:maint}
Recent NNLO and N$^3$LO determinations of the strong coupling $\alpha_s(M_Z)$ in DIS 
world data analyses. 
}
}}
\end{table}
}
\noindent
As is well known~\cite{Blumlein:1996gv}, though also questioned~\cite{stanote},
the fits at NLO exhibit scale uncertainties for both the renormalisation and 
factorisation scales of $\Delta_{r,f} \alpha_s(M_Z^2) \sim 0.0050$, which are too 
large to cope with the experimental precision of $O(1\%)$. Therefore, NNLO analyses 
are mandatory. In Table~1 recent NNLO results are summarised. NNLO non-singlet data 
analyses have been performed in~\cite{Blumlein:2006be,Gluck:2006yz}. 
The analysis~\cite{Blumlein:2006be}
is based on an experimental combination of flavour non-singlet data referring to 
$F_2^{p,d}(x,Q^2)$ for $x < 0.35$ and using the respective valence approximations for 
$x > 0.35$. The $\overline{d} - \overline{u}$ distributions and the $O(\alpha_s^2)$ 
heavy flavour corrections were accounted for. 
The analysis could be extended to N$^3$LO effectively due to the dominance of the Wilson 
coefficient in this order \cite{Vermaseren:2005qc} if compared to the anomalous 
dimension, cf.~\cite{Blumlein:2006be,Baikov:2006ai}. This analysis led to an increase of
$\alpha_s(M_Z^2)$ by $+0.0007$ if compared to the NNLO value.

A combined singlet and non-singlet NNLO analysis based on the DIS world data, 
including the Drell-Yan and di-muon data, needed for a correct description of 
the sea-quark densities, was performed in \cite{Alekhin:2009ni}. In the fixed 
flavour number scheme (FFNS) the value of $\alpha_s(M_Z^2)$ is the same as in 
the non-singlet case \cite{Blumlein:2006be}. 
The comparison between the FFNS and the BMSN 
scheme \cite{Buza:1996wv} for the description of the heavy flavour contributions induces a 
systematic uncertainty $\Delta \alpha_s(M_Z^2) = 0.0006$. One should note that
also in the region of medium and lower values of $x$ higher twist terms have to be 
accounted for within singlet analyses to cover data at lower values of $Q^2$. 
Moreover, systematic errors quoted by the different experiments usually cannot be 
combined in quadrature with the statistical errors, but require a separate treatment.
The NNLO analyses~\cite{JimenezDelgado:2008hf} are statistically compatible 
with the results of \cite{Blumlein:2006be,Gluck:2006yz,Alekhin:2009ni}, while those of 
\cite{Martin:2009bu} yield a higher value.

In~\cite{Alekhin:2010iu} the combined H1 and ZEUS data were accounted for 
in an NNLO analysis for the first time, which led to a shift of $+0.0012$. However,
running quark mass effects \cite{Alekhin:2010sv}
and the account of recent $F_L$ data reduce this value 
again to 
the NNLO value given in \cite{Alekhin:2009ni}. Other recent NNLO
analyses of precision data, as the measurement of $\alpha_s(M_Z^2)$ using thrust 
in high energy
$e^+e^-$ annihilation data  \cite{Gehrmann:2009eh,Abbate:2010xh}, result in 
$\alpha_s(M_Z^2) = 0.1153 \pm 0.0017 \pm 0.0023$, resp.
$0.1135 \pm 0.0011 \pm 0.0006$. Also the latter values are
lower than the 2009 world average \cite{Bethke:2009jm} based on NLO, NNLO and 
N$^3$LO results. 
%

Recent studies have found that $\alpha_s(M^2_Z)$ obtained from DIS data 
is closer to the world average than indicated by the large spread of values shown in Table~\ref{tab:maint}.
It is found to be necessary to perform global fits which include a careful treatment 
of the Tevatron jet data, since, at present, 
these data are the main constraint on the high $x$ gluon PDF. 
Note that the value of $\alpha_s$ is {\it anticorrelated} with the low $x$ gluon through the scaling violations of the HERA data. 
Thus $\alpha_s$ is {\it correlated} with the high $x$ gluon through the momentum sum rule. 
 As a consequence, the values of $\alpha_s$ found including a careful treatment of jets 
by MSTW08, NNPDF1.2 and CT10.1 give the most reliable determinations. 
Also HERAPDF gives a compatible value of $\alpha_s$ when jets are included, see Table 4.4.  
Ref. \cite{Thorne:2011kq} gives detailed reasons why the low values of $\alpha_s$ in Table 4.3 are questionable.  
For the reasons given in Section\,\ref{sec:gluon}, the LHeC will be able to considerably 
improve the gluon PDF at large $x$ (as well as at low $x$) and hence help to obtain the dramatic improvement in the determination of $\alpha_s$ from DIS.
%
%
\subsection{Simulation of $\alpha_s$ determination}
Since nearly twenty years, the $\alpha_s$ 
determination in DIS is dominated by the most precise data from
the BCDMS Collaboration, which hint to particularly low 
values of $\alpha_s (M_Z) \simeq 0.113$~\cite{Virchaux:1991jc}
and exhibit some peculiar systematic error effects,
when compared to the SLAC data and in the pQCD analyses
as are discussed in \cite{Adloff:2000qk,Twallny}. Recent analyses
seem to indicate that the influence of the BCDMS data is 
limited, which, however, is possible only when jet and 
nuclear fixed target data, extending to very low $Q^2$, are used.
Jet data sometimes tend to increase the value of
$\alpha_s$ and certainly introduce extra theoretical problems
connected with hadronisation effects in non-inclusive measurements.
The use of fixed target data poses problems due to the
uncertainty of corrections from higher twists and from nuclear effects,
because what is required is an extraordinary precision if indeed
one wants to unambiguously determine the strong coupling 
constant in DIS. These problems have been discussed in
detail above, and recently also in presentations by MSTW~\cite{alan}
and in a phenomenological study of the NNPDF group~\cite{Lionetti:2011pw}.

The question, of how large $\alpha_s$ is, remains puzzling, as
has been discussed at a recent workshop~\cite{mpialfa} and requires
a qualitatively and quantitatively new level of experimental input
if one wants to progress in DIS. 

Following the description of the simulated LHeC data 
(Sec.\,\ref{sec:simNC}) and the QCD fit
technique (Sec.\,\ref{sec:qcdfita}) a dedicated study has been
performed to estimate the precision of an $\alpha_s$ measurement
with the LHeC. In the fits, for the central values of the LHeC
data, the SM expectation is used smeared 
within the above uncertainties assuming their Gaussian distribution
and taking into account correlated uncertainties as well.

The QCD fit results are summarised in Tab.\,\ref{tab:alfa}.
The first two lines give the result of a fit to the HERA I data. One
observes  that the inclusion of DIS jet data reduces the
uncertainty, by a factor of two, but it also increases the central value
by more than the uncertainty. The LHeC alone, using only inclusive DIS,
reaches values of better than $0.2$\,\% which when complemented
with HERA data reaches a one per mille precision. From inspecting 
the results one finds that enlarging the $Q^2$ minimum still
leads to an impressive precision, as of two per mille in the LHeC plus HERA
case, at values which safely are in the DIS region.  A $Q^2$ cut
of for example $10$\,GeV$^2$  excludes also the lowest $x$ region
in which non-linear gluon interaction effects may require
changing the evolution equations. 
\begin{table}
\begin{center}
\begin{tabular}{|l|c|lc|c|}
\hline
case & cut [$Q^2$ (GeV$^2$)] & $\alpha_S$ & uncertainty & relative precision (\%)\\
\hline
HERA only (14p)& $Q^2>3.5$ & 0.11529 & 0.002238 &  1.94  \\
HERA+jets (14p)& $Q^2>3.5$  & 0.12203 & 0.000995 & 0.82 \\
\hline
LHeC only (14p)& $Q^2>3.5$  & 0.11680 & 0.000180  & 0.15  \\
LHeC only (10p)& $Q^2>3.5$  & 0.11796 & 0.000199   & 0.17 \\
LHeC only (14p)& $Q^2>20.$  & 0.11602 & 0.000292 &  0.25 \\
\hline
LHeC+HERA (10p)& $Q^2>3.5$  & 0.11769 & 0.000132 & 0.11 \\
LHeC+HERA (10p)& $Q^2>7.0$  & 0.11831 & 0.000238   & 0.20 \\
LHeC+HERA (10p)& $Q^2>10.$  & 0.11839 & 0.000304 & 0.26  \\
\hline
\end{tabular}
\caption{Results of NLO QCD fits to HERA data (top, without and with jets)
to the simulated LHeC data alone and to their combination.
Here $10$p or $14$p denotes two different sets of parameterisations,
one, with $10$ parameters, the minimum parameter set used
in~\cite{:2009wt} and the other one with four extra parameters added
as has been done for the HERAPDF1.5 fit. The central values 
of the LHeC based results are obviously of no interest. The result
quoted as relative precision includes all the statistical and the
systematic error sources taking correlations as from the energy scale
uncertainties into account. 
}
\label{tab:alfa}
\end{center}
\end{table}

It is clear from Table~\ref{tab:alfa} 
that the LHeC will give an enormous improvement in the experimental 
error on $\alpha_s$ from the evolution of structure functions,  and possibly
other processes including jets.
However, there is also the theory uncertainty to consider. It will be a great 
challenge to QCD theory to reduce this uncertainty, so as to make the most use of such 
results. This will require to study the effect of non-linear terms and additional ln$(1/x)$ 
contributions in DGLAP evolution at low $x$,  to control the parameterisations
and contributions of all PDFs much better than hitherto
and to have an accurate knowledge of the heavy quark contributions
as may be measured by the uncertainty 
of the charm quark mass, required to better than $10$\,MeV
for a knowledge of $\alpha_s$ to one per mille. 
Also one may have to include the QED corrections in the evolution. 
However, these limitations will be automatically improved by the LHeC itself.
As an important example, this is demonstrated for the determination of
$m_c$ in Section\,\ref{sec:mc}, which can be as accurate as about $5$\,MeV
based on the NC, CC cross sections and a precision measurement of $F_2^{cc}$.  
Then, to  reduce the uncertainty due to the choice of renormalisation and factorisation scales, 
it appears to be necessary, for the expected precision, to work at higher-order than NNLO.

From an experimental and
phenomenological point of view it appears extremely exciting that
with the LHeC  the $\alpha_s$ determination in DIS will be put on
much more solid ground, by the high precision and unprecedented
kinematic range. It has been a problem of continuous concern
that often crude parameterisations of PDFs are used. Assumptions,
like the link of the strange density to the anti-down, have so far
masked some of the genuine uncertainties on PDFs, as has been 
illustrated above. The LHeC for the first time in DIS offers the
prospect of obtaining a really complete set of  parton distributions,
of light and heavy quarks, often by direct measurements.
This can not only be expected to lead to much improved
precision, it also may result in surprises in a field which sometimes
and wrongly is considered to be solved.

In view of the importance of this result, the $\alpha_s$
simulation and QCD  analysis has been performed
independently twice, with separately generated NC and CC pseudodata
under somewhat different assumptions, albeit
using the same simulation program, and using 
different versions of the QCD fit program.
The results obtained before~\cite{tkluge} are 
in good agreement with the numbers presented here.

It is finally worth noting that there is an interest to measure
$\alpha_s$ also based on non-singlet quantities. The LHeC data
provide high precision information both on the valence quarks
and also on the proton-neutron structure function difference.
The precision expected from such measurements has not been estimated.

%% file: physics/deuterons.tex
%
%
The structure of the deuteron and of the neutron
are experimental unknowns over most of the kinematic region
of deep inelastic scattering. 
The last time lepton-deuteron scattering was measured occurred
in the fixed target $\mu D$ experiments at 
CERN~\cite{Benvenuti:1989gs,Benvenuti:1989fm,Aubert:1987da},
while it had only been considered at
 HERA~\cite{edH1,edAl,Greenshaw:2002wu}. The
LHeC extends the range of these measurements by nearly
four orders of magnitude in $Q^2$ and $1/x$, which
gives rise to a most exciting  programme in QCD and in 
experimental physics.
\subsubsection{DIS and partons} 
Electron-deuteron scattering complements $ep$ scattering in that it
makes possible accurate measurements of neutron structure
in the new kinematic range accessed by the LHeC.    In a collider
configuration, in which the hadron ``target"  has momentum
much larger than the lepton probe,
the spectator proton can be tagged\footnote{Such an eD experiment with
 tagged protons has been successfully carried out at the Jefferson 
laboratory~\cite{Baillie:2011za}, but at much lower energies and with 
much less statistics.} and its momentum measured
with high resolution~\cite{edH1}. The resulting neutron structure
function data are then free of  nuclear corrections which
have plagued the interpretation of deuteron data, especially
at larger $x$, until now~\cite{Schienbein:2009kk}.
At low $x$, for the first time, since diffraction is related to shadowing,
one will be able to control the shadowing corrections~\footnote{
For light nuclei, nuclear shadowing is dominated by the scattering off two nucleons. 
Since the probability of such double collisions  is primarily determined
by nuclear geometry, the $A$-dependence (though not the absolute value) 
of shadowing in light nuclei  ($A \le 12$)  is not sensitive
to details of the dynamics. Consequently, one can extract the nuclear 
shadowing correction  for  electron-deuteron scattering with a 
small uncertainty  (well below $1$\,\%) from say the ratio of 
the electron-carbon and electron-deuteron cross sections~\cite{Frankfurt:2006am}.
}
 at the per cent level 
of precision as is also discussed below.

\begin{figure}[htbp]
\hspace{3.0cm}
\includegraphics[clip=,width=0.5\textwidth]{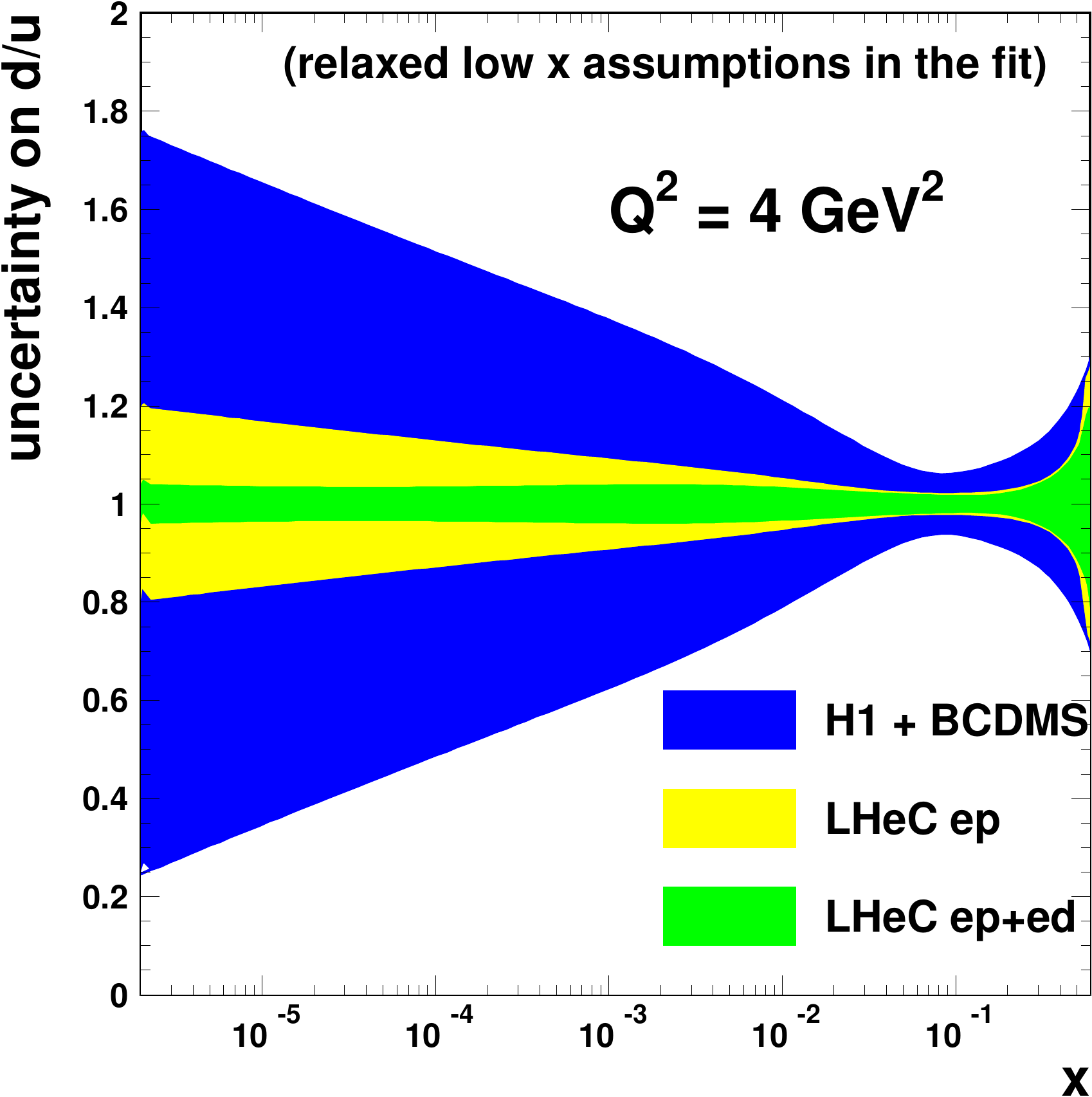}
\caption{Uncertainty of the $d/u$ ratio as a function of $x$
from a QCD fit to  H1 and BCDMS data (outer band, blue),
to the LHeC proton data (middle band, yellow) and
the combined simulated proton and deuteron data
from the LHeC (inner band, green). In these fits
the constraint of $u$ and $d$ to be the same at
low $x$ has been relaxed. 
}
   \label{fig:dudeut}
\end{figure}

Accurate $en$ cross section measurements will resolve the quark 
flavour decomposition of the
sea, i.e.  via isospin symmetry, unfolding $\bar{u}$ from $\bar{d}$
contributions to the rise of $F_2^p \propto x(4 \bar{u} + \bar{d})$
towards low $x$. From Fig.\,\ref{fig:dudeut} one can
see that a combination of H1 and BCDMS (proton and deuteron data
at larger $x$) leaves a very large uncertainty to the 
ratio of the light sea quarks at low $x$ if, as is done in this fit,
the conventional relation of $(\bar{u}-\bar{d}) \rightarrow 1$ for low $x$
is relaxed. In $ep$ at the LHeC,
it is mainly the charged current high statistics $ep$ data 
which constrain the $d/u$ ratio at lower $x$. In Fig.\,\ref{fig:dudeut} 
this may be recognised to be subject
to parameterisation effects to some extent because these
 mimic a reasonable precision
down to low $x < 10^{-5}$, although the LHeC CC data are limited
to $x \geq 10^{-4}$.   The light quark sea gets fully resolved
when one has $ep$ and $en$ data as this measures the orthogonal
combinations of $4u + d$ and $u+4d$.
%

For the study of the parton evolution with $Q^2$, the measurement
of $F_2^N=(F_2^p+F_2^n)/2$ is also crucial since it
disentangles the evolution of the non-singlet and the singlet contributions. 
Down to $x$ of about $10^{-3}$ the $W^{\pm}$ boson LHC data
will also provide  information on the up-down quark distributions,
albeit at high $Q^2$. With $ep$, $eD$ and $W^+/W^-$ data,
the low $x$ sea will be resolved for the first time, as all the low $x$
light quark information from HERA has been restricted to $F_2^p$ only.

A special interest in high precision neutron data at high $Q^2$
arises from the question of whether charge symmetry holds
at the parton level, as has been discussed 
recently~\cite{Hobbs:2011vy}. It may be studied in the charged
current $ep$ and $eD$ reactions, using both electrons and positrons,
by measuring the asymmetry ratio
\begin{equation}
R^- = 2 \frac{W_2^{-D} - W_2^{+D}} {W_2^{-p} + W_2^{+p}},
\end{equation}
which is directly sensitive to differences of up and down quark 
distributions in the proton and neutron, respectively, which
conventionally are assumed to be equal. With the prospect
of directly measuring the strange and anti-strange quark
asymmetry in $e^{\pm}p$ CC scattering and 
of tagging the spectator proton and thus eliminating
the Fermi motion corrections in $eD$, such a measurement
becomes feasible at the LHeC. It requires high luminosity of order
$1$\,fb$^{-1}$ in $eD$ scattering.

\subsubsection{QED corrections and photon PDFs of the proton and neutron}
The LHeC offers the unique opportunity to include ${\cal O}(\alpha)$
corrections to parton evolution by measuring the photon parton
distributions, $\gamma^{p,n}(x, Q^2)$, of the proton and the
neutron. The most direct measurement is to observe wide-angle
scattering of the photon by the electron beam. To be specific, the
processes $eN\to e\gamma X$ where the final state electron and photon
are produced with equal and opposite large transverse momentum. The
subprocess is then simply QED Compton scattering, $e\gamma \to
e\gamma$, and the cross sections are obtained by the
convolution \cite{Martin:2004dh}
$$ \frac{d\sigma(eN \to e\gamma X)}{
dx^\gamma}= \;\gamma^{p,n}(x^\gamma,\mu^2)\; \hat{\sigma}(e\gamma \to
e\gamma). $$ If the photon is produced with transverse energy
$E_T^\gamma$ and pseudorapidity $\eta^\gamma$ in the LHeC laboratory
frame, then
$$ x^\gamma=\frac{E_T^\gamma E_e{\rm exp}(\eta^\gamma
)}{2E_pE_e-E_T^\gamma E_p{\rm exp}(-\eta^\gamma )},$$ where $E_e$ and
$E_p$ are the energies of the electron and proton beams
respectively. At HERA only a single measurement of the $ep \to e\gamma
X$ cross section was made (for $x_\gamma \sim 0.005$), with a large
uncertainty \cite{Chekanov:2004wr}. Also, a first estimate of
$\gamma^{p,n}(x,Q^2)$ PDFs was performed in \cite{Martin:2004dh}.

Such measurements at the LHeC will be considerably more precise and
will allow an investigation of whether the ${\cal O}(\alpha)$
contributions have a sizeable effect, in comparison to the ${\cal
O}(\alpha_s^2)$ NNLO QCD terms, in a complete QED-modified DGLAP
evolution, including QED terms in the input. Even if they are found to
have a small effect, they necessarily lead to a precise determination
of the isospin violations $u^p \ne d^n$ and $u^n \ne d^p$. Recall that
it was these isospin violations, together with $s\ne {\bar s}$, which
explained away the NuTeV sin$^2 \Theta$ anomaly. Of course, ideally,
for precision physics we should anyway use QED-modified partons which
include $\gamma^{p,n}(x,Q^2)$.

\subsubsection{Hidden colour}
In nuclear physics, nuclei are simply the composites of nucleons.
 However, QCD provides a new 
perspective~\cite{Brodsky:1976rz,Matveev:1977xt}. Six quarks
 in the fundamental
$3_C$ representation of $SU(3)$ colour can combine into five 
different colour-singlet combinations, only one of which 
corresponds to a proton and
neutron.  The deuteron wavefunction is a proton-neutron bound 
state at large distances, but as the quark separation becomes 
smaller,
QCD evolution due to gluon exchange introduces four 
other ``hidden colour" states into the deuteron
wavefunction~\cite{Brodsky:1983vf}.  The normalisation 
of the deuteron form factor observed at large 
$Q^2$~\cite{Arnold:1975dd}, as well as the
presence of two mass scales in the scaling behaviour of 
the reduced deuteron form factor~\cite{Brodsky:1976rz},
 suggest sizeable hidden-colour
Fock state contributions  in the deuteron
wavefunction~\cite{Farrar:1991qi}.
The hidden-colour states of the deuteron can be 
materialised at the hadron level 
as   $\Delta^{++}(uuu)\Delta^{-}(ddd)$ and other novel quantum
fluctuations of the deuteron. These dual hadronic 
components become important as one probes the deuteron 
at short distances, such
as in exclusive reactions at large momentum transfer.  
For example, the ratio  
${{d \sigma/ dt}(\gamma d \to \Delta^{++} \Delta^{-})/{d\sigma/dt}(\gamma d\to n p) }$ 
is predicted to 
increase to a fixed ratio $2:5$ with increasing transverse
 momentum $p_T.$
Similarly, the Coulomb dissociation of the deuteron into
 various exclusive channels 
$e d \to e^\prime + p n, p p \pi^-, \Delta \Delta, \cdots$
will have a changing composition as the final-state hadrons 
are probed at high transverse momentum, reflecting the onset 
of hidden-colour
degrees of freedom.
The hidden colour of the deuteron can be probed at the LHeC 
in electron deuteron collisions by studying reactions such 
as $\gamma^* d \to n p X$ where the proton and neutron 
emerge in the target fragmentation region at high and 
opposite $p_T$.   In principle, one can also study DIS
 reactions $e d \to e^\prime X$ at very high $Q^2$ 
where $x > 1$.
  The production of high $p_T$ anti-nuclei 
at the LHeC is also sensitive to hidden colour-nuclear components.

%% file: physics/qce_hfl.tex
\subsection{Introduction and overview of expected highlights}
\label{sec:hfl_intro}
In this section it is shown that the measurements of charm and beauty
production at LHeC provide high precision pQCD tests 
and are crucial to improve the knowledge 
of the proton structure.
Historically the HERA charm and beauty studies  
extended by a large amount 
results from previous fixed target experiments.
%
This allowed a great advancement in the understanding 
of the dynamics of heavy quark production.
The LHeC is the ideal machine for a further extension 
of similar historic importance because
a higher centre of mass energy and a much larger
integrated luminosity compared to HERA are available.
On top of this the 
heavy flavour measurements will greatly benefit
from the advanced detector design at LHeC,
which will be well equipped with high precision Silicon trackers (see Section\,\ref{LHEC:MainDetector:tracking}).
At HERA the tagging was restricted to central rapidities and
effective efficiencies\footnote{The effective efficiency takes the background
pollution into account. It is defined as the efficiency of 
an equivalent background free sample with the same signal precision
as that obtained in the data.}
of only 0.1\% (1\%) for charm (beauty)
were reached.
At LHeC efficiencies of 10\% (50\%) should be possible for charm (beauty)
and a large rapidity range can be covered
from the very backward to the very forward regions.
In the following, 
the main heavy quark production processes are first introduced, 
together with the relevant pQCD theoretical schemes 
and some related open questions.
Afterwards, the exciting measurement prospects for heavy flavours 
at the LHeC are further elucidated.

\begin{figure}[h!b]
\centerline{
\includegraphics[clip=,width=0.4\textwidth]{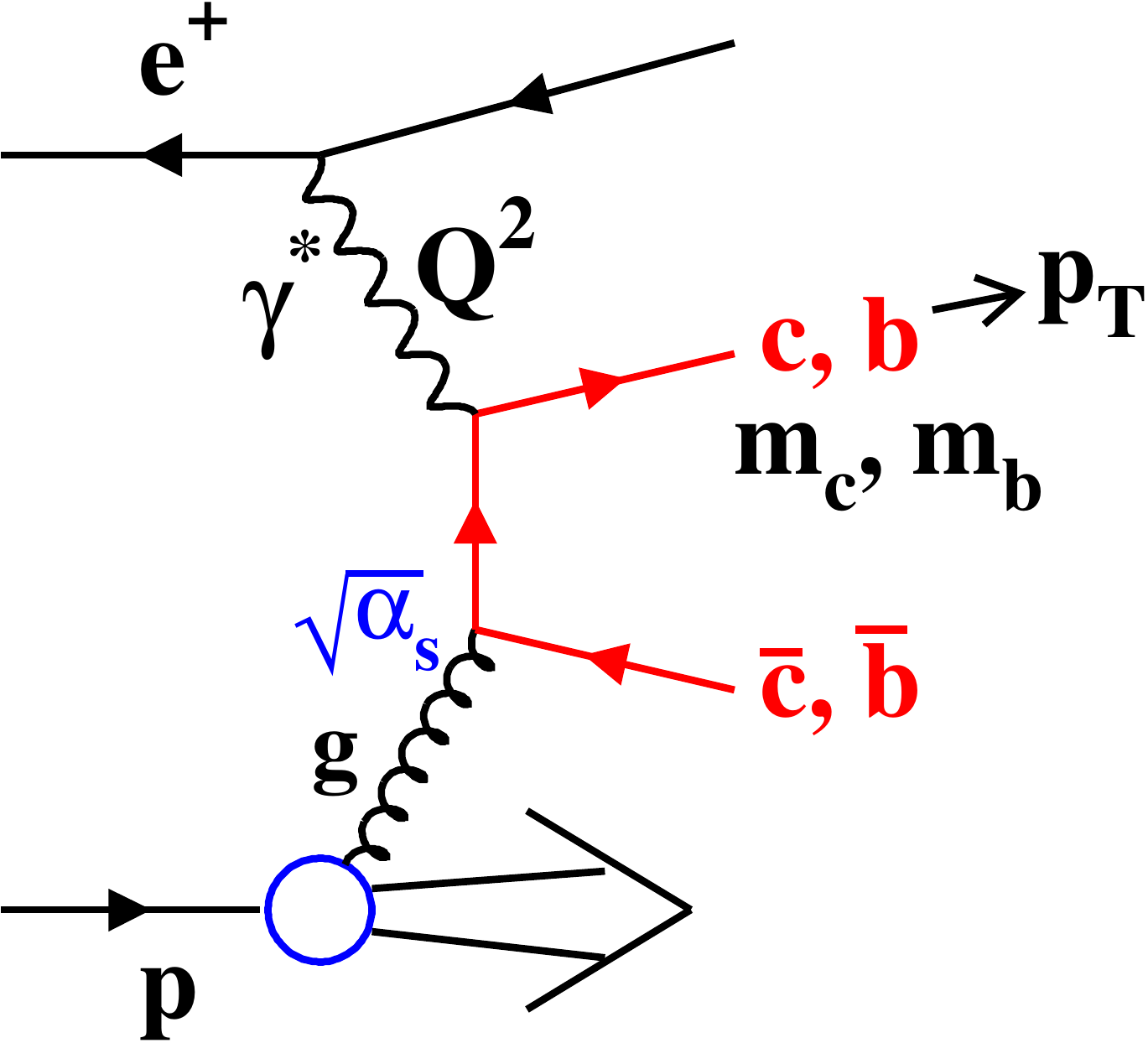}
\includegraphics[clip=,width=0.4\textwidth]{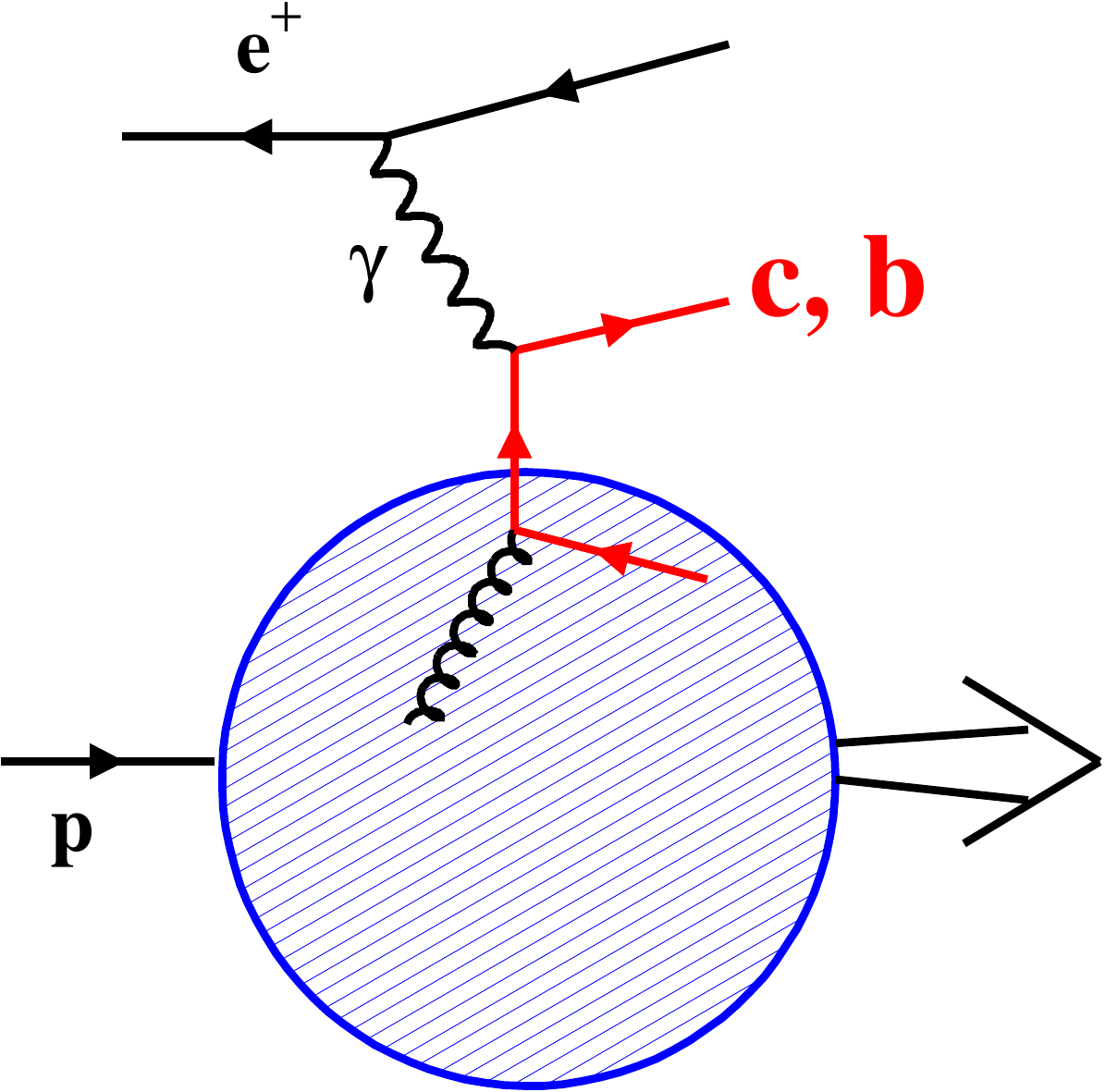}
}
\caption{Left: 
Leading order Boson Gluon Fusion (BGF) diagram for charm and beauty 
production in $ep$-collisions.
Right: Sketch of the leading order process in the massless approach
where charm and beauty quarks are treated as massless sea quarks in the proton.
}
\label{fig:hfl_bgf}
\end{figure}

In leading order, heavy quarks are produced in $ep$ collisions
via the Boson Gluon Fusion (BGF) process shown in Figure \ref{fig:hfl_bgf} on the left.
%
%
%
%
%
This process provides direct access to the gluon density in the proton.
BGF type processes dominate DIS scattering towards lower $x$,
due to the large gluon density.
In the high $Q^2$ limit, the  
events with charm and beauty quarks 
are expected to account for $\sim$ 36\% and $\sim 9\%$
of the BGF processes and hence contribute significantly
to inclusive DIS.
On the theoretical side, the description of heavy quark
production in the framework of perturbative QCD is complicated due to
the presence of several large scales
like the heavy quark masses, the transverse momentum
$p_T$ of the produced quarks and the momentum transfer $Q^2$.
Different calculation
schemes have been developed to obtain predictions from pQCD.
At low scales $p_T$ (or $Q^2$),  the fixed-flavour number scheme (FFNS) 
\cite{Harris:1997zq,Frixione:1994dv,Frixione:1995qc}
is expected to be most appropriate, where the quark masses 
are fully accounted for.
%
At very high scales the NLO FFNS scheme predictions are expected to break down
since large logarithms $\ln(p_T^2/m^2)$ are neglected that represent collinear gluon
radiation from the heavy quark lines.
These logarithms can be
resummed to all orders in the 
alternative zero-mass variable flavour number (ZM-VFNS) 
\cite{Binnewies:1998vm,Binnewies:1997gz,Kniehl:1996we,Cacciari:1995fs}
schemes.
%
%
Here the charm and beauty quarks are treated
above kinematic threshold as massless and appear also as active sea quarks 
in the proton, as depicted in figure~\ref{fig:hfl_bgf} in the sketch on the right.
Most widespread in use nowadays are the so-called
generalised variable flavour number schemes (GM-VFNS) 
\cite{Kramer:2003jw,Kniehl:2004fy}.
%
These mixed schemes converge to the massive and massless schemes at low and high
kinematic scales, respectively, and apply a suitable interpolation in
the intermediate region.
However, the exact modelling of the interpolation 
and in general the treatment of mass dependent 
terms in the perturbation series are still a highly
controversial issue among the various theory groups.
The different treatments
have profound implications for global PDF fits
and influence the fitted densities of 
gluons and other quark flavours in the proton.
%
This has direct consequences for many important cross section
predictions at LHC, for instance for Z and W production.
%
The value of the mass of the charm quark is also an important uncertainty in the predictions. 
In the determinations of $m_c$ we have to distinguish between the pole mass and the running mass. 
Fits to the present data have been performed using both as free parameters. 
First, Ref. \cite{Martin:2010db} used the pole mass as a free parameter and finds $m_c=1.45$ GeV at NLO and 1.26 GeV at NNLO. 
Alternatively, Ref.  \cite{Alekhin:2010sv} use the running mass and finds $m_c(m_c)=1.26$ GeV at NLO and 1.01 GeV at NNLO.  
Typically the uncertainties quoted in these results are about $\pm 10\%$. 
After the conversion from the pole to the running mass these values 
obtained by the two analyses are quite compatible with each other.  
Clearly, LHeC data are required to improve the perturbative stability 
and to increase the precision in our knowledge of $m_c$.

%
%

The following main physics highlights are expected for heavy quark
production measurements at LHeC:
\begin{itemize}
\item {\em Massive vs Massless scheme:}
At HERA the charm and beauty production data were found to be 
well described by the NLO FFNS scheme calculations
over the whole accessible phase space, up to the highest $p_T$ and $Q^2$ scales.
%
An LHeC collider would allow to extend these studies
to a much larger kinematic phase space, with much greater precision,
and thus precisely map the expected transition to the massless regime.
%
%
\item {\em Gluon density determination:}
At HERA the recorded charm data already provide 
some interesting sensitivity to the gluon density in the proton.
However due to the small tagging efficiencies the precision is far below
that obtained from the scaling violations of $F_2$ or those from jet data.
At LHeC this situation will greatly improve and it will be possible
to probe the gluon density via the BGF process 
down to proton momentum fractions $x_g \le 10^{-5}$, 
where it is currently not well known.
%
%

At such low values of $x_g$ a fixed-order perturbative computation becomes
unreliable. It is then necessary to resum both evolution equations and
hard matrix elements. In fact, heavy quark production is the first process
for which all-order small $x$ resummed terms were
computed, and  the high-energy
factorisation, on which the whole of perturbative small-$x$ resummation is based,
was proven in this context~\cite{Catani:1990xk,Catani:1990eg}.
Heavy quark production at the LHeC, with its high precision, energy and extended
kinematic coverage, would thus provide an ideal setting for tests of high-energy
factorisation and small $x$ resummation.

In this context it is also interesting to note that in the BGF process
one can reach for charm production much smaller 
$x_g$ values than with flavour inclusive jets 
since experimentally one can tag
charm quarks with small transverse momenta.
The studies of heavy flavour production sensitive to the
gluon density can be done both in DIS and in the photoproduction
kinematic regime.
\item {\em Charm and beauty densities in the proton:}
In general the measurements 
of the structure functions $F_2^{cc}$ and $F_2^{bb}$
are of the highest interest for theoretical analyses of
heavy flavour production in $ep$ collisions.
These structure functions describe the parts of $F_2$
which are due to events with charm or beauty quarks in the final state.
At sufficiently high $Q^2\gg m_c^2,m_b^2$,  the two structure functions 
can be directly related to effective densities of charm and beauty quarks in the proton.
%
This can be used for predictions of many interesting processes at LHC with charm
or beauty quarks in the initial state.
For instance, as discussed in~\cite{Belyaev:2005nu},
in the minimal supersymmetric extension of the standard model
the production of the neutral Higgs boson $A$ is driven by $b\bar{b} \rightarrow A$
and for the calculation of this process the PDF uncertainties
dominate over the theoretical uncertainties of the perturbative calculation.
At HERA the measurements of $F_2^{bb}$ barely reached the
necessary high $Q^2$ regime and only with modest precision.
Huge phase space extensions and improvements in precision
will be possible at LHeC.
%
%
\item {\em Constraining VFN parameters:}
The treatment of heavy quarks in QCD fits is subject to uncertainties,
both theoretically, as several schemes exist for describing the onset
of heavy quarks (thus far only for charm and beauty but with the LHeC 
also for top), and phenomenologically as the charm mass enters
as an external parameter. The LHeC precision NC and CC measurements,
combined with precision data on $F_2^{cc}$ are estimated to
determine this parameter to better than $5$\,MeV. This will resolve
the issue of the influence of $m_c$ on the determination of the
strong coupling constant and it will also clarify the theoretical 
treatment of heavy flavour in pQCD. 
\item {\em Intrinsic charm component:}
For a long time it has been suggested 
\cite{Brodsky:1980pb,Brodsky:1984nx,Harris:1995jx,Franz:2000ee} that 
the proton wave function might contain an intrinsic
charm component $uudc\bar{c}$.  
This would show up mainly at large $x>0.1$.
%
Unfortunately at HERA this large $x$ region could not be studied mainly due to the
limited detector acceptance in the forward region.
Due to the even larger boost in the forward direction at LHeC 
the situation is also not easy there.
However, with a forward tracking acceptance down to small polar angles 
there could be a chance to study this effect, in particular with the planned
low energy proton runs.
%
%
\item {\em Strange/antistrange densities:}
Events with charm quarks in the final state
can also be used as a tool for other purposes.
The strange and antistrange quark densities in the proton can be analysed
via the charge current process $sW \rightarrow c$, where the charm quark
is tagged in the event.
At HERA this was impossible due to the small cross sections, but at LHeC the
cross sections for CC reactions are much higher and as noted 
before the other experimental conditions (luminosities, detector) 
will greatly improve.
        This leads to the first and precise measurement of
         both the strange and the anti-strange quark densities
         as is demonstrated in Sect.\,\ref{sec:dirpart}.
\item {\em Electroweak physics:}
There are intriguing possibilities for LHeC electroweak physics studies   
with charm and beauty quarks in the final state.
For example one should be 
able to do a lepton beam polarisation asymmetry measurement for neutral current events,
where the scattered quark is tagged as a beauty quark.
This will provide direct access to the axial and vector couplings of the beauty quark
to the Z boson.
Similar measurements are possible for charm.
%
\end{itemize}

In summary the measurements of charm and beauty at an LHeC will be extremely
useful for high precision pQCD tests, in particular 
for the understanding of the treatment of mass terms in pQCD,
 to improve the knowledge of the proton PDFs: 
directly for g, c, b, s, $\bar{\mbox{s}}$\/ densities
and indirectly also for u and d.
Furthermore they provide a great potential for electroweak physics.
At the time when the LHeC will be operated, 
the pQCD theory calculations are expected to have advanced considerably.
In particular there is hope that full massive scheme NNLO calculations of order 
$o(\alpha_s^3)$ will be available by then.
These will allow theory to data comparisons for heavy flavour
production in $ep$ collisions with unprecedented precision.

In the following subsections several dedicated simulation studies are presented
which illustrate some of the expected highlights. 
First total cross sections are  presented for various processes involving
charm, beauty and also top quarks in the final state, showing that
LHeC will be a genuine {\em multi heavy flavour factory}.
Then the expected measurements of the structure functions
$F_2^{cc}$ and $F_2^{bb}$ are discussed and compared to the existing
HERA data.
Next a study is presented of the possibility to measure intrinsic charm 
with dedicated low proton energy runs.
Finally predictions for differential charm hadron production
cross sections in the photoproduction kinematic regime are 
presented and compared to HERA, demonstrating the large phase space extension.
\subsection{Total production cross sections for charm, beauty and top quarks}
\begin{figure}[h!]
\begin{center}
\includegraphics[clip=,width=1.0\textwidth]{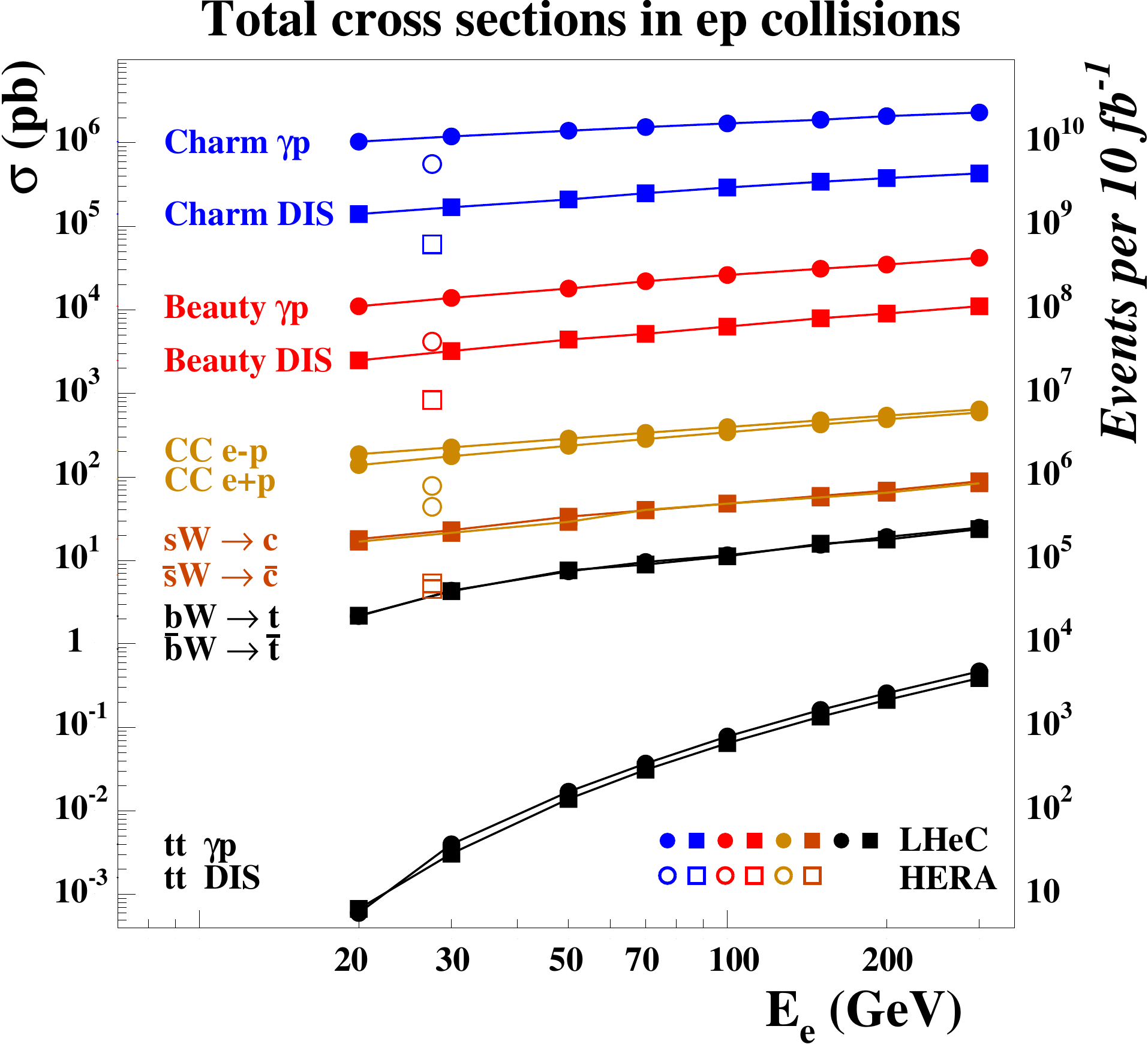}
\end{center}
\caption{Total production cross section predictions 
for various heavy quark processes at the LHeC (with 7 TeV proton energy),
as a function of the lepton beam energy.
The following processes
are covered: charm and beauty production in photoproduction ($Q^2<1\;\mbox{GeV}^2$) 
and DIS ($Q^2>2\;\mbox{GeV}^2$), 
the charged current processes
$sW\rightarrow c$ and $bW\rightarrow t$ and top pair production in 
photoproduction and DIS.
The flavour inclusive charged current total cross section is also shown. 
All predictions are taken from Monte Carlo simulations, 
some details can be found in Table \ref{tab:hfl_proctot}.
For comparison also the predicted cross sections at HERA 
(with 920 GeV proton energy) are shown.
}\label{fig:hfl_proctot}
\end{figure}

This section presents total cross sections for various 
heavy quark processes
at LHeC (with 7 TeV proton beam energy) as a function of the lepton beam energy.
Predictions are obtained for:
charm and beauty production in photoproduction and DIS, 
the charged current processes
$sW\rightarrow c$ and $bW\rightarrow t$ and top
quark pair production in 
photoproduction and DIS.
For comparison 
the flavour inclusive charged current total cross section is also shown. 
Table \ref{tab:hfl_proctot} lists the generated processes, 
the used Monte Carlo generators and
the selected parton distribution functions.
%
%
%
\begin{table}[h]
\begin{center}
\begin{tabular}{|l|l|c|}
\hline
Process & Monte Carlo & PDF \\
\hline 
Charm $\gamma p$ & PYTHIA6.4~\cite{Sjostrand:2006za}  
& CTEQ6L~\cite{Pumplin:2002vw} \\
Beauty $\gamma p$ &   &   \\
tt $\gamma p$ &  & \\
\hline
Charm DIS &  RAPGAP3.1~\cite{Jung:1993gf}
& CTEQ5L~\cite{Lai:1999wy} \\
Beauty DIS &  &   \\
tt DIS & & \\
\hline
CC $e^{+}p$ & LEPTO6.5~\cite{Ingelman:1996mq}
& CTEQ5L  \\
CC $e^{-}p$ &   &    \\
$sW \rightarrow c$ & &   \\
$\bar{s}W \rightarrow \bar{c}$ &  &   \\
$bW \rightarrow t$ &  &  \\
$\bar{b}W \rightarrow \bar{t}$ & &   \\
\hline
tt DIS & RAPGAP 3.1  & CTEQ5L  \\
\hline
\end{tabular}
\caption{Used generator programmes for the predictions of total cross sections
at LHeC, shown in Figure \ref{fig:hfl_proctot}. 
For all processes with top quarks the top mass was
set to a value of 170 GeV.
For both photoproduction (labelled as $\gamma p$) and DIS 
only direct photon processes were generated
and no reactions with resolved photons.
The $Q^2$ ranges of the generated data are 
$Q^2<1\;\mbox{GeV}^2$ for photoproduction with PYTHIA,
$Q^2>2\;\mbox{GeV}^2$ for DIS with RAPGAP and 
$Q^2>4\;\mbox{GeV}^2$ for the processes with LEPTO. 
}
\label{tab:hfl_proctot}
\end{center}
\end{table}
%
%
The resulting cross sections are shown in Figure \ref{fig:hfl_proctot}.
%
%
%
%
%
For comparison also the predicted cross sections for the HERA collider 
(with 920 GeV proton energy)
are presented.
The cross sections at LHeC are typically about one order of magnitude larger
compared to HERA.
Attached to the right of the plot are 
the number of events that are produced per
$10\;\mbox{fb}^{-1}$ of integrated luminosity.
For instance for charm more than 10 billion events
are expected in photoproduction and for beauty 
more than 100 million events. 
In DIS the numbers are typically 
a factor of five smaller.
The strange and antistrange densities can be probed
with some hundred thousands of charged current
events with charm 
in the final state.
The top quark production is dominated by the single production 
in the charged current reaction with beauty in the initial
state and about one hundred thousands tops and a similar 
number of antitops are expected.
In summary 
the LHeC will be the first $ep$ 
collider which provides access to all quark flavours
and with high statistics.
%
\subsection{Charm and beauty production in DIS}
This section presents predictions for 
charm and beauty production in neutral current DIS, 
for $Q^2$ values of at least a few $\mbox{GeV}^2$. 
The predictions are given for the 
structure functions $F_2^{c\bar{c}}$ and $F_2^{b\bar{b}}$
which denote the contributions from charm and beauty
events to $F_2$.
As explained in section \ref{sec:hfl_intro} the two structure functions are
of large interest for theoretical analyses. 
%
Experimentally they are obtained by determining
the total charm and beauty cross sections in 
two-dimensional bins of $x$ and $Q^2$.
The LHeC projections  shown here
were obtained with the Monte Carlo programme RAPGAP~\cite{Jung:1993gf}
%
which generates charm and beauty production 
with massive leading order matrix elements supplemented
by parton showers.
The proton Parton Distribution Function set CTEQ5L~\cite{Lai:1999wy}
were used and the heavy-quark masses were set to $m_{c}=1.5\;$GeV and
$m_{b}=4.75\;$GeV, respectively.
In general at HERA the RAPGAP predictions are known to provide 
a reasonable description of the measured charm and beauty DIS 
production data.
The RAPGAP data
were generated for an LHeC collider scenario
with 100 GeV electrons colliding with 7 TeV protons.
The statistical uncertainties have been evaluated
such that they correspond to an 
integrated data luminosity of $10\;\mbox{fb}^{-1}$.
All studies were done at the parton level, hadronisation effects were not
taken into account.
Tagging efficiencies of 10\% for charm quarks
and 50\% for beauty quarks have been assumed, respectively.
These efficiencies are about a factor 100 larger compared to 
the effective efficiencies 
(including the dilution due to background pollution)
at HERA which may look surprisingly but is explainable.
At HERA the charm quarks were tagged either with full
charm meson reconstruction or with inclusive secondary 
vertexing of charm hadron decays.
The first method suffered from very small branching ratios of 
suitable decay channels.
The second technique which was also used for the beauty tagging was 
affected by a large pollution
from light quark background events due to the limited detector
capabilities to separate secondary from primary vertices.
At LHeC one can expect 
a much better secondary vertex identification 
and thus a very strong background
reduction.
It is difficult to predict exactly how much background 
pollution will remain at LHeC, 
so for the purpose of this simulation study it was 
completely neglected.
Systematic uncertainties were  neglected for
the illustrations presented here, but an estimate was provided for the
subsequent investigation of the determination of the charm mass.
%
%
%

\begin{figure}[h!]
\begin{center}
\includegraphics[clip=,width=1.00\textwidth]{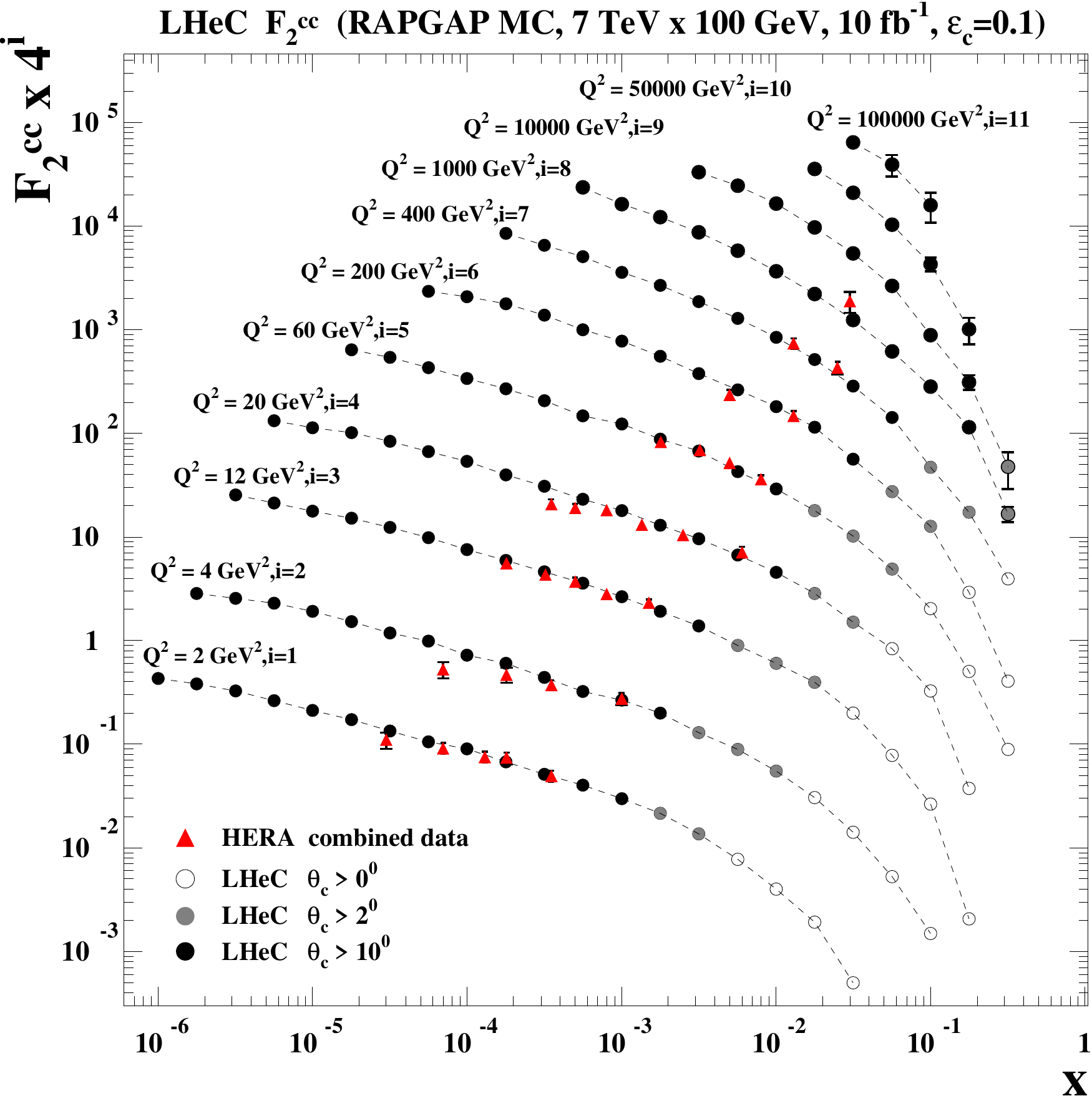}
\end{center}
\caption{$F_2^{cc}$ projections for LHeC compared 
to HERA data~\cite{h1zeusf2cc}, shown as a function of $x$ for various
$Q^2$ values. 
The expected LHeC results 
obtained with the RAPGAP MC simulation are shown as points
with error bars representing the statistical uncertainties.
The dashed lines are interpolating curves between the points.
For the open points 
the detector acceptance is assumed to cover the 
whole polar angle range.
For the grey shaded and black points events are only accepted
if at least one charm quark is found with polar angles
$\theta_c > 2^{0}$ and $\theta_c > 10^{0}$, respectively.
For further details of the LHeC simulation see the main text.
The combined HERA results from H1 and ZEUS
are shown as triangles with error bars representing
their total uncertainty.
}
\label{fig:hfl_f2cc_1}
\end{figure}

\begin{figure}[h!]
\begin{center}
\includegraphics[clip=,width=1.0\textwidth]{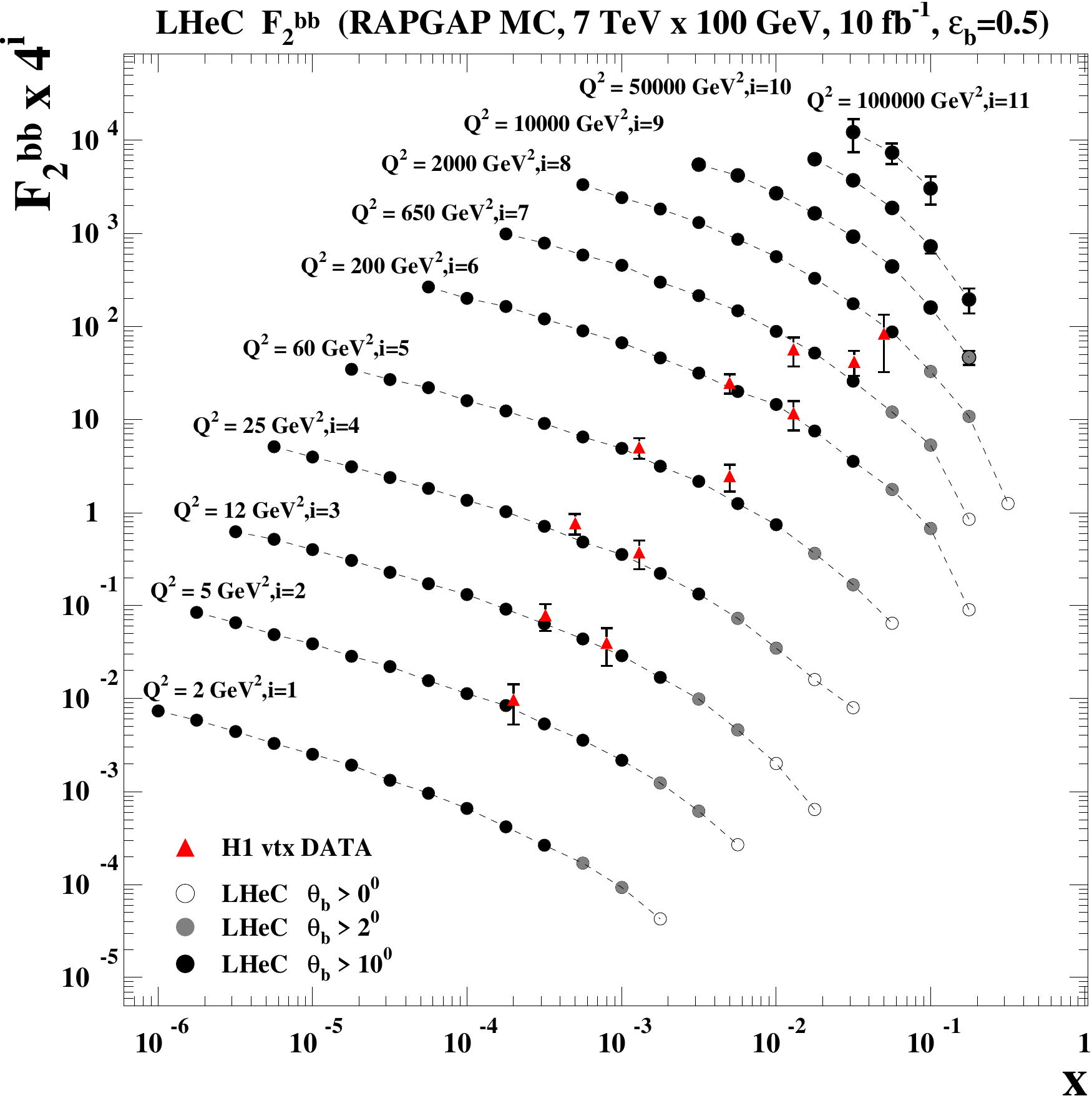}
\end{center}
\caption{
$F_2^{bb}$ projections for LHeC compared 
to HERA data~\cite{:2009ut} from H1, shown as a function of $x$ for various
$Q^2$ values. 
The expected LHeC results 
obtained with the RAPGAP MC simulation are shown as points
with error bars representing the statistical uncertainties.
The dashed lines are interpolating curves between the points.
For the open points 
the detector acceptance is assumed to cover the 
whole polar angle range.
%
%
For the grey shaded and black points events are only accepted
if at least one beauty quark is found with polar angles
$\theta_b > 2^{0}$ and $\theta_b > 10^{0}$, respectively.
For further details of the LHeC simulation see the main text.
The HERA results from H1
are shown as triangles with error bars representing
their total uncertainty.}
\label{fig:hfl_f2bb_1}
\end{figure}

%
Figures \ref{fig:hfl_f2cc_1} and \ref{fig:hfl_f2bb_1} 
show the resulting RAPGAP predictions at LHeC for 
the structure functions $F_2^{cc}$ and $F_2^{bb}$, respectively,
compared to recent measurements~\cite{h1zeusf2cc} from HERA.
%
%
%
%
%
%
%
%
The data are shown as a function of $x$ for various
$Q^2$ values.
The $Q^2$ values were chosen such that they cover a large fraction of
the specific values for which HERA results are available. 
Some further values demonstrate the phase space extensions at LHeC.
%
%
%
The projected LHeC data are presented as points with error bars 
which (where visible) indicate the estimated statistical uncertainties. 
For the open points 
the detector acceptance is assumed to cover the 
whole polar angle range.
For the grey shaded and black points events are only accepted
if at least one charm quark is found with polar angles
$\theta_c > 2^{0}$ and $\theta_c > 10^{0}$, respectively.
The selected results from HERA 
are shown as triangles with error bars indicating
the total uncertainty.
The HERA $F_2^{cc}$ results in Figure \ref{fig:hfl_f2cc_1} 
are those of a recent weighted average \cite{h1zeusf2cc} 
of almost all available
measurements from H1 and ZEUS.
In a large part of the covered phase space these
results are already rather accurate,
with precisions between 5\% and 10\%. 
The overlaid LHeC projections show 
a vast phase space increase to lower and larger
$x$ and also to much higher $Q^2$ values.
In the kinematic overlap region the 
expected statistical precision at LHeC is typically
a factor $\sim 40$ better than at HERA which can be easily
explained by the 20 times larger integrated luminosity
and the $\sim 100$ times better tagging efficiency.
For the smaller $x$ not covered by HERA the 
precision even improves at LHeC due to the growing
cross sections driven by the rise of the gluon density.
The best statistical precision in the LHeC simulation
is observed at smallest $x$ values and small $Q^2$
and reach down to 0.01\%.
As seen in the simulation (not shown here) the LHeC 
$F_2^{cc}$ data provide access to the
the gluon density in the BGF process 
down to proton momentum fractions $x_g \sim 10^{-5}$. 
The LHeC data can also provide a substantial extension
to higher $x$ compared to HERA 
where the measurements reached $x$ values of a few percent.
As evident from the simulated 
points with different polar angle cuts 
this necessitates an excellent forward tagging
of charm quarks.
In any case values of $x>0.1$ should be
accessible in the medium and large $Q^2$ domain.

Figure \ref{fig:hfl_f2bb_1} 
show the RAPGAP predictions at LHeC for $F_2^{bb}$.
Also shown are the results from the 
H1 analysis \cite{:2009ut} 
based on inclusive secondary vertex tagging.
Clearly these results and similar ones (not shown) from ZEUS
are not very precise, the typical total uncertainties are
20-50\%.
Again, the LHeC $F_2^{bb}$ projections demonstrate a vast phase space
increase, similar as for charm.
The best statistical precision obtained at LHeC for $F_2^{bb}$ is
seen in the simulation towards low $x$ and small and medium $Q^2$
and reach down to 1 per mille.
The measurements at LHeC will enable a precision mapping of 
beauty production from kinematic threshold to large $Q^2$.
In the context of the generalised variable flavour number schemes (GM-VFNS)
this will allow to study in detail the onset of the beauty quark density in
the proton and to compare it to the charm case.
%
%
As mentioned in Section\,\ref{sec:hfl_intro},
for high $Q^2 \gg m_b^2$ the $F_2^{bb}$ results 
can be directly interpreted in terms of an effective beauty
density in the proton.
The measurement of this density 
is of large interest because it can be used to predict
beauty quark initiated processes at the LHC.
As visible in the figure, HERA covers only a small phase space
in this region and with moderate precision.
However, at LHeC the prospects 
for measuring $F_2^{bb}$ in this region
are very good.
\subsection{Determination of the charm mass parameter in VFN schemes}
\label{sec:mc}
%
%
A quantitative understanding of proton structure, as has been mentioned above,
requires to correctly and precisely describe the contributions of the
heavy quarks. A quantity of particular concern is the charm-quark mass,
$m_c$, which formally enters as a parameter 
 the calculations of photon-gluon fusion into
$c \overline{c}$, with different meanings in
different variable flavour number schemes. Heavy quark densities
and specifically this parameter 
can be constrained with high precision inclusive and charm
production cross section measurements. 
The value of $m_c$ is directly related to the value of the strong
coupling constant, an uncertainty of $\delta m_c =100$\,MeV corresponding
to a relative uncertainty on $\alpha_s$ of about half a per cent,
as obtained by  H1~\cite{Adloff:2000qk}. 
The LHeC prospect of measuring $\alpha_s$ to per mille precision
requires to control $m_c$ to better than $10$\,MeV. Some of the observed
differences of recent $\alpha_s$ determinations in DIS can be
correlated with different assumptions on $m_c$.
The value of $m_c$ and the treatment of heavy flavour contributions
has similarly significant implications for the predictions of the $W$ and $Z$ boson
cross sections at the LHC.

A study is performed to estimate  the sensitivity of the charm 
quark production at the LHeC to the  $m_c$ parameter which enters the QCD fits.
As input the NC and CC pseudodata are used with their uncertainties
as described in Section~\ref{sec:simNC}. In addition data of
the charm structure function are simulated for
a luminosity of $10$\,fb$^{-1}$.  The assumed measurement method is
the impact parameter tagging technique as has been used by H1. 
The statistical uncertainty is 
scaled according to the charm tagging efficiency, 
assumed to be $10$\,\%, and a light 
quark background, of $1$\,\%. The dominating   systematic error 
comprises the correlated DIS cross section errors and an extra systematic
uncertainty of $2$\,\%.

A first study  uses the inclusive CC and NC 
cross section data from HERA, to which in a further step the 
combined H1 and ZEUS $F_2^{cc}$ data are added. In a second step the
LHeC pseudo-NC and CC data are represented by the
QCD fit central values  with their simulated uncertainties.  
Finally, the expected, simulated precision measurement of $F_2^{cc}$ with
the LHeC is added. In each case variations of $m_c$ in small steps
from $1.2$ to $1.8$\,GeV are considered and parabola fits made
to $\chi^2 (m_c)$. The resulting experimental uncertainties are summarised
in Table\,\ref{tab:errmc}.  It can be seen that the inclusive LHeC data
improve the uncertainty from the inclusive HERA data by a factor of $4$.
A genuine high precision measurement of $m_c$ can be obtained
from the simulated $F_2^{cc}$ data expected from the LHeC, with its
much improved range and precision based on a smaller beam spot
and dedicated vertex detector technology. The value obtained of $3$\,MeV
is an example for the huge potential for precision QCD physics of 
the LHeC. It specifically suggests that any uncertainty effect on the
measurement of $\alpha_s$ connected with the charm treatment
will be negligible. 

It is finally worth noting that at such a high level
of precision an improved determination of the beauty mass parameter
will become relevant. This was not studied numerically. From the
simulated $F_2^{bb}$ measurement, however, one can deduce
that $m_b$ would be determined very precisely as well.
The improvement with respect to HERA should be even more dramatic because,
unlike for charm,
the $F_2^{bb}$ data of HERA have been of limited accuracy, of order
$20$\,\%, only.
\begin{table}[h]
  \centering
  \begin{tabular}{|l|c|}
    \hline
Data input & Experimental uncertainty on $m_c$ [MeV] \\ \hline
HERA: NC+CC &  100  \\
HERA: NC+CC+ $F_2^{cc}$ & 60 \\ \hline
LHeC: NC+CC & 25 \\
LHeC: NC+CC+ $F_2^{cc}$ & 3 \\
 \hline
  \end{tabular}
\caption{Experimental (statistical and systematic) uncertainty
on the charm mass parameter, $m_c$, in NLO QCD analyses of
the HERA neutral (NC) and charged (CC) current cross section
data complemented by the HERA $F_2^{cc}$ data (top) and
the corresponding results estimated for the LHeC (bottom), see text.
}
\label{tab:errmc}
\end{table}
\subsection{Intrinsic heavy flavour}
It is usually assumed, for example in fits of parton distributions, 
that the charm and bottom quark distributions in
the proton structure only arise from gluon splitting
 $g \to Q \bar Q.$  However, the proton light-front wavefunction
contains {\it ab initio } intrinsic heavy quark Fock state
components such as
$|uud c \bar
c>$~\cite{Brodsky:1980pb,Brodsky:1984nx,Harris:1995jx,Franz:2000ee}.
Intrinsic charm and
bottom  may explain the origin of high $x_F$ open-charm and open-bottom
hadron production, as well as the single and double $J/\psi$
hadroproduction cross sections observed at high $x_F$.   The
factorisation-breaking nuclear $A^\alpha(x_F)$ dependence  of
hadronic $J/\psi$ production cross sections may  also be explained.

Some past phenomenological studies~\cite{Pumplin:2007wg} have shown
that at large $x$ and low scale (just above charm threshold) the
intrinsic component might be several times larger than the
intrinsically generated one. Neglecting a  
significant large $x$ intrinsic
component may also lead to an incorrect assessment of the large $x$
gluon distribution.\footnote{In~\cite{Brodsky:2006wb}  a novel mechanism for inclusive and diffractive
Higgs production $pp \to p H p $ is proposed, in which the Higgs boson carries
a significant fraction of the projectile proton momentum. The production
mechanism is based on the subprocess $(Q \bar Q) g \to H $ where
the $Q \bar Q$ in the $|uud Q \bar Q>$ intrinsic heavy quark Fock state
of the colliding proton has approximately $80\%$ of the projectile
protons momentum.   A similar mechanism could produce the
Higgs at large $x_F \sim 0.8$ in $\gamma p \to H X$ at the
LHeC based on the mechanism  $\gamma (Q \bar Q)  \to H $
since the heavy quarks typically each carry light-cone
momentum fractions $x \sim 0.4$ when they arise from the
intrinsic heavy quark Fock states $|uud Q \bar Q>$ of the proton.}

The LHeC could establish the phenomenology of intrinsic heavy
flavours, and in particular charm, at large $x$.    
In addition to DIS measurements, one
can test the charm (and bottom) distributions at the LHeC
by measuring reactions such as $\gamma p \to c X$ where the
charm jet is produced at high $p_T$ in the reaction $\gamma c \to c g$.

\begin{figure}[h!tbp]
\centerline{\includegraphics[width=1.0\textwidth]{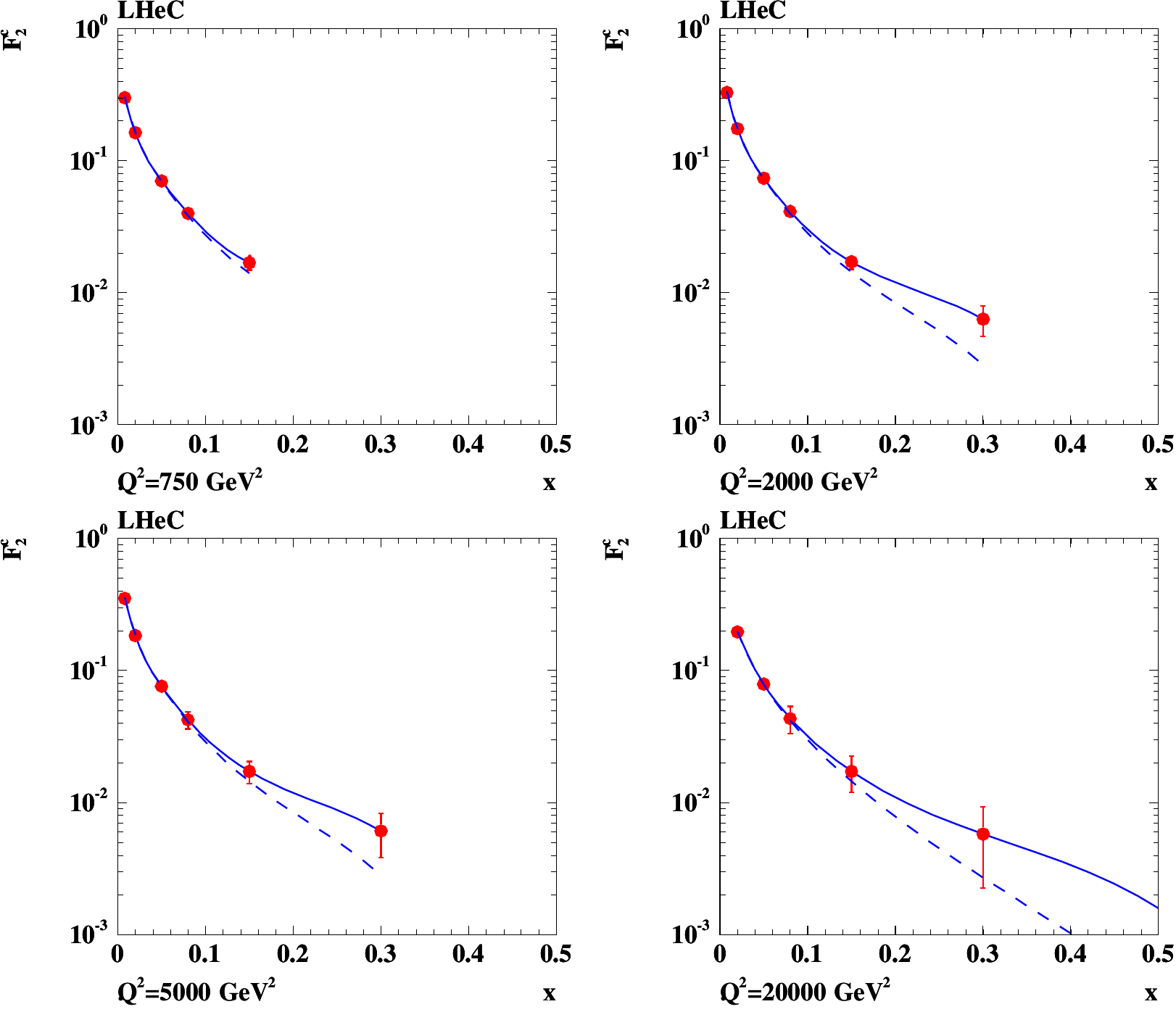}}
\caption{Simulation of measurement
of the charm structure function at large $x$, see text.
The errors are statistical, taking tagging and background
efficiencies into account. The tagging efficiency for charm
quarks was assumed to be 10\% and the amount of background
was estimated to be $0.01 \cdot  N_{ev}$, where $N_{ev}$ refers
to the total number of expected NC events in the respective 
$(Q^2,x)$ bin. 
Solid line: CTEQ66c predictions, including an intrinsic charm component,
dashed line: ordinary CTEQ6m.}
   \label{fig:chaintr}
\end{figure}

In order to access the charm and bottom distributions towards larger
Bjorken $x$, it is required to tag heavy flavour production in the
forward direction. As this is difficult in the asymmetric electron-proton
beam energy configuration such a measurement can favourably be
done with a reduced proton beam energy. Approximately,
as may be derived from Eq.\,\ref{xmax}, the small hadronic scattering
angle, $\theta_h$, is obtained from the relation,
$\theta_h^2 \simeq 2 \sqrt{Q^2}/E_p x$. Therefore a reduction by a factor
of $7$ of the proton beam energy $E_p$ enhances $x$ by $7$
at fixed $Q^2$ and $\theta_h$. One also notices that large $x$ is
reached at fixed $\theta_h$ and $E_p$ only at high $Q^2$.
The attempt to access maximum $x$ thus requires to find an
optimum of high luminosity, to reach high $Q^2$, and low proton
beam energy, to access large $x$. 
%
Fig.\,\ref{fig:chaintr} shows a simulated
measurement of the charm structure function for $E_p = 1$\,TeV
and a luminosity of $1$\,fb$^{-1}$. The two curves illustrate the
difference between  CTEQ66 PDF sets with and without an intrinsic
charm component, based on~\cite{Pumplin:2007wg}.
The actual amount of intrinsic charm may be larger than in the
CTEQ attempt, it may also be smaller.
One so finds
that a reliable detection of an intrinsic heavy charm component
at the LHeC may be possible, but will be a challenge for forward
charm detection and requires high luminosity.
The result yet may be rewarding
as it would have quite some theoretical consequences as sketched
above. It would be obtained in a region of high enough $Q^2$ to be
able to safely neglect any higher twist effects which may mimic
such an observation at low energy experiments.

\subsection{$D^{*}$ meson photoproduction study}
%
%
A study is presented of 
$D^{*}$ meson photoproduction at the LHeC.
It illustrates the large phase space extension to higher charm quark 
transverse momenta at LHeC compared to HERA; 
this will allow stringent tests
of the treatment of heavy quark mass dependent terms in pQCD.
The study is based on NLO predictions in the 
so-called general-mass variable-flavour-number scheme (GM-VFNS)
~\cite{Kramer:2003jw,Kniehl:2004fy}
for 1-particle inclusive heavy-meson production.
Both direct and resolved photon contributions are taken into account.
The cross section for direct photoproduction 
is a convolution of the proton PDFs, the cross section for the 
hard scattering process and 
the fragmentation functions FF for the transition of a 
parton to the observed heavy meson.
 For the resolved contribution, an additional 
convolution with the photon PDFs has to be performed. 
For the photoproduction predictions at the 
$ep$-colliders HERA and LHeC, the calculated photon proton cross
sections are convoluted with the photon flux using the Weizs\"{a}cker-Williams approximation.

In the GM-VFNS approach the large logarithms $\ln(p_T^2/m^2)$,
which appear due to the collinear mass singularities in the initial
and final state, are factorised into the PDFs and the FFs and summed by
the well known DGLAP evolution equations. The factorisation is
performed following the usual $\overline{\rm MS}$ prescription which
guarantees the universality of both PDFs and FFs. At the same time,
mass-dependent power corrections are retained in the hard-scattering
cross sections, as in the FFNS. 
For the photon PDF the parameterisation of Ref.\ \cite{Aurenche:2005da} 
with the standard 
set of parameter values is used and for the proton PDF the 
parameterisation CTEQ6.5 \cite{Tung:2006tb} of the CTEQ group.
For the FFs the set Belle/CLEO-GM of Ref.\ \cite{Kneesch:2007ey} is chosen. 
%
%
%
\begin{figure}[h!]
\begin{center}
\includegraphics[width=0.49\textwidth]{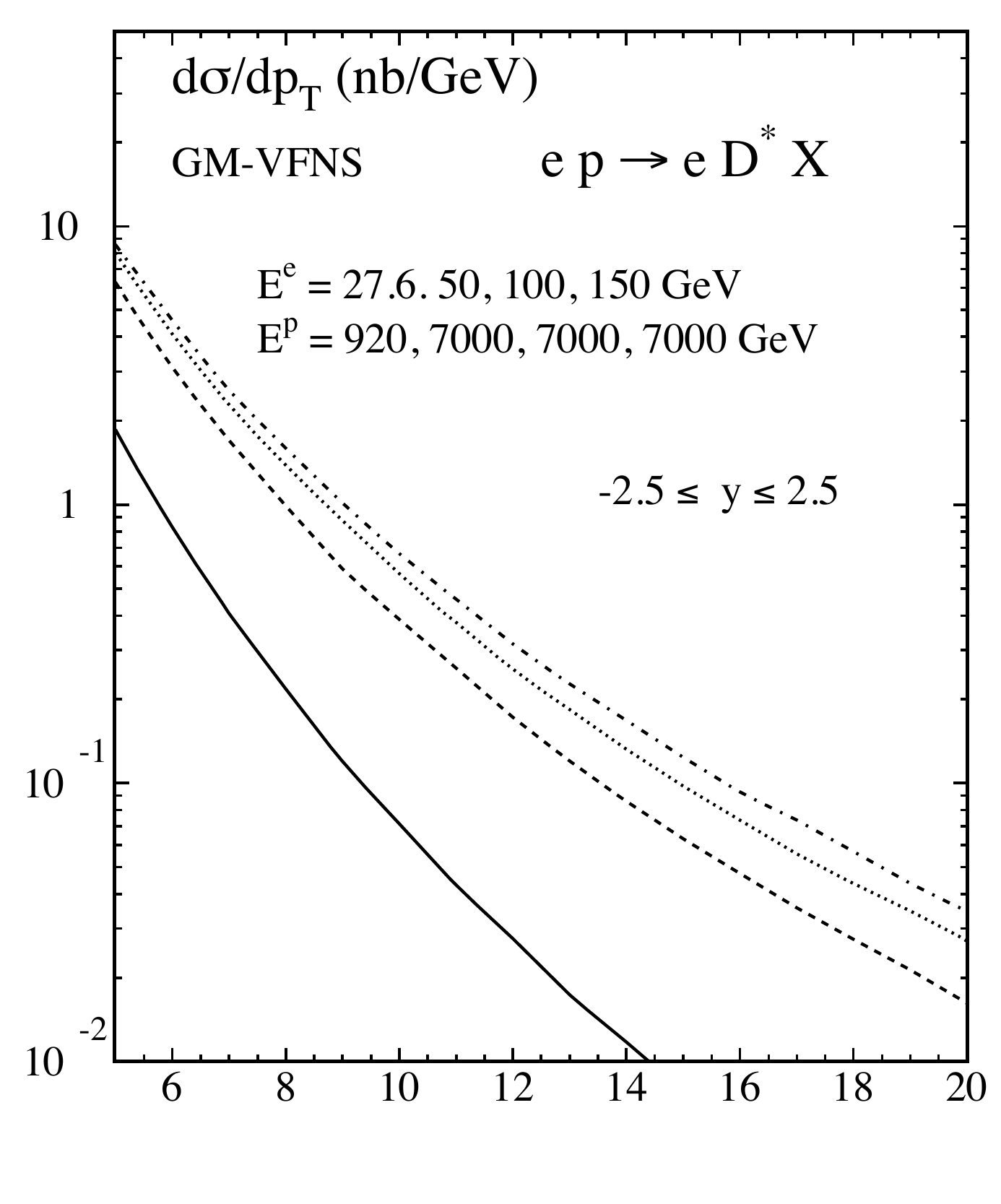}
\includegraphics[width=0.49\textwidth]{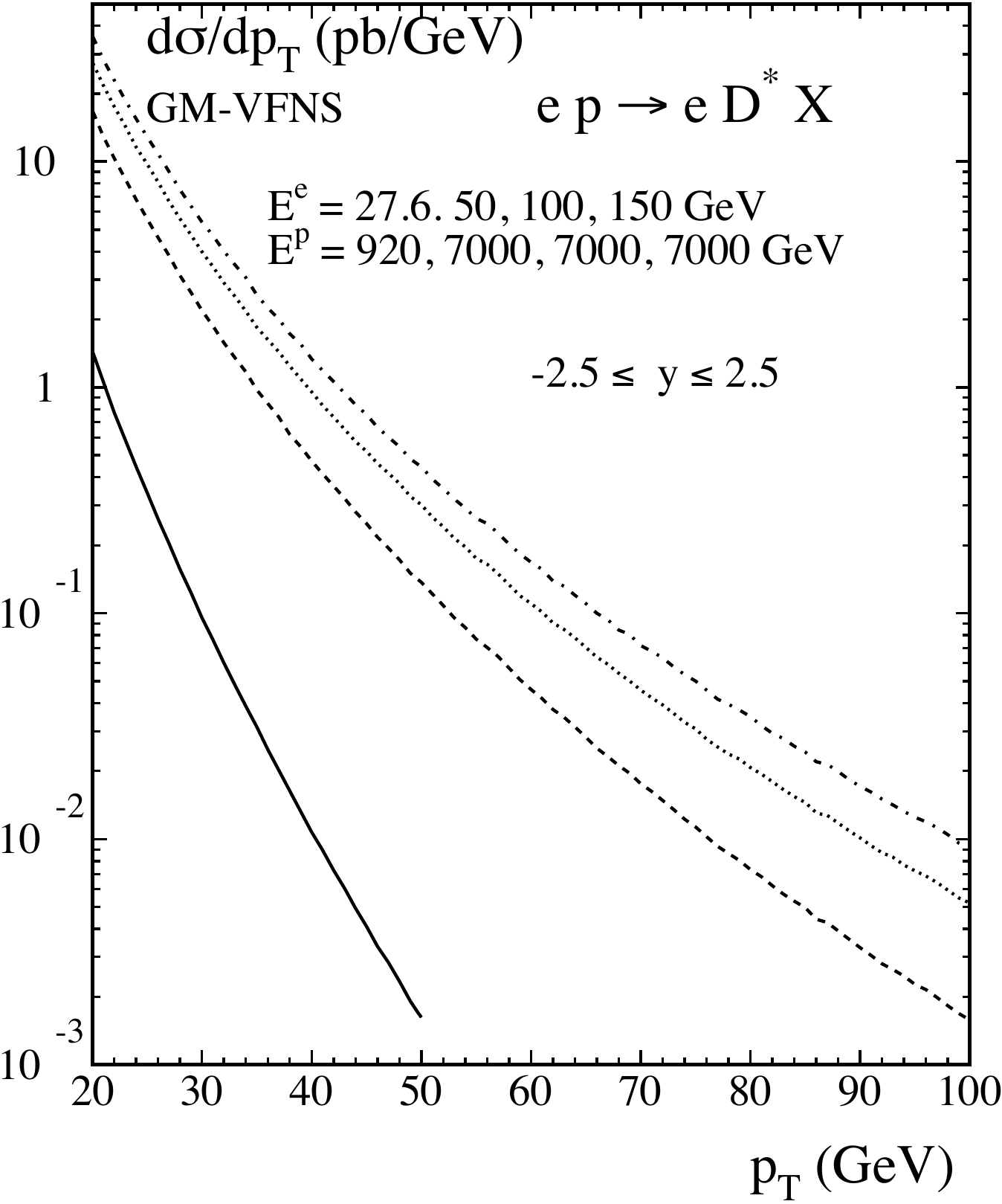}
\end{center}
\caption{\label{fig:dst_pt} The $p_T$-differential cross section 
for the production of $D^{\ast}$ mesons at LHeC for different beam 
energies integrated over rapidities $|\eta| \leq 2.5$, 
for the low-$p_T$ range 5 GeV$ \leq p_T 
\leq 20$ GeV (left) and for the high-$p_T$ range  20 GeV$ \leq p_T 
\leq 50$ GeV (right). The curves from 
bottom to top correspond to the combinations of beam energies as 
indicated in the figure. The lowest curves are showing the cross sections
at the HERA beam energies.}
\end{figure}
\begin{figure}[h!]
\begin{center}
\includegraphics[width=0.49\textwidth]{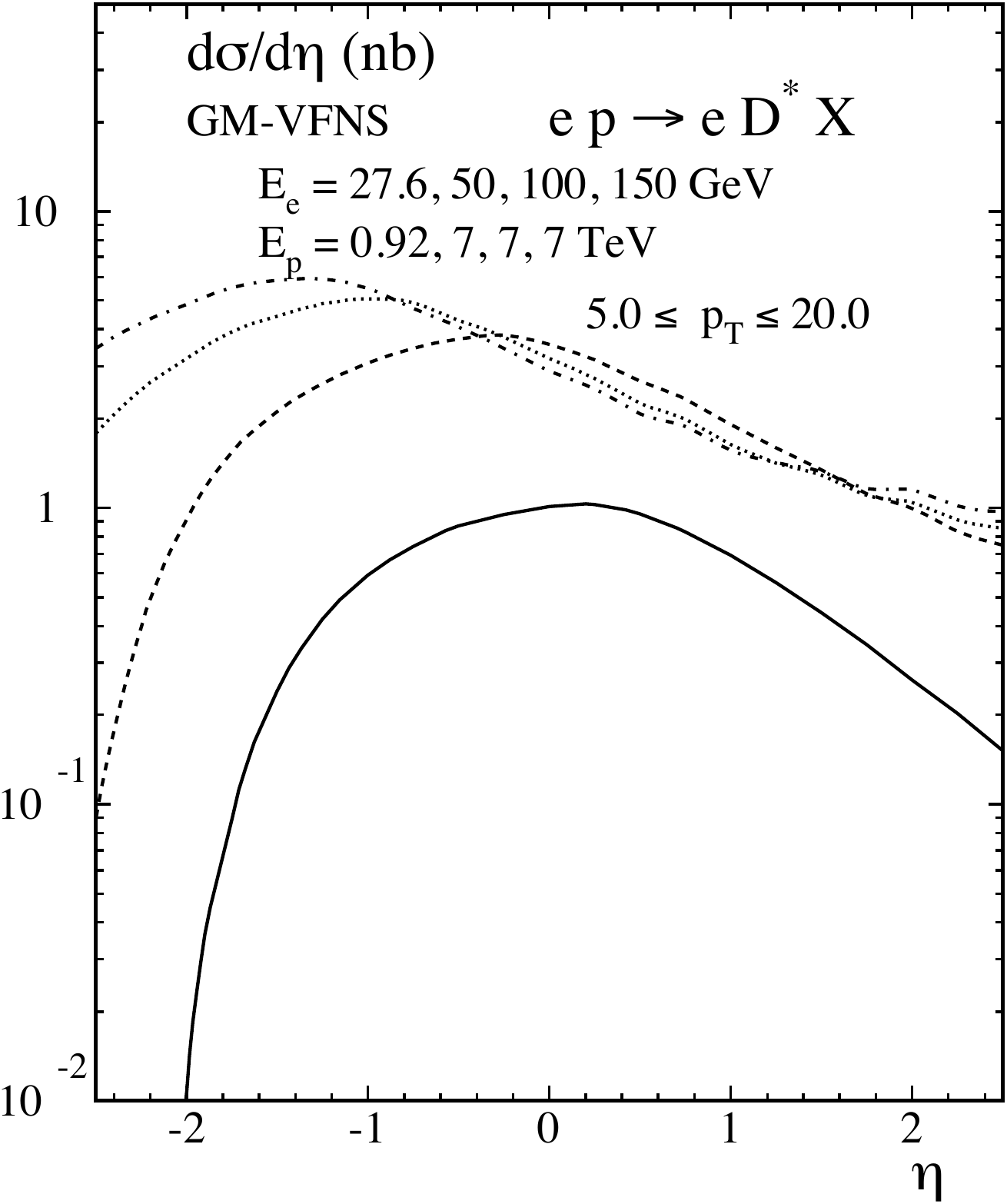}
\includegraphics[width=0.49\textwidth]{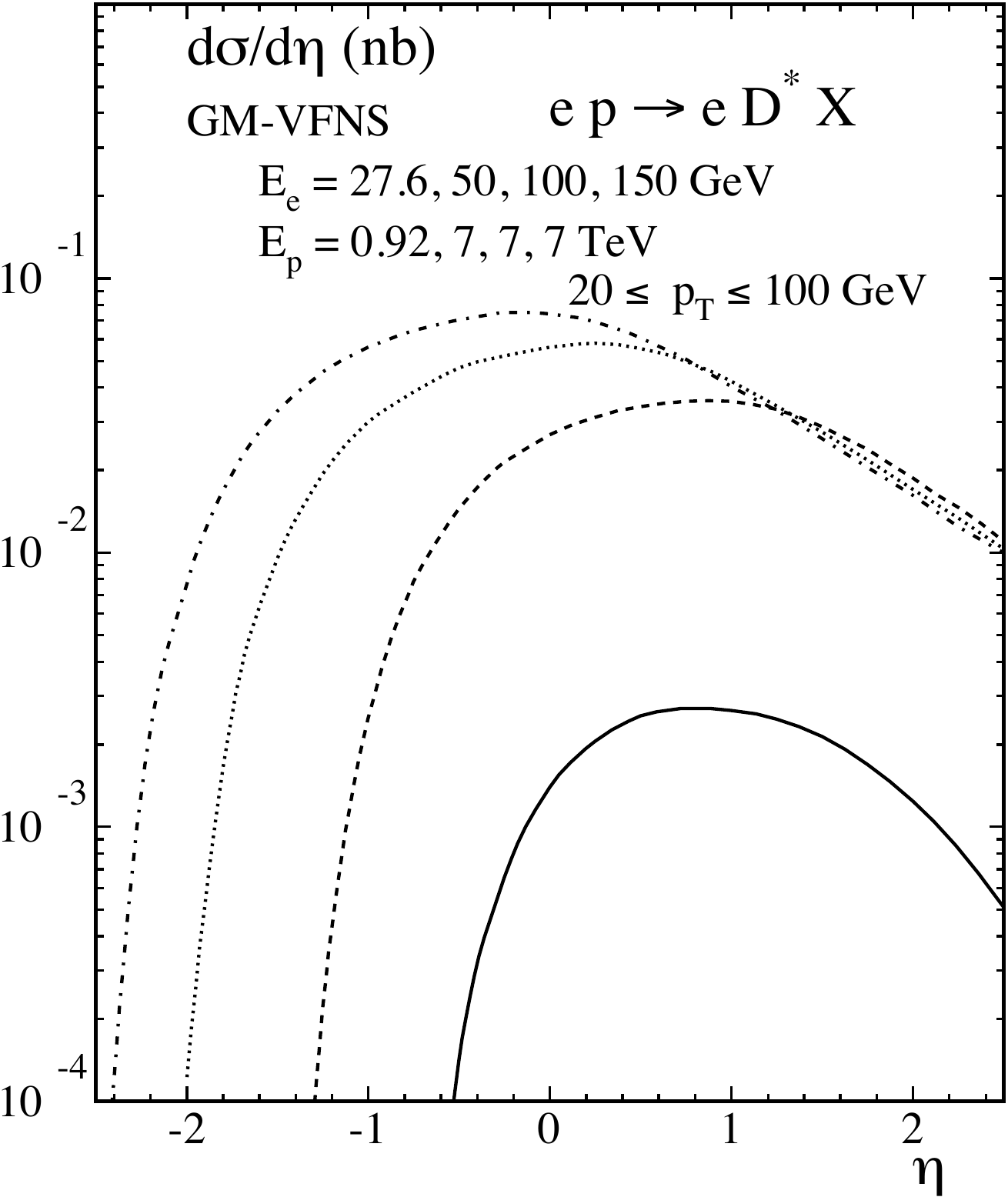}
\end{center}
\caption{\label{fig:dst_eta} Rapidity distribution of the cross 
section for the production of $D^{\ast}$ mesons at LHeC for different 
beam energies integrated over the low-$p_T$ range 5 GeV$ \leq p_T 
\leq 20$ GeV (left) and the high-$p_T$ range  20 GeV$ \leq p_T 
\leq 50$ GeV (right). The curves from bottom to top correspond to the 
combinations of beam energies as indicated in the figure.
The lowest curves are showing the cross sections at the HERA beam energies.}
\end{figure}
%
Various combinations of beam energies are studied.
To compare with the situation at HERA, as a reference, the 
values $E^p = 920$ GeV and $E^e = 27.5$ GeV for proton and electron 
energies, respectively, are also included. 
%
%
%
Numerical results of the study are shown in Fig.\ \ref{fig:dst_pt}.
%
The higher centre-of-mass energies 
available at the LHeC lead to a considerable increase of the cross 
sections as compared to HERA. Obviously one can expect 
an increase in the precision of corresponding measurements and much 
higher values of $p_T$, as well as higher values of the rapidity 
$\eta$, will be accessible. Since theoretical predictions also become 
more reliable at higher $p_T$, measurements of heavy quark production 
constitute a promising testing ground for perturbative QCD. One may 
expect that the experimental information will contribute to an 
improved determination of the (extrinsic and intrinsic) charm content 
of the proton and the charm fragmentation functions. 

%% file: physics/qce_jets.tex
\subsection{Jets in $ep$} 
\label{gammap:jets}
%
%

The study of the jet final states in lepton-proton collisions 
allows the determination of aspects of the nucleon structure 
which are not accessible in inclusive scattering. Moreover, jet production allows for 
probing predictions of QCD to a high accuracy. Depending on the virtuality of the exchanged 
photon, one distinguishes processes in photoproduction (quasi-real photon)
and deep inelastic scattering.

The photoproduction cross section for di-jet final states can be studied in different 
kinematic regions, thereby covering a wide spectrum of physical phenomena, and probing 
the structure of the proton and the photon. Two-jet production in deep inelastic scattering is  a
particularly sensitive probe of the gluon distribution in the proton and of the strong 
coupling constant $\alpha_s$. Both processes allow the study of potentially large 
enhancement effects in di-jet and multi-jet production. 

Jet production in photoproduction proceeds via 
the direct processes, in which the quasi-real photon interacts as a 
point-like particle with the partons from the proton, and the resolved 
processes, in which the quasi-real photon interacts with the partons 
from the proton via its partonic constituents. The parton distributions in the 
quasi-real photon are constrained mostly from the study of processes at $e^+e^-$ 
colliders, and are less well-determined than their counterparts in the proton. 
In both the direct and the resolved process, there are 
two jets in the final state at lowest-order QCD. The jet production 
cross section is given in QCD by the convolution of the flux of 
photons in the electron (usually estimated via the Weizs\"{a}cker-Williams 
approximation), the parton densities in the photon, the parton densities 
in the proton and the partonic cross 
section (calculable in pQCD). Therefore, the measurements of jet cross 
sections in photoproduction provide tests of perturbative QCD and the 
structure of the photon and the proton. 

Owing to the large size of the cross section, photoproduction of 
di-jets can be used for precision physics in QCD. A measurement at LHeC could 
improve upon previous HERA results and enter into a much larger kinematic 
region. In measurements made by the ZEUS collaboration, the available 
photon-proton centre-of-mass energy ranged from 142 to 293 GeV, and 
 jets of a transverse energy of up to 90 GeV could  be observed. 
By comparing the measured cross section with the theoretical prediction 
in NLO pQCD, a value 
of $\alpha_s(M_Z)$ was extracted with a total uncertainty of $\pm 3$\% and 
the running of $\alpha_s$ was tested over a wide range of $E_t^{\rm jet}$ 
in a  single measurement. The limiting factors in this measurement were 
the theoretical uncertainty inherent to the NLO prediction (which could be 
improved by computing  NNLO  corrections to jet photoproduction) and 
the experimental systematic uncertainty in the detector energy calibration. 
\begin{figure}[th]
\begin{center}
\includegraphics[width=0.6\textwidth]{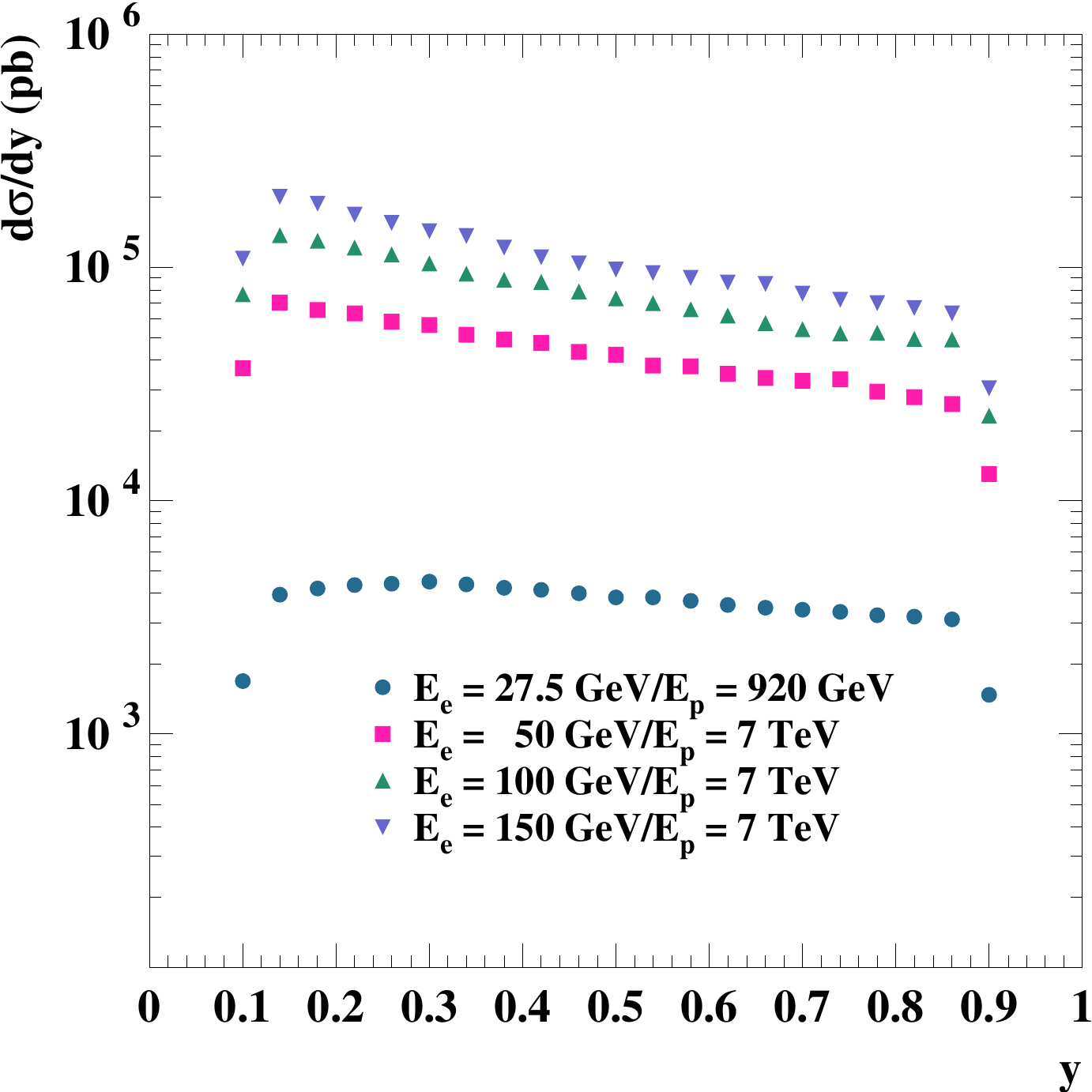}
\end{center}
\caption{\label{fig:phoprod1} 
PYTHIA predictions for photoproduction cross section at HERA and for three
LHeC scenarios.}
\end{figure}

Another motivation for making new photoproduction experiments is to 
improve the knowledge of the parton content of the photon. At present, 
most information on the photon structure is inferred from the collision 
of quasi-real photons with electrons at $e^+e^-$ colliders, resulting in 
a decent determination of the total (charge weighted) quark content of 
the quasi-real photon. Its gluonic content, and the quark flavour decomposition are 
on the other hand only loosely constrained. Improvements to the photon structure are 
of crucial importance to physics studies at a future linear $e^+e^-$ collider like the 
ILC or CLIC. Such a collider, operating far above the $Z$-boson resonance, 
will face a huge background from photon-photon collisions. This background can be 
suppressed only to a certain extent by kinematic cuts. Consequently, accurate 
predictions of it (which require an improved knowledge of the photon's parton content) 
are mandatory for the reliable interpretation of hadronic final states at the ILC or CLIC.
Several parameterisations of the parton distributions in the photon are available. They differ 
especially in the gluon content of the photon. For the studies presented here, 
the GRV-HO parameterisation~\cite{Gluck:1991jc} is used as default.

The photoproduction studies performed at LHeC were done for three 
different electron energy scenarios: $E_e$=50, 100 and 150 GeV. In all 
cases, the proton energy was set to 7 TeV. PYTHIA MC samples of 
resolved and direct processes were generated for these three 
scenarios. Jets were searched using the $k_t$-cluster algorithm in the 
kinematic region of $0.1<y<0.9$ and $Q^2<1$~GeV$^2$. Inclusive jet cross 
sections were done for jets of $E_t^{\rm jet}>15$~GeV and $-3<\eta^{\rm jet}<3$. 
Figure~\ref{fig:phoprod1} shows the PYTHIA MC cross sections as functions of $y$ for the three 
scenarios plus the corresponding cross section for the HERA 
regime. It can be seen that the LHeC cross sections are one to two orders of magnitude 
larger than the cross section at HERA.

The full study was complemented with fixed-order QCD 
calculations at order $\alpha_s$ and $\alpha_s^2$ using the program by Klasen 
et al.~\cite{Klasen:1996it} with the CTEQ6.1 sets for the proton PDFs, 
GRV-HO sets for the photon PDFs, $\alpha_s(M_Z) = 0.119$ and
 the renormalisation and factorisation scales 
were set to the transverse energy of each jet. 
\begin{figure}[t]
\begin{center}
\includegraphics[width=0.48\textwidth]{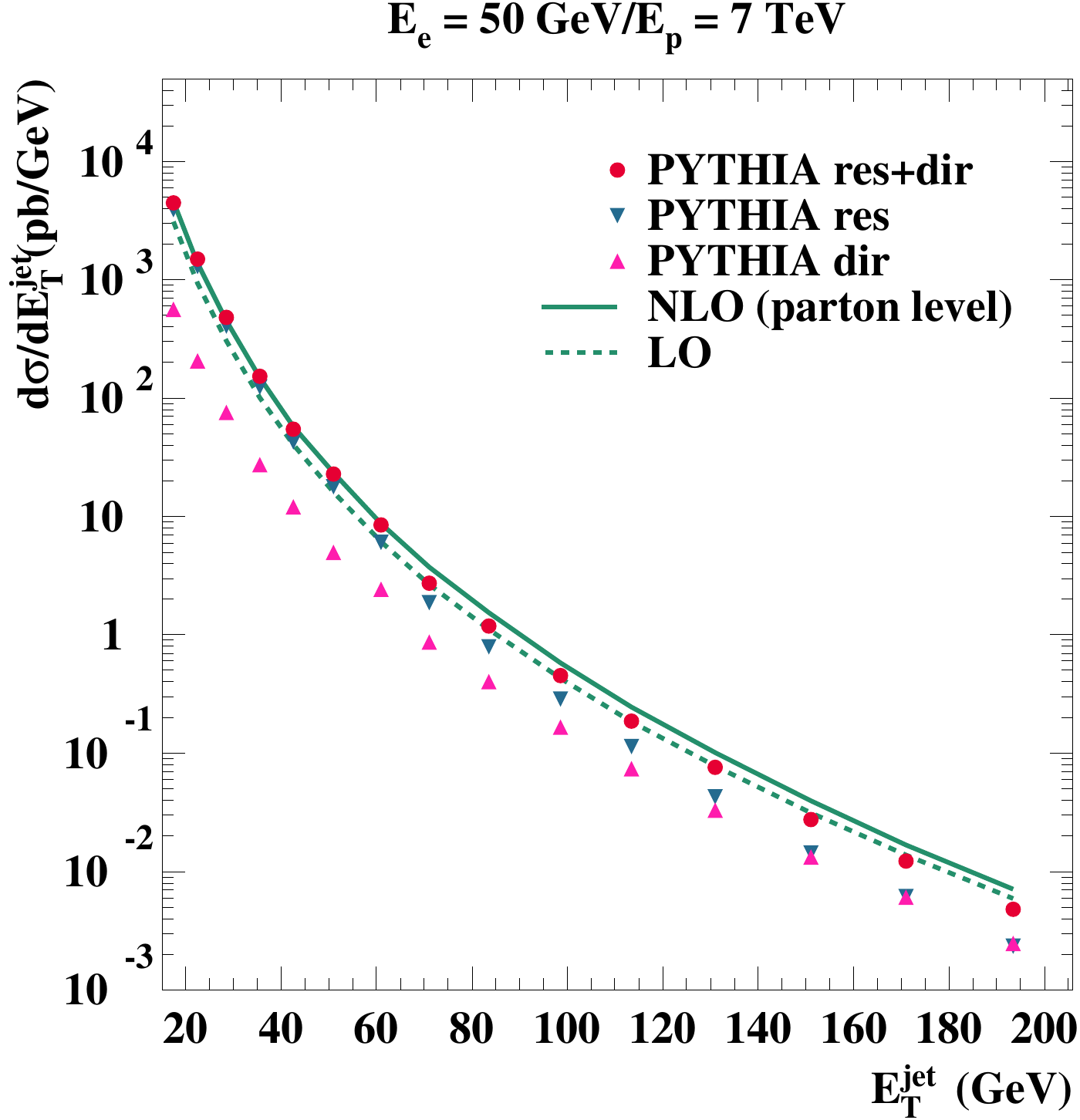}
\includegraphics[width=0.48\textwidth]{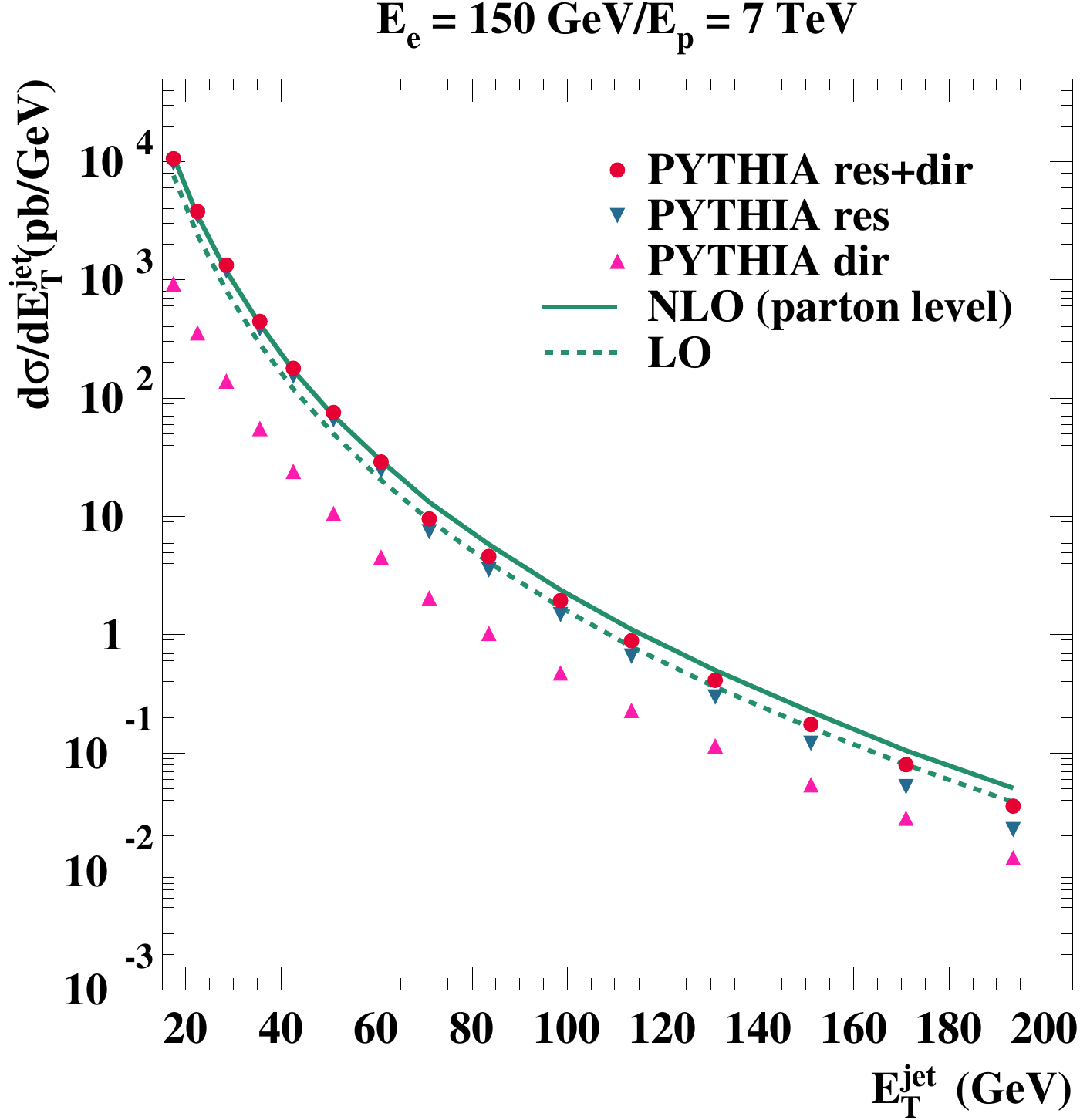}
\end{center}
\caption{\label{fig:etjet} 
Parton level predictions for the inclusive transverse energy distribution in photoproduction.}
\end{figure}
\begin{figure}[t]
\begin{center}
\includegraphics[width=0.48\textwidth]{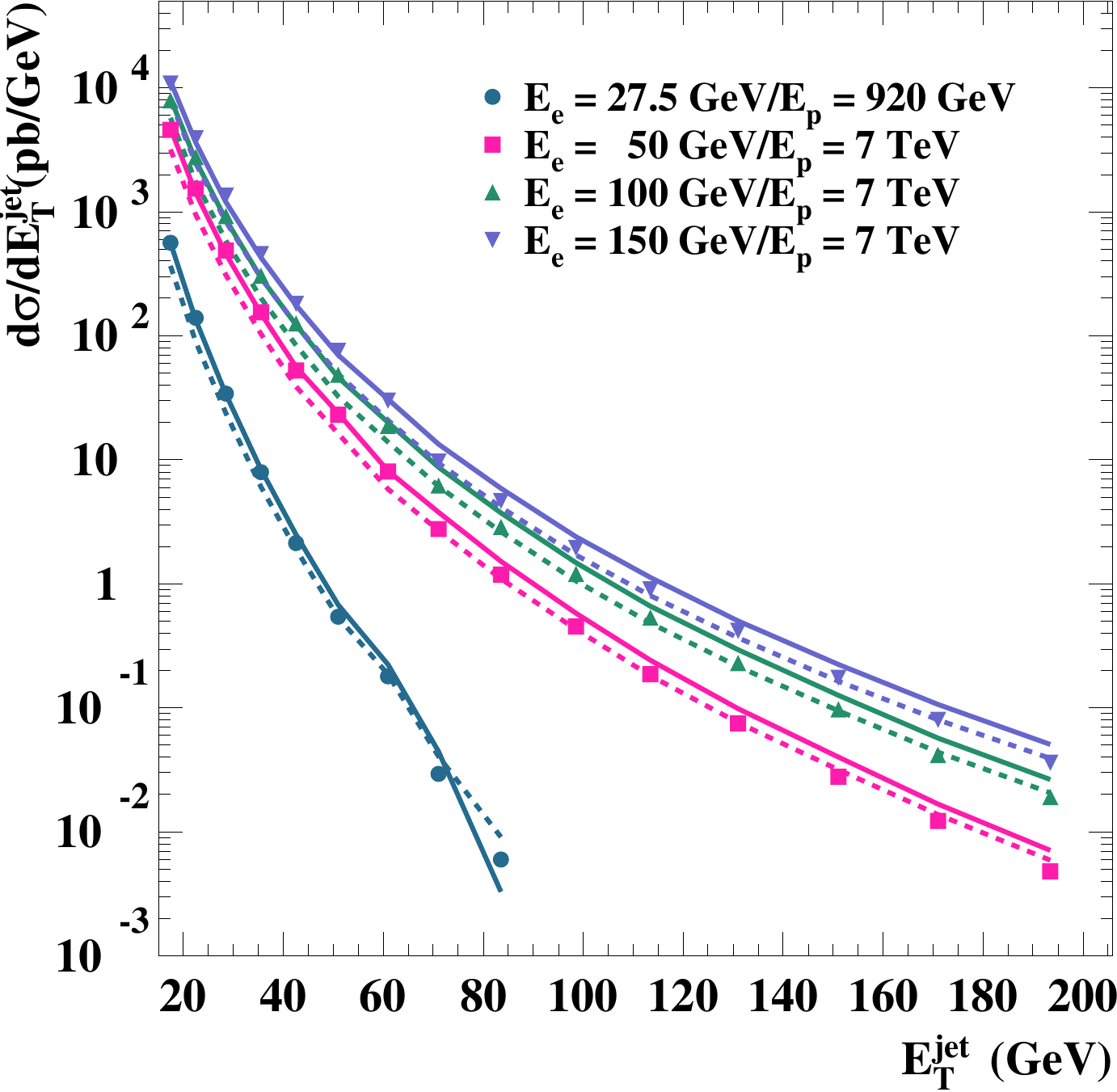}
\includegraphics[width=0.48\textwidth]{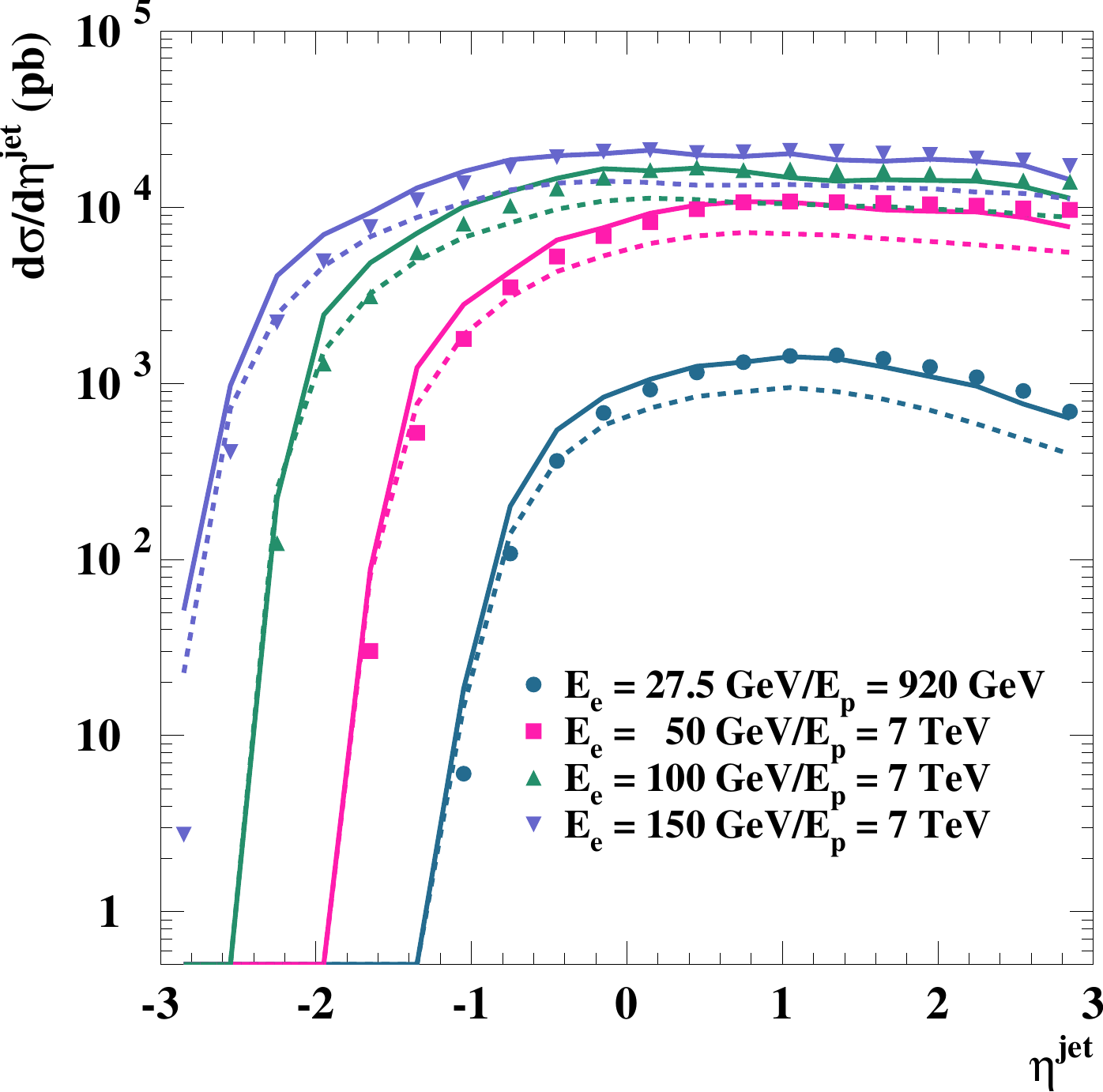}
\end{center}
\caption{\label{fig:jetdist} 
Dijet distributions in photoproduction as function of the jet transverse energy (left) and of
the jet rapidity (right) for different LHeC energies compared to the HERA kinematic range. }
\end{figure}

Figure~\ref{fig:etjet} shows the inclusive jet cross sections at parton level as 
functions of $E_t^{\rm jet}$ for the three energy scenarios for the PYTHIA 
res+dir (red dots), PYTHIA resolved (blue triangles) and PYTHIA direct 
(pink triangles) together with the predictions from the NLO (solid 
curves) and LO (dashed curves) QCD calculations. The calculations 
predict a sizeable rate for Etjet of at least up to 200 GeV. Resolved 
processes dominate at low $E_t^{\rm jet}$, but the direct processes become 
increasingly important as $E_t^{\rm jet}$ increases. The PYTHIA cross 
sections (which have been normalised to the NLO integrated cross 
section) agree well in shape with 
the NLO calculations. Investigating the $\eta^{\rm jet}$ distribution, one
finds that resolved processes dominate in the forward region, while direct processes 
produce more central jets.

Figure~\ref{fig:jetdist} show the inclusive jet cross sections at parton level as 
functions of $E_t^{\rm jet}$ (on the left) and $\eta^{\rm jet}$ (on the right) for the PYTHIA 
resolved+direct (symbols) and the predictions from the NLO (solid 
curves) and LO (dashed curves) QCD calculations together for the three 
energy scenarios. For comparison, the calculations for the HERA regime 
are also included. It is seen that the cross sections at fixed $E_t^{\rm jet}$ 
increase and that the jets tend to go more backward as the collision  energy
increases. The much larger photon-proton centre-of-mass energies that 
could be available at LHeC provide a much wider reach in $E_t^{\rm jet}$ and 
$\eta^{\rm jet}$ compared to HERA. 

Hadronisation corrections for the cross sections shown were 
investigated.  The corrections are predicted to be 
quite small, below +5\% for the chosen scenarios.
Since the hadronisation corrections are very small, the 
features observed at parton level remain unchanged.

Inclusive-jet and dijet measurements in deep-inelastic scattering (DIS) have
for a long time been a tool to test concepts and predictions of perturbative
QCD. Especially at HERA, jets in DIS have been thoroughly studied, and the
results have provided deep insights, giving for example precise values for the
strong coupling constant, $\alpha_s$ and providing constraints for the proton
PDFs. 

An especially interesting region for such studies has been the regime of large
(for HERA) $Q^2$ values of, for example, $Q^2 >$~125~${\rm GeV^2}$. In this
regime, the theoretical uncertainties, especially those due to the unknown
effects of missing higher orders in the perturbative expansion, are found to
be small. Recently, both the H1 and ZEUS collaborations have 
published measurements of inclusive-jet
and dijet events in this kinematic regime.

An extension of such measurements to the LHeC is interesting for two reasons:
First, the provided high luminosity will allow measurements in already
explored kinematic regions with still increased experimental
precision. Second, the extension in centre-of-mass energy, $\sqrt{s}$, and
thus in boson virtuality, $Q^2$, and in jet transverse energy, $E_{T,jet}$,
will potentially allow to study pQCD at even higher scales, extending the
scale reach for measurements of the strong coupling or the precision of the
proton PDFs at large values of $x$. 

To explore the potential of such 
a measurement, DIS jet production was investigated for the 
following LHeC scenario: proton beam energy 7~TeV, electron beam energy
70~GeV and integrated luminosity   10~${\rm fb^{-1}}$. The study concentrates 
on the phase space of high boson virtualities $Q^2$, with event selection 
cuts  100~$< Q^2 <$~500\,000~${\rm GeV^2}$ and  0.1~$< y <$~0.7, where $y$ 
is the inelasticity of the event. Jets are reconstructed using the $k_T$ 
clustering algorithm in the longitudinally invariant inclusive mode in the 
Breit reference frame. Jets were selected by requiring: 
a jet pseudorapidity in the laboratory of -2~$< \eta_{lab} <$~3, 
 a jet transverse energy in the Breit frame of $E_{T,jet}^{Breit} >$~20~GeV
 for the inclusive-jet measurement and
jet transverse energies in the Breit frame of 25(20)~GeV for the 
leading and the second-hardest jet in the case of the dijet selection. 

For inclusive-jet production cross sections were studied in the indicated kinematic regime as 
functions of $Q^2$, $x_{Bj}$, $E_{T,jet}^{Breit}$ and $\eta_{jet}^{lab}$, the jet pseudorapidity in 
the laboratory frame. For dijet production, studies are presented
as functions of $Q^2$, the logarithm of the proton momentum fraction $\xi$, $\log_{10}\xi$, 
the invariant dijet mass $M_{jj}$, the average transverse energy of the two jets in the Breit frame, 
$\overline{E_{T,jet}^{Breit}}$, and of  half of the absolute difference of the two jet 
pseudorapidities in the laboratory frame,  $\eta'$.

For the binning of the observables shown here, the statistical uncertainties
for the indicated LHeC integrated luminosity can mostly be neglected, even at
the highest scales. The systematic uncertainties were assumed to be dominated
by the uncertainty on the jet energy scale which was assumed to be known to
1\% or 3\% (both scenarios are indicated with different colours in the
following plots), leading to typical effects on the jet cross sections between
1 and 15\%. 
A further relevant uncertainty is the acceptance correction that is applied to
the data which was assumed to be 3\% for all observables.  

The theoretical calculations where performed with the {\textsc{disent}}
program~\cite{Catani:1996vz} using the CTEQ6.1 proton PDFs~\cite{Pumplin:2002vw,Stump:2003yu}. The 
central default squared
renormalisation and factorisation scales were set to $Q^2$.  
The theory calculations for the LHeC scenario were corrected for the effects
of hadronisation and $Z^0$ exchange  using Monte Carlo data samples simulated
with the $\textsc{lepto}$ program~\cite{Ingelman:1996mq}.

Theoretical uncertainties were assessed by varying the 
renormalisation scale up and down by a factor 2
(to estimate the potential effect of contributions beyond 
NLO QCD), by using the 40 error 
sets of the CTEQ6.1 parton distribution functions, and by varying 
$\alpha_s$ using the  
CTEQ6AB  PDF~\cite{Pumplin:2005rh}. The dominant theory uncertainty turned out to be 
due to the scale variations, resulting in effects of 
a few to up to 20\% or more,
  for example for low values of $Q^2$ or, for the case of the dijet
  measurement, for low values of the invariant dijet mass, $M_{jj}$, or the
  logarithm of momentum fraction carried into the hard scattering,
  $\log_{10}\xi$.  

Note that for the inclusive-jet results also the predictions for a HERA
scenario with almost the same selection are shown in order to indicate the
increased reach of the LHeC with respect to HERA. The only change is a
reduction in centre-of-mass energy to 318~GeV and a reduced $Q^2$ reach,
125~$< Q^2 <$~45\,000~${\rm GeV^2}$. The HERA predictions shown were also
corrected for hadronisation effects and the effects of $Z^0$ exchange.  
\begin{figure}[t]
\begin{center}
\includegraphics[width=0.48\textwidth]{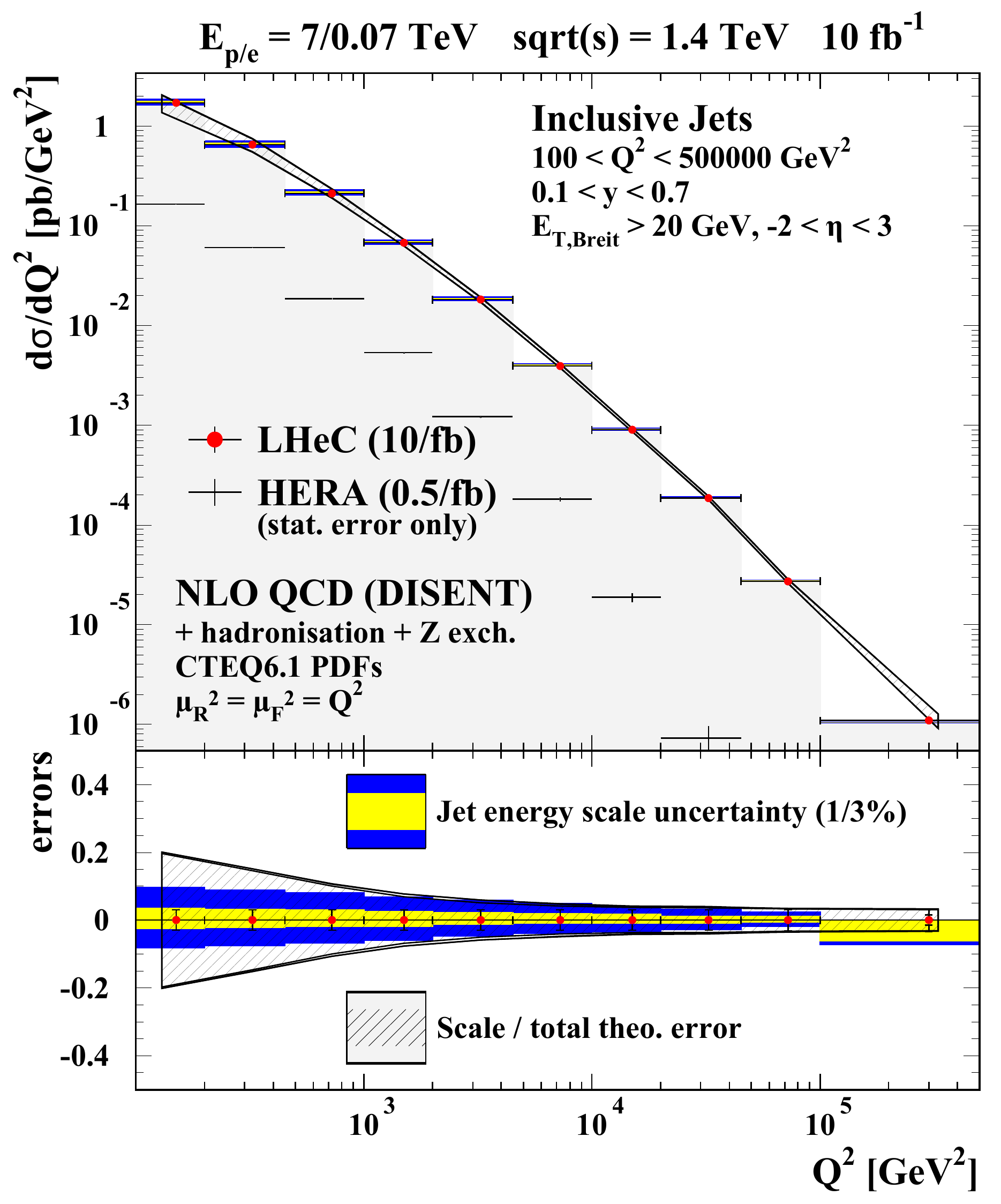}
\includegraphics[width=0.48\textwidth]{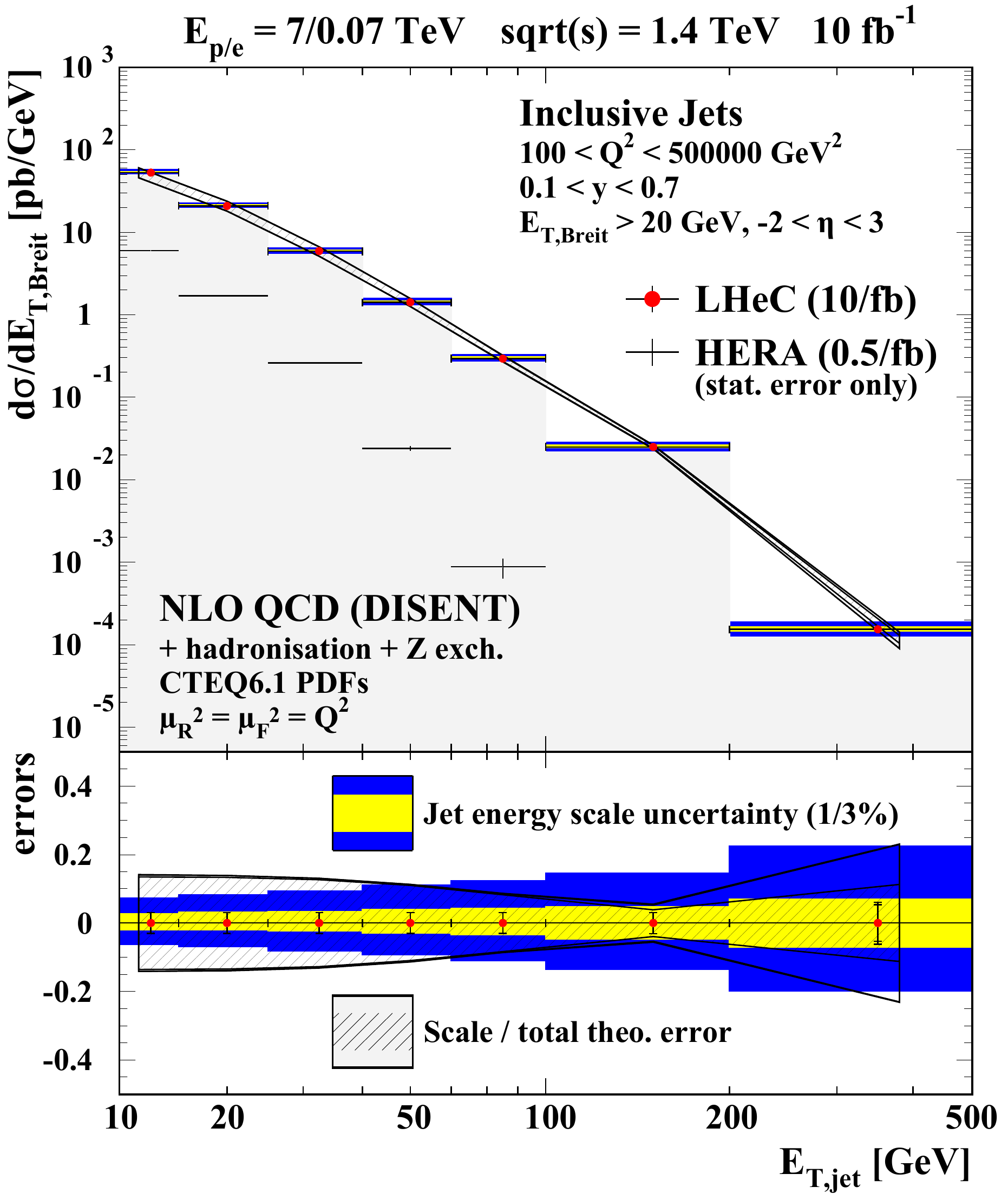}
\caption{\label{fig:dis.incjet}
Predicted LHeC results for inclusive jet production as function of $Q^2$ and of $E_{T}$ in the Breit frame. Predictions for HERA results
are also shown.}
\end{center}
\end{figure}

\begin{figure}[t]
\begin{center}
\includegraphics[width=0.48\textwidth]{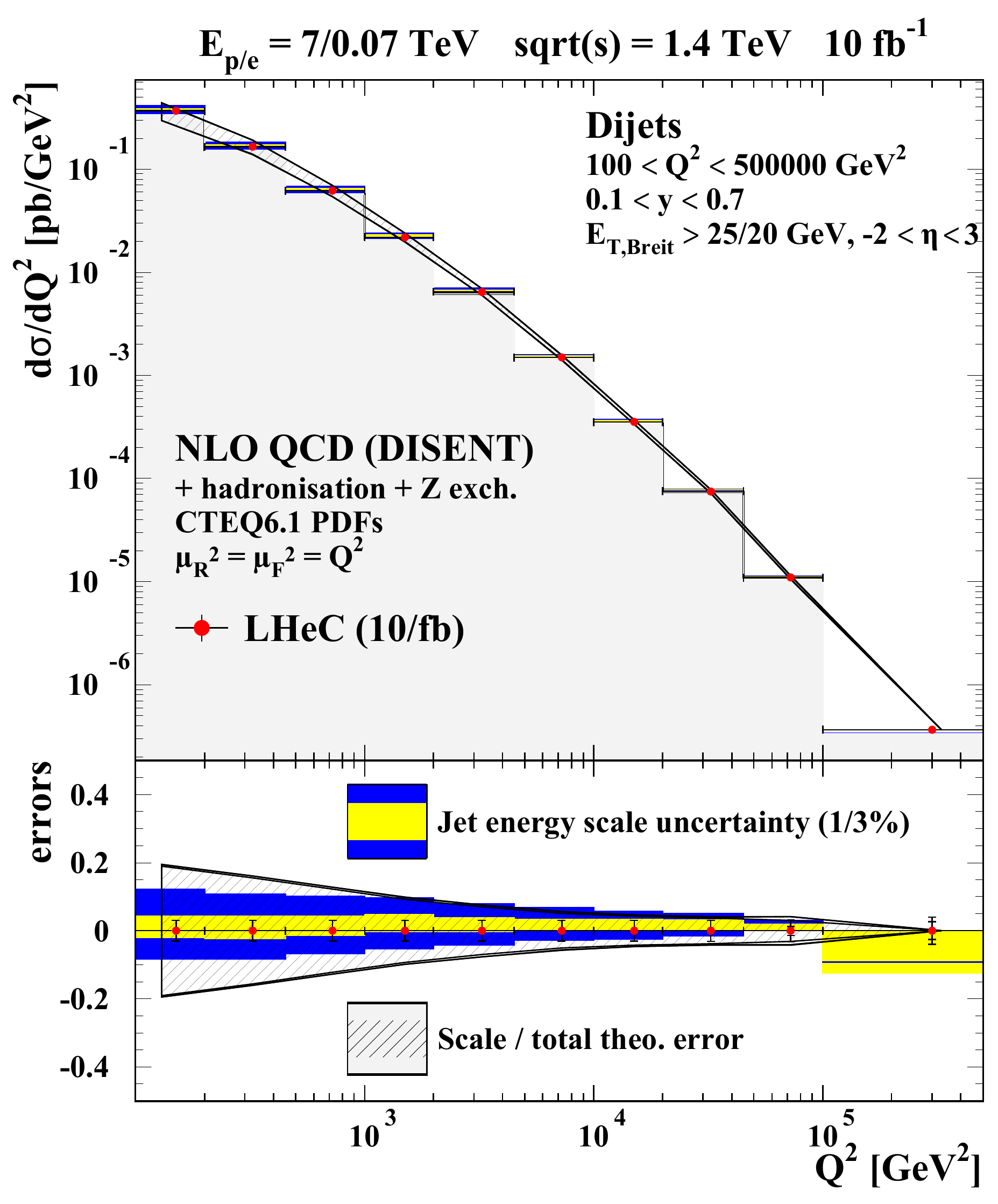}
\includegraphics[width=0.48\textwidth]{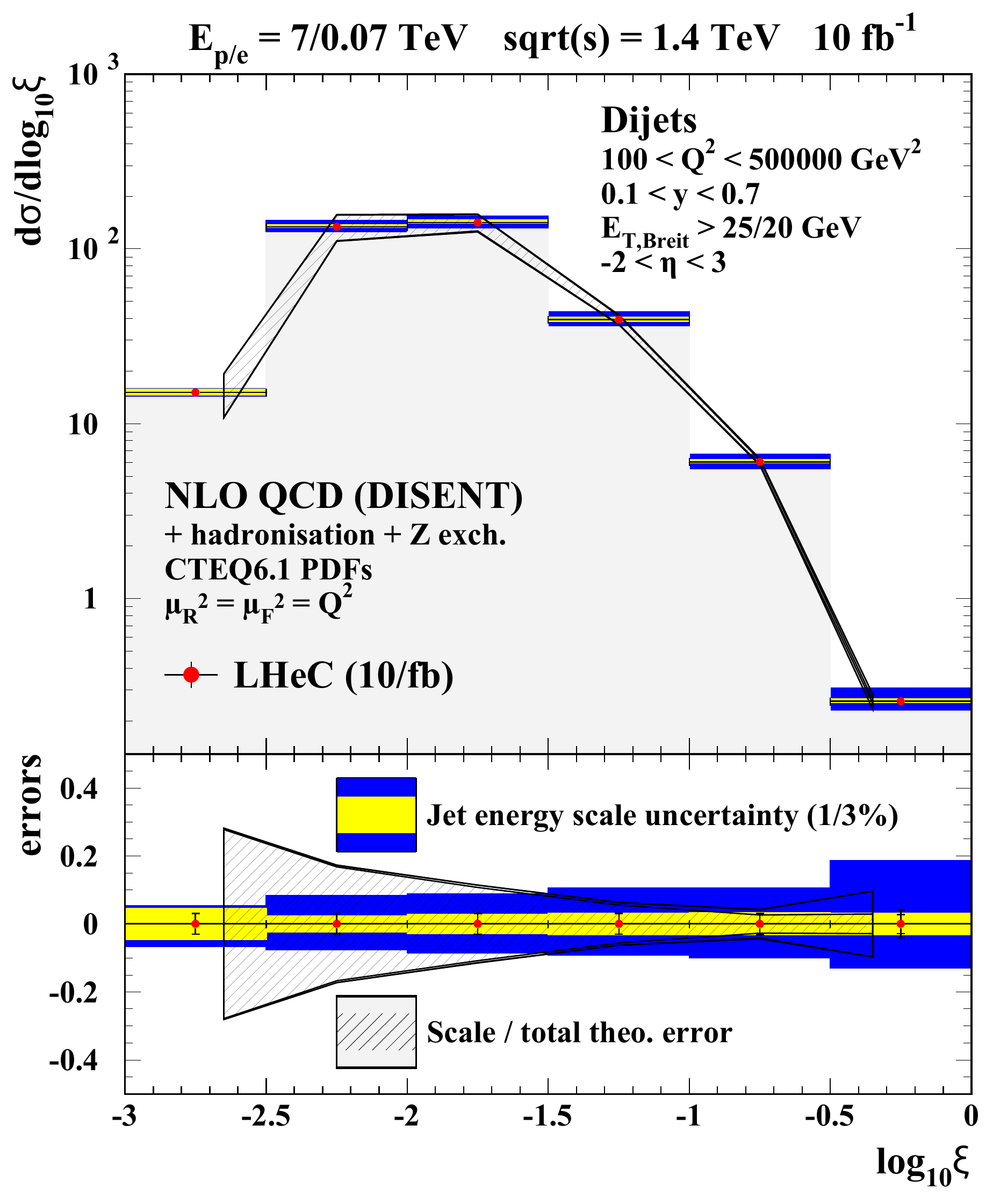}
\caption{\label{fig:dis.dijet}
Predicted LHeC results for dijet production as function of $Q^2$ and of $\xi$.}
\end{center}
\end{figure}

Figure~\ref{fig:dis.incjet} shows the inclusive jet cross section 
as function of $Q^2$ and of the jet transverse energy in the Breit frame,
while Figure~\ref{fig:dis.dijet} shows the dijet cross section as 
function of $Q^2$ and of $\xi = x_{Bj}(1+M^2_{jj}/Q^2)$. 
The top parts of the figures show
the predicted cross sections together with the expected statistical and
(uncorrelated) experimental systematic uncertainties as errors bars. The
correlated jet energy scale uncertainty is indicated as a coloured band; the
inner, yellow band assumes an uncertainty of 1\%, the 
outer, blue band one of 3\%. Also shown as a thin hashed area are the
theoretical uncertainties; the width of the band indicates the size of the combined
theoretical uncertainty. In case of inclusive-jet production, also the
predictions for HERA are indicated as a thin line.  

The bottom parts of the figures show the relative uncertainties due to the jet
energy scale (yellow band for 1\%, blue band for 3\%), the statistical and
uncorrelated experimental systematic uncertainties as inner / outer error
bars, and the combined theoretical uncertainties as hashed band. The inner
part of this band indicates the uncertainty due to the variation of the
renormalisation scale.  

The inclusive-jet cross section as function of $Q^2$ shows a typical picture:
In most region of the phase space, the uncertainties are dominated by the
theory uncertainties, and here mainly by the renormalisation scale
uncertainty. The typical size of experimental uncertainties is of the order of
10\%, with larger values in regions with low relevant scales --- i.e.\ low
invariant dijet masses, low jet transverse energies or low $Q^2$ values. The
theoretical uncertainties are typically between 5 and 20\%, with partially
strong variations over the typical range of the observable in question.  

A comparison with the HERA predictions for inclusive-jet production shows that
the LHeC cross sections is typically larger by 1 to 3 orders of magnitude. 
The dijet final state allows for a full reconstruction of the partonic 
kinematics, and can thus be used to probe the parton distribution functions 
in $Q^2$ and $\xi$. It can be seen that a measurement at LHeC
covers a large kinematic range down to $\xi \approx 10^{-3}$ and 
up to $Q^2= 10^5$~GeV$^2$. Potentially limiting factors in an extraction 
of parton distribution functions are 
especially the jet energy scale uncertainty 
on the experimental side and missing higher order (NNLO) corrections 
on the theory side. The jet energy scale uncertainty can be addressed 
by the detector design and by the experimental setup of the measurement. 
NNLO corrections to dijet production in deep inelastic scattering are 
already very much demanded by the precision of the HERA data, their 
calculation is currently in progress~\cite{Gehrmann:2009vu,Daleo:2009yj}.

In summary, jet final states in photoproduction and 
deep inelastic scattering at the LHeC promise a wide spectrum of 
new results on the partonic structure of the photon and the proton. They 
allow for precision tests of QCD by independent determinations of the 
strong coupling constant over a kinematic range typically one to 
two orders of magnitude larger than what was accessible at HERA. The 
resulting parton distributions will have a direct impact for precision 
predictions at the LHC and a future linear collider. 





%
%
%

\subsection{Jets in $\gamma$A} 
\label{gammap:jetseA}
\input{physics/tex/jetsgA}

%% file: physics/tex/jetsgA.tex
For photoproduction in $e$A collisions, jets provide 
an  abundant yield of high-energy probes 
of the nuclear medium. The expected cross sections 
have been computed using the calculations in \cite{Frixione:1995ms,Frixione:1997np}, for an electron beam of 50 GeV colliding with the LHC beams. 
For the nuclear case the same integrated luminosity (2 fb$^{-1}$) was 
assumed per nucleon as for $ep$. Only jets with $E_{Tjet}>20$ GeV are considered, and for the distribution in $E_{Tjet}$ the pseudorapidity acceptance is $|\eta_{jet}|<3.1$, corresponding to $5^{\rm o}<\theta_{jet}<175^{\rm o}$ 
in polar angle. 
The simulations use
the Weizs\"acker-Williams photon flux from the electron with 
the standard option in \cite{Frixione:1995ms,Frixione:1997np}.
The chosen photon, proton and nuclear modified PDFs are taken from 
GRV-HO \cite{Gluck:1991ee}, CTEQ6.1M \cite{Stump:2003yu} and
EPS09 \cite{Eskola:2009uj}, respectively - see Subsec. \ref{sec:nucleartargets} for explanations on the nuclear modifications of PDFs.
The renormalisation and factorisation scales are taken to be 
$\mu_R=\mu_F=\sum_{jets} E_{Tjet}/2$ and
the inclusive $k_T$ jet algorithm \cite{Ellis:1993tq} 
is used with $D=1$.
The statistical uncertainty in the computation (i.e. in the Monte Carlo integration) is smaller than 10 \% for all results shown.
This large statistical uncertainty is reached only for the largest $E_{Tjet}$,
with much smaller uncertainties at lower values of $E_{T}$.
No attempt has been made to estimate the uncertainties due 
to the choices of photons flux, photon or proton parton densities, scales 
or jet algorithms (see \cite{Adloff:2003nr,Frixione:1997ks} for 
such considerations at HERA). The issues of background subtraction, 
experimental efficiencies in the jet reconstruction or energy calibration 
have also yet to be addressed. The only uncertainty studied thus far 
is that due to the nuclear parton densities, which is extracted in 
the EPS09 framework \cite{Eskola:2009uj} using the Hessian method.

\begin{figure}
\begin{center}
\centerline{ \includegraphics[clip=,width=1.0\textwidth]{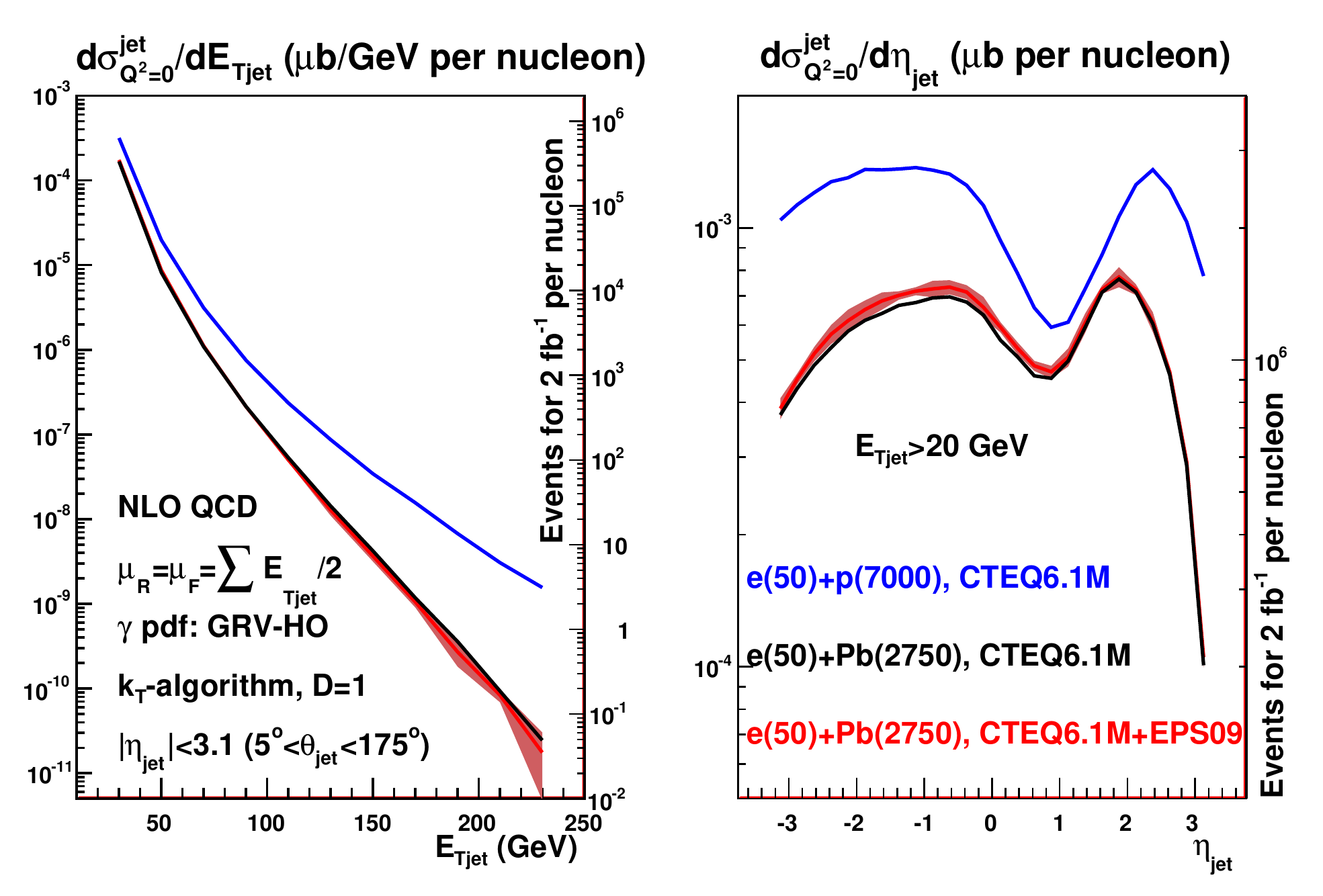}}
\vskip -0.5cm
\caption{Predictions for the inclusive jet distribution in photoproduction,
differential in $E_{Tjet}$ (left) and $\eta_{jet}$ (right)
for $e$(50)+$p$(7000) (blue,top lines), $e$(50)+Pb(2750) without nuclear modification
 of the parton densities (black lines), and $e$(50)+Pb(2750) with EPS09 nuclear
 modification of the parton densities (red lines for the central value and bands
 for the uncertainty coming from the nuclear modification factors). See the text
 and the legends on the plots for further details of the calculations and 
kinematic cuts. In both plots, the axis on the left corresponds to the cross
 section in $\mu$b, while the axis on the right provides the number of jets 
expected for an integrated luminosity of 2 fb$^{-1}$ per nucleon, per unit 
of $E_{Tjet}$ ($\eta_{jet}$) in the plot on the left (right).}
\label{Fig:gAjets}
\end{center}
\end{figure}

The results are shown in Fig. \ref{Fig:gAjets}. One observes that
yields of around $10^3$ jets per GeV are expected  with $E_{Tjet}\sim 95$ (80) GeV in $ep$ ($e$Pb), for $|\eta_{jet}|<3.1$ and the considered integrated luminosity of 2 fb$^{-1}$ per nucleon. The 
effects of the nuclear modification of parton densities and their uncertainties are smaller than 10 \%. The two-peak structure in the $\eta_{jet}$-plot results from the sum of the direct plus resolved contributions, each of 
which produce a single maximum, located in opposite hemispheres. 
Positive $\eta_{jet}$ values are dominated by direct photon interactions, 
whereas negative $\eta_{jet}$ values are dominated by 
contributions from resolved photons.

%% file: physics/tex/gpcs.tex
Due to the $1/Q^4$ propagator term, the LHeC $ep$ cross section is
dominated by
very low $Q^2$ quasi-real photons. With a knowledge of
the effective photon flux \cite{wwa}, measurements in this
kinematic region can be used to obtain real photoproduction ($\gamma$p)
cross sections. The real photon has a dual nature, 
sometimes interacting in a point-like manner and sometimes interacting
through its effective partonic structure, resulting from 
$\gamma \rightarrow q \bar{q}$ and higher multiplicity splittings
well in advance of the target \cite{gammap:review1,gammap:review2}, the 
details of
which are fundamental to the understanding of QCD evolution. 


The behaviour of the total photoproduction cross section at high energy is a topic of a major interest. It is now firmly established experimentally that all 
hadronic cross sections rise with centre of mass energy for large energies. The Froissart-Martin bound has been derived for hadronic probes. It therefore remains to be seen whether this bound is applicable to $\gamma p$ scattering. For example in Refs.
\cite{Frankfurt:2001nt,Rogers:2005qe} it has been argued that the bound for 
real photon-hadron interactions should be of a different functional form, namely $\ln ^3 s$. This would imply that 
the universality of the asymptotic behaviour of hadronic cross sections does not hold.
Therefore the measurement of the total photoproduction cross section at high energies  will bring an important insight into the  problems of universality of hadronic cross
sections, unitarity constraints, the role of diffraction 
and the interface between hard and soft physics.

\begin{figure}[h]
  \begin{center}
\includegraphics[width=0.7\textwidth]{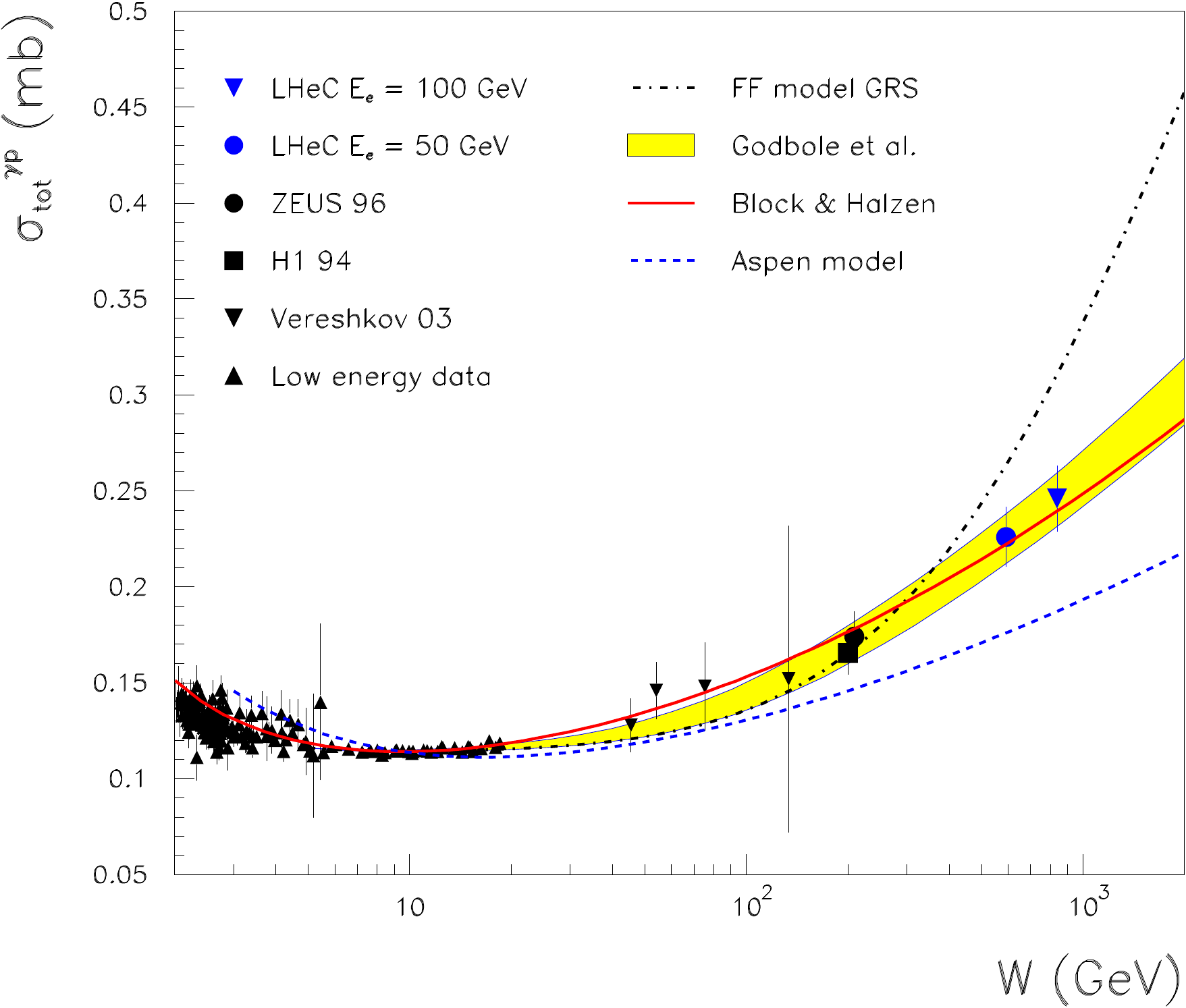}
  \end{center}
  \caption[]{Simulated LHeC measurements of the total photoproduction
cross section with $E_e = 50 \ {\rm GeV}$ or $E_e = 100 \ {\rm GeV}$,
compared with previous data and a variety of models (see text for details).
This is derived from a similar figure in \cite{Godbole:2008ex}.}
\label{gammap:sigmatot}
\end{figure}

In Fig.~\ref{gammap:sigmatot}, available data on the total cross
section are shown \cite{Chekanov:2001gw,Aid:1995bz,Vereshkov:2003cp,Nakamura:2010zzi}\footnote{The recent results by ZEUS \cite{Collaboration:2010wxa} refer only to the energy behaviour of the cross section in the range 194 $ <W<$ 296 GeV, but do not provide absolute values.}, together with a variety of models. More specifically, the dot-dashed black line labelled `FF model GRS' is a minijet model 
\cite{Godbole:2008ex}, the yellow band labelled `Godbole et al.' is an eikonalised minijet model with soft gluon resummation \cite{Godbole:2008ex} with the band defined by different choices of the parameters in the model, the red solid line labelled `Block \& Halzen' is based on a low energy
parameterisation of resonances joined with Finite
Energy Sum Rules and asymptotic
$\ln^2 s$-behaviour
 \cite{Block:2004ek,Block:2005pt}, and the dashed blue line labelled `Aspen model' is a QCD inspired model \cite{Block:1998hu}. 
 

The theoretical predictions
diverge at energies beyond those constrained by HERA data, where cross sections
were obtained by tagging and measuring the energies of 
electrons scattered through very small angles in dedicated calorimeters
located well down the beam pipe in the outgoing electron 
direction \cite{Chekanov:2001gw,Aid:1995bz}. 
As discussed in Chapter~\ref{detector:fwdbwd}, the most promising location for
similar small angle electron detectors at the LHeC is in the region
around $62 \ {\rm m}$ from the interaction point, which could be used
to tag scattered electrons in events with $Q^2 < 0.01 \ {\rm GeV^2}$ and 
$y \sim 0.3$. This naturally leads to measurements of the total photoproduction
cross section at $\gamma p$ centre-of-mass energies $W \sim 0.5 \surd{s}$.
The measurements would be strongly limited by systematics. In the absence of a 
detailed simulation of an LHeC detector these uncertainties are hard to
estimate. For the simulated data in Fig.~\ref{gammap:sigmatot}, uncertainties of
$7 \%$ have been assumed, matching the precision of the H1 and ZEUS data. 
This would clearly be more than adequate to distinguish between many of
the available models. The HERA uncertainties were dominated by the invisible
contributions from diffractive channels in which the diffractive masses were
too small to leave visible traces in the main detector. If 
detector acceptances to
$1^\circ$ are achieved at the LHeC, better precision 
is expected to be possible.

%% file: physics/electroweak.tex
%
%
%
Electroweak physics stands at a crossroad: 
if the tantalising hints from ATLAS and CMS for the
existence of the SM Higgs boson near to $125$\,GeV mass
get confirmed,  the mechanism for
the breaking of the electroweak symmetry and for assigning mass
to the weak intermediate bosons is likely discovered. Its investigation
will take a generation of precision electroweak measurements
focused around the mass, the couplings and CP properties of the
Higgs boson. The direct contributions to Higgs physics with the
LHeC are discussed in Section~\ref{sec:higgs}. For polarised
$e^-p$ scattering at the LHeC one can expect an estimated
number of $400$ well reconstructed   $H \rightarrow b \bar{b}$  events 
over a small background in CC, for $100$\,fb$^{-1}$
luminosity at $E_e=60$\,GeV.  These investigations are to be 
accompanied by a new level and generation of precision
electroweak measurements because it will be crucial to
verify the SM character of the electroweak sector and/or
to assign the observation to new physics phenomena and
their compliance with or violation of the Standard Model.
Similarly, if the SM Higgs boson is excluded,
the question of the relation of the basic electroweak parameters
at the quantum level, as of the top and $W$ boson mass,
will remain to be of high interest, while at maximum energy, at the
LHC, one will need to establish, with higher priority than 
otherwise, the damping of the $W_L W_L$
cross section as is predicted to avoid violation of unitarity.

The LHeC is a unique electroweak machine because
the $Q^2$ values exceed by far the masses, squared,
of $Z$, $W$
and also of $t$ and $H$, should that exist. It reaches a new
level of precision because of that coverage but also due
to the high luminosity and the special accuracy of measurements
in DIS. At the same time it provides a new level of high 
precision QCD measurements, of all PDFs in particular,
which will become crucial at the next level of precision
and the interpretation of subtle electroweak phenomena,
especially in connection with the LHC.

The following presents first a brief 
introduction to the context, mainly of previous electroweak
measurements. There follow
two simulations and analyses, which have been
undertaken to illustrate the very high precision one can obtain
with electroweak measurements at the LHeC, using
as suitable examples
the determination of the weak neutral current couplings
of light quarks and the evaluation of the scale dependence
of the weak mixing angle, $\sin^2 \Theta = 1 - (M_W/M_Z)^2$.
\subsection{Context}
Precision electroweak measurements at low energy have played a central role in 
establishing the Standard Model (SM) as the theory of fundamental interactions. 
Measurements at LEP, SLD, and the Tevatron have confirmed the 
SM at the quantum level, verifying the existence of its higher-order loop
contributions.  The sensitivity of these contributions
to virtual heavy particles has allowed for an estimate of the mass of the top
quark  prior to its actual discovery in 1995 by the CDF and D{\O} Collaborations.
Now that  the determination of the top mass at the Tevatron has become 
quite accurate, reaching the $1$\,\% level, and $M_W$ is known with an 
error of nearly $20$\,MeV, 
electroweak precision measurements have started to narrow the
range of  the mass  of the SM Higgs boson, see e.g.  \cite{Flacher:2008zq,Erler:2010wa}.
If the Brout-Englert-Higgs prediction, taken into the SM,
is correct, the SM scalar boson has a mass below 
$155$\,GeV, at $95$\,\% CL,  and it should 
appear measurable at the LHC.

Electroweak precision measurements are also a means to 
constrain  possible extensions of the SM.
Although the observed good quality of the SM fit disfavours new physics 
at an energy scale of O$(100~{\rm GeV})$ there are a few peculiarities worth
noting:
 a significant  tension exists between the forward-backward
asymmetry  of $Z\to b\bar{b}$, measured at LEP, 
which favours a heavy Higgs, and the left-right asymmetry in $Z\to \ell\bar{\ell}$
and the $W$ mass, which both favour a very light Higgs. 
The current prediction
of $M_H$ involves such conflicting information, the origin
of which may be  
statistical but could also be rooted in new physics~\cite{Gambino:2003xc}. A further 
$\sim 3\sigma$ hint for  physics beyond the SM, without such Higgs implications, 
is the deviation of the measured magnetic anomalous moment of the muon 
from its SM prediction~\cite{Davier:2010nc}.

Considerable efforts are ongoing to improve the precision and
to extend the reach of electroweak parameter measurements.
The Tevatron and subsequently the LHC will improve the 
current precision on the top mass.
A high precision measurement of the $W$ mass at the LHC will
require a corresponding new level of control of the PDFs~\cite{Krasny:2010vd},
for which the LHeC provides the ideal basis~\footnote{One may argue that 
this is for a long time hence, however, it is not new that precision
measurements in particle physics have a long duration, as can be
exemplified with the efforts to measure $\sin^2 \Theta$ in
neutrino fixed target experiments in the seventies until LEP $\sim 30$
years later, or the time it took from discovering the $W$ in the early eighties 
to its ongoing precision mass measurements.}. One notices the prospect for
LHCb to possibly achieve a good new measurement of 
$\sin^2\Theta$~\cite{Haywood:1999qg,Rabbertz:2010qi}.
 Two experiments at Jefferson Lab, Q-weak~\cite{Qweak}  and MOLLER \cite{moller},  
are to measure the  weak  mixing angle from parity violation in $ep$ and $e^- e^-$ scattering 
at low energy, which with high precision is important to verify the
scale dependence of $\sin^2\Theta$. This was recently much debated
when the NuTeV experiment claimed to have seen a too large angle,
which, however, lead to reanalyses of its other aspects, such as nuclear
and QED corrections and also PDFs. The LHeC will resolve
correlations between strong
and electroweak phenomena by providing PDFs free of nuclear
corrections and precise electroweak measurements.

The electroweak measurements possible at LHeC are principally of the kind
performed at HERA (see \cite{heraew,Zhang:2008aq} for an overview). However,
they will greatly benefit from the higher energy and larger luminosity, as well as from
highly polarised lepton beams, and therefore also include processes, such as
single standard model or anomalous top quark production, which were impossible
to study in $ep$ before.

%
\subsection{Light quark weak neutral current couplings}
\begin{figure}[th]
\begin{center}
\includegraphics[width=0.49\textwidth]{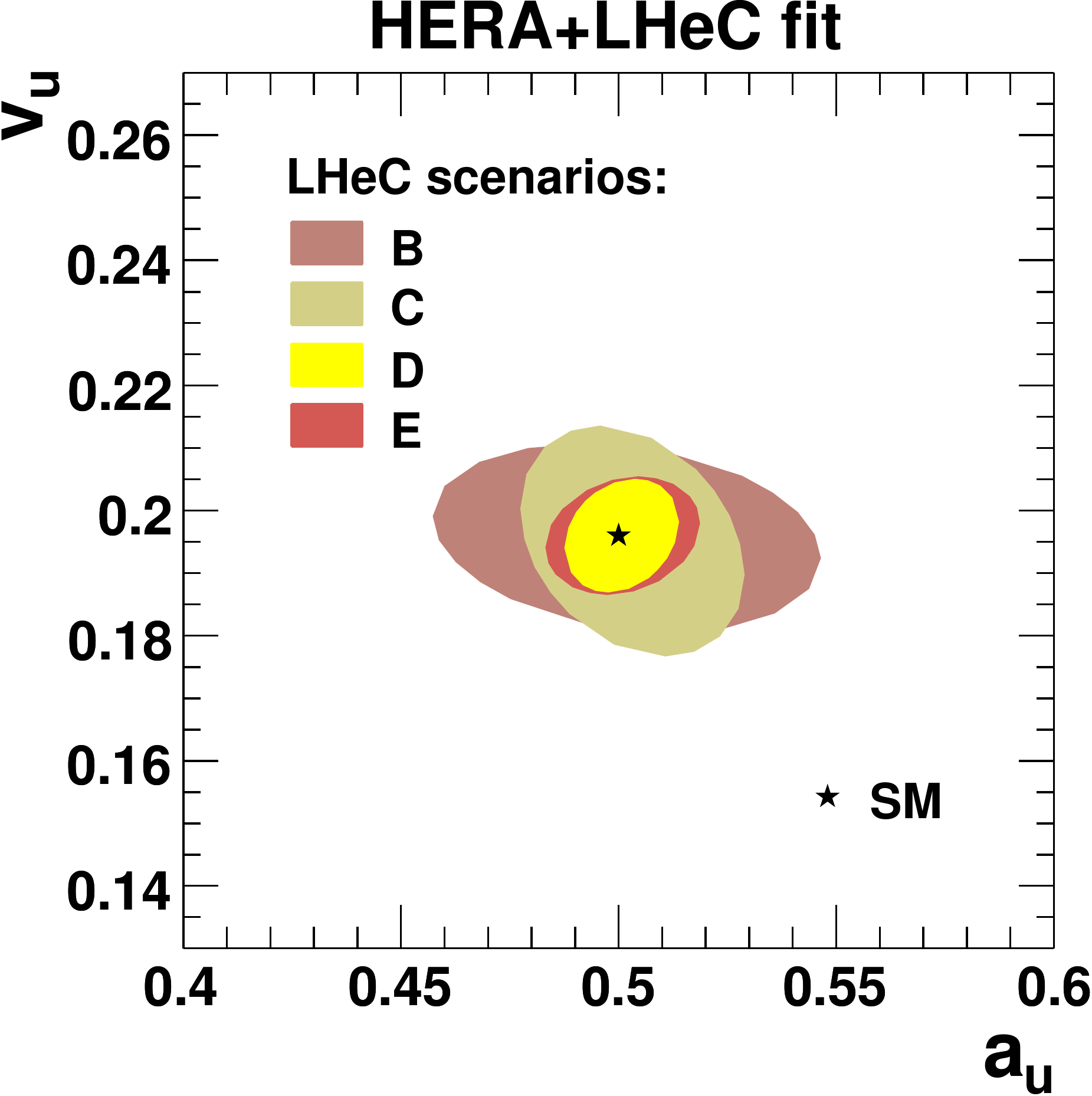} \includegraphics[width=0.49\textwidth]{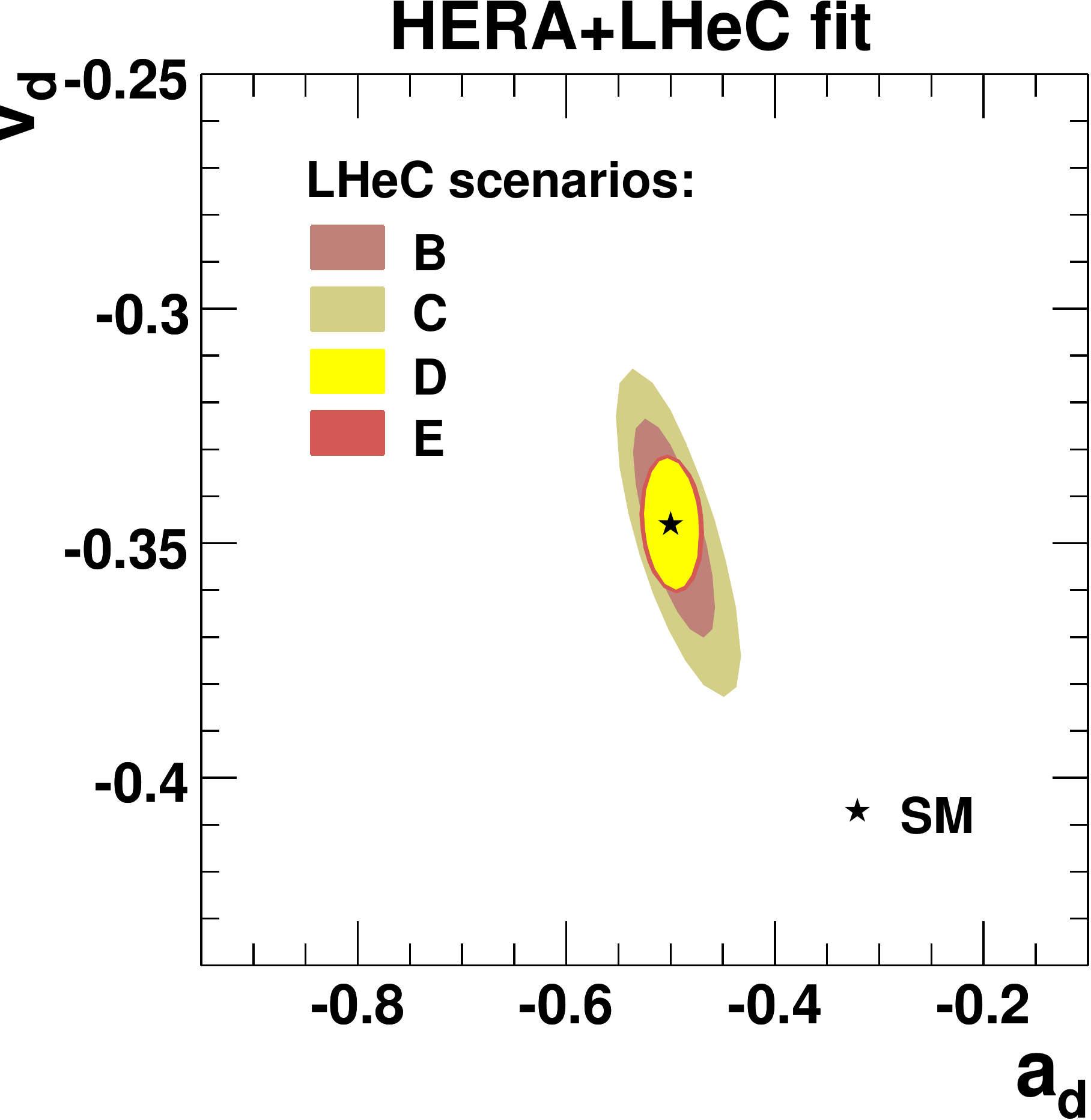}
\end{center}
\caption{ \label{Fig:ew:scenarios} Determination of the vector and axial-vector
weak neutral couplings of the light quarks at the LHeC, determined from a
joint NLO QCD and electroweak $\chi^2$ analysis of simulated NC and
CC cross section data using different beam scenarios as are summarised
in Table~\ref{tab:datasets}. The uncertainties comprise the full experimental
errors and consider their correlations.}
\end{figure}

The LHeC accesses with unprecedented precision
 the weak neutral current couplings which
enter the $\gamma Z$ interference and pure $Z$ exchange
parts of the NC cross sections, see Eq.\,\ref{ncfu}
in Section~\ref{sec:disformalism}.
As described in Section~\ref{sec:simNC}, a complete simulation
of DIS neutral and charged current inclusive cross
sections is performed including also their expected
uncorrelated and correlated systematic uncertainties.
The sensitivity of the LHeC to the light quark vector 
and axial-vector NC couplings ($v_q,~a_q$, with $q=u,~d$) 
is investigated with a  QCD fit to the simulated NC and CC data, in 
which the PDFs  and the  $v,~a$ couplings are simultaneously determined.
Various  beam conditions have been simulated,
which  are summarised in Table~\ref{tab:datasets}.
Figure~\ref{Fig:ew:scenarios} presents the precision with which
the up- and the down-quark couplings can be determined by the LHeC
as ellipses of uncertainties, which comprise the
statistical and the systematic uncertainties.
The experimental accuracy of the vector and axial vector couplings of the $u,~d$ quarks
amounts to a few \%, depending on the actual beam conditions.
%

\begin{figure}[t]
\begin{center}
\includegraphics[width=0.49\textwidth]{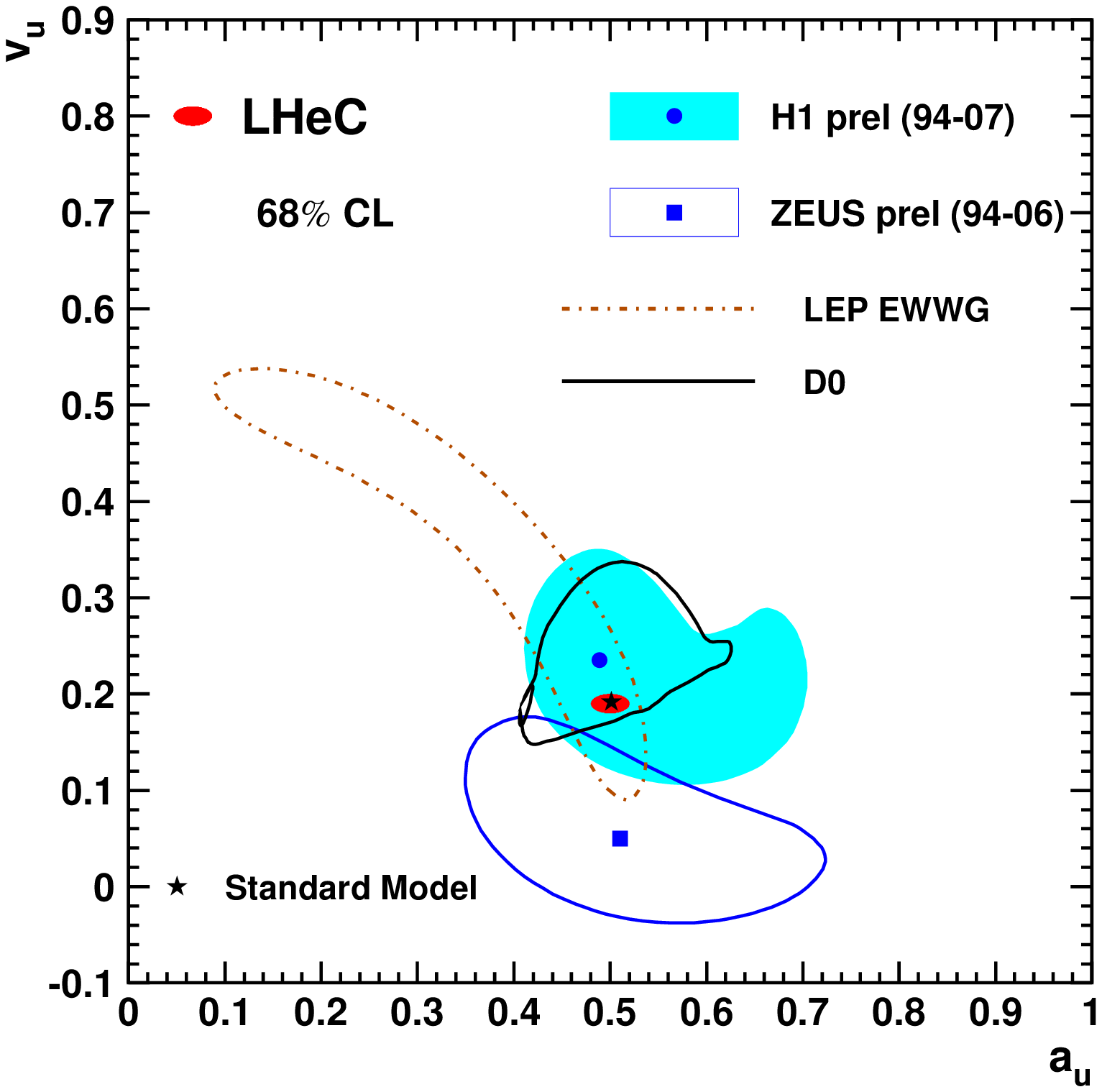}
\includegraphics[width=0.49\textwidth]{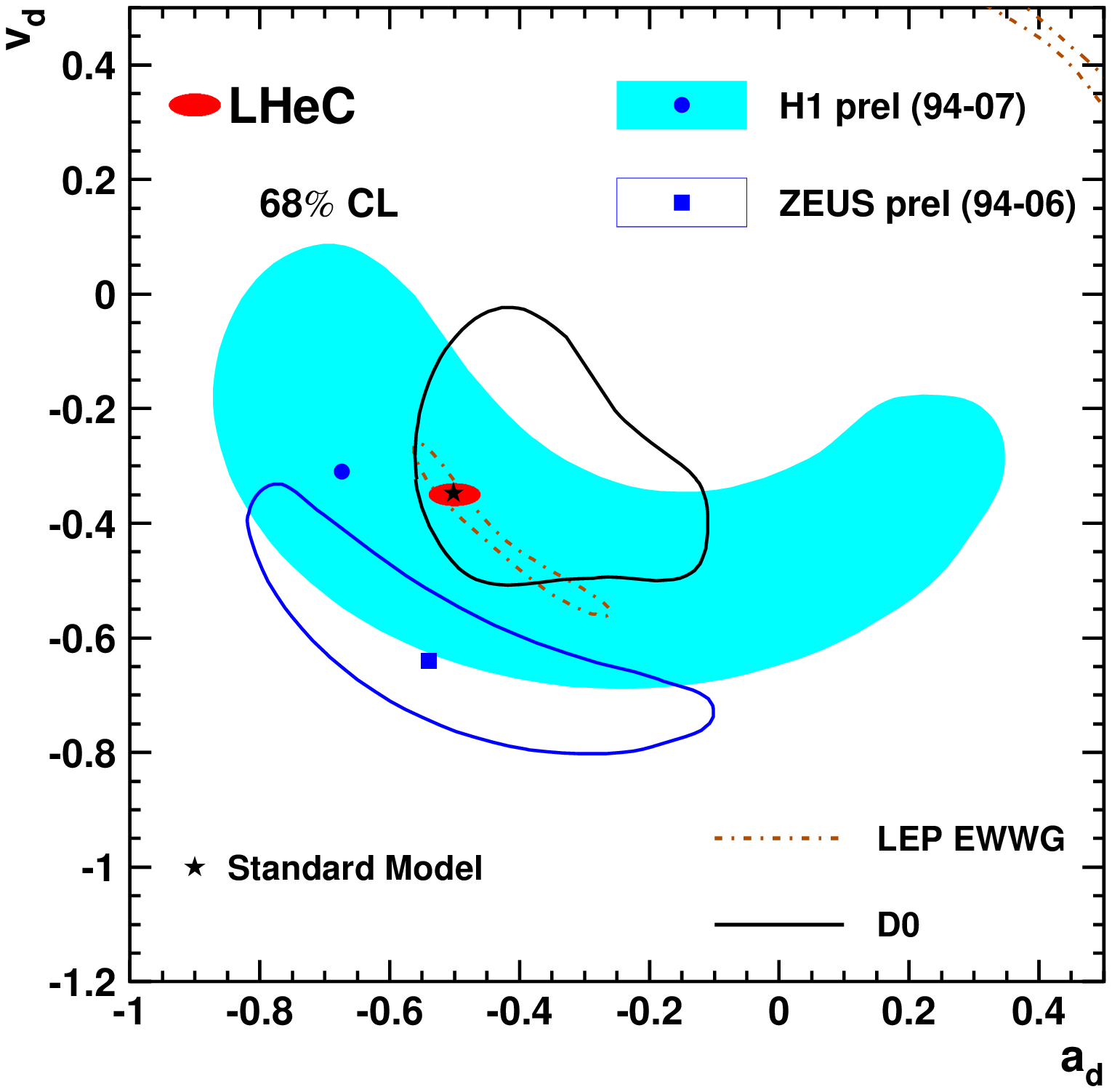}
\end{center}
\caption{ \label{Fig:ew:all} Determination of the vector and axial-vector
weak neutral current couplings of the light quarks by  LEP, D{\O}, H1 and 
ZEUS, compared with the simulated prospects for the LHeC.}
\end{figure}

The LHeC can completely
 disentangle the vector and axial-vector NC
couplings of up and down type light quarks with high precision. 
LEP has an ambiguity as it measures squares of the couplings
on the $Z$ pole while DIS and  Drell-Yan experiments 
access also their signs due to the $\gamma Z$ interference.
Recent results by ZEUS and H1 have already improved 
on the LEP determination in the case of  up quarks
while being less accurate for down quarks~\cite{Zhang:2008aq,Aktas:2005iv,Zhang:2010zz}.
The simultaneous determination of the four light quark couplings, based
on a luminosity of $5$\,fb$^{-1}$, by  the D{\O} experiment~\cite{Abazov:2011ws}
uses the $Z/\gamma^*$ forward-backward
asymmetry in the electron channel. It gives uncertainties of order $0.1-0.2$
which are an order of magnitude less precise than the expected DIS result at the LHeC.
This situation is illustrated in Fig.\ref{Fig:ew:all}. The LHeC determination, here 
drawn for scenario C,  of all four couplings is
shown as central ellipses around the SM prediction, and it is clearly of superior
quality.

%


The precise  determination of $v_{u,d}$ and $a_{u,d}$
will  constrain  new physics  models that modify significantly the light quark NC
couplings, without affecting the well-measured lepton and heavy quark couplings. 
It is not easy to realise such an exotic scenario in a natural way, although family
non-universal (leptophobic) Z' models 
(see for instance \cite{Salvioni:2009jp,Erler:1999nx}  and references therein), 
 R-parity violating supersymmetry 
(see \cite{Barbier:2004ez} for a review)  and 
leptoquarks \cite{Carpentier:2010ue} could be candidate theories.
 LHeC could therefore accurately test a spectrum of interesting 
new physics models. 
Anticipated results from the QWeak Collaboration 
\cite{Qweak}, when combined with existing precise measurements of Atomic Parity 
Violation and DIS experiments, 
could provide a per cent level determination of $v_u$ and $v_d$~\cite{Erler:2003yk} 
but it will not probe the axial-vector quark couplings.

%% file: physics/sinteta.tex
\subsubsection{Cross section asymmetries and ratios}
\begin{figure}
\centerline{\includegraphics[clip=,angle=0.,width=1.\textwidth]{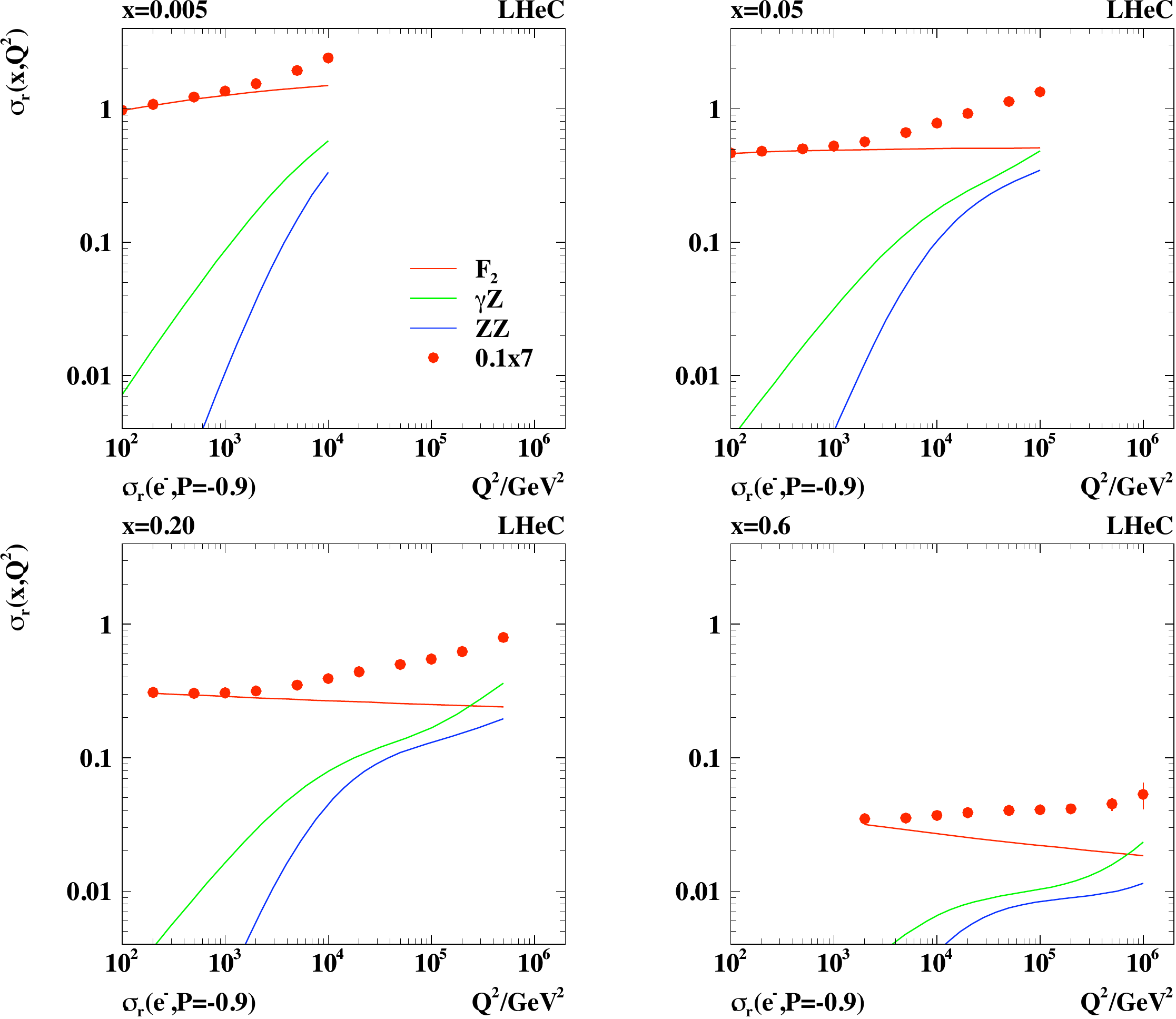}}
\caption{Simulated measurement of the neutral current DIS cross section (closed
points) with statistical errors for $10$\,fb$^{-1}$ shown as a function of
$Q^2$ for different values of Bjorken $x$. The different curves represent
the contributions of pure photon exchange (red), $\gamma Z$ interference (green)
and pure $Z$ exchange (blue) as prescribed in Eq.\,\ref{strf}. Note the
high precision of the reduced cross section measurement up to large $x$ 
and $Q^2$.
}
\label{fig:sigelw}
\end{figure}

The LHeC is a unique facility for electroweak physics because
of the very high luminosity, high measurement precision
and the extreme range of momentum
transfer $Q^2$. Fig.\,\ref{fig:sigelw} illustrates the reach and the
size of the electroweak effects in NC scattering.
Depending on the charge and polarisation of the electron beam,
the contributions from $\gamma Z$ interference and pure
$Z$ exchange become comparable to or even exceed the
photon exchange contribution, i.e. of $F_2$,
which has dominated hitherto all NC DIS measurements.
With the availability of two charge and two polarisation states, of neutral
and charged current measurements, proton and isoscalar targets,
a unique menu becomes available for testing the electroweak theory.
For example, one can very precisely measure light quark
weak neutral current couplings, discussed above.
One can also test the universality of $\gamma - g$ and $Z - g$ fusion
by extracting the heavy quark ($c,~b$)
contributions from $\gamma Z$ interference. A remarkable measurement
illustrated in the following regards
 the energy dependence of the weak mixing angle $\sin^2 \Theta$.

Tests of the electroweak theory in DIS require to simultaneously control
the parton distribution effects. With the outstanding data base from the
LHeC, joint QCD and electroweak fits become possible to high orders 
perturbation theory. Cross section asymmetries and ratios can also
be used to determine electroweak parameters. Particularly useful
examples are polarisation and charge asymmetries and also NC to
CC cross section ratios.
  
In NC scattering, the polarisation asymmetry 
\begin{equation} \label{apm}
A^\pm =  \frac{\sigma_{NC}^{\pm}(P_R) -\sigma_{NC}^{\pm}(P_L)}
                          {\sigma_{NC}^{\pm}(P_R) +\sigma_{NC}^{\pm}(P_L)}
\end{equation}
served for the decisive confirmation of the left handed weak neutral
current doublet structure as was predicted by the GWS theory in 
1979~\cite{Prescott:1979dh}. The size of the electroweak asymmetries
is given by the relative amount of $Z$ to photon exchange 
O($10^{-4} Q^2/ \rm{GeV}^2)$,
i.e. it becomes of order $1$ at high $Q^2$ at the LHeC.

To a  good approximation the asymmetry, normalised to the $L-R$
polarisation difference,
measures the structure function ratio  
\begin{equation}  \label{fgf}                                 
        \frac{2}{P_L-P_R} \cdot A^\pm      
        \simeq  \mp \kappa_Z   a_e \frac{F_2^{\gamma Z}}{(F_2 + \kappa_Z
         a_e Y_- xF_3^{\gamma Z}/Y_+)} 
         \simeq \mp  \kappa_Z a_e \frac{F_2^{\gamma Z}}{F_2}.
\end{equation}
Thus $A^+$ is expected to be about equal to $-A^-$
and to be only weakly dependent on the parton distributions.
The product of the axial coupling of the electron and the vector
coupling of the quarks, inherent in $F_2^{\gamma Z}$,
determines the polarisation asymmetry to be parity violating.
A measurement of $A^{\pm}$ provides a unique and precise
measurement of the scale dependence of
the weak mixing angle, as is discussed below (Sect.\,\ref{sec:sint}).
At large $x$  the polarisation asymmetry provides an NC
measurement of 
the $d/u$ ratio of the valence quark distributions, according to
\begin{equation} \label{doveru} 
          \frac{2}{P_L-P_R} \cdot A^\pm  \simeq  \pm  \kappa  \frac{1+d_v/u_v}{4+d_v/u_v}.
\end{equation}
Further asymmetries of NC cross sections have been discussed
in~\cite{Klein:1983vs}. 

The neutral-to-charged current cross section ratio
\begin{equation}
\label{RNCC}
R^{\pm} = \frac{\sigma_{NC}^{\pm}}{\sigma_{CC}^{\pm}}
= \frac{2}{ (1\pm P)  \kappa_W^2} \cdot \frac{ \sigma_{r,NC}^{\pm}}{\sigma_{r,CC}^{\pm}}
\end{equation}
is of interest for electroweak physics too as will be demonstrated below.
At very high $Q^2 \gg M_Z^2$ and neglecting terms in the NC
part proportional to $v_e$ it becomes approximately equal to
\begin{equation}
\label{Rhighq}
R^{\pm} \simeq  \frac{2 a_e^2}{ (1\pm P)  \cos^2 \Theta} 
\cdot \frac{ Y_+  F_2^{Z}  - Y_- P  xF_3^Z}{Y_+ W^{\pm}_2 + Y_- xW^{\pm}_3}
\end{equation}
which reveals the striking similarity of the neutral and charged weak interactions at high energies. 
One may further consider, for example,
a quantity which is the $eN$  analogue to the Paschos-Wolfenstein
relation~\cite{Paschos:1972kj} in $\nu N$ scattering
\begin{equation}
A_{NCC} =\frac{\sigma_{NC}^+ - \sigma_{NC}^-}{\sigma_{CC}^+ - \sigma_{CC}^-}.
\end{equation}
 
The  very high luminosity and $Q^2$ range of the
LHeC as compared even to HERA will open a completely new
era of electroweak physics in DIS.

\subsubsection{Measurement of the weak mixing angle}
\label{sec:sint}
Further tests of the SM at the quantum level and indirect
searches for new physics require ultimate precision.
Higher order corrections occur in the  factor $1- \Delta r$, see Eq.\,\ref{equG},
which depends on the top mass, logarithmically on the Higgs mass
and possibly on new, heavy particles.   
A measurement  of the weak mixing angle, $\sin^2 \Theta$,
to $0.01$\,\% precision should fix the Higgs mass to $5$\,\% accuracy.
The so far most precise measurements of  $\sin^2 \Theta$ have been performed
at the $Z$ pole in $e^+e^-$ scattering,  using the very high statistics, at LEP,
and in the case of the SLC, the large beam polarisation of $75$\,\% too.
The LHeC has the potential to measure weak asymmetries and 
cross section ratios at, below and beyond the $M_Z$ scale by
precisely measuring their dependence on $\sqrt {Q^2}$. 

The precision estimated for $\sin^2 \Theta$ depends on its definition.
Apart from the fermion and Higgs masses,
the electroweak theory has three independent parameters.
 For the subsequent study, as in a similar study of 
H1~\cite{Aktas:2005iv},
the values of $\alpha$ and $M_Z$ are fixed, which are best known,
$M_Z$ to $0.002$\,\%. For the estimate of the sensitivity to electroweak
effects, $\sin^2 \Theta$ is chosen here as the third parameter, which
is used together with $\alpha$ and $M_Z$ to calculate $G$ and $M_W$,
and also occurs in the weak neutral current couplings~\footnote{
An interesting test is also to fix $\alpha,~M_Z$ and the Fermi constant $G$ and to
determine derived electroweak parameters as $M_W$ or $\sin^2 \Theta$
for precision consistency checks in the search for deviations
from the SM. Such a study has not been undertaken so far for the LHeC.}.
 This way both
the NC and the CC cross sections are sensitive to $\sin^2 \Theta$.
Equivalently one could have expressed all parameters using 
$\alpha$, $M_Z$ and $M_W$, and determine $M_W$. Due to the relation
$\sin^2 \Theta = 1 - M^2_W/M^2_Z$, the error of such an 
indirect measurement of $M_W$ is
\begin{equation}
\Delta M_W = \frac{M_W \delta \sin^2 \Theta}{2 \sin^2 \Theta},
\end{equation} 
i.e. a one per mille precision on $\sin^2 \Theta$ corresponds 
to $\Delta M_W = 40$\,MeV.

\begin{figure}
\centerline{\includegraphics[clip=,angle=0.,width=0.9\textwidth]{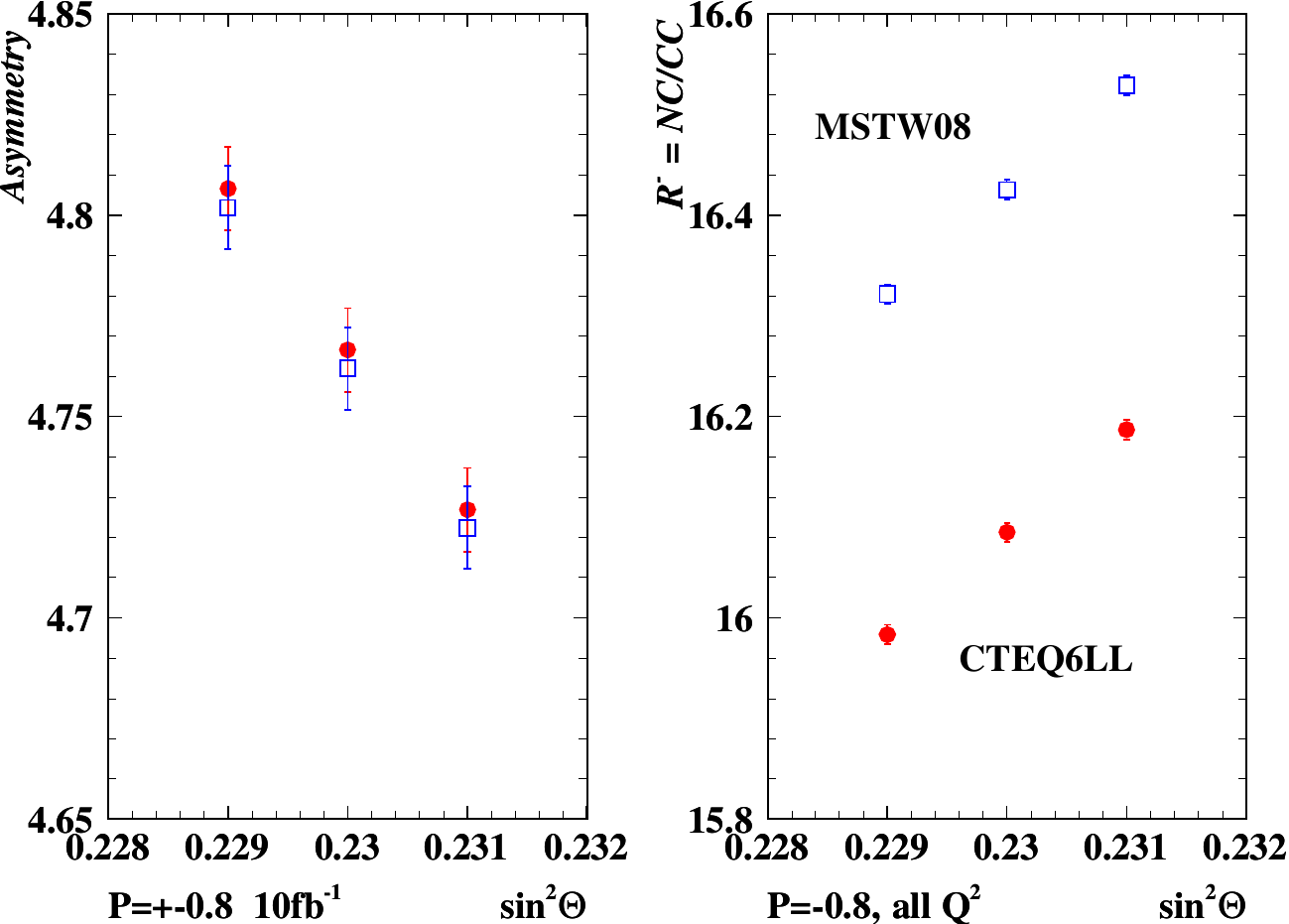}}
\caption{Simulated measurement of the polarisation NC cross
section asymmetry $A^-$ (left), in per cent for $P = \pm 0.8$,
and the ratio of neutral-to-charged current 
cross sections, $R=NC/CC$ (right), for $P=-0.8$,
for different values of $\sin^2 \Theta$.
The errors are statistical for
luminosities of $10$\,fb$^{-1}$ per beam for polarised electron scattering
for $E_e=60$\,GeV and the nominal $7$\,TeV proton beam.  The closed
(open) symbols show the simulation for the CTEQ6LL (MSTW08) leading
order parameterisations of the parton distributions. The average
$Q^2$ is $1300$\,GeV$^2$ for the NC asymmetry $A^-$,
while for the ratio $R$ the average CC $Q^2$ is about $9500$\,GeV$^2$.
Consequently, the mean $x$ in NC and CC differs by a factor of $6$,
which is at the origin of the large differences in $R$ between the
two PDF set predictions.
}
\label{fig:ARCMa}
\end{figure}

A simulation is done of the NC and CC cross sections depending on
the lepton beam charges and polarisations based on the 
formulae presented above. This allows to build a variety of
asymmetries and cross section ratios and derive their
sensitivity to the weak mixing angle. An example is illustrated
in Fig.\,\ref{fig:ARCMa}. Here the polarisation asymmetry (left)
and the NC/CC ratio (right) are calculated for different values
of $\sin^2 \Theta$ using two recent sets of leading order
parton distributions, CTEQ6LL and MSTW08. The measurement
precision of  $\sin^2 \Theta$ has a statistical, a polarisation,
a systematic and a PDF uncertainty. One derives that the
statistical precision is about $0.1$\,\% for the NC asymmetry $A^-$
and even $0.05$\,\% for the NC/CC ratio $R^-$ for $e^-p$ scattering
with an assumed  polarisation of $-0.8$ and a luminosity
of $10$\,fb$^{-1}$ for default beam energies.

At this early stage
of consideration one may not present a full error study.
However, a few first considerations are in order: 
The high luminosity and large $Q^2$ range move the electroweak
physics at this $ep$ machine to the level of highest precision
demands. Most of the systematic errors cancel in asymmetry
and ratio measurements. A $0.1$\,\% electron energy scale 
uncertainty, as has been achieved with H1, for example,
translates at the LHeC to a $0.15$\,\% change of $A^-$
and a negligible change of $R^-$. This measurement samples
data in a region of very high cross section accuracy and
can exclude the highest $x$ region where uncertainties 
grow like $1/(1-x)$.
The desired level of polarisation measurement is obviously
about a per mille, which seems to be possible
as is discussed in the detector chapter.

The requirements for $A^-$ and $R^-$ are different.
The asymmetry $A^-$ requires frequent changes of the
polarisation to control the time dependence of the measurement.
It measures essentially a ratio of the structure functions
$F_2^{\gamma Z}/F_2$ and therefore it is rather insensitive to 
uncertainties related to the parton distributions. In fact, one
observes in  Fig.\,\ref{fig:ARCMa} that the predictions of
the two PDF sets considered differ by less than the
statistical uncertainty for $A^-$. 
The NC/CC ratio $R$ is less sensitive to time drifts
as the NC and CC data are taken simultaneously. Its statistical
power is highest, as had already been noticed for HERA~\cite{Blumlein:1987fd}.
The present analysis indicates a large sensitivity to the PDFs,
which, however, is mainly related to the different mean
$Q^2$ values of the NC and CC samples. 

The high sensitivity of $R$ to the mixing angle can only be
employed when the PDFs are much better known than
so far. This, however, is one of the major goals of the LHeC
physics programme and large improvements are to be expected
as is discussed in Sec.\,\ref{sec:dirpart}.
The potential of measuring $\sin^2 \Theta$
from NC/CC ratios is observed to be particularly striking.
However, for the evaluation of the scale dependence
of $\sin^2 \Theta$ below, the results derived from $A^-$ 
are used due to its smaller PDF sensitivity, in this
first analysis.

\begin{table}[h]
{\small
  \centering
  \begin{tabular}{|c|r|r|r|r|r|r|r|}
    \hline
Type & $Q_1$  &  $P_1$ & $Q_2$  &  $P_2$ & $ \delta s (A_{12}) $ & $ \delta s (R_1) $ & $ \delta s (R_2)$ \\
\hline
e$^-$ Polarisation Conjugation  & -1.   & -0.8  &  -1.  & 0.8 &  0.00026 &  0.00009 &  0.00024 \\
e$^+$ Polarisation Conjugation &+1.  & -0.8  &  +1.  & 0.8 &  0.00027 &  0.00040 &  0.00015 \\
e$^-$ Low P Conjugation  & -1.   & -0.4  &  -1.  & 0.4 &  0.00052 &  0.00010 &  0.00015 \\
Charge Conjugation P=0 & +1.   & 0.  &  -1.  & 0. &  0.01600 &  0.00019 &  0.00012 \\ 
Charge Conjugation P=$\mp 0.8$  & +1.   & -0.8  &  -1.  & 0.8 &  --- &  0.00040 &  0.00024 \\ 
Charge Conjugation P=$\pm 0.8$  & +1.   & +0.8  &  -1.  & -0.8 &  0.00790 &  0.00015 &  0.00009 \\ \hline
e$^-$ PC Low $Q^2 \sim 300$\,GeV$^2$ & -1.   & -0.8  &  -1.  & 0.8 &  0.00068 &  0.00029 &  0.00083 \\
e$^-$ PC Med $Q^2 \sim 1500$\,GeV$^2$ & -1.   & -0.8  &  -1.  & 0.8 &  0.00027 &  0.00012 &  0.00029 \\
e$^-$ PC High $Q^2 \sim 22000$\,GeV$^2$ & -1.   & -0.8  &  -1.  & 0.8 &  0.00044 &  0.00071 &  0.00055 \\
e$^-$ PC  vHigh $Q^2 \sim 130000$\,GeV$^2$ &-1.   & -0.8  &  -1.  & 0.8 &  0.00170 &  0.00460 &  0.00200 \\ 
\hline
  \end{tabular}
}
\caption{Estimated precision of the weak mixing angle, $\delta s= \delta \sin^2 {\Theta}$,
 from simulated measurements of the NC cross section asymmetry, $A$,
 and the NC/CC cross section ratio, $R$,
 for different beam charge ($Q$) and polarisation ($P$) conditions.
}
\label{tab:sint}
\end{table}

\begin{figure}
\centerline{\includegraphics[clip=,angle=0.,width=1.\textwidth]{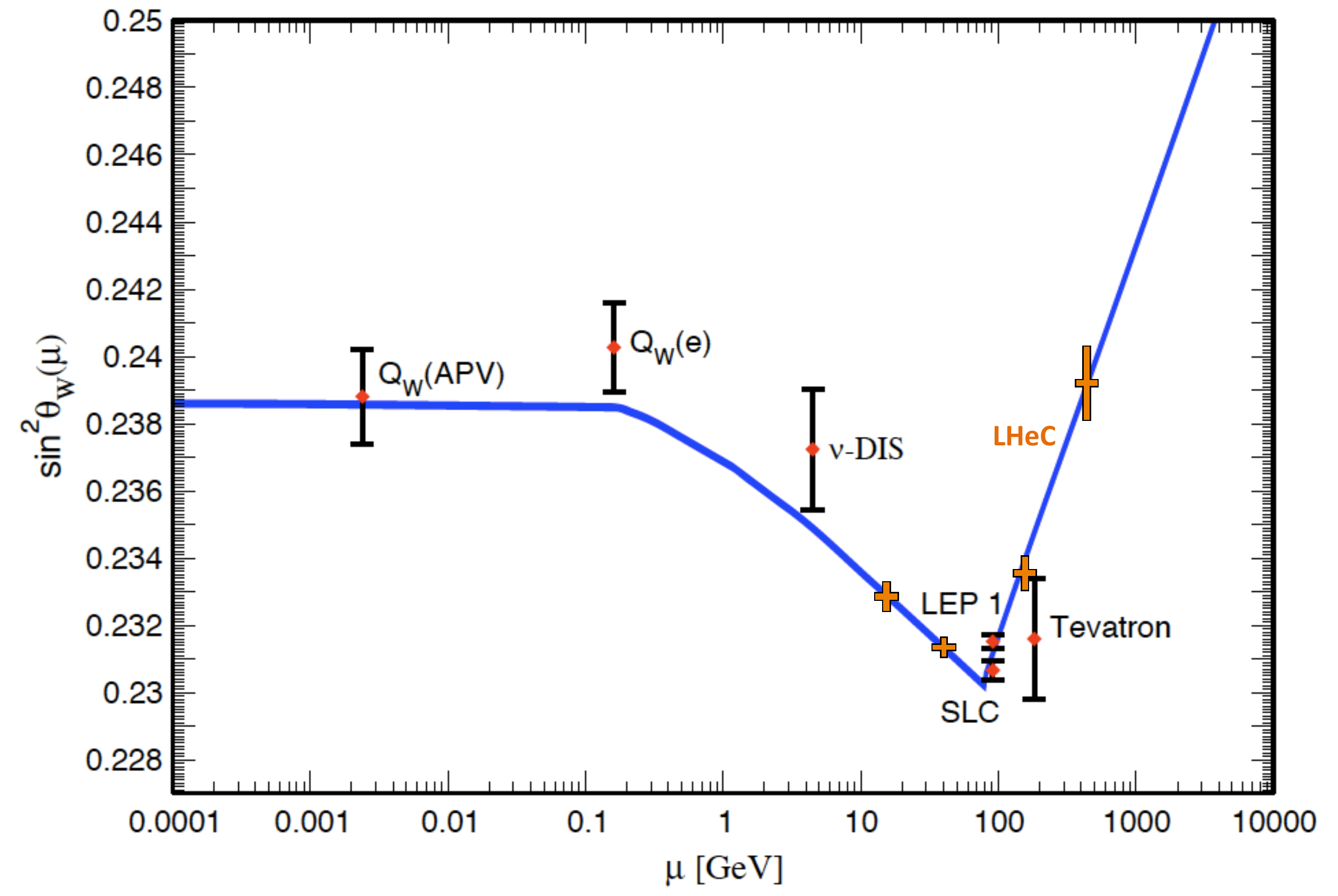}}
\caption{Dependence of the weak mixing angle 
 on the energy scale $\mu$,
taken from~\cite{Nakamura:2010zzi}. 
Four simulated points have been added based
on the estimated measurement accuracy using the polarisation asymmetry
$A^-$ binned in intervals of $\sqrt{Q^2}$, see text.
}
\label{fig:sintmu}
\end{figure}
The mixing angle
is predicted to vary strongly as a function of the scale $\mu$, which in
DIS is precisely known 
and given as $\sqrt{Q^2}$. This dependence results from higher order
loop effects as calculated in~\cite{Czarnecki:2000ic}.
Precise measurements to per mille uncertainty were performed
at the $Z$ pole by SLC and LEP experiments. Recent low energy
experiments have provided measurements of $\sin^2 \Theta$ at very low $Q^2$
as from the parity violation asymmetry due to polarisation conjugation
in M{\o}ller scattering at $Q^2 = 0.026$\,GeV$^2$ by the E158 experiment.
At scale values of about $5$\,GeV the NuTeV Collaboration has determined
the mixing angle which for some time created a substantial experimental
and theoretical effort when it appeared to be above the theoretical expectation
by a few standard  deviations. Explanations of this ``anomaly'' included 
variations of the strange quark density, effects from QED or nuclear corrections.
An ultra-precise measurement of $\sin^2 \Theta$ is envisaged,
yet still at $\mu = M_Z$, if a new $Z_0$ factory was built.

The current $\sin^2 \Theta$ measurements are summarised in
Fig.\,\ref{fig:sintmu}.  The plot also contains the projected
$\sin^2 \Theta$ uncertainty values from the LHeC for scales between
about $10$ and $400$\,GeV, as listed in Table\,\ref{tab:sint}, which
result from simulations of the parity violation asymmetry $A^-$ in
polarised $e^-p$ scattering~\footnote{It is to be noted that this comparison at the current stage
of the analysis is mainly illustrative. A quantitative comparison, in particular
with the LEP/SLD results, requires a more complete analysis including a
study of the systematic uncertainties, even when considered to be small in
the asymmetry measurement, the
 effect of higher-order corrections and the scheme dependence
of the result.}.
Due to the high statistics
nature of the DIS NC process, the variation of $\sin^2 \Theta$
as a function of $\sqrt{Q^2}$ can be measured for a large range of $\sqrt{Q^2}$.
At low scales the range is
limited by the sensitivity to the $Z$ exchange effects and at high
scales by the kinematic limit and luminosity.  It may deserve a study
to understand how low in $Q^2$ the asymmetry
$A^-$ can be determined in a meaningful measurement, which
is related to time drifts, polarisation flip times etc. and likely
can only be answered with real data. It is to be noted
that previous and planned fixed target experiments measure
this asymmetry at extremely small values of $Q^2$ as
compared to the range of the LHeC.


%% file: physics/CDR_lowx.tex



In Chapter\,\ref{chapter:qce}, the opportunities offered by the LHeC to 
perform precision QCD studies were discussed in detail. Such studies have 
been done, until now, within the framework of standard, 
fixed-order perturbation theory and collinear factorisation, which is
valid when momentum scales are sufficiently hard and when 
the hadron can be described as a dilute set of partons. 
On the other hand, the parton densities extracted from HERA data
exhibit a strong rise towards  
low $x$ at fixed $Q^2$, indicating that the proton becomes increasingly
densely packed. There are also compelling theoretical reasons 
to believe that collinear factorisation should break down
with increasing energies and sizes of the hadron. The low $x$
regime of proton structure thus represents an exciting and largely
unexplored territory whose dynamics are those of a densely 
packed partonic system. From very general considerations, it 
is clear that the increasing parton densities cannot continue untamed
throughout the region of LHeC sensitivity. Non-linear evolution 
must eventually become relevant and the parton densities must `saturate'. 
The LHeC offers the unique possibility of
observing these highly non-perturbative dynamics at sufficiently large 
$Q^2$ values for weak coupling methods to be applied, suggesting the
exciting possibility of a 
parton-level understanding of the collective properties of QCD. 
In this chapter we explore these possibilities in detail, addressing 
possible methods by which
LHeC data might be used to establish the existence of this 
new high parton density regime of QCD and
to explore its properties.

%

\section{Physics at small $x$}

\subsection{High energy and density regime of QCD}

\label{sec:lowxoverview}

\subsubsection{Introduction}
\input{physics/tex/lowxoverview}

\subsubsection{Beyond DGLAP evolution}
\input{physics/tex/dglaptonl}

\subsubsection{Resummation at low $x$}
\input{physics/tex/introresum}

\subsubsection{Saturation in perturbative QCD}
\input{physics/tex/cgc}

\subsubsection{The importance of diffraction}
\input{physics/tex/diffrintro}

\subsubsection{The importance of nuclei}
\input{physics/tex/nucintro}

\subsection{Status following HERA data}
\label{sec:statusheraintro}
\input{physics/tex/statusheraintro}

\subsubsection{Dipole models}
\input{physics/tex/dipolemodels}

\subsubsection{Hints of deviations from fixed-order linear DGLAP evolution in inclusive HERA data}
\input{physics/tex/sathera}

\subsubsection{Linear resummation schemes}
\input{physics/tex/resum}

\subsection{Low-$x$ physics perspectives at the LHC}
\label{sec:lowxlhc}

\input{physics/tex/LHC_lowx_3rv}

\subsection{Nuclear targets}
\label{sec:nucleartargets}

\input{physics/tex/npdfs}

\subsubsection{Requirements for the ultra-relativistic heavy ion programs at RHIC and the LHC}
\input{physics/tex/eA-hip}

\section{Prospects at the LHeC}

\subsection{Strategy: decreasing {\boldmath $x$} and increasing 
{\boldmath $A$}}
\label{sec:sxstrategy}
\input{physics/tex/sxstrategy}

\subsection{Inclusive measurements}
\label{sec:epincl}

\subsubsection{Predictions for the proton}
\input{physics/tex/predep}

\subsubsection{Constraining small-$x$  dynamics}
\input{physics/tex/testingnl}

\subsubsection{Predictions for nuclei: impact on nuclear parton distribution functions}
\input{physics/tex/predea}


\subsection{Exclusive production}
\label{sec:vm}
\input{physics/tex/excldiffintro}

\subsubsection{Exclusive production formalism in the dipole approach}
\input{physics/tex/vm}

\input{physics/tex/gpds_lhec}
\label{sec:dvcs}

\input{physics/tex/vmeA}

\subsection{Inclusive diffraction}

\label{sec:diffpdfs}
\input{physics/tex/ddisep}


\subsubsection{Predicting nuclear shadowing from inclusive diffraction in ep}
\input{physics/tex/dep2ns}


\subsubsection{Predictions for inclusive diffraction on nuclear targets}

\input{physics/tex/ddisea}




\subsection{Jet and multi-jet observables, parton dynamics and fragmentation}
\label{sec:jetspartdyn}

\subsubsection{Introduction}
\input{physics/tex/introjets}

\subsubsection{Unintegrated PDFs}
\input{physics/tex/updfs}

\input{physics/tex/dijets}

\subsubsection{Forward observables}
\input{physics/tex/forwardjets}


\subsubsection{Perturbative and non-perturbative aspects of final state radiation and hadronisation}
\input{physics/tex/fsr}


\subsection{Implications for ultra-high energy neutrino interactions and detection}
\label{sec:uhenu}
\input{physics/tex/neutrino}

%
%







%% file: physics/tex/lowxoverview.tex
Quantum Chromodynamics \cite{Fritzsch:1973pi}
is the fundamental theory of 
strong interactions
and has been extensively
tested in the last 39 years.  Still, many open questions remain to be solved.
One of them, which can be addressed at high energies, is the transition
between the regimes in which the strong coupling constant is either large  or
small - the so-called {\it strong } and {\it weak coupling} regimes. In the former,
standard perturbation theory techniques are not applicable  and
exact analytical results are not yet within the reach of current knowledge.
Therefore various models, {\it
effective} theories, whose parameters cannot yet be derived from  QCD, or numerical lattice computations, have to be employed.
 One example of such an effective theory which has been used through the years and
actually predates QCD, is Regge-Gribov
\cite{Regge:1959mz,Gribov:1968fc,Abarbanel:1975me} theory.


The weak coupling  regime has been well tested in high-energy experiments
through a selected class of measurements - often referred to as {\it hard
processes} - where weak and strong coupling effects can be cleanly  separated.
There exists a well-defined theoretical concept which has been derived  from
first principles and probed in the weak coupling  regime, namely the collinear
factorisation theorem (for a comprehensive review see \cite{Collins:1989gx} and
references therein). It allows a separation of the cross sections involving hadrons into: (i) parts that can be computed within perturbation theory, corresponding to the cross section for parton scattering, and (ii) pieces which cannot be calculated using weak coupling techniques, but whose 
evolution 
with momentum scales
 is still perturbative. The latter are universal, process-independent distributions that either characterise the partonic content of the hadron - {\it parton densities} on which we will mainly focus the discussion - 
or the eventual projection of partons onto hadrons. Together with their
corresponding (DGLAP) linear evolution equations
\cite{gribovlipatov,Altarelli:1977zs,Dokshitzer:1977sg}, they have been used to describe
experimental data to a high accuracy.  Examples include 
total DIS cross sections, the production of jets
with large transverse momenta and final states with heavy quarks,
see the analysis and discussion in Chapter\,\ref{chapter:qce}.

In recent years high-energy experiments have become sensitive to
kinematic regions in which the coupling is small but the factorisation
assumption may no longer be valid.  
We will refer to this region as the high parton density domain,
or simply the dense regime.
As an example, several HERA DIS measurements at
small longitudinal
momentum fractions $x$, 
where parton densities are large, indicate deviations from the
behaviour expected  with standard collinear factorisation.  Similarly,
hadronic or nuclear collisions involving partons with small 
values of $x$ may also show such deviations. At the same time, 
cross sections grow rapidly with decreasing $x$, so 
contributions from these
regions dominate hadronic cross sections in sufficiently 
high-energy scattering.
Experiments sensitive to this kinematic region thus provide a way to
test QCD in the new regime where the parton densities
become very large and highly novel effects are expected.
As has historically always been the case for the exploration of parton 
densities, the most promising approach is lepton-nucleon scattering,
exploiting the point-like, non-strongly interacting nature of the lepton
probe to take `snapshots' of the hadronic structure with 
deeply sub-femtoscopic
resolution.


From a theoretical viewpoint, this situation 
offers both opportunities and challenges. The fact that, at small-$x$, there is no abrupt transition between the dilute and dense regimes,
allows the use of techniques which, while still being weak coupling, go beyond those employed
in the dilute limit. The
usual parton multiplication processes have to be supplemented by processes in
which partons recombine - thus adding non-linear terms to the evolution
equations \cite{Gribov:1984tu}.  There are deep theoretical 
questions arising in this
new dense partonic 
regime of QCD. At high energies the scattering
amplitudes are close to the unitarity limit.
Unitarity is violated when the linear regime is extrapolated to very high energies, so the dynamics of QCD beyond the linear dilute regime has to be such that unitarity is fulfilled. The generic expectations are that the dynamical mechanism responsible for the fulfilment of unitarity is that accountable for the taming of parton densities due to recombination effects - this phenomenon is generically referred to as parton {\it saturation}.
Theoretical calculations
\cite{Mueller:1989st,JalilianMarian:1997gr,Balitsky:1995ub,Kovchegov:1999yj} 
in the limit of high energies support these expectations.
 Furthermore, the
experimental exploration of this transition region where the standard
perturbative description based on collinear factorisation and linear evolution equations requires large corrections, provides new possibilities
of further understanding the strong coupling regime.

Deep inelastic lepton-hadron scattering has 
already been shown to address these questions in
a very efficient manner. It provides the cleanest way of measuring the parton
densities, including the small-$x$ region in which
the transition between the
dilute and dense regimes of QCD should occur within the weak coupling region
where calculations can be done.
Approaching this transition region 
from the dilute side by decreasing $x$ or by increasing the 
number of nucleons in the target, one
should observe features which cannot be understood within the framework of
linear QCD evolution equations but, using more elaborate tools (non-linear
evolution equations) can still be analysed in terms of weak coupling
techniques.
Within the standard framework of leading-twist linear QCD
evolution equations (DGLAP) the parton densities are predicted to rise at small
$x$, and this rise has been seen 
very clearly at HERA. 
This rise should eventually be tamed by the novel, 
nonlinear effects leading to parton saturation. 
In hadron-hadron scattering, the growth
of total cross sections with energy is limited by unitarity bounds. As a result, 
according to Froissart and
Martin \cite{Froissart:1961ux,Martin:1962rt}, total cross sections satisfy
\begin{equation}
\sigma_{\rm tot} \le {\rm const.} \ln^2 s/s_0 \; ,
\label{eq:froissartmartin}
\end{equation}
where $s_0$ is a typical hadronic scale,
and the dimensionful 
coefficient `const.' is governed by the range of the strong interaction.
This bound comes from two fundamental assumptions. 
The first is that the amplitude
for the scattering at fixed value of impact 
parameter\footnote{The impact parameter in a scattering 
process between two particles with parallel trajectories is the perpendicular distance
between the centres of the particles.}
is bounded by unity
and the second is the finite range of the strong interaction.
The bound on the amplitude has a simple physical interpretation 
in terms of a situation where the
probability for the interaction becomes very high, so the target (or more
precisely the interaction region) 
becomes completely absorptive. This situation is
usually referred to as a {\it black disk} regime.  The description of this
regime is very challenging theoretically and it is expected that new phenomena
will occur which are direct manifestations of a new state of QCD which is
characterised by a high parton 
density \cite{Gribov:1968gs,Frankfurt:2001av}. 
The LHeC will uniquely offer the possibility of exploring
the transition towards this
new state of dense QCD matter, as it can pursue a two-pronged approach: high
centre-of-mass energy, extending the kinematic range to lower $x$, 
and the possibility of deep inelastic scattering off
heavy nuclei. 

In the rest of this introductory section, we will present 
different approaches that are currently under discussion to describe the high-energy regime of QCD. We will recall the ideas that lead from linear evolution equations to non-linear ones.
In the linear case we will discuss evolution equations computed within fixed
order perturbation theory (the DGLAP equations) as well as ones including some kind of
resummation - thus going beyond any fixed order in the perturbative expansion in the QCD coupling constant, the most famous example of which is
the Balitsky-Fadin-Kuraev-Lipatov
(BFKL) equation \cite{Kuraev:1977fs,Balitsky:1978ic}. Non-linear evolution leads to the phenomenon of saturation of partonic densities in the hadron or nucleus. We will briefly review the realisations of saturation of parton densities both at strong coupling and, mainly, at weak coupling. We will end by discussing the importance of diffractive observables and of the use of nuclear targets for the investigation of the small-$x$ behaviour of the hadron or nucleus wave function.

%% file: physics/tex/dglaptonl.tex
 In DIS the structure function $F_2(x,
Q^2)$ is proportional to the total cross section $\sigma_{\rm tot}$ for the
scattering of a virtual photon on a hadron $h$, $\gamma^* h\to X$.  The growth of
$F_2$ at small $x$ translates into the rise of $\sigma_{\rm tot}$	 as a
function of the energy of the virtual photon-hadron system. Although the
Froissart-Martin bound, derived for hadron-hadron scattering, cannot be applied
to a process involving a virtual photon, direct calculations
based on the evaluation of the QCD diagrams demonstrate unambiguously that, at
small $x$, large corrections exist and need to be resummed. 
These corrections suppress the leading-twist results and there is no doubt that, for $F_2$, the
rise with $1/x$ predicted by DGLAP is modified by contributions which are not
included in the framework of leading-twist linear evolution equations.
 The corrections which become numerically important in
the small-$x$ limit are also important for the restoration of the unitarity
bound, as mentioned previously.  As a result of these modifications parton saturation is reached for
sufficiently large energies or small values of Bjorken-$x$.

\begin{figure}
\centerline{ \includegraphics[clip=,width=0.6\textwidth,angle=0]{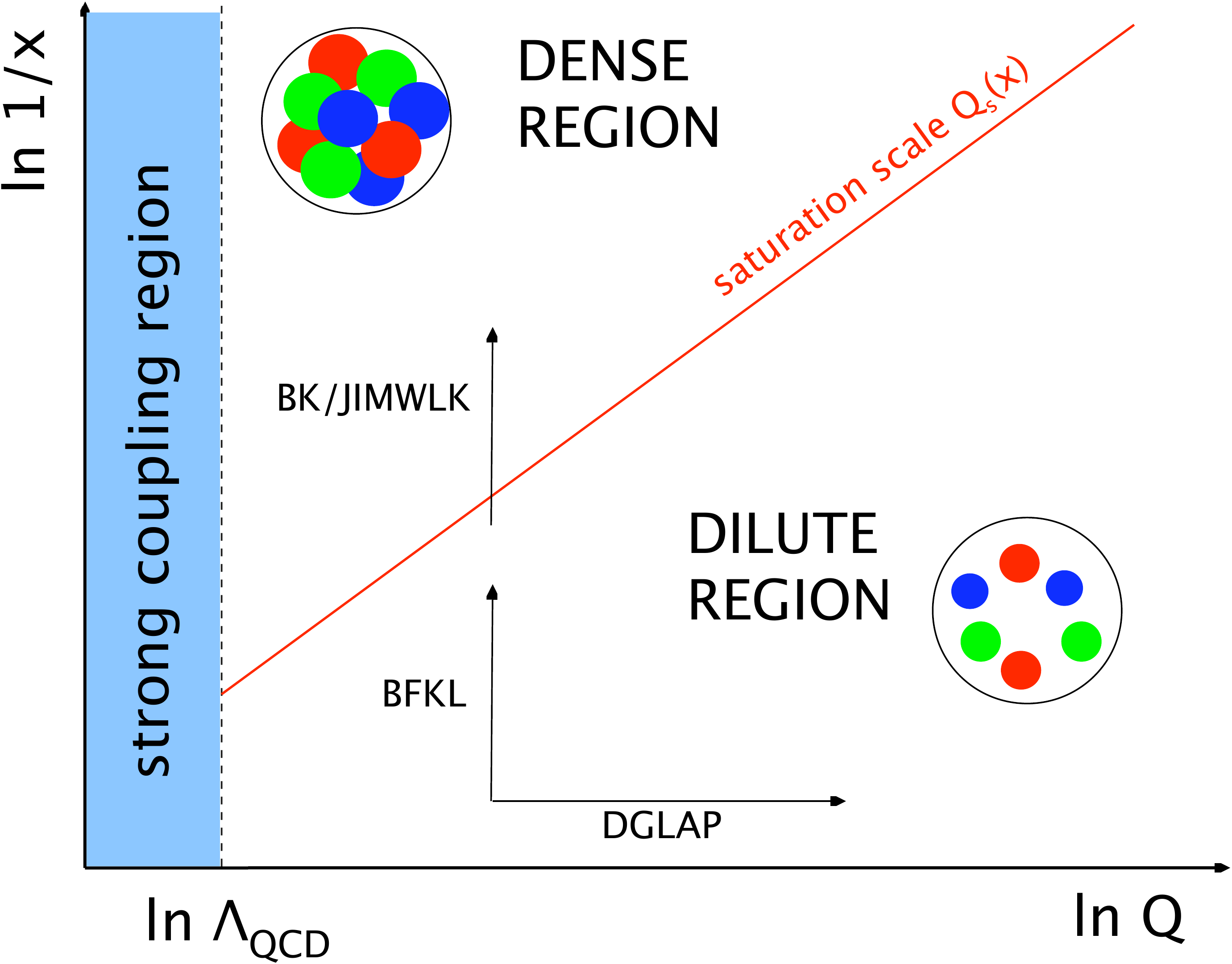}}
\caption{Schematic view of the different regions for the parton densities in the $\ln Q^2 - \ln1/x$ plane. See the text for comments.}
\label{Fig:satplane}
\end{figure}

 In deep inelastic electron-proton scattering, the virtual photon emitted by
the incoming electron interacts with partons inside the proton whose properties
are specified by the kinematics of the photon. In particular, the 
effective transverse
size of the partons is (roughly) inversely proportional to the square root of
the virtuality of the photon, $\langle r^2_T \rangle \sim 1/Q^2$.  The deep
inelastic cross section, parameterised through parton densities, 
thus {\it counts} the numbers of quarks and gluons 
per unit of phase space. For
sufficiently large photon virtualities $Q^2$ 
and not too small $x$, the improved QCD parton
model works well
because the partons forming the hadron, on the distance scale
defined by the small photon, are in a dilute regime, and they interact
only weakly.
This is a direct consequence of the
property of asymptotic freedom, which makes the strong coupling constant
small. This diluteness condition is not satisfied if the density of partons
increases. This happens if either the number of partons increases (large
structure function) or the interaction between the partons becomes strong
(large $\alpha_s$). The former situation is realised at small $x$, 
the latter for small photon virtuality
$Q^2$ which sets the scale of the strong coupling $\alpha_s(Q^2)$.  This simple
qualitative argument shows that corrections to the standard QCD parton picture
can be described in terms of quarks and gluons and their interactions as long
as $Q^2$ is not too small ($\alpha_s(Q^2) \ll 1$) and the gluon density is large
(small $x$). Combining these two conditions one arrives at the picture shown in
Fig.~\ref{Fig:satplane}: there is an approximately 
diagonal line in the $\ln Q^2 - \ln1/x$ plane below
which the parton distributions are dilute, and the standard QCD parton picture
applies.  In this regime linear evolution equations provide the correct
description of parton dynamics.  In the vicinity of the line, non-linear
QCD corrections become important, and above the line partons are in a
high-density state.
The division between the two
regimes is usually defined in terms of a
dynamically generated
`saturation scale', 
growing with decreasing $x$ and, 
in the case of nuclei, with increasing mass number.  Within this picture one
easily understands which type of corrections can be expected. Once the density
of gluons increases sufficiently, 
it becomes probable that, prior to their interaction with
the photon, gluons undergo recombination processes.

%% file: physics/tex/introresum.tex
As already mentioned in Sec.~\ref{sec:hfl_intro}, 
the generic challenges that the small-$x$ region bears in QCD are
inherently related to the divergence of the gluon number density with
decreasing values of $x$.  It is well known that the deep-inelastic partonic cross
sections and parton splitting functions receive large corrections in the
small-$x$ limit due
to the presence of powers of $[\alpha_s\log x]$ to all orders in
the perturbative
expansion~\cite{Kuraev:1977fs,Balitsky:1978ic,gribovlipatov,Catani:1990eg,Catani:1994sq}.
This suggests dramatic effects from logarithmically enhanced
corrections, so the success of
fixed order NLO perturbation theory at HERA has been very hard
to explain in regions where $x$ becomes small. 
Recently, hints have been found that indeed the 
quality of the DGLAP fits
tends to deteriorate systematically in the region of small $x$ and $Q^2$
\cite{Caola:2009iy,:2009wt}.  Direct calculations at next-to-leading logarithmic
accuracy in the BFKL framework were performed
\cite{Fadin:1998py,Ciafaloni:1998gs}, and showed a slow convergence of the
perturbative series in the high-energy, or small-$x$ regime.  Therefore,
generically one expects deviations from fixed-order DGLAP evolution in
the small-$x$ and small-$Q$ regime which call for a resummation of higher orders
in perturbation theory.

Extensive analyses have been performed in the last few years
~\cite{Altarelli:2003hk,Altarelli:2005ni,Altarelli:2008aj,Ciafaloni:2003rd,Ciafaloni:2003kd,Ciafaloni:2007gf},
which indeed point to the importance of resummation to all orders.  Resummation
should embody important constraints like kinematic effects, momentum sum rules
and running coupling effects. 

Several important questions arise here, such as the relation and interplay  of
the resummation and the non-linear effects, and possibly the role of
resummation in the transition between the perturbative and non-perturbative
regimes
in QCD.  Precise experimental measurements in extended kinematic regions are
needed to explore the deviations from standard DGLAP evolution and to
quantify the role of the resummation at small $x$.

%% file: physics/tex/cgc.tex
The original approach to implement unitarity and rescattering effects in high-energy hadron scattering was developed by Gribov \cite{Gribov:1968fc,Gribov:1968jf,Gribov:1968gs}. Models based on this non-perturbative Regge-Gribov framework  are
quite successful in describing existing data on inclusive and diffractive ep
and eA scattering (see e.g. \cite{Armesto:2010ee,Armesto:2010kr} and references
therein). 
However, they lack solid theoretical foundations within QCD.

On the other hand, attempts have been going on for the last 30 years to
implement parton rescattering or 
recombination\footnote{Note that the rescattering and recombination
concepts correspond to the same physical mechanism viewed in the rest frame  
and the infinite momentum frame of the hadron, respectively.} 
in perturbative QCD in order to
describe its high-energy behaviour.  In the pioneering work in
\cite{Gribov:1984tu,Mueller:1985wy}, a non-linear evolution equation in $\ln
Q^2$ was proposed to provide the  first correction to the linear equations. A
non-linear term appeared, which was proportional to the local density of colour
charges seen by the probe (the virtual photon). 

An alternative, independent approach was developed in \cite{Bartels:1994jj}, where the amplitudes for diffractive processes in the triple Regge limit were calculated. This resulted in the extraction of the triple Pomeron vertex in QCD at small $x$, 
which is responsible for the non-linear term in the evolution equations.

Later on these ideas were 
further 
developed to include all
corrections enhanced by the local 
parton 
density, to constitute what is called the
Colour Glass Condensate (CGC)
\cite{Mueller:1989st,McLerran:1993ni,McLerran:1993ka,McLerran:1994vd,JalilianMarian:1997gr,JalilianMarian:1997dw,Kovner:2000pt,Weigert:2000gi,Iancu:2000hn,Ferreiro:2001qy,Balitsky:1995ub,Kovchegov:1999yj}
(see also the most recent developments 
in \cite{Altinoluk:2009je,Gelis:2010nm,Kovchegov:2006vj,Balitsky:2008zza}).
The CGC provides a non-perturbative, but weak-coupling, realisation of 
parton saturation ideas within QCD.
The
linear limit of the basic CGC equation is the  
BFKL equation, which is the linear
evolution equation derived in  the high-energy limit. 
As illustrated in 
Fig. \ref{Fig:satplane}, the evolution in the $\ln Q^2 - \ln1/x$ plane is 
driven by both linear equations: along $\ln Q^2$ 
for DGLAP and along $\ln1/x$ for BFKL.

\begin{figure}
\centerline{ \includegraphics[clip=,width=0.6\textwidth]{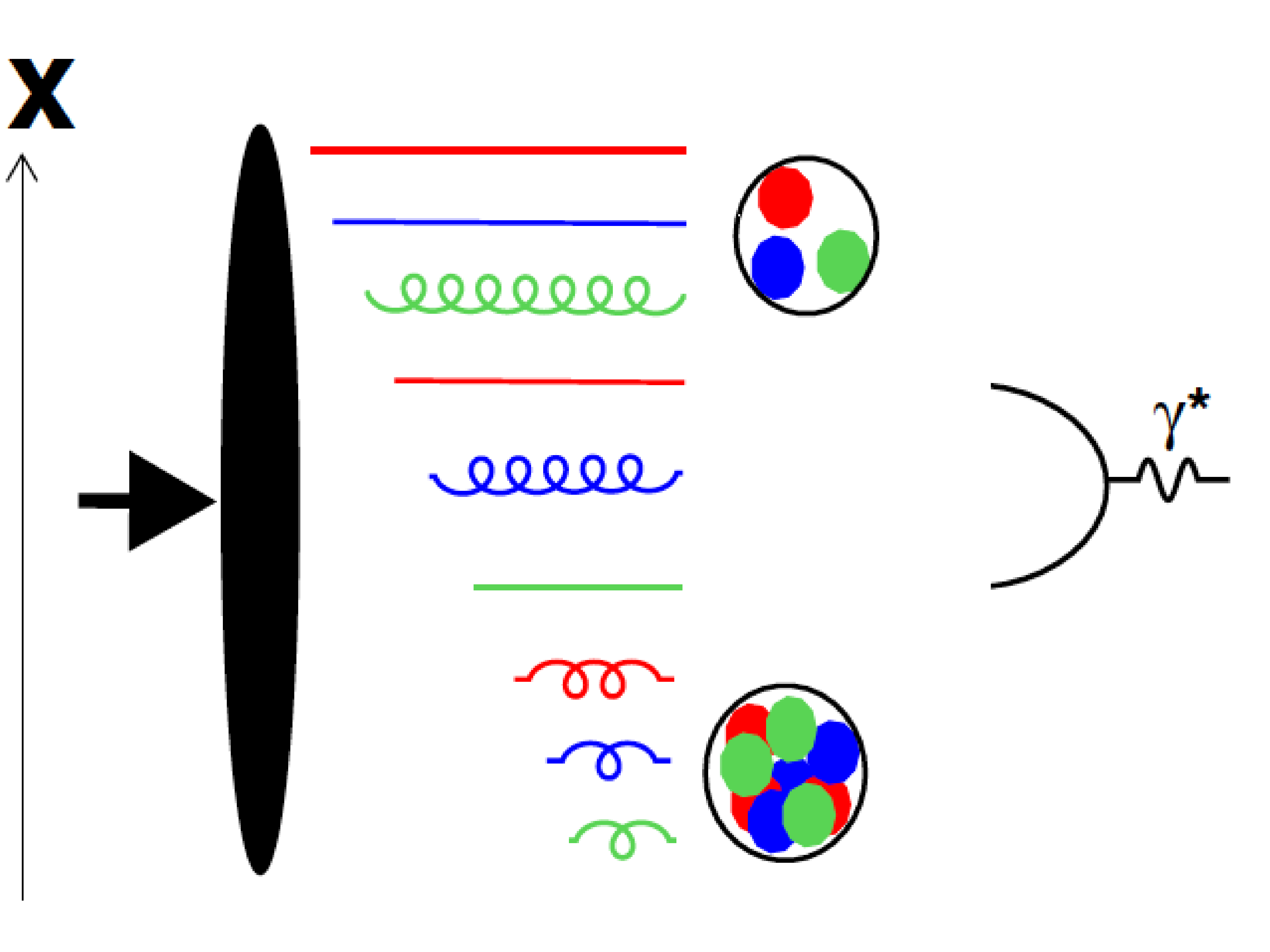}}
\caption{Illustration of saturation ideas. The hadron is moving very fast to the right, and its wave function contains many partonic components. Specifically, it includes partons with both large and small fractions of its longitudinal momentum $x$. The former are in a dilute regime and their lifetimes are very large, while the latter become densely packed due to multiple splitting and are short-lived (the length of the horizontal lines represents the extent of the lifetimes of the hadron fluctuations). Thus, the hard large $x$ 
partons act as a frozen source for the dynamics of the soft ones.
The photon with virtuality $Q^2$ 
is moving to the left and it constitutes a probe of the hadron wave 
function with a spatial resolution proportional to $1/Q$.}
\label{Fig:cgc}
\end{figure}

The basic framework in which saturation ideas are discussed is illustrated in
Fig. \ref{Fig:cgc}. 
One is considering the hadron wave
function at high energy. Its partonic components can be separated 
into those partons with a large momentum fraction $x$ and 
those with small $x$.  The large-$x$ components form dilute systems 
and provide
colour sources for the corresponding small-$x$ components. 
Due to multiple
splittings of the small-$x$ gluons, a dense system is eventually formed.  One
can then construct within this formalism an evolution equation for the gluon
correlators in the hadron wave function which is a renormalisation group
equation with respect to the rapidity separating large- and small-$x$ partons. This
renormalisation procedure assumes perturbative gluon emissions from the large-$x$
partons which imply a redefinition of the source at each step in rapidity.


The mean field version of the CGC evolution equations, the Balitsky-Kovchegov (BK) equation
\cite{Balitsky:1995ub,Kovchegov:1999yj}, provides a non-linear evolution
equation for the so-called unintegrated gluon densities. These distributions, unlike the standard integrated densities,  contain the information about the transverse momenta of the partons.
They naturally appear in the theoretical formulations of small-$x$ physics. A detailed description of these distributions as well as the prospects of their precise determination at the LHeC through a variety of  processes are discussed in Subsec.~\ref{sec:jetspartdyn}.

 It turns out that the BK approach
results in a gluon density which, for a fixed resolution of the probe, is
saturated for small longitudinal momentum fractions $x$, whereas
 at large values of $x$, the non-linear term is negligible.
The separation between these two limits
is given by a dynamically generated saturation
momentum $Q_s(x)$  which increases with decreasing 
$x$ (c.f. Fig.~\ref{Fig:satplane}), and therefore
saturation is determined by the condition $Q<Q_s(x)$.
Then, for large energies or small $x$, the system is in a dense regime of high gluon fields (thus non-perturbative) but the typical gluon momentum, $\sim Q_s$, is large (thus the coupling constant which determines gluon interactions is weak).
The qualitative behaviour
of the saturation scale with energy 
and nuclear size can be argued as follows.
The transition from a dilute to a dense regime 
occurs when the packing factor
(in this case, the product of the density of gluons per unit transverse area
times the gluon-gluon cross section) becomes of order unity i.e.
\begin{equation}
\frac{A\times xg(x, Q_s^2)}{\pi A^{2/3}} \times \frac{\alpha_s(Q_s^2)}{Q_s^2} \sim 1  \; \Longrightarrow \; Q_s^2 \sim A^{1/3}Q^2_0 \left(\frac{1}{x}\right)^\lambda,
\label{eq:qsat}
\end{equation}
where the growth of the gluon density at small $x$ 
in the dilute system has been approximated by a
power law, $xg(x, Q^2) \sim x^{-\lambda}$, logarithms are neglected and the
nucleus is considered a simple superposition of 
independent nucleons.  The exponent
$\lambda \simeq 0.3$ can be derived from QCD and is broadly consistent
with data from HERA. The scale $Q^2_0$ can only be
determined by experiment.

The BK equation was derived under several simplifying assumptions 
such as the 
scattering of a dilute projectile on a dense target, 
a large number of QCD colours and the absence of correlations in the target. 
At  present, the  discussion is concentrated on how to overcome these difficulties \cite{Iancu:2004es,Altinoluk:2009je,Kovchegov:2008mk}.
Possible phenomenological 
implications \cite{Dumitru:2010ak,Marquet:2010cf,Hatta:2006hs} are 
being considered. Also, the proposed relation between high-energy QCD and Statistical Mechanics \cite{Iancu:2004es,Munier:2009pc} is under investigation.

In the CGC formalism, the resummed terms are those enhanced by the
energy and by the local density of partons, and the saturation scale depends on
the matter (colour charge) 
density at the impact parameter probed by the virtual
photon. For a nucleus, 
the nuclear size plays the role of an enhancement factor, 
see Eq.~(\ref{eq:qsat}), in a manner which is
analogous to impact parameter scanning. 
Therefore, it is expected that when scanning the
impact parameter from the centre to the periphery of the 
hadron at high energy, one should go
from a non-linear to a linear regime. Analogously, non-linear effects will
become more important for large nuclei than for smaller ones or for nucleons.
Thus, a study of the variation of parton densities with impact parameter
and with the nuclear size, will 
provide an exacting test of our ideas on parton saturation.

%% file: physics/tex/diffrintro.tex
It was observed at HERA that a substantial fraction, about $10\%$,  
of deep
inelastic interactions are diffractive events of the type 
$ep \rightarrow eXp$. 
These are events in which the interacting proton 
stays intact, despite the inelasticity
of the interaction. 
Moreover, the proton appears  well separated from the rest of the
hadronic final state $X$
by a large rapidity gap. The events otherwise
look similar to normal deep inelastic events. 

Diffraction has been extensively analysed  at HERA, with a variety of
measurements as functions of $x$, $Q^2$ and the fractional proton
energy loss $\xpom$, as well as  more differential analyses
which include the dependence on the squared
four-momentum transfer $t$. Physically, for the
diffractive event to occur, there must be an exchange of a coherent,  colour
neutral  cluster of partons (a quasi-particle) which leaves the 
interacting proton intact.  This
colour neutral cluster is often called the {\it pomeron}, and it can be
characterised 
via a factorisation theorem \cite{Collins:1997sr} 
by a set of partonic densities analogous to
those for the proton or nucleus. At lowest order, the QCD realisation  
of the pomeron is a pair of gluons \cite{Low:1975sv,Nussinov:1975mw},
which leads to enhanced sensitivity to saturation phenomena compared to
the single gluon exchange in the bulk of non-diffractive processes.

There are strong theoretical indications that diffraction is closely linked
with the phenomenon of partonic saturation.  From a wide range of calculations,
mostly based on the so-called dipole model, see for example
\cite{GolecBiernat:1998js,GolecBiernat:1999qd},  it is known that 
diffractive DIS events involve softer 
effective scales than non-diffractive events at the same $Q^2$.
Thus, the exploration of diffractive phenomena offers a 
unique window to analyse both the relevance of non-linear effects and the
transition between perturbative and non-perturbative dynamics in QCD. 
 
The LHeC will provide a widely extended kinematic coverage for diffractive
events.
In addition to the enhanced sensitivity to saturation effects through the
basic 2-gluon exchange, 
their study at the LHeC will allow the extraction of
diffractive parton densities for a larger
range in $Q^2$ than at HERA, and will thus provide crucial tests of parton
dynamics and flavour decomposition
in diffraction as well as of the factorisation theorems. The high energy
involved also enables the production of diffractive states with large masses
which could include $W$ and $Z$ bosons as well as states with heavy 
flavours or
even exotic states with quantum numbers $1^-$.
 
\begin{figure}
\centerline{ \includegraphics[clip=,width=0.4\textwidth]{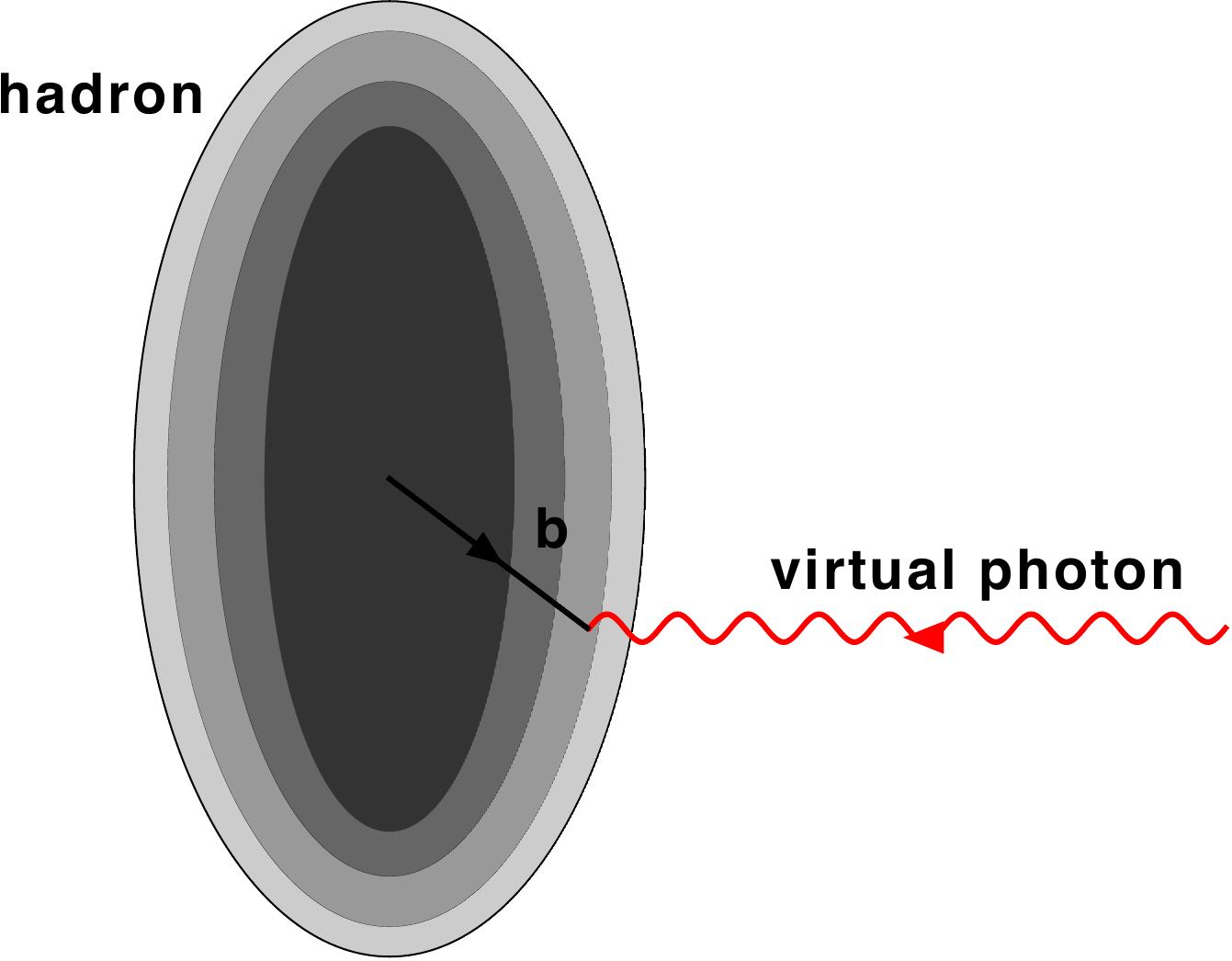}}
\caption{Illustration of the transverse profile of the hadron as explored by a virtual photon at impact parameter $b$.}
\label{Fig:transprof}
\end{figure}

Of particular importance is the exclusive diffractive production
of vector mesons, for which differential measurements 
as a function of squared four-momentum transfer, $t$, are most easily
performed.  It has been demonstrated that in this case, information about
the momentum transfer of the cross section can be translated into the
dependence of the scattering amplitude on impact parameter. As a result, a
profile in impact parameter of the 
interaction region, illustrated in 
Fig. \ref{Fig:transprof}, can be extracted.  The
precise determination of the dynamics governing the high parton density
regime
requires a detailed picture 
of the spatial distribution, in impact parameter space, of partons in
the interaction region.  By
selecting small impact parameter values (large $t$),
it is possible to probe the regions of 
highest parton density, where the onset of 
saturation phenomena should most readily occur. 
One can then extract the value of the saturation scale as a function of
energy and impact parameter. 

Even less differential measurements of the diffractive production of vector
mesons can provide valuable information about 
parton dynamics and non-linear effects. For
example, the measurement of the energy dependence of the diffractive cross
section for the photoproduction 
of $J/\psi$ mesons at the LHeC can distinguish between
different scenarios for parton evolution and thus explore 
parton saturation to a greater accuracy than ever before.

%% file: physics/tex/nucintro.tex
\begin{figure}
\centerline{ \includegraphics[clip=,width=0.8\textwidth,angle=0]{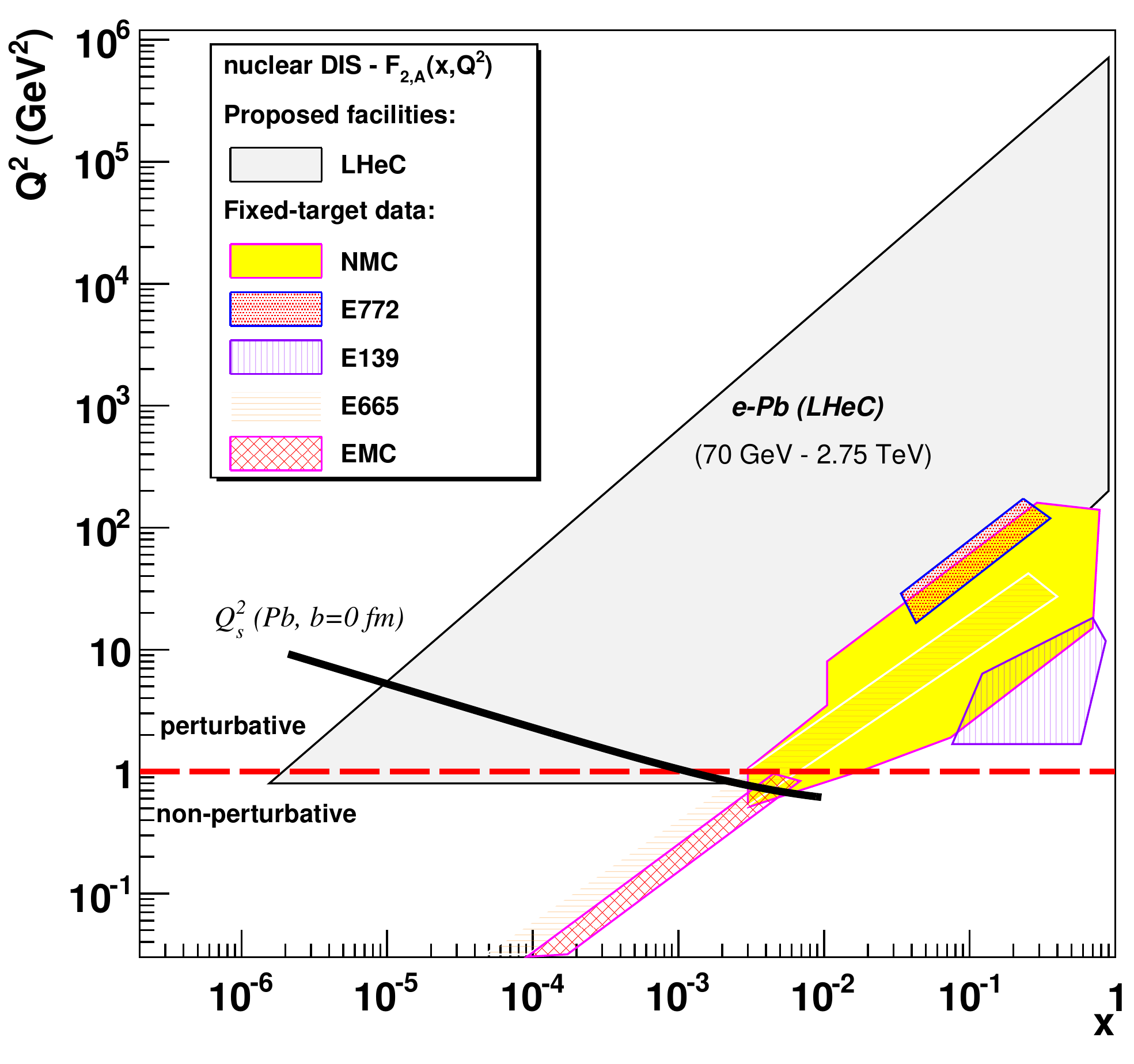}}
\caption{Kinematic coverage of the LHeC in the $\ln Q^2 - \ln 1/x$ plane for nuclear beams, compared with existing nuclear DIS and Drell-Yan experiments.}
\label{Fig:eAkinplane}
\end{figure}

Studying lepton-nucleus collisions is an important ingredient of the 
LHeC low $x$ programme for several reasons. Most obviously, as discussed in
Sections~\ref{sec:nucleartargets} and \ref{sec:epincl}, the nuclear structure
functions and parton densities are basically unknown at small $x$.
This is an issue which is becoming increasingly problematic in interpreting 
ultra-relativistic 
heavy ion collision data from RHIC and the LHC, 
as discussed in Subsec. \ref{sec:nucleartargets}.
The main
reason for this lack of knowledge comes from the rather small area in the $\ln
Q^2 - \ln 1/x$ plane covered by presently available experimental data, see Fig.
\ref{Fig:eAkinplane}.  Current theoretical  and phenomenological analyses
\cite{Armesto:2006ph} point to the importance of non-linear dynamics in DIS
off nuclei  at small and moderate $Q^2$ and small $x$, which needs to be tested
experimentally.  In this respect, a relation exists, as reviewed in
Sec.~\ref{sec:diffpdfs}, between diffraction in lepton-proton collisions and
the small-$x$ behaviour of nuclear structure functions. 
This relation relies on only 
basic properties of 
Quantum Field Theory and its verification provides stringent tests
of our understanding of the strong interaction.

\begin{figure}
\centerline{ \includegraphics[clip=,width=0.6\textwidth,angle=0]{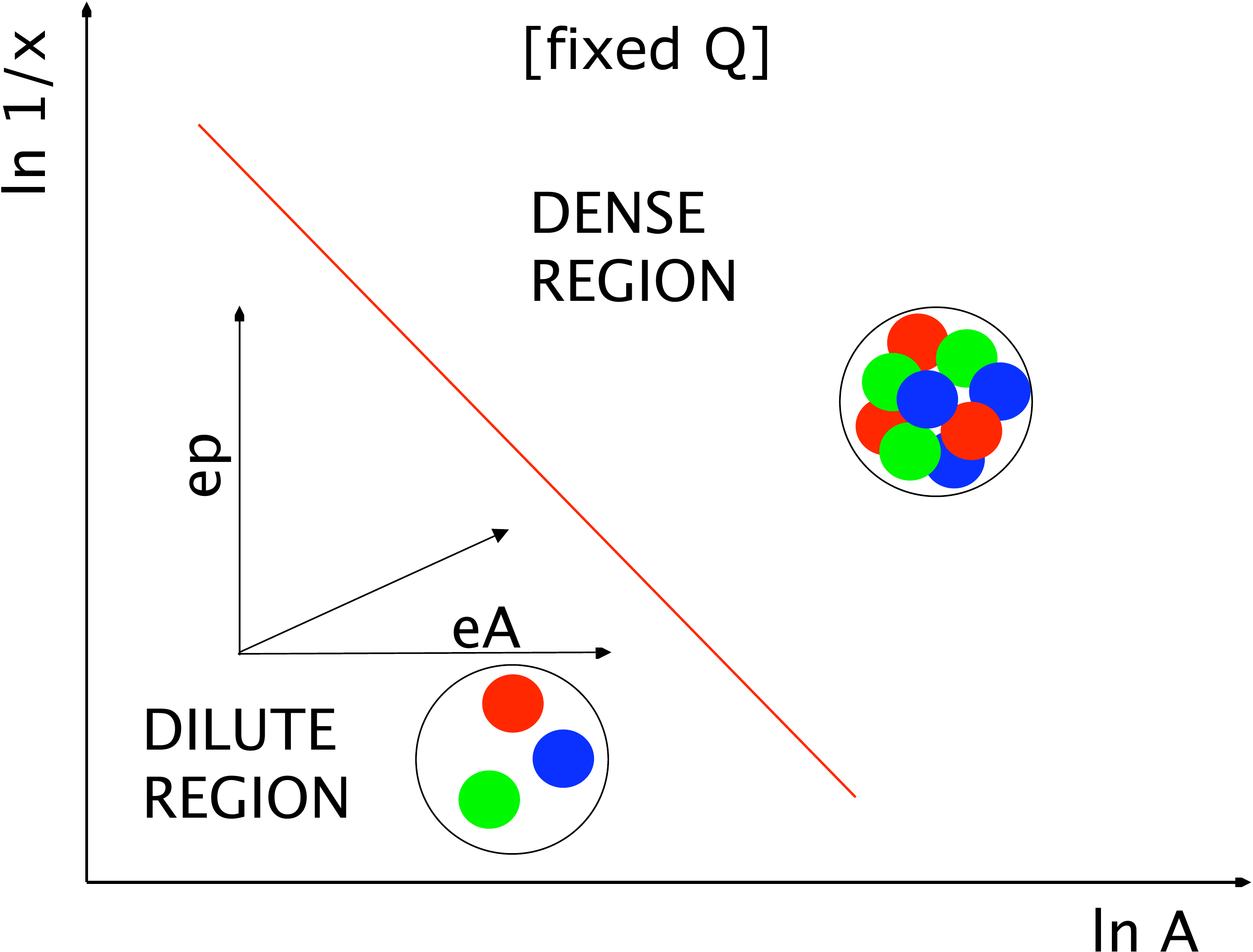}}
\caption{Schematic view of the different regions for the parton densities in the $\ln 1/x - \ln A$ plane, for fixed $Q^2$. 
Lines of constant occupancy of the hadron are parallel to the diagonal 
line shown. See the text for further comments.}
\label{Fig:satplaneeA}
\end{figure}

Non-linear effects in parton evolution are enhanced by  increasing
the density of partons. Such an 
increase can be achieved (see Fig. \ref{Fig:satplaneeA})  
either by increasing the energy of the collision
(decreasing $x$), or by increasing the 
nuclear mass number $A$.  The latter can be
accomplished by either using the largest nuclei possible,
or by selecting subsets of
collisions 
with small impact parameters $b$ (i.e. more central collisions)
between the relatively light nuclei and the virtual photon, such that
more nucleons are involved. The ideal situation would be to 
map out the dependence  
of the saturation scale on $x$, $b$ and
$A$ as fully as possible (see Eq. (\ref{eq:qsat})). This is
a key observable
in formulations which resum multiple interactions and
result in parton saturation.  As such it must be checked in experiment in order
to clearly settle the mechanism underlying non-linear parton dynamics.


Beyond inclusive variables, measurements of diffractive 
observables in lepton scattering from nuclei have never been 
obtained previously and the uncertainties in current theoretical predictions
are very large. Inclusive and exclusive diffraction 
measurements in 
lepton-nucleus collisions at the LHeC will offer 
a completely new testing ground for our ideas on nuclear 
structure at small $x$ and on parton saturation and non-linear 
dynamics in QCD.

%% file: physics/tex/statusheraintro.tex

As discussed in the previous section,
in the low-$x$ region a high parton density can be achieved in DIS and 
various novel phenomena are predicted.
Ultimately, unitarity constraints become important and a 
`black disk' limit is
approached \cite{Gribov:1968jf}, in which the cross section reaches the
geometrical bound given by the transverse proton or nucleus size. When $\alpha_s$ is small
enough for quarks and gluons to be the right degrees of freedom,
parton saturation
effects are therefore expected to occur within the theoretically controllable weak coupling regime.  In this small-$x$ limit, many striking
observable effects are predicted, such as $Q^2$ dependences of the cross sections which differ
fundamentally from the usual logarithmic variations, and diffractive cross
sections approaching $50\%$ of the total \cite{Frankfurt:2005mc}. This fairly
good phenomenological understanding of the onset of unitarity effects is,
unfortunately, not very quantitative. In particular, the precise location of
the saturation scale line in the DIS kinematic plane
(see Fig.~\ref{Fig:satplane}) is to be determined experimentally. The search for
parton saturation effects has  
therefore been a major issue throughout the lifetime of
the HERA project.

Although no conclusive saturation signals have been observed in parton density
fits to existing HERA data, various hints have been obtained, for example, by studying
the change in fit quality as low-$x$ and
$Q^2$ data are progressively omitted,  in the NNPDF \cite{Caola:2009iy,Albacete:2012rx} and HERAPDF \cite{:2009wt} analyses (see below).

A more common approach is to fit the data to dipole models
\cite{GolecBiernat:1998js,GolecBiernat:1999qd,Iancu:2003ge,Forshaw:2004vv},
which are applicable at very low $Q^2$ values beyond the range in which quarks
and gluons can be considered to be good degrees of freedom.  The typical
conclusion \cite{Forshaw:2004vv} is that HERA data in the perturbative regime
exhibit at best weak 
evidence for saturation.  However, when data in the $Q^2 < 1
\ {\rm GeV^2}$ region are included, models which include saturation effects are quite
successful in the description of the wide variety of experimental data.

\begin{figure}[h] \unitlength 1mm
  \begin{center}
      \includegraphics[width=0.49\textwidth]{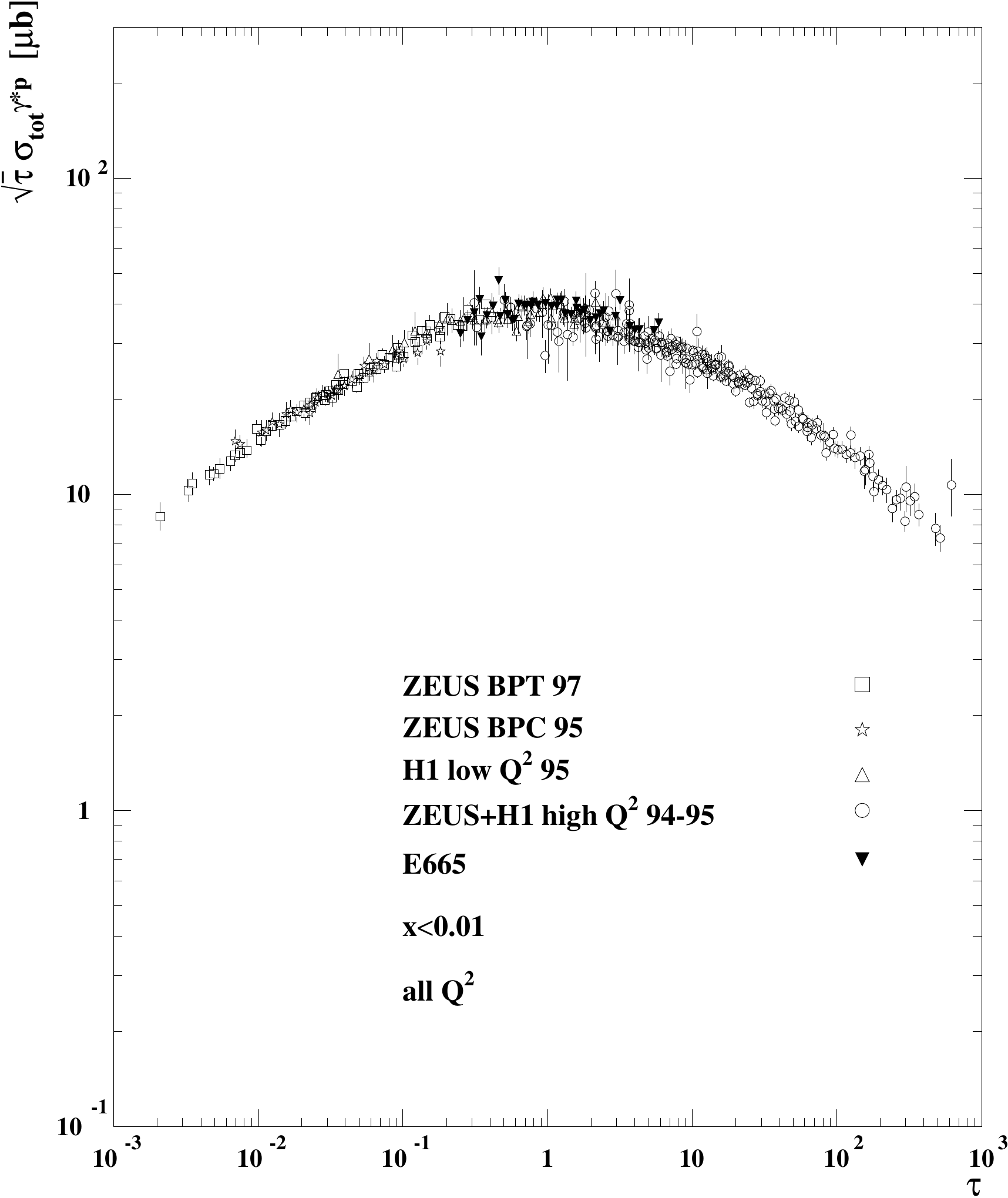}
      \includegraphics[width=0.49\textwidth]{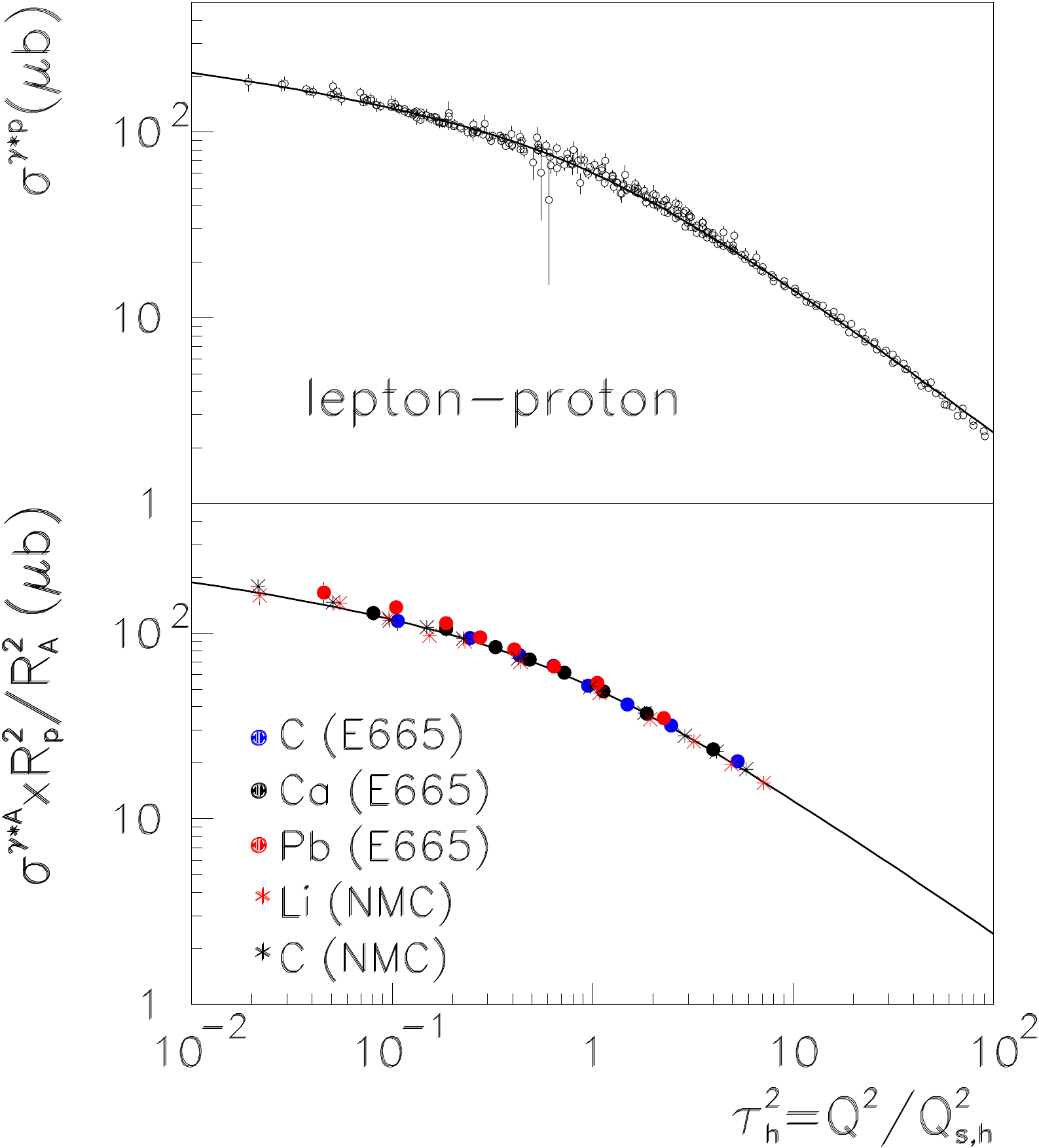}
  \end{center}
  \caption[]{(left) Geometric scaling plot \cite{Stasto:2000er}, in which
low $x$  data on the $\gamma^* p$ cross section
from HERA and E665 are plotted as a function of the 
dimensionless variable $\tau$ (see text). The cross sections are scaled by
$\surd \tau$ for visibility. (right) Geometric scaling plot showing cross sections
for electron scattering off nuclei as well as off protons \cite{Armesto:2004ud}.}
\label{fig:geoscale}
\end{figure}

The `geometric scaling' \cite{Stasto:2000er} feature of the HERA data
(Fig.~\ref{fig:geoscale}left) reveals that, to a good approximation, the low-$x$
cross section is a function of a single combined variable $\tau = Q^2 / Q^2_s(x)$, where
$Q^2_s = Q_0^2 \ x^{- \lambda}$ is the saturation scale, see Eq.
(\ref{eq:qsat}).  This parameterisation works well for scattering off both
protons and ions, as shown in Fig.~\ref{fig:geoscale}right
\cite{Stasto:2000er,Armesto:2004ud}.
Geometric scaling is observed not only for the total $\gamma^* p$ cross section, but also for other, more exclusive observables in  $\gamma^* p$ collisions \cite{Marquet:2006jb,Goncalves:2003ke} 
and even in hadron production in proton-proton collisions at the 
LHC \cite{McLerran:2010ex} 
and nucleus-nucleus collisions at RHIC \cite{Armesto:2004ud}.
This feature supports the view
(Subsec.~\ref{sec:lowxoverview}) of the cross section as being invariant along
lines of constant `gluon occupancy'.  When viewed in detail
(Fig.~\ref{fig:geoscale}), there is a change in behaviour in the geometric
scaling plot near $\tau = 1$, which has been interpreted as a transition to the
saturation region shown in Fig.~\ref{Fig:satplane}. However, data with $\tau
< 1$ exist only at very low, non-perturbative, $Q^2$ values to date, precluding
a partonic interpretation. Also, the fact that the scaling extends to large
values of $\tau$ which characterise the dilute regime, 
has prompted theoretical explanations
of this phenomenon which do not invoke the physics of saturation
\cite{Caola:2008xr}.

%% file: physics/tex/dipolemodels.tex

  As mentioned previously, one of the interesting observations at HERA is the
success of the description of many aspects of the experimental data within the
framework of the so-called dipole picture
\cite{Nikolaev:1990ja,Nikolaev:1991et,Mueller:1989st}  with models that include
unitarisation or saturation effects \cite{Mueller:1994jq,Mueller:1994gb}.
These models 
are based on the assumption that the relevant degrees of freedom at high energy
are colour dipoles. Dipole models in  DIS are closely related to
the Good-Walker picture \cite{Good:1960ba} previously developed
for soft processes in hadron-hadron collisions. In DIS, 
dipoles are shown to be the eigenstates of high-energy scattering in QCD, and
the photon wave function can be expanded onto the dipole basis.

\begin{figure}[ht]
\begin{center}
\includegraphics[width=0.50\textwidth]{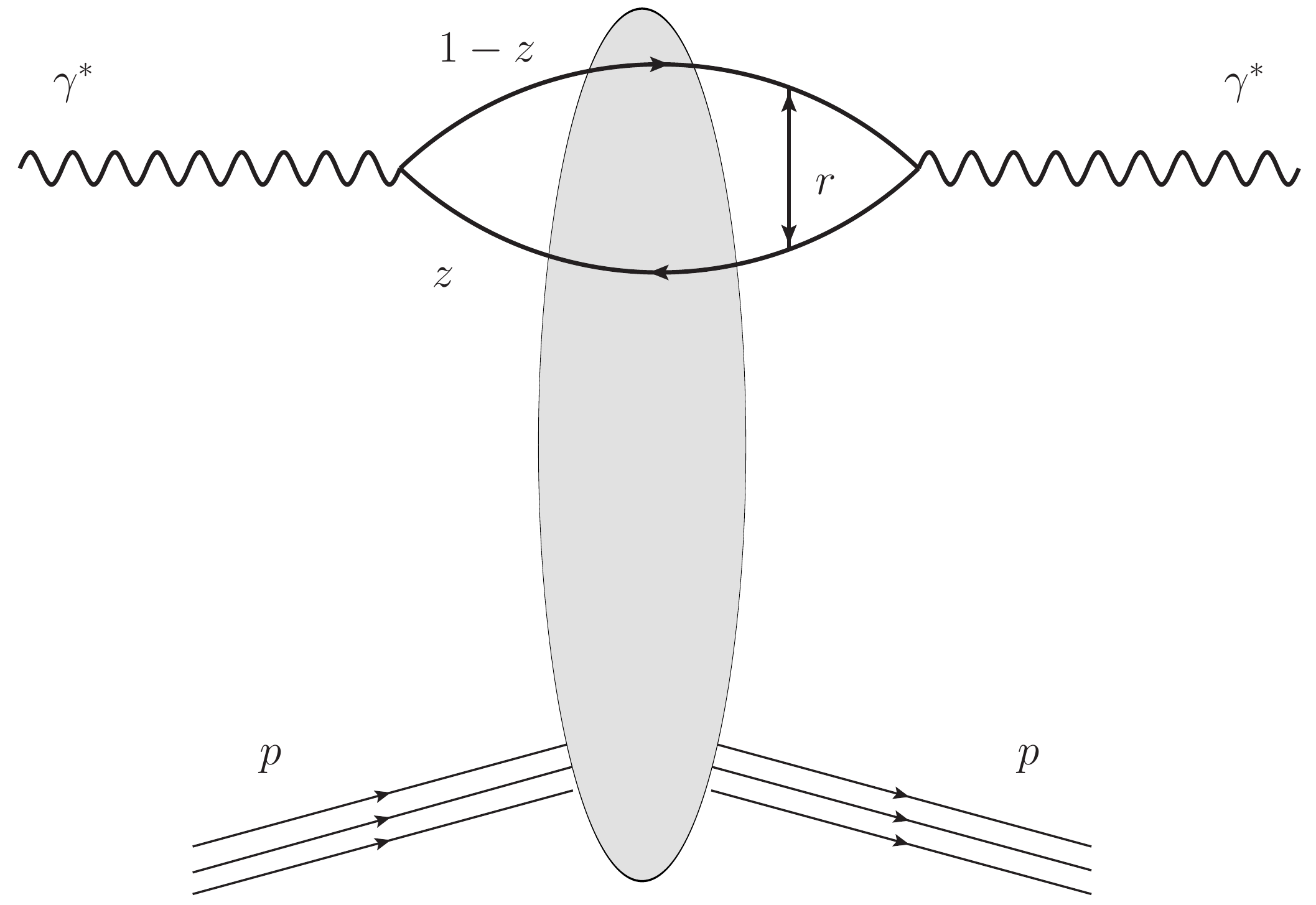}
\end{center}
\caption{ Schematic representation of dipole factorisation at small $x$ in DIS. The virtual photon fluctuates into a quark-antiquark pair and subsequently interacts with the target. All the details of the dynamics of the interaction are encoded in the dipole scattering amplitude.\label{fig:dipole}}
\end{figure}

The dipole factorisation for the inclusive cross section in DIS is
illustrated in Fig.~\ref{fig:dipole}. It differs from the usual picture of the
virtual photon probing the parton density of the target in that here the
partonic structure of the probed hadron is not evident. Instead, one
chooses a particular Lorentz frame where the photon  fluctuates
into a quark-antiquark pair with a transverse separation $r$ and at impact
parameter $b$ with respect to the target.  For sufficiently small
$x \ll (2m_NR_h)^{-1}$, with $m_N$ the nucleon mass and
$R_h$ the hadron or nuclear radius,
 the lifetime
of the $q\bar{q}$ fluctuation is much longer than the typical time for
interaction with the target.  The interaction of the $q\bar{q}$ dipole
with the hadron or nucleus is
then described by a scattering matrix $S(r,b;x)$ such that
$|S(r,b;x)|<1$.  The unitarity constraints can be incorporated  naturally in
this picture \cite{Mueller:2001fv} by the requirement that $|S(r,b;x)| \ge 0$,
with $S(r,b;x)=0$ corresponding to the black disk limit. Integrating
$1-S(r,b;x)$ over the impact parameter $b$ one obtains the dipole cross section
$\sigma^{q\bar{q}}(r,x)$, which depends on the dipole size and the energy
(through the dependence on $x=x_{\rm Bj}$).
The transverse size of the partons probed in this process is roughly proportional to the inverse of the virtuality of the photon $Q^2$. This statement 
is most accurate in the case of a longitudinally polarised photon, while in the case of a transversely polarised one, the distribution of the probed transverse sizes of dipoles is broadened due to the so-called aligned jet configurations.

At small values of the dipole size, such that $r \ll 1/Q$, the dipole cross
section can be shown to be related to the integrated gluon distribution
function
\begin{equation}
\sigma^{q\bar{q}}(r,x) \sim r^2 \, \alpha_s(C/r^2)\, xg(x,C/r^2) \; ,
\label{eq:dipolegluon}
\end{equation}
where $C$ is a constant.  In this regime, where $r$ is small, the dipole cross
section is small and consequently the amplitude is far from the unitarity
limits. With increasing energy the dipole cross section grows  and 
saturation corrections must be taken into account in order to guarantee the
unitarity bound on $S(r,b;x)$. The transition region between the two limits is
characterised by the saturation scale $Q_s(x)$. Several models
\cite{GolecBiernat:1998js,Bartels:2002cj,Iancu:2003ge} 
have been proposed
which successfully describe the  HERA data on the structure function $F_2$.

Once the dipole cross section has been constrained by the data on the inclusive
structure functions, it can be used to predict, with almost no 
additional parameters, the cross sections for diffractive production at
small $x$.  Inclusive diffraction has been computed within the dipole
picture in \cite{GolecBiernat:1999qd}, and exclusive diffraction of 
vector mesons in \cite{Kowalski:2003hm,Kowalski:2006hc}. One of the interesting
aspects of these models is that  they naturally lead to a constant ratio
of the diffractive to total cross sections as a function of energy
\cite{GolecBiernat:1999qd}. In models with saturation this is related to the
fact that the saturation scale provides a natural $x$-dependent cut-off and
gives the same leading-twist behaviour for inclusive and diffractive cross
sections.  As a result the ratio of inclusive to diffractive cross sections is
almost constant as a function of the energy.

In spite of the fact that this approach has been able to successfully describe inclusive data and  predict
diffraction at small values of $x$, there is still important conceptual
progress to be made. Certainly there are important hints from dipole models about the nature of the perturbative--non-perturbative transition in QCD. Nevertheless, dipole models should be rather regarded as effective phenomenological approaches. 
As such they only parameterise the essential dynamics at small $x$. 
For instance, the transverse impact parameter dependence
of the dipole scattering amplitude $S(r,b;x)$ is very poorly constrained.
Indeed, it is possible simultaneously to 
describe $F_2$ and $F_2^D$ with
a rather wide range of impact parameter dependences.
On the theoretical side, it has not been possible so far to 
fully predict the realistic  profile of the interaction region
in transverse size.
  It is therefore of vital importance to measure accurately the
$t$-dependencies of the
diffractive cross sections in an extended kinematic range to pin
down the impact parameter distribution of the proton  at high
energies.

%% file: physics/tex/sathera.tex

As discussed in previous sections, the experimental data on the inclusive
structure functions $F_2$ and $F_L$ measured at HERA have been successfully described  - with $\chi^2/d.o.f. \sim 1$ - by fits which use
linear fixed-order DGLAP evolution, see e.g. \cite{Lai:1999wy,Martin:2002aw,
Pumplin:2002vw,
Martin:2007bv,Nadolsky:2008zw,Watt:2008hi,Martin:2009iq,Lai:2010nw,Ball:2010de,:2009wt,Collaboration:2010ry}.  The
current status of the calculations is fixed order at next-to-next-to-leading accuracy.
On the other hand, see Subsec. \ref{sec:lowxoverview},
there are several theoretical reasons to expect that 
at small $x$ and/or at small
$Q^2$ the fixed-order DGLAP framework needs to be extended.
Possible relevant phenomena predicted by
perturbative QCD are linear small-$x$ resummation,
non-linear evolution and parton saturation or other higher-twist effects.  
Although the exact kinematic regime in which 
these effects should become important remains unclear,
it is evident that at some point they will lead to deviations from
fixed-order DGLAP evolution. Therefore, an important question 
is whether these
deviations are already present in HERA data. Several analyses have been
performed which aimed to address this question. 

In one analysis \cite{Forshaw:2004vv}, HERA $F_2(x, Q^2)$ 
data are subjected to three fits in the framework of a dipole model. 
In one of the fits, the parameterisation of the dipole cross
section does not contain saturation properties, 
whereas in the other two, saturation effects are included using
two rather different models \cite{Forshaw:2004vv,Iancu:2003ge}. All three
dipole fits are able to describe the HERA data adequately in the perturbative
region $Q^2 \geq 2 \ {\rm GeV^2}$. However, a clear preference for the models
containing saturation effects becomes evident when data in the range $0.045 <
Q^2 < 1 \ {\rm GeV^2}$ are added \cite{Forshaw:2004vv}. Similar conclusions
are drawn when the same dipole cross section
models are applied to various less
inclusive observables at HERA \cite{Forshaw:2006np}.
These observations provide an intriguing hint that saturation
effects may already be present in HERA data. However,  
due to the
non-perturbative nature of the low $Q^2$ kinematic region
in which the effects appear, there is no clear
interpretation in terms of perturbative QCD degrees of freedom 
and firm conclusions cannot be drawn 
on the existence and nature of parton recombination effects.

In another analysis \cite{Caola:2009iy}, possible indications of 
deviations from linear DGLAP evolution were discussed. It was based on an
unbiased PDF analysis of the inclusive HERA data. Here we present briefly an
updated version of this study which uses the most precise 
inclusive DIS data to date, the combined HERA--I
dataset~\cite{:2009wt} in the framework of the global NNPDF2.0
fitting framework.
%
The key idea 
is to perform global fits only in the large-$x$, large-$Q^2$
region, where NLO DGLAP is expected to be reliable. This way one can determine {\it
safe} parton distributions which are not contaminated by possible non-DGLAP
effects. These PDFs are then evolved backwards into the potentially {\it
unsafe} low-$x$ and low-$Q^2$ kinematic region, and 
are used to compute physical
observables, which are compared with data. A deviation between the predicted
and observed behaviour in this region can then provide a signal for effects
beyond NLO DGLAP.

\begin{figure}
\centering
\includegraphics[width=0.44\textwidth]{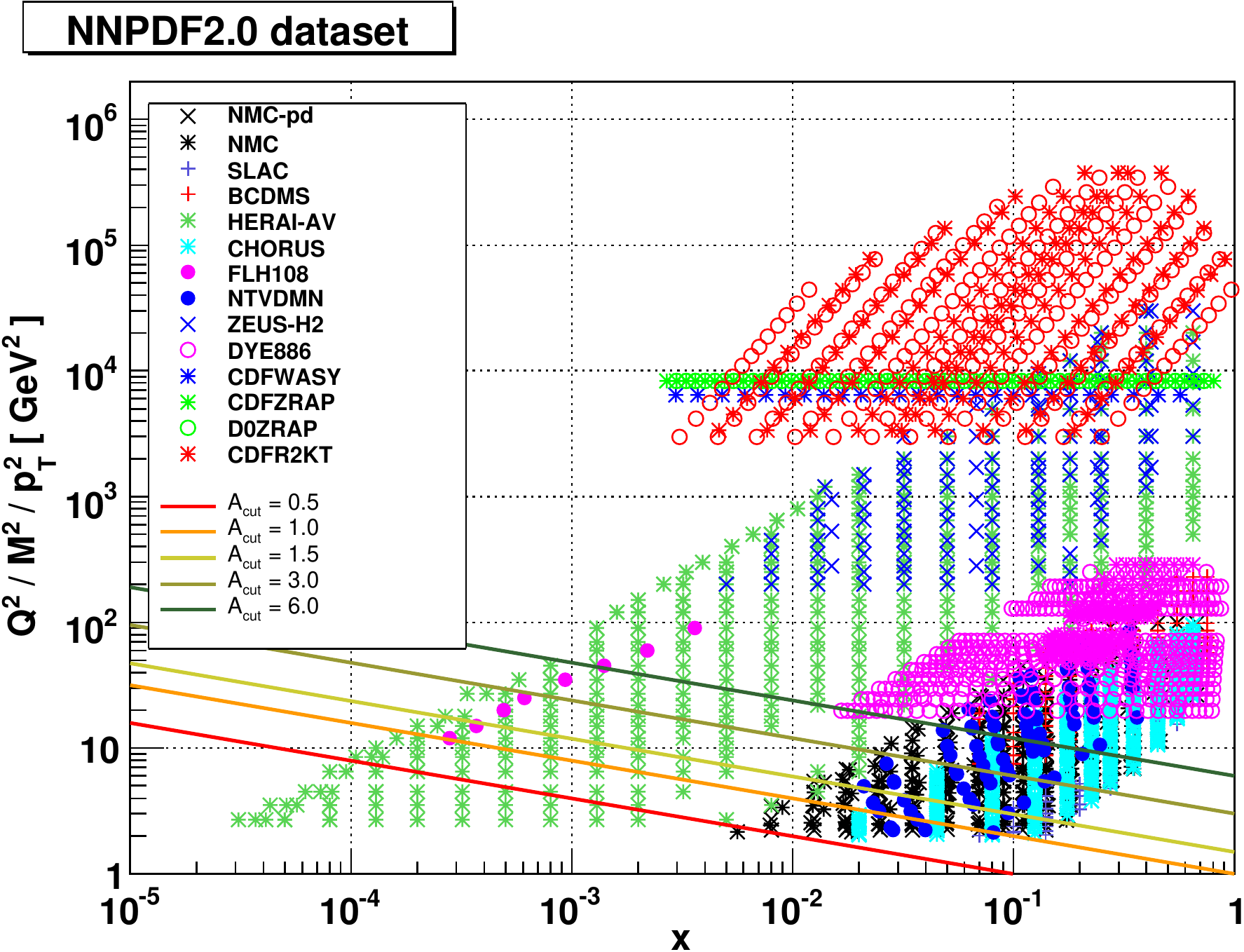}
\includegraphics[width=0.55\textwidth]{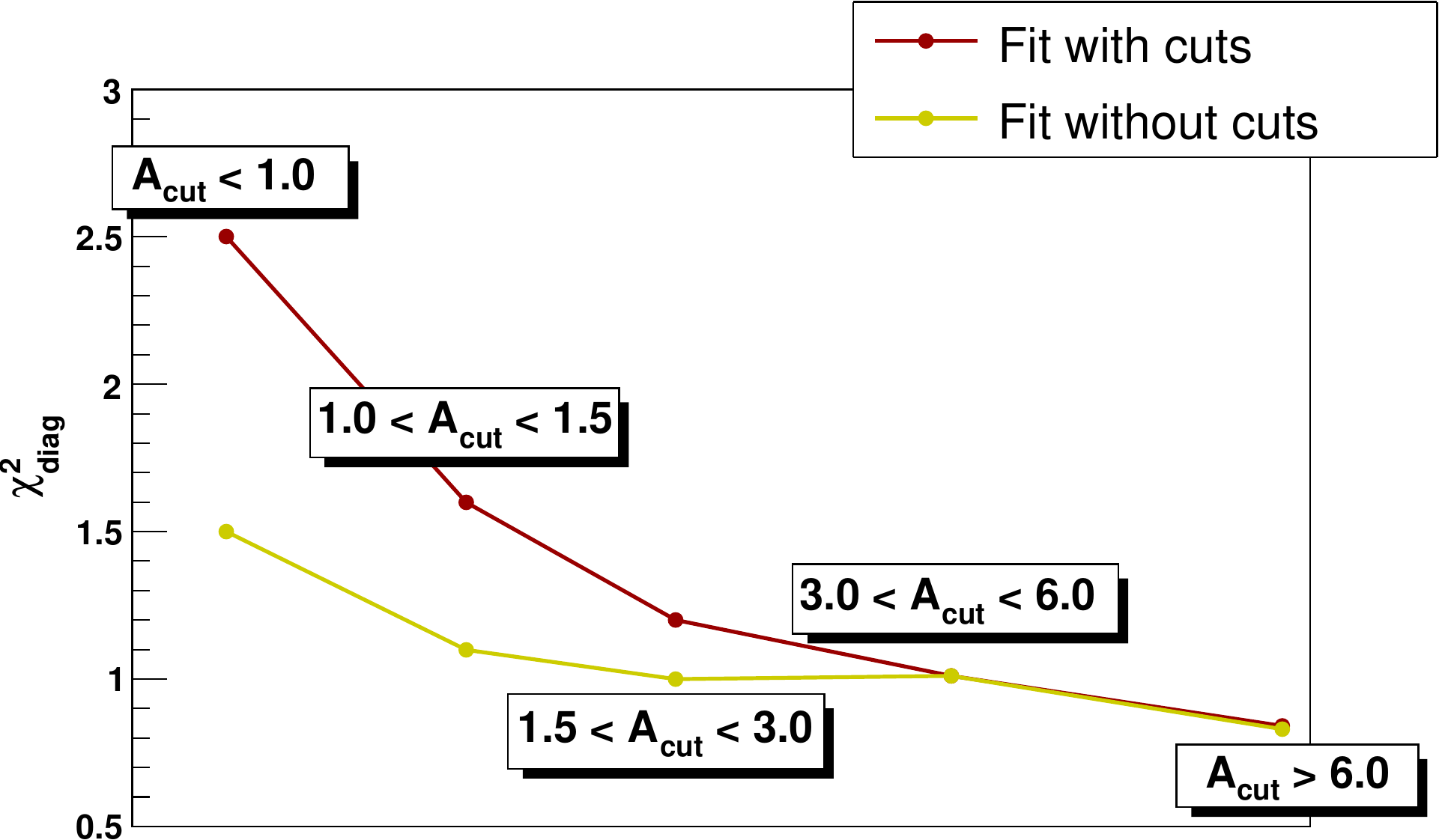}
\caption{\small Left plot: the kinematic
coverage of the data used in the NNPDF2.0 analysis, indicating
the different choices of $A_{\rm cut}$ used to probe
deviations from DGLAP. Right plot: the diagonal 
$\chi^2_{\rm diag}$ evaluated in 
kinematic slices corresponding to the different $A_{\rm cut}$ cuts, where 
 $\chi^2_{\rm diag}$ has been computed using both
the reference NNPDF2.0 fit
without kinematic cuts (yellow line) and the NNPDF2.0 with the 
maximum $A_{\rm cut}=1.5$ cut (red line).}\label{plot_chi2}
\end{figure}

\begin{figure}
\begin{center}
\includegraphics[width=0.49\textwidth]{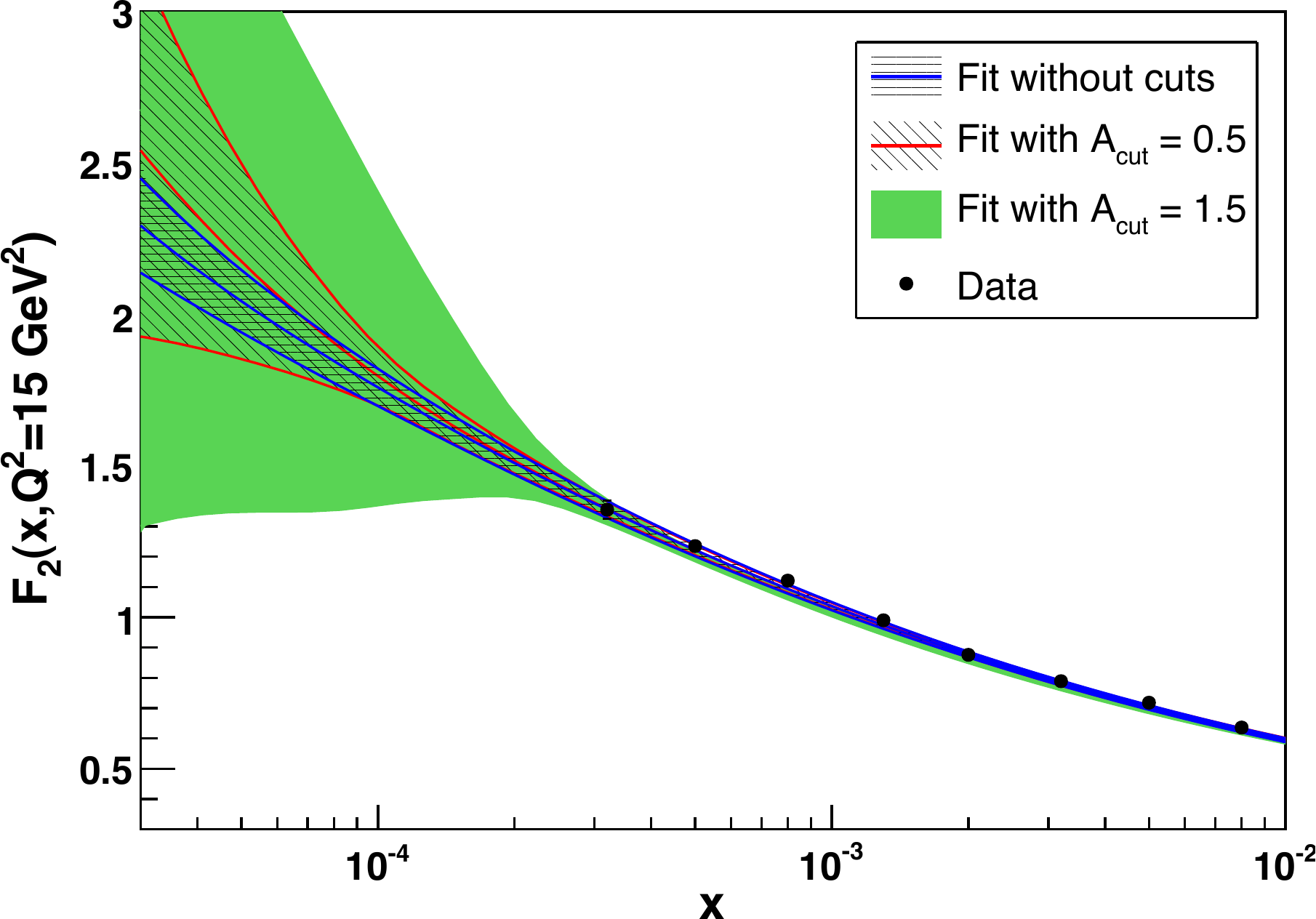}
\includegraphics[width=0.49\textwidth]{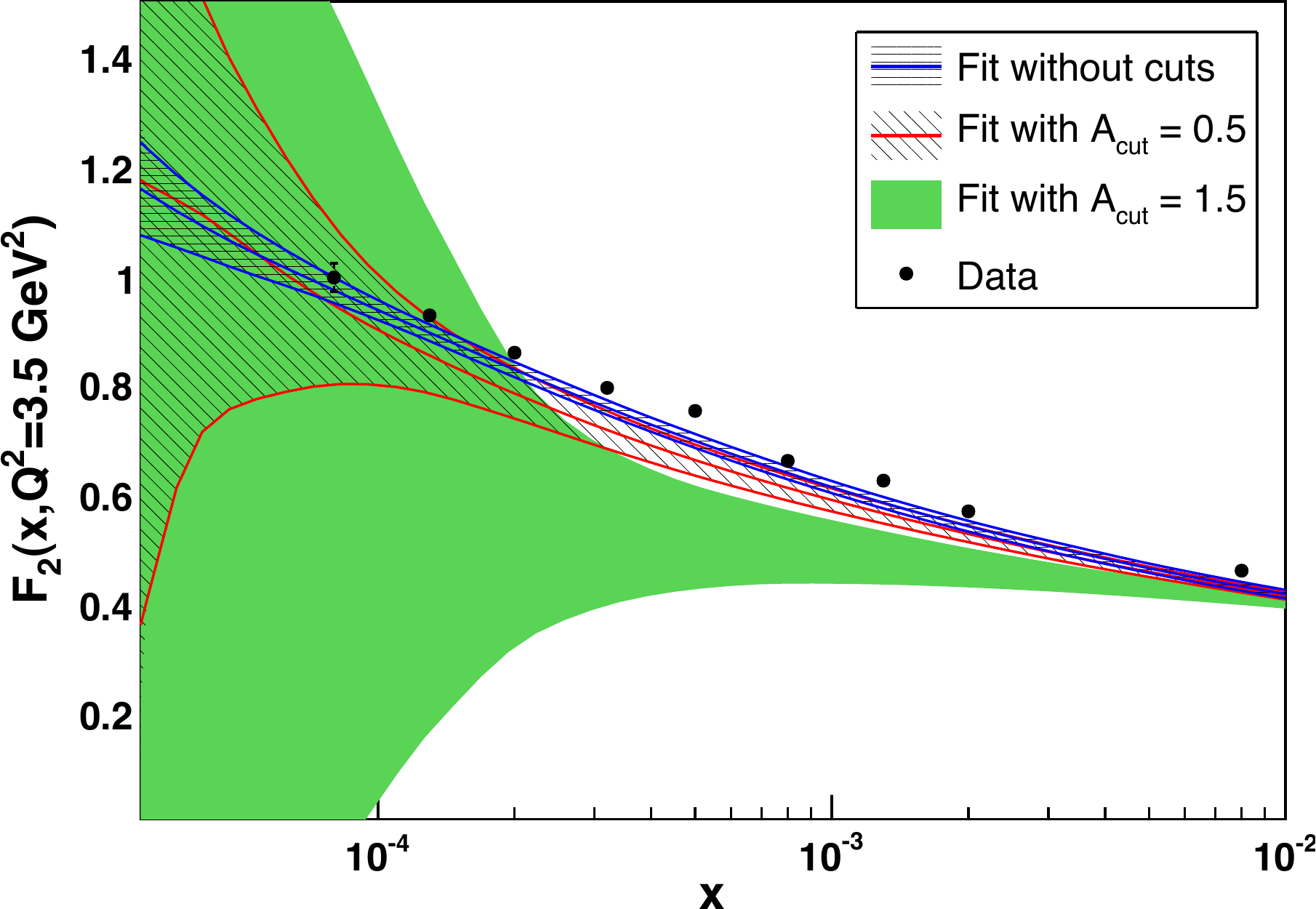}
\end{center}
\caption{\small 
Left: the proton structure function $F_2(x,Q^2=15~$GeV$^2)$ 
at small-$x$, computed from PDFs obtained from the
NNPDF2.0 fits with different 
values of $A_{\rm cut}$. Right: the same
but at a lower $Q^2=3.5$~GeV$^2$ scale.}\label{f2_35_pic}
\end{figure}

The PDFs
were determined within the {\it safe} kinematic region in which
$Q^2 \ge A_{\rm cut} \cdot x^{-\lambda}$,
where $\lambda=0.3$ and $A_{\rm cut}$ is a variable parameter
(see the left plot in
Fig.~\ref{plot_chi2} and \cite{Caola:2009iy} for details on the procedure).
The NNPDF2.0 analysis \cite{Ball:2010de} was 
repeated for different choices of the kinematic
cuts, one for each choice of $A_{\rm cut}$, and 
the results were compared with 
experimental data.  
As shown in Fig.~\ref{f2_35_pic}, at high
$Q^2=15$~GeV$^2$ one does not see any significant deviation from NLO
DGLAP. In this region all PDF sets agree with data and 
with one another, the only
difference between them being 
that as $A_{\rm cut}$ increases the PDF
uncertainty bands grow as expected due to the
experimental information removed by the cuts. The 
situation is  different at a lower
$Q^2=3.5$~GeV$^2$: the prediction obtained from the backwards evolution of the
data above the cut exhibits a systematic downward trend,
becoming more evident with increasing $A_{\rm cut}$. 
These results are indicative of deficiencies in the 
description of HERA data at 
low-$x$ and low-$Q^2$ by NLO DGLAP evolution\footnote{This problem cannot be solved by NNLO corrections which work in the opposite direction, see in this respect \cite{Martin:2009iq}. Also, in the HERAPDF framework \cite{:2009wt,Collaboration:2010ry} the fit quality tends to worsen when low-$Q^2$ data are included. See \cite{Albacete:2012rx} for a recent discussion and comparison with models containing non-linear dynamics.}. Specifically,  
the NLO DGLAP approach suggests a faster evolution with $Q^2$ than
is present in the data. 
To be sure that one is observing a genuine small-$x$ effect, one needs
to check that it becomes less and less relevant as $x$ and $Q^2$ increase. To
this aim  the diagonal 
$\chi^2_{\rm diag}$ 
was computed, see the right plot in Fig.~\ref{plot_chi2}, in different kinematic slices,
both from the fit without cuts and from that with the maximum cut 
$A_{\rm cut}=1.5$.  The expectation 
is that at larger $x$ and $Q^2$ the difference
between the two fits becomes smaller, as deviations from NLO DGLAP
should become negligible. The data support this expectation:
the contribution to the $\chi^2$ from the region 
with $A_{\rm cut} \ge 3$ is comparable for the fits with and without
cuts, in contrast to the lower $x$ and $Q^2$
region, where the $\chi^2$ is substantially larger in the version of
the fit with cuts applied. 
Nevertheless, it should be noted that there is
no general consensus on the origins of these effects. e.g. 
in \cite{Lai:2010vv} it is suggested that their origin 
lies in bias due to the chosen initial conditions for DGLAP evolution 


In summary, there are hints that the low-$Q^2$--low-$x$ region
covered by HERA may exhibit deviations from fixed-order linear evolution.
These hints are obtained from the success of dipole models with saturation features to describe the experimental data in this region, and 
from the fact that the 
quality of fixed-order DGLAP fits seems to deteriorate there. 
However, the region in which such effects may be present 
corresponds to rather small $Q^2$, preventing a clear
interpretation in terms of perturbative QCD degrees of freedom. 
In addition, the overall quality of the fixed-order DGLAP 
fits to HERA data remains high. It is therefore premature to draw 
any firm conclusion on the failure of fixed-order linear evolution 
as the appropriate tool to describe all HERA data. 
In any case, it is clear that the methods discussed in this
subsection should be used to 
analyse LHeC inclusive structure
function data, and would allow a detailed characterisation of any new
high-energy QCD dynamics unveiled by the LHeC. If the hints in the HERA
data are correct, the novel phenomena should appear at the LHeC 
in a higher $Q^2$ perturbative region where they can be established cleanly
and
understood in terms of parton dynamics.  

%% file: physics/tex/resum.tex
The deviations from DGLAP evolution could be caused by higher order effects at
small $x$ and small $Q$ which need to be resummed to all orders of perturbation
theory.  As mentioned previously, the problem of resummation at small $x$ has
been extensively studied in recent years, see for example
~\cite{Altarelli:2003hk,Altarelli:2005ni,Altarelli:2008aj,Ciafaloni:2003rd,Ciafaloni:2003kd,Ciafaloni:2007gf}.
It has been demonstrated that the small-$x$ resummation framework accounts for
running coupling effects, kinematic constraints, gluon exchange symmetry
and other physical constraints.  The results were shown to be very robust with
respect to scale changes and different resummation schemes.  As a result,
the effect of the resummation of terms which are enhanced at small $x$ is
perceptible but moderate - comparable in size to typical NNLO fixed order
corrections in the HERA region. 

\begin{figure}[ht]
\begin{center}
\includegraphics[width=0.49\textwidth]{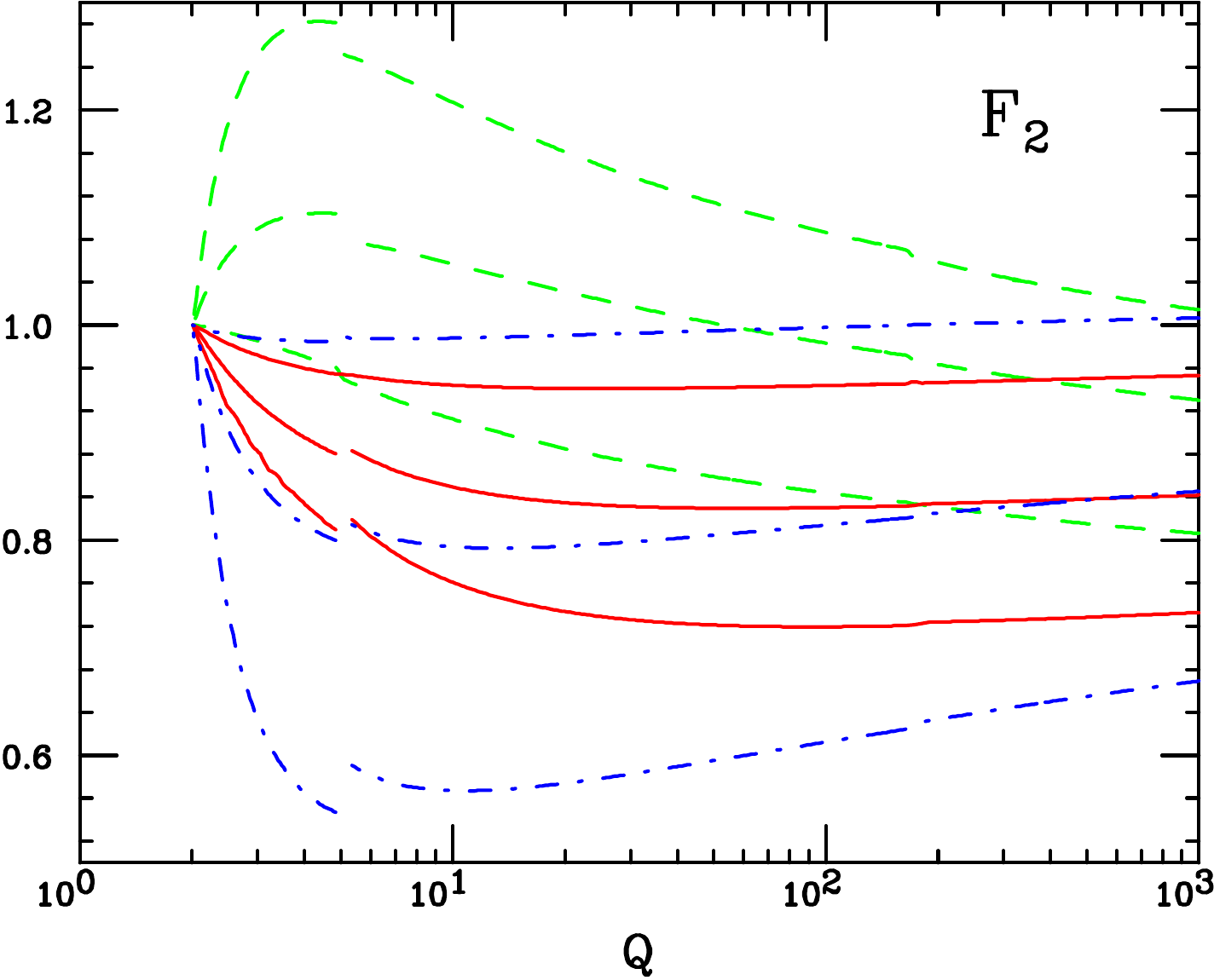}
\includegraphics[width=0.49\textwidth]{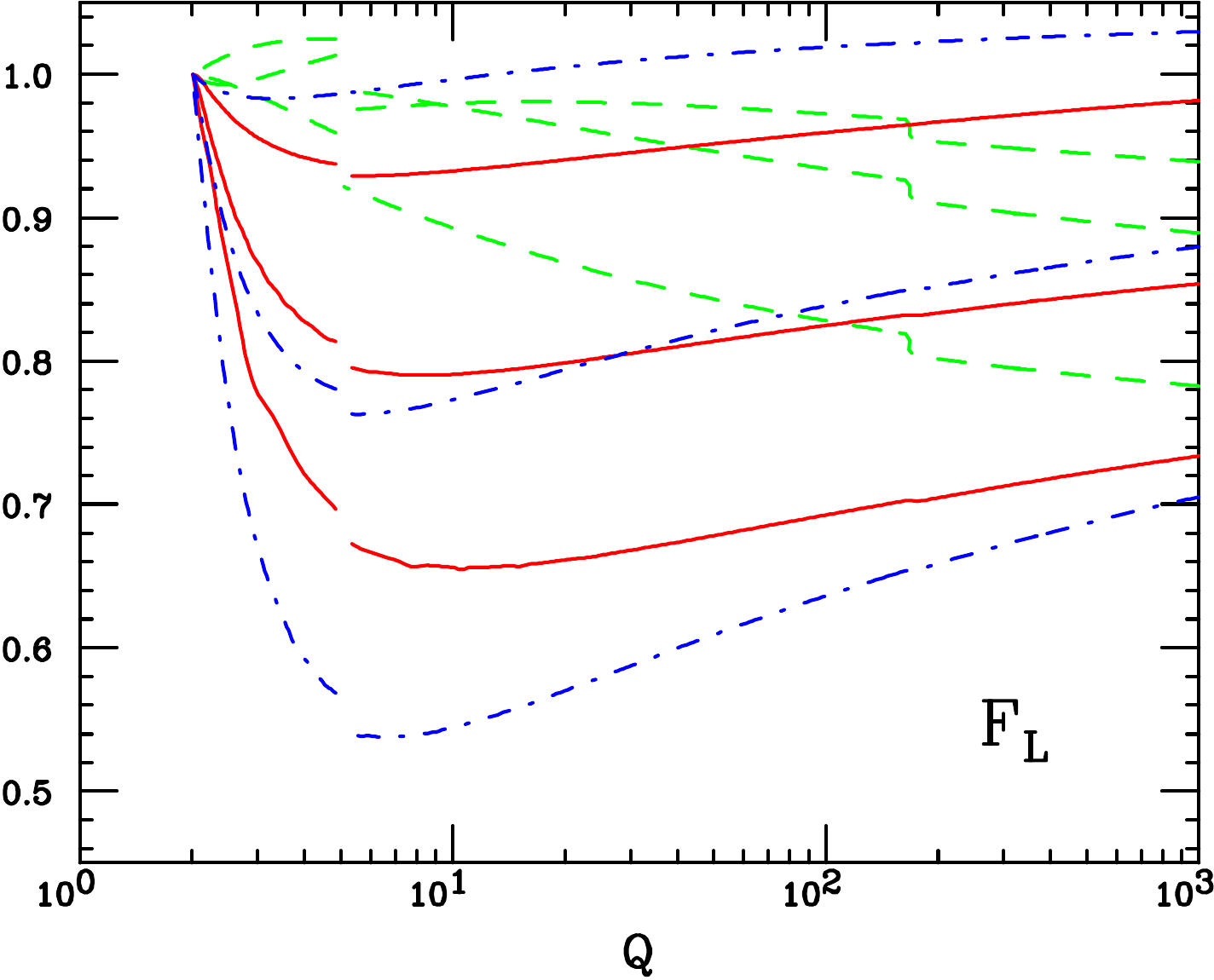}
\end{center}
\caption{The $K$-factors, defined as the ratio of the fixed-order NNLO
  or resummed calculation to the NLO fixed-order results for the
  singlet $F_2$ and $F_L$ structure functions, with $F_2$ and $F_L$
  kept fixed for all $x$ at $Q_0=2$~GeV. Results are shown at fixed
  $x=10^{-2},~~10^{-4}~{\rm or}~10^{-6}$ as a function of $Q$ in the
  range $Q=2-1000$~GeV with \as running and $n_f$ varied in a
  zero--mass variable flavour number scheme.  The breaks in the curves
  correspond to the $b$ and $t$ quark thresholds. The curves are:
  fixed order perturbation theory NNLO (green, dashed); resummed NLO
  in the \QMS\ scheme (red, solid), resummed NLO in the \MS\ scheme
  (blue, dot-dashed). Curves with decreasing $x$ correspond to those
  going from bottom to top for NNLO and from top to bottom in the
  resummed cases.}
\label{Fig:plot_f2resum}
\end{figure}

A major development for high--energy resummation was presented in
\cite{Altarelli:2008aj}, where the full small-$x$ resummation of
deep-inelastic scattering (DIS) anomalous dimensions and coefficient functions
was obtained including 
the quark contribution. This allowed for the first time a
consistent small-$x$ resummation of DIS structure functions. These results are
summarised in Fig.~\ref{Fig:plot_f2resum}, taken from
Ref.~\cite{Altarelli:2008aj}, where the $K$-factors for $F_2$ and $F_L$ for the
resummed results are compared.  As is evident  from this figure, resummation
is quite important in the region of low $x$ for a
wide range of $Q^2$ values.
One observes, for example, that the fixed order NNLO contribution leads to an
enhancement of $F_2$ with respect to NLO, whereas the resummed calculation
leads to a suppression. This means that a
truncation at any fixed order is very
likely to be insufficient for the description of the LHeC data and therefore
the fixed-order perturbative expansion becomes unreliable in the low-$x$ region,
which calls for the resummation.  Furthermore, the resummation of hard partonic
cross sections has been performed for several LHC processes such as heavy quark
production~\cite{Ball:2001pq}, Higgs
production~\cite{Marzani:2008az,Marzani:2008ih},
Drell-Yan~\cite{Marzani:2008uh,Marzani:2009hu} and prompt photon
production~\cite{Diana:2009xv,Diana:2010ef}.
The LHC is thus likely to provide a testing ground in the near future.

We refer to the recent review in Ref.~\cite{Forte:2009wh} as well as to the
HERA-LHC workshop proceedings~\cite{Dittmar:2009ii} for a more detailed summary
of recent theoretical developments in high-energy resummation.

To summarise, small-$x$ resummation is becoming a very important component for
precision  LHC physics, and will become a crucial ingredient of the LHeC
small-$x$ physics program~\cite{Rojo:2009us,Rojo:2009ut}. The LHeC extended
kinematic range will enhance the  differences between  the resummed
predictions and fixed-order DGLAP calculations.

%% file: physics/tex/LHC_lowx_3rv.tex
The low-$x$ regime of QCD can also be analysed in hadron and nucleus collisions at the LHC.
The experimentally accessible values of $x$ range from $x\sim 10^{-3}$ to $x\sim 10^{-6}$
for central and forward rapidities respectively. The estimates  for the corresponding saturation scale at $x\sim 10^{-3}$,
based on Eq.~(\ref{eq:qsat}), result in $Q_s^2\approx$ 1   GeV$^2$ for proton and  $Q_s^2\approx$ 5   GeV$^2$ for lead.

\begin{figure}
\begin{center}
\includegraphics[width=0.32\textwidth,height=5.5cm]{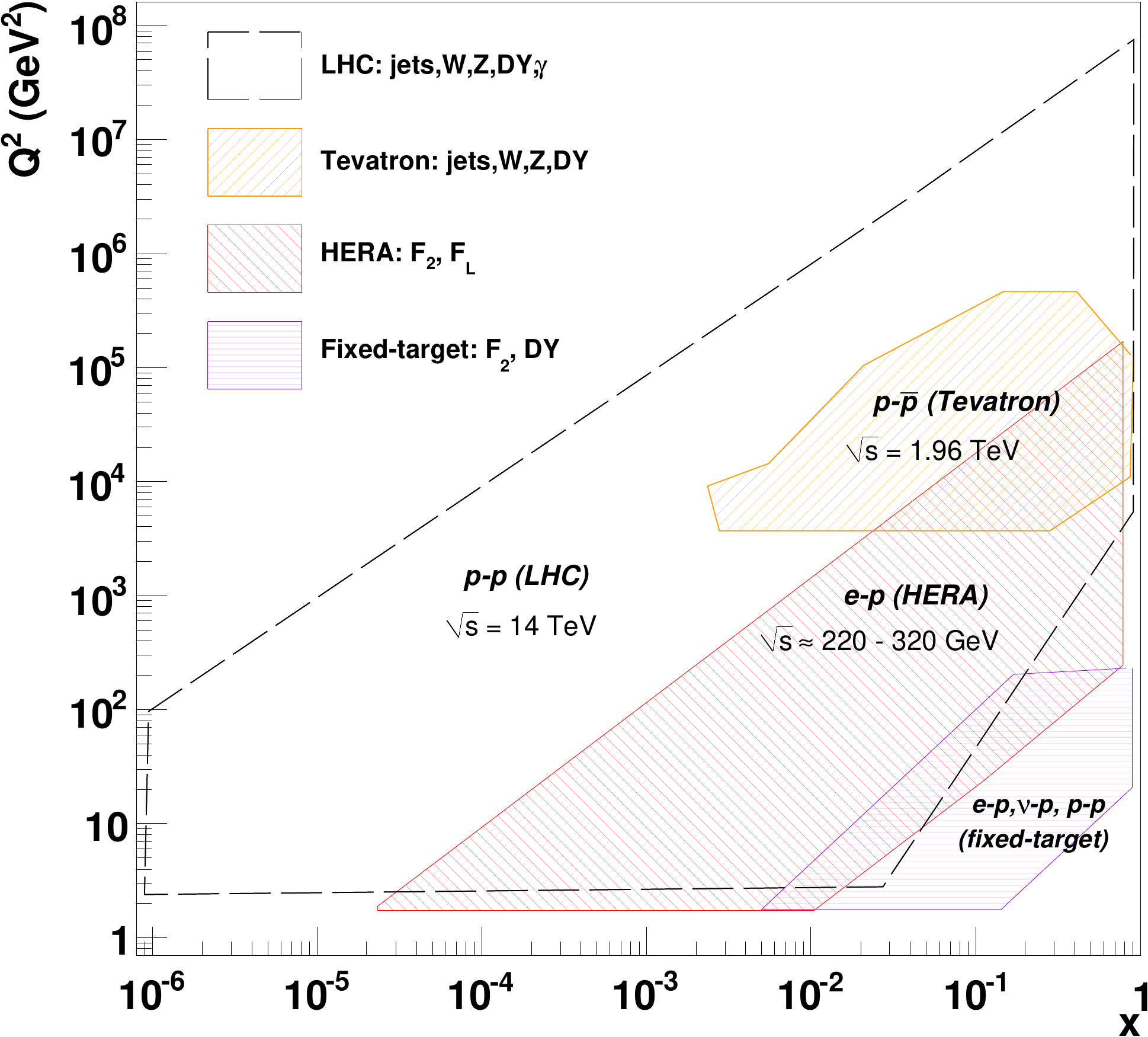}
\includegraphics[width=0.32\textwidth,height=5.8cm]{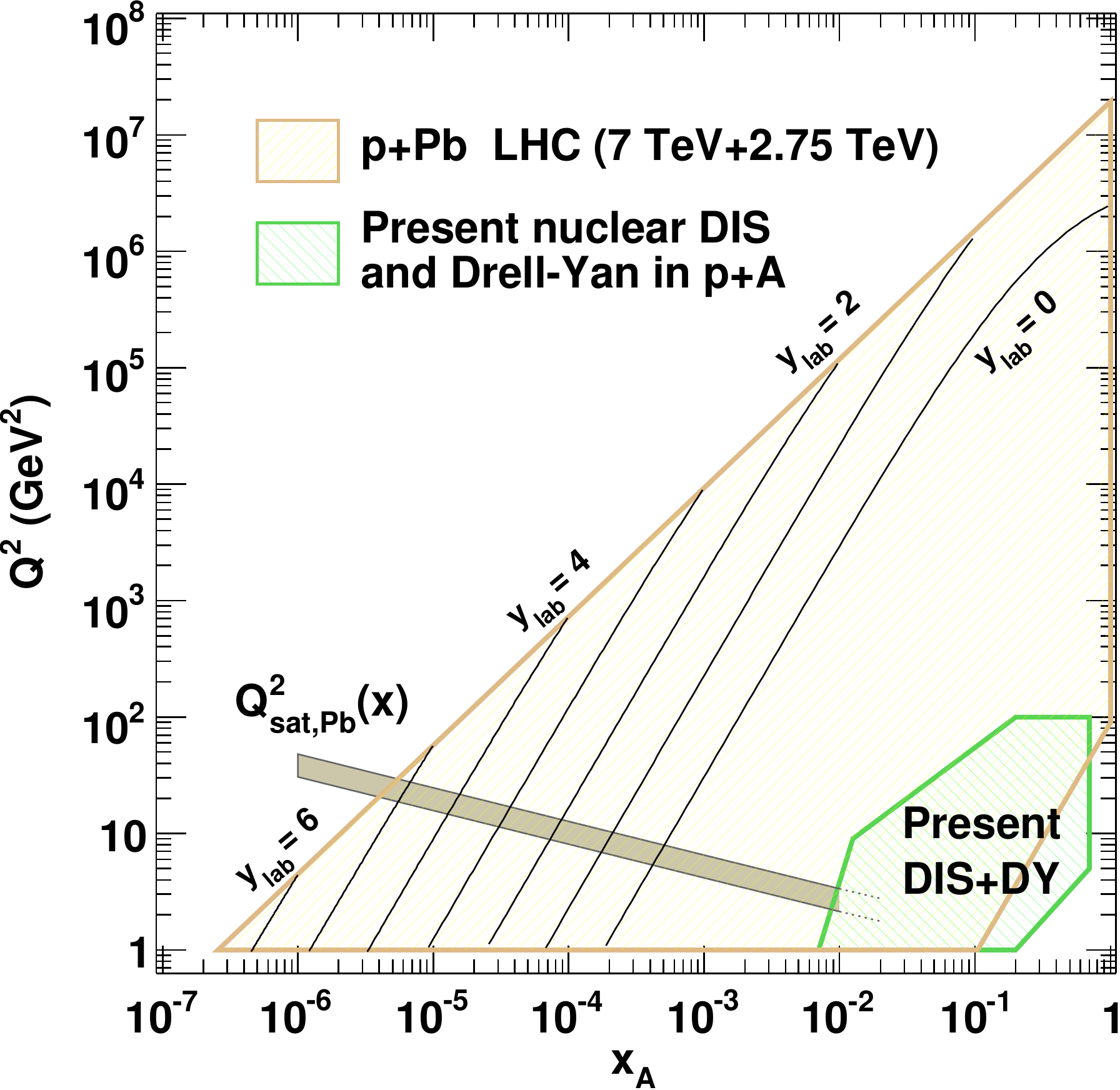}
\includegraphics[width=0.32\textwidth,height=5.5cm]{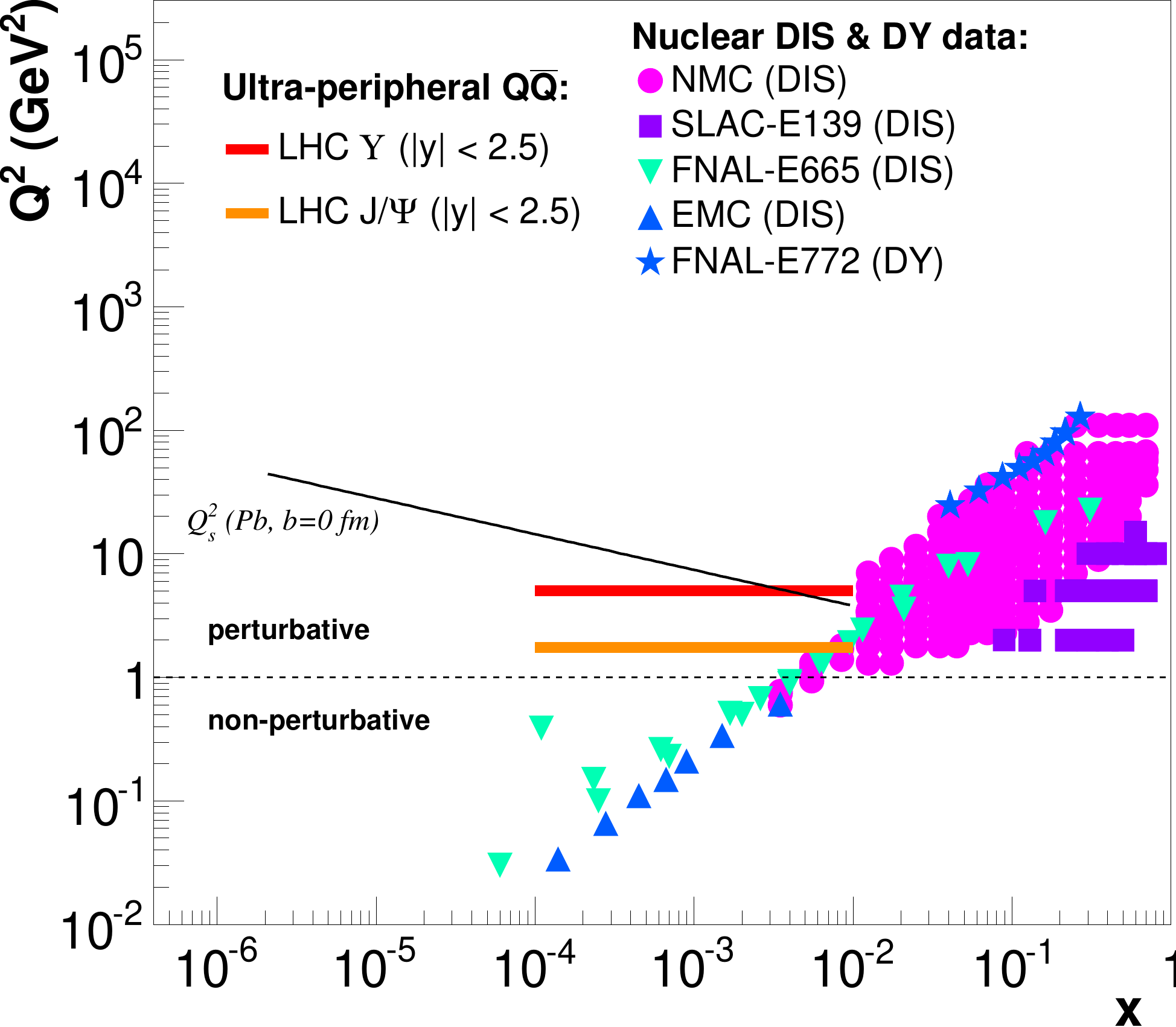}
\end{center}
\caption{Kinematic reaches in the ($x,Q^2$) plane covered in proton-proton (left),
proton-nucleus (centre)~\protect\cite{pAatLHC} and ultra-peripheral nucleus-nucleus 
(right)~\protect\cite{d'Enterria:2007pk} collisions at the LHC.
Also shown are the regions studied so far 
in collider and fixed-target experiments.
Estimates of the saturation scale for lead are also shown.} 
\label{fig:kinreachLHC}
\end{figure}

The significant increase in the centre-of-mass energy and the excellent rapidity coverage 
of the LHC detectors will extend the kinematic reach in the $x$--$Q^2$ plane 
by orders of magnitude compared to previous measurements at fixed-target and collider energies
(see Fig.~\ref{fig:kinreachLHC}).
Such measurements are particularly important in the nuclear case since, due to the scarcity of nuclear DIS data, 
the gluon PDF in the nucleus is virtually unknown at fractional momenta below $x\approx$~10$^{-2}$~\cite{Eskola:2009uj}.
In addition, due to the dependence of the saturation scale on the hadron transverse size, 
non-linear QCD phenomena are expected to play a central role in the phenomenology of collisions 
involving nuclei.
We succinctly review here the experimental possibilities for studying saturation physics
in $pp$, $p$A and AA collisions at the LHC.


\subsubsection{Low-$x$ studies in proton-proton collisions}

The LHC experiments feature 
detection capabilities at forward rapidities ($|\eta|\gtrsim$~3), 
which will allow measurements of various perturbative processes 
sensitive to the underlying parton structure and its dynamical evolution in the proton.
The {\it minimum} parton momentum fractions probed in a $2\rightarrow 2$ 
process with a particle of momentum $p_T$ produced at pseudo-rapidity $\eta$ is
\begin{equation}
x_{min} = \frac{x_T\,e^{-\eta}}{2-x_T\,e^{\eta}}\;\;,\;\;\; \mbox{ where } \;\; x_T=2p_T/\sqrt{s}\,,
\label{eq:x2_min}
\end{equation}
i.e. $x_{min}$ decreases by a factor $\sim$10 every 2 units of rapidity. The extra $e^\eta$ 
lever-arm motivates the interest in {\it forward} particle production measurements to study the PDFs
at small values of $x$. From Eq.~(\ref{eq:x2_min}) it follows that
the measurement at the LHC 
of particles with transverse momentum $p_{T}$~=~10~GeV 
at rapidities $\eta\approx$~5  
probes $x$ values as low as $x\approx 10^{-5}$ 
(Fig.~\ref{fig:kinreachLHC}, left).
Various experimental 
measurements have been proposed at forward rapidities at the LHC 
to constrain the low-$x$ PDFs in the proton and to look for 
possible evidence for non-linear 
QCD effects. These include forward jets and Mueller-Navelet dijets in ATLAS and CMS~\cite{d'Enterria:2009fa}; 
and forward isolated photons~\cite{Ichou:2010wc} and Drell-Yan (DY)~\cite{deLorenzi:2010} in LHCb.



\subsubsection{Low-$x$ studies in proton-nucleus collisions}

Until an electron-ion collider becomes available,
proton-nucleus collisions will be the best available tool 
to study 
small-$x$ physics in a nuclear environment without 
the strong influence of the
final-state medium as expected in the AA case. 
Though proton-nucleus collisions at the LHC are only scheduled to start in late 2012, detailed feasibility studies
exist~\cite{Jowett:2008hb} and strategies to define the accessible physics programme are being developed~\cite{pAatLHC}.
The $p$A
programme at the LHC serves a dual purpose~\cite{pAatLHC}: to provide ``cold QCD matter''
benchmark measurements for the physics measurements of the AA programme without significant final-state
effects, and to study the nuclear wavefunction in the small-$x$ region. 
In Fig.~\ref{fig:kinreachLHC} (centre) we show how dramatically the LHC will extend the region of phase space 
in the
$(x,Q^2)$ plane\footnote{Asymmetric colliding systems imply a rapidity shift in the two-in-one magnet design of the LHC. 
This shift has been taken into account in the figure: the quoted $y$ values are those in the laboratory
frame.} by orders of magnitude compared with those studied at 
present. The same figure also shows the scarcity of nuclear DIS and DY measurements and, correspondingly,
the lack of knowledge of nuclear PDFs in 
the regions needed to constrain the initial state for the AA programme 
- there is almost no information at present in the region $x\lesssim 10^{-2}$~\cite{Eskola:2009uj}.\\
Nuclear PDF constraints, checks of factorisation (universality of PDFs) 
and searches for saturation of partonic 
densities will be performed in $p$A collisions at the LHC by studying different production cross sections 
for e.g. inclusive light hadrons~\cite{QuirogaArias:2010wh}, 
heavy flavour particles~\cite{Eskola:2001gt}, isolated
photons~\cite{Arleo:2007js}, electroweak bosons~\cite{Paukkunen:2010qg} and jets.
Additional opportunities also appear in the so-called ultra-peripheral collisions in which the coherent 
electromagnetic field created by the proton or the large nucleus 
effectively acts as one of the colliding 
particles with photon-induced collisions at 
centre of mass energies higher than those reached 
in photoproduction at the HERA collider~\cite{Baltz:2007kq} 
(see next subsection).

At this point it is worth mentioning that particle production in the forward (proton) rapidity region in dAu collisions at RHIC shows features suggestive of saturation effects, although no consensus has been reached so far, see \cite{Arsene:2004ux,Kopeliovich:2005ym,Braidot:2010zh,Frankfurt:2007rn,Albacete:2010pg,Adare:2011sc,Stasto:2011ru} and references therein. The measurements at RHIC suffer from the limitation of working at the edge of the available phase space in order to study the small-$x$ region in the nuclear wave function. This limitation will be overcome by the much larger available phase space at the LHC.

\subsubsection{Low-$x$ studies in nucleus-nucleus collisions}

Heavy-ion ($AA$) 
collisions at the LHC 
aim at the exploration of collective partonic behaviour both in the initial
wavefunction of the nuclei as well as in the final produced matter, 
the latter being a hot and dense 
QCD medium (see the discussions in Section\,\ref{sec:nucleartargets}). The nuclear PDFs at small $x$ define the number of parton scattering centres and thus
the initial conditions of the system which then thermalises. 

A possible means of obtaining direct information on the nuclear parton 
distribution functions is through the 
study of final state particles which do not interact strongly 
with the surrounding medium, such as
photons~\cite{Arleo:2004gn} or 
electroweak bosons~\cite{Paukkunen:2010qg}.
Beyond this, 
global properties of the collision such as the 
total multiplicities or the existence of long-range 
rapidity structures (seen in AuAu collisions at RHIC \cite{Abelev:2009jv}
and in $pp$ and PbPb collisions at the 
LHC \cite{Khachatryan:2010gv,Chatrchyan:2011eka}) 
are sensitive to the saturation momentum 
which at the LHC is expected to be
well within 
the weak coupling regime \cite{Dumitru:2010iy}, $Q_{\rm sat,Pb}^2 \approx$ 5 -- 10 GeV$^2$.
CGC predictions for charged hadron multiplicities in central 
Pb-Pb collisions
at 5.5 TeV per nucleon
are $\dNdeta\approx$ 1500--2000~\cite{Armesto:2009ug}. (Note that the predictions done before the start of RHIC in 2000 were 3 times higher).
Recent data from ALICE \cite{Aamodt:2010pb} give $\dNdeta\approx$ 1600 in central Pb-Pb at 2.76 TeV per nucleon, in rough agreement with CGC expectations.

As already noted for the $p$A case, 
one of the cleanest ways to study the low-$x$ 
structure of the Pb nucleus at the LHC may be
via ultra-peripheral collisions (UPCs)~\cite{Baltz:2007kq} in which the strong electromagnetic fields 
(the equivalent flux of quasi-real photons) generated by the colliding nuclei can be used for photoproduction 
studies at maximum energies $\sqrtsgN\approx$ 1 TeV, that is 3--4 times larger than at HERA.
In particular, exclusive quarkonium 
photoproduction offers an attractive opportunity to constrain 
the low-$x$ gluon density at moderate virtualities, since in such processes the gluon 
couples {\em directly} to the $c$ or $b$ quarks 
and the cross section is proportional to the gluon density {\em squared}. 
The vector meson mass $M_V$ introduces a relatively large 
scale, amenable to a perturbative QCD treatment.
In $\gamma {\rm A}\rightarrow \jpsi \,(\ups)\,{\rm  A}^{(*)}$ processes at the LHC, the gluon 
distribution can be probed at values as low as 
$x=M_V^2/W_{\gamma {\rm A}}^2 e^{y}\approx 10^{-4}$, where $W_{\gamma {\rm A}}$
is the $\gamma$A centre of mass energy  
(Fig.~\ref{fig:kinreachLHC} right).
Full simulation studies~\cite{Nystrand:2008az,d'Enterria:2007pk} of 
quarkonium photoproduction
tagged with very-forward neutrons, show that ALICE and CMS can carry out 
detailed $p_T$,$\eta$ measurements in the dielectron and dimuon decay channels. 


In summary, $pp$, $p$A and AA collisions at the LHC have access to the small-$x$ regime, and will certainly help  to unravel the complex parton dynamics in this region. However, the excellent precision of a high energy electron-proton (ion) collider cannot be matched in hadronic collisions. The deep inelastic scattering process is  much cleaner experimentally and under significantly better theoretical control. The description of hadron-hadron and heavy ion collisions in the regime of small $x$ suffers from a variety of uncertainties, such as the 
question of the appropriate factorisation, if any, and the large indeterminacy of fragmentation functions in the relevant kinematic region.  Thus, 
the precise measurement of physical observables and parton densities and 
their interpretation in terms of QCD dynamics 
is only possible at an electron-hadron (ion) collider.

%% file: physics/tex/npdfs.tex
As discussed in Section\,\ref{sec:lowxoverview}, the use of 
nuclei offers a means of modifying the parton density 
both through colliding different nuclear species and by 
varying the impact parameter of the collision. Therefore, the study of DIS on nuclear targets is of 
the utmost importance for our understanding of the dynamics which control 
the behaviour of hadron and nuclear wave functions at small $x$. On the other hand, the characterisation of parton 
densities inside nuclei and the study of other aspects of lepton-nucleus collisions 
such as particle production, are of strong interest both fundamentally and because they are crucial for a correct interpretation of the experimental results from ultra-relativistic ion-ion collisions. In the rest of this section we focus on these last two aspects.

Additionally, nuclear effects have to be better understood in order to
improve the constraints on nucleon PDFs in analyses which include 
DIS data with neutrino beams (e.g. \cite{Martin:2009iq,Ball:2010de}). 
Due to the smallness of the cross section, such neutrino 
experiments use nuclear targets, so corrections for nuclear effects 
are a significant source of uncertainty in the extraction of 
parton densities even for the proton.

\subsubsection{Comparing nuclear parton density functions}

The nuclear modification of structure functions has been extensively studied since the early 70's \cite{Arneodo:1992wf,Geesaman:1995yd}. 
It is usually characterised through the so-called nuclear modification factor which, for a given structure function or parton density $f$, reads
\begin{equation}
R_f^A(x,Q^2)=\frac{f^A(x,Q^2)}{A\times f^N(x,Q^2)} \; .
\label{nmf}
\end{equation}
In this equation, the superscript $A$ refers to a nucleus of mass number $A$, while $N$ denotes the nucleon (either a proton or a neutron, or 
their average as obtained using deuterium). 
The absence of nuclear effects would result in $R=1$.

The nuclear modification factor for $F_2$ shows a rich structure: an enhancement ($R>1$) at large $x>0.8$, a suppression ($R<1$) for $0.3<x<0.8$, an enhancement for $0.1<x<0.3$, and a suppression for $x<0.1$ where isospin effects can be neglected. The latter effect is called shadowing \cite{Armesto:2006ph}, and is the dominant phenomenon at high energies (the kinematic region $x<0.1$ will determine particle production at the LHC, see Sec.~\ref{sec:lowxlhc} and \cite{Accardi:2004be}).

The modifications in each region are believed to be of different dynamical origin. In the case of shadowing, the explanation is usually given in terms of a coherent interaction involving several nucleons, 
which reduces the nuclear cross section from the totally incoherent situation, $R=1$, towards a region of total coherence. In the region of very small $x$, small-to-moderate $Q^2$ and for large nuclei, the unitarity limit of the nuclear scattering amplitudes is expected to be approached and some mechanism of 
unitarisation such as multiple scattering should come into play.
Therefore, in this region nuclear shadowing is closely related to the onset of the  unitarity limit in QCD and the transition  from coherent scattering of the probe off  a single parton to coherent scattering off many partons. 
The different dynamical mechanisms proposed to deal with this problem should offer a quantitative explanation for shadowing, with the nuclear size playing the role of a density parameter in the way discussed in Section\,\ref{sec:lowxoverview}.

At large enough $Q^2$ the generic expectation is that the parton system becomes dilute and the usual leading-twist linear DGLAP evolution equations should be applicable to nuclear PDFs. 
In this framework, global analyses of nuclear parton densities (in exact analogy to those of proton and neutron parton densities) have been developed up to NLO accuracy \cite{deFlorian:2003qf,Hirai:2007sx,Eskola:2009uj,deFlorian:2011fp}.
In these global analyses, the initial conditions for DGLAP evolution are parameterised by flexible functional forms but they lack theoretical motivation
in terms of e.g. the dynamical mechanisms for unitarisation mentioned above.
On the other hand, the relation between diffraction and nuclear shadowing \cite{Gribov:1968jf,Gribov:1968gs} can in principle be employed to 
constrain the initial conditions for DGLAP evolution, as has 
been explored previously at both LO \cite{Armesto:2010kr} and 
NLO \cite{Guzey:2009jr}\footnote{In the approach in \cite{Guzey:2009jr} predictions are provided only for sea quarks and gluons, with the valence taken from the analysis in \cite{Eskola:1998df}.} accuracy, see Subsec. \ref{sec:diffpdfs}.
All nuclear PDF analyses \cite{deFlorian:2003qf,Hirai:2007sx,Eskola:2009uj,deFlorian:2011fp} include data from NC DIS and DY experiments, \cite{Eskola:2009uj,deFlorian:2011fp} also 
use particle production data
at mid-rapidity in deuterium-nucleus collisions at RHIC, and \cite{deFlorian:2011fp} CC DIS data from neutrino experiments. Error sets obtained through the Hessian method are provided in \cite{Eskola:2009uj,deFlorian:2011fp}. Note that CC DIS data have been considered only recently \cite{Kovarik:2010uv,Paukkunen:2010hb,deFlorian:2011fp}\footnote{The analyses in \cite{Paukkunen:2010hb,Eskola:2009uj,deFlorian:2011fp} show the compatibility of the nuclear corrections as extracted from NC DIS, DY and particle production in dAu at RHIC, with CC DIS data on nuclear targets, while in \cite{Kovarik:2010uv} some tension is found between NC and CC DIS data.} in this context. 

\begin{figure}
\centerline{\includegraphics[width=\textwidth,angle=0]{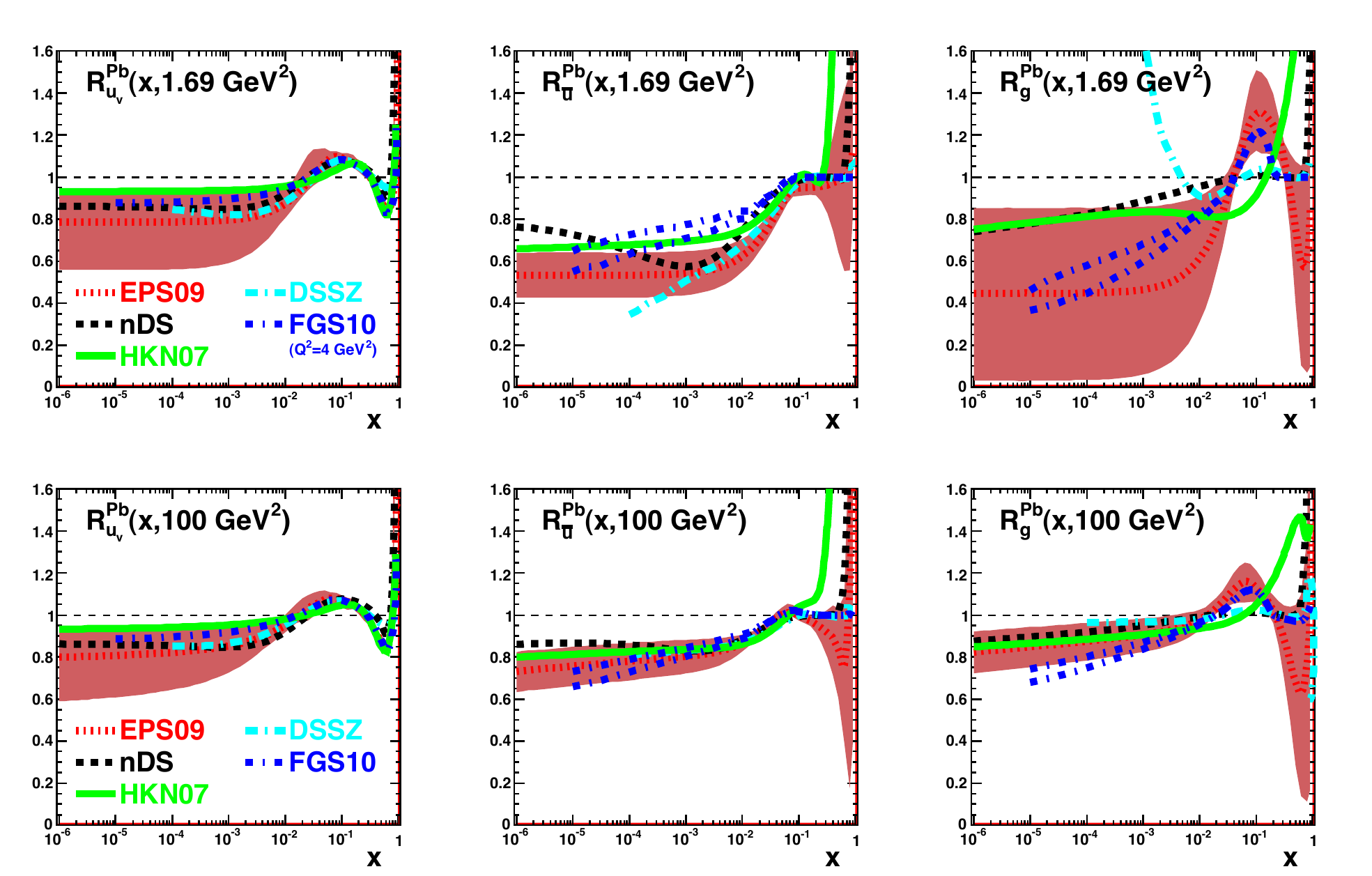}}
\caption{Ratio of parton densities in a bound proton in Pb to those in a 
free proton scaled by $A = 207$, for valence $u$ (left), $\bar u$ (middle) and $g$ (right), at $Q^2=1.69$ (top) and 100 (bottom) GeV$^2$. 
Results are shown from {\protect{\cite{deFlorian:2003qf}}} (nDS, black dashed), {\protect{\cite{Hirai:2007sx}}} (HKN07, green solid), {\protect{\cite{Eskola:2009uj}}} (EPS09, red dotted), {\protect{\cite{Guzey:2009jr}}} (FGS10, blue dashed-dotted; in this case the lowest $Q^2$ is 4 GeV$^2$ and two lines are drawn reflecting the uncertainty in the predictions) and \cite{deFlorian:2011fp} (DSSZ, cyan dashed-dotted). The red bands indicate 
the uncertainties according to the EPS09 analysis {\protect{\cite{Eskola:2009uj}}}.}
\label{Fig:npdfs}
\end{figure}

Results from different nuclear PDF analyses performed at NLO accuracy are shown in Fig. \ref{Fig:npdfs}, with the band indicating the uncertainty obtained using the error sets in \cite{Eskola:2009uj}. 
In addition to the discrepancies concerning the existence of an enhancement/suppression at large $x$, the different approaches lead to clear differences 
at small $x$, both in magnitude and in shape\footnote{The increasing shape of the gluon ratio with decreasing $x$ at small $x$ and $Q^2$ in \cite{deFlorian:2011fp}, is due to the fact that in this analysis the proton parton densities MSTW2008 \cite{Martin:2009iq}, in which the gluon distribution becomes negative in that kinematic region, are used.}, usually within the
large uncertainty band shown. With nuclear effects 
vanishing logarithmically in the DGLAP analysis, the corresponding differences and uncertainties 
diminish, although they remain sizeable until rather large $Q^2$.

These large uncertainties are due to the lack of experimental data on nuclear structure functions for $Q^2 > 2$ GeV$^2$ and $x$ 
smaller than a few times $10^{-2}$.
The constraints on the small-$x$ gluon are particularly poor.
Particle production data at mid-rapidity coming from deuterium-nucleus collisions at RHIC offer an indirect constraint on the small-$x$ sea and glue \cite{Eskola:2009uj,deFlorian:2011fp}, but these data are bound to contain sizeable uncertainties intrinsic to particle production in hadronic collisions at small and moderate scales. Therefore, only high-accuracy data on nuclear structure functions at smaller $x$, 
with a large lever arm in $Q^2$, as achievable at the LHeC, will be able to substantially reduce the uncertainties and clearly distinguish between the different approaches.

%% file: physics/tex/eA-hip.tex
The LHeC will offer 
extremely valuable information on several aspects of high-energy hadronic and nuclear collisions. On the one hand, it will characterise hard scattering processes in nuclei through a precise determination of the initial state. On the other hand, it will provide quantitative constraints on theoretical descriptions of initial particle production in ultra-relativistic nucleus-nucleus collisions and the subsequent evolution into the quark-gluon plasma, the deconfined partonic state of matter whose production and study offers key information about confinement. Such knowledge will complement that coming from pA collisions and self-calibrating hard probes in nucleus-nucleus collisions (see \cite{pAatLHC,Accardi:2004be,Accardi:2004gp,Bedjidian:2004gd,Arleo:2004gn}) regarding the correct interpretation of the findings of the heavy-ion programme at RHIC  (see e.g. \cite{Gyulassy:2004zy,d'Enterria:2006su} and refs. therein) and at the LHC. Beyond the qualitative interpretation of such findings, the LHeC will greatly improve the quantitative characterisation of the properties of QCD extracted from such studies. The relevant information can be classified into three items:

\begin{itemize}

\item[a.] \underline{Parton densities inside nuclei}:

The knowledge of parton densities inside nuclei is an essential piece of information for the analysis of the medium created in ultra-relativistic heavy-ion collisions using hard probes, i.e. those observables whose yield in nucleon-nucleon collisions can be predicted in pQCD (see \cite{Accardi:2004be,Accardi:2004gp,Bedjidian:2004gd,Arleo:2004gn}). The comparison between the expectation from 
an incoherent superposition of nucleon-nucleon collisions and the measurement in nucleus-nucleus collisions characterises the nuclear effects. 
However, we need to disentangle those effects which originate from the creation of a hot medium in nucleus-nucleus collisions, from effects arising only from differences in the partonic content between
 nucleons and nuclei.

Our present knowledge of parton densities inside nuclei is clearly insufficient in the kinematic regions of interest for RHIC and, above all, for the LHC (see \cite{Accardi:2004be} and Section\,\ref{sec:lowxlhc}). Such ignorance is reflected in uncertainties larger than a factor $3-4$ for the calculation of different cross sections in nucleus-nucleus collisions at the LHC (see Fig. \ref{Fig:npdfs} and \cite{QuirogaArias:2010wh}), thus weakening strongly the possibility of extracting quantitative characteristics of the produced hot medium. While the pA program at the LHC will offer new constraints on the nuclear parton densities (e.g. \cite{pAatLHC,QuirogaArias:2010wh}), measurements at the LHeC would be far more constraining and 
would reduce the uncertainties in nucleus-nucleus cross sections to less than a factor two.

\item[b.] \underline{Parton production and initial conditions for a heavy-ion collision}:

The medium produced in ultra-relativistic heavy-ion collisions develops very early a collective behaviour, usually considered as that of a thermalised medium and describable by relativistic hydrodynamics. The initial state of a heavy-ion collision for times prior to its eventual thermalisation, and the thermalisation or isotropisation mechanism, play a key role in the description of the collective behaviour. This initial condition for hydrodynamics or transport is presently modelled and fitted to data but
should eventually be
determined from a theoretical description of particle production within a saturation
framework embodying both aspects: parton
fluxes inside nuclei - discussed in the previous item, and particle production and evolution, eventually leading to isotropisation.

The CGC offers a well-defined framework in which 
the initial condition and thermalisation mechanism can be computed from QCD, see Section\,\ref{sec:lowxoverview} and e.g. \cite{Lappi:2009fq} and refs. therein. Although our theoretical knowledge is still incomplete, electron-nucleus 
collisions offer a setup, considerably less complex than nucleus-nucleus collisions,
in which these CGC-based calculations already exist and can be tested. In this way, electron-ion collisions offer a testing ground for ideas on parton production in a dense environment, which is required for a first principles calculation of the initial conditions for the collective behaviour in ultra-relativistic heavy-ion collisions. The LHeC offers the possibility of studying particle production  in the kinematic region relevant for experiments at RHIC and the LHC.

\item[c.] \underline{Parton fragmentation and hadronisation inside the nuclear medium}:

The mechanism through which a highly virtual parton evolves from an off-shell coloured state to a final state 
consisting of colourless hadrons, is still subject to great uncertainties. Electron-ion experiments offer a testing ground for our ideas and understanding of such phenomena, see \cite{Accardi:2009qv} and refs. therein, with the nucleus being a medium of controllable extent and density which modifies the radiation and hadronisation processes. 

The LHeC will have capabilities for particle identification and jet reconstruction for both nucleon and nuclear targets. Its kinematic reach will allow the study of partons travelling through the nucleus from low energies, for which hadronisation is expected to occur inside the nucleus, to high energies with hadronisation outside the nucleus. Therefore the modification of the yields of energetic hadrons, observed at RHIC\footnote{LHC experiments have already observed the jet quenching phenomenon both at the level of particle spectra \cite{Aamodt:2010jd,:2012nt,ALICE:2012ab,Aamodt:2011vg}
and through the study of jets \cite{Aad:2010bu,Chatrchyan:2011sx,:2012ni,Chatrchyan:2012gt}, which will play a central role in heavy-ion physics at these energies.} and usually attributed to in-medium energy loss - the so-called jet quenching phenomenon - will be investigated. With jet quenching playing a key role in the present discussions on the production and characterisation of the hot medium produced in ultra-relativistic heavy-ion collisions, the LHeC will offer most valuable information on effects in cold nuclear matter of great importance for clarifying and reducing the existing uncertainties.

\end{itemize}

%% file: physics/tex/sxstrategy.tex
As discussed previously, in order to analyse the 
regime of high parton densities at small $x$,
we propose a two-pronged approach which 
 is illustrated in Fig.~\ref{Fig:satplaneeA}. To reach an interesting novel regime of QCD one can either decrease $x$ by increasing the centre-of-mass energy or increase the matter density by increasing the mass number $A$ of the nucleus. In addition, we will see that diffraction, and especially
exclusive diffraction, will play a special role in unravelling the new dense partonic regime of QCD. 



The LHeC will offer a huge lever arm in $x$ and  also a possibility of changing the matter density at fixed values of $x$. This will allow us to pin down and compare the small $x$ and saturation phenomena both in protons and nuclei and 
will offer an excellent testing  ground for theoretical predictions. Thus, in the following, LHeC simulations of electron-proton collisions
are paralleled by those in electron-lead wherever possible.
For a complementary perspective on the opportunities 
for novel QCD studies offered by the 
LHeC, see \cite{stanote}.

%% file: physics/tex/predep.tex
The LHeC is expected to provide measurements of 
the structure functions of the proton with unprecedented precision,
which will allow detailed studies
of small-$x$ QCD dynamics. In particular, it will be highly sensitive to
departures of the inclusive observables $F_2$ and $F_L$ 
from  the  fixed-order  DGLAP framework, in the region of small $x$ and $Q^2$.
These deviations are expected by several theoretical arguments, as previously discussed in detail.

\begin{figure}
\begin{center}
\includegraphics[width=0.49\textwidth]{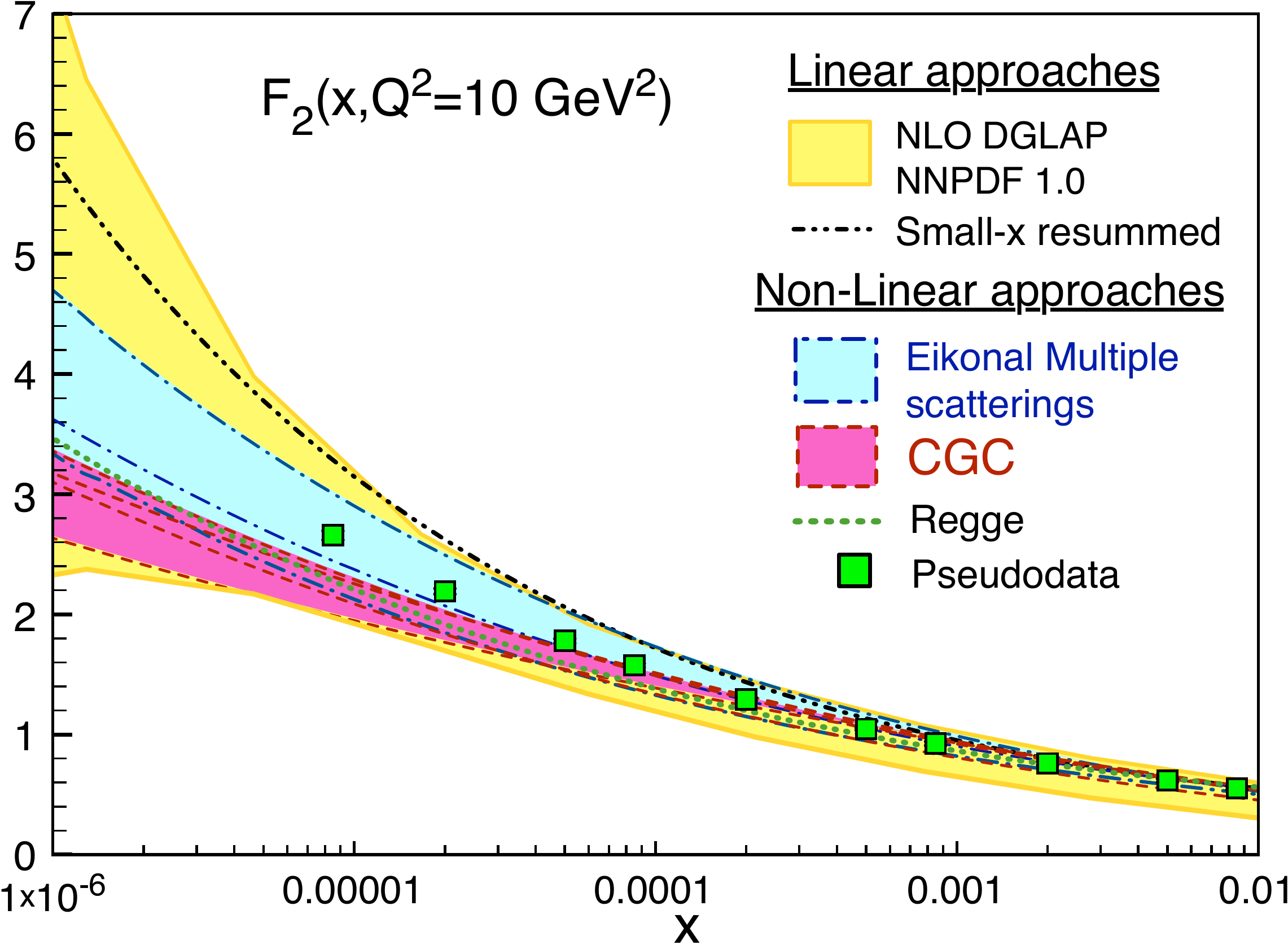}
\includegraphics[width=0.49\textwidth]{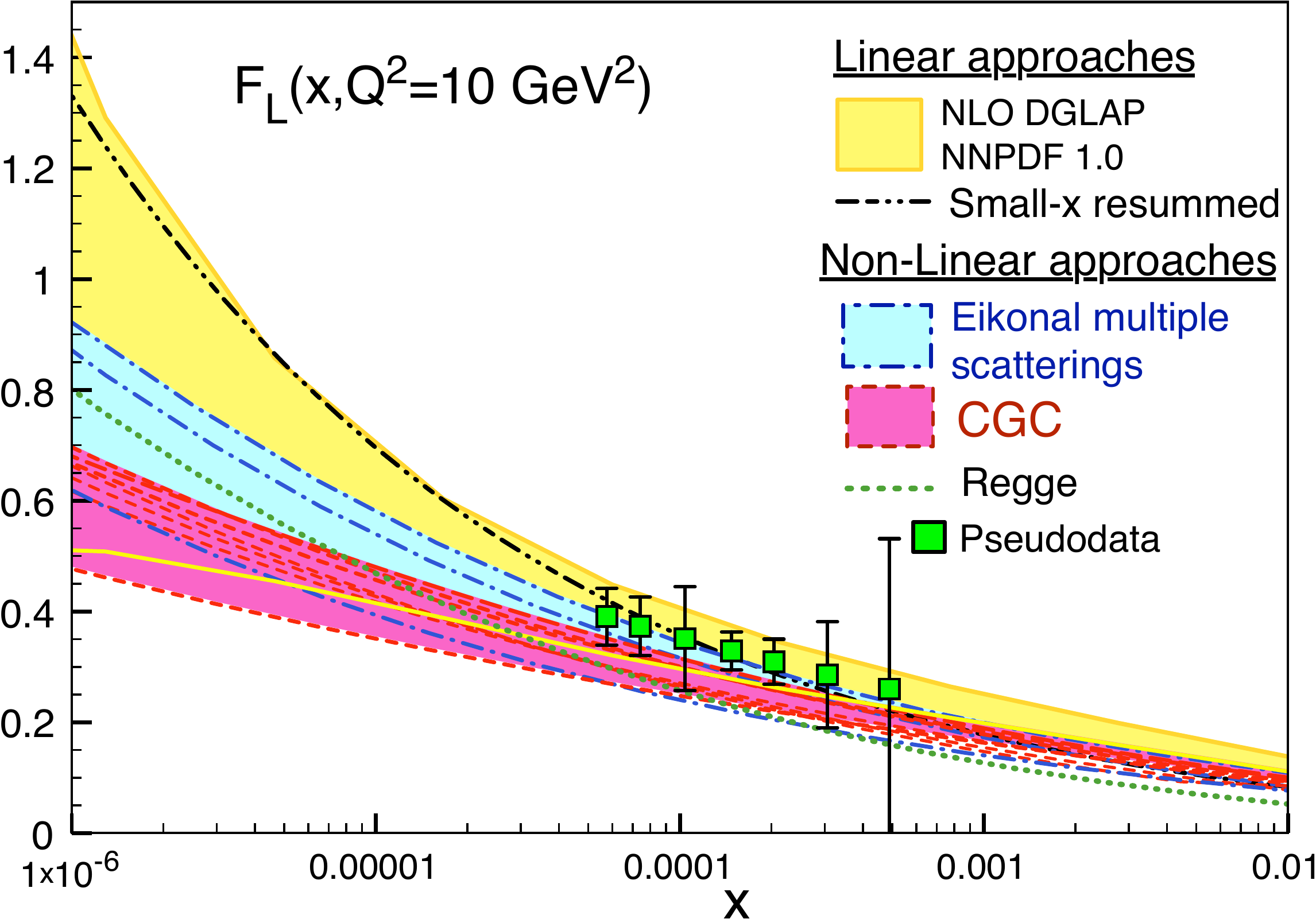}
\caption{Predictions from different models for $F_2(x,Q^2=10\ {\rm GeV}^2)$ (plot on the left) and $F_L(x,Q^2=10\ {\rm GeV}^2)$ (plot on the right) versus $x$, together with the corresponding pseudodata. See the text for explanations.}
\label{Fig:f2models}
\end{center}
\end{figure}

In Fig. \ref{Fig:f2models} we show some predictions for the proton structure functions, $F_2$ and $F_L$, in $e$p collisions at $Q^2=10$ GeV$^2$ and for $10^{-6}\leq x \leq 0.01$. The different curves correspond to the extrapolation of  models that correctly reproduce the available HERA data for the same observables in the small-$x$ region. They are of two types: those based on linear evolution approaches and those that include non-linear small-$x$ dynamics. Among the linear approaches we include extrapolation from the NLO DGLAP fit as performed by the NNPDF collaboration \cite{Ball:2009mk} (solid yellow bands) and the results from a combined 
DGLAP/BFKL approach, which includes resummation of small-$x$ effects \cite{GolecBiernat:2009be} (black dashed-dotted-dotted lines). The non-linear calculations shown here are all formulated within the dipole model. We distinguish two categories: those based on the eikonalisation of multiple scatterings together with DGLAP evolution of the gluon distributions \cite{Kowalski:2003hm,Bartels:2002cj} (blue dashed-dotted lines) and those relying in the Colour Glass Condensate effective theory of high-energy QCD scattering (red dashed lines). The latter include calculations based on solutions of the running coupling Balitsky-Kovchegov equation \cite{Albacete:2009fh} and other more phenomenological models of the dipole amplitude without \cite{Iancu:2003ge}, or with \cite{Kowalski:2006hc}
impact parameter dependence. Finally, we also include a hybrid approach, where initial conditions based on Regge theory and including non-linearities are evolved in $Q^2$ according to linear DGLAP evolution \cite{Armesto:2010ee} (green dotted line). In all cases the error bands are generated by allowing variations of the free parameters in each subset of models. The green filled squares correspond to the subset of the simulated LHeC pseudodata at 
$Q^2 = 10 \ {\rm GeV^2}$ (see Section~\ref{sec:simNC}).

Clearly, the accuracy
of the LHeC data will provide powerful discrimination between the different
models and constraints on the dynamics
underlying the small-$x$ region.

%% file: physics/tex/testingnl.tex

The potential impact of the LHeC on low $x$ parton densities within 
the framework of an NLO DGLAP analysis is assessed by adding
the pseudodata introduced in Section~\ref{sec:simNC} into the 
NNPDF fitting analysis. The pseudodata are first generated at the
extrapolated central values according to the existing NNPDF fits.

\begin{figure}[h]
\begin{center}
\includegraphics[width=0.49\textwidth]{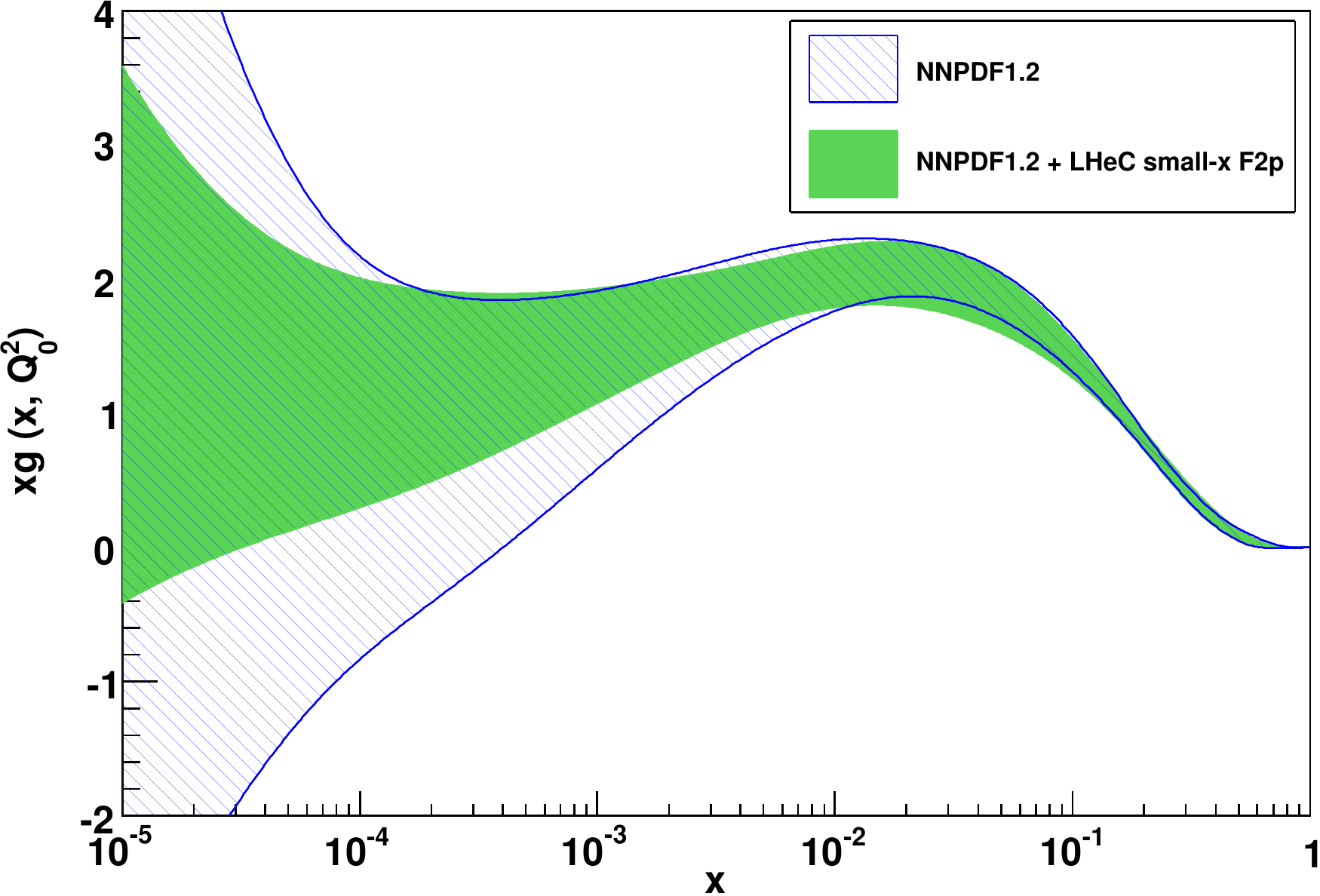}
\includegraphics[width=0.49\textwidth]{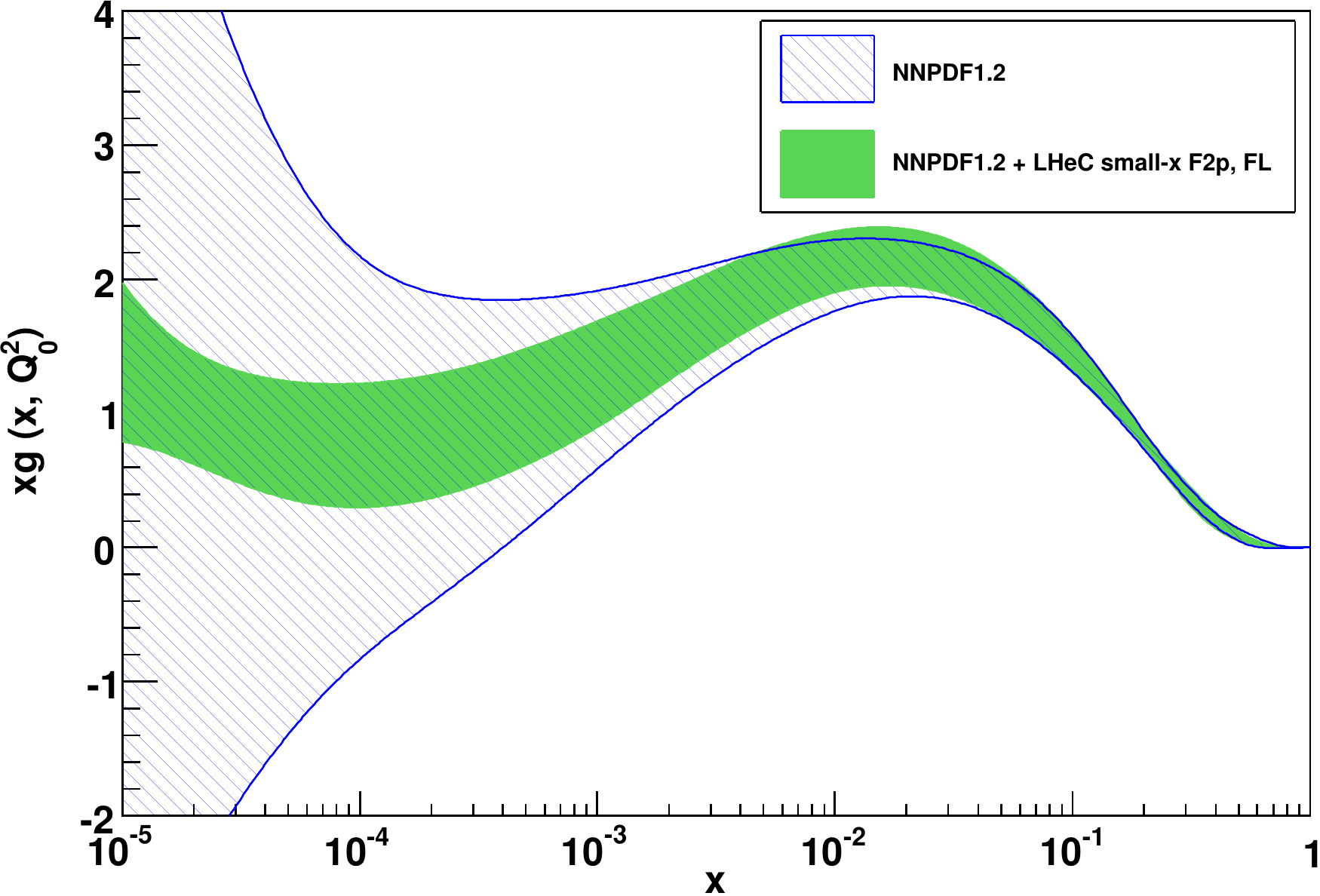}
\end{center}
\caption{\small The results for the gluon distribution in the 
standard NNPDF1.2 DGLAP fit \cite{Ball:2009mk}, together with the 
results when additionally including 
LHeC pseudodata  for $F_2$ (left) and for both $F_2$ and $F_L$ (right).
The results are shown at the starting scale for DGLAP evolution, 
$Q_0^2=2$ GeV$^2$.}
\label{fig:nnpdflhec}
\end{figure}
The extrapolated NNPDF1.2 
gluon density and its uncertainty band are shown 
at the starting scale for QCD evolution, $Q_0^2 = 2 \ {\rm GeV^2}$ in 
Fig. \ref{fig:nnpdflhec}, where it can be seen that the lack of
experimental constraints for 
$x \stackrel{<}{_{\sim}} 10^{-4}$ leads to an 
explosion in the uncertainties. When the LHeC $F_2$ pseudodata are
included in addition, the uncertainties improve considerably, but
remain rather large at the lowest $x$ values, 
due to the lack of a large lever-arm in $Q^2$ to constrain the
evolution. 
However, when the LHeC pseudodata on the
longitudinal structure function $F_L$ are included in addition, 
the additional constraints lead to a much more
substantial improvement in the uncertainties on the gluon density. 
\begin{figure}[h]
\begin{center}
\includegraphics[width=0.49\textwidth]{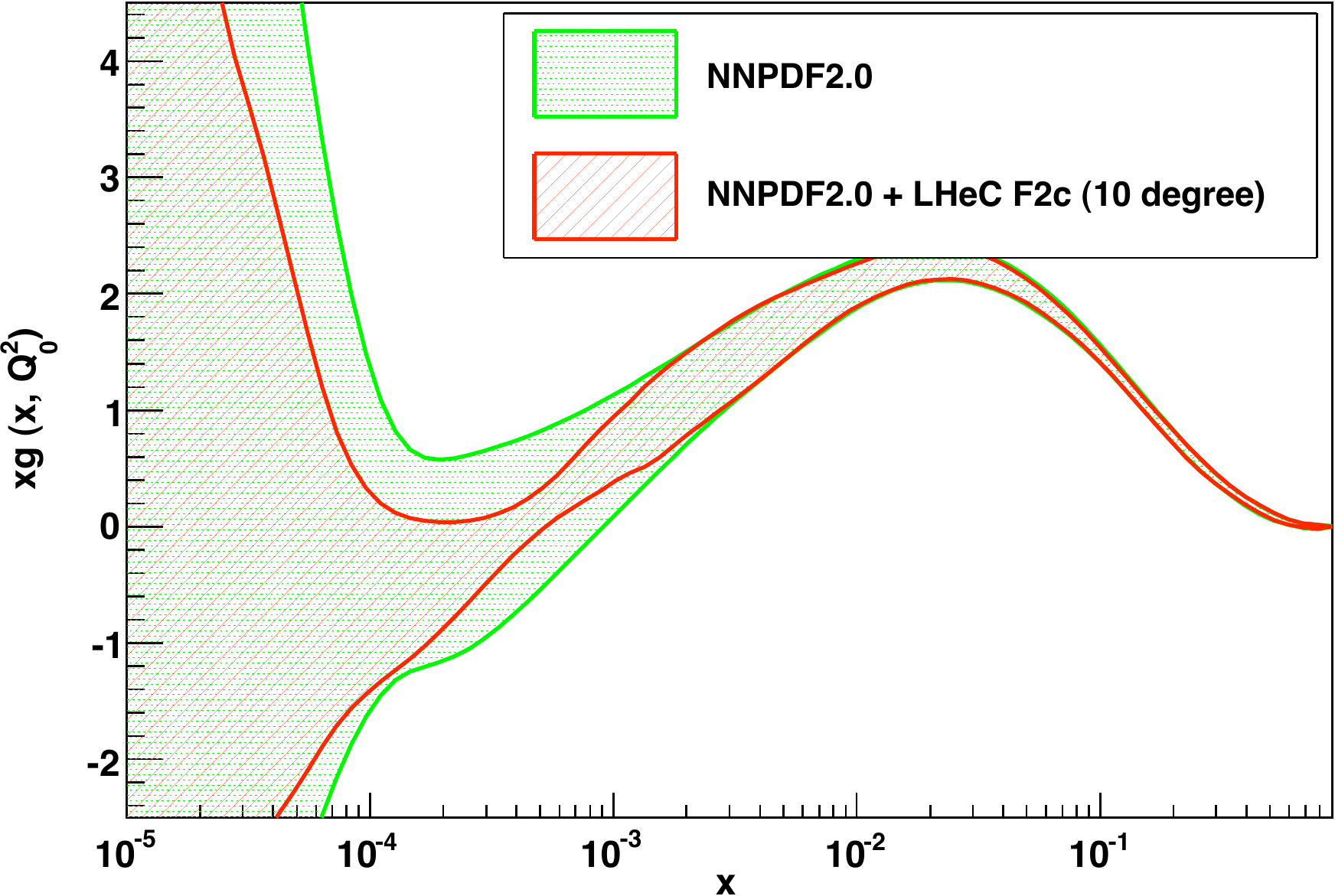}
\includegraphics[width=0.49\textwidth]{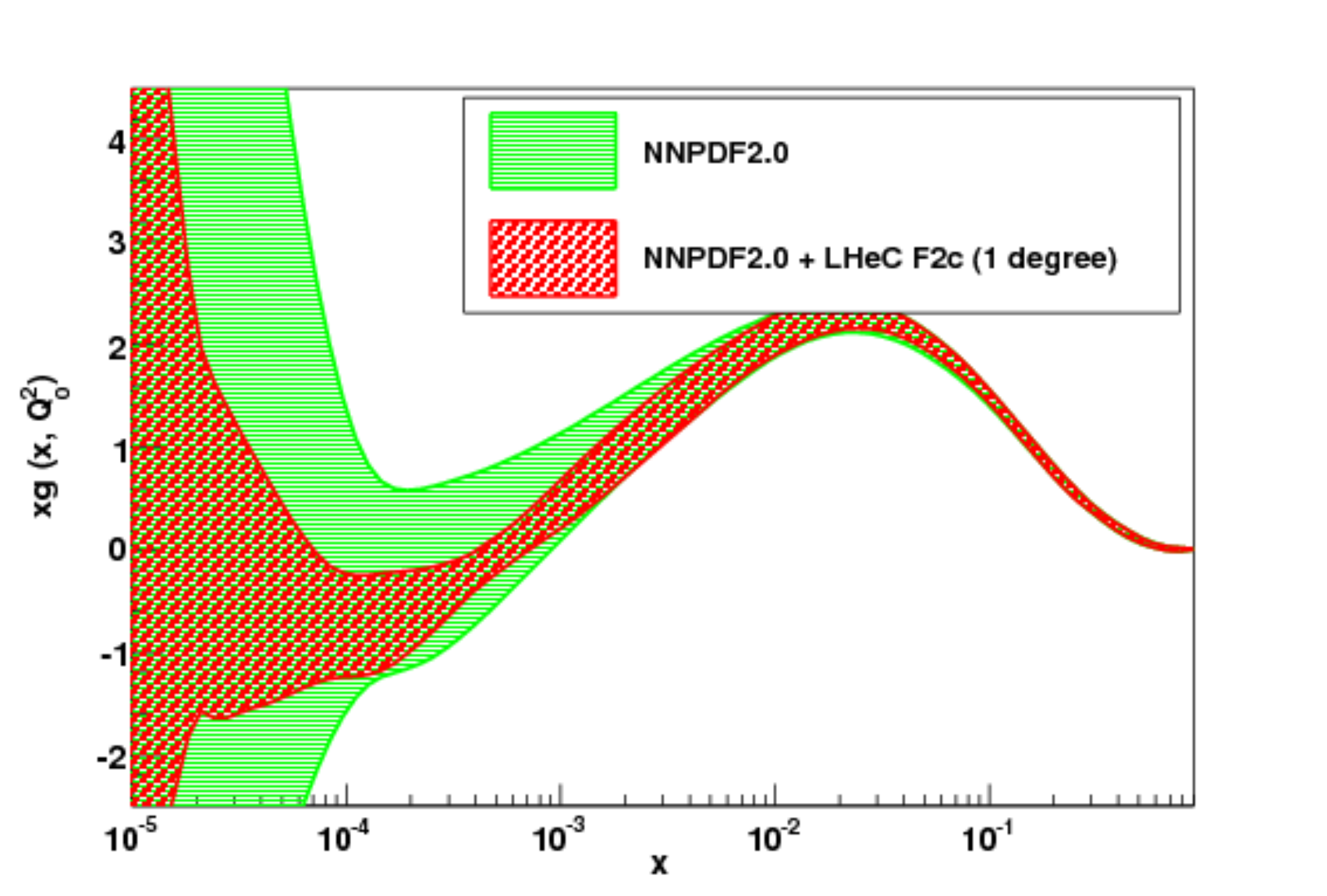}
\end{center}
\caption{\small The effect on the extracted gluon distribution function
of the inclusion of the 
LHeC pseudodata on the charmed structure function
in the NNPDF global analysis. Left plot: scattered electron
acceptance extending to within $10^\circ$ of the beam pipe. 
Right plot: $1^\circ$ acceptance.
The results are shown at the starting scale for DGLAP evolution, 
$Q_0^2=2$ GeV$^2$.} \label{f2cconstrain}
\end{figure}


As is well known from experience at HERA, 
the measurement of the longitudinal structure function 
presents many experimental challenges and involves possibly undesirable
modifications to the beam energies.
An alternative constraint on the gluon density from 
the charmed structure function $F_2^c$
has therefore also been investigated.
As discussed in detail in Subsec.~\ref{sec:hfl_intro}, the LHeC  will offer  
unique precision in the 
determination of the charm and beauty structure functions,
extending to very small $x$.

In Fig.~\ref{f2cconstrain} the gluon distribution function is shown,
as obtained from the NNPDF2.0 analysis. The green band corresponds to the 
standard analysis. The red band shows the 
modified analysis where additionally 
$F_2^c$ pseudodata from the LHeC are included, 
using a novel technique based on Bayesian reweighting \cite{nnpdf:2010gb}.
It is observed that the charmed structure function 
considerably improves the constraints on the gluon density at small 
values of $x$, especially between $3\times 10^{-5}-10^{-2}$, provided that the
scattered electron acceptance extends to within
around $1^\circ$ of the beam pipe. 
With a sufficiently good theoretical understanding, 
heavy flavour production data from the LHeC may 
thus offer an alternative to $F_L$ for precision constraints on
the gluon density at all but the lowest $x$ values. 

\begin{figure}[ht]
\begin{center}
\includegraphics[width=0.70\textwidth]{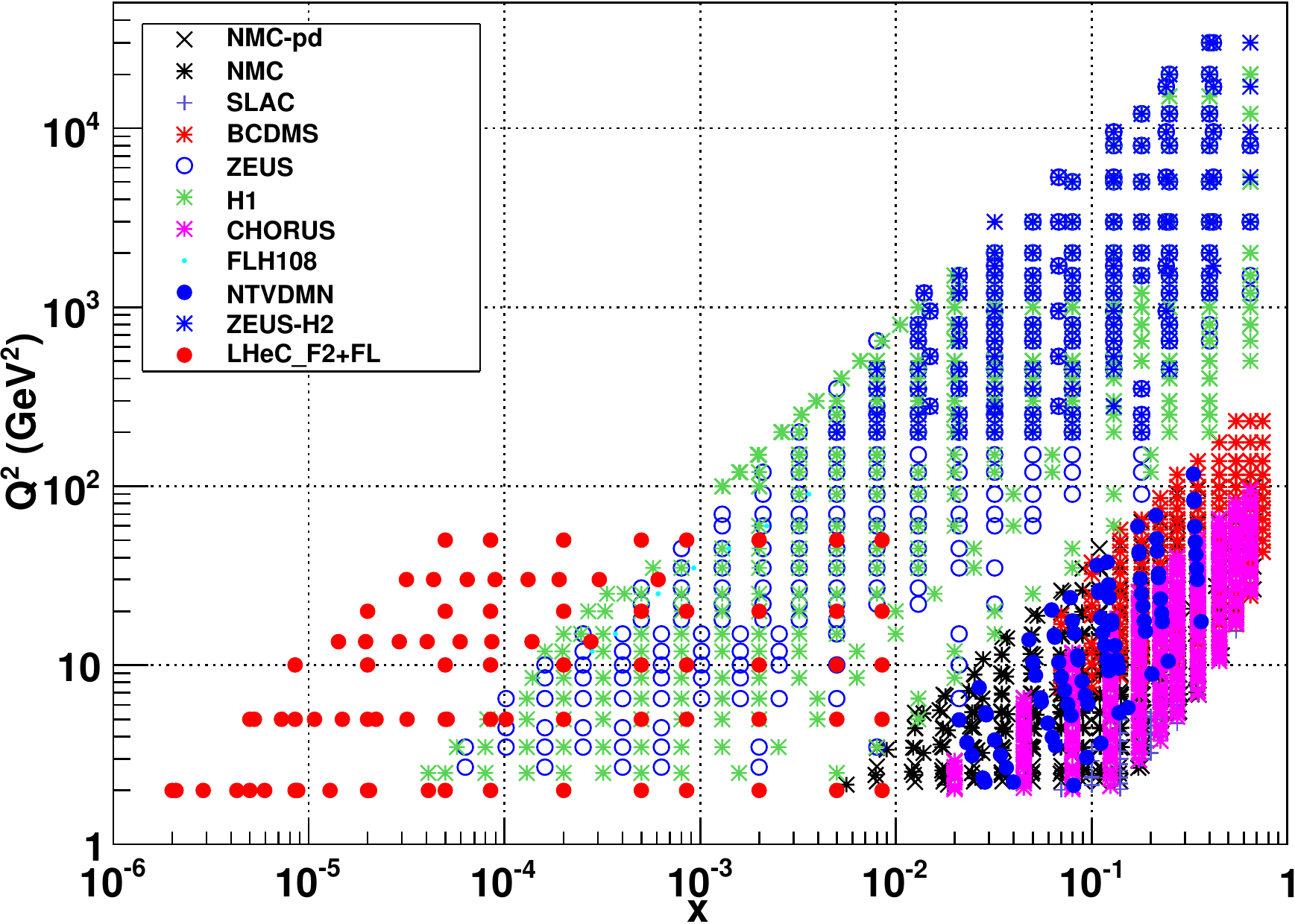}
\end{center}
\caption{The kinematic coverage of the
LHeC pseudodata used in the present studies, together
with the data already included in the 
reference NNPDF1.0 dataset. \label{fig:kin}}
\end{figure}

Given that for all models 
considered in Fig.~\ref{Fig:f2models} 
there are significant flexibilities in the initial parameterisations, 
it is conceivable that upon suitable 
changes of parameters it would be possible to obtain satisfactory fits of 
a wide range of models to the LHeC data.  
It is therefore essential to analyse in more detail the 
ability of the LHeC to distinguish 
unambiguously between different evolution dynamics.
With this aim, a PDF analysis is performed 
including LHeC pseudodata which are generated using different scenarios
 for small-$x$ QCD dynamics.
Pseudodata for $F_2(x,Q^2)$ and $F_L(x,Q^2)$
at small $x$ are considered in a scenario in which the LHeC
machine has
electron energy $E_e=70$ GeV  and electron
acceptance for $\theta_e\le 179^\circ$, for an integrated luminosity of
 $1 \ {\rm fb^{-1}}$. 
The study is carried out in the framework of the 
NNPDF1.0 analysis \cite{Ball:2008by} and includes all HERA and
fixed target data used in that analysis, in addition to LHeC
pseudodata. 
The kinematics of the LHeC pseudodata included in the fit
(together with other data included in the 
original NNPDF1.0 analysis) are shown 
in Fig.~\ref{fig:kin}. 
In order to avoid correlations between low $x$ and 
high $x$ data e.g. through the momentum sum rule constraint, 
only LHeC pseudodata with $x < 10^{-2}$ are considered. 
The average total uncertainty of the simulated
$F_2$ pseudodata is $\sim 2\%$, while that of 
$F_L$ is $\sim 8\%$.

For the NNPDF fits, the
input LHeC pseudodata are
generated not within the DGLAP framework, but rather 
using two
different models which
include saturation effects in the gluon density:
the AAMS09 model~\cite{Albacete:2009fh}, which is based
on non-linear Balitsky-Kovchegov evolution with a running coupling, and the 
FS04 dipole model~\cite{Forshaw:2004vv}. Both of these models deviate 
significantly from linear DGLAP evolution in the LHeC regime.


The global fit using the NNPDF1.0 framework with 
fixed-order DGLAP evolution is repeated, now
including LHeC pseudodata generated using the scenarios including 
saturation effects.
By assessing the quality of the fit with saturated LHeC pseudodata included,
this study tests the sensitivity to parton dynamics beyond fixed-order DGLAP.
The conclusions are the same
for both the AAMS09 and the FS04 models. The DGLAP analysis
yields an acceptable fit
when only the $F_2(x,Q^2)$ LHeC pseudodata are included.
This implies that although
the underlying physical theories are different,
the small-$x$ extrapolations
of AAMS09 and FS04 for $F_2$ are sufficiently similar
to DGLAP-based extrapolations for the differences to be 
absorbed as modifications to the shapes of the non-perturbative
initial conditions for the PDFs
at the starting scale $Q_0^2$ for DGLAP evolution.
More sophisticated analyses,
based for example on sequential kinematic cuts and backwards DGLAP
evolution, as presented in Subsec.~\ref{sec:statusheraintro}, 
could still be applied. However, it seems likely that it 
will not be possible 
unambiguously to establish non-linear effects using LHeC
data on $F_2$ alone.

\begin{figure}[ht]
\begin{center}
\includegraphics[width=0.95\textwidth]{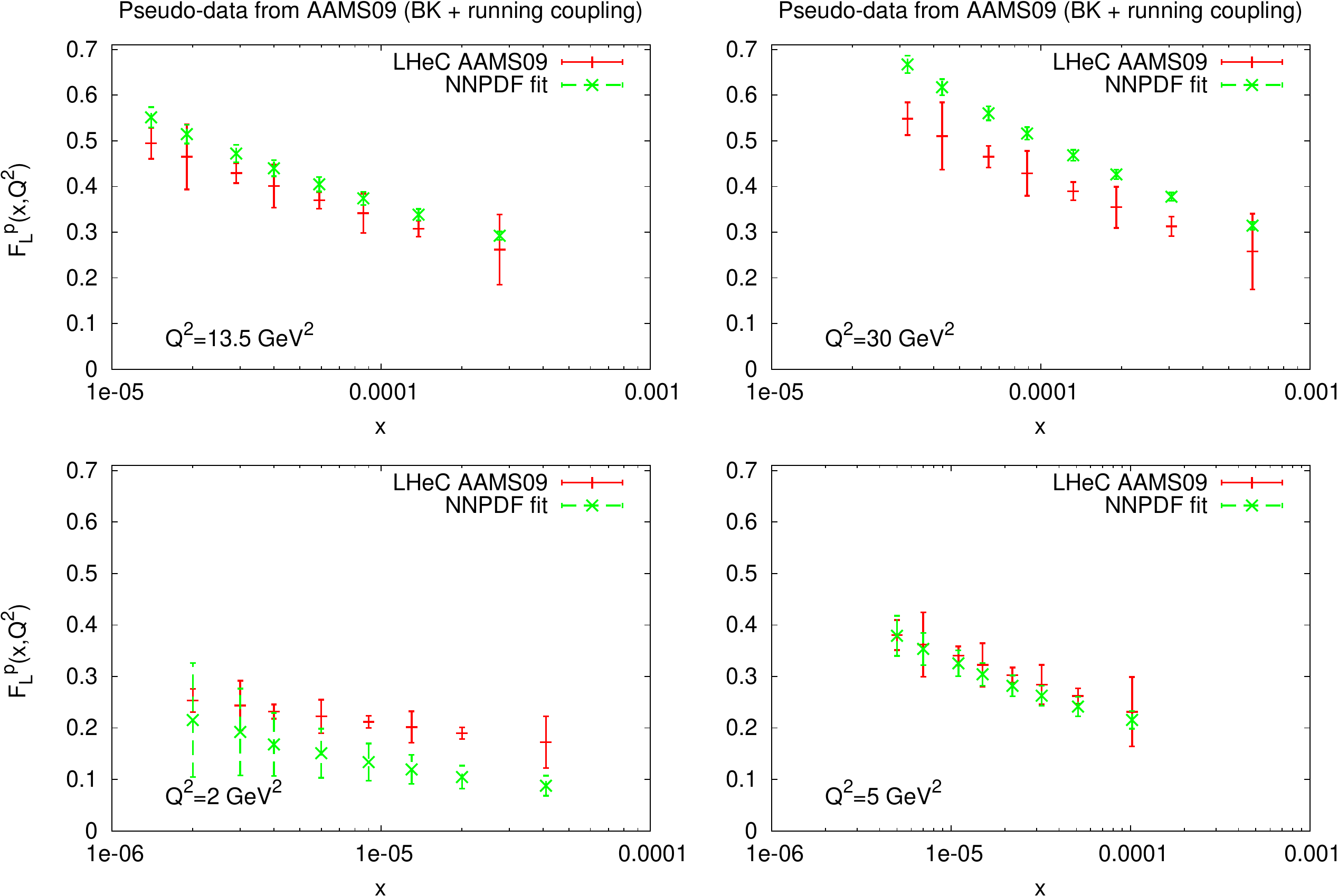}
\end{center}
\caption{\small The results for $F_L$ obtained from the best 
NLO DGLAP fit
to the standard NNPDF1.2 data set, 
together with the
LHeC pseudodata for $F_2(x,Q^2)$ and $F_L(x,Q^2)$
generated with the (saturating)
AAMS09 model. The fit results are compared with the 
input AAMS09 $F_L$ pseudodata. \label{aams}}
\end{figure}

The situation is very different
when data on the longitudinal
structure function $F_L(x,Q^2)$ are included in the NNPDF fit, 
provided the lever-arm in $Q^2$ is large enough
for the gluon sensitivity through the $Q^2$ evolution
of $F_2$ to conflict with that through $F_L$.
The  analysis 
based on linear 
DGLAP evolution fails to reproduce simultaneously $F_2$ and $F_L$ in all the
$Q^2$ bins, and thus the overall $\chi^2$ is very large.  
The effect is illustrated in Fig.~\ref{aams}, where
the best fits from the NNPDF DGLAP analysis are compared with the 
LHeC $F_L$ pseudodata generated from the AAMS09 model. 
This is a  clear
signal for a departure from fixed-order DGLAP of the simulated
pseudodata. 
This analysis
shows that the combined use of $F_2$ and $F_L$ 
data is a very sensitive probe of novel
small-$x$ QCD dynamics, and that their measurement would 
be very likely to
discriminate between different theoretical scenarios.
Using $F_2^{ \rm c}$ data in place of $F_L$ may  
offer a similarly  powerful means of establishing 
deviations from fixed-order linear DGLAP evolution at small $x$.

%% file: physics/tex/predea.tex
The LHeC, as an electron-ion collider in the TeV regime,  will have an enormous potential for measuring the nuclear parton distribution functions at small $x$.
Let us start by a brief explanation of how the pseudodata for inclusive observables in $e$Pb collisions are obtained: To simulate an LHeC measurement
of 
$F_2$ in electron-nucleus collisions, the points $(x,Q^2)$,
generated for $e$(50) + $p$(7000) collisions for a high acceptance, low luminosity scenario, as explained in Section\,\ref{sec:simNC}, are considered. Among them, we keep only those points at small $x\le 0.01$ and
not too large $Q^2<1000$ GeV$^2$ with $Q^2\le sx$, for a Pb beam energy of 2750 GeV per nucleon\footnote{In this document we have restricted the discussion and results to Pb because it is the presently accelerated ion at the LHC. But simulations also exist for a Ca nucleus of 3500 GeV per nucleon, and they can be easily produced for other nuclei as Ar (3150 GeV per nucleon), whose acceleration at the LHC has been discussed as part of the $AA$ program \cite{Jowett:2008hb}.}. Under the assumption that
the instantaneous
luminosity per nucleon is the same in $ep$ and $e$A (see Sections~\ref{sec:eAJowett} and \ref{sec:eALRJowett}), the number of
events is scaled
by a factor $1/(5\times 50\times A)$, with 50 coming from the transition from a
high luminosity to a low luminosity scenario, and 5 being a
crudely estimated 
reduction factor accounting for the shorter running time for ions than for
proton.

At each point of the grid, $\sigma_r$ and $F_2$ are generated using the dipole model of
\cite{GolecBiernat:1998js,Armesto:2002ny} to get the central value. Then, for
every point, the statistical error in $ep$ is scaled by the 
previously mentioned factor
$1/(5\times 50\times A)$, and corrected for the difference in $F_2$ or
$\sigma_r$ between the (Glauberized) 5-flavor GBW model \cite{Armesto:2002ny}
and the model used for the $ep$ simulation. The fractional systematic errors
are taken to be the same as for $ep$ - as
has been achieved in
previous DIS experiments on nuclear 
targets\footnote{A significant difference in the
systematics may eventually come from the different size of the QED radiative
corrections for protons and nuclei, an important point which remains to be
addressed in future studies.}. An analogous procedure is applied 
when obtaining
the nuclear
pseudodata for $F_{2}^c$ and $F_{2}^b$, considering the same tag and
background rejection efficiencies as in the $ep$ simulation.

To generate LHeC $F_L$ pseudodata 
for a heavy ion target, a dedicated 
simulation of $e$ + $p$(2750) collisions has been
performed, at three different energies: $10$, $25$ and $50$ GeV for the electron, with assumed
luminosities 5, 10 and 100 pb$^{-1}$ respectively, see Subsec.~\ref{sec:flong}.  Then, for each point in the simulated grid, $F_L$ values for
protons and nuclei are generated using the (Glauberized) 5-flavor GBW model
\cite{Armesto:2002ny}. The relative uncertainties are taken to be exactly the
same as in the $e$p simulation, as explained above.

\begin{figure}[htb]
\centerline{ \includegraphics[clip=,width=0.5\textwidth,angle=0]{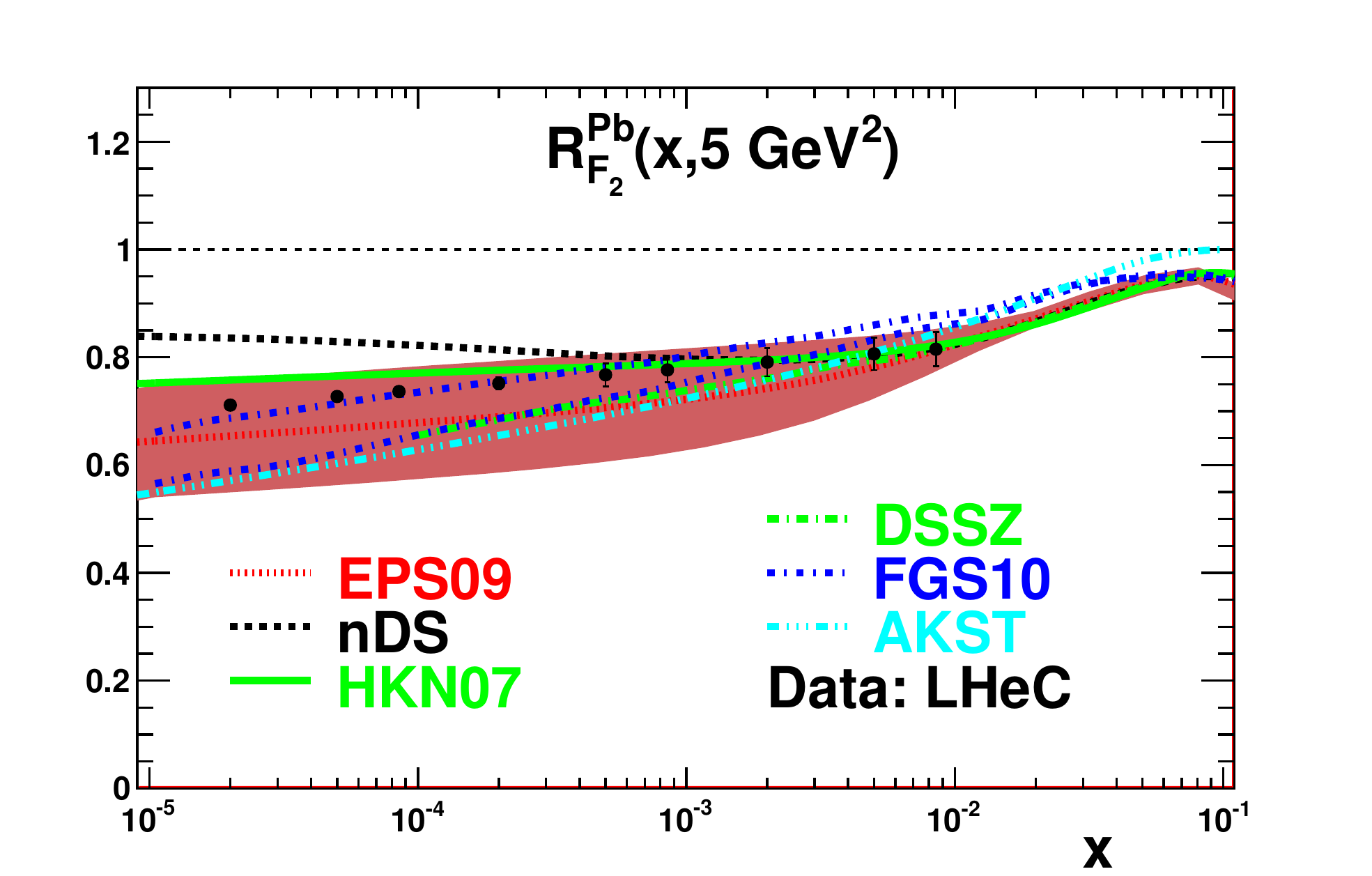}\includegraphics[clip=,width=0.5\textwidth,angle=0]{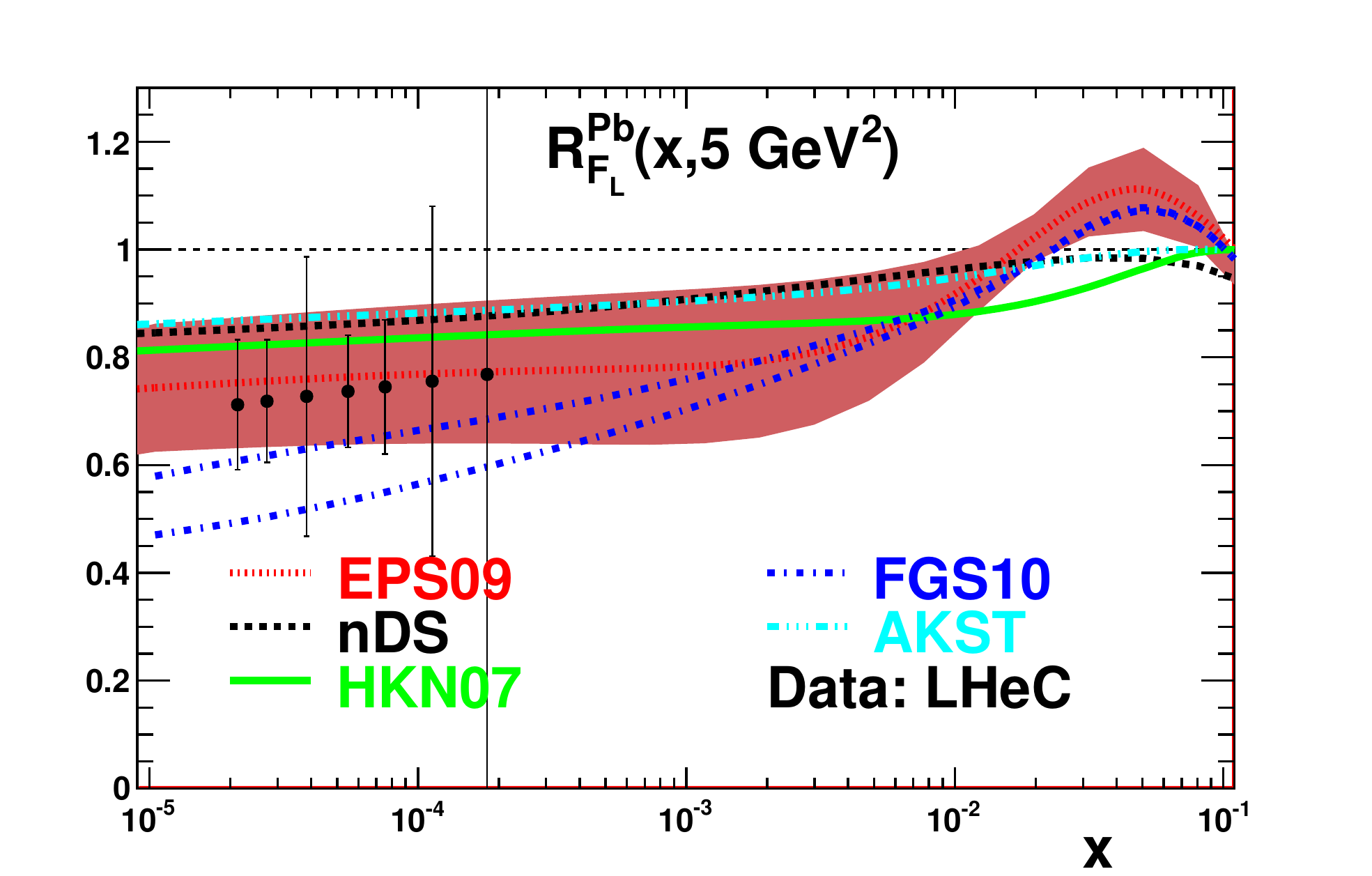}}
\caption{Predictions from different models for the nuclear modification factor, Eq. (\ref{nmf}) for Pb with respect to the proton, for $F_2(x,Q^2=5\ {\rm GeV}^2)$ (plot on the left) and $F_L(x,Q^2=5\ {\rm GeV}^2)$ (plot on the right) versus $x$, together with the corresponding 
LHeC pseudodata. Dotted lines correspond to the nuclear PDF set EPS09 \cite{Eskola:2009uj}, dashed ones to nDS \cite{deFlorian:2003qf}, solid ones to HKN07 \cite{Hirai:2007sx}, dashed-dotted ones to FGS10 \cite{Guzey:2009jr}, dashed-dotted-dotted ones to AKST \cite{Armesto:2010kr} and long dashed-dotted ones to DSSZ \cite{deFlorian:2011fp} (only for $F_2$). The band 
corresponds to the uncertainty in the Hessian analysis in EPS09 \cite{Eskola:2009uj}.}
\label{Fig:famodels}
\end{figure}

In Fig. \ref{Fig:famodels} we show several predictions for the nuclear suppression factor, Eq. (\ref{nmf}), with respect to the proton, for the total and longitudinal structure functions, $F_2$ and $F_L$ respectively, in $e$Pb 
collisions at an example $Q^2=5$ GeV$^2$ and for   $10^{-5} < x < 0.1$. 
Predictions based on global DGLAP analyses 
of existing data at NLO: nDS, HKN07, EPS09 and DSSZ  \cite{deFlorian:2003qf,Hirai:2007sx,Eskola:2009uj,deFlorian:2011fp}, plus those from models using the relation between diffraction and nuclear shadowing, AKST and FGS10 \cite{Armesto:2010kr,Guzey:2009jr}, are shown together with the LHeC pseudodata. Brief explanations on the different models can be found in Subsec.~\ref{sec:nucleartargets}.
Clearly, the accuracy of the data at the LHeC will offer huge possibilities for discriminating between different models and for constraining the dynamics underlying nuclear shadowing at small $x$.

In order to better quantify how the LHeC would improve the present situation concerning nuclear PDFs in global DGLAP analyses (see the uncertainty band in Fig. \ref{Fig:npdfs}), nuclear LHeC pseudodata have been included in the global EPS09 analysis \cite{Eskola:2009uj}. The DGLAP evolution was carried 
out at NLO accuracy, in the variable-flavor-number scheme (SACOT prescription) with the CTEQ6.6 \cite{Nadolsky:2008zw} 
set for free proton PDFs as a baseline. See \cite{Eskola:2009uj} and 
references therein for further details. The only
difference compared with the original EPS09 setup is that one additional gluon parameter, $x_a$, has been varied (this parameter was originally frozen in EPS09), and the only additionally 
weighted data set was the PHENIX data on $\pi^0$ production at mid-rapidity
\cite{Adler:2006wg} in dAu collisions at RHIC.

\begin{figure}[htb]
\centerline{ \includegraphics[clip=,width=0.85\textwidth,angle=0]{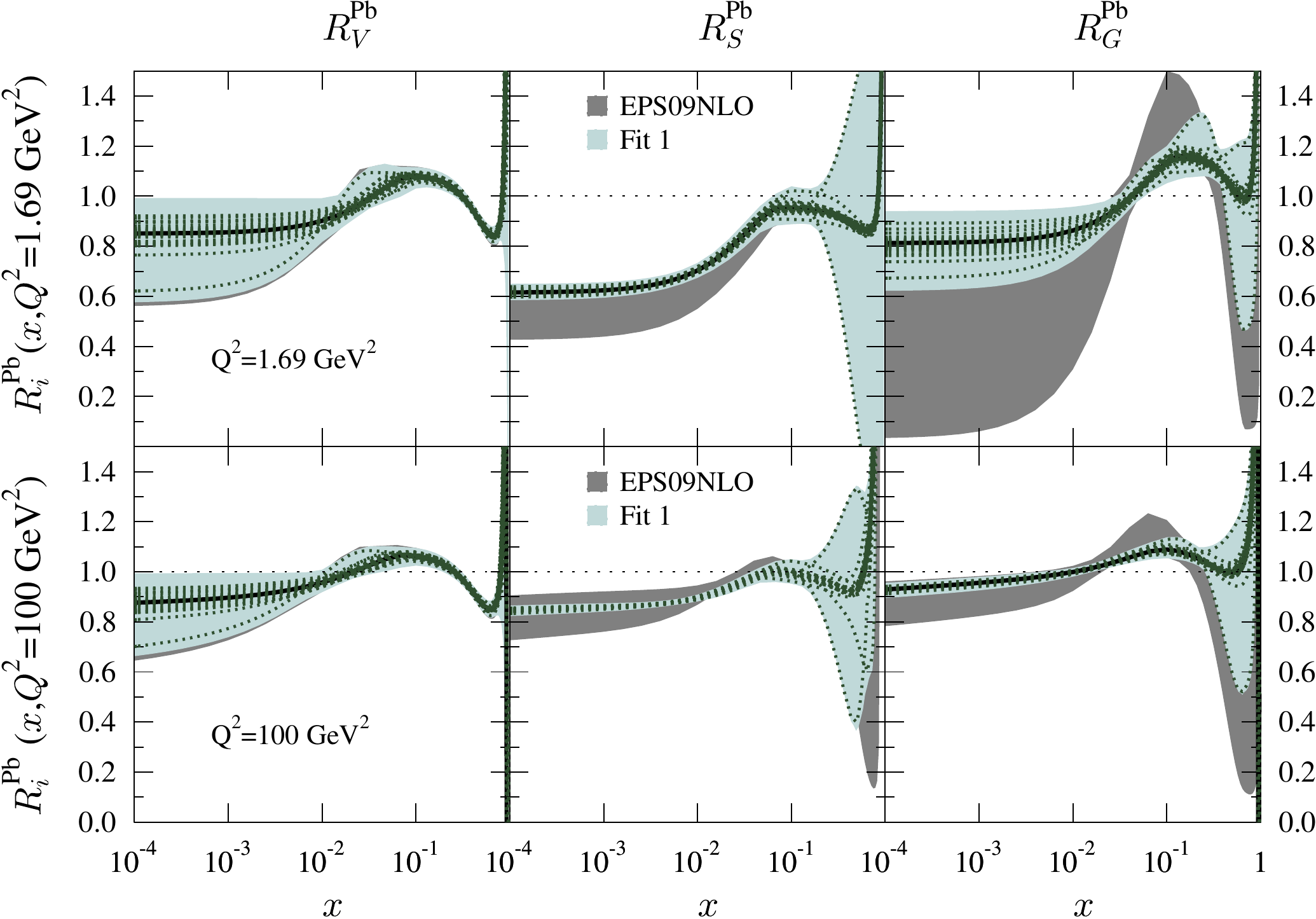}}
\caption{Ratio of parton densities for protons bound in Pb to those in a free proton, for valence $u$ (left), $\bar u$ (middle) and $g$ (right), at $Q^2=1.69$ (top) and 100 (bottom) GeV$^2$. The dark grey band corresponds to the uncertainty band using the Hessian method in the original EPS09 analysis \cite{Eskola:2009uj}, while the light blue band corresponds to the uncertainty obtained after including nuclear LHeC pseudodata on the total reduced cross sections (Fit 1). The dotted lines indicate the values corresponding to the different nPDF sets in the EPS09 analysis \cite{Eskola:2009uj}.}
\label{Fig:ndglap}
\end{figure}

Two different fits have been performed: the first one (Fit 1) includes pseudodata on the total reduced cross section. The results of the fit 
are shown in Fig. \ref{Fig:ndglap} in terms of the nuclear modification
factors for the parton densities. A large improvement in the 
determination of sea quark and gluon densities at small $x$ is evident.

\begin{figure}[htb]
\centerline{ \includegraphics[clip=,width=0.6\textwidth,angle=0]{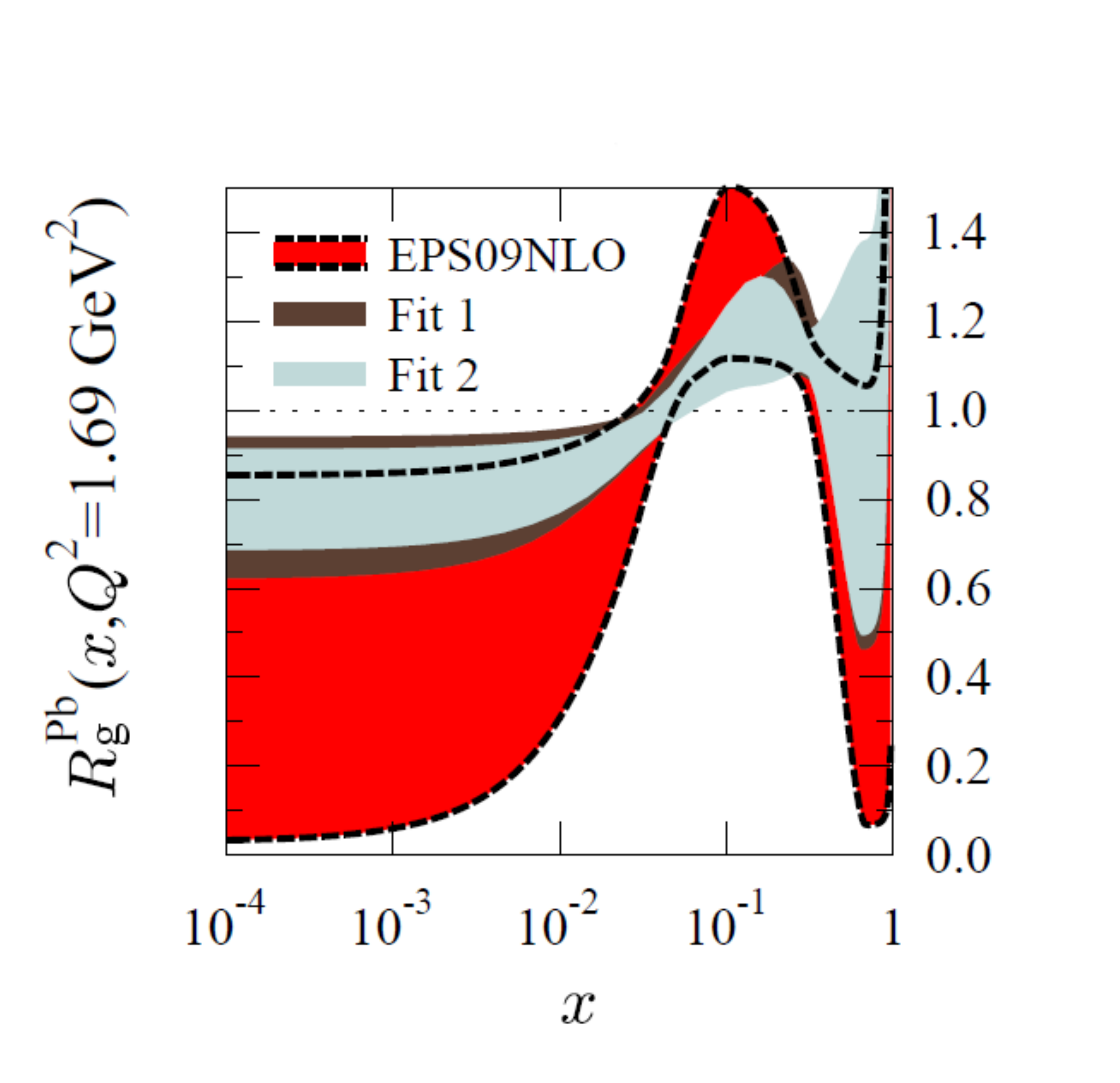}}
\caption{Ratio of the gluon density for protons bound in 
Pb to that of a free proton at $Q^2=1.69$  GeV$^2$. The red band corresponds to the uncertainty using the Hessian method in the original EPS09 analysis \cite{Eskola:2009uj}, while the dark brown band corresponds to the uncertainty obtained after including nuclear LHeC pseudodata on the total reduced cross sections (Fit 1), and the light blue band shows the uncertainty obtained after further including pseudodata on charm and beauty reduced cross sections (Fit 2).}
\label{Fig:ndglap2}
\end{figure}

The second fit (Fit 2) includes not only nuclear LHeC pseudodata on
the total reduced cross section but also on its charm and beauty
components. These data provide direct
information on the nuclear effects on charm and beauty parton
densities, which are generated mainly dynamically from the gluons through DGLAP evolution. Thus, the inclusion of such pseudodata further improves the
determination of the nuclear effects on the gluon at small $x$, as
illustrated in Fig. \ref{Fig:ndglap2}.

In both Figs. \ref{Fig:ndglap} and \ref{Fig:ndglap2} a sizeable reduction of the uncertainties in the sea quark and gluon nuclear parton distributions at large $x>0.1$ can also be observed. This improvement is basically 
due to the constraints imposed by sum rules and to the fact that DGLAP evolution links large and small $x$. Although the study of parton distributions at large $x$ is not the subject of this chapter, it is worth commenting that $F_2$ could be measured in $e$A collisions at the LHeC with a statistical accuracy better than a few percent up to $x\sim 0.6$ but for large $Q^2>1000$ GeV$^2$. On the other hand, flavor decomposition will only be accessible for $x<0.1$. Therefore, the LHeC will provide additional information on the antishadowing ($R>1$, $0.1<x<0.3$) and - with less precision - on the EMC-effect ($R<1$, $0.3<x<0.8$) regions. The latter is valence-dominated and there exist data from fixed target experiments, though at much smaller $Q^2$, so at the LHeC the validity of leading-twist DGLAP evolution will be tested.

Furthermore, the large lever-arm in $Q^2$ opens the possibility of measuring CC events in electron scattering on nuclear targets, thus helping to improve the loose constraints on the flavour decomposition of the nuclear parton densities coming from existing DIS and DY data. In this respect (see the comments in Subsec. \ref{sec:nucleartargets}) the LHeC may help to clarify the issue of the compatibility of the nuclear corrections extracted in neutrino-nucleus collisions with those coming from electron- or muon-nucleus collisions\footnote{Note that the nuclear modifications of the structure function $F_2$ in these two types of process are expected to differ due to the different coupling to quarks \cite{Brodsky:2004qa}.}.

In conclusion, the precision and large lever-arm in $x$ and $Q^2$ of the nuclear data at the LHeC will offer huge possibilities for discriminating different models and for constraining the parton densities in global DGLAP analyses. Besides measurements of the reduced cross section, data on its charm and bottom components and on $F_L$ will help to constrain the nuclear effects on PDFs, see e.g. the recent work in \cite{Cazaroto:2008qh,Armesto:2010tg}.

%% file: physics/tex/excldiffintro.tex
\subsubsection{Introduction}


\begin{figure}
 \centering
  \includegraphics[width=0.70\textwidth]{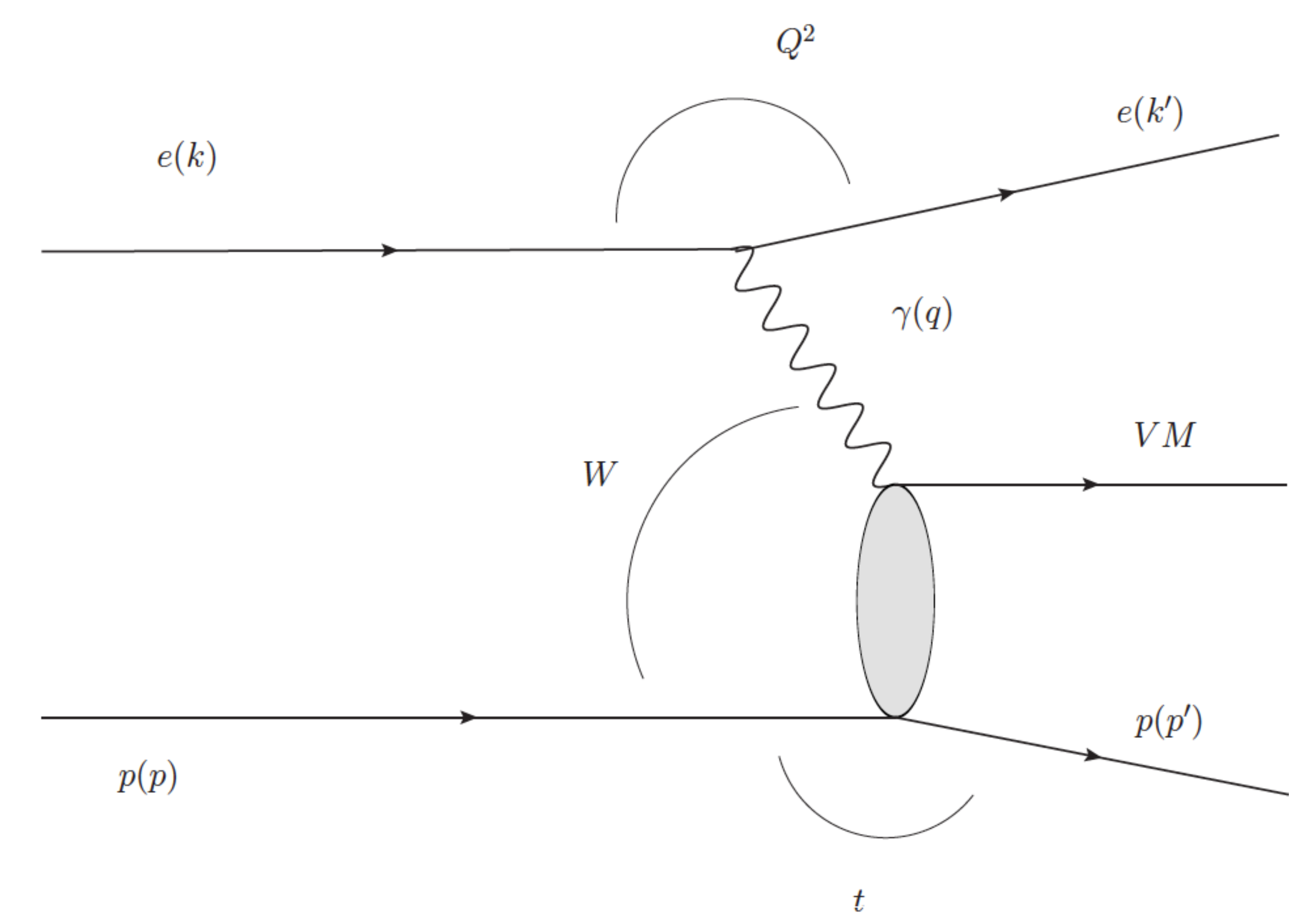}
  \caption{Schematic illustration of the exclusive vector meson
production process and the kinematic variables used to describe it
in photoproduction ($Q^2 \rightarrow 0$) and DIS (large $Q^2$). The
outgoing particle labelled `VM', may be either a vector meson 
with $J^{PC} = 1^{--}$ or a photon.}
\label{vmschematic}
\end{figure}

Exclusive processes such as the electroproduction of vector mesons 
and photons, $\gamma^\ast N \rightarrow VN (V = \rho^0, \phi, \gamma)$, 
or photoproduction of heavy quarkonia, $\gamma N \rightarrow VN 
(V = J/\psi, \Upsilon )$ - see Fig.~\ref{vmschematic} - 
provide information on nucleon structure
and small-$x$ dynamics which is  complementary to that obtained in inclusive
measurements \cite{Frankfurt:2005mc}. 
The exclusive production of $J/\psi$ and $\rho$ mesons in $ep$ collisions
and Deeply-Virtual Compton Scattering
(DVCS, $ep \rightarrow e \gamma p$),
have been particularly prominent in the 
development of our understanding of HERA physics \cite{marage}.

Diffractive channels such as these 
are favourable, since the underlying exchange 
crudely equates to a pair of gluons, making the process sensitive to the 
square of  the gluon density \cite{Martin:2007sb}, 
in place of the linear dependence 
for $F_2$ or $F_L$.
With a sufficiently good theoretical understanding of the exclusive
production mechanism,  
this may enhance substantially the sensitivity to 
non-linear evolution and saturation phenomena.    
As already shown at HERA, 
$J/\Psi$ production in particular is a 
potentially very clean probe of the gluonic structure 
of the hadron \cite{Kowalski:2006hc,Martin:2007sb}.  
The same exclusive processes can be measured in deep inelastic scattering off nuclei, 
where the gluon density is 
modified by nuclear effects \cite{Caldwell:2010zza}.
In addition, exclusive processes give access to the spatial 
distribution of the gluon density, parameterised by the impact 
parameter \cite{Munier:2001nr} of the collision.  
The correlations between the gluons coupling to
the proton contain information on the three-dimensional structure of the
nucleon or nucleus, which is encoded in the Generalised Parton 
Densities (GPDs). The GPDs combine aspects of parton densities and elastic 
form factors and have emerged as a key concept for describing nucleon structure 
in QCD (see \cite{Goeke:2001tz,Diehl:2003ny,Belitsky:2005qn}
for a review). 

Exclusive processes can be treated conveniently 
within the dipole picture described in Subsec.~\ref{sec:statusheraintro}.
In this framework, the cross section can be
represented  as a product of three factorisable terms: the splitting
of an incoming photon into a $q \bar{q}$ dipole; the `dipole'
cross section for the interaction of this $q \bar{q}$ pair with
the proton and, in the case of vector mesons, a wave function term for
the projection of the dipole onto the meson.
As discussed in Subsec.~\ref{sec:statusheraintro} the dipole 
formalism is particularly convenient
since saturation effects can be easily incorporated.
 
\subsubsection{Generalised parton densities and spatial structure}


At sufficiently 
large $Q^2$ the exclusively produced
meson or photon is in a configuration of transverse 
size much smaller than the typical hadronic size, 
$r_\perp \ll R_{\rm hadron}$. As a result its interaction with the target can 
be described using perturbative QCD \cite{Brodsky:1994kf}.
A QCD factorisation theorem \cite{Collins:1996fb} states that the 
exclusive amplitudes in this regime can be factorised into a perturbative QCD 
scattering process and certain universal process-independent 
functions describing the emission and absorption of the active partons 
by the target, the generalised parton distributions (GPDs).


Let us briefly review  (see \cite{Goeke:2001tz,Diehl:2003ny,Belitsky:2005qn}
for details) the definition of GPDs and their relation to the ordinary parton densities discussed in detail in Chapter\,\ref{chapter:qce}.
The parton distributions of the proton (or any other hadron) are given by the diagonal matrix elements
$\langle P,\lambda|\hat{O} | P,\lambda\rangle$, where $P$ and $\lambda$ are the 4-momentum and helicity of the proton, and $\hat{O}$ is a twist-2
quark or gluon operator. However, there is new information in the
GPDs defined in terms of the off-diagonal matrix elements
$\langle P^\prime,\lambda^\prime|\hat{O} |  P,\lambda\rangle$. Unlike the diagonal PDFs, the GPDs cannot be regarded as parton densities,
but are to be interpreted as probability amplitudes.

The physical significance of GPDs is best seen using light-cone coordinates, $z^\pm = (z^0 \pm
z^3)/\sqrt{2}$, and in the light-cone gauge, $A^+ = 0$. It is conventional to define the generalised quark
distributions in terms of quark operators at light-like separation, resulting in
\begin{equation}
\label{eq:gpddef}
F_q(x, \xi, t) 
=\frac{1}{2\bar P^+}\left[ H_q((x, \xi, t)\bar{u}(P^\prime) \gamma^+ u(P)
+  E_q((x, \xi, t)\bar{u}(P^\prime) \frac{i\sigma^{+\alpha}\Delta_\alpha}{2m} u(P)   \right]
\end{equation}
with $\bar P = (P +P^\prime)/2$ and $\Delta= P^\prime - P$, and where we have suppressed the helicity labels of the
protons and spinors. We now have two extra kinematic variables:
$t = \Delta^2$, $\xi= - \Delta^+/(P + P^\prime)^+$.
We see that $-1\le\xi\le 1$. Similarly, we may define GPDs $\tilde H_q$ and $\tilde E_q$ with an additional $\gamma_5$
between the quark operators in Eq.~(\ref{eq:gpddef}); and also an analogous set of gluon GPDs, $H_g$, $E_g$, $\tilde H_g$ and
$\tilde E_g$. These definitions correspond to helicity-conserving GPDs. Analogous definitions exist for helicity-flip (transversity), chiral-odd GPDs $H_T$, $E_T$, $\tilde{H}_T$, $\tilde{E}_T$~\cite{Diehl:2001pm}.

For $P^\prime = P$, $\lambda^\prime =\lambda$  the matrix elements reduce to the ordinary PDFs:
\begin{eqnarray}
&H_q(x, 0, 0) = q(x),\ \  H_q( - x, 0, 0) = -\bar q(x),\ \  H_g(x, 0, 0) = xg(x),&\nonumber \\
&\tilde H_q(x, 0, 0) = \Delta q(x),\ \  \tilde H_q( - x, 0, 0) = \Delta \bar q(x),\ \  \tilde H_g(x, 0, 0) = x\Delta g(x),& \nonumber \\
& H_T(x,0,0) =\Delta_T q(x),&\label{eq:gpdtopdf}
\end{eqnarray}
where $\Delta q$ ($\Delta_T q(x)$) is the difference between quark densities with opposite helicities (transversities). No corresponding relations exist for $E$, $\tilde E$, $E_T$, $\tilde H_T$, $\tilde E_T$ as they decouple
in the forward limit, $\Delta = 0$.
For properties of all these distributions, see the reviews \cite{Goeke:2001tz,Diehl:2003ny,Belitsky:2005qn}.

For the evolution of the GPDs, there are two types of domain: (i) the time-like domain, with $|x| < |\xi|$, where
the GPDs describe the wave functions of a t-channel $q\bar q$ (or gluon) pair and evolve according to
modified ERBL equations \cite{Efremov:1979qk,Lepage:1980fj}; (ii) the space-like domain, with $|x| > |\xi|$, where
the GPDs generalise the familiar $q$, $\bar q$ (and gluon) PDFs and describe DVCS and exclusive vector meson production, and evolve according to modified DGLAP equations.
The splitting functions for the evolution of GPDs are known to NLO \cite{Belitsky:2000yn}.

The GPDs contain new information about proton structure and should be determined from
experiment. We can parameterise them in terms of 'double distributions' \cite{Radyushkin:1998es,Radyushkin:1998bz}, which reduce to
diagonal PDFs as $\xi \to 0$. With an additional physically reasonable 'Regge' assumption of no
extra singularity at $\xi= 0$, GPDs at low $\xi$ are uniquely given in terms of diagonal PDFs to $\cal{O}(\xi)$
\cite{Martin:2009zzb}. Alternatively, flexible $SO(3)$-based parameterisations have been used to determine GPDs
from DVCS data \cite{Kumericki:2009uq}.


The Fourier transform of the GPDs with respect to the transverse 
momentum transferred to the nucleon describes the transverse spatial 
distribution of partons (illustrated in Fig. \ref{Fig:transprof}) with a given longitudinal momentum fraction 
$x$ \cite{Burkardt:2000za,Diehl:2002he,Ralston:2001xs}. 
The transverse spatial distributions of quarks and gluons are fundamental 
characteristics of the nucleon, which reveal the size of the configurations
in its partonic wave function and allow the study of the non-perturbative 
dynamics governing their change with $x$, such as Gribov diffusion,
chiral dynamics, and other phenomena. The nucleon transverse
gluonic size is also an essential input in studies of saturation
at small $x$. It determines the initial conditions of the non-linear
QCD evolution equations and thus directly influences the impact
parameter dependence of the saturation scale for the 
nucleon \cite{Kowalski:2003hm,Rogers:2003vi}, which in turn 
predicates its nuclear enhancement \cite{Kowalski:2007rw}. 
Information on the nucleon transverse quark and gluon distributions 
is further required in the phenomenology of high-energy $pp$ collisions
with hard processes, including those with new particle production,
where it determines the underlying event structure (centrality dependence) 
in inclusive scattering \cite{Frankfurt:2003td} and the rapidity gap 
survival probability in hard single diffraction \cite{Aaron:2010su}
and central exclusive 
diffraction \cite{Frankfurt:2006jp,Deile:2010mv}. In view of its 
considerable interest, the transverse quark/gluon imaging of the nucleon with 
exclusive processes has been recognised as an important objective of 
nucleon structure and small-$x$ physics.

Mapping the transverse spatial distribution of quarks and gluons 
requires measurement of the 
$t$-dependence of hard 
exclusive processes up to large values of  $|t|$, of the order of 
$1 \, {\rm GeV}^2$.
Studies of the $Q^2$-dependence and comparisons between different
channels provide crucial tests of the reaction mechanism and the
universality of GPDs. Vector meson production at small $x$ and 
heavy quarkonium photoproduction at high energies probe the 
gluon GPD of the target, while real photon production (DVCS) involves the singlet quark as well as the 
gluon GPDs. Measurements of exclusive $J/\psi$ photo/electroproduction 
\cite{Chekanov:2004mw,Aktas:2005xu}
and $\rho^0$ and $\phi$ electroproduction at HERA have confirmed the 
applicability of the factorised QCD description through several 
model-independent tests, and have provided basic information on the 
nucleon gluonic size in the region $10^{-4} < x < 10^{-2}$ and its 
change with $x$ \cite{Frankfurt:2005mc}. Measurements of 
DVCS at HERA \cite{Aaron:2007cz,Chekanov:2008vy} hint that the transverse 
distribution of singlet quarks may 
extend further than that of gluons. 
While these experiments have given important insight 
into transverse nucleon structure, 
the interpretation of the HERA data is limited by the low statistics
which preclude a fully differential analysis. 
A major source of systematic uncertainty at larger $t$ arises from the lack
of a complete separation between elastically scattered protons and
proton excitations, illustrating the importance of good scattered
proton detection at the LHeC. 

As discussed in the following, 
the LHeC would enable a comprehensive program of gluon and singlet quark
transverse imaging through exclusive processes, with numerous 
applications to nucleon structure and small-$x$ physics.
The high statistics would permit fully differential measurements
of exclusive channels, as needed to 
understand the reaction mechanism.
For example, measurements of the $t$-distributions for fixed $x$
differentially in $Q^2$ are needed to confirm the dominance of small-size
configurations. The LHeC would also push such measurements
to the region $Q^2 \sim {\rm few} \times 10 \, {\rm GeV}^2$ 
where finite-size (higher-twist) effects are small and the
effects of QCD evolution can be cleanly identified. 
Measurements of gluonic exclusive channels ($J/\psi, \phi, \rho^0$)
at the LHeC would provide gluonic transverse images of the nucleon 
down to $x \sim 10^{-6}$ with unprecedented accuracy, testing
theoretical ideas about diffusion dynamics in the wave function.
Because exclusive cross sections are proportional to the
square of the gluon GPD (i.e. the gluon density),
such measurements would also offer new insight into non-linear
effects in QCD evolution, and enable new tests of the approach
to saturation by measuring the impact parameter dependence of
the saturation scale.
Along these lines, saturation effects in the 
exclusive vector meson production on protons and nuclei have been studied in \cite{Marquet:2009vs,Caldwell:2010zza,Lappi:2010dd,Horowitz:2011jx}.
Furthermore, measurements of DVCS would provide 
additional information on the nucleon singlet quark size
and its dependence on $x$. Besides its intrinsic interest for
nucleon structure and small-$x$ physics, this information would
greatly advance our theoretical understanding of the transverse 
geometry of high-energy $pp$ collisions at the LHC. We note that 
these exclusive measurements at the LHeC would complement similar 
measurements at moderately small $x$ ($0.003 < x < 0.2$) with 
the COMPASS experiment at CERN and in the valence region $x > 0.1$ with 
the JLab 12 GeV Upgrade, providing a comprehensive picture of the
nucleon spatial structure.

Further interesting information comes from hard exclusive measurements
accompanied by the diffractive dissociation of the nucleon,
$\gamma^\ast N \rightarrow V + Y$ ($Y = $ low-mass 
proton dissociation state).
The ratio of inelastic to elastic diffraction in these processes 
provides information on the quantum fluctuations of the gluon
density, which reveals the quantum-mechanical nature of the 
non-perturbative colour fields in the nucleon and can be related
to dynamical models of low-energy nucleon 
structure \cite{Frankfurt:2008vi}. HERA results 
are in qualitative agreement with such model predictions but do
not permit a quantitative analysis.
These measurements of exclusive diffraction at the LHeC, and similar ones for $e$A collisions, would allow for detailed quantitative studies of all these new aspects of nucleon and nuclear structure.

%% file: physics/tex/vm.tex

\begin{figure}
  \begin{minipage}[t]{0.5\textwidth}
    (a)\hfill$\,$\\
    \includegraphics[width=\textwidth]{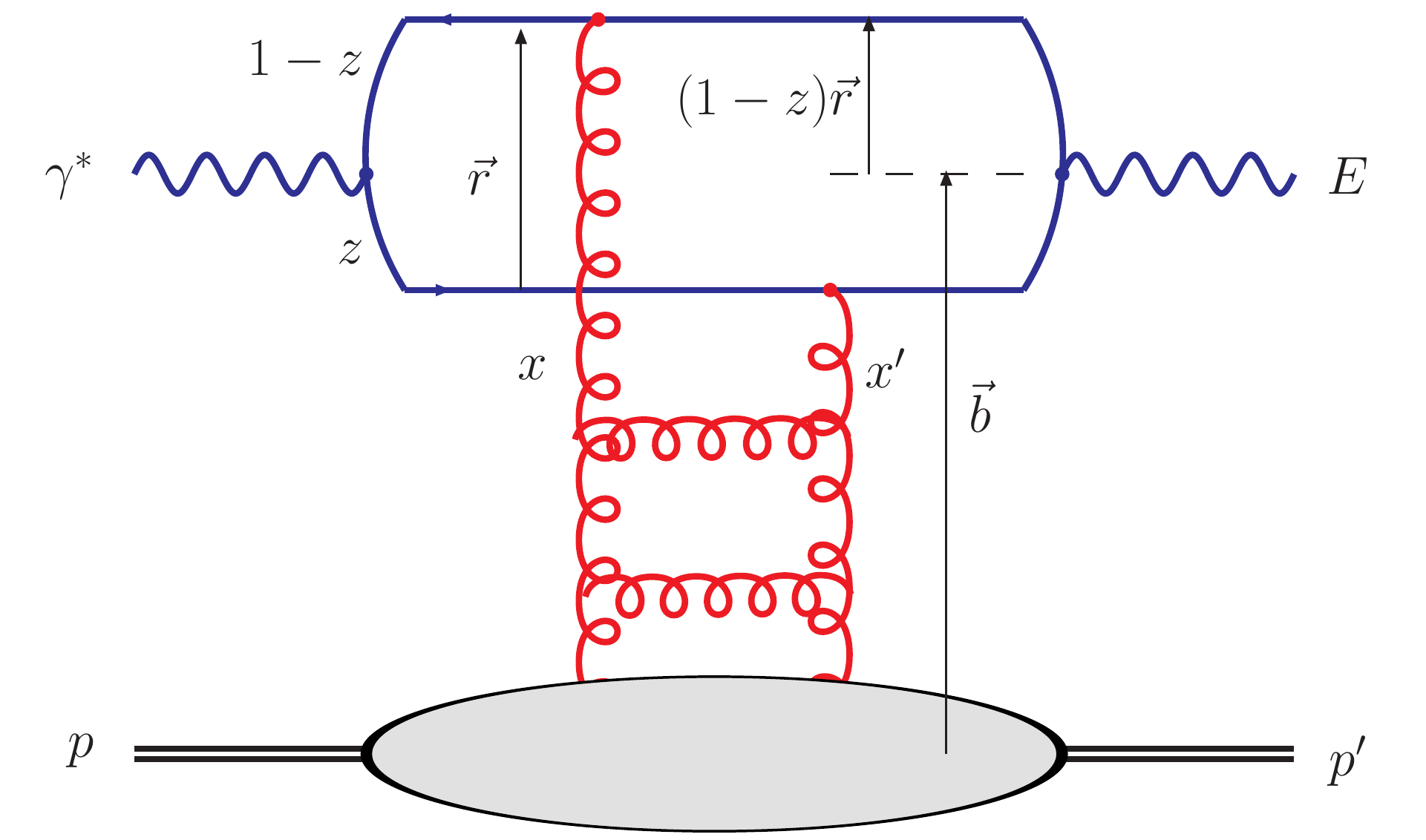}
  \end{minipage}\hfill
  \begin{minipage}[t]{0.5\textwidth}
    (b)\hfill$\,$\\
    \includegraphics[width=\textwidth]{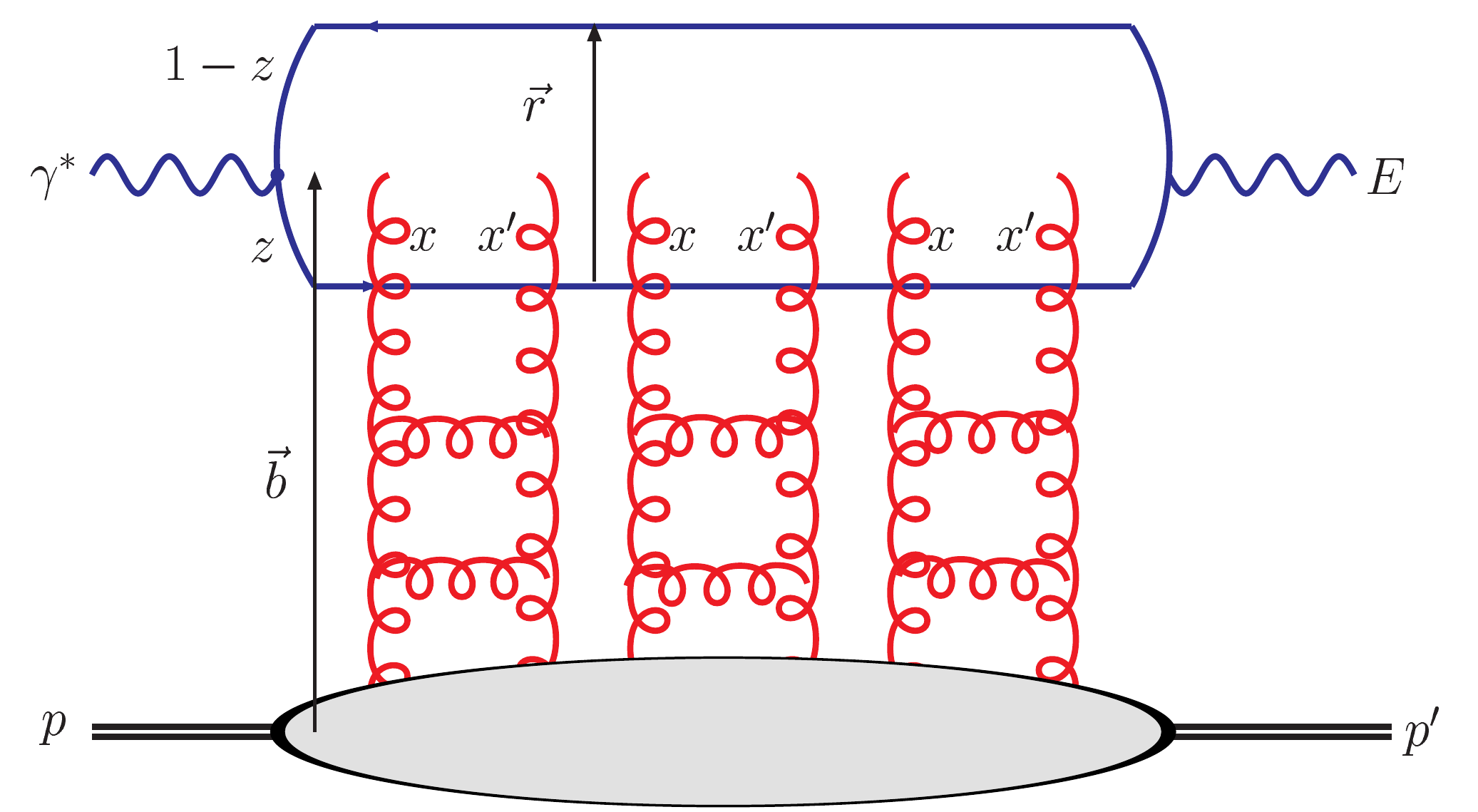}
  \end{minipage}
    \caption{Parton level diagrams representing the $\gamma^*p$ scattering amplitude proceeding via (a)~single-Pomeron and (b)~multi-Pomeron exchange, where the perturbative QCD Pomeron is represented by a gluon ladder.  For exclusive diffractive processes, such as vector meson production ($E=V$) or DVCS ($E=\gamma$), we have $x^\prime\ll x\ll 1$ and $t=(p-p^\prime)^2$.  
These diagrams are related through the optical theorem to inclusive DIS, 
where $E=\gamma^*$, $x^\prime=x\ll 1$ and $p^\prime=p$.}
  \label{fig:diagrams}
\end{figure}

For the exclusive production of vector mesons, a QCD factorisation theorem has been demonstrated (for $\sigma_L $) in \cite{Brodsky:1994kf}.  The dipole model follows from this QCD factorisation theorem in the LO approximation. 
Within the dipole model, see Subsec.~\ref{sec:statusheraintro}, the amplitude for the exclusive diffractive production of a particle $E$, $\gamma^* p\to Ep$, shown in Fig.~\ref{fig:diagrams}(a), can be expressed as
\begin{equation} \label{eq:exclamp}
  \mathcal{A}^{\gamma^* p\rightarrow E+p}_{T,L}(x,Q,\Delta) = \mathrm{i}\,\int\!\mathrm{d}^2\boldsymbol{r}\int_0^1\!\frac{\mathrm{d}{z}}{4\pi}\int\!\mathrm{d}^2\boldsymbol{b}\;(\Psi_{E}^{*}\Psi)_{T,L}\;\mathrm{e}^{-\mathrm{i}[\boldsymbol{b}-(1-z)\boldsymbol{r}]\cdot\boldsymbol{\Delta}}\;\frac{\mathrm{d}\sigma_{q\bar q}}{\mathrm{d}^2\boldsymbol{b}} \; .
\end{equation}
Here  $E=V$ for  vector meson production, or $E=\gamma$ for  deeply virtual Compton scattering (DVCS).
In Eq. (\ref{eq:exclamp}), $z$ is the fraction of the photon's light-cone momentum carried by the quark, $r=|\boldsymbol{r}|$ is the transverse size of the $q\bar{q}$ dipole, while $\boldsymbol{b}$ is the impact parameter, that is, $b=|\boldsymbol{b}|$ is the transverse distance from the centre of the proton to the centre-of-mass of the $q\bar{q}$ dipole; see Fig.~\ref{fig:diagrams}(a).  The transverse momentum lost by the outgoing proton, $\boldsymbol{\Delta}$, is the Fourier conjugate variable to the impact parameter $\boldsymbol{b}$, and $t\equiv(p-p^\prime)^2=-\Delta^2$.  The forward overlap function between the initial-state photon wave function and the final-state vector meson or photon wave function in Eq.~\eqref{eq:exclamp} is denoted $(\Psi_E^*\Psi)_{T,L}$, while the factor $\exp[\mathrm{i}(1-z)\boldsymbol{r}\cdot\boldsymbol{\Delta}]$ originates from the non-forward wave function~\cite{Bartels:2003yj}.  The differential cross section for an exclusive diffractive process is obtained from the amplitude, Eq.~\eqref{eq:exclamp}, by
\begin{equation} \label{eq:dsdt}
  \frac{\mathrm{d}\sigma^{\gamma^* p\rightarrow E+p}_{T,L}}{\mathrm{d}t} = \frac{1}{16\pi}\left\lvert\mathcal{A}^{\gamma^* p\rightarrow E+p}_{T,L}\right\rvert^2,
\end{equation}
up to corrections from the real part of the amplitude and from 
skewedness ($x^\prime\ll x \ll 1$ for the variables shown in 
figure~\ref{fig:diagrams}a).  Taking the imaginary part of the forward scattering amplitude immediately gives the formula for the total $\gamma^*p$ cross section (or equivalently, the proton structure function $F_2=F_T+F_L$) via the
optical theorem:
\begin{equation} \label{eq:inclamp}
  \sigma^{\gamma^* p}_{T,L}(x,Q) = {\rm Im}\,\mathcal{A}^{\gamma^* p\rightarrow \gamma^*p}_{T,L}(x,Q,\Delta=0) = \sum_f \int\!\mathrm{d}^2\boldsymbol{r} \int_0^1\!\frac{\mathrm{d} z}{4\pi}(\Psi^{*}\Psi)_{T,L}^f\,\int\!\mathrm{d}^2\boldsymbol{b}\;\frac{\mathrm{d}\sigma_{q\bar q}}{\mathrm{d}^2\boldsymbol{b}} \; .
\end{equation}
The dipole picture therefore provides a unified description of both exclusive diffractive processes and inclusive DIS at small $x$.

The unknown quantity common to Eqs.~\eqref{eq:exclamp} and \eqref{eq:inclamp} is the $b$-dependent dipole--proton cross section, 
\begin{equation} \label{eq:dsdb}
  \frac{\mathrm{d}\sigma_{q\bar q}}{\mathrm{d}^2\boldsymbol{b}} = 2\;\mathcal{N}(x,r,b) \; ,
\end{equation}
where $\mathcal{N}$ is the imaginary part of the dipole--proton scattering amplitude, which can vary between zero and one, with $\mathcal{N}=1$ corresponding to  the unitarity (``black disk'') limit.  The scattering amplitude $\mathcal{N}$ encodes the information about the details of the strong interaction between the dipole and the target (proton or nucleus). It is generally parameterised according to some theoretically-motivated functional form, with the parameters fitted to data.  Most dipole models assume a factorised $b$ dependence, $\mathcal{N}(x,r,b) = T(b)\,\mathcal{N}(x,r)$, with $\mathcal{N}(x,r)\in[0,1]$ and, for example, $T(b)=\Theta(R_p-b)$, so that the $b$-integrated $\sigma_{q\bar q}=(2 \pi R_p^2)\,\mathcal{N}(x,r)$.  
However, the ``saturation scale'' is strongly dependent on impact 
parameter and the chosen of $b$-dependence must be made consistent 
with the $t$-dependence of exclusive diffraction at HERA. 
This matching is complicated by the non-zero effective ``Pomeron slope'' 
$\alpha_{\mathbb{P}}^\prime$ measured at HERA, which implies a correlation 
between the $x$- and $b$- dependences of $\mathcal{N}(x,r,b)$.  
Therefore, for accurate results, $\mathcal{N}(x,r,b)$ should be determined from the simultaneous description of inclusive DIS and exclusive diffractive processes.

An impact-parameter-dependent saturation (``b-sat'') model~\cite{Kowalski:2003hm,Kowalski:2006hc} has been shown to describe very successfully a broad range of HERA data on exclusive diffractive vector meson ($J/\psi$, $\phi$, $\rho$) production and DVCS (see also the rather different approach in \cite{Marquet:2007qa}), including almost all aspects of the $Q^2$, $W$ and $t$ dependence with the exception of $\alpha_{\mathbb{P}}^\prime$, together with the inclusive structure functions $F_2$, $F_2^{c\bar{c}}$, $F_2^{b\bar{b}}$ and $F_L$.
The ``b-Sat'' parameterisation is based on LO DGLAP evolution of an initial gluon density, $xg(x,\mu_0^2) = A_g\,x^{-\lambda_g}\,(1-x)^{5.6}$, with a Gaussian 
impact parameter dependence, $T(b) \propto \mathrm{exp}(-b^2/2B_G)$.  The dipole scattering amplitude is parameterised as
\begin{equation} \label{eq:bsat}
  \mathcal{N}(x,r,b) = 1-\exp\left(-\frac{\pi^2}{2N_c}r^2\alpha_S(\mu^2)\,xg(x,\mu^2)\,T(b)\right),
\end{equation}
where the scale $\mu^2=4/r^2+\mu_0^2$, $B_G=4$ GeV$^{-2}$ was fixed from the $t$-slope of exclusive $J/\psi$ photoproduction at HERA, and the other three parameters ($\mu_0^2=1.17$~GeV$^2$, $A_g=2.55$, $\lambda_g=0.020$) were fitted to ZEUS $F_2$ data with $x_{\mathrm{Bj}}\le 0.01$ and $Q^2\in[0.25,650]$ GeV$^2$~\cite{Kowalski:2006hc}.  The eikonalised dipole scattering amplitude of Eq.~\eqref{eq:bsat} can be expanded as
\begin{equation} \label{eq:expand}
  \mathcal{N}(x,r,b) = \sum_{n=1}^{\infty}\;\frac{(-1)^{n+1}}{n!}\;\left[\frac{\pi^2}{2N_c}r^2\alpha_S(\mu^2)\,xg(x,\mu^2)\,T(b)\right]^n,
\end{equation}
where the $n$-th term in the expansion corresponds to $n$-Pomeron exchange; for example, the case $n=3$ is illustrated in Fig.~\ref{fig:diagrams}(b).  The terms with $n>1$ are necessary to ensure unitarity.

\subsubsection{Simulations of LHeC elastic 
{\boldmath $J/\psi$} and {\boldmath $\Upsilon$} production}



Due to the extremely
clean final states produced, 
the relatively low effective $x$-values ($x_{\rm eff} \sim (Q^2 + m_V^2) / (Q^2 + W^2)$) 
and scales ($Q^2_{\rm eff} \sim (Q^2 + m_V^2) / 4$)
accessed \cite{Ryskin:1992ui,Martin:2007sb}, and the experimental
possibility of varying both $W$ and $t$ over wide
ranges, 
$J/\psi$ photoproduction ($Q^2 \rightarrow 0$)
may offer the cleanest 
available signature to study 
the transition between the dilute and dense regimes of small-$x$ partons.
It should be possible to detect the muons
from $J/\psi$ or $\Upsilon$ decays with 
acceptances extending to
within $1^\circ$ of the beam pipe
with dedicated muon chambers on the outside of the 
experiment. Depending on the electron
beam energy, 
this makes invariant photon-proton masses $W$ of well beyond
$1 \ {\rm TeV}$ accessible. 

For the analysis presented here we concentrate on the 
photoproduction limit, where the HERA data are most precise 
due to the large cross sections and where unitarity effects 
are most important. 
Studies have also been made at larger $Q^2$ \cite{PRN:EDS}, 
where the extra hard scale 
additionally allows a perturbative treatment of exclusive light 
vector meson (e.g.~$\rho$, $\omega$, $\phi$) production.  Again, 
perturbative unitarity effects are expected to be important for 
light vector meson production when $Q^2\gtrsim1$~GeV$^2$ is not too large.

LHeC pseudodata for elastic $J/\psi$ and $\Upsilon$ photoproduction
and electroproduction
have been generated 
using the DIFFVM Monte Carlo generator \cite{diffvm}
under the assumption of $1^\circ$ acceptance
and a variety of luminosity scenarios. The DIFFVM generator involves 
a simple Regge-based parameterisation of the dynamics and a 
full treatment of decay angular distributions. Statistical 
uncertainties are estimated for each data point. Systematic
uncertainties are hard to estimate without a detailed simulation
of the muon identification and reconstruction 
capabilities of the detector, but are likely to be at least as good as the 
$10 \%$ measurements typically achieved for the elastic 
$J/\psi$ at HERA.
   
\begin{figure}
\begin{center}
  \includegraphics[width=0.7\textwidth]{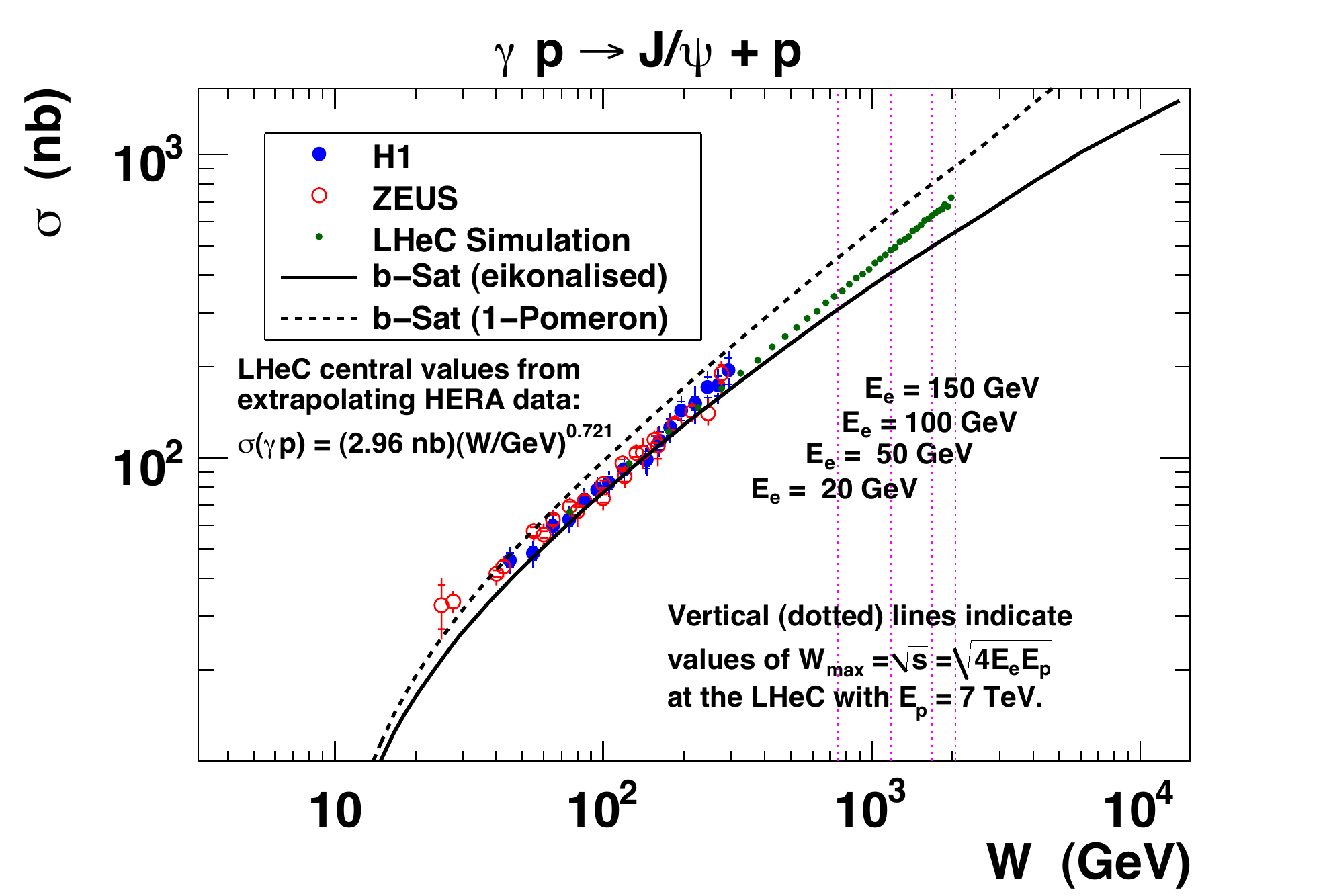}
  \includegraphics[width=0.7\textwidth]{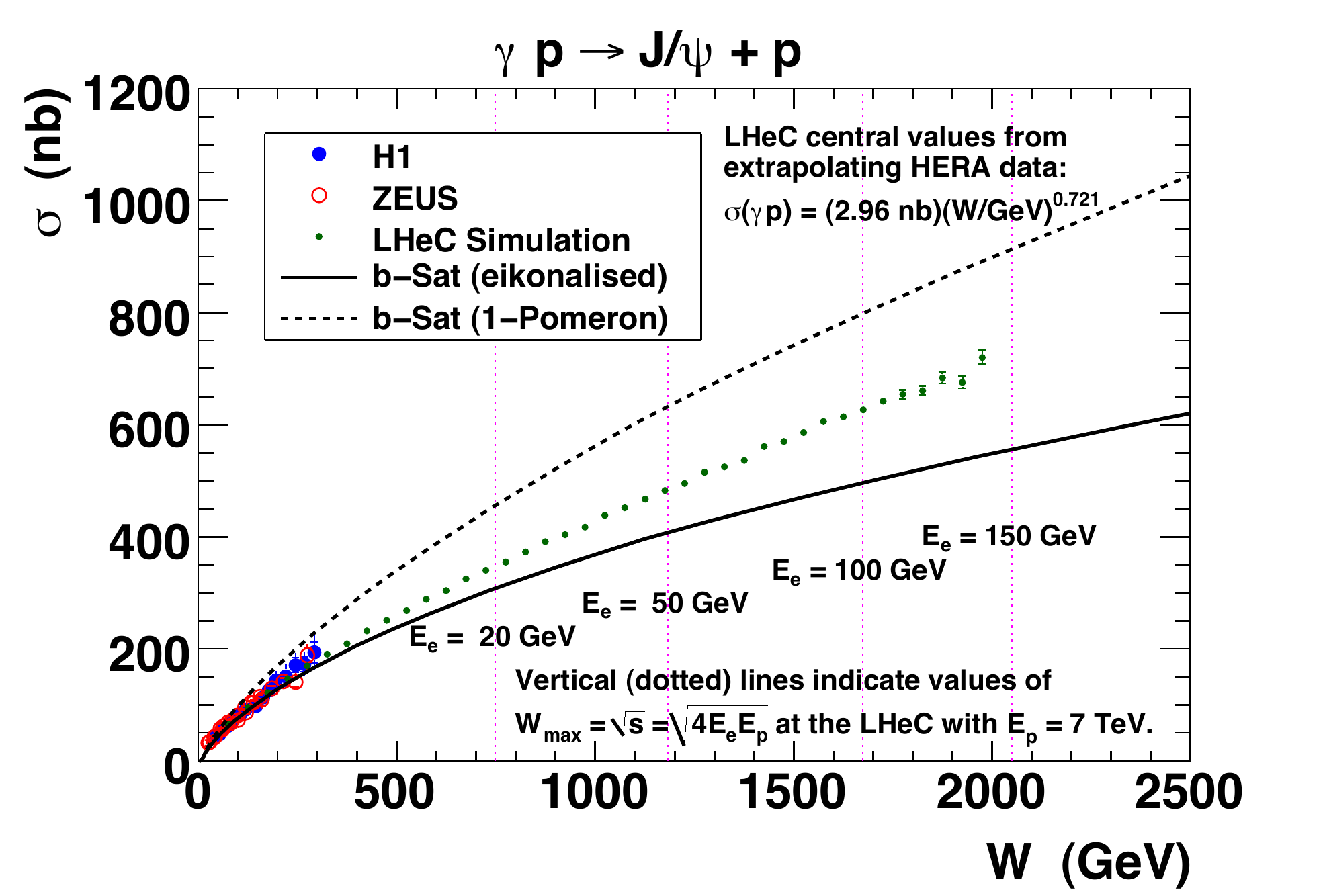}
\end{center}
  \caption{LHeC exclusive $J/\psi$ photoproduction pseudodata, as a function of the $\gamma p$ centre-of-mass energy $W$, plotted on a (top)~log--log scale and (bottom)~linear--linear scale.  The difference between the solid and dashed curves indicates the size of unitarity corrections according to the b-Sat dipole model.}
  \label{fig:jpsi}
\end{figure}

The plots in Fig.~\ref{fig:jpsi} show $t$-integrated predictions for exclusive $J/\psi$ photoproduction ($Q^2=0$) obtained from Eqs.~\eqref{eq:exclamp} and \eqref{eq:dsdt}, using the eikonalised ``b-Sat'' dipole scattering amplitude given in Eq.~\eqref{eq:bsat} together with a ``boosted Gaussian'' vector meson wave function~\cite{Forshaw:2003ki,Kowalski:2006hc}.  Also shown is the single-Pomeron exchange contribution obtained by keeping just the first ($n=1$) 
term in the expansion of Eq.~\eqref{eq:expand}, 
such that the scattering amplitude 
is linearly dependent 
on the gluon density, without refitting any of the input parameters.  %
%

The difference between the ``eikonalised'' and ``1-Pomeron'' predictions therefore indicates the importance of unitarity corrections, which increase significantly with rising $\gamma p$ centre-of-mass energy $W$.  The maximum kinematic limit accessible at the LHeC, $W=\sqrt{s}$, is indicated for different options for electron beam energies ($E_e$) and not accounting for the angular acceptance of the detector.  The 
most precise HERA data~\cite{Chekanov:2002xi,Aktas:2005xu} are overlaid, together with sample LHeC pseudodata points,
assuming $1^\circ$ muon acceptance, 
with the errors (statistical only) given by an LHeC simulation with $E_e=150$~GeV. The central values of the LHeC pseudodata points were obtained from a Gaussian distribution with the mean given by extrapolating a power-law fit to the HERA data~\cite{Chekanov:2002xi,Aktas:2005xu} and the standard deviation given by the statistical errors from the LHeC simulation.  The plots in Fig.~\ref{fig:jpsi} show that the errors on the LHeC pseudodata are much smaller than the difference between the ``eikonalised'' and ``1-Pomeron'' predictions.  Therefore, exclusive $J/\psi$ photoproduction at the LHeC may be an ideal observable for investigating unitarity corrections at a perturbative scale provided by the charm-quark mass.

\begin{figure}
\begin{center}
  \includegraphics[width=0.7\textwidth]{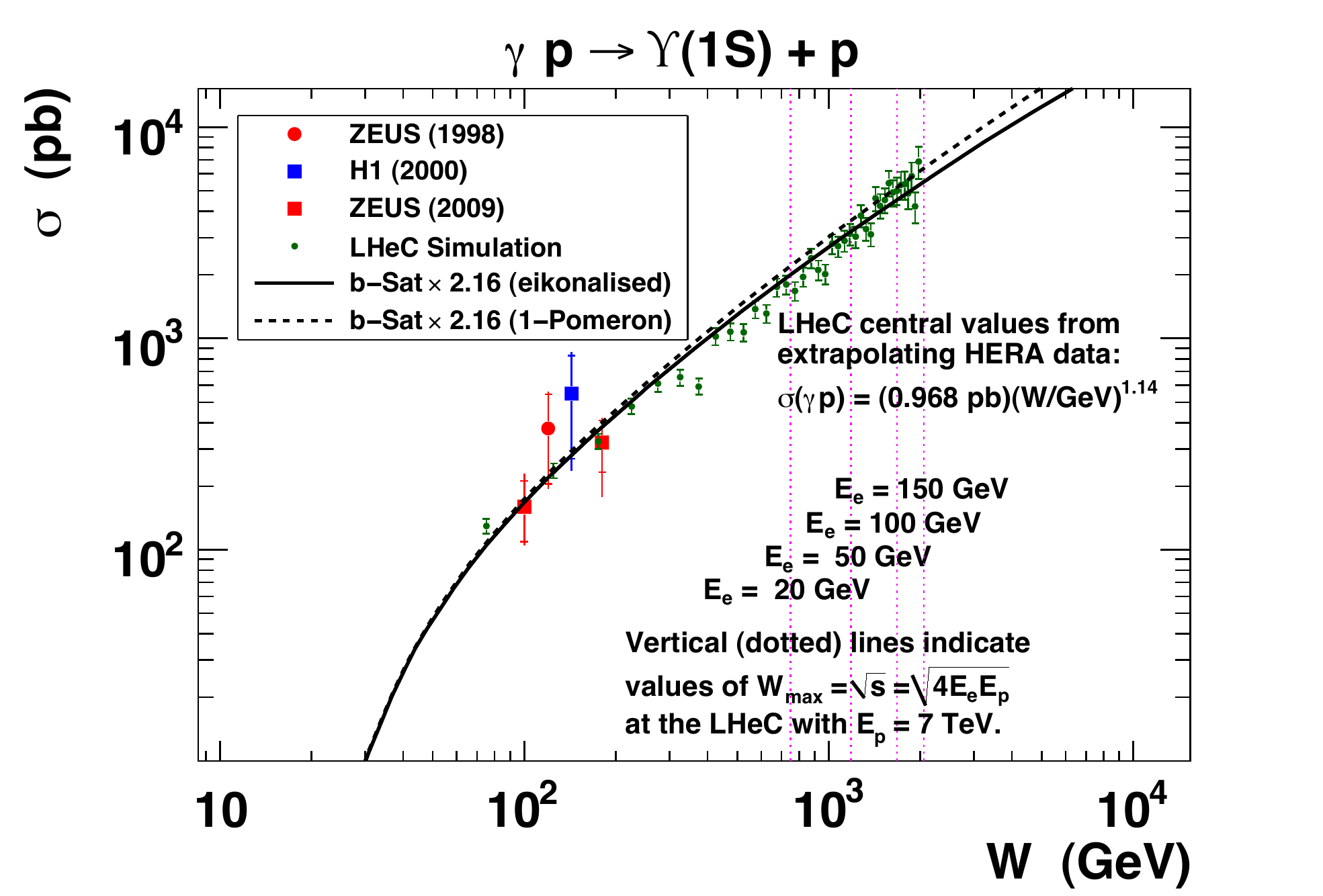}
\end{center}
  \caption{LHeC exclusive $\Upsilon$ photoproduction pseudodata, 
as a function of the $\gamma p$ centre-of-mass energy $W$, plotted on a
log--log scale.
The difference between the solid and dashed curves indicates the size of unitarity corrections according to the 
b-Sat model.  The b-Sat theory predictions have been scaled by a factor 2.16 to best-fit the existing HERA data.}
  \label{fig:upsilon}
\end{figure}

Similar plots for exclusive $\Upsilon$ photoproduction are shown in Fig.~\ref{fig:upsilon}.  Here, the unitarity corrections are smaller than for $J/\psi$ production due to the larger scale provided by the bottom-quark mass and therefore the smaller typical dipole sizes $r$ being probed.  The simulated LHeC pseudodata points also have larger statistical errors than for $J/\psi$ production due to the much smaller cross sections. Nonetheless, the simulations indicate that
a huge improvement in kinematic range and precision is possible compared
with the very sparse $\Upsilon$ data from 
HERA~\cite{Breitweg:1998ki,Adloff:2000vm,Chekanov:2009zz}.

In order to achieve a satisfactory description of the experimental data on exclusive $\Upsilon$ photoproduction, an additional normalisation factor of $\sim 2$ has to be included in the dipole calculation (a similar factor is required for other calculations using the dipole model, see 
for example Ref.~\cite{Cox:2009ag}). This normalisation factor does not arise from any theoretical considerations. Therefore, the dipole model prediction for the $\Upsilon$ in diffractive exclusive processes in DIS still poses significant theoretical questions which cannot be resolved without LHeC data.


%

%

\begin{figure}
  \centering
  \begin{minipage}[t]{0.6\textwidth}
    (a)\hfill$\,$\\
    \includegraphics[width=0.9\textwidth]{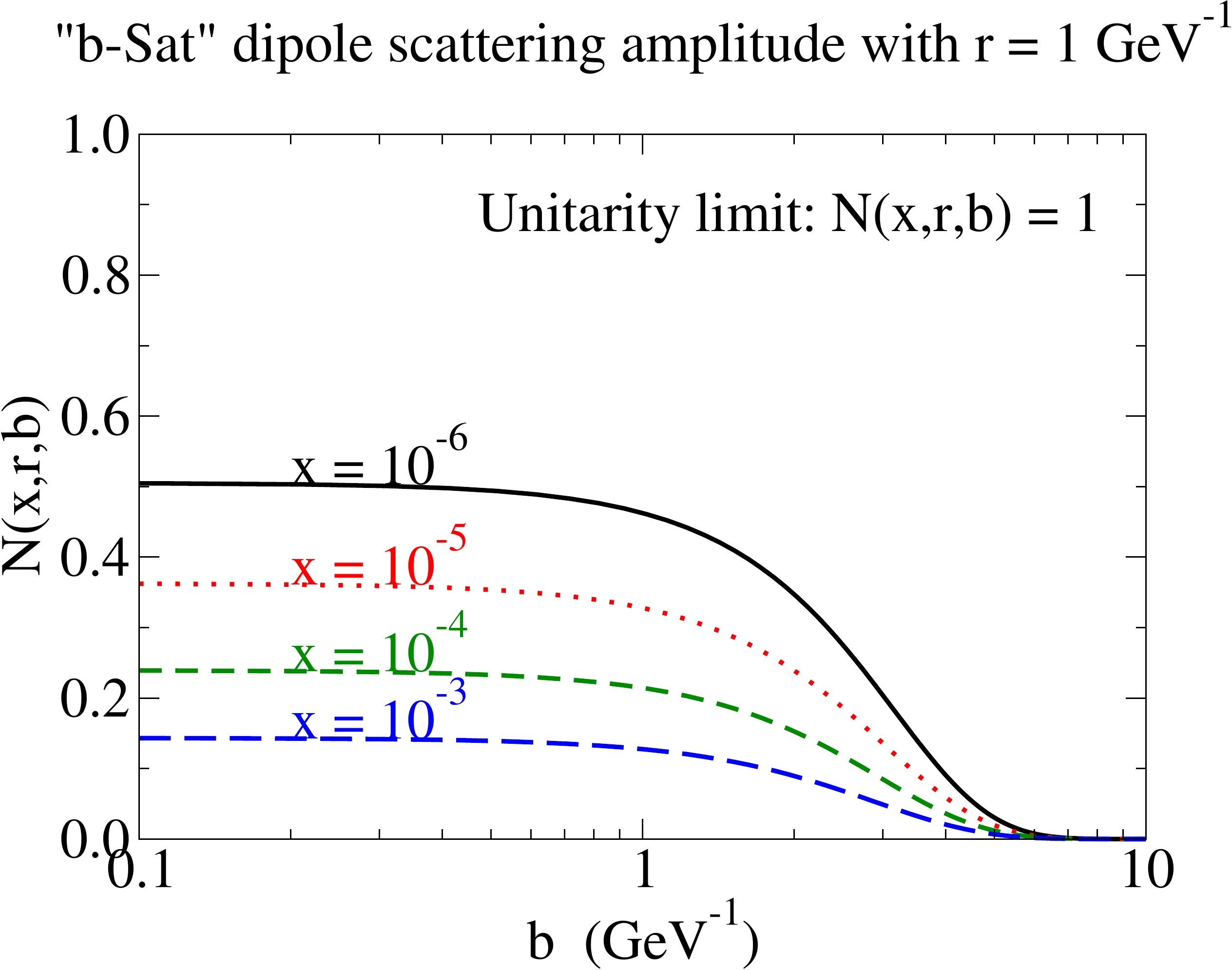}
  \end{minipage}\hfill
  \begin{minipage}[t]{0.35\textwidth}
    (b)\hfill$\,$\\
    \includegraphics[width=\textwidth,clip]{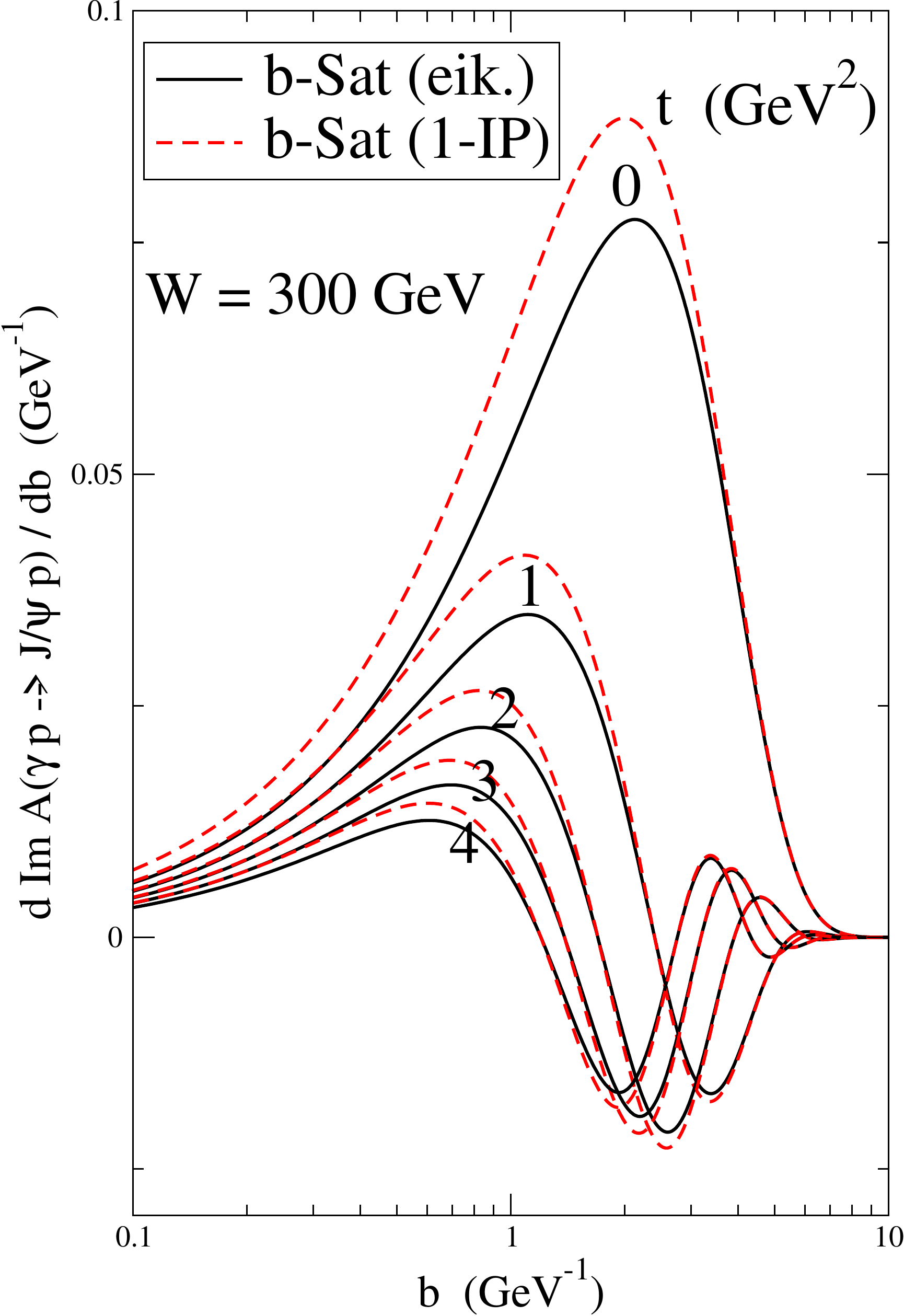}
  \end{minipage}
  \caption{(a)~The imaginary part of the dipole scattering amplitude, $\mathcal{N}(x,r,b)$, as a function of the impact parameter $b$, for fixed values of dipole size $r=1$~GeV$^{-1}$ (typical for exclusive $J/\psi$ photoproduction) and different $x$ values.  (b) The ($r$-integrated) amplitude - the integrand of Eq. (\ref{eq:exclamp}) - for exclusive $J/\psi$ photoproduction as a function of $b$, for $W=300$ GeV and $|t|=0,1,2,3,4$~GeV$^2$.}
  \label{fig:bdep}
\end{figure}

The cross sections shown in Figs.~\ref{fig:jpsi} and \ref{fig:upsilon} 
are integrated over $t\equiv(p-p^\prime)^2=-\Delta^2$, where 
$\boldsymbol{\Delta}$ is the Fourier conjugate variable 
to the impact parameter $\boldsymbol{b}$.  
One expects that at high centre-of-mass energies (small $x$), 
saturation effects are most important close to the centre of the proton 
(small $b$), where the interaction region is densest. 
This is illustrated   in Fig.~\ref{fig:bdep}(a)  where 
the b-Sat model dipole scattering amplitude is shown as a function of $b$ for various $x$ values.
By measuring exclusive diffraction in bins of $|t|$ one can extract the impact parameter profile of the interaction region.  This is illustrated in Fig.~\ref{fig:bdep}(b) where the integrand
of Eq.~(\ref{eq:exclamp}) is shown for different values of $t$ as a function of impact parameter. Clearly for large values of $|t|$, small values of $b$ 
are probed in the impact parameter profile, corresponding to the
most densely populated region, where saturation effects should be most clearly
visible.  Indeed, the eikonalised dipole model of Eq.~\eqref{eq:bsat} leads to ``diffractive dips'' in the $t$-distribution of exclusive $J/\psi$ photoproduction at large $|t|$ 
(reminiscent of the dips seen in the $t$-distribution of the  
proton-proton elastic cross section), departing from the exponential fall-off in the $t$-distribution seen with single-Pomeron exchange~\cite{Kowalski:2003hm}.  The HERA experiments have only been able to make precise measurements of exclusive $J/\psi$ photoproduction at relatively small $|t|\lesssim1$~GeV$^2$, and no significant departure from the exponential fall-off, $\mathrm{d}\sigma/\mathrm{d}t\sim \exp(-B_D|t|)$, has been observed.  

\begin{figure}
 \centering
  \includegraphics[width=0.6\textwidth]{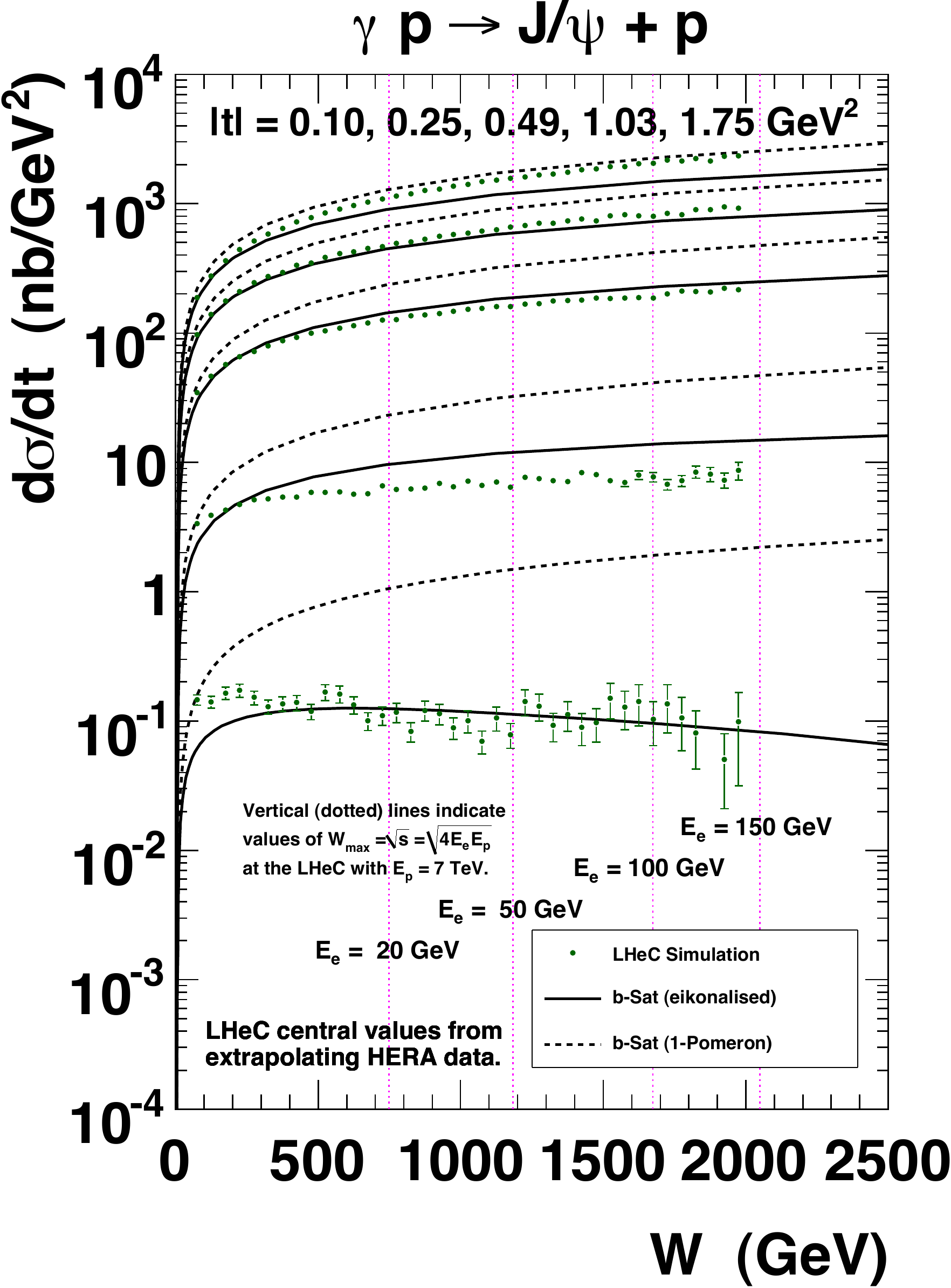}
  \caption{Simulated LHeC measurements of the $W$-dependence of exclusive 
$J/\psi$ photoproduction at the LHeC, differentially in bins  of 
$|t|=0.10, 0.20, 0.49, 1.03, 1.75\;  {\rm GeV}^2$.  The difference between the solid and dashed curves indicates the size of unitarity corrections 
according to the b-Sat dipole model.
The central values of the LHeC pseudodata points were obtained from a Gaussian distribution with the mean given by extrapolating a parameterisation of HERA data and the standard deviation given by the statistical errors from the LHeC simulation with $E_e=150$~GeV.  The $t$-integrated cross section ($\sigma$) as a function of $W$ for the HERA parameterisation was obtained from a power-law fit to the data from both ZEUS~\cite{Chekanov:2002xi} and H1~\cite{Aktas:2005xu}, then the $t$-distribution was assumed to behave as $\mathrm{d}\sigma/\mathrm{d}t = \sigma\cdot B_D\exp(-B_D|t|)$, with $B_D=[4.400+4\cdot 0.137\,\log(W/90{\rm~GeV})]$~GeV$^{-2}$ obtained from a linear fit to the values of $B_D$ versus $W$ given by both ZEUS~\cite{Chekanov:2002xi} and H1~\cite{Aktas:2005xu}.}
\label{plots:graeme2}
\end{figure}

In Fig.~\ref{plots:graeme2}, LHeC pseudodata on 
the differential cross section 
${\rm d} \sigma / {\rm d} t$ is shown as a function of the energy $W$ in different bins of $t$ for the case of exclusive $J/\Psi$ production.
Again two different b-Sat model scenarios 
are shown, with unitarisation effects and with 
single Pomeron exchange. Already for small values of $|t|\sim 0.2 \; {\rm GeV}^2$  and low values of electron energies there is a large discrepancy between the models. The LHeC simulated data still have very small errors in this regime, and can clearly distinguish between the different models.
The differences are of course amplified for large $t$ and large 
electron beam energies.
However the precision of the data deteriorates at large $t$.

Summarising, it is clear that the precise measurements of large-$|t|$ exclusive $J/\psi$ photoproduction at the LHeC would have significant sensitivity to unitarity effects.

%% file: physics/tex/gpds_lhec.tex
\subsubsection{Simulations of deeply virtual compton scattering at the LHeC}


Simulations of the DVCS measurement possibilities with the 
LHeC
have been made using the Monte Carlo generator MILOU \cite{Perez:2004ig}, in the 
`FFS option', for which the DVCS cross section is estimated 
using the model of Frankfurt, Freund and Strikman \cite{Frankfurt:1997at}.
A $t$-slope of $B=6$ GeV$^{-2}$ is assumed. 

The $ep\to e \gamma p$ DVCS cross section is estimated in various
scenarios for the electron beam energy and the  statistical
precision of the measurement is estimated for different integrated luminosity
and detector acceptance choices. 
Detector acceptance cuts at either $1^\circ$ or $10^\circ$
are placed on the polar angle of the final state electron and photon. 
Based on experience with controlling backgrounds in HERA DVCS 
measurements \cite{Aaron:2007cz,Chekanov:2008vy,:2009vda}, an additional
cut is placed on the 
transverse momentum $P_T^\gamma$ of the final state photon.


\begin{figure}
 \begin{center}
\centerline{ \includegraphics[clip=,width=0.9\textwidth]{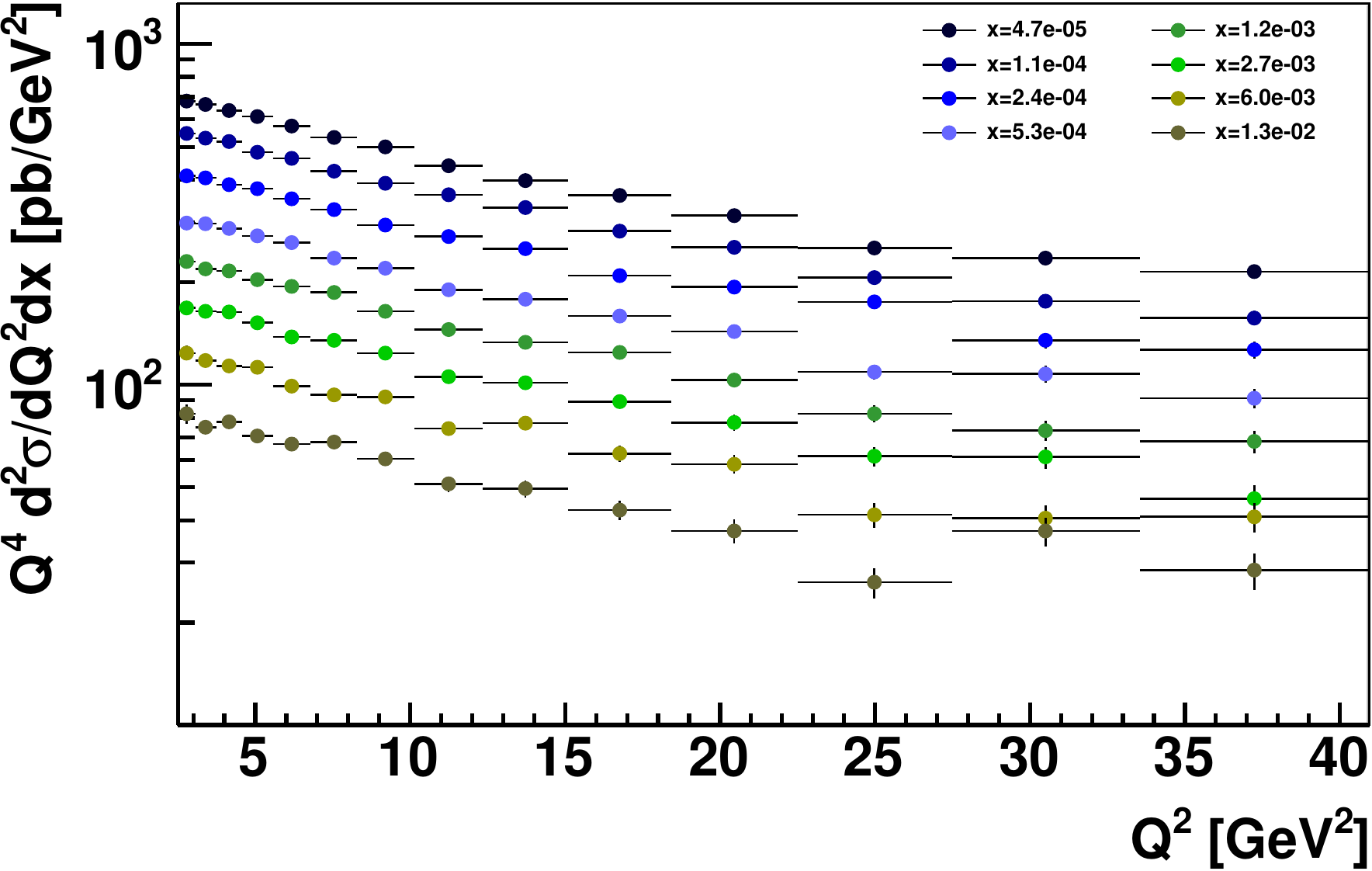}}
  \caption{Simulated LHeC measurement of the
DVCS cross section multiplied by $Q^4$ 
for different
$x$ values for a luminosity of $1 \ {\rm fb^{-1}}$, with 
$E_e = 50 \ {\rm GeV}$, and electron and photon 
acceptance extending to within $1^\circ$ of the beam pipe with a
cut at $P_T^\gamma = 2 \ {\rm GeV}$. Only statistical uncertainties
are considered.}
  \label{fig:RR2}
 \end{center}
\end{figure}

The kinematic limitations due to the scattered electron acceptance
follow the 
same patterns as for the inclusive cross section (see Section~\ref{sec:epincl}).
The photon $P_T^\gamma$ cut is found to be a further important
factor in the $Q^2$ acceptance, with measurements at 
$Q^2 < 20 \ {\rm GeV^2}$ almost completely impossible for a 
cut at $P_T^\gamma > 5 \ {\rm GeV}$, even in the scenario with 
detector acceptances reaching $1^\circ$. 
If this cut is relaxed to $P_T^\gamma > 2 \ {\rm GeV}$,
it opens the available phase space towards the lowest $Q^2$ and $x$ 
values permitted by the electron 
acceptance.

A simulation of a
possible LHeC DVCS measurement double differentially in $x$ and
$Q^2$ is shown in Fig.~\ref{fig:RR2} for a very modest luminosity
scenario ($1 \ {\rm fb^{-1}}$) 
in which the electron beam energy is $50 \ {\rm GeV}$, the 
detector acceptance extends to $1^\circ$ and photon 
measurements are possible down to $P_T^\gamma = 2 \ {\rm GeV}$. 
High precision is possible throughout the region 
$2.5 < Q^2 < 40 \ {\rm GeV^2}$
for $x$ values extending down to $\sim 5 \times 10^{-5}$. The need to
measure
DVCS therefore places constraints on the detector performance for low 
transverse momentum photons, which in practice translates into 
the electromagnetic calorimetry noise conditions and response linearity 
at low energies.

\begin{figure}
 \begin{center}
\centerline{ \includegraphics[clip=,width=0.9\textwidth]{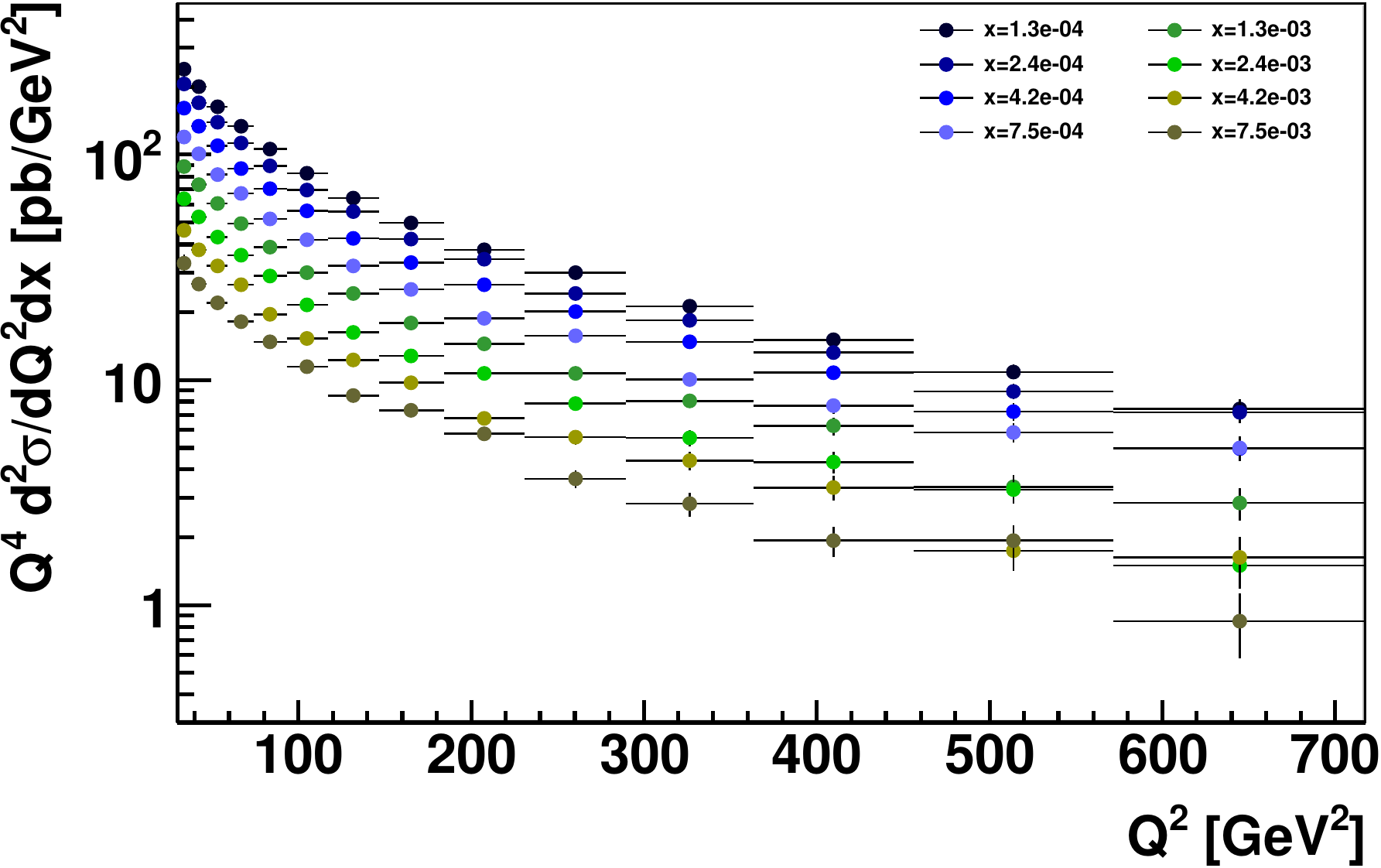}}
  \caption{Simulated LHeC measurement of the
DVCS cross section multiplied by $Q^4$ 
for different
$x$ values for a luminosity of $100 \ {\rm fb^{-1}}$, with 
$E_e = 50 \ {\rm GeV}$, and electron and photon 
acceptance extending to within $10^\circ$ of the beam pipe with a
cut at $P_T^\gamma = 5 \ {\rm GeV}$. Only statistical uncertainties
are considered.}
  \label{fig:RR3}
 \end{center}
\end{figure}

If the detector acceptance extends to only $10^\circ$, the
$P_T^\gamma$ cut no longer plays such an important role.
Although the low $Q^2$ acceptance is lost in this scenario, the 
larger luminosity will allow precise measurements for $Q^2 \stackrel{>}{_{\sim}} 50 \ {\rm GeV^2}$, a region which is not well covered in the 
$1^\circ$ acceptance scenario due to the small cross section.
In the 
simulation shown in Fig.~\ref{fig:RR3}, a factor of 100 increase in
luminosity is considered, resulting in precise measurements extending
to $Q^2 > 500 \ {\rm GeV^2}$, well beyond the range explored for DVCS
or other GPD-sensitive processes to date. 

Maximising the lepton beam energy potentially gives access to the largest
$W$ and smallest $x$ values, provided the low $P_T^\gamma$ region can be
accessed. However, the higher beam lepton energy 
boosts the final state photon in the
scattered lepton direction, resulting in an additional acceptance
limitation. 

Further studies of this process will require a better understanding of
the detector in order to estimate systematic uncertainties. A particularly
interesting extension would be to investigate 
possible beam charge \cite{Aaron:2007cz,:2009vda} and polarisation
asymmetry measurements
at lower $x$ or larger $Q^2$ than 
was possible at HERA. With the addition of such information,
a full study of the potential of the LHeC to constrain GPDs could be
performed. 

\subsubsection{Accessing chiral-odd transversity GPDs in diffractive processes}


Transversity quark distributions in the nucleon remain among the most unknown leading-twist hadronic observables.   The four chiral-odd transversity GPDs~\cite{Diehl:2001pm}, denoted  $H_T$, $E_T$, $\tilde{H}_T$, $\tilde{E}_T$, offer a new way to access the transversity-dependent quark content of the nucleon. The factorisation properties of exclusive amplitudes apply in principle both to chiral-even and to chiral-odd sectors. However, one photon or one meson electroproduction leading-twist amplitudes are insensitive to the latter~\cite{Diehl:1998pd,Collins:1999un}.
At leading twist, they can be accessed experimentally through the quasi-forward exclusive electro- or photoproduction of a vector meson pair with a large invariant mass ~\cite{Ivanov:2002jj,Enberg:2006he}. 
\begin{figure}[htb]
\centerline{  \includegraphics[width=0.4\textwidth]{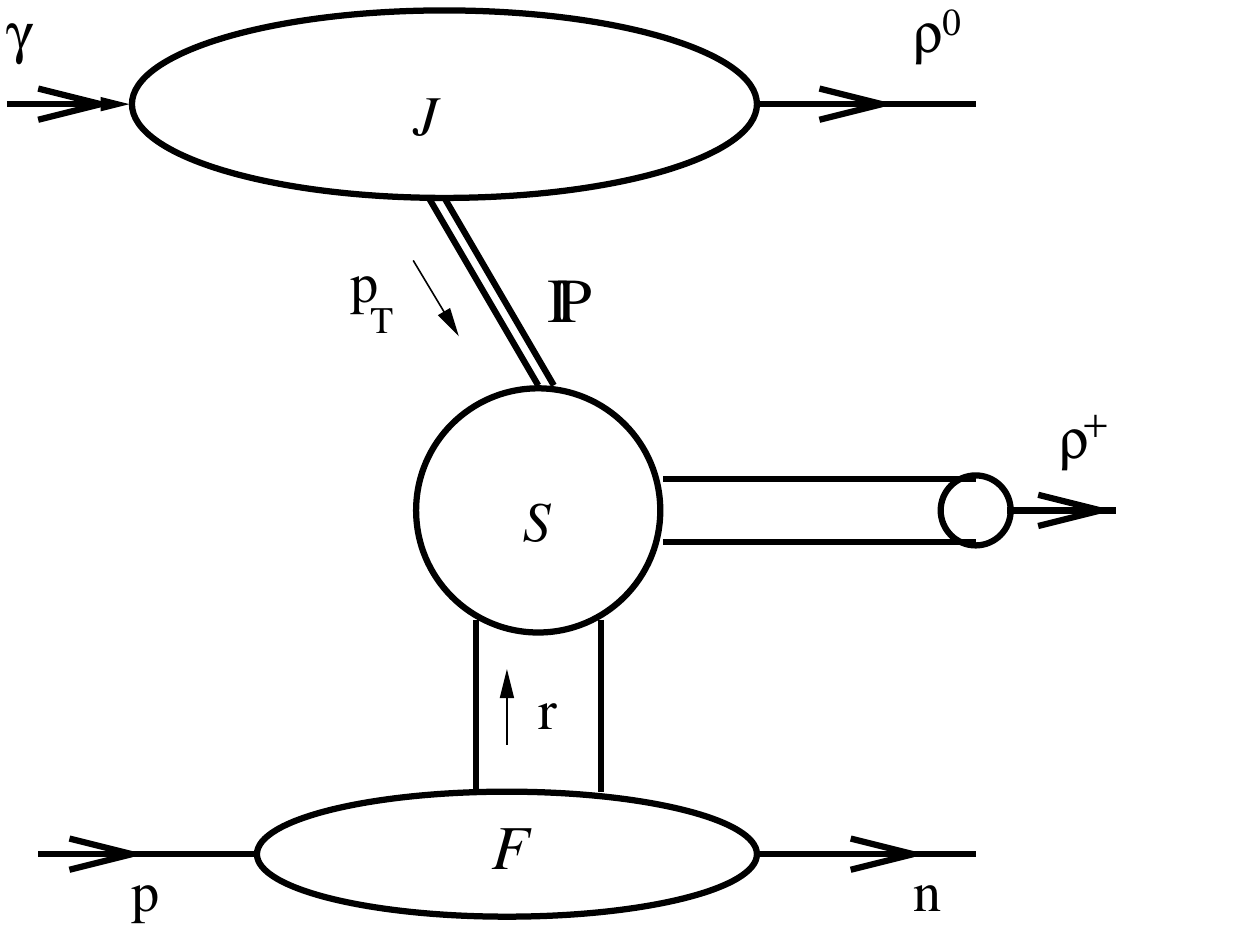}}
  \caption{Factorisation of the process  $ \gamma^{(*)} p \to \rho \rho N' $ in the asymmetric kinematics discussed in the text. $\cal{P}$ is the hard Pomeron modelled by two gluon exchange.}
\label{figPro}
\end{figure}
In analogy with the virtual photon exchange occurring in the deep inelastic electroproduction of a meson, one considers the subprocess:
\begin{equation}
{\cal P} (q_P)\;\; p (p_2) \to  \rho_{T}(p_\rho)\;N'(p_{2'})\;,
\end{equation}
of almost forward scattering of a  virtual  Pomeron  on a nucleon, the hard scale being the virtuality $- q_P^2$ of this Pomeron. 
The choice of a transversely polarised vector meson $\rho_{T}(p_\rho)$  involves at leading twist a chiral-odd distribution amplitude (DA), which in turn selects the chiral-odd GPDs. Let us stress that the target need not be polarised for the amplitude to contain the transversity GPD.
This subprocess is at work in the diffractive process
\begin{equation}
\label{2mesongen}
e p (p_2) \to e' \gamma^{(*)}_{L/T} (q)\;\; p (p_2) \to  e' \rho_{L,T}^0(q_\rho)\;\; \rho_{T}(p_\rho)\;
N'(p_{2'})\;,
\end{equation}
 shown in  Fig.~\ref{figPro}.   The final state may be either $\rho^0 \rho^0 p$ or  $\rho^0 \rho^+ n$.  We consider the kinematics where  the energy of the system 
($\rho_T (p_\rho)\; N'$) is smaller
than the energy of the system ($\rho_{L,T} \; \rho_T$) but still large enough to
justify a factorised approach  (in particular much larger than baryonic resonance masses).
In  this regime, the amplitude  is calculable consistently within
the collinear factorisation method, as an
integral (over  the longitudinal momentum fractions of the quarks)  of
the product of two amplitudes: the first one 
(the {\em impact factor} $ J^{\gamma \to \rho^0}$) describes 
the transition $\gamma^{(*)} \to \rho_{L,T}^0$ in 
the Born approximation via two gluon exchange and
the second one  describes the subprocess
${\cal P}\;p\;\to \;\rho_T\;N'$. The fact that this latter process  is
closely related to the electroproduction process 
$\gamma^*\,p \to\rho\,N'$
allows the separation of its long distance dynamics  expressed
through the GPDs from a perturbatively calculable coefficient function. 
The skewness parameter $\xi$ is related in the usual way ($\xi \approx x_B/(2-x_B)$) to the  Bjorken variable
defined by the Pomeron momentum $x_B = -q_P^2/(2q_P\cdot p_2)$. 

The resulting scattering amplitude ${\cal M}^{\gamma^*\,p\,\to \rho^0\, \rho_T\,p}$ then receives contributions from the four chiral-odd GPDs $H_T, \tilde H_T,E_T $ and $\tilde E_T$,  but  only the first contribution does not vanish kinematically in the forward direction. Thus, assuming  that the Mandelstam variable $-t= -(p_2-p_{2'})^2$ is sufficiently small,
the transversity GPD $H_T$ contribution dominates the amplitude which reads in the $\rho^0 \rho^+_T$ case:
\begin{eqnarray}
{\cal M}^{\gamma \,p\,\to \rho^0\, \rho^+_T\,n}
&=& \sin \theta \;16\pi^2 W^2 \alpha_s f_\rho^T \xi \sqrt\frac{1-\xi}{1+\xi}
\frac{C_F}{N_c\,(p_T^{\;2})^2}
\label{CON} \\
&\times&
\int\limits_0^1
\frac{\;du\;\phi_\perp(u)}{ \,u^2 \bar u^2 }
 J^{\gamma \to \rho^0}(up_T,\bar up_T)
 \frac{H^{ud}_T(\xi(2u-1),\xi,t)}{\sqrt 2},
\nonumber
\end{eqnarray}
with $ H_T^{ud}= H_{T}^u- H_T^{d}$, $f_\rho$ the $\rho$ decay constant, $\phi_\perp(u)$ the DA of 
the $\rho_T$ meson, $W^2=(q+p_2)^2$,  $\theta$ the angle between the transverse polarisation vector of the target $\vec{n}$ and the polarisation vector
$\vec{\epsilon}_T$ of the produced $\rho_T-$meson, and $p_T$  the transverse momentum of the $\rho^0$ meson (see \cite{Ivanov:2002jj,Enberg:2006he}).
 Note that the squared amplitude averaged over the nucleon polarisations does not cancel,
leading to the remarkable feature that these exclusive unpolarised reactions
are sensitive to the transversity GPDs.

\begin{figure}[htb]
\centerline{    \includegraphics[width=0.5\textwidth]{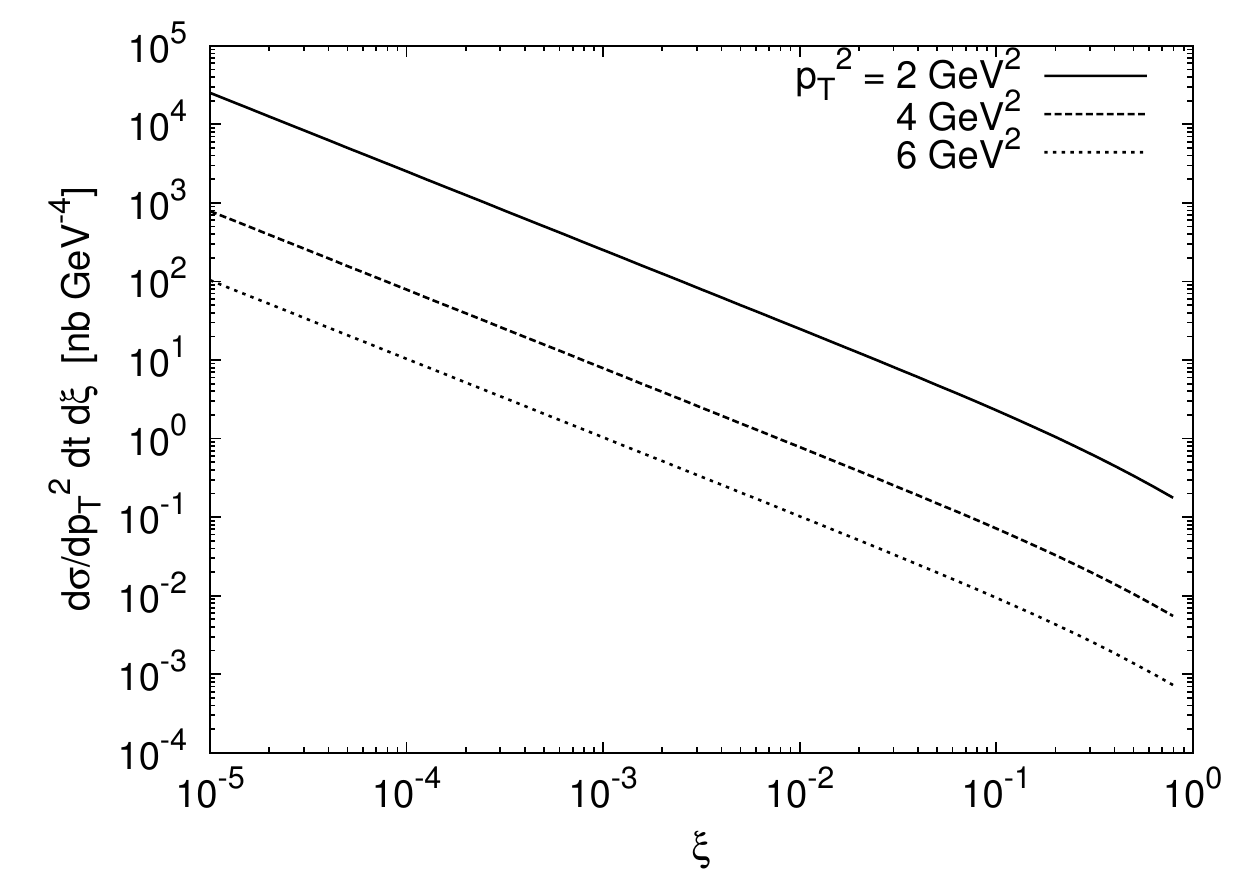}
 \includegraphics[width=0.5\textwidth]{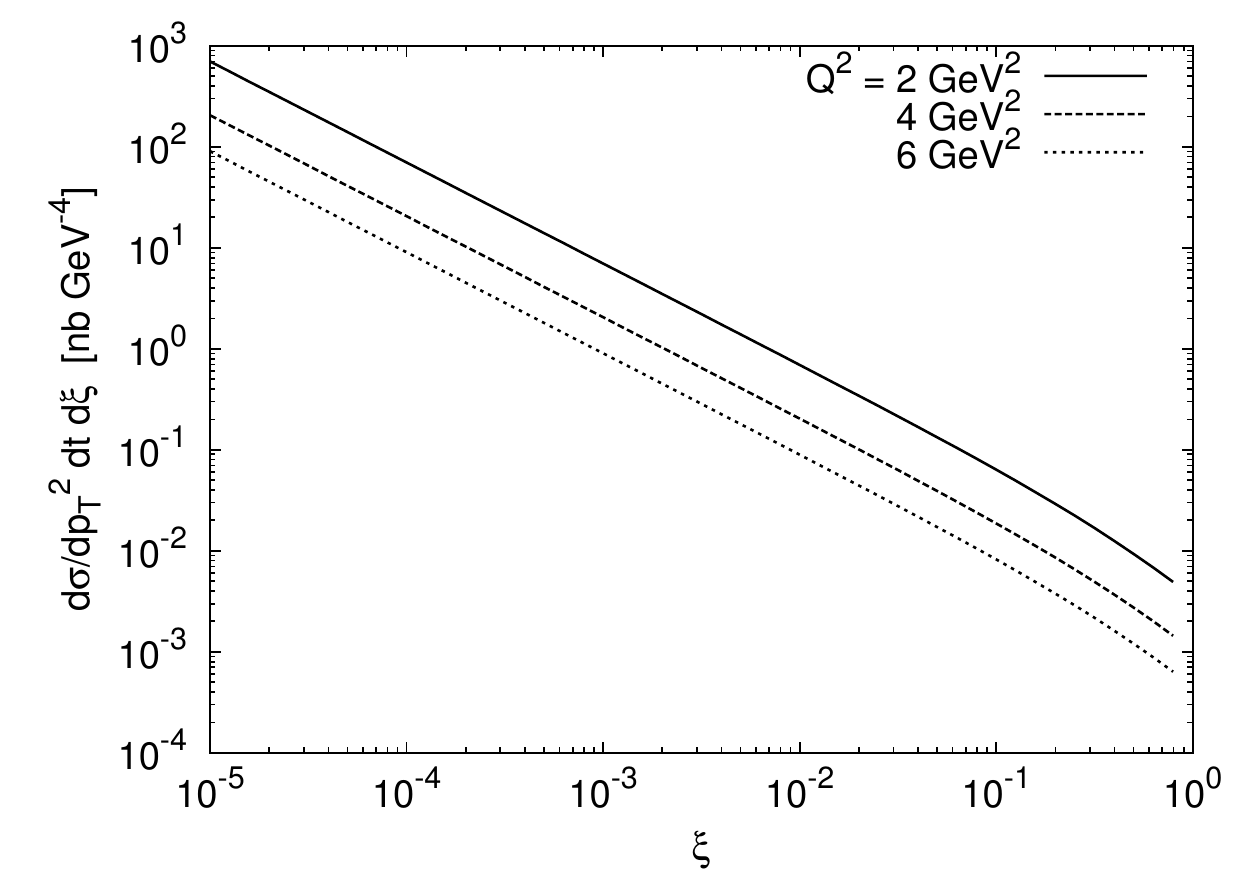}}
  \caption{The differential cross section for the photoproduction (a) and electroproduction (b) of the meson pair
$\rho^0_T \, \rho^+_T$  as a function of $\xi$ for  (a)  $p_T^2 =2,\,4\,$ and $6$ GeV$^2$ and for (b)    $p_T^2 =2 $ GeV$^2$ and $Q^2=2,\,4\,$ and $6$ GeV$^2$. The cross sections for the production of the meson pair
$\rho^0_T \, \rho^0_T$  are two times smaller.}
\label{transGPDRes}
\end{figure}
To get an estimate of the differential cross section of this process, we use a simple meson pole model for the transversity GPD $H_T^q(x,\xi,t)$ 
 starting with the effective interaction Lagrangian 
$
{\cal L_{ANN}}= 
 \frac  {g_{A\, NN}}{2M}\bar N \sigma_{\mu\nu}\gamma_5
\partial^\nu A^\mu N$.
This yields, identifying  the axial meson as
$A=b_1(1235)$, 
\begin{equation}
\label{Hpole}
H^{ud}_T(x,\xi,0)=\frac{g_{b_1 NN}f_{b_1}^{T}\langle k_\perp^2\rangle }{
2\sqrt{2}M_N\,m_{b_1}^2}\,
\frac{\phi^{b_1}_\perp\left(\frac{x+\xi}{2\xi}\right)}{2\xi}\;,
\end{equation}
with
 the average of the intrinsic transverse  momentum of the quarks
$\langle k_\perp^2 \rangle \approx 0.8$ GeV$^2$. 
The resulting cross sections estimated in the 
approximation where the Pomeron is modelled by a two gluon exchange 
do not depend on the variable $W^2$, but on the variable $\xi$. They
are shown in Fig. \ref{transGPDRes} as 
a function of  $\xi$ for various values of $p_T^2 $ and $Q^2$.
The rise at small $\xi$ comes mostly from the phase space factor. 
NLO corrections for this amplitude are as yet unknown. 
The cross sections look reasonably large.
Studies into the prospects for detection of the final states and
of the accessible kinematic range are left for the future.

%% file: physics/tex/vmeA.tex
\subsubsection{Diffractive vector meson production off nuclei}


Exclusive diffractive processes are 
similarly promising as a source of information on the gluon density 
in the nucleus \cite{Caldwell:2010zza}. 
Quasi-elastic scattering of photons from nuclei at small $x$ 
can be treated within the same dipole model 
framework as for $ep$ scattering, making the comparisons with the proton case relatively straightforward. The interaction of the dipole with the nucleus can be viewed as a sum of dipole scatterings off the nucleons forming the nucleus. Nuclear effects can be incorporated into the dipole cross section by  modifying  the transverse gluon distribution 
and adding the corrections due to Glauber rescattering from
multiple nucleons \cite{Kowalski:2003hm,Caldwell:2010zza}.  
Previous experimental data on exclusive production from 
nuclei exist \cite{Sokoloff:1986bu,Adams:1994bw}, but 
are limited in both kinematic range and precision. 

\begin{figure}
\centerline{ \includegraphics[clip=,width=0.4\textwidth]{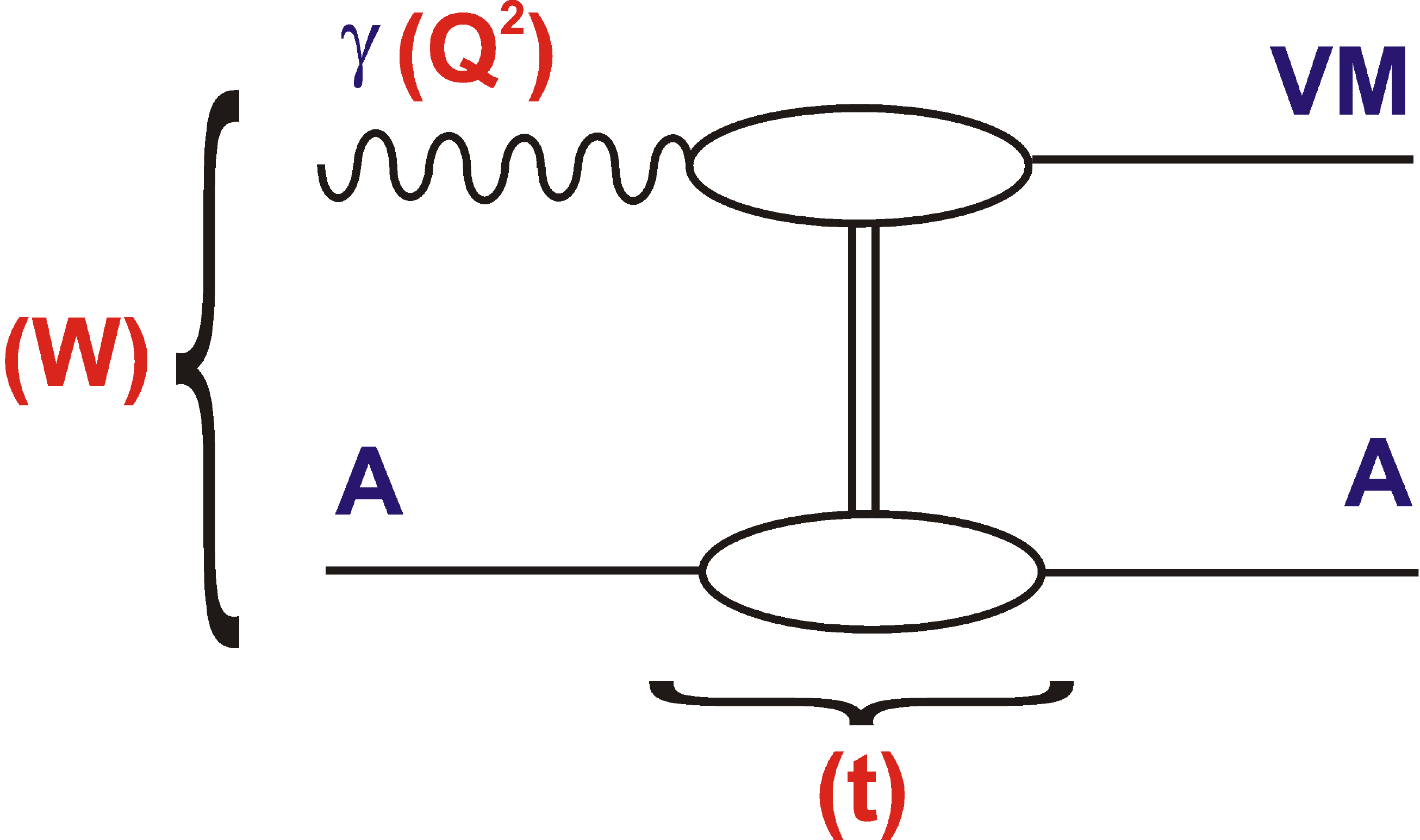} \hfill \includegraphics[clip=,width=0.5\textwidth]{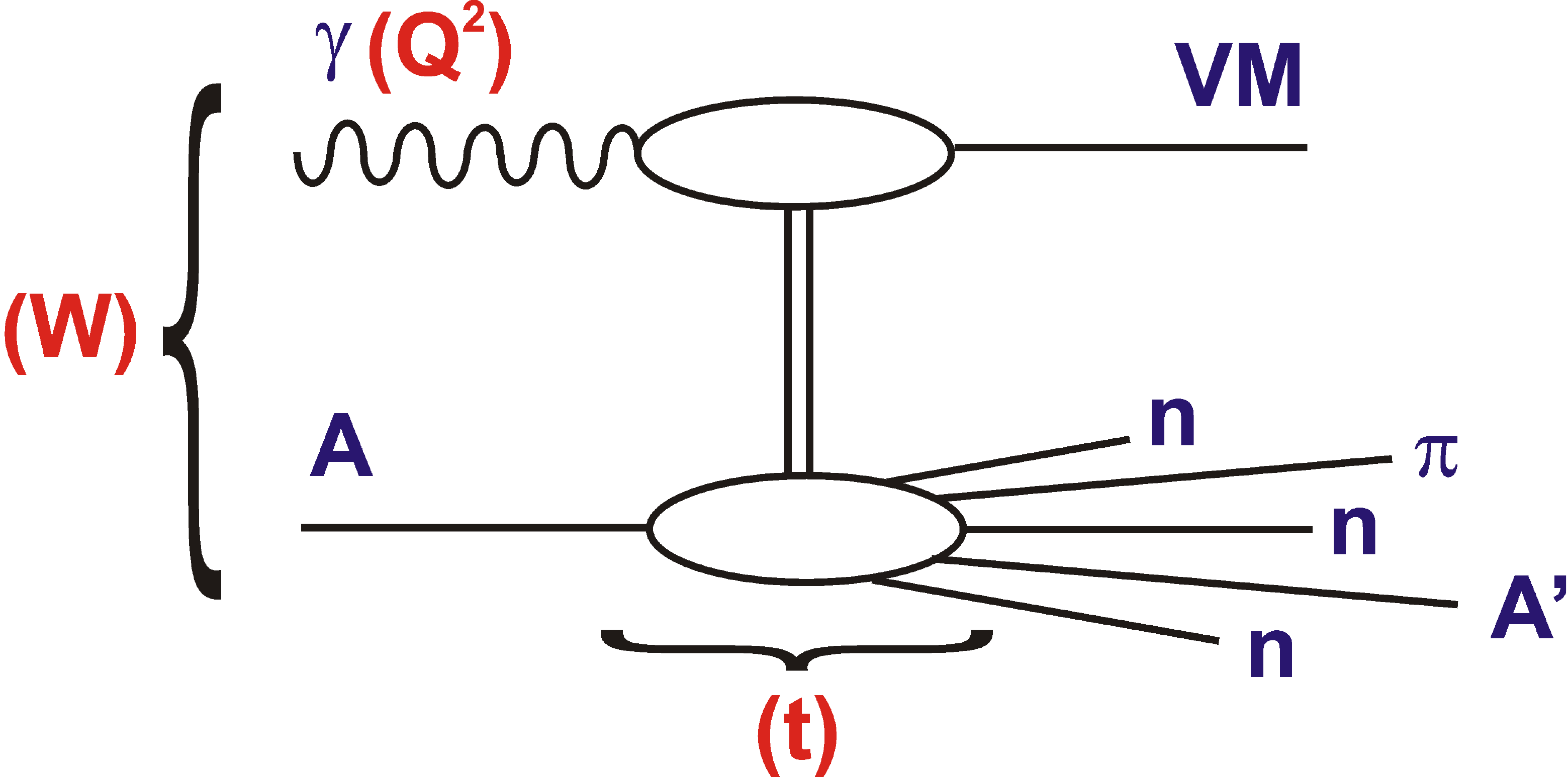}}
\caption{Diagrams illustrating the different types of exclusive 
diffraction in the nuclear case: coherent (plot on the left) and incoherent (plot on the right). While the diagrams have been drawn for the case of exclusive vector meson production, they equally apply to an arbitrary diffractively produced state.}
\label{fig:coherent}
\end{figure}

There is one aspect of diffraction which is specific to nuclei.
The structure of incoherent diffraction with nuclear break-up 
($e$A$\to$$eXY$) is more complex than with a proton target, and it can also be 
more informative. In the case of a target nucleus, we expect the following qualitative changes in the $t$-dependence. First, the low-$|t|$ regime of coherent diffraction illustrated in Fig.~\ref{fig:coherent} left, in which the nucleus scatters elastically and remains in its ground state, will be 
dominant up to a smaller value of $|t|$ (about $|t|=0.05$ GeV$^2$) 
than in the proton case, reflecting the larger size of the nucleus. 
The nuclear 
dissociation regime (incoherent case), see Fig.~\ref{fig:coherent} right, will consist of two parts: an intermediate regime in momentum transfer up to 
perhaps $|t| = 0.7$ GeV$^2$, 
where the nucleus will predominantly break up into its constituent 
nucleons, and a large-$|t|$ regime where the nucleons inside the nucleus will also break up, implying - for instance - pion production in the $Y$ system. While these are only qualitative expectations, it is crucial to study this aspect of diffraction quantitatively in order to 
complete our understanding of the transverse structure of nuclei.

\begin{figure}[htb]
 \centering
  \includegraphics[width=0.57\textwidth]{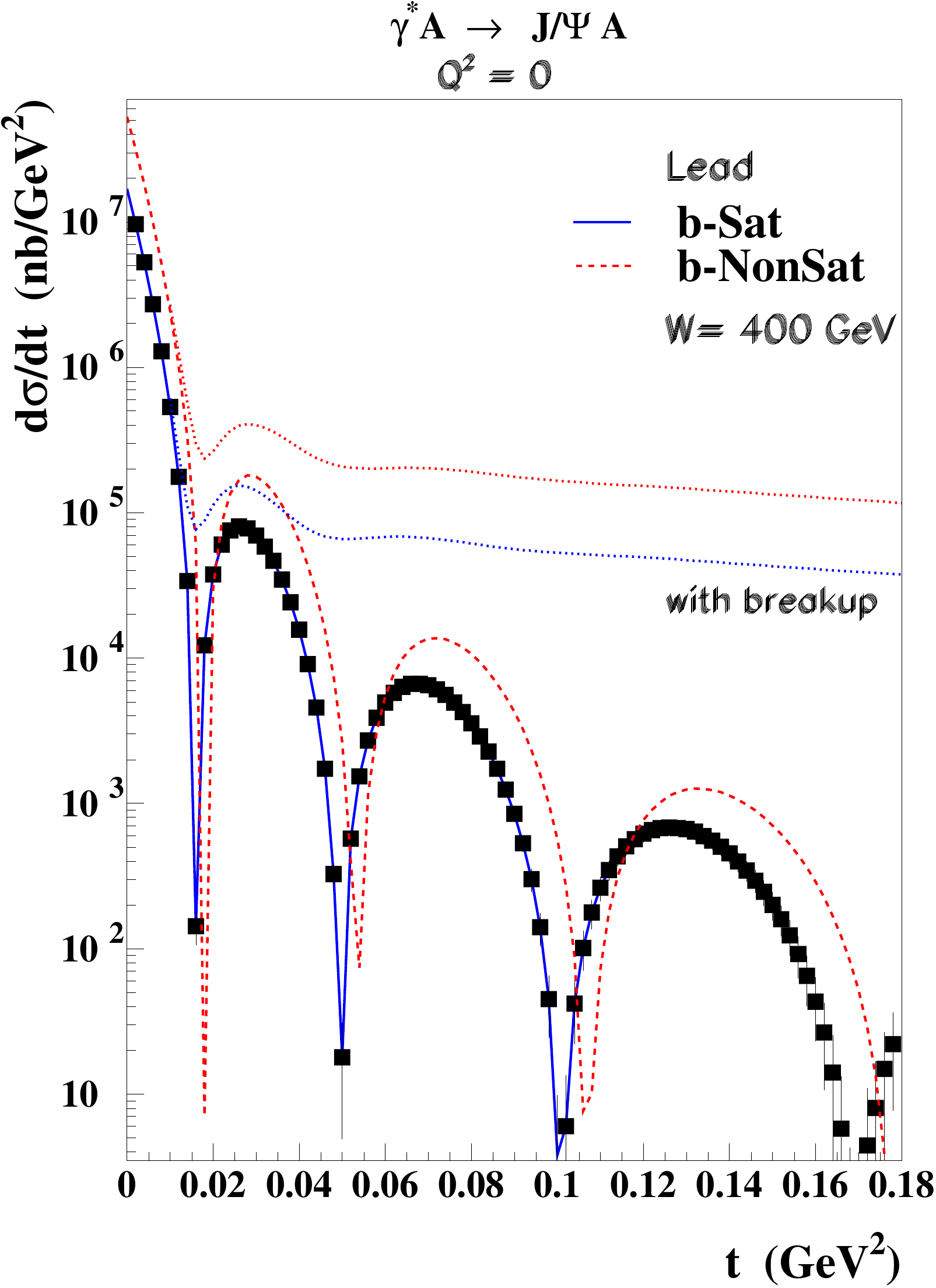}
  \caption{Differential cross section for the diffractive production of $J/\Psi$ on a lead nucleus, as a function of the momentum transfer $|t|$. 
The dashed-red and solid-blue lines 
correspond to the b-Sat model predictions for 
coherent production without and with saturation effects, respectively. 
The dotted lines correspond to the predictions for the incoherent case.
The pseudodata shown for the coherent case are explained in the text.}
\label{plots:henri1}
\end{figure}

Fig.~\ref{plots:henri1} shows the diffractive cross sections for 
exclusive $J/\Psi$ production off a lead nucleus with (b-Sat) and without (b-NonSat) saturation effects. 
The figure shows both the coherent and incoherent cross sections. 
According to both models shown, 
the cross section for $t \sim 0$  is dominated by coherent production,
whereas the nuclear break-up contribution becomes dominant for
$|t| \stackrel{>}{_{\sim}} 0.01 \ {\rm GeV^2}$, leading to a relatively
flat $t$ distribution. The coherent cross section exhibits a characteristic
multiple-dip structure at these relatively large $t$ values, the details
of which are sensitive to gluon saturation effects. Resolving these
dips requires a clean separation between the coherent and nuclear break-up
contributions, which may be possible with sufficient forward instrumentation. 
In particular, preliminary studies suggest that the detection 
of neutrons from the nuclear break-up
in the Zero Degree Calorimeter (Subsec.~\ref{LHEC:Detector:zdc}) reduces the 
incoherent backgrounds dramatically. Assuming that it is possible
to obtain a relatively clean sample of 
coherent nuclear diffraction, resolving the rich structure at large $t$
should be possible based on the measurement of the transverse momentum
of the elastically produced $J/\psi$ according to 
$t = - p_T^2(J/\psi)$. The resolution on the $t$ measurement is thus
related to that on the $J/\psi$ by 
$\Delta t = 2 \sqrt{-t} \ \Delta p_T(J/\psi)$, amounting to 
$\Delta t < 0.01 \ {\rm GeV^2}$ throughout the range shown in 
Fig.~\ref{plots:henri1} 
assuming $\Delta p_T(J/\psi) < 10 \ {\rm MeV}$,
as has been achieved at HERA. The pseudodata 
for the coherent process shown in the figure
are consistent with this resolution and correspond to a
modest integrated luminosity of order $10 \ {\rm pb^{-1}}$.

Independently of the large $|t|$ behaviour, important information 
can be obtained from the low $|t|$ region alone.  
Coherent production for $t \sim 0$
can easily be related to the properties of dipole-nucleon interactions,  
because all nuclear effects can be absorbed into the nuclear wave functions, 
such that only the average gluon density of the nucleus enters
the calculation. For this forward 
cross section, the exact shape of the nuclear wave function 
is not important, in contrast to what happens at larger $|t|$
where the distribution reflects the functional form of the 
nuclear density.

\begin{figure}[htb]
 \centering
  \includegraphics[width=0.57\textwidth]{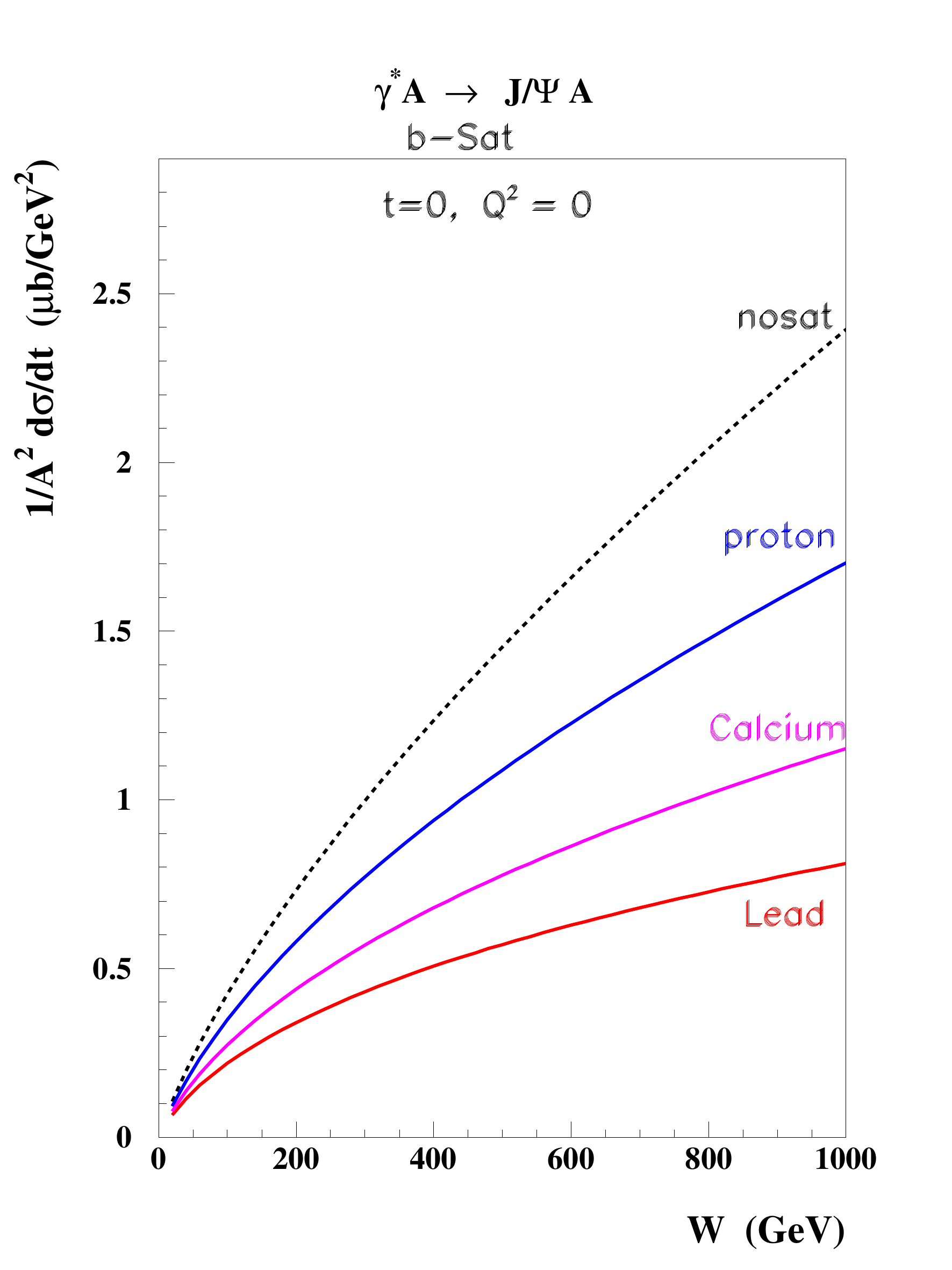}
  \caption{Energy dependence of the coherent photoproduction of the $J/\Psi$ on a proton and different nuclei in the forward case $t=0$
according to the b-Sat model. The cross sections are normalised by a factor $1/A^2$, corresponding to the dependence on the gluon density squared if no nuclear effects are present.}
\label{plots:henri2}
\end{figure}

\begin{figure}[htb]
 \centering
  \includegraphics[width=0.9\textwidth]{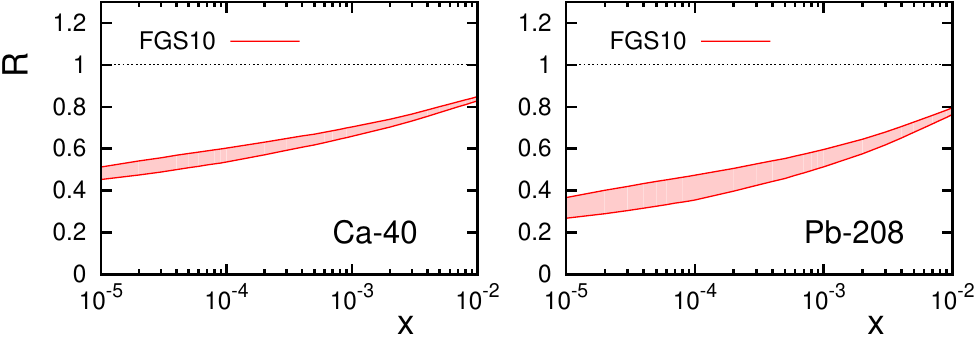}
  \caption{The $x$ dependence of the nuclear modification 
ratio for the gluon density squared, from nuclei to protons (re scaled by $A^2$), for 
  the scale corresponding to the exclusive production of the $J/\Psi$. 
The results have been obtained from the model described in \cite{Frankfurt:2011cs}.}
\label{plots:vadim}
\end{figure}

Saturation effects can be studied in a very clean way using the
$t$-averaged gluon density obtained in this way from
the forward coherent cross section. Fig. \ref{plots:henri2}  
shows this cross section for $J/\Psi$ production 
as a function of $W$ for different nuclei. The cross section 
varies substantially as a function of the $\gamma^* p$ centre of mass energy $W$ and the nuclear mass number $A$. It is also very sensitive to shadowing or saturation effects due to the fact that the differential cross section at $t=0$ has a quadratic dependence on the gluon density and  $A$. 
Due to this fact,  
the ratios of the cross sections for nuclei and protons are roughly proportional
to the ratios of the gluon densities squared. This has been exploited in the calculation \cite{Frankfurt:2011cs} presented in
Fig.~\ref{plots:vadim}, where the nuclear modification factor 
$R$ for the square of the gluon density is shown. The predictions are 
consistent with those obtained from the b-Sat model (Fig.~\ref{plots:henri2}).
Therefore, a precise measurement of the
$J/\psi$ 
cross section around $t=0$ is an invaluable source of information 
on the gluon density and in particular on non-linear effects.

Another region of interest is the measurement at larger $|t|$, 
$|t| \stackrel{>}{_{\sim}} 0.15\ {\rm GeV}^2$. Here the reaction is fully dominated by the incoherent processes in which the nucleus breaks up. 
The shadowing or saturation effects should be stronger in this region than in the coherent case \cite{Kowalski:2007rw} and the shape of the diffractive cross section should be only weakly sensitive to 
nuclear effects \cite{Caldwell:2010zza}. Finally, 
the intermediate region between $|t| \sim 0.01 \ {\rm GeV^2}$ and 
$|t| \sim 0.1 \ {\rm GeV^2}$ 
is also very interesting because 
here the barely known gluonic nuclear effects can be studied.

\subsubsection{Searching for the Odderon}

Exclusive processes in photoproduction and DIS offer 
unique sensitivity to rare exchanges in QCD. 
One prominent example is that of exclusive pseudoscalar 
meson production, which could proceed via the exchange of the 
Odderon. 
The Odderon is the postulated Reggeon which is the C-odd partner  
of the Pomeron. The exchange of an Odderon should 
contribute with different signs to particle-particle and 
particle-antiparticle scattering. Therefore, in the case of 
hadron-hadron collisions it could lead, 
via the optical theorem, to a  
difference between proton-proton and proton-antiproton total 
cross sections at high energies, provided the intercept of the 
Odderon is close to unity.  Despite many searches, no evidence 
for Odderon exchange has been found so far, see for 
example \cite{Nicolescu:1999qi}. Nevertheless, the existence 
of the Odderon is a firm prediction of high-energy QCD, 
for a comprehensive review see \cite{Ewerz:2003xi}. At lowest 
order in perturbation theory it can be described as a system 
of three non-interacting gluons. In the leading logarithmic 
approximation in $x$ its evolution is governed by the 
Bartels-Kwieci\'nski-Prasza\l{}owicz (BKP) 
equations \cite{Bartels:1978fc,Bartels:1980pe,Kwiecinski:1980wb}. 
Up to now, two solutions to the BKP equations are known, one with 
intercept slightly below one \cite{Janik:1998xj} and the 
other with intercept exactly equal to one \cite{Bartels:1999yt}.

Several channels involving Odderon exchange are possible at the LHeC,
leading to the exclusive production of pseudoscalar mesons,
$\gamma^{(\star)} p \rightarrow Cp$, where
$C = \pi^0,\eta,\eta',\eta_c \dots$ 
Searches for the Odderon in the reaction $ep\rightarrow e\pi^0 N^*$ 
were performed by the H1 collaboration at HERA \cite{Adloff:2002dw} at an 
average $\gamma p$ c.m.s energy $\langle W \rangle= 215 \; {\rm GeV}$. 
No signal was found and an upper limit on the cross section was derived, 
$\sigma(ep\rightarrow e\pi^0 N^*,\; 0.02 < |t| < 0.3 \; {\rm GeV}^2) <49 \; {\rm nb}$ at the $95\; \%$ confidence level. 
Although the predicted cross sections for processes governed by Odderon
exchange are rather small, they are not suppressed with increasing
centre-of-mass energy and the large luminosities offered by the 
LHeC may be exactly what is required for a discovery. 
In addition to $\pi^0$ production, Odderon searches at the LHeC could be 
based on other exclusive channels, for example with heavier 
mesons $\eta_c,\eta_b$ \cite{Czyzewski:1996bv}. 

\begin{figure}[htb]
  \centerline{\includegraphics[width=0.9\textwidth]{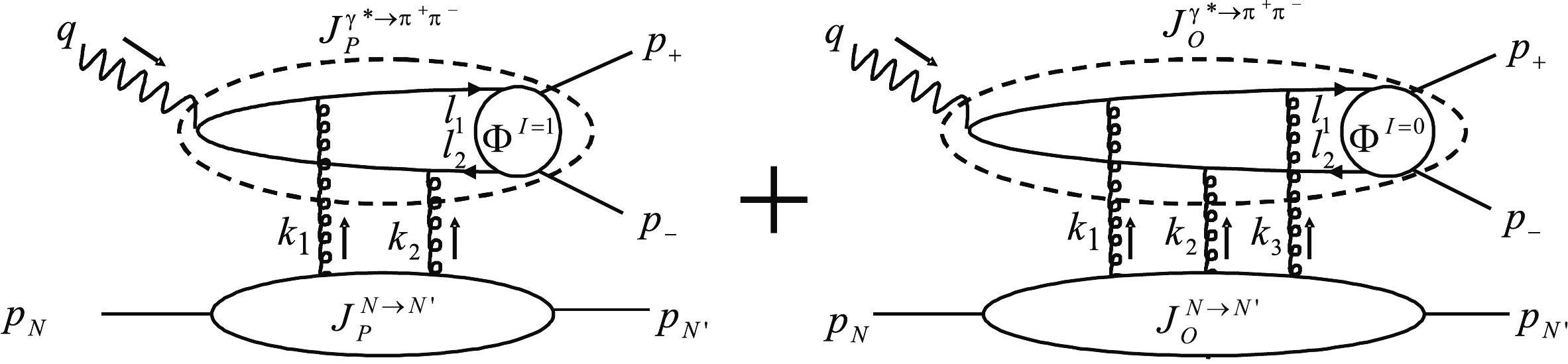}}
  \caption{Feynman diagrams describing $\pi^+ \pi^-$ electroproduction in the Born approximation.}
\label{FigOdderon}
\end{figure}

It has been advocated \cite{Brodsky:1999mz} that one could devise more sensitive tests of the existence 
of the Odderon exchange by searching for interference effects between Pomeron and Odderon exchange amplitudes. Such an observable is the measurement of the  difference between charm and anti-charm angular or 
energy distributions in $\gamma^*p\rightarrow c\bar{c}N^*$. Another channel is the exclusive photo or electroproduction of two pions \cite{Hagler:2002nh, Ginzburg:2002zd ,  Hagler:2002nf}. Indeed a $\pi^+ \pi^-$ pair may be produced  both as a charge symmetric $C^+$ and a charge antisymmetric $C^-$ state. The Pomeron exchange amplitude will contribute to the $C^-$ $\pi^+ \pi^-$ state, the Odderon exchange  amplitude will contribute to the $C^+$ $\pi^+ \pi^-$ state.  A (mesonic) charge antisymmetric observable will select the interference of these two amplitudes. In the hard electroproduction case, one may estimate the effect through a lowest order calculation where Pomeron (Odderon) exchange is calculated through
the exchange of two (three) non-interacting gluons in a colour singlet state in the $t$-channel, as shown in Fig. \ref{FigOdderon}. 

The impact representation of the amplitude has the form of an integral over the 2-dimensional transverse momenta $k_i$ of the $t$-channel gluons:
\begin{eqnarray}
\label{pom}
 {\cal M}_P &=& -i\,W^2\,\int\;\frac{d^2 k_1 \; d^2 k_2 \;
  \delta^{(2)}(k_1 +k_2-{p}_{2\pi})}{(2\pi)^2\,k_1^2\,k_2^2}
 J_P^{\gamma^* \rightarrow \pi^+\pi^-}\cdot
J_P^{N \rightarrow N'}\;,\label{odd}
\\
\nonumber
 {\cal M}_O &=&-\frac{8\,\pi^2\,W^2}{3!}\int\;\frac{d^2 k_1 \; d^2k_2 d^2 k_3\;
  \delta^{(2)}(k_1 +k_2 +k_3-{p}_{2\pi})}{(2\pi)^6\,k_1^2\,k_2^2\,k_3^2}
 J_O^{\gamma^* \rightarrow \pi^+\pi^-}\cdot
J_O^{N \rightarrow N'}\;,
\end{eqnarray}
where $J_{P/O}^{\gamma^* \rightarrow \pi^+\pi^-}$
is the impact factor
 for the transition 
$\gamma^* \to \pi^+\ \pi^-$
  and   $J_{P/O}^{N \rightarrow N'}$ is the impact factor
 for the transition of the nucleon in the initial state $N$ into the nucleon in the
 final state $N'$.

\begin{figure}[htb]
 \centering
  \includegraphics[width=0.6\textwidth]{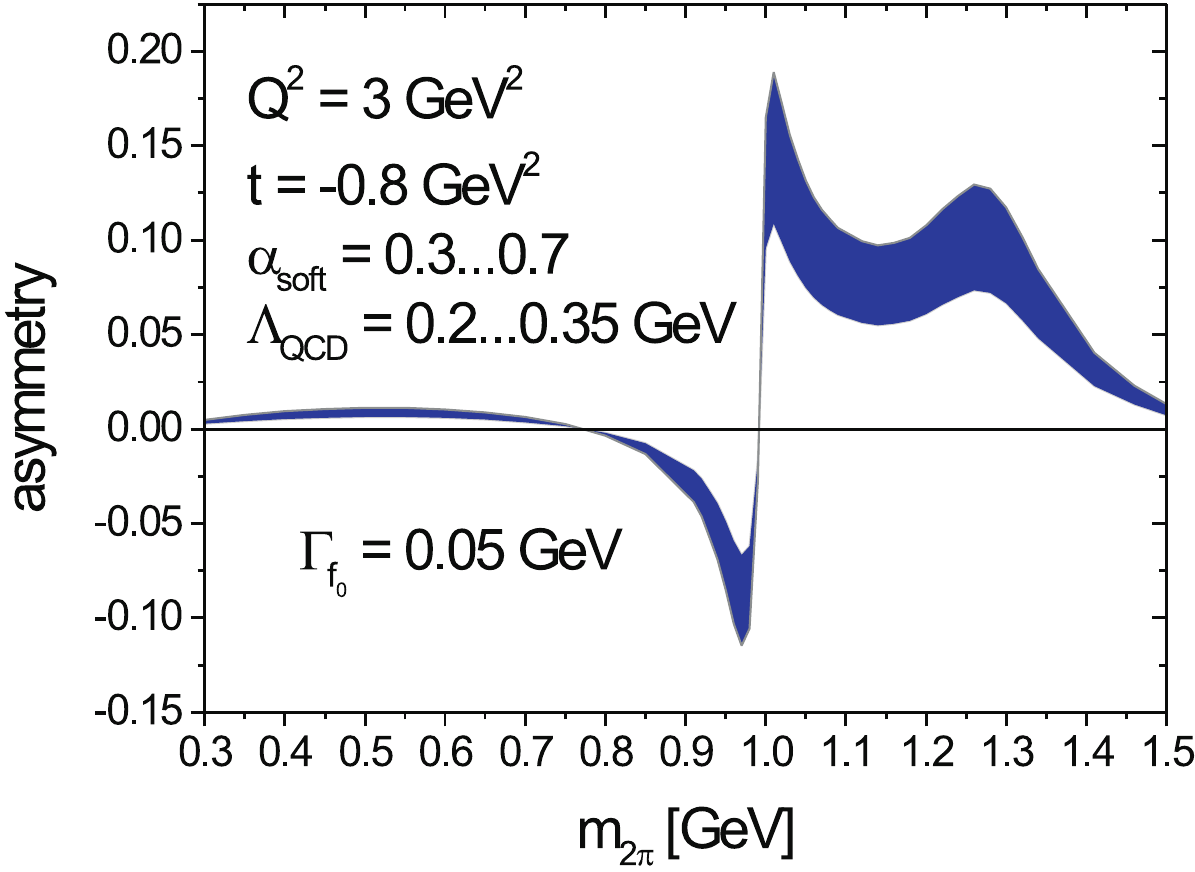}
  \caption{The charge asymmetry defined in Eq. (\ref{asym}) as a function of the $\pi^+\pi^-$ invariant mass $m_{2\pi}$.}
\label{FigOdderonRes}
\end{figure}

The impact factors are calculated by  standard methods. An important feature of the $J_{P/O}^{\gamma^* \rightarrow \pi^+\pi^-}$  impact factors is the presence of the appropriate two-pion generalised distribution amplitude (GDA) \cite{Diehl:1998dk,Polyakov:1998td,Polyakov:1998ze}:
\begin{equation}
\label{lP}
J_P^{\gamma^*_L \rightarrow \pi^+\pi^-}(k_1,k_2) =
-\frac{i\,e\,g^2\,\delta^{ab}\,Q}{2\,N_C}\;
\int_0^1\,dz\,z{\bar z}\,P_P(k_1,k_2)\,
\Phi^{I=1}(z,\zeta,m_{2\pi}^2)\;,
\end{equation}
\begin{equation}
\label{lO}
J_O^{\gamma^*_L \rightarrow \pi^+\pi^-}(k_1,k_2,k_3) =
-\frac{i\,e\,g^3\,d^{abc}\,Q}{4\,N_C}\;
\int_0^1\,dz\,z{\bar z}\,P_O(k_1,k_2,k_3)\,
\frac{1}{3}\Phi^{I=0}(z,\zeta,m_{2\pi}^2)\;,
\end{equation}
where $P_P$ and $P_O$ are known perturbatively calculated functions. $\zeta$ is the light-cone momentum fraction of the  $\pi^+$ in  the two pion system of invariant mass $m_{2\pi}$, which is related to the   polar decay angle $\theta$ of the $\pi^+$ in the rest
frame of the two pion system.
The GDAs $\Phi^I(z,\zeta,m_{2\pi}^2)$ are non-perturbative matrix elements containing the full strong interactions between the two pions. They are universal quantities  much related to GPDs in the meson. One must distinguish the GDA $\Phi^{I=0}$ where the pion pair is in an isosinglet state from the GDA $\Phi^{I=1}$ where it is in an isovector state. The charge conjugation parity of the exchanged particle selects the charge parity, hence the isospin of the emerging two-pion state: the Pomeron (Odderon) exchange process involves the
production of a pion pair in the $C$-odd (even) channel which corresponds to
odd(even) isospin. In the numerical studies we   use a
simple ansatz \cite{Diehl:2000uv} for the generalised distribution amplitudes $\Phi^I(z,\zeta,m_{2\pi}^2)$.
A crucial point  is the choice of the parameterisation of the phases in the
  GDA's since, through interference effects, the rapid variation of a phase
shift leads to a characteristic  $m_{2\pi}$-dependence of the asymmetry. We show on Fig. \ref{FigOdderonRes} the resulting estimate for the charge asymmetry defined as
\begin{eqnarray}
\label{asym}
A(Q^2, t, m_{2\pi}^2)
=\frac{ \int \cos \theta \,d \sigma
(W^2,Q^2,t,m_{2\pi}^2,\theta)}{
\int d \sigma (W^2,Q^2,t,m_{2\pi}^2,\theta)}
=\frac{ \int_{-1}^{1}\,\cos \theta\,d\cos \theta \,2\ \mbox{Re}
\left[{\cal M}_P^{\gamma^*_L}({\cal M}_O^{\gamma^*_L})^*  \right]}{
\int_{-1}^{1}\,d\cos \theta \left[ |{\cal M}_P^{\gamma^*_L}|^2
+ |{\cal M}_O^{\gamma^*_L}|^2  \right]}\;,
\end{eqnarray}
where $\theta$ is the  polar decay angle of the $\pi^+$ in the rest
frame of the two pion system.
In order to visualise a rather large uncertainty in our modelling we present
our results with an error band dominated by the value of the soft coupling constant ${\alpha}_{soft}$ which we vary in the interval of ${\alpha}_{soft} = 0.3 - 0.7$ (see Ref.\cite{Hagler:2002nf} for details).
While detailed studies on the possibilities for detection of the final states are left for the future, this estimate demonstrates that the presence of the perturbative Odderon may be discovered in two pion electroproduction at high energy (note that the asymmetry (\ref{asym}) is independent of $W^2$).

%% file: physics/tex/ddisep.tex
\subsubsection{Introduction to diffractive deep inelastic scattering}

\begin{figure}[h] \unitlength 1mm
  \begin{center}
\includegraphics[width=0.35\textwidth]{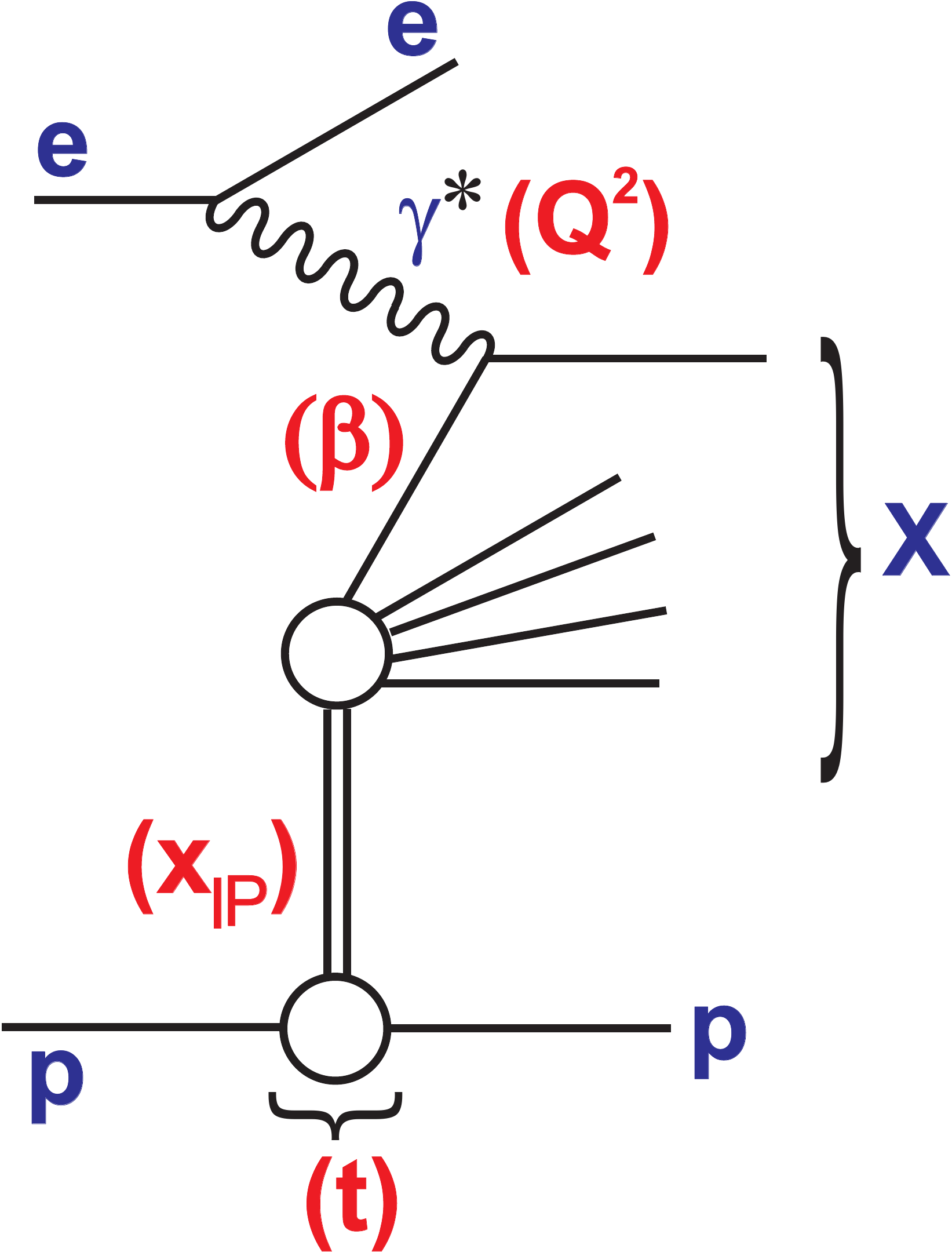}
  \end{center}
  \caption[]{Illustration of the kinematic variables used to describe the
inclusive diffractive DIS process $ep \rightarrow eXp$.}
\label{ddis:feynman}
\end{figure}

Approximately 10\% of low-$x$ DIS events are of the diffractive type, 
$ep \rightarrow eXp$, with the proton surviving the collision intact
despite the large momentum transfer from the 
electron (Fig.~\ref{ddis:feynman}). This process is usually
interpreted as the diffractive dissociation of the exchanged virtual
photon to produce any hadronic 
final state system $X$ with mass much smaller than $W$
and the same net quantum numbers as the exchanged photon
($J^{PC} = 1^{--}$).
Due to the lack of colour flow,
diffractive DIS events are characterised by 
a large gap in the rapidity distribution of final state hadrons between
the scattered proton and the diffractive final state $X$.

As discussed in Section~\ref{sec:vm},
similar processes exist in electron-ion scattering, 
where they
can be sub-divided into fully coherent diffraction, where the nucleus 
stays intact ($eA \rightarrow eXA$) and
incoherent diffraction, where the nucleons within the nucleus are
resolved and the nucleus breaks up 
($eA \rightarrow eXY$, $Y$ being a system produced via
nuclear or nucleon excitation, with the 
same quantum numbers as $A$).

Theoretically, rapidity gap production is usually described in terms of
the exchange of a net colourless object in the $t$-channel,
which is often referred to as a 
pomeron \cite{Kaidalov:1979jz,Goulianos:1982vk}. In the
simplest models \cite{Ingelman:1984ns,Donnachie:1987xh}, this pomeron has a 
universal structure and its vertex couplings factorise, such that 
it is applicable for example to proton-(anti)proton
scattering as well as DIS. One of the main achievements at HERA
has been the development of an understanding of diffractive DIS
in terms of parton dynamics and QCD \cite{Wolf:2009jm}. 
Events are selected using the experimental signatures
of either a leading proton \cite{Aktas:2006hx,Chekanov:2008fh,Aaron:2010kz} 
or the presence of a large rapidity gap \cite{Aktas:2006hy,Chekanov:2008fh}. 
The factorisable pomeron picture has proved remarkably successful for 
the description of most of these data.

The kinematic variables used to describe diffractive DIS are illustrated in 
Fig.~\ref{ddis:feynman}. In addition to $x$, $Q^2$ and the squared
four-momentum transfer $t$, the mass $M_X$ 
of the diffractively produced final state provides a further
degree of freedom. In practice, 
the variable $M_X$ is often replaced by
\begin{equation}
\beta \;  = \; \frac{Q^2}{Q^2+M_X^2-t}\ .
\end{equation}
Small values of $\beta$ refer to events with diffractive masses much bigger than the photon virtuality, while values of $\beta$ close to unity 
are associated with small $M_X$ values.
In models based on a factorisable pomeron, $\beta$ may be interpreted as
the fraction of the pomeron longitudinal momentum 
which is carried by the struck parton.
The variable
\begin{equation}
\xpom=\frac{x}{\beta}=\frac{Q^2+M_X^2-t}{Q^2+W^2-M^2}\ ,
\end{equation}
with $M$ the nucleon mass, is then interpreted as the longitudinal 
momentum fraction of the Pomeron with respect to the 
incoming proton or ion.
It also characterises the 
size of the rapidity gap as  $\Delta\eta\simeq\ln(1/\xpom)$.


\subsubsection{Measuring diffractive deep inelastic scattering at the LHeC}

Diffractive DIS (DDIS) can be studied in a substantially increased kinematic
range at the LHeC, which will allow a whole new level of 
investigations of the factorisation properties of inclusive diffraction,
will lead to new insights into low-$x$ dynamics and will provide a 
subset of final states with known quantum numbers for use in searches
for new physics and elsewhere.

As shown in \cite{Collins:1997sr}, 
collinear QCD factorisation holds in the leading-twist approximation 
in diffractive DIS and can be used to 
define diffractive parton distribution functions for the proton or ion.  
That is, within the collinear framework, the diffractive structure 
functions \cite{Blumlein:2001xf} 
can be expressed as convolutions of the appropriate 
coefficient functions with diffractive quark and gluon 
distribution functions,
which in general depend on all of $\beta$ , $Q^2$, $\xpom$ and $t$.
The diffractive parton distribution functions (DPDFs) are physically 
interpreted as probabilities for finding a 
parton with a small fraction of 
the proton momentum $x = \beta \xpom$, 
under the condition that the proton stays intact with a final
state four-momentum which is specified 
up to an azimuthal angle by $\xpom$ and $t$.
The DPDFs may then be evolved in $Q^2$ with the DGLAP evolution 
equations, 
with $\beta$ playing the role of the 
Bjorken-$x$ variable. 
The other two variables $\xpom$ and $t$ play the role of external parameters to the DGLAP evolution.

In various extractions using HERA
DDIS data \cite{Aktas:2006hy,Aktas:2007bv,Chekanov:2009qja,Martin:2005hd}
the DPDFs have been found to 
be dominated by gluons. Proton vertex factorisation holds to good 
approximation, such that the DPDFs vary only in normalisation
with the four-momentum of the final state proton, the normalisation 
being well modelled using Regge phenomenology \cite{Goulianos:1982vk}. 

\begin{figure}
\begin{center}
\includegraphics[clip=,width=0.6\textwidth,angle=0]{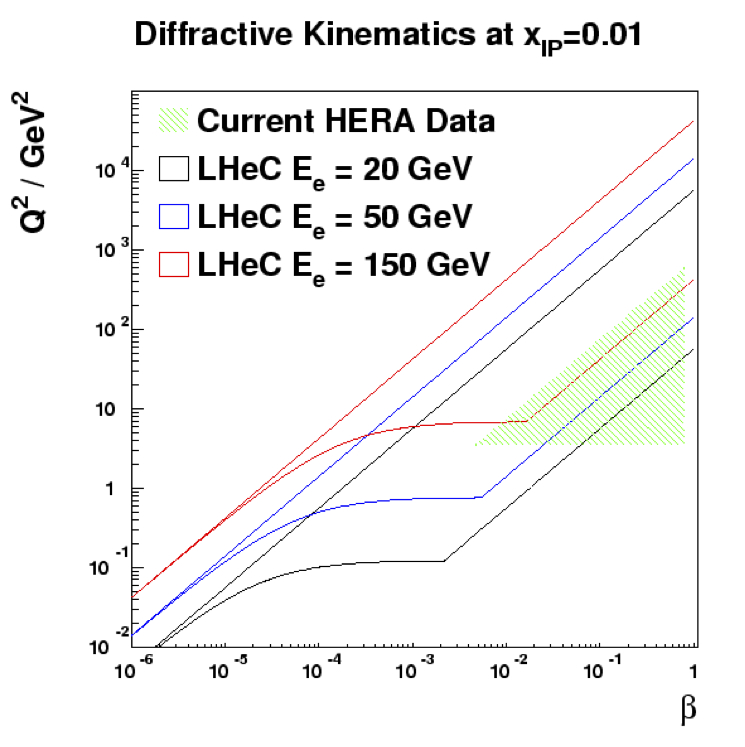}
\includegraphics[clip=,width=0.6\textwidth,angle=0]{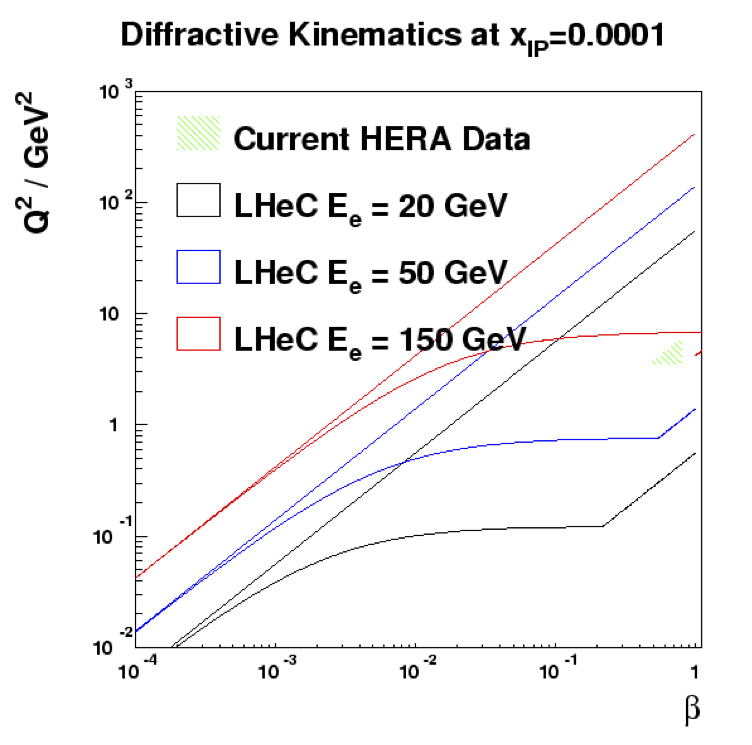}
\end{center}
\caption{Diffractive DIS kinematic ranges in $Q^2$ and $\beta$ of HERA  
and of the LHeC for different electron 
energies $E_e=20,50,150 \; {\rm GeV}$ at $\xpom=0.01$ (upper plot), 
and $\xpom=0.0001$ (lower plot). In both cases, $1^{\rm o}$ acceptance 
is assumed for the scattered electron and the typical experimental
restriction $y > 0.01$ is imposed. No rapidity gap restrictions are applied.}
\label{Fig:diffrangekin}
\end{figure}

The LHeC will offer the opportunity to study diffractive DIS in an 
unprecedented kinematic range.
The diffractive kinematic plane is illustrated in Fig.~\ref{Fig:diffrangekin} 
for two different values of the Pomeron momentum fraction, $\xpom=0.01$ 
and $\xpom=0.0001$. In each plot, accessible kinematic ranges are
shown for
three different electron energies in collision with the $7 \ {\rm TeV}$
proton beam. 
Figure~\ref{Fig:diffrangekin}a corresponds to the coverage that
will be possible based on leading
proton detection (see Chapter~\ref{detector:fwdbwd}).
Figure~\ref{Fig:diffrangekin}b is more representative of the possibilities
using the large rapidity gap technique (see the following).
It is clear that the LHeC will have a much increased reach compared 
with HERA 
towards low values of $\xpom$, 
where the interpretation of diffractive events is
not complicated by the presence of sub-leading 
meson exchanges, rapidity gaps are large and diffractive
event selection systematics are correspondingly small.
The range in the fractional struck quark
momentum $\beta$ extends by a 
factor of around 20 below that 
accessible at HERA. 

\begin{figure}[htbp] \unitlength 1mm
  \begin{center}
\includegraphics[width=0.7\textwidth]{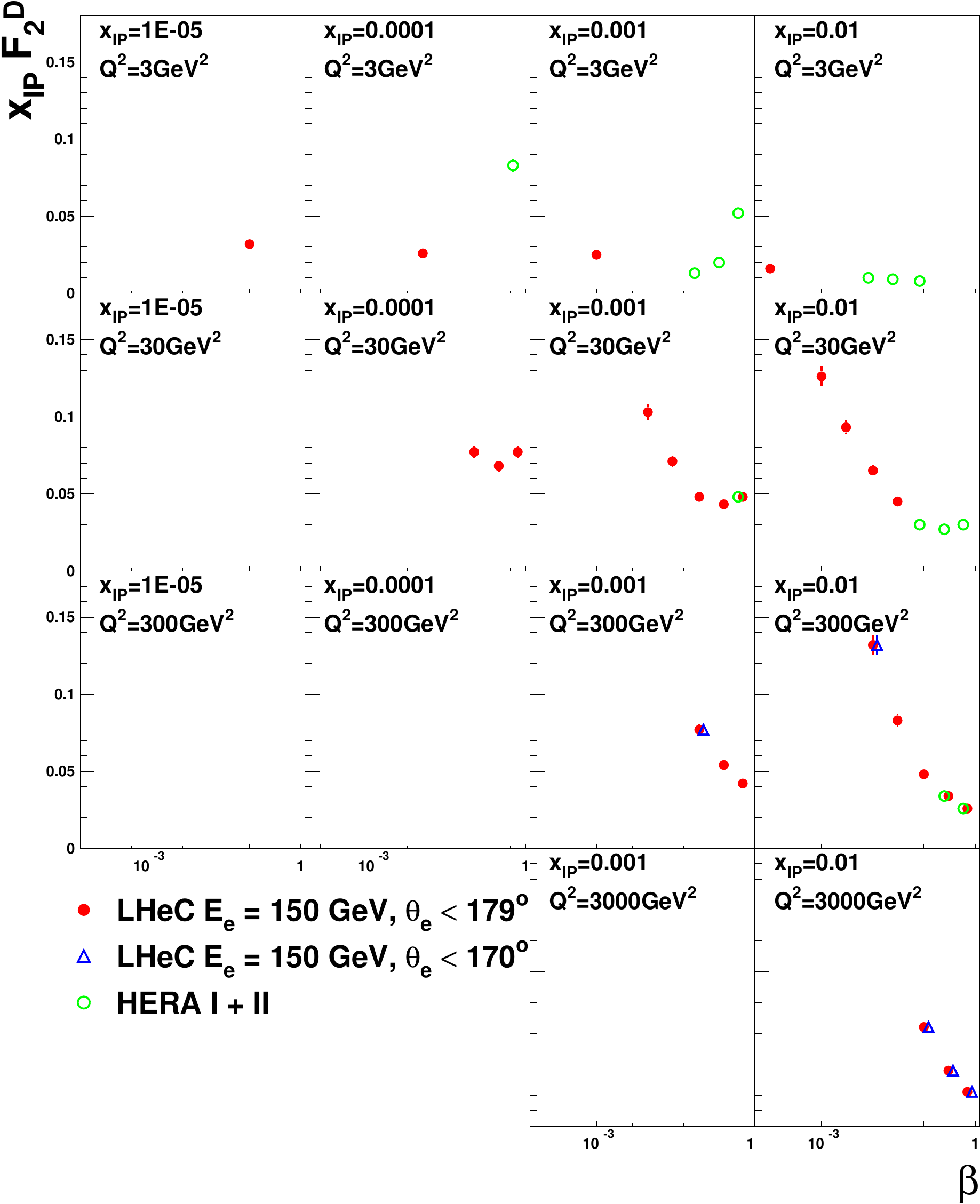}
  \end{center}
  \caption[]{Simulation of a possible LHeC measurement of the 
diffractive structure function, $F_2^D$ using a
$2 \ {\rm fb^{-1}}$ sample, compared with an estimate
of the optimum results achievable at HERA using the full
luminosity for a single experiment ($500 \ {\rm pb^{-1}}$). 
The loss of kinematic
region if the LHeC scattered electron acceptance extends to within $10^\circ$
of the beam-pipe, rather than $1^\circ$ is also illustrated.}
\label{ddis:f2d150}
\end{figure}

Figure~\ref{ddis:f2d150} further illustrates the achievable kinematic range 
of diffractive DIS measurements at the LHeC for the example of a 
$150 \ {\rm GeV}$ electron beam combining large rapidity gap
and proton tagging acceptance, compared with an estimation of the
final HERA performance. For ease of illustration, a binning scheme is
chosen in which the $\beta$ dependence is emphasised and very large
bins in $\xpom$ and $Q^2$ are taken. 
There is a large difference between the 
kinematically accessible ranges with
backward acceptance cuts of $1^\circ$ and $10^\circ$.
Statistical 
uncertainties are typically much smaller
than 1\% for a luminosity of $2 \ {\rm fb^{-1}}$, so a much finer
binning is possible, as required. The data points are plotted according to
the H1 Fit B DPDF predictions \cite{Aktas:2006hy}, which amounts to
a crude extrapolation based on dependences in the HERA range.

Systematic uncertainties are difficult to estimate without a 
detailed knowledge of the forward detectors and their acceptances. At HERA,
sub-5\% systematics have been achieved in the bulk of the phase
space and it is likely that the LHeC could do at least as well. 

\begin{figure}[htbp] \unitlength 1mm
  \begin{center}
\includegraphics[width=0.6\textwidth]{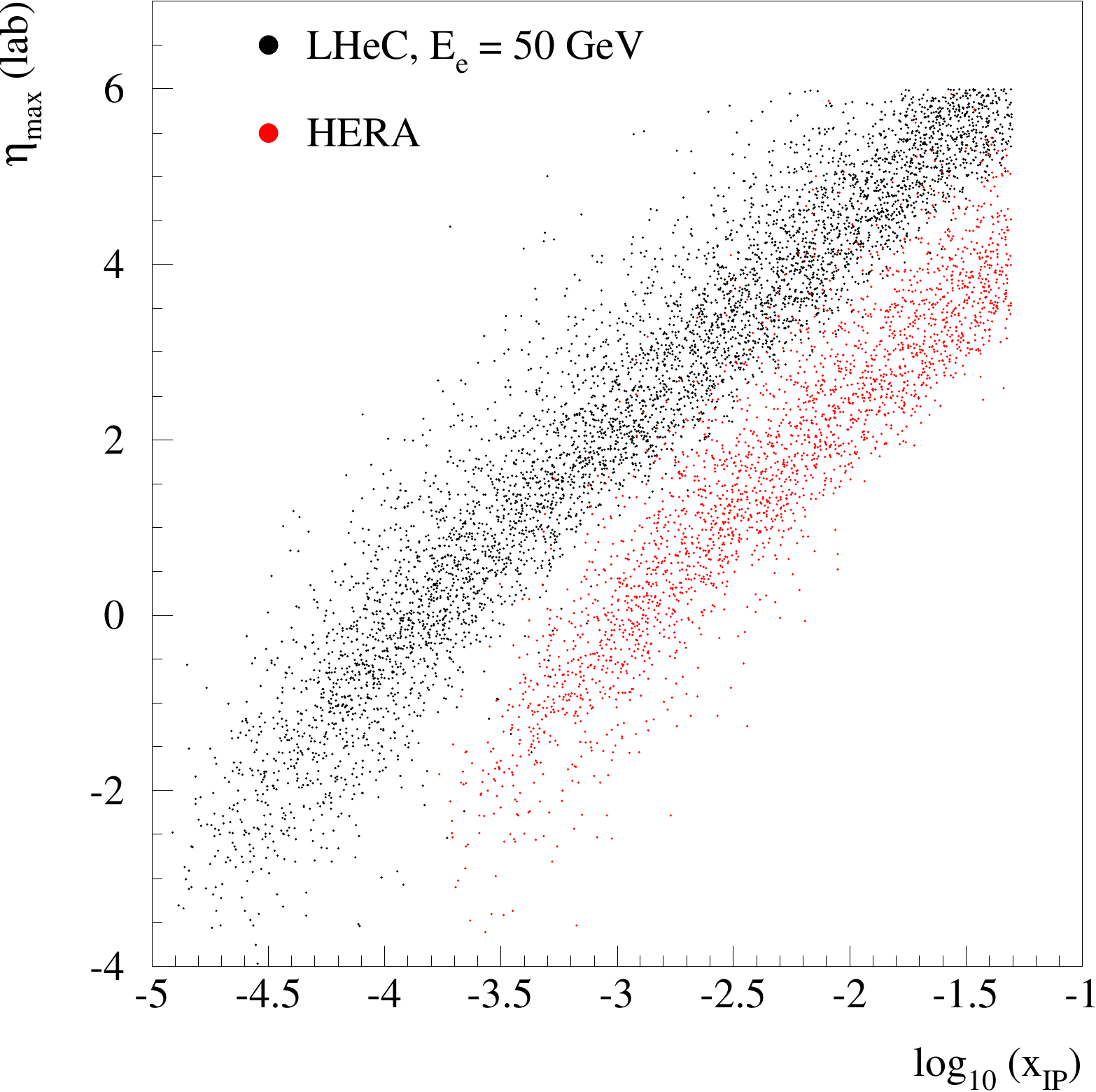}
  \end{center}
  \caption[]{Comparison of the correlation between the rapidity gap selection
variable, $\eta_{\rm max}$ and $\xpom$ at HERA and at the LHeC, using
events simulated with the RAPGAP Monte Carlo generator.}
\label{ddis:scatter}
\end{figure}

The limitations in the kinematic range accessible with the 
large rapidity gap technique are investigated in Fig.~\ref{ddis:scatter}.
This shows the correlation between $\xpom$ and the pseudorapidity
$\eta_{\rm max}$ of the most forward particle in the hadronic 
final state system $X$, in simulated samples with LHeC and HERA
beam energies, according to the RAPGAP event 
generator \cite{Jung:1993gf}.
This correlation depends
only on the proton beam energy and is thus the same for all LHeC
running scenarios. 
At HERA, a cut at $\eta_{\rm max} \sim 3.2$ has been used to select
diffractive events.  
Assuming LHeC forward instrumentation 
extending to around $\theta = 1^\circ$, a
cut at $\eta_{\rm max} = 5$ may be possible, which
would allow measurements to be made comfortably
up to $\xpom \sim 0.001$, with some 
limited sensitivity at larger $\xpom$, a region where the proton
tagging acceptance takes over (see Chapter~\ref{detector:fwdbwd}). The two methods are thus complementary,
and offer some common acceptance in an overlap region of
$\xpom$. This redundancy could be used for cross-calibration of
the two methods and their systematics, as has been done at HERA.

\subsubsection{Diffractive parton densities and final states}

\begin{figure}
\begin{center}
\includegraphics[width=0.49\textwidth,angle=0]{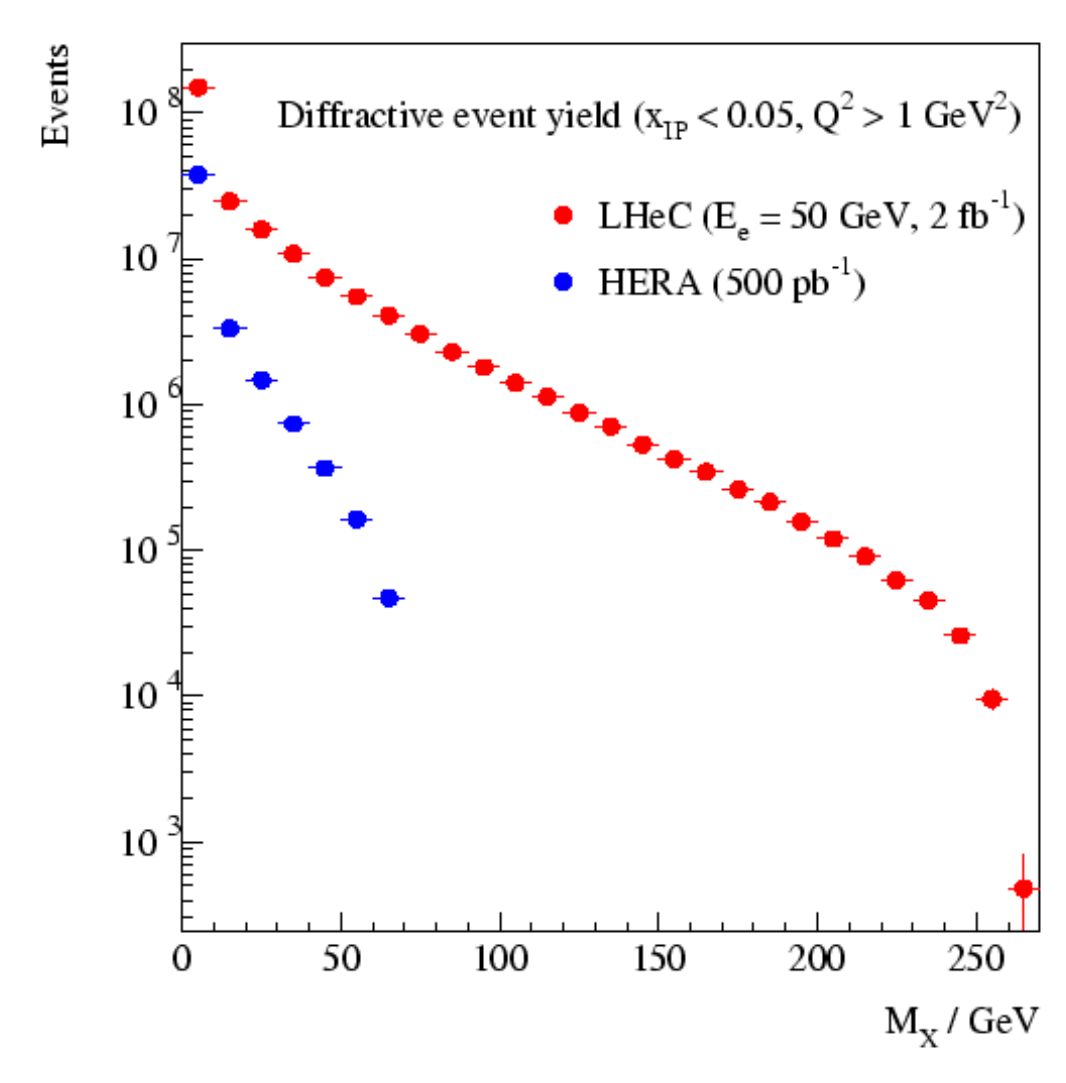}
\includegraphics[width=0.49\textwidth,angle=0]{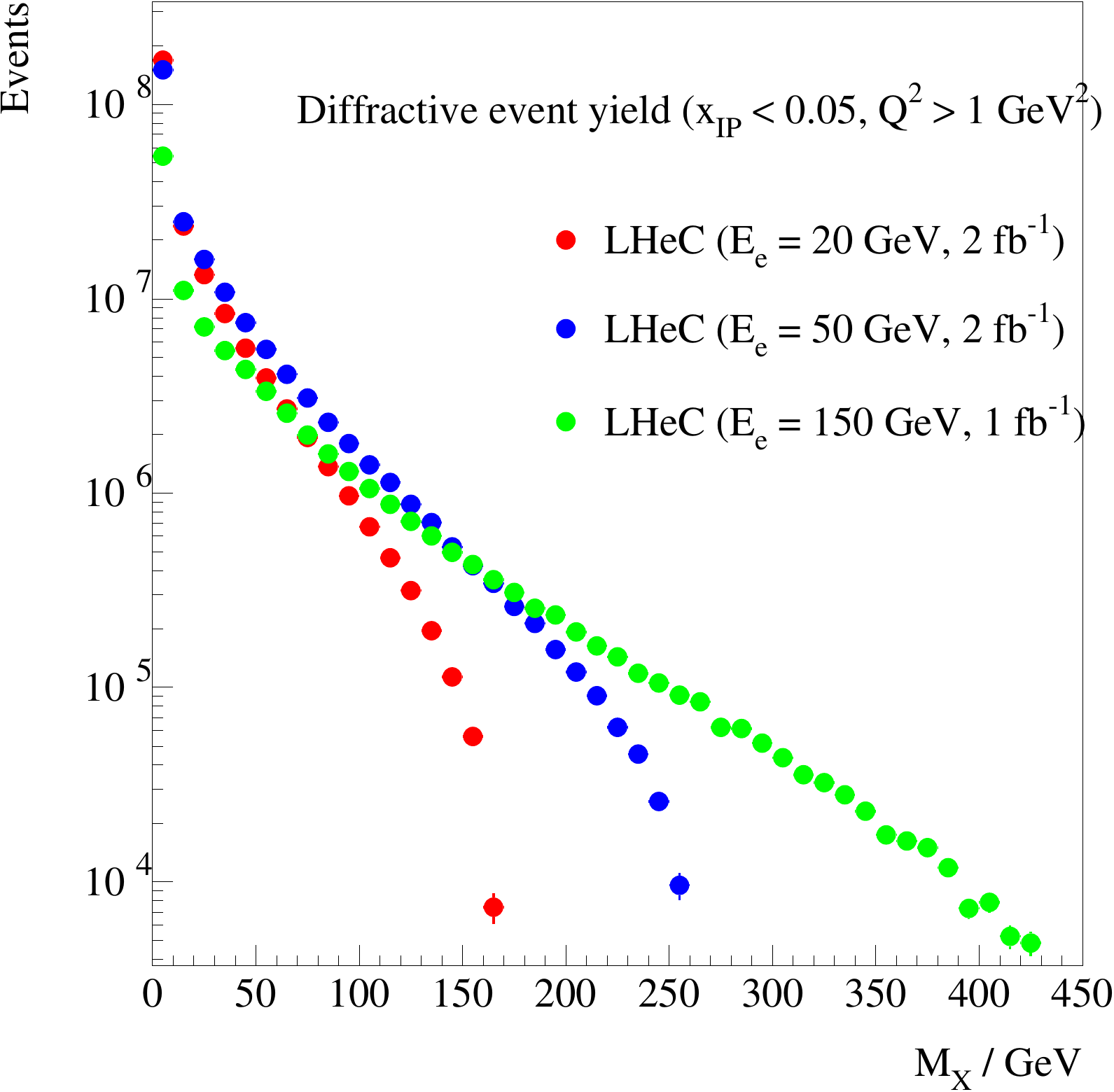}
\end{center}
\caption{Simulated distributions in the invariant mass $M_X$
according to the RAPGAP Monte Carlo model for samples of events
obtainable with $\xpom < 0.05$ 
Left: one year of high acceptance LHeC running at $E_e = 50 \ {\rm GeV}$ 
compared with 
HERA (full luminosity for a single
experiment). Right: comparison between three different high acceptance 
LHeC luminosity and $E_e$ scenarios.}
\label{Fig:diffyield}
\end{figure}

The previously unexplored
diffractive DIS region of very low $\beta$ is of particular 
interest.  Here, diffractively produced systems will be created
with unprecedented invariant masses.
Figure~\ref{Fig:diffyield} left
shows a comparison between HERA and the LHeC in
terms of the 
$M_X$ distribution which could be 
produced in diffractive processes with $x_{_{I\!\!P}} < 0.05$
(using the RAPGAP Monte Carlo model \cite{Jung:1993gf}).
Figure~\ref{Fig:diffyield} right compares the expected
$M_X$ distributions for one year of running at three 
LHeC electron beam energy choices. 
Diffractive masses
up to several hundred ${\rm GeV}$ are accessible
with reasonable rates, such that diffractive final
states involving beauty quarks and $W$ and $Z$ bosons, or even exotic
states with $1^-$ quantum numbers, could be
produced. 

Large improvements in DPDFs
are likely to be possible from NLO DGLAP fits to 
LHeC diffractive structure 
function data. 
In addition to the extended phase space in $\beta$, 
the extension of the kinematic range towards
larger $Q^2$ increases the
lever-arm for extracting the diffractive gluon density and opens the
possibility of significant weak gauge boson exchange, which would allow a
quark flavour decomposition for the first time. 



Proton vertex
factorisation can be tested precisely by comparing the
$\beta$ and $Q^2$ dependences at the LHeC at different small $\xpom$ values
in their considerable regions of overlap. 
The production of dijets or heavy quarks
as components of the diffractive system $X$ will allow precise
 testing of QCD collinear factorisation. These processes
are driven by boson-gluon fusion ($\gamma^* g \rightarrow
q \bar{q}$) and thus provide complementary sensitivity to the 
diffractive gluon density to be compared with that from the scaling
violations of the inclusive diffractive cross section. 

\begin{figure}
\begin{center}
\includegraphics[width=0.65\textwidth,angle=0]{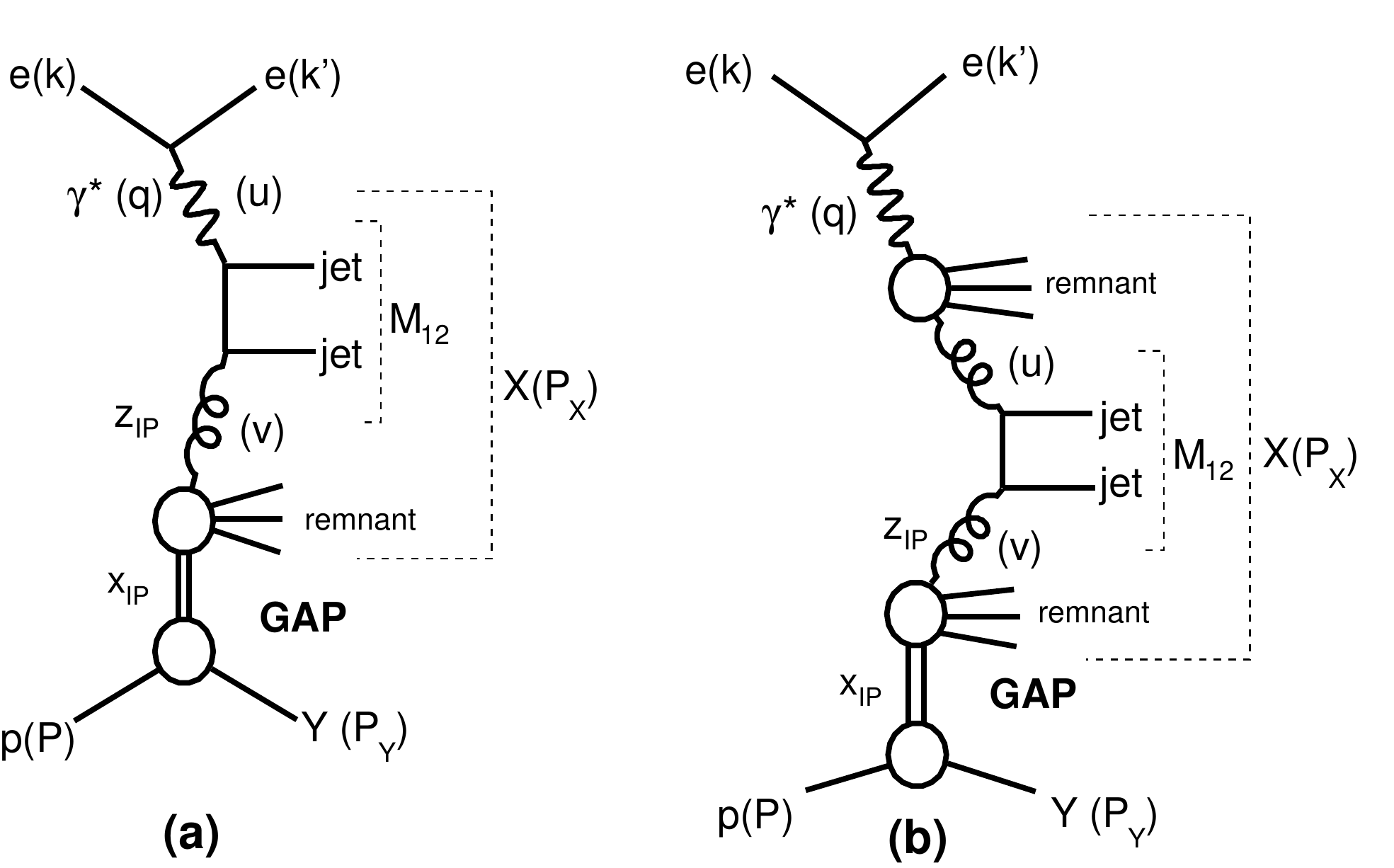}
\end{center}
\caption{Leading order diagrams for diffractive dijet photoproduction. 
Diagrams (a) and (b) are examples of direct and resolved photon 
interactions, respectively.
}
\label{Fig:diffjetfeyn}
\end{figure}

Diffractive final states containing charm signatures or 
relatively high transverse
momentum dijets have been analysed in detail 
at HERA. In the DIS regime, the cross sections for these processes
are reproduced within uncertainties by calculations based on NLO DPDFs
extracted from inclusive diffractive data 
for both the 
dijet \cite{Aktas:2007bv,Chekanov:2007aa,Aktas:2007hn,Aaron:2011mp} and 
charm \cite{Aktas:2006up,Chekanov:2003gt} cases. 
By far the limiting factor in 
the precision of these tests is the large scale uncertainty
on the theoretical predictions, 
due to the strong kinematic limitations on the accessible 
jet transverse energies in diffraction at HERA.
The situation from HERA photoproduction data is more complex
and is usually divided into direct and resolved photon 
contributions (figures~\ref{Fig:diffjetfeyn}a and~\ref{Fig:diffjetfeyn}b,
respectively).
In the direct photon case, where the highly 
virtual photon has a point-like coupling,
the process is driven by photon-gluon fusion and 
at the current level of precision, 
cross sections are well predicted using DPDFs extracted in fits to inclusive
diffractive data \cite{Aktas:2007hn,Aaron:2010su,Chekanov:2007rh}.
In contrast, the resolved photon case
introduces sensitivity to the rich partonic structure of the 
quasi-real photon. It is these partons which participate in 
the hard scattering sub-process producing the dijets, 
in a manner which resembles the situation in hadron-hadron scattering. 
In this case, the possibility of additional 
rescatterings between the hadronic remnants leads to a non-unit 
`survival
probability' for the rapidity 
gap \cite{Dokshitzer:1991he,Bjorken:1992er,Gotsman:1993vd}
and a breakdown of factorisation. 
Factorisation tests 
have been carried out 
on several occasions in diffractive dijet photoproduction
at HERA, resulting in a somewhat  
confused situation on the size of the 
gap destruction effects \cite{Aaron:2010su,Chekanov:2007rh}
and the roles of resolved and direct contributions.
Data in which the parton entering the hard scattering
carries a lower fraction $x_\gamma$ of the photon momentum are
required to clarify the situation, both experimentally and theoretically.  

\begin{figure}
\begin{center}
\includegraphics[width=0.65\textwidth,angle=0]{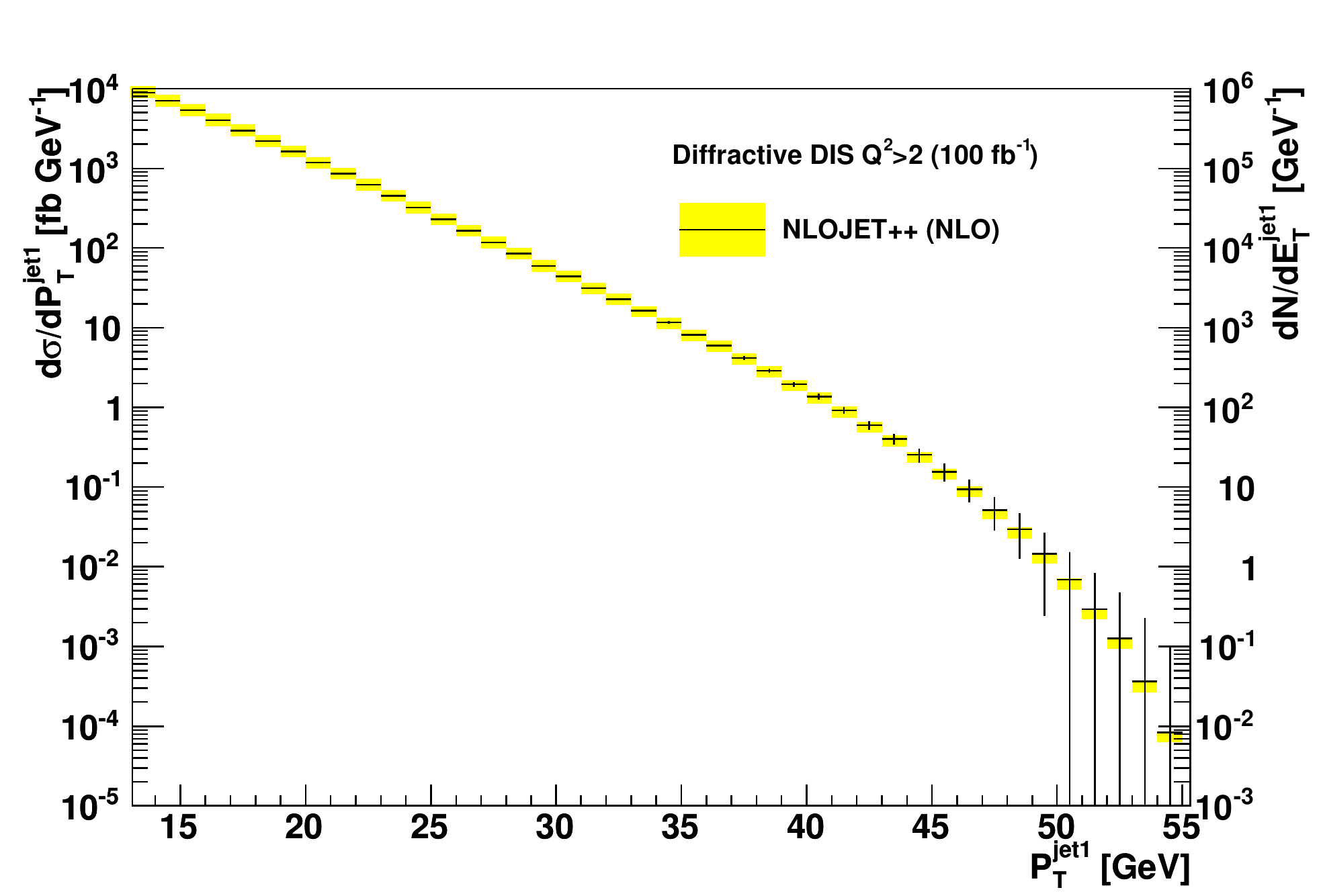}
\end{center}
\caption{Simulated transverse momentum distribution of the jets in diffractive dijet production in DIS ($Q^2 > 2$ GeV$^2$).  The simulation was performed using NLOJET++, assuming integrated luminosity of $100 \; {\rm fb}^{-1}$ and high acceptance for the scattered electron ($1^\circ$).  Scale uncertainties are illustrated by varying the factorisation scale in the range $(0.25 \mu^2,4 \mu^2)$.
}
\label{Fig:diffdijets}
\end{figure}

At the LHeC, much larger diffractive jet transverse momenta
are measurable ($p_T \stackrel{<}{_{\sim}} M_X / 2$) in both photoproduction
and DIS. An example study is
shown in Fig. ~\ref{Fig:diffdijets},
where the diffractive DIS dijet
cross section is simulated for the LHeC kinematics 
and acceptance, using NLOJET++ \cite{Nagy:2003tz}, 
with the H1 2006 Fit B DPDFs \cite{Aktas:2006hy}. 
Kinematic cuts of $\xpom<0.01$, $Q^2 > 2 \; {\rm GeV}^2$,
$0.1<y<0.7$ and $\theta_e> 1^\circ$, matching the expected LHeC detector
geometry and ensuring good containment for the jets and the scattered
electron.  
Jets were reconstructed using the $k_T$ algorithm with  
$R=1$ and an 
integrated luminosity of
$100 \; {\rm fb}^{-1}$ is assumed. 
The statistical precision remains excellent up to jet $p_T$ values of
around $40 \ {\rm GeV}$, with measurements possible up to 
around $50 \ {\rm GeV}$.   
Theory scale variations in the range of $(0.25 \mu^2,4 \mu^2)$ 
lead to much smaller uncertainties than is the case in the HERA data.

Diffractive dijet photoproduction at the LHeC 
is expected to be dominated by the resolved photon 
contribution.
A range of transverse momenta similar to the DIS case 
is accessible in photoproduction,
assuming tagging of electrons scattered through small angles as
described in Section~\ref{sec:gammaptot}. 
Fractional
DPDF momenta $z_{I\!\!P}$, and in the resolved photoproduction case, $x_\gamma$
values, between one and two orders of magnitude smaller than 
at HERA are typically accessible. All of these improvements will 
lead to a new level of precision in
tests of factorisation 
and constraints on the diffractive gluon density 
in new kinematic regions from diffractive jet production at the 
LHeC \cite{Newman:2009mb}.


The simulated measurement of the longitudinal proton structure 
function, $F_L$ described in Section~\ref{sec:flong}, could also
be extended to extract the diffractive analogue, $F_L^D$. 
At small $\beta$, where the cross section for longitudinally polarised
photons is expected to be dominated by a leading twist contribution,
an $F_L^D$ measurement provides further complementary
constraints on the role of gluons in the diffractive PDFs. 
As $\beta \rightarrow 1$, a higher twist contribution from 
longitudinally polarised photons, closely related to that 
driving vector meson electroproduction, dominates the diffractive
cross section in many models \cite{Bartels:1998ea} and a measurement to even
modest precision would give considerable insight. 
A first measurement of this quantity has recently been reported by
the H1 Collaboration \cite{Collaboration:2011ij}, 
though the precision is strongly limited by statistical
uncertainties. The LHeC provides the opportunity to explore it in
much finer detail.

In contrast to leading proton production, the production of leading
neutrons in DIS ($ep \rightarrow eXn$)
requires the exchange of a net isovector system. Data
from HERA have supported the view that this process is driven
dominantly by charged pion exchange over a wide range of neutron
energies \cite{Aaron:2010ze}. With the planned
emphasis on zero degree calorimetry for leading neutron 
measurements (see Chapter~\ref{detector:fwdbwd}), 
LHeC data will thus constrain the 
structure of the pion at much lower $x$ and larger $Q^2$ values
than has been possible hitherto.
Note also that the combination of rapidity gap detection and zero degree calorimetry offers the possibility of disentangling coherent from incoherent nuclear diffraction.

\subsubsection{Diffractive DIS, dipole models and sensitivity to non-linear effects}


Diffractive DIS at the LHeC will provide an opportunity to test the 
predictions of collinear factorisation and the possible onset of 
non-linear or higher-twist effects in the evolution. Of particular 
importance is the semi-hard regime $Q^2<10$ GeV$^2$ and 
$x$ as small as possible. It is possible that the non-linear saturation
regime will be easier to reach with diffractive than with inclusive 
measurements, since diffractive processes are mostly sensitive to 
quantum fluctuations in the proton wave function that have a virtuality 
of order of the saturation scale $Q_s^2$, instead of $Q^2$. As a result, 
power corrections (not the generic $\Lambda_{QCD}^2/Q^2$ corrections, but 
rather the sub-class of them of order $Q_s^2/Q^2$) 
are expected to come into play starting from a higher value of $Q^2$ 
in diffractive than in inclusive DIS. Indeed, there is already a hint of
this at HERA: collinear factorisation starts to fail below about 3 GeV$^2$ 
in the case of $F_2$ \cite{:2009wt}, while it breaks down 
already around 8 GeV$^2$ in the case of $F_2^D$ \cite{Aktas:2006hy}.
This fact can alternatively be observed in the feature that models 
which in principle should only work for small $Q^2$, can in practice be 
used up to larger $Q^2$ for diffractive than  for inclusive 
observables (see e.g. \cite{Armesto:2010ee}).

\begin{figure}[ht] \unitlength 1mm
  \begin{center}
\includegraphics[width=0.8\textwidth]{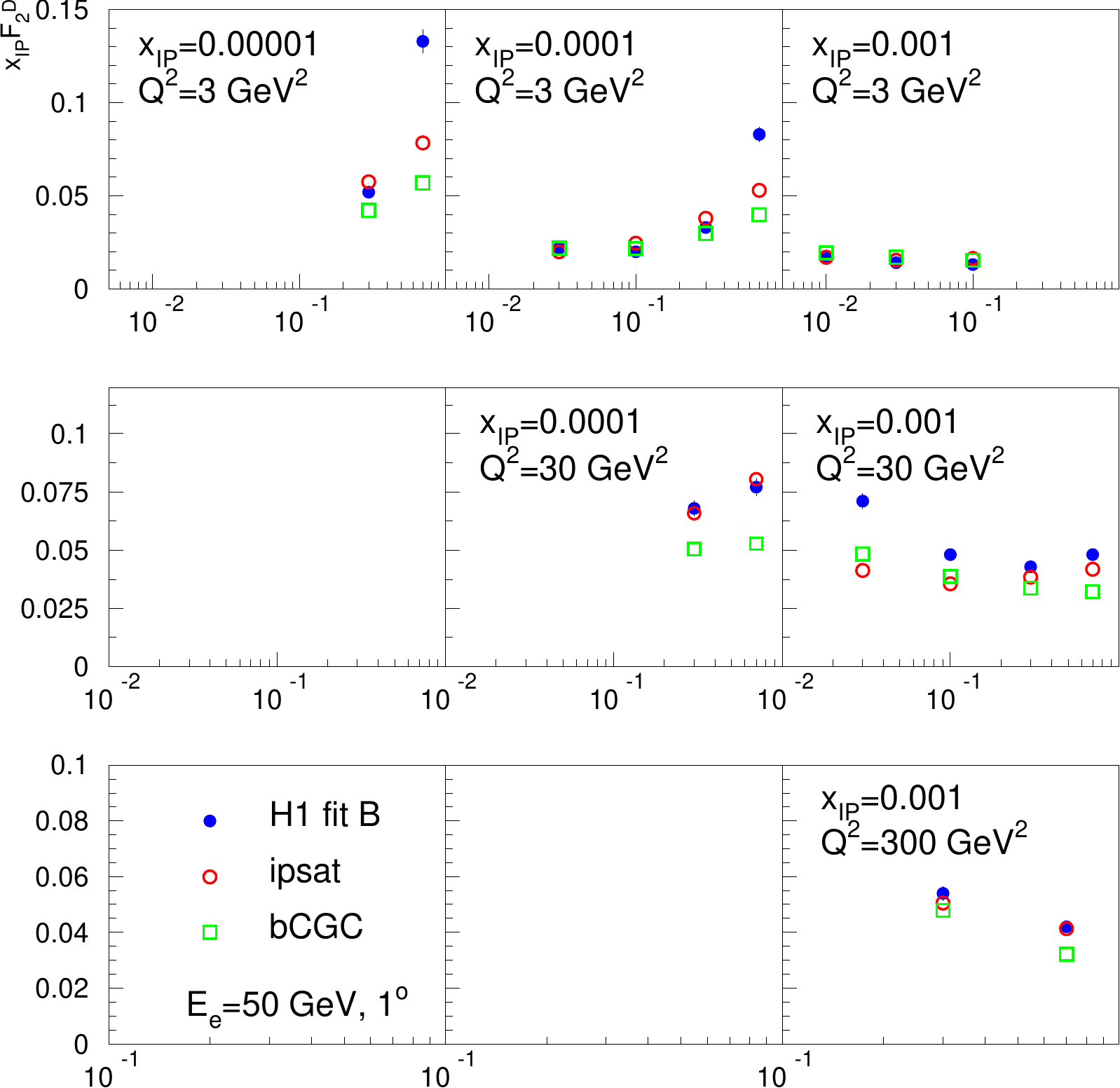}
  \end{center}
  \caption[]{Simulated $F_2^D$ measurements in selected $\xpom$, $\beta$
and $Q^2$ bins. An extrapolation of the H1 Fit B DPDF fit to HERA data
is compared with two different implementations of the dipole model,
both of which contain saturation effects and 
include $q \bar{q} g$ photon fluctuations in addition to 
$q \bar{q}$ ones.}
\label{ddis:tuomas}
\end{figure}

With the sort of measurement precision for $F_2^D$ achievable at the LHeC, 
it ought to be possible
to distinguish between different models, as illustrated in 
Fig.~\ref{ddis:tuomas}. For the simulated data shown here, a 
conservative situation is
assumed, in which the electron beam energy
is $50 \ {\rm GeV}$ and only the rapidity gap 
selection method is used, such that the highest  
$\xpom$ bin is at $0.001$. H1 Fit B \cite{Aktas:2006hy} extrapolations 
(as in Fig.~\ref{ddis:f2d150}) are compared with the 
``b-sat" \cite{Kowalski:2003hm,Kowalski:2006hc}
and bCGC \cite{Watt:2007nr} 
dipole models.
As has been found to be necessary to describe HERA data, 
photon fluctuations to $q \bar{q} g$ states are included in addition
to the usual $q \bar{q}$ dipoles used to describe inclusive 
and vector meson cross
sections.
Both dipole models differ substantially from the H1 Fit B extrapolation.
The LHeC simulated precision and kinematic range are sufficient
to distinguish between a range of models with and without saturation effects,
and also between different models which incorporate saturation.

%% file: physics/tex/dep2ns.tex
The connection between nuclear shadowing and diffraction was established a long time ago by Gribov \cite{Gribov:1968jf}.  Its key approximation is that the nucleus can be described as a dilute system of nucleons in the nucleus rest frame.
The accuracy of this approximation for hadron-nucleus interactions is on the level of a few \%, which reflects the small admixture of non-nucleonic degrees of freedom in nuclei and the small off-shellness of the nucleons in nuclei as compared to the soft strong interaction scale.
Gribov's result can be derived using  the AGK cutting rules \cite{Abramovsky:1973fm} and hence it is a manifestation of unitarity \cite{Frankfurt:1998ym,Frankfurt:2003zd}.  
The formalism can be used to calculate directly  cross sections of  $\gamma (\gamma^*)$-nucleus scattering for the interaction with $N=2$ nucleons, 
but has to be supplemented by additional considerations to account for the contribution of the interactions with $N\ge 3$ nucleons.
 
In this context, nuclear PDFs  at small $x$ can be calculated \cite{Frankfurt:1998ym,Frankfurt:2003zd} combining
unitarity relations for different cuts of the shadowing  diagrams corresponding to diffractive and inelastic final states, with the QCD factorisation theorem for hard diffraction \cite{Collins:1997sr}. A {\it model-independent}  expression for the nuclear PDF at fixed impact parameter $b$, valid for the case $N=2$   \cite{Frankfurt:1998ym}, reads:
\begin{eqnarray}
\Delta \left[ x f_{j/A}(x,Q^2,b)\right]&=&  x f_{j/N}(x,Q^2,b) -x f_{j/A}(x,Q^2,b)   \nonumber \\
&=&  8 \pi A(A-1) \Re e \left[\frac{(1-i\eta)^2}{1+\eta^2}  \int^{0.1}_x d \xpom
\beta f_j^{D(4)}(\beta,Q^2,\xpom,t_{{\rm min}})\right. \nonumber \\ 
&\times& \left.\int^{\infty}_{-\infty}d z_1  \int^{\infty}_{z_1}  d z_2 \, 
\rho_A(\vec{b},z_1) \rho_A(\vec{b},z_2) 
e^{i (z_1-z_2) \xpom m_N}\right],
\label{eq:m12}
\end{eqnarray}
where $ f_{j/A}(x,Q^2)$ and $f_{j/N}(x,Q^2)$ are nuclear and nucleon PDFs, $ f_j^{D(4)}(\beta,Q^2,\xpom,t_{{\rm min}})$ are diffractive nucleon PDFs, $\eta =\Re e \,A^{diff}/ \Im m \,A^{diff}\approx 0.17$, $  \rho_A(r)$ is the nuclear matter density, and $t_{{\rm min}}=-m_N^2\xpom^2$ with $m_N$ the nucleon mass. 
Eq. (\ref{eq:m12}) satisfies the QCD evolution equations to all orders in $\alpha_s$.
Numerical studies indicate that the dominant contribution to the shadowing probed by present experiments - corresponding to not very small $x$ - comes from the region of relatively large $\beta$,
for which small-$x$ approximations which involve resummation of  $\ln x$ terms are not important.

In Eq. (\ref{eq:m12}), the interaction of different configurations of the hard probe (e.g. $q\bar q$, $q\bar q g$, vector meson resonances,$\dots$) are encoded in $ f_j^{D(4)}(\beta,Q^2,\xpom,t_{{\rm min}})$. 
For the case of more than $N=2$ nucleons, there are two or more intermediate nucleon diffractive states which may be different and thus result in a different interaction between the the virtual photon and the nucleus. Therefore the interaction of the hard probe with $N \ge 3$ nucleons is sensitive to finer details of  the diffractive dynamics, namely the interplay 
between the interactions of the hard probe with $N$ nucleons
with different cross sections.
This (colour) fluctuation effect  is  analogous to the inelastic shadowing phenomenon for the scattering of hadrons from nuclei, with the important difference that 
the dispersion of the interaction cross sections for the configurations  in the projectile  is much smaller in the hadronic case than in DIS. 
 
In order to estimate this effect, one should note that,
experimentally, the energy dependence of hard diffraction is close to that 
observed for soft Pomeron dynamics (the soft Pomeron intercept $\alpha_{\mathbb P}\approx 1.11$)   with  
the  hard Pomeron contribution ($\alpha_{\mathbb P}\approx 1.25$) being a small correction.
%
This fact indicates that  hadron-like (aligned jet) configurations \cite{Abramowicz:1995hb}, evolved via DGLAP evolution to large $Q^2$,  dominate hard diffraction in DIS, while
 point-like  configurations  give an important, and increasing with $Q^2$, contribution to small-$x$ PDFs.
 This  reduces the uncertainties in the treatment of $N\ge 3$ contributions \cite{Guzey:2009jr,Frankfurt:2011cs}.  Calculations show that the difference between two extreme scenarios of colour fluctuations   is  $\le  20\% $ for $A\sim 200$ and much smaller for lighter nuclei, see the  two FGS10 curves in Figs. \ref{Fig:npdfs} and \ref{Fig:famodels}. Besides, fluctuations tend to reduce 
the shadowing somewhat compared with the approximations neglecting 
them \cite{Armesto:2010kr,Frankfurt:1998ym,Armesto:2003fi,Tywoniuk:2007xy} 
(compare the FGS10 results in Fig. \ref{Fig:famodels} left with those 
labelled AKST). The gluon density is more sensitive to the
magnitude of fluctuations than $F_2$, as can be inferred from
Figs. \ref{Fig:npdfs} and \ref{Fig:famodels} right.

Finally, the AGK technique also allows 
the calculation of the nuclear diffractive PDFs, see below, and  fluctuations of multiplicity in non-diffractive DIS \cite{Frankfurt:1998ym,Frankfurt:2011cs, Frankfurt:2003gx}. Both observables turn out to be  sensitive to the pattern of colour fluctuations.

%% file: physics/tex/ddisea.tex

Inclusive diffraction was first measured in  DIS events  in $ep$  collisions at the HERA collider.
LHeC would be the first electron-ion collider machine, and therefore 
 DDIS off nuclei at this machine will be a completely unexplored
territory throughout the whole kinematic domain accessed. This  implies  a huge discovery potential
in this field.

Despite this lack of experimental information on DDIS off nuclei, we have expectations, based on our current understanding of QCD, of how it should look.  For instance, the theory of nuclear shadowing allows us to construct nuclear diffractive PDFs for large $Q^2$ (see the previous item) while, within the Colour Glass Condensate framework, nuclear diffractive structure functions can be predicted at small $x$. Depending on kinematics and the heavy ion species, different patterns of nuclear shadowing or antishadowing are expected as a function of $\beta$ and $\xpom$. This is just one of many examples of what should be checked with an $e$A collider. Others are the impact parameter dependence introduced in the models, or
the relation between nuclear shadowing and diffraction in $ep$ which relies on what we know on DDIS from HERA. Therefore, in the larger kinematic domain accessible at the LHeC there are many things to discover about the structure of nuclei with diffractive measurements.

One of the main issues which needs to be established is whether the collinear, leading twist, factorisation of inclusive diffraction, proved for protons, 
is applicable for scattering off nuclei, and the region of its applicability. An important question arises as to where the factorisation would  break down, i.e. for which values of $Q^2$ and $W$, and whether it depends on the mass number, which would provide most important information on the role of the higher twists in different nuclei.
A related issue is whether the factorisation of the hadron vertex which is used in the proton case also holds in the nuclear case. In the analysis of the diffractive structure functions, the Regge-type factorisation is usually assumed. This factorisation states that the diffractive structure function is written as a product of the two factors: one of them is the Pomeron structure function that depends on $\beta$ and $Q^2$, and the other is the Pomeron flux factor that is a function of $t$ and $\xpom$. The latter one is usually parameterised using a Regge form with a Pomeron intercept
being close to, albeit slightly higher than, the value obtained from  soft interactions. It is currently unclear whether such factorisation would still hold in the nuclear case, and this is one of the issues that can be tested at the LHeC. Also  the range of possible parameters, like  the Pomeron intercept,
extracted from such analysis, would provide important details on the nuclear dynamics.

\begin{figure}
 \centering
  \includegraphics[width=0.8\textwidth]{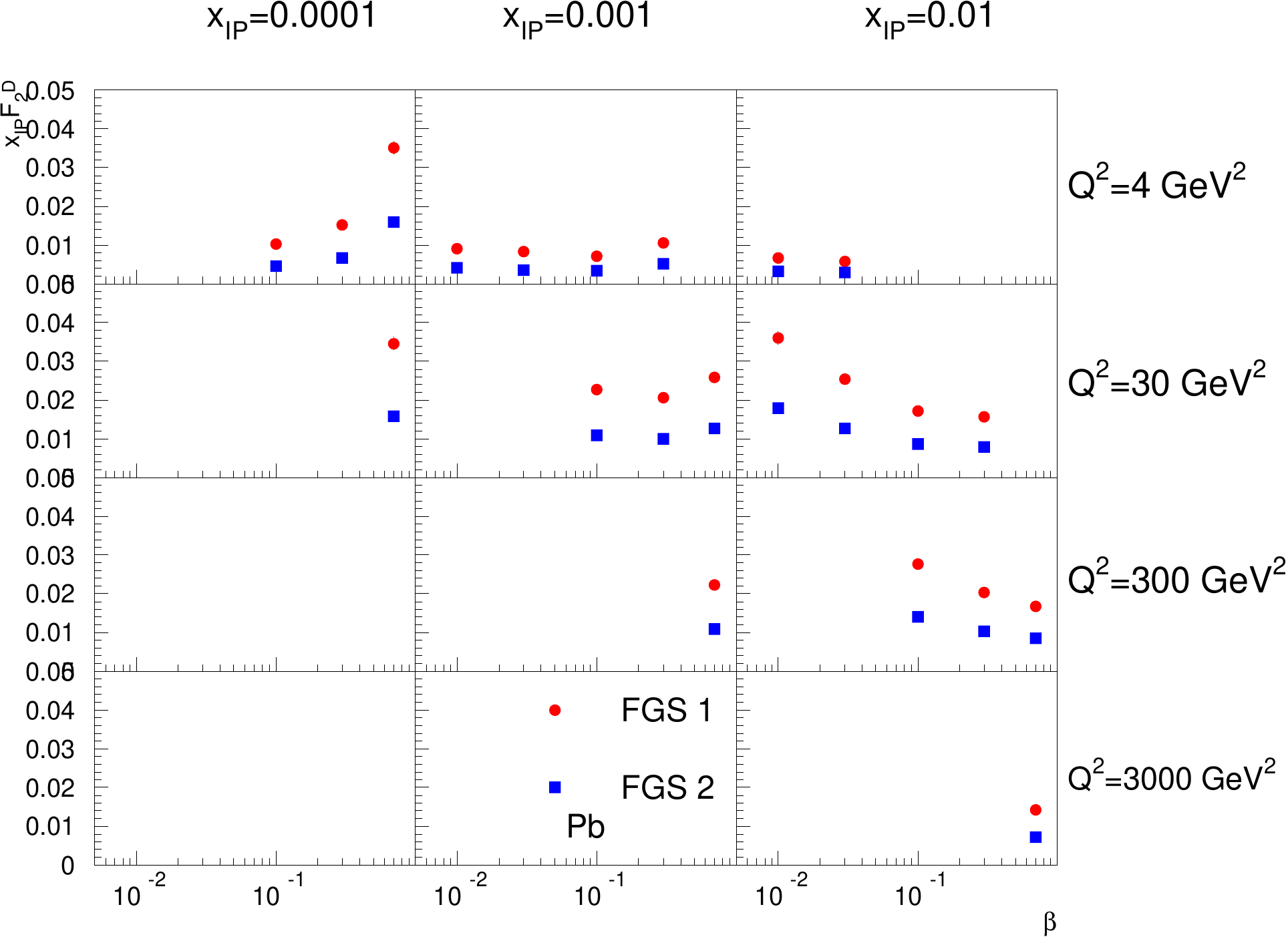}
  \caption{Diffractive structure function $\xpom F_2^D$ for Pb in bins of $Q^2$ and $\xpom$ as a function of $\beta$. Model calculations are taken from \cite{Frankfurt:2011cs}.}
\label{fig:fgsdiff10}
\end{figure}

\begin{figure}
 \centering
  \includegraphics[width=0.7\textwidth]{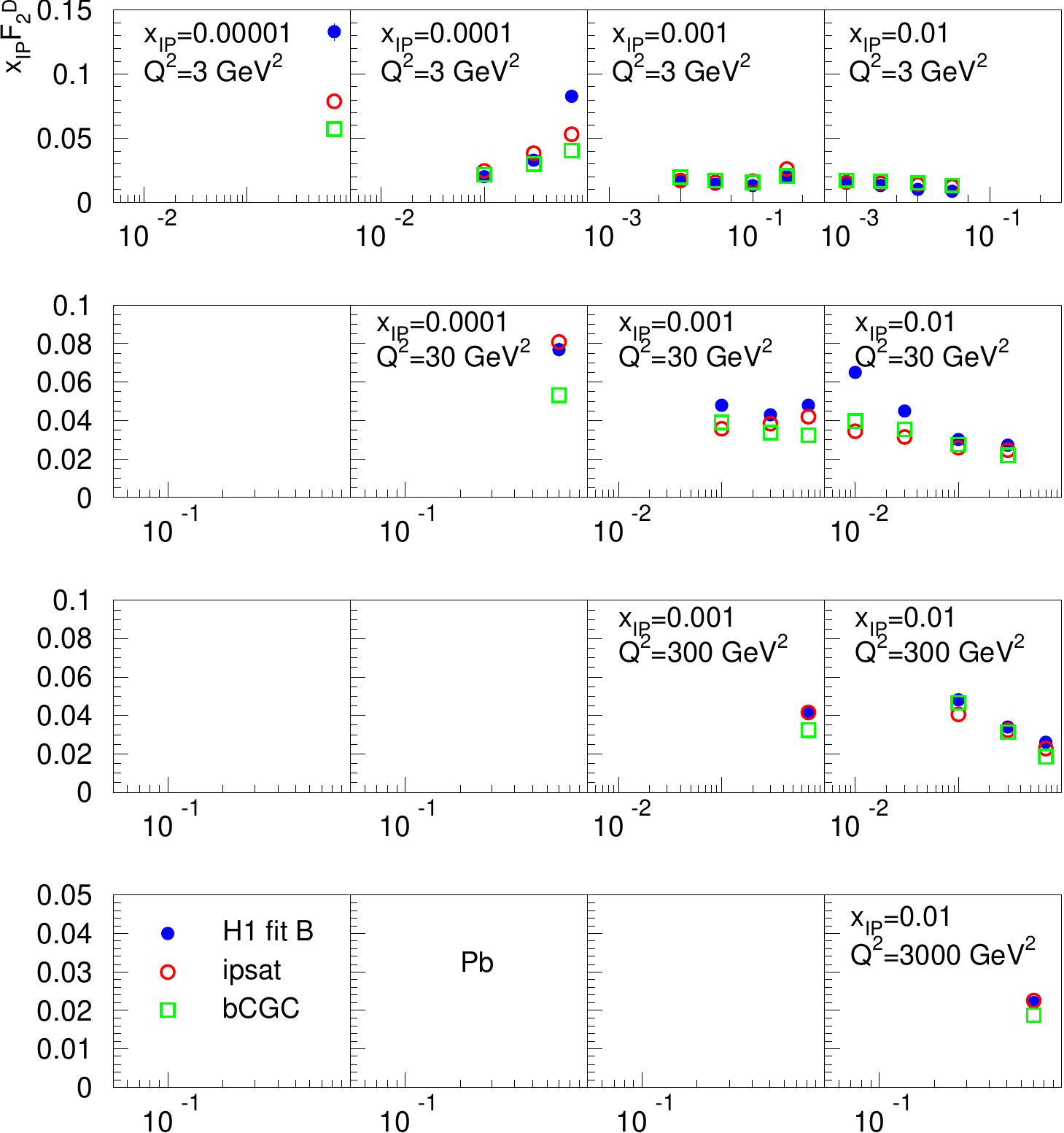}
  \caption{Diffractive structure function $\xpom F_2^D$ for Pb in bins of $Q^2$ and $\xpom$ as a function of $\beta$. Model calculations are based on the dipole framework \cite{Marquet:2007nf,Kowalski:2008sa}. }
\label{fig:lappidiff}
\end{figure}

Predictions from a variety of models for nuclear coherent 
diffraction (see comments on the different types of diffractive processes 
on nuclei in Section\,\ref{sec:vm}), are shown in Figs.~\ref{fig:fgsdiff10} and \ref{fig:lappidiff}. 
The chosen models here are FGS10 \cite{Frankfurt:2011cs} and  KLMV \cite{Marquet:2007nf,Kowalski:2008sa}. Both plots show 
selected LHeC pseudodata for
$\xpom F_2^D$ as a function of $\beta$ in bins of $Q^2$ and $\xpom$. Statistical and systematic errors are added in quadrature, with systematic errors estimated to be at the level of $5\%$. The models give very different predictions both in absolute value and in their detailed dependence on $\xpom$ and $Q^2$, which cannot be resolved without LHeC 
data.

\begin{figure}
 \centering
  \includegraphics[width=0.6\textwidth]{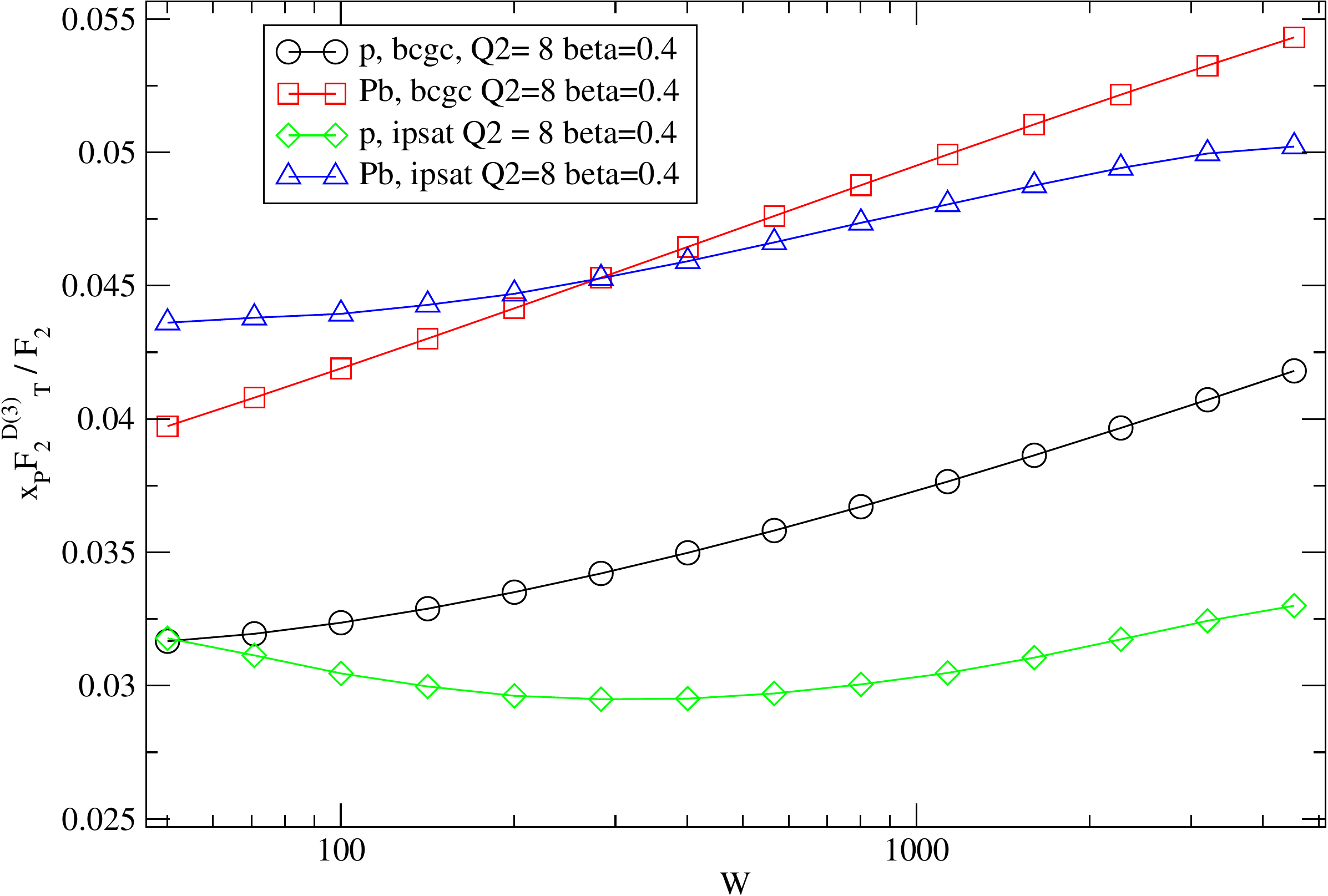}
  \caption{Ratio of the transversely polarised photon contribution to the 
diffractive structure function $\xpom F_2^D$ to the inclusive structure function $F_2$ in $p$ and Pb for fixed values of $Q^2$ and $\beta$ as a function of the energy $W$.
   Model calculations are based on the dipole framework \cite{Marquet:2007nf,Kowalski:2008sa}.}
\label{fig:ratiodiff}
\end{figure}

Also shown in Fig.~\ref{fig:ratiodiff} are 
predicted diffractive-to-total ratios of the structure function $F_2$ as a function of $W$.  It was demonstrated in \cite{GolecBiernat:1999qd} that the constancy with $W$ of this ratio for the proton - approximately shown by HERA data - can be naturally explained in models which include saturation effects, because in the black disk regime the ratio of diffractive-to-total cross sections tends to a constant value. It has been predicted that in the black disk regime this ratio (for coherent diffraction) may grow as large as $50\%$ \cite{Nikolaev:1995xu}. In reality, it could be smaller due to the density distribution  in impact parameter.
Within the given energy range the models shown in Fig.~\ref{fig:ratiodiff}
predict a slight variation with energy. Note however the rather substantial difference between predictions coming from the different models as well as the fact that the plot shows the ratio of structure functions for given $\beta$ and $\xpom$ and not integrated cross sections.
 The uncertainty 
in modelling the impact parameter is one of the main sources of the discrepancies between the models. Precise LHeC data
are required for clarifying these aspects. 

Finally we note that, if the scattering on a nucleus at small $x$ is dominated almost entirely by the so-called black disk regime, then in principle dramatic effects are expected  that can be revealed by studying the final states in diffractive events \cite{Frankfurt:2001nt}.  As demonstrated in  \cite{Gribov:1968gs}, the total virtual photon-nucleus  cross section in the black disk  limit reads simply
\begin{equation}
\sigma_{\gamma^* A} \; = \; 2 \pi R_A^2 \, (1-Z_3) ,
\label{eq:gribovbdl}
\end{equation}
where $R_A$ is the nuclear radius and $Z_3$ the charge renormalisation constant due to hadrons. The coefficient $1-Z_3$ can be computed in terms of the hadronic components of the photon wave function and related to the cross section for the annihilation of electron-positron pairs into hadrons. Since the elastic part (i.e. that due to diffraction) is half the total cross section in this regime, one can obtain from eq.~(\ref{eq:gribovbdl}) a spectrum of the diffractive masses \cite{Frankfurt:2001nt} that, in the centre-of-mass of the diffractively produced system, should be the same as in $e^+e^-$ annihilation. A similar analysis for exclusive processes in this limit shows that the exclusive diffractive production cross sections on nuclei (see section\,\ref{sec:vm})  would exhibit a $1/Q^2$ behaviour instead of the $1/Q^6$ behaviour expected from pQCD. This is due to the fact that a factor $1/Q^4$ which comes from  the square of the cross section of the interaction of a small dipole with the target disappears in the black disk limit.

%% file: physics/tex/introjets.tex


Inclusive measurements provide essential information about the
integrated distributions of partons in a proton. However, as was
discussed in previous sections, more exclusive measurements are needed
to pin down the essential details of the small-$x$ dynamics.  For
example, a central prediction of the BFKL framework at small $x$ is the
diffusion of the transverse momenta of the emitted partons between the
photon and the proton. In the standard collinear approach with
integrated parton densities the information about the transverse
momentum is not accessible. However, it can be recovered within a
different framework which utilises unintegrated parton
distribution functions, dependent on parton transverse momentum as
well as $x$ and $Q^2$. 
Unintegrated PDFs are  natural in the BFKL approach to
small-$x$ physics. A general, fundamental expectation is that
as $x$ decreases, the distribution in transverse momentum
of the emitted partons broadens, resulting in diffusion.

The specific
parton dynamics can be tested by a number of exclusive measurements.
These in turn can provide valuable information about the distribution
of transverse momentum in the proton.  
As discussed in
\cite{Collins:2005uv}, for  many inclusive observables the
collinear approximation with integrated PDFs
is completely insufficient, and even just including parton transverse
momentum effects by hand may not be sufficient to
describe many observables.  
In DIS, for example, processes needing unintegrated distributions
include the transverse momentum distribution of heavy quarks.
Similar problems are encountered in hadron
collisions when studying heavy quark and Higgs production. The natural
framework using unintegrated PDFs
gives a much more reliable description.  
Furthermore, lowest-order calculations in
the framework with unintegrated PDFs 
provide a much more realistic description of
cross sections concerning kinematics. This may well lead to NLO and
higher corrections being much smaller numerically than they typically
are at present in standard collinear factorisation, since the LO
description is better.

This approach, however, calls for precise measurements of a variety of
relatively exclusive processes in a wide kinematic range.  
As discussed below,
measurements of dijets, forward jets and particles, as well as transverse
energy flow, are 
required to constrain the unintegrated PDFs and
will give valuable information about  parton dynamics at small $x$.
While we will discuss the case of DIS on a proton, all conclusions can be paralleled for DIS on nuclei.

%% file: physics/tex/updfs.tex


\newcommand\BIBjour[1]{\emph{#1}}
\newcommand\BIBvol[1]{\textbf{#1}}

\newcommand{\eqdef}{\stackrel{\textrm{def}}{=}}
\newcommand{\eqprelim}{\stackrel{\textrm{prelim}}{=}}
\newcommand{\eqbad}{\stackrel{\textrm{?}}{=}}

\newcommand\3[1]{\boldsymbol{#1}}

\newcommand{\T}[1]{\boldsymbol{#1}_{\perp}}
\newcommand{\Tj}[2]{\boldsymbol{#1}_{#2\,\perp}}
\newcommand{\Tsc}[1]{#1_{\perp}}

\DeclareRobustCommand{\MSbar}{\ensuremath{ \overline{\rm MS} }}

\DeclareRobustCommand*\diff[2][]{%
   \mathop{\mathrm{d}^{#1}{}#2}\nolimits
}


The standard integrated parton densities are functions of the
longitudinal momentum fraction of a parton relative to its parent
hadron, with an integral over the parton transverse momentum.  In
contrast, unintegrated, or transverse-momentum-dependent (TMD), parton
densities depend on both parton longitudinal momentum fraction and parton
transverse momentum.  Processes for which unintegrated densities are
natural include the Drell-Yan process (and its generalisation to Higgs
production), and semi-inclusive DIS (SIDIS).  In SIDIS, we need TMD
fragmentation functions as well as TMD parton densities.

In the literature there are several apparently different approaches to
TMD parton densities, with varying degrees of explicitness in the
definitions and derivations.
\begin{itemize}
\item The CSS approach \cite{Collins:1981uk, Collins:1981uw,
    Collins:2003fm, Collins:2008ht} and some further developments
  \cite{Ji:2004wu}.
\item The CCFM approach
  \cite{Ciafaloni:1987ur,Catani:1989yc,Catani:1989sg,Marchesini:1994wr}
  for small $x$.
\item Related BFKL associated works 
  \cite{Balitsky:2008zza,Balitsky:2001gj}.
\end{itemize}

Central to this subject is the concrete definition of TMD densities,
and complications arise because QCD is a gauge theory.  A natural
initial definition uses light-front quantisation: the unintegrated
density of parton $j$ in hadron $h$ would be
\begin{equation}
  \label{eq:int.pdf.def1}
  f_{j/h}(x, \T{k}) \eqbad
         \frac{1}{ 2x (2\pi)^3 }
         \sum_\lambda 
         \frac{ \braket{P,h| b_{k,\lambda,j}^{\dag} b_{k,\lambda,j} |P,h}_{\rm c} }
              { \braket{P,h|P,h} }\, ,
\end{equation}
where $b_{k,\lambda,j}$ and $b_{k,\lambda,j}^{\dag}$ are light-front annihilation
and creation operators, $j$ and $\lambda$ label parton flavor and helicity,
while $k=(k^+,\T{k})$ is its momentum, and only connected graphs `c'
are considered.  The `?' over the equality sign warns that the formula
does not apply literally in QCD.  Expressing $b_{k,\lambda,j}$ and
$b_{k,\lambda,j}^{\dag}$ in terms of fields gives the TMD density as the
Fourier transform of a light-front parton correlator.  For example, for a
quark
\begin{equation}
  \label{eq:TMD.pdf.def1a}
  f_j(x,\T{k}) 
  \eqbad
  \int \frac{ \diff{w^-} \diff[2]{\T{w}} }{ (2\pi)^3 } \,
    e^{ -ix P^+w^- + i \T{k}\cdot\T{w} }
    \braket{P| \,
            \overline{\psi}_j(0,w^-,\T{w}) \,
            \frac{\gamma^+}{2} \,
       \psi_j(0) \,
    |P}_{\text{c}} \,.
\end{equation}
One can similarly define a TMD fragmentation function \cite{Collins:1981uw}
$d_{h/j}(z,\T{p})$, for the probability density of final-state hadron
$h$ in an outgoing parton $j$.

\begin{figure}
  \centering
  \begin{tabular}{c@{\hspace*{2cm}}c}
    \includegraphics[scale=0.8]{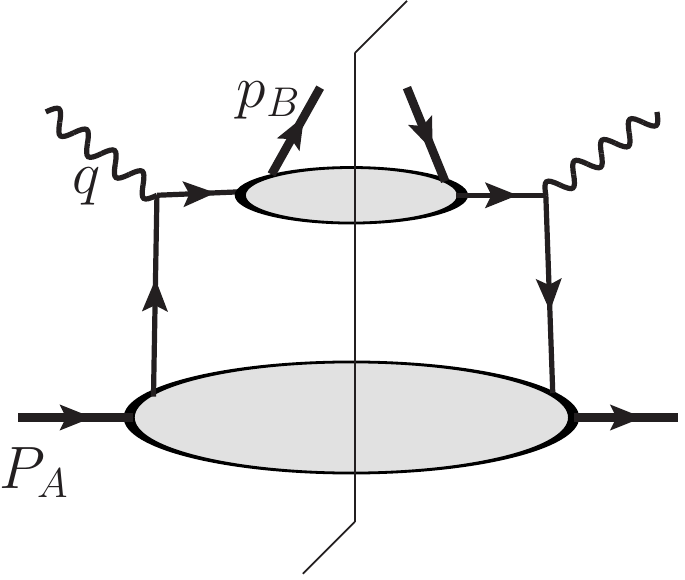} 
  &
    \includegraphics[scale=0.8]{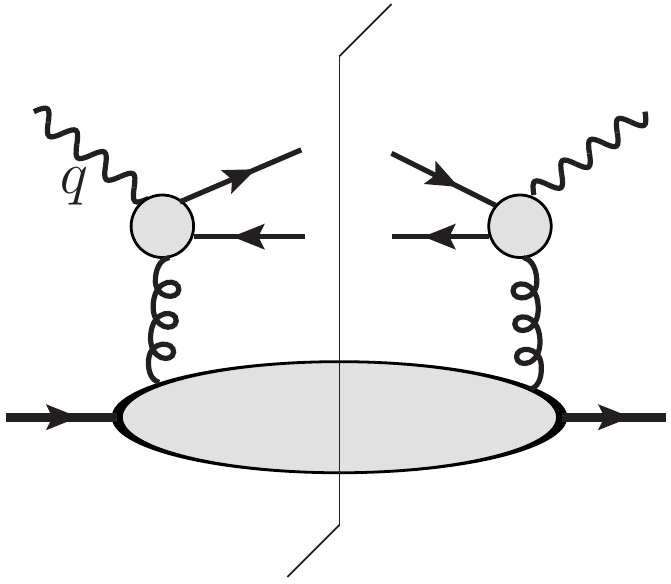}
  \\
    (a) & (b)
  \end{tabular}
  \caption{(a) Parton model factorisation for a SIDIS cross section. 
           (b) Factorisation for high-energy $q\bar{q}$ photoproduction.} 
  \label{fig:small.x.fact}
\end{figure}

The corresponding factorisation formula for SIDIS $e+A(P_A) \to
e+B(p_B)+X$ is \cite{Ji:2004wu}
\begin{equation}
\label{eq:SIDIS.fact}
  \frac{ \diff{\sigma} }
       { \diff{x} \diff{Q^2} \diff{z} \diff[2]{\Tj{P}{B}} }
  = \sum_j \int \diff[2]{\T{k}} H_j f_{j/A}(x, \T{k}) d_{B/j}(z,\Tj{p}{B}+z\T{k}),
\end{equation}
where $z$ and $\Tj{P}{B}$ are the fractional longitudinal momentum
and the transverse momentum of the detected hadron relative to the
simplest parton-model calculation of the outgoing jet, while $H_j$ is
the hard-scattering factor for electron-quark elastic scattering;
see Fig.\ \ref{fig:small.x.fact}(a).  
In the fragmentation function $d_{B/j}$
in Eq.\ (\ref{eq:SIDIS.fact}), the use
of $z\T{k}$ with its factor of $z$ is because the transverse-momentum
argument of the fragmentation function is a transverse momentum of the
outgoing hadron relative to the parton initiating the jet, whereas
$\T{k}$ is the transverse momentum of a parton relative to a hadron.

The most obvious way of applying (\ref{eq:TMD.pdf.def1a}) in QCD is to
define the operators in light-cone gauge $A^+=0$, or, equivalently, to
attach Wilson lines to the quark fields with a light-like direction
for the Wilson lines.  One minor problem in QCD is that, because the 
wave function is infinite (see below), the exact probability interpretation of parton
densities cannot be maintained.

A much harder problem occurs because QCD is a gauge theory.
Evaluating TMD densities defined by (\ref{eq:TMD.pdf.def1a}) in
light-cone gauge gives divergences where internal gluons have
infinite negative rapidity \cite{Collins:1981uk}.  These cancel only
in the integrated density.  The physical problem is that any coloured
parton entering (or leaving) the hard scattering is accompanied by a
cloud of soft gluons, and the soft gluons of a given transverse
momentum are distributed uniformly in rapidity.  A parton density
defined in light-cone gauge corresponds to the asymptotic situation of
infinite available rapidity.

A quark in a realisable hard scattering can be considered as having a
transverse recoil against the soft gluons, but with a physically
restricted range of rapidity.  So a proper definition of a TMD density
must implement a rapidity cut-off in the gluon momenta.  Evolution
equations must take into account the rapidity cut-off.  The CSS
formalism \cite{Collins:1981uk} has an explicit form of the rapidity
cut-off and an equation for the dependence of TMD functions on the cut-off.
But in any alternative formalism the need in the definitions for a
cut-off to avoid rapidity divergences is non-negotiable.

Parton densities and fragmentation functions are only useful because
they appear in factorisation theorems, so a useful definition must
allow useful factorisation theorems to be formulated and derived.  An
improved definition involving Wilson line operators has recently been
given in \cite{Collins:2011qcdbook}; see also \cite{Aybat:2011zv}.

A second train of argument leads to a related kind of factorisation
(the so-called $\Tsc{k}$-factorisation) for processes at small $x$
\cite{Catani:1990eg}.  A classic process is photo- or
electro-production of charm pairs $\gamma(p_1)+h(p_2) \to Q(p_3) +
\bar{Q}(p_4) +X$, for which $\Tsc{k}$-factorisation has the form
\begin{equation}
  \label{eq:small.x.fact}
  4M^2 \sigma_{\gamma g}(\rho,M^2/Q_0^2)
  = \int \diff[2]{\T{k}} \int_0^1 \frac{\diff{z}}{z}
     \hat{\sigma}(\rho/z,\T{k}^2/M^2) f_{g/h}(x, \T{k}),
\end{equation}
see Fig.\ \ref{fig:small.x.fact}(b).  Here $\rho=M^2/(p_1+p_2)^2 \ll 1$,
and $M$ is the mass of the heavy quark.  The corresponding definition
of the TMD gluon density \cite{Ciafaloni:1987ur} is said to use
light-cone gauge, but there is in fact a hidden rapidity cut-off
resulting from the use of the BFKL formalism. 

Although both (\ref{eq:SIDIS.fact}) and (\ref{eq:small.x.fact}) use
$\Tsc{k}$-dependent parton densities, there are important differences.
In (\ref{eq:small.x.fact}), the hard scattering 
cross section $\hat{\sigma}$ has the
incoming gluon \emph{off}-shell, whereas in (\ref{eq:SIDIS.fact}), the
hard scattering $H_j$ uses on-shell partons. This is associated with a
substantial difference in the kinematics.  In \eqref{eq:SIDIS.fact}
for SIDIS, the transverse momenta of the partons relative to their
hadrons are less than $Q$, which allows the neglect of parton
virtuality in the hard scattering.  This approximation fails at large
partonic transverse momentum, $\T{k}\sim Q$, but ordinary
collinear factorisation is valid in that region.  
So the factorisation formula is
readily corrected, by adding a suitable matching term
\cite{Collins:1981uk}.

In contrast, in the small-$x$ formula \eqref{eq:small.x.fact}, the
gluon transverse momentum is comparable with the hard scale $M$.  So
it is not appropriate to neglect $\T{k}$ with respect to $M$, and the
hard scattering is computed with an off-shell gluon.  Factorisation is
actually obtained from BFKL physics, where the gluons in Fig.\
\ref{fig:small.x.fact}(b) couple the charm quark subgraph to a
subgraph where the lines have much larger rapidity.  

The evolution equation of the CS-style TMD functions used in
(\ref{eq:SIDIS.fact}) gives the dependence of the TMD functions on the
rapidity difference between the hadron and the virtual photon momenta.
The results for TMD functions and for the cross sections can finally
be obtained \cite{Ji:2004wu} in terms of (a) ordinary integrated
parton densities and fragmentation functions, (b) perturbatively
calculable quantities, and (c) a restricted set of non-perturbative
quantities.  The most important of these non-perturbative quantities is
the distribution in recoil transverse momentum per unit rapidity
against the emission of the soft interacting gluons, 
which is exponentiated after
evolution.  Importantly, it is independent of $x$ and $z$, and it is
universal between processes \cite{Collins:2004nx}, and different only
between gluons (colour octet) and quarks (colour triplet).  There is
also what can be characterised as a non-perturbative intrinsic
transverse momentum distribution in both parton densities and
fragmentation functions.  In the quark sector, all but the
fragmentation function are well measured in Drell-Yan processes
\cite{Landry:2002ix}.

On the other hand, evolution for the small-$x$ formalism in \eqref{eq:small.x.fact} is
given by the BFKL method.

The avenues for further improvement on this subject are both
theoretical and experimental. On the theory side, these concern the
relation between different formalisms for evolution
\cite{Collins:1981uk, Ji:2004wu, Balitsky:2008zza, Balitsky:2001gj,
  Collins:1984kg}, the extension of factorisation theorems to a larger
number of particles in the final state, and the matching to Monte
Carlo generators.  On the experimental side, the sensitivity to TMD
functions is linked to a sensitivity to parton transverse momentum.
This is the case of SIDIS at low transverse momentum. Another
interesting process which would enable the TMD gluon functions to be
probed is $ep \to e \pi\pi X$, with the pions being in different directions
(different jets), but such that they are close to back-to-back in the
$(q,p_i)$ (the so-called brick wall) frame.

%% file: physics/tex/dijets.tex
Finally, measuring SIDIS and dijet production off protons or nuclei at the LHeC will allow detailed investigations of non-linear parton evolution in QCD. In this respect, the SIDIS cross section \cite{Marquet:2009ca} and dihadron production \cite{Dominguez:2010xd} have been studied in the CGC framework. It turns out that, for small $x$, one is sensitive to the saturation regime of the target (proton or nucleus) wave function if the transverse momentum of the produced hadron is of the order of the saturation momentum.

\subsubsection{Dijet production and  angular decorrelation}

\begin{figure}[htbp]
\begin{center}
\includegraphics[width=0.6\columnwidth]{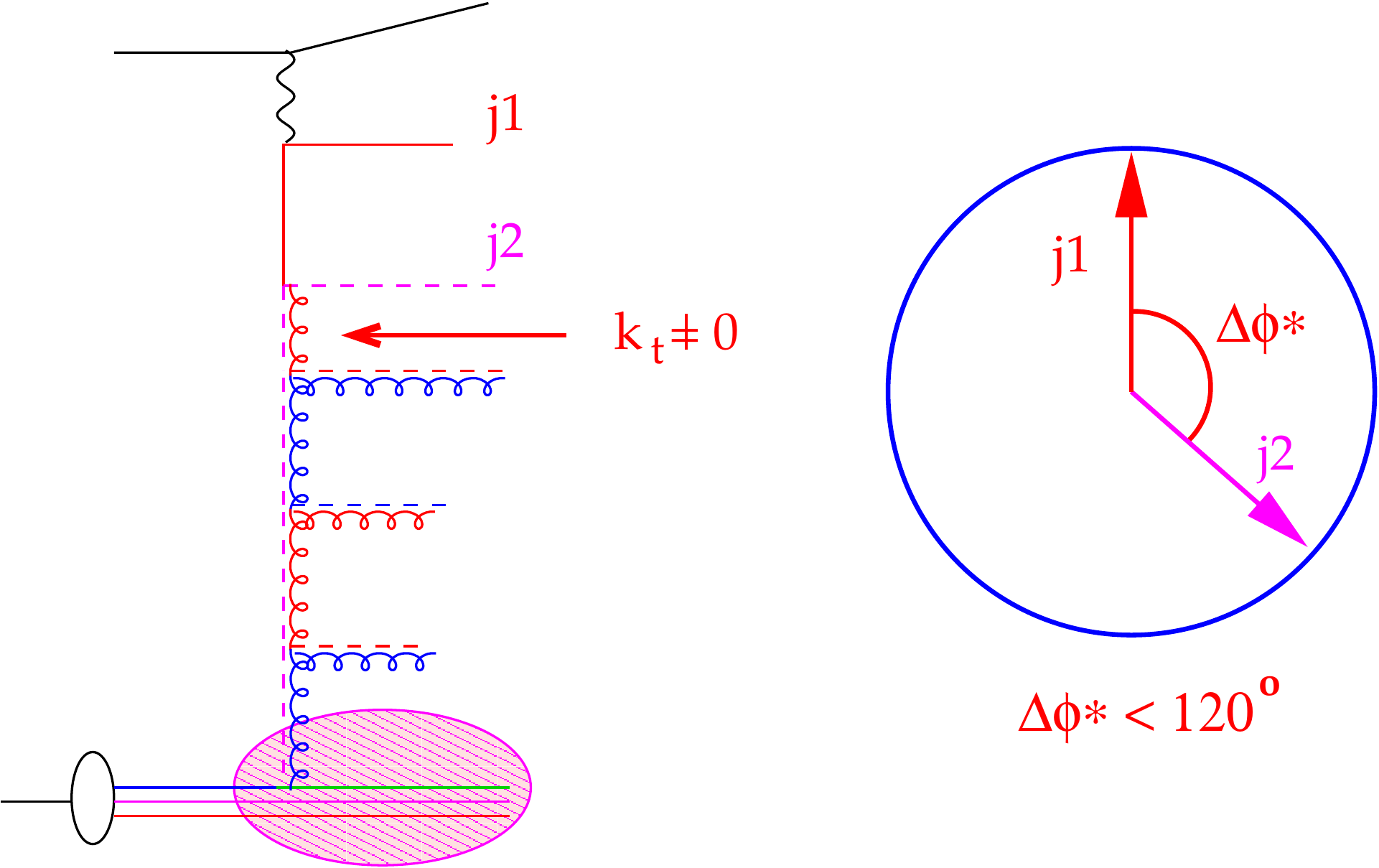}
\caption{Schematic representation of the production of a system of two jets in the process of virtual photon-gluon fusion.  The incoming 
gluon has non-vanishing transverse momentum $k_T\neq 0$ which leads to the decorrelation of the jets. $\Delta \phi$  is the angle between two jets.}
\label{fig:dijet-diagram}
\end{center}
\end{figure}

Dijet production in high energy deep inelastic electron-proton scattering is 
a very valuable process for the study of the     
small-$x$ behaviour in QCD. The dominant process is illustrated in Fig.~\ref{fig:dijet-diagram}, which is that of the   $\gamma^* g \rightarrow q\bar{q} \rightarrow$~dijet production.
The       
incoming gluon can have sizeable transverse momentum accumulated from diffusion       
in $k_T$ along the gluon chain.
As Bjorken-$x$ becomes smaller, and therefore the longitudinal momentum of the gluon also decreases, larger values of the transverse momentum $k_T$ can be sampled. This will lead to an 
 azimuthal decorrelation between the jets which increases with    
decreasing $x$. The definition of $\Delta \phi$ is indicated in Fig.~\ref{fig:dijet-diagram}. That is, the jets are no longer back-to-back since they must balance    
the sizeable transverse momentum $k_T$ of the incoming virtual gluon.  



This picture of dijet production is to be contrasted with the conventional 
picture which uses integrated parton distributions, and 
typically leads to a narrow distribution about the back-to-back
jet configuration. Higher orders usually broaden the distribution. However, as shown by direct measurements of DIS dijet data  \cite{Aktas:2003ja}, NLO DGLAP calculations are not able to accommodate the pronounced effect of the decorrelation.

Explicit calculations for HERA kinematics show that the models which include the resummation of powers of $\log 1/x$  compare favourably with the experimental data \cite{Askew:1994is,Kwiecinski:1999wj,Szczurek:2000pj,Hansson:2007de,Hautmann:2008vd}.
The proposal and calculations to extend such studies to diffractive DIS also exist \cite{Bartels:1996tc,Bartels:1999tn}.

\begin{figure}[htbp]
\begin{center}
\includegraphics[width=0.8\columnwidth]{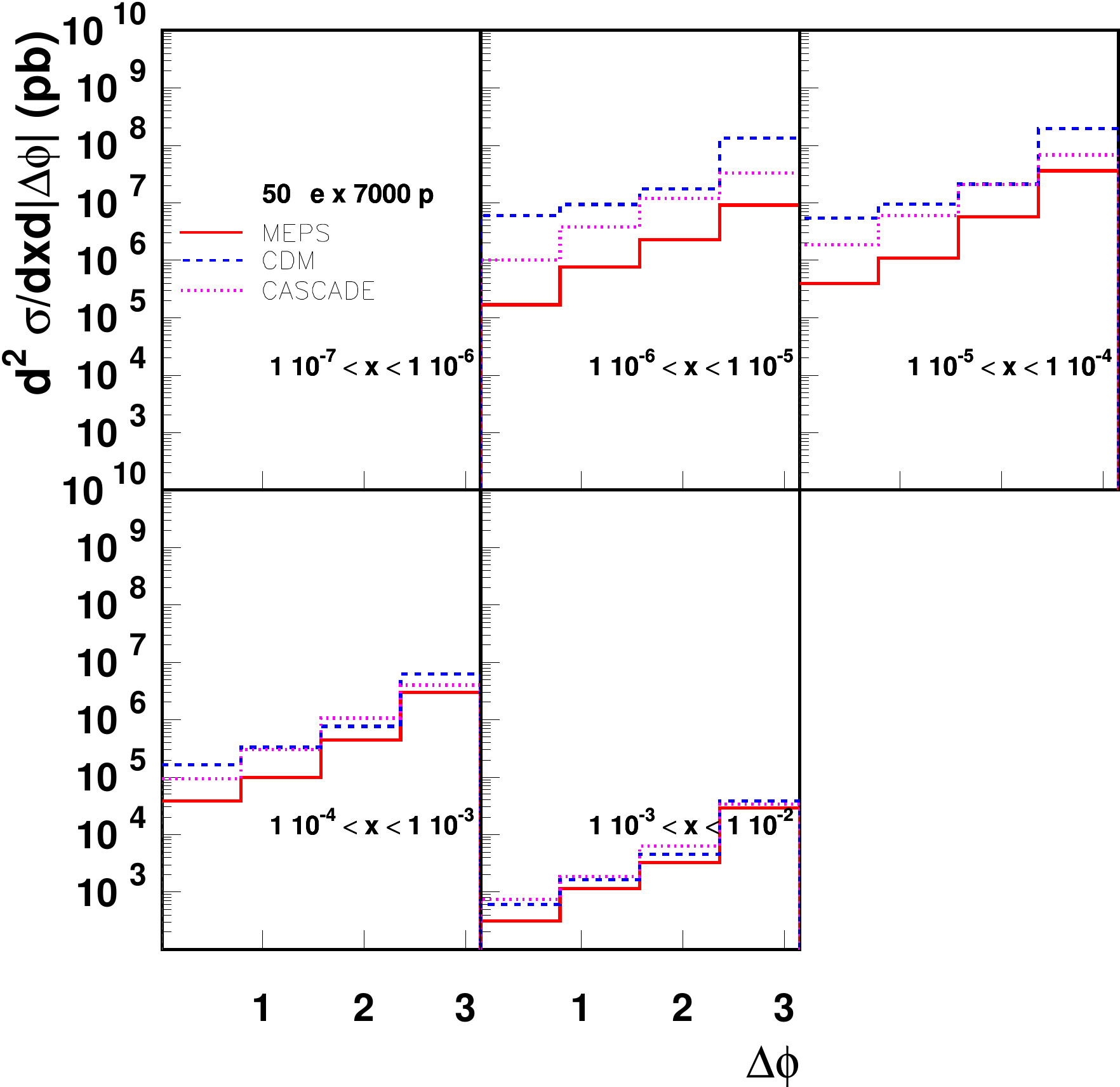}
\caption{Differential cross section for dijet production
as a function of the azimuthal separation 
$\Delta \phi$ for dijets with $E_{{\rm T, jet 1}}  > 7$~GeV  and 
$E_{{\rm T, jet 2}} > 5$~GeV.}
\label{fig:dijethannes}
\end{center}
\end{figure}

In Fig.~\ref{fig:dijethannes} we show the differential cross section as a function of $\Delta \phi$ for jets in the region
$-1 < \eta_{jet} <2.5$ with $E_{{\rm T, jet 1}}  > 7$~GeV  and $E_{{\rm T, jet 2}} > 5$~GeV found with the $k_t$ jet algorithm in the kinematic range $Q^2 > 5$~GeV, $0.1<y<0.6$ for different regions in $x$. 
The `MEPS' prediction comes from a Monte Carlo generator \cite{Jung:1993gf} 
using ${\cal O}(\alpha_s)$ matrix elements with a DGLAP-type parton 
shower. The `CDM' prediction uses the same generator \cite{Jung:1993gf},
but with higher order parton radiation simulated with the 
Colour Dipole Model~\cite{Lonnblad:1992tz}, thus effectively including some
$k_t$ diffusion. Finally, the 
CASCADE Monte Carlo prediction ~\cite{Jung:2010si}, uses off-shell matrix 
elements 
convoluted with an unintegrated gluon distribution (CCFM set A), with 
subsequent parton showering according to the CCFM evolution equation. 

At large $x$ all predictions agree reasonably well, in
both shape and normalisation. At smaller $x$ the $\Delta \phi$-distribution becomes flatter for CDM and CASCADE, indicating higher order effects leading to a larger decorrelation of the produced jets. Whereas a decorrelation is observed, its size depends on the details of the parton evolution and thus a measurement of the $\Delta \phi$ cross section provides a direct measurement of higher order effects which need to be taken into account at small $x$.

Thus, in principle, a measurement of the azimuthal dijet distribution offers
a direct determination of the $k_T$-dependence of the unintegrated gluon distribution.
 When additionally  supplemented
by inclusive measurements, it  can serve as an important  constraint for the precise determination of the fully unintegrated parton distribution, with the transverse momentum dynamics in the proton completely unfolded.

\subsubsection{Dihadron correlations}

\begin{figure}
\centerline{ \includegraphics[width=0.6\textwidth,angle=0]{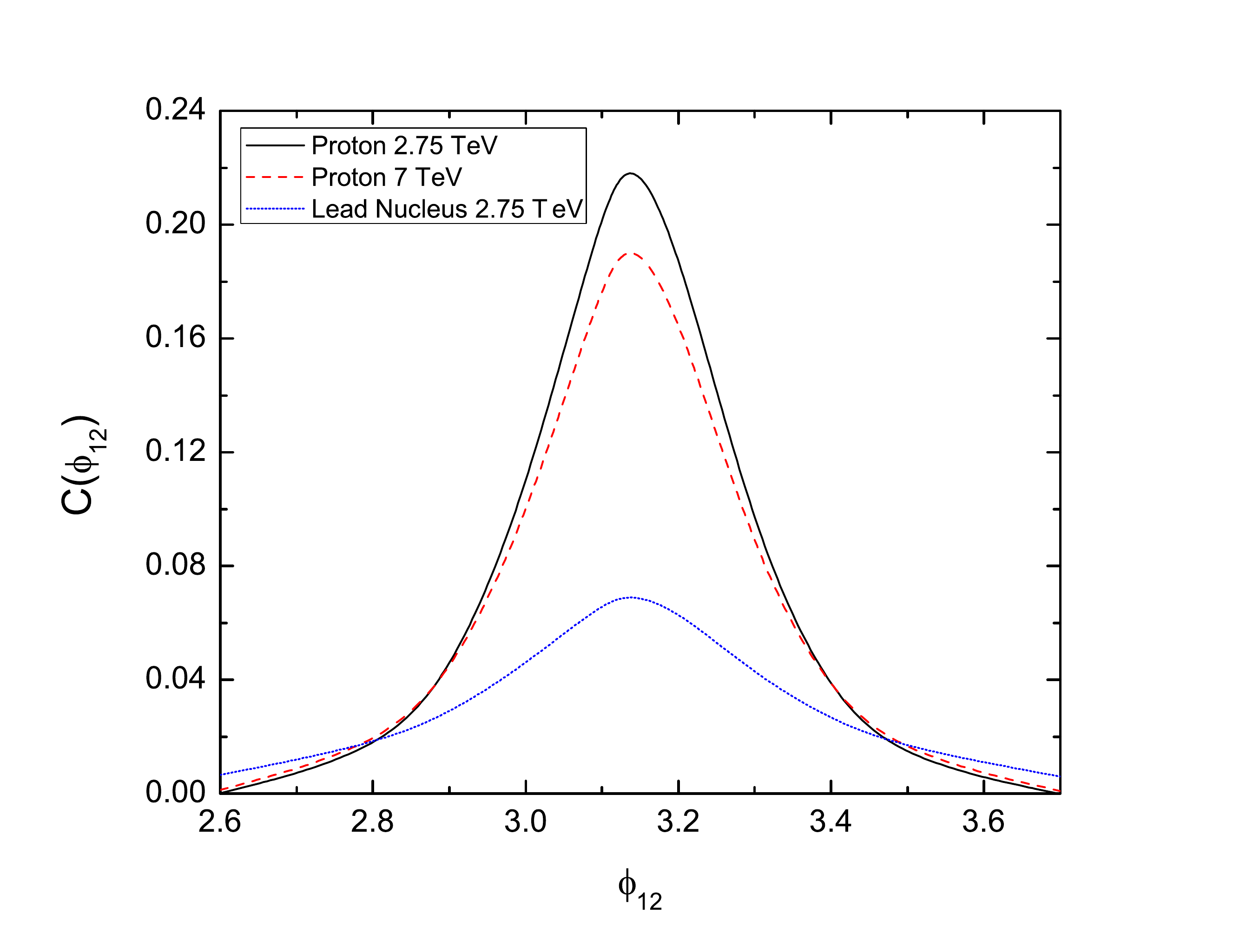}
}
\caption{Di-hadron correlation function  for the case of the scattering off the proton (red-dashed and black-solid lines) compared to the $eA$ case (blue-dotted line). The energy of the electron is assumed to be equal $E_e=50 \  {\rm GeV}$. The observed hadrons are pions. }
\label{fig:dihadron}
\end{figure}

Another interesting observable which is directly sensitive to the transverse momentum dependence of the 
parton distribution in the proton or nucleus is the process of two hadron production\footnote{This observable is currently discussed in the forward (proton) rapidity region in dAu collisions at RHIC and it shows features suggestive of physics beyond standard collinear factorisation, although no consensus has been reached so far, see \cite{Braidot:2010zh,Frankfurt:2007rn,Albacete:2010pg,Adare:2011sc,Stasto:2011ru} and references therein.}. Instead of two jets, one observes semi-inclusively two hadrons with certain transverse momentum.  One can define the function which describes the angular correlation of the two produced hadrons in the following way:
\begin{equation}
C(\phi_{12}) = \frac{1}{\frac{d\sigma(\gamma^* N\rightarrow h_1 X)}{dz_{h1}}}\frac{d\sigma^{\gamma^* N\rightarrow h_1 h_2+X}}{dz_{h1}dz_{h2} d\phi_{12}} \;.
\end{equation}
In the above formula $z_{h1},z_{h2}$ are the longitudinal momentum fractions of the two produced hadrons w.r.t. the photon momentum and $\phi_{12}$ is the azimuthal angle between them. The quantity $\frac{d\sigma(\gamma^* N\rightarrow h_1 X)}{dz_{h1}}$ is the single inclusive cross section. In Fig.~\ref{fig:dihadron} we show the results of the calculation using the formalism presented in \cite{Marquet:2009ca}. The gluon density was evaluated using the GBW model \cite{GolecBiernat:1998js} for the proton and a modified version of the same model for the nucleus.  The electron energy is assumed to be $E_e=50$ GeV, the proton energy is $7 \  {\rm TeV}$ and the nucleus energy is $2.75  \ {\rm TeV}$.
Also for the direct comparison with the nuclear case the curve with  proton energy of $2.75 \ {\rm TeV}$ is shown. The transverse momenta of the produced pions are integrated over, it is assumed that the leading particle has a minimum transverse momentum of $p_T=3 \  {\rm GeV}$ and the associated particle $p_T=2\  {\rm  GeV}$. The photon virtuality is $Q^2=4 \  {\rm GeV}^2$, $y=0.7$
and the fractions of the longitudinal momenta of the produced pions are fixed to be equal to $z_{1h}=z_{2h}=0.3$.
One clearly sees that the correlation function is wider for a larger target (nucleus) than for the proton. This suppression of the peak in the correlation function can be  interpreted in this model  as the effect of the stronger saturation in the gluon density for the nucleus than for the proton.
We also see that the correlation function varies mildly with the available energy for the same target (i.e. proton). One observes stronger de-correlation of the produced hadrons with a higher energy or at smaller values of $x$ which is indicative of the importance of the $\ln 1/x$ effects for this observable. Therefore the measurement of the dihadron correlation provides another way of constraining the unintegrated gluon distribution. In particular, measuring the dihadron correlations in DIS provides with a unique opportunity\cite{Dominguez:2010xd,Dominguez:2011wm} to directly study the so-called Weizs\"{a}cker-Williams unintegrated gluon distribution.

%% file: physics/tex/forwardjets.tex

\begin{figure}
 \centering
  \includegraphics[width=0.45\textwidth]{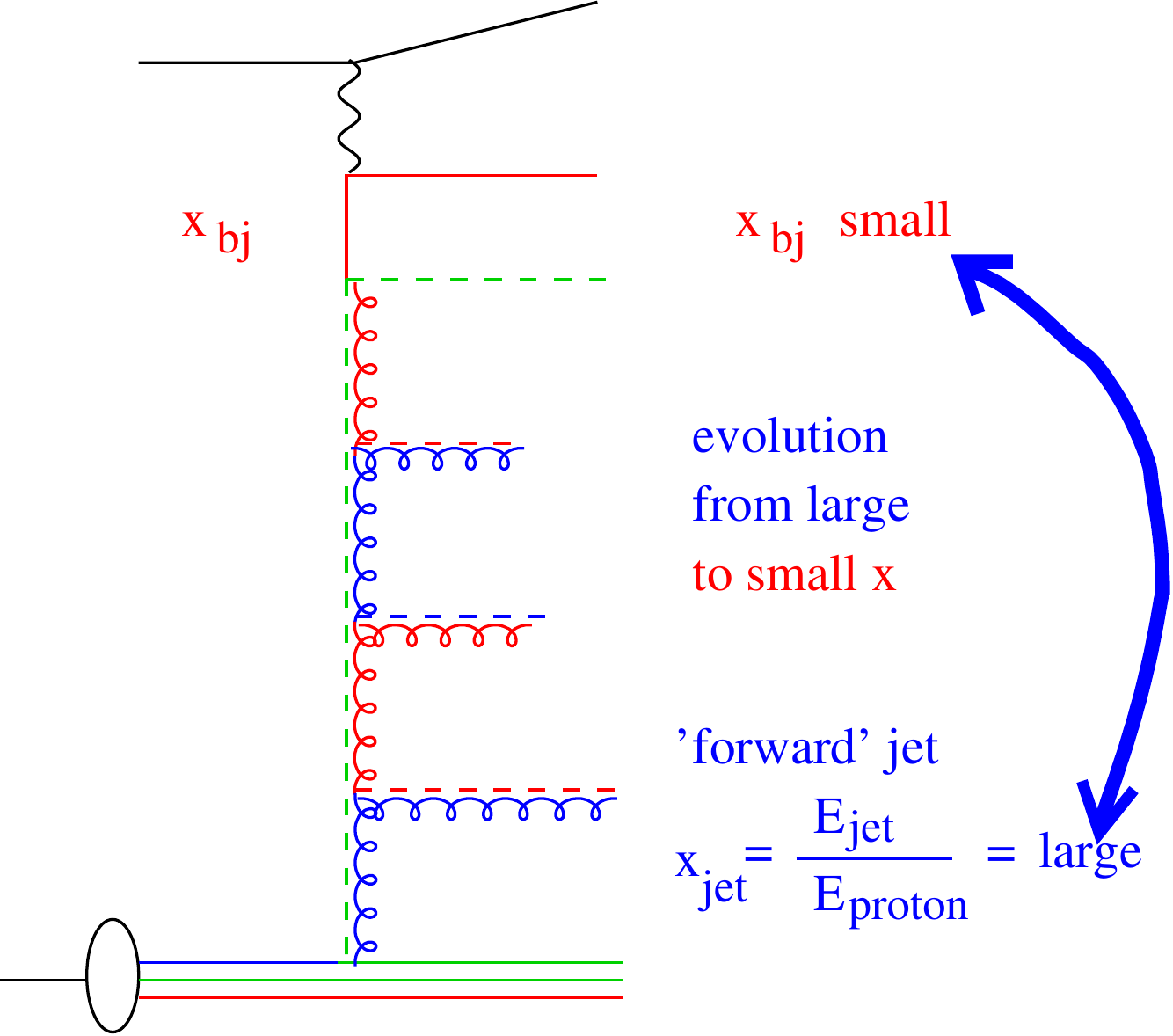}
  \caption{Schematic representation of the production of 
a high transverse momentum forward jet in DIS.}
\label{fig:forjet}
\end{figure}

It was proposed some time ago \cite{Mueller:1990er,Mueller:1990gy} that a 
process which would be very sensitive to the parton dynamics and the transverse momentum distribution was the production of forward jets in DIS. According to \cite{Mueller:1990er,Mueller:1990gy}, DIS events containing identified forward jets provide a particularly clean window on small-$x$ dynamics. The schematic view of the process is illustrated in Fig.~\ref{fig:forjet}. The 
forward jet transverse momentum provides the second hard scale $p_T$. Hence one has a process with two hard scales: the photon virtuality $Q$ and the transverse momentum of the forward jet $p_T$. As a result  the collinear (DGLAP) configurations (with no diffusion and strongly ordered transverse momenta) can be eliminated by choosing the scales to be of comparable size, $Q^2 \simeq p_T^2$. Additionally, the jet is required to be produced in the forward direction
by demanding that $x_J$, the longitudinal momentum fraction of the produced jet,  is as large as possible, and $x/x_J$ 
is as small as possible. This requirement selects events with a large sub-energy between the jet and the virtual photon, such that the BFKL framework should be applicable.
There have been dedicated measurements of forward jets at HERA \cite{Aid:1995we,Adloff:1998fa,Aktas:2005up,Breitweg:1998ed,Breitweg:1999ss,Chekanov:2005yb}, which demonstrated that DGLAP dynamics at NLO are indeed incompatible with the experimental measurements. 
On the other hand, calculations based on resummations of powers of $\log 1/x$ (BFKL and others) \cite{Kwiecinski:1997jk,Kwiecinski:1999hv,Bottazzi:1998rs,Jung:1998mi,Jung:1999yi,Jung:2000hk,Kepka:2006cg} are consistent with the data. The azimuthal dependence of forward jet production has also been studied \cite{Bartels:1996wx,Vera:2007dr} as a sensitive probe of the small-$x$ dynamics.


Another observable that provides a valuable insight into the features of small-$x$ physics is the  transverse energy ($E_T$-flow) 
accompanying DIS events at small $x$.
The diffusion of the transverse momenta in this region 
leads to a strongly enhanced distribution of $E_T$ at
small $x$. As shown in \cite{Kwiecinski:1994zs,GolecBiernat:1994fw}, 
small-$x$
evolution results in a  broad Gaussian $E_T$-distribution as a function of rapidity. This should be
contrasted with the much smaller $E_T$-flow obtained assuming strong $k_T$-ordering as in DGLAP-based approaches, which give
an $E_T$-distribution that narrows with decreasing $x$, for fixed $Q^2$.

The first experimental measurements of the $E_T$-flow in small-$x$ DIS events indicate that there is significantly more $E_T$ than is given by conventional QCD cascade models based on DGLAP evolution. Instead we find that they are in much better agreement with estimates which incorporate dynamics beyond fixed-order DGLAP \cite{Brook:1998jd,Lonnblad:1992tz,Jung:2000hk} 
such as BFKL evolution. The latter dynamics are characterised by an increase of the $E_T$-flow in the central region with decreasing $x$.

However, the experimental data from HERA do not enable a detailed analysis due
to their constrained kinematics.  At the LHeC one could perform measurements
with large separations in rapidity and for different selections of the scales
$(Q,p_T)$. In particular, there is a possibility of varying scales to test
systematically the parton dynamics from the collinear (strongly ordered) regime
$Q^2 \gg p_T^2$ to the BFKL (equal scale, Regge kinematics) regime $Q^2 \simeq
p_T^2$. 
Measurements of the energy flow in different $x$-intervals, in the small-$x$
regime, should therefore allow a definitive check of the applicability of BFKL
dynamics and of the eventual presence of more involved, non-linear effects.

\begin{figure}[htbp]
\begin{center}
\includegraphics[width=0.45\columnwidth]{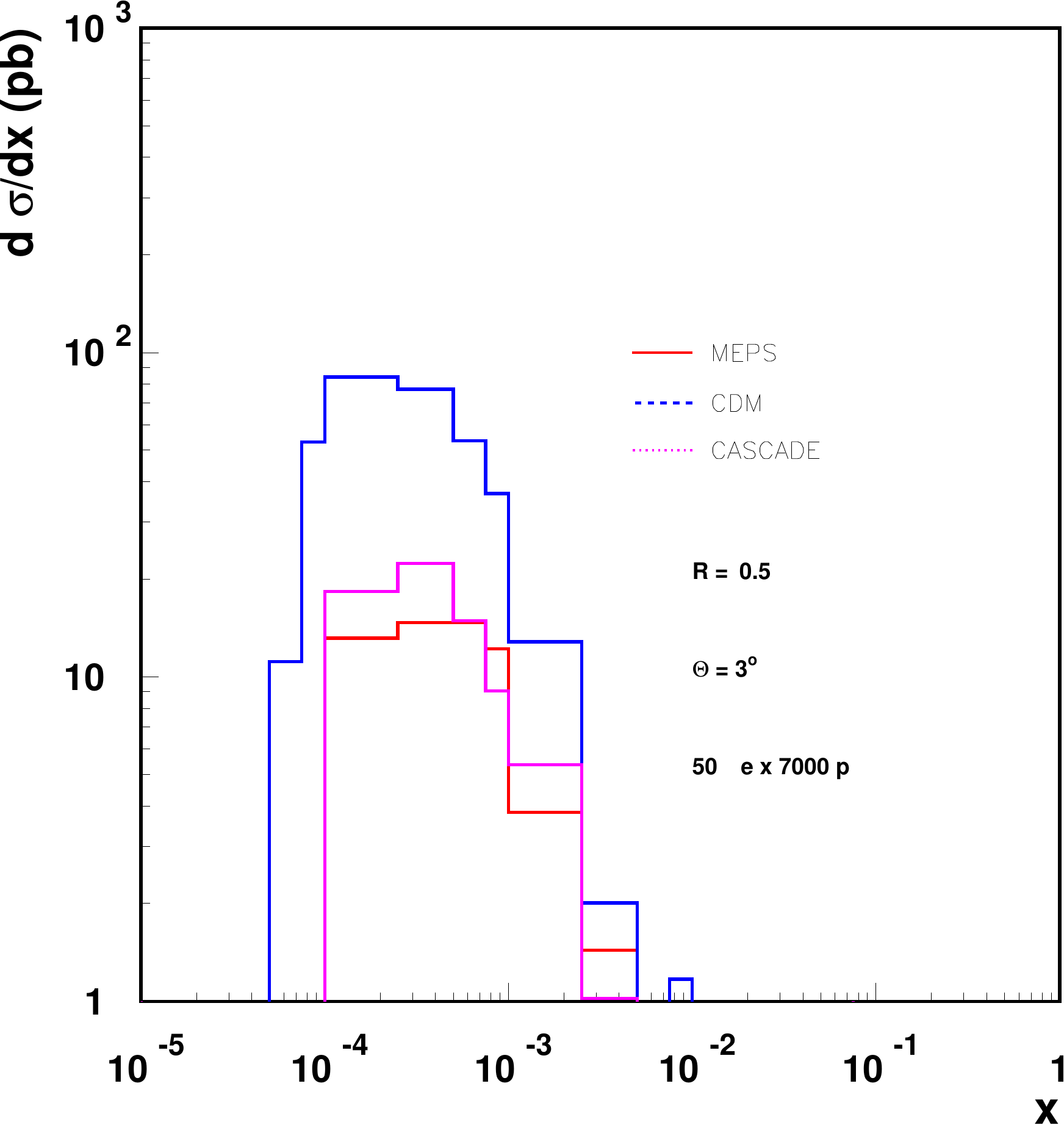}
\includegraphics[width=0.45\columnwidth]{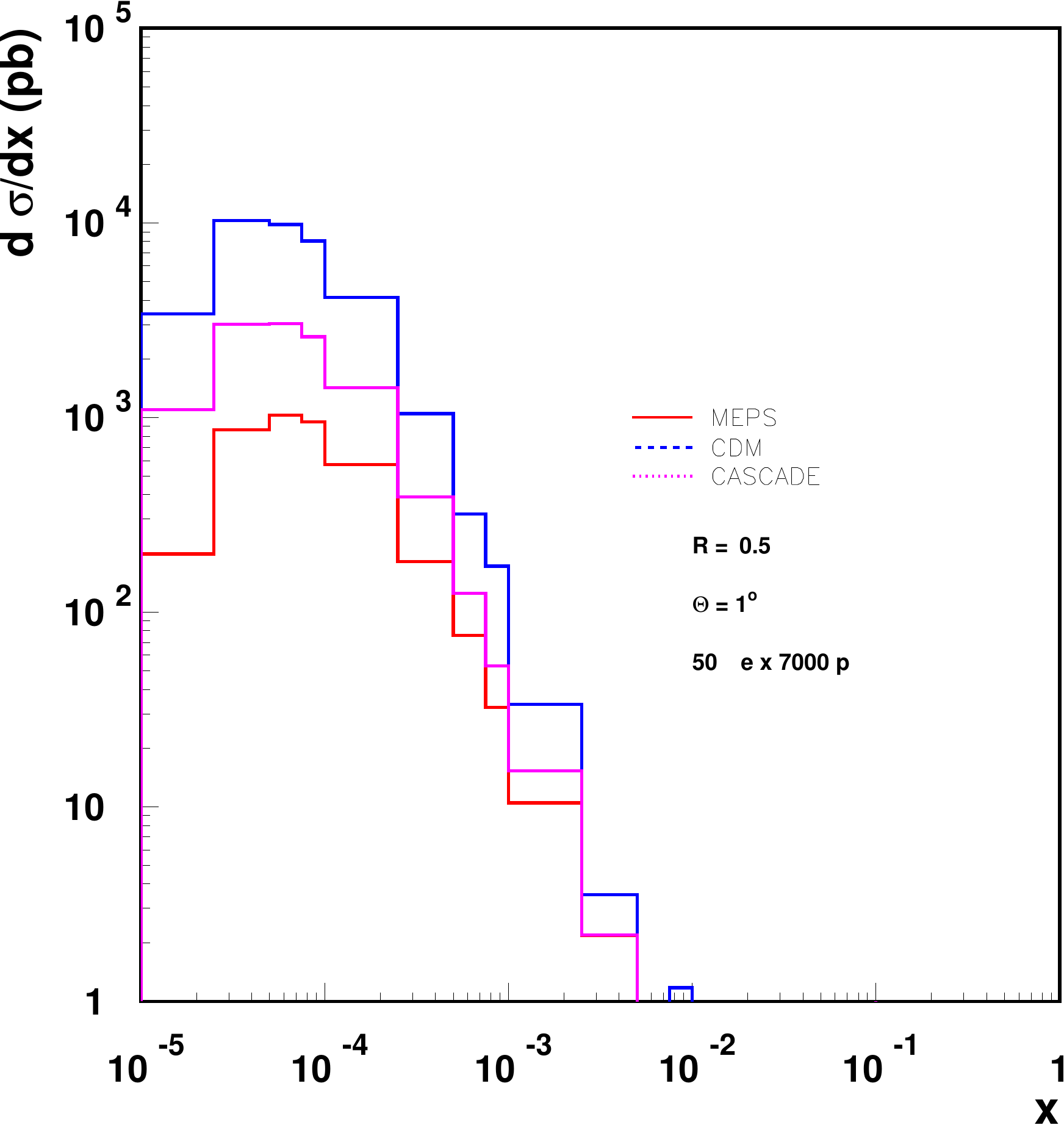}
\caption{Cross section for forward jets with $\Theta_{jet} >3^\circ$ (left) and  $\Theta_{jet} > 1^\circ$ (right).
Predictions from MEPS, CDM and CASCADE are shown. 
Jets are found with the SISCone algorithm using $R=0.5$.}
\label{fwdjets-0.5}
\end{center}
\end{figure}

\begin{figure}[htbp]
\begin{center}
\includegraphics[width=0.45\columnwidth]{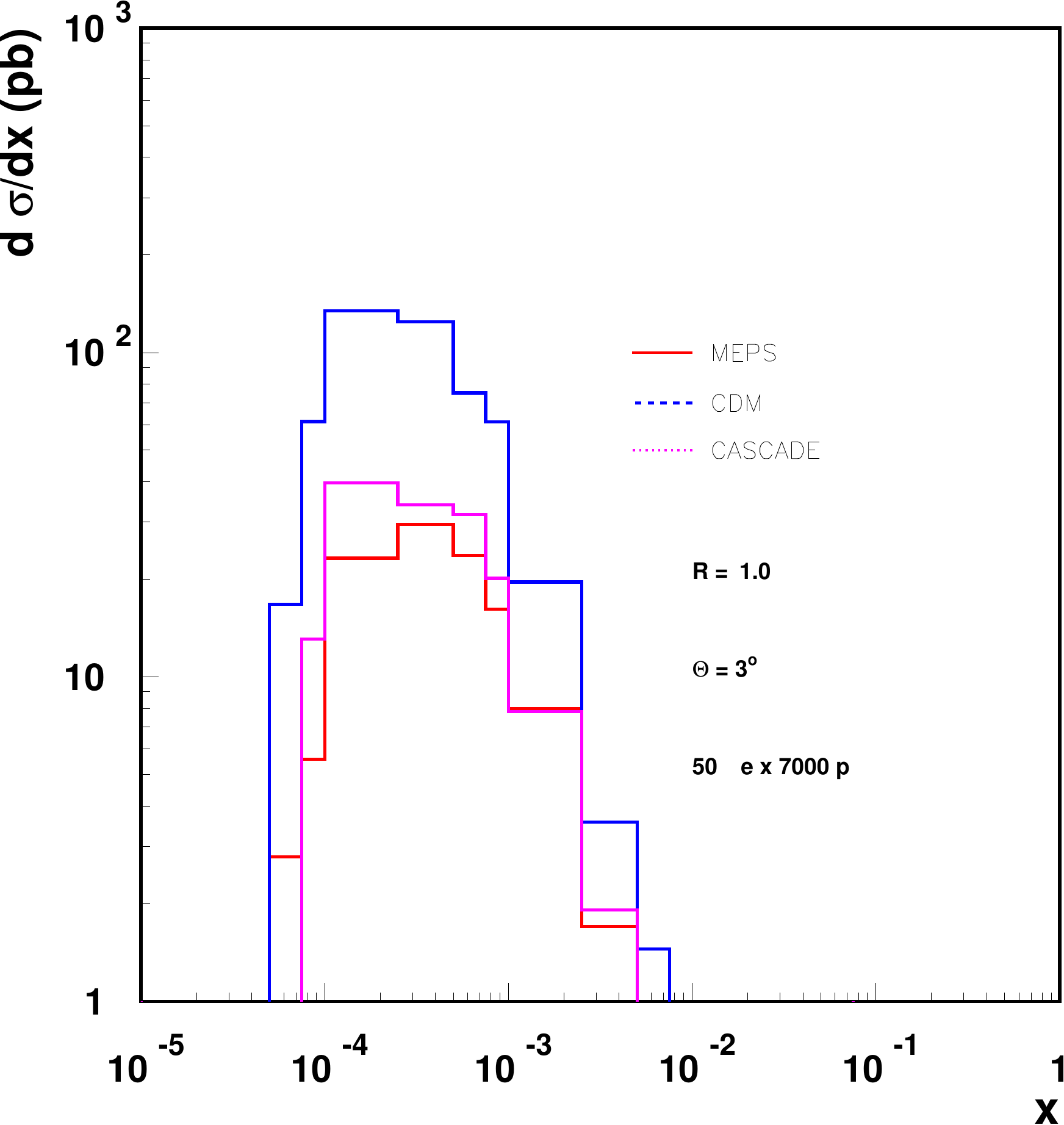}
\includegraphics[width=0.45\columnwidth]{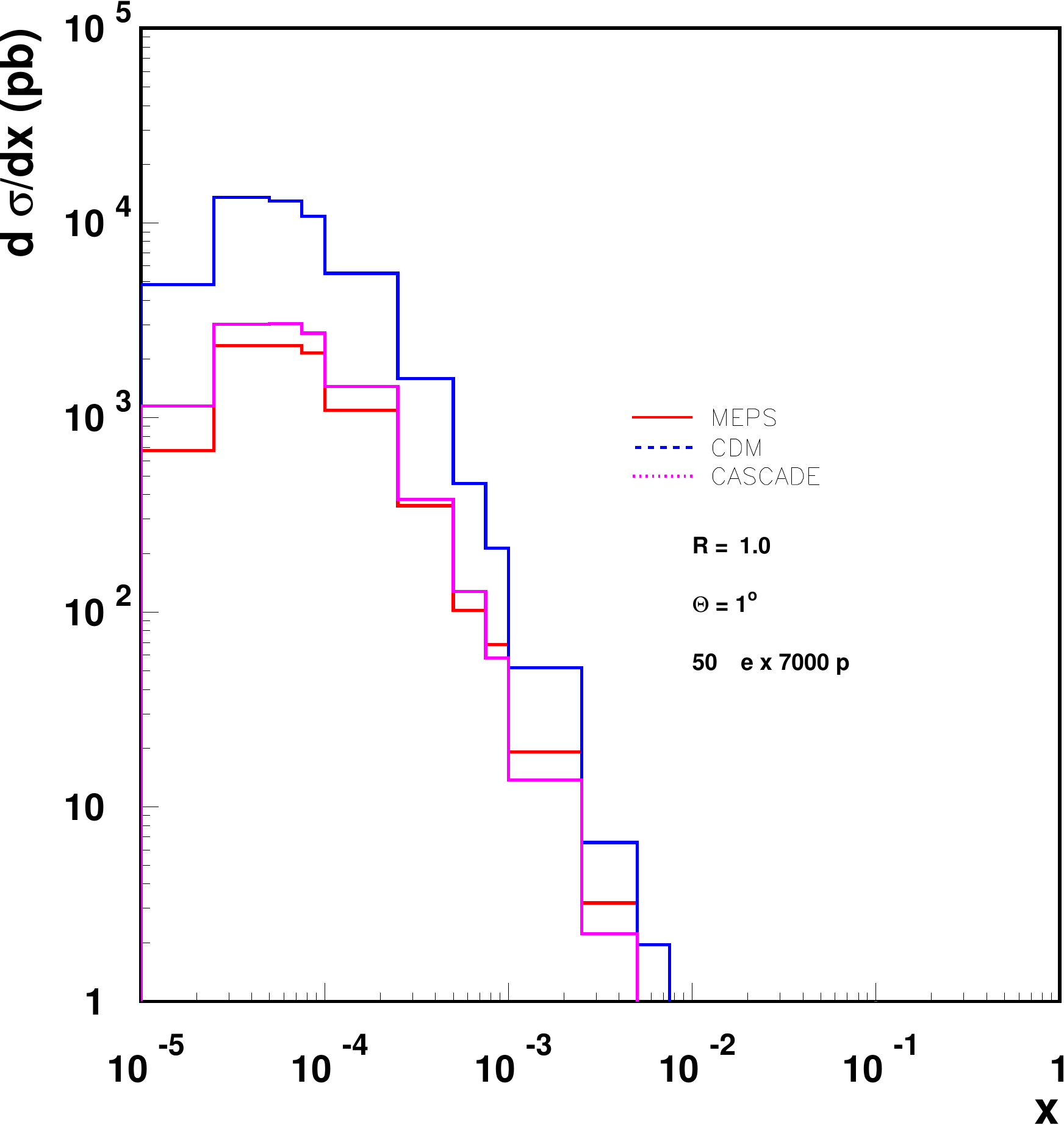}
\caption{Cross section for forward jets with $\Theta_{jet} >3^\circ$ (left) and  $\Theta_{jet} > 1^\circ$ (right).
Predictions from MEPS, CDM and CASCADE are shown. 
Jets are found with the SISCone algorithm using $R=1.0$.}
\label{fwdjets-1.0}
\end{center}
\end{figure}

A simulation of forward jet production at the LHeC is shown in Figs.~\ref{fwdjets-0.5} and \ref{fwdjets-1.0}. The jets are required to have $E_T > 10$~GeV with a polar angle $\Theta_{jet} >  1^\circ$ or $3^\circ$ in the laboratory frame. Jets are found with the SISCone jet-algorithm \cite{Salam:2007xv}. The DIS phase space is defined by $Q^2 > 5$~GeV, $0.05<y<0.85$.

In Fig.~\ref{fwdjets-0.5} the differential cross section is shown 
as a function of Bjorken $x$ for an electron energy of $E_e=50$~GeV.
The calculations are obtained 
from the MEPS \cite{Jung:1993gf},  
CDM \cite{Lonnblad:1992tz} and 
CASCADE \cite{Jung:2000hk} Monte Carlo models, as described in the 
previous section.
Predictions for $\Theta_{jet} > 3^\circ$ and $\Theta_{jet} > 1^\circ$ are shown. One can clearly see that the small-$x$ range is explored 
in detail with the small angle scenario. In Fig.~\ref{fwdjets-1.0} the forward jet cross section is shown when using $R=1$ instead of $R=0.5$ (Fig.~\ref{fwdjets-0.5}). 
It is important to note that good forward acceptance of the detector is crucial for the measurement of forward jets. The dependence of the cross section on the acceptance angle is very strong as is evident from 
comparisons between the cross sections for different $\Theta_{jet}$
cuts in Figs.~\ref{fwdjets-0.5} and \ref{fwdjets-1.0}. 

A complementary reaction to that of forward jets is the production of forward $\pi^0$ mesons in DIS. Despite having a lower rate,  this process offers some advantages over forward jet production. By looking onto single particle production the dependencies on the jet finding algorithms can be eliminated. Also, the non-perturbative hadronisation effects can be effectively encompassed into fragmentation functions \cite{Kwiecinski:1999hv}.

%% file: physics/tex/fsr.tex
\begin{figure}
\centerline{ \includegraphics[clip=,width=0.95\textwidth,angle=0]{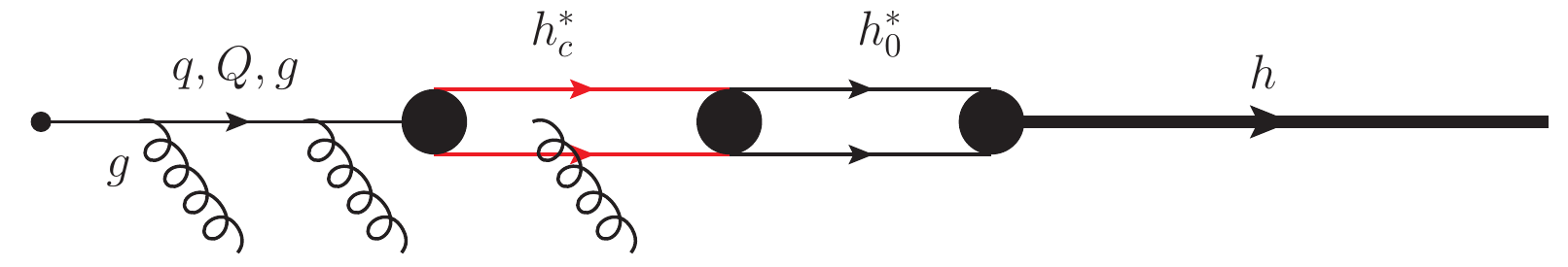}}
\caption{Sketch of the different postulated stages 
in the hadronisation of a highly virtual parton.
From left to right: radiating parton;
radiating coloured pre-hadron, colourless pre-hadron and final 
state hadron.}
\label{Fig:fsr}
\end{figure}

The mechanism through which a highly virtual parton produced in a hard
scattering gets rid of its virtuality and colour and
finally projects onto an observable final state hadron, is unknown to a great
extent (see \cite{Accardi:2009qv} and references therein). 
The different postulated stages 
of the process are illustrated 
in  Fig.~\ref{Fig:fsr}. The coloured parton undergoes
QCD radiation before forming first a coloured excited bound state
(pre-hadron), then a colourless pre-hadron and ultimately a 
final state hadron. These sub-processes are characterised by
different time scales. While the first stage can be described in perturbative
QCD \cite{Dokshitzer:1991wu}, subsequent ones require models (e.g. the QCD
dipole model for the pre-hadron stages) and non-perturbative information.

The LHeC offers great opportunities to study these aspects and improve
our understanding of all of them. The energy of
the parton which is struck by the virtual photon implies a Lorentz dilation
of the time scales for each stage of the radiation and
hadronisation processes. All of them are influenced 
by the fact that they do not take place in the vacuum, but within the
QCD field created by the other components of the hadron or
nucleus. While at fixed target SIDIS or 
DY experiments,
the lever arm in energy is relatively small (energy transfer
to the struck parton in its rest frame $\nu< 100$ GeV), at the
LHeC this lever arm will be huge ($\nu< 10^5$ GeV; see also in Subsec. \ref{gammap:jetseA} the abundant yield of expected high transverse momentum jets in photoproduction), implying that the
different stages can be considered to happen in or out of the hadron field 
depending on the parton energy.
Furthermore, the fact that we can  introduce a piece of coloured matter of
known length and density - a nucleus - by doing $e$Pb collisions at different
centralities, allows a controllable 
variation of the contribution of the different processes. The
induced differences in the final distributions of hadrons, 
both in terms of their momenta and
of their relative abundance, will provide 
important information about the
time scales and the detailed physical mechanisms at work in each stage.
Dramatic effects are predicted in some models \cite{Frankfurt:2001nt}, 
with a significant suppression of the forward hadron spectra due to the  existence of a dense partonic system.
Note that SIDIS experiments already provide information for the
determination of standard fragmentation functions (see
\cite{deFlorian:2007aj,deFlorian:2007hc} for a recent
analysis). The other pieces of
information, coming mainly
from $e^+e^-$ experiments, will not be improved until next-generation linear
colliders become available.

Furthermore, these studies will shed light on two aspects already discussed in
Subsec.~\ref{sec:nucleartargets}, related to 
the study of ultra-relativistic heavy-ion collisions: the characterisation of the medium created in
such collisions through hard probes, and the details of
particle production in a dense situation which will define the initial conditions
for the collective behaviour of this medium.
Concerning the latter, our theoretical tools for
computing particle production in $e$A collisions 
are more advanced e.g. within the CGC framework, and on a safer ground 
than in
nucleus-nucleus collisions (see Subsec. \ref{sec:lowxoverview} and e.g.
\cite{Lappi:2009fq} and refs. therein).
The possibility of disentangling the 
different mechanisms through which the factorisation that is
used in dilute systems - collinear factorisation \cite{Collins:1989gx} -
becomes broken by density effects (e.g. initial and final state energy loss
or
final state absorption) will be possible at the LHeC and will complement
existing studies done at much smaller energies
in fixed target SIDIS and DY experiments \cite{Accardi:2009qv}.

\begin{figure}
\centerline{ \includegraphics[clip=,width=0.45\textwidth,angle=0]{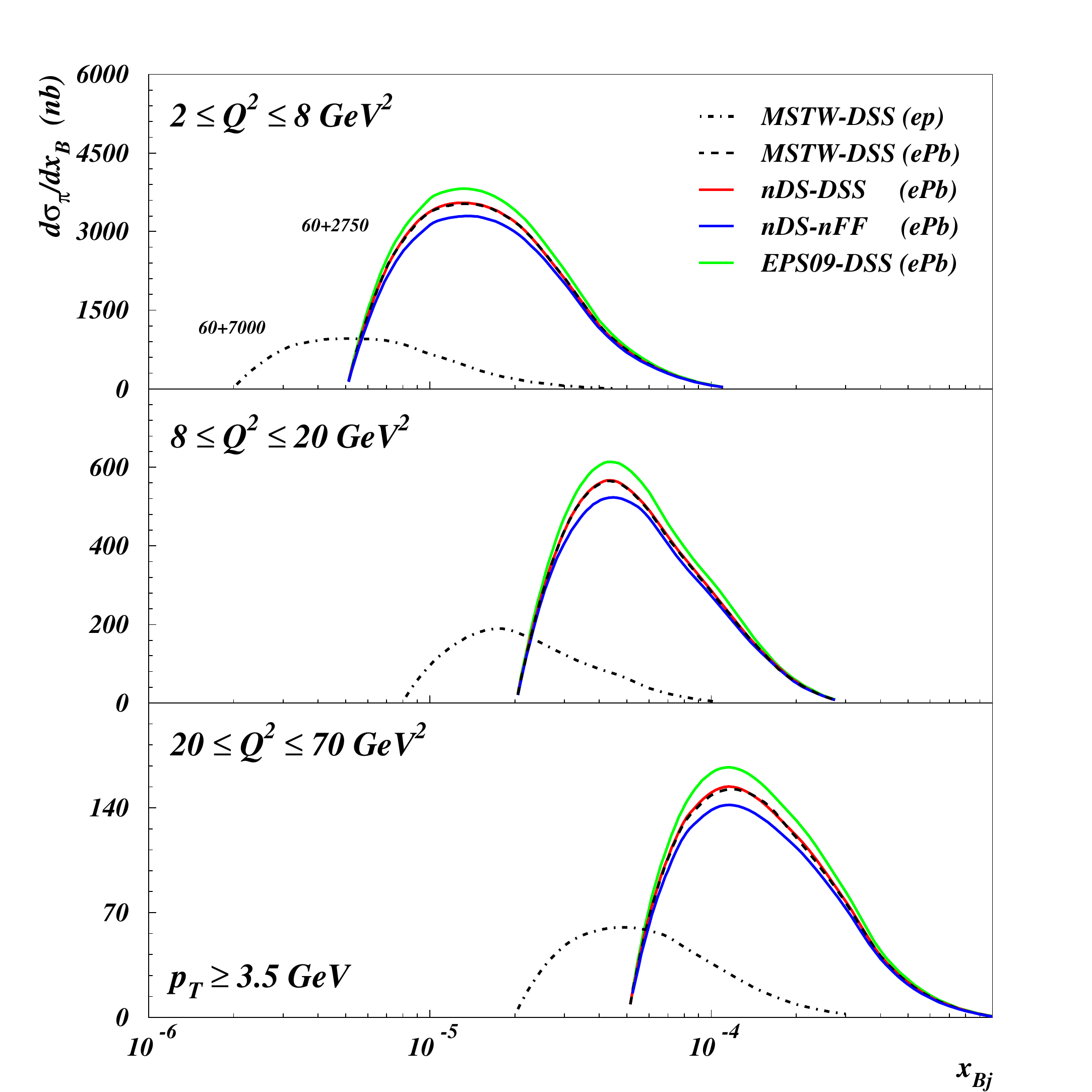}\includegraphics[clip=,width=0.45\textwidth,angle=0]{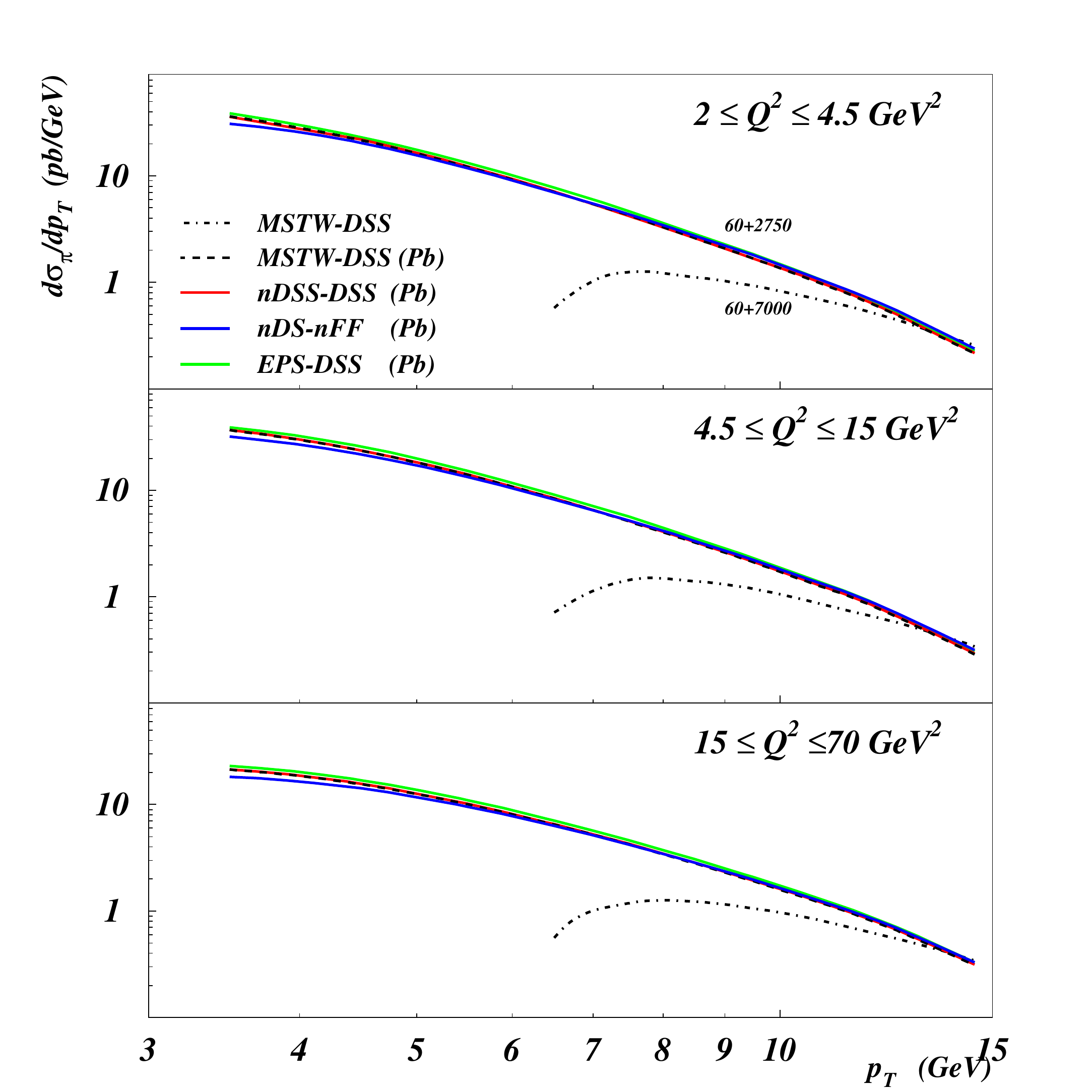}}
\caption{Cross section for inclusive $\pi^0$ production versus Bjorken $x_{Bj}$ for $p_T>3.5$ GeV/c (left) and versus $p_T$ (right), computed in NLO QCD \cite{Daleo:2004pn}. Dashed-dotted black lines refer to $ep$ collisions. All other line types refer to $e$Pb collisions: dashed black ones to standard nucleon PDFs \cite{Martin:2009iq} and fragmentation functions \cite{deFlorian:2007aj,deFlorian:2007hc}, solid red (green) ones to nuclear PDFs \cite{deFlorian:2003qf} (\cite{Eskola:2009uj}) and nucleon fragmentation functions, and solid blue ones to nuclear PDFs \cite{deFlorian:2003qf}  and nuclear fragmentation functions \cite{Sassot:2009sh}. All cross sections are given per nucleon i.e. divided by 208 for Pb. Cuts: $\theta_\pi \in [5^\circ,25^\circ]$, $x_\pi=E_\pi/E_p>0.01$, have been applied. See the text for further explanations.}
\label{Fig:newpi0}
\end{figure}

In order to quantify the possibilities for SIDIS studies, we first show the expected cross sections for $\pi^0$ production in $ep$ and $e$Pb collisions at the LHeC for $E_e=60$ GeV, see Fig. \ref{Fig:newpi0}.
There the calculations are done at NLO \cite{Daleo:2004pn}, using as nucleon PDFs those from \cite{Martin:2009iq} and, in order to illustrate their effect, different nuclear PDFs \cite{deFlorian:2003qf,Eskola:2009uj} and both ordinary \cite{deFlorian:2007aj,deFlorian:2007hc} and modified \cite{Sassot:2009sh}\footnote{In this reference, fragmentation functions in nuclear matter are extracted in a DGLAP analysis at LO and NLO.} fragmentation functions. Cuts have been applied as in the H1 study \cite{Aktas:2004rb}\footnote{Studies with looser cuts - a more realistic situation at the LHeC, and of the achievable resolution in $x$ and $p_T$, are left for the future.} whose data are well reproduced by the NLO calculation: angle of the $\pi^0$ from the proton  in the laboratory $\theta_\pi \in [5^\circ,25^\circ]$, pion energy fraction $x_\pi=E_\pi/E_p>0.01$ and pion transverse momentum $2.5<p_T<15$ GeV/c. All scales in the calculation have been fixed to $(Q^2+p_T^2)/2$ ($K$-factors and the scale dependence of the results are discussed in  \cite{Daleo:2004pn}).
From the plots in the figure, it becomes clear that even for these very restrictive cuts and for a modest integrated luminosity of 1 fb$^{-1}$, a large number of pions will be produced with relatively large transverse momentum. The nuclear effects on PDFs and on fragmentation require measurements with good statistic and systematic precision in order to be disentangled. 

The results with looser cuts: $\theta_\pi \in [1^\circ,25^\circ]$, $x_\pi=E_\pi/E_p>0.005$ that could be achieved at the LHeC, have also been studied.
Their effect is an increase of the cross section by a factor $\sim 3$ with respect to the results with the more restrictive H1 cuts.


SIDIS also offers the possibility to measure the nuclear effects on fragmentation functions through the double ratio for nucleus $A$ and particle $k$:
\begin{equation}
R_{A}^{k}(\nu,z,Q^2)={\frac{1}{N_A^e}\frac{dN_A^k}{d\nu dz}}\biggm/{\frac{1}{N_p^e}\frac{dN_p^k}{d\nu dz}}\,,
\label{Eq:ratsidis}
\end{equation}
with $N^e$ the number of scattered electrons at a given $\nu$ and $Q^2$ i.e. the DIS cross section. At LO and for a single quark flavour, this double ratio becomes the ratio of fragmentation functions in $e$A over $ep$, see \cite{Accardi:2009qv}. Usually, the energy of the lepton-hadron/nucleus collisions are the same in numerator and denominator, and the collisions in the denominator are $e$D in order to suppress isospin effects as much as possible.

In order to estimate the nuclear modifications of fragmentation functions for the case of the LHeC, we compute this double ratio. For the numerator, we consider $e$Pb collisions at 60+2750 GeV while for the denominator we take $ep$ collisions at 60+7000 GeV.
We follow the model in
 \cite{Arleo:2003jz} which considers the energy loss of the parent parton though radiative processes\footnote{For this, we use the quenching weights in \cite{Salgado:2003gb} instead of the simplified expressions employed in  \cite{Arleo:2003jz}.} plus formation time arguments which make the effective length of traversed nuclear matter $L$ smaller at small $\nu$ than the geometrical one $L_{max}$. We use the LO nucleon PDFs in  \cite{Martin:2009iq} and the nucleon fragmentation functions in  \cite{deFlorian:2007aj,deFlorian:2007hc}, and also considered the nuclear modification of PDFs in \cite{Eskola:2009uj}. We employ a value of the transport coefficient characterising the strength of the interaction of a quark with nuclear matter $\hat q=0.7$ GeV$^2$/fm \footnote{This value is larger than the one used in \cite{Arleo:2003jz}. We have checked
that the model reproduces fixed target data on the $\nu$ dependence of the ratio (\ref{Eq:ratsidis}) for pion production on Kr over D in \cite{Airapetian:2007vu} using this value of $\hat q$ without formation time considerations.}.

\begin{figure}[h!]
\centerline{ \includegraphics[clip=,width=0.32\textwidth,angle=0]{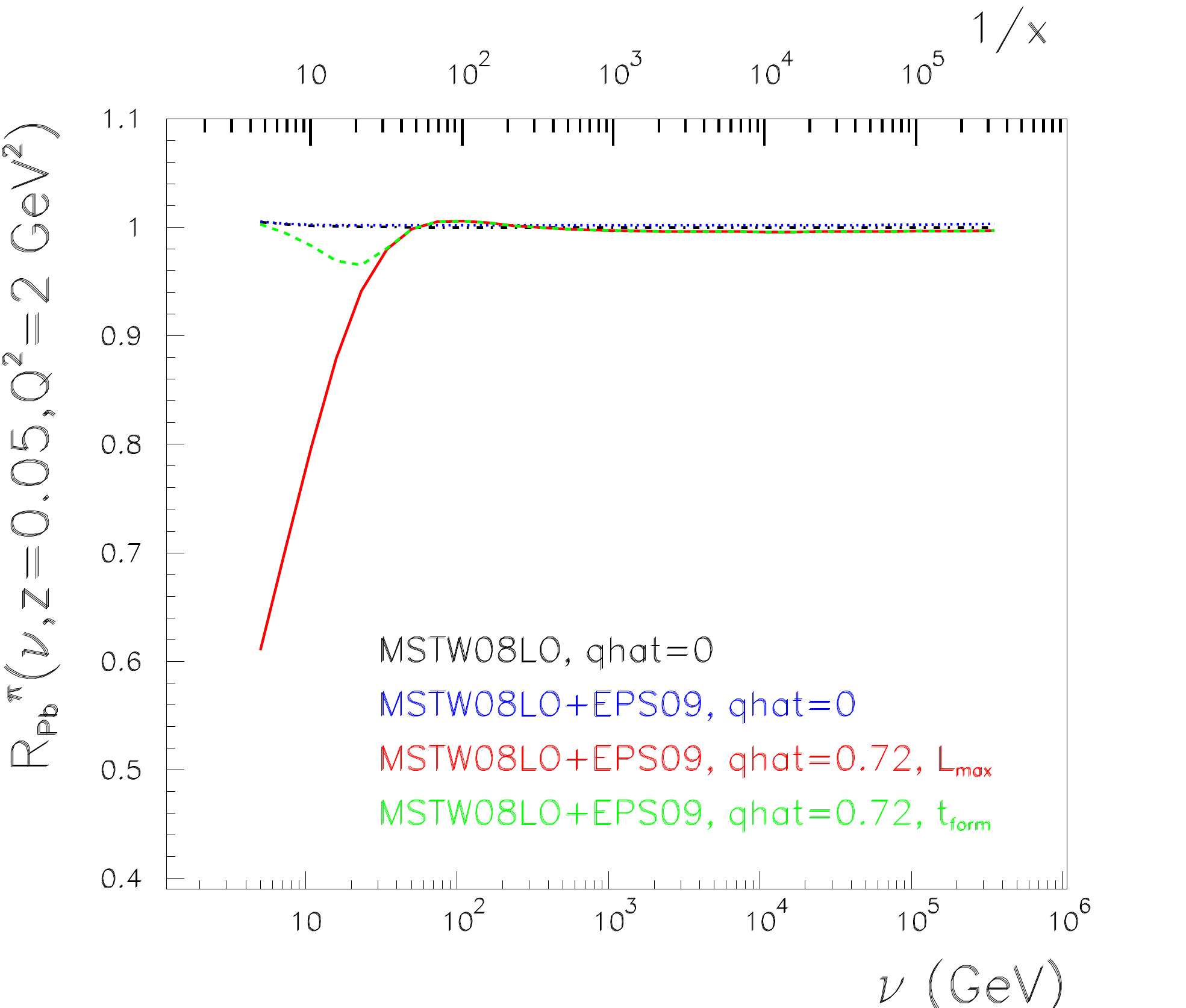} \includegraphics[clip=,width=0.32\textwidth,angle=0]{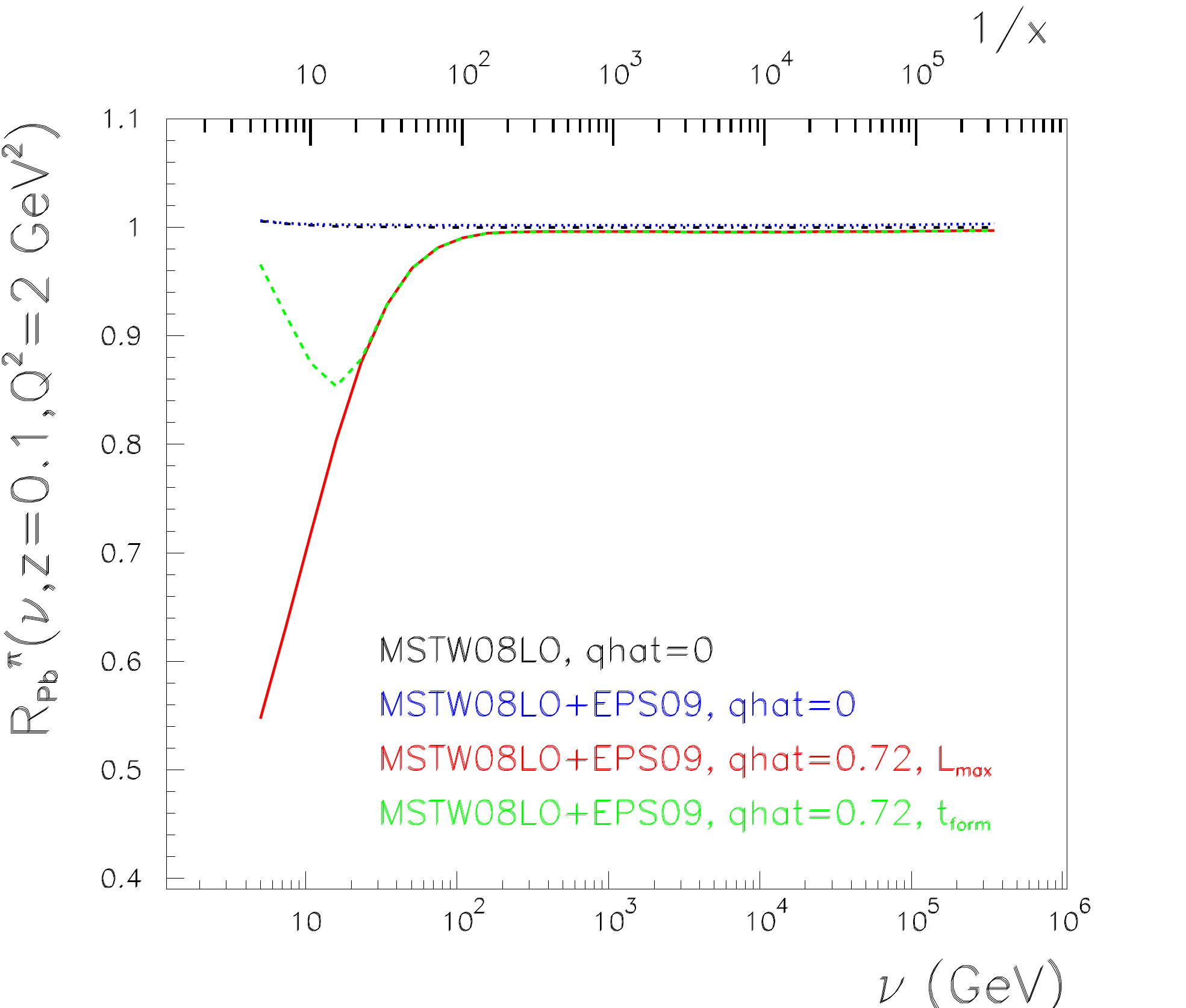} \includegraphics[clip=,width=0.32\textwidth,angle=0]{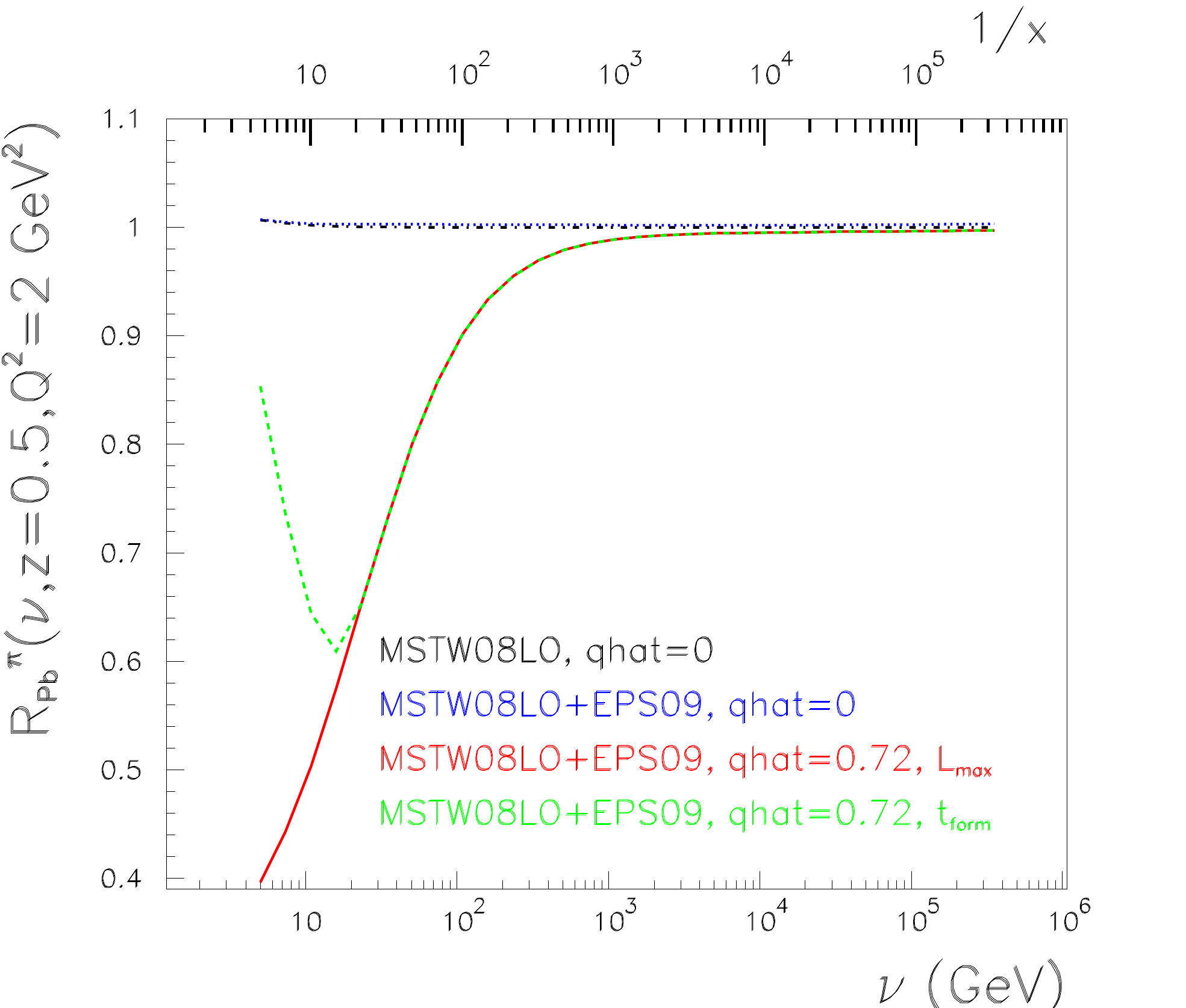}}
\centerline{ \includegraphics[clip=,width=0.32\textwidth,angle=0]{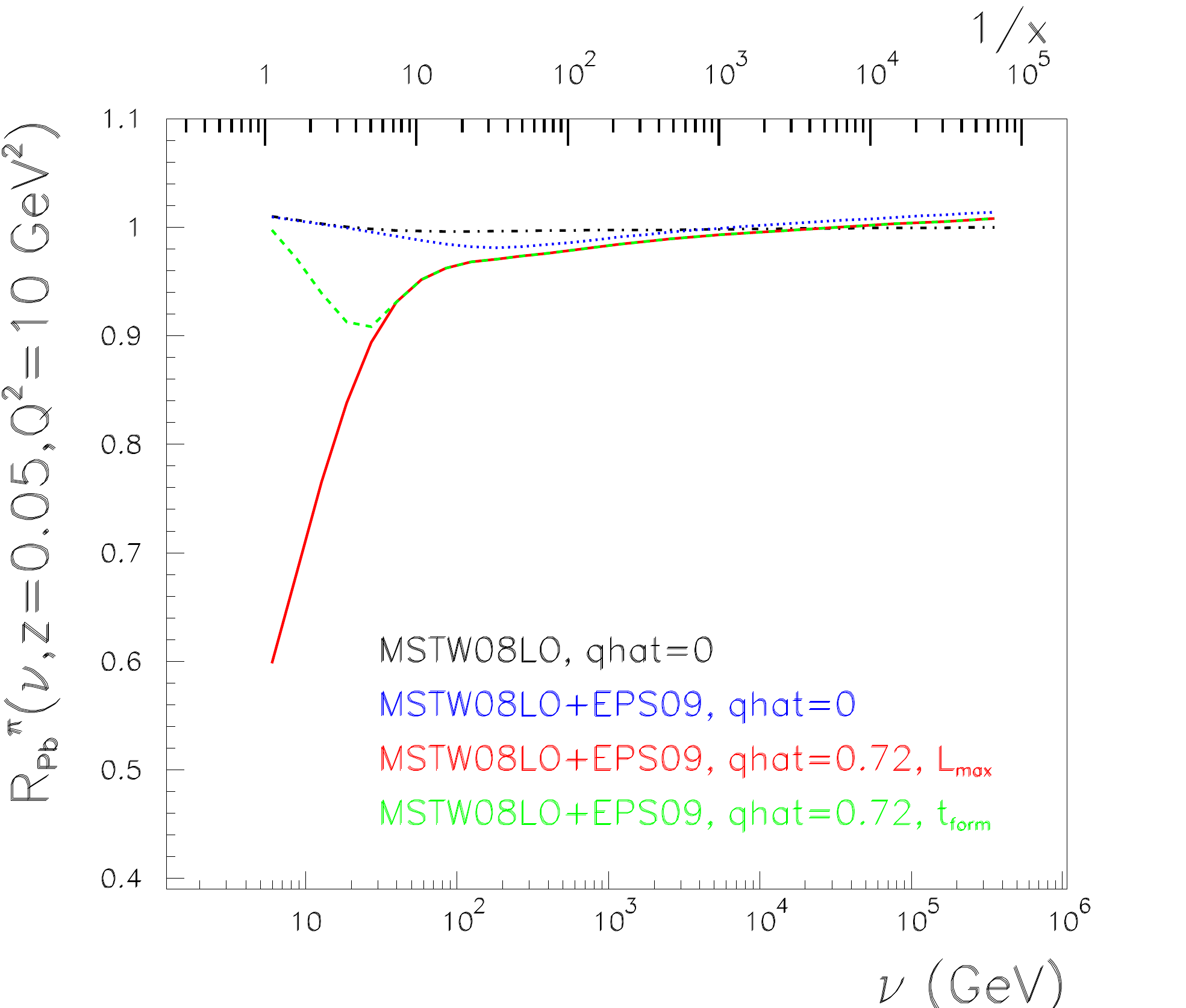} \includegraphics[clip=,width=0.32\textwidth,angle=0]{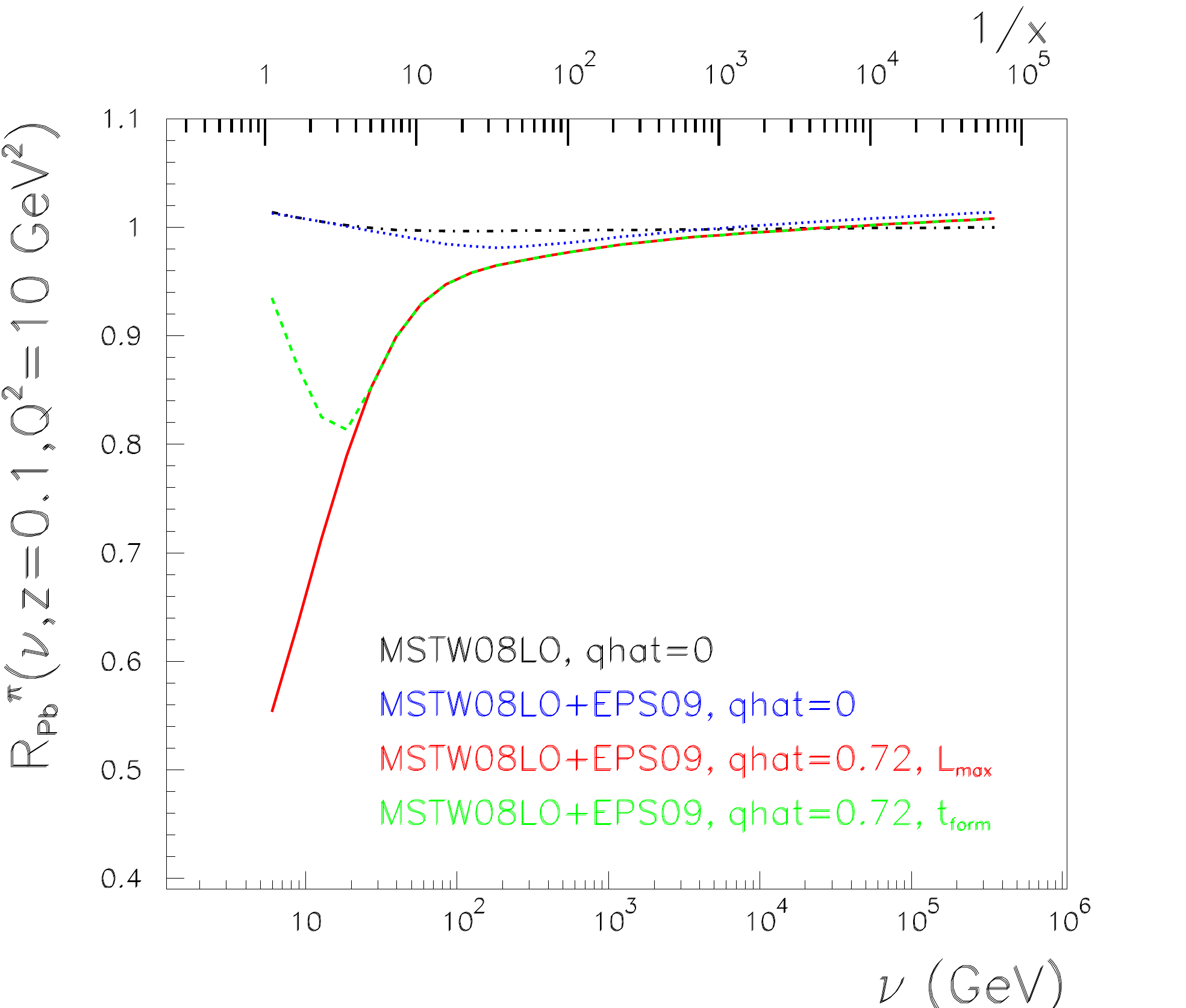} \includegraphics[clip=,width=0.32\textwidth,angle=0]{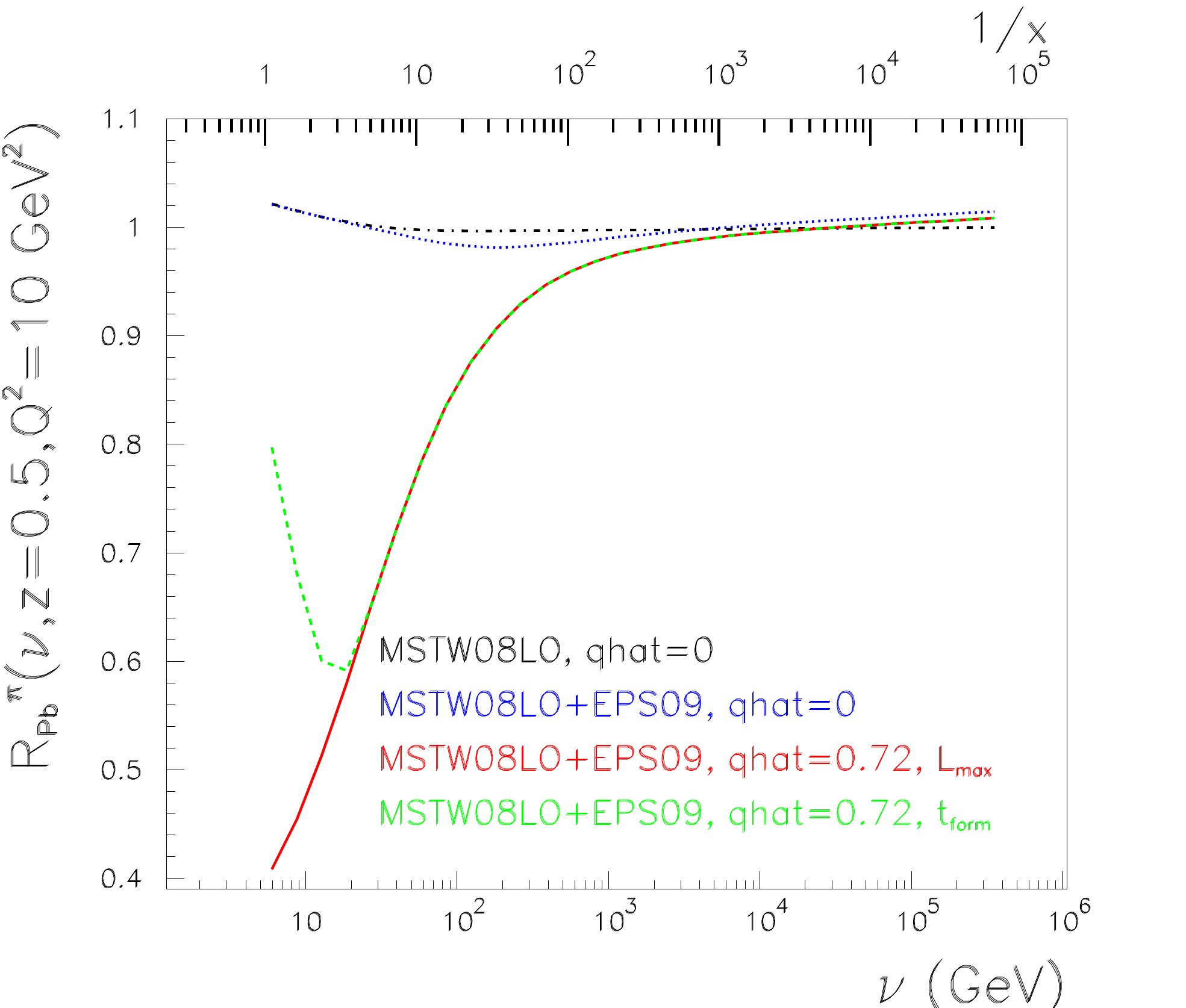}}
\centerline{ \includegraphics[clip=,width=0.32\textwidth,angle=0]{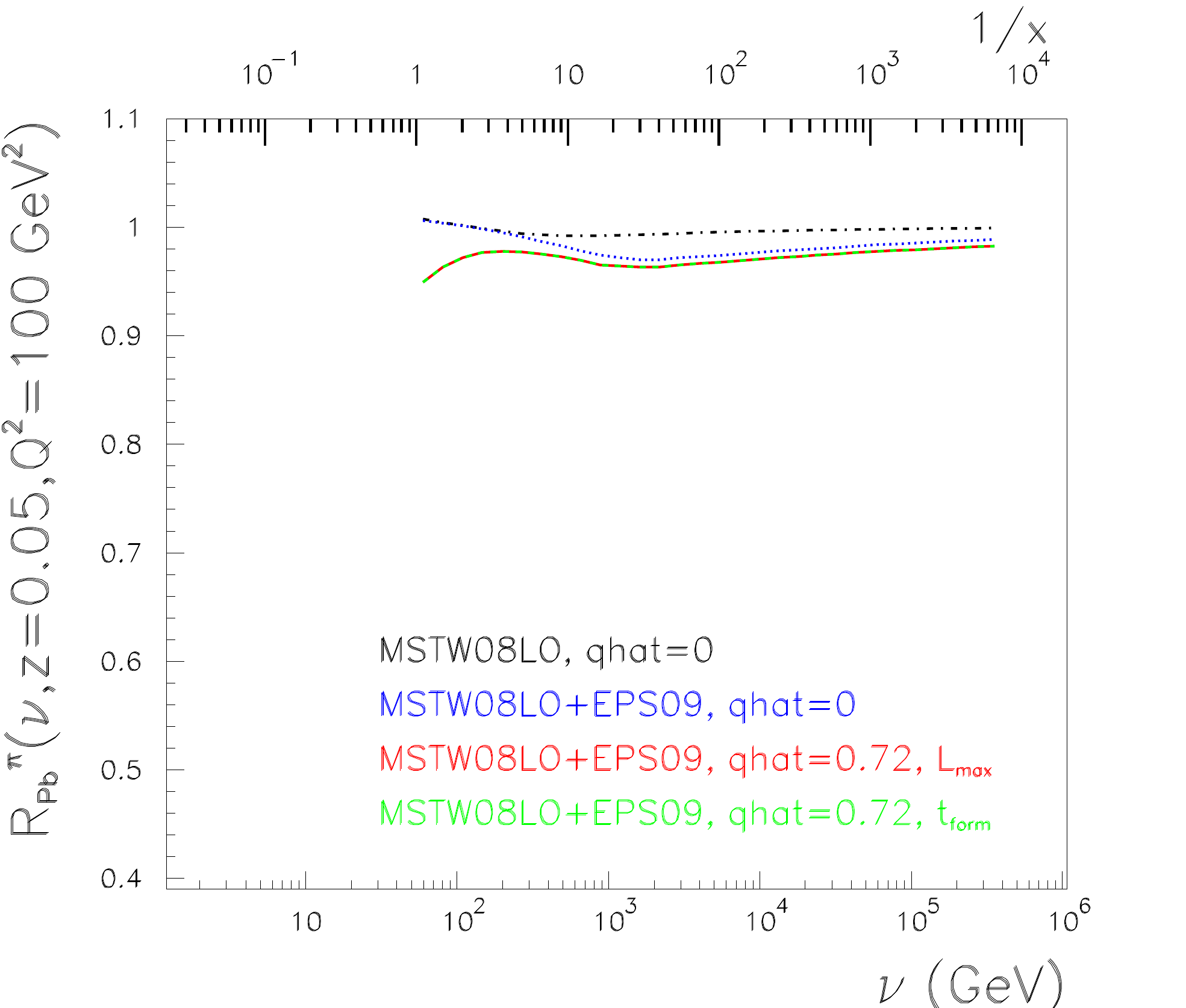} \includegraphics[clip=,width=0.32\textwidth,angle=0]{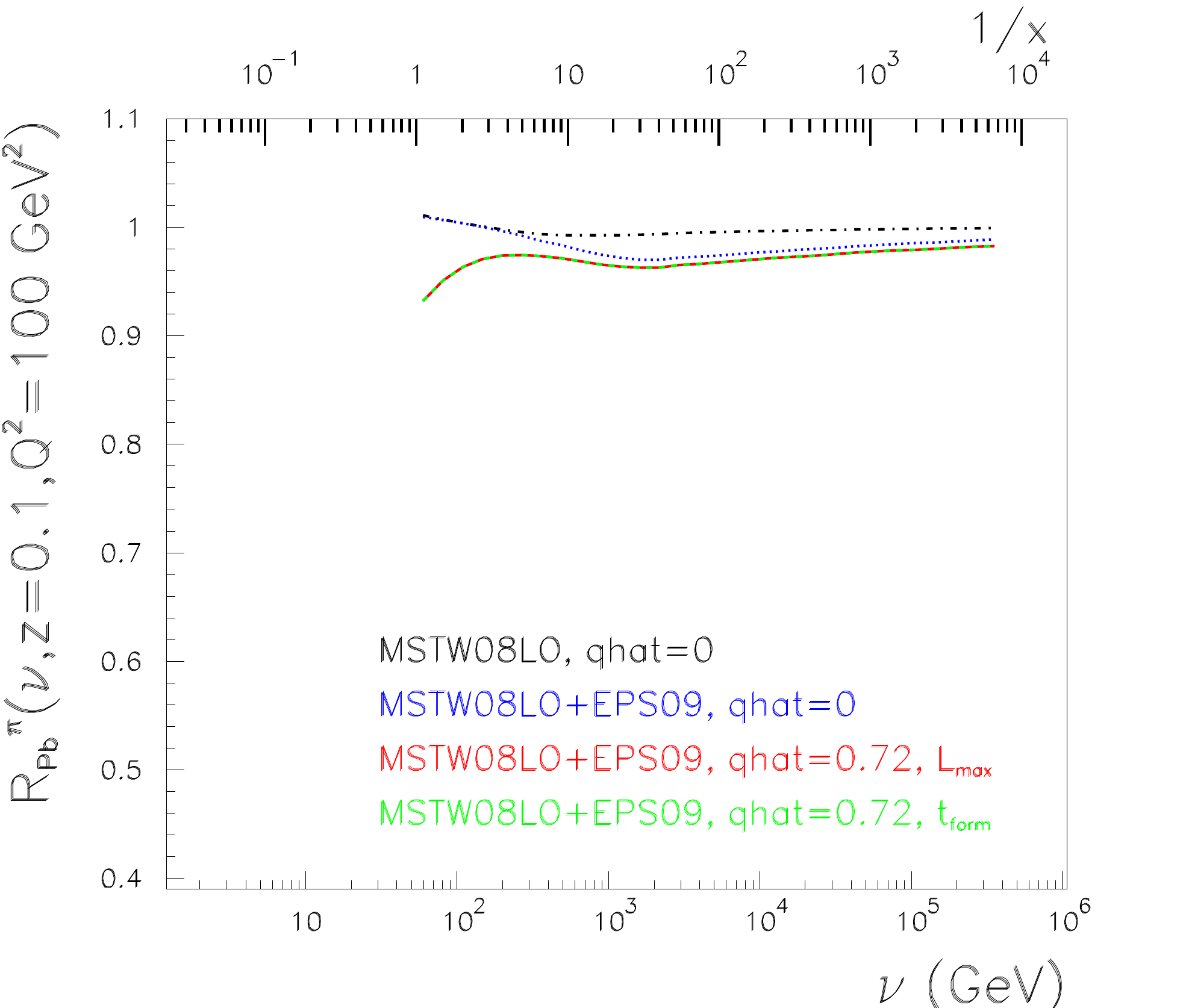} \includegraphics[clip=,width=0.32\textwidth,angle=0]{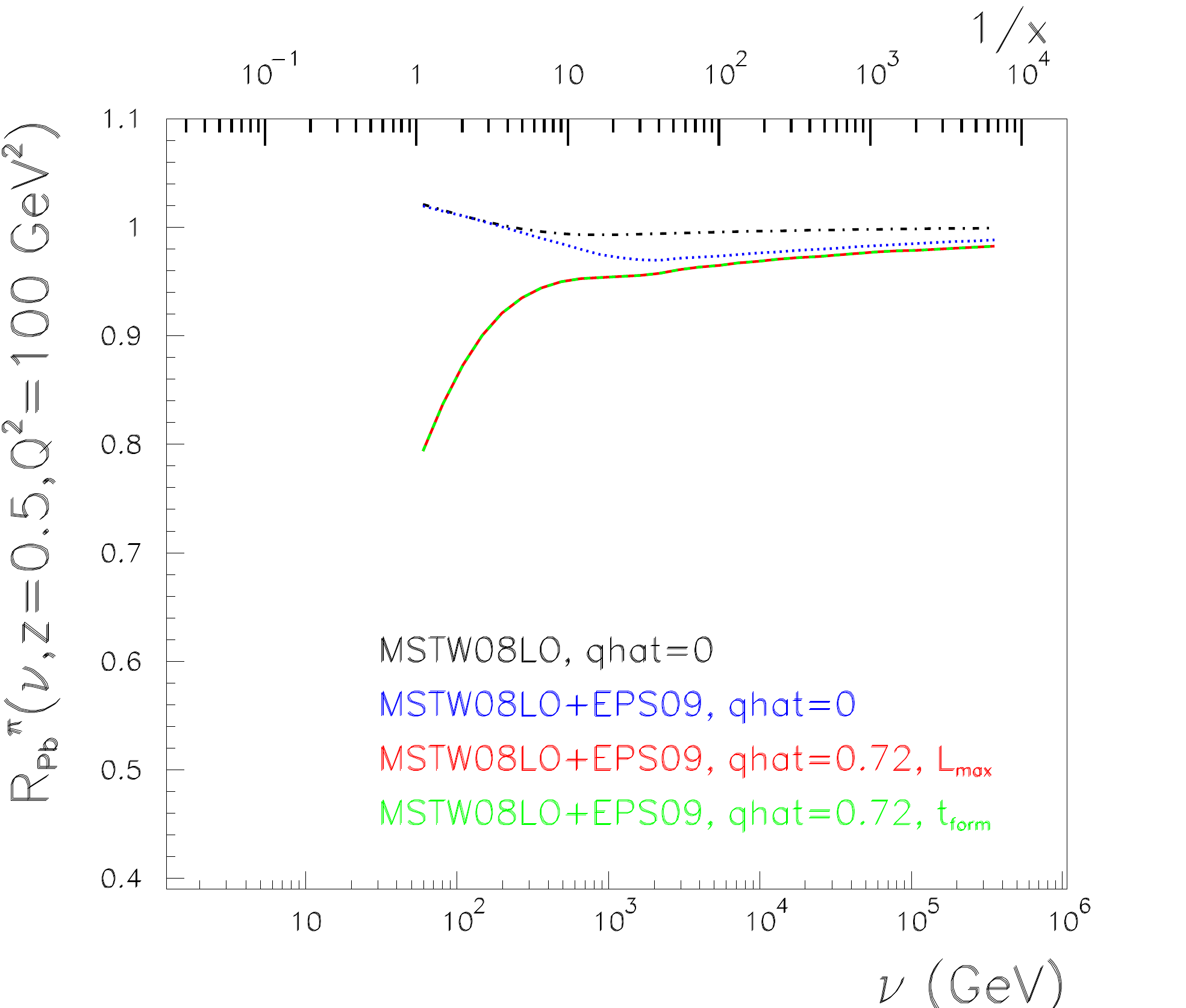}}
\caption{Ratio $R_{\rm Pb}^{\pi^0}(\nu,z,Q^2)$, Eq. (\ref{Eq:ratsidis}), versus $\nu$ (lower horizontal axes) or $1/x$ (upper horizontal axes) in $e$Pb over $ep$ at the LHeC, for $z=0.05$, 0.1 and 0.5 (from left to right) and $Q^2=2$, 10 and 100 GeV$^2$ (from top to bottom). Dashed-dotted black lines show the results without any nuclear effect but isospin, dotted blue ones further include the nuclear modification of PDFs  \cite{Eskola:2009uj}, solid red ones the effect of parton energy loss with a geometrical length, and dashed green include formation time considerations. See the text and \cite{Arleo:2003jz} for details of the calculation.}
\label{Fig:sidiseA}
\end{figure}

The results for $\pi^0$ production are shown in Fig. \ref{Fig:sidiseA}.
Several conclusions can be drawn. First, the effect of the difference in energy between numerator and denominator, and of isospin, are very small. Second, nuclear effects on fragmentation are larger for smaller $\nu$, as expected in a model in which the energy loss becomes energy-independent \cite{Arleo:2003jz,Salgado:2003gb}. 
Third, the nuclear suppression is larger for larger $z$ and it decreases with increasing $Q^2$, both effects due to the steepness of the fragmentation function and its evolution with $Q^2$. Finally, formation time limitations are only sizeable for small $\nu$, as naively expected due to the possibility of hadron formation inside the nucleus in this kinematic region, see \cite{Arleo:2003jz}.

From these results we conclude that the study of SIDIS at the LHeC looks very promising.
Still, extensive analyses at detector level are required in order to establish the accessible kinematic regions and to further explore the possibilities for particle identification.

%% file: physics/tex/neutrino.tex

The stringent constraints of the parton distributions at very small $x$ from
a future LHeC will
have important implications for
neutrino astronomy. Ultra-high energy neutrinos can provide important
information about distant astronomical objects and the origin of the
Universe.
%
They have attracted a lot of
attention during recent  years, see the reviews \cite{Stasto:2003xq,Becker:2007sv}. Neutrino astronomy has many advantages over
conventional photon astronomy.  This is due to the fact that neutrinos,
unlike photons, interact only weakly, so they can travel long distances  being
practically undisturbed. The typical interaction lengths for neutrinos and
photons at energy $E \sim 1 \  {\rm TeV}$ are about 
$$
{\cal L}_{int}^{\nu} \sim 250 \times 10^9 \ {\rm g/cm^2} \; , \hspace*{2cm}
{\cal L}_{int}^{\gamma} \sim 100 \ {\rm g/cm^2} \; .
$$
Thus, very energetic photons with energy bigger than $ \sim 10 \  {\rm TeV}$ cannot reach the Earth from the very distant corners
of our Universe without being rescattered. 
In contrast, neutrinos can travel very long distances without
interacting. They are also not deflected by  galactic magnetic fields, and therefore at ultra-high energies the angular distortion of the neutrino trajectory is very small.
As a result, highly energetic neutrinos reliably point back to their sources.
The interest in the  neutrinos at these high energies has led to the development
of several neutrino observatories, see \cite{Becker:2007sv} and references therein.

For reliable observations based on neutrino detection,
precise knowledge about their production rates and interactions is
essential to estimate the background, the expected fluxes and the detection
probabilities.
Even though neutrinos interact only weakly with other particles, strong interactions play an essential role in the calculations of their production
rates and interaction cross sections. This is due to the fact that neutrinos 
are produced in the decays of various mesons such as $\pi, K, D$ and
even $B$, which are produced in high-energy proton-proton (or proton-nucleus or nucleus-nucleus) collisions. These hadronic processes occur mainly in the 
atmosphere
though possibly also in the accretion discs of remote
Active Galactic Nuclei. Further, the interactions of highly energetic neutrinos with matter are dominated by the deep inelastic cross section with nucleons
or nuclei.
Hence, low-$x$ information from high-energy collider experiments such as HERA, Tevatron, LHC and, most importantly, the future LHeC, is
invaluable.

One of the main uncertainties (if not the dominant one) in the current limits
on high-energy neutrino production is due to the neutrino-nucleon (nucleus) 
cross section.
In fact, event rates are proportional to the neutrino cross section in many experiments.
This cross section involves
the gluon distribution probed at very small
values of Bjorken $x$,  down to even $  \sim 10^{-9}$, which corresponds to a very high centre of mass energy. 

\begin{figure}
\centering
\includegraphics[width=0.47\textwidth]{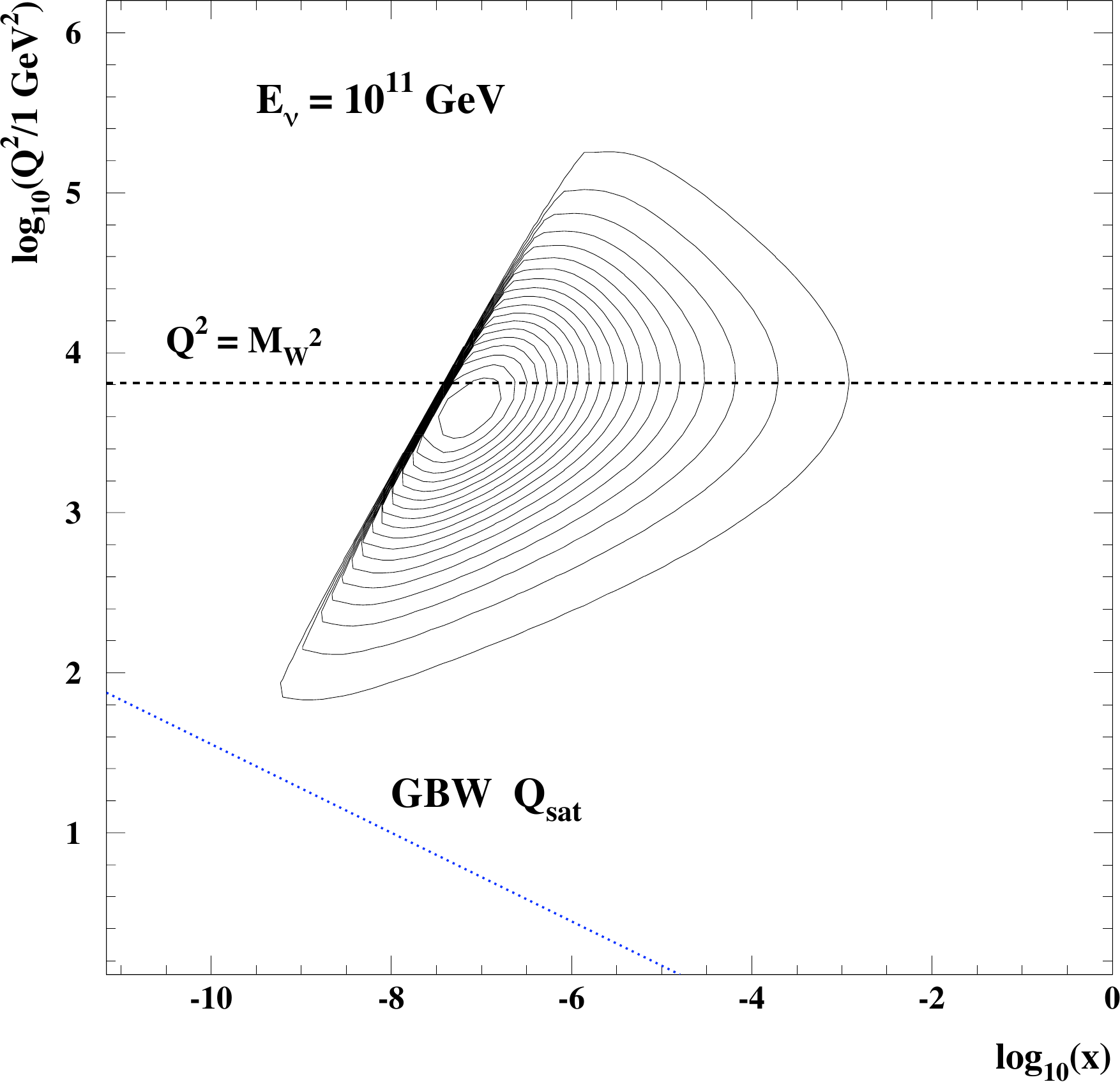}
\hspace*{0.3cm}
\includegraphics[width=0.47\textwidth]{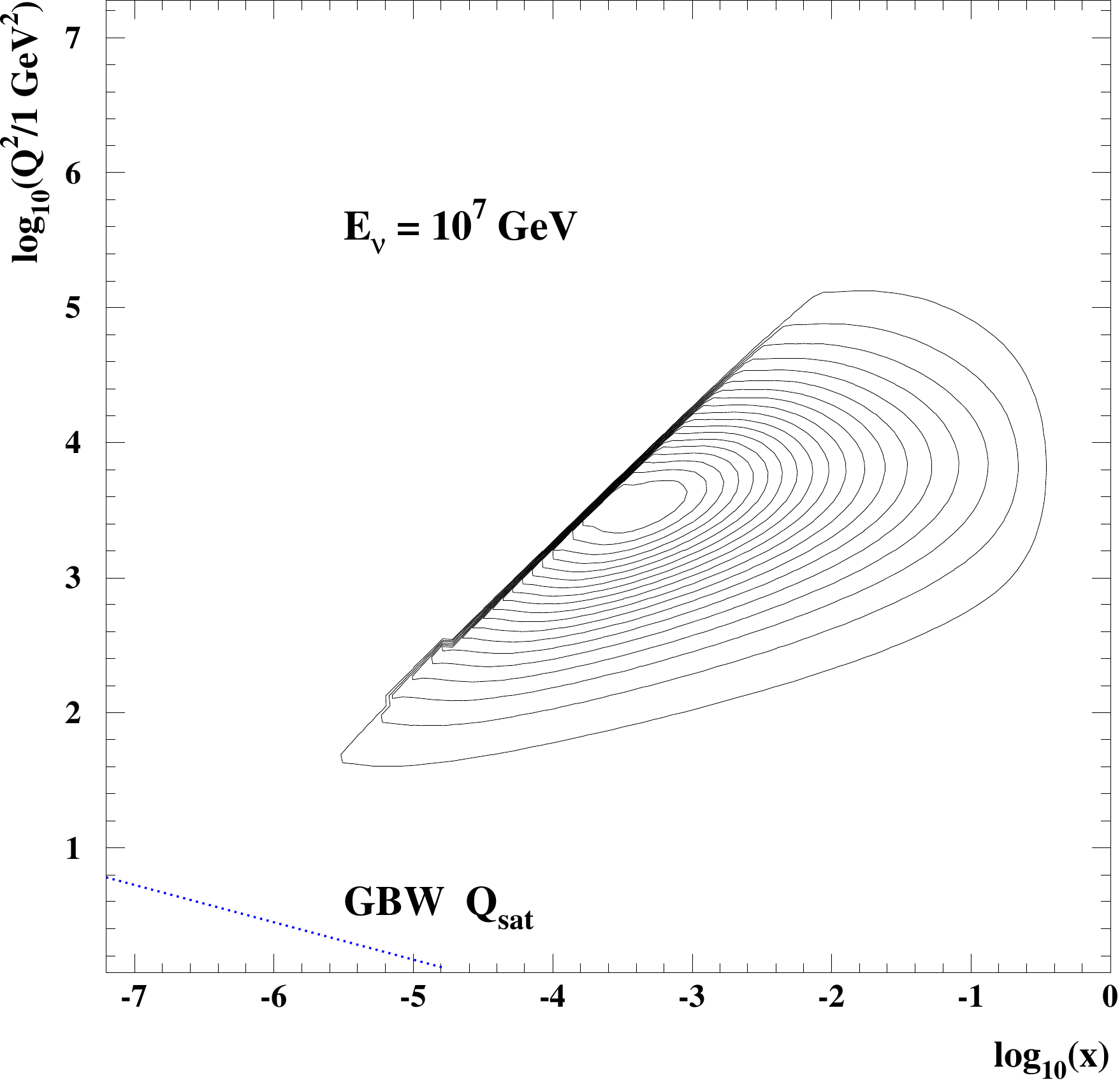}
\caption{Contour plot showing the $x, Q^2$ domain of the dominant 
contribution to the differential cross section
${\rm d} \sigma/ {\rm d} \ln (1/x) d \log Q^2$ for the total $\nu$-nucleon 
interaction at neutrino laboratory energies of $E_\nu = 10^{11}$ GeV (left plot) and $E_\nu = 10^{7}$ GeV (right plot).  The 20 contours                      
enclose contributions of 5, 10, 15 $\cdots$ 100 \% of the 
cross section. The saturation scale 
according to the model in \cite{GolecBiernat:1998js} is shown 
as a dashed line. See the text for further explanation.}
\label{fig:contouruhe}
\end{figure}

To visualise the kinematic regime probed in ultra-high energy neutrino-nucleon
interactions, contour plots of the
differential cross section  $\frac{d^2\sigma}{d\ln 1/x d\ln Q^2/\Lambda^2}$ 
in the
$(x,Q^2)$ plane are shown
in Fig.~\ref{fig:contouruhe}. 
The contours enclose regions with different contributions
to the total cross section $\sigma(E_{\nu})$.  For very high
energy $E_\nu =10^{11}\ {\rm GeV}$ 
the dominant contribution comes from the domain
$Q^2 \simeq M_W^2$ and $x_{\rm min} \simeq M^2_W /(2M_N E) \sim
10^{-8}-10^{-7}$ where $M_N$ is the nucleon mass, inaccessible to
any current or proposed accelerators. However, at lower neutrino 
energy $E_\nu =10^7 \, {\rm GeV}$ the relevant 
domain of $(x,Q^2)$ could be very well covered by the LHeC, 
thus providing 
important new constraints on the neutrino-nucleon cross section.

\begin{figure}[ht]
\begin{center}
\includegraphics[width=0.49\textwidth]{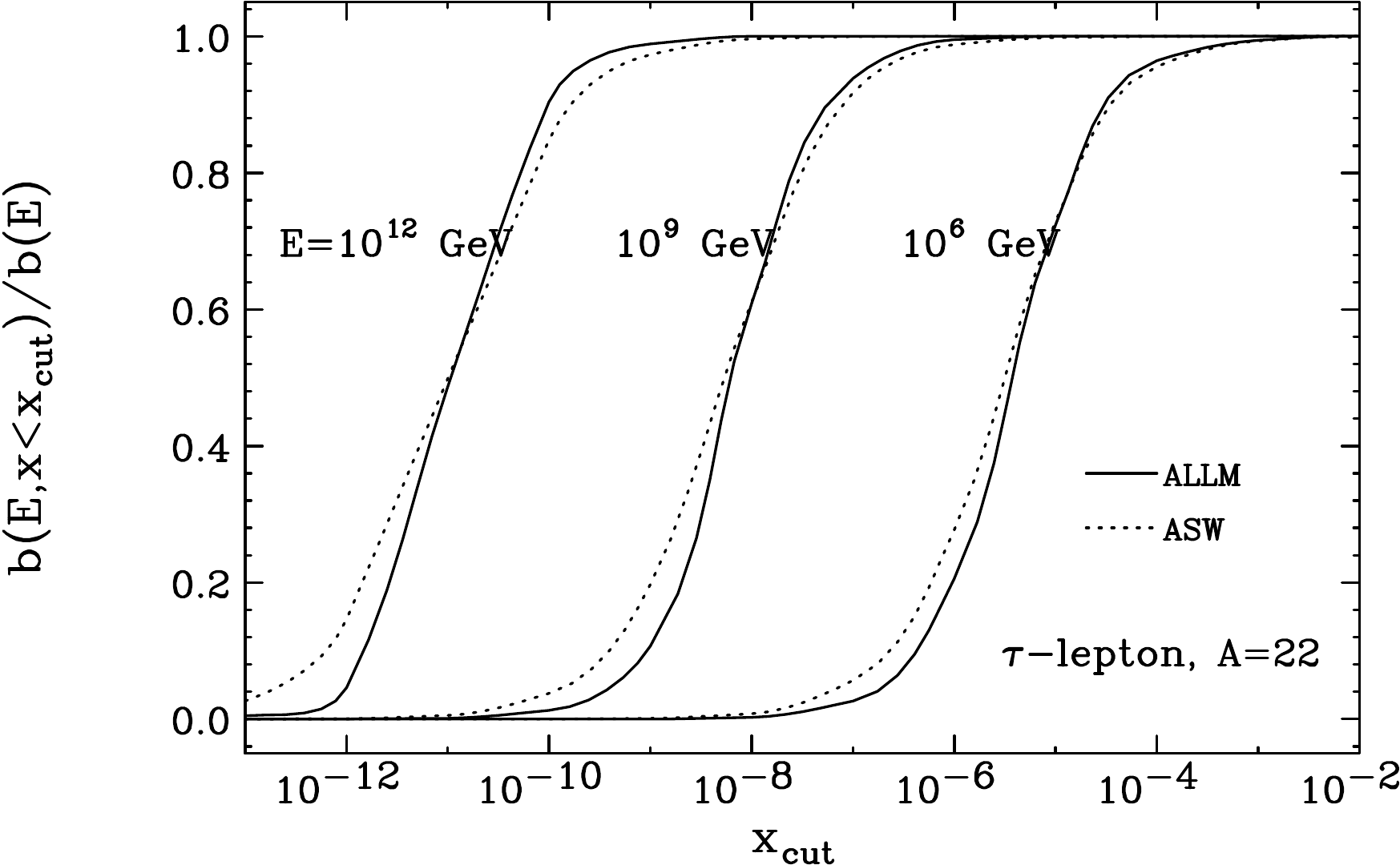}
\includegraphics[width=0.49\textwidth]{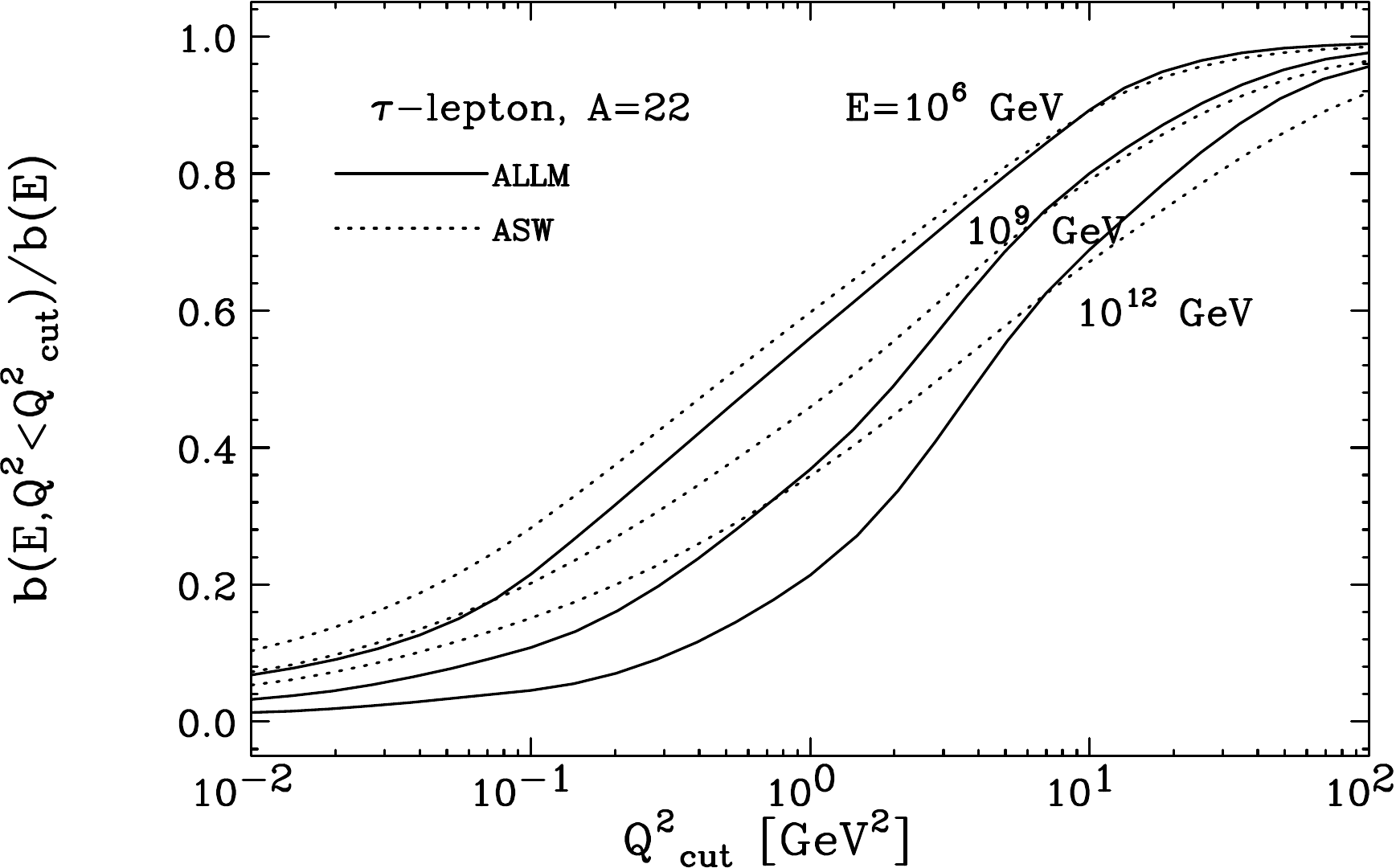}
\end{center}
\caption{The relative contribution of $x<x_{cut}$ (plot on the left)
and of $Q^2<Q^2_{cut}$ (plot on the right)
to the photo-nuclear energy loss rate, $b(E)$, for different neutrino energies $E=10^6$, $10^9$ and $10^{12}$ GeV, in two different models for the extrapolation of structure functions to very small $x$. See the text and \cite{Armesto:2007tg} - from which these plots were taken - for explanations.}
\label{fig:xq2taueloss} 
 \end{figure}

On the other hand, another process that has been proposed for neutrino
detection
comes from the discovery of neutrino flavor oscillations,
which makes it possible that 
high rates of $\tau$ neutrinos reach the Earth
despite being  
heavily suppressed in most postulated production mechanisms. 
The possibility to search for $\nu_\tau$'s
by looking for $\tau$ leptons that exit the Earth, 
Earth-skimming neutrinos, has been 
shown to be particularly advantageous to detect neutrinos of energies 
in the EeV ($10^{18} \ {\rm eV}$) range \cite{Zas:2005zz}.
The short lifetime of a $\tau$ lepton originating a neutrino
charged current interaction allows the $\tau$ to decay in flight while
still close to the Earth's surface, 
producing an outgoing air shower, detectable in
principle by various techniques.
This channel suffers from negligible contamination for
other neutrino flavors. 
The sensitivity to $\nu_\tau$'s through the Earth-skimming channel
directly depends both on the neutrino charged current cross section
and on the $\tau$ range (the energy loss)
which is determined by the amount of 
matter with which the neutrino has to interact to produce an emerging
$\tau$. It turns out that the $\tau$ energy loss is also determined by the behaviour of the proton and nucleus structure functions at very small values of $x$, see e.g. \cite{Armesto:2007tg}.
The average energy loss per unit depth, $X$, 
is conveniently represented by:
\begin{equation}
- \left<\frac{dE}{dX}\right> = a(E) + b(E) E , \ \ b(E) = 
\frac{N_A}{A} \int dy \; y \int dQ^2 \frac{d\sigma^{lA}}{dQ^2 dy} \; ,
\label{eq:taueloss}
\end{equation}
where the $a(E)$ term is due to ionisation, $b(E)$ is the sum of
fractional losses due to $e^+ e^-$ pair production,
bremsstrahlung and photo-nuclear interactions,
$N_A$ is Avogadro's number and $A$ is the mass number.
The parameter $a(E)$ is nearly constant and the term $b(E) E$ 
dominates the energy loss above a critical energy that for $\tau$ 
leptons is a few TeV, with
the photo-nuclear interaction being dominant for $\tau$ energies exceeding
$E=10^7$~GeV (as already assumed in Eq. (\ref{eq:taueloss})).
In Fig. \ref{fig:xq2taueloss} the relative contribution to $b(E)$ of different
$x$ and $Q^2$ regions is shown. It can be observed that the energy loss is
dominated by very small $x$ and, in contrast to the case of the neutrino cross
section, by small and moderate $Q^2 \lsim m_\tau^2$.

As the LHeC will be able to explore a new  regime of low $x$ and moderate-to-high $Q^2$, and
constrain the parton distributions, the measurements performed at this collider
will be invaluable for the precise evaluation of the neutrino-nucleon (or
nucleus) scattering cross sections  and $\tau$ energy loss
necessary for ultra-high energy neutrino astronomy.

%% file: physics/bsm.tex
\newcommand{\Rp}{\mbox{$\not \hspace{-0.15cm} R_p$}}
\newcommand{\MeV}{\mbox{\rm MeV}}
\def\GeV{\hbox{$\;\hbox{\rm GeV}$}}
\def\TeV{\hbox{$\;\hbox{\rm TeV}$}}
\newcommand{\pb}{\mbox{{\rm ~pb}}}
\newcommand{\fb}{\mbox{{\rm ~fb}}}

\input{physics/bsm_intro.tex}

\input{physics/bsm_highq2.tex}

\input{physics/bsm_Leptoquarks.tex}

\input{physics/bsm_ExcitedFermions.tex}
\input{physics/bsm_newquarks_gammaq.tex}

\input{physics/bsm_higgs.tex}

\input{physics/bsm_HWW_azimuthal.tex}

\clearpage

%% file: physics/bsm_intro.tex
The LHC is the primary  machine to search for physics
beyond the Standard Model at the TeV energy scale.
The role of the LHeC, which is projected to operate when the
LHC begins its high luminosity phase, is to complement
and possibly resolve the observation of new phenomena
based on the specifics of deep inelastic $ep$ scattering
at energies extending to beyond a TeV. At the LHC,
it will not always be
possible to measure with precision the parameters of the new physics.
In this section, it is shown that in several cases
the LHeC can probe in detail deviations from the expected electroweak
interactions shared by leptons and quarks, thus adding essential information
on the new  physics. Previous
studies~\cite{Bagger:1984su,Altarelli:1984rn,Cashmore:1985xn,Jarlskog:1990dv}
of the potential of high-energy $ep$ colliders for the discovery of
exotic phenomena have considered a number of 
processes, most of which are reviewed here.
At the time this report is completed, the only sign for new physics at the
LHC, apart from new $b$ quark states and a plethora of more and more
stringent limits on the mass of new particles and their existence,
higher symmetries or extra dimensions, consists in the still
tentative observation of a new state at about $125$\,GeV mass, which
may be associated to the long searched for SM Higgs boson.
This section therefore concludes with a study of the
Higgs production at the LHeC in the rather clean 
$WW \rightarrow H \rightarrow b \overline{b}$ channel.

%% file: physics/bsm_highq2.tex
\section{New physics in inclusive DIS at high {\boldmath{$Q^2$}}}
\label{sec:NewPhysicsAtHighQ2}


The LHeC collider
would enable the study of deep inelastic neutral current scattering at
very high squared momentum transfers, $Q^2$, thus probing the structure
of electron-quark ($eq$) interactions at very short distances. At these small scales new
phenomena not directly detectable may yet become observable as deviations
from the Standard Model predictions. A convenient tool to assess the
experimental sensitivity beyond the maximal available centre of mass
energy and to parameterise indirect signatures of new physics is the
concept of an effective four-fermion contact interaction.  If the
contact terms originate from a model where fermions have a
substructure, a compositeness scale can be related to the size of the
composite object. If they are due to the exchange of a new heavy
particle, such as a leptoquark, the effective scale is related to the
mass and coupling of the exchanged boson.  Contact interaction
phenomena are best observed as a modification of the expected $Q^2$
dependence and all information is essentially contained in the
differential cross section $\mathrm d\sigma / \mathrm d Q^2$.
An alternative way to parameterise the effects of fermion substructure
makes use of form factors, which would also lead to deviations of
$\mathrm d\sigma / \mathrm d Q^2$ with respect to the SM prediction.
As a last example, low scale quantum gravity effects, which may be 
mediated via gravitons coupling to SM particles and propagating into 
large extra spatial dimensions, could also be observed as a 
modification of $\mathrm d\sigma / \mathrm d Q^2$ at highest $Q^2$.
These possible manifestations of new physics in inclusive DIS are addressed
in this section.

\subsection{Quark substructure}

The remarkable similarities in the electromagnetic and weak
interactions of leptons and quarks in the Standard Model, and
their anomaly cancellations in the family structure, strongly suggest
a fundamental connection. It would therefore be natural to conjecture
that they could be composed of more fundamental constituents, or that
they form a representation of a larger gauge symmetry group than that
of the Standard Model, in a Grand Unified Theory.  

A possible method to investigate fermion substructures is to assign a
finite size of radius $R$ to the electroweak charges of leptons and/or quarks
while treating the gauge bosons $\gamma$ and $Z$ still as point-like
particles~\cite{Kopp:1994qv}.
A convenient parameterisation is to introduce `classical'
form factors $f(Q^2)$ at the gauge boson--fermion vertices,
which are expected to diminish the Standard Model cross section at high
momentum transfer:
\begin{eqnarray}
  f (Q^2) & = & 1 - \frac{1}{6}\, \langle r^2 \rangle \,Q^2 \ , \\[.4em]
  \frac{d\sigma}{dQ^2} & = &
  \frac{d\sigma^{SM}}{dQ^2} \, f^2_e(Q^2)\,f^2_q(Q^2) \ .
\end{eqnarray}

The form factor $f(Q^2)$ is related to the Fourier transform of the
electroweak charge distribution within the fermion.
The square root of the mean-square radius of this distribution,
$R = \sqrt{\langle r^2 \rangle}$, is taken as a measure of the particle size.
Since the point-like nature of the electron/positron is already established down 
to extremely low distances in $e^+\,e^-$ and $(g - 2)_e$ experiments,
only the quarks are allowed to be extended objects, i.e. the form factor
$f_e$ can be set to unity in the above equation.

\begin{figure}[bhtp]
\begin{center}
\includegraphics[width=0.6\columnwidth]{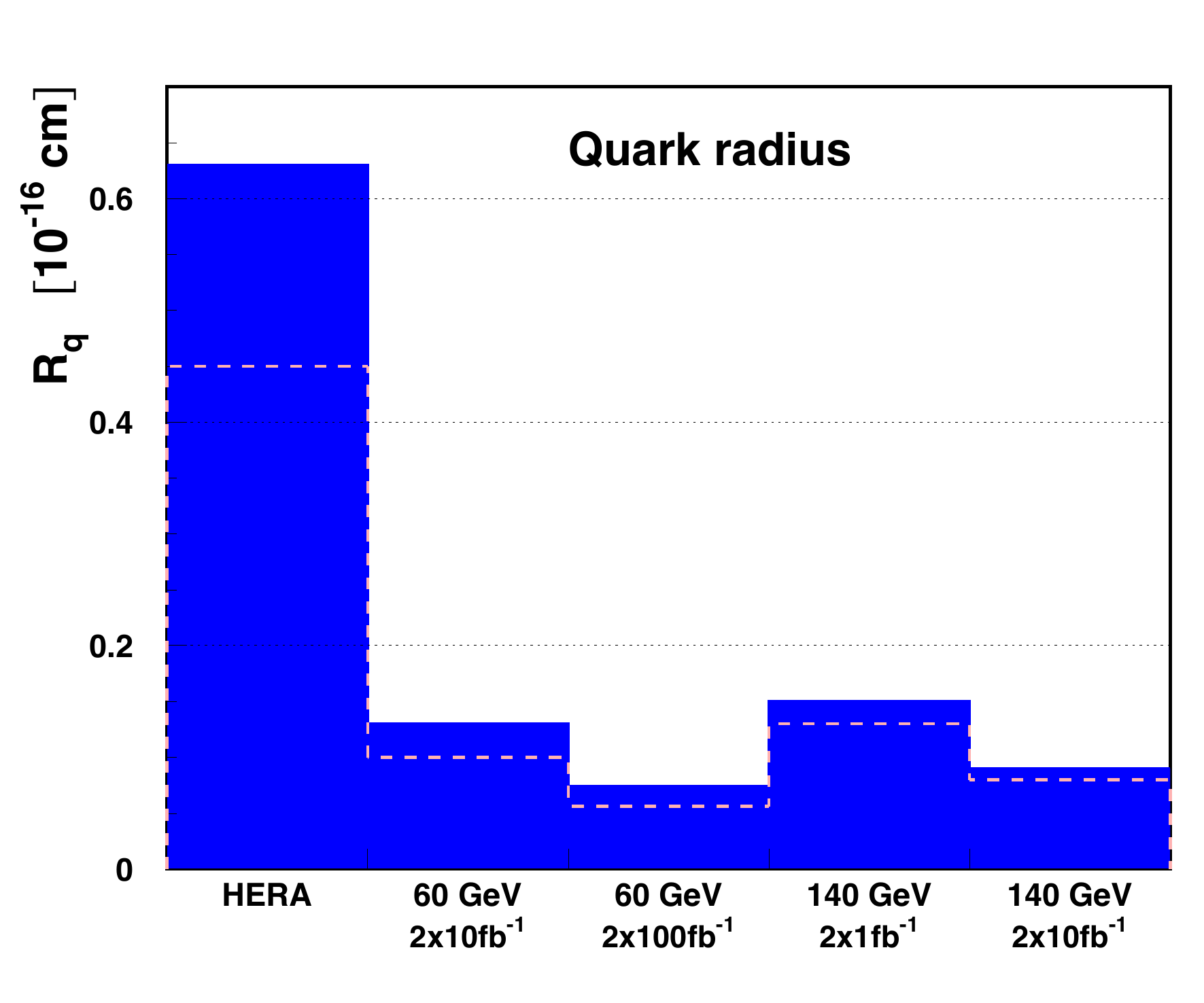}
\caption{Sensitivity ($95 \%$ confidence level limits) of an LHeC collider
to the effective quark radius.
The dashed lines show the sensitivity when systematic uncertainties are neglected,
while a systematic uncertainty of $5 \%$ is accounted for when calculating the sensitivities
shown as the full histograms. 
}
\label{fig:quark_radius}
\end{center}
\end{figure}
Figure.\ref{fig:quark_radius} shows the sensitivity that the LHeC
could reach on the ``quark radius"~\cite{Zarnecki:2008cp}. Two
beam energy configurations have been studied
($E_e = 60$~GeV and $E_e = 140$~GeV), and two values of the integrated
luminosity, per charge, have been assumed in each case.
A sensitivity to quark radius below $10^{-19}$~m could be reached, about one order
of magnitude better than the current constraints.

At the LHC, quark compositeness can be investigated by studying the properties
of dijet events, in particular their mass
spectrum together with angular distributions. 
This is usually done in the context of four-quark contact interactions (CI),
defined similarly to the $eeqq$ contact interactions that are considered in the next paragraph
(see Eq.~\ref{lcontact} and Eq.~\ref{etacoeff}).
With the statistics collected in 2011 at $\sqrt{s} = 7$~TeV, the ATLAS experiment
rules out four-quark contact interaction scales lower than $7.8$~TeV~\cite{ATLAS-CONF-2012-038}.
This is not directly related to the quark radius considered above, the latter being defined
from the distribution of electroweak charge within the quark. Dijet production at the LHC
is largely dominated by strong interactions, and a deviation from the SM of the electroweak 
production of dijets would lead to a very small effect in the total dijet production cross section.
From a naive scaling of the CI contribution by $ ( \alpha_{em} / \alpha_S)^2$, the current
bound would translate into an upper limit of $7 \cdot 10^{-19}$~m on the quark radius.
With $300$~fb$^{-1}$ of LHC data at $14$~TeV, a factor of about $4$ could be gained on
this sensitivity.


\subsection{Contact interactions}

New currents or heavy bosons may produce indirect effects
through the exchange of a virtual particle interfering with
the $\gamma$ and $Z$ fields of the Standard Model.
For particle masses and scales well above the available energy,
$\Lambda \gg \sqrt{s}$,
such indirect signatures may be investigated by searching for
a four-fermion point-like $(\bar{e}\,e)(\bar{q}\,q)$ contact interaction.
The most general chiral invariant 
Lagrangian for neutral current vector-like contact interactions
can be written in the form~\cite{Eichten:1983hw, Ruckl:1983ag, Haberl:1991vu}
\begin{eqnarray}
  {\cal L}_V  &=&\sum_{q \, = \, u,\, d}\left\{\eta^q_{LL}\,
   (\bar{e}_L\gamma_\mu e_L)(\bar{q}_L\gamma^\mu q_L)
   +\eta^q_{LR}\,
   (\bar{e}_L\gamma_\mu e_L)(\bar{q}_R\gamma^\mu q_R)\right.
 \nonumber \\
   &&\ \ \ \ \left.+\;\eta^q_{RL}\,(\bar{e}_R\gamma_\mu e_R)
   (\bar{q}_L\gamma^\mu q_L) +\eta^q_{RR}\,
   (\bar{e}_R\gamma_\mu e_R)(\bar{q}_R\gamma^\mu q_R)\right\} \; ,
 \label{lcontact}
\end{eqnarray}
where the indices $L$ and $R$ denote the left-handed and right-handed
fermion helicities and the sum extends over {\em up}-type and
{\em down}-type quarks and antiquarks $q$.
In deep inelastic scattering at high $Q^2$ the contributions from the
first generation $u$ and $d$ quarks  dominate and contact terms
arising from sea quarks $s$, $c$ and $b$ are strongly suppressed.
Thus, there are eight independent effective coupling coefficients,
four for each quark flavour
\begin{eqnarray}
  \eta_{ab}^q & \equiv & \epsilon\frac{ 4 \pi}{\Lambda^{q \ 2}_{ab}} \ ,
  \label{etacoeff}
\end{eqnarray}
where $a$ and $b$ indicate the $L,\ R$ helicities,
$\Lambda^q_{ab}$ is a scale parameter
and $\epsilon$ is  often set to $\epsilon = \pm 1$,
which determines the interference sign with the Standard Model currents.
The ansatz eq.~(\ref{lcontact}) can be easily applied to any new
phenomenon, {\em e.g.} $(e q)$ compositeness, leptoquarks or new gauge bosons,
by an appropriate choice of the coefficients $\eta_{ab}$.
Scalar and tensor interactions of dimension~6 operators
involving helicity flip couplings  are strongly suppressed
at {\sc Hera}~\cite{Haberl:1991vu} and therefore not considered.

\begin{figure}[htbp]
\begin{center}
\includegraphics[width=0.49\columnwidth]{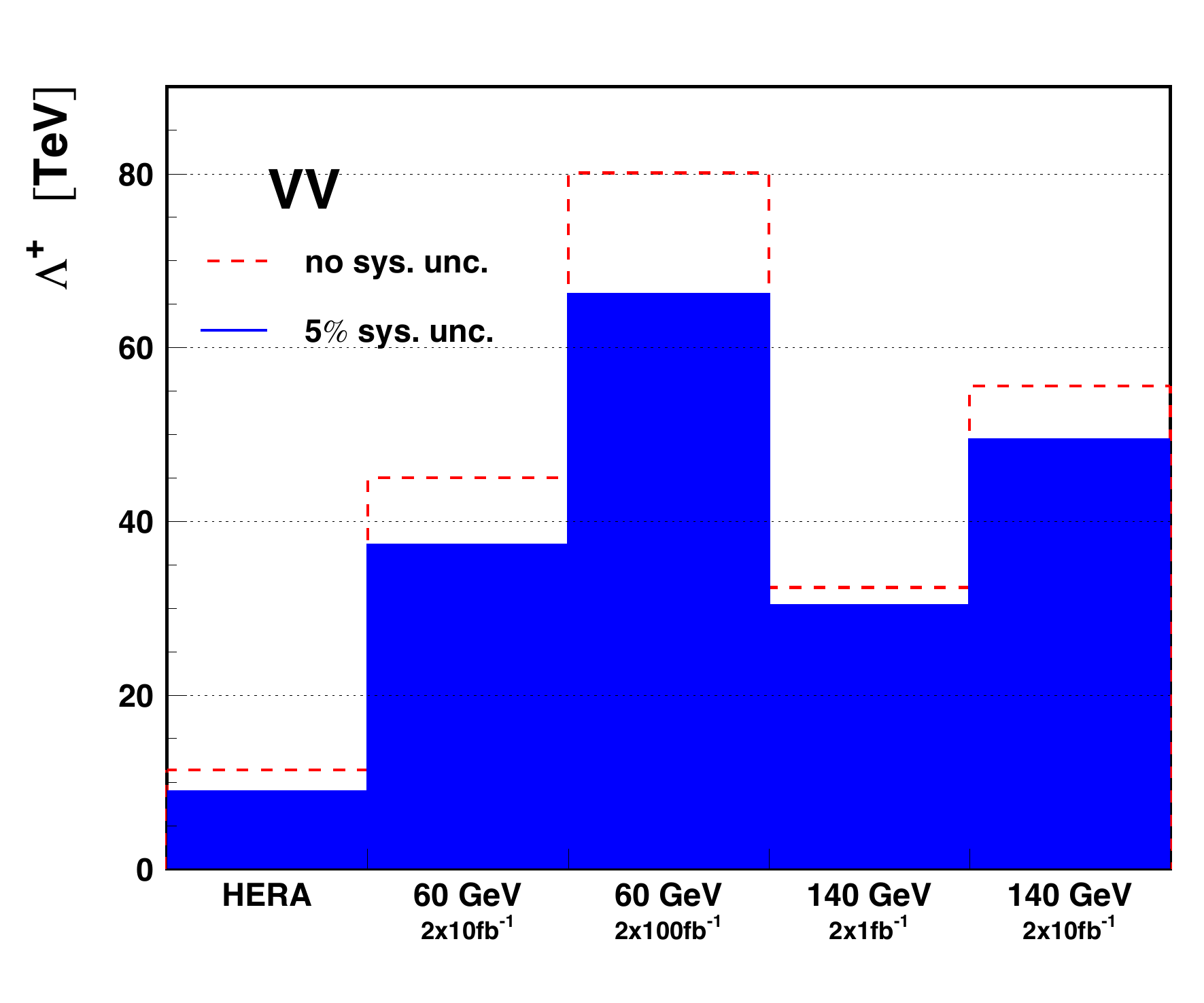}
\includegraphics[width=0.49\columnwidth]{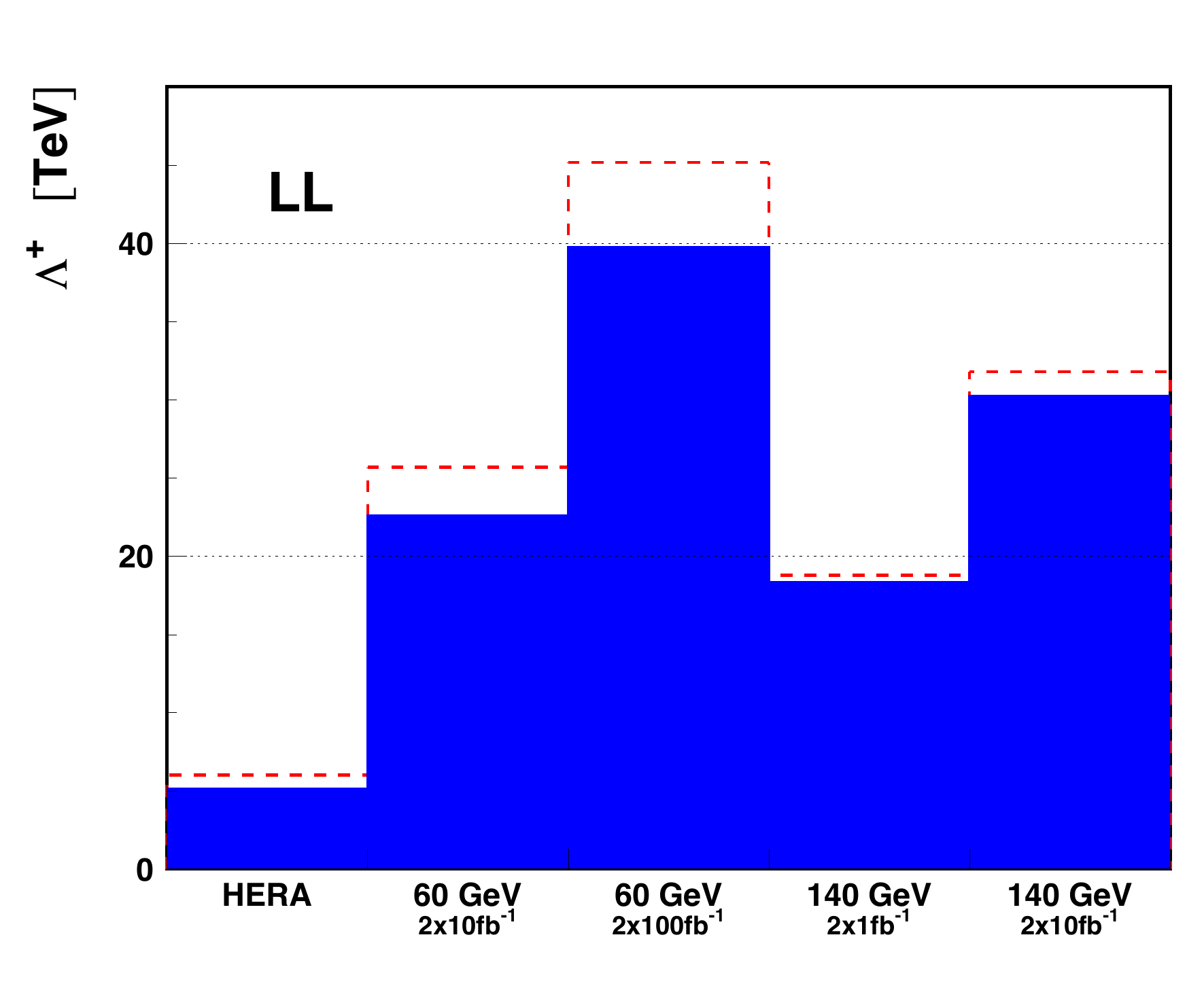}
\end{center}
\caption{Sensitivity ($95 \%$ confidence level limits) on the scale $\Lambda$
for two example contact interactions.
The dashed lines show the sensitivity when systematic uncertainties are neglected,
while a systematic uncertainty of $5 \%$ is accounted for when calculating the sensitivities
shown as the full histograms.
}
\label{fig:ci_limits}
\end{figure}

Figure~\ref{fig:ci_limits} shows the sensitivity that the LHeC could
reach on the scale $\Lambda$, for two example cases of 
contact interactions~\cite{Zarnecki:2008cp}.
In general, with $10$ fb$^{-1}$ of data, LHeC would probe scales between
$25 \TeV$ and $45 \TeV$, depending on the model. The ultimate sensitivity of LHC
to such $eeqq$ interactions, which would affect the di-electron Drell-Yan (DY) spectrum
at high masses, is similar.
With $\sim 1$~fb$^{-1}$ of data at $\sqrt{s} = 7$~TeV, the ATLAS and CMS experiments rule out 
$eeqq$ contact interactions with a scale below $\sim 10$~TeV. The sensitivity will extend to
typically $30$~TeV with $100$~fb$^{-1}$ of data at $\sqrt{s} = 14$~TeV.


\begin{figure}[t]
\centerline{\includegraphics[width=0.55\columnwidth]{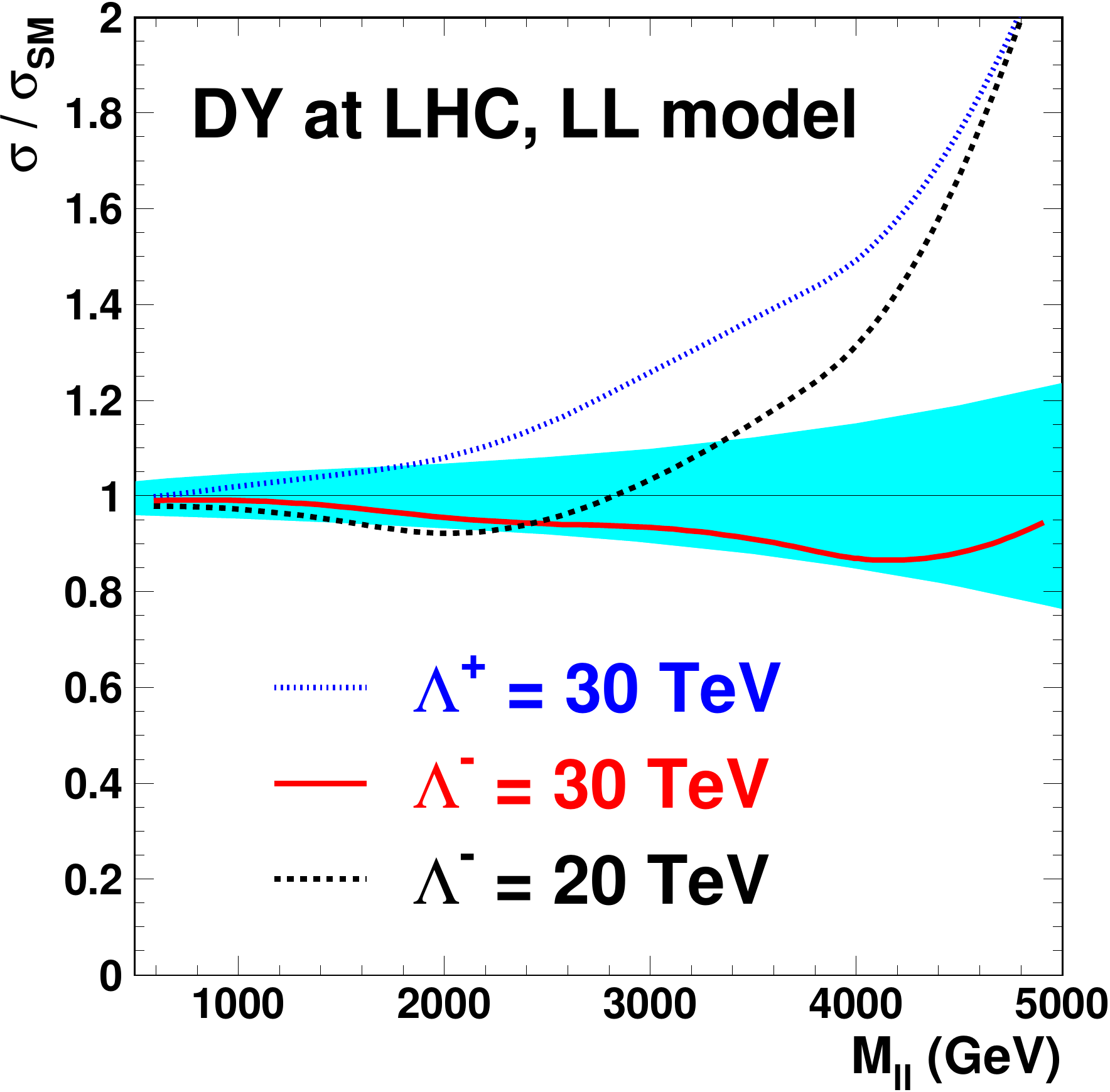}}
\caption{Example deviations, from its SM value, of the Drell-Yan cross section at LHC as a function of the
dilepton mass, in the presence of an $eeqq$ contact interaction. The blue band shows the relative
 uncertainty of the predicted SM cross sections due to the current uncertainties of the
 parton distribution functions, as obtained from the CTEQ 6.1 sets. With a luminosity of
$300$~fb$^{-1}$, the statistical uncertainty of the measurement would be about $20 \%$
($60 \%$) at $M_{ll} = 3$~TeV ($M_{ll} = 4$~TeV). }\label{Fig:CI_DY}
\end{figure}

\begin{figure}[htbp]
 \centerline{\includegraphics[width=0.7\columnwidth]{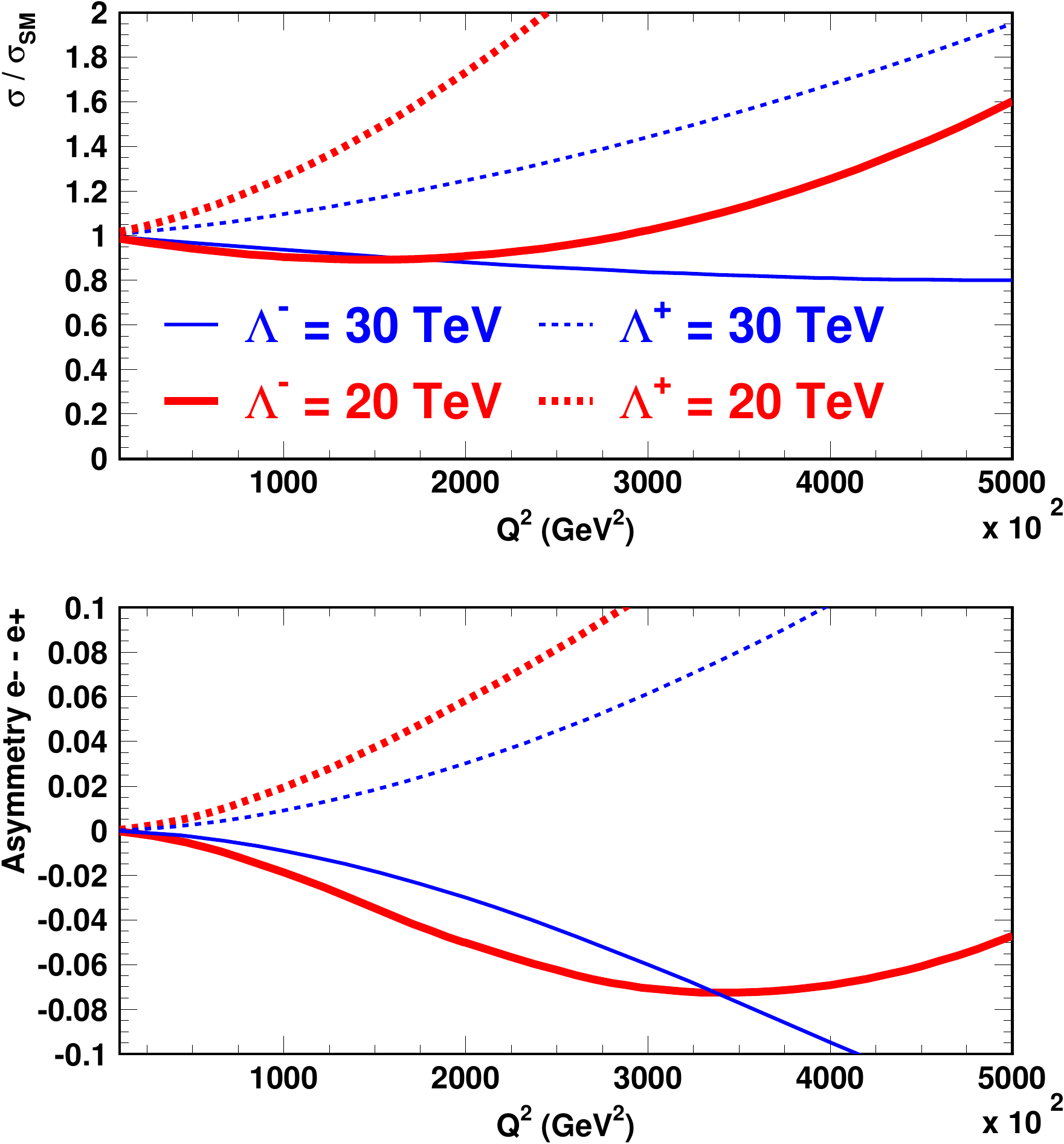}}
 \caption{(top) Example deviations of the $e^- p$ DIS cross section at LHeC, in the presence of an $eeqq$
 CI, for $E_e = 70$~GeV. The ratio of the ``measured" to the SM cross sections, $r = \sigma / \sigma_{SM}$, is shown. 
 The cross sections would be measured with a 
 statistical (systematic) accuracy of 3\,(1)$\%$ at $Q^2 = 2 \cdot 10^5$~GeV$^2$
 and of  10\,(2)$\%$ at $Q^2 = 4 \cdot 10^5$~GeV$^2$ 
 for an assumed integrated luminosity of $10$\,fb$^{-1}$.
 (bottom) Asymmetry $\frac{r(e^-) - r(e^+)}{ r(e^-) + r(e^+)}$  between $e^- p$ and $e^+ p$
 measurements of $\sigma / \sigma_{SM}$.}\label{Fig:CI_LHeC}
\end{figure}
 
Figure~\ref{Fig:CI_DY} shows how the DY cross section at the
LHC would deviate from the SM value, for
three examples of $eeqq$ contact interactions. In the ``LL" model considered here,
the sum in eq.~(\ref{lcontact})
only involves left-handed fermions and all amplitudes have the same phase $\epsilon$.
With only $pp$ data, it will be difficult to
determine simultaneously the size of the contact interaction scale $\Lambda$ and the sign
of the interference of the new amplitudes with respect to the SM ones: for example, for $\Lambda = 20 \TeV$
and $\epsilon = -1$,
the decrease of the cross section with respect to the SM prediction for di-electron masses
below $\sim 3 \TeV$, which is characteristic of a negative interference, is too small to
be firmly established when uncertainties due to parton distribution functions are taken into account.
Angular distributions and forward-backward asymmetries can help in principle to disentangle
between the various possible CI scenarios. However, the statistical uncertainties
expected for dilepton masses above $\sim 2.5 - 3$~TeV limit the power of these variables
to scales well below the sensitivity limit. A similar conclusion was reached in~\cite{Rizzo:2009pu}
in a study of the indirect effects of a very heavy $Z'$ boson on dilepton events at the LHC.

For the same ``LL" model, the sign of this interference can be unambiguously determined
at LHeC from the asymmetry of $\sigma / \sigma_{SM}$ in $e^+ p$ and $e^- p$ data, as shown
in Fig.~\ref{Fig:CI_LHeC}. \\

Moreover, with a polarised lepton beam, $ep$ collisions would help determine the chiral structure
of the new interaction.
More generally, it is very likely that both $pp$ and $ep$ data would be necessary to
underpin the structure of new physics which would manifest itself as an $eeqq$ contact interaction.
Such a complementarity of $pp$, $ep$ (and also $ee$) data was studied in~\cite{Zarnecki:1999je} in the
context of the Tevatron, HERA and LEP colliders.

\subsection{Kaluza-Klein gravitons in extra-dimensions}

In some models with $n$ large extra dimensions, the SM particles reside
on a four-dimensional ``brane", while the spin 2 graviton propagates into the 
extra spatial dimensions and appears in the four-dimensional world as a tower 
of massive Kaluza-Klein (KK) states. The summation over the enormous number of Kaluza-Klein 
states up to the ultraviolet cut-off scale, taken as the Planck scale $M_S$ in 
the $4+n$ space, leads to effective contact-type interactions $f f f' f'$  
between two fermion lines, with a coupling $\eta = O(1) / M_S^4$.
In $ep$ scattering, the exchange of such a tower of Kaluza-Klein gravitons would
affect the $Q^2$ dependence of the DIS cross section $\mathrm d\sigma / \mathrm d Q^2$.
At LHeC, such effects could be observed as long as the scale $M_S$ is below
$4-5 \TeV$. While at the LHC, virtual graviton exchange may be observed for
scales up to $\sim 10 \TeV$, and the direct production of $KK$ gravitons, for
scales up to $5-7 \TeV$ depending on $n$, would allow this phenomenon to be
studied further, LHeC data may determine that the new interaction is universal
by establishing that the effect in the $e q \rightarrow eq$ cross section is
independent of the lepton charge and polarisation, and, to some extent, of
the quark flavour.

%% file: physics/bsm_Leptoquarks.tex
\section{Leptoquarks and leptogluons}

\label{sec:leptoquarks}

The high energy of the LHeC extends the kinematic range of DIS physics
to much higher values of electron-quark mass $M=\sqrt{sx}$, beyond those of HERA.
By providing both baryonic and leptonic quantum numbers in the
initial state, it is ideally suited to a study of the properties of new
bosons possessing couplings to an electron-quark pair in this new mass
range.  
Such particles can be squarks in supersymmetric models with $R$-parity
violation (\Rp), or first-generation leptoquark (LQ) bosons which
appear naturally in various unifying theories beyond the Standard
Model (SM) such as: $E_6$~\cite{Hewett:1988xc}, where new fields can
mediate interactions between leptons and quarks; extended
technicolor~\cite{Farhi:1980xs,Hill:2002ap}, where leptoquarks result from
bound states of technifermions; the Pati-Salam
model~\cite{Pati:1974yy}, where the leptonic quantum number is a
fourth colour of the quarks or in lepton-quark compositeness models.
They are produced as single $s-$channel resonances via the fusion of
incoming electrons with quarks in the proton.  
They are generically
referred to as ``leptoquarks" in what follows.  The case of
``leptogluons", which could be produced in $ep$ collisions as a fusion
between the electron and a gluon, is also addressed at the end of this
section.

\subsection{Phenomenology of leptoquarks in $\bold{ep}$ collisions}



\begin{figure}[htb]
 \begin{center} 
   \epsfig{figure=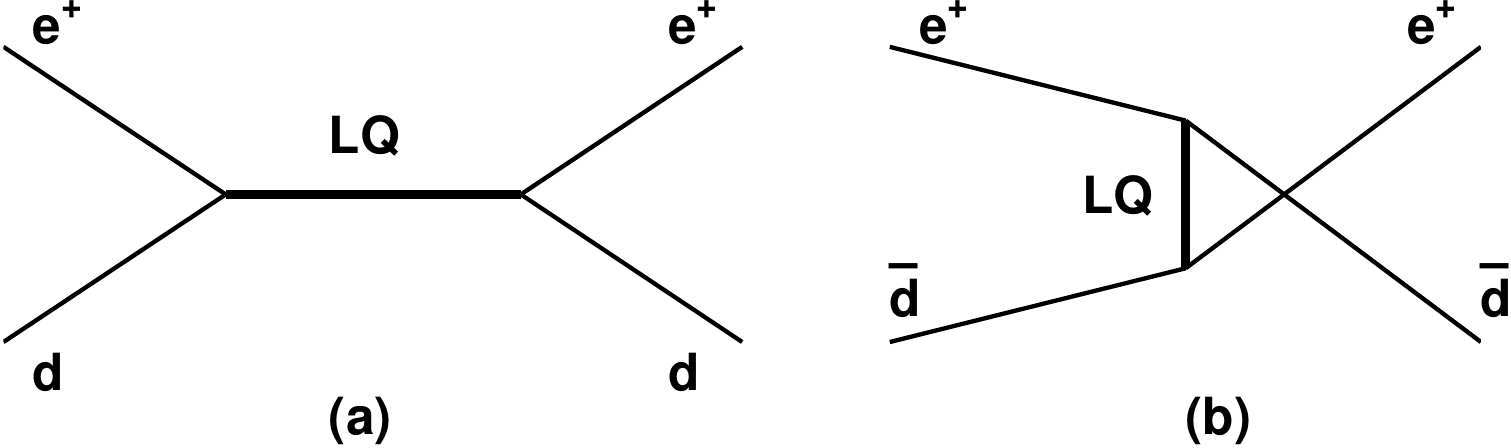,width=0.8\textwidth}
  \caption{\label{fig:lqdiag}
            Example diagrams for resonant production in the $s$-channel
            (a) and exchange in the $u$-channel (b) of a LQ
            with fermion number $F=0$.
            The corresponding diagrams for $|F|=2$ LQs are obtained
            from those depicted by exchanging the quark and antiquark. }
 \end{center}
\end{figure}

In $ep$ collisions, LQs may be produced resonantly up to the
kinematic limit of $\sqrt{s}$
via the fusion of the incident lepton with a quark or
antiquark coming from the proton, or
exchanged in the $u$ channel, as illustrated in
Fig.~\ref{fig:lqdiag}.
%
The coupling $\lambda$ at the $LQ-e-q$ vertex is
an unknown parameter of the model.

%
In the narrow-width approximation, the resonant production
cross section is proportional to $\lambda^2 q(x)$
where
$q(x)$ is the density of the struck parton in the incoming proton.

The resonant production or $u$-channel exchange of a leptoquark
gives $e + q$ or $\nu + q'$ final states leading to
individual events indistinguishable from SM NC and CC DIS respectively.
For the process $e q \rightarrow LQ \rightarrow eq$, the distribution
of the transverse energy $E_{T,e}$ of the final state lepton shows a Jacobian
peak at $M_{LQ} / 2$, $M_{LQ}$ being the LQ mass.
Hence the strategy to search for a LQ signal in $ep$ collisions is
to look, among high $Q^2$ (i.e. high $E_{T,e}$) DIS event candidates, for a peak
in the invariant mass $M$ of the final $e-q$ pair.
Moreover, the significance of the LQ signal over the SM DIS background
can be enhanced by exploiting the specific angular distribution of
the LQ decay products (see spin determination, below).

\subsection{The Buchm\"uller-R\"uckl-Wyler Model}

A reasonable phenomenological framework to study
first generation LQs is provided by
the BRW model~\cite{Buchmuller:1986zs}. This model is based on
the most general Lagrangian that is invariant under
$SU(3) \times SU(2) \times U(1)$, respects lepton and baryon number
conservation, and incorporates dimensionless
family diagonal couplings of
LQs to left- and/or right-handed fermions.
Under these assumptions LQs can be classified according to their
quantum numbers into
10 different LQ isospin multiplets (5 scalar and 5 vector),
half of which carry a vanishing fermion number $F=3B+L$
($B$ and $L$ denoting the baryon and lepton number respectively)
and couple
to $e^- + \bar{q}$ while the other half carry $|F|=2$ and couple to
$e^- + {q}$.
These are listed in Table~\ref{tab:brwscalar}.
%
%
\begin{table*}[htb]
  \renewcommand{\doublerulesep}{0.4pt}
  \renewcommand{\arraystretch}{1.2}
 \vspace{-0.1cm}

\begin{center}
    \begin{tabular}{|c|c|c||c|c|c|}
      \hline
       $F=2$ & Prod./Decay & $\beta_e$
              & $F=0$ & Prod./Decay & $\beta_e$  \\

      \hline
%
     \multicolumn{6}{|c|}{Scalar Leptoquarks} \\ \hline
    $^{1/3}S_0$     & $e^-_L {u}_L\rightarrow e^- {u}$ & $1/2$
  & $^{5/3}S_{1/2}$ & $e^-_L \bar{u}_L \rightarrow e^- \bar{u}$            & $1$  \\
                          & $e^-_R {u}_R\rightarrow e^- {u}$ & $1$
  &                       & $e^-_R \bar{u}_R \rightarrow e^- \bar{u}$            & $1$ \\
      \cline{1-3}
      $^{4/3}\tilde{S}_0$
        & $e^-_R {d}_R\rightarrow e^- {d}$ & $1$
  & $^{2/3}S_{1/2}$ & $e^-_R \bar{d}_R \rightarrow e^- \bar{d}$            & $1$ \\
      \hline
      $^{4/3}S_1$
        & $e^-_L {d}_L \rightarrow e^- {d}$
         & $1$
  & $^{2/3}\tilde{S}_{1/2}$ & $e^-_L \bar{d}_L \rightarrow e^- \bar{d}$ & $1$ \\
      $^{1/3}S_1$
        & $e^-_L {u}_L \rightarrow e^- {u}$
         & $1/2$
             & & &  \\
      \hline
%
     \multicolumn{6}{|c|}{Vector Leptoquarks} \\ \hline
    $^{4/3}V_{1/2}$ & $e^-_R {d}_L\rightarrow e^-  {d}$ & $1$
  & $^{2/3}V_{0}$   & $e^-_R \bar{d}_L \rightarrow e^- \bar{d}$              & $1$ \\
                          & $e^-_L {d}_R\rightarrow e^-  {d}$ & $1$
  &                       & $e^-_L \bar{d}_R \rightarrow e^- \bar{d}$              & $1/2$ \\
      \cline{4-6}
    $^{1/3}V_{1/2}$ & $e^-_R {u}_L\rightarrow e^-  {u}$ & $1$
  & $^{5/3}\tilde{V}_0$
        & $e^-_R \bar{u}_L \rightarrow e^- \bar{u}$ & $1$ \\
      \hline
    $^{1/3}\tilde{V}_{1/2}$
        & $e^-_L {u}_R\rightarrow e^- {u}$ & $1$
  & $^{5/3}V_{1}$    & $e^-_L \bar{u}_R \rightarrow e^- \bar{u}$              & $1$ \\
                          &                                            &
  & $^{2/3}V_{1}$    & $e^-_L \bar{d}_R \rightarrow e^- \bar{d}$              & $1/2$ \\
      \hline
      \hline
    \end{tabular}
    \caption {\small \label{tab:brwscalar}
               Leptoquark isospin families in the Buchm\"uller-R\"uckl-Wyler
               model.
               For each leptoquark, the superscript corresponds to its
               electric charge, while the subscript denotes its weak
               isospin.
               $\beta_e$ denotes the branching ratio of the
               LQ into $e + q$. }

\end{center}
\end{table*}

We use the nomenclature of~\cite{Schrempp:1991vp} to label the different
LQ states.
In addition to the underlying hypotheses of BRW, we restrict
LQs couplings to only one chirality state of the lepton,
given that deviations from lepton universality in helicity suppressed
pseudoscalar
meson decays have not been observed~\cite{Davidson:1993qk,Leurer:1993em}.

In the BRW model, LQs decay exclusively into $eq$ and/or $\nu q$
and the branching ratio $\beta_e = BR (LQ \rightarrow eq)$ is fixed
by gauge invariance to 0.5 or 1 depending on the
LQ type.

%

\subsection{Phenomenology of leptoquarks in $\bold{pp}$ collisions}

\paragraph{Pair production}

In $pp$ collisions leptoquarks would be mainly pair-produced via $gg$
or $qq$ interactions.
As long as the coupling
$\lambda$ is not too strong (e.g. $\lambda \sim 0.3$ or below, 
corresponding to a strength similar to or lower than that of the electromagnetic
coupling, $\sqrt{ 4 \pi \alpha_{em}}$), the production cross section is essentially 
independent of $\lambda$. At the LHC, LQ masses up to about $1.2$ (scalar LQs) and $1.5 \TeV$
(vector LQs)
will be probed~\cite{Belyaev:2005ew}, independently of the coupling $\lambda$. 
However, the determination of the quantum
numbers of a first generation LQ in the pair-production mode is not possible (e.g. for the fermion
number) or ambiguous and model-dependent (e.g. for the spin). Single LQ production is
much better suited for such studies.

\paragraph{Single production}

\begin{figure}[htb]
\begin{center}
\begin{tabular}{cc}
\includegraphics[height=3.5cm]{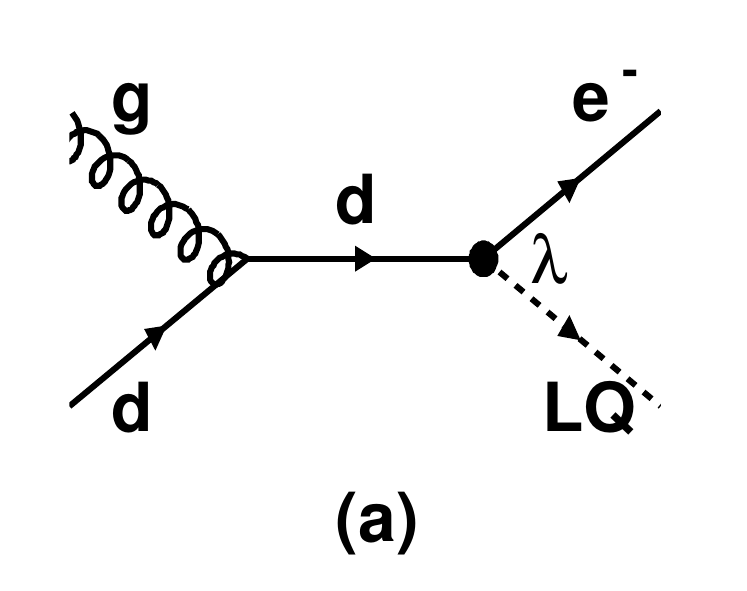} &
\includegraphics[height=3.5cm]{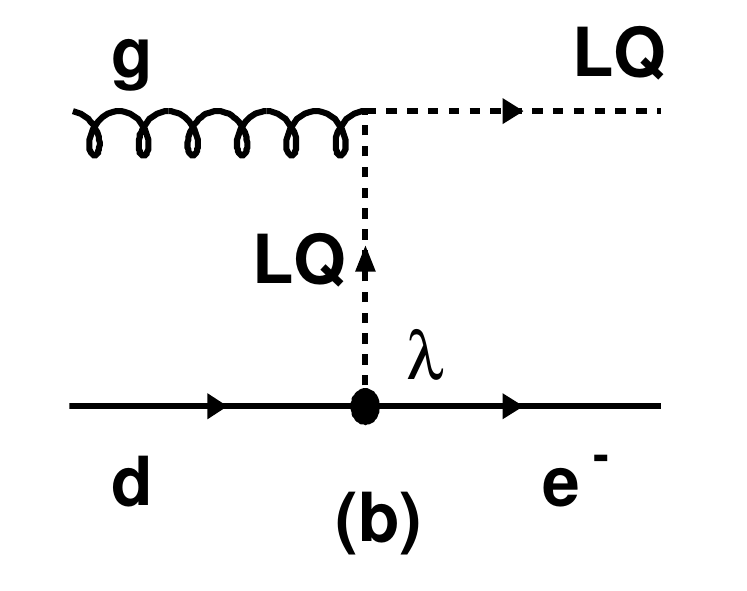}
\end{tabular}
\begin{tabular}{ccc}
\includegraphics[height=3.5cm]{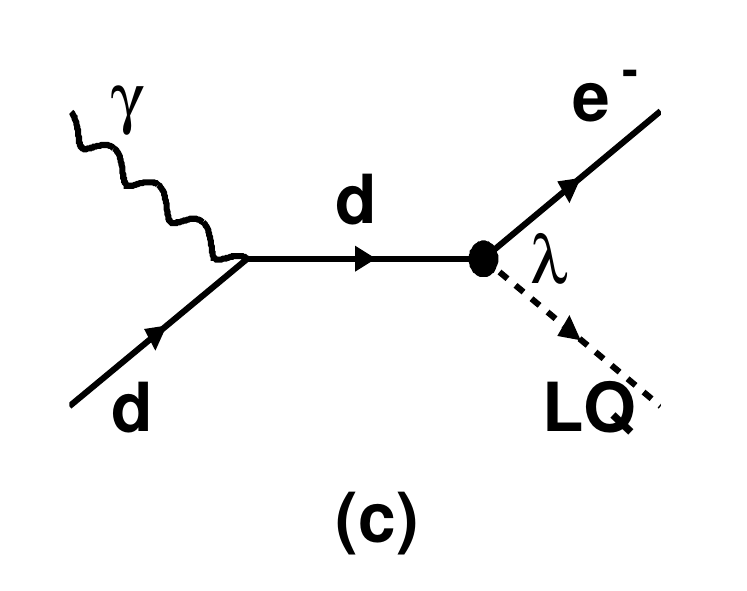} &
\includegraphics[height=3.5cm]{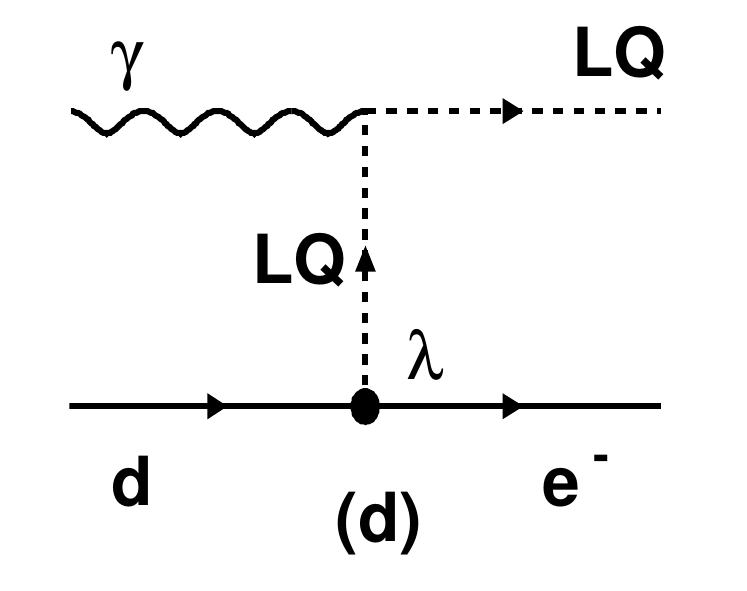} &
\includegraphics[height=3.5cm]{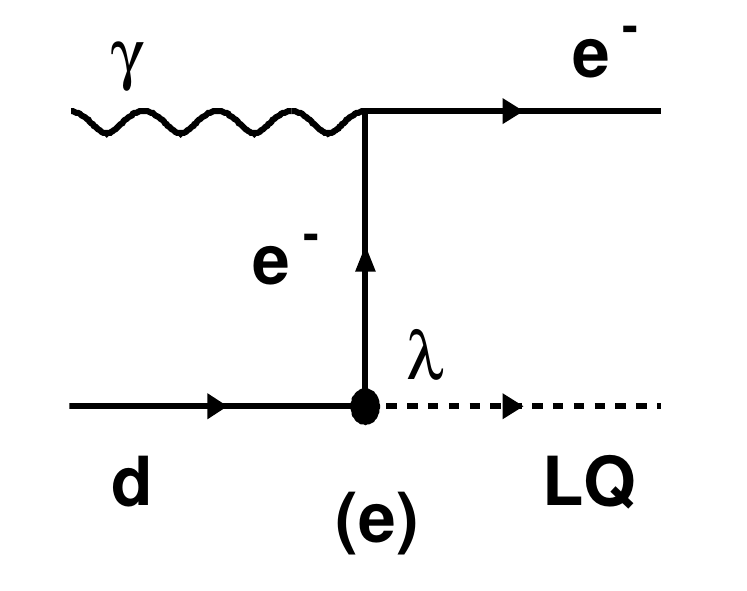}
\end{tabular}
  \caption{\label{fig:LQsingle_pp_diags}
    Diagrams for single LQ production in $pp$ collisions, shown for the example case
    of the $\tilde{S}_{1/2}^{L}$ scalar leptoquark. The production may occur via
    $qg$ interactions (a and b), or via $q \gamma$ interactions (c, d and e). In the
    latter case, the photon can be emitted by the proton (elastic regime) or by a quark
    coming from the proton (inelastic regime). }
\end{center}
\end{figure}

\begin{figure}[htb]
\begin{center}
  \includegraphics[width=0.49\textwidth]{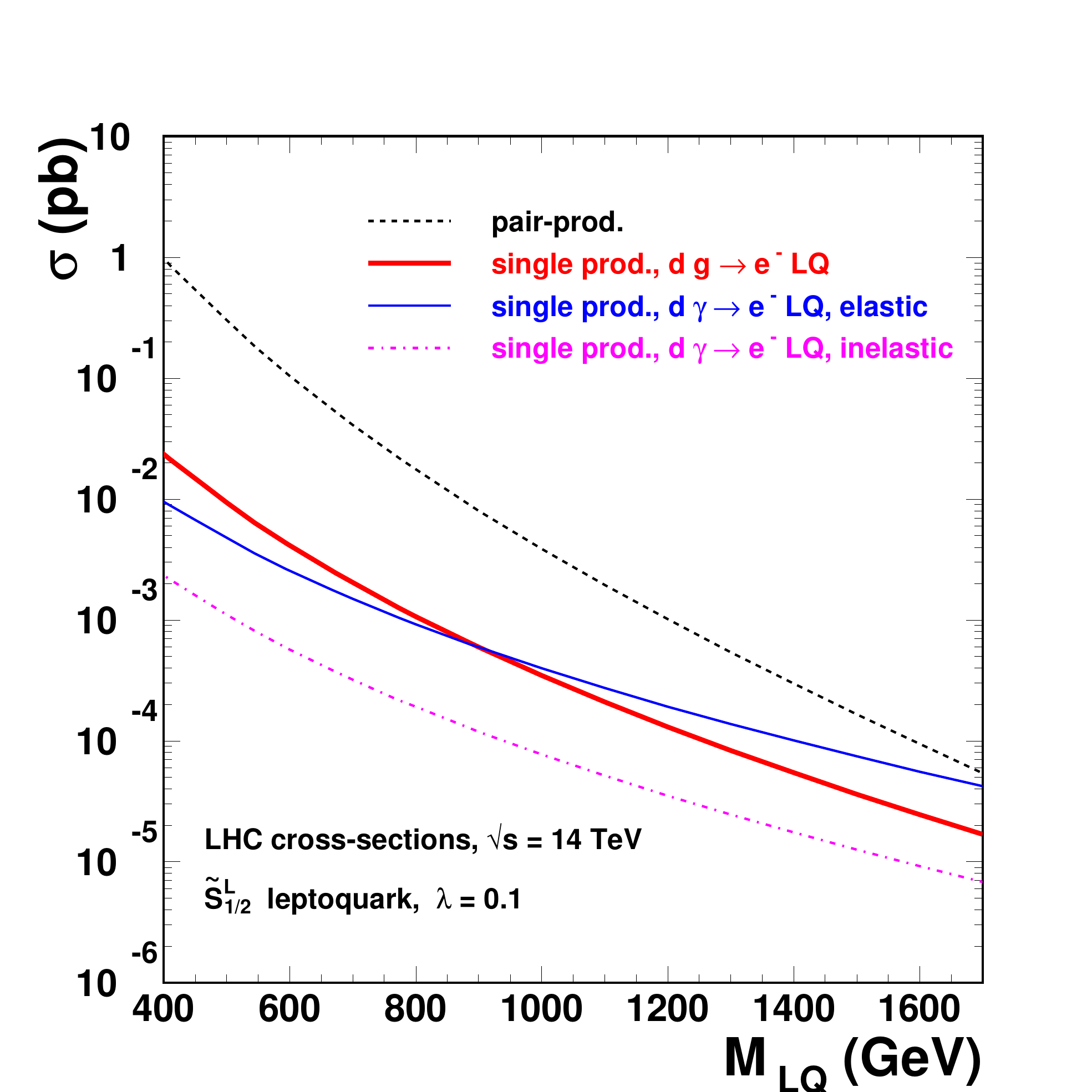}
  \includegraphics[width=0.49\textwidth]{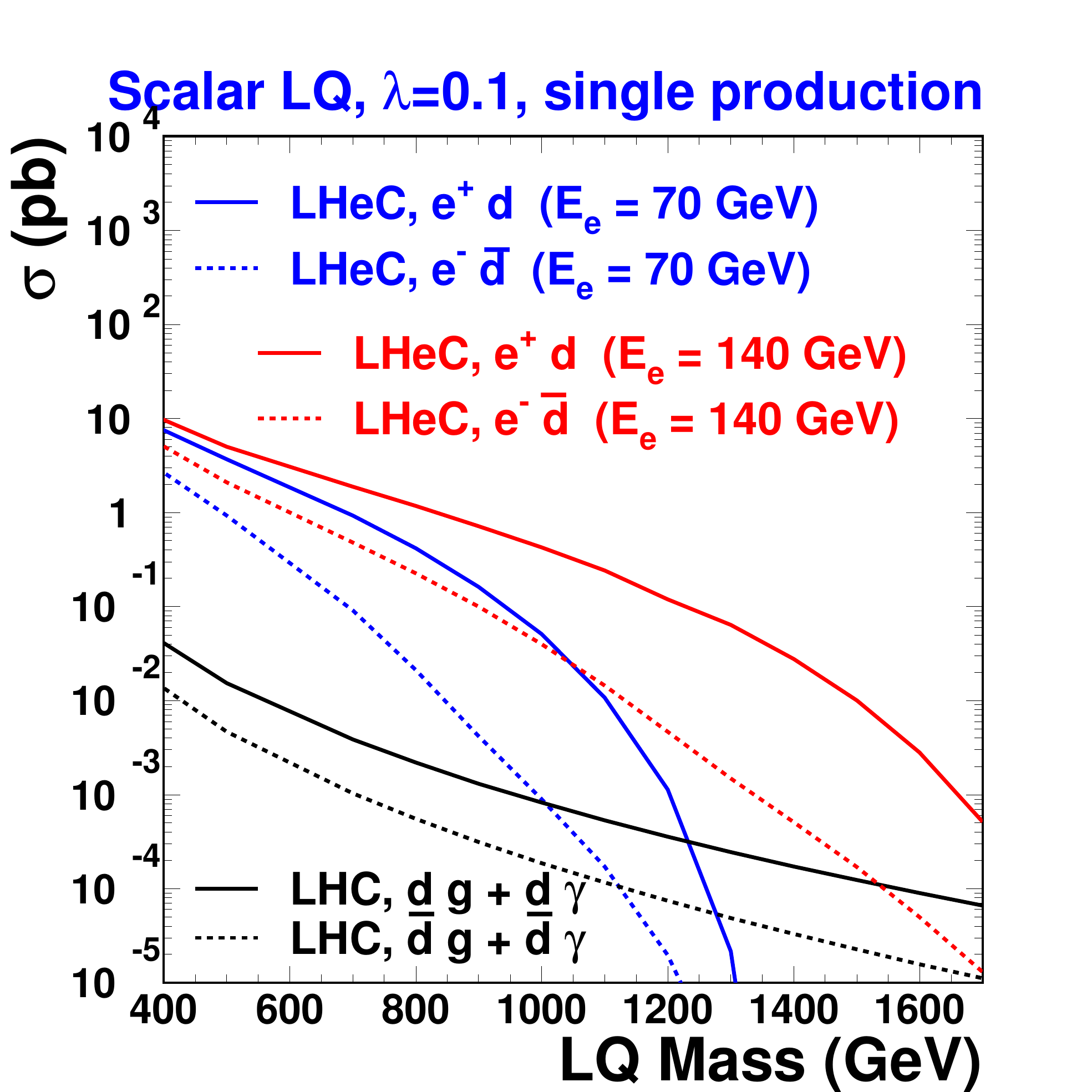}
  \caption{\label{fig:lq_pp_singlexsec}
     left: Single LQ production cross section at the LHC.
     right: comparison of the cross section for single LQ production,
     at LHC and at LHeC.
   }
\end{center}
\end{figure}

Single LQ production at the LHC is also possible. So far, only the production
mode $g q \rightarrow e + LQ$ (see example diagrams in Fig.~\ref{fig:LQsingle_pp_diags}a and b)
has been considered in the literature (see e.g.~\cite{Belyaev:2005ew}).
In the context of this study, the additional production mode
$\gamma q \rightarrow e + LQ$ has been considered as well (see example diagrams in
Fig.~\ref{fig:LQsingle_pp_diags}c, d and e). This cross section has been calculated by taking into
account:
\begin{itemize}
\item the inelastic regime, where the photon virtuality $q^2$ is large enough
  and the proton breaks up in a hadronic system with a mass well above the proton
  mass. In that case, the photon is emitted by a parton in the proton, and
  the process $q q' \rightarrow q + e + LQ$ is calculated.
\item the elastic regime, in which the proton emitting the photon remains intact.
 This calculation involves the elastic form factors of the proton.
\end{itemize}
Similarly to resonant LQ production in $ep$ collisions, the cross section of single $LQ$ 
production in $pp$ collisions approximately scales with the square of the coupling,
$\sigma \propto \lambda^2$.
Figure~\ref{fig:lq_pp_singlexsec} (left) shows the cross section for single $LQ$ production
at the LHC as a function of the LQ mass, assuming a coupling $\lambda = 0.1$.
While the inelastic part of the $\gamma q$ cross section can be neglected, the elastic
production (which often yields an associated electron in the forward direction)
plays an important role at high masses; its cross section is larger than
that of LQ production via $g q$ interactions for masses above $\sim 1 \TeV$.
However, the cross section for single LQ
production at LHC is much lower than that at LHeC, in $e^+ p$ or $e^- p$ collisions, as shown
in Fig.\ref{fig:lq_pp_singlexsec} (right).

\paragraph{LQ exchange in the t-channel}

In $pp$ collisions, the $t$-channel exchange of first generation LQs would lead to di-electron events,
$q \bar{q} \rightarrow e^+ e^-$. The squared amplitude of that process is proportional to
$\lambda^4$ and its interference with the standard Drell-Yan production scales as $\lambda^2$.
Hence, its effect is sizeable only for large values of the coupling $\lambda$. It can
be used to explore part of the very high mass domain, beyond the discovery reach offered by pair-production.

\subsection{Contact term approach}

For LQ masses far above the kinematic limit, the contraction
of the propagator in the $eq \rightarrow eq$ and $qq \rightarrow ee$ amplitudes
leads to
a four-fermion interaction. 
%
%
Such interactions are studied in the context of general contact terms,
which can be used to parameterise any new physics process with a
characteristic energy scale far above the kinematic limit.
%
%

In $ep$ collisions, Contact Interactions would interfere with
NC DIS processes and lead to a distortion of the $Q^2$ spectrum
of NC DIS candidate events.
The results presented in Section~\ref{sec:NewPhysicsAtHighQ2}
can be re-interpreted into expected sensitivities on high mass leptoquarks.

\begin{figure}[htb]
\begin{center}
  \includegraphics[width=0.6\textwidth]{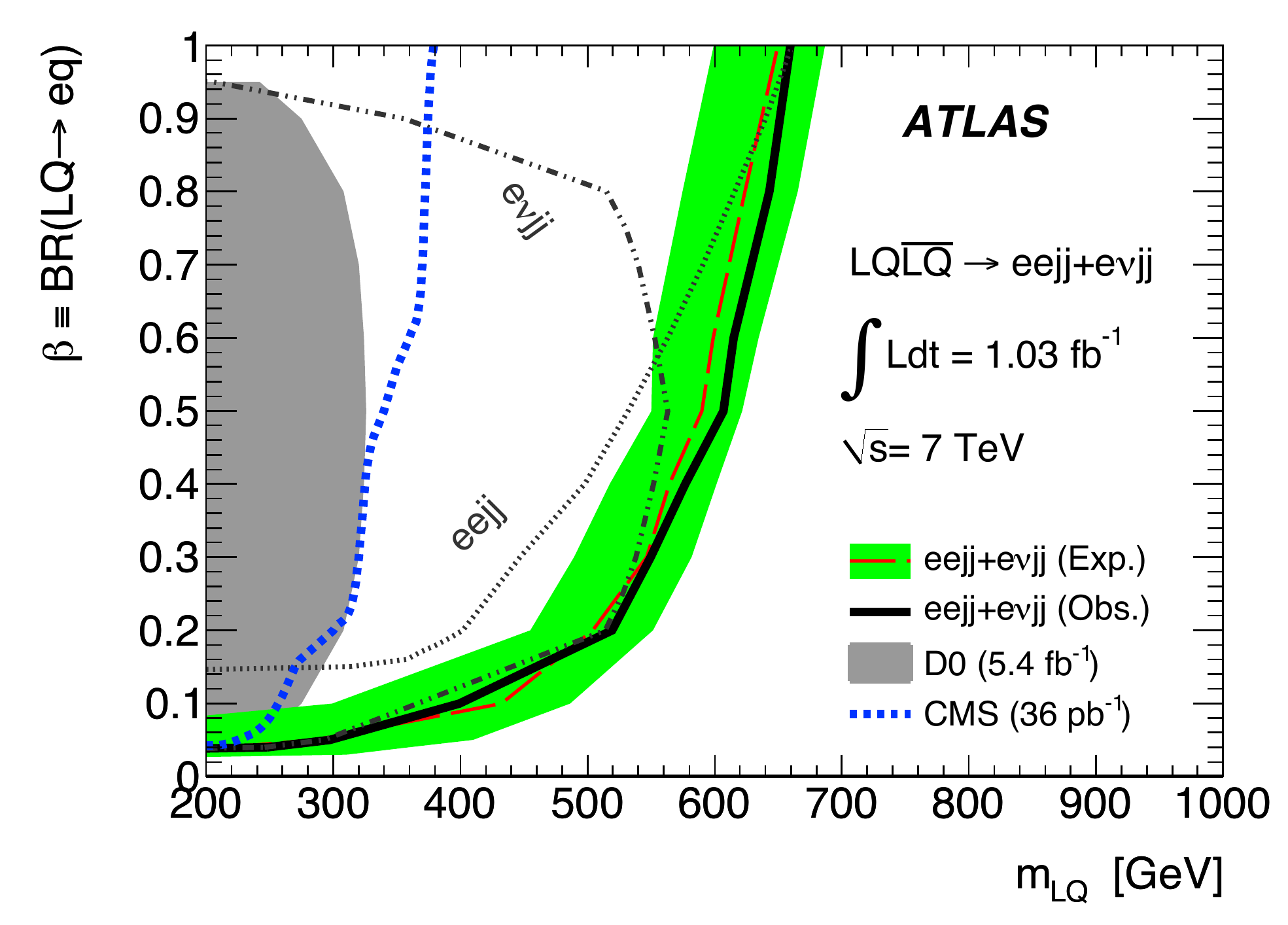}
  \caption{
     \it Constraints on first generation scalar leptoquarks obtained by the ATLAS experiment
         with $1$~fb$^{-1}$ of data taken at $\sqrt{s} = 7$~TeV. From~\cite{Aad:2011ch}.}
    \label{fig:CMS_LQs}
\end{center}
\end{figure}

\subsection{Current status of leptoquark searches}

The H1 and ZEUS experiments at the HERA $ep$ collider have constrained the coupling
$\lambda$ to be smaller than the electromagnetic coupling 
($\lambda < \sqrt{4 \pi \alpha_{em}} \sim 0.3$) for first generation
LQs lighter than $300$~GeV.
The D0 and CDF experiments at the Tevatron $pp$ collider set constraints on
first-generation LQs that are independent of the coupling $\lambda$, by looking
for pair-produced LQs that decay into $eq$ ($\nu q$) with a branching ratio
$\beta$ ($1 - \beta$). For a branching fraction $\beta = 1$, masses below $299$~GeV
are excluded by the D0 experiment~\cite{:2009gf}.
The CMS and ATLAS experiments have recently set tighter constraints~\cite{Aad:2011ch, Chatrchyan:2011ar}.
The most recent published result is illustrated in Fig.~\ref{fig:CMS_LQs}. With $\sim 1$~fb$^{-1}$ of data 
taken in 2011 at $\sqrt{s} = 7$~TeV, the ATLAS experiment rules out scalar LQ masses below $660$~GeV ($607$~GeV) for
$\beta = 1$ ($\beta = 0.5$).


\subsection{Sensitivity on leptoquarks at LHC and at LHeC}


Leptoquark searches at the LHC will greatly benefit from the increased centre of mass energy,
which was already raised to $8$~TeV for the $2012$ data taking.
Assuming that $2 \times 25$~fb$^{-1}$ of data can be collected by the end of 2012, combining
the results from ATLAS and CMS should allow scalar LQ masses up to nearly $900$~GeV to be probed.
A similar sensitivity would be obtained, per experiment, with $10$~fb$^{-1}$ of data at $14$~TeV.
With $100$~fb$^{-1}$ the mass domain below $1$~TeV should be fully covered, and with $300$~fb$^{-1}$
the sensitivity could reach about $1.1$ to $1.2$~TeV.

\begin{figure}[htb]
\begin{center}
  \includegraphics[width=0.6\textwidth]{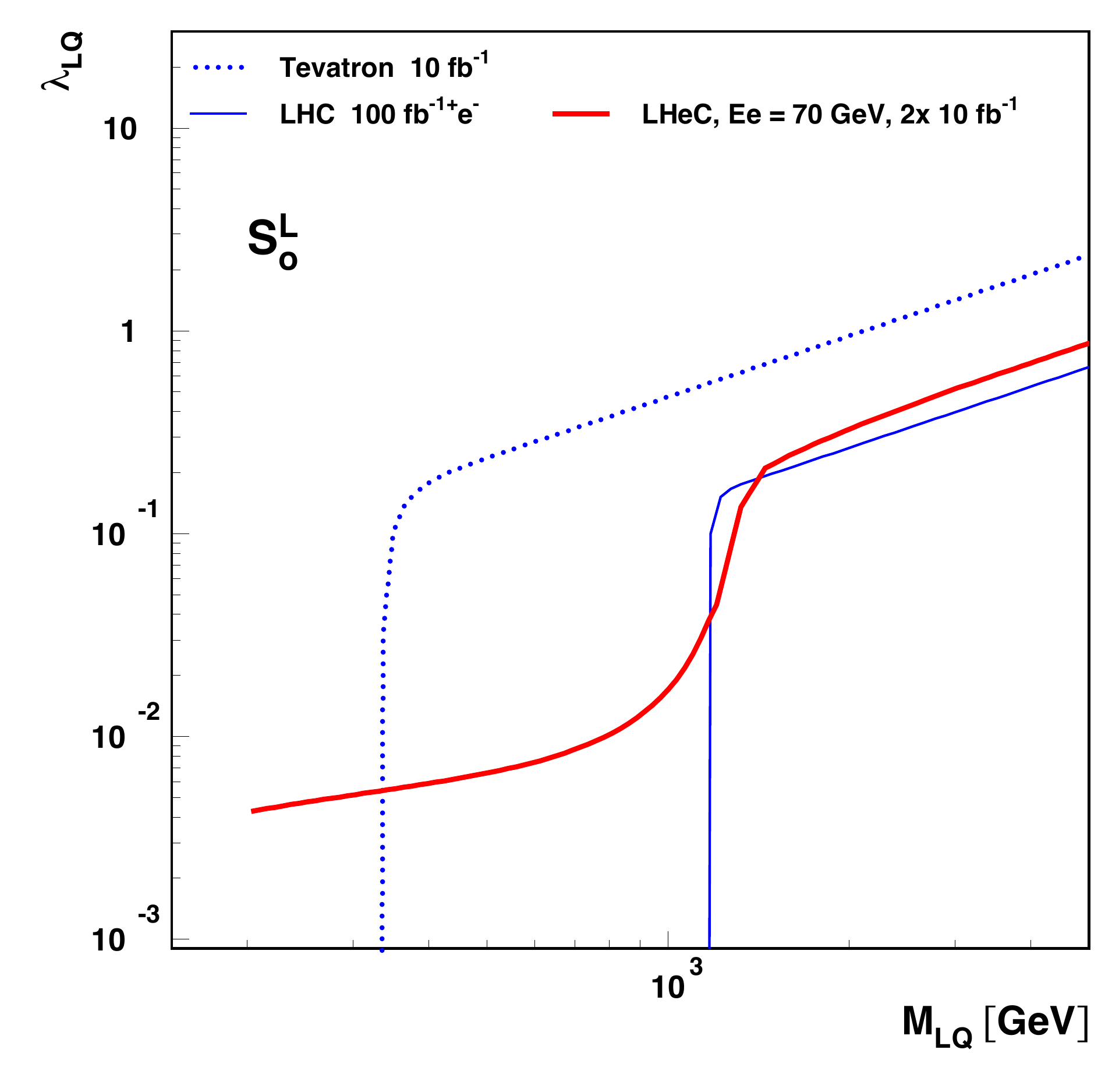}
  \caption{
    \it  Mass-dependent upper bounds on the LQ coupling
     $\lambda$ as expected at LHeC for a luminosity of $10 \fb^{-1}$
     per lepton charge
     (full red curve) and at the LHC for $100 \fb^{-1}$ (full blue curve).
     These are shown for an example scalar LQ coupling to $e^- u$.
     The LHC curve shows the sensitivity expected from LQ pair-production, that is
     insensitive to the value of the coupling $\lambda$; beyond that limit, the curve
     shows the sensitivity expected from $t$-channel exchange.
     } \label{fig:lq_lambda}
\end{center}
\end{figure}

Figure~\ref{fig:lq_lambda} shows the expected sensitivity \cite{Zarnecki:2008cp} of the LHC and LHeC colliders
for scalar leptoquark production. 
For a coupling $\lambda$ of ${\cal{O}}(0.1)$, LQ masses up to about $1 \TeV$ could
be probed at the LHeC.
In $pp$ interactions at the LHC, such leptoquarks would be
mainly pair-produced. Beyond the mass domain that can be probed via pair-production, independently of
the coupling $\lambda$, the LHC curve in Fig.~\ref{fig:lq_lambda}
shows the sensitivity expected from $t$-channel exchange.


\subsection{Determination of LQ properties}

In $ep$ collisions LQ production can be probed in detail, taking advantage of the formation and
decay of systems which can be observed directly as a combination of jet and lepton invariant mass in
the final state. It will thereby be possible at the LHeC to probe directly
and  with high precision  the
perhaps complex structures which will result in the lepton-jet system
and to determine the quantum numbers of new states.
Examples of the sensitivity of high energy $ep$ collisions to the properties
of LQ production follow. In particular, a quantitative comparison of the
potential of LHC and LHeC to measure the fermion number of a LQ, and
the flavour of the quark it couples to, is given. \\

\begin{figure}[htb]
\begin{center}
\includegraphics[height=4cm]{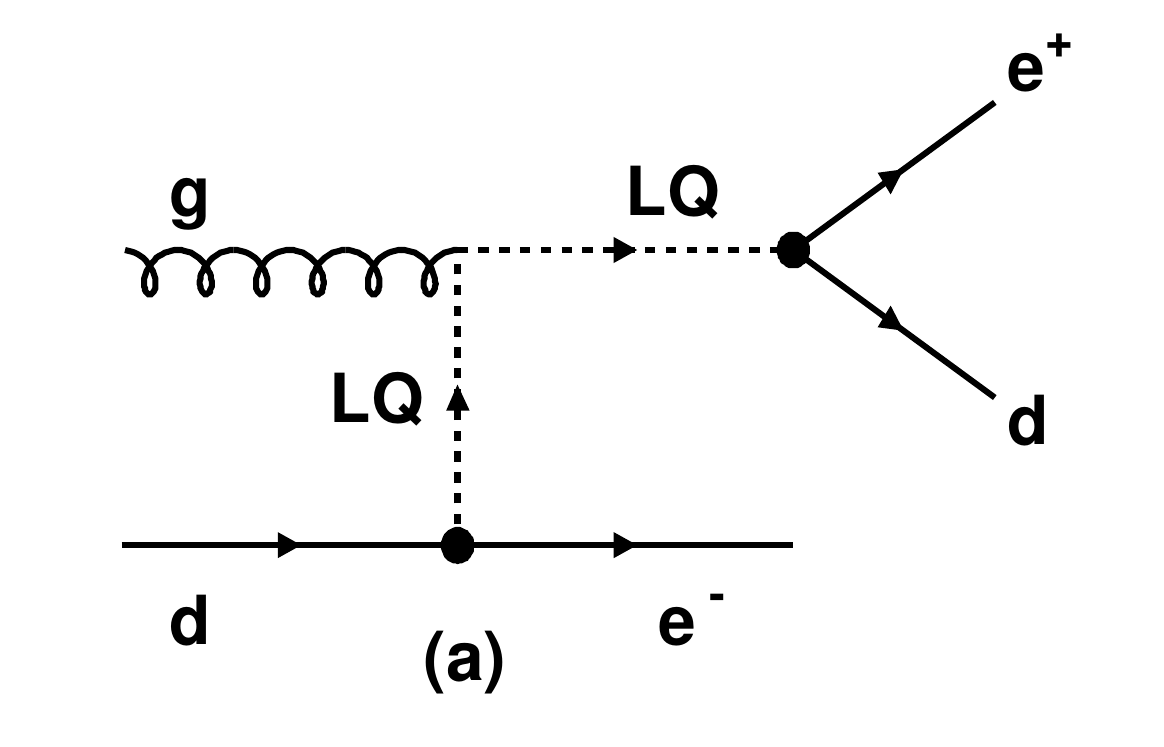}
\includegraphics[height=4cm]{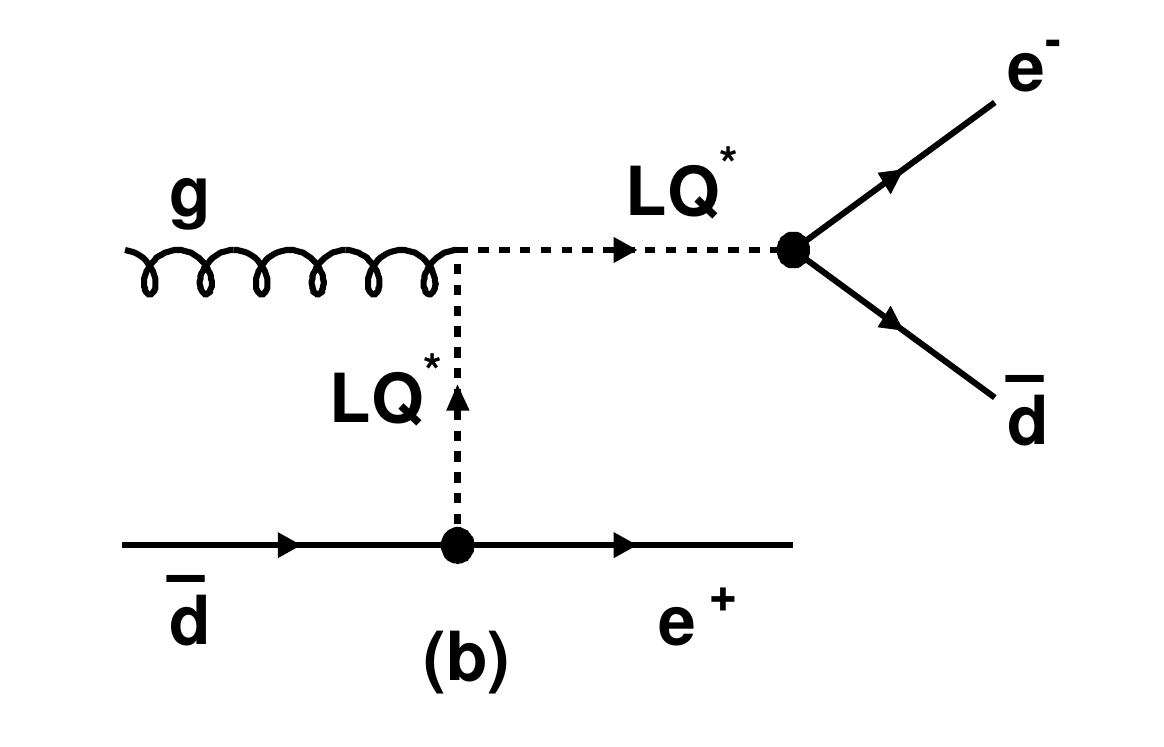}
  \caption{\label{fig:lq_lhc_asym_method}
  \it Single production of a $F=0$ leptoquark decaying (a) into a positron and a jet and (b) into
  an electron and a jet. In (a) (resp. (b)), the jet comes from a quark (an antiquark); conservation of the
  baryon number implies that the parton involved in the initial state is a quark (an antiquark).
  }
\end{center}
\end{figure}

\begin{figure}[htb]
\begin{center}
\includegraphics[width=0.49\textwidth]{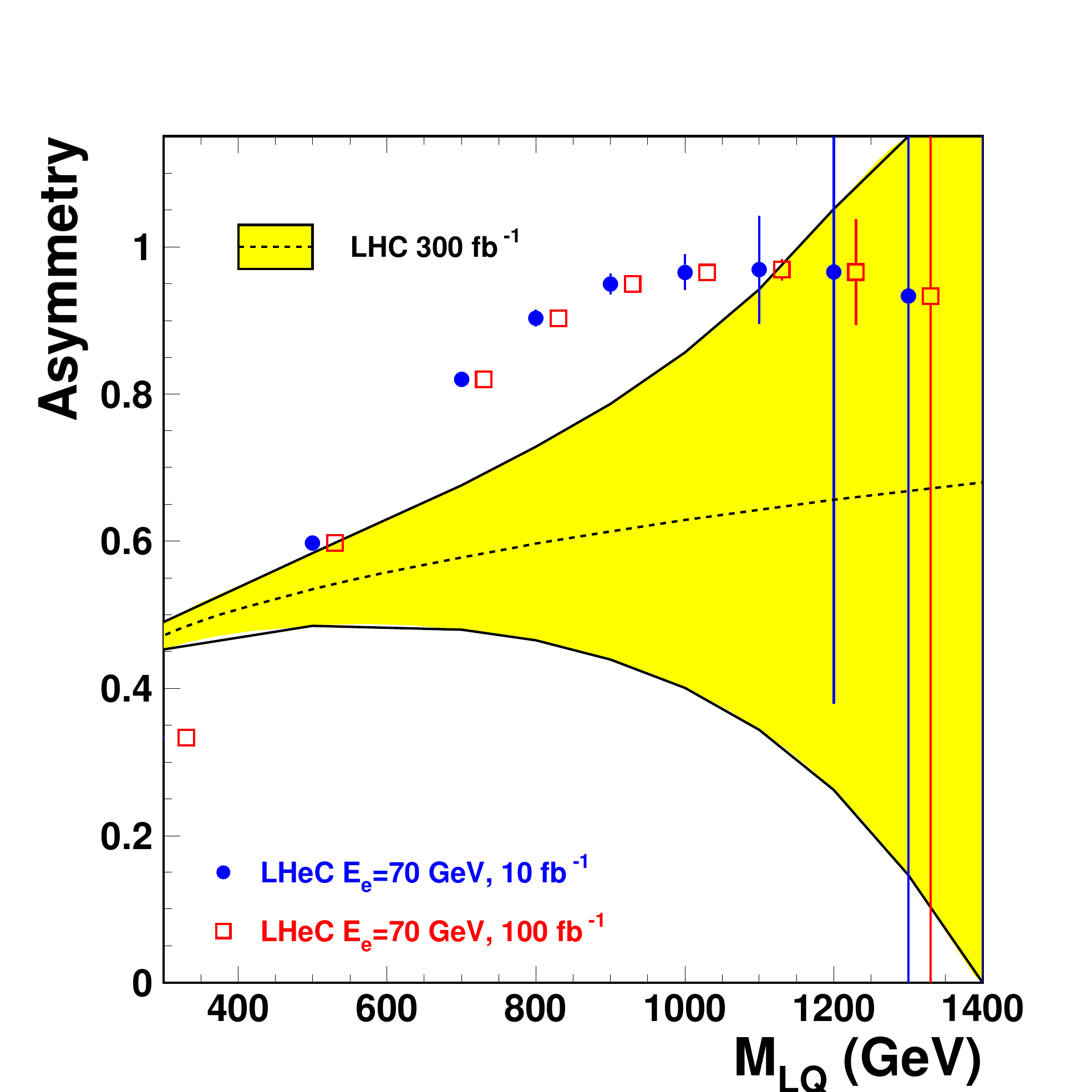}
\includegraphics[width=0.49\textwidth]{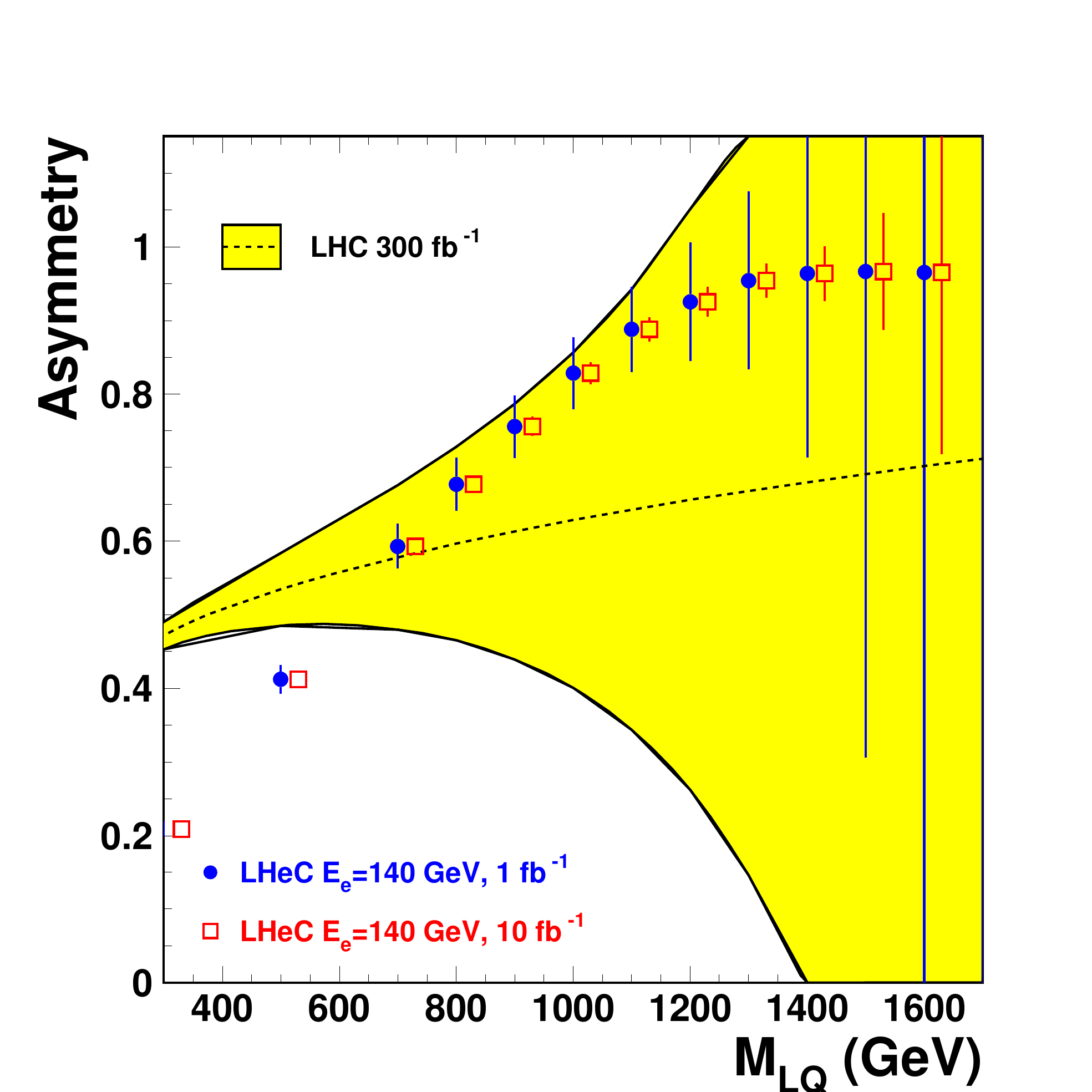}
\end{center}
  \caption{\label{fig:lq_single_asym} \it
     Asymmetries which would determine the fermion number $F$ of
     a LQ and the flavour of the quark the LQ couples to.
     The sign of the asymmetry is the relevant quantity to determine $F$.
     The dashed curve shows the asymmetry that could be measured at the LHC; the yellow
     band shows the statistical uncertainty of this quantity, assuming an
     integrated luminosity of $300 \fb^{-1}$.
     The red and blue symbols, together with their error bars, show the asymmetry
     that would be measured at LHeC, assuming $E_e = 70 \GeV$ (left) or
     $E_e = 140 \GeV$ (right). Two values of the integrated luminosity have been
     assumed.
     These determinations correspond to the $\tilde{S}_{1/2}^{L}$ (scalar
      LQ coupling to $e^+ + d$), with a coupling of $\lambda=0.1$.
     }
\end{figure}

\begin{figure}[htb]
\begin{center}
\includegraphics[height=8.3cm]{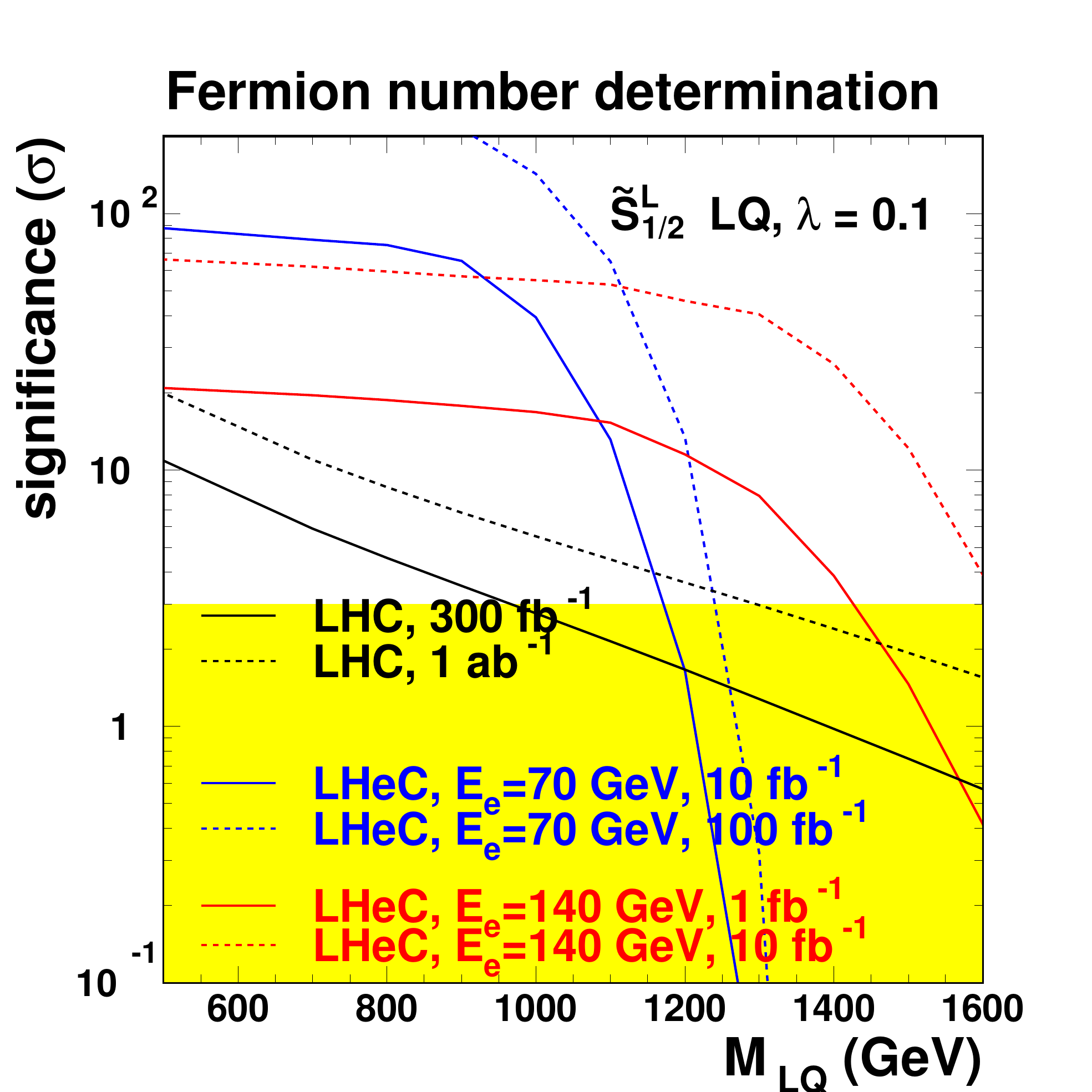}
  \caption{\label{fig:lq_single_asym_significance}
    Significance of the determination of the fermion number of a LQ, at the
    LHC (black curve) and at the LHeC (blue and red curves). This corresponds
    to a $\tilde{S}_{1/2}^{L}$ leptoquark, assuming a coupling
    of $\lambda=0.1$. 
    }
\end{center}
\end{figure}

\paragraph{Fermion number ($F$)}
 Since the parton densities for $u$ and $d$ at high $x$ are much larger than
 those for $\bar{u}$ and $\bar{d}$, the production cross section at LHeC of an
 $F=0$ ($F=2$) LQ is much larger in $e^+ p$ ($e^- p$) than in
 $e^- p$ ($e^+ p$) collisions. A measurement of the asymmetry between the $e^+p$
 and $e^- p$ LQ cross sections,
$$ {\cal{A}}_{ep} = \frac {\sigma_{prod}(e^+ p ) - \sigma_{prod}(e^- p) } { \sigma_{prod}(e^+ p ) + \sigma_{prod}(e^- p) } $$
 thus determines, via its sign, the fermion number
 of the produced leptoquark. 
 Pair production of first generation LQs at the LHC will not allow
 this determination.  Single LQ production at the LHC, followed
 by the LQ decay into  $e^{\pm}$ and  $q$ or $\bar{q}$, could determine $F$
 by comparing the signal cross sections with an $e^+$
 and an $e^-$ coming from the resonant  state.
 Indeed, for a $F=0$ leptoquark, the signal observed when the resonance is made by a positron
and a jet corresponds to diagrams involving
a {\it{quark}} in the initial state (see Fig.\ref{fig:lq_lhc_asym_method}a).
Hence the corresponding cross section, $\sigma(e^+_{out} j)$ is larger than that 
of the signal observed when the resonance is made by an electron and a jet, 
$\sigma(e^-_{out} j)$, since a high $x$ {\it{antiquark}} is involved in that latter case (see Fig.\ref{fig:lq_lhc_asym_method}b).
In contrast, for a $F = 2$ LQ, $\sigma(e^+_{out} j)$ is smaller than $\sigma(e^-_{out} j)$.
The measurement of (the sign of) the asymmetry
$$ {\cal{A}}_{pp} = \frac { \sigma(e^+_{out} j) - \sigma(e^-_{out} j) }  { \sigma(e^+_{out} j) + \sigma(e^-_{out} j) }$$
should thus provide a determination of the LQ fermion number.
%
%
 However, the single  LQ production cross section
 at the LHC is two orders of magnitude lower than at the LHeC (Fig.\,\ref{fig:lq_pp_singlexsec}),
 so that the asymmetry ${\cal{A}}_{pp}$ measured at the LHC will suffer from statistics
 in a large part of the parameter space. 
 For a LQ coupling to $ed$ and $\lambda = 0.1$, no information on $F$ can be extracted from
 $300$~fb$^{-1}$ of LHC data for a LQ mass above $\sim 1 \TeV$, while the LHeC can determine 
 $F$ for LQ masses up to $1.5 \TeV$ (Fig.\,\ref{fig:lq_single_asym} and
Fig.~\ref{fig:lq_single_asym_significance}). 
Details of the determination of ${\cal{A}}_{pp}$ at the LHC are given in the next paragraph. \\



\par An estimate of the precision with which 
the asymmetry ${\cal{A}}_{pp}$ can be measured at the LHC was
obtained from a Monte Carlo simulation. First, using the
model~\cite{Belyaev_LQ} implemented in CalcHep~\cite{Pukhov:2004ca},
samples were generated for the processes $g~u \to e^+ e^- u $ and
$g~\bar u \to e^+ e^- \bar u $, keeping only diagrams involving the
exchange of a scalar LQ of charge 1/3, isospin 0 and fermion
number 2.  This leptoquark ($^{1/3}S_0$ in the notation of
Table~\ref{tab:brwscalar}) couples to $e_R^- u_R$. Assuming that it is chiral, 
only right-handed coupling was allowed. The $^{1/3}S_0$ leptoquark was also assumed to
couple only to the first generation. Masses of 500 GeV, 750 GeV and 1 TeV were considered. The
renormalisation and factorisation scales were set at $Q^2 = m_{LQ}^2$
and the coupling parameter $\lambda = 0.1$. A centre
of mass energy of 14 TeV was assumed at the LHC.

High statistics background samples, corresponding to 150 fb$^{-1}$
were also produced by generating the same processes $p p \to e^+ e^- + \mathrm{jet}$, including all
diagrams except those involving the exchange of leptoquarks.
Kinematic preconditions were applied at the generation level to both
signals and background: (i)$p_T({\mathrm jet}) >$ 50 GeV, (ii) $p_T(e^\pm) > 20$ GeV,
(iii) invariant mass of jet-$e^+-e^-$ system $>$ 200 GeV.
The cross sections for the signals and backgrounds under these
conditions are: 19.7 fb, 3.4 fb and
0.87 fb for LQ's of mass 500 GeV, 750 GeV and 1 TeV respectively, and
1780 fb for the background.  These events were subsequently passed to Pythia~\cite{Sjostrand:2006za}
to perform parton showering and hadronisation, then processed
through Delphes~\cite{Ovyn:2009tx} for a fast simulation of the ATLAS
detector.  Finally, considering events with two reconstructed electrons 
of opposite sign and,
assuming that the leptoquark has already been discovered (at the LHC),
the combination of the highest $p_T$ jet with
the reconstructed $e^-$ or $e^+$ with a mass closest to the known leptoquark
mass is chosen as the LQ candidate. The following cuts for $m_{LQ}=$ 500, 750
and 1000 GeV, respectively, are applied:
\begin{itemize}
 \item dilepton invariant mass $m_{ll} >$ 150, 200, 250 GeV. This cut
rejects very efficiently the $Z$+ jets background.
 \item $p_T(e_1) >$ 150, 200, 250 GeV and $p_T(e_2) >$ 75, 100, 100 GeV, 
where $e_1$ is the reconstructed $e^\pm$ with higher $p_T$ and $e_2$ the lower $p_T$ electron.
 \item $p_T(j_1) >$ 100, 250, 400 GeV, where $j_1$ is the
reconstructed jet with highest $p_T$, used for the reconstruction of
the LQ.
\end{itemize}

Table~\ref{tab:LQ_F} summarises the results of the simulation for an
integrated luminosity of 300 fb$^{-1}$.  The expected number of signal
events shown in the table is then simply the number of events due to
the leptoquark production and decay, falling in the resonance peak
within a mass window of width (60, 100, 160 GeV) for the three cases
studied, respectively.  Although this simple analysis can be improved by
considering other less dominant backgrounds and by using 
optimised selection criteria, it should give a good estimate of the
precision with which the asymmetry can be measured. This precision
falls rapidly with increasing mass and, above $\sim $ 1 TeV, it
becomes impossible to observe simultaneously single production of both
$^{1/3}S_0$ and $^{1/3}\bar{S_0}$. It must be noted that the asymmetry
at the LHC will be further diluted by the abundant leptoquark pair
production, not taken into account here.

\begin{table}[h]
  \centering
  \begin{tabular}{|c|c|c|c|c|c|}
    \hline
  LQ mass   &   \multicolumn{2}{c|}{$^{1/3}S_0 \to e^+\bar u$} &  \multicolumn{2}{c|}{$^{1/3}{\bar S_0} \to e^- u$} &  Charge Asymmetry \\ \cline{2-5}
  (GeV)     &      Signal        &      Background            &      Signal        & Background                  &                   \\ \hline
   500      &        121         &      431                   &       771          & 478                         & $ 0.73 \pm 0.05$ \\ 
   750      &        18.3        &      137                   &       132          & 102                         & $ 0.76^{+0.16}_{-0.14}$ \\
  1000      &        4.9         &      57                    &        44          &  42                         & $ 0.77^{+0.23}_{-0.24}$  \\ \hline
  \end{tabular}
\caption{Estimated number of events of signal and background, and the charge asymmetry measurement with 300 fb$^{-1}$ at
the LHC, for $\lambda=0.1$.}
\label{tab:LQ_F}
\end{table}

\begin{figure}[htbp]
\centering \includegraphics[width=0.5\textwidth]{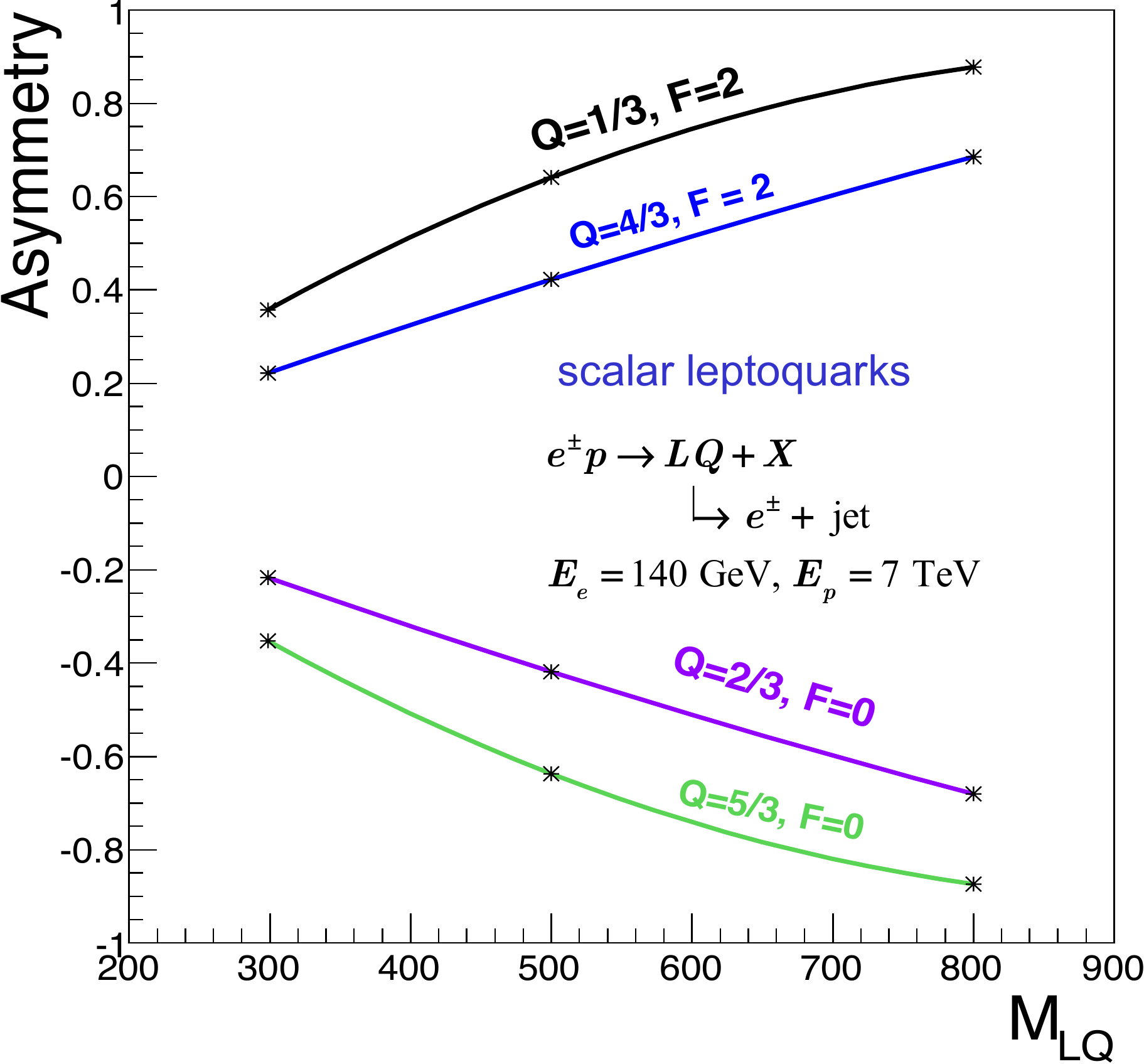}\put(-65,-10){{LQ Mass (GeV)}}
\caption{ Charge asymmetry ${\cal{A}}_{ep}$ for different types of
  scalar LQs as a function of the LQ mass.}
\label{fig:ChargeAsymmetry}
\end{figure}

\begin{figure}[htbp]
\centering \includegraphics[width=0.7\textwidth]{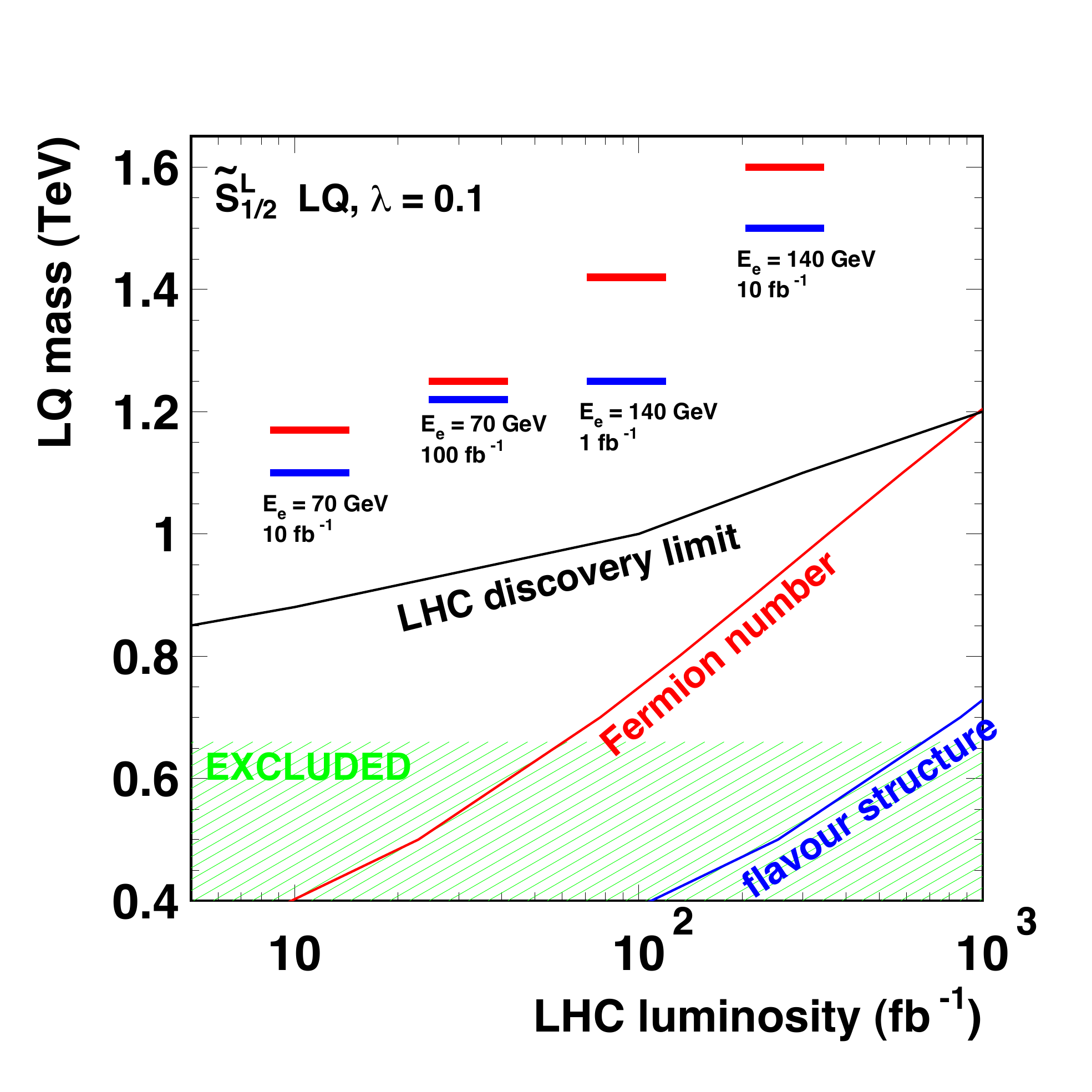}
\caption{ The mass domain over which the LHC could discover a leptoquark (upper black curve), determine
its fermion number (middle red curve), and determine the flavour of the quark it couples to (lower blue curve),
as a function of the integrated luminosity. The mass ranges where the LHeC could determine the
LQ fermion number (the quark flavour it couples to)
are shown in the top part of the figure, as the horizontal upper red (lower blue) lines, for two values of 
the lepton beam energy and two values of the integrated luminosity.
The hatched area corresponds to the mass domain that is already ruled out by the LHC experiments.
 }
\label{fig:lqsummary}
\end{figure}

\paragraph{Flavour structure of the LQ coupling} More generally, the same charge asymmetry observables
are sensitive to the flavour of the quark the LQ couples to, through the dependence
on the parton distribution functions of the interacting quark in the proton. For example, Fig.~\ref{fig:ChargeAsymmetry}
shows the calculated asymmetry ${\cal{A}}_{ep}$ that could be measured at LHeC, for scalar LQs. Provided that the coupling $\lambda$ is not too small,
the accuracy of the measurement of ${\cal{A}}_{ep}$ at LHeC (see Fig.~\ref{fig:lq_single_asym}) would
allow the various LQ types to be disentangled, as different LQs lead to values of ${\cal{A}}_{ep}$ 
that differ by typically $20-30 \%$.
A similar measurement at the LHC would be possible only in a very limited part of the
phase space (low masses and large couplings), where the statistics would be large enough to
yield an accuracy of less than $\sim 10 \%$ on the measured asymmetry ${\cal{A}}_{pp}$.
This is illustrated in Fig.~\ref{fig:lqsummary} which shows, as a function of the integrated luminosity, the
mass range where the LHC experiments could discover a leptoquark, determine its fermion number,
and determine the flavour of the quark it couples to. This is shown for an example LQ type,
the scalar $\tilde{S}_{1/2}^L$, and an example value of the coupling, $\lambda = 0.1$.
The mass range where the LHeC could make the same measurements is also depicted, for
four LHeC configurations.
The LHeC
would be able to determine these properties over
the full mass range where the LHC could discover a leptoquark,
even in a configuration where $E_e \sim 70$~GeV provided that the integrated luminosity is large enough.
On the other hand, the LHC will not deliver any information on the flavour structure of a 
leptoquark, unless it is discovered with a mass very close to the current limit and the
experiments collect a very large amount of luminosity.

 \paragraph{Spin} At the LHeC, the angular distribution of the
 LQ decay products is unambiguously related to its spin.
Indeed, scalar LQs produced in the $s$-channel decay isotropically in their rest
frame leading to a flat ${\rm d} \sigma\,/\,{\rm d}y \;$ spectrum where
$y= \frac{1}{2}\left(1+\cos{\theta^*}\right)$
is the
Bjorken scattering variable in DIS and $\theta^*$ is the decay polar
angle of the lepton relative to the incident proton in the LQ centre of
mass frame.
In contrast, events resulting from the production and decay of
vector LQs would be distributed according to
${\rm d} \sigma\,/\, {\rm d} y \propto \, (1-y)^2$.
These $y$ spectra
from scalar or vector LQ production are markedly
different from the ${\rm d} \sigma\,/\, {\rm d} y \propto \,y^{-2}$
distribution expected at fixed $M$ for the dominant $t$-channel
photon exchange in neutral current DIS
events\footnote{At high momentum transfer, $Z^0$ exchange is no longer
          negligible and contributes to less pronounced differences
          in the $y$ spectra between LQ signal and DIS background.}.
Hence, a LQ signal in the NC-like channel will be statistically most prominent
at high $y$. 

 The spin determination will be much more complicated,
 even possibly ambiguous, if only the LHC
 leptoquark pair production data are available.  
 Angular distributions for vector LQs depend 
 strongly on the structure of the $g \, LQ \, {\overline{LQ}}$ coupling, 
 i.e. on possible anomalous couplings.
 For a structure similar to that of the $\gamma W W$ vertex, vector LQs produced via
 $q \bar{q}$ fusion are unpolarised and, because both LQs are produced with the
 same helicity, the distribution of the LQ production angle will be similar to that
 of a scalar LQ. The study of LQ spin via single LQ production at the LHC
 will suffer from the relatively low rates and more complicated backgrounds.

\paragraph{Neutrino decay modes}
 At the LHeC, there is similar sensitivity 
 for LQ decay into both $eq$ and $\nu q$.  At the LHC, in $pp$ collisions,
 LQ decay into neutrino-quark final states is plagued by huge QCD
background. At the LHeC, production through $eq$ fusion with subsequent
$ \nu q$ decay is thus very important if the complete pattern of LQ decay couplings
is to be determined. 

\paragraph{Coupling $\lambda$} 
The intrinsic width of a leptoquark, which depends on the coupling $\lambda$ and on the LQ mass,
is expected to be small. For example, for a scalar LQ of $1$~TeV and $\lambda = 0.1$, the width is
below $0.2$~GeV, smaller than the experimental mass resolution. Hence, the coupling $\lambda$ cannot 
be extracted from a measurement of the intrinsic width of the leptoquark.

 However, the production cross section of a LQ in $ep$
 collisions can be written, in the narrow-width approximation, as :
  $$ \sigma_{prod}  = \frac{\lambda^2}{ 16 \pi} q(x = M^2 / s_{ep})  \qquad ( J=0 ) \qquad {\mbox{ or}} \qquad \sigma_{prod}  = \frac {\lambda^2}{ 8 \pi} q(x = M^2 / s_{ep})  \qquad ( J=1 ) $$
depending on its spin $J$.
Hence, at LHeC, the determination of:
\begin{itemize}
 \item the $LQ$ spin, via the analysis of the angular distribution of its decay products;
 \item the flavour of the quark $q$ involved in the $e-q-LQ$ vertex, via the charge
  asymmetry described above;
 \item the production cross section, via the cross sections measured in the $eq$ and
    $\nu q$ decay modes
\end{itemize}
allows the value of the coupling $\lambda$ to be determined, from the above formula.

\paragraph{Chiral structure of the LQ coupling}  Chirality is central to the SM Lagrangian.
 Polarised electron and positron  beams
 at the LHeC will shed light on the chiral structure of the 
 LQ-e-q couplings. The asymmetry between the production cross sections
 measured in $e^-_L p $ and $e^-_R p$ collisions would determine whether a $ |F| = 2$ leptoquark
 couples to $e^-_L$ or to $e^-_R$. 
 For a LQ of $F=0$, a polarised positron beam would be needed to make this determination
over the full mass range (some information could also be obtained from polarised
electrons, but in a smaller mass-coupling range).
Measurements of a similar nature at LHC are impossible. \\

\par
In summary, if a first generation leptoquark were to exist in the TeV mass range with a
coupling $\lambda$ of ${\cal{O}}(0.1)$, the LHeC would allow a rich program of
``spectroscopy" to be carried out, resulting in the determination of most of
the LQ properties. 



\subsection{Leptoquarks as R-parity violating squarks}

As already mentioned, squarks in R-parity violating supersymmetric 
models\footnote{The potential of LHeC to observe supersymmetric particles in models where the
$R$-parity is conserved has been studied as well. However, the leading process
of squark-selectron pair production would have a sizeable cross section only when the sum of the
masses of the produced sparticles is below $\sim 1$~TeV. The constraints
on squarks already set by the LHC experiments using the data taken in $2011$ largely rule out this possibility.
} could be an
example of ``leptoquark" scalar bosons. While the LHC experiments already constrain the
squark masses to be above $\sim 1$~TeV in the case of five or four degenerate squarks,
the limits are much weaker on a stop or a sbottom that would be much lighter than the other squarks,
this possibility being well motivated theoretically.
Should the light stop or sbottom possess sizeable R-parity violating couplings to first generation
leptons, the constraints shown in Fig.~\ref{fig:CMS_LQs} would apply, as well as the general
discussion presented above.
In addition, the $R$-parity conserving decay modes of this squark, if not negligible, could be
studied as well at LHeC. The relatively clean environment may allow, for example, mass measurements
to be performed with an interesting precision. This possibility has not been investigated yet.

\subsection{Leptogluons}

While leptoquarks and excited fermions are widely discussed in the literature,
leptogluons have not received the same attention. However, they are predicted in all models with coloured 
preons \cite{Harari:1979gi,Fritzsch:1981zh,Greenberg:1980ri,Barbieri:1981cy,Baur:1985ud,Celikel:1998dj}. 
For example, in the framework of fermion-scalar
models, leptons would be bound states of a fermionic preon and
a scalar anti-preon $l=(F\bar{S})=1\oplus8$ (both F and S are colour
triplets), and each SM lepton would have its own colour octet partner~\cite{Celikel:1998dj}.

A study of leptogluons production at LHeC is presented in~\cite{LHEC_Note_leptogluons_Sultansoy}.
It is based on the following Lagrangian:
\begin{equation}
  L=\frac{1}{2\Lambda}\underset{l}{\sum}\left\{ \bar{l}_{8}^{\alpha}g_{s}G_{\mu\nu}^{\alpha}\sigma^{\mu\nu}(\eta_{L}l_{L}+\eta_{R}l_{R})+h.c.\right\} 
\end{equation}
where $G_{\mu\nu}^{\alpha}$ is the field strength tensor for gluon,
index $\alpha=1,2,...,8$ denotes the colour, $g_{s}$ is gauge coupling,
$\eta_{L}$ and $\eta_{R}$ are the chirality factors, $l_{L}$ and
$l_{R}$ denote left and right spinor components of lepton, $\sigma^{\mu\nu}$
is the anti-symmetric tensor and $\Lambda$ is the compositeness scale.
The leptonic chiral invariance implies $\eta{}_{L}$$\eta_{R}=0$.

\begin{figure}[htb]
\begin{center}
\includegraphics[width=0.49\textwidth]{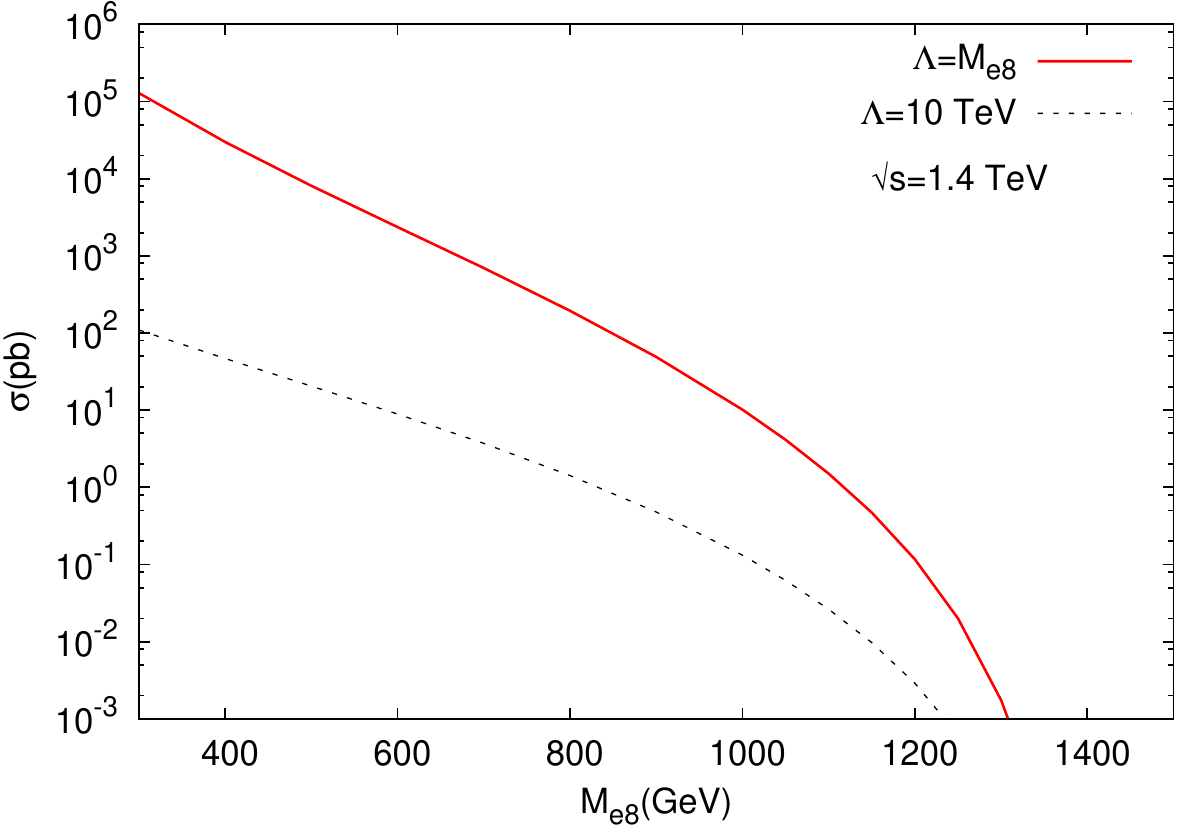}
\includegraphics[width=0.49\textwidth]{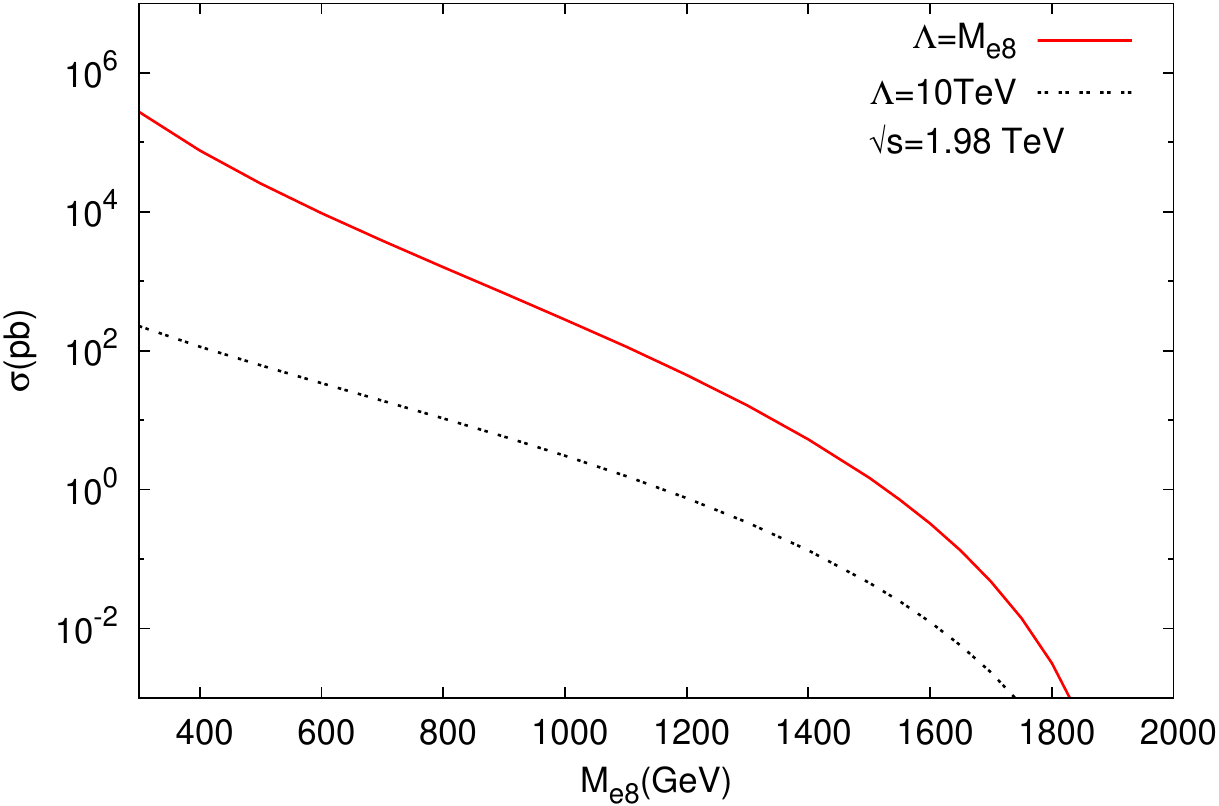}
\end{center}
\caption{Resonant $e_{8}$ production at the LHeC, for two values of the centre-of-mass energy.}
\label{fig:leptogluon_xsec}
\end{figure}

The phenomenology of leptogluons at LHC and LHeC is very similar to that of leptoquarks,
despite their different spin (leptogluons are fermions while leptoquarks are bosons) and
their different interactions. Figure~\ref{fig:leptogluon_xsec} shows typical cross sections
for single leptogluon production at the LHeC, assuming $\Lambda$ is equal to the leptogluon mass.
It is estimated that, for example, a sensitivity up to a compositeness scale
of 200 TeV, at $3\sigma$ level can be achieved with LHeC having $E_e = 70$ GeV and with 1 fb$^{-1}$.
The mass reach for $M_{e8}$ is 1.1 TeV for $\Lambda = 10$ TeV.

Similarly to leptoquarks, should leptogluons be discovered at the LHC, LHeC data would be of
the highest value for the determination of the properties of this new particle.

%% file: physics/bsm_ExcitedFermions.tex
\section{Excited leptons and other new heavy leptons}

The three-family structure and mass hierarchy of the known fermions is
one of the most puzzling characteristics of the Standard Model (SM) of
particle physics. Attractive explanations are provided by models
assuming composite quarks and leptons~\cite{Harari:1982xy}.  The
existence of excited states of fermions ($F^{*}$) is a natural
consequence of compositeness models.  More generally, various models predict the existence of fundamental new heavy leptons,
which can have similar experimental characteristics as excited
leptons. They could, for example, be part of a fourth Standard model family. 
They arise also in Grand Unified Theories, and appear as colourless fermions in technicolor models. 

New heavy leptons could be pair-produced at the LHC for masses up to
${\cal{O}}(300)$~GeV. As for the case of leptoquarks, $pp$ data from pair-production 
of new leptons may not allow for a detailed study
of their properties and couplings. Single production
of new leptons is also possible at the LHC, but is expected to have a larger cross section 
at LHeC, via $e \gamma$ or $e W$ interactions. The case of excited electrons
is considered in the following, with more details being given in~\cite{LHEC_Note_estar_Sauvan} together
with a similar study of the production of excited neutrinos.
The production of new leptons from a fourth generation ($l_4, \nu_4$) via magnetic interactions mixing
the first and fourth generation is very similar and was studied in~\cite{LHEC_Note_fourth_family_Ciftci}.


Single production of excited leptons at the LHC ($\sqrt{s}$ up to
$14$~TeV) may happen via the reactions
$pp{\rightarrow}e^{\pm}e^{*}{\rightarrow}e^+e^-V$ and
$pp{\rightarrow}\nu{e^{*}}+\nu^{*}{e^{\pm}}{\rightarrow}e^{\pm}\nu{V}$. 
The LHC should be able 
to tighten considerably the current constraints on these possible new states~\cite{Eboli:2001hi}. 

Recent results of searches for excited leptons~\cite{Aaron:2008cy, :2008xe, Aaron:2009iz}
at HERA using all data collected by the H1 detector have demonstrated that $ep$ colliders are very 
competitive to $pp$ or $e^+ e^-$ colliders. 
Indeed limits set by HERA extend at high mass beyond the kinematic reach of LEP 
searches~\cite{Abbiendi:2002wf, Abdallah:2004rc} and to higher compositeness scales
than those obtained at the Tevatron~\cite{:2008hw} using $1$~fb$^{-1}$ of data.
Therefore a future LHeC machine, with a centre of mass energy of $1-2$~TeV,
much higher than at the HERA $ep$ collider, 
should provide a good environment to search for and study excited leptons.

\subsection{Excited fermion models}

Compositeness models attempt to explain the hierarchy of masses in the SM by the existence 
of a substructure within the fermions. Several of these 
models~\cite{Hagiwara:1985wt, Boudjema:1992em, Baur:1989kv} predict excited 
states of the known fermions, in which excited fermions  are assumed to have 
spin $1/2$ and isospin $1/2$ in order to limit the number of parameters of the 
phenomenological study. They are expected to be grouped into both left- 
and right-handed weak isodoublets with vector couplings. The existence of 
the right-handed doublets is required to protect the ordinary light fermions 
from radiatively acquiring a large anomalous magnetic moment via $F^{*}FV$ 
interaction (where V is a $\gamma, Z$ or $W$).

Interactions between excited and ordinary fermions may be mediated by gauge bosons, as described by the effective Lagrangian:
\begin{equation}
\label{eq:lagrangianGM}
{\cal L}_{GM} = \frac{1}{2\Lambda} \bar{F^*_R}\,\sigma^{\mu\nu} \left[ g\,f\frac{\vec{\tau}}{2}\,\vec{W_{\mu\nu}}+g'\,f'\,\frac{Y}{2}\,B_{\mu\nu}+g_s\,f_s\,\frac{\vec{\lambda}}{2}\,\vec{G_{\mu\nu}} \right] F_L~+~h.c.,
\end{equation}
where $Y$ is the weak hypercharge, $g_s,~g=\frac{e}{\sin{\theta_W}}$ and $g'=\frac{e}{\cos{\theta_W}}$ are the strong and electroweak gauge couplings, where $e$ is the electric charge and $\theta_W$ is the weak mixing angle; $\vec{\lambda}$ and $\vec{\tau}$ are the Gell-Mann matrices and the Pauli matrices, respectively. $G_{\mu\nu},~W_{\mu\nu}$ and $B_{\mu\nu}$ are the field strength tensors describing the gluon, the $SU(2)$, and the $U(1)$ gauge fields;
$f_s,~f$ and $f'$ are factors multiplying the coupling constants associated to each gauge field. 
They depend on the composite dynamics. The parameter $\Lambda$ has units of energy and can be regarded 
as the compositeness scale which reflects the range of the new confinement force.

In addition to gauge mediated (GM) interactions, a new interaction could take place at the scale of the
binding energy of the constituents of quarks and leptons. It would result in  new interaction terms
between excited fermions and ordinary fermions, that can be described by an effective four-fermion
Lagrangian~\cite{Baur:1989kv}:\begin{equation}
\label{eq:CILagrangian}
{\cal L}_{CI} = \frac{ g^2_*}{2\Lambda^2}j^{\mu}j_{\mu}\,,
\end{equation}
where $g_*$ denotes the strength of the new interaction and $j_{\mu}$ is the fermion current
\begin{equation}
j_{\mu} = \eta_L\bar{F}_L\gamma_{\mu}F_L + \eta'_{L}\bar{F^*}_L\gamma_{\mu}F^*_L + \eta"_L\bar{F^*}_L\gamma_{\mu}F_L + h.c. + (L{\rightarrow}R).\end{equation}
In the following, we set $g_* = 0$ and only consider the ``gauge" terms of Eq.\ref{eq:lagrangianGM}.
Thus, the results presented below are independent of the strength of the new interaction, 
and can be generically applied to the production of any new lepton
coupling to an electron-photon pair via the standard electromagnetic interaction.

\subsection{Simulation and results}

In the following study, excited electron ($e^{*}$) production and decays via GM interactions are considered.
The $e^{*}$ production cross section under the assumption $f=-f'$ becomes much smaller than for $f=+f'$ and therefore only the case $f=+f'$ is studied.

\begin{figure}[htbp]
\begin{center}
\includegraphics[width=0.6\textwidth]
{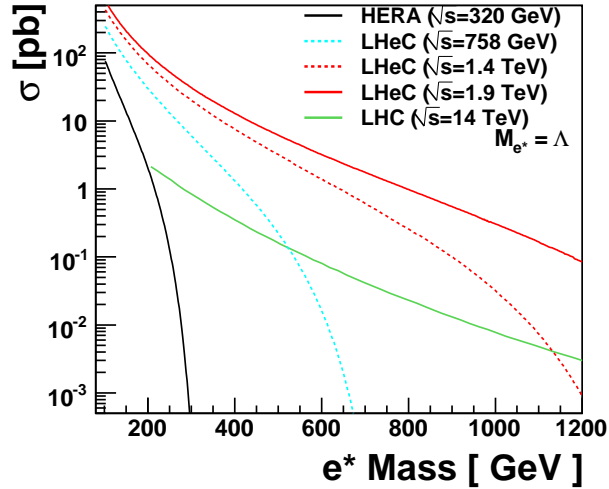}
\caption{The $e^{*}$ production cross section via gauge mediated interactions,
for different design scenarios of the LHeC electron-proton collider, compared to the cross sections at HERA and at the LHC. The cross sections shown correspond to the choice $f = f' = 1$. }
\label{fig:Xsection_GM}
\end{center}
\end{figure}

Excited electrons could be produced in $ep$ collisions at the LHeC via a $t$-channel $\gamma$ or $Z$ boson exchange. The Monte Carlo (MC) event generator COMPOS~\cite{Kohler:1991yu} 
is used for the calculation of the $e^{*}$ production cross section and 
the simulation of signal events. 
%
The resulting cross sections for several LHeC configurations, assuming $f=+f' = 1$ and $M_{e^{*}}=\Lambda$, 
are shown in Fig.~\ref{fig:Xsection_GM}, together with
the corresponding production cross section at HERA and at the LHC~\cite{Eboli:2001hi}.
In the mass range accessible by the LHeC, the $e^{*}$ production cross section via GM interactions is clearly much  higher than at the LHC. 

\begin{figure}[htbp]
\begin{center}
\includegraphics[width=0.6\textwidth]{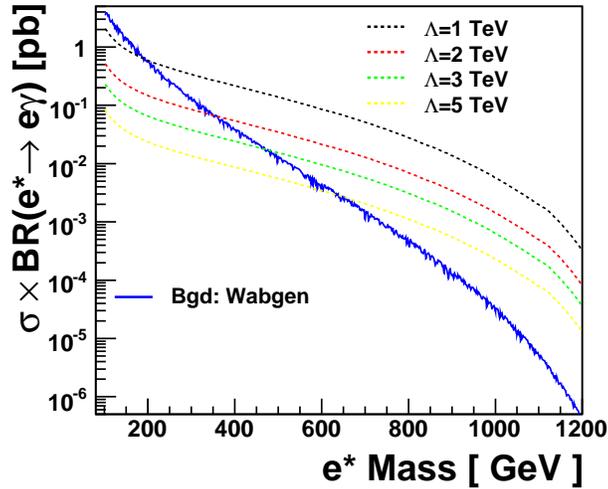}
\caption{Electromagnetic production cross section for $e^{*}$ ($e^{*}\rightarrow e\gamma$) for different values of $\Lambda$,
  together with the cross section from background processes.}
\label{fig:Brackground}
\end{center}
\end{figure}

In order to estimate the sensitivity of excited electron searches at the LHeC, the $e^{*}$ production followed by its decay in the channel $e^{*}{\rightarrow}e\gamma$ is considered.
This is the key channel for excited electron searches in $ep$ collisions as it provides a very clear signature and has a large branching ratio.
The main sources of backgrounds from SM processes are considered here, namely neutral currents (NC DIS) and QED-Compton ($e\gamma$) events. Other possible SM backgrounds are negligible.
The MC event generator WABGEN~\cite{Berger:1998kp} is used to generate these background events.
Figure~\ref{fig:Brackground} compares the  $e^{*}$ production cross section to the total cross section of SM backgrounds. Background events dominate in the low $e^{*}$ mass region.
Hence to enhance the signal, candidate events are selected with two isolated electromagnetic clusters with a polar angle between $5^{\circ}$ and $145^{\circ}$ and transverse energies greater than $15$~GeV and $10$~GeV, respectively.

To translate the results into exclusion limits, expected upper limits on the coupling $f/\Lambda$ are derived at $95\%$ Confidence Level (CL) as a function of excited electron masses.

\begin{figure}[htbp]
\begin{center}
\includegraphics[width=0.6\textwidth]{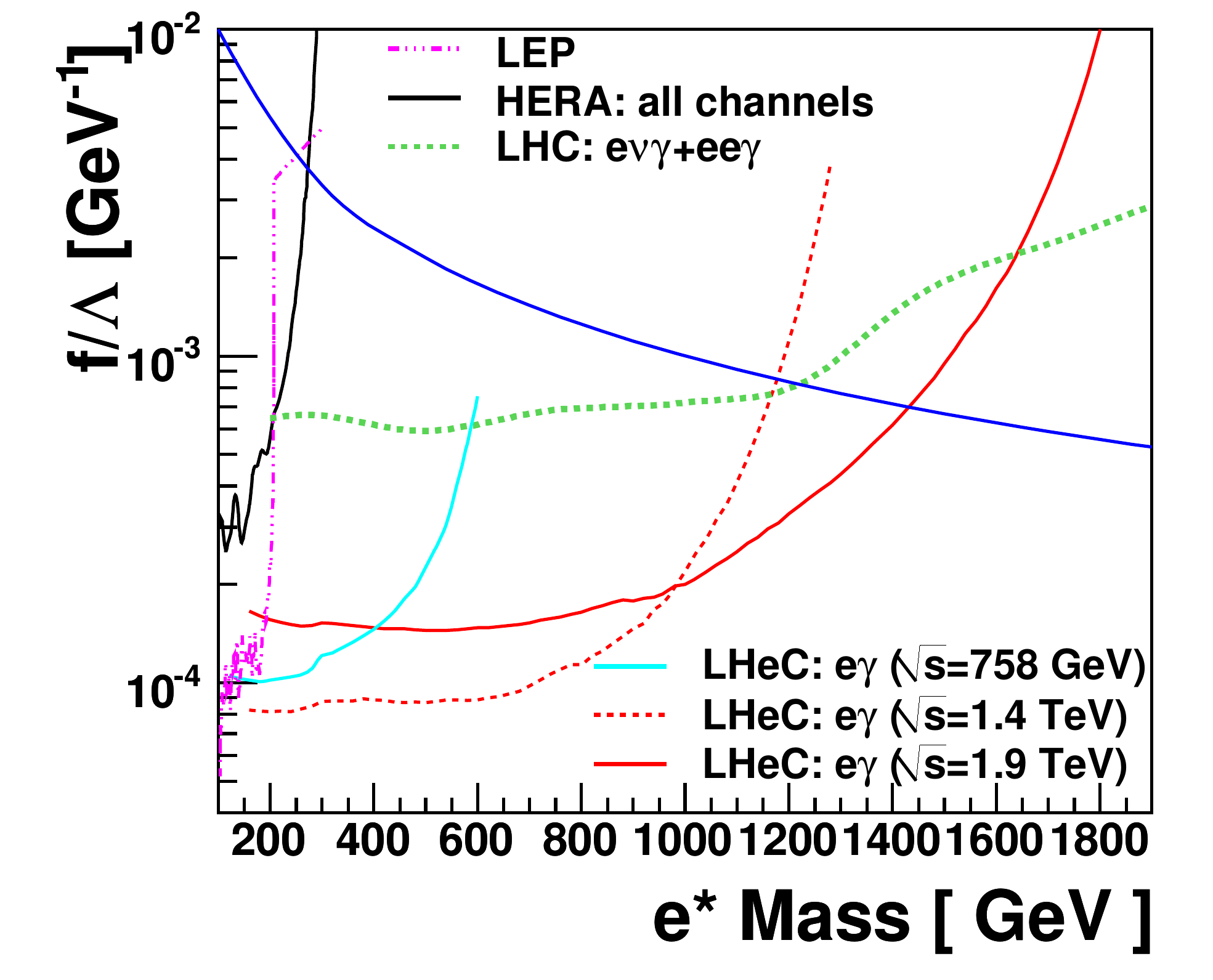}\put(-160,150){{$\color{blue}{f/\Lambda=1/M_{e^{*}}}$}}
\caption{Sensitivity to excited electron searches for different design scenarios of the LHeC electron-proton collider, compared to the expected sensitivity of the LHC ($\sqrt{s}=14$~TeV, $L=100$~fb$^{-1}$). Different integrated luminosities at the LHeC  ($L=10$~fb$^{-1}$ for $\sqrt{s}$ up to $1.4$~TeV and $L=1$~fb$^{-1}$ for $\sqrt{s}$ up to $2$~TeV) are assumed. The curves present the expected exclusion limits on the coupling $f/\Lambda$ at $95\%$ CL as a function of the mass of the excited electron with the assumption $f=+f'$. Areas above the curves are excluded. Present experimental limits obtained at LEP and HERA are also represented.}
\label{fig:Limit_GM}
\end{center}
\end{figure}

The attainable limits at the LHeC on the ratio $f/\Lambda$ are shown in figure~\ref{fig:Limit_GM} for excited electrons, for the hypothesis $f=+f'$ and different integrated luminosities $L=10$~fb$^{-1}$ for $\sqrt{s}$ up to $1.4$~TeV and $L=1$~fb$^{-1}$ for $\sqrt{s}$ up to $2$~TeV.
They are compared to the upper limits obtained at 
LEP~\cite{Abbiendi:2002wf, Abdallah:2004rc}, HERA~\cite{Aaron:2008cy} and also to the 
expected sensitivity of the LHC~\cite{Eboli:2001hi}.
Considering the assumption $f/\Lambda=1/M_{e^{*}}$ and $f=+f'$, excited electrons with masses up to 1.2(1.5)~TeV, corresponding to centre of mass energies of $\sqrt{s}=1.4(1.9)$~TeV of the LHeC, are excluded.
Under the same assumptions, LHC ($\sqrt{s}=14$~TeV) could exclude $e^{*}$ masses up to $~1.2$~TeV 
for an integrated luminosity of $100$~fb$^{-1}$. In the accessible mass range of LHeC, the LHeC would be able to probe smaller values of
the coupling $f / \Lambda$ than the LHC.
Similarly to leptoquarks (see Section\,\ref{sec:leptoquarks}), if an excited electron is observed at the LHC with a mass
of ${\cal{O}}(1 \TeV)$, the LHeC would be better suited to study the properties of this particle, thanks to the larger
single production cross section (see Fig.~\ref{fig:Xsection_GM}).

%

The ATLAS and CMS experiments have carried out a search for excited leptons using data taken in $2011$ at
$\sqrt{s} = 7$~TeV~\cite{ATLAS:2012af, Chatrchyan:2011pg}.
These analyses assume that the production of excited electrons via $qq e e^*$ contact interactions
is dominant, by setting $g^2_* = 4 \pi$ in Eq.~\ref{eq:CILagrangian}. This is markedly different from the conservative
hypothesis made here, $g_* = 0$, where the production of excited electrons is dominated by gauge interactions such
that the results are independent of the strength of the new interaction $g_*$.
Under the assumption that $g^2_* = 4 \pi$, the ATLAS experiment rules out $e^*$ masses below
$1.87$~TeV for $f = f' = 1$ and $\Lambda = M_{e^*}$. 
This rules out, in this specific model and for the couplings assumed, the observability of an excited
electron at the LHeC.
Lighter $e^*$ with lower couplings, for which the LHC has no sensitivity yet, may be observed
and studied at the LHeC.

%% file: physics/bsm_newquarks_gammaq.tex
\section{New physics in boson-quark interactions}

Several extensions of the Standard Model predict new phenomena that would
be directly observable in boson-quark interactions. For example, the top quark
may have anomalous couplings to gauge bosons, leading to Flavour Changing Neutral
Current (FCNC) vertices $tq \gamma$, where $q$ is a light quark. Similarly,
excited quarks ($q^*$) or quarks from a fourth generation ($Q$) could be produced via
$\gamma q \rightarrow q*$ or $\gamma q \rightarrow Q$.
The transitions $\gamma q \rightarrow t, q^*, Q$ can be studied in $ep$ collisions
at the LHeC, but a much larger cross section would be achieved at a $\gamma p$ collider,
due to the much larger $\gamma p$ centre-of-mass energy.
The single production of $q^*$, $Q$ or of a top quark via anomalous couplings is also
possible at the LHC, but it involves an anomalous coupling together with an
electroweak coupling and the main background processes involve the strong interaction.
The signal to background ratio will thus be much more challenging at the LHC, and 
any constraints on anomalous couplings would therefore be obtained from the decay channels of these quarks.
The example of anomalous single top production is detailed in the following.

\subsection{An LHeC-based $\bold{\gamma p}$ collider}

The possibility to operate the LHeC as a $\gamma p$ collider is described in~\ref{gammap}.
If the electron beam is accelerated by a linac, it can be converted into a beam of
high energy real photons, by backscattering off a laser pulse. The energy of these photons
would be about $80 \%$ of the energy of the initial electrons.


\subsection{Anomalous single top production at a $\gamma$p collider}

\vspace{0.2cm}
\noindent

%
%
The top quark is expected to be most sensitive to physics
beyond the Standard Model (BSM) because it is the heaviest available
particle of the Standard Model (SM). A precise measurement
of the couplings between SM bosons and fermions provides a powerful
tool for the search of  BSM physics allowing a possible detection of 
deviations from SM predictions~\cite{AguilarSaavedra:2008zc}.
Anomalous $tqV$ ($V=g,\gamma,Z$
and $q=u,c$) couplings can be generated through dynamical mass
generation~\cite{Fritzsch:1999rd}, sensitive
to the mechanism of dynamical symmetry breaking. They have a similar chiral structure as the
mass terms, and the presence of these couplings would be interpreted
as signals of new interactions. This motivates the study of top quark
flavour changing neutral current (FCNC) couplings at present and future
colliders.

Current experimental constraints at 95\% C.L. on the anomalous top
quark couplings are~\cite{Nakamura:2010zzi}: $BR(t\rightarrow\gamma u)<0.0132$
and $BR(t\rightarrow\gamma u)<0.0059$ from HERA; $BR(t\rightarrow\gamma q)<0.041$
from LEP and $BR(t\rightarrow\gamma q)<0.032$ from CDF. The HERA
experiments have a 
much higher sensitivity to $u\gamma t$ than $c\gamma t$ due
to more favourable parton density, and provide the best constraint to date on $BR(t\rightarrow\gamma u)$.
The ZEUS experiment also considered an anomalous (vector) coupling $tuZ$, but the cross section is much suppressed
due to the $Z$ boson mass in the $t$-channel exchange, and the resulting constraints were not competitive
with those obtained at LEP or at the Tevatron.
In this section, the possibility to study anomalous couplings $t u \gamma$ at the LHeC is addressed.

The top quarks will be copiously produced at the LHC, allowing for detailed studies of their properties.
For a luminosity of $1$~fb$^{-1}$ ($100$~fb$^{-1}$)
the expected ATLAS sensitivity to the top quark FCNC decay is $BR(t\to q\gamma)\sim10^{-3}(10^{-4})$
~\cite{Aad:2009wy,:1999fr}. The production of top quarks by FCNC interactions at
hadron colliders has been studied in~\cite{Han:1996ce,Malkawi:1995dm,Tait:1996dv,Han:1998tp,
Tait:2000sh,Liu:2004bb,Liu:2005dp,Cao:2007bx,Cao:2007dk,Ferreira:2008cj,Yang:2008sb,
Han:2009zm,Cao:2008vk}, $e^{+}e^{-}$colliders
in~\cite{Fritzsch:1999rd,Obraztsov:1997if,Han:1998yr,Cao:2002si,AguilarSaavedra:2004wm} 
and lepton-hadron collider in ~\cite{Fritzsch:1999rd,Alan:2002wv,Ashimova:2006zc,Aaron:2009vv}. LHC will
give an opportunity to probe $BR(t\rightarrow ug)$ down to $5\times10^{-3}$
~\cite{Cakir:2005rf}; ILC/CLIC has the potential to probe $BR(t\rightarrow q\gamma)$
down to $10^{-5}$ ~\cite{MoortgatPick:2005cw}.

The potential of the LHeC to search for anomalous top quark interactions in $ep$ collisions 
was studied in~\cite{GerhardAtDivonne} and the sensitivity on a coupling $t u \gamma$ was shown
to be lower than what could be probed at the LHC.
In contrast, operating the LHeC as a $\gamma p$ collider offers interesting possibilities
to study anomalous top quark interactions. 
These have been investigated in~\cite{Cakir:2009rq}
and are summarised here.
The effective Lagrangian involving anomalous $t\gamma q$ $(q=u,c)$ interactions
is given by:
\begin{equation}
L=-g_{e}\sum_{q=u,c}{\displaystyle Q_{q}\frac{\kappa_{q}}{\Lambda}\bar{t}\sigma^{\mu\nu}(f_{q}+h_{q}\gamma_{5})qA_{\mu\nu}+h.c.}\label{eq:anotop}\end{equation}
 where $A_{\mu\nu}$ is the usual photon field tensor, $\sigma_{\mu\nu}=\frac{i}{2}(\gamma_{\mu}\gamma_{\nu}-\gamma_{\nu}\gamma_{\mu})$,
$Q_{q}$ is the quark charge, in general $f_{q}$ and $h_{q}$ are
complex numbers, $g_{e}$ is the electromagnetic coupling constant, $\kappa_{q}$
is a real and positive anomalous FCNC coupling constant and $\Lambda$ is the
new physics scale. The neutral current magnitudes in the Lagrangian
satisfy $|(f_{q})^{2}+(h_{q})^{2}|=1$ for each term. The anomalous
decay width can be calculated as

\begin{equation}
\Gamma(t\rightarrow q\gamma)=(\frac{\kappa_{q}}{\Lambda})^{2}\frac{2}{9}\alpha_{em}m_{t}^{3}\label{eq:2}\end{equation}

Taking $m_{t}=173$ GeV and $\alpha_{em}=0.0079$, the anomalous decay
width $\approx9$ MeV for $\kappa_{q}/\Lambda=1$ TeV$^{-1}$
while the SM decay width is about 1.5 GeV. 

\begin{figure}[htb]
\centering{\includegraphics[clip=,width=0.60\textwidth]{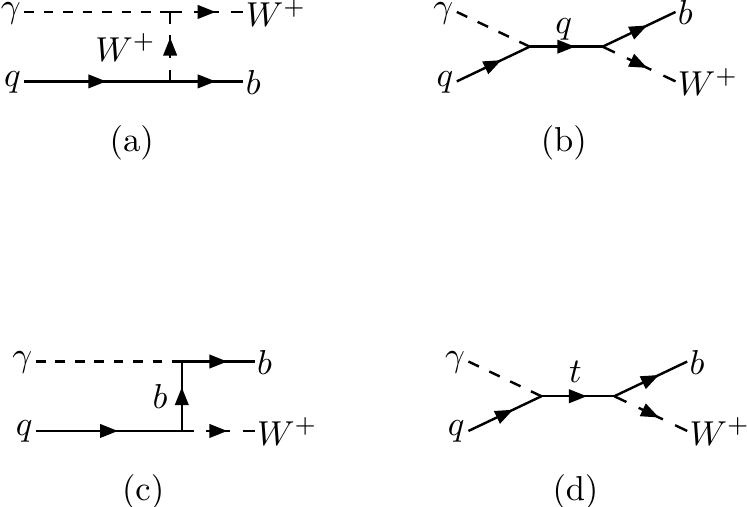}\caption{Feynman diagrams for $\gamma q\rightarrow W^{+}b$, where $q=u,c$.\label{fig:anotop1}}}
\end{figure}

For numerical calculations
anomalous interaction vertices are implemented into the CalcHEP package~\cite{Pukhov:2004ca} 
using the CTEQ6M ~\cite{Pumplin:2002vw} parton distribution functions. The Feynman diagrams for the
subprocess $\gamma q\rightarrow W^{+}b$ are shown
in Fig.~\ref{fig:anotop1}, where $q=u,c$. 
The first three diagrams correspond to irreducible
backgrounds and the last one to the signal. The main background comes from associated production of $W$ boson
and the light jets. 

\begin{figure}[htb]
\centering{\includegraphics[clip=,width=0.60\textwidth,scale=0.6]{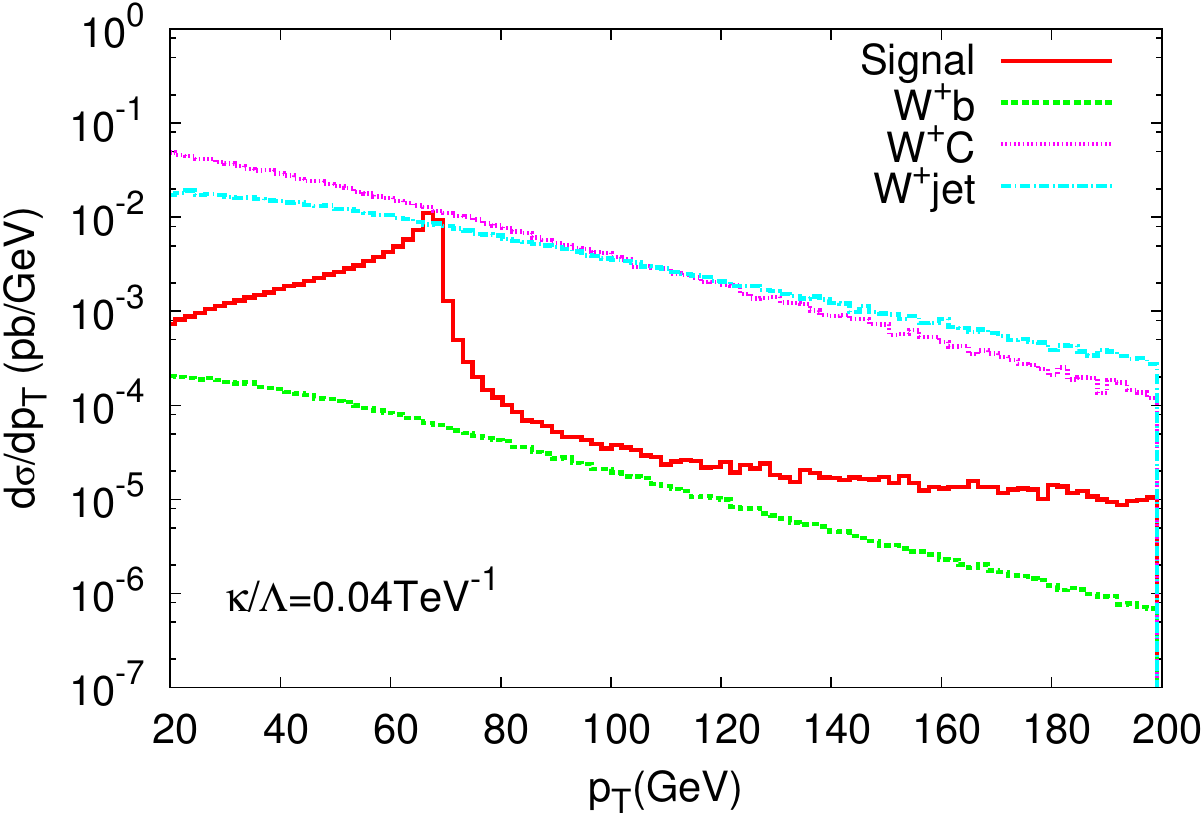}\caption{The transverse momentum distribution of the final state jet for the
signal and background processes. The differential cross section includes
the b-tagging efficiency and the rejection factors for the light jets.
The centre of mass energy $\sqrt{s_{ep}}=1.4$ TeV and $\kappa/\Lambda=$0.04
TeV $^{-1}$\label{fig:anotop2}. }}

\end{figure}

The differential cross sections for the final state jets are given
in Fig. \ref{fig:anotop2} ($\kappa/\Lambda=0.04$ TeV$^{-1}$) for $E_{e}=70$
GeV and $E_{p}=7000$ GeV assuming $\kappa_{u}=\kappa_{c}=\kappa$.
It is seen that the transverse momentum distribution of the signal
has a peak around 70 GeV.

Here, b-tagging efficiency is assumed to be 60\% and
the mistagging factors for light ($u,d,s$) and $c$ quarks are taken
as 0.01 and 0.1, respectively.
A $p_{T}$ cut reduces the signal ( by $\sim30\%$ for $p_{T}>50$
GeV), whereas the background is essentially suppressed (by a factor 4-6) .
In order to improve the signal to background ratio further, one can
apply a cut on the invariant mass of $W+jet$ around top mass. In Table \ref{anotop_tab:1},
the cross sections for signal and background processes are given after
having applied both a $p_{T}$ and an invariant mass cuts ($M_{Wb}=150-200$ GeV).

\begin{table}[ht]
\begin{center}
\begin{tabular}{|c|c|c|c|}
\hline
$\kappa/\Lambda=0.01$ TeV$^{-1}$  & $p_{T}>20$ GeV  & $p_{T}>40$ GeV  & $p_{T}>50$ GeV\tabularnewline
\hline
\hline
Signal  & $8.86\times10^{-3}$  & $7.54\times10^{-3}$  & $6.39\times10^{-3}$\tabularnewline
\hline
Background: $W^{+}b$  & $1.73\times10^{-3}$  & $1.12\times10^{-3}$  & $7.69\times10^{-4}$\tabularnewline
\hline
Background: $W^{+}c$  & $3.48\times10^{-1}$  & $2.30\times10^{-1}$  & $1.63\times10^{-1}$\tabularnewline
\hline
Background: $W^{+}jet$  & $1.39\times10^{-1}$  & $9.11\times10^{-2}$  & $6.38\times10^{-2}$\tabularnewline
\hline
\end{tabular}
\end{center}
\caption{The cross sections (in pb) according to the $p_{T}$ cut and invariant
mass interval ($M_{Wb}=150-200$ GeV) for the signal and background
at $\gamma p$ collider based on the LHeC with $E_{e}=70$ GeV and
$E_{p}=$7000 GeV. \label{anotop_tab:1}}
\end{table}

In order to calculate the statistical significance (\emph{SS}) we
use following formula ~\cite{Ball:2007zza} :

\begin{equation}
SS=\sqrt{2\left[(S+B)\ln(1+\frac{S}{B})-S\right]}\label{eq:3}\end{equation}
 where $S$ and $B$ are the numbers of signal and background events,
respectively. Results are presented in Table \ref{anotop_tab:2} for different
$\kappa/\Lambda$ and luminosity values. It is seen that even with
$2$ fb$^{-1}$ the LHeC based $\gamma p$ collider will provide $5\sigma$
discovery for $\kappa/\Lambda=0.02$ TeV$^{-1}$.

\begin{table}[ht]
\begin{center}
\begin{tabular}{|c|l|c|}
\hline
$SS$  & $L=2$ fb$^{-1}$  & $L=10$ fb$^{-1}$\tabularnewline
\hline
\multicolumn{1}{||c||}{$\kappa/\Lambda=0.01$ TeV$^{-1}$} & 2.6 (2.9)  & 5.8 (6.5)\tabularnewline
\hline
$\kappa/\Lambda=0.02$ TeV$^{-1}$  & 5.3 (5.9)  & 11.8 (13.3)\tabularnewline
\hline
\end{tabular}
\caption{The signal significance ($SS$) for different values of $\kappa/\Lambda$
and integral luminosity for $E_{e}=70$ GeV and $E_{p}=$7000 GeV
(the numbers in parenthesis correspond to $E_{e}=140$ GeV). \label{anotop_tab:2}}
\end{center}
\end{table}

\begin{figure}[htb]
\begin{center}
\includegraphics[clip=,width=0.49\textwidth]{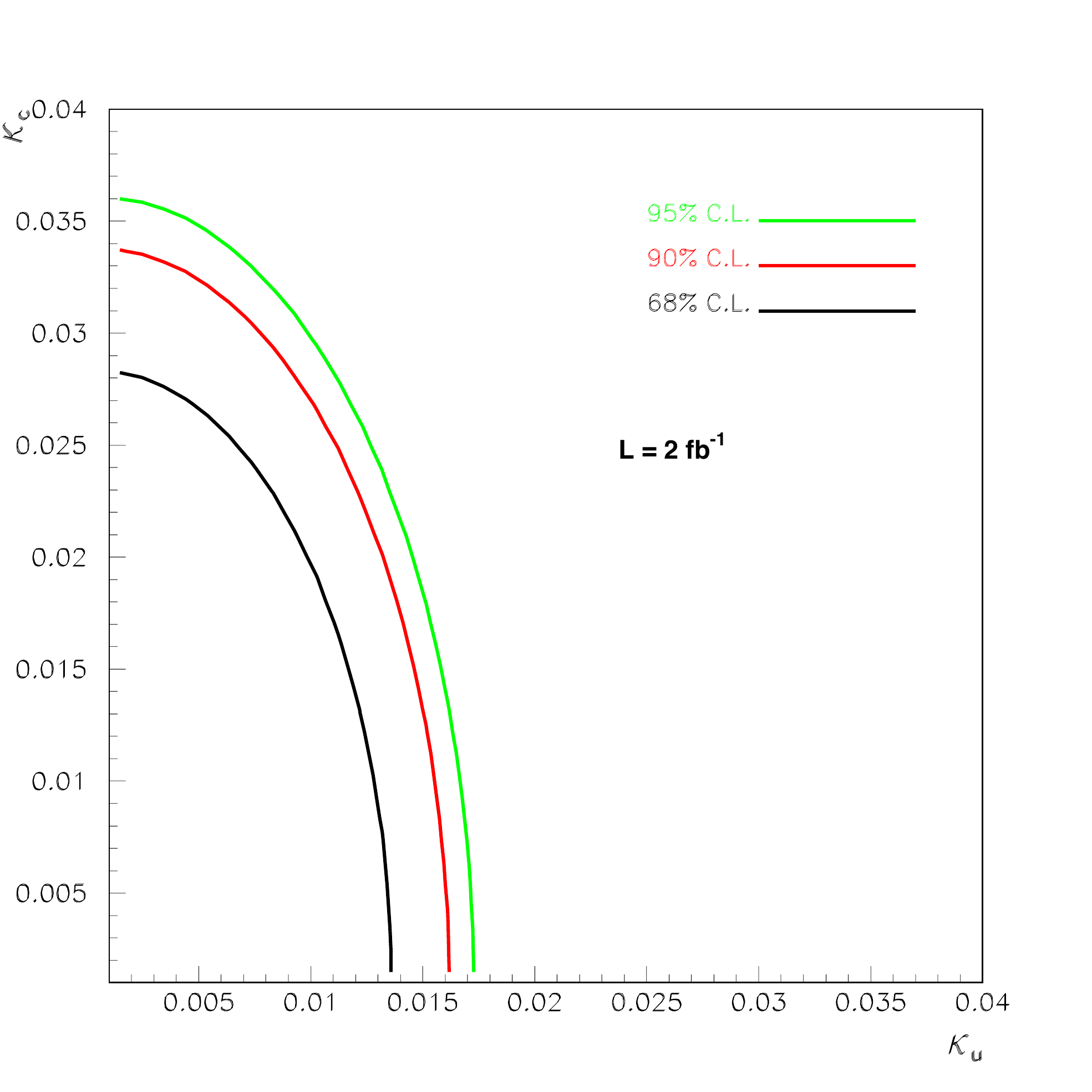}
\includegraphics[clip=,width=0.49\textwidth]{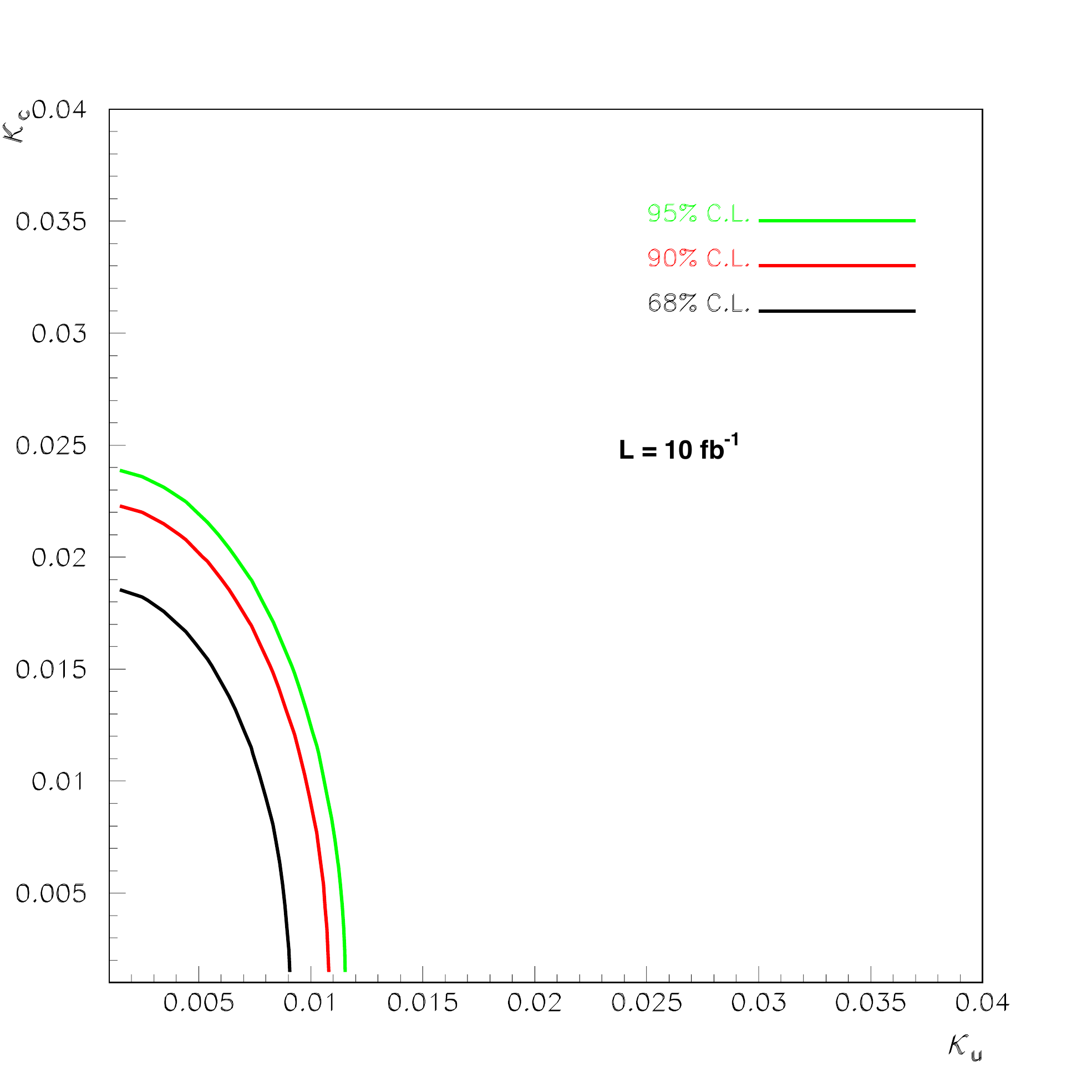}
\caption{\label{fig:anotop3-4} Contour plot for the anomalous couplings reachable at the LHeC based
$\gamma p$ collider with the $ep$ centre of mass energy $\sqrt{s_{ep}}=1.4$
TeV and integrated luminosity of $L_{int}=2$ fb$^{-1}$ (left) or $L_{int}=10$ fb$^{-1}$ (right)}
\end{center}
\end{figure}

Up to now, we have assumed $\kappa_{u}=\kappa_{c}=\kappa$. However, it
would be interesting to analyse the case $\kappa_{u}\neq\kappa_{c}$.
Indeed, at HERA, valence $u$-quarks dominate whereas at LHeC energies the $c$-quark
and $u$-quark contributions become comparable.
Therefore, the sensitivity to $\kappa_{c}$ will be enhanced at LHeC
comparing to HERA. In Fig.~\ref{fig:anotop3-4} contour
plots for anomalous couplings in $\kappa_{u}-\kappa_{c}$ plane
are presented. For this purpose, a $\chi^{2}$ analysis was performed with
\begin{eqnarray}
\chi^{2} & = & \sum_{i=1}^{N}\left({\scriptstyle {\textstyle \frac{\mbox{\ensuremath{\sigma}}_{S+B}^{i}-\mbox{\ensuremath{\sigma}}_{B}^{i}}{\Delta\sigma_{B}^{i}}}}\right)^{2}\label{eq:4}\end{eqnarray}
 where $\mbox{\ensuremath{\sigma}}_{B}^{i}$ is the cross section
for the SM background in the $i^{th}$ bin, including both $b$-jet
and light-jet contributions with their corresponding efficiency factors.
In the $\sigma_{S+B}$ calculations, we take into account the different values for $\kappa_{u}$
and $\kappa_{c}$ as well as the signal-background interference.
Fig.~\ref{fig:anotop3-4} shows that the sensitivity is enhanced by
a factor of 1.5 when the luminosity changes from 2 fb$^{-1}$ to 10
fb$^{-1}$. Concerning the energy upgrade, increasing electron energy
from 70 GeV to 140 GeV results in 20$\%$ improvement for $\kappa_{c}$~\cite{Cakir:2009rq}.
Increasing the electron energy further (energy frontier $ep$ collider)
does not give an essential improvement in the sensitivity to anomalous couplings~\cite{Cakir:2003cg}.

Table~\ref{anotop_tab:2} shows that a sensitivity to anomalous coupling $\kappa/\Lambda$
down to 0.01 TeV$^{-1}$ could be reached.
Noting that the value of $\kappa/\Lambda=0.01$ TeV$^{-1}$ corresponds
to $BR(t\rightarrow\gamma u)\approx2\times10^{-6}$ which is two orders
smaller than the LHC reach with 100 fb$^{-1}$, it is obvious that
even an upgraded LHC will not be competitive with LHeC based $\gamma p$
collider in the search for anomalous $t\gamma q$ interactions. Different
extensions of the SM (SUSY, technicolor, little Higgs, extra dimensions
etc.) predict branching ratio $BR(t\rightarrow\gamma q)$=$O$(10$^{-5}$),
hence the LHeC will provide an opportunity to probe these models. 

\subsection{Excited quarks in $\gamma$p collisions at the LHeC}

Excited quarks will have vertices with SM quark and gauge bosons (photon, gluon, Z or W bosons). They can be produced at $ep$ and $\gamma p$ colliders via quark photon fusion. Interactions involving excited quarks are described by the Lagrangian of eq.~\ref{eq:lagrangianGM} (where $F$ is now a quark $q$).

\begin{figure}[htb]
\begin{center}
{\includegraphics[width=0.49\textwidth]{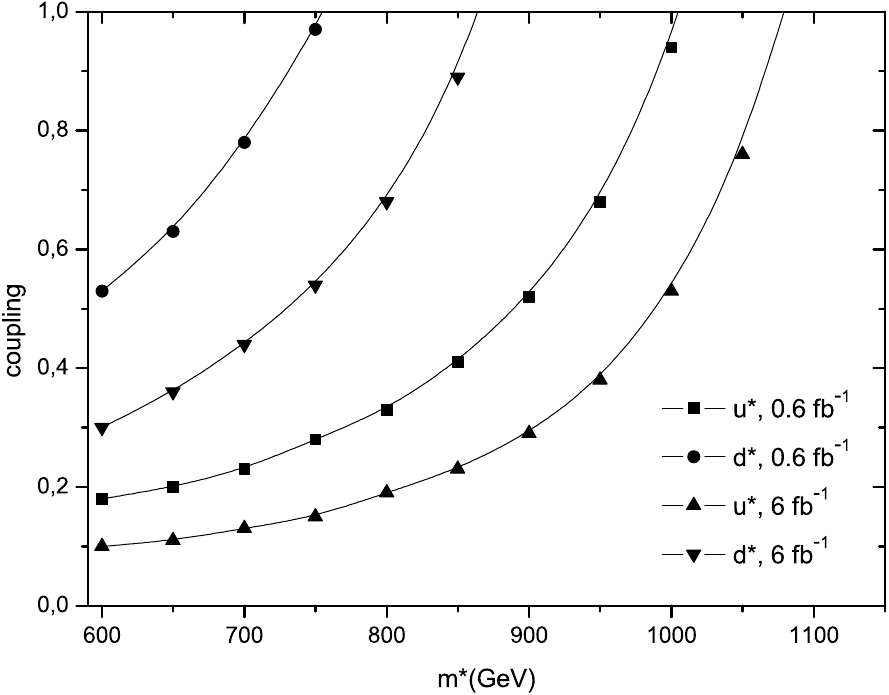}} 
{\includegraphics[width=0.49\textwidth]{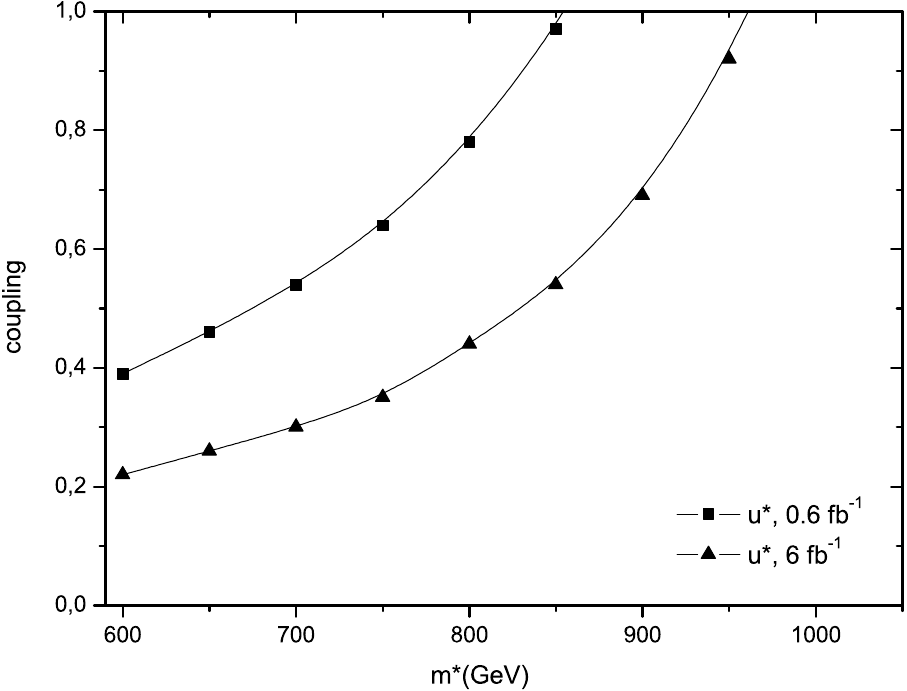}}
\end{center}
\caption{Observation reach at $3\sigma$ for coupling and excited quark mass at a $\gamma p$ collider 
with $\sqrt{s}$ = 1.27 TeV from an analysis of (left) the $jj$ channel and (right) the $\gamma j$ channel.
\label{fig:bsm_ciftci_1star_fig1}}
\end{figure}

A sizeable $f_s$ coupling would allow for resonant $q^*$ production at the LHC via quark-gluon fusion.
In that case, the LHC would offer a large discovery potential for excited quarks and would be well suited to
study the properties and couplings of these new quarks. 
With $1$~fb$^{-1}$ of data collected at $\sqrt{s} = 7$~TeV, the ATLAS collaboration already rules out
$q^*$ with masses below $\sim 3$~TeV for $f_s = 1$ and $\Lambda = M(q^*)$~\cite{Aad:2011fq}.
However, if the coupling of excited quarks to $gq$
happens to be suppressed, the LHC would mainly produce $q^*$ via pair-production and would have little sensitivity to
couplings $f / \Lambda$ or $f' / \Lambda$. Such couplings would be better studied, or probed down to
much lower values, via single-production of $q^*$ at the LHeC.
A study of the LHeC potential for excited quarks is presented in~\cite{LHEC_Note_qstar_Ciftci}.
An example of the $3 \sigma$ discovery reach, assuming $f = f' = f_s$ and setting $\Lambda$ to be equal to the
$q^*$ mass, is given in Fig.~\ref{fig:bsm_ciftci_1star_fig1}. Both decays $q^* \rightarrow q \gamma$ and 
$q^* \rightarrow q g$ have been considered here.


\subsection{Quarks from a fourth generation at LHeC}

\begin{figure}[htb]
\begin{center}
{\includegraphics[width=0.45\textwidth]{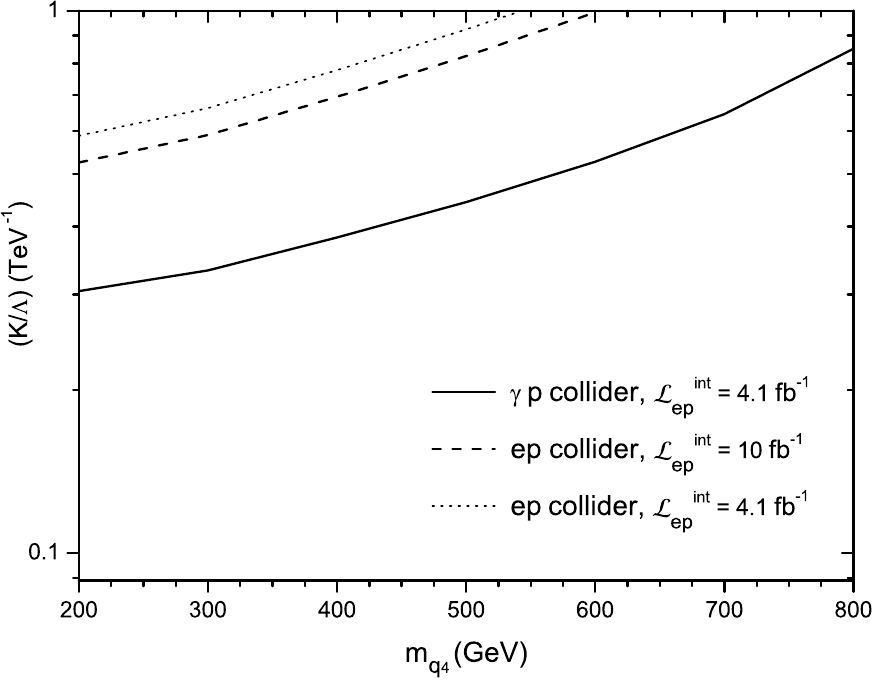}}
{\includegraphics[width=0.45\textwidth]{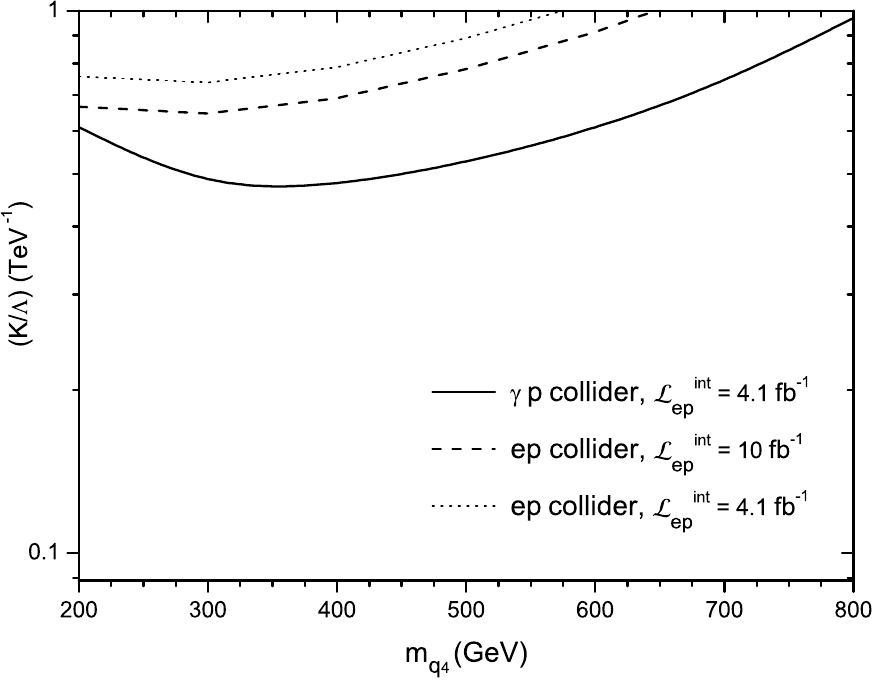}}
{\includegraphics[width=0.45\textwidth]{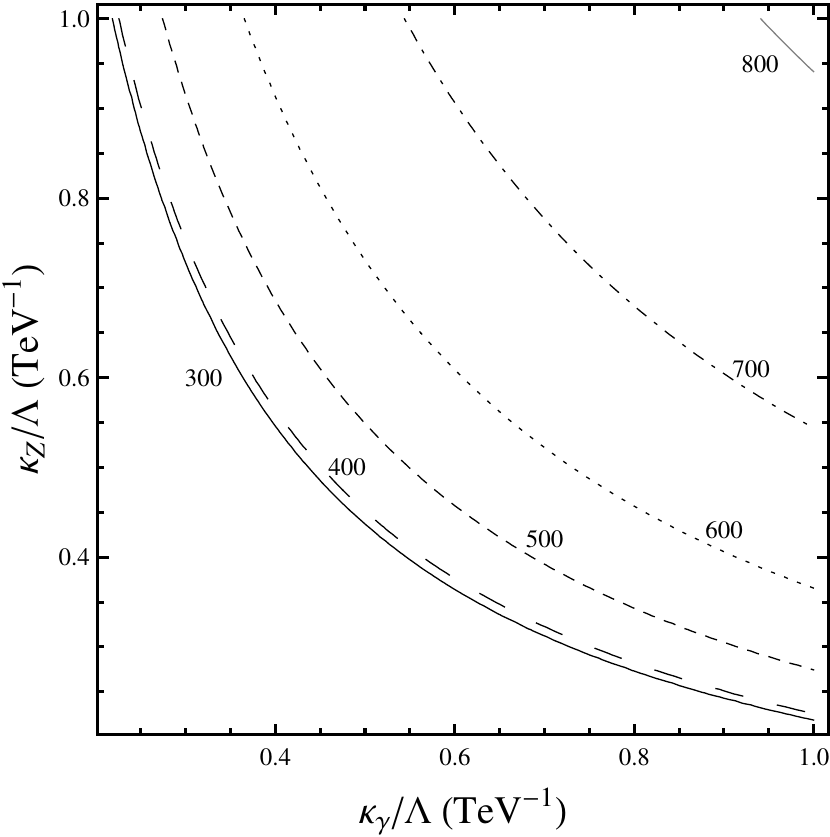}}
\end{center}
\caption{The achievable values of the anomalous coupling strength at $ep$ and $\gamma p$ colliders for a) $q_4 \rightarrow \gamma q$ anomalous process and (b) $q_4 \rightarrow Z q$ anomalous process as a function of the $q_{4}$ mass; (c) the reachable values of anomalous photon and Z couplings with $L_{int}=4.1$ fb$^{-1}$. 
\label{fig:bsm_ciftci_fourth_fig3}}
\end{figure}

The case of fourth generation quarks with magnetic FCNC interactions to gauge bosons
and standard quarks,
\begin{equation}
 {\cal L}=\left(\frac{\kappa^{q_4 q_{i}}_{\gamma}}{\Lambda}\right) e_{q}g_{e}\bar{q}_{4}\sigma_{\mu\nu}q_{i} F^{\mu\nu}+\left(\frac{\kappa^{q_4 q_{i}}_{Z}}{2\Lambda}\right) g_{Z}\bar{q}_{4}\sigma_{\mu\nu}q_{i} Z^{\mu\nu}+\left(\frac{\kappa^{q_4 q_{i}}_{g}}{\Lambda}\right) g_{s}\bar{q}_{4}\sigma_{\mu\nu}T^{a}q_{i} G^{\mu\nu}_{a}+h.c. 
\label{eq:q4_anomalous}
\end{equation}
is very similar to that of excited quarks.
A $\gamma p$ collider based on LHeC would have a better sensitivity than LHC to anomalous couplings
$\kappa_{\gamma}$ and $\kappa_Z$. A detailed study is presented in~\cite{LHEC_Note_fourth_family_Ciftci}
and example results are shown in Fig.~\ref{fig:bsm_ciftci_fourth_fig3}. These figures also show the clear
advantage of a $\gamma p$ collider compared to an $ep$ collider, for the study of new physics in
$\gamma q$ interactions.

\subsection{Diquarks at LHeC}

The case of diquark production at LHeC has been studied in~\cite{LHEC_Note_diquark_Cakir}. 
The production cross section can be sizeable at a high energy $ep$ machine, especially when operated 
as a $\gamma p$ collider. The measurement of the $\gamma p \rightarrow DQ + X$ 
cross section, for a diquark $DQ$ of known mass and known coupling to the diquark pair\footnote{The LHC
would observe diquark as di-jet resonances, and could easily determine its mass, width and
coupling to the quark pair.} would 
provide a measurement of the electric charge of the diquark. It would thus be complementary
to the $pp$ data, which offer no simple way to access the $DQ$ electric charge.
However, the diquark masses and couplings that could be accessible at LHeC appear to
be already excluded by the recent search for dijet resonances at the LHC~\cite{Khachatryan:2010jd}.

\subsection{Quarks from a fourth generation in $Wq$ interactions}

In case fourth generation quarks do not have anomalous interactions as in Eq.~\ref{eq:q4_anomalous},
they (or vector-like quarks coupling to light generations~\cite{Atre:2008iu,Atre:2011ae}) could be produced in $ep$ collisions by $Wq$ interactions provided that the $V_{Qq}$ elements of the extended
CKM matrix are not too small, via the usual vector $W q Q$ interactions.
An example of the sensitivity that could be reached at LHeC is presented in~\cite{LHEC_Note_fourth_Cakir},
assuming some values for the $V_{Qq}$  parameters. Measurements of single $Q$ production at LHeC
would provide complementary information to the LHC data, that could help in
determining the extended CKM matrix.

%% file: physics/bsm_higgs.tex
\section{Sensitivity to a Higgs boson}
\label{sec:higgs}
\begin{figure}[h!]
\begin{center}
\includegraphics[width=0.6\columnwidth]{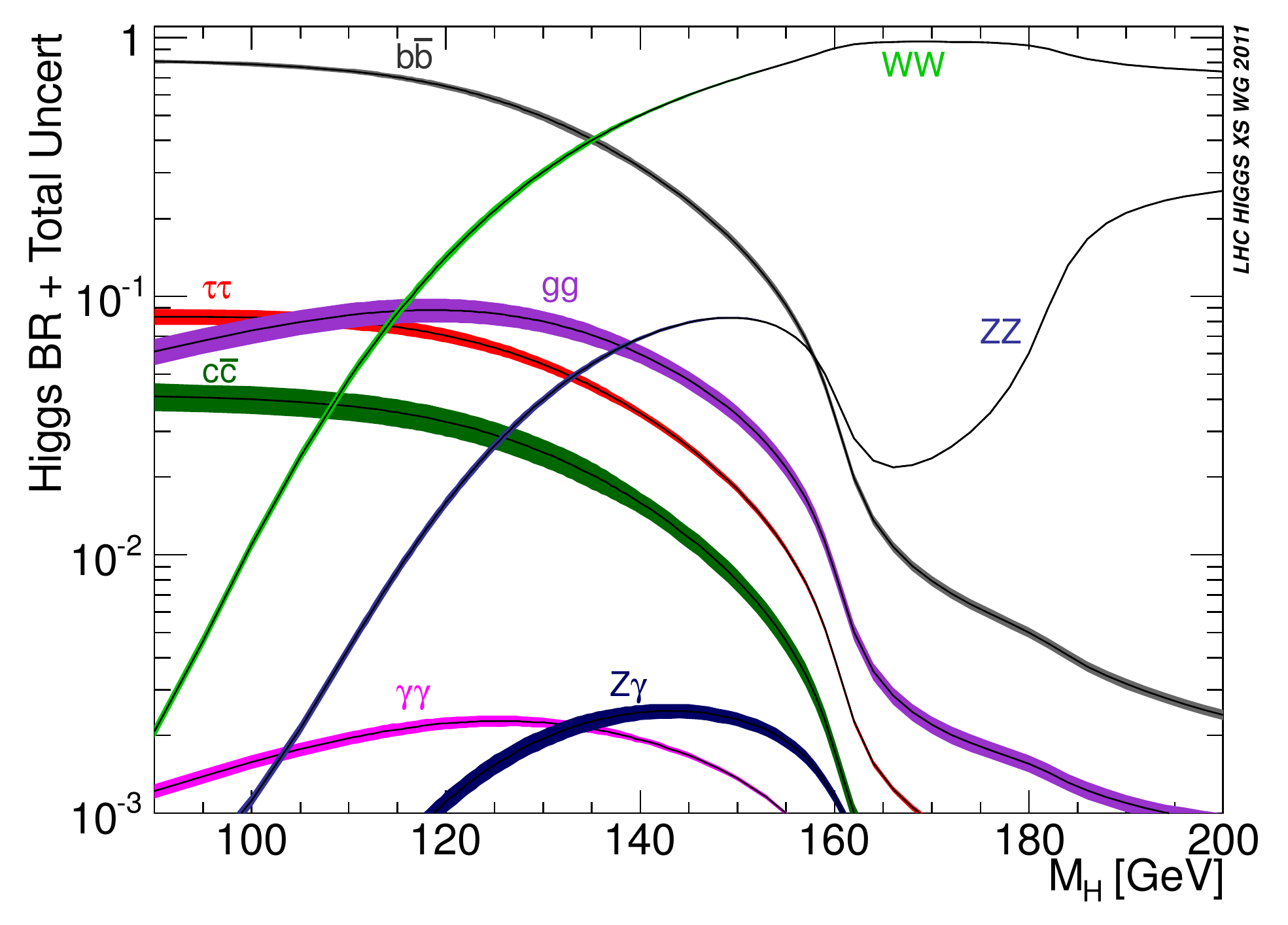}
\caption{Branching fractions of the SM scalar boson as a function of its mass.
The analysis presented in this report has solely considered the
$H$ to $b \bar{b}$ decay.}
\label{fig:hbranch}
\end{center}
\end{figure}

Unlike HERA, the LHeC has an exciting sensitivity to the Higgs boson,
should it exist, because of the increase in energy and luminosity.
It is cleanly produced via either $ZZ$ or $WW$ fusion and is thus
complementary to the dominant $gg$ fusion in $pp$ scattering.
The final state in $ep$ scattering is also cleaner than in $pp$, which can
be exploited to identify complex final states. As an example, this section
describes first considerations on the Higgs at the LHeC, the reconstruction
of its dominant decay channel, into $b \bar{b}$, and the determination of its
CP properties, based on its uniquely identifiable production via
 $WW \rightarrow H$ fusion in CC scattering. The results are encouraging as
they point to a $\sim5$\,\% precision determination of the $WWHb\bar{b}$
coupling, with the default $60$\,GeV energy electron beam and
for $100$\,fb$^{-1}$ of integrated luminosity. 
In future  studies much can be done to develop this further,
using a dedicated simulation of an optimised $ep$ detector,
refined analysis techniques such as those which are often employed at the LHC now,
by considering neutral current or photoproduction
of the Higgs and also including further
final states, such as $WW,~ZZ$ and $c \bar{c}$
as are illustrated in Figure\,\ref{fig:hbranch}. If indeed the Higgs
particle exists at $125$\,GeV, this will undoubtedly strongly motivate 
the LHeC design to go beyond the 
$10^{33}$\,cm$^{-2}$s$^{-1}$ luminosity considered as baseline in this design 
concept report.
\subsection{Introductory remarks}
Understanding the mechanism of electroweak symmetry breaking is a key
goal of the LHC physics programme. In the SM, the symmetry breaking is
realised via a scalar field (usually known as the Higgs field) which, at the minimum of
the potential, develops a non-zero vacuum expectation value. The
breaking of the $SU(2)_L \times U(1)_Y $ symmetry gives mass to the
electroweak gauge bosons via the Brout-Englert-Higgs mechanism while the fermions
obtain their mass via Yukawa couplings with the Higgs field.  The LHC
experiments should be able to discover a SM scalar boson (Higgs boson) within the full
allowable mass range up to about $1$~TeV. Following its possible 
discovery at the LHC, it will be crucial to measure the
couplings of the Higgs boson to the SM particles, in particular to
the fermions, in order to:
\begin{itemize}
 \item establish that the Higgs field is indeed responsible for the
  fermion masses, via Yukawa couplings $ y_f H \bar{f} f$;
 \item distinguish between the SM and (some of) its possible extensions. For
  example, despite the richer content of the Higgs sector in the Minimal Supersymmetric
  Standard Model, only the light SUSY Higgs boson $h$ may be observable at the LHC
  in certain regions of parameter space. Its properties are very similar to 
  those of the SM Higgs $H$, and precise measurements of ratios
  $ BR ( \Phi \rightarrow VV) / BR (\Phi \rightarrow f \bar{f}) $  
  will be essential in determining whether or not the observed boson, $\Phi$,  is the SM Higgs scalar.
\end{itemize}
The LEP experiments have ruled out a SM boson lighter than $114.5$~GeV, and
electroweak precision measurements suggest that the SM Higgs boson should be light.
Latest results from Higgs searches at the LHC constrain the SM Higgs mass
to lie within $117.5  - 127.5$~GeV or above about 
$600$~GeV~\cite{ATLAS-CONF-2012-019, CMS-PAS-HIG-12-008}
with $5$~fb$^{-1}$
of data collected at $\sqrt{s} = 7$~TeV by  both the ATLAS and the CMS experiments.
In the allowed low mass range, the Higgs would predominantly decay into a $b \bar{b}$ pair
with a branching ratio of about $60$\%, but a measurement of the $H b \bar{b}$ coupling
will be challenging at the LHC~\cite{Ball:2007zza, Aad:2009wy,Duhrssen:2005sp} 
and a direct observation of $H \rightarrow b \bar{b}$ in the inclusive production mode 
is made very difficult by the overwhelming QCD background. A possible search channel
would be associated $WH$ and $ZH$ production, with highly boosted Higgs, 
leading to a high mass jet with substructure~\cite{Butterworth:2008iy}. 
The observability of the signal
in the $t \bar{t} H$ production mode may also suffer from a large background,
including background of combinatorics origin, and from experimental systematic uncertainties.
 The signal $H \rightarrow b \bar{b}$ may be observed at the LHC
in the exclusive production mode, thanks to the much cleaner environment in a
diffractive process. However,
the production cross section in this mode is expected to be small and predictions
 suffer from large theoretical uncertainties,
such that this measurement, if feasible at all, is not expected to translate into a precise
measurement of the $H b \bar{b}$ coupling.

At the LHeC, a light Higgs boson could be produced via
 weak vector boson fusion (WBF) with a sizeable cross section.
This section focuses on the 
observability of the signal $e p \rightarrow H (\rightarrow b \bar{b}) + X $
at LHeC, which may deliver a clear observation of the $H \rightarrow b \bar{b}$ decay.
The studies have been performed using the nominal $7$~TeV LHC
 proton beams  and electron and positron beam energies in the range of $50$ to  
$150$~GeV, i.e. only the lepton beam energies will be specified in the following.
A similar study,  using parton-level events only,  can be found in~\cite{Han:2009pe}.


\subsection{Higgs production at the LHeC}
\begin{figure}[h!]
\begin{center}
\includegraphics[width=0.3\columnwidth]{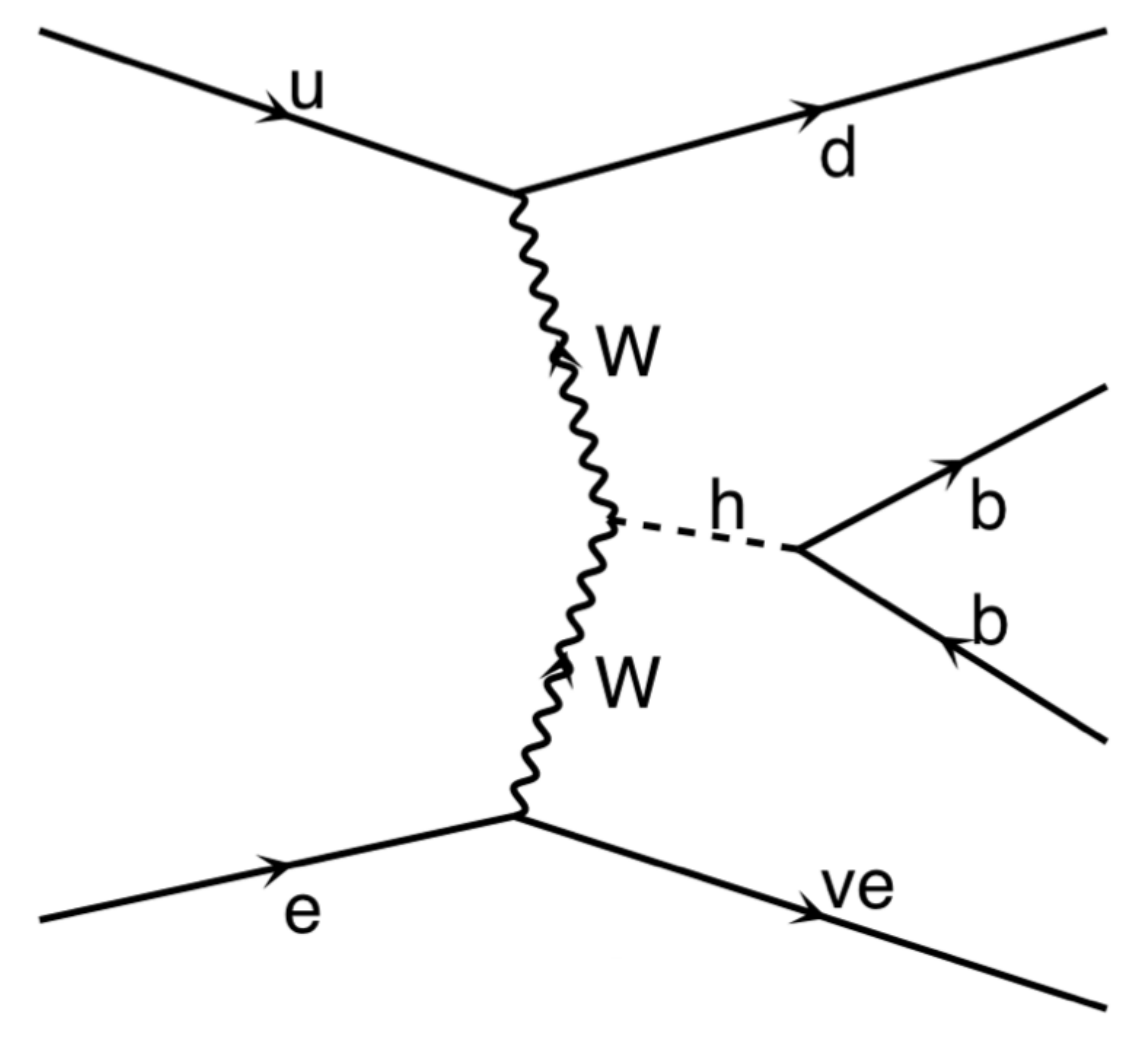}
\hspace{5mm}
\includegraphics[width=0.3\columnwidth]{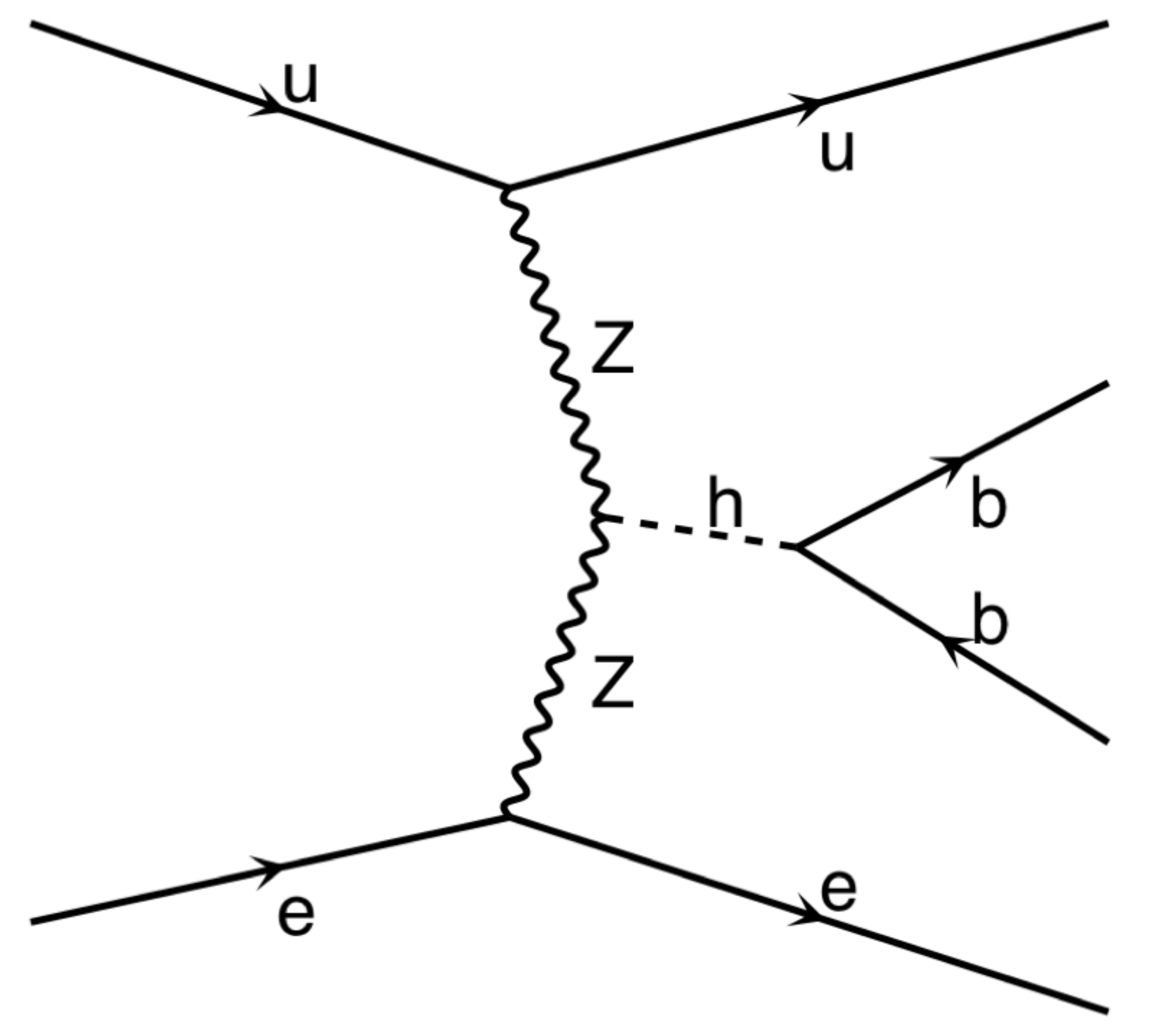}
\caption{Feynman diagrams for CC (left) and NC (right) Higgs production in
leading order QCD at the LHeC. Diagrams produced using MadGraph.}
\label{fig:higgs_diagrams}
\end{center}
\end{figure}

\begin{figure}[h!]
\begin{center}
\includegraphics[width=0.81\columnwidth]{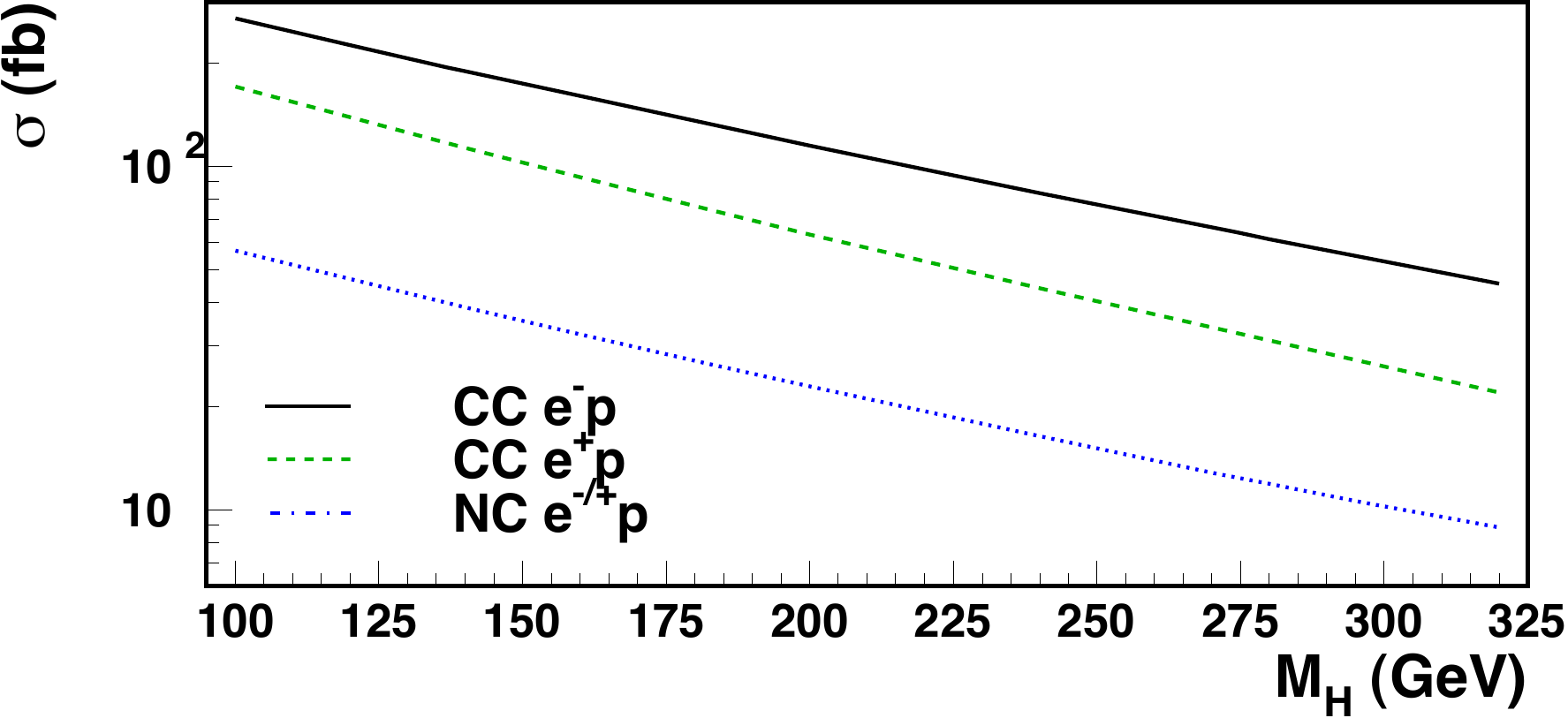}
\caption{Total production cross section of a SM Higgs boson in $e^{\pm}p$ 
collisions with $E_e$=140~GeV and $E_p$=7~TeV, as a function of the Higgs mass.}
\label{fig:higgs_xsection}
\end{center}
\end{figure}

In $ep$ collisions, the Higgs boson would be cleanly 
produced in neutral current (NC) interactions via
the $ZZH$ coupling, and in charged current 
(CC) interactions via the $WWH$ coupling. 
The production mechanism therefore excludes the
gluon-gluon fusion which determines the Higgs
production at the LHC. The
corresponding leading order 
diagrams are shown in Fig.~\ref{fig:higgs_diagrams}. 
The total Higgs production
cross sections for CC and NC $e^{\pm}p$ scattering, as a function of the Higgs mass,
are displayed in Fig.~\ref{fig:higgs_xsection}.
%
The $WWH$ production dominates the total cross section.  As is the case for the
inclusive CC DIS interactions, the cross section is much larger in $e^- p$ collisions than
in $e^+ p $ collisions, due to the more favourable density of the valence quark that
is involved ($u$ in $e^- p$, $d$ in $e^+ p$), and to the more favourable helicity factors.
Table~\ref{tab:higgs_xsec} shows the total
 Higgs production cross section (at leading order $\alpha_S$) via
CC interactions in $e^- p$ collisions, for various values of the Higgs mass and three example
values of the electron beam energy.  If the input Higgs mass is changed, the electroweak 
parameters are recalculated according to the SM expectations.
The renormalisation and factorisation scales are set to the partonic 
centre-of-mass energy  which gives an about 10\% smaller cross section
prediction than using scales fixed to the input Higgs mass. 
%
This ${\cal{O}}(10 \%)$  uncertainty is well covered by the expected size 
of   leading QED corrections~\cite{Jager:2010zm,Blumlein:1992eh} and 
next-to-leading order QCD corrections~\cite{Blumlein:1992eh}.  
Both effects are expected to be small, i.e. moderately affecting  the
shape of some kinematic distributions in the range of 5\% to
 ${\cal{O}}(20 \%)$. However, those estimates may deserve further 
study of their dependence on phase space requirements.   
Remaining NNLO QCD contributions can be expected to modify the cross section
to the $1$\,\% level, which is not important for the present study.
%
\begin{table*}[h!]
\begin{center}
 \begin{tabular}{|c|c|c|c|c|c|c|}
 \hline
        &  $M_H=$100  GeV & 120  GeV &  160  GeV & 200 GeV & 240  GeV & 280  GeV  \\  \hline
 $E_e = 50$~GeV  &   102   &  81  & 50   & 32  & 20   & 12    \\
 $E_e = 100$~GeV  & 201  & 165  & 113  & 79  &  55  &  39   \\
 $E_e = 150$~GeV   & 286  & 239  & 170   & 123  & 90  &  67  \\
 \hline \hline
 \end{tabular}
 \caption{ \label{tab:higgs_xsec} Total production cross sections in fb of a SM Higgs 
boson with masses in the range of 100 to 280 GeV via charged current
    interactions in $e^- p$ collisions, for three example values of the electron beam 
energy of 50, 100 and 150 GeV, using the program package 
MadGraph~\cite{Alwall:2007st} and the CTEQ6L1 parton distribution functions.}
\end{center}
\end{table*}
%
%

%
%


\subsection{Observability of the signal}
\label{sec:bsm_higgs_observability}
%
\begin{figure}[h!]
\begin{center}
\includegraphics[width=0.81\columnwidth]{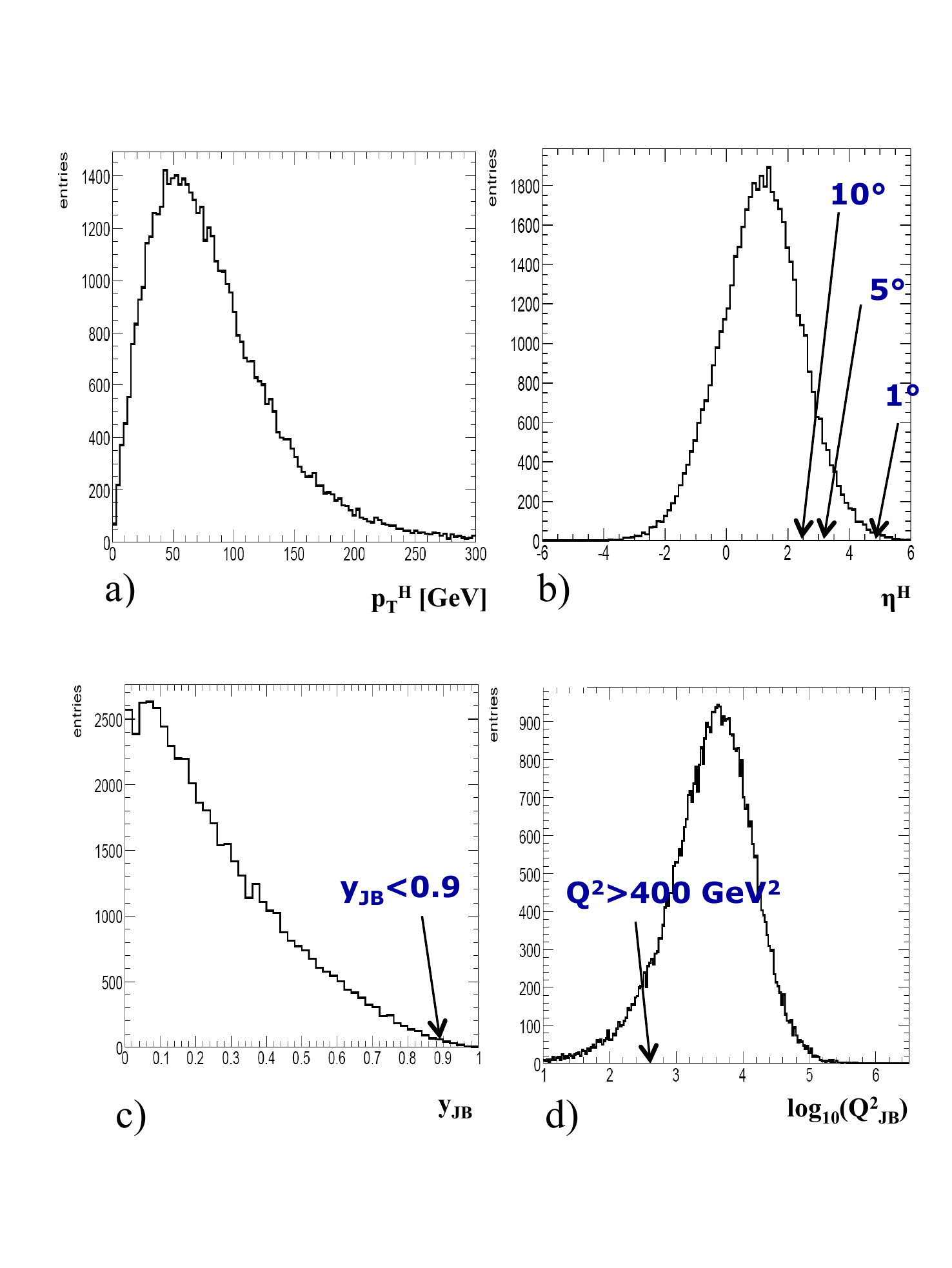}
\caption{Generated (a) transverse momentum and (b) pseudorapidity 
distributions of a 120 GeV SM Higgs boson using a 150 GeV electron beam 
and a CC selection, see text.  Indicated are also typical values for detector 
acceptance in the polar scattering angle. Reconstructed (c)  inelasticity, $y_{JB}$, 
and (d) negative four-momentum transfer, $Q^2_{JB}$ distributions where 
in both cases the applied selections are shown.}
\label{fig:higgs_rec}
\end{center}
\end{figure}
A first, baseline feasibility study  is performed using charged current
DIS events produced in $e^-p$ collisions
which provide the  largest expected Higgs cross section, see Fig.~\ref{fig:higgs_xsection}. 
MadGraph~\cite{Alwall:2007st} has been used to generate SM Higgs production, 
CC and NC DIS background events.
An  electron beam energy of 150\,GeV and a  proton beam energy of 7\,TeV is used 
as the reference beam configuration and a 120\,GeV SM Higgs boson mass in the MC simulation study.
Results are also obtained with a different electron beam energy and also Higgs masses.

Calculations of cross sections and generation of final states of 
outgoing particles are performed by MadGraph using the
chosen  beam parameters and considering the  dominant  SM tree-level Feynman diagrams.
Typical kinematic distributions obtained for a generated $120$~GeV SM Higgs boson 
are shown in Figs.~\ref{fig:higgs_rec} a) and b). The average polar scattering angle 
of the Higgs boson is forward at about $40^o$ and a pseudorapidity of 
about $1$, respectively.

Fragmentation and hadronisation processes are simulated using
PYTHIA~\cite{Sjostrand:2006za} with custom modifications to apply for $ep$ collisions.
In the absence of a completed detector design and simulation at the time
of this investigation, the particles were passed through a generic, LHC-style 
detector using the PGS~\cite{PGS} fast detector simulation tool.  The 
 tracking coverage is assumed to extend to pseudorapidities of
$|\eta|<3$. The calorimeter coverage is assumed to be  $|\eta| < 5$ 
with an electromagnetic calorimeter resolution of $5$\,\%$/\sqrt{E({\rm GeV})}$ 
(plus $1$\,\% of constant term)
and a hadronic calorimeter resolution of 60\,\%$/\sqrt{E({\rm GeV})}$.
Jets are reconstructed by a cone algorithm with a cone size of $\Delta R = 0.7$.
The efficiency of b-flavour tagging is assumed to be 60\,\% and 
flat within the tracking coverage, whereas mistagging probabilities 
of $10$\,\% and $1$\,\% for charm-quark jets and for
light-quark jets, respectively, are taken into account.

The dominating source of background at large missing transverse energy is coming
from multi-jet production in CC DIS interactions. In particular, a 
good rejection of the background coming from single top
production ($e^- b \rightarrow \nu t$), where 
the top decays hadronically, puts  constraints on the acceptance, the
resolution and the b-tagging performance of the detector, as will be seen below.
The background due to multijet production in NC interactions is also considered.

In the case of simulating NC DIS background events, since the cross 
section is very high (diverging at $Q^2 \sim 0 $ values), only 
processes producing two or more b quarks
are generated in order to have sufficient
 MC statistics after the selection. Using an artificially increased mistag 
probability, it could be verified that the remaining 
NC background is indeed due to events with two  b-quark jets in the final state.

The following selection criteria are applied, based on observable
 variables reconstructed by the PGS detector simulation, to 
distinguish $H \rightarrow b \bar{b}$ from the CC and NC DIS backgrounds.
\begin{itemize}
\item {\bf cut (1): Primary cuts}
\begin{itemize}
\item Exclude electron-tagged events
\item $E_{T,miss} > 20\,{\rm GeV}$
\item $N_{jet} (P_{T,jet}>20\,{\rm GeV}) \geq 3$
\item $E_{T, total} > 100\,{\rm GeV}$
\item $y_{JB} < 0.9$, where $y_{JB} = \Sigma (E-p_z) / 2E_e$, as shown in Fig.~\ref{fig:higgs_rec} c)  
\item $Q_{JB}^{2} > 400\,{\rm GeV}$, where $Q^2_{JB} = E_{T,miss} ^2 / (1 - y_{JB})$, as shown in  Fig.~\ref{fig:higgs_rec} d) 

\end{itemize}
\item {\bf cut (2): b-tag requirement}
\begin{itemize}
\item $N_{b\mathchar`-jet} (P_{T, jet} >20\,{\rm GeV}) \geq 2$, where b-jet means a b-tagged jet
\end{itemize}
\item {\bf cut (3): Higgs invariant mass cut}
\begin{itemize}
\item $90 < M_{H} < 120\,{\rm GeV}$; due to the energy carried by the neutrino from $b$ decays, the mass peaks are slightly lower than the true Higgs mass
\end{itemize}
\end{itemize}
\begin{figure}[htbp]
\begin{center}
\includegraphics[width=0.49\columnwidth]{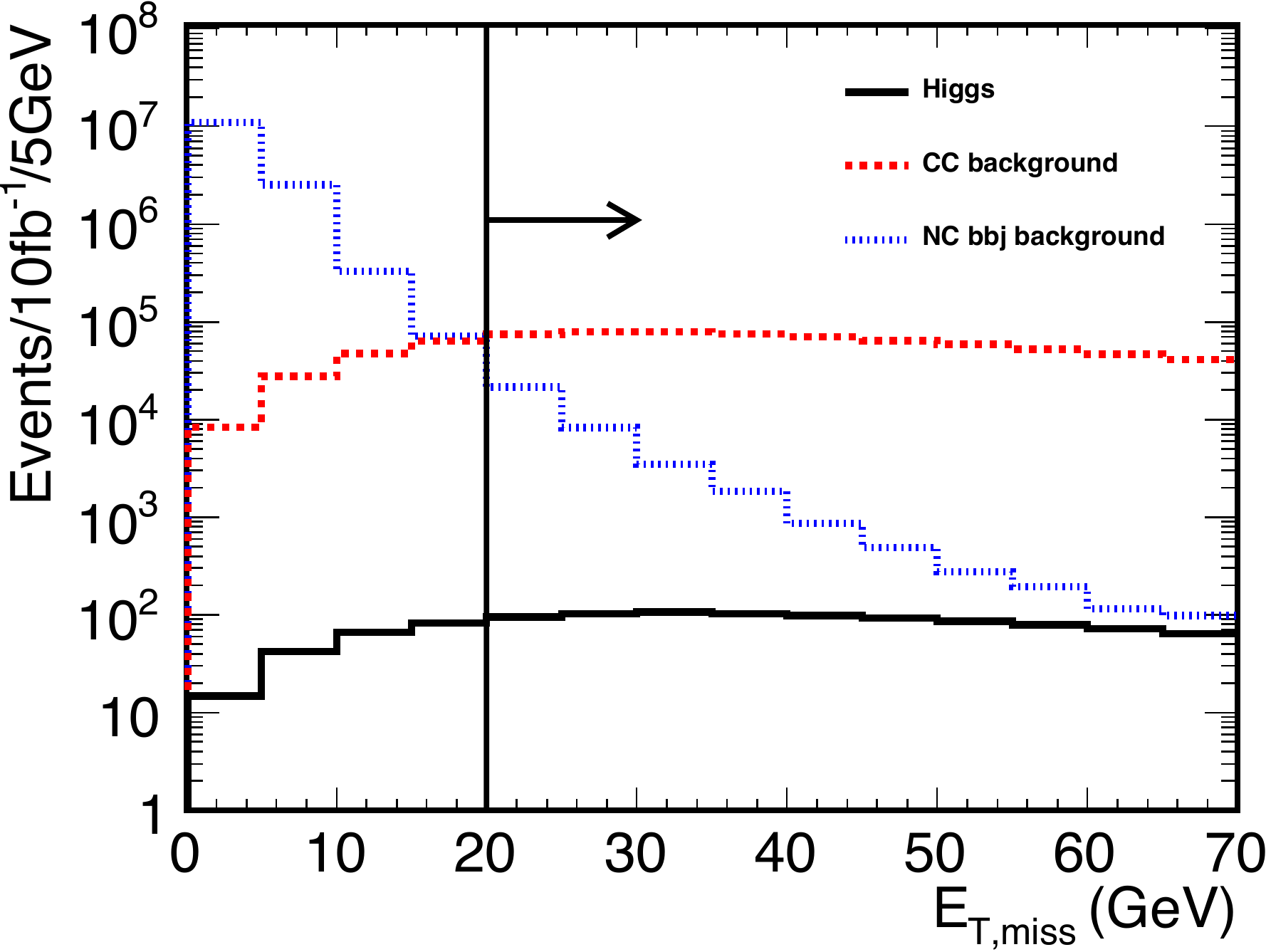}
\includegraphics[width=0.49\columnwidth]{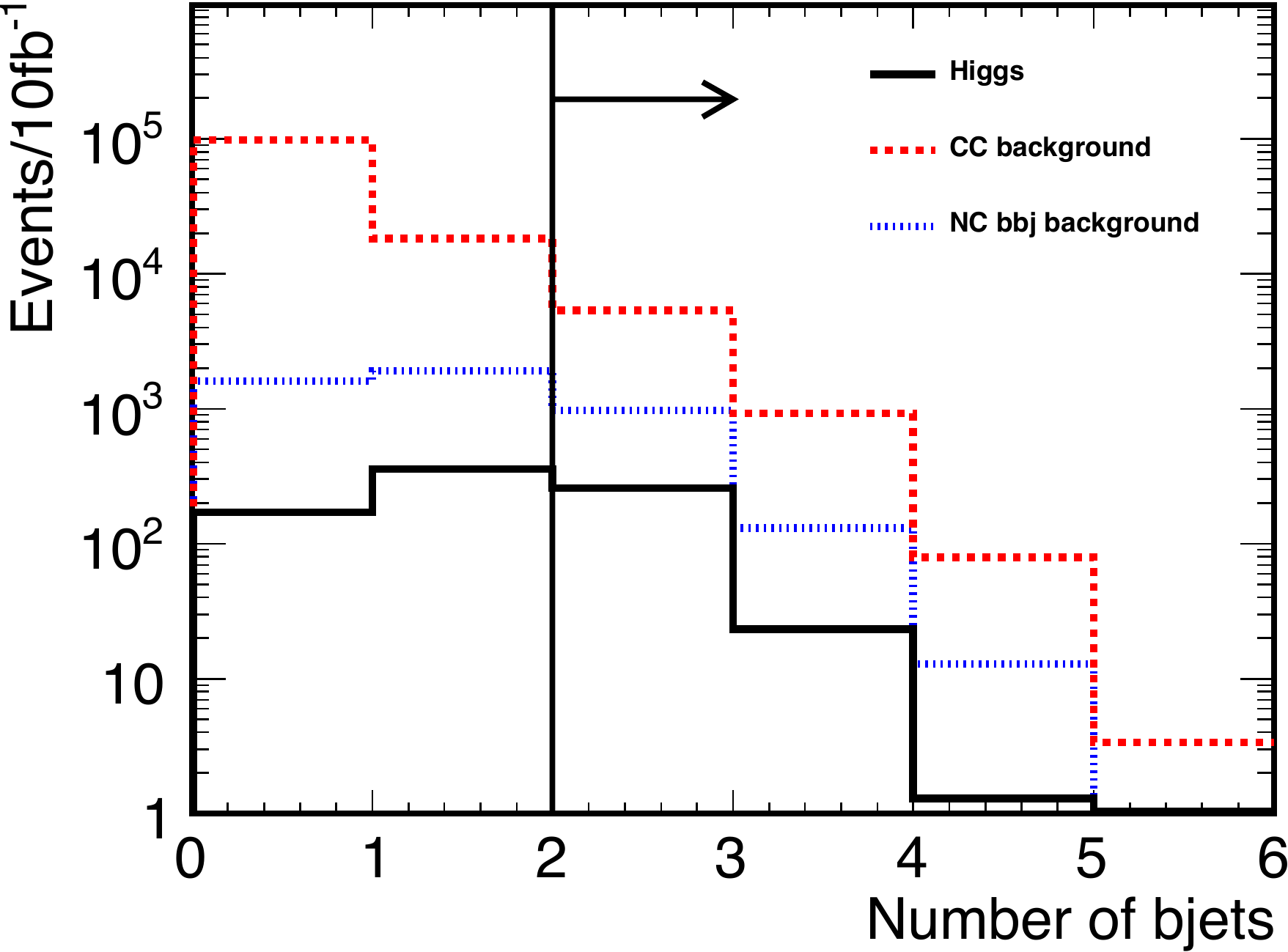}
\caption{Missing $E_{T}$ (left) and number of b-tagged jets (right). Solid (black), dashed (red) and dotted (blue) histograms show $H \rightarrow b \bar{b}$, CC and NC DIS multi-jet background events, respectively.
The right plot is for events passing cut (1), see text.}
\label{fig:higgs_cut1}
\end{center}
\end{figure}
Fig.~\ref{fig:higgs_cut1} shows the missing energy, $E_{T, miss}$,
 and number of b-tagged jets for $H \rightarrow b \bar{b}$ events 
together with the CC and NC DIS background.
The NC background is strongly suppressed by the missing $E_{T,miss}$ cut and electron-tag requirement.
Requiring at least two b-tagged jets, the Higgs invariant mass is reconstructed 
using the  b-tagged jets which are most central, i.e. the ones with the lowest and 
second lowest pseudorapidity values, $\eta$.
After cuts (1) to (3) are applied, about 45\,\%  
of the remaining CC background is due to single top production 
where  light-quark jets can be misidentified as b-tagged jets.
The single top background is further reduced by 
the requirements as follows.
\begin{itemize}
\item {\bf cut (4): Rejection of single top production}
Single top events result in a final state with one 
 b-jet and a W boson decaying into two light-quark jets. The following cuts are found to 
be efficient in suppressing this background. 
\begin {itemize}
\item $M_{jjj, top} > 250\,{\rm GeV}$, where the three-jet invariant mass ($M_{jjj,top}$) 
is reconstructed from three mainly centrally produced jets using two b-tagged jets with the 
lowest $\eta$ and any third jet with the lowest $\eta$ (b-tag not required for the third jet)
\item $M_{jj, W} > 130\,{\rm GeV}$, where the di-jet invariant mass ($M_{jj, W}$) is 
reconstructed from one b-tagged jet with the lowest $\eta$ and any second jet with the 
lowest $\eta$ regardless of b-tag but excluding the second lowest $\eta$ b-jet
\end{itemize}
\item {\bf cut (5): Forward jet tagging}
\begin{itemize}
\item $\eta_{jet} > 2$ for the jet with the lowest pseudorapidity (lowest-$\eta$ jet) but  
excluding the two b-tagged jets used to reconstruct the Higgs boson candidate
\end{itemize}
\end{itemize}
\begin{figure}[htbp]
\begin{center}
\includegraphics[width=0.49\columnwidth]{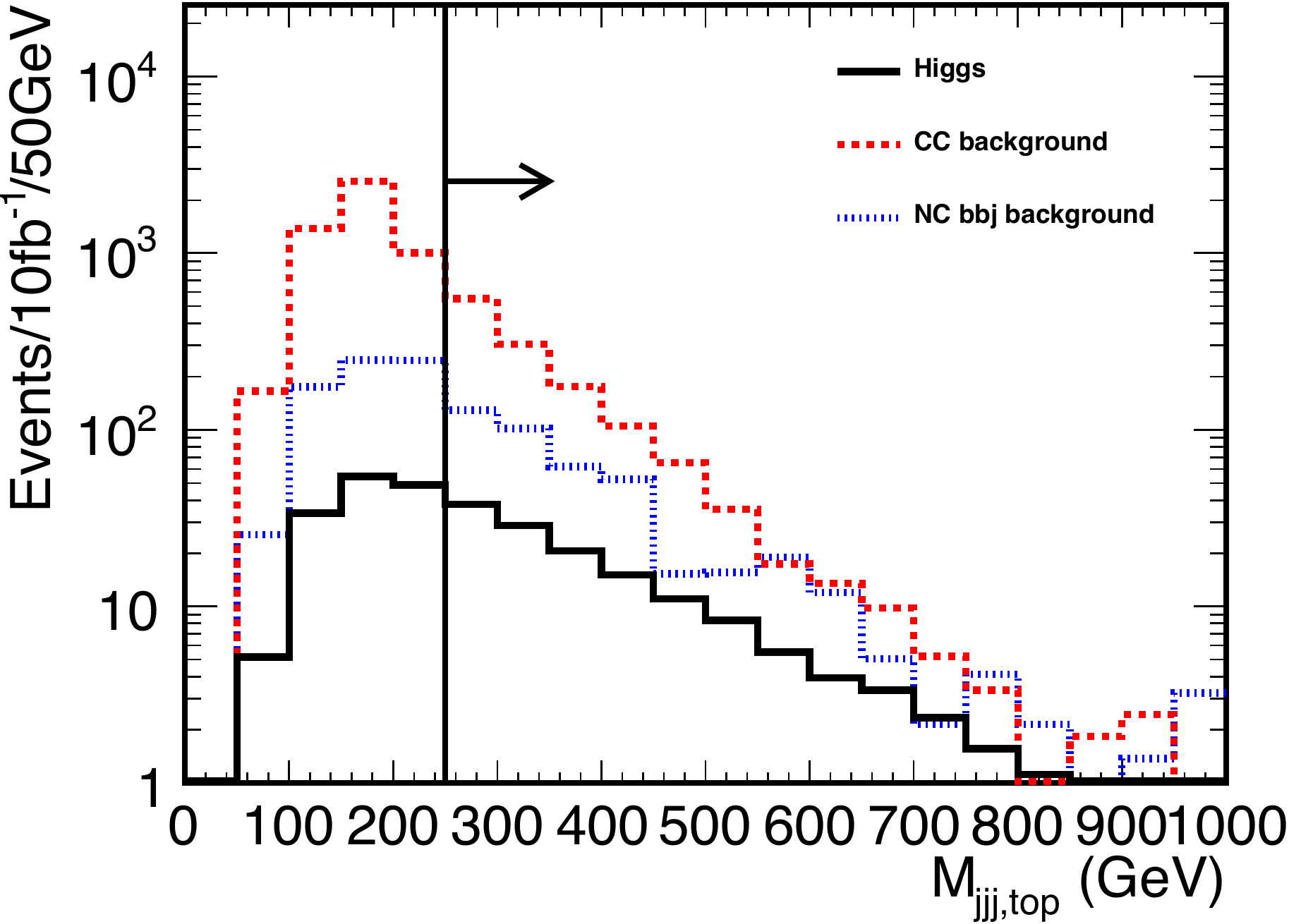}
\includegraphics[width=0.49\columnwidth]{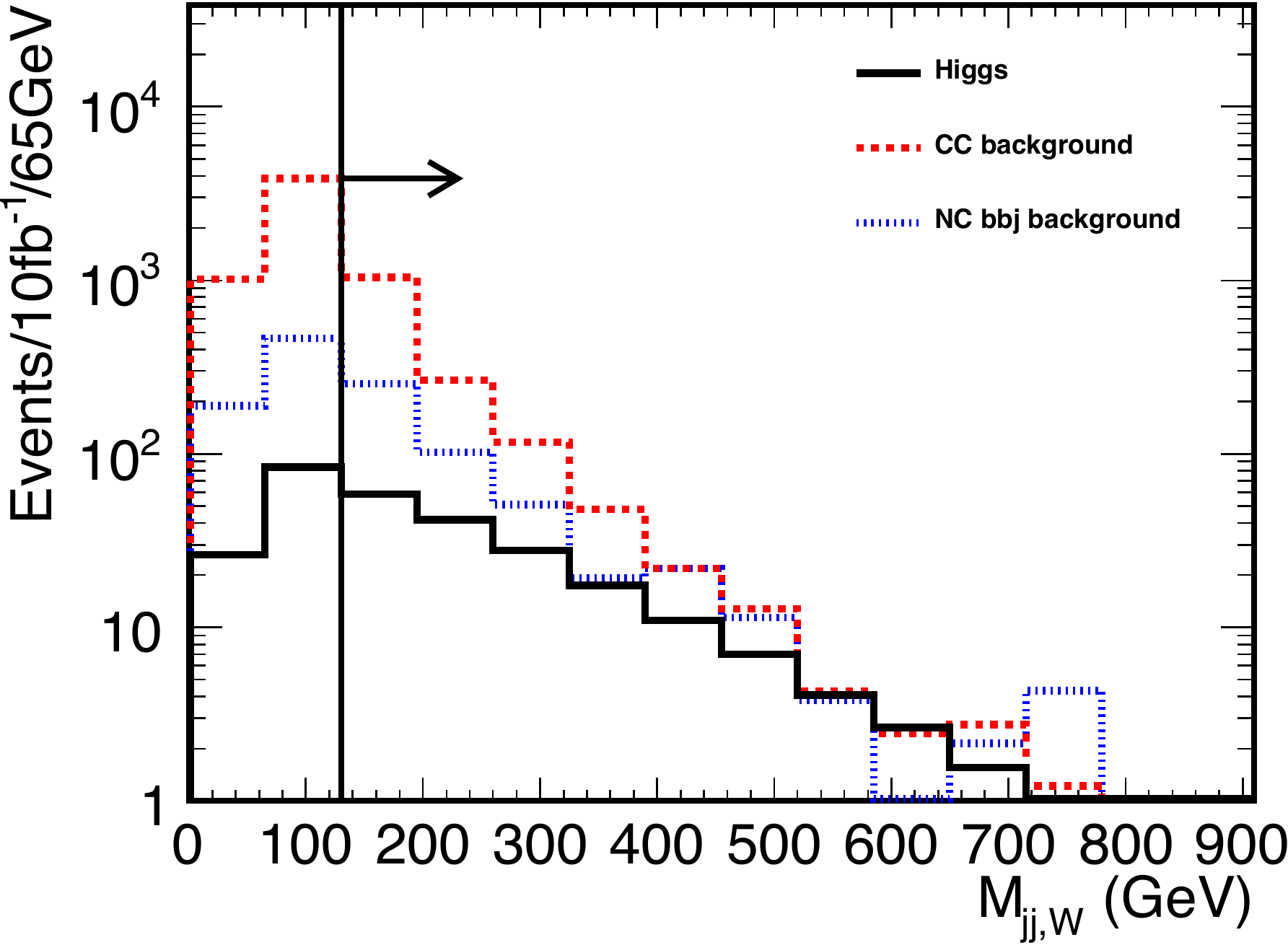}
\caption{Invariant mass distributions for (left) three-jet, $M_{jjj, top}$,  and (right) di-jet, $M_{jj, W}$, candidates.    The solid (black), dashed (red) and dotted (blue) histograms show the $H \rightarrow b \bar{b}$ signal events, and the  CC and NC DIS background events, respectively.}
\label{fig:higgs_cut2}
\end{center}
\end{figure}
\begin{figure}[htbp]
\begin{center}
\includegraphics[width=0.55\columnwidth]{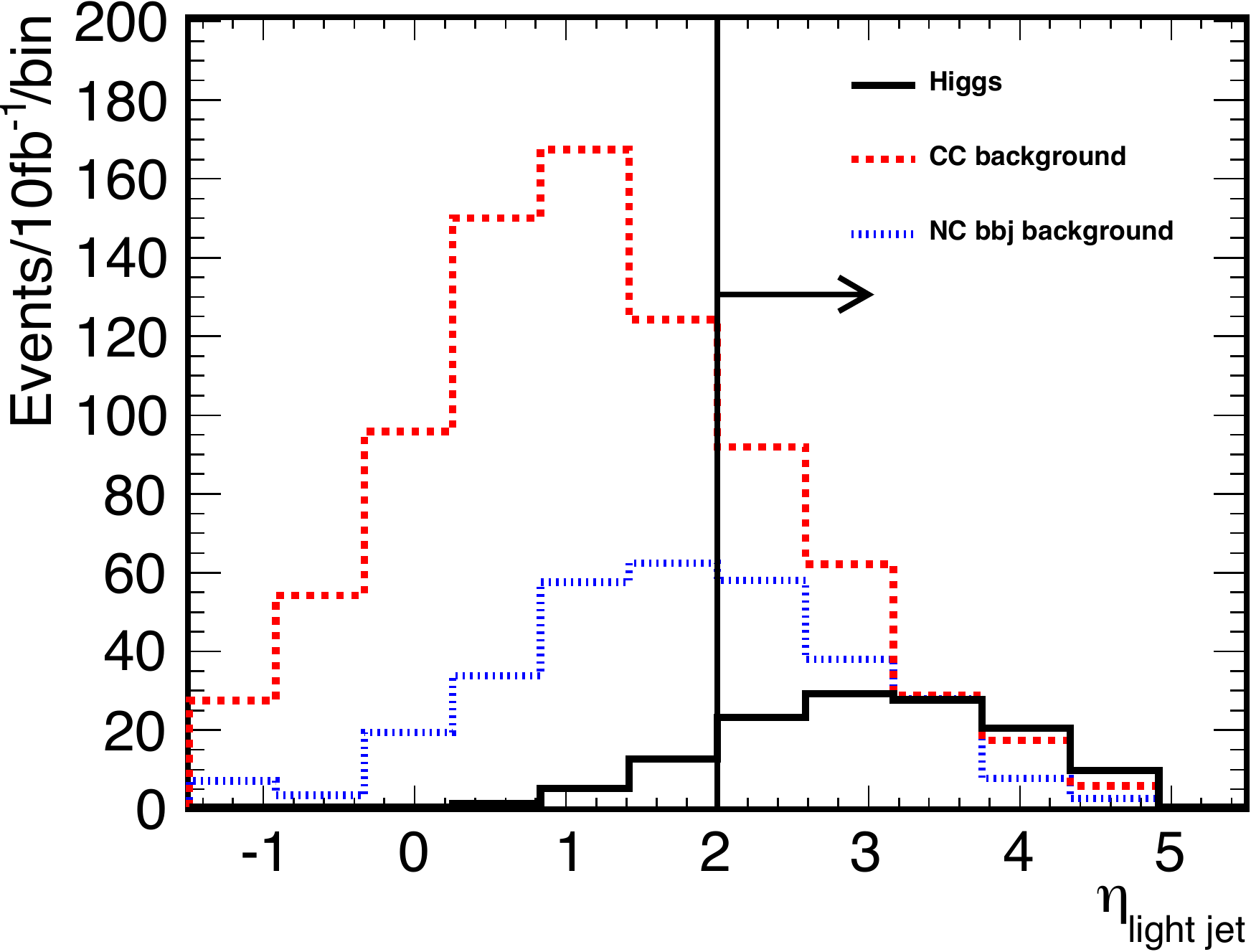}
\caption{ Jet pseudorapidity, $\eta_{jet}$, distribution for the lowest-$\eta$ jet excluding the two $b$-tagged jets used for the reconstruction of the Higgs boson candidate.
The solid (black), dashed (red) and dotted (blue) histograms show the $H \rightarrow b \bar{b}$ signal events, and the  CC and NC DIS background events , respectively.}
\label{fig:higgs_forwjet}
\end{center}
\end{figure}
\begin{figure}[htbp]
\begin{center}
\includegraphics[width=0.55\columnwidth]{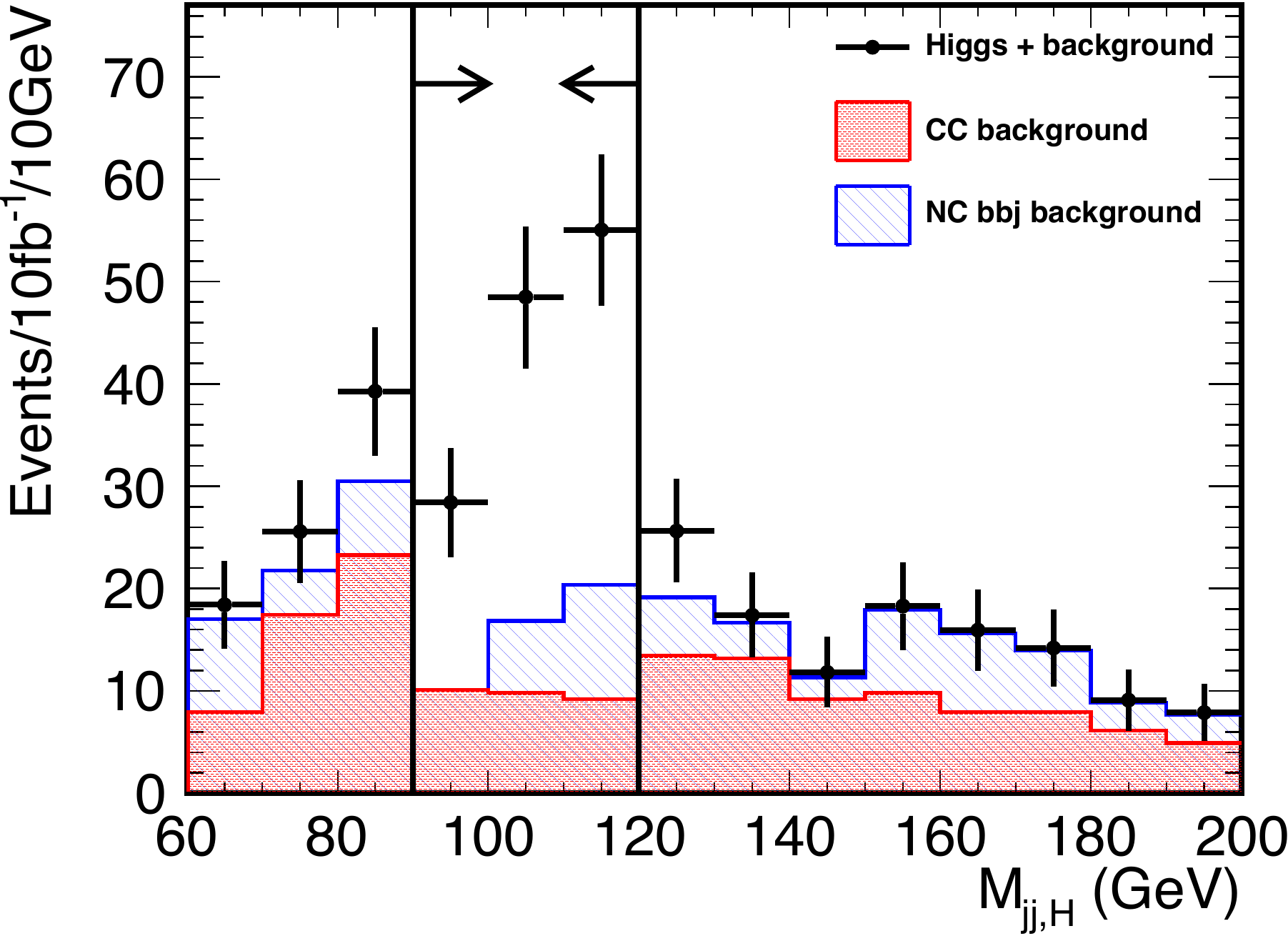}
\caption{Reconstructed invariant Higgs boson mass after all selection criteria, except for the invariant mass cut, have been applied. Points with error bars (black) show the $H \rightarrow b \bar{b}$ signal added to the CC (red histogram) and NC (hatched blue histogram) DIS background for an integrated luminosity of 10\,fb$^{-1}$.}
\label{fig:higgs_final_mass}
\end{center}
\end{figure}
Fig.~\ref{fig:higgs_cut2} shows the reconstructed three-jet ($M_{jjj, top}$) and di-jet
 ($M_{jj, W}$) invariant mass distributions after cuts (1) and (2) are applied. 
 For the simulated CC multi-jet
background, the former distribution peaks at the top mass and the latter one peaks at the $W$ mass.
The final cut is motivated by the fact that the jet from a light quark participating in the CC reaction
for the signal-type events is kinematically boosted to forward rapidities (in the proton beam direction),
as shown in Fig.~\ref{fig:higgs_forwjet}.

Fig.~\ref{fig:higgs_final_mass} shows the reconstructed Higgs mass distribution 
for an integrated luminosity of $10\,{\rm fb^{-1}}$,  after all selection criteria 
except for the final invariant mass cut have been applied.
The results are summarised in Table~\ref{tab:higgs_selection}.
After the selection, 85 $H \rightarrow b \bar{b}$ events are expected for 
$10$\,fb$^{-1}$ luminosity with a $150$\,GeV electron beam.
The signal to background ratio is $1.79$ and the significance 
of the signal $S/\sqrt{N}$ is $12.3$.
\begin{table*}[htb]
\begin{center}
 \begin{tabular}{|c||c|c|c|c|c|}
 \hline
     &  Higgs production & CC DIS  &  NC $bbj$  & $S/N$ & $S/\sqrt{N}$   \\ 
 \hline
 cut (1) & 816  & 123000  & 4630 & $6.38\times10^{-3}$ &  2.28  \\
cut (1) to (3)  & 178  & 1620  & 179  & $9.92\times 10^{-2}$  &   4.21  \\
All cuts  & 84.6  & 29.1  & 18.3   & 1.79  & 12.3  \\
 \hline \hline
 \end{tabular}
 \caption{ \label{tab:higgs_selection} Expected $H \rightarrow b \bar{b}$ signal and 
background events with 150\,GeV electron beam for an integrated luminosity of 
$10$\,fb$^{-1}$. Contents of the cuts are listed in text.}
\end{center}
\end{table*}
For a higher Higgs boson mass, $m_H=150$\,GeV, which is already excluded
for SM couplings, the production cross 
section and the $b\bar b$ branching ratio both decrease. 
The expected number of signal events becomes $25$ and
$S/N$ and $S/\sqrt{N}$  are $0.52$ and $3.60$, respectively.

Promising results are also obtained with a $60$\,GeV electron beam, for an integrated 
luminosity of $100$\,fb$^{-1}$:
for a $120$\,GeV, SM Higgs boson, $250$ $H \rightarrow b \bar{b}$ signal events 
are expected after the same cuts have been applied.
Considering the CC and NC DIS background, the $S/N$ and $S/\sqrt{N}$  
are $1.05$ and $16.1$, respectively.

The results shown here are consistent with the recent parton-level study 
in~\cite{Han:2009pe} for the signal. That study 
does not include e.g. b-quarks produced in the parton showering and thus overestimates
the signal-to-background expectation, by an estimated
 about a factor of five.
The estimation of the background rejection remains subject to large uncertainties 
and is sensitive to details of the detector design where the hadronic energy 
resolution and the b-tagging capabilities are crucial.
As mentioned above, the  large NC background cross section at 
forward lepton scattering angles (low $Q^2$ values) makes it very challenging to simulate a sufficient 
number of events to limit the Monte Carlo statistical uncertainty using the current set-up. 
A conservative estimate of the background evaluation presented here, where 
only events with at least two b quarks have been simulated, indicates an uncertainty 
of about a factor 3. With a full simulation, it can be expected to 
become negligible when the true measurement is realised. Neglecting
therefore this source of uncertainty, the expected dominant systematic errors , 
 besides theoretical estimates of signals and backgrounds 
which may improve, also with LHeC QCD measurements, are
from
instrumental effects, related to the efficiency and acceptance of lepton and jet
reconstruction (hadronic energy resolution) 
and b-tagging and mistagging probabilities. These are
difficult to estimate without real data or a more realistic
detector simulation. The
statistical uncertainty on the cross section can, however, be
estimated: $15$\,\% for the  case of $150$\,GeV $\times$ $7$\,TeV beams
and a Higgs of mass $120$\,GeV, which reduces to $6$\,\% for the
default $60$\,GeV electron beam energy scenario.
 This measurement represents a direct determination of 
$g^2_{Hbb} \cdot g^2_{HWW} / \Gamma_H$,
where $g_{Hbb}$ and $g_{HWW}$ denote the $Hbb$ and $HWW$ couplings and
$\Gamma_H$ is the total width of the Higgs.

In addition to providing a constraint on the $Hbb$ coupling, this measurement,
combined with the measurements of (products of) couplings expected from
the LHC~\cite{ThesisOfDuehrssen},
would also provide an interesting consistency check of the $HWW$ coupling.
However, this extraction requires a few assumptions
to be made, in particular relating the $HZZ$ and $HWW$ couplings.
The LHeC provides the unique opportunity to select experimentally the $HWW$ coupling
in Higgs production via weak boson fusion, in contrast to WBF production at the LHC
where the contributions from the $HZZ$ and $HWW$ couplings can not be disentangled.
Hence the LHeC could probe the $HWW$ coupling without any assumption on the $HZZ$ coupling.
This is of particular interest since these couplings could receive sizeable anomalous
contributions from physics beyond the Standard Model. This possibility is further
explored in the following.

%% file: physics/bsm_HWW_azimuthal.tex
\subsection{Probing anomalous $HWW$ couplings at the LHeC}
A measurement of the $HWW$ vertex provides  insights into the nature
 of the coupling of a scalar field to a heavy vector boson.
Parameterising the $H(k)-W^+_\mu(p)-W^-_\nu(q)$ 
vertex in the form $i\Gamma^{\mu\nu}(p,q)\ \epsilon_\mu(p)\ 
\epsilon^\ast_\nu(q)$, any deviations from the simple SM formula
$\Gamma_{\rm (SM)}^{\mu\nu}(p,q) = gM_W\,g^{\mu\nu}$
at a level incompatible with SM loop corrections would immediately 
indicate the presence of new physics. Following Ref.~\cite{Plehn:2001nj},
 these deviations can be parameterised using two dimension-5 operators
\begin{equation}
\Gamma^{\rm (BSM)}_{\mu\nu}(p,q) = \frac{-g}{M_W}\left[ 
\lambda \left(p.q\, g_{\mu\nu} - p_\nu q_\mu  \right)
+ i\, \lambda^\prime\ \epsilon_{\mu\nu\rho\sigma}p^\rho q^\sigma
\right]
\label{eqn:BSMvertex}
\end{equation}
where $\lambda$ and $\lambda^\prime$ are, respectively, effective coupling 
strengths for the $CP$-conserving and the $CP$-violating parts. 

An unambiguous determination of the CP property of the Higgs boson, 
particularly to test if it is a CP eigenstate, should optimally employ 
its coupling to the heavy fermions, 
mainly via $H t \bar t$ production~\cite{Godbole:2011hw}.
Similarly one may use
 the $HVV$ coupling which is expected to be more easily accessible.
The above parameterisation of anomalous $HWW$ and similar couplings illustrates 
the important point that the CP properties of the Higgs boson are rather 
difficult to measure directly. 
Information on the couplings $\lambda$ and $\lambda'$ to any degree of
certainty can throw light on the CP property of the Higgs.
Several suggestions have been made on how this 
can be done at colliders, using angular correlations  between the 
final state particles as well as other kinematic 
quantities~\cite{Godbole:2004xe,Accomando:2006ga}. An additional 
complication arises, however, because most of the 
observables studied so far in the context of the LEP, Tevatron
and LHC machines are dependent on more than one of these 
couplings~\cite{Hankele:2006ma}, barring the case of $HZZ$ coupling. 
At the $e^+e^-$ colliders the Higgstrahlung process and at the LHC the
decay $H \rightarrow Z Z^{(*)}$ offer the chance to study the same
quite cleanly.
If the 'hints' for a light Higgs should be confirmed, the 
$H \rightarrow  ZZ^{*}$
would offer a chance to establish the CP property of the Higgs if it is a CP
eigenstate and possibility to explore the anomalous $HZZ$ coupling ~\cite{Godbole:2007cn,DeRujula:2010ys};
the case for  the $HWW$ vertex may be less clear though.
Further, even at the ILC, 
a determination of an anomalous $HWW$ vertex will still be contaminated by 
the $HZZ$ vertex~\cite{Biswal:2005fh}.

An $ep$ collider has a unique 
advantage in the fact that the $HWW$ vertex gives rise to the process
$e + p \to \nu_e + X + H(b\bar{b})$ through the single Feynman diagram shown 
in Figure~\ref{fig:higgs_diagrams}(left), with no "pollution" from the $HZZ$ coupling. 
Other advantages, with respect to the $pp$ environment, include:
\begin{itemize}
\item Very good signal to background ratio, see Sect.~\ref{sec:bsm_higgs_observability}.
\item The Higgs boson signal does not have contamination from other production mechanisms, such as gluon-gluon fusion.
\item As opposed to the LHC,  at LHeC the forward and backward directions 
can be disentangled because the
direction of the missing neutrino and the struck quark, respectively,
is well defined, a feature which could be exploited in further studies.
\item Since $ep$ cross sections are much smaller than those in $pp$,
even at maximum luminosity there shall be no pile-up of events
which deteriorates the event selection, resolution and missing energy 
reconstruction in $pp$.
\end{itemize}

The final state has
missing transverse energy (MET) and three jets $J_1$, $J_2$ and $J_3$, 
of which two (say $J_2$ and $J_3$) are tagged as $b$-jets. It can be 
shown~\cite{Plehn:2001nj} that in the limit when there is practically no energy transfer 
to the $W$ boson and the final states are very forward, 
the $CP$-conserving ($CP$-violating) coupling $\lambda$ ($\lambda^\prime$) 
contributes to the matrix element for this process a term of the form
which goes through zero when the missing transverse momentum is perpendicular to the
$p_T$ of the jet:
\begin{equation}
{\cal M} \sim +\lambda \, \not{\!\vec{p}}_T.\vec{p}_T^{J_1}
\qquad\qquad
\widetilde{\cal M} \sim -\lambda^\prime \, \not{\!\vec{p}}_T.\vec{p}_T^{J_1} \ .
\end{equation}

\begin{figure}
\begin{center}
\includegraphics[width=0.6\columnwidth]{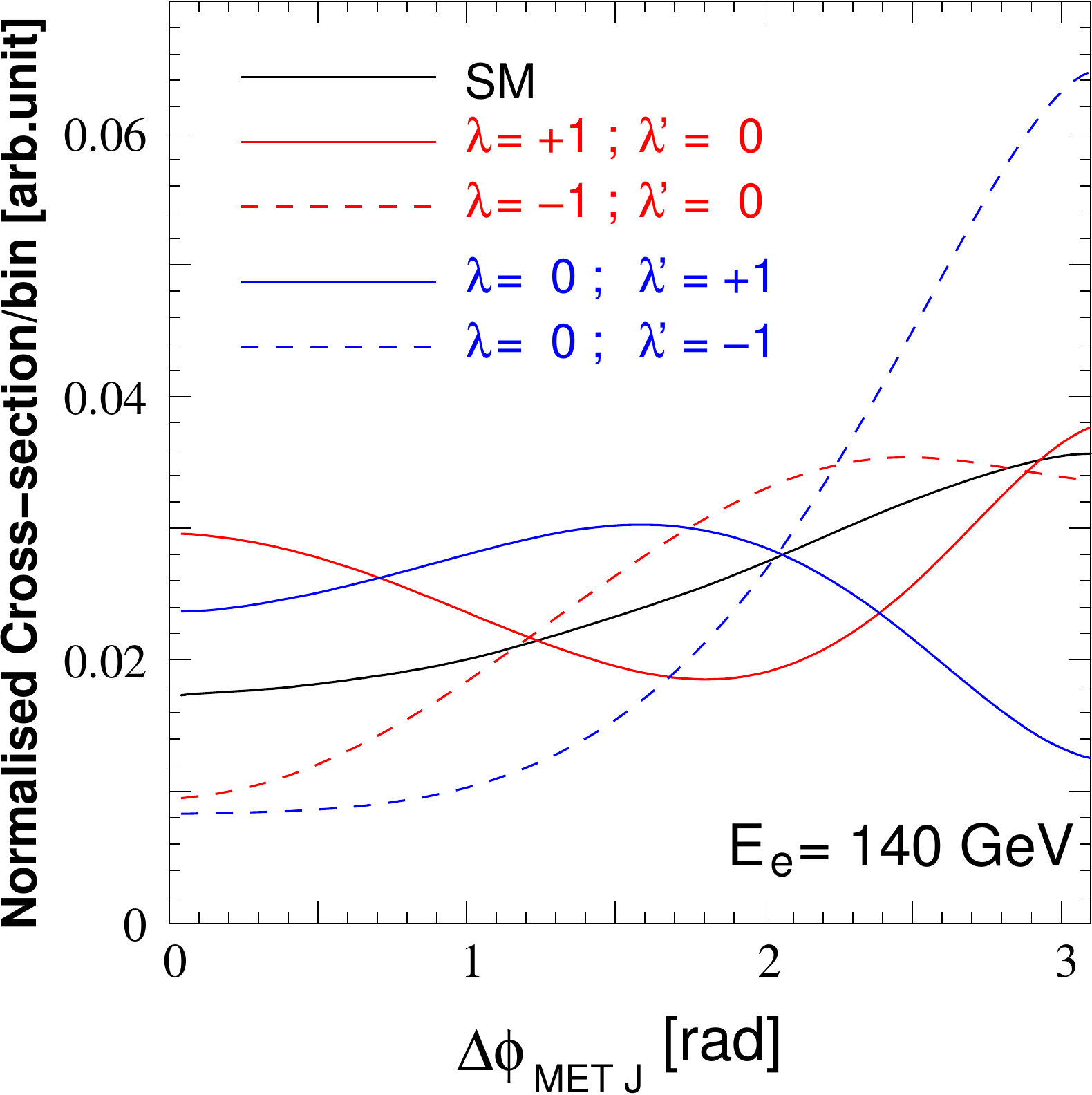}
\caption{ Illustrating the SM distribution in azimuthal
angle and deviations therefrom which are due to anomalous $HWW$ couplings.}
\label{fig:HWW_theory}
\end{center}
\end{figure}

This explains the general trend  illustrated  in 
Figure~\ref{fig:HWW_theory}, for an exact calculation of the 2 $\to$ 3 process 
$eq \to \nu_e q' H$ at the parton level, with parton density functions from the CTEQ-6L1 set~\cite{Pumplin:2002vw}.
In the case considered, 140~GeV electrons collide with
7~TeV protons and the Higgs boson mass is set to 120 GeV.

\begin{figure}[h]
\begin{center}
\includegraphics[width=\columnwidth]{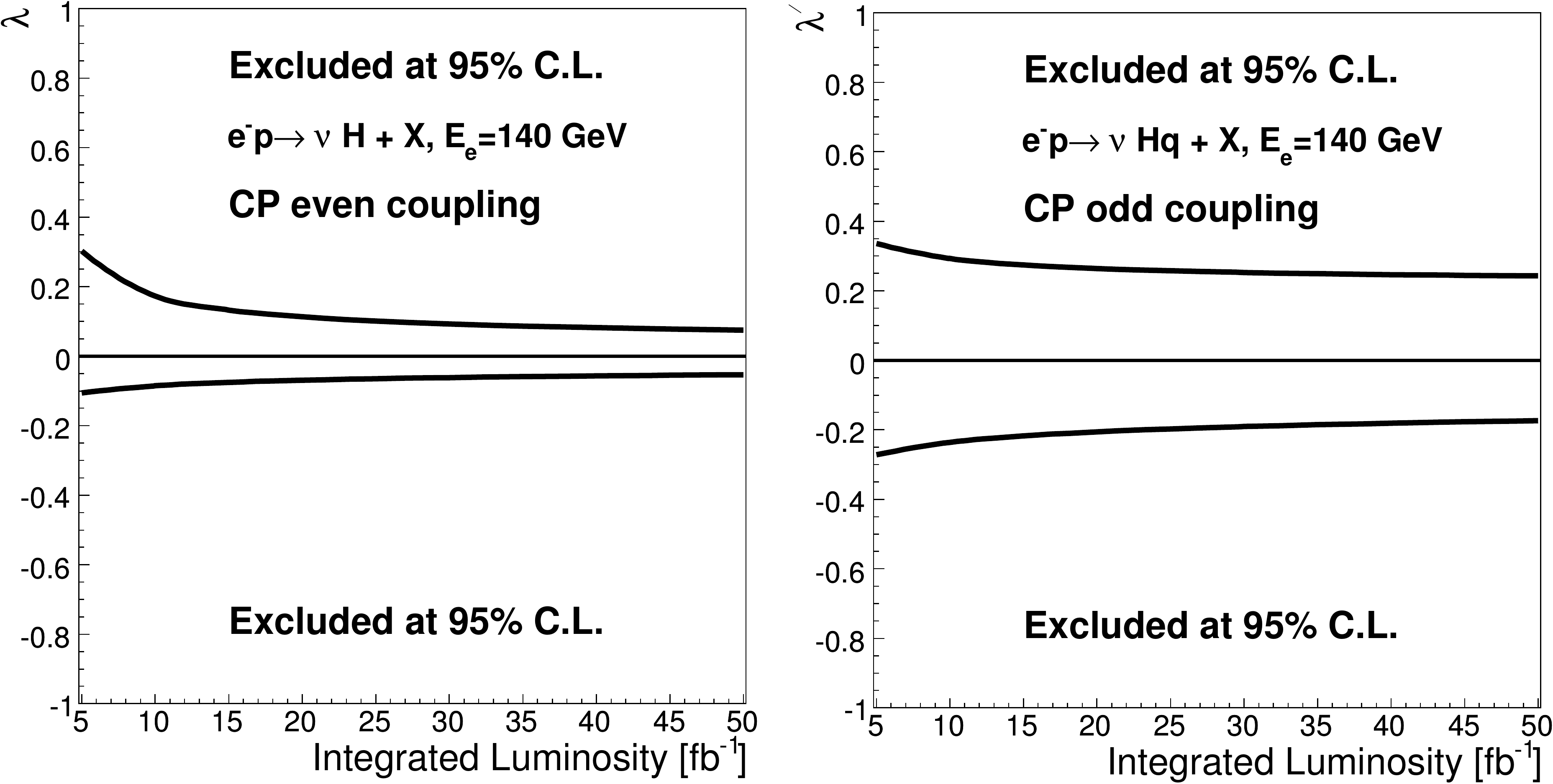}
\caption{Exclusion plots obtainable by a
study of the azimuthal angle distributions at the LHeC for the 
$CP$-even coupling $\lambda$ and the $CP$-odd coupling $\lambda^\prime$ . Note that this study is for $M_H = 120$ GeV.}
\label{fig:HWW_simul}
\end{center}
\end{figure}

The analysis is based on the kinematic cuts and efficiencies adopted 
in~\cite{Han:2009pe}. The azimuthal distribution has been simulated in $10$
equidistant bins and the 
signal and SM backgrounds have been calculated in each bin using the 
same formulae used to create Figure~\ref{fig:HWW_theory}, followed by
a detailed simulation of fragmentation, jet identification and detector
effects. In addition, the number of expected background events has been  varied according to the values reported in Sect.~\ref{sec:bsm_higgs_observability}. Assuming statistical
errors dependent on the integrated luminosity $L$, the sensitivity, 
for a given $L$, of the experiment to $\lambda, \lambda^\prime$
is determined with a log-likelihood analysis. 
The results are shown in 
Figure~\ref{fig:HWW_simul}, where a 95\% exclusion limit is indicated for the
$\lambda$ and $\lambda^\prime$ couplings as a function of $L$. It is 
clear from this figure that by the time the LHeC has collected 10~fb$^{-1}$
of data,  anomalous $HWW$ couplings to the level of
0.3 or lower could be excluded. The experimental set-up is somewhat more sensitive to the 
$CP$-even coupling, as evidenced by the narrower inaccessible region as shown in Fig.~\ref{fig:HWW_simul}, left. 
This study is further detailed in Ref.~\cite{Biswal:2012mp}. 

While keeping the energy of the proton beam constant the acceptance increases slightly for electron beam energies above 100\,GeV. For energies below 100\,GeV the loss of acceptance becomes significant. The acceptance of the Higgs boson signal for 50\,GeV decreases by 25\% with respect to that of 100\,GeV. Most of the acceptance loss stems from the requirement of two $b$-tagged jets. Part of the acceptance can be recovered by extended the tracking and calorimeter coverage  
further into the forward direction.

\vspace{0.4cm}
To summarise, the LHeC in its configuration as presented in this report allows for
important investigations of  the Higgs boson, its $HWW$, $Hbb$ couplings and CP properties. The initial studies presented here are to be pursued further as the design of the apparatus and its simulation proceed. Clearly, the study of light Higgs boson properties demands excellent detector capabilities such as efficient b-tagging, missing energy reconstruction and very good hadronic
energy resolution, better than the $60$\,\% considered here and linked to the
tracker. With a luminosity enhancement to the $10^{34}$\,cm$^{-2}$s$^{-1}$
level, the LHeC can become a Higgs machine of striking potential. Therefore,
if a light Higgs is confirmed then the LHeC design, of both machine and detector,
will to an important extent be steered towards an optimum investigation
of the electroweak symmetry breaking mechanism.

%% file: machine/ring.tex
\chapter{Ring-Ring Collider}
\input{machine/rintro}
\input{machine/fitterer}
\clearpage
\input{machine/holzer_thompson}
\input{machine/bernard_rr}

\input{machine/herr}

\input{machine/eAJowett}

\input{machine/barber_wienands}
\input{machine/mess}

\input{machine/Burkhardt}

%% file: machine/rintro.tex
\section{Baseline parameters and configuration}
Intense electron-proton beam interactions in the LHC tunnel can be realised with an electron storage ring and the LHC proton beams, as has been discussed already at the Lausanne workshop back in 1984.  This solution was revived~\cite{Dainton:2006wd} when it had been seen that a hundred fold higher luminosity
can be achieved than with HERA,  owing to the intense proton beams available with the LHC. With an electron beam energy 
set between about $50$ and $100$\,GeV and the $7$\,TeV proton beam energy one can realise a new $ep$ collider
of cms energy, $\sqrt{s} = 2 \sqrt{E_e E_p}$ beyond $1$\,TeV. The advantages of a ring-ring (RR)
configuration are that one uses known technology, with much experience from HERA and LEP, and that
intense beams of both lepton charges can be generated without technical difficulty.

For the present design study, the electron beam energy has been set to $60$~GeV as discussed above, Sect.~\ref{sect:elener}. With extra efforts and higher investments one may double that energy, as had been achieved for LEP~\cite{lep2design}, should there be strong physics requirements. One yet has to consider that power losses vary $\propto E_e^{-4}$ and much higher synchrotron radiation occurs, which causes the operation and technical conditions to be increasingly demanding as $E_e$ increases. A $60$~GeV the $e^{\pm}$ beam may be polarised while, following the calculations presented below, that becomes questionable when $E_e$ increases.

Due to the smallness of the $ep$ tune-shift, synchronous $pp$ and $ep$ interactions can be realised with the LHC and the LHeC. This requires to bypass the active $pp$ experiments with separate tunnels which, in adjacent caverns, can house the RF. Excavation of such tunnels could proceed in parallel to LHC operation, similar to the CMS cavern which was excavated while LEP ran. Due to machine hardware placements or unfortunate geological conditions, none of the 4 machine points (3,4 and 6,7) could house the LHeC interaction region. For the present study IP2 was chosen as the $ep$ IR, currently housing ALICE, and bypasses were considered for ATLAS and CMS.

Maximum luminosity can be achieved with focusing magnets placed close to the interaction point. This limits, however, the polar angle acceptance. Two principal interaction optics solutions have been developed, the high luminosity optics, with acceptance down to about $8^{\circ}$, and the large acceptance optics, covering polar angles down to $1^{\circ}$. As is shown below, there is only a factor of $4$ difference in the product of the $\beta$~functions. It therefore would be possible to only consider the large acceptance solution, avoiding large delays in rearranging the IR as has happened during the HERA luminosity upgrade in 2000-2003. Nevertheless,  both configurations are  documented here, including options of the detector with and without focusing magnets placed close to the interaction point. 

A complete lattice has been designed for the new ring. This takes into account some peculiarities due to the LHC. In particular, an asymmetric FODO cell, of half the LHC FODO cell length, had to be designed to account for LHC service modules and the DFBs. Similarly, a non-standard solution for the dispersion matching had to be developed, using eight individually powered quadrupoles instead of regulating the position of dipoles which is too constrained by the LHC.

A further baseline parameter is the injection energy. The LHeC electron storage ring differs from LEP in its bunch structure. The LHeC has a maximum of about $2 \cdot 10^{10}$ electrons per bunch in a much higher repetition rate than LEP, which had a bunch intensity of $4 \cdot 10^{11}$. The smaller intensity allows to inject directly from a Linac without accumulation, which, in turn, suggests an injection at low energy so that no additional circular injector is required. For the current design a new injector is considered, using linac technology with high frequency cavities, of energy as low as $10$\,GeV. This poses constraints on the quality of the main dipole magnets, which have to ensure a magnetic field reproducibility of about $10^{-4}$. Therefore dipole prototypes had been designed and built: $C$- (and $H$) shape prototype magnets have been developed, built and successfully tested at BINP Novosibirsk. Alternative magnets have also been built and were successfully tested at CERN. Besides the magnetic field properties, attention was given in both designs to small outer dimensions (of about $35\times 35$\,cm$^2$, compared to  $50\times 50$\,cm$^2$ at LEP), and to a reduction of the weight (from $800$\,kg/m at LEP to $260$\,kg/m for the LHeC) in order to facilitate the installation. 
The total number of magnets is in the order of $4000$. Such an amount is large, but it could be obtained within a few years of production time, 
following $1:1$ prototyping within the technical design phase.

The key question for the storage ring is its possible installation in the LHC tunnel without posing too harsh constraints on the LHC operation schedule. A first inspection was made of the various elements of concern, as described below, with the conclusion that installation of the LHeC was possible but very demanding and would take a few years of shutdown of the LHC.
For a TDR of the ring-ring solution, a detailed 3D CAD integration study of both accelerators would be mandatory. 

The subsequent chapter describes the studies dedicated to characterise the RR option. The most important parameters are listed for a better overview in Table~\ref{tabparring}, \ref{tabcompring} and \ref{tabirring}. It is followed by  a similar chapter on the LR option. Much of the system hardware is common or similar and thus it is contained in a following
chapter. From today's perspective both options may be realised within the coming ten years, albeit the differences which distinguish them. It had been part of the referee process to understand the relative merits in terms of physics, technical aspects, operation, infrastructure and future developments. 
The next phase of prototyping and the technical design will be developed for only one of them.

\begin{table}[hbt]
   \centering
   \begin{tabular}{|l|c|}
       \hline
      electron beam $60$ GeV & \\
\hline
$e^-$ ($e^+$)  per bunch $N_e$ [$10^{10}$]  &  $ 1.97~(1.97) $ \\ 
$e^-$ ($e^+$) polarisation [\%]& $40~(40)$ \\
bunch spacing [$\rm{ns}$]& $25$  \\
bunch length [mm]  &  $ 6 $\\ 
transverse emittance at IP $\gamma \epsilon^e_{x,y}$ [ mm] & $ 0.59,~0.29 $  \\ 
beam current [mA] & $ 100 $ \\
total wall plug power [MW] & $100$  \\ 
syn rad power [MW] & $44$  \\ 
\hline
       proton beam 7 TeV & \\
       \hline
protons per bunch $N_p$ [$10^{11}$] & $1.7$        \\  
transverse emittance $\gamma \epsilon^p_{x,y}$ [$\rm{\mu m}$ & $3.75$  \\ 
\hline
   \end{tabular}
   \caption{Parameters of the RR configuration.}
   \label{tabparring}
\end{table}

\begin{table}[hbt]
   \centering
   \begin{tabular}{|l|c|}
       \hline
      magnets  &  \\
\hline
number of dipoles  &  $ 3080 $    \\ 
dipole field [T] & $0.013-0.076$  \\
number  of quadrupoles  &  $ 968 $    \\ 
\hline
      RF and cryogenics &  \\
\hline
number of cavities & $112$  \\
gradient [MV/m]  & $11.9$ \\
cavity voltage  [MV]  & $5$  \\  
cavity $R/Q$ [$\Omega$] & 114  \\ 
cooling power [kW]  & $5.4$@$4.2$ K  \\ 
\hline
   \end{tabular}
   \caption{Components of the electron accelerators.}
   \label{tabcompring}
\end{table}

\begin{table}[hbt]
   \centering
   \begin{tabular}{|l|c|c|}
       \hline
     &  HA & HL    \\
       \hline
      electron beam $60$ GeV & &\\
\hline
IP $\beta$ function $\beta^*_{x,y}$ [m] & $ 0.4,~0.2 $ & $ 0.18,~0.1 $\\
syn rad power (interaction region) [$\rm{kW}$] &  $ 51 $  &  $ 33 $\\
critical energy [$\rm{keV}$] &  $ 163 $ &  $ 126 $\\
\hline
       proton beam 7 TeV & & \\
       \hline
IP $\beta$ function $\beta^*_{x,y}$ [m] & $ 4.0,~1.0 $ & $ 1.8,~0.5 $\\
\hline
      collider & & \\
\hline
Lum $e^-p$ ($e^+p$) [$10^{32}$cm$^{-2}$s$^{-1}$]& $ 9~(9) $ & $ 18~(18) $ \\
rms beam spot size $\sigma_{x,y}$ [$\rm{\mu m}$] & $ 45,~22 $ & $ 30,~16 $\\ \hline
crossing angle $\theta$ [mrad] & \multicolumn{2}{c|}{$ 1 $ }  \\\hline
$L_{ep}(\theta)$ [$10^{32}$cm$^{-2}$s$^{-1}$] & $ 7.3~(7.3) $ & $ 13~(13) $ \\\hline
$L_{eN}=A~L_{eA}$ [$10^{32}$cm$^{-2}$s$^{-1}$] & \multicolumn{2}{c|}{$ 0.45 $ }\\
\hline
   \end{tabular}
   \caption{Parameters of the RR interaction region.}
   \label{tabirring}
\end{table}

\newpage

%% file: machine/fitterer.tex
\section{Geometry}
\label{lat}
All lattice descriptions in this chapter are based on LHeC lattice Version 1.1.
\subsection{General layout}
\label{lat:1}
The general layout of the LHeC consists of eight arcs, six straight sections and two bypasses around the experiments in Point 1 and Point 5. The e-p collision experiment is assumed to be located in Point 2, the only interaction point of the beams. All straight sections except those in the bypasses have the same length as the LHC straight sections: $538.8 \ \mathrm{m}$ at even points and $537.8 \ \mathrm{m}$ at odd points.

\begin{figure}[hb]
\centerline{\includegraphics[clip=,width=1.0\textwidth]{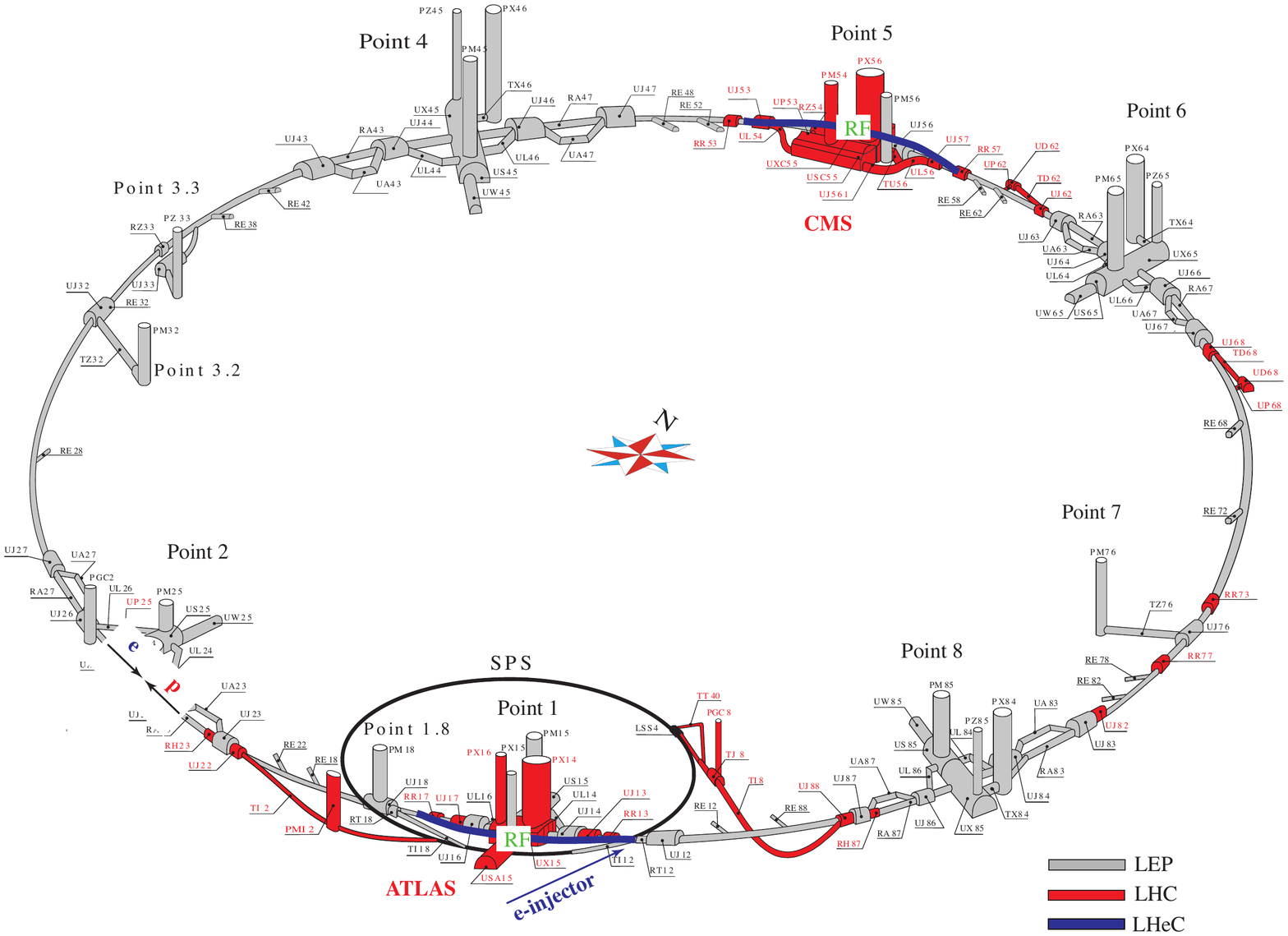}}
\caption{Schematic Layout of the LHeC: In grey the LEP tunnel now used for the LHC, in red the LHC extensions. The two LHeC bypasses are shown in blue. The RF is installed in the central straight section of the two bypasses. The bypass around Point 1 hosts in addition the injection.} \label{lat:fig:1:1}
\end{figure}

The insertions shared with the LHC are already used for the experiments or for LHC equipment. Therefore the RF for the electron ring is installed in the straight sections of the bypasses (see Section $\ref{RR-RF-section}$). For the same reason the beam is injected in the bypass around Point 1. Point 1 is preferred over Point 5 for geological and infrastructural reasons. The overall layout of the LHeC is shown in Fig. \ref{lat:fig:1:1}.

\subsection{Electron ring circumference and e-p synchronisation}
\label{lat:2}
The LHeC electron beam collides only in one point (Point 2) with the protons of the LHC. This leaves the options to either exactly match the circumferences of the proton and electron rings or to allow a difference of a multiple of the LHC bunch spacing. In the case of different circumferences the proton beam could become unstable due to beam-beam interactions with the electrons \cite{Hirata1990156}, \cite{myersbb}. To avoid this possible effect in the LHeC, the electron ring circumference is matched exactly to the proton ring circumference.

The circumference can be adjusted in two ways:
\begin{enumerate}
\item Different bypass designs, e.g. inner and outer bypass, which compensate each other in length.
\item Radial displacement of the electron ring to the inside or outside of the LHC in the places where the two rings share the same tunnel to compensate for the path length difference caused by the bypasses.
\end{enumerate}
The various design possibilities for the bypasses are discussed in Sec. \ref{lat:4}. Considering their characteristics, the best choice seems to be outer bypasses around both experiments.

In general synchronisation between the e- and p-beam could arise from small differences in the circumferences of the central orbits. Both beams could be synchronised by adjusting the RF frequency of the electron or proton beam accordingly \cite{jowettlhec2009}. The feasibility of this method was demonstrated with proton lead in the LHC \cite{jowettpAsynch} and also for electrons and protons in Hera \cite{herasync}.


\subsection{Idealised ring}
\label{lat:3}
In the following the average between LHC Beam~1 and Beam~2 is taken as reference geometry for the LHC. 
\subsubsection{General layout}
\label{lat:3:1}
To compensate the path length differences from the bypasses, the electron ring is placed on average 61~cm to the inside of the LHC in the sections where both rings share the tunnel. For this a complete ring with an ideally constant radial offset of  61~cm to the LHC was designed. In the following we refer to this ring as the \emph{Idealised Ring}.

\begin{figure}[hb]
\centerline{\includegraphics[clip=,width=0.7\textwidth]{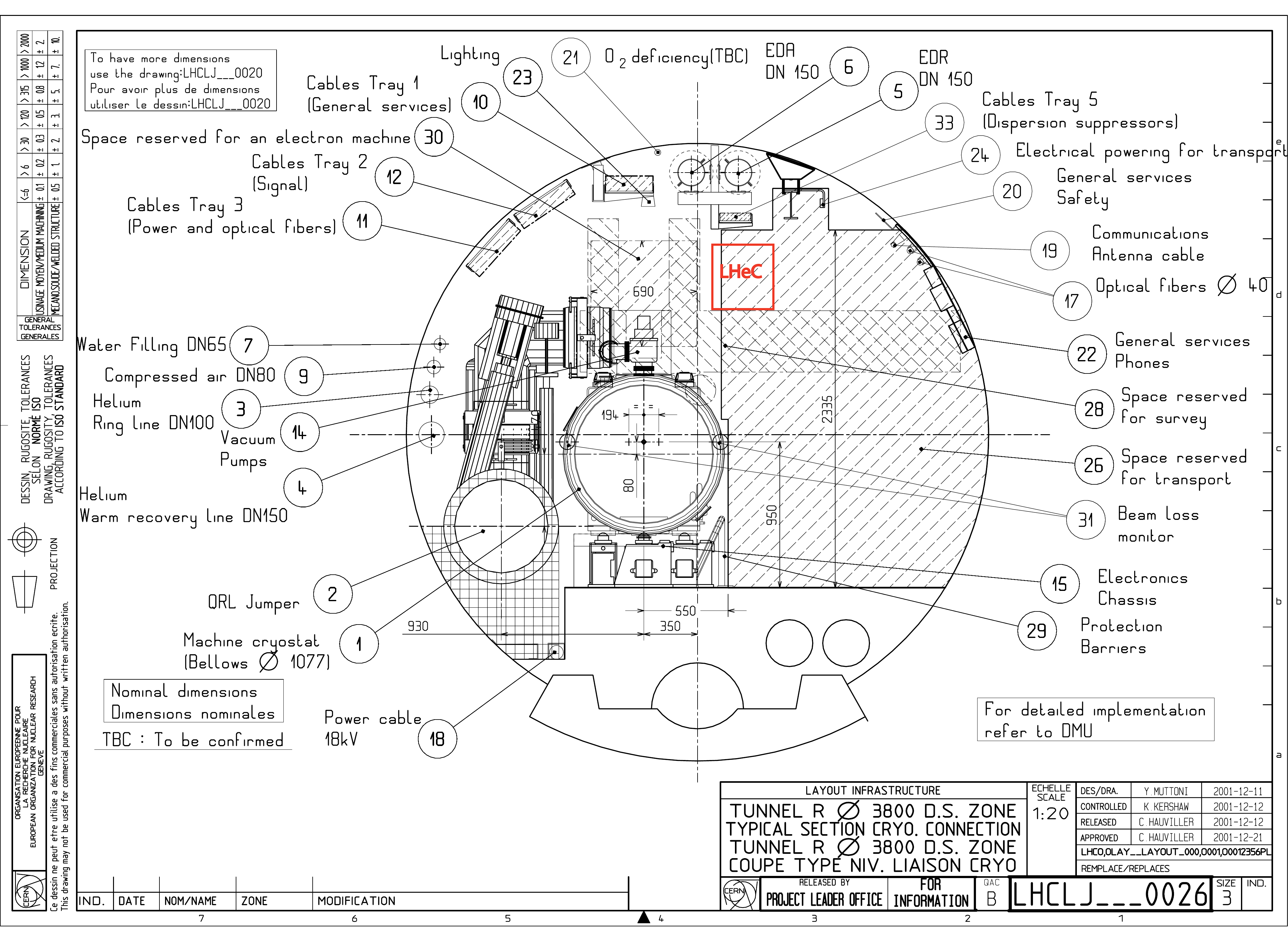}}
\caption{Representative cross section of the LHC tunnel. The location of the electron ring is indicated in red.}\label{lat:fig:1:2}
\end{figure}

In addition to the horizontal displacement, the electron ring is set  1~m above the LHC in order to minimise the interference with the LHC elements. The main remaining conflict in the arc are then the service modules as shown in Fig. \ref{fg:mess_LHCtunnel3_4} and the DFBs in the insertions (see Section $\ref{sec:Space requirements}$). A representative cross section of the LHC tunnel is shown in Fig. \ref{lat:fig:1:2}.

In the main arcs the service modules have a length of 6.62~m and are installed at the beginning of each LHC arc cell. The insertions host a different number of DFBs with a varying placement and length. The idealised ring lattice is designed to avoid overlaps of magnet elements with all service modules in the main arcs. In order to show that it is possible to design an optics with no e-ring elements at any DFB positions in the insertions, the dispersion suppressors of the even and odd insertions were adapted to the DFB positions and lengths in IR2 and IR3 respectively. For simplicity all straight sections are filled with a regular FODO cell structure. 

\subsubsection{Geometry}
\label{lat:3:2}
To adjust the beam optics to the regular reappearance of the service modules at the beginning of each LHC arc cell it was suggested to use a multiple $n$ or sub-multiple $1/n$ ($n \in \mathbb{N}$) of the LHC arc cell length as LHeC FODO cell length. Beside the integration constraints, the cell has to provide the right emittance. Taking half the LHC arc cell length as LHeC FODO cell length already fulfils this second criterion (Sec. \ref{opt:1}). 

\begin{figure}[hb]
\centerline{\includegraphics[clip=,width=0.7\textwidth]{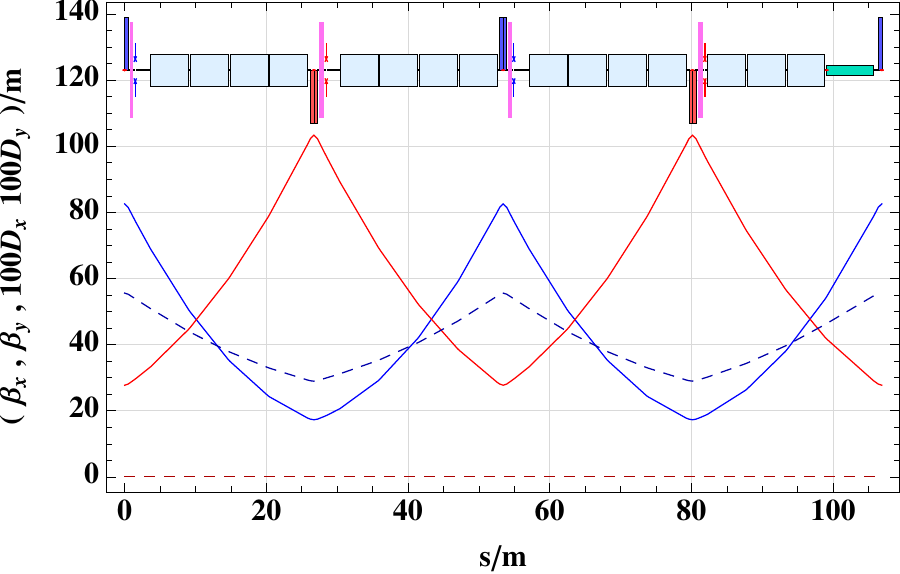}}
\caption{Electron ring arc cell optics. One arc cell consists of two FODO cells symmetric in the placement of the quadrupoles and asymmetric for the dipoles.}\label{lat:fig:3:2:1}
\end{figure}

As the LHC arc cell is symmetric, the best geometrical alignment with the LHC main arc would be achieved, if the LHeC cell also had a symmetrical layout. Because of the service modules, no elements can be placed in the first 6.9~m of two consecutive cells. If all cells had the same layout, another 6.9~m would be lost in the second FODO cell. This would result in additional unwanted synchrotron radiation losses as the energy loss in a dipole magnet is proportional to the inverse length of the dipole
\begin{equation}\label{lat:eqn:3:2:1}
U_{\mathrm{dipole}}=\frac{C_{\gamma}}{2\pi}E_0^4\frac{\theta^2}{l} \ , \ C_{\gamma}= \frac{4\pi}{3} \frac{r_e}{(m_e c^2)^3}
\end{equation}
where $\theta$ is the bending angle, $l$ the length of the dipole and $E_0$ the beam energy. In order to avoid this, the LHeC arc cell is a double FODO cell, symmetric in the positioning of the quadrupoles but asymmetric in the placement of the dipoles (Fig. \ref{lat:fig:3:2:1}). 

The bending angle in the arc cells and also in the DS is determined by the LHC geometry. In the following we refer to the LHC DS as the section from the end of the arc to the beginning of the LSS. With this definition the LHC DS consists of two cells. Keeping the same conversion rule as in the arc (one LHC FODO cell corresponds to two LHeC FODO cells), the LHeC DS would then ideally consist of 4 equal cells. For consistency the ratio between the LHeC DS and arc cell lengths is the same as between the LHC DS and arc cell. For the LHC this ratio is $2/3$. This leaves the following choices for the number of dipoles in the arc and DS cell:
\begin{equation}\label{lat:eqn:3:2:2}
 N_{\textnormal{Dipole, arc cell}}=\frac{3}{2} N_{\textnormal{Dipole, DS cell}}=3,6,9,12,15 \ \ldots
\end{equation}
A good compromise between a reasonable dipole length and optimal use of the available space for the bending is $15$ dipoles per arc cell. The dipoles are then split up in packages of $3+4+4+4$ in one arc cell and $2+3$ in one DS cell. 

Beside the bending angle, the module length of the electron ring has to be matched to the LHC geometry. As the electron ring is radially displaced to the inside of the proton ring, all e-ring modules are slightly shorter than their proton ring equivalents (Table \ref{lat:tab:3:2:1}).

\begin{table}[h]
  \centering
  \begin{tabular}{|l|c|c|}
    \hline
 & Proton Ring & Electron Ring \\ \hline \hline
Arc Cell Length & $106.9 \ \mathrm{m}$ & $106.881 \ \mathrm{m}$\\ \hline
DSL Length (even points) & $172.80 \ \mathrm{m}$ & $172.78 \ \mathrm{m}$ \\ \hline
DSR Length (even points) & $161.60 \ \mathrm{m}$ & $161.57 \ \mathrm{m}$ \\ \hline
DSL Length (odd points) & $173.74 \ \mathrm{m}$ & $173.72 \ \mathrm{m}$ \\ \hline
DSR Length (odd points) & $162.54 \ \mathrm{m}$ & $162.51 \ \mathrm{m}$ \\ \hline
  \end{tabular}
\caption{Proton and Electron-Ring Module Lengths. DSL=Dispersion Suppressor Left side, DSR=Dispersion Suppressor Right side}
\label{lat:tab:3:2:1}
\end{table} 

The above considerations already fix the bending angle of the dipoles, which leaves only position and length as free parameters. Ideally the dipole length would be chosen as long as possible, but because of the asymmetry of the arc cell, the dipoles have to be shortened and moved to the right in order to fit the LHC geometry.

\begin{figure}[hb]
\centerline{\includegraphics[clip=,width=0.7\textwidth]{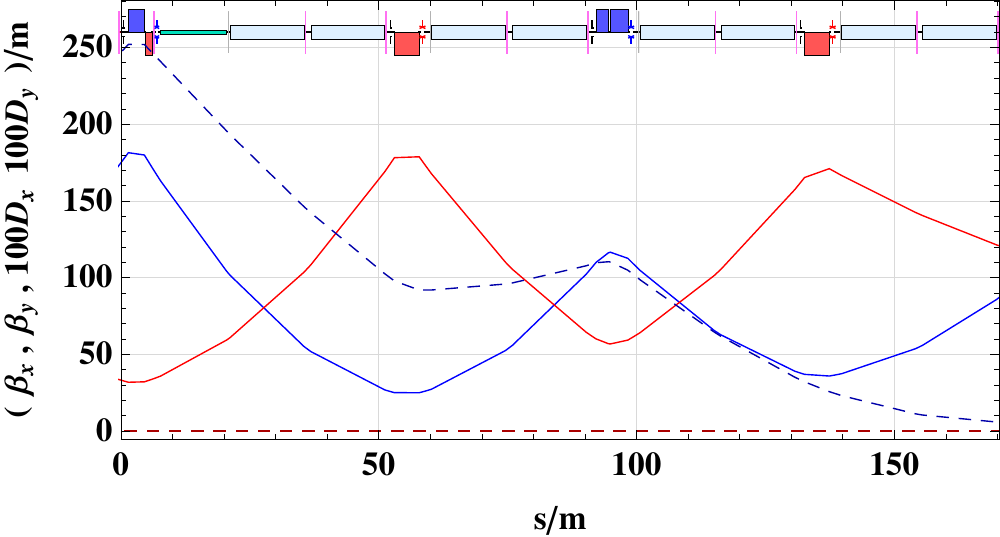}}
\caption{LHC DS on the left side of IP2.}\label{lat:fig:3:2:2}
\end{figure}

\begin{figure}[hb]
\centerline{\includegraphics[clip=,width=0.7\textwidth]{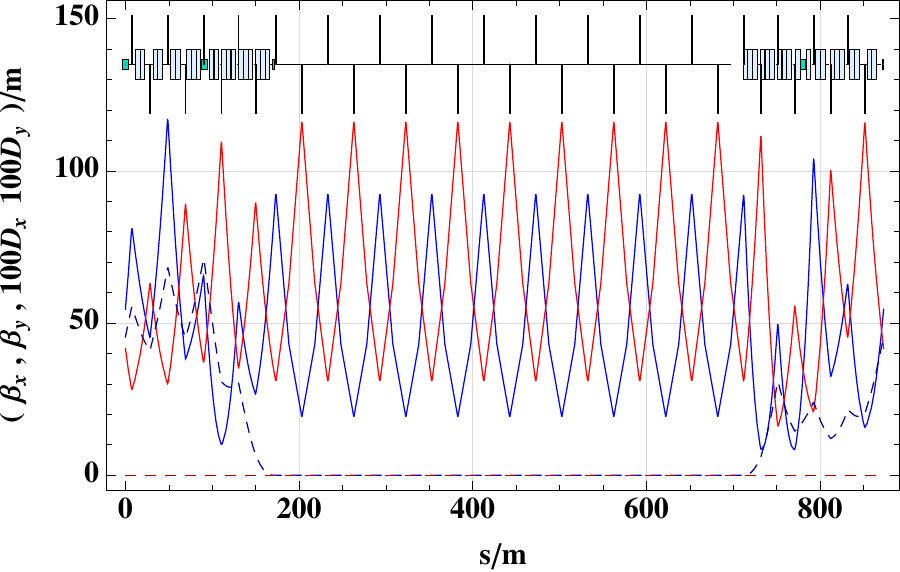}}
\caption{LHeC IR for even IRs, based on the DFB configuration in Point 2.}\label{lat:fig:3:2:3}
\end{figure}

\begin{figure}[hb]
\centerline{\includegraphics[clip=,width=0.7\textwidth]{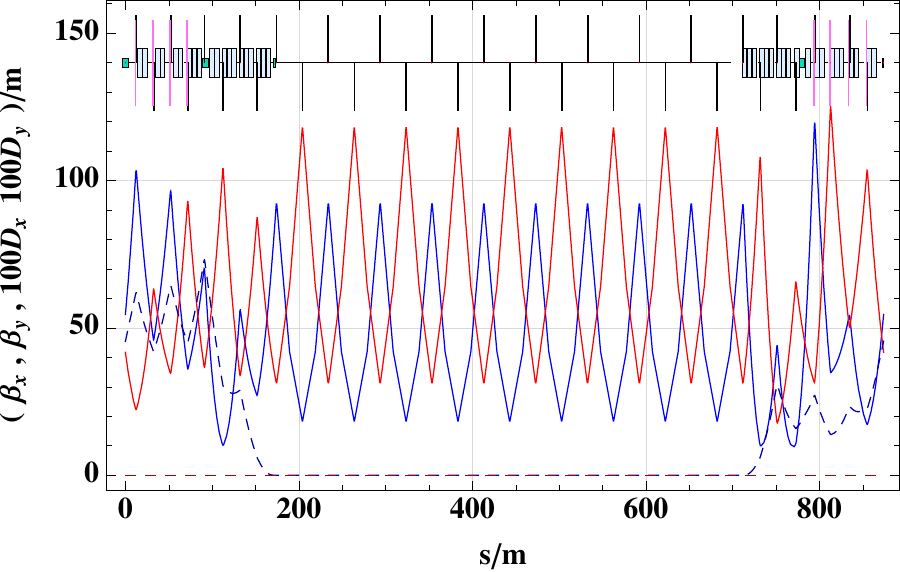}}
\caption{LHeC IR for odd IRs, based on the DFB configuration in Point 3.}\label{lat:fig:3:2:4}
\end{figure}

\begin{figure}[hb]
\centerline{\includegraphics[clip=,width=0.7\textwidth]{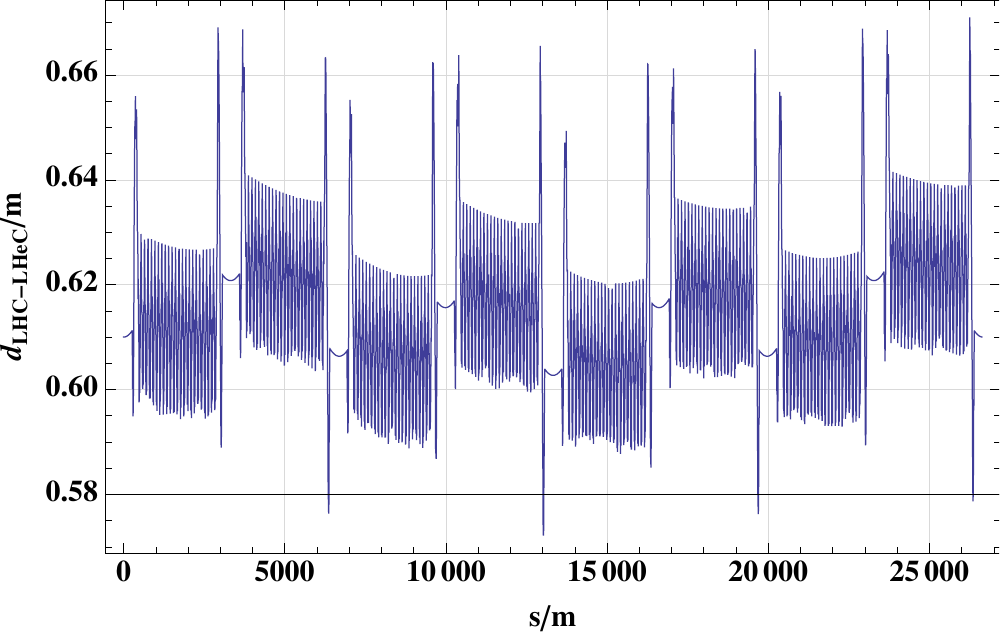}}
\caption{Radial distance between the idealised electron ring and the proton ring}\label{lat:fig:3:2:6}
\end{figure}

\begin{figure}[hb]
\centerline{\includegraphics[clip=,width=0.7\textwidth]{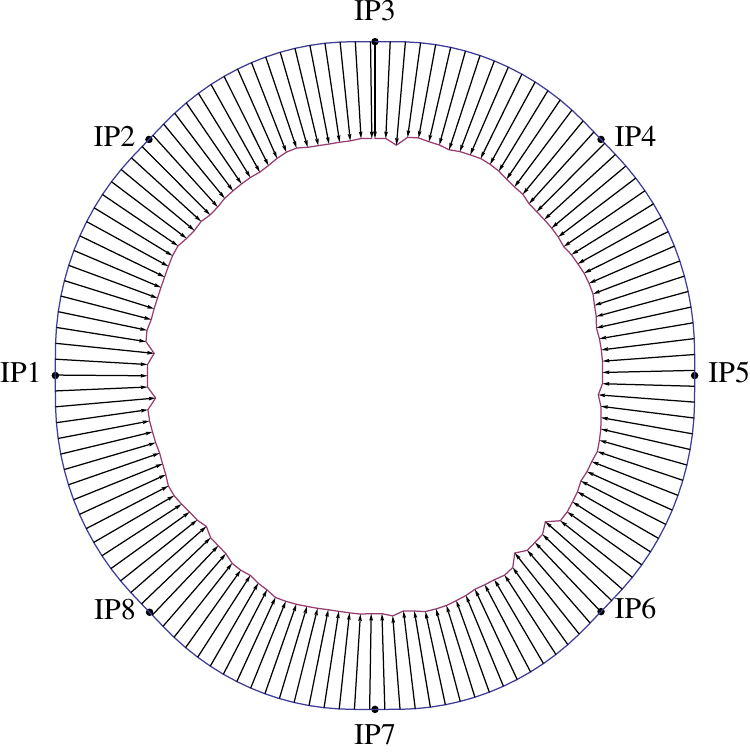}}
\caption{LHC and LHeC. The distance between the two rings is exaggerated by a factor 2000.}\label{lat:fig:3:2:7}
\end{figure}

The LHeC DS layout would ideally be similar to the LHC DS layout (Fig. \ref{lat:fig:3:2:2}), but has to be modified in order to leave space for the DFBs in the DS region. In the final design the dipoles are placed as symmetrically as possible between the regular arrangement of the quadrupoles (Fig. \ref{lat:fig:3:2:3}, \ref{lat:fig:3:2:4}).
The difference between the LHC proton ring and the idealised LHeC electron
ring is shown in Fig. \ref{lat:fig:3:2:6} and \ref{lat:fig:3:2:7}.


\subsection{Bypass options}
\label{lat:4}
In the design of the e-ring geometry, it is foreseen to bypass the LHC experiments at Point 1 and Point 5. The main requirements for both bypasses are that all integration constraints are respected, synchrotron radiation losses are not significantly increased and that the change in circumference can be compensated by increasing or decreasing the radius of the ring. 

Three different options are considered as basic bypass designs:
\begin{description}
\item \textbf{Vertical Bypass:} A vertical bypass would have to be a vertically upward bypass as downward would imply crossing the LHC magnets and other elements. For this a separation of about $20$ to $25 \ \mathrm{m}$ is required \cite{sylweisz}. This can only be achieved by strong additional vertical bending. In general a vertical bypass would therefore be rather long, increase the synchrotron radiation due to the additional vertical bends and decrease the polarisation compared to a horizontal bypass. A vertical bypasses is therefore only considered as an option if horizontal bypasses are not possible.
\item \textbf{Horizontal Inner Bypass:} A horizontal inner bypass can be constructed by simply decreasing the bending radius of the main bends. Consequently the synchrotron radiation losses for an inner bypass are larger than for a comparable outer bypass. The advantage of an inner bypass is, if used in combination with an outer one, that it reduces the circumference and the two bypasses could compensate each other's path length differences.
\item \textbf{Horizontal Outer Bypass:} A horizontal outer bypass uses the existing curvature of the ring instead of additional or stronger dipoles and consequently does not increase the synchrotron radiation losses. In general this is the preferred option.
\end{description}

\subsection{Bypass point 1}
\label{lat:5}
\begin{figure}[hb]
\centerline{\includegraphics[clip=,width=0.7\textwidth]{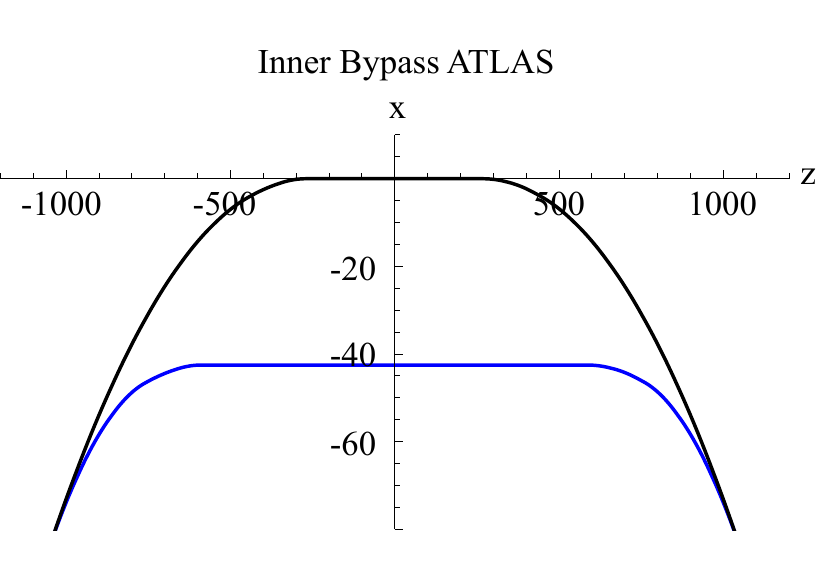}}
\caption{Example of an inner Bypass around Point 1. The Bypass is shown in blue, The LHC proton ring in black.}\label{lat:fig:5:1}
\end{figure}

The cavern in Point 1 reaches far to the outside of the LHC, so that a separation of about  $100 \ \mathrm{m}$ would be necessary in order to fully bypass the experimental hall. For a bypass on the inside, a smaller separation of about  $39 \ \mathrm{m}$ would be required. For an inner bypass with minimal separation, the bending strength in three normal arc cells would have to be doubled resulting in a bypass of more than $2 \ \mathrm{km}$ length. A sketch of such an inner bypass is shown in Fig. \ref{lat:fig:5:1}.

\begin{figure}[hb]
\centerline{\includegraphics[clip=,width=0.7\textwidth]{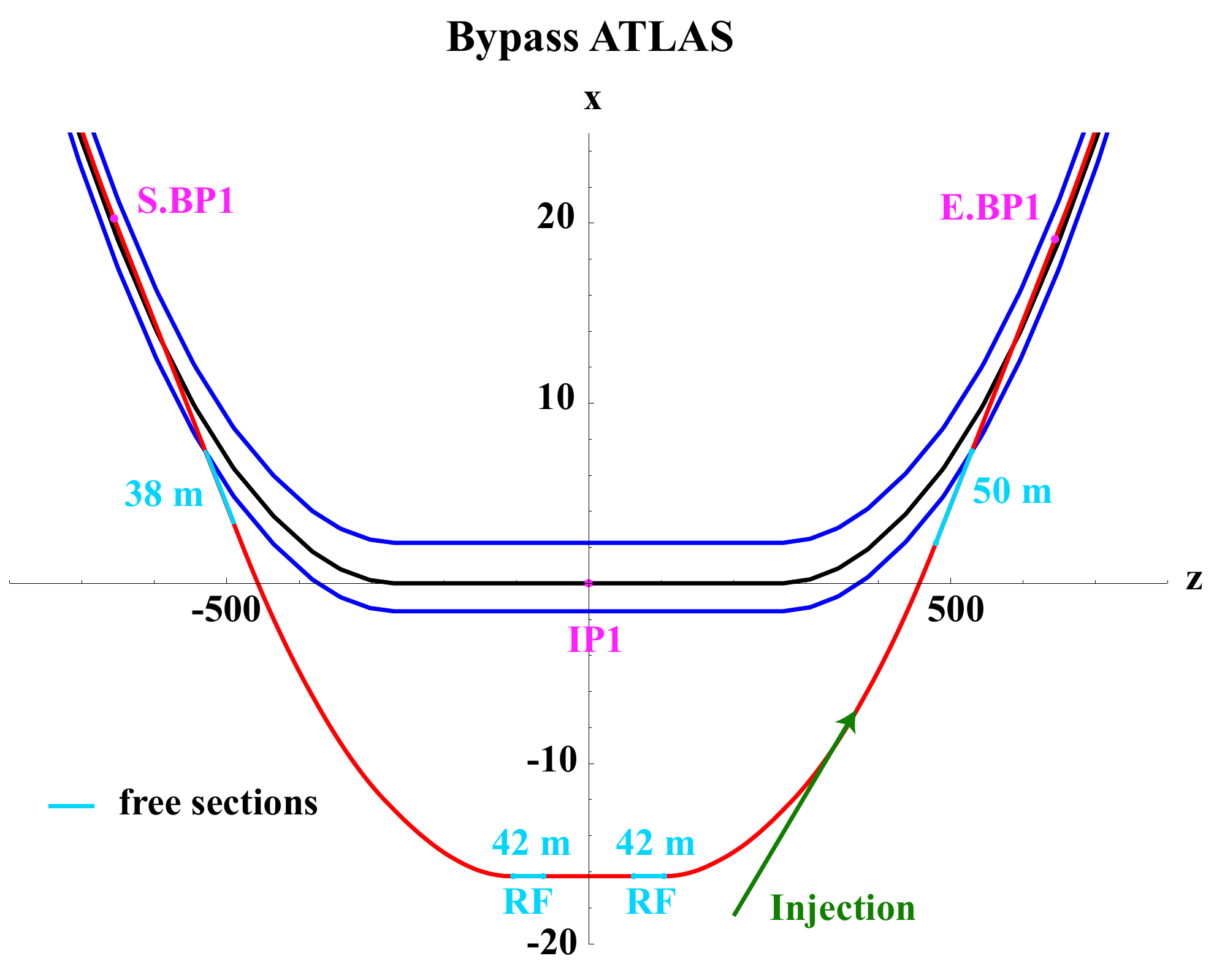}}
\caption{Final bypass design using the survey gallery in Point 1. The LHC proton ring is shown in black, the electron ring in red and the tunnel walls in blue. Dispersion free sections reserved for the installation of RF, wiggler(s), injection and other equipment are marked in light blue. The injection is marked in green and is located in the right arc of the bypass. Beginning and end of the bypass are marked with S.BP1 and E.BP1}\label{lat:fig:5:2}
\end{figure}

Instead of a long inner bypass, an outer bypasses using the existing survey gallery is chosen as final design. With this design the separation is brought down to $16.25 \ \mathrm{m}$. The RF is installed in the straight section next to the straight section of the proton ring. The electron beam is injected into the arc on the right side of the bypass. The design is shown in Fig. \ref{lat:fig:5:2}.

\subsection{Bypasses point 5}
\label{lat:6}
\begin{figure}[hb]
\centerline{\includegraphics[clip=,width=0.7\textwidth]{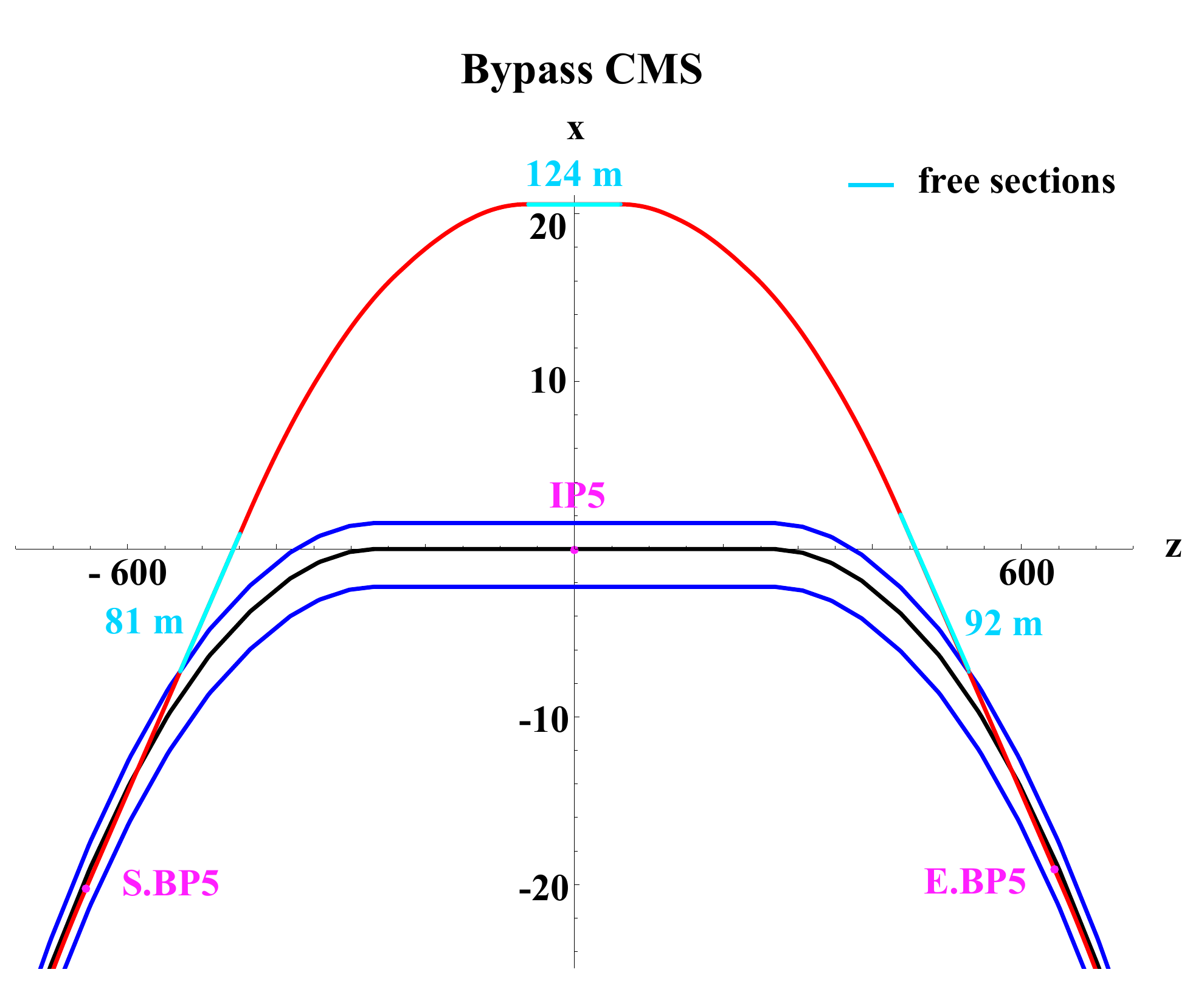}}
\caption{Horizontal outer bypass in Point 5. The LHC proton ring is shown in black, the electron ring in red and the tunnel walls in blue. Dispersion free sections reserved for the installation of RF, wiggler(s), injection and other equipment are marked in light blue. Beginning and end of the bypass are marked with S.BP5 and E.BP5}\label{lat:fig:6:1}
\end{figure}

Due to the compact design of the cavern in Point 5 a separation of only about $20 \ \mathrm{m}$ is needed to completely bypass the experiment on the outside (Fig. \ref{lat:fig:6:1}). The separation in the case of an inner horizontal bypass or a vertical bypass would be the same or larger and therefore, as in the case of Point 1, the horizontal outer bypass is preferred over an inner or vertical one. The RF is installed in the centre straight section parallel to the proton ring.

\subsection{Matching proton and electron ring circumference}
\label{lat:7}
\begin{figure}[hb]
\centerline{\includegraphics[clip=,width=0.7\textwidth]{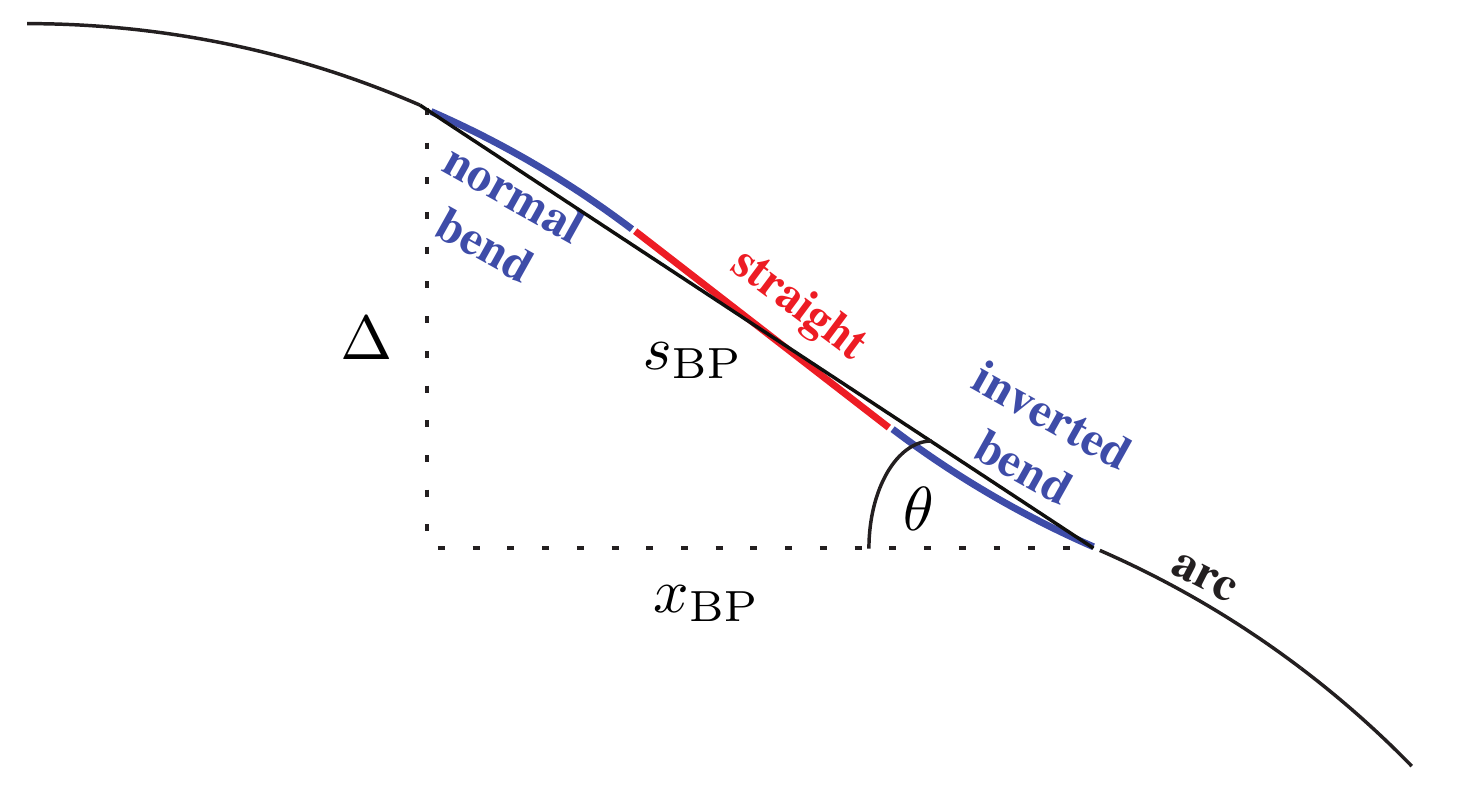}}
\caption{Outer bypass: a straight section is inserted to obtain the required separation. A larger separation could be achieved by inserting inverted bends.}\label{lat:fig:7:1}
\end{figure}

Both bypasses in Point 1 and Point 5 require approximately the same separation and a similar design was chosen for both. To obtain the necessary separation $\Delta_{\mathrm{BP}}$ a straight section of length $s_{\mathrm{BP}}$ is inserted into the lattice of the idealised ring (Sec. \ref{lat:3}) in front of the last two arc cells. The separation $\Delta_{\mathrm{BP}}$, the remaining angle $\theta_{\mathrm{BP}}$ and the inserted straight section $s_{\mathrm{BP}}$  are related by (Fig. \ref{lat:fig:7:1}):
\begin{equation}\label{lat:eqn:7:1}
 \Delta_{\mathrm{BP}}=s_{\mathrm{BP}}\sin{\theta_{\mathrm{BP}}}
\end{equation}
As indicated in Fig. \ref{lat:fig:7:1} the separation could be increased by inserting a S-shaped chicane including negative bends. The advantage of additional bends would be the faster separation of the electron and proton ring. On the other hand the additional bends would need to be placed in the LHC tunnel, the straight sections of the bypass would be reduced and the synchrotron radiation losses increased. Hence this is not the preferred solution.

In the following, estimates for the current bypass design, which does not include any extra bends, are presented. Given the separation, angle and length of the inserted straight section, the induced change in circumference is then:
\begin{equation}\label{lat:eqn:7:2}
 \Delta s_{\mathrm{BP}}=s_{\mathrm{BP}}-x_{\mathrm{BP}}=2 \Delta_{\mathrm{BP}} \tan{(\frac{\theta_{\mathrm{BP}}}{2})}
\end{equation}
This change can be compensated by a change in radius of the idealised ring by:
\begin{equation}\label{lat:eqn:7:3}
 \Delta s_{\mathrm{BP}}=2 \pi \Delta R
\end{equation}

Taking the change in radius into account, the separation $\Delta_{\mathrm{BP}}$ has to be substituted by $\Delta_{\mathrm{BP,tot}}:=\Delta_{\mathrm{BP}}+\Delta R$. The radius change and the total separation are then related by: 
\begin{equation}\label{lat:eqn:7:4}
 \Delta R=\frac{\Delta_{\mathrm{BP}}}{\pi\cot{\left(\frac{\theta_{\mathrm{BP}}}{2}\right)}-2} \ ,\quad \textnormal{with} \ \Delta_{\mathrm{BP}}=\Delta_{\mathrm{BP1}}+\Delta_{\mathrm{BP5}}
\end{equation}
As the bypass in Point 1 passes through the existing survey gallery, the geometry and with it the separation in Point 1, cannot be changed. The bypass in Point 5, on the other hand, is fully decoupled from the existing LHC cavern and tunnel and is therefore used for the fine adjustment of the circumference. The design values of both bypasses are summarised in Table~\ref{lat:tab:7:1}.
\begin{table}[h]
  \centering
  \begin{tabular}{|l|c|c|}
    \hline
 & Point 1 & Point 5 \\ \hline \hline
Total bypass length & $1303.3 \ \mathrm{m}$ & $1303.7 \ \mathrm{m}$\\ \hline
Separation & $16.25 \ \mathrm{m}$ & $20.56 \ \mathrm{m}$ \\ \hline
Dispersion free straight section & $172 \ \mathrm{m}$ & $297 \ \mathrm{m}$ \\ \hline
Ideal radius change of the idealised ring & \multicolumn{2}{c|}{$61 \ \mathrm{cm}$}  \\ \hline
  \end{tabular}
\caption{Lengths characterising the bypasses.}
\label{lat:tab:7:1}
\end{table} 

\section{Layout and optics}
\label{opt}
Throughout the whole electron ring lattice, the choice of the optics is strongly influenced by the geometrical constraints and shortage of space in the LHC tunnel. The main interference with the LHC beside Point 1 and Point 5, which have to be bypassed, are the service modules and DFBs in the tunnel, where no electron ring elements can be placed. 
\subsection{Arc cell layout and optics}
\label{opt:1}
The LHC service modules are placed at the beginning of each LHC main arc cell. In order to obtain a periodic solution of the lattice, the electron ring arc cell length can only be a multiple or $1/n$th, $n \in \mathsf{N}$, of the LHC FODO cell length. Given the same phase advance and bending radius, the emittance increases with increasing cell length $L$ of a FODO cell. In the case of the LHeC electron ring a FODO cell length corresponding to half the LHC FODO cell length delivers an emittance close to the design value of $\epsilon_{\mathrm{rms},x/y}=5.0/2.5 \ \mathrm{nm}$. The emittance of a cell with the full LHC FODO cell length is about a factor of $4$ too large.

Choosing half the LHC FODO cell length divides the arc into $23$ equal double FODO cells with a symmetric configuration of the quadrupoles and an asymmetric distribution of the dipoles, precisely $8$ dipoles in the first FODO cell and $7$ in the second. The dipole configuration is asymmetric in order to use all available space for the bending of the e-beam and consequently minimise the synchrotron radiation losses. With a phase advance of $180^\circ$ horizontally and $120^\circ$ vertically over the complete double FODO cell, which corresponds to a phase advance of  $90^\circ/60^\circ$ per FODO cell, the horizontal emittance lies with $3.96 \ \mathrm{nm}$ well below the design value of $5 \ \mathrm{nm}$. The optics of one arc cell is shown in Fig. \ref{lat:fig:3:2:1} and the parameters are listed in Table~\ref{opt:tab:1:1}.
\begin{table}[h]
  \centering
  \begin{tabular}{|l|l|}
    \hline
Beam Energy & $60 \ \mathrm{GeV}$ \\ \hline
Phase Advance per Cell & $180^{\circ}/120^{\circ}$ \\ \hline
Cell length & $106.881 \ \mathrm{m}$ \\ \hline
Dipole Fill factor & $0.75$ \\ \hline
Damping Partition $J_x/J_y/J_e$ & $1.5/1/1.5$ \\ \hline
Coupling constant $\kappa$ & $0.5$ \\ \hline
Horizontal Emittance (no coupling)& $3.96 \ \mathrm{nm}$ \\ \hline
Horizontal Emittance ($\kappa=0.5$)& $2.97 \ \mathrm{nm}$ \\ \hline
Vertical Emittance ($\kappa=0.5$)& $1.49 \ \mathrm{nm}$ \\ \hline
  \end{tabular}
\caption{Optics Parameters of one LHeC arc cell with a phase advance of $90^{\circ}/60^{\circ}$ per half cell.}
\label{opt:tab:1:1}
\end{table}

\subsection{Insertion layout and optics}
\label{opt:2}
For simplicity all even and all odd insertions of the electron ring have the same layout as described in Sec. \ref{lat:1}. Each insertion is divided in three parts: the dispersion suppressor on the left side (DSL), the straight section and the dispersion suppressor on the right side (DSR).

\subsubsection{Dispersion suppressor}
\label{opt:2:2}
Various well known standard DS designs like the missing bend or half bend scheme exist, but they are all based on specific placement of the dipoles. In the case of the LHeC the position of the dipoles is strongly determined by the LHC geometry and does not match any of the standard schemes. Therefore the dispersion matching is achieved by 8 individually powered quadrupoles and not with the positioning of the dipoles. The DS on the left side is split into two DS sections, reaching from the first DFB to the second and from the second to the beginning of the straight section. In the DSL the quadrupoles are distributed equally in each section. In the DSR they are placed with equal distances from each other throughout the complete DS. This layout turned out to be better for the right side due to the different arrangement of the DFBs. The DSs of the even and odd points differ slightly in their length but have the same general layout. The lengths of the DSs are listed in Table~\ref{lat:tab:3:2:1}. The DS optics are shown in Fig. \ref{lat:fig:3:2:3} and \ref{lat:fig:3:2:4}.

\subsubsection{Straight section}
\label{opt:2:3}
For simplicity the straight sections consist of a regular FODO lattice with a phase advance of $90^{\circ}/60^{\circ}$ except the straight section at Point~3 and Point~7 where the phase advance of the FODO cells is used for the adjustment of the working point. In a later stage the lattice and optics of the straight sections will have to be adjusted to the various insertions.

\subsection{Bypass layout and optics}
\label{opt:3}
\begin{figure}[hb]
\centerline{\includegraphics[clip=,width=0.7\textwidth]{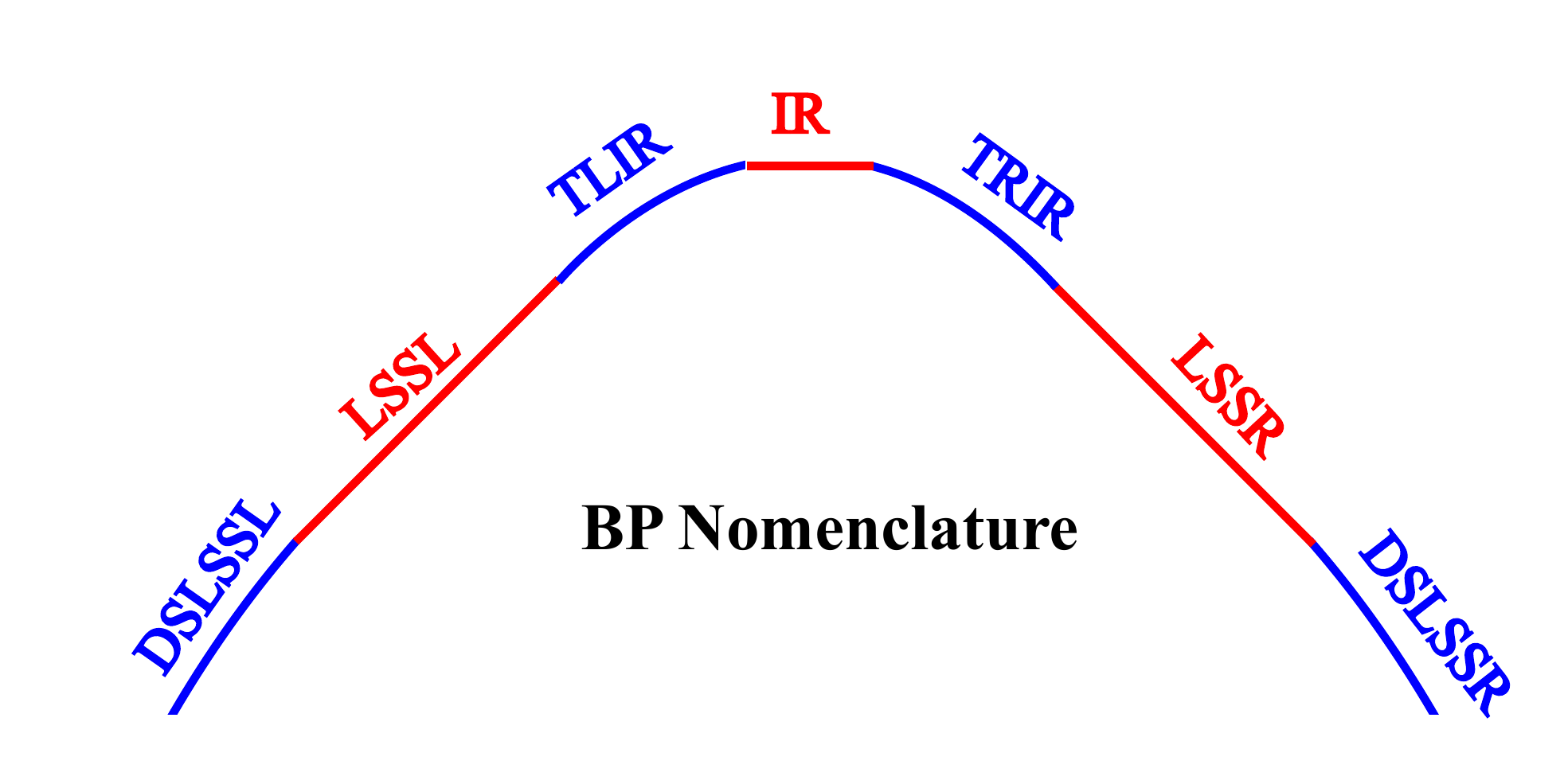}}
\caption{Bypass layout and nomenclature.}\label{opt:fig:3:1}
\end{figure}

The general layout and nomenclature of the bypasses is illustrated in Fig. \ref{opt:fig:3:1}. The straight sections LSSL, LSSR and IR are dispersion free sections reserved for the installation of RF, wiggler(s), injection etc. Two normal arc cells (4 FODO cells) with 8 individual quadrupoles are used as dispersion suppressor before the first straight section LSSL and after the last straight section LSSR. In the sections TLIR and TRIR the same configuration of dipoles is kept as in the idealised lattice for geometric reasons. Among this fixed arrangement of dipoles 14 matching quadrupoles per side are placed as equally as possible. 

\begin{figure}[hb]
\centerline{\includegraphics[clip=,width=0.9\textwidth]{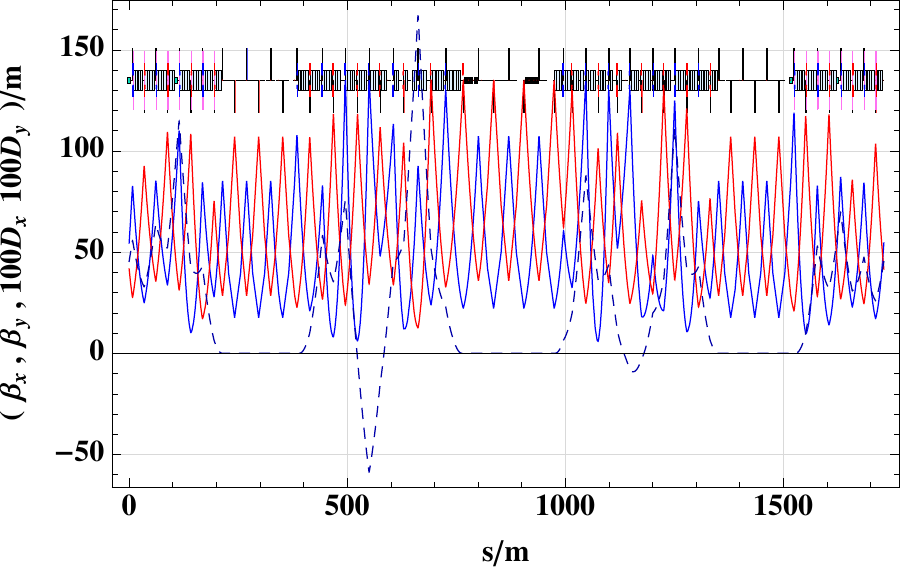}}
\caption{Bypass optics Point 1.}\label{opt:fig:3:2}
\end{figure}

\begin{figure}[hb]
\centerline{\includegraphics[clip=,width=0.9\textwidth]{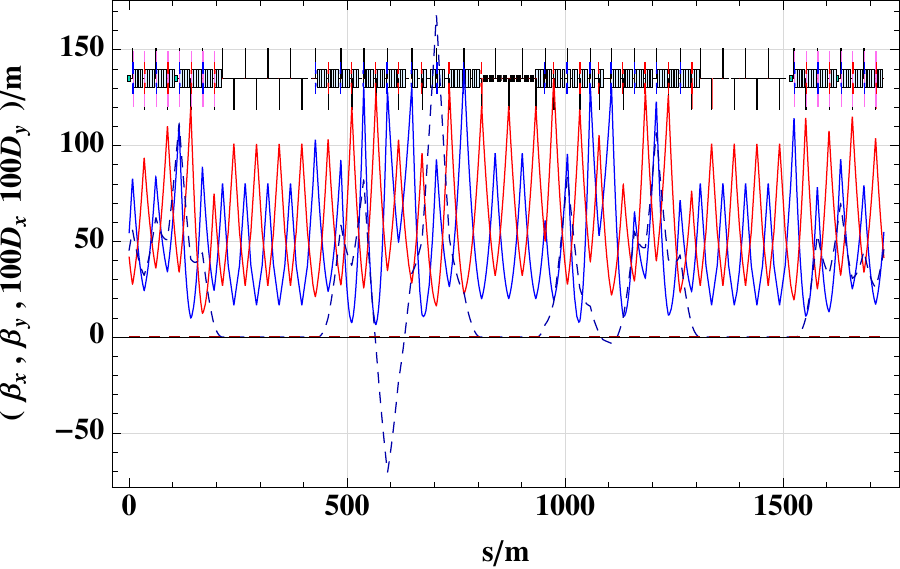}}
\caption{Bypass Optics Point 5.}\label{opt:fig:3:3}
\end{figure}

The straight sections consist of a regular FODO lattice with a phase advance of $90^{\circ}/60^{\circ}$.\\
The complete bypass optics in Point 1 and Point 5 are shown in Fig. \ref{opt:fig:3:2} and \ref{opt:fig:3:3}.

\subsection{Chromaticity correction}
\label{opt:4}
\begin{figure}[hb]
\centerline{\includegraphics[clip=,width=0.9\textwidth]{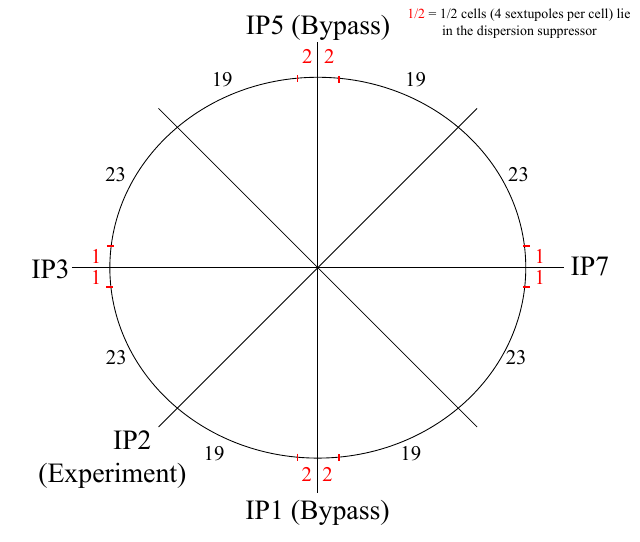}}
\caption{LHeC Sextupole Scheme for a phase advance of $90^{\circ}/60^{\circ}$ with sextupoles also placed in the dispersion suppressor.}\label{opt:fig:4:1}
\end{figure}

The phase advance of one LHeC FODO cell of $90^{\circ}/60^{\circ}$ suggests a chromaticity correction with in total 5 interleaved sextupole families, 2 horizontal and 3 vertical. In order to reduce the chromatic stopband and the off momentum beta beating each arc contains an equal number of sextupoles per family, so $\mathrm{n \cdot 2}$ horizontal and $\mathrm{m \cdot 3}$ in the vertical. Further to reduce the sextupole strength and therefore the excitation of resonances, the  families are completed by placing sextupoles also in the dispersion suppressors. This yields a sextupole scheme as illustrated in Fig.~\ref{opt:fig:4:1}. A large part of the total natural chromaticity usually comes from the experiments due to their large $\beta$-functions and magnet strength in the final focus quadrupoles.  This is only true for the vertical plane of the HA optics. In the case of the HL option and the horizontal plane of the HA optics, all insertions including the experimental insertion in Point 2 contribute more or less equally to the chromaticity. This suggests a global correction of the chromaticity with 2 sextupoles for the horizontal and 3 for the vertical plane for the HL option. For the HA option a local correction of the off-momentum beta-beating with the two arcs adjacent to IP2 could be considered instead of a simple global correction \cite{ipac12mflhec}. The contribution of the different insertions to the total chromaticity is listed in Table~\ref{opt:tab:4:1} and Table~\ref{opt:tab:4:2}. 
\begin{table}[H]
    \centering
    \begin{tabular}{|l|l|l|}
    \hline
& $-\mathrm{d} Q_{x/y}$ & $-(\mathrm{d}Q_{x/y}/\mathrm{d}Q_{x/y, \mathrm{tot}})\cdot 100$ \\ \hline \hline
full sequence & 142.1/115.6 &  100/100\\ \hline
IR 1 & 9.6/8.2&  6.8/7.1\\ \hline
IR 2 & 4.6/3.8&  3.2/3.3\\ \hline
IR 3/7 & 4.5/3.6&  3.2/3.1\\ \hline
IR 4/6/8 & 4.6/3.8&  3.2/3.3\\ \hline
IR 5 & 10.0/7.8&  7.0/6.7\\ \hline
  \end{tabular}
\caption{Contribution of the insertions to the natural chromaticity for the HL Option}
\label{opt:tab:4:1}
\end{table}

\begin{table}[H]
    \centering
    \begin{tabular}{|l|l|l|}
    \hline
& $-\mathrm{d} Q_{x/y}$ & $-(\mathrm{d}Q_{x/y}/\mathrm{d}Q_{x/y, \mathrm{tot}})\cdot 100$ \\ \hline \hline
full sequence & 144.1/136.2&  100/100\\ \hline
IR 1 & 9.9/7.5&  6.7/5.5\\ \hline
IR 2 & 7.5/25.0&  5.2/18.3\\ \hline
IR 3/7 & 4.7/3.7&  3.2/2.7\\ \hline
IR 4/6/8 & 4.6/3.7&  3.2/2.7\\ \hline
IR 5 & 10.2/7.8&  7.0/5.7\\ \hline
  \end{tabular}
\caption{Contribution of the insertions to the natural chromaticity for the HA Option}
\label{opt:tab:4:2}
\end{table}
In general the chromaticity correction is expected to be rather unchallenging.  

\subsection{Working point}
\label{opt:5}
\begin{figure}[hb]
\centerline{\includegraphics[clip=,width=0.9\textwidth]{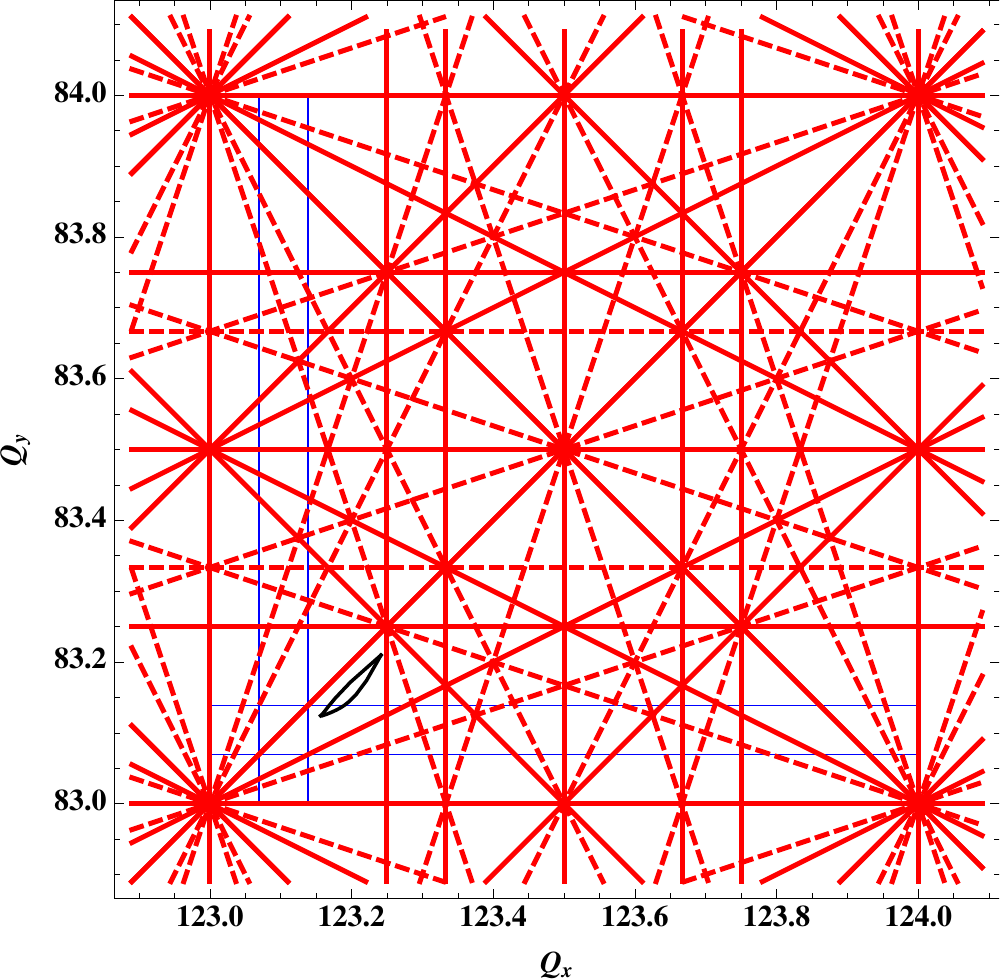}}
\caption{Working Point for the HA optics. The dashed lines are the coupling resonances up to $4$th order, the solid lines the constructive resonances up to $4$th order. The black line indicates the working point without beam-beam tune shift, while the blue lines indicate the working point with beam-beam tune shift.}\label{opt:fig:5:1}
\end{figure}

\begin{figure}[hb]
\centerline{\includegraphics[clip=,width=0.9\textwidth]{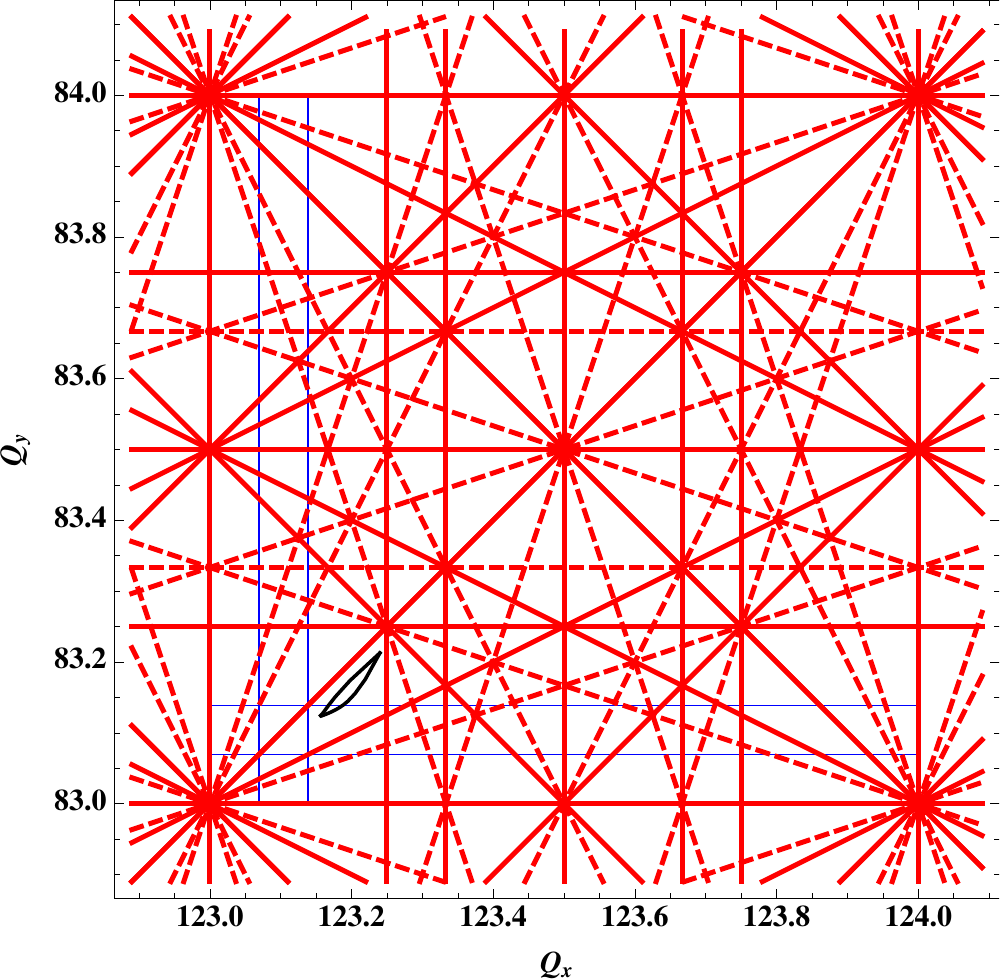}}
\caption{Working Point for the HL optics. The dashed lines are the coupling resonances up to $4$th order, the solid lines the constructive resonances up to $4$th order. The black line indicates the working point without beam-beam tune shift, while the blue lines indicate the working point with beam-beam tune shift.}\label{opt:fig:5:2}
\end{figure}

Because of the bypasses and the single interaction region, the LHeC lattice has no reflection or rotation symmetry. As 50\% emittance ratio is required, betatron coupling resonances may be excited and must be taken into account for the choice of the working point. In addition the beam will suffer a maximum beam-beam tune shift of $0.087$ in both planes in the case of the HA option and $0.085$ in the horizontal and $0.090$ in the vertical plane in the case of the HL option. Besides the systematic resonances also the first synchrotron sidebands of at least the integer resonances have to be avoided. Taking the beam-beam tune shift and the detuning with amplitude from head-on interactions into account a possible working point could be $Q_x=123.155/Q_y=83.123$ for the HA as well as for the HL option. The working point diagrams for both cases are shown in Figs. \ref{opt:fig:5:1} and \ref{opt:fig:5:2}.

\subsection{Aperture}
\label{opt:6}
The current LHeC e-ring magnet apertures (see Sec.~$\ref{arcmagnets}$) are based on the experience from LEP \cite{GreenBooksLEP3} applied on the LHeC arc cells. They correspond to minimum 36.2~$\sigma$~hor./39.9~$\sigma$~ver. in the arc dipoles, 32.9~$\sigma$~hor./59~$\sigma$~ver. in the arc quadrupoles, 14.7~$\sigma$~hor./35.9~$\sigma$~ver. in the insertion dipoles and 14.6~$\sigma$~hor./51.6~$\sigma$~ver. in the insertion quadrupoles. In the estimate all insertions were included whereas for the IP (Point 2) the values were only calculated for the HA option. All values are summarised in Table~\ref{opt:tab:6:1}, \ref{opt:tab:6:2}, \ref{opt:tab:6:3}, \ref{opt:tab:6:4}. The hor. aperture in the insertion dipoles and quadrupoles is slightly too tight, but as the gradients are small, it can be easily increased by around 5 to 7~mm without changing considerably the magnet design. In all calculations a Gaussian beam profile in all three dimensions was assumed and the maximum beam size is consequently given by:
\begin{equation}\label{opt:eqn:6:1}
\sigma_{x,y}=\sqrt{\beta_{x,y} \epsilon_{x,y}+D_{x,y}^2\sigma_E^2}
\end{equation}
where $\epsilon_{x,y}$ are the design emittances of 5 and 2.5 nm respectively.
\begin{table}[H]
 \begin{minipage}{.45\textwidth}
    \centering
    \begin{tabular}{|l|l|}
    \hline
Hor. Half Apert. Dip. & $30 \ \mathrm{mm}$ \\ \hline
Ver. Half Apert. Dip. & $20 \ \mathrm{mm}$ \\ \hline
Max. Hor. Beam Size & $0.82 \ \mathrm{mm}$ \\ \hline
Max. Ver. Beam Size & $0.50 \ \mathrm{mm}$ \\ \hline
Hor. Apert./Max. Beam Size  & $36.2$ \\ \hline
Ver. Apert./Max. Beam Size  & $39.9$ \\ \hline
  \end{tabular}
\caption{Aperture and beam sizes for the arc dipoles}
\label{opt:tab:6:1}
 \end{minipage}
 \hspace{0.5cm}
 \begin{minipage}{.45\textwidth}
    \centering
  \begin{tabular}{|l|l|}
    \hline
Hor. Half Aperture Dipole & $30 \ \mathrm{mm}$ \\ \hline
Ver. Half Aperture Dipole & $20 \ \mathrm{mm}$ \\ \hline
Max. Hor. Beam Size & $2.04 \ \mathrm{mm}$ \\ \hline
Max. Ver. Beam Size & $0.56 \ \mathrm{mm}$ \\ \hline
Hor. Aperture/Max. Beam Size  & $14.7$ \\ \hline
Ver. Aperture/Max. Beam Size  & $35.9$ \\ \hline
  \end{tabular}
\caption{Aperture and beam sizes for the insertion dipoles including Point 2 (HA~Option)}
\label{opt:tab:6:2}
 \end{minipage}
\end{table}

\begin{table}[H]
 \begin{minipage}{.45\textwidth}
    \centering
\begin{tabular}{|l|l|}
    \hline
Apert. Radius Arc Quad. & $30 \ \mathrm{mm}$ \\ \hline
Max. Hor. Beam Size & $0.91 \ \mathrm{mm}$ \\ \hline
Max. Ver. Beam Size & $0.51 \ \mathrm{mm}$ \\ \hline
Hor. Apert./Max. Beam Size & $32.9$ \\ \hline
Ver. Apert./Max. Beam Size & $59.0$ \\ \hline
  \end{tabular}
\caption{Aperture and beam sizes for the arc quadrupoles}
\label{opt:tab:6:3}
 \end{minipage}
 \hspace{0.5cm}
 \begin{minipage}{.45\textwidth}
    \centering
  \begin{tabular}{|l|l|}
    \hline
Apert. Radius Quad.& $30 \ \mathrm{mm}$ \\ \hline
Max. Hor. Beam Size & $2.06 \ \mathrm{mm}$ \\ \hline
Max. Ver. Beam Size & $0.58 \ \mathrm{mm}$ \\ \hline
Hor. Apert./Max. Beam Size & $14.6$ \\ \hline
Ver. Apert./Max. Beam Size & $51.6$ \\ \hline
  \end{tabular}
\caption{Aperture and beam sizes for the insertion quadrupoles including Point 2 (HA~Option)}
\label{opt:tab:6:4}
 \end{minipage}
\end{table}

\newpage

%% file: machine/holzer_thompson.tex
%
%
\section{Interaction region layout}
\label{LHEC:machine:RR-IR-layout}
The design of the Interaction Region (IR) of the LHeC is particularly challenging as it 
has to consider boundary conditions from 
\begin{itemize}
\item{The lattice design and beam optics of the electron and proton beams}
\item{The geometry of the LHC experimental cavern and the tunnel} 
\item{The beam separation scheme which is determined by the bunch pattern of the LHC standard proton operation and related to this
 the optimisation of the synchrotron light emission and collimation}
\item{The technical feasibility of the hardware.} 
\end{itemize}
Therefore the IR has to be optimised with respect to a well matched beam optics that adapts the optical parameters from the new electron-proton interaction point to the standard LHC proton beam optics in the arc and to the newly established beam optics of the electron ring. At the same time the two colliding beams as well as the non-colliding proton beam of LHC have to be separated efficiently and guided into their corresponding magnet lattices. As a general rule that has been established in the context of this study any modification in the standard LHC lattice and any impact on the LHC proton beam parameters had to be chosen moderately to avoid detrimental effects on the performance of the LHC proton-proton operation.
\par
The layout and parameters of the new e/p interaction point are defined by the particle physics requirements. At present the physics program that has been proposed for the LHeC ~\cite{Det_Acceptance} follows two themes - a high luminosity, high Q$^2$ program requiring a forward and backward detector acceptance of around 10$^{\circ}$ and a low x, low Q$^2$ program, which requires an increased detector acceptance  in forward and backward direction of at least 1$^{\circ}$ and could proceed with reduced luminosity. Accordingly two machine scenarios 
have been studied for the interaction region design. Firstly, a design that has been optimised for high luminosity with an acceptance of 10$^{\circ}$ and secondly, a high acceptance design that allows for a smaller opening angle of the detector. In both cases the goal for the machine luminosity is in the range of 10$^{33}\,\mathrm{cm}^{-1}\,\mathrm{s}^{-1}$ but the layouts differs in the magnet lattice, the achievable absolute luminosity and mainly the synchrotron radiation that is emitted during the beam separation process. Both options will be presented here in detail and the corresponding design luminosity, the technical requirements and the synchrotron radiation load will be compared. In both cases however, a well matched spot size of the electron and proton beam  had to be established at the collision point: Experience in SPS and 
HERA  ~\cite{Bieler:1999iz}, ~\cite{Zimmermann:1992da} showed that matched beam cross sections have to be established between the two colliding beams to guarantee stable beam conditions. Considering the different nature of the beams, namely the emittances of the electron beam in the two transverse planes, the interaction region design has to consider this boundary condition and the beam optics has to be established to achieve equal beam sizes  $ \sigma _x (p) = \sigma _x (e) $, $ \sigma _y (p) = \sigma _y (e)$  at the IP. 

The basic beam parameters however like energy, particle intensity and beam emittances  are identical for both designs, determined by the electron and proton ring lattices and the pre-accelerators. They are summarised in Table~\ref{tab:parameters}.

\begin{table}[h]
\begin{center}
\begin{tabular}{|l|c|c|c|}
\hline
Quantity                      &  unit   &     e    &   p        \\
\hline
Beam energy                   &  GeV    &    60    &  7000      \\
Total beam current            &   mA    &   100    &   860      \\
Number of bunches             &         &  2808    &  2808      \\
Particles/bunch $N_b$         &$10^{10}$&  2.0    &   17       \\
Horiz.\ emittance             &   nm    &   5.0    &  0.5       \\
Vert.\ emittance              &   nm    &   2.5    &  0.5       \\
\hline
Bunch distance                &   ns    & \multicolumn{2}{c|}{25}     \\
\hline
\end{tabular}
\caption{Main parameters for e/p collisions.}
\label{tab:parameters}
\end{center}
\end{table}

Colliding two beams of different characteristics, the luminosity obtained is given by the equation
\begin{equation}
L=\sum_{i=1}^{n_b} ( I_e  I_p)\frac{1}{ e^2 f_0 2\pi \sqrt{ \sigma_{xp}^2 + \sigma_{xe}^2} \sqrt{ \sigma_{yp}^2 + \sigma_{ye}^2} },
\end{equation}
where $\sigma_{x,y}$  denotes the beam size of the electron and proton beam in the horizontal and vertical plane and $I_e$, $I_p$  the electron and proton single bunch currents. In all IR layouts the electron beam size at the IP is matched to the proton beam size in order to optimise the delivered luminosity and minimise detrimental beam beam effects. 
\par

The main difference of the IR design for the electron proton collisions with respect to the existing LHC interaction regions is the fact that the two beams of LHeC cannot be focused and / or guided at the same time: The different nature of the two beams, the fact that the electrons emit synchrotron radiation and mainly the large difference in the particle momentum make a simultaneous focusing of the two beams impossible. The strong gradients of the proton quadrupoles in the LHC triplet structure cannot be tolerated nor compensated for the electron lattice and a stable optical solution for the electrons is not achievable under the influence of the proton magnet fields. The electron beam therefore has to be separated from the proton beam after the collision point  before any strong `` 7~TeV like''  magnet field is applied. 

In order to obtain still a compact design and to optimise the achievable luminosity of the new e/p interaction region, the beam separation scheme has to be combined with the electron mini-beta focusing structure.
 
\begin{figure}
\centerline{\includegraphics[clip=,width=0.60\textwidth]{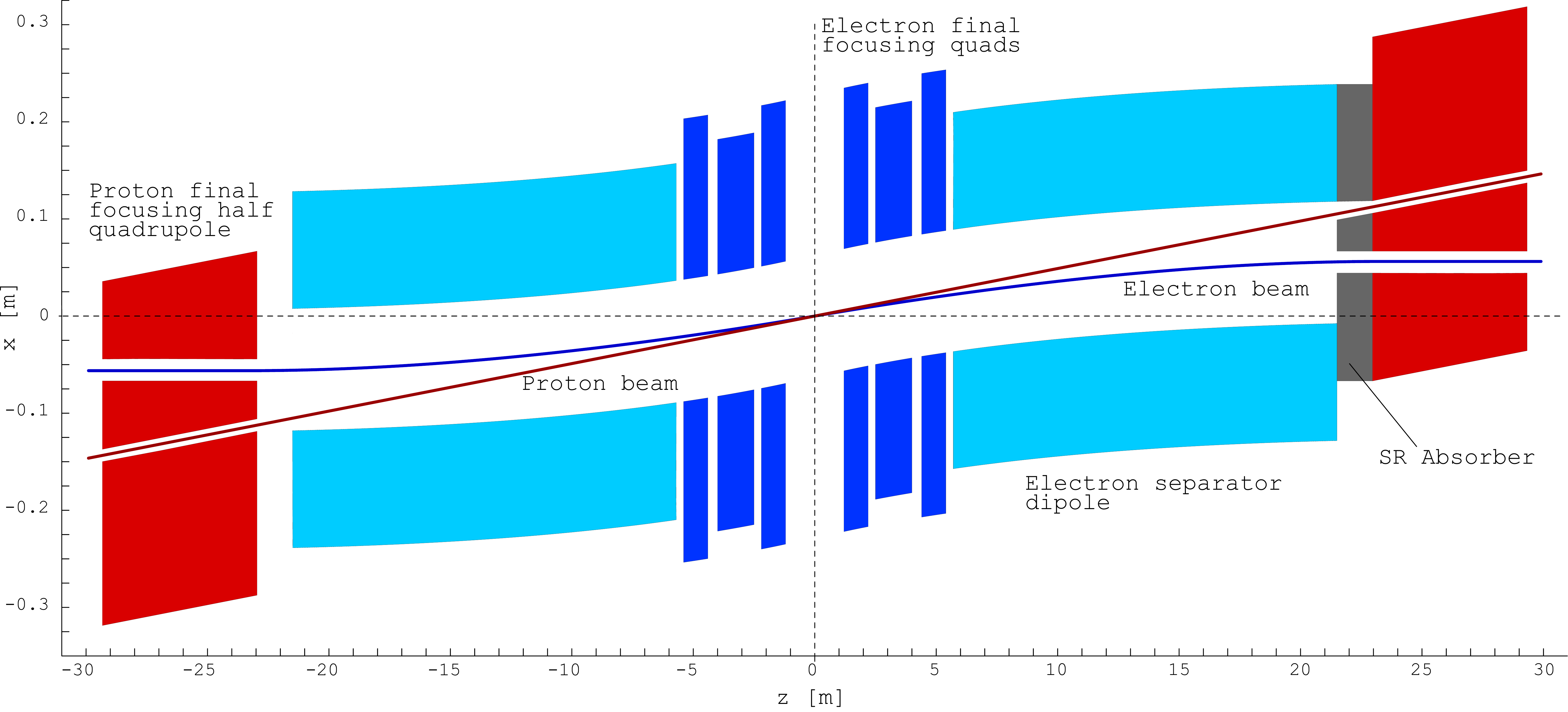}}
\caption{Schematic layout of the LHeC 10 Degree interaction region}
\label{Fig:IR_gross}
\end{figure}

Figure~\ref{Fig:IR_gross} shows a schematic layout of the interaction region. It refers to the 10 Degree option and shows a compact triplet structure that is used for early focusing of the electron beam. The electron mini beta quadrupoles are embedded into the detector opening angle and in order to obtain the required separation effect they are shifted in the horizontal plane and  act effectively as combined function magnets: Thus focusing and separation of the electron beam are combined in a very compact lattice structure, which is the prerequisite to achieve luminosity values in the range of $10^{33}$ $\mathrm{cm}^{-2}\mathrm{s}^{-1}$. 
 
\subsection{Beam separation scheme}
The separation scheme of the two beams has to be optimised with respect to an efficient (i.e. fast) beam separation and a synchrotron radiation power and critical energy of the emitted photons that can be tolerated by the absorber design. Two main issues have to be accomplished: a sufficient  horizontal distance between the beams has to be generated at the position of the first proton (half) quadrupole, located at a distance of $s=$~23~m from the interaction point (the nominal value of the LHC proton lattice). In addition to that, harmful beam beam effects have to be avoided at the first   parasitic bunch encounters which will take place at $s=$~3.75~m, as the nominal bunch distance in LHC corresponds to $\Delta t=25$~ns. These so-called parasitic bunch crossings have to be avoided as they would lead to intolerable beam-beam effects in the colliding beams. As a consequence the separation scheme has to deliver a sufficiently large horizontal distance between the two counter rotating bunches at these locations. 
\par
To achieve the first requirement a separation effect is created inside the mini beta quadrupoles of the electron beam:  The large momentum difference of the two colliding beams provides a very elegant way to separate the lepton and the hadron beams:  Shifting the mini-beta quadrupoles of the electron beam and installing a 15.8~m long, but weak separator dipole  magnet close to the IP provides the gentle separation that is needed to keep the synchrotron radiation level in the IR within reasonable limits. 

\begin{figure}
\centerline{\includegraphics[clip=,width=0.55\textwidth]{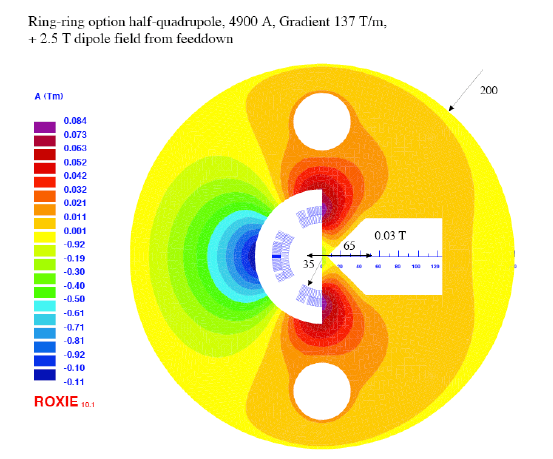}}
\caption{Super conducting half  quadrupole  in the proton lattice: The electron beam will pass on the right hand side of the mirror plate in 
a quasi field free region (see Sec.~\ref{tripletmagnets}).}
\label{Fig:half_quad}
\end{figure}

The nearest proton quadrupole to the IP is designed as a half-quadrupole to ease the extraction of the outgoing electron beam. At this location (at $s=$~23~m) a minimum separation of $ \Delta x=$~55~mm  is needed to guide the electron beam along the mirror plate of a sc. proton half quadrupole (see Sec.~\ref{tripletmagnets}). A first layout of this magnet is sketched in figure \ref{Fig:half_quad}

The horizontal offsets of the mini beta lenses are chosen individually in such a way that the resulting bending strength in the complete separation scheme (quadrupole triplet / doublet and separator dipole) is constant. In this way a moderate separation strength is created with a constant bending radius of $ \rho=6757$~m for the 10 Degree option. In the case of the 1 Degree option the quadrupole lenses of the electron lattice cannot be included inside the detector design as the opening angle of the detector does not provide enough space for the hardware of the electron ring lattice. Therefore a much larger distance between the IP and the location of the first electron lens had to be chosen ($\Delta s=$6.2~m instead of $\Delta s=$~1.2~m). As a consequence - in order to achieve the same overall beam separation - stronger magnetic separation fields have to be applied resulting in a  bending radius of  $ \rho=$~4057~m in this case. In both cases the position of the electron quadrupoles is following the design orbit of the electron beam to avoid local strong bending fields and keep the synchrotron radiation power to a minimum. This technique has already been successfully applied at the layout of the HERA electron-proton collider \cite{Rossbach:1985ve}. 
 
Still the separation at the location of the first proton magnet is small and  a half quadrupole design for this super conducting magnet has  been chosen at this point. The resulting beam parameters - including the expected  luminosity for this Ring-Ring option - are summarised in Table~\ref{tab:IR_parameters}.

\begin{table}
\begin{center}
\begin{tabular}{|l|c|c|c|c|c|}
\hline
Detector Option       &            &      \multicolumn{2}{c|} { 1$^{\circ}$  }  &    \multicolumn{2}{c|} {10$^{\circ}$ }    \\ \hline
Quantity              &     unit   &       electrons      &    protons           &     electrons      &            protons    \\
\hline
Number of bunches     &            &           \multicolumn{4}{c|} {2808}                           \\\hline
Particles/bunch $N_b$ & $10^{10}$  &          1.96        &         17            &           1.96    &                 17    \\
Horiz. beta-function  &   m        &           0.4        &        4.0            &           0.18    &                1.8    \\
Vert. beta-function   &   m        &           0.2        &        1.0            &            0.1    &                0.5    \\
Horiz.\ emittance     &  nm        &           5.0        &        0.5            &            5.0    &                 0.5   \\
Vert.\ emittance      &  nm        &           2.5        &        0.5            &            2.5    &                 0.5   \\
\hline
Distance to IP        &   m        &           6.2        &         22            &            1.2    &                  22   \\ \hline
Crossing angle        &  mrad      &           \multicolumn{2}{c|} {1.0}         &      \multicolumn{2}{c|} {1.0}           \\ \hline
Synch. Rad. in IR     &  kW        &           \multicolumn{2}{c|} {51}          &      \multicolumn{2}{c|} {33}            \\ \hline
absolute Luminosity   &m$^{-2}$ s$^{-1}$&      \multicolumn{2}{c|} {$8.54*10^{32}$ }&      \multicolumn{2}{c|} {$1.8*10^{33}$} \\  \hline
Loss-Factor S         &            &        \multicolumn{2}{c|} {0.86}           &      \multicolumn{2}{c|} {0.75}          \\  \hline
effective Luminosity  &m$^{-2}$ s$^{-1}$&      \multicolumn{2}{c|} {$7.33*10^{32}$ }&      \multicolumn{2}{c|} {$1.34*10^{33}$}\\  \hline   
\end{tabular}
\caption{Parameters of the mini beta optics for the 1$^{\circ}$ and 10$^{\circ}$ options of the LHeC Interaction Region.}
\label{tab:IR_parameters}
\end{center}
\end{table}

It has to be pointed out in this context that the arrangement of the off centre quadrupoles as well as the strength of the separator dipole depend on the beam optics of the electron beam. The beam size at the parasitic crossings and at the proton quadrupole will determine the required horizontal distance between the electron and proton bunches.  The strength and position of these magnets however will determine the optical parameters, including the dispersion function that is created during the separation process itself. Therefore a self-consistent layout concerning optics, beam separation and geometry of the synchrotron light absorbers has to be found. 

It is obvious that these boundary conditions have to be fulfilled not only during luminosity operation of the e/p rings. During injection and the complete acceleration procedure of the electron ring the influence of the electron quadrupoles on the proton beam has to be compensated with respect to the proton beam orbit (as a result of the separation fields) as well as to the proton beam optics: The changing deflecting fields and gradients  of the electron magnets will require correction procedures in the proton lattice that will compensate this influence at any moment.  

\subsection{Crossing angle} 
\begin{figure}
\centerline{\includegraphics[clip=,width=0.55\textwidth]{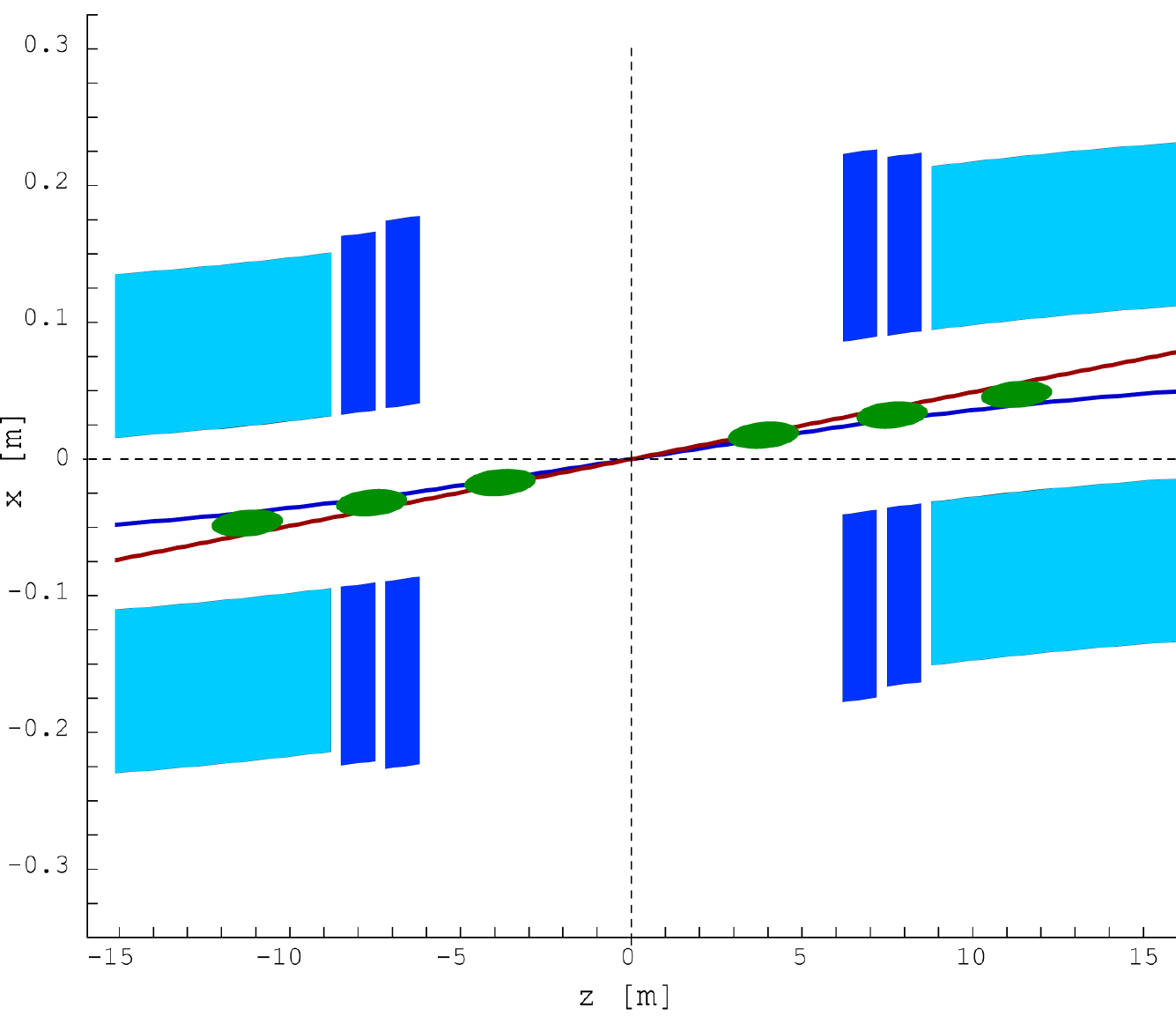}}
\caption{LHeC interaction region including the location of the first parasitic bunch encounters where 
a sufficient beam separation is achieved by a crossing angle of 1 mrad. The location of the parasitic encounters is indicated by green ovals.}
\label{Fig:IR_zoom_bunche}
\end{figure}

A central aspect of the LHeC IR design is the beam-beam interaction of the colliding electron and proton bunches. The bunch structure of the electron beam will match the pattern of the LHC proton filling scheme for maximal luminosity, giving equal bunch spacing of 25~ns  to both beams. The IR design therefore is required to separate the bunches as quickly as possible to avoid additional bunch interactions at these positions and limit the beam-beam effect to the desired  interactions at the IP. The design bunch distance in the LHC proton bunch chain corresponds to $ \Delta t=$~ 25~ns  or $\Delta s=$~7.5~m. The counter rotating bunches therefore meet after the crossing at the interaction point at additional, parasitic collision points in  a distance $s=$~3.75~m from the IP.  To avoid detrimental effects from these parasitic crossings the above mentioned separation scheme has to be supported by a crossing angle that will deliver a sufficiently large horizontal distance between the bunches at the first parasitic bunch crossings. 
This technique is used in all LHC interaction points. In the case of the LHeC however, the crossing angle is determined by the emittance of the electron beam and the resulting beam size which is considerably larger than the usual proton beam size in the storage ring. In the case of the LHeC IR a crossing angle of $\theta=$~1~mrad is considered as sufficient in the 1$^{\circ}$ as well as in the 10$^{\circ}$ option to avoid beam-beam effects from this parasitic crossings. Figure~\ref{Fig:IR_zoom_bunche} shows the position of the first possible parasitic encounters and the effect of the crossing angle to deliver a sufficient separation at these places. 

The detailed impact of one beam on another is evaluated by a dedicated beam-beam interaction study which is included in this report, based on a minimum separation of 5$\sigma_e + 5\sigma_p$ at every parasitic crossing node. Due to the larger electron emittance the separation is mainly dominated by the electron beam parameters, and as a general rule it can be stated that the rapid growth of the $\beta$-function in the drift around the IP,
\begin{equation}
\beta(s)=\beta^* + \frac{s^2}{\beta^*},
\end{equation}
makes it harder  to separate the beams if small $\beta^*$ and a large drift space $s$ is required in the optical design.  

In any design for the LHeC study, a crossing angle is used to establish an early beam separation, reduce the required strength in the separation magnets and minimise the synchrotron radiation power that is created inside the interaction region. 

As a draw back however  the luminosity is reduced  due to the fact that the bunches will not collide anymore head on. This reduction is expressed in a geometric luminosity reduction factor ``S'', that depends on the crossing angle $ \theta$, the length of the electron and proton bunches $ \sigma_{ze} $ and $ \sigma_{zp} $ and the transverse beam size in the plane of the bunch crossing $ \sigma^*_x $:

\begin{equation}
S(\theta) = \left[1+\left(\frac{\sigma^2_{sp}+\sigma^2_{se}}{2\sigma^{*2}_x}\right)\tan^2\frac{\theta}{2}\right]^{-\frac{1}{2}}\ .
\label{eqn:S}
\end{equation}\\

Accordingly, the effective luminosity that can be expected for a given IR layout is obtained by 
\begin{equation}
L = S(\theta) * L_0
\end{equation}\\
  
For the two beam optics that have been chosen for this design study (the  1$^{\circ}$ and the  10$^{\circ}$ option) and a  crossing angle of $ \theta $ = 1mrad  the loss factor amounts to 
$S=86 \%$ and $S=75 \%$ respectively.

\subsection{Beam optics and luminosity}

\begin{figure}
\centerline{\includegraphics[clip=,width=0.65\textwidth]{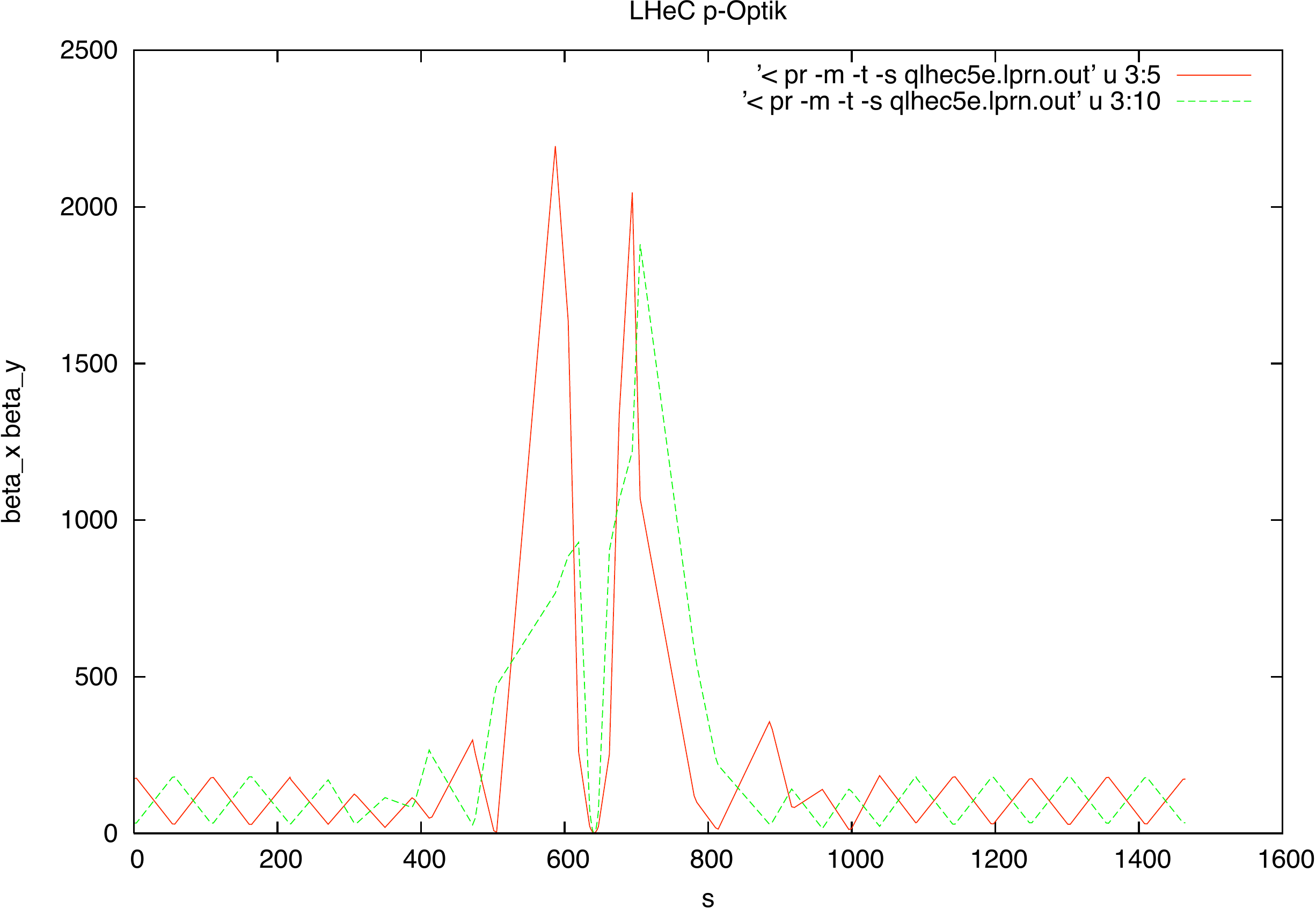}}
\caption{Proton optics for the LHeC interaction region. The gradients of the antisymmetric triplet lattice in the standard LHC have been modified to adopt for the requirements of the LHeC flat beam parameters.}
\label{Fig:lhec_p_optik}
\end{figure}

A special boundary condition had to be observed in the design of the proton beam optics of the LHeC: For the layout of the four present proton-proton interaction regions in the LHC machine an anti-symmetric option had been chosen: A solution that is appropriate for a round beam optics ( $ {\sigma_x}^* = {\sigma_y}^* $ ). An optimised design for collisions with the flat e$^{\pm} $  beams however requires unequal $ \beta $ -functions for the hadron beam at the IP and the existing LHC optics can no longer be maintained. Therefore the optical layout of the existing triplet structure in the LHC had to be modified to match the required beta functions ( $ \beta_x=$~1.8~m, $\beta_y=$~0.5~m) at the IP  to the regular optics of the FODO structure in the arc (Figure~\ref{Fig:lhec_p_optik}).

In the case of the electron beam optics, two different layouts of the interaction region are considered: One optical concept for highest achievable luminosity and a solution for maximum detector acceptance. In the first case an opening angle of  10$^{\circ}$ is available inside the detector geometry and  allows to install an embedded magnet structure where the first electron quadrupole lenses can be placed as close as $s=$~1.2~m  from the IP. This early focusing scheme leads to moderate values of the $\beta$ function inside the mini beta quadrupoles and therefore allows for a smaller spot size at the IP and larger luminosity values can be achieved. Still however the quadrupoles require a compact design: While the gradients required by the optical solution are small (for a super conducting magnet design) the outer radius of the first electron quadrupole has been limited to $ r_{max}$ = 210~mm.

\begin{figure}
\centerline{\includegraphics[clip=,width=0.65\textwidth]{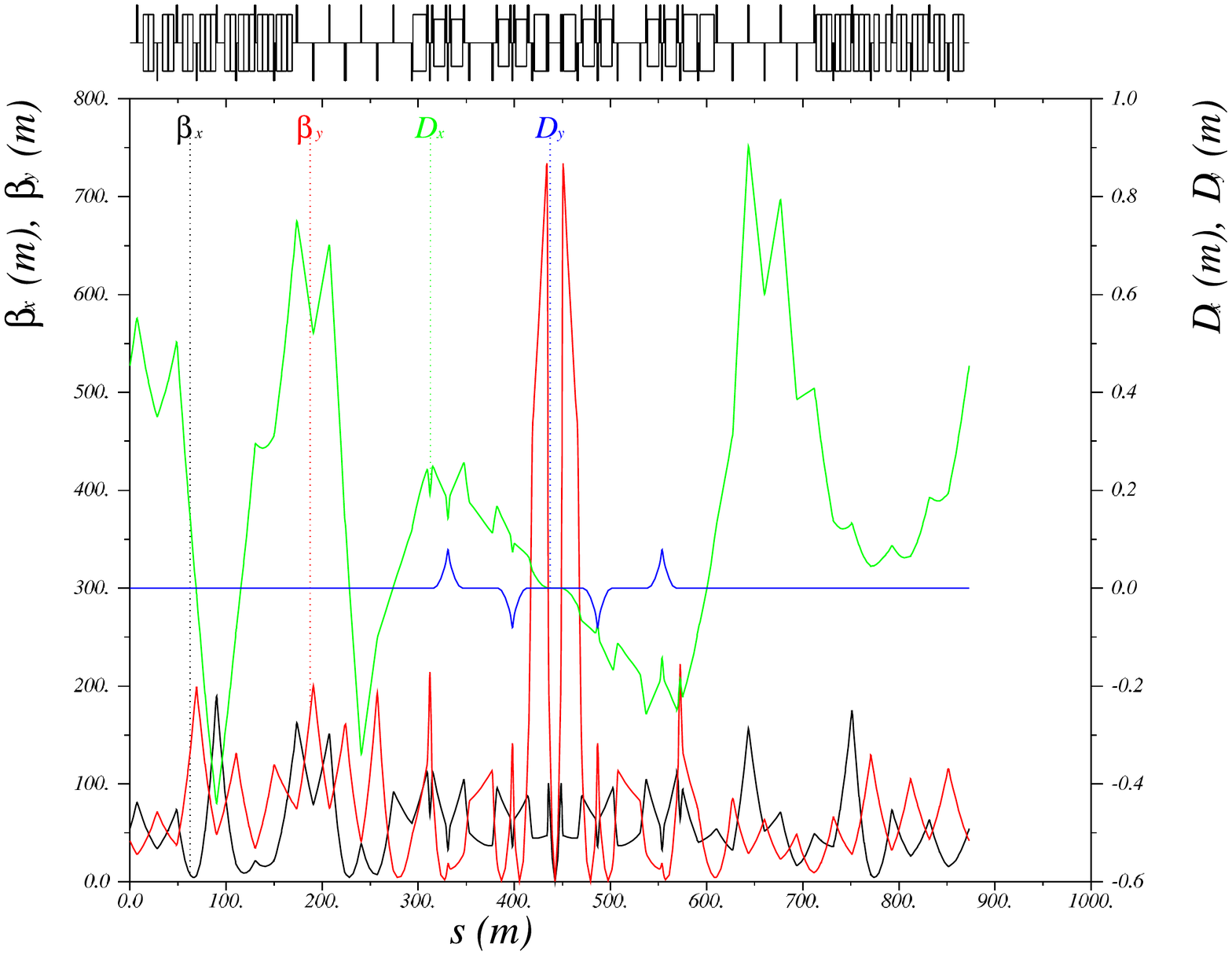}}
\caption{Electron optics for the LHeC interaction region. The plot corresponds to the 1 Degree option where a doublet structure combined with a 
separation dipole has been chosen to separate the two beams.}
\label{Fig:lhec_e_optik}
\end{figure}

In the case of the 1$^{\circ}$ option the detector design is optimised for largest detector acceptance. Accordingly the opening angle of the detector hardware is too small to deliver space for accelerator magnets. The mini beta quadrupoles therefore have to be located outside the detector, and  a distance $s=$~6.2~m from the IP had to be chosen in this case. Even if the magnet dimensions are not limited by the detector design in this case, the achievable luminosity is about a factor of two smaller than in the 10$^{\circ}$ case.  

The two beam optics that are based on these considerations are discussed in detail in the next chapter of this report. In the case of the 10$^{\circ}$ option a triplet structure has been chosen to allow for moderate values of the beta functions inside the mini beta quadrupoles. As a special feature of the optics that is shown in Figure~\ref{Fig:lhec_e_optik} the focusing effect of the first quadrupole magnet is moderate: Its gradient has been limited as it has to deliver mainly the first beam separation.  Table~\ref{tab:IR_parameters} includes as well the overall synchrotron radiation power that is produced inside the IR. Due to the larger bending radius (i.e. smaller bending forces) in the case of the  10$^{\circ}$ option the produced synchrotron radiation power is limited to about 30~kW, while the alternative - high acceptance - option has to handle 50~kW of synchrotron light.

The details of the synchrotron light characteristics   are covered in the next chapters of this report for both cases, including the critical energies and the  design of the required absorbers. 

For the 1$^{\circ}$ option the mini beta focusing is based on a quadrupole doublet as the space limitations in the transverse plane are much more relaxed compared to the alternative option and the main issue here was to find  a compact design in the longitudinal coordinate: Due to the larger distance of the focusing and separating magnets from the IP the magnet structure has to be more compact and the separating field stronger to obtain the required horizontal beam distance at the location $s=$~23m of the first proton quadrupole. The corresponding beam optics for both options are explained in full detail below.  

\section{Design requirements}
\label{IR.Design}
\subsection{Detector coverage and acceptance}
\label{IR.Design.MA}
\begin{figure*}[!h]
\centerline{\includegraphics[scale=0.5]{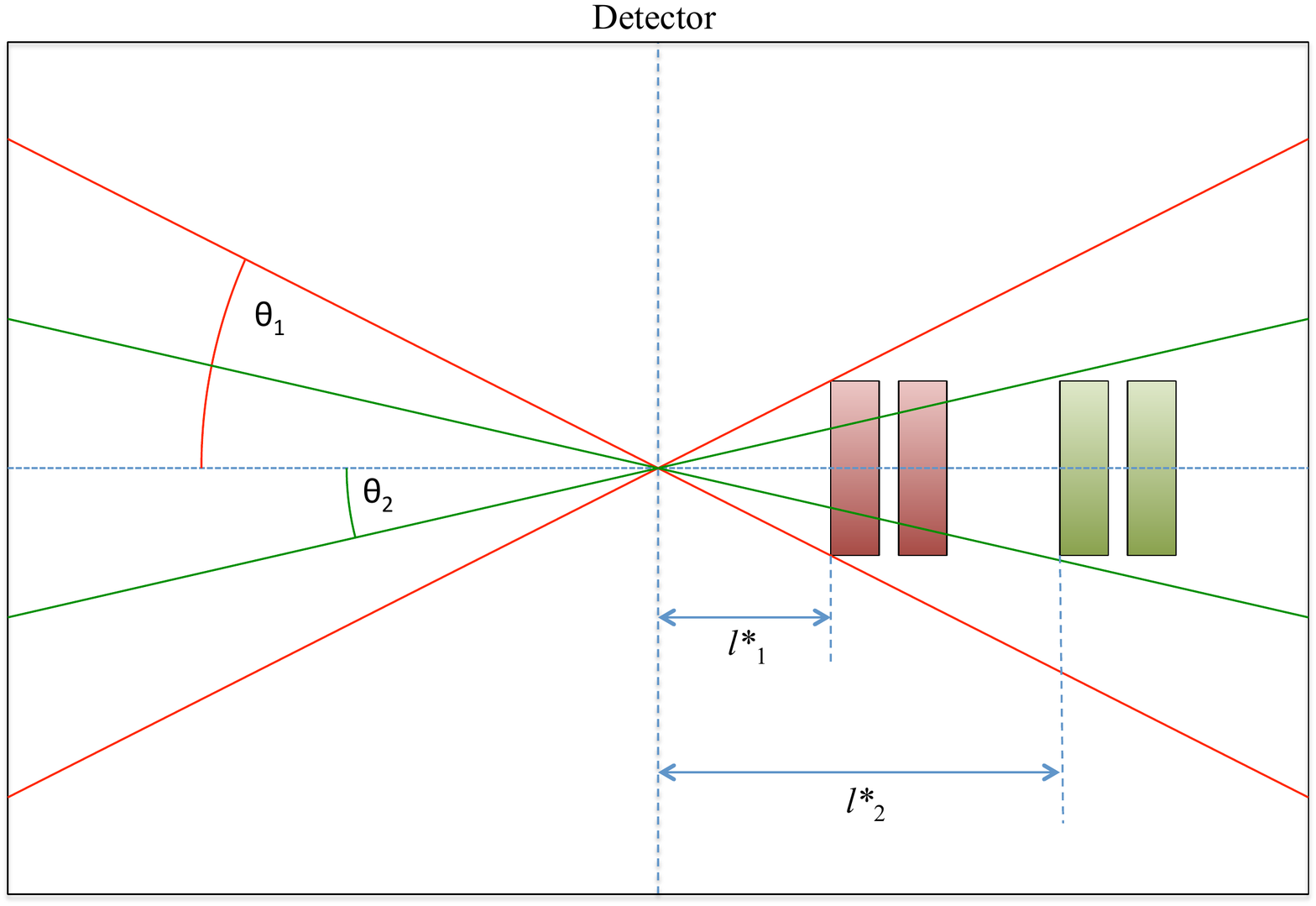}}
\caption{Graphical representation of acceptance. $\theta_1$ shows a lower acceptance cone, while $\theta_2$ shows a higher acceptance cone. For machine elements of constant diameter, higher acceptance increases $l$*.}
\label{fig:IR.MA.MA}
\end{figure*}

Acceptance describes the amount of angular obstruction of the detector due to the presence of machine elements, as shown in Figure~\ref{fig:IR.MA.MA}. For example, an acceptance of 10\textdegree~implies a protrusion of machine elements into the detector such that a cone of 10\textdegree~half-angle along the beam axis is blocked. The detector is thus unable to see particles emitted at less than this angle, and event data is lost at high pseudo-rapidities. Accordingly larger detector opening angles denote lower acceptance 
but allows to position machine elements at a smaller distance to the IP.

Since $\beta$ grows quadratically with distance, a smaller $l$* generally allows stronger focusing of a beam and thus higher luminosity. While there is no direct relationship between $l$* and luminosity, a balance must be found to optimise both luminosity and acceptance. Two IR designs are proposed as solutions to the balance between luminosity and acceptance. Both designs aim to achieve a luminosity in the range of $\sim$10$^{33}$~cm$^{-2}$s$^{-1}$.

\begin{enumerate}
\item High Luminosity Layout (HL)
	\begin{itemize}
		\item 10\textdegree~acceptance
		\item Higher luminosity
	\end{itemize}
\item High Acceptance Layout (HA)
	\begin{itemize}
		\item 1\textdegree~acceptance
		\item Lower luminosity
	\end{itemize}
\end{enumerate}

\noindent In concert with these designs, two plans are proposed for running LHeC. One option is to run with the HL layout, then switch to the HA layout during a shutdown. The second option is to optimise the HA layout for sufficient luminosity to replace the HL layout entirely.

\subsection{Lattice matching and IR geometry}
\label{IR.Design.LM}
The principle layout and requirements of the beam separation scheme have been described above. A minimum separation of 5$\sigma_e$ + 5$\sigma_p$ is specified at each parasitic node. In addition an overall distance between the proton and electron beam of  55~mm at the location of the first proton magnet, $s=$~23~m, has been chosen as an attainable target from optical, radiation (see Sec.~\ref{sec:NATHAN}) and magnet design (see Sec.~\ref{tripletmagnets}) standpoints. 

Once the beams are separated into independent beam pipes, the electron beam must be transported into the ring lattice. Quadrupoles are used in the long straight section (LSS) of the electron machine  to transport the beam from the IP to the dispersion suppressor and match the twiss parameters at either end. Space must be available to insert dipoles and further quadrupoles to allow the orbit of the beam to be designed with regard to the physical layout of the ring and the IR.

The IR and LSS geometries must be designed around a number of further constraints. In addition to the beam separation required to avoid parasitic bunch encounters, the electron beam must be steered from the electron ring into the IR and back out again. The colliding proton beam must be largely undisturbed by the electron beam. The non-colliding proton beam must be guided through the IR without interacting with either of the other beams.

\section{High luminosity IR layout}
\label{IR.HL}

\subsection{Parameters}
\label{IR.HL.P}
Table \ref{tab:IR.HL.P.Params} details the interaction point parameters and other parameters for this design. To optimise for luminosity, a small $l$* is desired. An acceptance angle of 10\textdegree~is therefore chosen, which gives an $l$* of 1.2~m for final focusing quadrupoles of reasonable size.

\begin{table}[!h]
\begin{centering}
\begin{tabular}{|l|l|}
	\hline
	$L(0)$&1.8$\times 10^{33}$\\ \hline
	$\theta$&1$\times 10^{-3}$\\ \hline
	$S(\theta)$&0.746\\ \hline
	$L(\theta)$&1.34$\times 10^{33}$\\ \hline
	$\beta_x*$&0.18 m\\ \hline
	$\beta_y*$&0.1 m\\ \hline
	$\sigma_x*$&3.00$\times 10^{-5}$ m\\ \hline
	$\sigma_y*$&1.58$\times 10^{-5}$ m\\ \hline
	SR Power&33 kW\\ \hline
	$E_c$&126 keV\\ \hline
\end{tabular}
\caption{Parameters for the HL IR. Note that the geometric luminosity reduction factor, S, is calculated using the LHC ultimate bunch length of 75 mm.}
\label{tab:IR.HL.P.Params}
\end{centering}
\end{table}

\noindent SR calculations are detailed in section (see Sec.~\ref{sec:NATHAN}). The total power emitted in the IR is similar to that in the HERA-2 IR\cite{Schneekloth:1998kh} and as such appears to be reasonable, given enough space for absorbers.

\subsection{Layout of the electron lattice}
\label{IR.HL.L}
\begin{figure*}[!h]
\centerline{\includegraphics[scale=0.6]{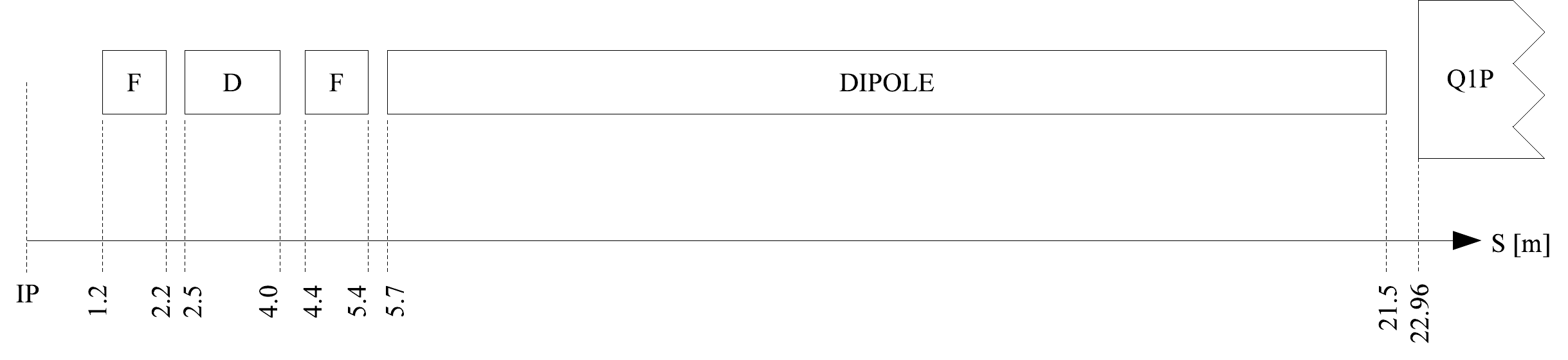}}
\caption{Layout of machine elements in the HL IR. Note that the left side of the IR is symmetric.}
\label{fig:IR.HL.L.Layout}
\end{figure*}

A symmetric final quadrupole triplet layout followed by a long weak dipole magnet has been chosen for this design, due to the relatively round beam spot aspect ratio of 1.8:1. Figure~\ref{fig:IR.HL.L.Layout} and table \ref{tab:IR.HL.L.Layout} detail the layout.

\begin{table}[!h]
\begin{centering}
\begin{tabular}{|l|l|l|l|l|l|l|}
	\hline
	Element&S$_{\mathrm{entry}}$ [m]&L [m]&Gradient [T/m]&Dipole Field [T]&Offset [m]\\ \hline\hline
	BS.L&-21.5&15.8&-&-0.0296&-\\ \hline
	Q3E.L&-5.4&1.0&89.09229&-0.0296&-3.32240$\times 10^{-4}$\\ \hline
	Q2E.L&-4&1.5&-102.2013&-0.0296&2.89624$\times 10^{-4}$\\ \hline
	Q1E.L&-2.2&1.0&54.34071&-0.0296&-5.44711$\times 10^{-4}$\\ \hline
	IP&0.0&-&-&-&-\\ \hline
	Q1E.R&1.2&1.0&54.34071&0.0296&5.44711$\times 10^{-4}$\\ \hline
	Q2E.R&2.5&1.5&-102.2013&0.0296&-2.89624$\times 10^{-4}$\\ \hline
	Q3E.R&4.4&1.0&89.09229&0.0296&3.32240$\times 10^{-4}$\\ \hline
	BS.R&5.7&15.8&-&-0.0296&-\\ \hline
\end{tabular}
\caption{Machine elements for the HL IR. S$_{\mathrm{entry}}$ gives the leftmost point of the idealised magnetic field of an element. Note that S is relative to the IP.}
\label{tab:IR.HL.L.Layout} 
\end{centering}
\end{table}

\noindent The distance of the first electron magnet from the IP,  $l$* of 1.2 m, allows both strong focusing of the beam, and constant bending of the beam from \mbox{s=1.2~m} to \mbox{21.5~m}. This is achieved with offset quadrupoles and a separation dipole.

\begin{figure*}[!h]
\centerline{\includegraphics[scale=0.6,angle=90]{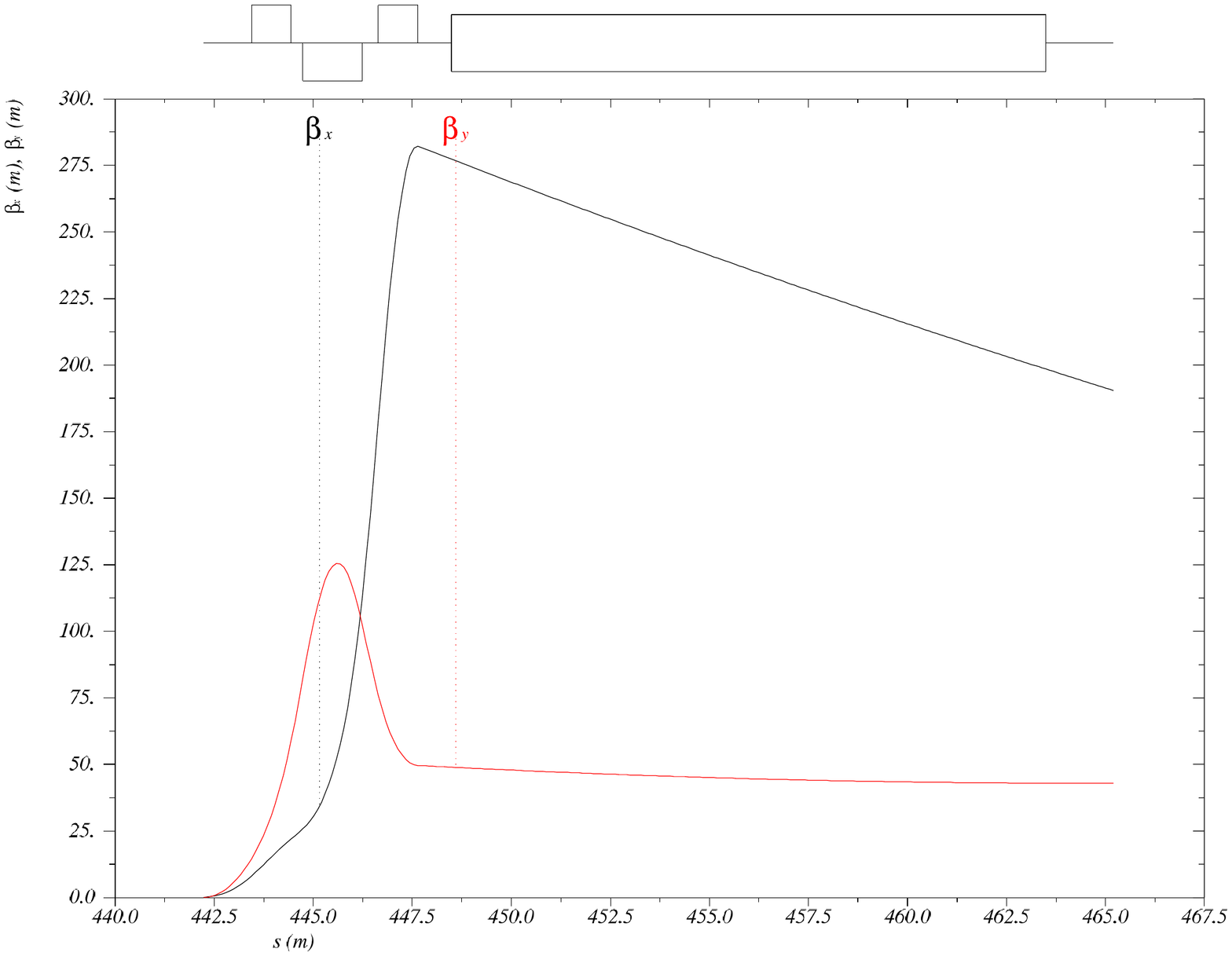}}
\caption{$\beta$ functions in both planes for the HL IR layout, from the IP to the face of the final proton quadrupole at s=23~m. Note that s is relative to the ring, which begins at the left side of the left dispersion suppressor of IP2.}
\label{fig:IR.HL.L.Twiss}
\end{figure*}

Figure~\ref{fig:IR.HL.L.Twiss} shows the $\beta$ functions of the beam in both planes from the IP to the face of the final proton quadrupole at s=23~m.

\subsection{Separation scheme}
\label{IR.HL.S}
The electron triplet is powered in FDF mode generating a large peak in $\beta_x$, but is designed such that the peak is between parasitic crossings. The first F quadrupole reduces $\beta_x$ at \mbox{s=3.75~m} compared to an initial D quadrupole. The third F quadrupole then reduces $\beta_x$  sufficiently to avoid large beam-beam interactions at the second parasitic crossing, $s=$~7.5~m.

This is aided by the bending provided by the offset quadrupoles, and also the IP crossing angle of 1~mrad. These elements ensure that the separation between the beams, normalised to the beam size, increases at each parasitic crossing. Note that 1~mrad is not a minimum crossing angle required by beam-beam interaction separation criteria but is  a chosen balance between luminosity loss and minimising bend strength. In theory, this layout could support an IP with no crossing angle; however the bend strength required to achieve this would generate an undesirable level of SR power.

\section{High acceptance IR layout}
\label{IR.HA}

\subsection{Parameters}
\label{IR.HA.P}
Table \ref{tab:IR.HA.P.Params} details the main parameters for this design. The chosen acceptance for this layout is 1\textdegree. For final electron focusing magnets of reasonable strength this places all elements outside the limits of the detector, at $s=±6.2$~m. Due to the small crossing angle the first electron magnets have to be placed beyond this distance. As such, the actual acceptance of the layout is limited by the beam pipe diameter rather than the size of machine elements. This also gives further flexibility in the strengths and designs of the final focusing quadrupoles.

\begin{table}[!h]
\centering
\begin{tabular}{|l|l|}
	\hline
	$L(0)$&8.54$\times 10^{32}$\\ \hline
	$\theta$&1$\times 10^{-3}$\\ \hline
	$S(\theta)$&0.858\\ \hline
	$L(\theta)$&7.33$\times 10^{32}$\\ \hline
	$\beta_x*$&0.4 m\\ \hline
	$\beta_y*$&0.2 m\\ \hline
	$\sigma_x*$&4.47$\times 10^{-5}$ m\\ \hline
	$\sigma_y*$&2.24$\times 10^{-5}$ m\\ \hline
	SR Power&51 kW\\ \hline
	$E_c$&163 keV\\ \hline
\end{tabular}
\caption{Parameters for the HA IR. Note that the geometric luminosity reduction factor, S, is calculated using the LHC ultimate bunch length of 75 mm.}
\label{tab:IR.HA.P.Params}
\end{table}

\noindent SR calculations are detailed in Sec.~\ref{sec:NATHAN}. Again, the total power emitted in the IR is similar to that in the HERA-2 IR\cite{Schneekloth:1998kh} and as such appears to be reasonable, given enough space for absorbers. However it is significantly higher than that in the HL layout. As discussed in Sec.~\ref{sec:NATHAN}, an option exists to reduce the total SR power by including a dipole field in the detector, thus mitigating the limitation imposed on dipole length by the larger $l$*.

\subsection{Layout}
\label{IR.HA.L}
\begin{figure*}[!h]
\centerline{\includegraphics[scale=0.6]{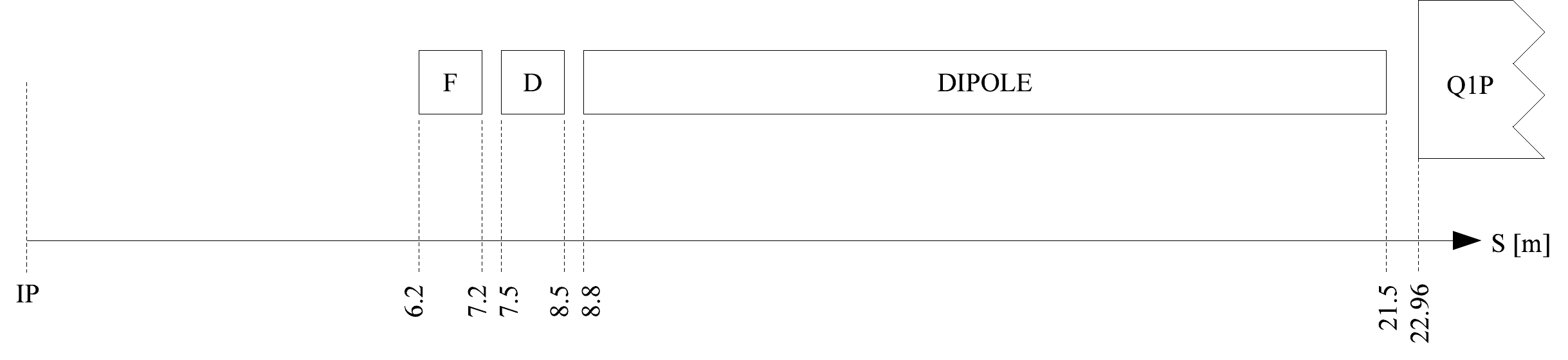}}
\caption{Layout of machine elements in the HA IR. Note that the left side of the IR is symmetric.}
\label{fig:IR.HA.L.Layout}
\end{figure*}

A symmetric final quadrupole doublet layout has been chosen for the electron lattice in this design. The beam spot aspect ratio of 2:1 is marginally flatter than the HL layout, and as such a triplet is less suitable. Figure~\ref{fig:IR.HA.L.Layout} and table \ref{tab:IR.HA.L.Layout} summarise the details of the layout.

\begin{table}[!h]
\centering
\begin{tabular}{|l|l|l|l|l|l|l|}
	\hline
	Element&S$_{\mathrm{entry}}$ [m]&L [m]&Gradient [T/m]&Dipole Field [T]&Offset [m]\\ \hline\hline
	BS.L&-21.5&12.7&-&-0.0493&-\\ \hline
	Q2E.L&-8.5&1.0&-77.30906&-0.0493&6.37700$\times 10^{-4}$\\ \hline
	Q1E.L&-7.2&1.0&90.38473&-0.0493&-5.45446$\times 10^{-4}$\\ \hline
	IP&0.0&-&-&-&-\\ \hline
	Q1E.R&6.2&1.0&90.38473&0.0493&5.45446$\times 10^{-4}$\\ \hline
	Q2E.R&7.5&1.0&-77.30906&0.0493&-6.37700$\times 10^{-4}$\\ \hline
	BS.R&8.8&12.7&-&0.0493&-\\ \hline
\end{tabular}
\caption{Machine elements for the HA IR. S$_{\mathrm{entry}}$ gives the leftmost point of the idealised magnetic field of an element. Note that S is relative to the IP.}
\label{tab:IR.HA.L.Layout} 
\end{table}

\noindent The $l$* of 6.2m imposes limitations on focusing and bending in this case. Focusing is limited by quadratic $\beta$ growth through a drift space, which is increased for smaller $\beta$*. As such, the achievable luminosity is smaller than in the HL design lattice.

Again offset quadrupoles are used to separate the beams. However this layout has less total dipole length available. Additionally, the first parasitic crossing occurs before the location of the  first electron quadrupole. This further limits final focusing as the beam cannot be permitted to grow too large by this time. Due to the reduced effective length for focusing and beam separation, stronger bending must be applied to obtain the overall separation of 55 mm at the place of the first proton quadrupole. Accordingly higher synchrotron radiation power is generated in this design.

\begin{figure*}[!h]
\centerline{\includegraphics[scale=0.6,angle=90]{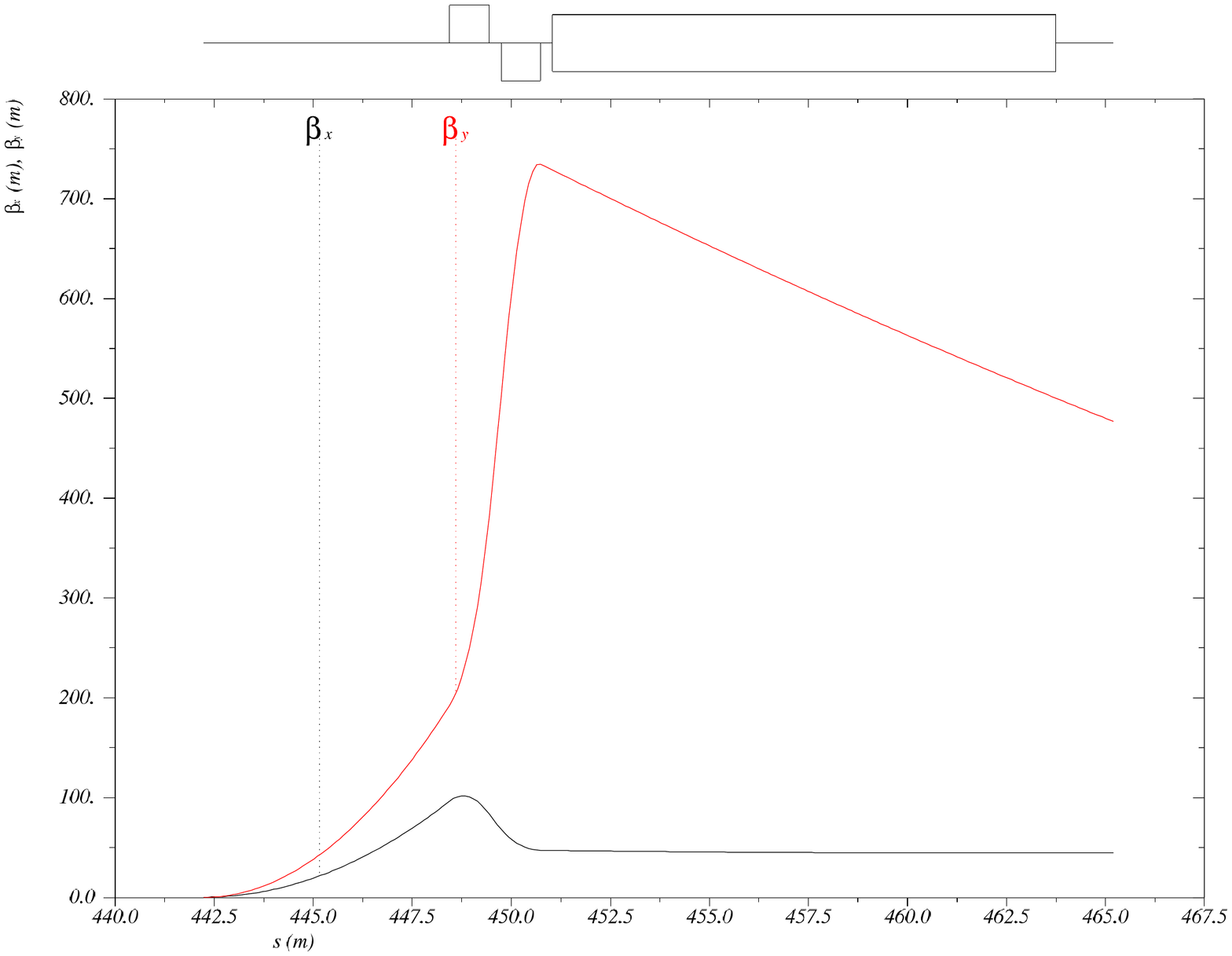}}
\caption{$\beta$ functions in both planes for the HA IR layout, from the IP to the face of the final proton quadrupole at \mbox{s=23~m}. Note that s is relative to the ring, which begins at the left side of the left dispersion suppressor of IP2.}
\label{fig:IR.HA.L.Twiss}
\end{figure*}

Figure~\ref{fig:IR.HA.L.Twiss} shows the $\beta$ functions of the beam in both planes from the IP to the face of the final proton quadrupole at \mbox{s=23~m}.

\subsection{Separation scheme}
\label{IR.HA.S}
The final electron doublet is optimised to limit the peak in $\beta_x$ on the cost of higher $\beta_y$.  Unlike the HL layout, the first parasitic crossing is reached before focusing begins. As such  a minimum crossing angle of roughly 0.7~mrad is required, which is dependent solely upon $\beta$ growth in the drift space. As a balance between luminosity loss and SR power generation, and aiding comparison with the HL layout, a crossing angle of 1~mrad has been chosen.

\section{Comparison of the two layouts}
\label{IRs.C}
Table \ref{tab:IR.Comp.Params} shows a direct comparison of various parameters of the two layouts.

\begin{table}[!h]
\centering
\begin{tabular}{|l|l|l|}
	\hline
	Parameter&HL&HA\\ \hline \hline
	$L(0)$&1.8$\times 10^{33}$&8.54$\times 10^{32}$\\ \hline
	$\theta$&1$\times 10^{-3}$&1$\times 10^{-3}$\\ \hline
	$S(\theta)$&0.746&0.858\\ \hline
	$L(\theta)$&1.34$\times 10^{33}$&7.33$\times 10^{32}$\\ \hline
	$\beta_x*$&0.18 m&0.4 m\\ \hline
	$\beta_y*$&0.1 m&0.2 m\\ \hline
	$\sigma_x*$&3.00$\times 10^{-5}$ m&4.47$\times 10^{-5}$ m\\ \hline
	$\sigma_y*$&1.58$\times 10^{-5}$ m&2.24$\times 10^{-5}$ m\\ \hline
	SR Power&33 kW&51 kW\\ \hline
	$E_c$&126 keV&163 keV\\ \hline
\end{tabular}
\caption{Parameter comparison for the HL and HA layouts.}
\label{tab:IR.Comp.Params}
\end{table}

\begin{figure*}[!h]
\centerline{\includegraphics[scale=0.6]{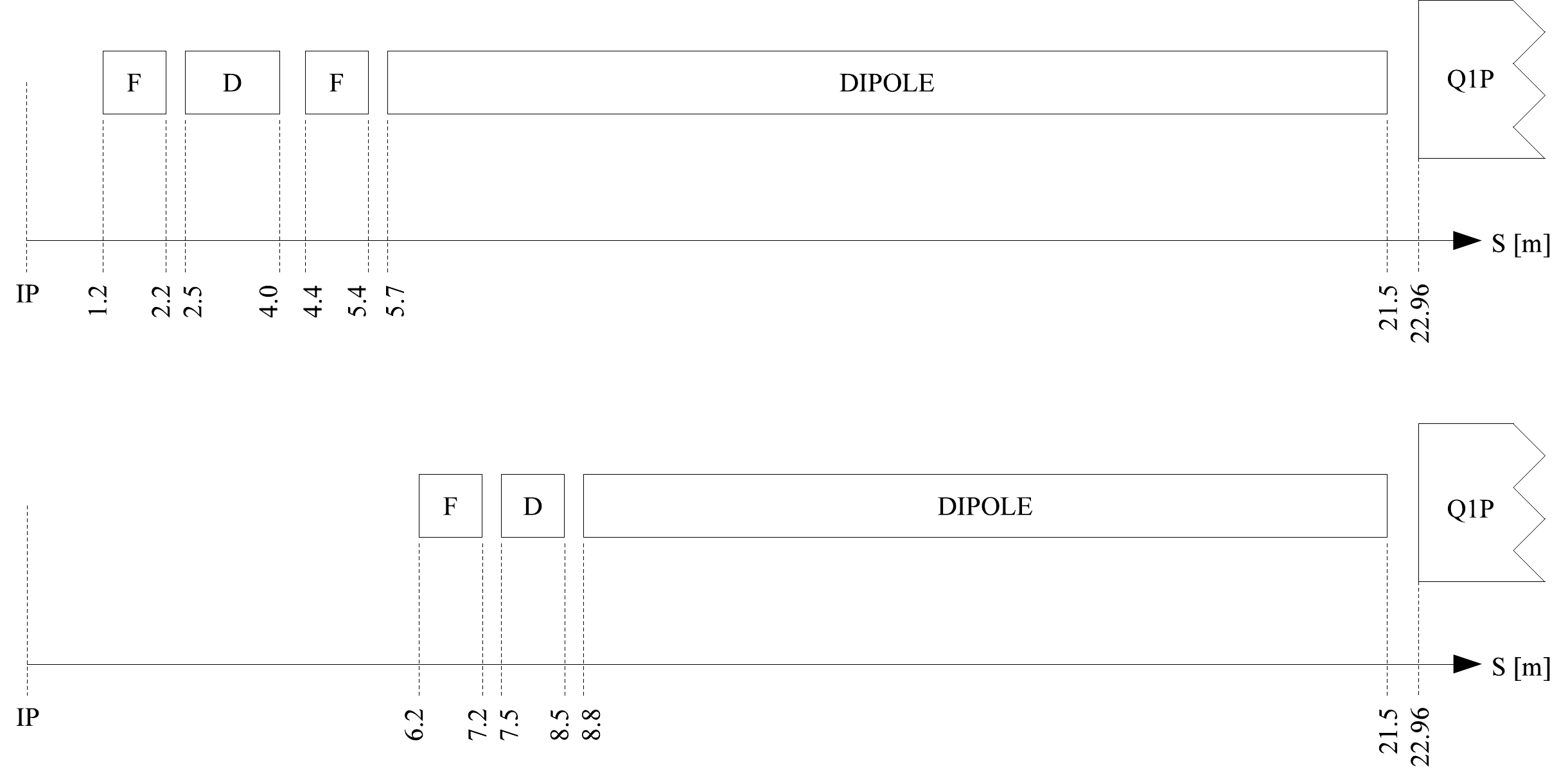}}
\caption{Scale comparison of the layouts for the HL and HA designs. Note the large difference in $l$*.}
\label{fig:IR.Comp.Layouts}
\end{figure*}

\noindent The difference in luminosity after considering losses due to the crossing angle is a factor of 1.8. However it should be noted that this design strives for technical feasibility and both layouts could potentially be squeezed further to decrease $\beta$* in both planes. The HL layout could likely be squeezed further than the HA layout due to the large difference in $l$*, as shown in Figure~\ref{fig:IR.Comp.Layouts} which compares the two IR layouts. At this stage both designs deliver their required IP parameters of luminosity and acceptance and appear feasible.

The HA design on the other side generates more SR power. This appears to be within reasonable limits and is discussed in Sec.~\ref{sec:NATHAN}. Furthermore, an option is discussed to install a dipole magnet in the detector. This early separation would reduce the required strength of the dipole fields in the IR, significantly reducing total SR power.

\subsection{Crab cavities}
\label{IRs.C.Crabs}
Both IR designs incorporate a crossing angle of 1mrad to facilitate fast beam separation. As discussed this introduces a luminosity loss factor S. The crossing angle is optimised to balance separation, SR power and luminosity. The loss factor is greater for the HL layout (0.746) than the HA (0.858) due to the smaller beam spot. However both are moderate, and as such a need for crab cavities is not foreseen.

Crab cavities rotate the bunch locally to the IP to counteract the effect of the crossing angle. They present a significant technical challenge, although feasibility has been demonstrated at KEKB\cite{Abe:2007bj}. It is preferred to avoid their necessity. However, their use remains a possibility if needs arise. For example, if designs for the proton half-quadrupoles prove to require larger beam separation than expected, increasing the crossing angle is likely the best option, as increased bending would quickly generate unfeasible levels of SR power. In this case, crab cavities would need to be considered to recover luminosity.

\section{Long straight section}
\label{IRs.LSS}
The Long Straight Section (LSS) geometrically and optically matches the IR to the rest of the LHeC ring lattice. For the purposes of this report, the LSS is defined from the start of the left dispersion suppressor (DS) to the end of the right DS. This is due to the need to alter the DS's optically and geometrically from the nominal design to obtain a valuable solution.

The LSS geometry for the electron ring uses a complex bending scheme in the horizontal and vertical plane to satisfy the various constraints. These include the 0.6~m radial offset of the LHeC ring as mentioned in Sec.~\ref{lat}, the 1~m vertical offset, and the IR separation geometry. The resulting small path length difference must be compensated elsewhere in the ring, nominally in the bypasses. 

It has to be be noted that in the current LSS design there are some conflicts between placements of the magnets for the LSS layout of the LHeC and standard LHC rings. The aim has been to design a self-consistent LHeC solution, and then iterate upon this to eliminate these conflicts. Future plans are discussed later in this section. It should also be noted that the solution presented is only matched for the HA IR layout. However generating a similar solution for the HL layout presents no additional challenges.

\subsection{Dispersion}
\label{IRs.LSS.D}
\begin{figure*}[!h]
\centerline{\includegraphics[scale=0.6]{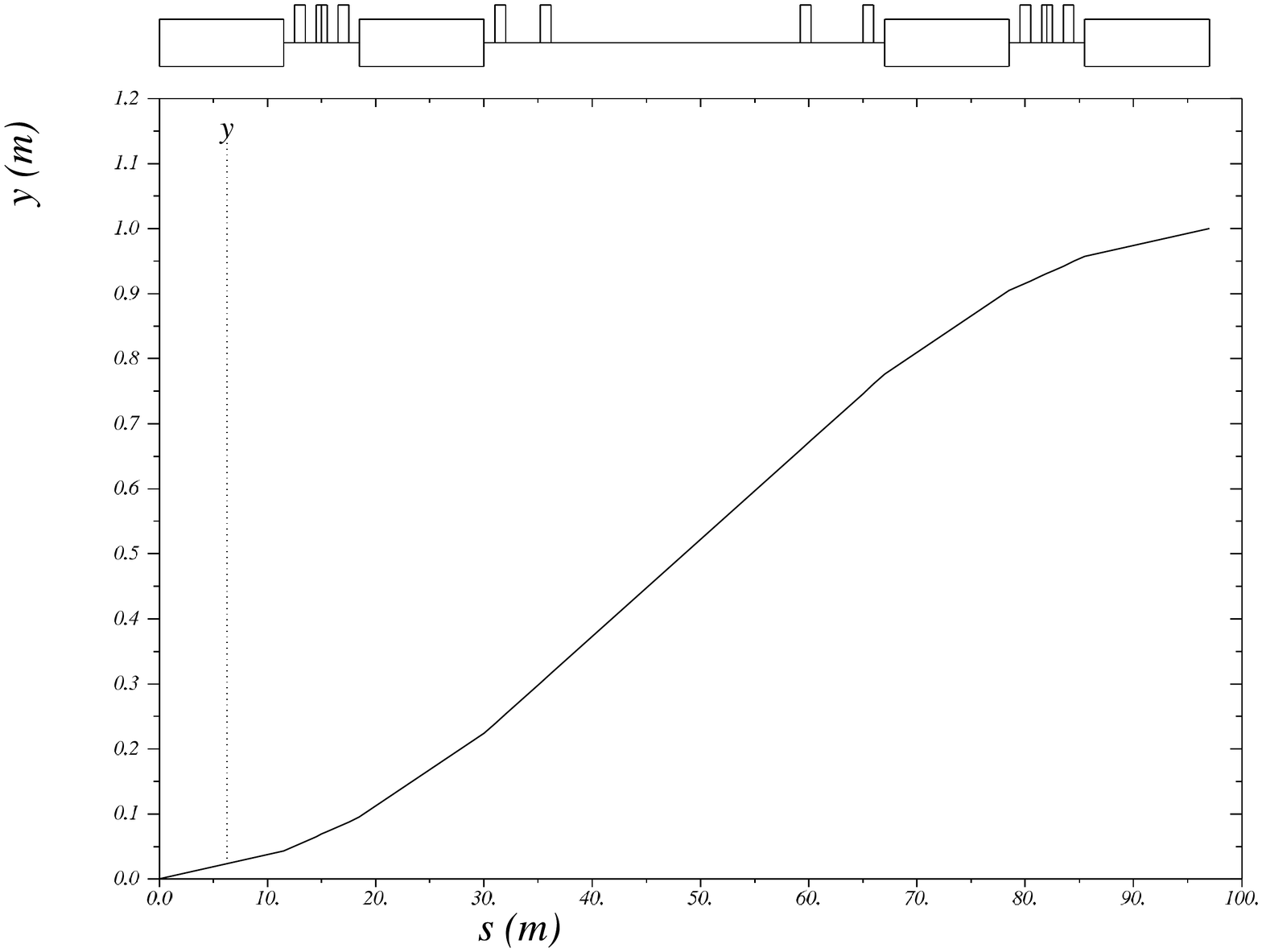}}
\caption{Geometry plot for a DBA dogleg pair in the HA LSS design.}
\label{fig:IR.LSS.D.DBAGeom}
\end{figure*}

\begin{figure*}[!h]
\centerline{\includegraphics[scale=0.6]{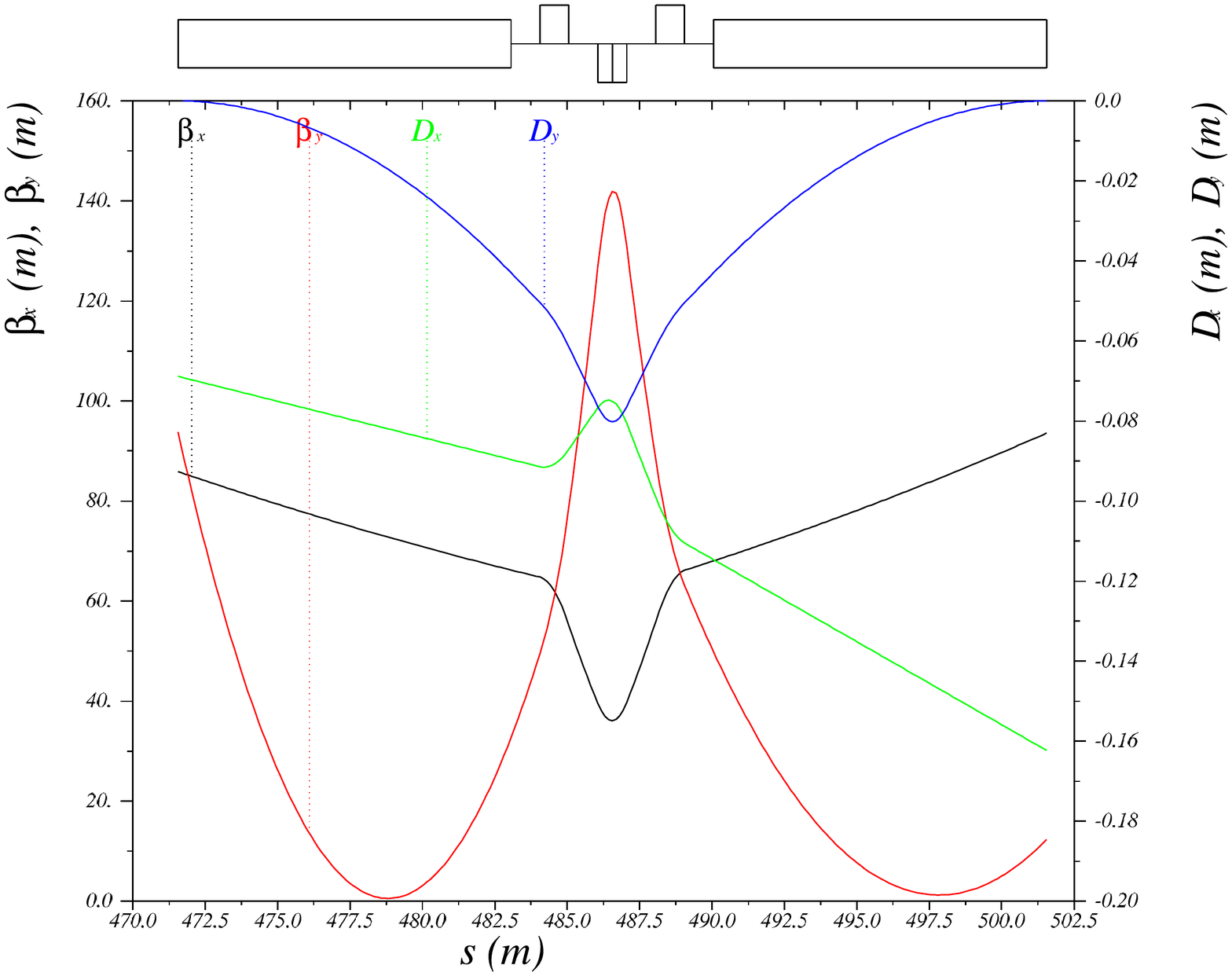}}
\caption{Optics plot for a single DBA module in the HA LSS design. Note waists and peaks in $\beta_y$.}
\label{fig:IR.LSS.D.DBAOptics}
\end{figure*}

A key constraint coupled to optics and geometry is dispersion. Since dispersion is an optical quantity generated by the deflecting fields, this becomes a challenge for the complex LSS bending scheme. The LHeC DSs are designed to match horizontal dispersion from the LSS to the arc. There is no equivalent scheme to deal with large vertical dispersion. Therefore an achromatic vertical separation scheme is proposed. Two vertical double bend achromat (DBA) sections on either side of the IR form doglegs while generating no vertical dispersion outside this region.  Figures~\ref{fig:IR.LSS.D.DBAGeom} and \ref{fig:IR.LSS.D.DBAOptics} detail the geometry and optics of the DBA sections used in the LSS.

\subsection{Geometry}
\label{IRs.LSS.Geom}
\begin{figure*}[!h]
\centerline{\includegraphics[scale=0.6]{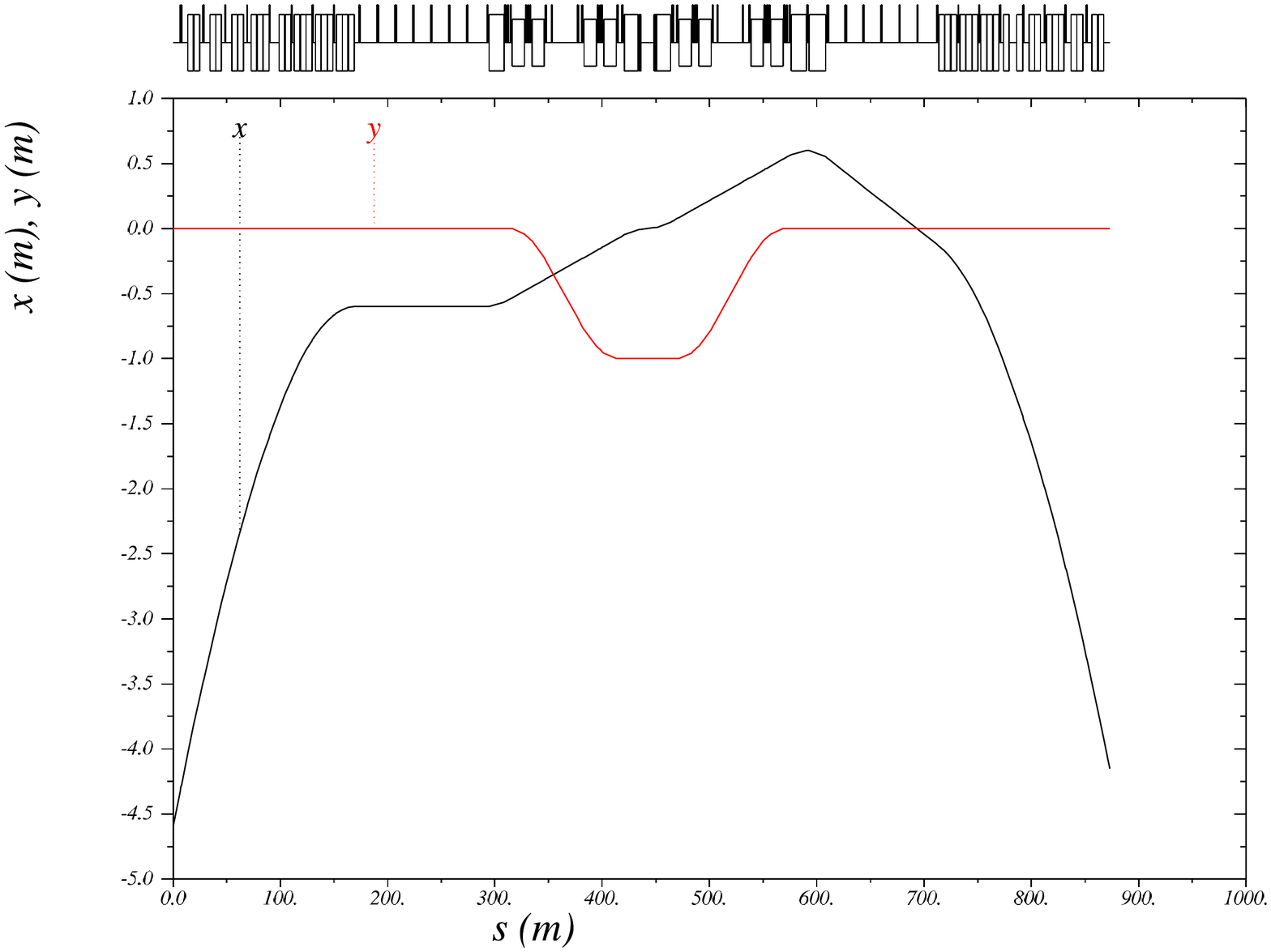}}
\caption{Geometry of the LSS design. Due to small angles involved, the s axis approximates the z axis well, and is used to allow MADX to display lattice elements.}
\label{fig:IR.LSS.Geom.LSSGeom}
\end{figure*}

\begin{figure*}[!h]
\centerline{\includegraphics[scale=0.6]{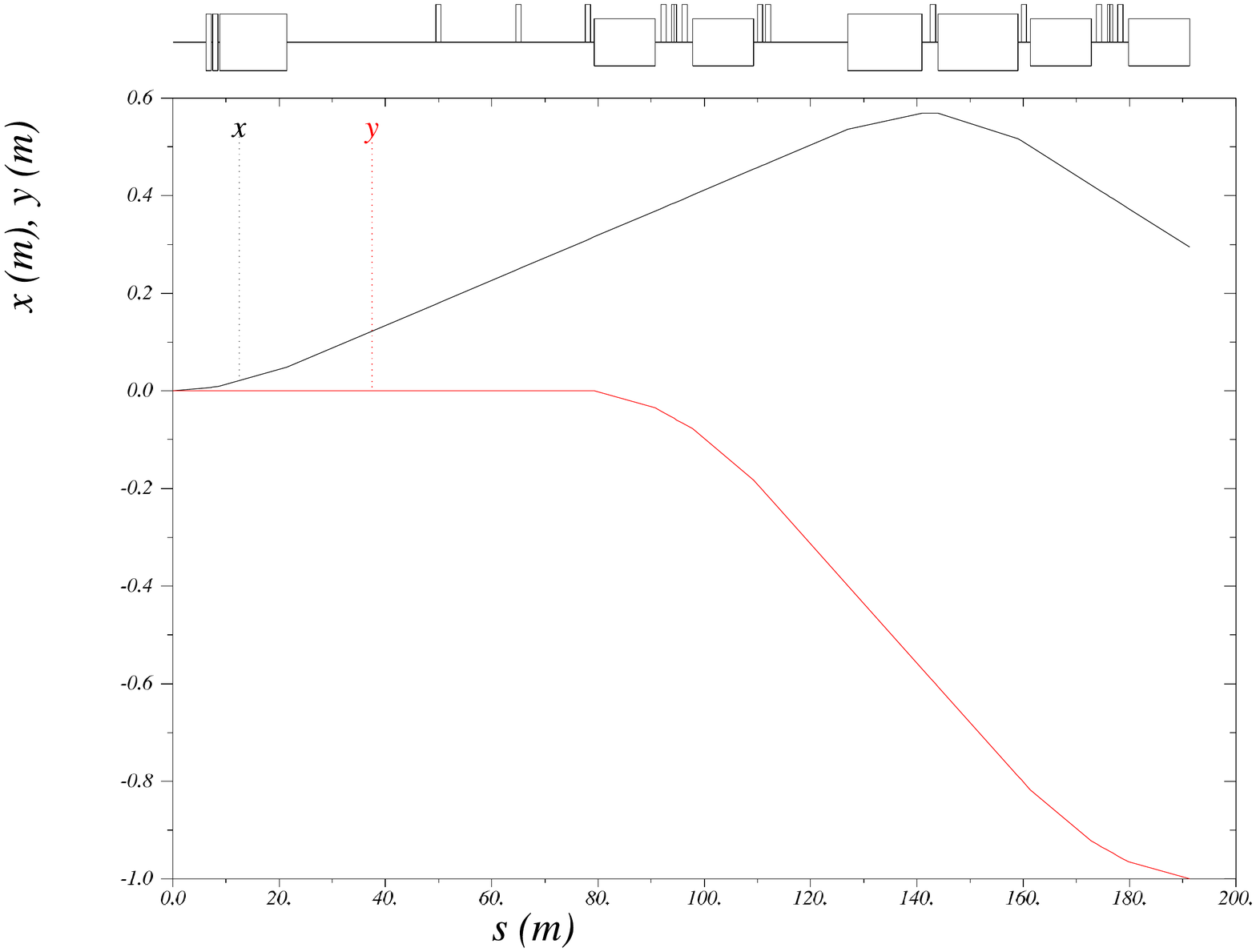}}
\caption{Example of geometry of a design with coupled horizontal and vertical bends. Interleaving bends in this way generates roll around s axis. The IP is at zero in both axes.}
\label{fig:IR.LSS.Geom.LSSCoupled}
\end{figure*}

Figure~\ref{fig:IR.LSS.Geom.LSSGeom} shows the geometry of the LSS solution on a larger scale. Note that the vertical doglegs are placed between the two horizontal dipole sets. To maximise use of space, schemes were explored with interleaved horizontal and vertical bends, as shown in Figure~\ref{fig:IR.LSS.Geom.LSSCoupled}. This allows increased bend length and distance between the bending magnets to reduce the SR power. However this coupled bending generates rotation of the beam around the s axis, effectively causing all subsequent quadrupoles to have a skew component.

\noindent Note that the left DS has nominal bend strength, while the right DS dipoles are weakened to accommodate the 1.2~m horizontal separation. Note also that future iterations of the LSS will include changes to accommodate the solution for the non-colliding proton beam detailed in Sec.~\ref{IRs.NCBeam}. In practise this simply manifests as a rotation of the IR section, and no complex changes are required.

\subsection{Electron optics in the LSS}
\label{IRs.LSS.Optics}
Placement of quadrupole elements is constrained by LSS geometry requirements, and by the LHC lattice, although this constraint is ignored for this iteration. While the LSS horizontal dipoles alone do not significantly constrain space, the combination of these and the vertical DBA scheme takes up large amounts of space.

\begin{figure*}[!h]
\centerline{\includegraphics[scale=0.6]{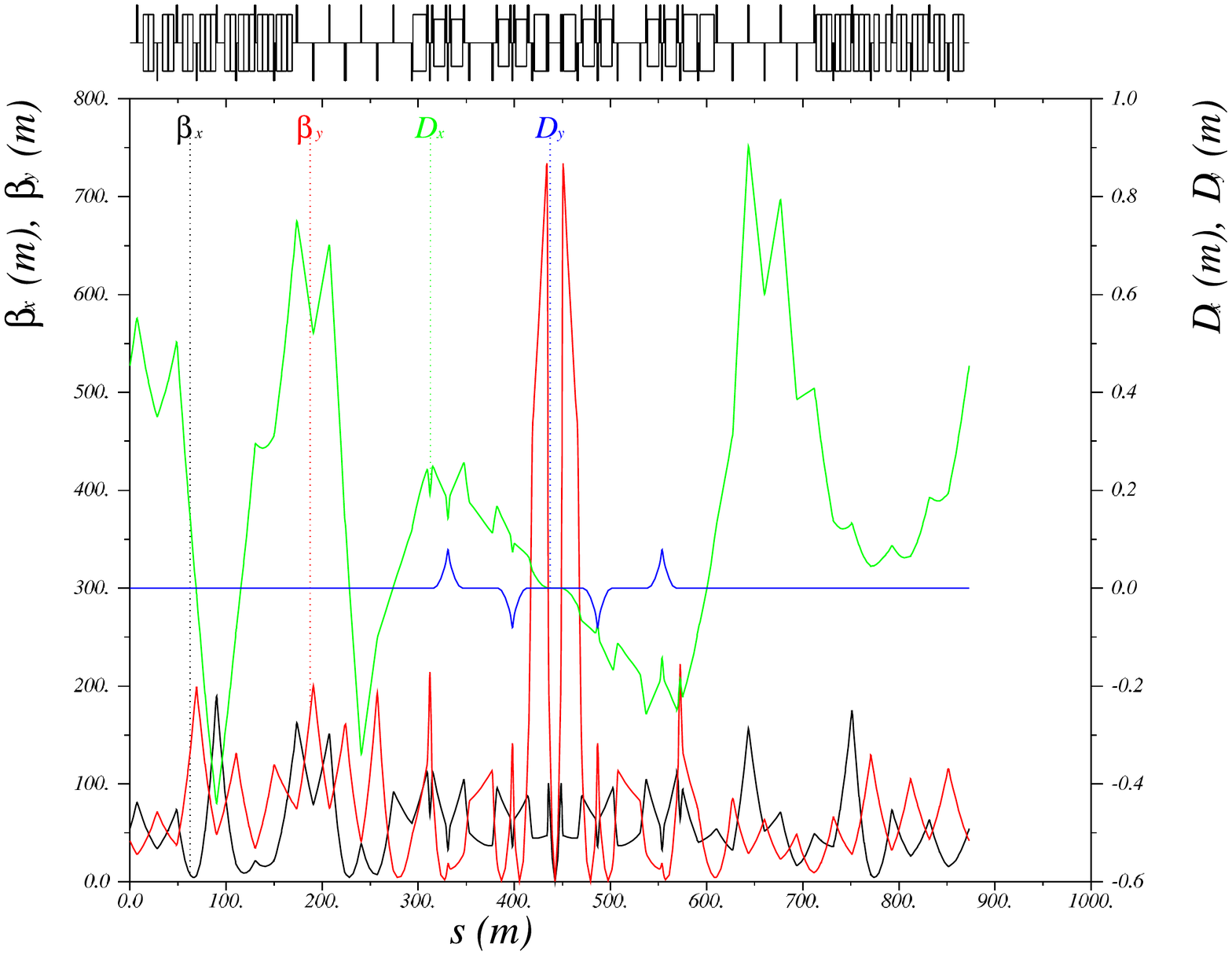}}
\caption{Optics plot for the HA LSS design.}
\label{fig:IR.LSS.Optics.Optics}
\end{figure*}

To gain sufficient matching flexibility, quadrupole triplets are used in the centre of the DBAs. The triplet DBA  generates a characteristic beta function shape, resulting in peaks and waists which make matching more challenging but feasible.  Figure~\ref{fig:IR.LSS.Optics.Optics} shows the beta and dispersion functions of the LSS optics.

\subsection{Synchrotron radiation}
\label{IRs.LSS.SR}
While detailed simulations have not yet been run, a simple analytical calculation of SR generated by the dipoles in the LSS has been performed, giving an initial estimate of $\sim$1.4~MW. Note that this includes the left and right DS sections. This is manageable considering the $\sim$50~MW estimate for the rest of the ring.

\subsection{LHC integration}
\label{IRs.LSS.LHC}
Currently, the DBA modules and quadrupoles near the IP conflict with the LHC proton triplet. After sufficient horizontal and/or vertical separation electron elements may be placed arbitrarily. Work is in progress on an updated design which moves vertical separation outward from the IP, after horizontal separation. In this case, no quadrupoles are required until $\sim$75~m from the IP, leaving space for the proton triplet. This geometry also successfully incorporates the solution for the non-colliding proton beam. However at the time of writing, optical matching is not yet finalised.

This "late vertical separation" scheme changes optical constraints. In the current "early" vertical separation scheme, limited space between the IR and the DBA decreases matching flexibility. In the "late" design, flexibility between the IR and DBA increases, but decreases correspondingly between the DBA and the DS.

Note that it is to some degree possible to reduce a bending scheme's space requirements arbitrarily, at the cost of more SR power.

\section{The non-colliding proton beam}
\label{IRs.NCBeam}
In both IRs, a solution must be found for dealing with the second proton beam. The second beam must not collide with either of the other two beams, or generate significant beam-beam effects. Also, detector designs strongly prefer for the second beam to occupy the same central beam pipe as the other two beams, rather than allowing space through the detector for a second pipe.

\subsection{Design elements}
\label{IRs.NCBeam.Design}
To avoid collisions and beam-beam effects, the bunches of the non-colliding (NC) beam will be shifted in time by half a bunch distance. This prevents proton-proton collisions at the IP, and allows the NC beam to overlap with the co-rotating electron beam.

Proton-proton interactions at the parasitic encounters however and accordingly beam-beam effects can still occur. To minimise these, the NC beam is left unsqueezed, and a proton-proton crossing angle is implemented which generates sufficient separation at these locations. For the unsqueezed optics, the so-called LHC alignment optics \cite{alignment} is modified for use on the NC beam only. The same scenario is proposed in the linac-ring design in Sec.~\ref{sec:LR-IR}.

\begin{figure*}[!h]
\centerline{\includegraphics[scale=0.5]{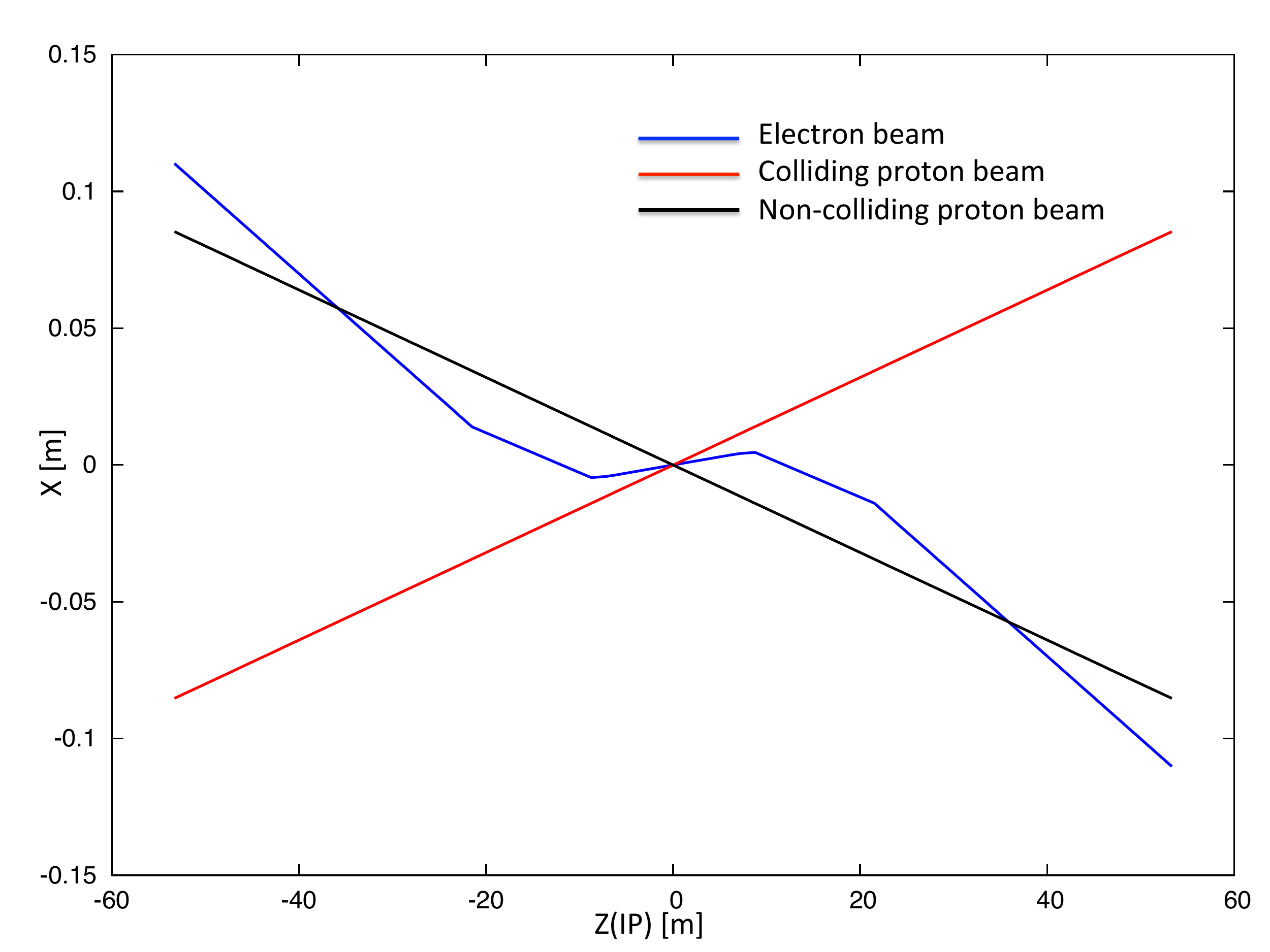}}
\caption{Trajectories of the three beams in the HA interaction region design. Note that in this plot the beams are reversed compared to the LSS plots.}
\label{fig:IR.NCBeam.Design.3Beam}
\end{figure*}

The required  crossing angle for the second proton beam is generated by changing the LHC separator dipoles D1 and D2. Figure~\ref{fig:IR.NCBeam.Design.3Beam} shows the trajectories of the three beams for the HA design. The proton final triplet is rotated in the horizontal plane and moved to match the new trajectory of the colliding beam while its position in s stays constant.

\noindent Note that the electron trajectory is rotated as well to match the colliding proton beam, such that the electron-proton crossing angle of 1~mrad is kept constant. This requires a change to the LSS geometry and optics solution which has not yet been implemented. This will be included in the next iteration of the LSS design. No new issues are likely to be introduced. Note also that the electron IR itself is unchanged in both the HL and HA designs, so SR calculations and detector designs do not require updates.

\subsection{Solution}
\label{IRs.NCBeam.Soltn}
For the unsqueezed optics of the second proton beam, zero triplet strength is required. The triplet quadrupoles each have a single proton aperture and as such the proton beams cannot be focused differently if both pass through the main aperture. Therefore the NC beam is guided through the same aperture as the electron beam, and experiences effectively no focusing. The proton LSS matching quadrupoles, which are separately powered for each beam, are then used to implement the NC beam optics.

As shown in Sec.~\ref{tripletmagnets}, Q1 will be a half-quadrupole. A large field-free aperture accommodates the electron beam and the NC proton beam. Q2 and Q3 have standard designs which incorporate low-field pockets which will be used for the shared electron and NC proton apertures.

Aperture calculations are based on 15$\sigma$ proton envelopes and 20$\sigma$ electron envelopes. In both cases, the aperture need is driven by horizontal requirements, since the horizontal envelopes and horizontal separation dominate over the vertical electron envelope. Note that the Q2 and Q3 apertures are circular; aperture radius is thus determined by the larger dimension.

\subsubsection{High luminosity}
\label{IRs.NCBeam.Soltn.HA}
The proton-proton crossing angle is optimised to 3~mrad to minimise aperture requirements, by making the NC beam follow the electron beam closely. The electron trajectory is determined by the IR separation scheme. 

\begin{table}[!h]
\centering
\begin{tabular}{|l|l|l|}
\hline
Element&Ap Radius&Ap Centre\\
\hline
Q1&0.0311&-0.0666\\
\hline
Q2A&0.0274&-0.1001\\
\hline
Q2B&0.0259&-0.1251\\
\hline
Q3&0.0257&-0.1592\\
\hline
\end{tabular}
\caption{Proton triplet aperture requirements of the non-colliding proton beam for the HL layout.}
\label{tab:IR.NCBeam.Soltn.HL.Apertures}
\end{table}

\begin{figure*}[!h]
\centerline{\includegraphics[scale=0.5]{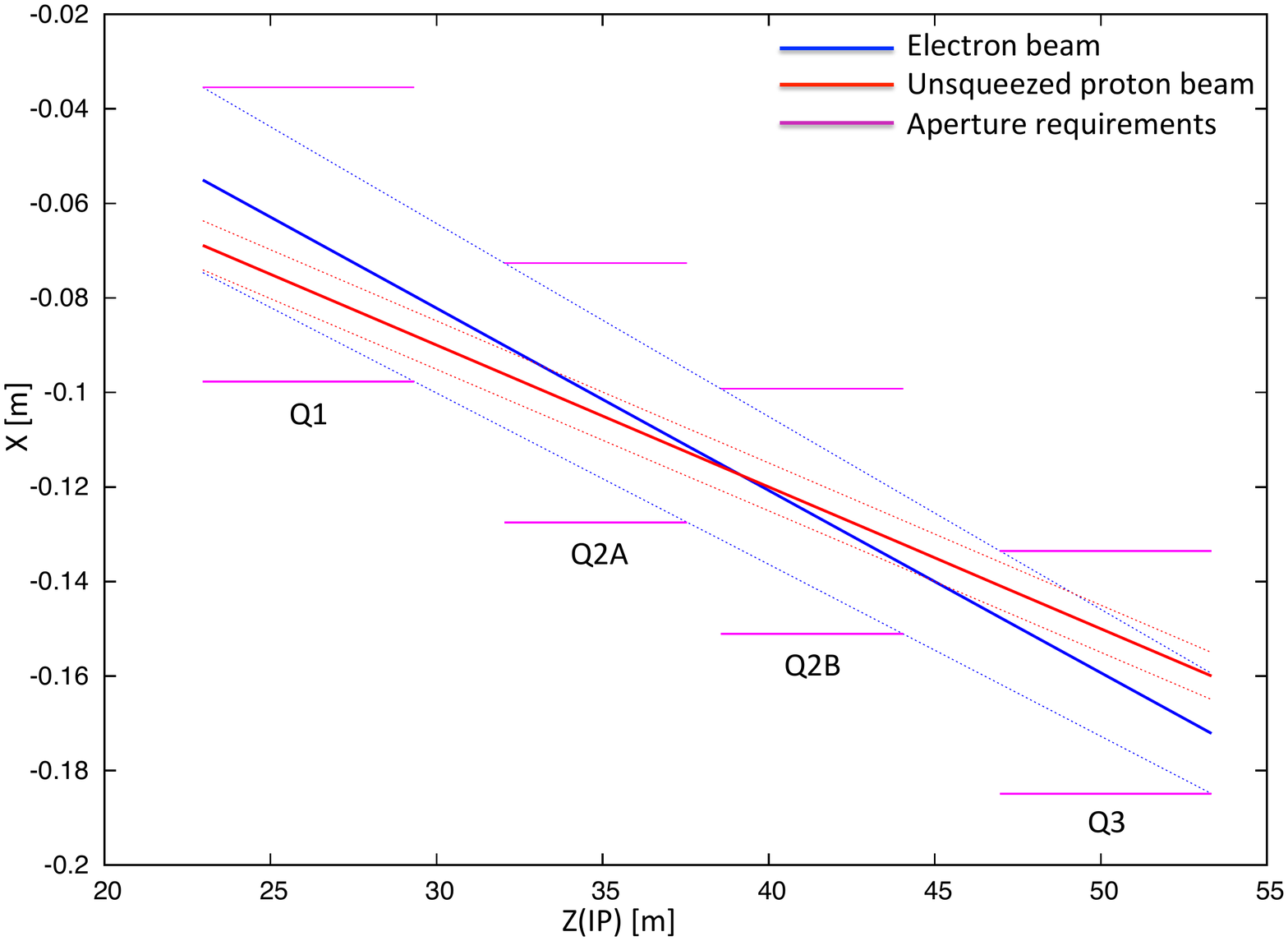}}
\caption{Proton triplet aperture requirements with trajectories and envelopes of the electron beam and NC proton beam for the HL layout. Note that in this plot the beams are reversed compared to the LSS plots.}
\label{fig:IR.NCBeam.Soltn.HL.Apertures}
\end{figure*}

\subsubsection{High acceptance}
\label{IRs.NCBeam.Soltn.HA}
\begin{figure*}[!h]
\centerline{\includegraphics[scale=0.5]{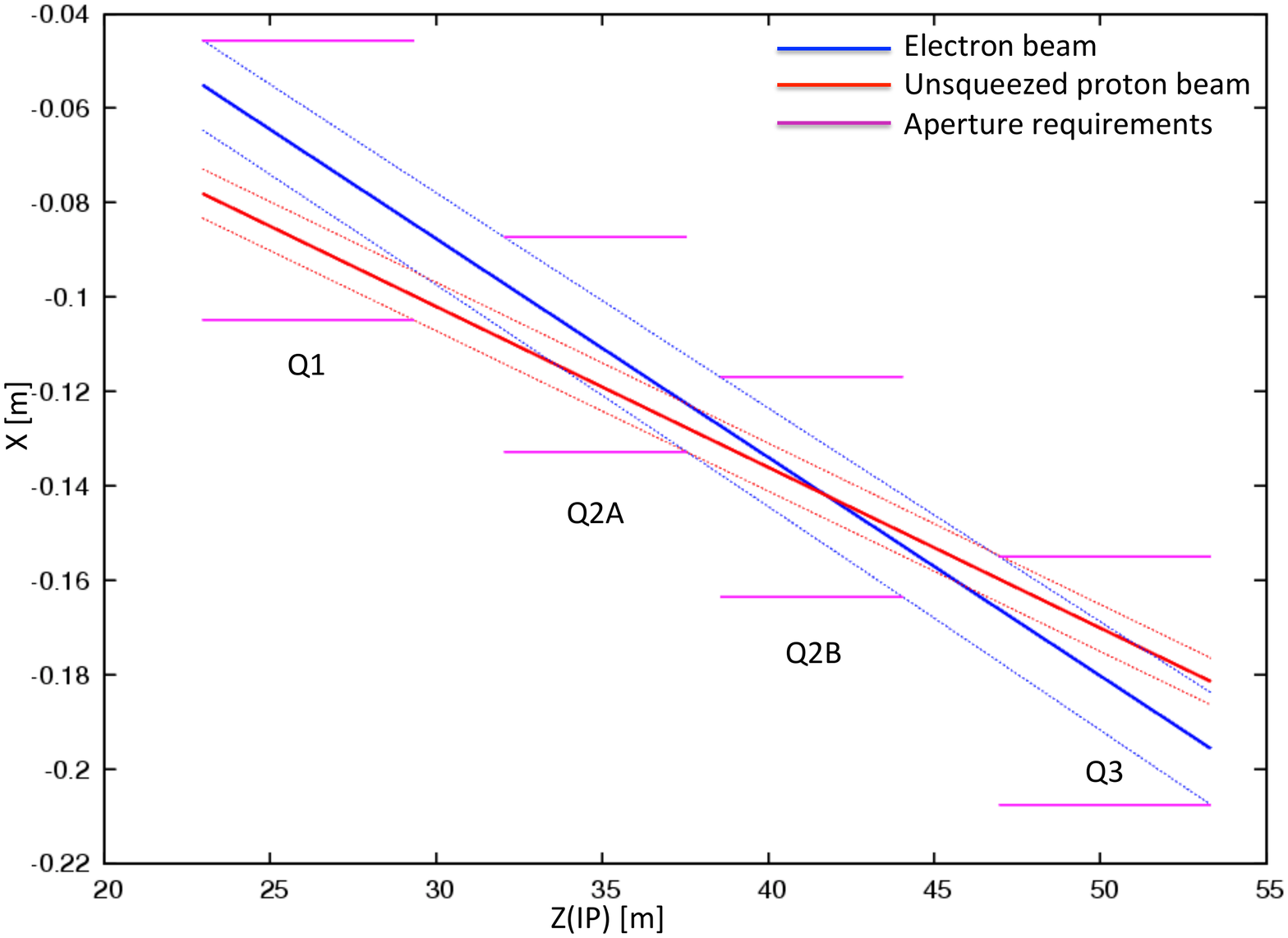}}
\caption{Proton triplet aperture requirements with trajectories and envelopes of the electron beam and NC proton beam for the HA layout. Note that in this plot the beams are reversed compared to the LSS plots.}
\label{fig:IR.NCBeam.Soltn.HA.Apertures}
\end{figure*}

In this case the proton-proton crossing angle is optimised to 3.4~mrad to minimise aperture requirements. Again the NC proton beam will follow closely the electron beam trajectory, which is determined by the IR separation scheme. The electron beam, having larger emittance, dominates aperture requirements. The separation between the electron beam and the NC proton beam is larger in the HA layout than in the HL layout, due to the later bending in the HA separation scheme. Table~\ref{tab:IR.NCBeam.Soltn.HA.Apertures} and figure~\ref{fig:IR.NCBeam.Soltn.HA.Apertures} show the required apertures.

\begin{table}[!h]
\centering
\begin{tabular}{|l|l|l|}
\hline
Element&Aperture Radius&Aperture Centre\\
\hline
Q1&0.0296&-0.0752\\
\hline
Q2A&0.0227&-0.1100\\
\hline
Q2B&0.0233&-0.1402\\
\hline
Q3&0.0264&-0.1811\\
\hline
\end{tabular}
\caption{Proton triplet aperture requirements of the non-colliding proton beam for the HA layout.}
\label{tab:IR.NCBeam.Soltn.HA.Apertures}
\end{table}

\subsection{Summary}
\label{IRs.NCBeam.Summary}

Aperture requirements for the HL layout are somewhat less demanding than for the HA layout, but both sets of requirements are feasible and do not present difficulties in magnet design using existing technology. The existing Q1 design is easily sufficient. Q2A and Q2B would ideally be two copies of the same yoke, requiring a larger hole in each. Q3 requires a larger yoke than the existing 200~mm radius design, but the tooling limit of 270~mm should be sufficient.

In both designs, the crossing angle may be increased if desired for beam-beam reasons. The existing Q1 design supports a crossing angle up to 4~mrad, but this would require significantly larger apertures in the other magnets.

%% file: machine/bernard_rr.tex

%
\section{Synchrotron radiation and absorbers}
\label{sec:NATHAN}
\subsection{Introduction}
	
	The synchrotron radiation (SR) in the interaction region has been analysed in three ways. The SR was simulated in depth using a program made with the GEANT4 (G4) toolkit. In addition a cross check of the total power and average critical energy was done in IRSYN, a Monte Carlo simulation package written by R. Appleby~\cite{rob}. A final cross check has been made for the radiated power per element using an analytic method. These other methods confirmed the results seen using G4. The G4 program uses Monte Carlo methods to create Gaussian spatial and angular distributions for the electron beam. The electron beam is then guided through vacuum volumes that contain the magnetic fields for the separator dipoles and electron final focusing quadrupoles. 
	
	The SR is generated in these volumes using the appropriate G4 process classes. The G4 SR class was written for a uniform magnetic field, and therefore the quadrupole volumes were divided such that the field remained approximately constant in each volume. This created agreement between upstream and downstream quadrupoles since for a downstream quadrupole the beta function at the entrance and exit are reversed from its upstream counterpart. This agreement confirms that the field was approximately constant in each volume. 
	
	The position, direction, and energy of each photon created is written as ntuples at user defined Z values. These ntuples are then used to analyse the SR fan as it evolves in Z. The analysis was done primarily through the use of MATLAB scripts. It was necessary to make two versions of this program. One for the high luminosity design and one for the high detector acceptance design.
	
	Before going further, some conventions used for this section will be explained. The electron beam is referred to as \emph{the beam} and the proton beams will be referred to as either the interacting or non interacting proton beams. The beam propagates in the -Z direction and the interacting proton beam propagates in the +Z direction. A right handed coordinate system is used where the X axis is horizontal and the Y axis is vertical. The beam centroid always remains in the Y = 0 plane.  The \emph{angle of the beam} will be used to refer to the angle between the beam centroid's velocity vector and the Z axis, in the Y = 0 plane. This angle is set such that the beam propagates in the -X direction as it traverses Z.
	
	 The SR fans extension in the horizontal direction is driven by the angle of the beam at the entrance of the upstream separator dipole. Because the direction of emitted photons is parallel to the direction of the electron that emitted it, the angle of the beam and the distance to the absorber are both greatest at the entrance of the upstream separator dipole and therefore this defines one of the edges of the synchrotron fan on the absorber. The other edge is defined by the crossing angle and the distance from the IP to the absorber. The S shaped trajectory of the beam means that the smallest angle of the beam will be reached at the IP. Therefore the photons emitted at this point will have the lowest angle and for this given angle the smallest distance to the absorber. This defines the other edge of the fan in the horizontal direction.
	 
	  The SR fans extension in the vertical direction is driven by the beta function and angular spread of the beam. The beta function along with the emittance defines the r.m.s. spot size of the beam. The vertical spot size defines the Y position at which photons are emitted. On top of this the vertical angular spread defines the angle between the velocity vector of these photons and the Z axis. Both of these values produce complicated effects as they are functions of Z. These effects also affect the horizontal extension of the fan however are of second order when compared to the angle of the beam. Since the beam moves in the Y = 0 plane these effects dominate the vertical extension of the beam. 
	  
	  The number density distribution of the fan is a complicated issue. The number density at the absorber is highest between the interacting beams. The reason for this is that although the separator dipoles create significantly more photons the number of photons generated per unit length in Z is much lower for the dipoles as opposed to the quadrupoles due to the high fields experienced in the quadrupoles. The position of the quadrupole magnets then causes the light radiated from them to hit the absorber in the area between the two interacting beams. 
	  
\subsection{High luminosity}	
\subsubsection{Parameters}

	\indent
	The parameters for the high luminosity option are listed in Table \ref{tab:Param10}. The separation refers to the displacement between the two interacting beams at the face of the proton triplet. 
	
\begin{table}[!htb]
  \centering	
\begin{tabular}{| c | c |}
  \hline
Characteristic & Value \\
\hline
\hline
Electron Energy [GeV] & 60  \\
\hline
Electron Current [mA] & 100  \\
\hline
Crossing Angle [mrad] & 1 \\
\hline
Absorber Position [m] & -21.5 \\
\hline
Dipole Field [T] & 0.0296  \\
\hline
Separation [mm] & 55 \\
\hline
$\gamma/s$ & $ 5.39 \times 10^{18}$\\
  \hline
\end{tabular}
\caption{High Luminosity: Parameters}
\label{tab:Param10}
\end{table}

	The energy, current, and crossing angle ($\theta_{c}$) are common values used in all RR calculations. The dipole field value refers to the constant dipole field created throughout all dipole elements in the IR. The direction of this field is opposite on either side of the IP. The quadrupole elements have an effective dipole field created by placing the quadrupole off axis, which is the same as this constant dipole field. The field is chosen such that 55 mm of separation is reached by the face of the proton triplet. This separation was chosen based on S. Russenschuck's SC quadrupole design for the proton final focusing triplet~\cite{schavannes}. The separation between the interacting beams can be increased by raising the constant dipole field. However, for a dipole magnet $P_{SR} \propto |B^2|$~\cite{nathan}, therefore an optimisation of the design will need to be discussed. The chosen parameters give a flux of $5.39 \times 10^{18}$ photons per second at Z = -21.5 m.

\subsubsection{Power and critical energy}
	Table \ref{tab:PCE10} shows the power of the SR produced by each element along with the average critical energy produced per element. This is followed by the total power produced in the IR and the average critical energy. Since the G4 simulations utilise Monte Carlo, multiple runs should be made with various seeds to get an estimate for the standard error.

\begin{table}[!htbp]
  \centering
\begin{tabular}{| c | c | c | }
  \hline
  Element &  Power [kW] & Critical Energy [keV]\\
\hline\hline
 DL & 6.4 &   71  \\
 \hline
  QL3 & 5.3 &   308   \\
 \hline
  QL2 & 4.3 &    218  \\
 \hline
  QL1 & 0.6 &   95   \\
 \hline
 QR1 & 0.6 &   95   \\
 \hline
  QR2 & 4.4 &   220  \\
 \hline
  QR3 & 5.2 &   310   \\
 \hline
 DR & 6.4 &   71   \\
 \hline\hline
 Total/Avg & 33.2 &  126  \\
\hline
\end{tabular}
\caption{High Luminosity: Power and Critical Energies as calculated with GEANT4.}
\label{tab:PCE10}
\end{table}

	 The power from the dipoles is greater than any one quadrupole however the critical energies of the quadrupoles are significantly higher than in the dipoles. It is expected that the dipole and quadrupole elements can create power on the same order however have very different critical energies. This is because the dipole is an order of magnitude longer than the quadrupole elements. Since the SR power created for both the quadrupole and dipoles are linearly dependent on length \cite{nathan} one needs to have a much higher average critical energy to create comparable amounts of power.  
	
\subsubsection{Comparison}
	The IRSYN cross check of the power and critical energies is shown in Table \ref{tab:GIRSYN10}. This comparison was done for the total power and the average critical energy. 
	
	\begin{table}[!htbp]
  \centering
  \begin{tabular}{| c | c | c | c | c |}
 \hline
 &   \multicolumn{2}{|c|}{Power [kW]} & \multicolumn{2}{|c|}{Critical Energy [keV]} \\
\hline\hline
 & GEANT4 & IRSYN & GEANT4 & IRSYN \\
 \hline
 Total/Avg & 33.2 &33.7 & 126 & 126 \\
  \hline
  \end{tabular}
\caption{High Luminosity: GEANT4 and IRSYN comparison}
\label{tab:GIRSYN10}
\end{table}
	
	A third cross check to the G4 simulations was made for the power as shown in Table \ref{tab:GA10}. This was done using an analytic method for calculating power in dipole and quadrupole magnets~\cite{nathan}. This was done for every element which provides confidence in the distribution of this power throughout the IR. 
	
	\begin{table}[!htbp]
  \centering
  \begin{tabular}{| c | c | c | }
 \hline
 &   \multicolumn{2}{|c|}{Power [kW]}  \\
\hline\hline
Element & GEANT4 & Analytic \\
\hline
 DL & 6.4 &   6.3   \\
 \hline
 QL3 & 5.3 &   5.4   \\
 \hline
 QL2 & 4.3 &   4.6  \\
 \hline
 QL1 & 0.6 &   0.6   \\
 \hline
 QR1 & 0.6 &   0.6   \\
 \hline
 QR2 & 4.4 &    4.6  \\
 \hline
 QR3 & 5.2 &   5.4   \\
 \hline
 DR & 6.4 &   6.3  \\
 \hline
 Total/Avg & 33.2 & 33.8 \\
  \hline
  \end{tabular}
\caption{High Luminosity: GEANT4 and Analytic method comparison}
\label{tab:GA10}
\end{table}

\subsubsection{Number density and envelopes}

	\begin{figure}[!htbp]
\centerline{\includegraphics[clip=,width=1.\textwidth]{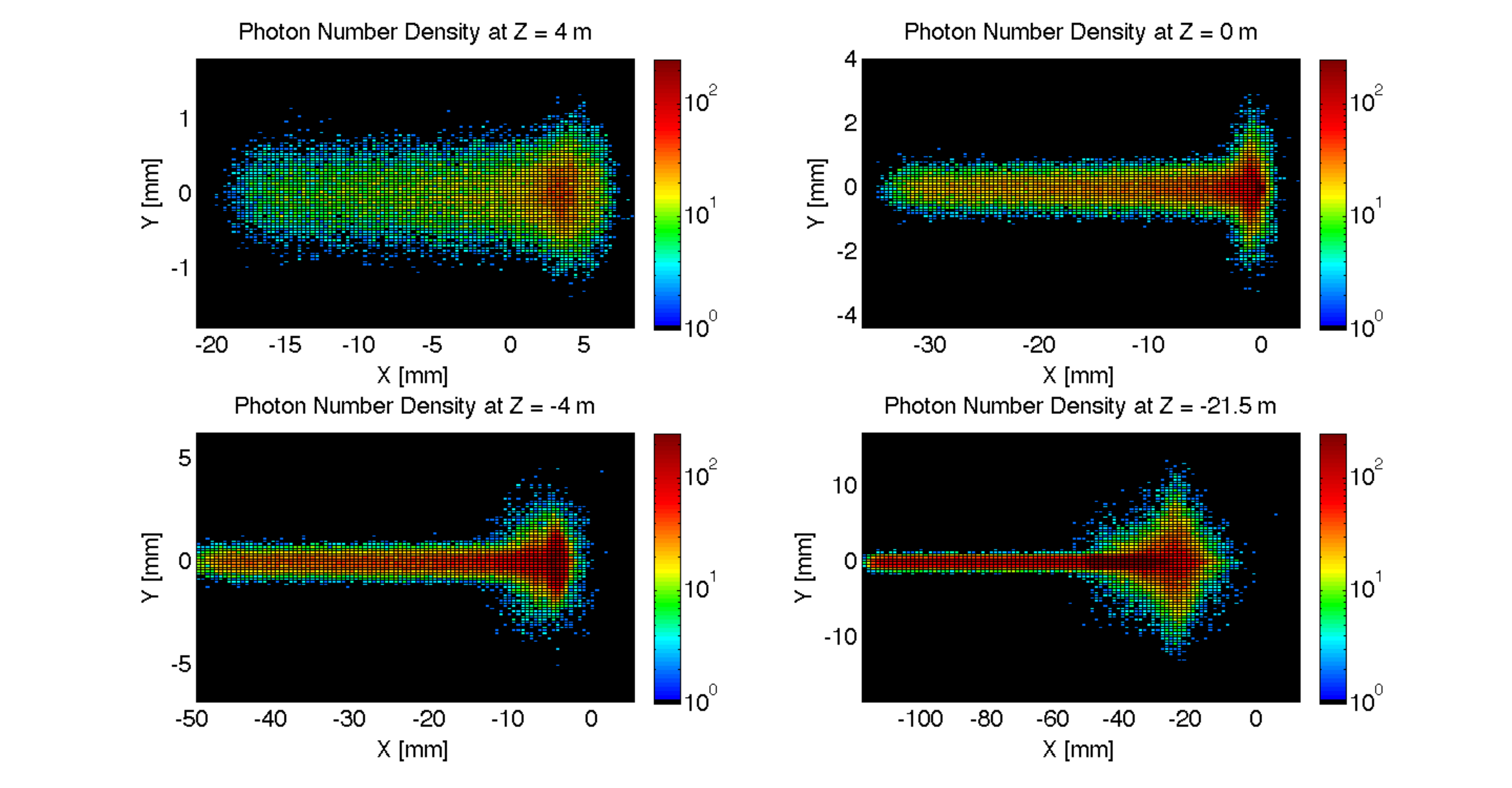}}
\caption{High Luminosity: Number Density Growth in Z}
\label{Fig:10degnumdens}
\end{figure}

	The number density of photons as a function of Z is shown in Figure \ref{Fig:10degnumdens}. Each graph displays the density of photons in the $Z = Z_o$ plane for various values of $Z_o$. The first three figures give the growth of the SR fan inside the detector area. This is crucial for determining the dimensions of the beam pipe. Since the fan grows asymmetrically in the -Z direction an asymmetric elliptical cone geometry will minimise these dimensions, allowing the tracking to be placed as close to the beam as possible. The horizontal extension of the fan in the high luminosity case is the minimum for the two Ring Ring options as well as the Linac Ring option, which is most important inside the detector region. This is due to the lower value of $l^*$. Because the quadrupoles are closer to the IP and contain effective dipole fields the angle of the beam at the entrance of the upstream dipole can be lower as the angle of the beam doesn't need to equal the crossing angle until $Z = l^*$. The number density of this fan appears as expected. There exists the highest density between the two beams at the absorber. 
	
\begin{figure}[!htbp]
\centerline{\includegraphics[clip=,width=1.\textwidth]{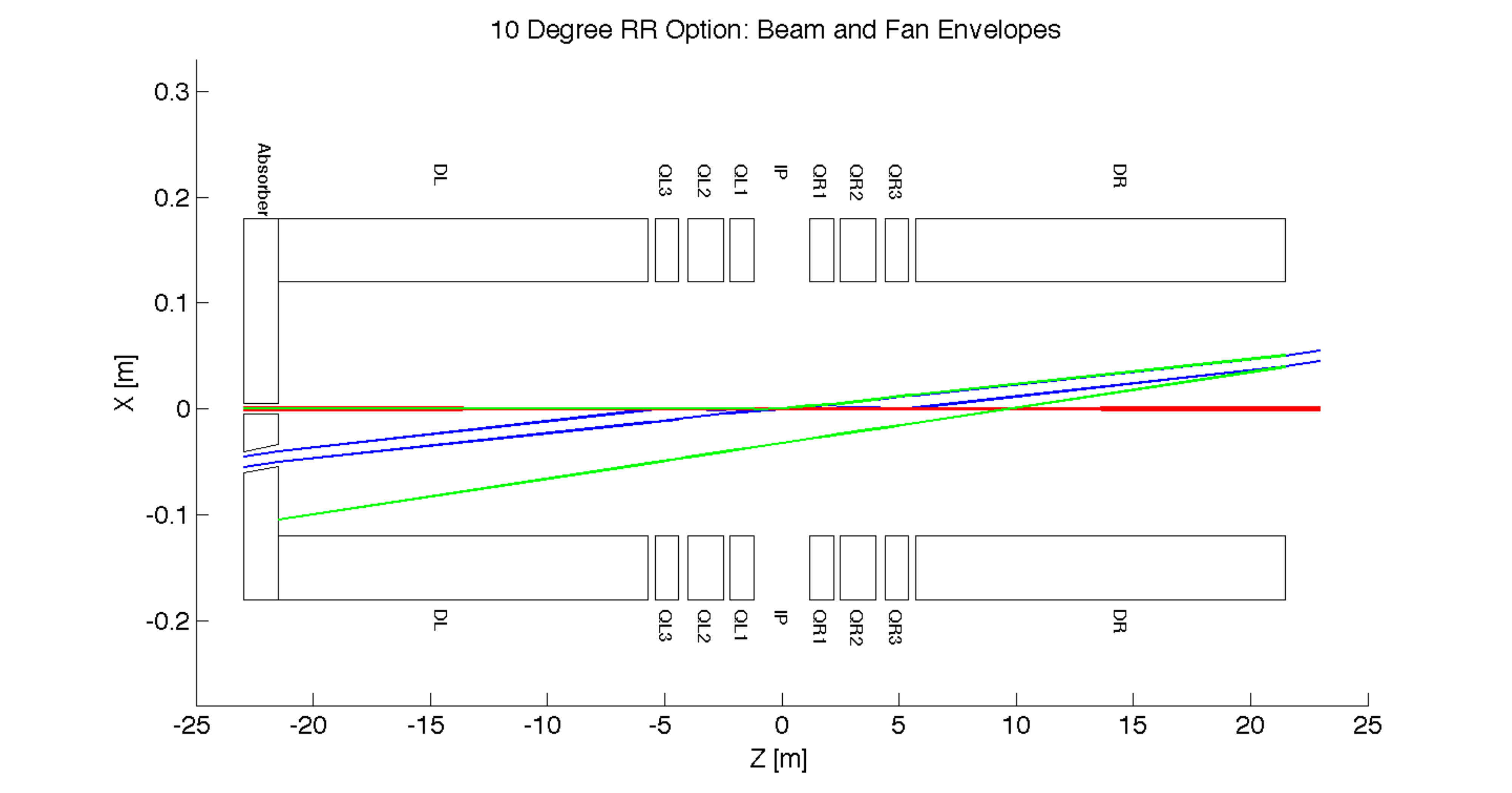}}
\caption{High Luminosity: Beam Envelopes in Z}
\label{Fig:10degenv}
\end{figure}  

	In Figure \ref{Fig:10degnumdens} the distribution was given at various Z values however a continuous envelope distribution is also important to see everything at once. This can be seen in Figure \ref{Fig:10degenv}, where the beam and fan envelopes are shown in the Y = 0 plane. This makes it clear that the fan is antisymmetric which comes from the S shape of the electron beam as previously mentioned.

	\subsubsection{Critical energy distribution}
	
\begin{figure}[!htbp]
\centerline{\includegraphics[clip=,width=1.\textwidth]{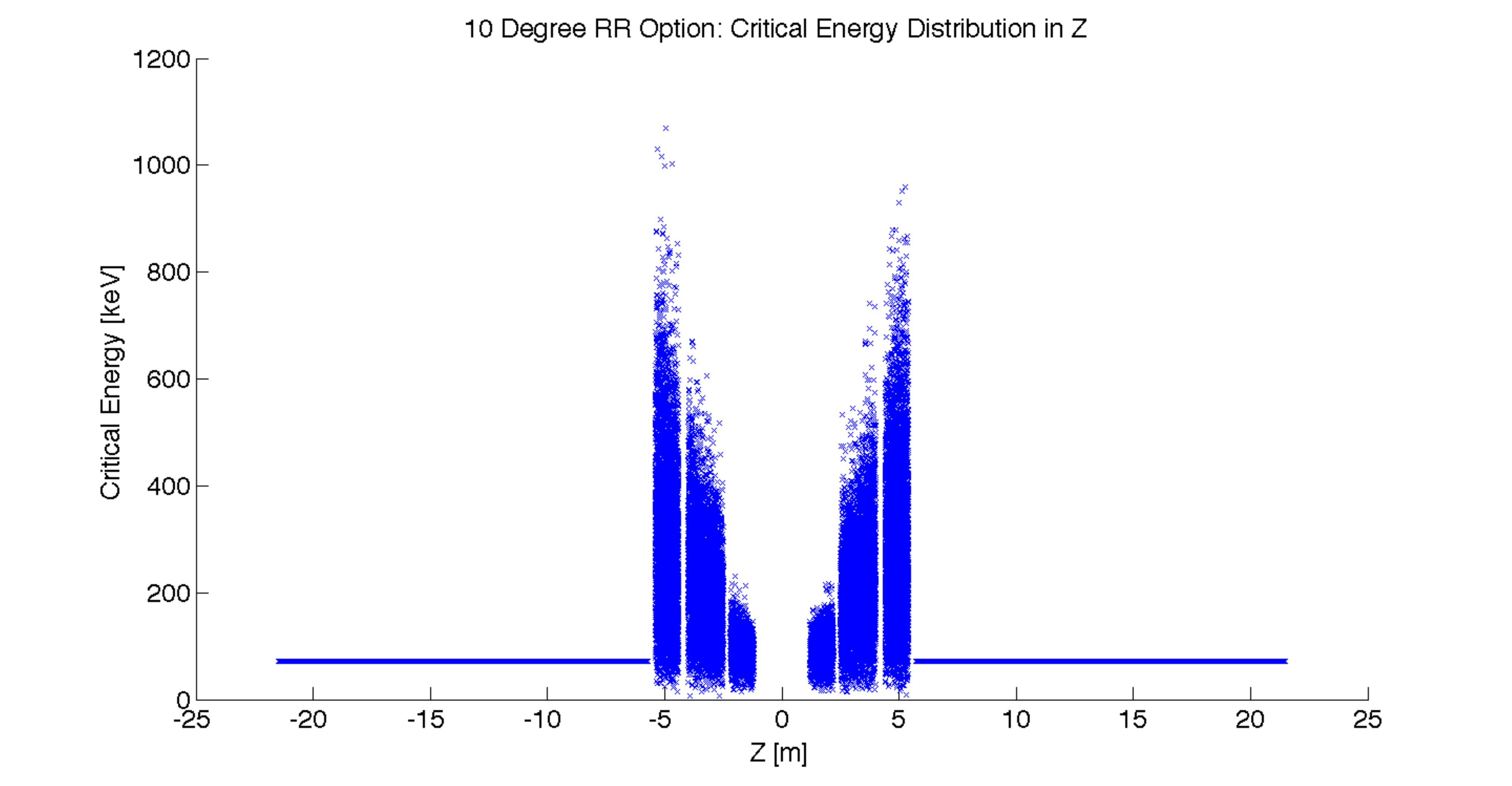}}
\caption{High Luminosity: Critical Energy Distribution in Z}
\label{Fig:10degCE}
\end{figure}

	The Critical Energy is dependent upon the element in which the SR is generated, and for the quadrupole magnets it is also dependent upon Z. This is a result of the fact that the critical energy is proportional to the magnetic field component that is perpendicular to the particle direction. i.e. $E_c \propto B_{\perp}$~\cite{SREcrit}. Since the magnitude of the magnetic field is dependent upon x and y, then for a Gaussian beam in position particles will experience different magnetic fields and therefore have a spectrum of critical energies. In a dipole the field is constant and therefore regardless of the position of the particles as long as they are in the uniform field area of the magnet they have a constant critical energy. Since the magnetic field is dependent upon x and y it is clear that as the r.m.s. spot size of the beam decreases there will be a decrease in critical energies. The opposite will occur for an increasing spot size. This is evident from Figure \ref{Fig:10degCE}.
	
	\subsubsection{Absorber}
	
	\begin{figure}[!htbp]
\centerline{\includegraphics[clip=,width=1.\textwidth]{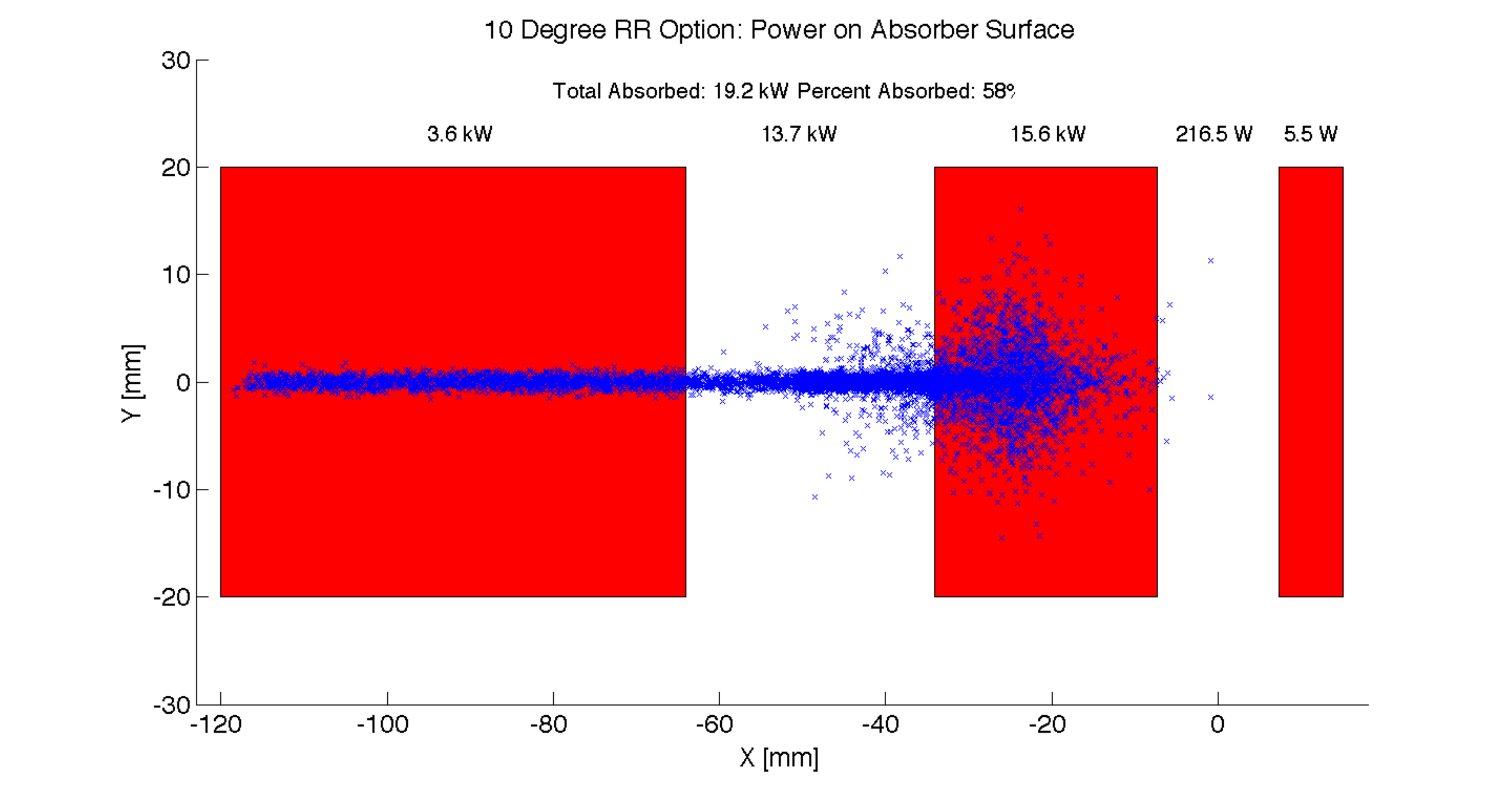}}
\caption{High Luminosity: Photon distribution on Absorber Surface}
\label{Fig:10degAbs}
\end{figure}
	
	The Photon distribution on the absorber surface is crucial. The distribution decides how the absorber must be shaped. The shape of the absorber in addition to the distribution on the surface then decides how much SR is backscattered into the detector region. In HERA backscattered SR was a significant source of background that required careful attention~\cite{zeus}. Looking at Figure \ref{Fig:10degAbs} it is shown that for the high luminosity option 19.2 kW of power from the SR light will fall on the face of the absorber which is $58\%$ of the total power. This gives a general idea of the amount of power that will be absorbed. However, backscattering and IR photons will lower the percent that is actually absorbed.

\subsubsection{Proton triplet}

	The super conducting final focusing triplet for the protons needs to be protected from radiation by the absorber. Some of the radiation produced upstream of the absorber however will either pass through the absorber or pass through the apertures for the two interacting beams. This is most concerning for the interacting proton beam aperture which will have the superconducting coils. A rough upper bound for the amount of power the coils can absorb before quenching is 100W~\cite{stephan}. There is approximately 217 W entering into the interacting proton beam aperture as is shown in Figure \ref{Fig:10degAbs}. This doesn't mean that all this power will hit the coils but simulations need to be made to determine how much of this will hit the coils. The amount of power that will pass through the absorber can be disregarded as it is not enough to cause any effects. The main source of power moving downstream of the absorber will be the photons passing through the beams aperture. This was approximately 13.7 kW as can be seen from Figure \ref{Fig:10degAbs}. Most of this radiation can be absorbed in a secondary absorber placed after the first downstream proton quadrupole. Overall protecting the proton triplet is important and although the absorber will minimise the radiation continuing downstream this needs to be studied in depth.

	\subsubsection{Backscattering}
	
\begin{figure}[!h!t!b]
\centerline{\includegraphics[clip=,width=1.\textwidth]{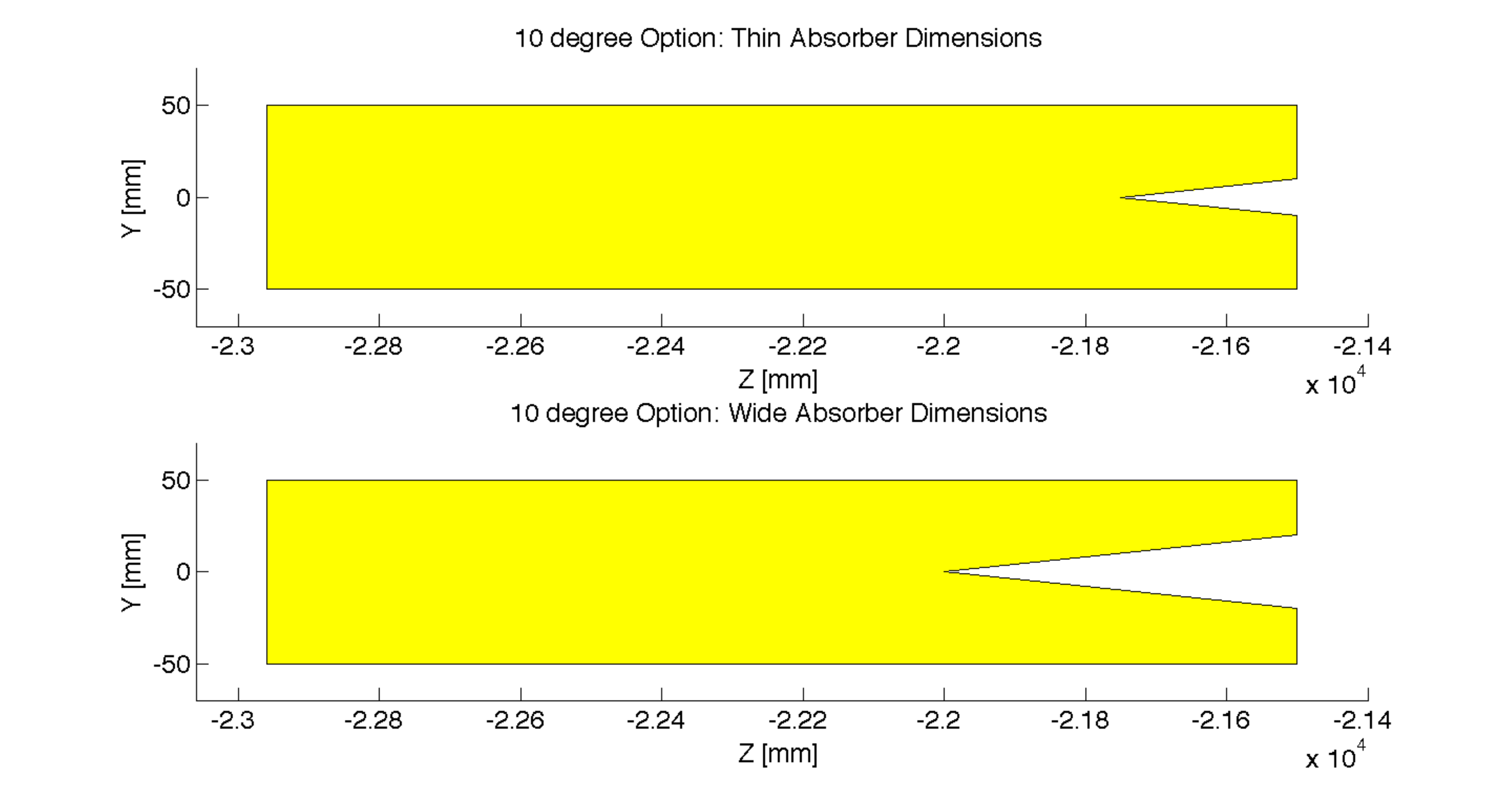}}
\caption{High Luminosity: Absorber Dimensions}
\label{Fig:10degAbsDim}
\end{figure}

	Another GEANT4 program was written to simulate the backscattering of photons into the detector region. The ntuple with the photon information written at the absorber surface is used as the input for this program. An absorber geometry made of copper is described, and general physics processes are set up. A detector volume is then described and set to record the information of all the photons which enter in an ntuple. The first step in minimising the backscattering was to optimise the absorber shape. Although the simulation didn't include a beam pipe the backscattering for different absorber geometries was compared against one another to find a minimum. The most basic shape was a block of copper that had cylinders removed for the interacting beams. This was used as a benchmark to see the maximum possible backscattering. In HERA a wedge shape was used for heat dissipation and minimising backscattering~\cite{zeus}. The profile of two possible wedge shapes in the YZ plane is shown in Figure \ref{Fig:10degAbsDim}. It was found that this is the optimum shape for the absorber. The reason for this is that a backscattered electron would have to have its velocity vector be almost parallel to the wedge surface to escape from the wedge and therefore it works as a trap. As can be seen from Table \ref{tab:BM10} utilising the wedge shaped absorber did not reduce the power by much. This appears to be a statistical limitation and needs to be redone with higher statistics to get a better estimate of the difference between the two geometries.
	
	After the absorber was optimised it was possible to set up a beam pipe geometry. An asymmetric elliptical cone beam pipe geometry made of beryllium was used since it would minimise the necessary size of the beam pipe as previously mentioned. The next step was to place the lead shield and masks inside this beam pipe. To determine placement a simulation was run with just the beam pipe. Then it was recorded where each backscattered photon would hit the beam pipe in Z. A histogram of this data was made. This determined that the shield should be placed in the Z region ranging from -20 m until the absorber (-21.5 m). The shields were then placed at -21.2 m and -20.5 m. This  decreased the backscattered power to zero as can be seen from Table \ref{tab:BM10}. Although this is promising this number should be checked again with higher statistics to judge its accuracy. Overall there is still more optimisation that can occur with this placement. 
	
		\begin{table}[!htbp]
  \centering
  \begin{tabular}{| c | c | }
 \hline
Absorber Type & Power [W] \\
\hline\hline
 Flat & 22  \\
 \hline
Wedge & 18.5 \\
 \hline
 Wedge \& Mask/Shield & 0  \\
\hline
  \end{tabular}
\caption{High Luminosity: Backscattering/Mask}
\label{tab:BM10}
\end{table}

\begin{figure}[!h!t!b]
\centerline{\includegraphics[clip=,width=1.\textwidth]{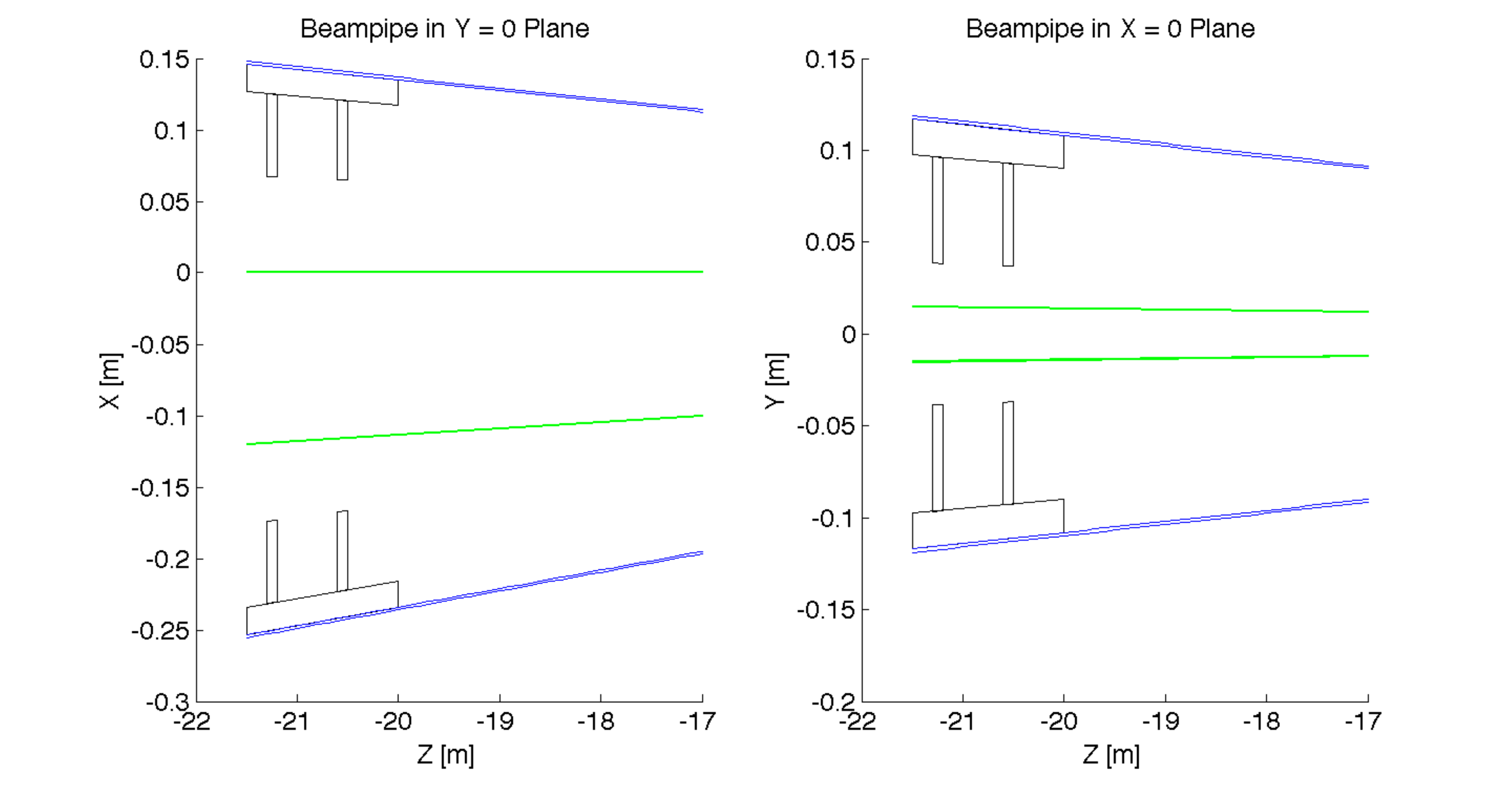}}
\caption{High Luminosity: Beam pipe Cross Sections}
\label{Fig:10degBP}
\end{figure}

Cross sections of the beam pipe in the Y = 0 and X = 0 planes with the shields and masks included can be seen in Figure \ref{Fig:10degBP}.

\subsection{High detector acceptance}

\subsubsection{Parameters}

\indent	
	For the Ring Ring high acceptance option the basic parameters are listed in Table \ref{tab:1degparam}. The separation refers to the displacement between the two interacting beams at the face of the proton triplet. 
	
\begin{table}[!htbp]
  \centering	
\begin{tabular}{| c | c |}
  \hline
Characteristic & Value \\
\hline
\hline
Electron Energy [GeV] & 60  \\
\hline
Electron Current [mA] & 100  \\
\hline
Crossing Angle [mrad] & 1 \\
\hline
Absorber Position [m] & -21.5 \\
\hline
Dipole Field [T] & 0.0493  \\
\hline
Separation [mm] & 55.16 \\
\hline
$\gamma/s$ & $ 6.41 \times 10^{18}$\\
  \hline
\end{tabular}
\caption{High Acceptance: Parameters}
\label{tab:1degparam}
\end{table}

The energy, current, and crossing angle ($\theta_{c}$) are common values used in all RR calculations. The dipole field value refers to the constant dipole field created throughout all dipole elements in the IR. The separation is the same as in the high luminosity case and can be altered for the same reasons with the same ramifications.The chosen parameters give a flux of $6.41 \times 10^{18}$ photons per second at Z = -21.5 m, which is slightly higher than in the high luminosity case. This is expected as the fields experienced in the high acceptance case are higher. 

\subsubsection{Power and critical energy}
	Table \ref{tab:PCE1} shows the power of the SR produced by each element along with the average critical energy produced per element. This is followed by the total power produced in the IR and the average critical energy. Since the G4 simulations utilise Monte Carlo, multiple runs should be made with various seeds to get an estimate for the standard error.

\begin{table}[!htbp]
  \centering
\begin{tabular}{| c | c | c | }
  \hline
  Element &  Power [kW] & Critical Energy [keV]\\
\hline\hline
 DL & 13.9 &   118   \\
 \hline
 QL2 & 6.2 &   318  \\
 \hline
 QL1 & 5.4 &   294   \\
 \hline
 QR1 & 5.4 &   293   \\
 \hline
 QR2 & 6.3 &    318  \\
 \hline
 DR & 13.9 &   118  \\
 \hline\hline
 Total/Avg & 51.1 &  163  \\
\hline
\end{tabular}
\caption{High Acceptance: Power and Critical Energies [GEANT4]}
\label{tab:PCE1}
\end{table}

	The distribution of power and critical energy over the IR elements is similar to that of the high acceptance option with the exception of the upstream and downstream separator dipole magnets. The power and critical energies are significantly higher than before. This is due to the higher dipole field and the quadratic dependence of power on magnetic field and linear dependence of critical energy on magnetic field~\cite{SREcrit}.
	
\subsubsection{Comparison}
	The IRSYN cross check of the power and critical energies is shown in Table \ref{tab:GIRSYN1}. This comparison was done for the total power and the critical energy. 
	
	\begin{table}[!htbp]
  \centering
  \begin{tabular}{| c | c | c | c | c |}
 \hline
 &   \multicolumn{2}{|c|}{Power [kW]} & \multicolumn{2}{|c|}{Critical Energy [keV]} \\
\hline\hline
 & GEANT4 & IRSYN & GEANT4 & IRSYN \\
 \hline
 Total/Avg & 51.1 & 51.3 & 163 & 162 \\
  \hline
  \end{tabular}
\caption{High Acceptance: GEANT4 and IRSYN comparison}
\label{tab:GIRSYN1}
\end{table}
	
	A third cross check to the G4 simulations was also made for the power as shown in Table \ref{tab:GA1}. This was done using an analytic method for calculating power in dipole and quadrupole magnets~\cite{nathan}. This comparison provides confidence in the distribution of the power throughout the IR.
	
	\begin{table}[!htbp]
  \centering
  \begin{tabular}{| c | c | c | }
 \hline
 &   \multicolumn{2}{|c|}{Power [kW]}  \\
\hline\hline
Element & GEANT4 & Analytic \\
\hline
 DL & 13.9 & 14      \\
 \hline
 QL2 & 6.2 &  6.2   \\
 \hline
 QL1 & 5.4 &  5.3   \\
 \hline
 QR1 & 5.4 &    5.3  \\
 \hline
 QR2 & 6.3 &  6.2   \\
 \hline
 DR & 13.9 & 14   \\
 \hline\hline
 Total & 51.1 &  51  \\
\hline
  \end{tabular}
\caption{High Acceptance: GEANT4 and Analytic method comparison}
\label{tab:GA1}
\end{table}

\subsubsection{Number density and envelopes}
	
	\begin{figure}[!htbp]
\centerline{\includegraphics[clip=,width=1.\textwidth]{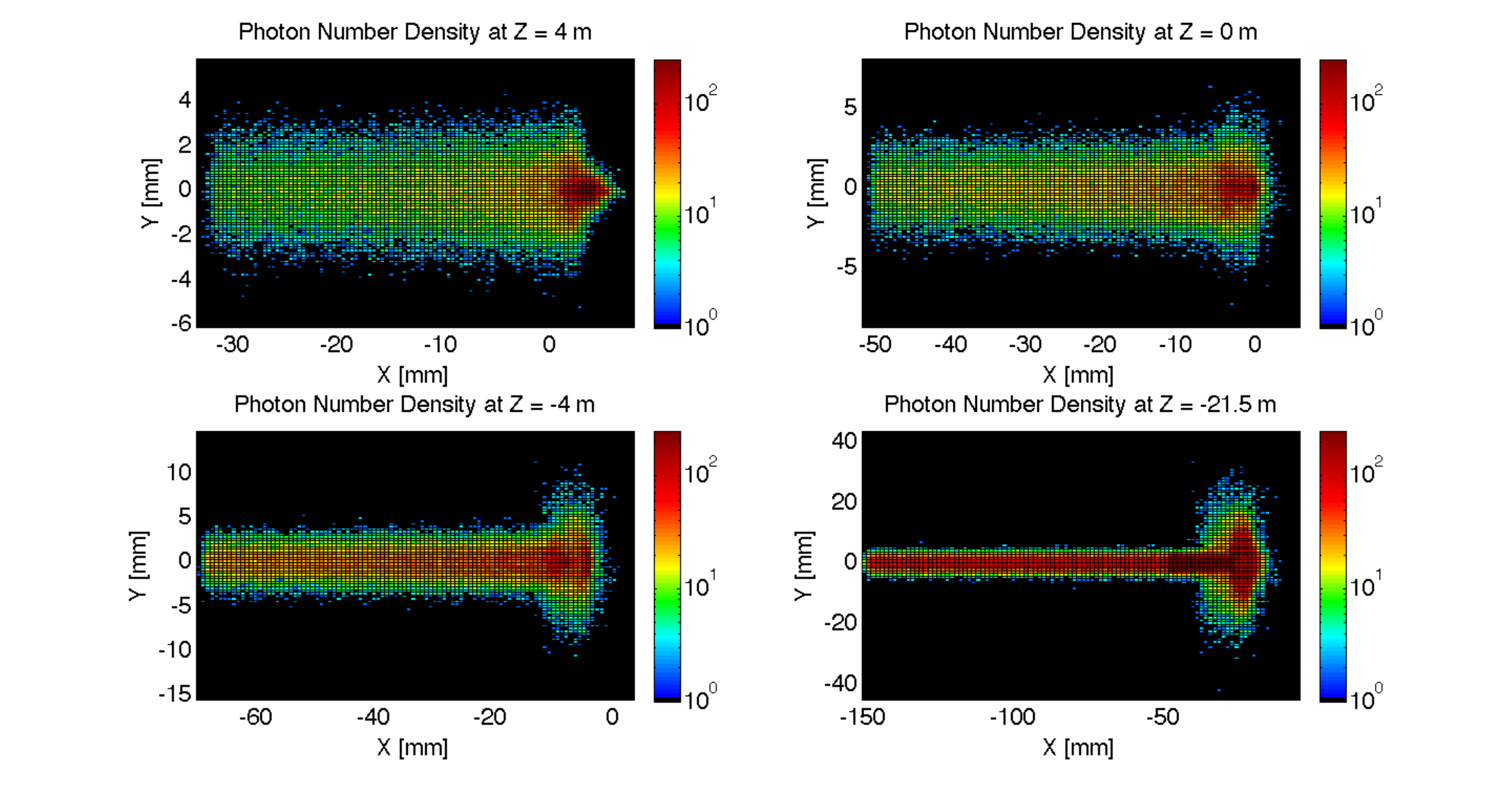}}
\caption{High Acceptance: Number Density Growth in Z}
\label{Fig:1degnumdens}
\end{figure}

	The number density of photons as a function of Z is shown in Figure \ref{Fig:1degnumdens}. The horizontal extension of the fan in the high acceptance case is larger than in the high luminosity case however still lower than in the LR option. Since the beam stays at a constant angle for the first 6.2 m after the IP it requires larger fields to bend in order to reach the desired separation. This means that an overall larger angle is reached near the absorber, and since the S shaped trajectory is symmetric in Z the angle of the beam at the entrance of the upstream quadrupoles is also larger and therefore the fan extends further in X. 
	
	\begin{figure}[!htbp]
\centerline{\includegraphics[clip=,width=1.\textwidth]{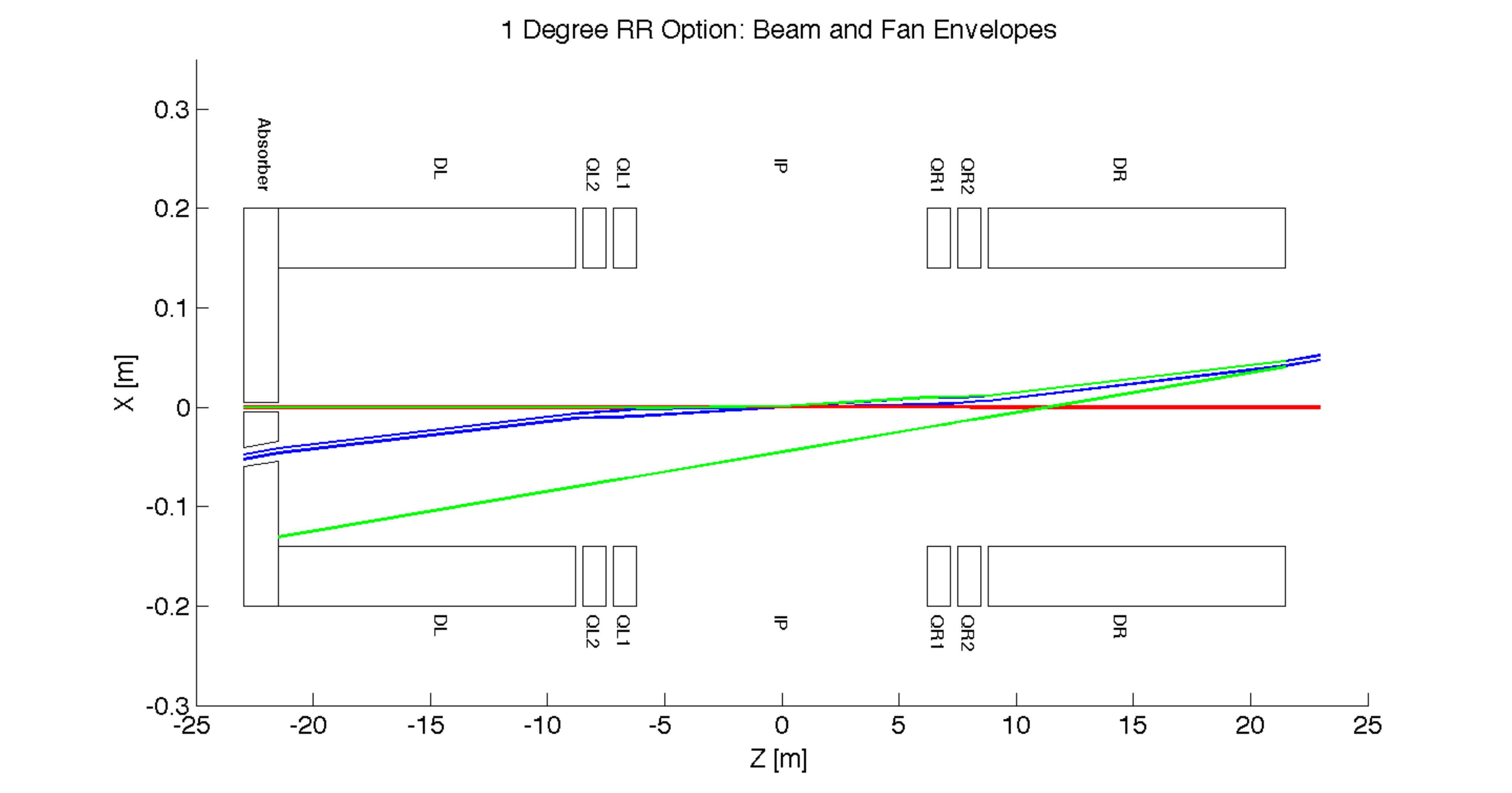}}
\caption{High Acceptance: Beam Envelopes in Z}
\label{Fig:1degenv}
\end{figure}
	
	The envelope of the SR fan can be seen in Figure \ref{Fig:1degenv}, where the XZ plane is shown at the value Y = 0. Once again the fan is antisymmetric due to the S shape of the electron beam.
	
	\subsubsection{Critical energy distribution}
	
		\begin{figure}[!htbp]
\centerline{\includegraphics[clip=,width=1.\textwidth]{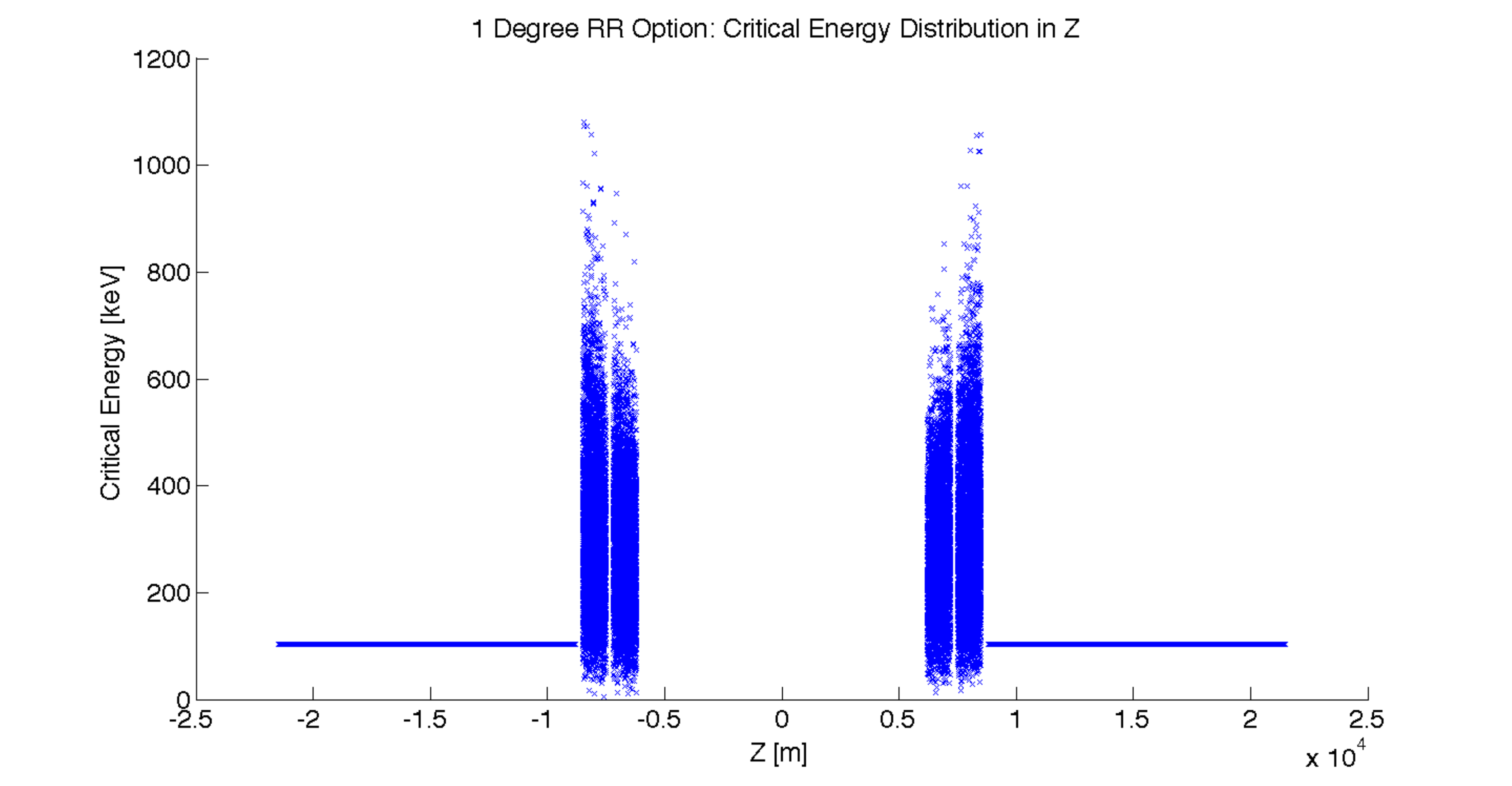}}
\caption{High Acceptance: Critical Energy Distribution in Z}
\label{Fig:1degCE}
\end{figure}
	
	The critical energy distribution in Z is similar to that of the high luminosity case. This is due to the focusing of the beam in the IR. This is evident from Figure \ref{Fig:1degCE}.
	
	\subsubsection{Absorber}
	
	\begin{figure}[!htbp]
\centerline{\includegraphics[clip=,width=1.\textwidth]{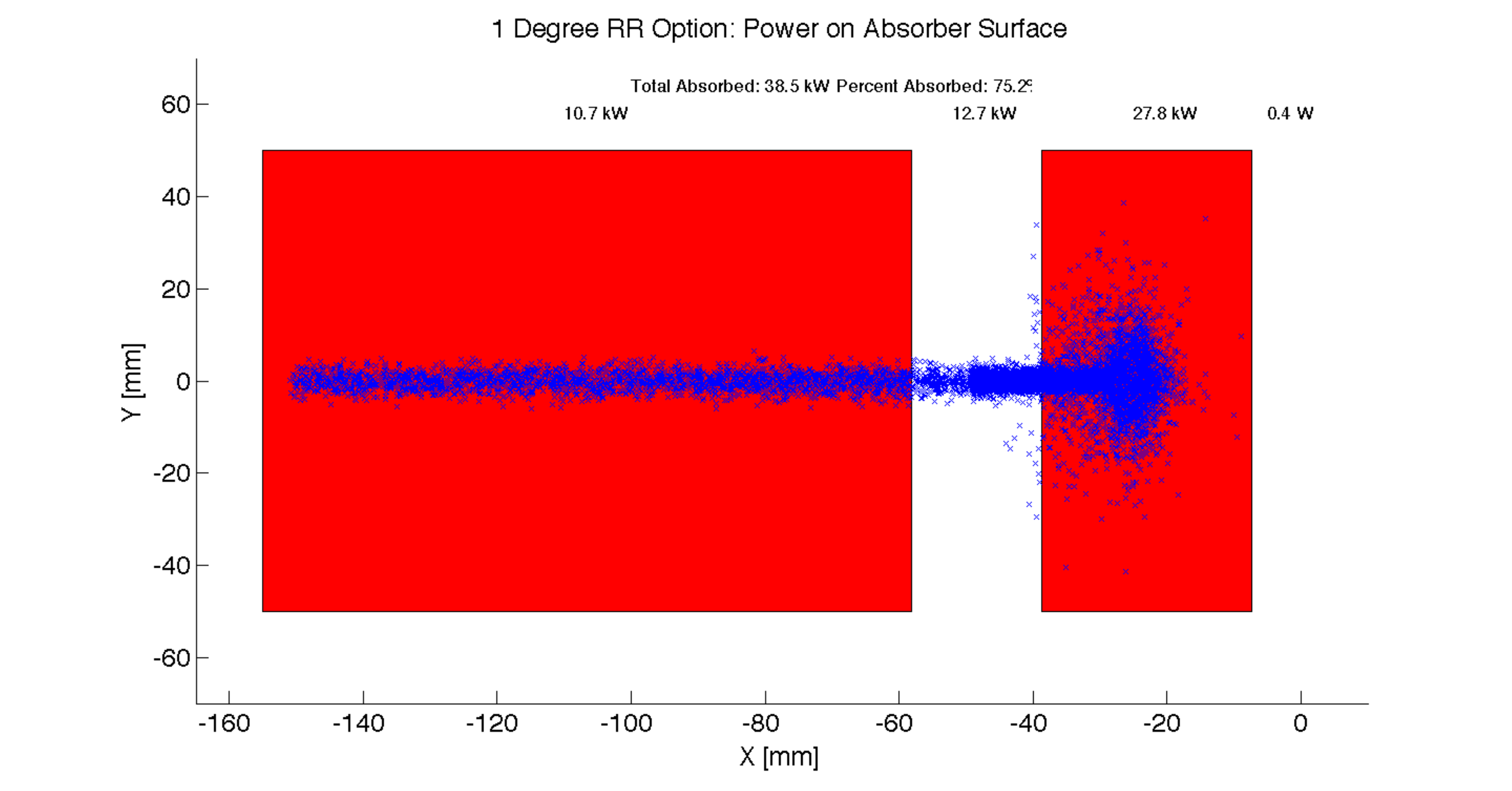}}
\caption{High Acceptance: Photon distribution on Absorber Surface}
\label{Fig:1degAbs}
\end{figure}
	
	 Looking at Figure \ref{Fig:1degAbs} it is shown that for the high acceptance option 38.5 kW of power from the SR light will fall on the face of the absorber which is $75\%$ of the total power. This gives a general idea of the amount of power that will be absorbed. However, backscattering and IR photons will lower the percent that is actually absorbed. 

\subsubsection{Proton triplet}

	The super conducting final focusing triplet for the protons needs to be protected from radiation by the absorber. Some of the radiation produced upstream of the absorber however will either pass through the absorber or pass through the apertures for the two interacting beams. This is most concerning for the interacting proton beam aperture which will have the superconducting coils. A rough upper bound for the amount of power the coils can absorb before quenching is 100 W~\cite{stephan}. In the high acceptance option there is approximately 0.4 W entering into the interacting proton beam aperture as is shown in Figure \ref{Fig:1degAbs}. Therefore for the high acceptance option this is not an issue. The amount of power that will pass through the absorber can be disregarded as it is not enough to cause any significant effects. The main source of power moving downstream of the absorber will be the photons passing through the beams aperture. This was approximately 12.7 kW as can be seen from Figure \ref{Fig:1degAbs}. Most of this radiation can be absorbed in a secondary absorber placed after the first downstream proton quadrupole. Overall protecting the proton triplet is important and although the absorber will minimise the radiation continuing downstream this needs to be studied in depth.

	\subsubsection{Backscattering}
	
\begin{figure}[!h!t!b]
\centerline{\includegraphics[clip=,width=1.\textwidth]{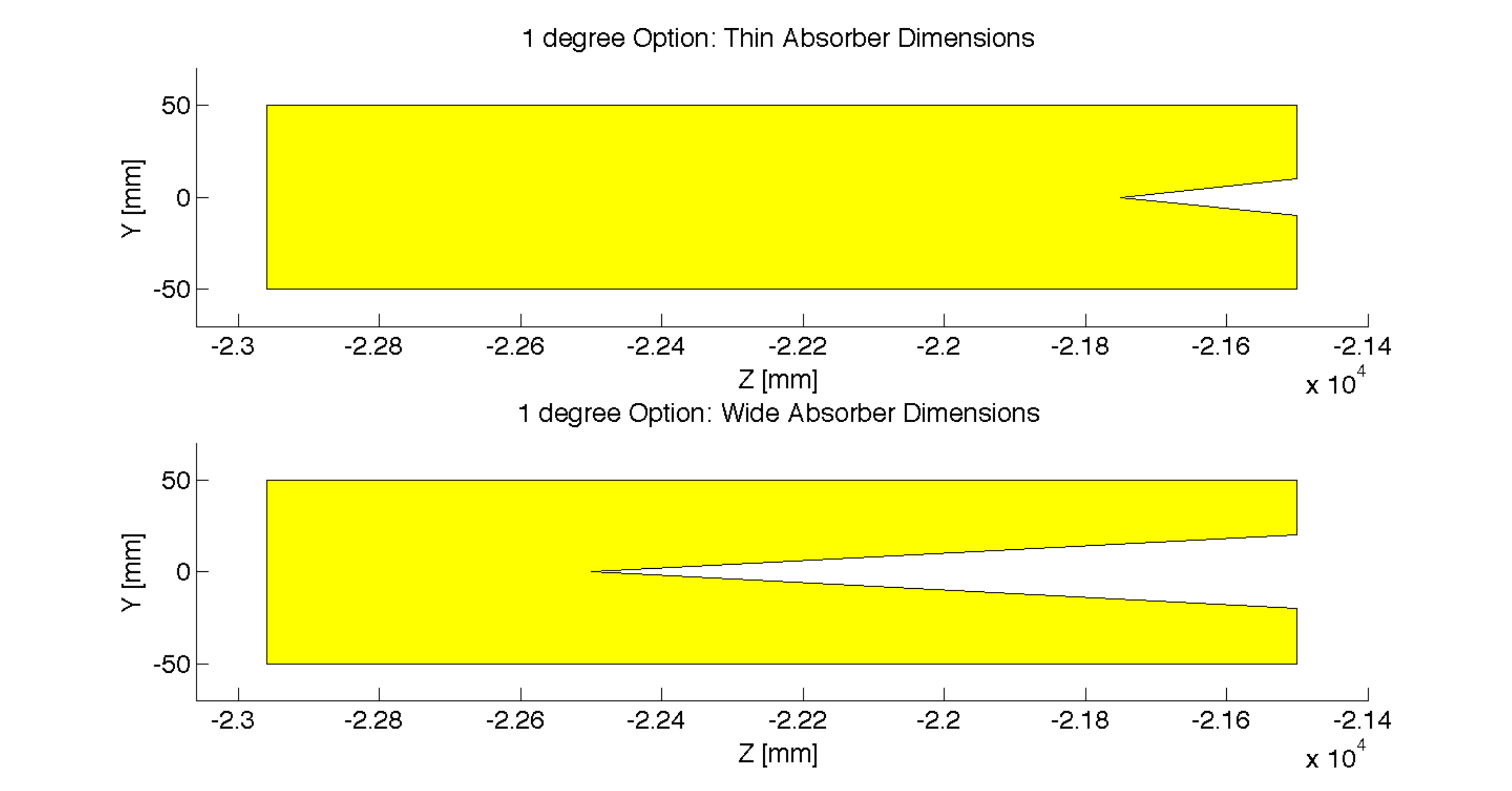}}
\caption{High Acceptance: Absorber Dimensions}
\label{Fig:1degAbsDim}
\end{figure}

	Another GEANT4 program was written to simulate the backscattering of photons into the detector region. The ntuple with the photon information written at the absorber surface is used as the input for this program. An absorber geometry made of copper is described, and general physics processes are set up. A detector volume is then described and set to record the information of all the photons which enter in an ntuple. The first step in minimising the backscattering was to optimise the absorber shape. Although the simulation didn't include a beam pipe the backscattering for different absorber geometries was compared against one another to find a minimum. The most basic shape was a block of copper that had cylinders removed for the interacting beams. This was used as a benchmark to see the maximum possible backscattering. In HERA a wedge shape was used for heat dissipation and minimising backscattering~\cite{zeus}. The profile of two possible wedge shapes in the YZ plane is shown in Figure \ref{Fig:1degAbsDim}. It was found that this is the optimum shape for the absorber. The reason for this is that a backscattered electron would have to have its velocity vector be almost parallel to the wedge surface to escape from the wedge and therefore it works as a trap. As can be seen from Table \ref{tab:BM1} utilising the wedge shaped absorber decreased the backscattered power by a factor of 9. 
	
	After the absorber was optimised it was possible to set up a beam pipe geometry. An asymmetric elliptical cone beam pipe geometry made of beryllium was used since it would minimise the necessary size of the beam pipe as previously mentioned. The next step was to place the lead shield and masks inside this beam pipe. To determine placement a simulation was run with just the beam pipe. Then it was recorded where each backscattered photon would hit the beam pipe in Z. This determined that the shield should be placed in the Z region ranging from -20 m until the absorber (-21.5 m). The shields were then placed at -21.2 m and -20.6 m. This  decreased the backscattered power to zero as can be seen from Table \ref{tab:BM1}. Although this is promising this number should be checked again with higher statistics to judge its accuracy. Overall there is still more optimisation that can occur with this placement. 
	
		\begin{table}[!htbp]
  \centering
  \begin{tabular}{| c | c | }
 \hline
Absorber Type & Power [W] \\
\hline\hline
 Flat & 91.1  \\
 \hline
Wedge & 10 \\
 \hline
 Wedge \& Mask/Shield & 0  \\
\hline
  \end{tabular}
\caption{High Acceptance: Backscattering/Mask}
\label{tab:BM1}
\end{table}

\begin{figure}[!h!t!b]
\centerline{\includegraphics[clip=,width=1.\textwidth]{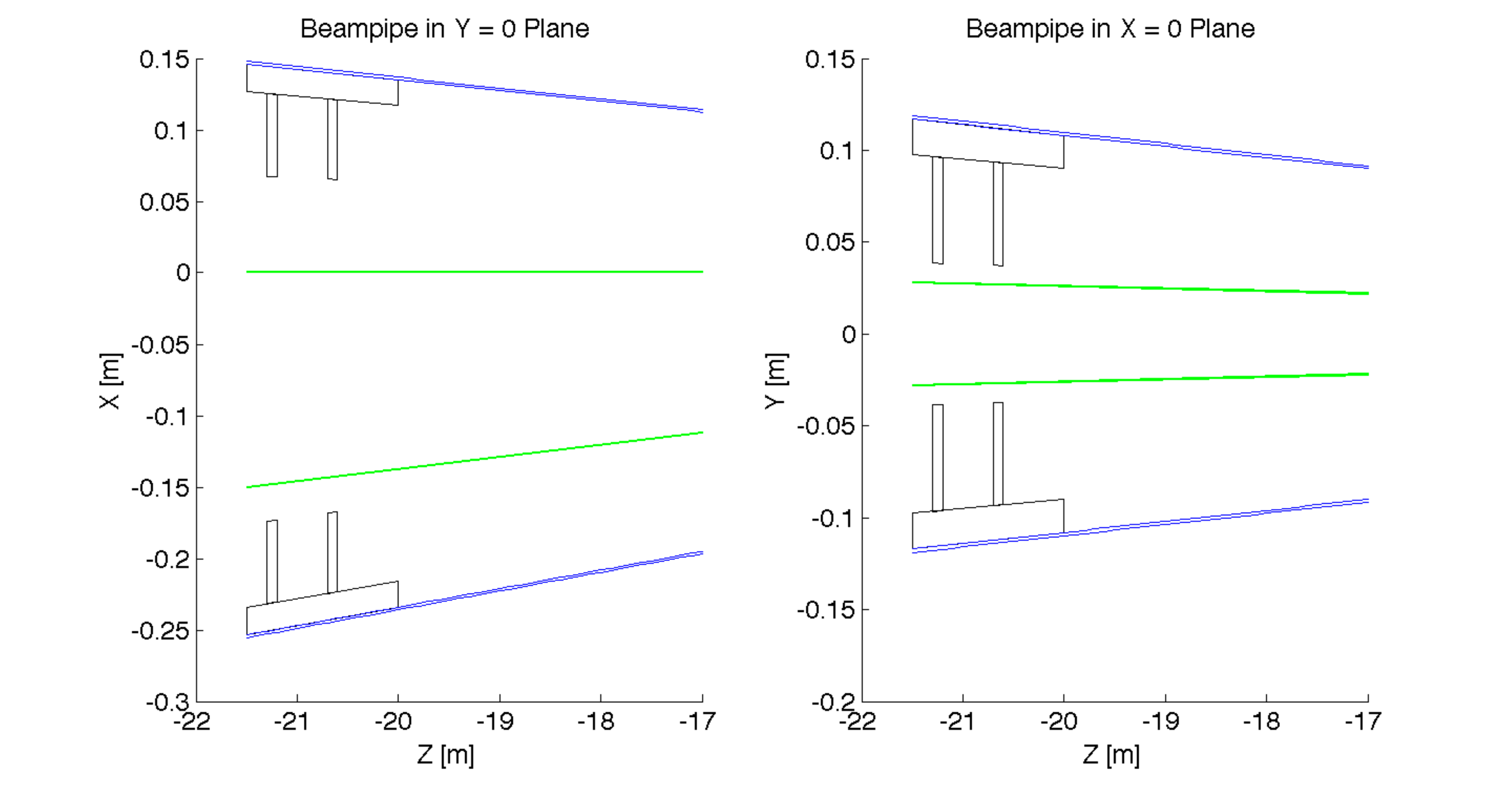}}
\caption{High Acceptance: Beam pipe Cross Sections}
\label{Fig:1degBP}
\end{figure}

Cross sections of the beam pipe in the Y = 0 and X = 0 planes with the shields and masks included can be seen in Figure \ref{Fig:1degBP}.

%% file: machine/herr.tex
\section{Beam-beam effects in the LHeC}
In the framework of the Large Hadron electron Collider a ring-ring option is considered where protons of one beam collide with the protons of the second proton beam as well as with leptons from a separate ring.  To deduce possible limitations the present knowledge of the LHC beam-beam effects from proton-proton collisions are fundamental to define parameters of an interaction point with electron-proton collisions.  From past experience it is known that the maximum achievable luminosity in a collider is limited by beam-beam effects.  These are often quantified by the maximum beam-beam tune shifts in each of the two beams.  An important aspect in electron-proton collisions is that the proton beam, more sensitive to transverse noise, could be perturbed by a higher level of noise in the electron beam.  In this section we will assess some limits to the possible tune shift achievable in collision based on experience from past colliders as CESR \cite{Young:1997zzc} and LEP \cite{Brandt:1999zs} and more recent ones like the LHC \cite{MDNoteLHC}.
\subsection{Head-on beam-beam effects}
A first important performance issue in beam-beam interaction comes from the restricted choice of the $\beta$-function at the interaction point to keep the transverse beam sizes equal for the two beams since proton and electron emittances are different. 
The choice of beta functions at the interaction point has to be different for the two beams in order to keep $\sigma_{x}^{e} = \sigma_{x}^{p}$ and  $\sigma_{y}^{e} = \sigma_{y}^{p}$ for the reasons explained in detail in \cite{malika}. 
In a mismatched collision the larger bunch may suffer more because a large part of the particle distribution will experience the non-linear beam-beam force of the other bunch. 
With this in mind it is preferable to keep the electron beam slightly larger than the proton beam 
since the electron beam may be less sensitive due to strong radiation damping. 
This matching implies that the electron emittances must be controlled during operation and kept 
as constant as possible (i.e. H/V coupling).
For the proton beam the beam-beam effects from the electron beam will be different for the two planes. Optical matching of the beam sizes at the IP is the first constraint for any interaction region layout proposed.
~\\

Another important issue is the achievable tune shift and how this relates to the linear beam-beam parameter which is normally the parameter used to evaluate the strength of the beam-beam interaction.
~\\

The linear beam-beam parameter is defined as $\xi_{bb}$ and is expressed for the case of round beams like in proton-proton collision at the LHC as:

\begin{equation}\label{eq:01}
\xi_{bb}~=~\frac{N r_{p} \beta^{*}}{4\pi\gamma\sigma^{2}}
\end{equation}
where $r_{p}$ is the classical proton radius, $\beta^{*}$ is the
optical amplitude function ($\beta$-function) at the interaction
point, $\sigma = \sigma_{x,y}$ is the transverse beam size in metres
at the interaction point, $N_{p}$ is the bunch intensity and $\gamma$
is the relativistic factor.  For proton-proton collisions where
$\xi_{bb}$ does not reach too large values and the operational tune is
far enough away from linear resonances, this parameter is about equal
to the linear tune shift $\Delta Q$ expected from the head-on
beam-beam interaction.  This is the case for the LHC proton-proton
collisions at IP1 and IP5 where the linear tune shift per IP is of the
order of 0.0034/0.0037 for nominal beam parameters as summarised in
Table~\ref{tab:01} and corresponds to the linear beam-beam parameter
$\xi_{bb}$.  This is in general not true for lepton colliders where
the operational scenario differs from hadron colliders and other
effects become dominant and have to be taken into account.
~\\
In the case of electron beams the transverse shape of the beams is
normally elliptical with $\sigma_{x} > \sigma_{y}$.  In this
configuration one can generalise the linear beam-beam parameter
calculation with the following formula \cite{Bassetti:1980by}:

\begin{equation}\label{eq:02}
{{\xi_{x,y}}}~=~\frac{N r_{e} \beta^{*}_{x,y}}{2\pi\gamma\sigma_{x,y}(\sigma_{x} + \sigma_{y})}
\end{equation}

with $r_{e}$ is the electron classical radius.

In the case of electron-proton collisions one has to also take into account the different species during collision and the beam-beam parameters become:

\begin{equation}
\label{eq:03}
{\xi_{(x,y),b_{1}}}~=~\frac{N_{b_{2}} r_{b_{1}} \beta^{*}_{(x,y),b_{1}}}{2\pi\gamma_{b_{1}}\sigma_{(x,y),b_{2}}(\sigma_{x,b_{2}} + \sigma_{y,b_{2}})}
\end{equation}
Here $b_{1}$ and $b_{2}$ refer to Beam1 and Beam2 respectively.
The linear beam-beam parameter $\xi$ is often used to quantify the strength of the
beam-beam interaction, however it does not reflect the non-linear
nature of the electromagnetic interaction.
Nevertheless, it can be used for comparison and as a scaling parameter. 
Since a general beam-beam limit cannot be found and will be different from one
collider to the next, the interpretation should be conservative.

\begin{table}
\begin{center}

\begin{tabular}{|l|c|c|c|} 
\hline 
Parameter & LEP & LHC (nominal) \\
\hline \hline
Beam sizes $\sigma_x/\sigma_y$ & 180~$\mu$m$/$7~$\mu$m & 16.6~$\mu$m$/$16.6~$\mu$m\\ \hline 
Intensity N & $4.0\times 10^{11}$/bunch & $1.15\times 10^{11}$/bunch\\ \hline
Energy & 100 GeV & 7000 GeV \\ \hline
$\beta_{x}^{*}/\beta_{y}^{*}$ &1.25 m$/$0.05 m & 0.55 m$/$0.55 m\\ \hline
Crossing angle $\theta_x/\theta_y$ & 0.0 &0~$\mu$rad/285~$\mu$rad\\ \hline
Beam-beam tune shift($\Delta Q_x/\Delta Q_y)$ & 0.0400/0.0400 & 0.0037/0.0034 \\ \hline
\end{tabular}
\caption{Comparison of parameters for the LEP collider and the LHC.}\label{tab:01}
\end{center}
\end{table}

In Table~\ref{tab:01} we compare LEP and LHC beam parameters and achieved linear beam-beam parameters.
Some of the differences are striking: while the beams in the LHC are round
at the interaction point, they are very flat in LEP. This is due to
the excitation of the beam in the horizontal plane by the strong
synchrotron radiation and damping in the vertical plane.
Another observation is the much larger beam-beam parameter in LEP.

One reason for the larger achievable beam-beam parameter in lepton colliders is due to a significant 
dynamic beta effect when operating at a working point close to integer tune. 
This is considered more difficult with proton beams. 
In Equation~\ref{eq:04} the perturbed $\beta^{*}$ is expressed as a function of the beam-beam parameter $\xi$ and the phase advance between two interaction points $2 \pi Q^{i}$. The tune shift $\Delta$Q becomes a function of the tune which can be chosen to keep the actual shift small. 

\begin{equation}
\label{eq:04}
{\beta^{*}}(Q, \xi)~=~\frac{\beta}{\sqrt{1 + 4 \pi \xi(cot(2\pi Q^{i})) - 4\pi^{2}\xi^{2}}}
\end{equation}

 From experience it is known that electrons have a bigger range for the linear head-on beam-beam parameter: 
LEP II has proved an unperturbed beam-beam parameter of 0.07 per interaction point corresponding to a measured $\Delta Q$ of 0.03 - 0.04 as also 
confirmed in other lepton colliders. 
The large difference between the beam-beam parameter and the achieved tune shift was due to the strong dynamic $\beta$ effect in LEP. CESR demonstrated the possibility to achieve tune shifts of the order of 0.09. A second and most important reason for a higher acceptable tune shift in lepton colliders is the synchrotron radiation damping. Furthermore, while for lepton colliders a clear indication for a ``beam-beam limit" exists, not
such criteria can be easily defined for hadron machines \cite{MDNoteLHC}. From these considerations we have to assume that  the choice of beam-beam parameters $\xi_{bb}$ of the proton beam is restricted.

The LHC as a proton-proton collider has confirmed previous experience from S$p\bar{p}$S and Tevatron that a total linear tune shift of 0.018 (0.006 per IP) is tolerable with neither important losses nor reduction of beam lifetime during normal operation. 
It is generally admitted that $\xi_{bb}$ could reach a value of 0.01 per interaction point.
Recent experiments at the LHC with very high intensity beams beyond ultimate and reduced transverse beam sizes 
demonstrated the possibility to reach head-on tune shifts well beyond the nominal values \cite{MDNoteLHC}. 
At the LHC tune shifts per IP close to 0.02 have been achieved. 
Total tune shifts exceeding 0.034 have also been achieved with stable beams for two symmetric crossings at IP1 and IP5. 
These latest experiments demonstrate the possibility to operate with larger than nominal beam-beam parameters.

The calculated beam-beam parameters for the electron and proton beams due to an electron-proton collision 
in the LHeC are summarised in Table~\ref{tab:02} for the two interaction region 
options (1~Degree option and 10~Degree option).

\begin{table}
\begin{center}
\begin{tabular}{|l|c|c|c|c|} 
\hline
IR Option     & \multicolumn{2}{c|}{1 Degree}  & \multicolumn{2}{c|}{10 Degree} \\
\hline \hline
Beams & Electrons & Protons & Electrons & Protons \\ \hline
Energy & 60 GeV & 7 TeV & 60 GeV & 7 TeV \\ \hline
Intensity & $2\times 10^{10}$ & $1.7\times 10^{11}$ & $2\times 10^{10}$ & $1.7\times 10^{11}$\\ \hline
$\beta_{x}^{*}$ & 0.4 m & 4.0 m & 0.18 m & 1.8 m\\  \hline
$\beta_{y}^{*}$ & 0.2 m & 1.0 m & 0.1 m & 0.5 m\\ \hline
$\epsilon_{x}$ & 5 nm & 0.5 nm & 5 nm & 0.5 nm\\ \hline
$\epsilon_{y}$ & 2.5 nm & 0.5 nm & 2.5 nm & 0.5 nm\\ \hline
$\sigma_{x}$  & \multicolumn{2}{c|}{45 $\mu$m} & \multicolumn{2}{c|}{30 $\mu$m}\\  \hline
$\sigma_{y}$ &\multicolumn{2}{c|}{22 $\mu$m} & \multicolumn{2}{c|}{15.8 $\mu$m}\\  \hline
Crossing angle & \multicolumn{2}{c|}{1 mrad} & \multicolumn{2}{c|}{1 mrad} \\ \hline
$\xi_{bb,x}$ & 0.086 & 0.00086 & 0.085 & 0.00085\\  \hline
$\xi_{bb,y}$ & 0.086 & 0.00043 & 0.089 & 0.00045\\  \hline
Luminosity &  \multicolumn{2}{c|}{$7.33\times 10^{32}$ cm$^{-2}$s$^{-1}$} &  \multicolumn{2}{c|}{$1.34\times 10^{33}$ cm$^{-2}$s$^{-1}$}\\
\hline
\end{tabular}
\caption{Beam parameters for the interaction region options and the linear beam-beam parameters $\xi$.}\label{tab:02}
\end{center}
\end{table}

\begin{table}
\begin{center}
\begin{tabular}{|l|c|c|c|c|} 
\hline
              & \multicolumn{2}{c|}{Nominal}  & \multicolumn{2}{c|}{Upgrade} \\  \hline \hline
      & Electrons & Protons & Electrons & Protons \\ \hline
$\xi_{bb,x}$ & 0.016 & 0.0013 & 0.027 & 0.0017\\  \hline
$\xi_{bb,y}$ & 0.018 & 0.0012 & 0.041 & 0.0005\\ \hline
\end{tabular}
\caption{Linear beam-beam parameters for HERA, nominal machine and upgrade parameters.}\label{tab:02a}
\end{center}
\end{table}

The two proposed interaction region options will give for the proton beam a maximum beam-beam parameter in the horizontal plane of about $8.5\times 10^{-4}$. This effect is in the shadow of the proton-proton collision at IP1 and IP5 which will give a beam-beam parameter of $5.5\times 10^{-3}$ per IP for nominal beam emittances and assuming intensities of $1.7\times 10^{11}$ protons/bunch, which was already exceeded during 2010 operation at the LHC with reduced emittances and nominal beam intensities. One should not expect detrimental effects of the head-on interactions with the electron beam apart from a potential coupling of noise from the electron into the proton beam. 

For the electron beam, on the contrary, the beam-beam parameter of  $8.6\times 10^{-2}$ is large and represents a value at the limit of what has been achieved so far in other lepton 
machines (LEP at 90 GeV energy achieved an unperturbed beam-beam parameter of 0.07, (with a maximum tune shift of 0.04) while  KEK and HERA achieved a maximum $\xi_{bb} = 0.04$ 
during operation, CESR achieved a beam-beam parameter of 0.09 for single IP but with lower luminosity). 
The beam-beam tune shifts achieved at HERA for the nominal and upgrade version are summarised in Table~\ref{tab:02a}
for comparison.
The foreseen beam-beam parameter of  $8.6\times10^{-2}$ is optimistic and a significant reduction due to dynamic beta and
the small number of interaction points could make it feasible.

\subsection{Long range beam-beam effects}
So far we have discussed head-on beam-beam interactions but an important issue are the long range interactions which will occur at the electron-proton collision and their interplay with the proton-proton crossings at IP1 and IP5. The two interaction points IP1 and IP5 will give up to 60 proton-proton long-range interactions which should be added to the two interaction region options which will give two additional parasitic encounters. 
The beam separation at this encounters should be as large as possible to reduce any non-linear perturbation. 
The parasitic encounters occur every 3.75~m from the interaction point for a bunch spacing of 25~ns. 
The proposed optics will then lead to parasitic beam-beam interactions which will occur at a 
transverse separation $d$ as:
\begin{equation}
\label{eq:05}
d(s)_{x,y} ~=~ \alpha \frac{s}{\sqrt{\epsilon_{x,y} \beta(s)_{x,y}}}
\end{equation}
with $\epsilon_{x,y}$ are the beam emittance in the separation plane and $\beta(s)$ is the betatron function at a distance s from the interaction point.
 
\begin{table}
\begin{center}
\begin{tabular}{|l|c|c|c|c|} 
\hline
IR Option     & \multicolumn{2}{c|}{1 degree}  & \multicolumn{2}{c|}{10 degree} \\
\hline  \hline
Beams & Electrons & Protons & Electrons & Protons \\ \hline
$\beta_{x}^{*}$ & 0.4 m & 4.0 m & 0.18 m & 1.8 m\\  \hline
$\beta_{y}^{*}$ & 0.2 m & 1.0 m & 0.1 m & 0.5 m\\ \hline
$\epsilon_{x}$ & 5 nm & 0.5 nm & 5 nm & 0.5 nm\\ \hline
$\epsilon_{y}$ & 2.5 nm & 0.5 nm & 2.5 nm & 0.5 nm\\ \hline
Crossing angle & \multicolumn{2}{c|}{1 mrad} & \multicolumn{2}{c|}{1 mrad} \\ \hline
$d_{x}$  & 90 $\sigma_{p}$ & 9.0 $\sigma_{e}$ & 60 $\sigma_{p}$ & 6.0 $\sigma_{e}$\\ \hline
\end{tabular}

\caption{Normalised beam separation $d_{x}$ at beam-beam long range encounters for the two interaction region options.}\label{tab:03}
\end{center}
\end{table}

In Table~\ref{tab:03} the distances of the parasitic encounters in units of the transverse beam sizes are shown for both interaction region layouts. 

The 1~Degree option gives long range interactions at larger separation with respect to the 10~Degree option 
which results in small separations of  $\approx$~6~$\sigma$ for the proton beam. 
Particles in the tail of the proton beam particles will experience the non linearity of the electron beam 
electromagnetic force. 
The presence of two long range at 6 $\sigma$ separation may be acceptable since it is shown experimentally  
that few encounters also at smaller separation do not affect the beams dramatically \cite{levelingLHC}. 
However, the interplay of these two encounters with the long-range interactions from IP1 and IP5  
should be studied in detail with numerical simulation to highlight possible limitations. 
In this framework future experiments at the LHC will help defining a possible beam parameters space for the 
control of the long-range effects from proton-proton collisions. 
If encounters at 6~$\sigma$ present a limitation to the collider performance then a possible cure to increase the 
long-range separation could be a further increase of the crossing angle and using crab cavities can recover 
the increased geometric luminosity reduction factor. 
In this case a study of the crab cavities effects on the proton beam would be essential to define the 
effects of transverse noise on colliding beams.
~~\\
For any reliable study of the LHeC project one has to address other possible beam-beam issues with extensive numerical simulations of the operational scenario of the LHeC. 
This is fundamental since there is no other possible simplification which can be adopted in evaluating 
the non-linear parts of the beam-beam forces. 
For this reason a detailed and full interaction layout with crossing schemes matched in thin lens version is needed. 
With the complete optic layout beam-beam effects which still need further studies by means of numerical 
simulation campaign are the following:
\begin{itemize}
\item[$\bullet$]Long-range tune shifts and orbit effects.
\item[$\bullet$]Self-consistent study of the proton-proton and electron-proton beam dynamics interplay.
\item[$\bullet$]Dynamic aperture tracking studies.
\item[$\bullet$]Multi-bunch effects.
\item[$\bullet$]Noise coupling from the electron to the proton beam.
\end{itemize}
The evaluation of the non-linear effects of the beam-beam interactions with self-consistent calculations will define 
a set of parameters for operation  \cite{LHCBBWebpage}.

%% file: machine/eAJowett.tex

\section{Performance as an electron-ion collider}
\label{sec:eAJowett}

\subsection{Heavy nuclei, e-Pb collisions}
With the first collisions of lead nuclei (\Pb) in 2010~\cite{Jowett:2008hb,JowettChamonix2011},
the LHC has already demonstrated its
capability as a heavy-ion collider and this naturally opens up the possibility of electron-nucleus
(e-A) collisions in the LHeC.

In order to avoid interference with the high luminosity proton-proton operation, this mode of operation would naturally be included in the  annually-scheduled ion operation period of the LHC.
In principle, the CERN complex could provide A-A (or even p-A) collisions to the LHC experiments while
the LHeC operates with e-A collisions.
The lifetime of the nuclear beam would depend mainly on
whether it was exposed to the losses from A-A luminosity in the LHC (in this case it would be at
least a few hours).

In the first decade or so of LHC operation, the ion injector chain is expected to provide mainly
\Pb, but also other species such as \Ar\ or \Xe, either to the LHC or from the SPS to fixed target
experiments in the North Area.
These beams could also be collided with electrons in the LHeC but
solid intensity estimates are not yet available for the lighter ions.
For simplicity, we shall
estimate LHeC performance in e-Pb collisions with the design performance values of the ion injector
chain as described in~\cite{Benedikt:2004wm} and the assumption of a single nuclear beam in one ring
of the LHC with parameters as recalled from~\cite{Bruning:2004ej} in Table~\ref{tab:Pbparameters}.
It
is assumed that present uncertainties about the Pb intensity limits at full energy in the LHC will
have been resolved, if necessary, by installation of new collimators in the dispersion suppressors
of the collimation insertions in the LHC.
This simplifies the discussion because the design
emittances of Pb and proton beams in the LHC are such that both species have the same geometric beam
sizes and considerations of optics and aperture can be taken over directly.
The ``Ultimate Pb'' value of the Pb single bunch intensity was
already attained in 2010~\cite{JowettChamonix2011}
using a
simplified injection scheme but not yet with the nominal filling scheme for 592 bunches;
it can be considered an optimistic goal.
At present, there are no prospects for increasing the number of bunches significantly.
Lower Pb emittances may be possible but would not increase e-Pb luminosity
unless matched with smaller optical functions or emittances for the electron beam.

\begin{table}[!h]
\begin{center}
\begin{tabular}{|l|c|c|c|}
\hline
& & Design Pb & Ultimate Pb \\\hline\hline
Energy & $E_{\text{Pb}}$ & \multicolumn{2}{c|}{574. TeV} \\
\hline
Energy per nucleon & $E_{N}$ & \multicolumn{2}{c|}{2.76 TeV} \\
\hline
No.\ of bunches & $n_b$ & \multicolumn{2}{c|}{592} \\
\hline
Ions per bunch & $N_{\text{Pb}}$  & \enum{7.}{7} &  \enum{1.2}{8} \\
\hline
Normalised emittance & $\varepsilon _n$ & \multicolumn{2}{c|}{\qty{1.5}{\mu m}}  \\
\hline
\end{tabular}
\end{center}
\caption{\label{tab:Pbparameters}
         Parameters for the \Pb\ beam according to
         Chapter~21 of~\cite{Bruning:2004ej}.
        }
\end{table}

   Assume that the injection system can create an electron bunch train matching the 592-bunch train
of Pb nuclei in the LHC so that every Pb bunch finds a collision partner in the electron beam.
Assuming further that the hadron optics can be adjusted to match the sizes of the electron and Pb
beams, the luminosity can be expressed in terms of the interaction point optical functions and
emittances of the electron beam.
Since the e-A physics is focused on low-$x$ these are taken from Table~\ref{tab:IR.HA.P.Params}
describing the Ring-Ring High Acceptance optics, which reduces the luminosity by a factor 2 as compared with
the High-Luminosity optics.

In e-p mode, the intensity of the 2808 electron  bunches,  $N_{\text{e}}$ is limited for the Ring-Ring version of the LHeC 
by the total RF power available to compensate the synchrotron radiation loss.  For the same power
(some 44~MW for  $N_{\text{e}}=\enum{2}{10}$ of Table~\ref{tab:IR_parameters}),
the intensity of the $n_b=592$~bunches required to collide with the Pb nuclei can be increased by a factor
$2808/592$ to
$N_{\text{e}}=\enum{9.5}{10}$.
Electron beam parameters for the LHeC Ring-Ring option other than the single bunch intensity can be taken from
Table~\ref{tab:IR_parameters}.
Present experience with beam-beam effects in the LHC suggests that the
additional electron  intensity would not present any problem for the proton beam.
The single-bunch
intensity is still well below that achieved in LEP although the feasibility of these values should
be confirmed by further analysis of the ring impedance and collective effects.

Neglecting the geometric reduction factor due to the crossing angle and the hourglass effect, the \emph{electron-nucleon}  luminosity,
$L_{eN}= A L_{eA}$,
is then given by
\begin{equation}
L_{eN}  = \frac{{n_b f_0 N_{\text{e}} (AN_{{\text{Pb}}} )}}
{{4\pi \sqrt {\beta _{xe}^* \varepsilon _x^{} } \sqrt {\beta _{ye}^* \varepsilon _y^{} } }} = \left\{ {\begin{array}{*{20}c}
   {2.6 \times 10^{31} {\text{ cm}}^{{- 2}} {\text{s}}^{{ - 1}} } & {{\text{(Nominal Pb)}}}  \\
   {4.5 \times 10^{31} {\text{ cm}}^{{- 2}} {\text{s}}^{{-1}} } & {{\text{(Ultimate Pb)}}}
\end{array} } \right.
\end{equation}
This gives an indication of the range of peak luminosities that can be expected.  A factor of
2~could be gained by switching to the high-luminosity interaction region optics.

By the time the LHeC
comes into operation, it is not unreasonable to hope that ways to increase the number of Pb bunches
and perhaps to reduce their emittance (by cooling) may be implemented.  Therefore, on an optimistic
view, the luminosity could be even higher than the value quoted here.

Finally, we note that the dependence of luminosity on electron beam energy ($\propto E_e^{-6}$) is
very strong at the power limit so that a trade-off between energy and luminosity
may be of interest.

\subsection{Electron-deuteron collisions}

As discussed in \cite{pAatLHC},
deuteron beams are not presently available in the CERN complex.
Meanwhile it has been clearly demonstrated~\cite{Stovall:2011wc}
that it would not be feasible to set up a
D$^{-}$ source and accelerate them via Linac4.
The present proton Linac2 is due to be shut down so the only
way to accelerate them would be via the heavy ion Linac3.
However this would require a new source, RFQ and switch-yard at the input to Linac3.
The  study of practical feasibility,  space limitations, design
and potential performance of
these modifications to the injector complex started in late 2011 with a view
to supplying 
light ions to fixed target experiments and the LHC in several years' time.

Assuming that a practical design can be implemented,
the intensity of bunches in the LHC ring can be estimated as follows.

The present GTS-LHC source delivers \isotope{208}{Pb}{29+} ions with a charge-to-mass ratio $Q/A=
1/7.2$. A safe estimate of the space-charge limit at the entrance of Linac3 is \qty{200}{\mu A}.
To accelerate deuterons with $Q/A=1/2$, all magnetic and electric fields would have to be
reduced by a factor 3.6, leading to a space-charge limited current of
\qty{55}{\mu A}.

However there is then a very comfortable margin in the electric and magnetic fields
and deuterons are not subject to the loss factors associated with the subsequent stripping stages for Pb.
If enough deuteron current is  available from the source (say \qty{5}{mA}), and one accepts losses
in the linac and a somewhat degraded beam quality at the end,
then a current in the range of
\qty{\text{200-500}}{\mu A} would probably be available at the end of the linac.

As a caveat, early measurements of poor transmission of helium ions in
Linac3~\cite{Hill:2001vt} should be mentioned.
However the explanation is unclear due to the lack of appropriate diagnostics.

The bunch number and filling pattern in the LHC would be similar to that of the Pb beam.
A naive transposition of the   scaling of the ratios   of Linac3 output current (\qty{50}{\mu A}) to
LHC bunch intensity (\enum{7}{7}) from Pb to deuterons would suggest that the
deuteron single-bunch intensity in the LHC could be
\(
N_{\text{D}}  \approx \enum{1.5}{10}
\).

However this does not consider the differences in performance of the remainder of the injector chain
(the LEIR cooling ring, PS and SPS synchrotrons).
A proper evaluation of these requires a more detailed study.
To be safe, we can apply a factor 5 reduction to this value.

Then, assuming that we collide
such a beam with the electron beam   described in the preceding sub-section,
we see that  \emph{electron-nucleon} luminosities of order
\(
L_{eN}  \gtrsim   10^{31} {\text{ cm}}^{-2} {\text{s}}^{{  - 1}}
\)
could be accessible in e-D collisions at the LHeC.

%% file: machine/barber_wienands.tex
\section{Spin polarisation -- an overview}
Before describing concepts for attaining electron and positron
spin polarisation for the ring-ring option of the LHeC we present a brief overview of the theory and 
phenomenology. We can then draw on this later as required.   
This overview is necessarily brief but more details can be found in
\cite{handbooka,mont98}. 

\subsection{Self polarisation}

The spin polarisation of an ensemble of spin--1/2 fermions with the same energies 
travelling  in the same direction is defined as 
\begin{eqnarray} 
\vec P = \langle \frac{2}{\hbar} \vec \sigma \rangle
\end{eqnarray}
where $\vec \sigma$ is the spin operator in the rest frame and 
$ \langle \  \rangle $
denotes the expectation value for the mixed spin state.
We denote the single-particle rest-frame expectation value
of $\frac{2}{\hbar} \vec \sigma$
by $\vec S$ and we call this the ``spin''.
The polarisation is then the average of $\vec S$ over an ensemble
of particles such as that of a bunch of particles.

Electrons and positrons circulating in the (vertical) guide
field of a storage ring emit synchrotron radiation and a tiny fraction
of the photons can cause spin flip from up to down and vice versa.
However, the up--to--down and down--to--up rates differ,  
with the result that in ideal circumstances the electron (positron) 
beam can become spin polarised anti-parallel (parallel) to the field,
reaching a maximum polarisation, $P_{\rm st}$, of $\frac{8}{5 \sqrt{3}} =
92.4\%$.  This, the Sokolov-Ternov (S-T) polarising process, is
very slow on the time scale of other dynamical phenomena occurring in
storage rings, and the inverse time constant for the exponential build
up is \cite{st64}:
\begin{eqnarray}
\tau_{\rm st}^{-1}= \frac{5\sqrt{3}}{8}
      \frac{r_{\rm e} \gamma^{5}\hbar}
           {m_{\rm e}  {{|\rho|}^{3} } }
\label{eq:ST}
\end{eqnarray}
where $r_{\rm e}$ is the classical electron radius,  $\gamma$ is the Lorentz 
factor,  $\rho$ is the radius of curvature
in the magnets and the other symbols have their usual meanings.
The time constant is usually in the range of a few minutes to a few hours.

However, even without radiative spin flip, the spins are not stationary
but precess in the external fields. In particular, the motion of $\vec S$ 
for  a charged particle
travelling in electric and magnetic fields is governed by the
Thomas-BMT equation 
$d{\vec S}/ds = \vec{\Omega} \times {\vec S}$
where $s$ is the distance around the ring \cite{mont98,jackbook}. The 
vector 
$\vec{\Omega}$ depends on the electric ($\vec E$) and magnetic ($\vec B$)
fields, the energy and the velocity (${\vec v}$) which evolves 
according to the Lorentz equation:
\begin{eqnarray}
   \vec{\Omega}
 =       \frac{e}{m_{\rm e} c}
            \left[
        -\left(\frac{1}{\gamma}+a\right) \vec{B}
        +\frac{a\gamma}{1+\gamma} \frac{1}{c^{2}}
     ({\vec{v}} \cdot \vec{B})
      {\vec{v}}
         + \frac{1}{c^2} \left(a+\frac{1}{1+\gamma}\right)
     ({\vec{v}}\times {\vec E})
                  \right]  \\
= \frac{e}{m_{\rm e} c}
            \left[
        -\left(\frac{1}{\gamma}+a\right) {\vec{B}}_{\perp}
        -\frac{g}{2 \gamma} {\vec{B}}_{\parallel}
         + \frac{1}{c^2} \left(a+\frac{1}{1+\gamma}\right)
     ({\vec{v}}\times {\vec E})
                  \right] \; .
\end{eqnarray}
Thus $\vec{\Omega}$ depends on $s$ and on the position of the particle 
$u \equiv (x, p_x, y, p_y, l, \delta) $ in the 
6-D phase space of the motion. The coordinate $\delta$ is the 
fractional deviation of the energy from the energy of a synchronous 
particle (``the beam energy'') and $l$
is the distance from the centre of the bunch. The coordinates $x$ and $y$ are 
the
horizontal and vertical positions of the particle relative to the 
reference trajectory and 
$p_x =  x', p_y = y'$ (except in solenoids) are their conjugate momenta.
The quantity $g$ is the appropriate gyromagnetic factor and $a=(g-2)/2$ is the
gyromagnetic anomaly.
For $e^{\pm}$,  $a \approx 0.0011596$.
${\vec{B}}_{\parallel}$ and ${\vec{B}}_{\perp}$ are the magnetic fields
parallel and perpendicular to the velocity.

In a simplified picture, the majority of the photons in the synchrotron 
radiation do not cause
spin flip but tend instead to randomise the $e^{\pm}$ orbital motion in
the (inhomogeneous) magnetic fields.  Then, if the ring is
insufficiently-well geometrically aligned and/or if it contains
special magnet systems like the ``spin rotators'' needed to produce
longitudinal polarisation at a detector (see below), the spin-orbit coupling
embodied in the Thomas-BMT equation can cause spin diffusion, i.e.
depolarisation.  Compared to the S-T polarising effect the
depolarisation tends to rise very strongly with beam energy.  The equilibrium
polarisation is then less than 92.4\% and will depend on the relative
strengths of the polarisation and depolarisation processes.
As we shall see later, even without depolarisation certain
dipole layouts can reduce the equilibrium polarisation to below 
92.4\%. 

Analytical estimates of the attainable equilibrium polarisation are best 
based on the
Derbenev-Kondratenko (D-K) formalism \cite{dk73,mane87a}. This 
implicitly asserts that
the value of the equilibrium polarisation in an $e^{\pm}$
storage ring is the same at all points in phase space and is given by

\begin{eqnarray}
 {P}_{\rm dk} &=& \mp \frac{8}{5\sqrt{3}}
     \frac{
   { \oint {ds} \left<  \frac{1}{|\rho(s)|^{3}}
              \hat{b} \cdot
      ( \hat{n}- \frac{\partial{\hat{n}}} {\partial{\delta}} )
          \right>_{s}  } }
          { {\oint {ds} \left< \frac{1}{|\rho(s)|^{3}}
          ( 1- \frac{2}{9} { ( \hat{n}\cdot\hat{s} )}^{2}
              +
      \frac{11}{18}
      |\frac{\partial{\hat{n}}} {\partial{\delta}}| ^{2} \, )
          \right>_{s} } }
\label{eq:PDK}
\end{eqnarray}
where $<\ >_{s}$ denotes an average over phase space at azimuth $s$,
$\hat s$ is the direction of motion and $\hat b = ({\hat s} \times
{\dot{\hat s}})/|{\dot{\hat s}}|$.  $\hat b$ is the magnetic field
direction if the electric field vanishes and the motion is
perpendicular to the magnetic field.  $\hat{n}(u; s) $ is a unit 3-vector
field over the phase space satisfying the Thomas-BMT equation along
particle trajectories $u(s)$ (which are assumed to be integrable),
and it is 1-turn periodic: $\hat{n}(u;
s + C ) = \hat{n}(u; s)$ where $C$ is the circumference of the ring.

The field $\hat{n}(u; s)$ is a key object for systematising spin dynamics in 
storage rings.
It provides a reference direction for spin at each point in phase space
and it is now called the {\em ``invariant spin field''}
\cite{mont98,hvb99a,spin2000}.
At zero orbital amplitude, i.e. on the periodic (``closed'') orbit,
the  $\hat{n}(0; s)$ is written as  $\hat{n}_{0}(s)$. 
For $e^{\pm}$ rings and away from spin-orbit resonances (see below),
$\hat{n}$  is normally at most a few milliradians away from $\hat{n}_{0}$. 

A central ingredient of the D-K formalism is the implicit assumption
that the $e^{\pm}$ polarisation at each point in phase space is
parallel to $\hat n $ at that point. In the approximation
that the particles have the same energies and are travelling in the same 
direction, the polarisation of a bunch measured in a
polarimeter at $s$ is then the ensemble average
\begin{eqnarray}
  { \vec  P}_{\rm ens,dk}(s)
              \ =\
     P_{\rm dk}~
     \langle \hat{n} \rangle_{s} \; .
\label{eq:pensdk}
\end{eqnarray}
In conventional situations in $e^{\pm}$ rings, $ \langle \hat{n}\rangle_{s}$
is very nearly aligned along ${{\hat n}_0}(s)$. 
The {\em value} of the ensemble average, ${P}_{\rm ens,dk}(s)$, is
essentially independent of $s$.

Equation \ref{eq:PDK} can be viewed as having three components. 
The piece 
\begin{eqnarray}
 {P}_{\rm bk} &=& \mp \frac{8}{5\sqrt{3}}
     \frac{
   { \oint {ds} \left<  \frac{1}{|\rho(s)|^{3}}
              \hat{b} \cdot\hat{n}   
          \right>_{s}  } }
          { {\oint {ds} \left< \frac{1}{|\rho(s)|^{3}}
          ( 1- \frac{2}{9} { ( \hat{n}\cdot\hat{s} )}^{2} ) \right>_{s} } }
\approx 
\mp \frac{8}{5\sqrt{3}}
     \frac{
   { \oint {ds} \frac{1}{|\rho(s)|^{3}}
              \hat{b} \cdot\hat{n}_0   
            } }
          { {\oint {ds} \frac{1}{|\rho(s)|^{3}}
     ( 1- \frac{2}{9} {n}_{0 s}^{2} ) } } \; .
\label{eq:PBK}
\end{eqnarray}
gives the equilibrium polarisation due to radiative spin flip.
The quantity ${n}_{0 s}$ is the component of $\hat{n}_{0}$ along the 
closed 
orbit.
The subscript ``bk'' is used here instead of ``st'' to reflect
the fact that this is the generalisation by Baier and Katkov 
\cite{bk68,bks70} of the original S-T expression  to cover the case of 
piece-wise  homogeneous fields. 
Depolarisation is then accounted for by including the term
with $\frac{11}{18}
  | \frac{\partial{\hat{n}}} {\partial{\delta}}| ^{2} $
in the denominator.
Finally, the term with $\frac{\partial{\hat{n}}} {\partial{\delta}}$ in the 
numerator
is the so-called kinetic polarisation term. This results from the 
dependence of the radiation power on the initial spin direction and
is not associated with spin flip. It can normally be neglected but is 
still of interest in rings with special layouts. 

In the presence of radiative depolarisation 
the rate in Eq.\ \ref{eq:ST} must be replaced by
\begin{eqnarray}
      \tau^{-1}_{\rm dk}
       &=&
      \frac{5\sqrt{3}}{8}
      \frac{r_{\rm e} \gamma^{5}\hbar}
           {m_{\rm e}}
      \frac{1}{C}
               \oint ds    
\left< \frac{1- \frac{2}{9} (\hat{n}\cdot \hat{s})^{2}
                             +
 \frac{11}{18}  |\frac{\partial{\hat{n}}}
                      {\partial{\delta}}| ^{2} }
                     {|\rho(s)|^{3}}  \right>_s   \; .
\end{eqnarray}
This can be written in terms of the spin-flip polarisation rate, 
$\tau^{-1}_{\rm bk}$, and the depolarisation rate, $\tau^{-1}_{\rm dep}$, 
as:
\begin{eqnarray}
      \frac{1}{{\tau}_{\rm dk}}
           &=&
      \frac{1}{\tau_{\rm bk}}
            +
      \frac{1}{\tau_{\rm dep}}\ ,
\label{eq:tau}
\end{eqnarray}
where
\begin{eqnarray}
      \tau^{-1}_{{\rm dep}} 
      &=&
      \frac{5\sqrt{3}}{8}
      \frac{r_{\rm e}\gamma^{5}\hbar}
           {m_{\rm e}}
      \frac{1}{C}
                 \oint ds\,
                            \left<\,
                \frac{
                        \frac{11}{18}\,
                | \frac{\partial{\hat{n}}}
                      {\partial{\delta}}| ^{2}}
                     {|\rho(s)|^{3}}
                                       \right>_s 
\label{eq:TDK}
\end{eqnarray}
and 
\begin{eqnarray}
      \tau^{-1}_{\rm bk}
       &=&
      \frac{5\sqrt{3}}{8}
      \frac{r_{\rm e} \gamma^{5}\hbar}
           {m_{\rm e}}
      \frac{1}{C}
               \oint ds    
\left< \frac{1- \frac{2}{9} (\hat{n}\cdot \hat{s})^{2} }
                             {|\rho(s)|^{3}}  \right>_s \; .
\label{eq:TBK}
\end{eqnarray}

The time dependence for build-up from an initial polarisation $P_0$ to
equilibrium is
\begin{equation}
P(t)\ =\
  {P}_{\rm ens,dk}
    \left[
          1- e^{-t/{\tau_{\rm dk}}}
              \right] +
                        P_0 e^{-t/{\tau_{\rm dk}}} \; .
\label{eq:timedep}
\end{equation}

In perfectly aligned $e^{\pm}$  storage rings containing just horizontal
bends, quadrupoles and accelerating cavities, there is no vertical betatron 
motion and ${\hat n}_0(s)$ is vertical. Since the spins do
not ``see'' radial quadrupole fields and since the electric fields in
the cavities are essentially parallel to the particle motion, ${\hat n}$ is
vertical, parallel to the guide fields and to ${\hat n}_0(s)$ at all $u$ and 
$s$.
Then the derivative $\frac{\partial{\hat{n}}} {\partial{\delta}}$ vanishes
and there is no depolarisation.
However, real rings have misalignments.
Then there is vertical betatron motion so that the spins also see radial 
fields which tilt them from the vertical. Moreover, ${\hat n}_0(s)$ is
also tilted and the spins can couple to vertical quadrupole fields too.
As a result ${\hat n}$ becomes dependent on $u$  and  
``fans out'' away from  ${\hat n}_0(s)$ by an amount
which usually increases with the orbit amplitudes. 
Then in general $\frac{\partial{\hat{n}}} {\partial{\delta}}$ no longer 
vanishes in the dipoles (where $1/|\rho(s)|^{3}$ is large) and
depolarisation occurs. In the presence of skew quadrupoles and solenoids and,
in particular, in the presence of spin rotators, 
$\frac{\partial{\hat{n}}} {\partial{\delta}}$ can be
non-zero in dipoles even with perfect alignment. 
The deviation of ${\hat n}$ from ${\hat n}_0(s)$, and the depolarisation,
tend to be particularly large near to the spin-orbit resonance condition 
 
\begin{equation}
       \nu_0
         \ =\
          k_{_{0}}
          +
          k_{_{I}}Q_{_{I}}
          +
          k_{_{II}}Q_{_{II}}
          +
          k_{_{III}}Q_{_{III}} \; .
\end{equation}
Here $k_{_{0}}, k_{_{I}}, k_{_{II}}, k_{_{III}}$
are integers, $Q_{_{I}}, Q_{_{II}}, Q_{_{III}}$ are the 
three tunes of the synchrobetatron
motion and $\nu_0$ is the spin tune on the closed orbit, i.e. 
the number of 
precessions around ${\hat n}_0(s)$ per turn, made by a spin on the closed
orbit
{\footnote{In fact the resonance condition should be more precisely
expressed in terms of the so-called amplitude dependent spin tune 
\cite{mont98,hvb99a,spin2000}. But for typical $e^{\pm}$ rings, the amplitude 
dependent spin tune differs only insignificantly from $\nu_0$.}}.
In the special case, or in the approximation, of no synchrobetatron 
coupling 
one can make the associations: $I \rightarrow x$, 
$II \rightarrow y$ and $III \rightarrow s$, where, here, the subscript 
$s$ labels the synchrotron mode.
In a simple flat ring with no closed-orbit distortion,
$\nu_0 = a \gamma$ where $\gamma$ is the Lorentz factor for 
the nominal beam energy.
For $e^{\pm}$, 
$a \gamma$ increments by 1 for every 441 MeV increase in beam energy. 
In the presence of misalignments and special elements like rotators, 
$\nu_0$ is usually still approximately proportional to the beam 
energy.  
Thus an energy scan will show peaks in $\tau^{-1}_{{\rm dep}}$ and dips in
${P}_{\rm ens,dk}(s)$, namely at around the resonances. Examples can be seen 
in figures \ref{fig:poln_067}  and \ref{fig:poln_1} below.
The resonance condition expresses the fact that the disturbance to spins
is greatest when the $|\vec{\Omega}(u; s) - \vec{\Omega}(0; s)|$ along a trajectory
is coherent (``in step'') with the natural spin precession. 
The quantity ($|k_{_{I}}| + |k_{_{II}}| + |k_{_{III}}|$) is called the order 
of the  resonance. Usually, the strongest resonances are those 
for which $|k_{_{I}}| + |k_{_{II}}| + |k_{_{III}}| = 1$, i.e., the first-order resonances. 
The next strongest are usually the so-called 
{\em ``synchrotron sideband resonances''}
of parent first-order resonances, i.e. resonances for which 
$\nu_0 = 
k_{_{0}} \pm Q_{_{I,II,III}} + {\tilde k}_{_{III}}Q_{_{III}}$
where ${\tilde k}_{_{III}}$ is an integer and mode $III$ is associated
with synchrotron motion.
All resonances are due to the non-commutation of successive 
spin rotations 
in 3-D and they therefore occur even with purely linear orbital motion.

We now list some keys points.
\begin{itemize}
\item
The approximation on the r.h.s. of Eq.\ \ref{eq:PBK} makes it clear that if 
there are  dipole magnets with fields not parallel to $\hat{n}_0$, 
as is the case, for example, when spin rotators are used, then 
$P_{\rm bk}$  can be lower than the 92.4\% attainable in the case of a simple
ring with no solenoids and where all dipole fields and  
$\hat{n}_0 (s)$  are vertical. 
\item
If, as is usual, the kinetic polarisation term makes just a small
contribution, the 
above formulae can be combined to give
\begin{eqnarray}
      P_{\rm ens,dk} &\approx& P_{\rm bk} \frac{\tau_{\rm dk}}{\tau_{\rm bk}} \; .
\label{eq:pmeas}
\end{eqnarray}
From Eq.\ \ref{eq:tau} it is clear that $\tau_{\rm dk} \le \tau_{\rm bk}$.

\item
The underlying rate of polarisation due to the S-T effect,
$\tau_{\rm bk}^{-1}$, increases with the fifth power of the energy  
and decreases with the third power of the bending radii.
\item
It can be shown that as a general rule the ``normalised''  strength of the 
depolarisation, 
$\tau_{\rm dep}^{-1}/\tau_{\rm bk}^{-1}$, increases with beam energy 
according to a tune-dependent polynomial in even powers of the beam energy.
So we expect that the attainable equilibrium polarisation decreases as the energy increases.
This was confirmed LEP, where with the tools available,
little polarisation could be obtained at 60 GeV \cite{AssmannPACorEPAC}.
\end{itemize}

\subsection{Suppression of depolarisation -- spin matching}
Although the S-T effect offers a convenient way to obtain stored
high energy $e^{\pm}$ beams, it is only useful in practice 
if there is not too much depolarisation.
Depolarisation can be significant if the ring is misaligned, if it 
contains spin rotators or if it contains uncompensated solenoids 
or skew quadrupoles. Then if  $P_{\rm ens,dk}$ and/or $\tau_{\rm dk}$
are too small, the layout and the optic must be adjusted 
so that $(|\frac{\partial{\hat{n}}} {\partial{\delta}}|)^2 $
is small where $1/|\rho(s)|^{3}$ is large.
So far it is only possible to do this within the linear approximation
for spin motion.  
This technique is called {\em ``linear spin matching''} and when successful,
as for example at HERA \cite{bar95a}, it immediately reduces the strengths of 
the first-order spin-orbit resonances.
Spin matching requires two steps: {\em ``strong synchrobeta spin matching''}
is applied to the optics and layout of the perfectly aligned ring 
and then {\em ``harmonic closed-orbit spin matching''} is applied to
soften the effects of misalignments. This latter technique aims to
adjust the closed orbit so as to reduce the tilt of ${\hat n}_0$ from the
vertical in the arcs. Since the misalignments
can vary in time and are usually not sufficiently well known,
the adjustments are applied empirically while the polarisation is being
measured.

Spin matching must be approached on a case--by--case basis. An overview can 
be found in \cite{handbooka}. 

\subsection{Higher order resonances}
Even if the beam energy is chosen so that first-order resonances
are avoided and in
linear approximation $P_{\rm ens,dk}$ and/or $\tau_{\rm dk}$ 
are expected to be large, it can happen that that beam energy corresponds
to a higher order resonance. 
As mentioned above, in practice the most intrusive higher order resonances are those for which 
$ \nu_0 = k_0 \pm Q_{k} + {\tilde k}_{{s}}Q_{{s}}$
 ($k \equiv I, II ~{\rm or}~ III$). These
synchrotron sideband resonances of the first-order parent resonances
are due to modulation by energy oscillations of the instantaneous rate of 
spin precession around 
$\hat{n}_{0}$. 
The depolarisation rates 
associated with sidebands of isolated parent resonances
($ \nu_0 = k_0 \pm  Q_{k}$)
are related to the depolarisation rates for the parent 
resonances. For example,  if the beam energy is such that the system is near 
to a dominant $Q_y$ resonance we can 
approximate $\tau_{\rm dep}^{-1}$ in the form
\begin{eqnarray}
\tau_{{\rm dep}}^{-1} \propto \frac{A_y}{{\left(\nu_0 - k_0 \pm  Q_y\right)}^2} \; .
\label{eq:firstorderres}
\end{eqnarray}
This becomes 
\begin{eqnarray}
\tau_{{\rm dep}}^{-1} \propto \sum_{{\tilde k}_s = -\infty}^{\infty}
\frac{A_y\,{B_y}(\zeta;{\tilde k}_s)} {{\left(\nu_0 - k_0 \pm Q_y \pm {\tilde k}_s \, Q_s \right)}^2}
\nonumber
\end{eqnarray}
if the synchrotron sidebands are included.
The quantity $A_y$ depends on the beam energy and the optics and is reduced 
by spin matching. 
The proportionality constants ${B_y}(\zeta;{\tilde k}_s)$ are called {\em enhancement factors},  
and they contain modified Bessel functions 
$I_{|{\tilde k}_s|}(\zeta)$ and $I_{|{\tilde k}_s|+1}(\zeta) $ which depend  
on $Q_s$ and the energy spread $\sigma_{\delta}$ through the  {\em modulation index} 
$\zeta = (a \gamma ~\sigma_{\delta}/ Q_s)^2$. More formulae can be found 
in \cite{mane90,mane92}.

Thus the effects of synchrotron sideband resonances can be reduced by
doing the spin matches described above.  Note that these formulae are just 
meant as a guide since they are approximate and explicitly
neglect interference between the first-order parent resonances.
To get a complete impression, the Monte-Carlo simulation mentioned
later must be used.
The sideband strengths generally increase with the energy spread and 
the beam energy and the sidebands are a major contributor to the increase of 
$\tau_{\rm dep}^{-1}/\tau_{\rm bk}^{-1}$ with energy.

\subsection{Calculations of the $e^{\pm}$ polarisation in the LHeC} 

As a first step towards assessing the attainable polarisation 
we have considered an early version of the LHeC lattice: 
a flat ring with no rotators, no interaction point and no bypasses.
The tunes are $Q_x = 123.83$ and $Q_y = 85.62$. The horizontal emittance is
$8$ nm.
The ring is therefore typical of the designs under consideration.
With perfect alignment, $\hat n_0$ is vertical everywhere and
there is no vertical dispersion. The polarisation will then reach 92.4\%.
At $\approx 60$ GeV, $\tau_{\rm bk} \approx60$ minutes.

For the simple flat ring these values can be obtained by hand from Eq.\ \ref{eq:PBK}
and Eq.\ \ref{eq:TBK}. However, in general, e.g., in the presence of misalignments
or  rotators, the calculation of polarisation requires special software and for this study, the
thick-lens code SLICKTRACK was used \cite{barslicktrack}. This essentially consists of four
sections which carry out the following tasks:
\begin{itemize}
\item [(1)] 
  Simulation of misalignments followed by orbit correction with correction coils.

\item [(2)] 
  Calculation of the optical properties of the beam and the beam sizes.

\item [(3)] 
  Calculation of $\partial {\hat n} / \partial \delta$ for
  linearised spin motion with the thick-lens version (SLICK
  \cite{bar82}) of the SLIM algorithm \cite{handbooka}.  

  The equilibrium polarisation is then obtained from Eq.\
  \ref{eq:PDK}.  This provides a first impression and only exhibits
  the first order resonances.

\item [(4)]
  Calculation of the rate of depolarisation beyond the linear approximation    
  of item 3. 

  In general, the numerical calculation of the integrand in Eq.\
  \ref{eq:TDK} beyond first order represents a difficult computational
  problem. Therefore a pragmatic approach is adopted, whereby the rate
  of depolarisation is obtained with a Monte-Carlo spin-orbit tracking
  algorithm which includes radiation emission. The algorithm employs
  full 3-D spin motion in order to see the effect of the higher order
  resonances.  The Monte-Carlo algorithm can also handle the effect on
  the particles and on the spins of the non-linear beam-beam
  forces.  An estimate of the equilibrium polarisation is then
  obtained from Eq.\ \ref{eq:pmeas}.

\end{itemize}

\begin{figure}[htb]
\begin{center}
\includegraphics[width=10.0cm]{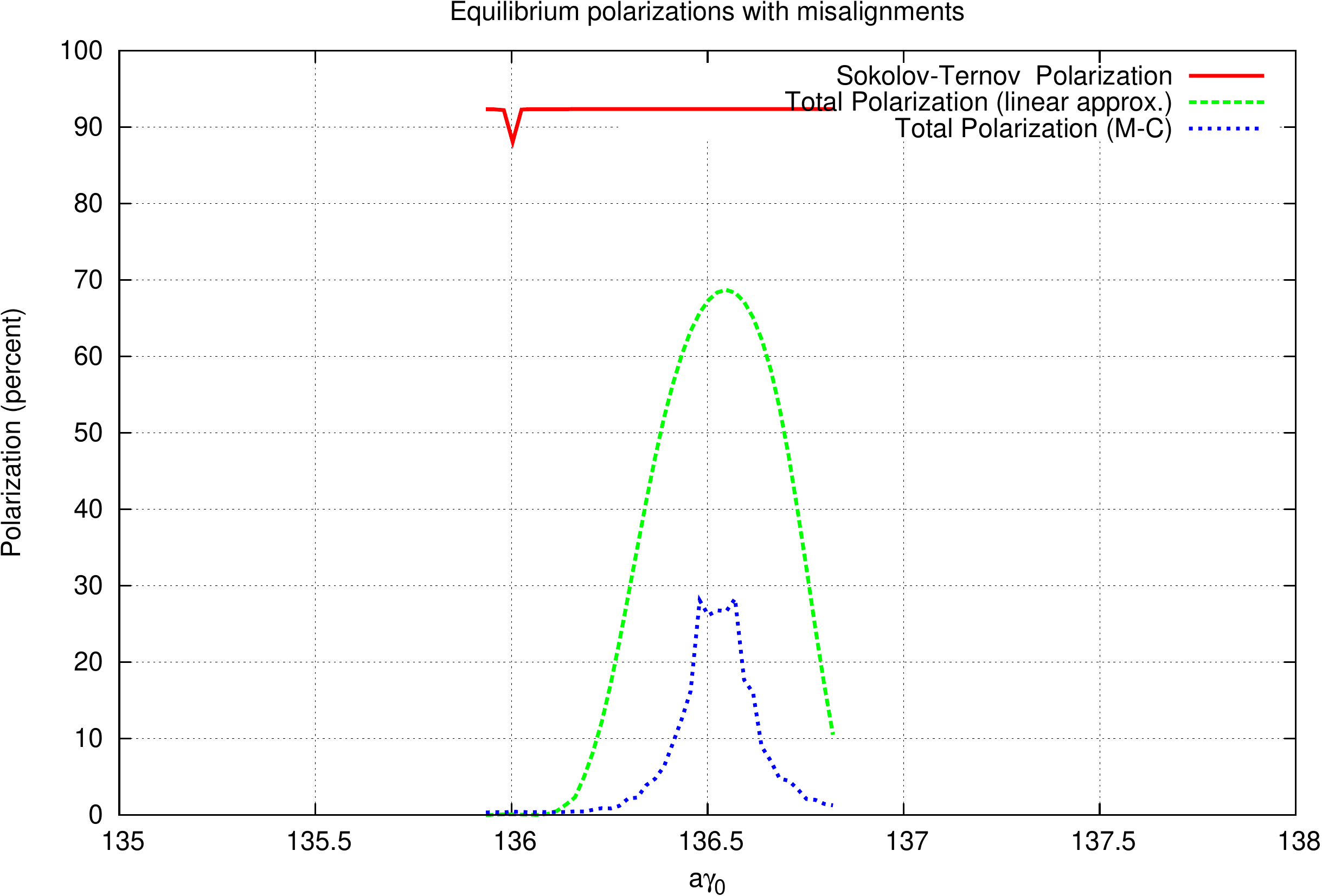}
\end{center}
\caption{Estimated polarisation for the LHeC without spin rotators, $Q_s = 0.06$.}
\label{fig:poln_067}
\end{figure}

\begin{figure}[htb]
\begin{center}
\includegraphics[width=10.0cm]{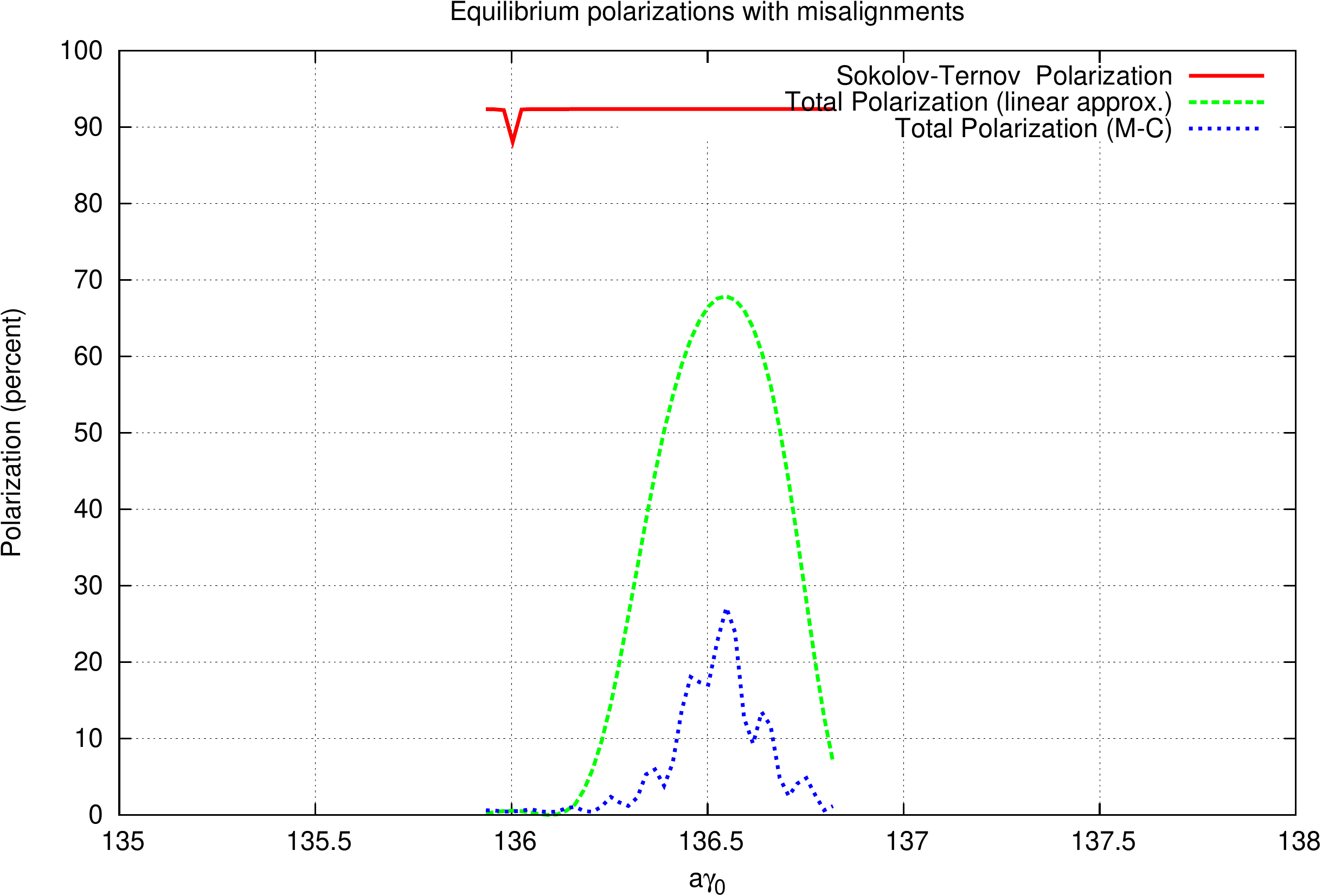}
\end{center}
\caption{Estimated polarisation for the LHeC without spin rotators, $Q_s = 0.1$.}
\label{fig:poln_1}
\end{figure}


Some basic features of the polarisation for the misaligned flat ring
are shown in figures \ref{fig:poln_067} and \ref{fig:poln_1} 
where polarisations are plotted against $a
\gamma$ around 60 GeV. In both cases the r.m.s. vertical closed-orbit 
deviation is about 75$\mu$m.  This is obtained
after giving the quadrupoles r.m.s. vertical misalignments of 150$\mu$m
and assigning a correction coil to every
quadrupole. The vector $\hat n_0$ has an r.m.s. tilt of about 4
milliradians from the vertical near $a \gamma = 136.5$. For figure \ref{fig:poln_067} the synchrotron tune,
$Q_s$, is 0.06 so that $\xi \approx 5$. For figure \ref{fig:poln_1},  $Q_s = 0.1$ so
that $\xi \approx 1.9$.

The red curves depict the polarisation due to the Sokolov-Ternov
effect alone.  The dip to below 92.4\% at $a \gamma = 136$ is due to
the characteristic very large tilt of $\hat n_0$ from the vertical at an
integer value of $a \gamma$. See \cite{handbooka}. 

The green curves depict the equilibrium polarisation after taking into
account the depolarisation associated with the misalignments and the
consequent tilt of $\hat n_0$.  The polarisation is calculated with
the linearised spin motion as in item 3 above. In these examples the
polarisation reaches about 68 \%. The strong fall off on each side of
the peak is mainly due to first-order ``synchrotron'' resonances $\nu_0
= k_0 \pm Q_s$.  Since $Q_s$ is small these curves are similar for the
two values of $Q_s$.

The blue curves show the polarisation obtained as in item 4 above.
Now, by going beyond the linearisation of the spin motion, the peak
polarisation is about 27 \%. The fall from 68 \% is mainly due to
synchrotron sideband resonances. With $Q_s = 0.06$ (Fig.~\ref{fig:poln_067}) the resonances are
overlapping. With $Q_s = 0.1$, (Fig.~\ref{fig:poln_1}) the sidebands begin to separate.  In
any case these curves demonstrate the extreme sensitivity  of the
attainable polarisation to small tilts of $\hat n_0$ at high energy. 
Simulations for $Q_s = 0.1$ with a series of differently misaligned rings, all with  
r.m.s. vertical closed-orbit distortions of about 75$\mu$m,  exhibit 
peak equilibrium polarisations ranging from about about 10 \% to about  40 \%.
Experience at HERA suggests that harmonic closed-orbit spin matching
can eliminate the cases of very low polarisation.

\begin{figure}[htbp]
\begin{center}
\includegraphics[width=10.0cm]{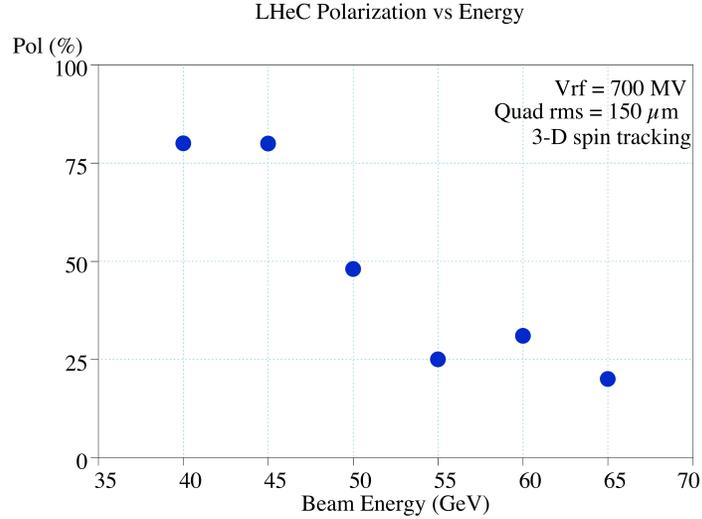}
\end{center}
\caption{Equilibrium polarisation {\em vs} ring energy, full 3-D spin
  tracking results}
\label{fig:poln_energy}
\end{figure}

Figure~\ref{fig:poln_energy} shows a typical energy dependence of the peak equilibrium
polarisation for a fixed RF voltage and for one of the misaligned rings. 
The synchrotron tune varies from $Q_s=0.093$ at 40 GeV to $Q_s=0.053$
at 65 GeV due to the change in energy loss per turn.
As expected the attainable polarisation falls steeply as the energy
increases.  However, although with this good alignment, a high
polarisation is predicted at 45 GeV, $\tau_{\rm bk}$ would be about 5
hours as at LEP.  A small $\tau_{\rm bk}$ is not only essential for a
programme of particle physics, but essential for the application of
empirical harmonic closed-orbit spin matching.

As mentioned above, it was difficult to get polarisation at 60 GeV at
LEP.  However, these calculations suggest that by adopting the levels
of alignment that are now standard for synchrotron-radiation sources
and by applying harmonic closed-orbit spin matching, there is reason
to hope that high polarisation in a flat ring can still be obtained.

\subsection{Spin rotator concepts for the LHeC}
The LHeC, like all analogous projects involving spin, needs
longitudinal polarisation at the interaction point. 
However, if the S-T effect is to be the means of producing  and maintaining the 
polarisation, 
then as is clear from Eq.\ \ref{eq:PBK}, ${\hat n}_0$ must be close to 
vertical in most of the dipoles. 
We have seen at Eq.\ \ref{eq:pensdk}
that the polarisation is essentially parallel to ${\hat n}_0$.
So to get longitudinal polarisation
at a  detector, it must be arranged that ${\hat n}_0$ is longitudinal
at the detector but vertical in the rest of the ring. This can be achieved
with magnet systems called spin rotators which rotate  ${\hat n}_0$ from
vertical to longitudinal on one side of the detector and back to vertical
again on the other side. 

Spin rotators use sequences of magnets which generate large spin
rotations around different axes and exploit the non-commutation of
successive large rotations around different axes.  According to the
T-BMT equation, the rate of spin precession in longitudinal fields is
inversely proportional to the energy.  However, for motion
perpendicular to a magnetic field spins precess at a rate essentially proportional
to the energy: $\delta {\theta}_{\rm spin} = (a \gamma + 1){\delta
  \theta}_{\rm orb}$ in obvious notation.  Thus for the high-energy
ring considered here, spin rotators should be based on dipoles as in
HERA \cite{bar95a}. 
In that case the rotators consisted of interleaved
horizontal and vertical bending magnets set up so as to generate
interleaved, closed, horizontal and vertical bumps in the design orbit.
The individual orbit deflections were small but the spin
rotations were of the order of a radian. The success in obtaining high
longitudinal polarisation at HERA attests to the efficacy of such rotators.

Eq.\ \ref{eq:PBK} shows that $P_{\rm bk}$
essentially scales with the cosine of the angle of tilt of ${\hat n}_0$ from the vertical
in the arc dipoles. Thus a rotation error resulting in a tilt of ${\hat n}_0$
of even a few degrees would not reduce $P_{\rm bk}$ by too much.
However, as was mentioned above,  a tilt of ${\hat n}_0$ in the arcs can lead 
to depolarisation. In fact the calculations show that at 60 GeV, tilts of more than a few milliradians cause 
significant depolarisation. Thus well-tuned rotators
are essential for maintaining polarisation.

Dipole rotators require a significant amount of space in the
ring. To minimise the power density as well as to preserve the polarisation, 
the amount of synchrotron radiation from the rotators needs to be kept to a
minimum, in direct conflict with the desire to keep the dipole magnets
as short as possible. In addition, longer dipole magnets lead to larger
orbit excursions. A numerical example for HERA-type spin rotators in the
LHeC with a bending radius of each dipole equal to that of the arc
dipoles yields a length of each spin rotator of about 170~m. The net
space appears to be available; the challenge being the integration of
the string of dipoles and the vertical magnet movers in an already
crowded area of the LHC tunnel. Note that the rotator incorporates a
certain amount of bending angle. The excursion away from the
nominal orbit is about 0.3~m.

A scheme using two Siberian Snakes has been considered by Derbenev and
Grote\cite{derbgrote} (see
below) that would integrate the IR rotators with the vertical dogleg
required to bring the beams into collision. For this the horizontal bends
are all of the same polarity and contribute to the overall $360^\circ$
bend so that the added dipole strength in the IR is minimised.

Table~\ref{tab:Rotatorparms} gives an indication of possible
parameters for LHeC spin rotators. These are subject to change as the
specific geometry in the IR is being further refined. Note that the
effect of these rotators on the degree of polarisation remains to be
evaluated (but see below for further comments on the Derbenev-Grote scheme).

\begin{table}[htb]
\centering
\begin{tabular}{|l|c|r|r|}
\hline
 Parameter & Unit & HERA-type &  Derbenev-Grote (IP only)\\
\hline\hline
No. of vertical dipole magnets & & 12 & 10 \\\hline
No. of horizontal dipole magnets & & 12 & 10\\\hline
Bending angle/magnet & $^\circ$ & 0.110 & 0.132 \\\hline
Length of magnet & m & 5.45 & 5.45 \\\hline
Total length of rotator & m & 170 & 80 \\\hline
Net bending angle & $^\circ$ & 0.66 & 1.32 \\\hline
Vertical offset & m & 0 & 1.25 \\\hline
\end{tabular}
\caption{Possible Parameters for LHeC Spin Rotators}
\label{tab:Rotatorparms}
\end{table}

\subsection{Further work}

We now list the next steps towards obtaining longitudinal polarisation at the interaction point. 

\begin{itemize}

\item [(1)] 
  A harmonic closed-orbit spin matching algorithm must be implemented
  for the LHeC to try to correct the remaining tilt of $\hat n_0$ and thereby
  increase the equilibrium polarisation.

\item [(2)]
  Practical spin rotators must be designed and appropriate strong
  synchrobeta spin matching must be implemented.  The design of the
  rotators and spin matching are closely linked.  Some preliminary
  numerical investigations (below) show, as expected, that without this spin
  matching, little polarisation will be obtained.

\item [(3)]
  If synchrotron sideband resonances are still overwhelming after items 1
  and 2 are implemented, a scheme involving Siberian Snakes could be
  tried.  Siberian Snakes are arrangements of magnets which manipulate
  spin on the design orbit so that the closed-orbit spin tune is
  independent of beam energy.  Normally the spin tune is then 1/2 and
  heuristic arguments suggest that the sidebands should be suppressed.
  However, the two standard schemes \cite{mont84} either cause
  $\hat n_0$ to lie in the machine plane (just one snake) or ensure
  that it is vertically up in one half of the ring and vertically down
  in the other half (two snakes). In both cases Eq.\ \ref{eq:PBK} shows that ${P}_{\rm bk}$ vanishes. 
  In principle, this problem can be overcome for two
  snakes by again appealing to Eq.\ \ref{eq:PBK} and having short strong dipoles in the
  half of the ring where $\hat n_0$ points vertically up and long
  weaker dipoles in the half of the ring where $\hat n_0$ points
  vertically down (or vice versa). Of course, the dipoles must be chosen so that the total bend  
  angle is $\pi$ in each half of the ring.  Moreover, Eq.\ \ref{eq:PBK} shows that the pure Sokolov-Ternov
  polarisation would be much less than 92.4\%. One version of this concept 
  \cite{derbgrote} uses a pair of rotators which together form a snake
  while a complementary snake is inserted diametrically opposite to
  the interaction point.  Each rotator comprises interleaved strings
  of vertical and horizontal bends which not only rotate the spins
  from vertical to horizontal, but also bring the $e^{\pm}$ beams down
  to the level of the proton beam and then up again.  However, the use of short
  dipoles in the arcs increases the radiation losses.
      
  Note that because of the energy dependence of spin rotations in the
  dipoles, $\hat n_0$ is vertical in the arcs at just one energy.
  This concept has been tested with SLICKTRACK but in the absence of a
  strong synchrobeta spin match, the equilibrium polarisation is very
  small as expected. Nevertheless the effects of misalignments and of the
  tilt of $\hat n_0$ away from design energy, have been isolated by
  imposing an artificial spin match using standard facilities in
  SLICKTRACK. The snake in the arc has been represented as a thin element that
  has no influence on the orbital motion. Then it looks as if the synchrotron sidebands are indeed
  suppressed in the depolarisation associated with tilts of $\hat n_0$.
  In contrast to the rotators in HERA, this kind of rotator allows only
  one helicity for electrons and one for positrons. 

\item [(4)] 
  If a scheme can be found which delivers sufficient longitudinal
  polarisation, the effect of non-linear orbital motion, the effect
  of beam-beam forces and the effect of the magnetic fields of the detector must then be studied.

\end{itemize}

\subsection{Summary}
We have investigated the possibility of polarisation in the LHeC
electron ring. At this stage of the work it appears  that a polarisation of
between 25 and 40\% at 60~GeV can be reasonably aimed for, assuming the efficacy of  harmonic closed-orbit spin matching. 
Attaining this degree
of polarisation will require precision alignment of the magnets to
better than $150\mu$m rms, a challenging but achievable
goal. The spin rotators necessary at the IP need to be properly
spin matched to avoid additional depolarisation and this work is in progress.
An interesting alternative involving the use of Siberian Snakes to try to 
avoid the depolarising synchrotron sideband resonances is being
investigated. At present, this appears to potentially yield a similar degree of
polarisation, at the expense of increased energy dissipation in the
arcs arising from the required differences of the bending radii in the two
halves of the machine.

%% file: machine/mess.tex
\section{Integration and machine protection issues}
\subsection{Space requirements}
\label{sec:Space requirements}
The integration of an additional electron accelerator into the LHC is a difficult task. Firstly, the LEP tunnel was designed for LEP and not for the LHC, which is now using up almost all space in the tunnel. It is not evident, how to place another accelerator into the limited space. Secondly, the LHC will run for several years, before the installation of a second machine can start. Meanwhile the tunnel will be irradiated and all installation work must proceed as fast as possible to limit the collective and individual doses. The activation after the planned high-luminosity-run of the LHC and after one month of cool-down is expected to be around $0.5 ... 1 \ \mu \mathrm{Sv/h}$ \cite{Forkel} on the proton magnets and many times more at exposed positions. Moreover the time windows for installation will be short and other work for the LHC will be going on, maybe with higher priority. Nevertheless, with careful preparation and advanced installation schemes an electron accelerator can be fitted in.

For the installation of the LHC machine proper, all heavy equipment had to pass the UJ2, while entering the tunnel. There the equipment had to be moved from TI2, which comes in from the outside, to the transport zone of LHC, which is on the inner side of the ring. Clearly, applying this procedure to the installation of the LHeC everything above the cold dipoles has to be removed. The new access shafts and the smaller size of the equipment for the electron ring may render this operation unnecessary.

\paragraph {General}
The new electron accelerator will be partially in the existing tunnel and partially in specially excavated tunnel sections and behind the experiments in existing underground areas. The excavation work will need special access shafts in the neighbourhood of the experiments from where the stub-tunnels can be driven. The connection to the existing LEP tunnels will be very difficult. The new tunnel enters with a very small grazing angle, which means over a considerable length. Very likely the proton installation will have to be removed while the last metres of the new tunnel is bored. 

\begin{figure}[ht]

\centerline{  \includegraphics* [clip=,,width=0.8\textwidth]{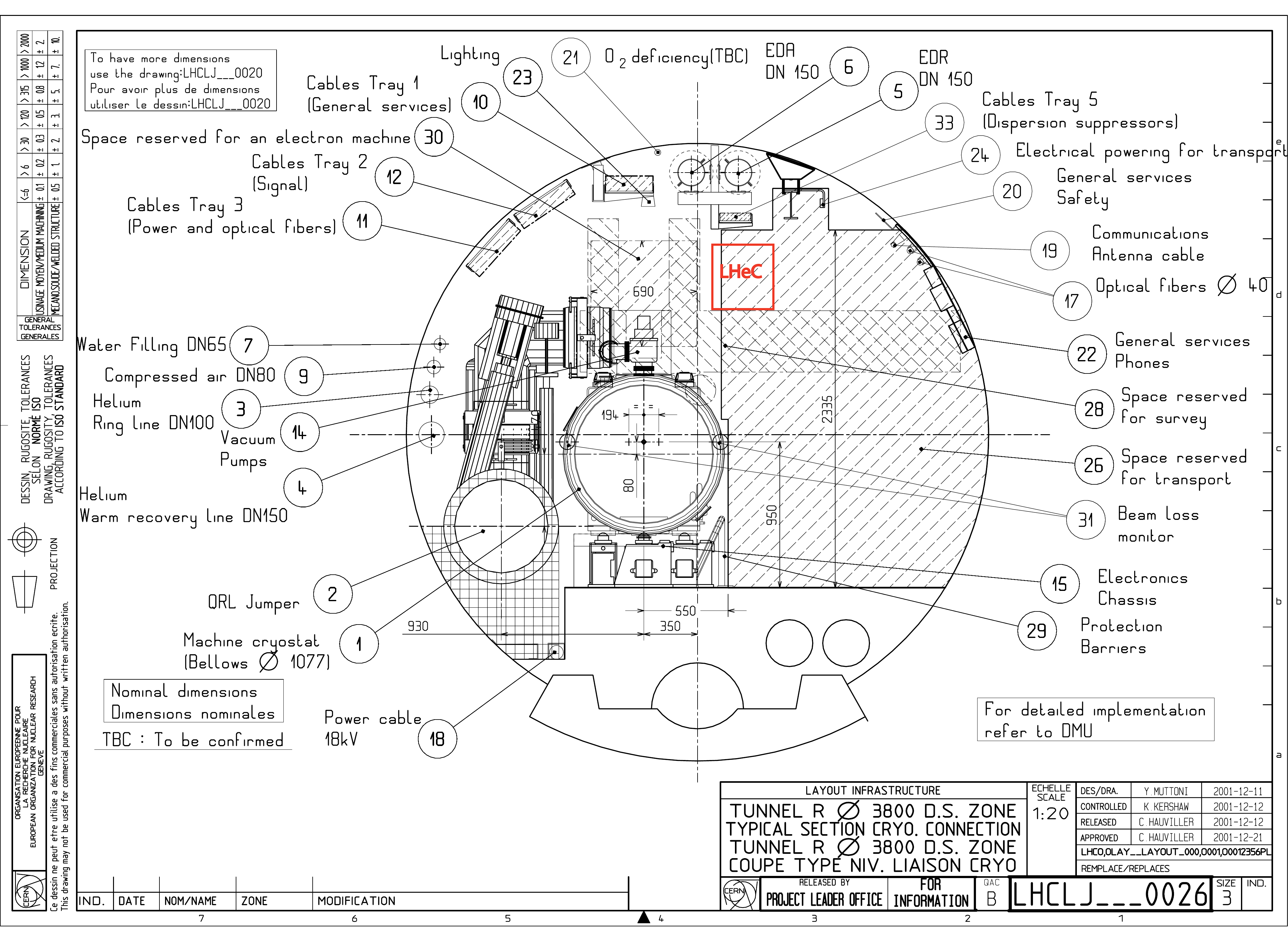}}
   \caption{Cross-section of the LHC tunnel with the original space holder for the electron beam installation directly above the LHC cryostat and the shifted new required space due to the additional bypass in IR1 and IR5 and the need to keep the overall circumference of the electron ring identical to that of the proton beams.}\label{fg:mess_3_13_tunnel1}
\end{figure}

Figure~\ref{fg:mess_3_13_tunnel1} shows a typical cross section of the LHC tunnel, where the two machines are together.  The LHC dipole dominates the picture. The transport zone is  indicated at the right (inside of the ring). The cryogenic installations (QRL) and various pipes and cable trays are on the left. The dipole cross section shows two concentric circles. The larger circle corresponds to the largest extension at the re-enforcement rings and marks a very localised space restriction on a  very long object. The inner circle is relevant for items shorter than about 10 m longitudinally.  A hatched square above the dipole labelled \it{30} \rm indicates the area, which was kept free in the beginning for an electron machine. Unfortunately, the centre of this space is right above the proton beam. Any additional machine will, however, have to avoid the interaction Points 1 and 5. In doing so additional length will be necessary, which can only be compensated for by shifting the electron machine in the arc about 60 cm to the inside (right), as indicated by the red square in Figure~\ref{fg:mess_3_13_tunnel1}. The limited space for compensation puts a constraint on the extra length created by the bypasses. The transport zone will, however, be affected. This requires an unconventional way to mount the electron machine. 
Nevertheless, there is clearly space to place an electron ring into the LHC, for most of the arc. 
\begin{figure}

\centerline{  \includegraphics* [clip=,,width=0.8\textwidth]{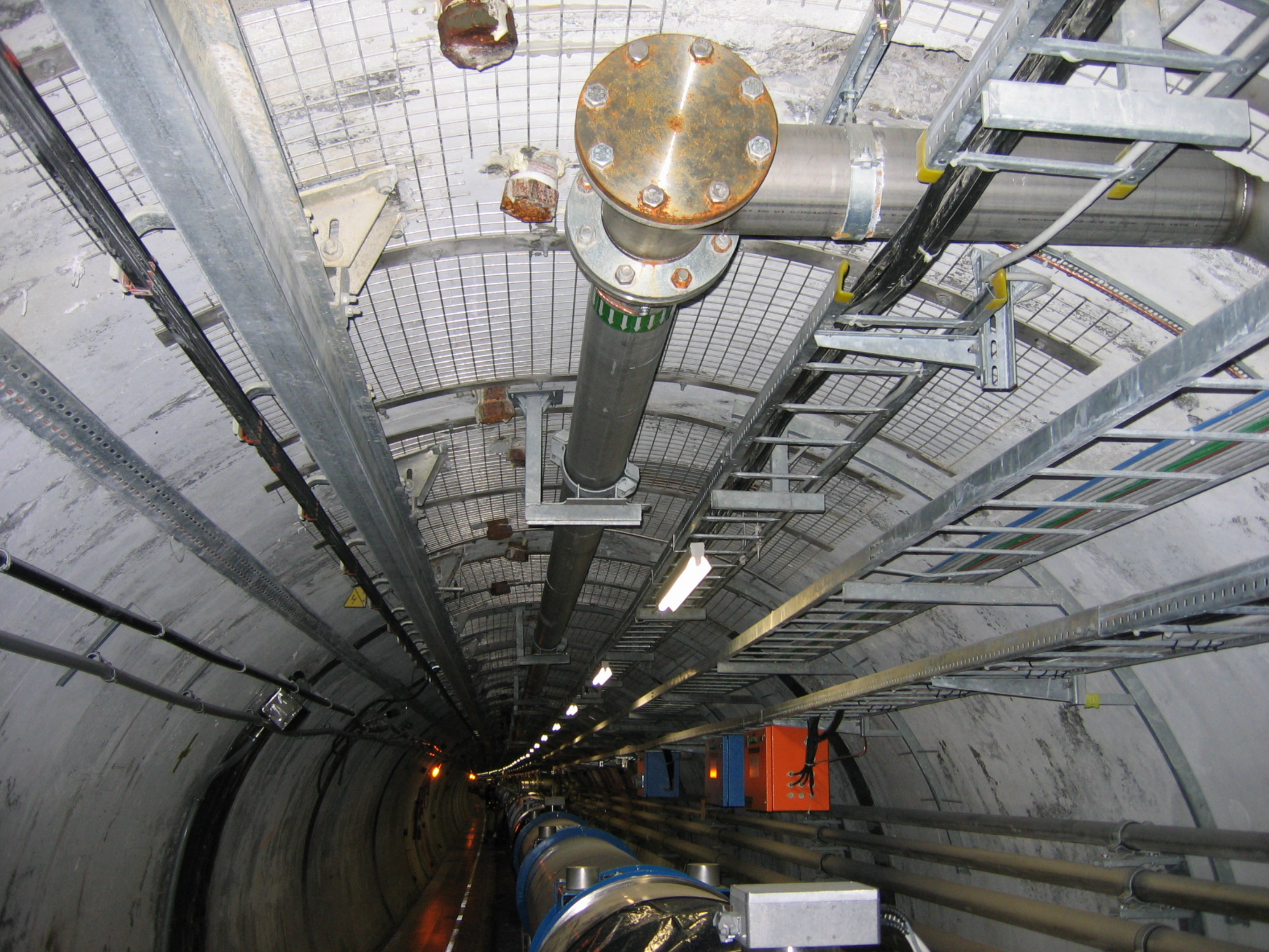}}
  \caption{View of sector 4 showing the chain of superconducting magnets in the arc.}\label{fg:mess_LHCtunnel3_4}
\end{figure}
Figure \ref{fg:mess_LHCtunnel3_4} gives the impression that the tunnel for most of its length is not too occupied.

\begin{figure}

\centerline{  \includegraphics* [clip=,,width=0.8\textwidth]{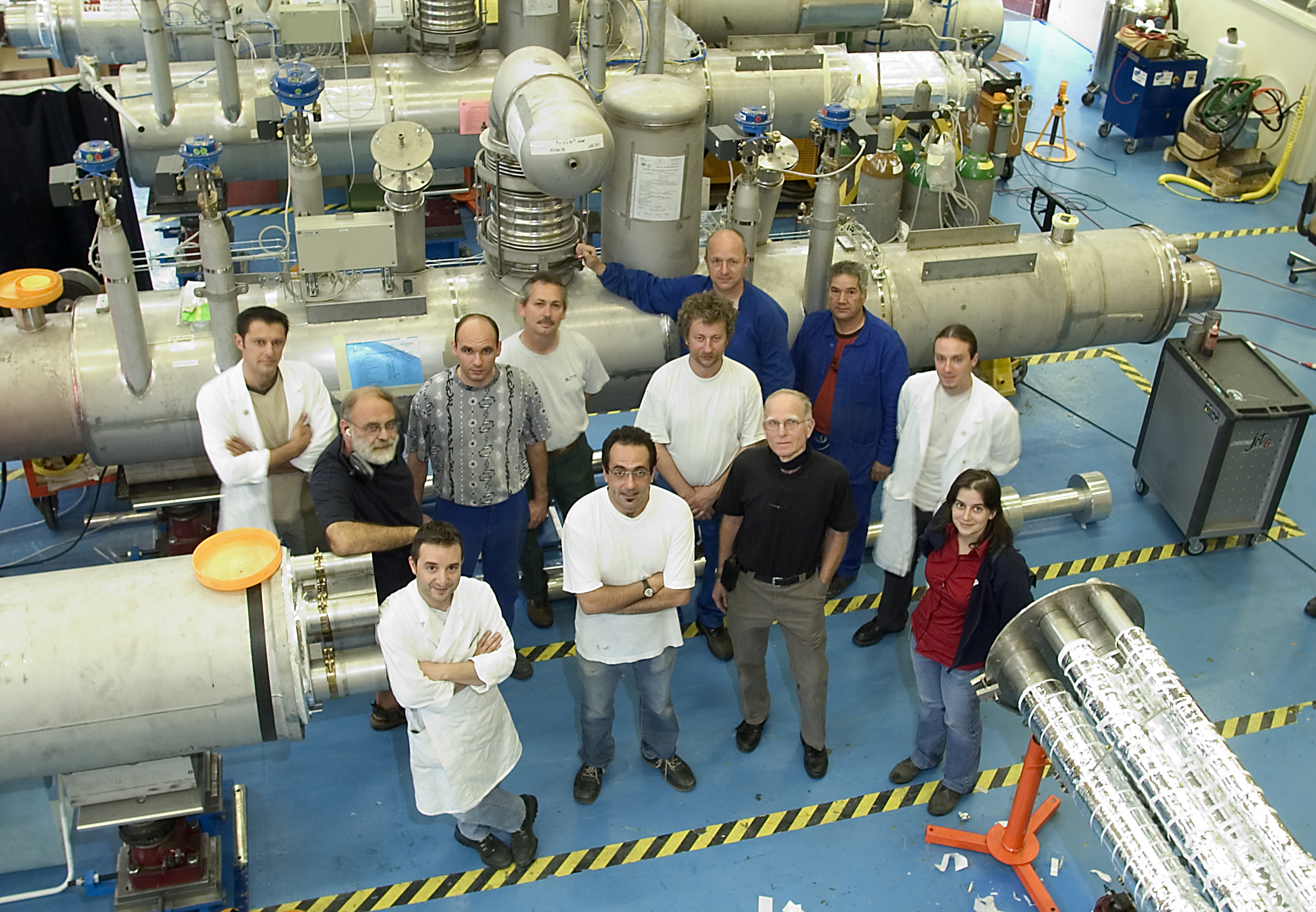}}
  \caption{Sideview of a QRL service module with the jumper that extends vertically above the LHC cryostat and the cryogenic distribution line.}\label{fg:mess_ServiceModul}
\end{figure}

\paragraph {In the arc}
In Fig. \ref{fg:mess_LHCtunnel3_4} one sees the chain of superconducting magnets and in the far distances the \it{QRL Service Module }\rm with its jumper, the cryogenic connection between the superconducting machine and the cryogenic distribution line. The service modules come always at the position of every second quadrupole and have a substantial length. The optics of the LHeC foresees no e-ring magnet at these positions. A photo of service modules in the workshop is shown in figure \ref{fg:mess_ServiceModul} (courtesy CERN) illustrating that the QRL extends substantially in the vertical direction above the LHC arc cryostat and cryo line. 
The picture \ref{fg:mess_LHCtunnel3_4}, taken in sector 3, shows also the critical tunnel condition in this part of the machine. Clearly, heavy loads cannot be suspended from the tunnel ceiling. The limit is set to 100 kg per metre along the tunnel. The e-ring components have to rest on stands from the floor wherever possible. Normally there is enough space between the LHC dipoles and the QRL to place a vertical 10 cm quadratic or rectangular support. Alternatively a steel arch bolted to the tunnel walls and resting on the floor  can support the components from above. This construction is required wherever the space for a stand is not available. 

The electron machine, though partially in the transport zone, will be high up in the tunnel.
The transport of cryogenic equipment may need the full height. Transports of that kind will only happen, when part of the LHC are warmed up. This gives enough time to shift the electron ring to the outside by 30 cm, if the stands are prepared for this operation. The outside movement causes also a small elongation of the inter-magnet connections. This effect is locally so small that the expansion joints, required anyway, can accommodate it. One could even think of moving large sections of the e-machine outwards in a semi-automatic way. Thus the time to clear the transport path can be kept in the shadow of the warm-up and cool-down times.

\paragraph {Dump area} 
The most important space constraints for the electron machine are in the proton dump area, the proton RF cavities, Point 3, and in particular the collimator sections.

\begin{figure} 
\centerline{\includegraphics[clip=,width=0.7\textwidth]{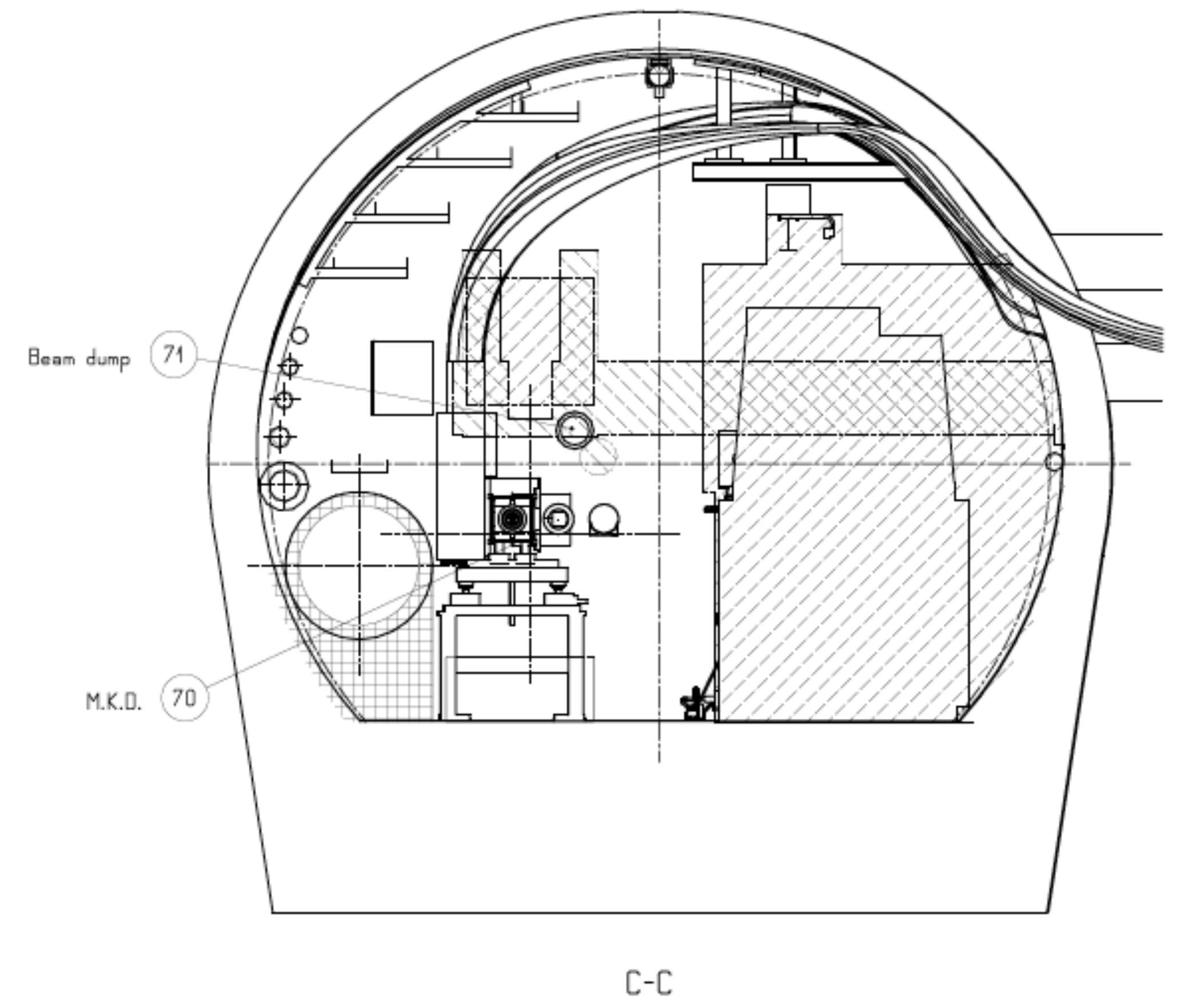}}
\caption{Dump kicker \cite{mess_3_13_c}}\label{fg:mess_RA63_dumpkicker}
\end{figure}

\begin{figure} 
\centerline{\includegraphics[clip=,width=0.8\textwidth]{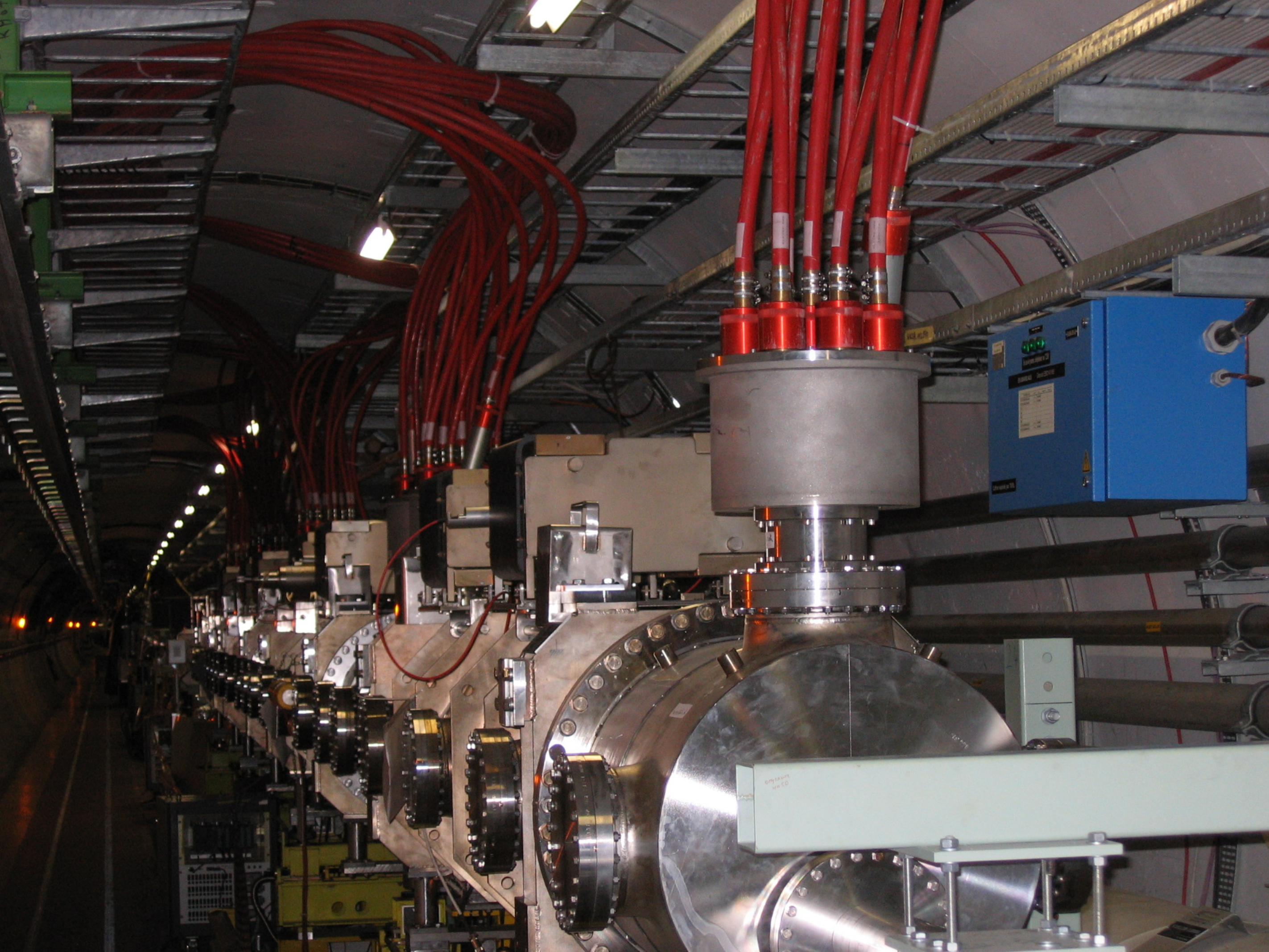}}
\caption{Dump kicker installation in IR6 for one of the two LHC proton rings.}\label{fg:mess_LHC_dump_kicker}
\end{figure}

\begin{figure}
\centerline{\includegraphics[clip=,width=0.8\textwidth]{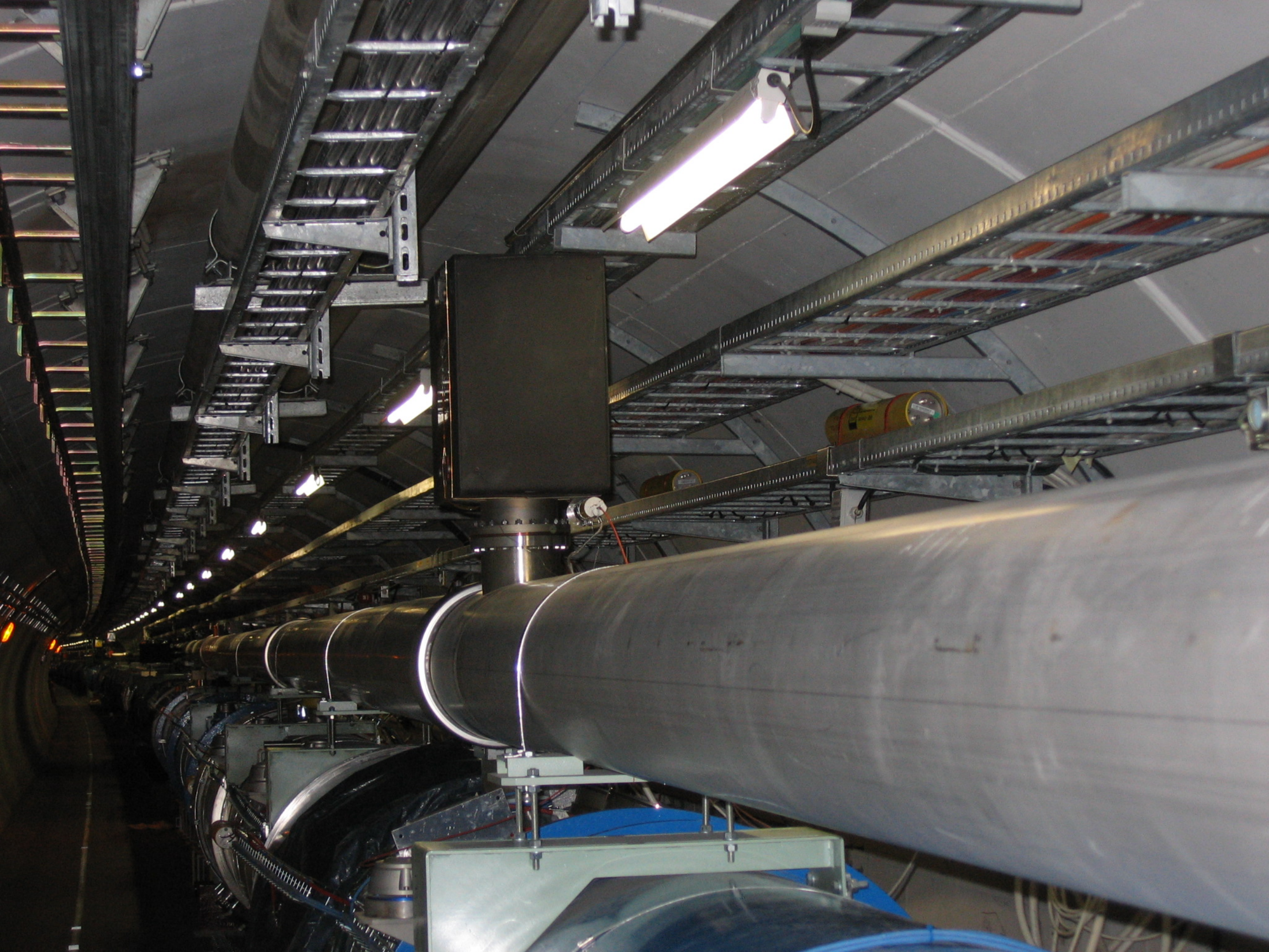}}
\caption{Dump line of one of the LHC proton rings.}\label{fg:mess_LHC_dump}
\end{figure}

Figure \ref{fg:mess_RA63_dumpkicker} \cite{mess_3_13_c} shows the situation at the dump kicker. The same area is also shown in a photo in Figure \ref{fg:mess_LHC_dump_kicker}, while Figure \ref{fg:mess_LHC_dump} shows one of the outgoing dump-lines. The installation of the e-machine requires the proper rerouting of cables (which might be damaged by radiation and in need of exchange anyhow), eventually turning of pumps by 90 degrees or straight sections in the electron optics to bridge particularly difficult stretches with a beam pipe only.


\begin{figure}h
\centerline{\includegraphics[clip=,width=0.8\textwidth]{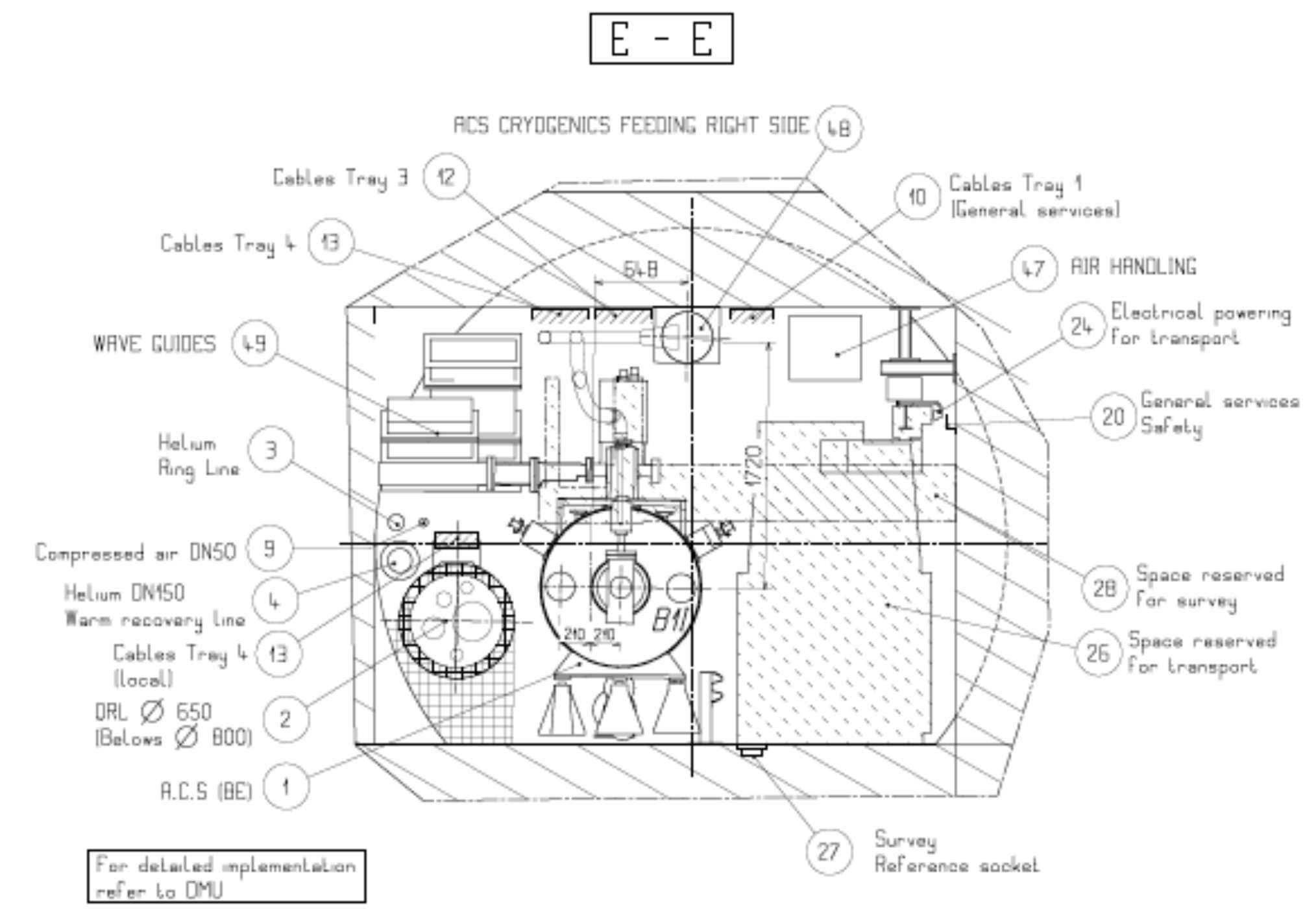}}
\caption{Schematic tunnel cross section with the LHC Proton Proton RF in Point 4 \cite{mess_3_13_d}.}\label{fg:mess_RUX45_pRF}
\end{figure}

\begin{figure*}[h!]
\centerline{\includegraphics[clip=,width=0.8\textwidth]{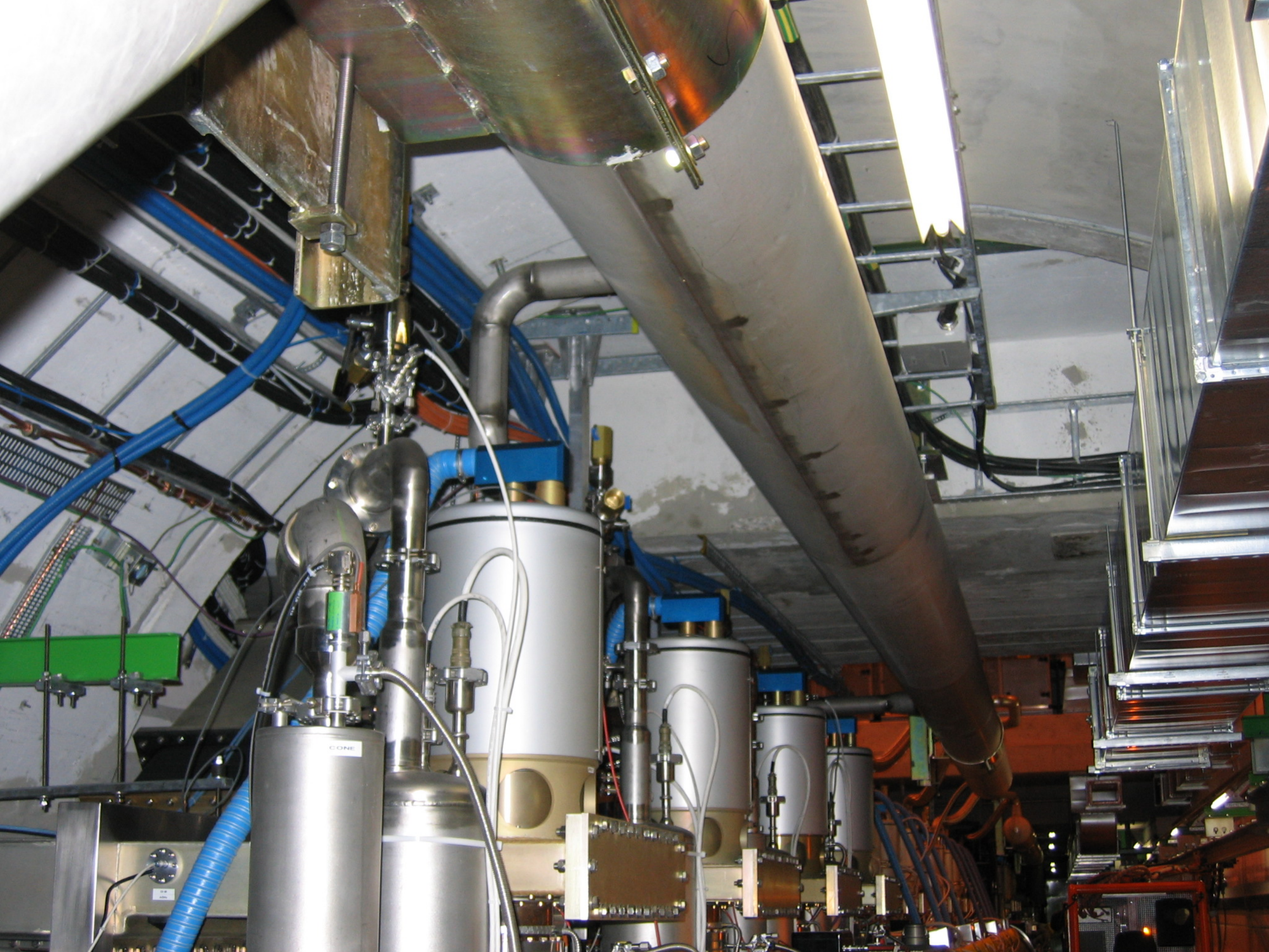}}
\caption{Tight space restriction in Point 4 due to the LHC proton RF installation.}\label{fg:mess_LHC_p_RF_2}
\end{figure*}

\paragraph {Point 4, proton RF}
The Figures \ref{fg:mess_RUX45_pRF} \cite{mess_3_13_d} and \ref {fg:mess_LHC_p_RF_2} illustrate the  situation at the Point 4, where the LHC RF is installed. Fortunately, the area is not very long. A short straight section could be created for the electron ring. This would allow to pass the area with just a shielded beam pipe.

\begin{figure*} [h!]
\centerline{\includegraphics[clip=,width=\textwidth]{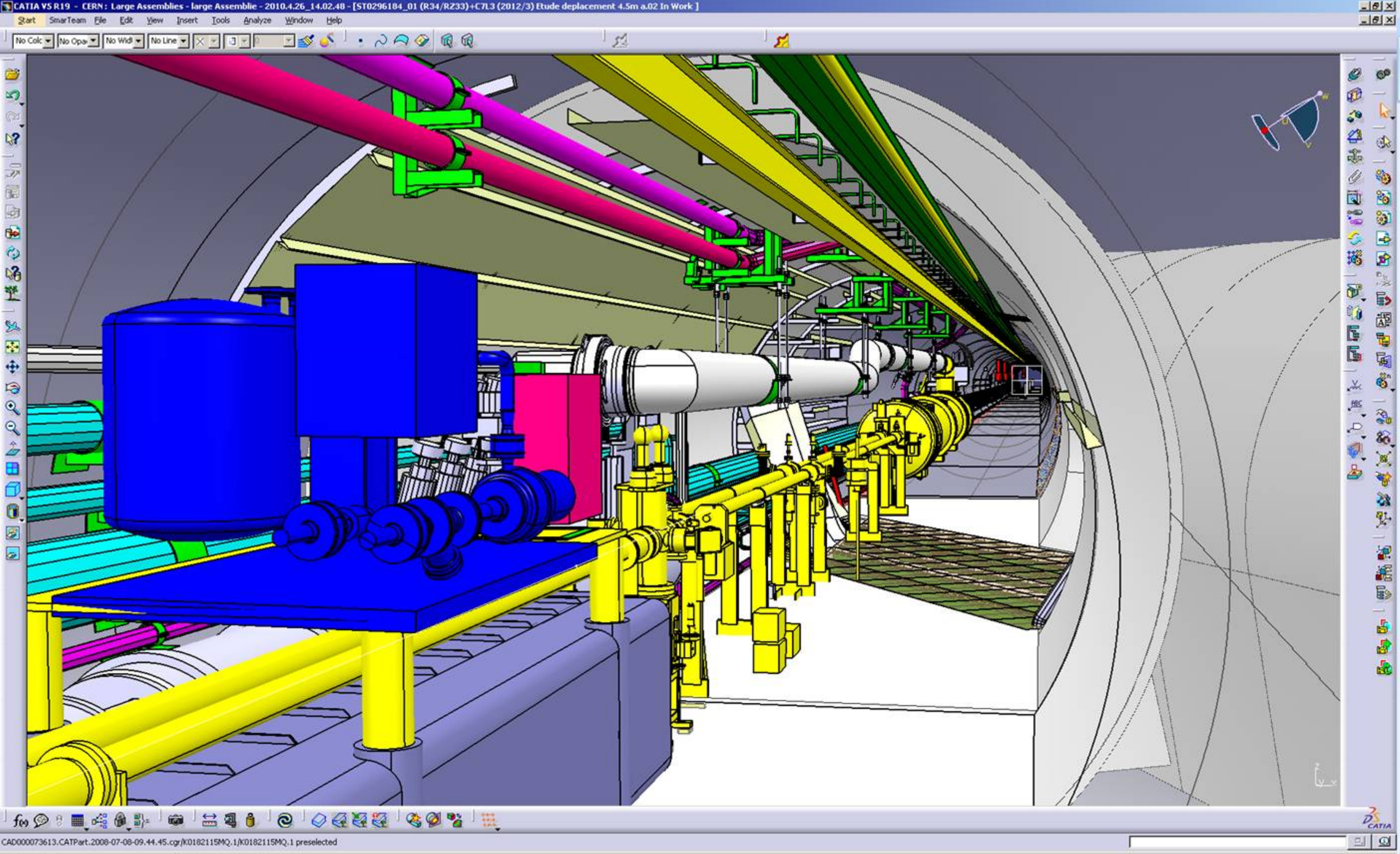}}
\caption{The cryogenic connection in Point 3}\label{fg:mess_LHC_CryLink1}
\end{figure*}

\begin{figure*}[h!]
\centerline{\includegraphics[clip=,width=\textwidth]{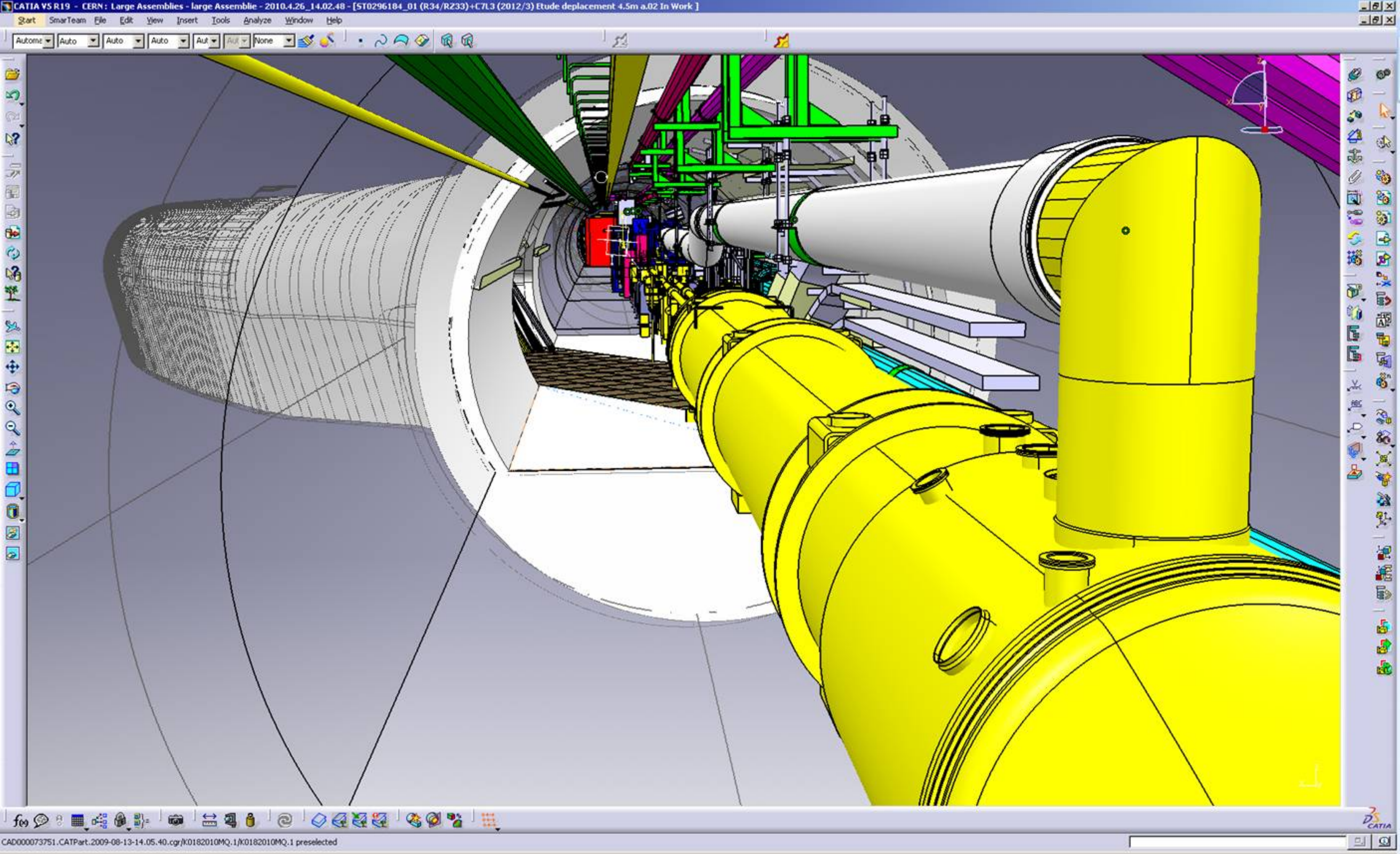}}
\caption{The cryogenic connection in Point 3 (grey tube passing above the two LHC proton beam vacuum tubes [yellow]).}\label{fg:mess_LHC_CryLink2}
\end{figure*}

\begin{figure*}[h!]
\centerline{\includegraphics[clip=,width=\textwidth]{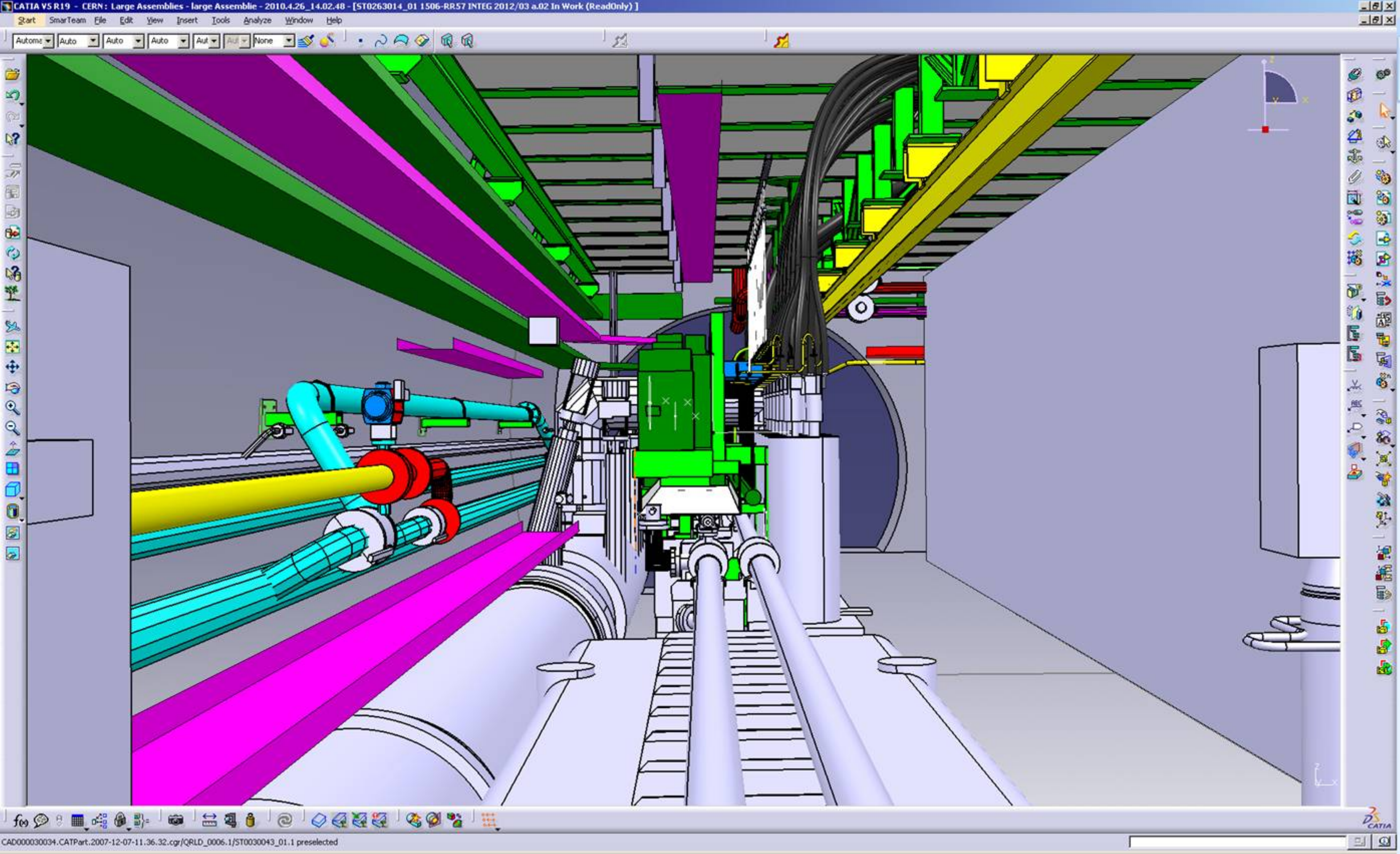}}
\caption{A typical big current feed-box (DFBA) on top of (green) and next to (grey shafts with black power lines) the two proton beam pipes.}\label{fg:mess_LHC_DFBA}
\end{figure*}

\begin{figure}[h!]
\centerline{\includegraphics[clip=,width=0.8\textwidth]{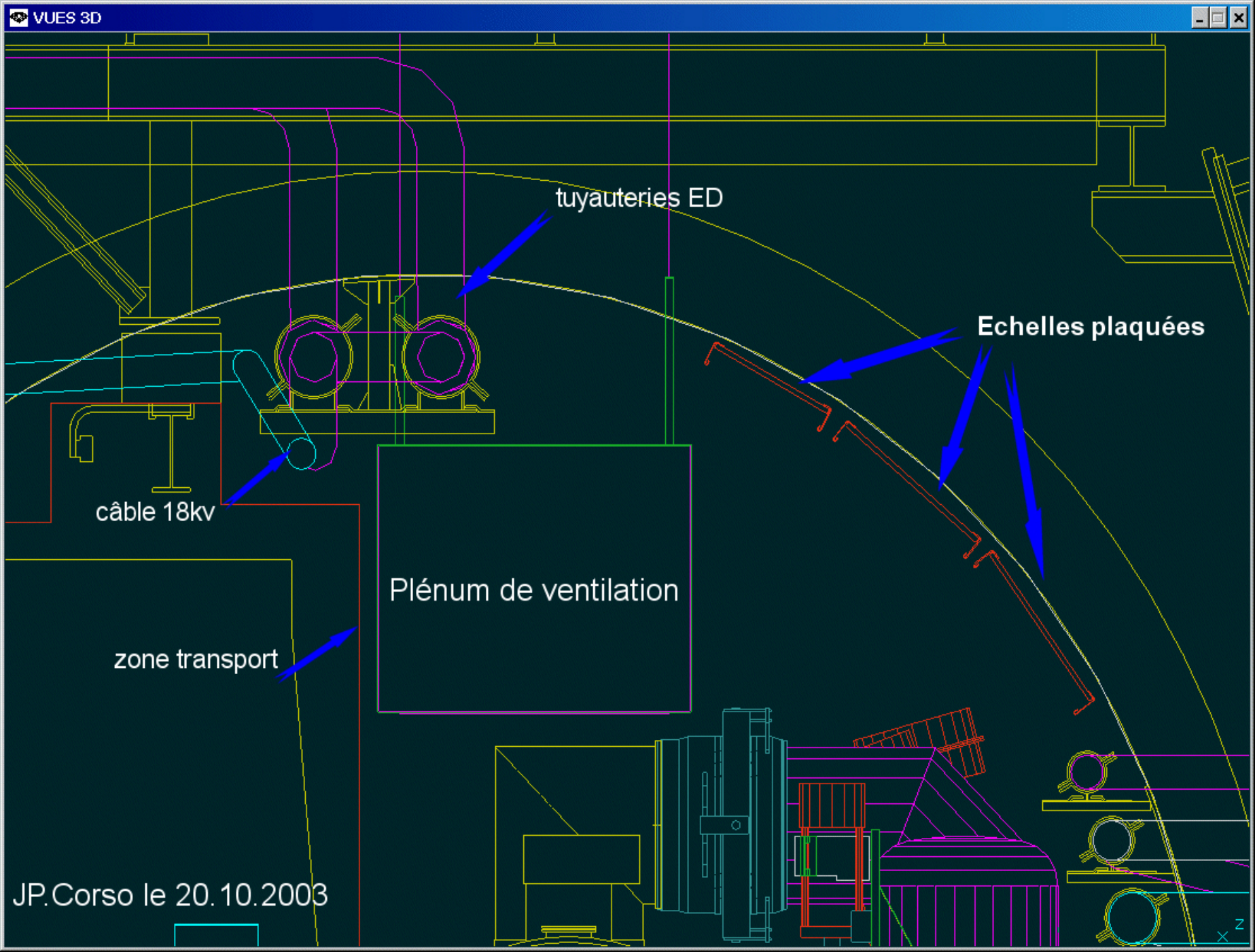}}
\caption{Air-duct in LSS7 indicated by the box labelled 'Plenum de ventilation' \cite{mess_3_13_d}.}\label{fg:mess_LHC_airduct}
\end{figure}

\paragraph {Cryolink in Point 3}
The geography around Point 3 did not permit to place there a cryoplant. The cryogenic cooling for the feedboxes is provided by a cryolink, as is shown in the figures \ref{fg:mess_LHC_CryLink1} and \ref{fg:mess_LHC_CryLink2}. In particular above the Q6 proton quadrupole changes have to be made.
There are other interferences with the cryogenics, as for example at the DFBAs (main feedboxes). An example is shown in figure \ref{fg:mess_LHC_DFBA}. Eventually  the electron optics has to be adapted to allow the beam pipe to pass the cables, which may have to be moved a bit.
\paragraph {Long straight section 7}
An extra air duct is mounted in the long straight section 7  (LSS7) as is indicated in Fig. \ref{fg:mess_LHC_airduct} (labelled Plenum de ventilation) avoiding the air pollution of the area above Point~7. The duct occupies the space planned for the electron machine. The air duct has to be replaced by a slightly different construction mounted further outside (to the right in the figure). There are also air ducts at Points~1 and 5, but they are not an issue. The electron ring is passing behind the experiments in these Points
\paragraph {Proton collimation}
The areas around Point 3 (-62...+177~m) and Point 7 (-149...+205~m) \cite {mess_3_13_coll} are heavily used for the collimation of the proton beam. The high dose rate in the neighbourhood of a collimator makes special precautions for the installation of new components or the exchange of a collimator necessary. Moreover, the collimator installation needs the full height of the tunnel. Hence, the electron ring installation has to be suspended from the re-enforced tunnel roof. The electron machine components must be removable and installable, easy and fast. The re-alignment must be well prepared and fast, possibly in a remote fashion. It is uncommon  to identify fast mounting and demounting as a major issue. However, with sufficient emphasis during the R\&D phase of the project, this problem can be solved.

\subsection{Impact of the synchrotron radiation on tunnel electronics}
\label{sec:SR}
It is assumed that the main power converters of the LHC will have been moved out of the RRs because of the single event upsets, caused by proton losses.

The synchrotron radiation has to be intercepted at the source, as in all other electron accelerators. A few millimetre of lead are sufficient for the relatively low (critical) energies around 100 to 200 keV. The K-edge of lead is at 88 keV, the absorption coefficient is above 80/cm at this energy \cite {mess_3_13_att}. One centimetre of lead is sufficient to suppress 300 keV photons by a factor of 100. Detailed calculations of the optics will determine the amount of lead needed in the various places. The primary shielding needs an effective water cooling to avoid partial melting of the lead. 

The electronics is placed below the proton magnets. Only backscattered photons with correspondingly lower energy will reach the electronics. If necessary, a few millimetre of extra shielding could be added here.

The risk for additional single event upsets due to synchrotron radiation is negligible.

\subsection{Compatibility with the proton beam loss system}
\label{sec:BLM}
The proton beam loss monitoring system works very satisfactory. It has been designed to detect proton losses by observing secondaries at the outside of the LHC magnets.  The sensors are ionisation chambers.
Excessive synchrotron radiation (SR) background will presumably trigger the system and dump the proton beam. The SR background at the monitors has to be reduced by careful shielding of either the monitors or the electron ring. Alternatively, the impact of the photon background can be reduced by using a new loss monitoring system which is based on coincidences (as was done elsewhere \cite{Wittenburg:2000rp}).

\subsection{Space requirements for the electron dump}
\label{sec:e-dump}
The electron beam of the LHeC installation requires a dedicated dump section. Potential interference of the losses during or after an electron beam dump with equipment of the LHC proton rings still needs to be studied and a suitable space still needs to be found in the LHC tunnel.

\subsection{Protection of the p-machine against heavy electron losses}
\label{sec:e-loss}
The existing proton loss detectors are placed, as mentioned above, at the LHC magnets. The trigger threshold requires certain number of detectors to be hit by a certain number of particles. The assumption is that the particles come from the inside of the magnets and the particle density there is much higher. Electron losses, creating a similar pattern in the proton loss detectors will result in a much lower particle density in the superconducting coils. Hence, still tolerable electron losses will unnecessarily trigger the proton loss system and dump the proton beam. 
The proton losses are kept at a low level by installing an advanced system of collimators and masks. Fast changes of magnet currents, which will result in a beam loss, are detected. A similar system is required for the electrons.  An electron loss detection system, like the one mentioned in Ref.\cite{Wittenburg:2000rp}, combined with the proton loss system can be used to identify the source of the observed loss pattern and to minimise the electron losses by improved operation. It seems very optimistic to think of a hardware discrimination system, which determines very fast the source of the loss and acts correspondingly. Such a system could be envisaged only after several years of running.   
\subsection{How to combine the machine protection of both rings?}
\label{sec:Machine-Protection}
The existing machine-protection system combines many different subsystems. The proton loss system, the quench detection system, cryogenics, vacuum, access, and many other subsystems may signal a dangerous situation. This requirement lead to a very modular architecture, which could be expanded to include the electron accelerator.

%% file: machine/Burkhardt.tex
\section{LHeC injector for the Ring-Ring option}
\subsection{Injector}

The LEP pre-injectors have been dismantled and the infrastructure re-used
for the CLIC test facility CTF3.
The RF cavities that accelerated leptons in
the SPS have been removed to reduce its impedance.
Re-installation of an injector chain similar to LEP's through the PS and
SPS would be costly and potentially limit the proton performance.

The LHeC e-ring therefore requires new lepton injectors.

In the 30 years from the design of the LEP injectors, there has been substantial progress in accelerator technology. This is particularly true in the field of superconducting radio frequency technology which was very successfully used for LEP2 on a large scale and which has been further developed for TESLA and the ILC. It makes it feasible to design a very compact and efficient 10\,GeV injector based on the principle of a recirculating LINAC and to take advantage of the studies for ELFE at CERN\,\cite{Aulenbacher:1999hu}.

\subsection{Required performance}

The main requirements for the LHeC ring-ring electron and positron injectors are summarised in Table\,\ref{tab:RReinjParm}.

\begin{table}[htdp]
\begin{center}
\begin{tabular}{|c|c|}
\hline
particle types   & e$^+$, e$^-$ \\\hline
polarised        & no \\\hline
injection energy & $E_b = \,10\,{\rm GeV}$ \\\hline
bunch intensity  & $2\times10^{10}\,e = 3.2\,{\rm nC}$ \\\hline
pulse frequency  & $\geq 5\,/{\rm s}$ \\\hline
\end{tabular}
\end{center}
\caption{Main parameters for the LHeC RR injector}
\label{tab:RReinjParm}
\end{table}%

Polarisation is not required from the ring injectors. It would be very difficult to maintain the polarisation during the acceleration in the main ring.
Instead, polarisation can be built up at top energy from synchrotron radiation.

The electron bunch intensity for nominal LHeC performance is $1.4\times10^{10}$.
The target intensity for the injector is taken as $2\times10^{10}$ which includes a safety factor and allows for losses at injection and during the ramp.
Higher single-bunch intensities may be useful, with a smaller number of bunches, for the e-A mode of operation.
LEP was operated with much higher bunch intensities up to $4\times 10^{11}$ limited by the transverse mode coupling instability (TMCI).
The TMCI threshold current can be estimated from\,\cite{Brandt:1994ze}
\begin{equation}
I_{th} = \frac{\omega_s E}{e  \sum  \beta \; k_\perp (\sigma_s)}\,
\label{eq:kperp}
\end{equation}
where $\omega_s = 2 \pi Q_s f_{\rm rev}$ is the synchrotron frequency, $e$ the elementary charge, $E$ is the beam energy, $\beta$ the
beta function value at the location of the impedance and $k_\perp$ the loss factor which accounts for the transverse impedance of the machine.
LEP had a design injection energy of 20\,GeV. It was raised to 22\,GeV to increase the TMCI threshold.

The relatively low bunch intensity required for the LHeC allows for direct injection without accumulation and for a lower injection energy compared to LEP.
The LHeC transverse impedance will be similar to LEP, with a smaller contribution from the reduced number of cavities and an increased impedance contribution from the more compact beam-pipe cross section. Lowering the beam energy results in weak bending fields and loss of synchrotron radiation damping. A beam energy of a few GeV may still be tolerable for transverse mode coupling but would not be practical for magnet stability and require strong wigglers to get a significant radiation damping (otherwise this requires a minimum beam energy of the order of 10\,GeV).

A pulse frequency of on average 5\,Hz is required, to fill the LHeC electron ring with 2808 bunches in 10\,minutes.

The injector requirements summarised in Table\,\ref{tab:RReinjParm} are within the reach of proven technology and concepts.
An example is the FACET facility at SLAC which provides $2\times10^{10}$ electrons of 23\,GeV energy at 30\,Hz repetition frequency\,\cite{FACET-PARM-2010}.

The intensities and repetition frequency required here match well with the performance of the LIL, the first part of the LEP pre-injectors, which we reconsider here for the source, positron accumulation and pre-acceleration to 0.6\,GeV.
For the acceleration to 10\,GeV we propose a new, superconducting recirculating LINAC.

\subsection{Source, accumulator and acceleration to 0.6\,GeV}

\begin{figure}
\begin{center}
\includegraphics[width=1.0\columnwidth]{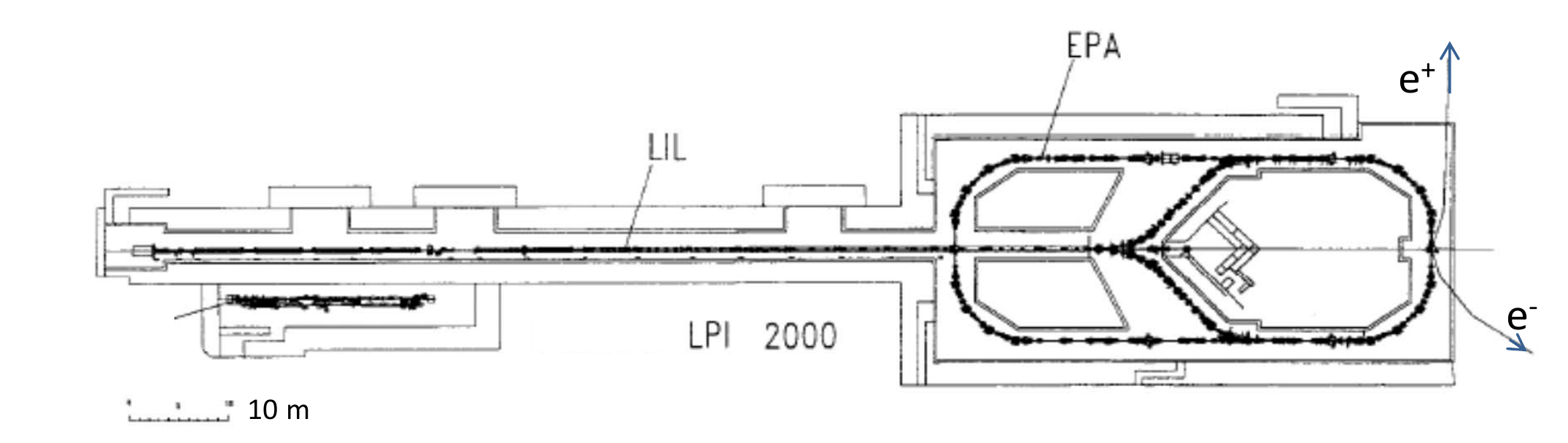}
\caption{Layout of the LPI in 2000.}
\label{Fig:LHeCInjector5}
\end{center}
\end{figure}

Figure $\ref{Fig:LHeCInjector5}$ shows the layout of the LPI (LEP Pre-Injector) as it was working in 2000.
The  LPI was composed of the LIL (LEP Injector Linac) and the EPA (Electron Positron Accumulator).

Table $\ref{tab:rinolfi6}$ gives the beam characteristics at the end of LIL.

\begin{table}[h]
\centering
\begin{tabular}{|l|l|}
\hline
Beam energy & $200$ to $700$ MeV \\ \hline
Charge & $5 \times 10^8$ to $2 \times 10^{10} e^-$ / pulse \\ \hline
Pulse length & $10$ to $40$ ns (FWHM) \\ \hline
Repetition frequency & $1$ to $100$ Hz \\ \hline
Beam sizes (rms) & $3$ mm \\ \hline
\end{tabular}
\caption{LIL beam parameters.}
\label{tab:rinolfi6}
\end{table}

Table $\ref{tab:rinolfi7}$ gives the electron and positron beam parameters at the exit of EPA.

\begin{table}[h]
\centering
\begin{tabular}{|l|l|}
\hline
Energy & $200$ to $600$ MeV \\ \hline
Charge & up to $4.5 \times 10^{11} e{\pm}$ \\ \hline
Intensity & up to $0.172$ A \\ \hline
Number of bunches & $1$ to $8$ \\ \hline
Emittance & $0.1$ mm.mrad \\ \hline
Tune & $Q_x = 4.537, Q_y = 4.298$ \\ \hline
\end{tabular}
\caption{The electron and positron beam parameters at the exit of EPA.}
\label{tab:rinolfi7}
\end{table}

With 8 bunches in the EPA for a 1.14\,s cycle, the 2808 electron bunches required for the LHeC could be filled in 6.7\,min which is perfectly adequate.
According to the original LEP injector design report \,\cite{GreenBooksLEP1,GreenBooksLEP2,GreenBooksLEP3} Vol.I, the cycle length for positrons is 11.22\,s which would allow the 2808 bunches to be filled in 66\,minutes. We conclude that the LIL+EPA performance is fully adequate for the LHeC. A reduction of the cycle length for positrons would be useful to reduce the filling time.

\begin{figure}[htpb] 
 \center{\includegraphics[width=0.6\textwidth]{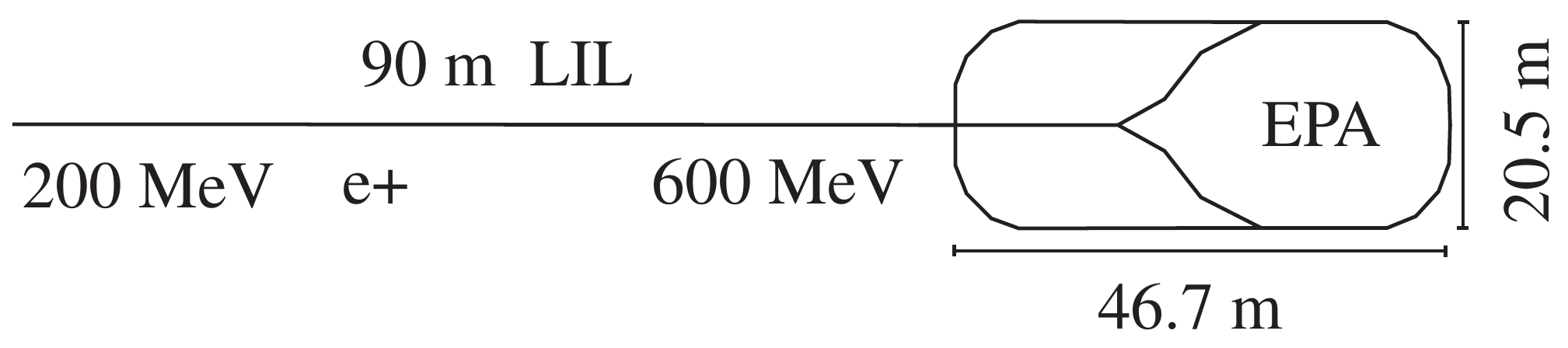}}
  \vspace{-2mm}\caption{LIL and EPA}
  \label{plot:EPA_layout}
\end{figure}

\subsubsection*{Timing considerations}

EPA was planned for 1 to 8\,bunches compatible with the LEP RF-frequency. The EPA circumference of 125.665\,m corresponds to $t_{\rm rev}= 419.173\,{\rm ns}$, which is $16.75\times 25\,{\rm ns}$ and would in theory allow for 16\,bunches spaced by 25\,ns as relevant for the LHeC. Injection in batches of 72\,bunches as possible for protons into the LHC would require
a five times larger damping ring which would be rather expensive.

EPA had an RF-frequency $f_{\rm RF} = 19.0852\,{\rm MHz}$. It will be increased to 40\,MHz to allow for a bunch spacing of 25\,ns.
For the injection into the LHC we propose a fast kicker system with a kicker rise-time below 25\,ns. 
This conserves the dimensions of EPA and gives full flexibility to place the bunches into the LHeC electron ring as required to collide with the proton or ion bunches\,\cite{Bruning:2004ej,Benedikt:2004wm}.

\subsection{10\,GeV injector}

For the acceleration to 10\,GeV we propose a re-circulating LINAC, designed as a downscaled, low energy version of the 25\,GeV ELFE at CERN design \,\cite{Aulenbacher:1999hu} using modern ILC-type RF-technology.

\begin{figure}[htpb] 
 \center{\includegraphics[width=0.9\textwidth]{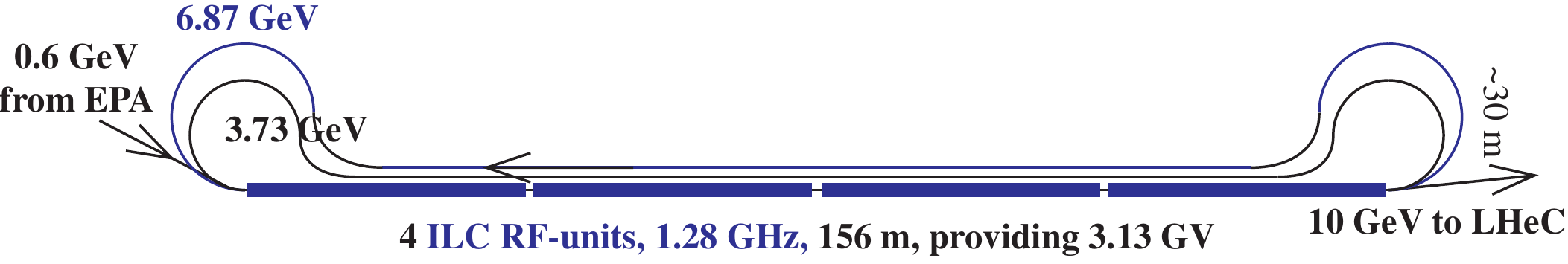} }
  \vspace{-2mm}\caption{Recirculator using 4 ILC modules.}
  \label{plot:Recirculator4ILCmodules}
\end{figure}

A sketch of the proposed machine is shown in Fig.\,\ref{plot:Recirculator4ILCmodules}.
The acceleration is provided by 4 RF-units of the ILC type, providing together 3.13\,GV acceleration.

The acceleration from 0.6\,GeV to 10\,GeV is achieved in three passages through the LINAC.
This requires only two re-circulation arcs which can be constructed in the horizontal plane.
The maximum energy in the last re-circulation arc is $10 - 3.13 = 6.87\,{\rm GeV}$.

For a beam energy $E$ and bending radius $\rho$, the energy loss $U_0$ by synchrotron radiation in the  single passage through a re-circulation arc is
\begin{equation}
U_0 = C_\gamma \frac{E^4}{\rho}
\end{equation}
where
\[ C_\gamma = \frac{e^2}{3\epsilon_0} \frac{1}{(m c^2)^4}
= 8.846\times 10^{-5} \mbox{ m {\rm GeV}}^{-3}\;.
\]
where $e$ is the elementary charge and $m$ the electron mass.
The relative energy spread is increased by the synchrotron radiation in a single passage by
\begin{equation}
\sigma_{e} = r_{\rm e} \, c_f \, \frac{\gamma^{5/2}}{\rho}
\end{equation}
where $r_e$ is the classical electron radius and
\begin{equation}
c_f = \frac{3}{2} \sqrt{\frac{55\pi}{27 \sqrt{3}\,\alpha }} = 33.75\,.
\end{equation}
A bending radius of $\rho=2\,{\rm m}$ at $E = 6.87\,{\rm GeV}$ would result in an energy loss by recirculation of $U_0 = 98\,{\rm MeV}$ 
and an energy spread of $10^{-3}$. This would both be tolerable, but require very strong superconducting 11\,tesla magnets for the 6.87\,GeV recirculation.

At this stage, we propose the use of warm 2\,tesla magnets, resulting in a bending radius of $\rho = 11.5\,{\rm m}$ for the $6.87\,{\rm GeV}$
recirculation and $\rho = 6.2\,{\rm m}$ for the 3.73\,GeV recirculation. The values for the energy loss and spread are listed in Table\,\ref{tab:Recirculator}.

\begin{table}[htdp]
\begin{center}
\begin{tabular}{|c|c|c|c|c|}
\hline
$E$ [GeV]  & $B$  [T]  & $\rho$ [m] & $U_0$ [MeV]& $\sigma_e$ \\ \hline \hline
6.87  & 2      & 11.45  & 17.1  & $1.7\times10^{-4}$ \\  \hline
3.73  & 2      &  6.23  &  2.8  & $7\times10^{-5}$ \\  \hline
\end{tabular}
\end{center}
\caption{Energy, bending field and radius, energy loss and energy spread in the recirculator magnets.}
\label{tab:Recirculator}
\end{table}%

To save space and allow for a single LINAC tunnel, we propose a dogbone-like shape for the recirculators as shown in Fig.\,\ref{plot:Recirculator4ILCmodules}.

%% file: machine/linac.tex
\chapter{Linac-Ring Collider}
\input{machine/zimmermann}
\input{machine/LR-IR}

\input{machine/bernard}

\input{machine/schulte}

\input{machine/eALRJowett}
\input{machine/rinolfilrem}

\input{machine/SpinRotator}

\input{machine/positrons}

%% file: machine/zimmermann.tex
\section{Basic parameters and configurations}

\subsection{General considerations}
A high-energy electron-proton collider can be realised by accelerating electrons (or positrons) in a linear accelerator (linac)
to 60--140 GeV and colliding them with the 7-TeV protons circulating in the LHC. 
Except for the collision point and the surrounding interaction region, the tunnel and the infrastructure for such a linac 
are separate and fully decoupled from the LHC operation, from the LHC maintenance work, 
and from other LHC upgrades (e.g., HL-LHC and HE-LHC). 

The technical developments required for this type of collider can both benefit from and be used for many future projects. 
In particular, to deliver a long or continuous beam pulse, as required for high luminosity, 
the linac must be based on superconducting (SC) radio frequency (RF) technology. 
The development and industrial production of its components can exploit synergies with numerous other advancing 
SC-RF projects around the world, such as the European XFEL at DESY, eRHIC, ESS, ILC, CEBAF 
upgrade, CESR-ERL, JLAMP, and the 
CERN HP-SPL.

For high luminosity operation at a beam energy of 50--70 GeV the linac should be operated in continuous wave (CW) mode, 
which restricts the maximum RF gradient through the associated cryogenics power, to a value of about 20 MV/m or less. 
In order to limit the active length of such a linac and to keep its construction and operating costs low, the linac should, 
and can, be recirculating. For the sake of energy efficiency and to limit the overall site power, while boosting the luminosity, 
the SC recirculating CW linac can be operated in energy-recovery (ER) mode.

Electron-beam energies higher than 70 GeV, e.g.~140 GeV, can be achieved by a pulsed SC linac, similar to the XFEL, ILC or SPL. 
In this case the accelerating gradient can be larger than for CW operation, i.e.~above 30 MV/m, which minimises the total length, 
but recirculation is no longer possible at this beam energy 
due to prohibitively high synchrotron-radiation energy losses in any return arc of reasonable dimension. 
As a consequence the standard energy recovery scheme using recirculation cannot be implemented and the luminosity of such a higher-energy lepton-hadron 
collider would be more than an order of magnitude lower than the one of the lower-energy CW ERL machine, at the same wall-plug power. 

For a linac it is straightforward to deliver a 80--90\% polarised electron beam.

The production of a sufficient number of positrons to deliver positron-proton collisions at a similar 
luminosity as for electron-proton collisions is challenging for a linac-ring collider\footnote{A review of 
linac-ring type collider proposals can be found in Ref.~\cite{akay}.}
A conceivable path towards decent proton-positron luminosities would include a recycling of the spent positrons, 
together with the recovery of their energy. 

The development of a CW SC recirculating energy-recovery linac (ERL) for LHeC would prepare the ground, 
the technology and the infrastructure for many possible future projects, e.g., for an International Linear Collider, 
for a Muon Collider\footnote{The proposed Muon Collider heavily relies 
on SC recirculating linacs for muon acceleration as well as on a SC-linac 
proton driver.}, for a neutrino factory, or for a proton-driven plasma wake field accelerator. 
A ring-linac LHeC would, therefore, promote any conceivable future high-energy physics project, 
while pursuing an attractive forefront high-energy physics programme in its own right.

\subsection{ERL performance and layout}
Particle physics imposes the following performance requirements. 
The lepton beam energy should be 60 GeV or higher and the electron-proton luminosity of order $10^{33}$~cm$^{-2}$s$^{-1}$. 
Positron-proton collisions are also required, with at least a few percent of the electron-proton luminosity.
Since the LHeC should operate simultaneously with LHC $pp$ physics, it should not degrade the $pp$ luminosity. 
Both electron and positron beams should be polarised.
Lastly, the detector acceptance should extend down to 1$^{\circ}$ or less.
In addition, the total electrical power for the lepton branch of the LHeC collider should stay below 100 MW. 

For round-beam collisions, the luminosity of the linac-ring collider \cite{pgw} is written as  
\begin{equation}
L = \frac{1}{4 \pi e} \frac{N_{b,p}}{\epsilon_{p}} \frac{1}{\beta^{\ast}_{p}} I_{e} H_{hg} H_{D}\; ,
\label{lumi}
\end{equation}
where $e$ denotes the electron charge, $N_{b,p}$ the proton bunch population, $\beta_{p}^{\ast}$
the proton IP beta function, $I_{e}$ the average electron beam current, $H_{hg}$ the
geometric loss factor arising from crossing angle and hourglass effect, and $H_{D}$ the disruption 
enhancement factor due to the electron pinch in collision, or luminosity reduction factor from the anti-pinch 
in the case of positrons.   
In the above formula, it is assumed that the electron bunch spacing is a multiple of the proton beam bunch spacing.
The latter could be equal to 25, 50 or 75 ns, without changing the luminosity value.

The ratio $N_{b,p}/\epsilon_{p}$ is also called the proton beam brightness.
Among other constraints, the LHC beam brightness is limited by the proton-proton beam-beam limit. 
For the LHeC design we assume the brightness value obtained for the ultimate bunch intensity,
$N_{p,p}=1.7\times 10^{11}$, and the nominal proton beam emittance, $\epsilon_{p}=0.5$~nm 
($\gamma \epsilon_{p}=3.75$~$\mu$m). This corresponds to a total $pp$ 
beam-beam tune shift of 0.01. More than two times higher values have already been
demonstrated, with good $pp$ luminosity lifetime, during initial LHC beam commissioning,
indicating a potential for higher $ep$ luminosity.

To maximise the luminosity the proton IP beta function is chosen as $0.1$~m.
This is considerably smaller than the 0.55 m for the $pp$ collisions of the 
nominal LHC.
The reduced beta function can be achieved by reducing the free length between
the IP and the  first proton quadrupole (10 m instead of 23 m), and 
by squeezing only one of the two proton beams, namely the one colliding with the leptons, 
which increases the aperture available for this beam in the last quadrupoles.
In addition, we assume that the final quadrupoles could be based on 
Nb$_{3}$Sn superconductor technology instead of Nb-Ti.
The critical field for Nb$_{3}$Sn is almost two times 
higher than for Nb-Ti, at the same temperature and current density, 
allowing for correspondingly larger aperture and higher quadrupole gradient. 
Nb$_{3}$Sn quadrupoles are presently under development 
for the High-Luminosity LHC upgrade (HL-LHC).

The geometric loss factor $H_{hg}$ needs to be optimised as well.
For round beams with $\sigma_{z,p}\gg \sigma_{z,e}$ (well fulfilled for 
$\sigma_{z,p}\approx 7.55$~cm, $\sigma_{z,e}\approx 300$~$\mu$m)
and $\theta_{c}\ll 1$,  it can be expressed as\footnote{The derivation of this formula 
is similar to the one for the LHC in Ref.~\protect\cite{frfz}, with the difference that here 
the two beams have different emittances and IP beta functions, and the electron bunch length is neglected.
Curves obtained with formula (\ref{hhg}) were first reported in \protect\cite{epac08}.} 
\begin{equation}
H_{hg} = \frac{\sqrt{\pi} z e^{z^{2}} {\rm erfc} (z)}{S}\; ,
\label{hhg} 
\end{equation}
where 
\begin{eqnarray}
z & \equiv & 2 \frac{(\beta^{\ast}_{e}/\sigma_{z,p})(\epsilon_{e}/\epsilon_{p})}{\sqrt{1+(\epsilon_{e}/\epsilon_{p})^{2}}} S  \nonumber \nonumber 
\end{eqnarray}
and 
\begin{eqnarray}
S & \equiv & \sqrt{1+\frac{\sigma_{x,p}^{2} \theta_{c}^{2}}{8\sigma_{p}^{\ast\; 2}}}\; . \nonumber
\end{eqnarray}
Luminosity loss from a crossing angle is avoided by head-on collisions.
The luminosity loss from the hourglass effect, due to the long proton bunches and 
potentially small electron beta functions, is kept small, thanks to a ``small'' linac electron beam emittance
of 0.43 nm ($\gamma \epsilon_{e}=50$~$\mu$m). 
We note that the assumed electron-beam emittance, though small when compared with a storage ring of comparable energy, 
is still very large by linear-collider standards. 

The disruption enhancement factor for electron-proton collisions is about $H_{D}\approx 1.35$,   
according to Guinea-Pig simulations \cite{schultechav} and a simple estimate based on
the fact that the average rms size of the electron beam during the collision approaches a value 
equal to $1/\sqrt{2}$ of the proton beam size.
This additional luminosity increase from disruption is not taken into account in the numbers given below. 
On the other hand, for positron-proton collisions the disruption of the positrons leads to a significant
luminosity reduction, by roughly a factor $H_{D}\approx 0.3$, 
similar to the case of electron-electron collisions \cite{ee}.

The final parameter determining the luminosity is the
average electron (or positron) beam current $I_{e}$.
It is closely tied to the total electrical power available (taken to be 100 MW).

\subsubsection{Crossing angle and IR layout}
\begin{figure}
\centerline{\includegraphics[angle=0,clip=,width=0.7\textwidth]{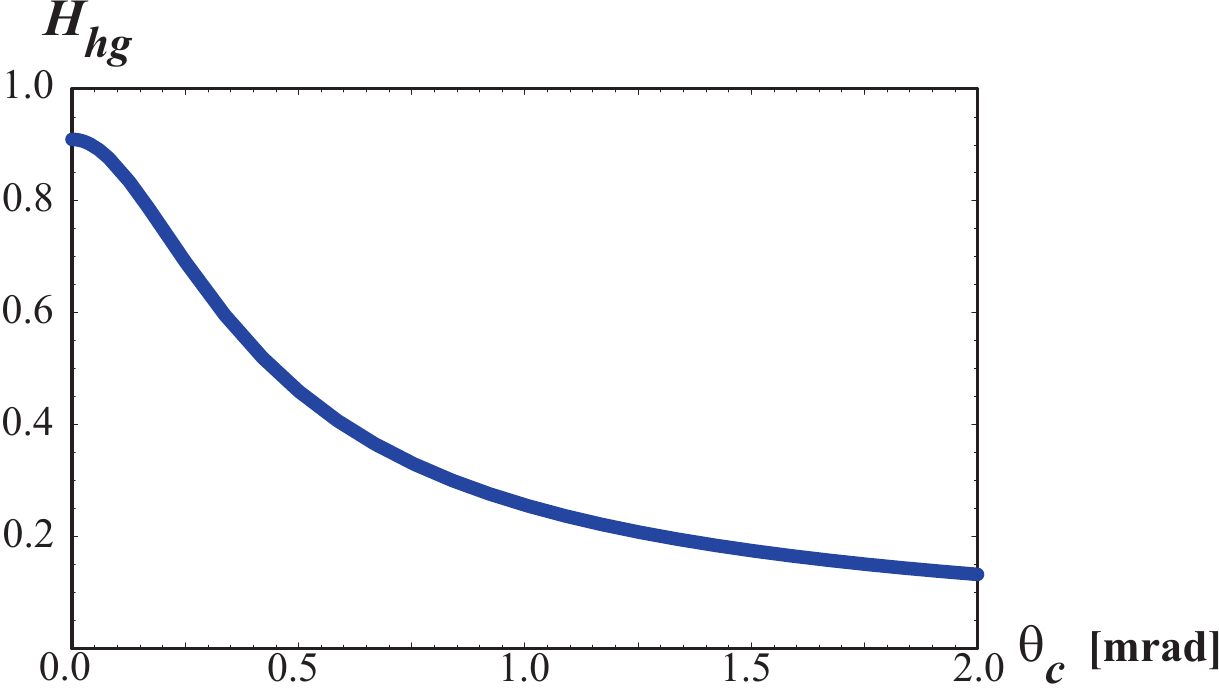}}
\caption{Geometric luminosity loss factor $H_{hg}$, (\protect\ref{hhg}), 
as a function of the total crossing angle}
\label{cross} 
\end{figure}

The colliding electron and proton beams need to be separated
by 7 cm at a distance of 10 m from the IP in order to enter through
separate holes in the first proton quadrupole magnet.
This separation could be achieved with a crossing angle of 7 mrad and crab cavities.
The required crab voltage would, however, need to be of order 200 MV, which is 
20--30 times the voltage needed for $pp$ crab crossing at the HL-LHC. 
Therefore, crab crossing is not considered an option for the L-R LHeC.
Without crab cavities, any crossing angle should be smaller than 0.3 mrad,  
as is illustrated in Fig.~\ref{cross}.
Such small a crossing angle is not useful,  
compared with the 7 mrad angle required for the separation.
The R-L interaction region (IR), therefore, uses detector-integrated dipole fields around
the collision point, to provide head-on $ep$ collisions ($\theta_{c}=0$ mrad)
and to separate the beams by the required amount.
A dipole field of about 0.3 T over a length of $\pm9$ m accomplishes these goals.

\begin{figure}
\centerline{\includegraphics[angle=0,clip=,width=0.7\textwidth]{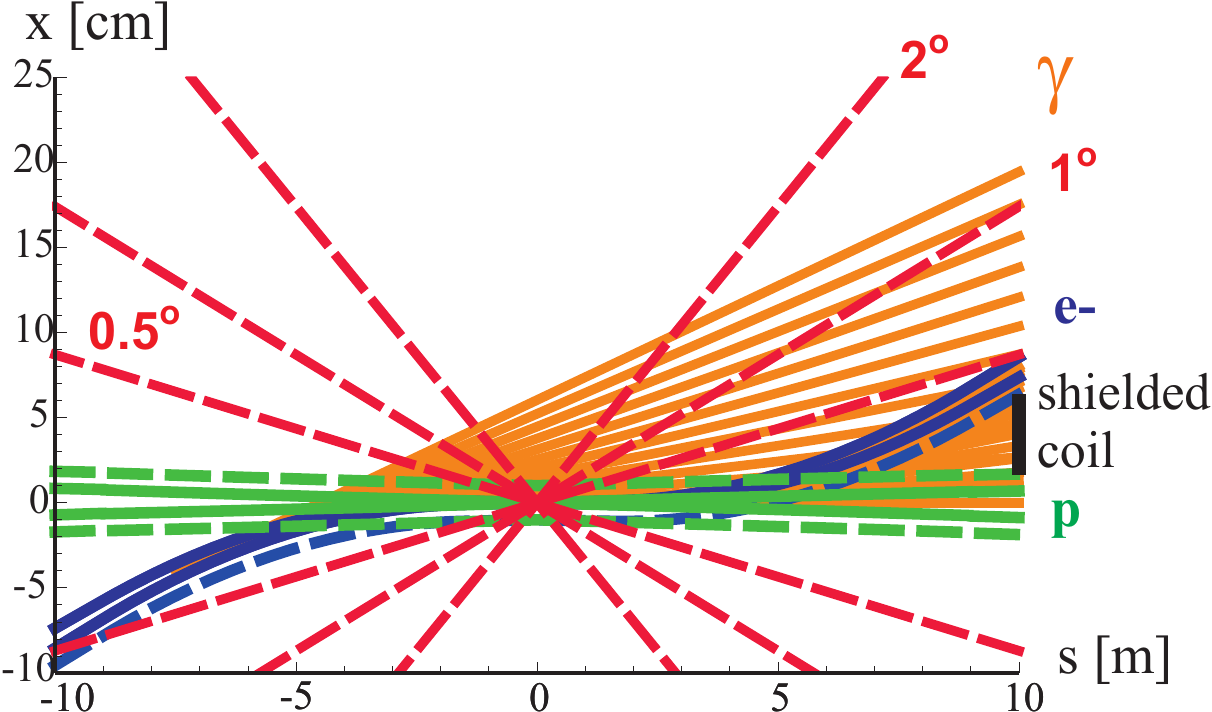}}
\caption{Linac-ring interaction-region layout. Shown are the beam envelopes of $10\sigma$ (electrons) [solid blue]
or 11$\sigma$ (protons) [solid green], the same envelopes with an additional constant 
margin of 10 mm [dashed], the synchrotron-radiation 
fan [orange], the approximate location of the magnet coil between
incoming protons and outgoing electron beam [black], and a ``1 degree'' line.}
\label{irlayout} 
\end{figure}

The IR layout with separation dipoles and crossing angle is sketched in
Fig.~\ref{irlayout}. Significant synchrotron radiation, with 48 kW average power, 
and a critical photon energy of 0.7 MeV, is emitted in the dipole fields.
A large portion of this radiation is extracted through the electron and proton beam pipes.
The SC proton magnets can be protected against the radiation heat load by 
an absorber placed in front of the first quadrupole and by a liner inside the beam pipe. 
Backscattering of synchrotron radiation into the detector is minimised by shaping the surface of
absorbers and by additional masking.

\begin{figure}
\centerline{\includegraphics[angle=0,clip=,width=0.7\textwidth]{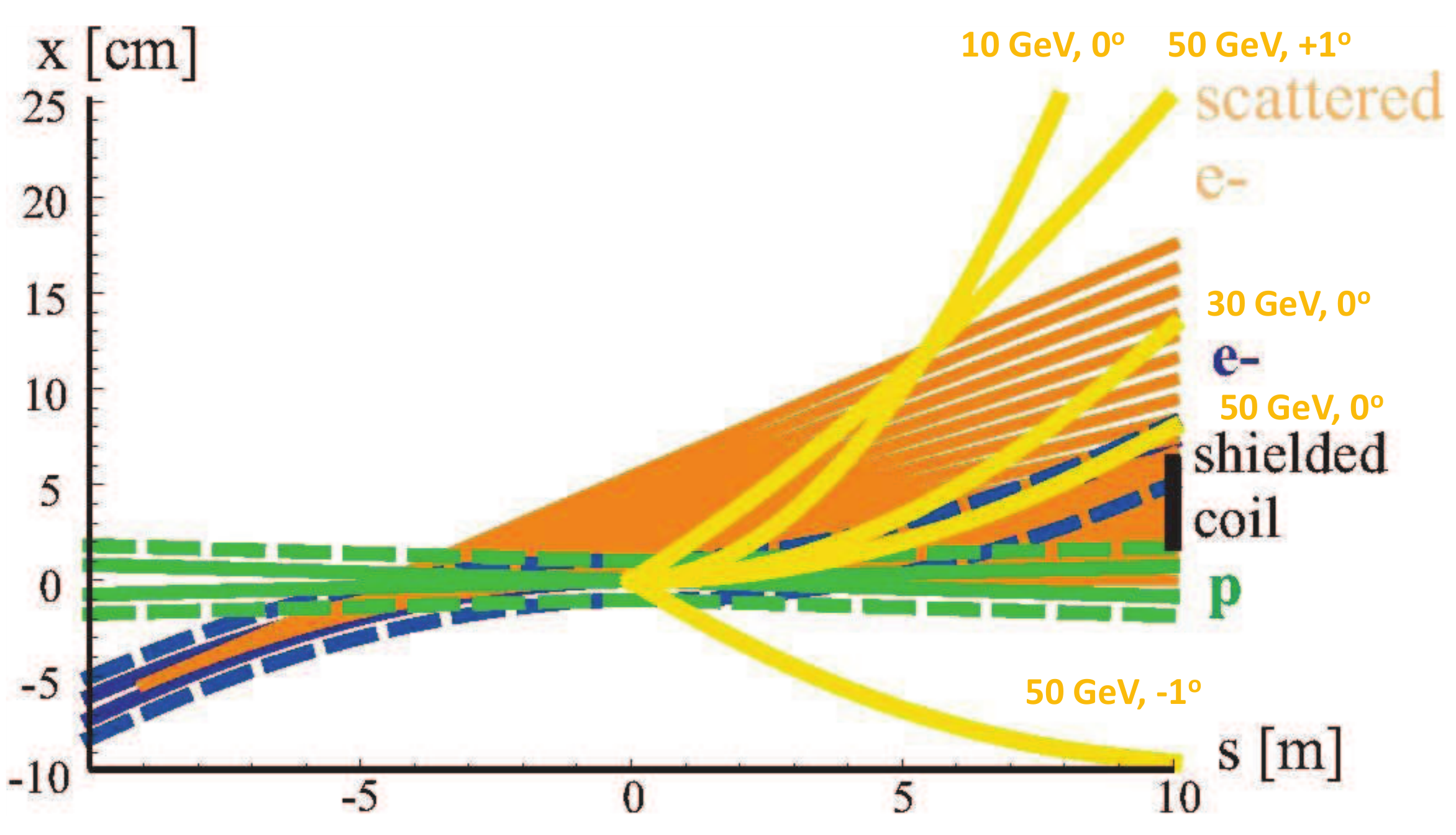}}
\caption{Example trajectories in the detector dipole fields for 
electrons of different energies and scattering angles,
demonstrating an enhancement of the detector acceptance by the dipoles.}
\label{iracceptance} 
\end{figure}

The separation dipole fields modify, and enhance, the geometric acceptance of the detector.
Figure \ref{iracceptance} illustrates that scattered electrons with energies of 10--50 GeV
might be detected at scattering angles down to zero degrees.

\subsubsection{Electron beam and the case for energy recovery}
\begin{figure}
\centerline{\includegraphics[angle=0,clip=,width=0.7\textwidth]{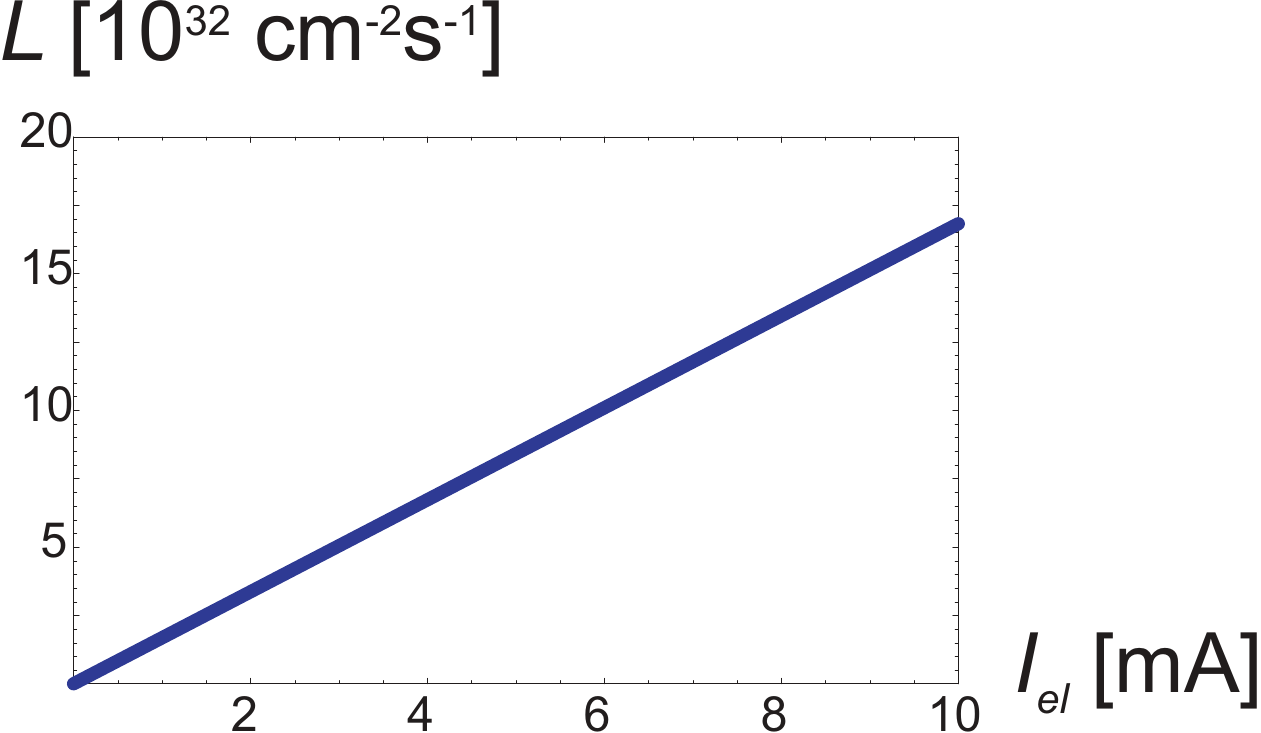}}
\caption{Linac-ring luminosity versus average electron beam current, according to (\protect\ref{lumi}).}
\label{lumiie} 
\end{figure}

The electron-beam emittance and the electron IP beta function are not critical,
since the proton beam size is large by electron-beam standards (namely about 7 $\mu$m rms
compared with nm beam-sizes for linear colliders).
The most important parameter for high luminosity is the average beam current, $I_{e}$,
which linearly enters into the luminosity formula (\ref{lumi}).
In addition to the electron beam current, also the bunch spacing (which should be a 
multiple of the LHC 25-ns proton spacing) and polarisation (80--90\% for the electrons) 
need to be considered. 
Having pushed all other parameters in (\ref{lumi}), Fig.~\ref{lumiie} illustrates that an average 
electron current of about 6.4 mA is required to reach the target luminosity 
of $10^{33}$~cm$^{-2}$s$^{-1}$.

For comparison, the CLIC main beam has a design average current of 0.01 mA \cite{clic2008}, 
so that it falls short by a factor 600 from the LHeC requirement.
For other applications it has been proposed to raise the CLIC beam power by
lowering the accelerating gradient, raising the bunch charge by a factor of two, 
and increasing the repetition rate up to three times, which raises 
the average beam current by a factor 6 to about 0.06 mA  
(this type of CLIC upgrade is described in \cite{Aksakal:2006fa}).
This ultimate CLIC main beam current is still a factor 100 below the LHeC target. 
On the other hand, the CLIC drive beam would have a sufficiently high current,
namely 30 mA, but at the low energy 2.37 GeV, which would not be 
useful for high-energy $ep$ physics.
Due to this low energy, also the drive beam power is still a factor 
of 5 smaller than the one required by LHeC.
Finally, the ILC design current is about 0.04 mA \cite{Phinney:2007gp}, which also falls more than 
a factor 100 short of the goal.

Fortunately, SC linacs can provide higher average current, e.g. by increasing the linac duty factor 
10--100 times, or even running in continuous wave (CW) mode, at lower accelerating gradient.
Example average currents for a few proposed designs illustrate this point:
The CERN High-Power Superconducting Proton Linac aims at about 1.5 mA average current (with 50 Hz pulse rate) \cite{hpspl},
the Cornell ERL design at 100 mA (cw) \cite{cornellerl}, and the eRHIC ERL at about 50 mA average current at
20 GeV beam energy (cw) \cite{erhicerl}.
All these designs are close to, or exceed, the LHeC requirements for average beam current
and average beam power (6.4 mA at 60 GeV).
It is worth noting that the JLAB UV/IR 4th Generation Light Source FEL is  
routinely operating with 10 mA average current (135 pC pulses at 75 MHz) \cite{neil}.
The 10-mA current limit in the JLAB FEL arises from well understood
beam break up \cite{bbughh} and significantly larger currents would be 
possible with suitably designed cavities.
It is, therefore, believed  that more than 6.4 mA for the LHeC ERL would be feasible. 

The target LHeC IP electron-beam power is 384 MW.
With a standard wall-plug-power to RF conversion efficiency around 50\%, this would imply
about 800 MW electrical power, far more than available. 
This highlights the need for energy recovery where the energy of the spent beam, after collision, is recuperated
by returning the beam 180$^{\circ}$ out of phase through the same RF structure that had earlier been used for
its acceleration, again with several recirculations.
An energy recovery efficiency $\eta_{\rm ER}$ reduces the electrical 
power required for RF power generation at a given beam current by a factor $(1-\eta_{\rm ER})$.
We need an efficiency $\eta_{\rm ER}$ above 90\% or higher to reach the beam-current goal 
of 6.4 mA with less than 100 MW total electrical power.

The above arguments have given birth to the LHeC Energy Recovery Linac high-luminosity baseline design,
which is being presented in this chapter.

\subsubsection{Choice of RF frequency}
Two candidate RF frequencies exist for the SC linac.
One possibility is operating at the ILC and XFEL RF frequency around 1.3 GHz,
the other choosing a frequency of about 720 MHz, close to the RF frequencies
of the CERN High-Power SPL, eRHIC, and the European Spallation Source (ESS).

The ILC frequency would have the advantage of synergy with the 
XFEL infrastructure, of profiting from the high gradients reached 
with ILC accelerating cavities, and of smaller structure size, 
which could reduce the amount of high-purity 
niobium needed by a factor 2 to 4.


Despite these advantages,  
the present LHeC baseline frequency is 720 MHz, or, more precisely, 721 MHz to be compatible with the LHC bunch spacing.
The arguments in favour of this lower frequency are the following:
\begin{itemize}
\item 
A frequency of 721 MHz requires less cryo-power (about two times less than at 1.3 GHz according to BCS
theory; the exact difference will depend on the residual resistance \cite{tuckmantel}).
\item
The lower frequency will facilitate the design and operation of high-power couplers \cite{napoly},
though the couplers might not be critical \cite{ciapala}. 
\item 
The smaller number of cells per module (of similar length) at lower RF frequency 
is preferred with regard to trapped  modes \cite{tuckmantel2}.
\item
The lower-frequency structures reduce beam-loading effects and 
transverse wake fields. 
\item
The project can benefit from synergy with SPL, eRHIC and ESS.
\item
Other projects, e.g.~low-emittance ERL light sources, can reduce the bunch charge by choosing a 
higher RF frequency. This is not the case for the LHeC, where the bunch distance is not determined by the 
RF frequency, but by the distance between proton bunches.  
\end{itemize}
In case the cavity material costs at 721 MHz would turn out to be a major concern, they could be reduced 
by applying niobium as a thin film on a copper substrate, rather than using bulk niobium. 
Establishing the necessary cavity performance with thin-film coating will require further R\&D. 
It is expected that the thin-film technology may also enhance the intrinsic cavity properties, e.g.~increase 
the $Q_{0}$ value. 

Linac RF parameters for both 720 MHz and 1.3 GHz in CW mode as well as 
for a pulsed 1.3-GHz option are compared in Table \ref{tablerf}.
The 721 MHz parameters are derived from eRHIC \protect\cite{lbz}.
Pulsed-linac applications for LHeC are discussed in Sections\,\ref{p140} 
and \ref{gammap}.

\begin{table}[htbp]
\begin{center} 
\begin{tabular}{|l|c|c|c|}
\hline
& ERL 721 MHz &  ERL 1.3 GHz & Pulsed \\ 
\hline
RF duty factor & CW & CW & 0.05 \\
RF frequency [GHz] & 0.72 & 1.3 & 1.3 \\
cavity length [m] & 1.04 & $\sim$1 & $\sim$1 \\
energy gain / cavity [MeV] & 20.8 & 20.8 & 31.5 \\
R/Q [$\text{circuit} \ \Omega$] & 285 & 518 & 518 \\
$Q_{0}$ [$10^{10}$] & 2.5 & 1 & 1 \\ 
power loss RF [W/cav.] & 30  & 42 
& 5  \\
``W per W'' (1.8 K to RT) & 700 & 700 & 700 \\
length / GeV [m] (filling=0.57)& 97 & 97 & 56 \\
\hline
\end{tabular}
\caption{Linac RF parameters 
for two different RF frequencies and two modes of operation.
The row ``W to W'' refers to the power needed at room temperature (RT)
to cool a heat unit at 1.8 K.
The numbers quoted for 721 MHz refelect the (measured) parameters 
of eRHIC prototype cavity BNL-I and an extrapolation to 
the improved cavity BNL-III \protect\cite{ilan}. 
The heat-load values at $20$\,MV/m indicated for 1.3 GHz 
have been extrapolated  
from \protect\cite{Phinney:2007gp}.
%
The additional static heat loss depends on the cryomodule
design and can be made small compared with the dynamic loss. 
} 
\label{tablerf}
\end{center} 
\end{table}

\subsubsection{ERL electrical site power}
The cryopower for two 10-GeV accelerating SC linacs is 28.9 MW, 
assuming 23 W/m heat load at 1.8 K
and 18 MV/m cavity gradient 
and 700 ``W per W'' cryo efficiency as for the ILC.
The RF power needed to control microphonics for the accelerating RF is estimated at 22.2 MW,
considering that 10 kW/m RF power may be required, as for eRHIC,
with 50\% RF generation efficiency. 
The electrical power for the additional RF 
compensating the synchrotron-radiation energy loss is 24.1 MW, with
an RF generation efficiency of 50\%. 
The cryo power for the compensating RF is 2.1 MW, provided in additional
1.44 GeV linac sections, and the microphonics control for the compensating RF requires
another 1.6 MW.
In addition, with an injection energy of 50 MeV, 6.4 mA beam current, and as usual
50\% efficiency, the electron injector consumes about 6.4 MW.
A further 3 MW is budgeted for the recirculation-arc magnets \cite{tommasini}.
Together this gives a grand total of 88.3 MW electrical power, some 25\% below the 
100 MW limit. 
The LHeC ERL power budget is summarised in Table \ref{tablepower}. 

\begin{table}[htbp]
\begin{center} 
\begin{tabular}{|l|c|}
\hline
Item &  Electrical Power [MW] \\
\hline
Main linac cryopower & 18.0 \\
Microphonics control & 22.2 \\
Extra RF to compensate SR losses & 24.1 \\
Extra-RF cryopower & 1.6 \\
Electron injector & 6.4 \\
Arc magnets & 3.0 \\
\hline
Total &  75.3 \\
\hline
\end{tabular}
\caption{ERL power budget.}
\label{tablepower}
\end{center} 
\end{table}



\subsubsection{ERL configuration}
\begin{figure}
\centerline{\includegraphics[angle=0,clip=,width=0.9\textwidth]{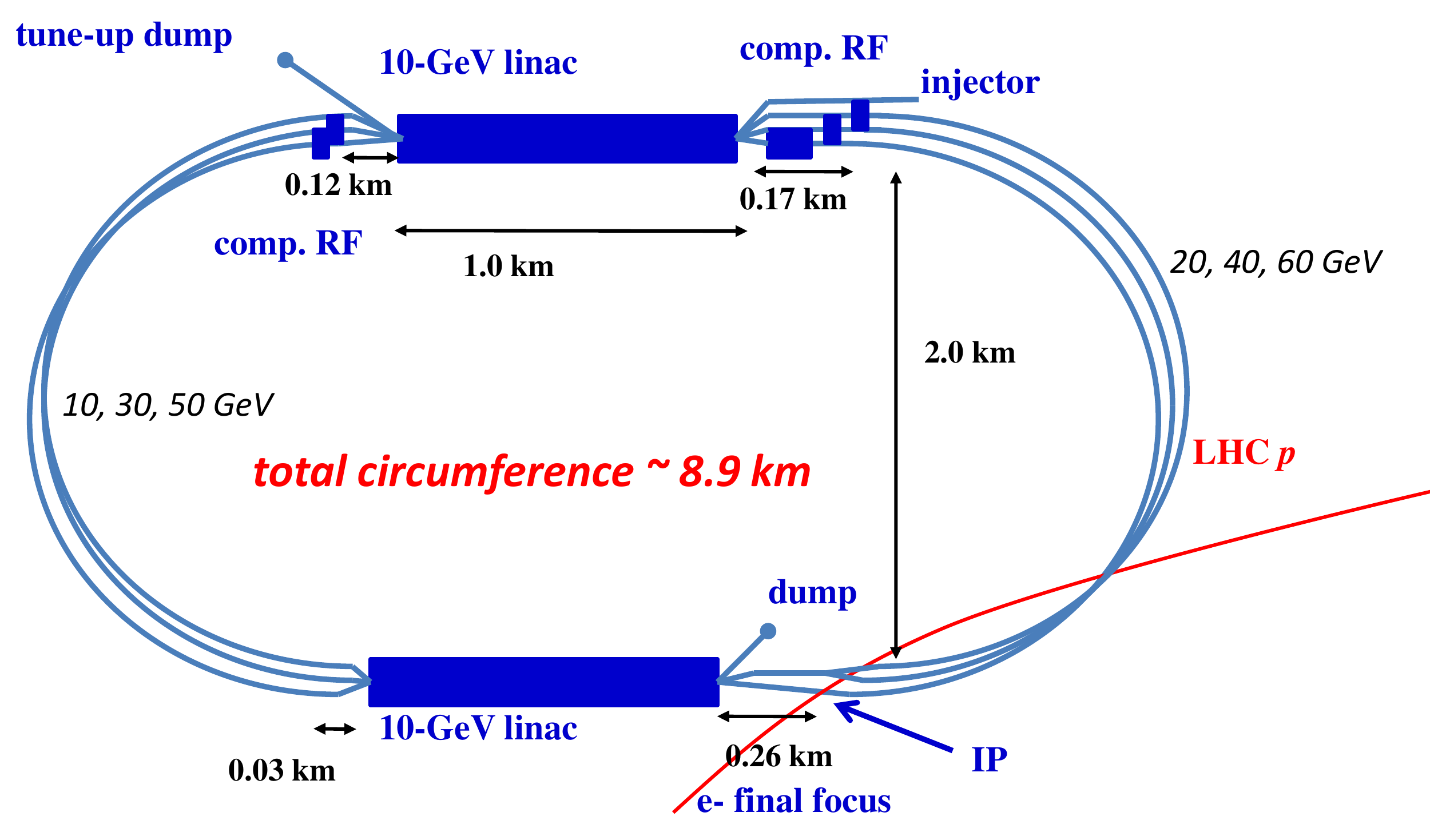}}
\caption{LHeC ERL layout including dimensions.}
\label{erllayout} 
\end{figure}

The ERL configuration is depicted in Fig.~\ref{erllayout}.
The shape, arc radius and number of passes have been optimised
with respect to construction cost and with respect to synchrotron-radiation
effects \cite{skrabacz}.

The ERL is of racetrack shape. 
A 500-MeV electron bunch coming from the injector is accelerated in each of the two 
10-GeV SC linacs during three revolutions, after which it has 
obtained an energy of 60 GeV. The 60-GeV beam is focused and collided with the 
proton beam. It is then bent by 180$^{\circ}$ in the highest-energy arc beam line
before it is sent back through the first linac, at a decelerating RF phase. 
After three revolutions with deceleration, re-converting the energy stored in the beam 
to RF energy, the beam energy is back at its original value of 500 MeV, 
and the beam is now disposed in a low-power 3.2-MW beam dump. 
A second, smaller (tune-up) dump could be installed behind the first linac.

Strictly speaking, with an injection energy into the first linac of 0.5 GeV, the energy gain in the 
two accelerating linacs need not be 10 GeV each, but about 9.92 GeV, in order to reach 60 GeV
after three passages through each linac. Considering a rough value of 10 GeV means that we overestimate the
 electrical power required by about 1\%. 

Each arc contains three separate beam lines
at energies of 10, 30 and 50 GeV on one side, and 20, 40 and 60 GeV on the other. 
Except for the highest energy level of 60 GeV, at which there is only one beam, 
in each of the other arc beam lines  
there always co-exist a decelerating and an accelerating beam. 
The effective 
arc radius of curvature is 1 km, with a dipole bending radius of 764 m \cite{bogacz}. 

The two straight sections accommodate the 1-km long SC accelerating linacs.
In addition to the 1km linac section, there is an additional space of 290 m in each straight section of the racetrack.
In one straight of the racetrack 
260 m of this additional length is allocated for the electron final focus (plus matching and splitting), 
the residual 30 m on the other side of the same straight allows for 
combining the beam and matching the optics into the arc.
In the second straight section of the racetrack the additional length of the straight sections houses the additional linacs for compensating the 1.88 GeV energy loss in the return arcs \cite{dschulte2}. 
For the highest energy, 60 GeV, there is a single beam and the compensating RF (750 MV) 
can have the same frequency, 721 MHz, as in the main linac \cite{dschulte2}.
For the other energies, a higher harmonic RF system, e.g. at 1.442 GHz, can compensate the energy loss
for both decelerating and accelerating beams, which are 180$^{\circ}$ out of phase at 721 MHz.
On one side of the second straight  
one must compensate a total energy loss of about 907 MeV per particle (=750+148+9 MeV, corresponding to the energy loss
at 60, 40 and 20 GeV, respectively), which 
should easily fit within a length of 170 m.
On the other side one has to compensate 409 MeV (=362+47 MeV), corresponding to SR energy losses at 50 and 30 GeV),
for which a length of 120 m is available. 

The total circumference of the ERL racetrack is chosen as 8.9 km, equal to one third of 
the LHC circumference. This choice has the advantage that one could introduce ion-clearing 
gaps in the electron beam which would match each other on successive revolutions 
(e.g. for efficient ion clearing in the linacs that are shared by six different 
parts of the beam) and which would also always coincide with the same proton bunch locations
in the LHC, so that in the latter a given proton beam would either always collide
or never collide with the electrons \cite{schulte}. Ion clearing may be necessary to suppress ion-driven
beam instabilities. The proposed implementation scheme would remove ions while minimising the
proton emittance growth which could otherwise arise when encountering
collisions only on some of the turns. 
In addition, this arrangement can be useful for comparing 
the emittance growth of proton bunches which are colliding with the electrons and those
which are not.

The length of individual components is as follows. 
The exact length of the 10-GeV linac is 1008 m.
The individual cavity length is taken to be 1 m. The optics consists of 56-m long FODO cells
with 32 cavities. The number of cavities per linac is 576. 
The linac cavity filling factor is 57.1\%.
The effective arc bending radius is set to be 1000 m.
The bending radius of the dipole magnets is 764 m, corresponding to a dipole
filling factor of 76.4\% in the arcs.
The longest SR compensation linac has a length of 84 m (replacing the energy lost by SR at 60 GeV).
Combiners and splitters between straights and arcs require about 20--30 m space each.
The electron final focus may have a length of 200--230 m.

\subsubsection{IP parameters and beam-beam effects}
Table \ref{ipparam} presents interaction-point (IP) parameters for the electron and
proton beams.

\begin{table}[htbp]
\begin{center} 
\begin{tabular}{|l|cc|}
\hline
& protons & electrons \\
\hline
beam energy [GeV] & 7000 & 60 \\
Lorentz factor $\gamma$ & 7460 & 117400 \\
normalised emittance $\gamma\epsilon_{x,y}$ [$\mu$m] & 3.75 & 50 \\
geometric emittance $\epsilon_{x,y}$ [nm] & 0.40 & 0.43 \\a
IP beta function $\beta_{x,y}^{\ast}$ [m] & 0.10 & 0.12 \\
rms IP beam size $\sigma_{x,y}^{\ast}$ [$\mu$m] & 7 & 7 \\
initial rms IP beam divergence $\sigma_{x',y'}^{\ast}$ [$\mu$rad] & 70 & 58 \\
beam current [mA] & $\ge$430 & 6.4 \\
bunch spacing [ns] & 25 or 50 &  (25 or) 50  \\
bunch population [ns] & $1.7\times 10^{11}$ & (1 or) $2\times 10^{9}$ \\
\hline
\end{tabular}
\caption{IP beam parameters}
\label{ipparam}
\end{center} 
\end{table}

Due to the low charge of the electron bunch, the proton head-on beam-beam 
tune shift is tiny, namely $\Delta Q_{p}=+0.0001$, 
which amounts to only about 1\% of the LHC $pp$ design tune shift (and is of opposite sign).
Therefore, the proton-beam tune spread induced by the $ep$ collisions is negligible.
In fact, the electron beam acts like an electron lens and could conceivable
increase the $pp$ tune shift and luminosity, but only by about 1\%.
Long-range beam-beam effects are equally 
insignificant for both electrons and protons, since  the detector-integrated dipoles separate 
the electron and proton bunches by about 36$\sigma_{p}$ at the first parasitic encounter,  
3.75 m away from the IP. 

One further item to be looked at is the proton beam emittance growth. 
Past attempts at directly simulating the emittance growth from $ep$ collisions 
were dominated by numerical noise from the finite number of macroparticles
and could only set an upper bound \cite{epac2004}, nevertheless indicating  
that the proton emittance growth due to the pinching electron beam 
might be acceptable for centred collisions. 
Proton emittance growth due to electron-beam position jitter 
and simultaneous $pp$ collisions is another potential concern.
For a 1$\sigma$ offset between the electron and proton orbit at the IP, 
the proton bunch receives a deflection of about 10 nrad (approximately $10^{-4}\sigma_{x',y'}^{\ast}$).
Beam-beam simulations for LHC $pp$ collisions have determined the acceptable 
level for random white-noise dipole excitation as $\Delta x/\sigma_{x}\le 0.1\%$ \cite{ohmipac07}.  
This translates into a very relaxed electron-beam random orbit jitter tolerance 
of more than 1$\sigma$.
The tolerance on the orbit jitter 
will then not be set by beam-beam effects, but by the luminosity loss resulting from off-centre collisions,
which, without disruption, scales as $\exp(-(\Delta x)^{2}/(4 \sigma_{x,y}^{\ast\; 2})$.
The random orbit jitter observed at the SLAC SLC 
had been of order 0.3--0.5$\sigma$ \cite{slcjitter,slcjitter2}.  
A $0.1\sigma$ offset at LHeC would reduce the luminosity by at most 0.3\%,
a $0.3\sigma$ offset by 2.2\%.
Disruption further relaxes the tolerance. 

The strongest beam-beam effect is encountered by the electron beam, which is heavily disrupted.  
The electron disruption parameter is $D_{x,y}\equiv N_{b,p}r_{e}\sigma_{z,p}/(\gamma_{e}\sigma^{\ast\; 2}) 
\approx 6$,
and the ``nominal disruption angle'' $\theta_{0}\equiv D\sigma^{\ast}/\sigma_{z,p}
=N_{b,p}r_{e}/(\gamma_{e}\sigma^{\ast})$ \cite{chenyokoya} is about 600 $\mu$rad (roughly 10$\sigma_{x',y'}^{\ast}$),
which is huge. Simulations show that the actual maximum angle of the disrupted electrons is less than half $\theta_{0}$.

\begin{figure}
\centerline{\includegraphics[angle=-0.5,clip=,width=0.45\textwidth]{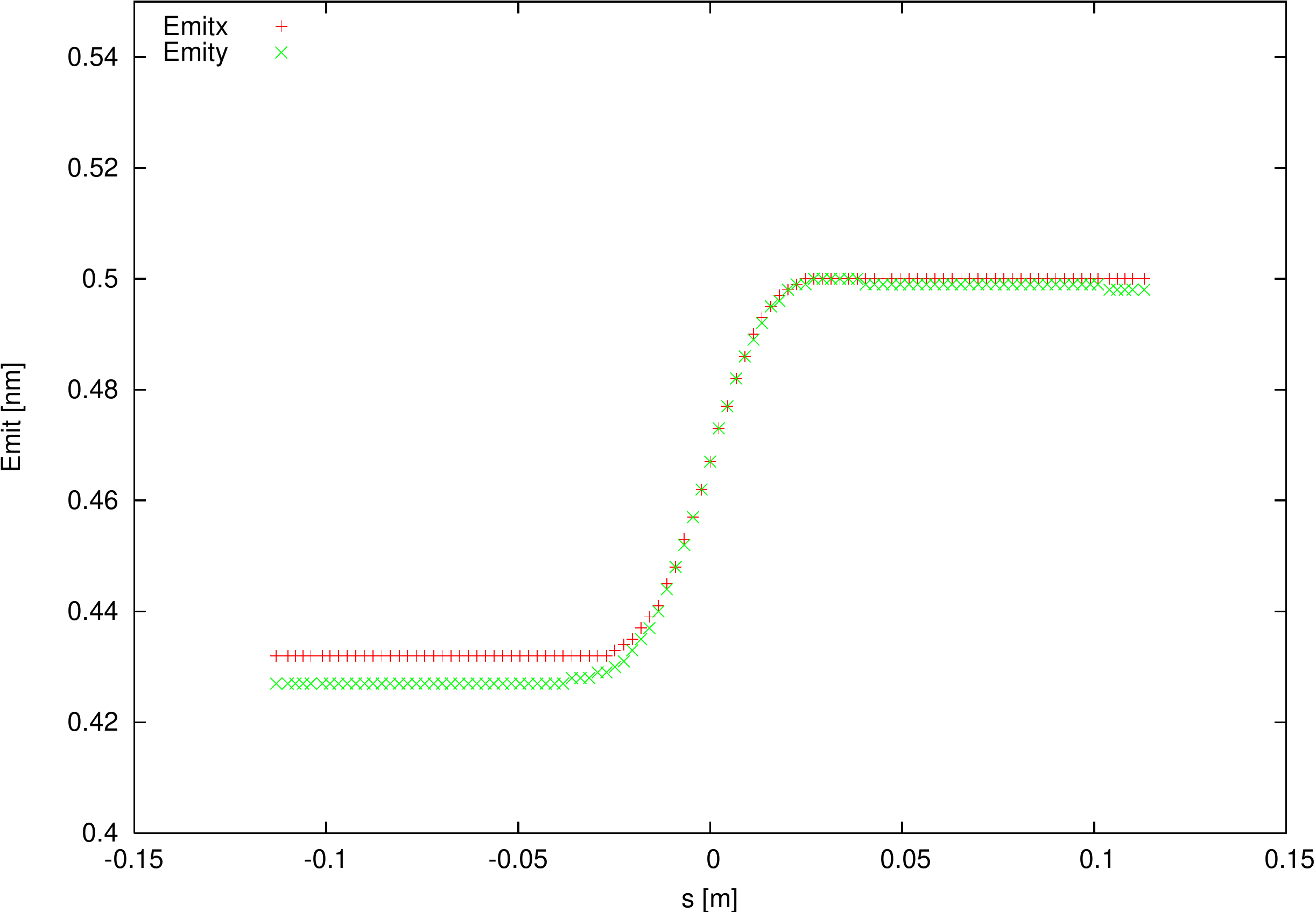}
\includegraphics[angle=-0.5,clip=,width=0.45\textwidth]{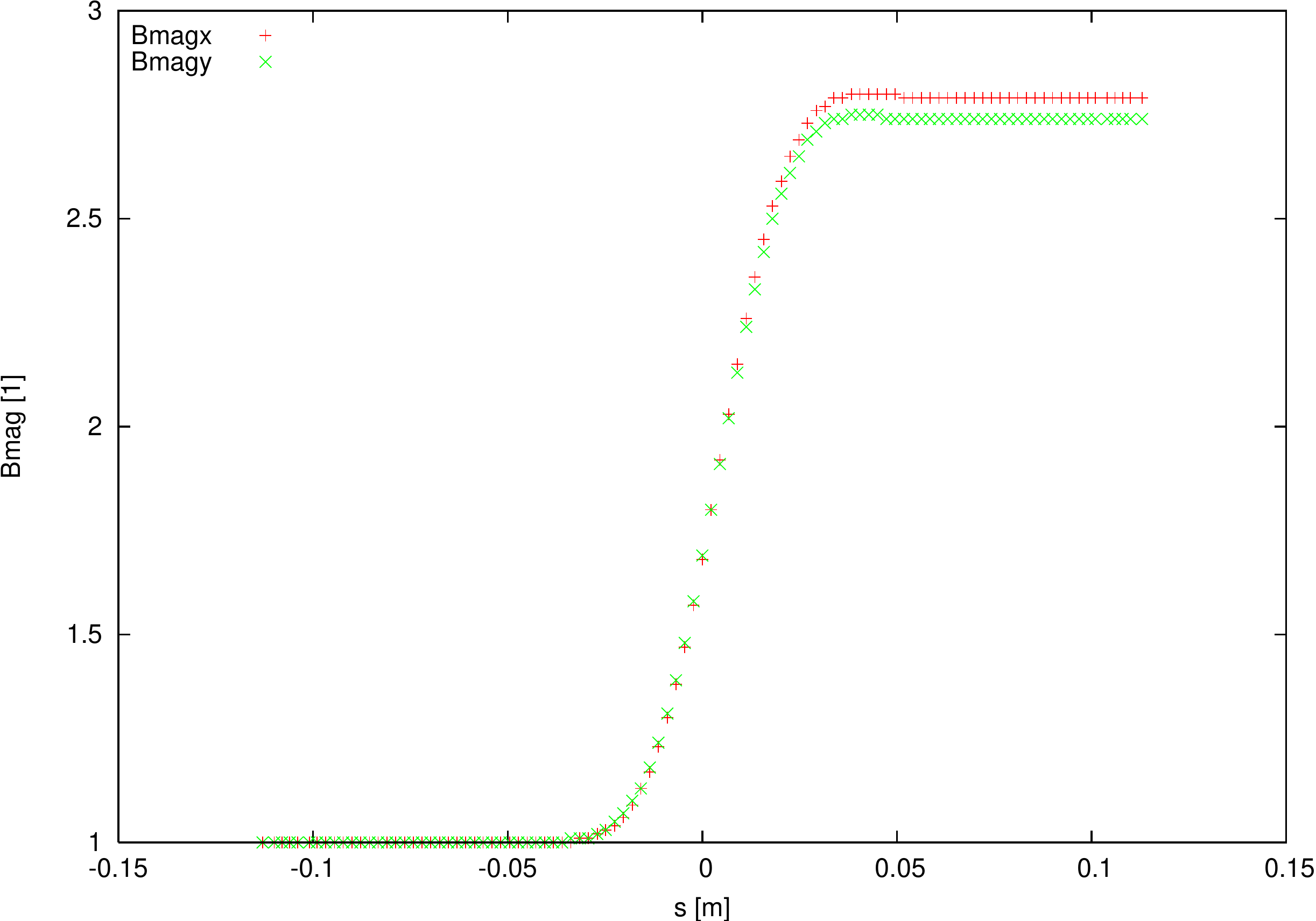}}
\centerline{\includegraphics[angle=-0.5,clip=,width=0.45\textwidth]{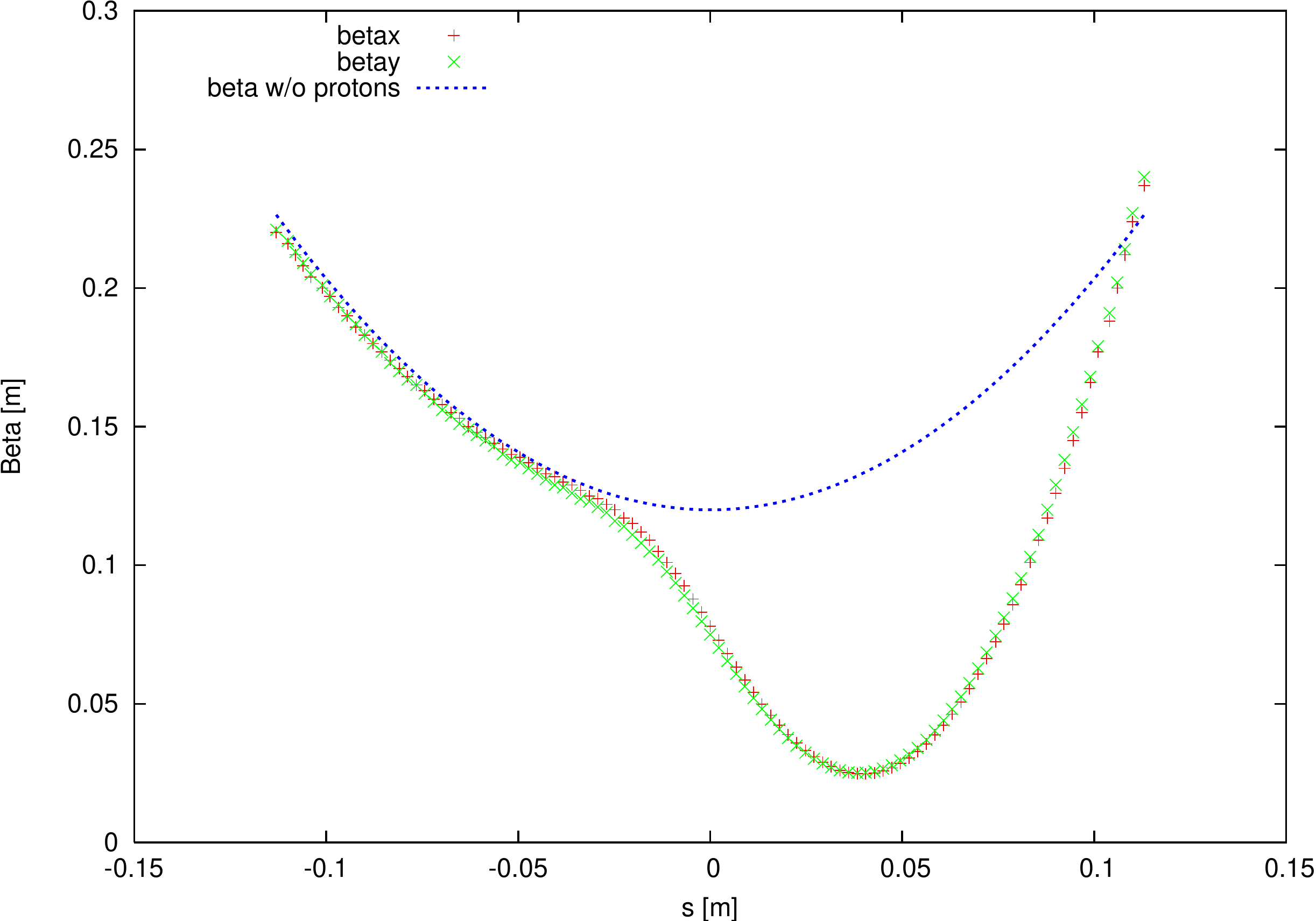}
\includegraphics[angle=-0.5,clip=,width=0.45\textwidth]{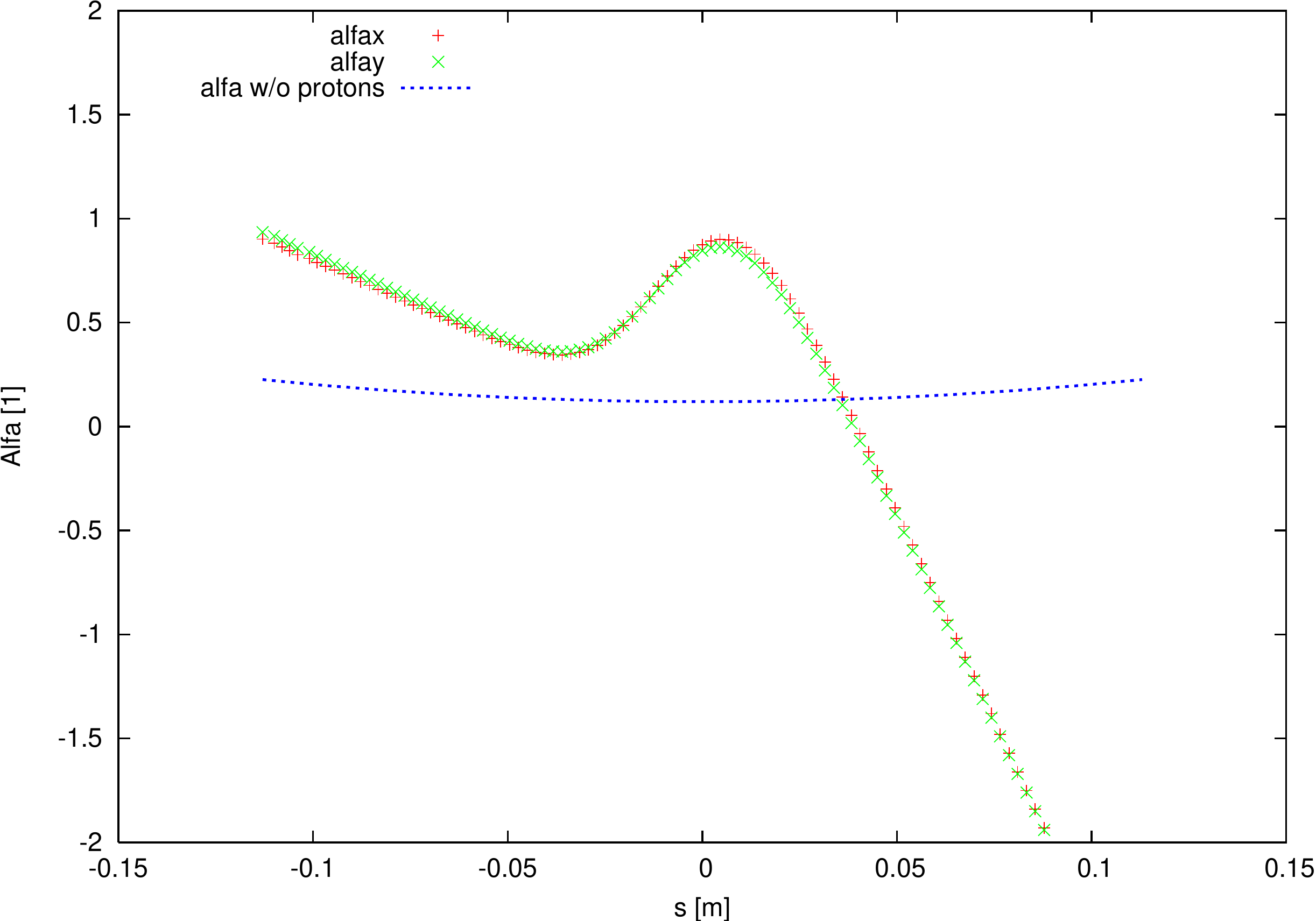}}
\caption{Simulated evolution of the 
electron beam emittance (top left), mismatch factor $B_{\rm mag}$ (top right)
beta function (bottom left) and alpha function (bottom right) during the collision
with a proton bunch, as a function of distance from the IP.}
\label{disr} 
\end{figure}

Figure \ref{disr} illustrates the emittance growth and optics-parameter change 
for the electron beam due to head-on collision with a ``strong'' proton bunch.
The intrinsic emittance grows by only 15\%, but there is a 180\% growth in the 
mismatch parameter ``$B_{\rm mag}$'' (defined as $B_{\rm mag}=(\beta \gamma_{0}-2 \alpha \alpha_{0}+\beta_{0}\gamma)/2$,
where quantities with and without subindex ``0'' refer to the optics without and with 
collision, respectively.
Without adjusting the extraction line optics to the parameters of the mismatched beam
the emittance growth will be about 200\%. This would be acceptable since the 
arc and linac physical apertures have been determined  
assuming up to 300\% emittance growth for the decelerating beam \cite{bogacz}. 
However, if the optics of the extraction line is rematched for the colliding 
electron beam (corresponding to an effective $\beta^{\ast}$ of about 3 cm rather than the nominal 
12 cm; see Fig.\ref{disr} bottom left), the net emittance growth can be much reduced, to only 
about 20\%.
The various optics parameters shown in Fig.~\ref{disr} 
vary by no more than 10--20\% for beam-beam orbit offsets up to 1$\sigma$.

\begin{figure}
\centerline{\includegraphics[angle=-0.5,clip=,width=0.6\textwidth]{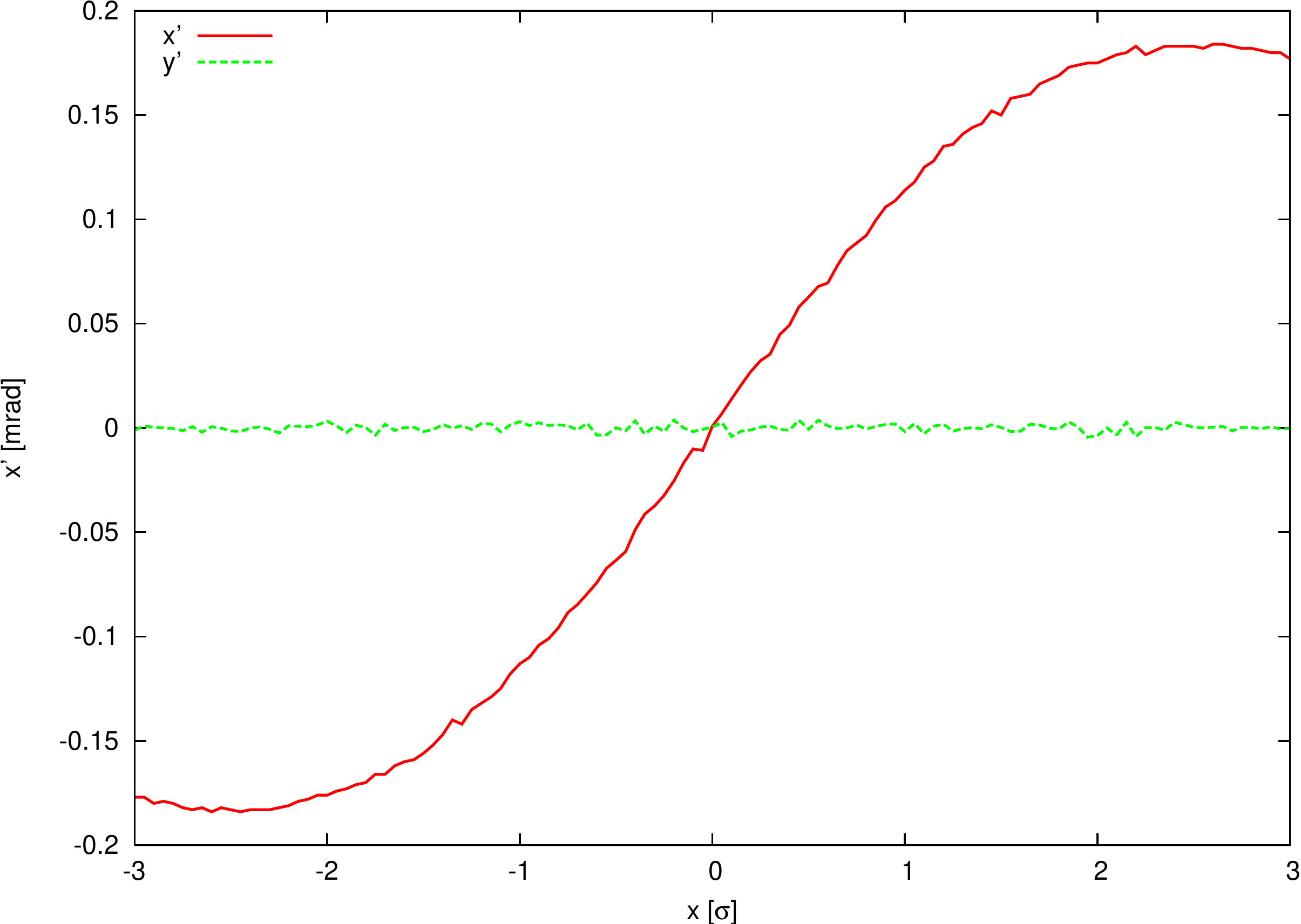}}
\caption{Simulated electron horizontal 
centre-of-mass deflection angle as a function of the 
horizontal beam-beam offset.}
\label{deflangle} 
\end{figure}

Figure \ref{deflangle} presents the average electron deflection angle as a function
of the beam-beam offset. 
The extraction channel for the electron beam must have sufficient aperture to accommodate
both the larger emittance due to disruption and the average trajectory change 
due to off-centre collisions. 

\subsection{Polarisation}
The electron beam can be produced from a polarised DC gun with about 90\% polarisation,
and with, conservatively, 10--50 $\mu$m  normalised emittance \cite{polarizedsource}.
Spin-manipulation tools and measures for preserving polarisation, like 
a Wien filter and/or spin rotators, and polarimeters should be included in the optics design
of the injector, the final focus, and the extraction line. 

As for the positrons, up to about 60\% polarisation can be achieved either 
with an undulator \cite{undulatorsource} or 
with a Compton-based e$^{+}$ source \cite{comptonsource1,comptonsource2}\footnote{The primary challenge for positrons is to produce them in sufficient
number and with a small enough emittance.}.

\subsection{Pulsed linacs}
\label{p140}
\begin{figure}
\centerline{\includegraphics[angle=0,clip=,width=0.9\textwidth]{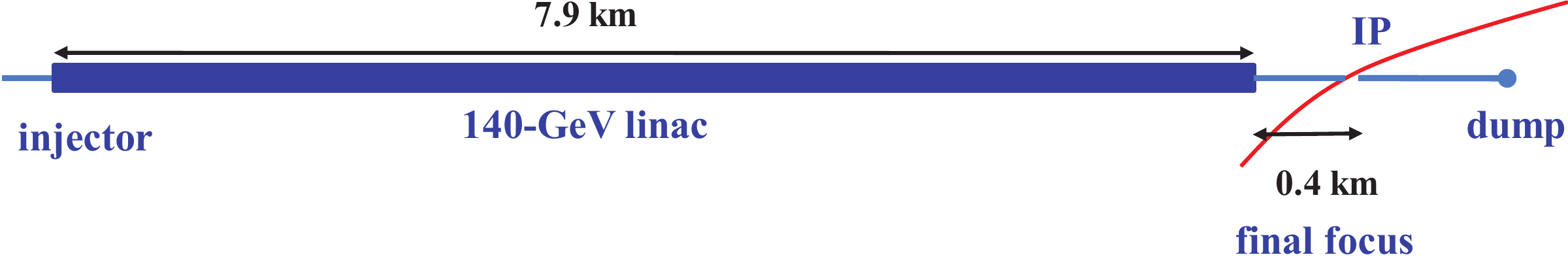}}
\caption{Pulsed single straight 140-GeV linac for higher-energy ep collisions.}
\label{pulsedlinac} 
\end{figure}

For beam energies above about 140 GeV, due to the growing impact of synchrotron radiation, 
the construction of  a single straight linac is cheaper than that of a recirculating  linac \cite{skrabacz}. 
Figure \ref{pulsedlinac} shows the schematic of an LHeC collider based on a 
pulsed straight 140-GeV linac, including injector, final focus, and beam dump.
The linac could be either of ILC type (1.3 GHz RF frequency) 
or operate at 721 MHz as the preferred ERL version.
In both cases, ILC values are assumed for the cavity gradient (31.5 MV/m) 
and for the cavity unloaded $Q$ value ($Q_{0}=10^{10}$).
This type of linac would be extendable to ever higher beam energies and could conceivably
later become part of a linear collider.
In its basic, simplest and conventional version no energy recovery is possible for this configuration, 
since it is impossible to bend the 140-GeV beam around.
The lack of energy recovery leads to significantly lower luminosity.
For example, with 10 Hz repetition rate, 5 ms pulse length (longer than ILC), a geometric reduction
factor $H_{g}=0.94$ and $N_{b,e}=1.5\times 10^{9}$ per bunch, 
the average electron current would be 0.27 mA and the luminosity $4\times 10^{31}$~cm$^{-2}$s$^{-1}$.

The construction of the 140-GeV pulsed straight linac 
could be staged, e.g.~so as to first feature a pulsed linac at 60 GeV,
which could also be used for $\gamma$-$p/A$ collisions (see Section\,\ref{gammap}). 
The linac length decreases directly in proportion to the beam energy.
For example, at 140-GeV the pulsed linac measures 7.9 km, while at 60 GeV its length would be 3.4 km. 
For a given constant wall-plug power, of 100 MW, both the average electron 
current and the luminosity scale roughly inversely with the beam energy.
At 60 GeV the average electron current becomes 0.63 mA and the 
pulsed-linac luminosity, without any energy recovery, would be more than  
$9\times 10^{31}$~cm$^{-2}$s$^{-1}$.

\subsection{Higher-energy LHeC ERL option}
\begin{figure}
\centerline{\includegraphics[angle=0,clip=,width=0.9\textwidth]{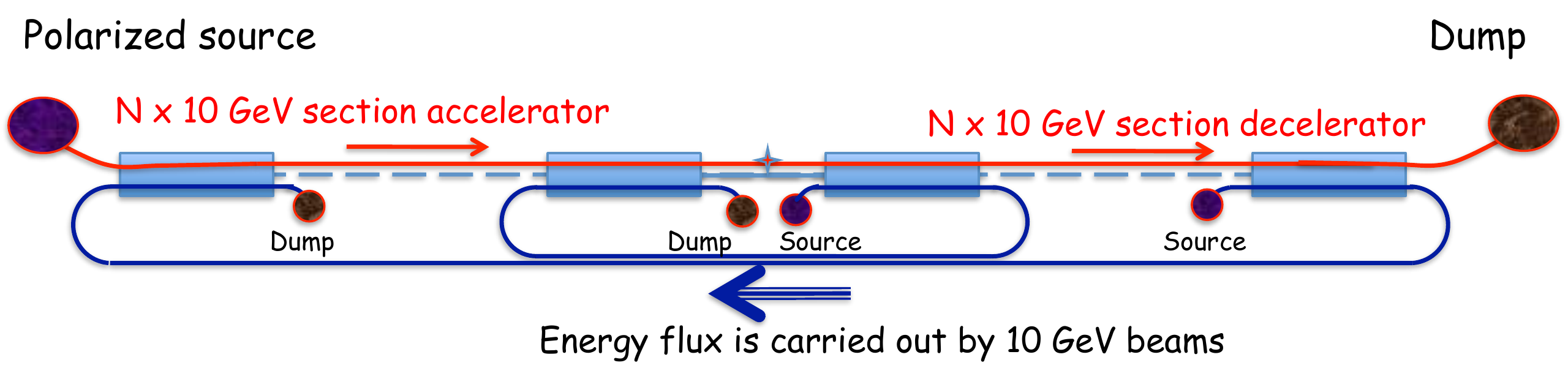}}
\caption{Highest-energy high-luminosity ERL option based on two straight linacs and 
multiple 10-GeV energy-transfer beams \protect\cite{litvinenkotwolinac}.} 
\label{erllinac} 
\end{figure}

The simple straight linac layout of Fig.~\ref{pulsedlinac} can be expanded as shown 
in Fig.~\ref{erllinac} \cite{litvinenkotwolinac}.
The main electron beam propagates from the left to the right.
In the first linac it gains about 150 GeV, then collides with the hadron beam,
and is then decelerated in the second linac. By transferring the 
RF energy back to the first accelerating linac, with the help of multiple, e.g.~15, 10-GeV ``energy-transfer beams,'' 
a novel type of energy recovery is realised without bending the spent beam. 
With two straight linacs facing each other this configuration could easily be converted into a 
linear collider, or vice versa, pending on 
geometrical and geographical constraints of the LHC site. 
As there are negligible synchrotron-radiation losses the energy recovery could be more efficient
than in the case of the 60-GeV recirculating linac. 
Such novel form of ERL could push the LHeC luminosity to the $10^{35}$~cm$^{-2}$s$^{-1}$ level. 
In addition, it offers ample synergy with the CLIC two-beam technology.

\subsection{$\gamma$-$p/A$ Option}
\label{gammap}
\begin{figure}
\centerline{\includegraphics[angle=0,clip=,width=0.9\textwidth]{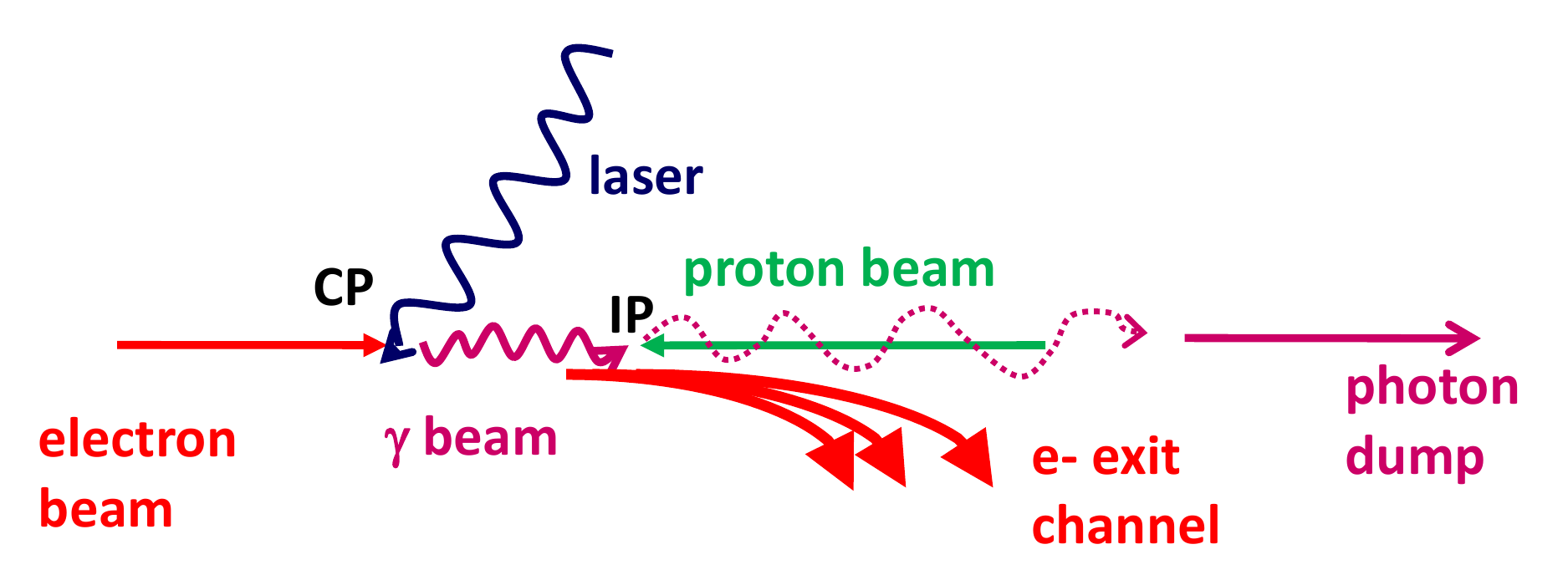}}
\caption{Schematic of $\gamma$-$p/A$ collision; prior to the photon-hadron interaction point (IP), 
the electron beam is scattered off a several-J laser pulse at the conversion point (CP).}
\label{gammap1} 
\end{figure}

In case of a (pulsed) linac without energy recovery the electron beam can be converted
into a high-energy photon beam, by backscattering off a laser pulse, as is illustrated in Fig.~\ref{gammap1}. 
The rms laser spot size at the conversion point should be similar to the size of the electron beam at this location, 
that is $\sigma_{\gamma}\approx 10 \mu$m. 

With a laser wavelength around $\lambda_{\gamma} \approx 250$ nm ($E_{\gamma,0}\approx 5$~eV),
obtained e.g.~from a Nd:YAG laser with frequency quadrupling, 
the Compton-scattering parameter $x$ \cite{telnov,burkhardt},  
\begin{equation}
x \approx 15.3\; \left[ \frac{E_{e,0}}{\rm TeV}\right]\; \left[ \frac{E_{\gamma,0}}{\rm eV}\; \right]\; , 
\end{equation}
is close to the optimum value 4.8 for an electron energy of 60 GeV 
(for $x>4.8$ high-energy photons get lost due to the creation of $e^+e^-$ pairs).
The maximum energy of the Compton scattered photons is given by
$E_{\gamma,{\rm max}} = x/(x+1) E_{0}$, which is larger than 80\% of 
the initial electron-beam energy $E_{e,0}$, for our parameters.
The cross section and photon spectra depend on the longitudinal electron
polarisation $\lambda_{e}$ and on the circular laser polarisation $P_{c}$.
With proper orientation ($2 \lambda_{e} P_{c}=-1$) the photon spectrum is
concentrated near the highest energy $E_{\gamma ,{\rm max}}$. 
 
The probability of scattering per individual electron is \cite{nlczdr}
\begin{equation}
n_{\gamma} = 1 - \exp (-q)
\end{equation}
with
\begin{equation}
q = \frac{\sigma_{c} A}{E_{\gamma,0} 2 \pi \sigma_{\gamma}^{2}}\; ,
\end{equation}
where $\sigma_{c}$ denotes the (polarised) Compton cross section
and $A$ the laser pulse energy. 
Using the formulae in \cite{telnov2}, the Compton cross section 
for $x=4.8$ and $2 \lambda_{e} P_{c}=-1$ is computed to be 
$\sigma_{c}=3.28\times 10^{-25}$~cm$^{2}$. 
The pulse energy corresponding to $q=1$, i.e.~to a conversion efficiency of 65\%, is estimated as
$A\approx E_{\gamma,0} 2 \pi \sigma_{\gamma}^{2}/\sigma_{c}\approx 16$~J.
To set this into perspective, for a $\gamma\gamma$ collider at the ILC, 
Ref.~\cite{klemz} considered a pulse energy of 9 J 
at a four times longer wavelength of $\lambda\approx 1\; \mu$m.


The energies of the leftover electrons 
after conversion extend from about 10 to 60 GeV. 
This spent electron beam, with its enormous energy spread, must be safely 
extracted from the interaction region. 
The detector-integrated dipole magnets will assist in this process. They 
will also move the scattered electrons away from the interaction point.
A beam dump for the high-energy photons should also be installed, 
behind the downstream quadrupole channel.

\begin{figure}
\centerline{\includegraphics[angle=0,clip=,width=0.43\textwidth]{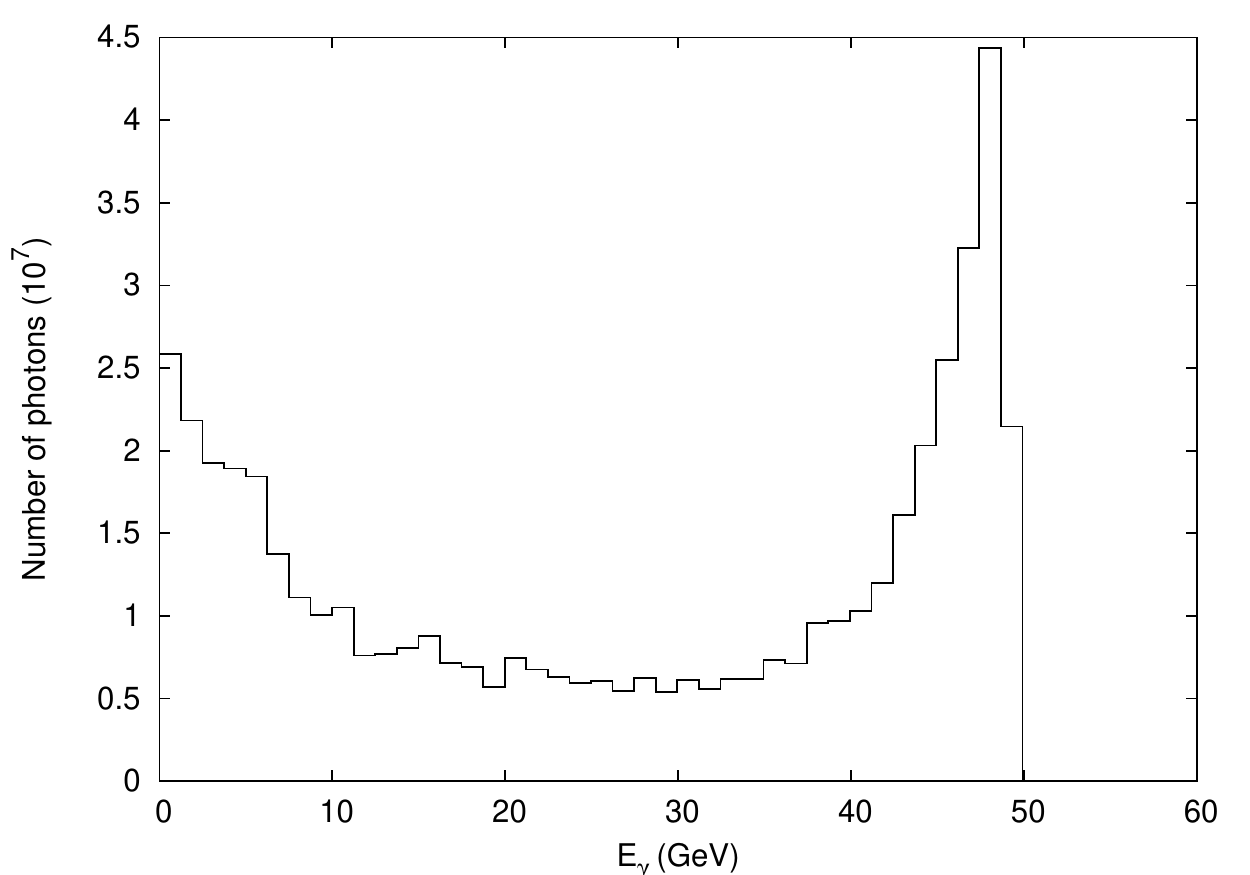} 
\includegraphics[angle=0,clip=,width=0.47\textwidth]{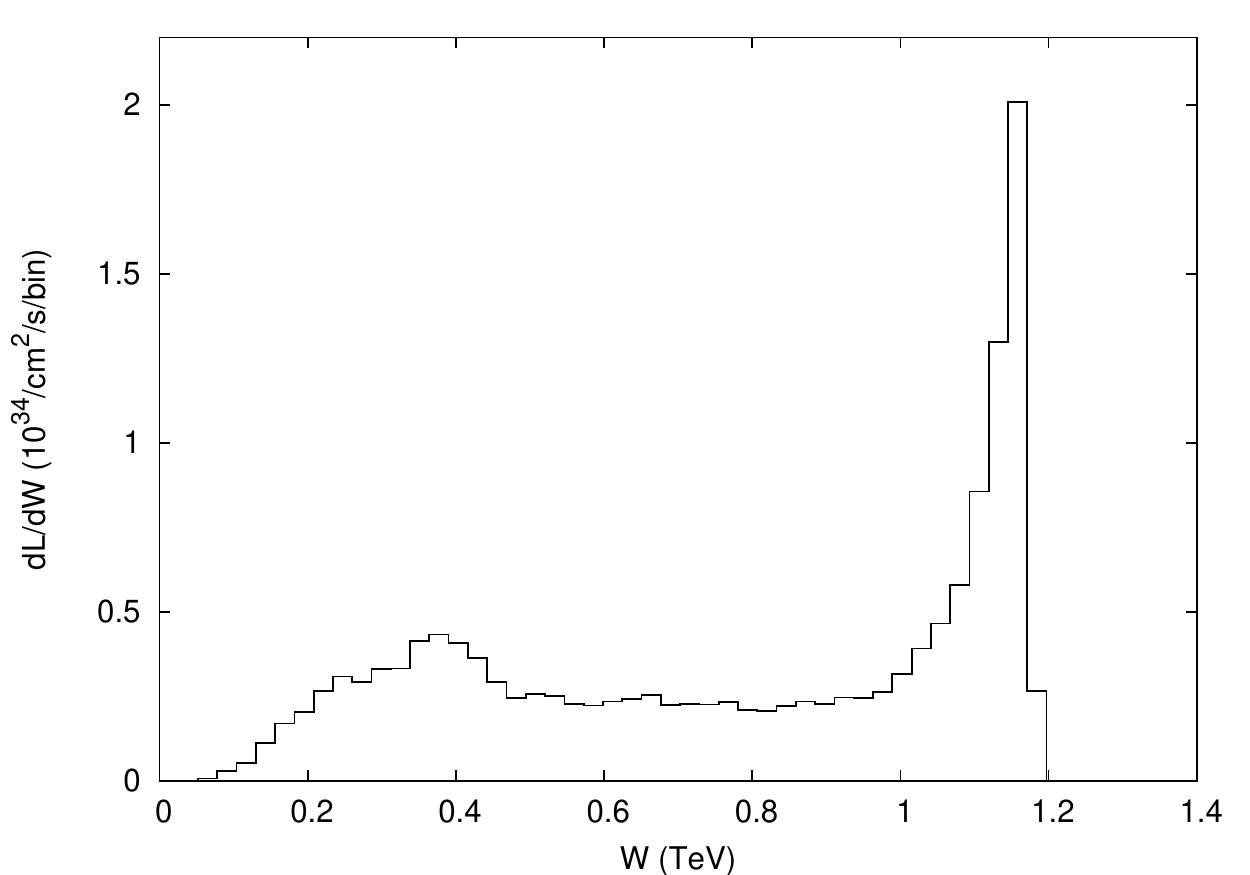}}
\caption{Simulated example photon spectrum after the conversion point (left) 
and  $\gamma$-$p$ luminosity spectrum 
\protect\cite{aksakalnergiz}.}
\label{gammap2} 
\end{figure}

Figure \ref{gammap2} presents the photon energy spectrum
after the conversion and the luminosity spectrum \cite{aksakalnergiz},
obtained from a simulation with the Monte-Carlo code CAIN \cite{cain}.

The much larger interaction-point spot size and the lower electron beam energy at the LHeC
compared with $\gamma\gamma$ collisions at a linear collider allow placing the  
conversion point at a much greater distance
$\Delta s \approx \beta^{\ast}\sim 0.1$~m from the interaction point, which could
simplify the integration in the detector, and is also necessary as otherwise, with e.g.~
a mm-distance between CP and IP, the conversion would take place inside the proton bunch.

\begin{figure}
\centerline{\includegraphics[angle=0,clip=,width=0.6\textwidth]{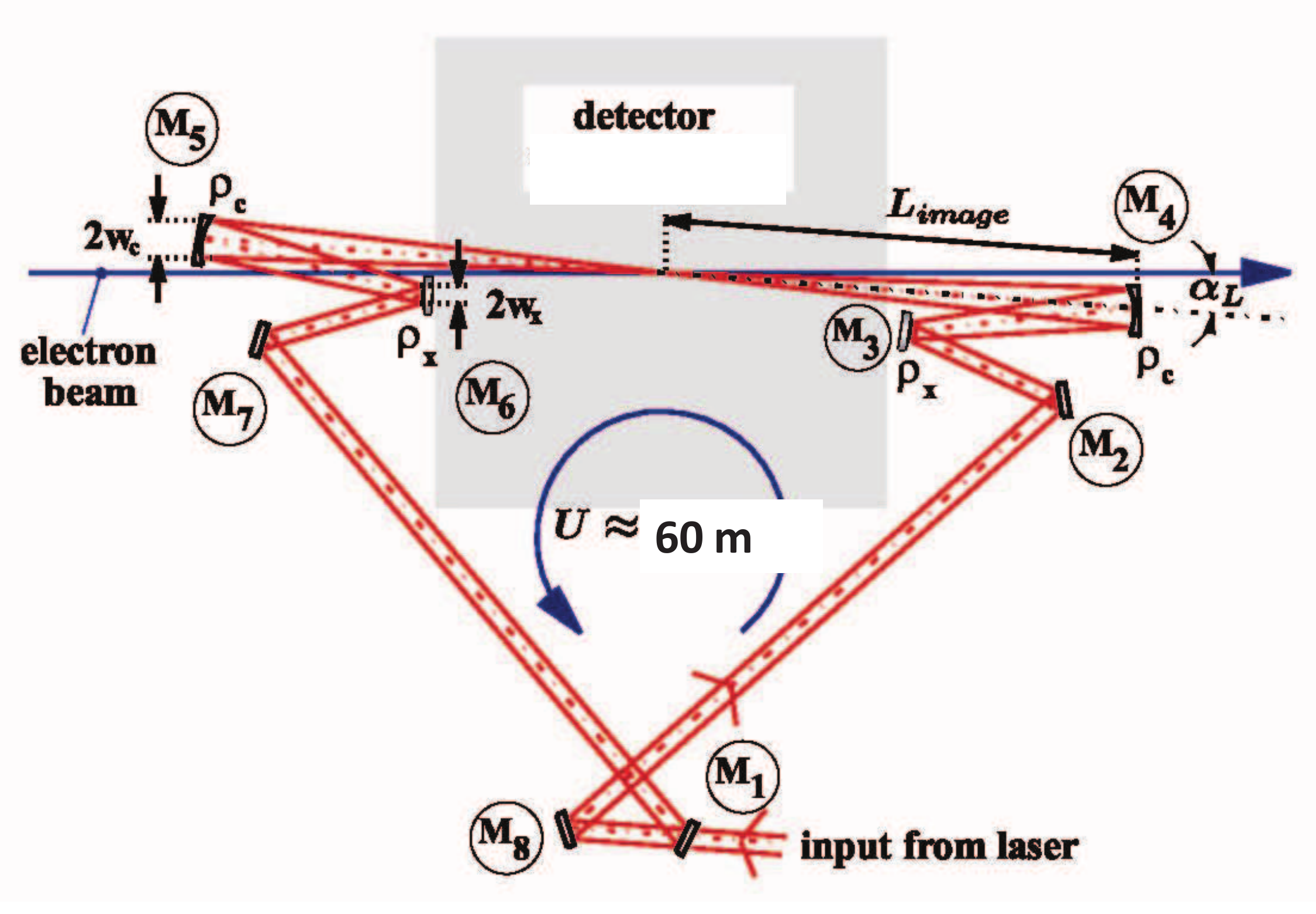}}
\caption{Recirculating mirror arrangement providing a laser-pulse path length 
of 60 m for pulse stacking synchronously with the arriving electron bunches 
(adapted from \protect\cite{klemz}).}
\label{gammap4} 
\end{figure}

To achieve the required laser pulse energy, external pulses can be stacked in a recirculating optical cavity.
For an electron bunch spacing of e.g.~200  ns, the path length of the recirculation could be 
60m. A schematic of a possible mirror system is sketched in Fig.~\ref{gammap4} (adapted from \cite{klemz}).

\subsection{Summary of basic parameters and configurations}
The baseline
 60-GeV ERL option presented here can provide a $ep$ luminosity of 10$^{33}$~cm$^{-2}$s$^{-1}$,
at less than 100 MW total electrical power for the electron branch of the collider, and
with less than 9 km circumference. The 21 GV of SC-RF installation represents its main hardware component. 

A pulsed 140-GeV linac, without energy recovery, could achieve a luminosity 
of $1.4\times 10^{31}$~cm$^{-2}$s$^{-1}$, at higher c.m.~energy, 
again with less than 100 MW electrical
power, and shorter than 9 km in length.
The pulsed linac can accommodate a $\gamma$-$p/A$ option.
An advanced, novel type of energy recovery, proposed for the single straight high-energy linac
case, includes a second decelerating linac, and multiple 10-GeV ``energy-transfer beams''.
This type of collider could potentially reach luminosities of $10^{35}$~cm$^{-2}$s$^{-1}$.

High polarisation is possible for all linac-ring options. 
Beam-beam effects are benign, especially for the proton beam, 
which should not be affected by the presence of the electron beam.

Producing the required number of positrons needed 
for high-luminosity proton-positron collisions 
is the main open challenge for a linac-ring LHeC. 
Recovery of the positrons together with their energy, as well as fast
transverse cooling schemes, are likely to be essential ingredients for any linac-based
high-luminosity $ep$ collider involving positrons. 

%% file: machine/LR-IR.tex
\section{Interaction region}
\label{sec:LR-IR}
This section presents a first conceptual design of the LHeC linac-ring 
Interaction Region (IR). The merits of the IR are a very low $\beta^*$
of 0.1m with proton triplets as close as possible to the IP
to minimise chromaticity. Head-on proton-electron collisions are
achieved by means of dipoles around the Interaction Point (IP).
The Nb$_3$Sn superconductor has been chosen for the proton triplets
since it provides the largest gradient. If this technology proves
not feasible in the timescale of the LHeC a new design of the
IR can be pursued using standard technology.

The main goal of this first design is to
evaluate potential obstacles, decide on the needs of special approaches
for chromaticity correction  and evaluate the impact of 
the IR synchrotron radiation.

\subsection{Layout}
\label{sec:Layout}
\begin{figure}
\centerline{\includegraphics[angle=0,clip=,width=0.8\textwidth]{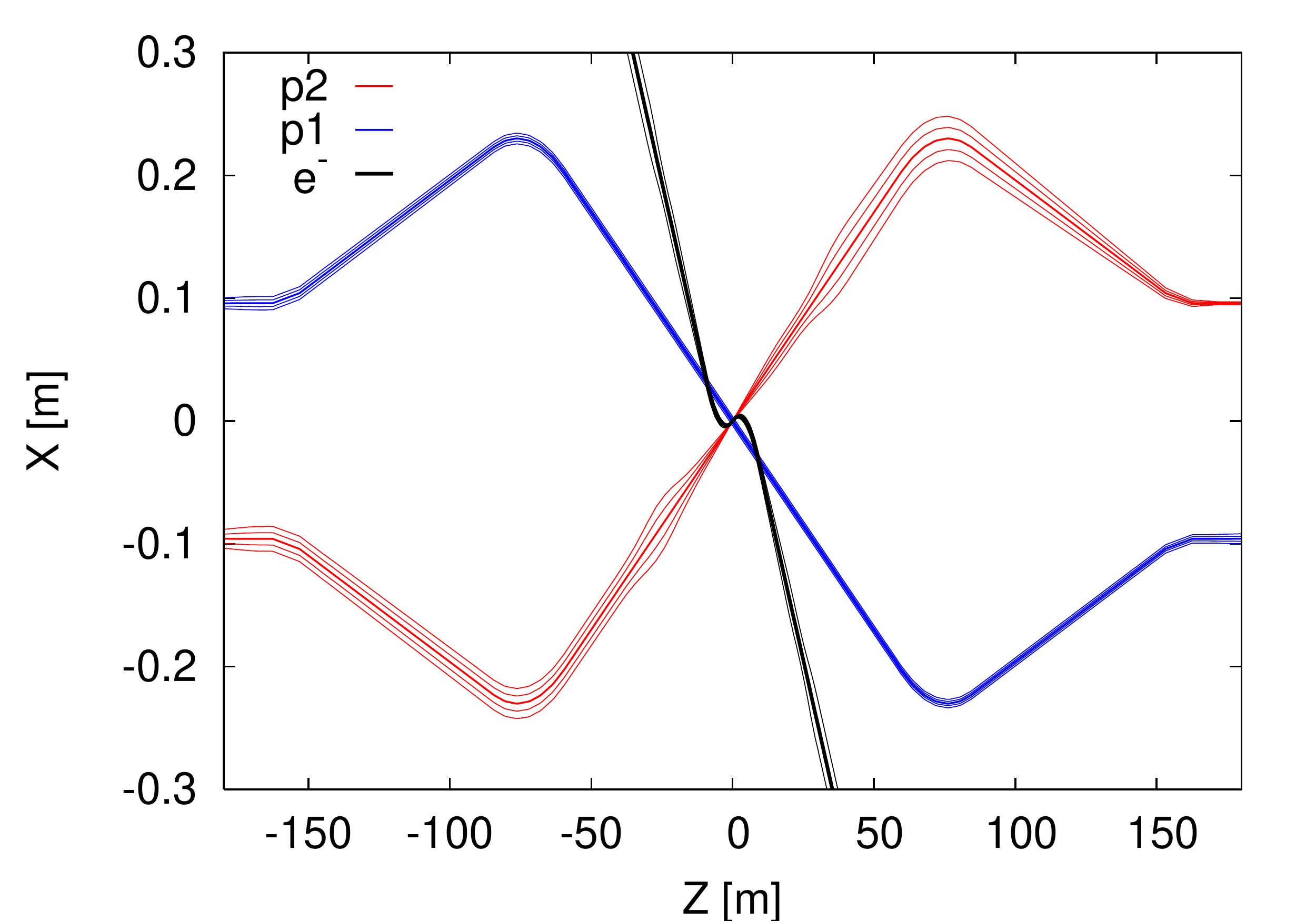}}
\caption{LHeC interaction region displaying the two proton beams
and the electron beam trajectories with 5$\sigma$ and 10$\sigma$ envelopes. }\label{Fig:IR}
\end{figure}

A crossing angle of 6.8~mrad between the non-colliding proton beams 
allows  enough separation to place the proton triplets. Only
the proton beam colliding with the electrons is focused.  A possible
configuration in IR2 could be to inject the electrons parallel
to the LHC Beam 1 and collide them head-on with Beam 2, see Fig.~\ref{Fig:IR}.
The signs of the separation and recombination dipoles (D1 and D2) have
to be changed to allow for the large crossing angle at the IP.
The new  D1 has one aperture per beam and
is 4.5 times stronger than the LHC design D1.
The new D2 is 1.5 times stronger than the LHC design D2.
Both dipoles feature about a 6~T field. 
The lengths of the nominal LHC D1 and D2 dipoles have been left unchanged, 23~m and 9~m, respectively.
However the final IR design will need to incorporate a escape line 
for the neutral particles coming from the IP, probably requiring to split D1 into two
parts separated by tens of metres.

%

\begin{figure}
\centerline{\includegraphics[angle=-90,clip=,width=0.8\textwidth]{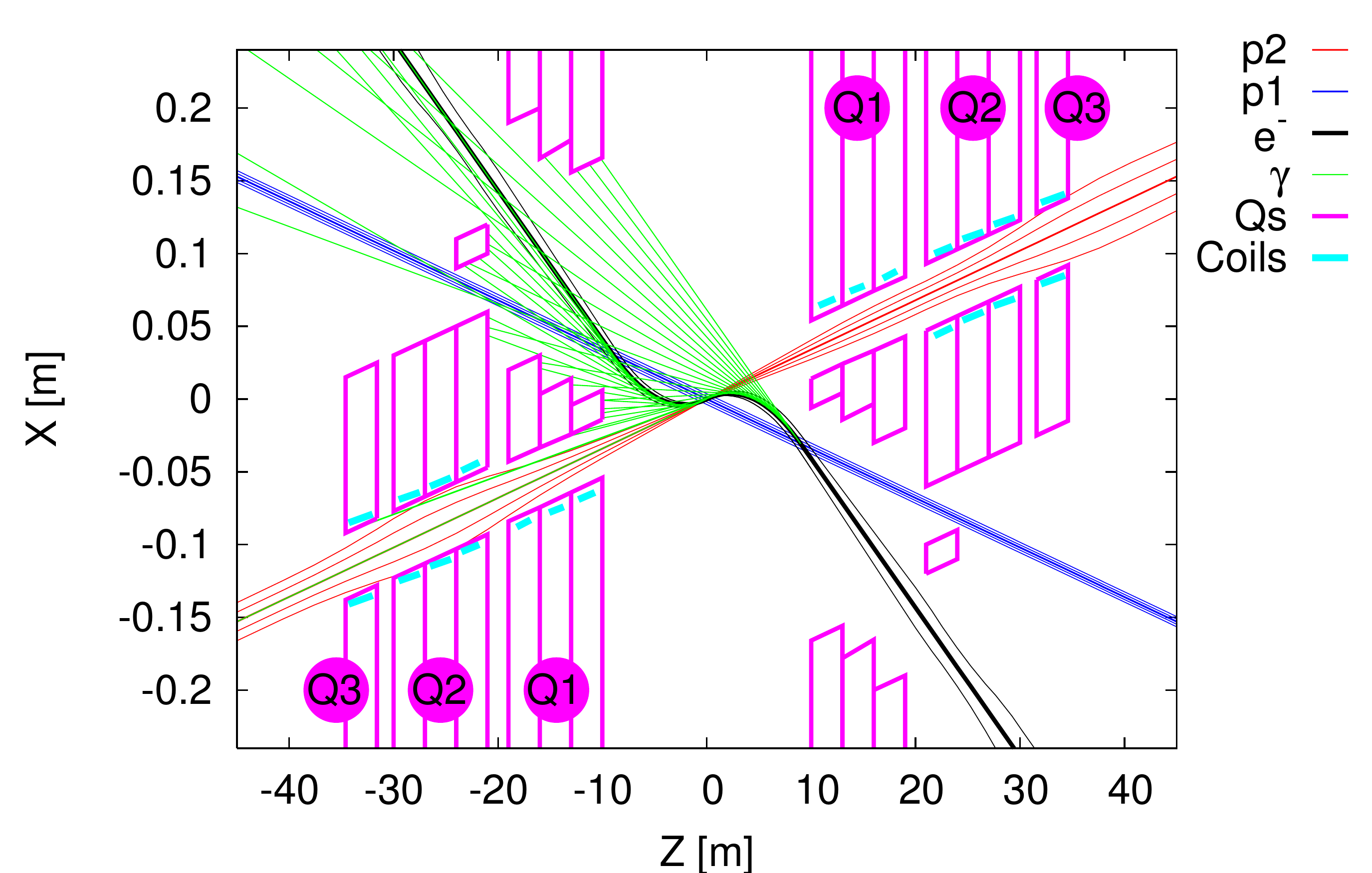}}
\caption{LHeC interaction region with a schematic view of synchrotron radiation. 
Beam trajectories with 5$\sigma$ and 10$\sigma$ envelopes are shown.
The parameters of the Q1 and Q2 quadrupole segments correspond to the Nb$_{3}$Sn half-aperture 
and single-aperture (with holes) quadrupole of Fig.~\ref{srfig4}.
}\label{Fig:IR-SR}
\end{figure}

Bending dipoles around the IP are used to make the electrons  collide head-on with  Beam2
and to safely extract the disrupted electron beam. The required field of these dipoles is
determined by the L$^*$ and the minimum separation of the electron and the focused beam 
at the first quadrupole (Q1). A 0.3~T field extending over 9~m allows for a beams separation
of 0.07~m at the entry of Q1. This separation distance is compatible with mirror quadrupole designs
 using  Nb$_3$Sn technology; see Section\,\ref{tripletmagnets}.
The electron beam radiates 48~kW in the IR dipoles. A sketch of the 3 beams, the synchrotron radiation
fan and the proton triplets is shown in Fig.~\ref{Fig:IR-SR}.

\subsection{Optics}
\label{sec:Optics}

\subsubsection{Colliding proton optics}

\begin{figure}
\centerline{\includegraphics[angle=0,clip=,width=0.8\textwidth]{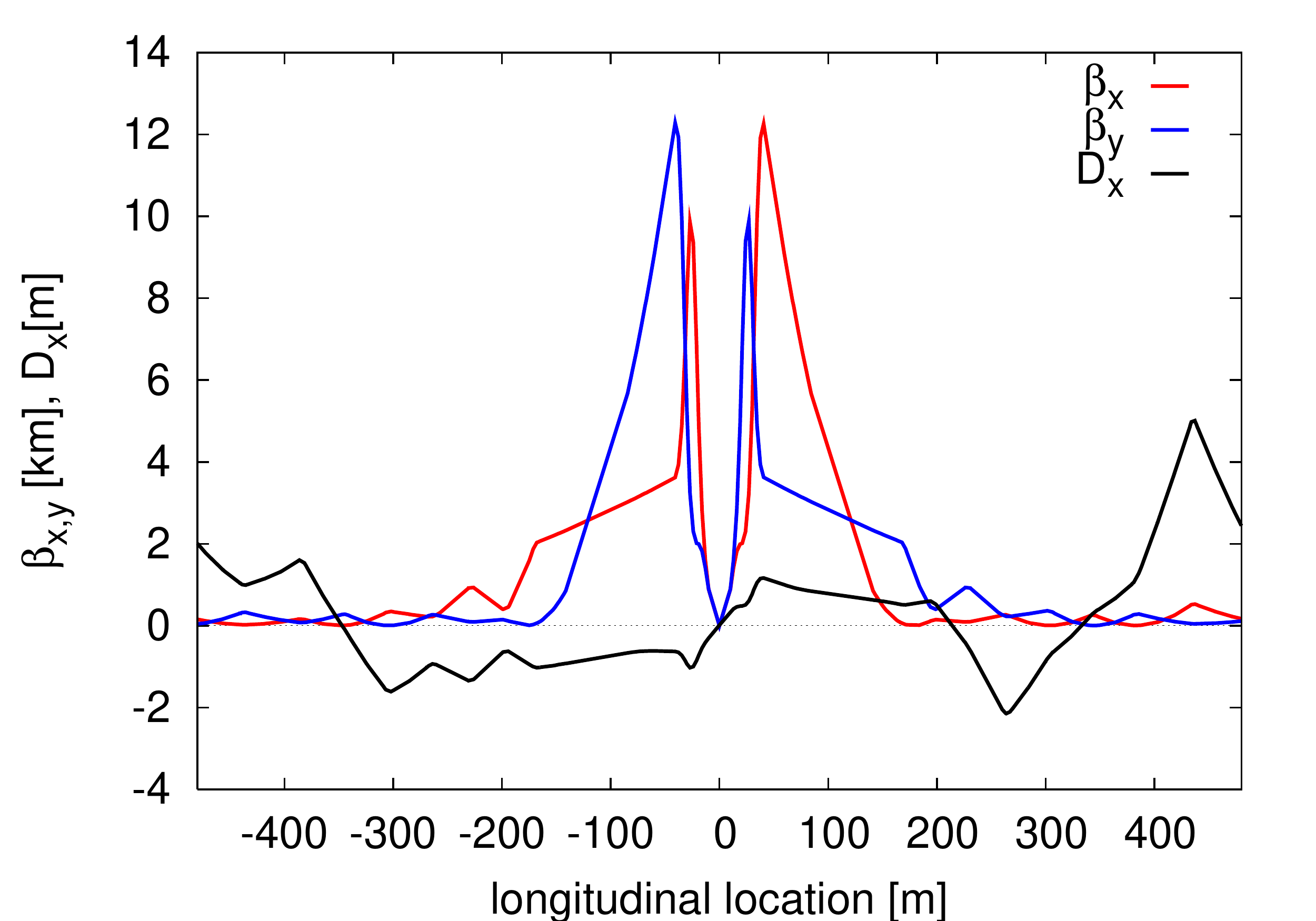}}
\caption{Optics functions for main proton beam.}\label{Fig:p1optics}
\end{figure}

The colliding beam triplet starts at L*=10m from the IP and it
consists of 3 quadrupoles, where the main parameters are given in
Table~\ref{tab:quads}.  The quadrupole aperture is computed as
11max($\sigma_x$,$\sigma_y$)$+5$~mm.  The 5~mm split into 1.5~mm for
the beam pipe, 1.5~mm for mechanical tolerances and 2~mm for the
closed orbit.  The magnet parameters for the first two quadrupoles
correspond to Nb$_3$Sn design described in Section\,
\ref{tripletmagnets}.  The total chromaticity from the two IP sides
amounts to 960 units.  The optics functions for the colliding beam are
shown in~Fig.~\ref{Fig:p1optics}

\begin{table}
\begin{center}
\begin{tabular}{|c|c|c|c|c|c|}\hline
Name & Gradient & Length & Radius & p1-p2 Sep. & ``Radius'' of Field-Free Hole \\
     & [T/m]  & [m]  & [mm] & [mm] & [mm] \\ \hline
Q1 & 187 & 9 & 22 & 63 & 40 \\ \hline 
Q2 & 308 & 9 & 30 & 87 & 26 \\ \hline
Q3 & 185 & 9 & 32 & -- & -- \\ \hline
\end{tabular}
\end{center}
\caption{Parameters of the proton triplet quadrupoles. The radius is computed as 
11max($\sigma_x$,$\sigma_y$)$+5$~mm. For Q2 the hole ``radius'' describes 
the distance from the closest aperture. ``p1-p2 Sep.'' refers to the distance 
between the two proton beams at the entrance of the quadrupole. 
}\label{tab:quads}
\end{table}

\begin{figure}
\centerline{\includegraphics[angle=0,clip=,width=0.8\textwidth]{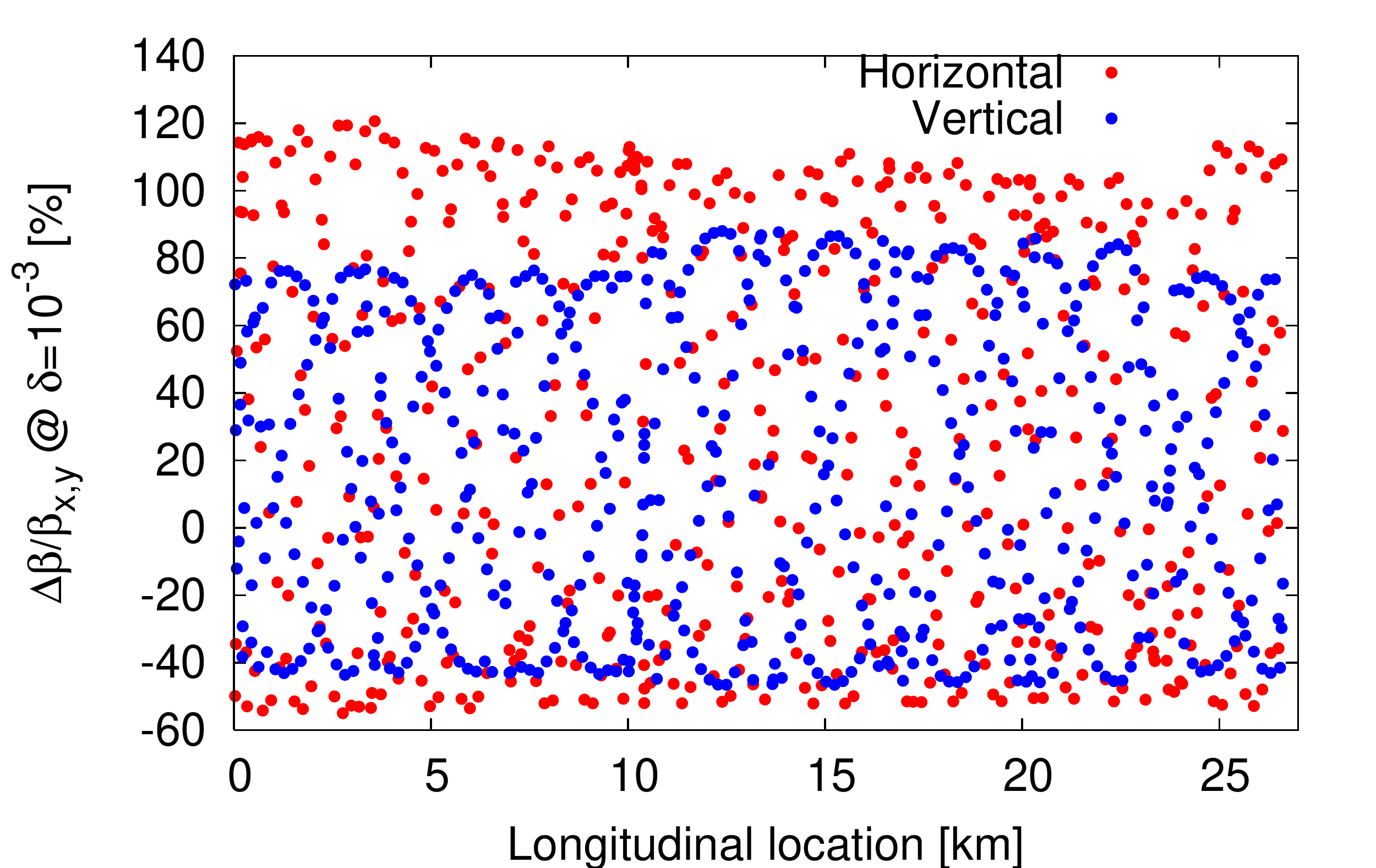}}
\caption{Chromatic beta-beating at dp/p=0.001.}\label{Fig1:p1chromoptics}
\end{figure}

It was initially hoped that a compact Nb$_3$Sn triplet with L$^*$=10m
would allow for a normal chromaticity correction using the
arc sextupoles. However after matching this triplet to the LHC
and correcting linear chromaticity
the chromatic $\beta$-beating at dp/p=0.001 is about 100\% (see Fig.~\ref{Fig1:p1chromoptics}).
This is intolerable regarding collimation and machine protection
issues. Therefore a dedicated chromaticity correction scheme
has to be adopted. A large collection of studies exist 
showing the feasibility of correcting even larger chromaticities 
in the LHC~\cite{Johnstone,Fartoukh38,Fartoukh49}.
Other local chromatic correction approaches as~\cite{pantaleo,jaipac12}, 
where quadrupole doublets are used to provide the strong focusing,
could also be considered for the LHeC.

Since LHeC anyhow requires a new dedicated chromaticity correction scheme,
current NbTi technology could be pursued instead of Nb$_3$Sn
and the L$^*$ could also be slightly increased. The same conceptual
three-beam crossing scheme as in~Fig.~\ref{Fig:IR} could be kept.

To achieve L$^*$ below 23~m requires a cantilever supported
on a large mass as proposed for the CLIC QD0~\cite{gaddi}
to provide sub-nanometre stability at the IP.
The LHeC vibration tolerances are much more relaxed, being on the sub-micrometre
level.

\subsubsection{Non-colliding proton optics}
\begin{figure}
\centerline{\includegraphics[angle=0,clip=,width=0.8\textwidth]{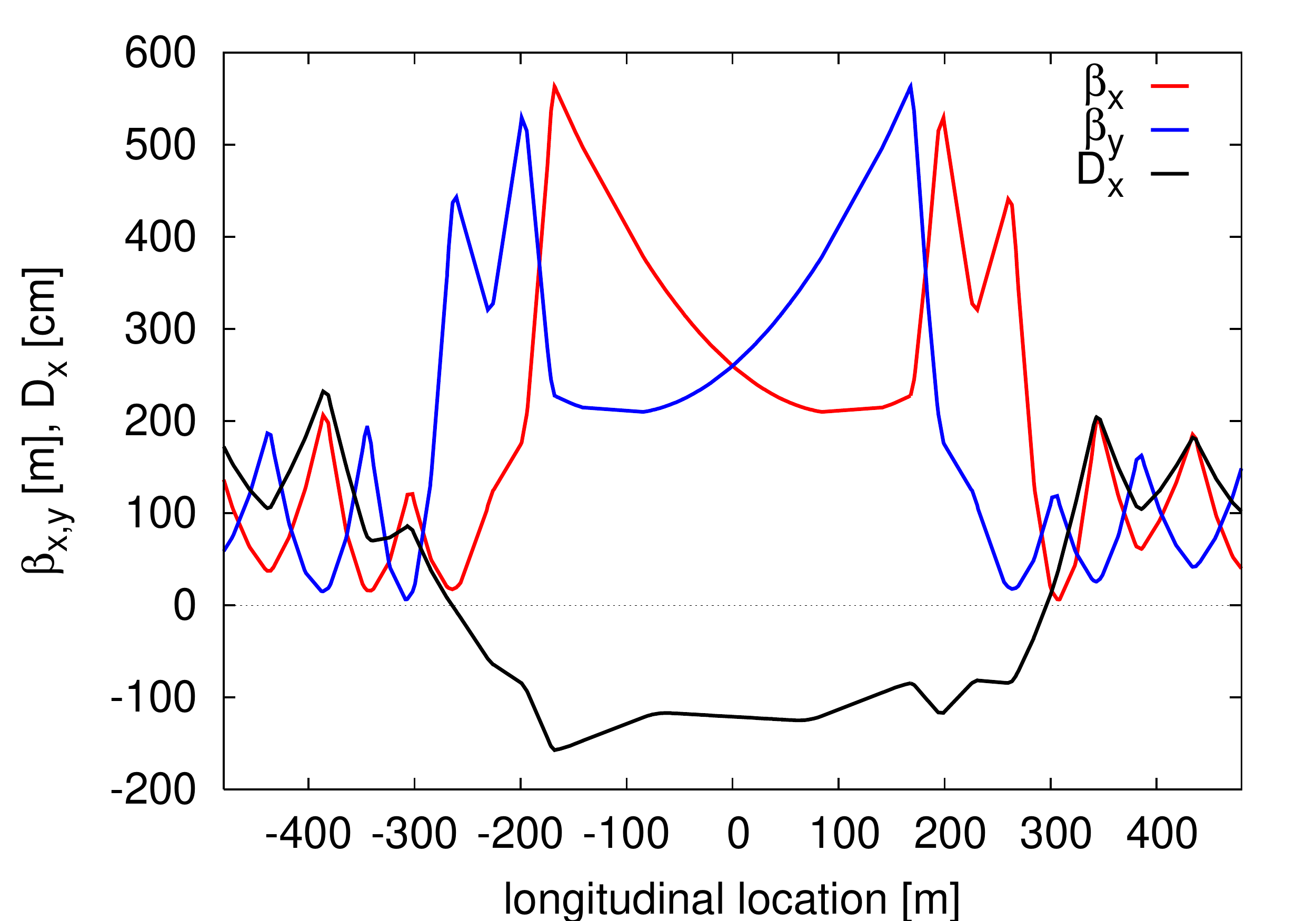}}
\caption{Optics functions for the non-colliding proton beam without triplets.}\label{Fig:p2optics}
\end{figure}

The non-colliding beam has no triplet quadrupoles since it does not
need to be focused. The LHC ``alignment optics''~\cite{alignment}
was used as a starting point. Figure~\ref{Fig:p2optics} shows the optics
functions around the IP.
The LHeC IP longitudinal
location can be chosen so as to completely avoid unwanted proton-proton collisions.
 
The non-colliding proton beam travels through dedicated holes in the proton triplet quadrupoles,
in Q1 together  with the electron beam. 
The Q1 hole dimensions are determined by the
electron beam, see below. By contrast, the non-colliding proton beam travels
alone through the first module of the Q2, requiring about 30~mm full
aperture. No fields are assumed in these apertures but the
possible residual fields could easily be taken into account
for the proton optics.

\subsubsection{Electron optics}
\begin{figure}
\begin{center}
\includegraphics[angle=0,clip=,width=0.59\textwidth]{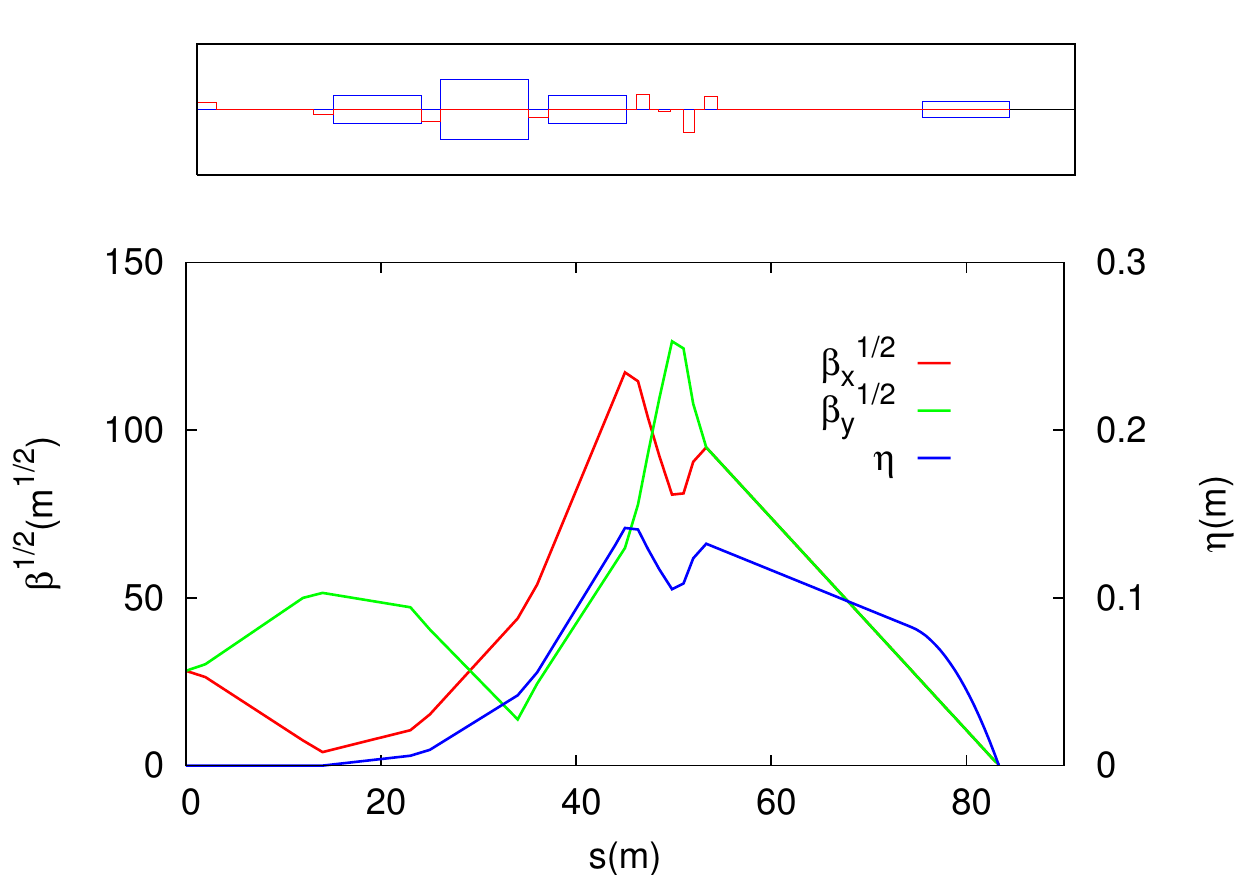}
\includegraphics[angle=0,clip=,width=0.59\textwidth]{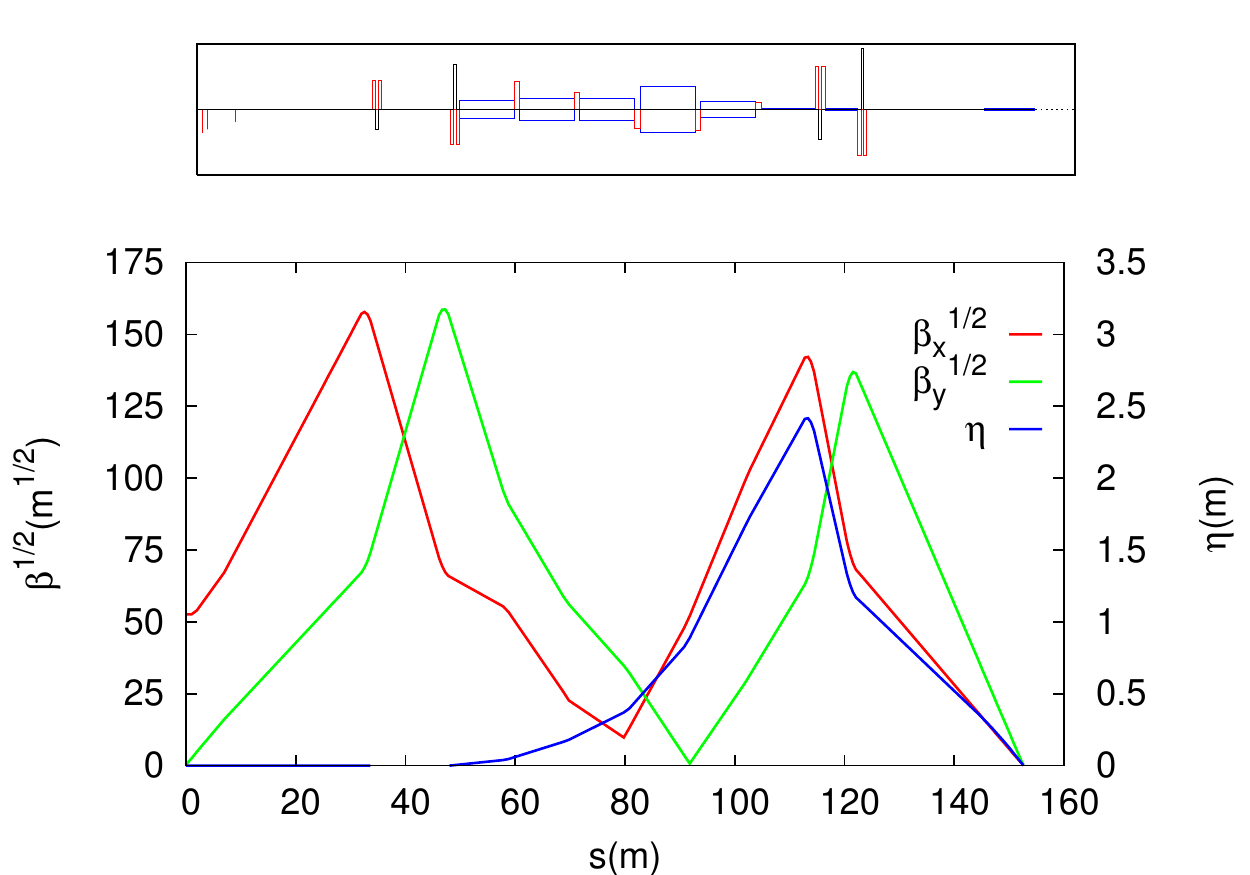}
\includegraphics[angle=0,clip=,width=0.6\textwidth]{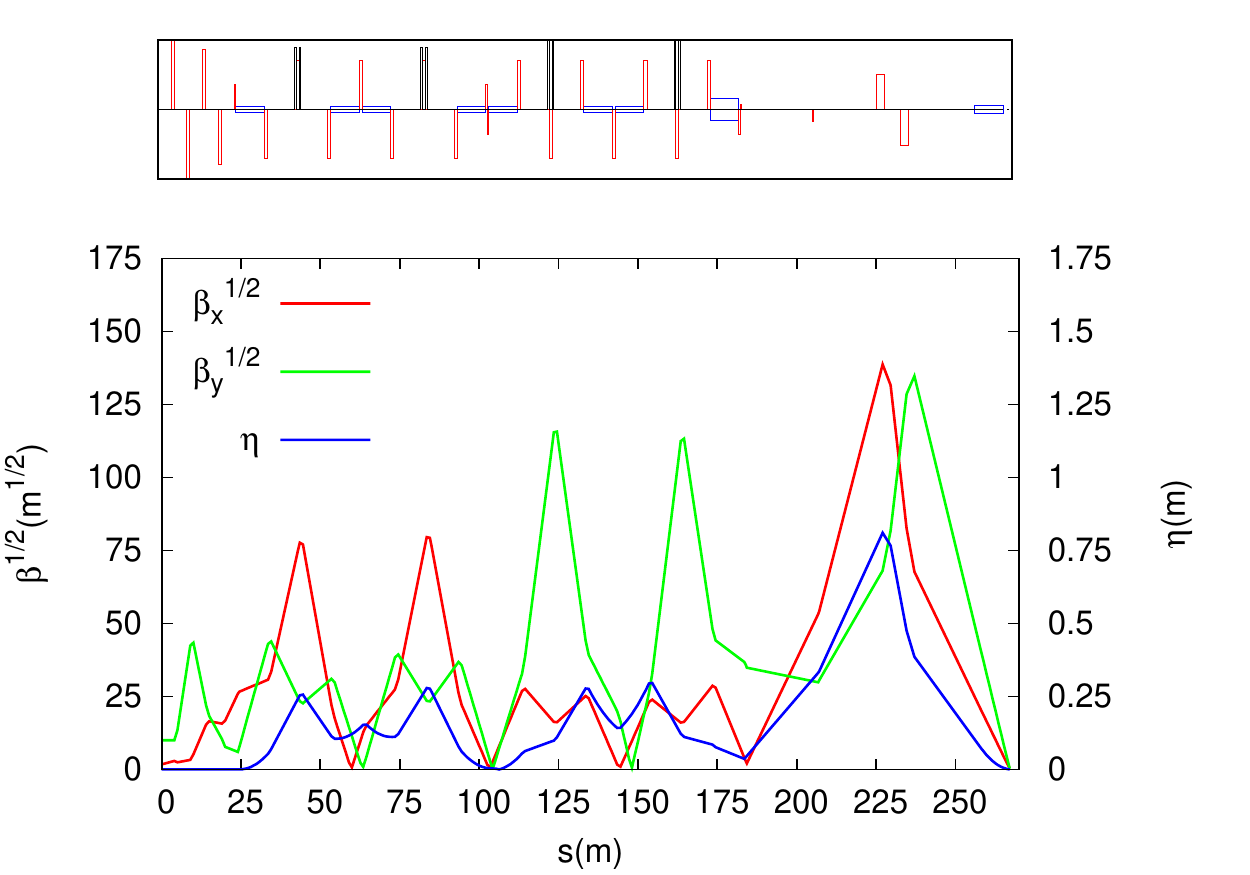}
\end{center}
\caption{Electron final focus optics for the three different options: triplet (top), doublet with local chromatic correction (middle) and doublet with traditional chromatic correction (bottom).}\label{Fig:e-optics}
\end{figure}

About 200~m are available between the exit of the linac and the IP, of which at least 40 m should be
allocated for matching, collimation and beam diagnostics.
On the IP side, a free length $L^*$ of 30~m is chosen to allow for enough separation
between the proton and the electron final focusing quadrupoles.
Respecting these length constraints three alternative 
final-focus optics for the electron beam have been developed.
They are illustrated in Fig.~\ref{Fig:e-optics}.

The first optics is a round-beam electron optics with $\beta_{e;x,y}^{*}=0.1$ m realised by a plain triplet
without any sextupoles (Fig.~\ref{Fig:e-optics} top picture). 
Upstream bending magnets complement the separation dipole so as to match the dispersion at the IP. 
The total length is 90~m. The SR power is small, about 25 kW on the incoming side 
of the IP, coming almost entirely from the separation dipole before the collision point.
Without any chromatic correction 
the IP beam size increase for an rms 
relative momentum spread of $3\times 10^{-4}$ is about 10\% horizontally and 21\%
vertically.

\begin{figure}
\begin{center}
\includegraphics[angle=0,clip=,width=0.45\textwidth]{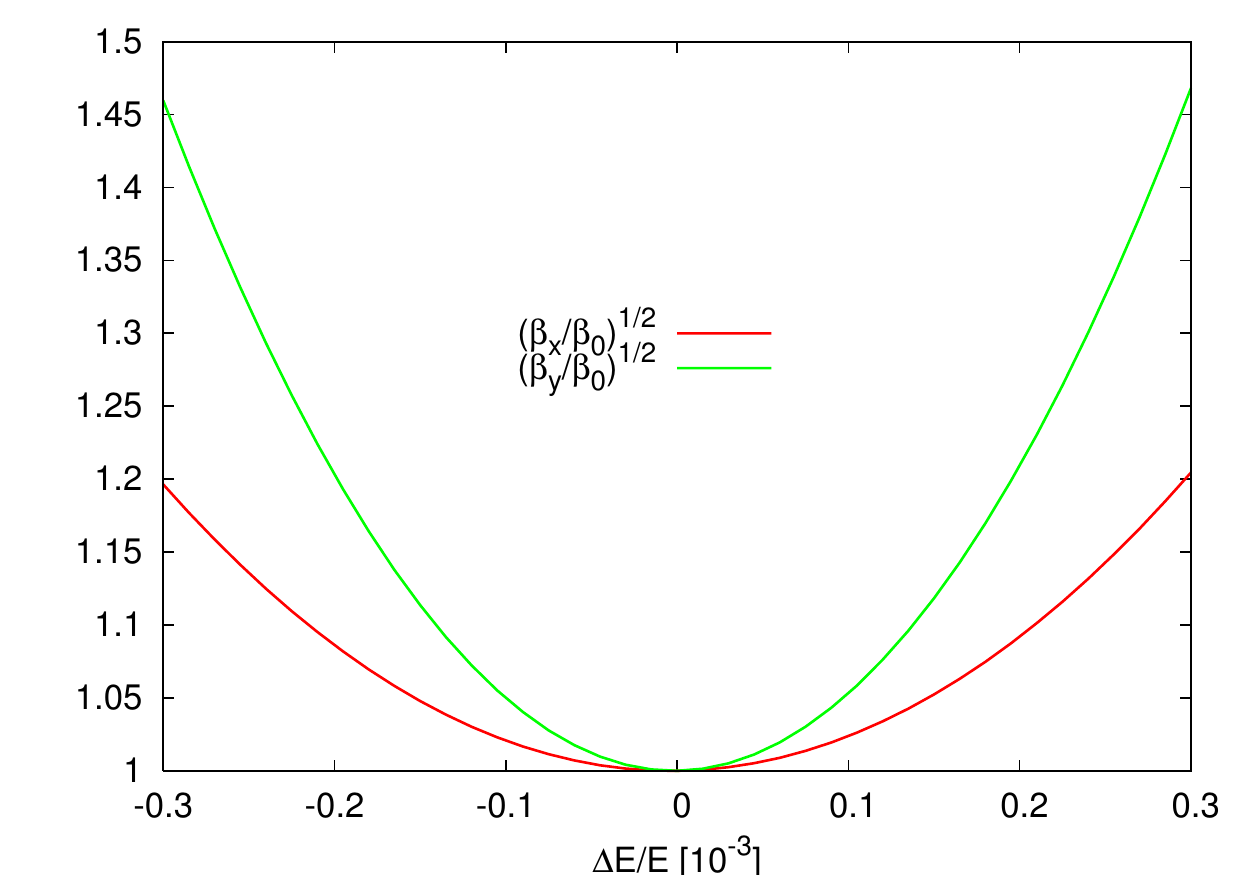}
\includegraphics[angle=0,clip=,width=0.45\textwidth]{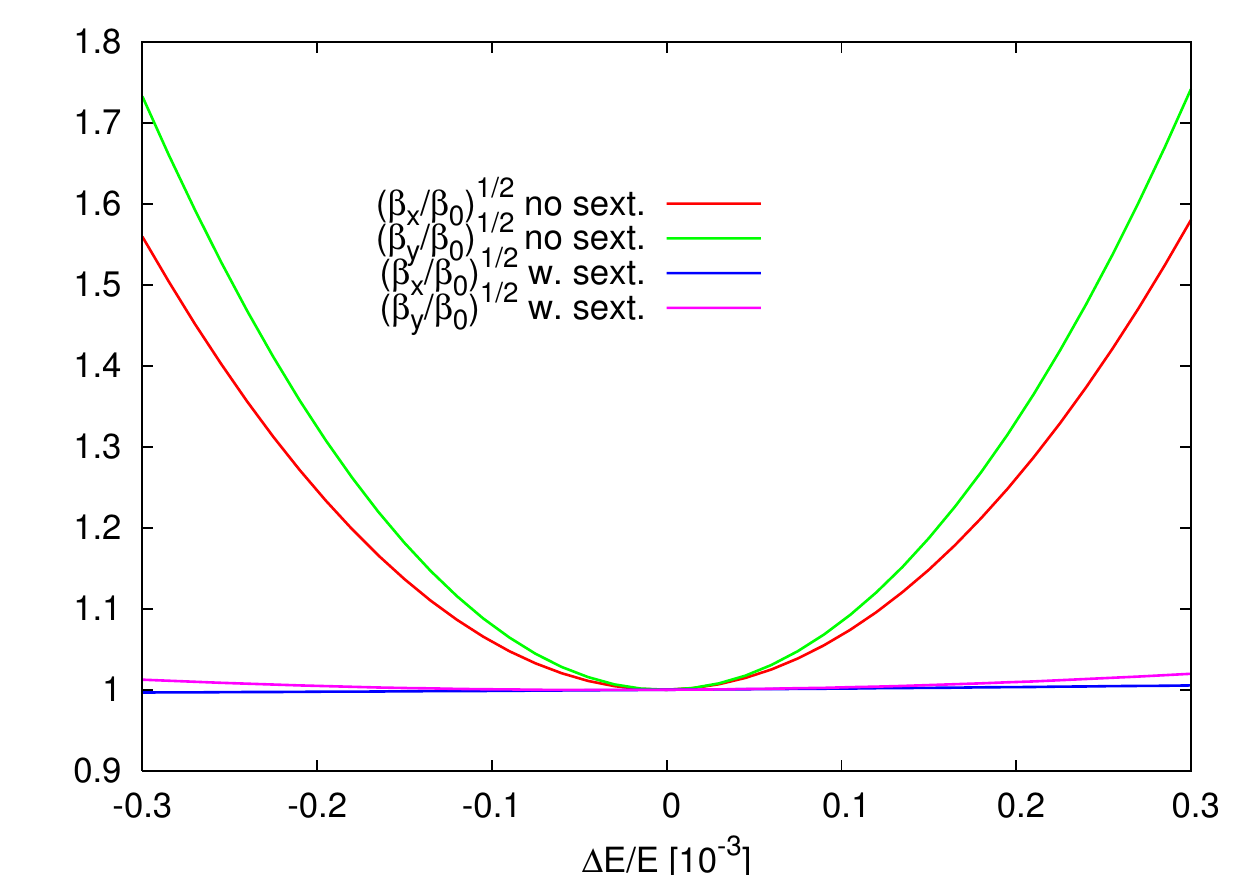}
\end{center}
\caption{Relative increase in the linear beam size ($\sqrt{\beta}$) as a function 
of beam energy error for the triplet and doublet-local options, as computed by MAD-X.}    
\label{Fig:linearbandwidth}
\end{figure}

The second optics~\cite{ipac11ja} employs a final quadrupole doublet with local chromatic correction 
using 4 sextupoles arranged according to the ``compact
final-focus'' scheme proposed for future linear colliders~\cite{pantaleo} (Fig.~\ref{Fig:e-optics} centre picture). 
It is optimised for unequal IP beta functions $\beta_{e;x}^{*}=0.2$ m and $\beta_{e;y}^{*}=0.05$ ,
which are more suitable for a final doublet.
In order to correct the chromaticity without generating unacceptable residual geometric
aberrations a sufficiently large dispersion is needed across the final quadrupoles. 
Achieving this without introducing too much synchrotron radiation requires a longer system.
The actual doublet optics has a length of 150~m. The SR power is 84~kW 
for the entire final focus on the incoming side of the IP, of which only about 
one third, 24 kW, is due to last separation dipole, with (at least) 
the same 24 kW again on the outgoing side. 
With this optics the IP beam size increase for an rms 
relative momentum spread of $3\times 10^{-4}$ is about 0.2\% 
horizontally and 1.3\% vertically, only due optical aberrations.
However, synchrotron radiation increases the horizontal beam size by 138\%.
A future optimisation of the location and strength of the bending magnets may 
improve this figure. 
The linear momentum bandwidths for the triplet and the doublet-local  optics are compared in
Fig.~\ref{Fig:linearbandwidth}. The bandwidth was computed by MAD-X
for a mono-chromatic beam with zero energy spread and varying 
offset from the design beam energy.  
These plots reveal the benefit of a chromatic correction.

The third optics~\cite{ipac12ja} employs a final quadrupole doublet with a traditional modular scheme
for the chromaticity correction (Fig.~\ref{Fig:e-optics} bottom picture). 
This implies having dedicated sections for the correction
of the horizontal and vertical chromaticities, thus requiring an even longer system.
The $\beta$-functions at the IP are $\beta_x^\ast=0.2$~m and $\beta_y^\ast=0.05$~m 
and the total length of the system is $L_\text{FFS}=267.1$~m. 
The linear spot size is $  \sigma_x^\ast=9.23\; \mu$m 
and  $\sigma_y^\ast=4.61\; \mu$m and including nonlinear effects and after correction the beam 
sizes are  $\sigma_x^\ast=10.48\; \mu$m and  
$\sigma_y^\ast=5.66 \mu$m.
In other words, the beam size increases by a $10\%$ in the horizontal plane 
and $25\%$ in the vertical plane due to the non-linearities. 
The compensation of the nonlinear effects is not optimum, 
because the strength of the dipoles was lowered in order to reduce the 
synchrotron-radiation effects, and the  
system was optimised  by finding the minimum beam size 
while varying the dispersion in the sextupoles. 
The final radiated power due to synchrotron radiation is $49$ kW. 
The radiation increases the 
horizontal spot size to $ \sigma_x^\ast=12.8\; \mu$m.

The optics of the three systems are shown in Fig.~\ref{Fig:e-optics},
already matched to the exit of the linac. 
The electron focusing quadrupoles feature moderately low gradients as shown in Table~\ref{tab:e-quads}. 

The higher-order aberrations for the three optics 
were analysed and minimised by applying a combination of the codes 
MAD-X/PTC and MAPCLASS \cite{mapclass} with and without 
the effect of synchrotron radiation. 
Table \ref{tab:e-aberrations} summarises the relative 
beam-size increase for the three optics together with an estimate of the luminosity loss
based on the geometric overlap of unequal beams.

\begin{table}
\begin{center}
\begin{tabular}{|c||c|c|c||c|c|c||c|c|c|}\hline
& \multicolumn{3}{c||}{triplet} & \multicolumn{3}{c|}{doublet - local}& \multicolumn{3}{c|}{doublet - traditional}
\\ \hline
Name & Grad. & Len. & Rad. & Grad. & Len. & Rad. & Grad. & Len. & Rad.
\\
& [T/m] & [m] & [mm] & [T/m] & [m] & [mm]& [T/m] & [m] & [mm]
\\\hline
Q1 & 19.7 & 1.34 &  20 & $-19.1$ & 1.1 & 36     & -20.54 & 2.5 &  36      \\\hline
Q2 & $-38.8$ & 1.18 &  32 & $17.7$ & 1.1 & 37   & 20.31  & 2.5 &  35        \\\hline
Q3 & $-3.46$ & 1.18 &  20 & $-14.7$ & 1.1 & 41  & -6.59  & 0.3 &  17      \\\hline
Q4 & 22.3 & 1.34 &  22 &  11.8 & 1.1 & 41       & 2.85   & 0.3 &  13    \\\hline
\end{tabular}
\end{center}
\caption{Final electron quadrupole parameters for the triplet and the 2 doublet optics.
The radius is computed as 11 max($\sigma_x$,$\sigma_y$)$+5$~mm.\label{tab:e-quads}
In the doublet solution the third and fourth quadrupole, Q3 and Q4, are located
further upstream.}
\end{table}

\begin{table}[h!]
\begin{center}
\vspace{0.1cm}
\begin{tabular}{|c|c|c|c|}\hline
~~~~~~~~~~~~~~~~~~~~~~~~~~~~~~~~~~~~~~~~ & triplet &   doublet - local & doublet - traditional    \\\hline
$\Delta{\sigma}_x/\sigma_{x,0}$, no SR &   9$\%$ &  1.5$\%$ &     5.75$\%$     \\\hline
$\Delta{\sigma}_y/\sigma_{y,0}$, no SR &   21$\%$ &  1.7$\%$ &     14.1$\%$     \\\hline
$\Delta{\sigma}_x/\sigma_{x,0}$, with SR &   10$\%$ &  141$\%$ &    39.3$\%$       \\\hline
$\Delta{\sigma}_y/\sigma_{y,0}$, with SR &   21$\%$ &  1.9$\%$ &   14.3$\%$       \\\hline
$\Delta L/L_0$, with SR                  &   $-14\%$ &    $-46\%$ &    $-23\%$   \\\hline
\end{tabular}
\label{tab:e-aberrations}
\end{center}
\caption{Relative IP electron beam-size increase
with respect to the linear spot size $\sigma_{0,x(y)}=\sqrt{\epsilon_{x(y)}\beta_{x(y)}^*}$
considering a Gaussian momentum distribution of $\delta_{\rm rms}=3\times 10^{-4}$.
An indication of the luminosity loss due to the geometric overlapping of unequal proton and
electron beams is also given.}
\vspace{-0.2cm}
\end{table}


The electrons share a hole with the non-colliding proton beam in the
first half-quadrupole, Q1, and then travel through a dedicated hole in the cryostat of Q2.
The common hole in the proton Q1 must have about 160~mm full 
horizontal aperture to allow for the varying separation between 
the electron and non-colliding proton orbit (120~mm) with 
the usual electron-beam aperture assumptions ($\pm$20~mm). 
First design of mirror magnets for Q1 feature a field of 0.5~T in the electron
beam pipe. This value is considered too large when compared to the IR dipole
of 0.3~T, but new designs with active isolation or dedicated coils
could considerably reduce this field. Migrating to NbTi technology
would reduce this field too.

\vspace{-0.1cm}
\subsubsection{Spent electron beam}
\begin{figure}
\centerline{\includegraphics[angle=0,clip=,width=0.8\textwidth]{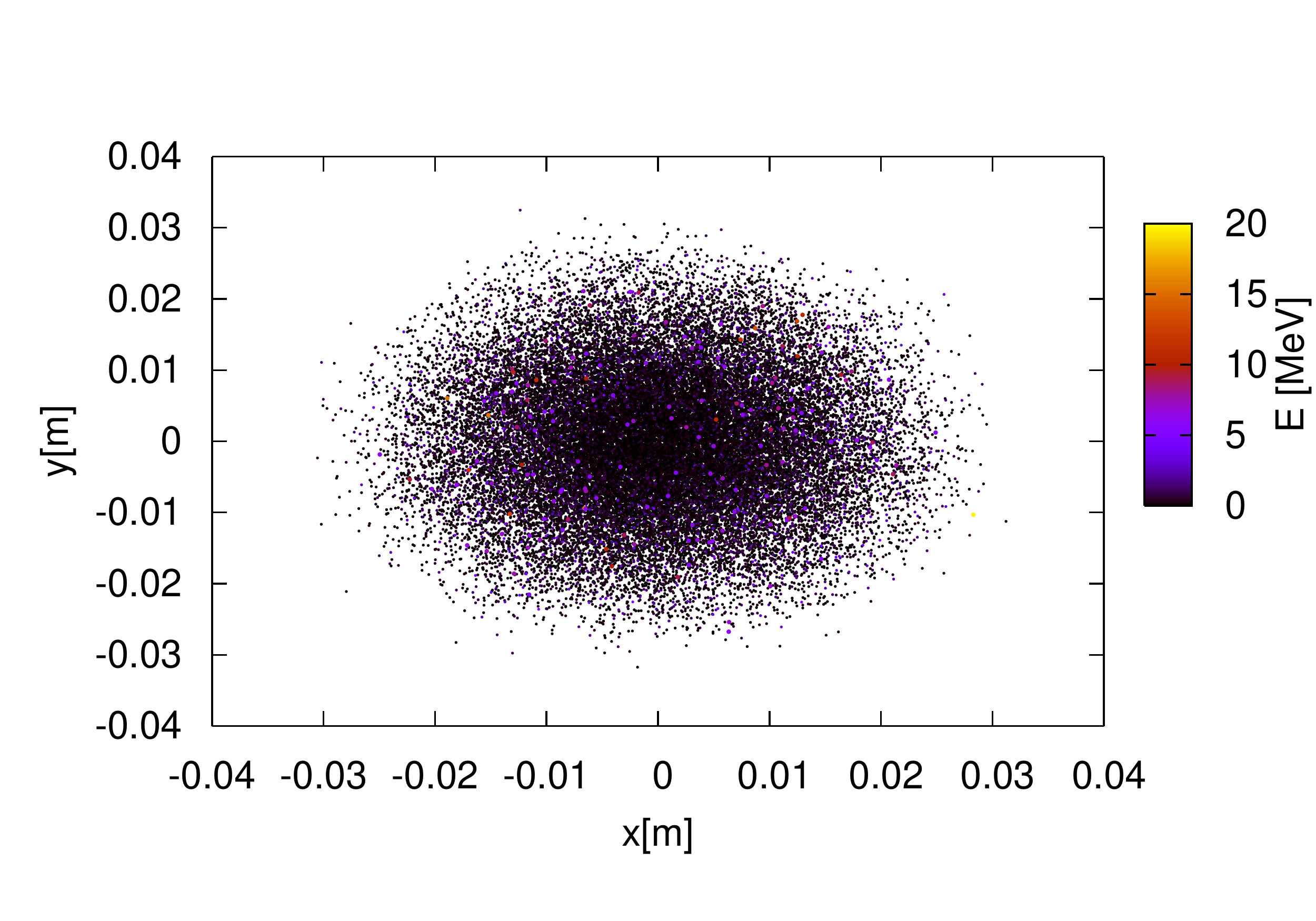}}
\caption{Distribution of the spent electron beam at 10~m from the IP. The Gaussian and rms sigmas are shown on the plot.}\label{Fig:e-spend}
\end{figure}

The electromagnetic field pf the proton beam during the collision
provides extra focusing for the electron beam. 
This increases the divergence of the spent electrons. 
Figure~\ref{Fig:e-spend} shows 
the horizontal distribution of the electrons at 10~m from the IP (entry of Q1)
as computed by GuineaPig~\cite{guinea}. The contribution of 
dispersion and energy spread to the transverse size of the exiting collided beam 
can be neglected. 
Therefore, it is possible to linearly scale the sigmas at 10~m to estimate
both the horizontal and vertical sigmas at any other longitudinal location.
The simulation used $10^{5}$ particles. No particles are observed beyond 4.5~mm
from the beam centroid at 10~m from the IP and beyond 9~mm at 20~m.
A radial aperture of 10~mm  has been reserved for the beam size at the incoming electron Q1 hole.
The same value of 
10~mm seem to be enough to also host the spent electron beams, although it might
be worth to allocate more aperture margin in the last block of Q1.

\subsection{Modifications for $\gamma$p or $\gamma$-A}
\label{sec:gammap} 
The electron beam can be converted into photons by Compton scattering off  
a high-power laser pulse, as discussed Section\,\ref{gammap}. 
For this option a laser path and high-finesse optical cavities 
must be integrated into the interaction region.
A multiple mirror arrangement has been sketched in Fig.~\ref{gammap4}.
The 0.3-T dipole field after the (now) $\gamma$-p interaction point 
will help to separate the Compton-scattered spent electron beam 
from the high-energy photons. 
The high-energy photons 
propagate straight into the direction of the incoming 
proton beam through the main openings of Q1 and Q2, while the 
spent electrons will be extracted through the low-field
exit holes shared with the non-colliding proton beam, as
for electron-proton collisions.

%% file: machine/bernard.tex
\subsection{Synchrotron radiation and absorbers}
\label{LHEC:machine:linac:bernard}

\subsubsection{Introduction}
	
The synchrotron radiation (SR) in the linac-ring interaction region
has been analysed by three different approaches.  The SR was simulated
using a program made with the GEANT4 (G4) toolkit.  In addition, a
cross check of the total power and average critical energy was done in
IRSYN, a Monte Carlo simulation package written by
R.~Appleby~\cite{rob}.  A final cross check of the radiated power has
been performed using an analytic method.  The latter two checks
confirmed the results obtained from G4.  The G4 program uses Monte
Carlo methods to create the desired Gaussian spatial and angular
distributions of an electron beam.  This electron beam distribution is
then transported through a ``vacuum system,'' including the magnetic
fields for the separator dipoles.  In a non-zero magnetic field SR is
generated using the appropriate G4 process classes.  The position,
direction, and energy of each photon emitted is written as ntuples at
user defined longitudinal positions ($Z$ values).  These ntuples are
then used to analyse the SR fan as it evolves in $Z$.  The latter
analysis was done primarily through MATLAB scripts.
	
	This section uses the following conventions. 
The electron beam is being referred to as \emph{the beam} 
and the proton beams will be called either the interacting or non interacting proton beams. 
The (electron) beam propagates in the $-Z$ direction and the interacting proton beam propagates 
in the $+Z$ direction. At the collision point both beams propagate in the straight $Z$ (or $-Z$) direction.
A right-handed coordinate system is used where 
the $X$ axis is horizontal and the $Y$ axis vertical. 
The beam centroid always remains in the $Y = 0$ plane.  
The \emph{angle of the beam} will be used to refer 
to the angle between the beam centroid's direction 
and the $Z$ axis, in the $Y = 0$ plane. 
This angle is defined such that the beam propagates in the $-X$ 
direction when it passes through the dipole field 
as it moves along $Z$.
	
	 The SR fan's extension in the horizontal direction 
is determined by the angle of the beam at the entrance of the upstream separator dipole. 
Because the direction of the photons is parallel to the direction 
of the electron from which it is emitted, 
the angle of the beam and the $X$-distance to the interacting proton beam at 
the $Z$ location of the last proton quadrupole
are both greatest for photons generated 
at the entrance of the upstream separator dipole and, therefore, 
this angle defines one of the edges of the synchrotron fan on the absorber in front of the proton quadrupole. 
The other edge is defined by the crossing angle, which is zero for the linac-ring option. 
The S shaped trajectory of the beam means that the smallest angle of the beam will be reached at the IP. 
Therefore, the photons emitted at this point will move exactly along the $Z$ axis. 
This defines the other edge of the fan in the horizontal direction.
	 
	  The SR fan's extent  
in the vertical direction is determined by the beta function and angular spread of the beam. 
The beta function along with the emittance defines the local rms beam size. 
The vertical rms beam size characterises the range of $Y$ positions at which photons are emitted.
Possibly more importantly, the vertical angular spread defines the angle 
between the velocity vector of these photons and the $Z$ axis. 
Both of these dependencies are functions of $Z$. 
Similar effects also affect the horizontal extension of the SR fan, however, 
in the horizontal plane they are of second order when compared to the 
horizontal deflection angle in the strong dipole field. 
	
	The number density distribution of the SR fan is inferred from the simulations.  
The number density at the location of the absorber is highest in the region between the two interacting beams. 
This is due to the S shaped trajectory of the beam.

\subsubsection{Parameters}

	\indent
	The parameters for the Linac Ring option are listed in Table \ref{tab:ParamLR}. 
The separation refers to the displacement between the two interacting beams at the face of the proton triplet. 
	
\begin{table}[h]
  \centering	
\begin{tabular}{| c | c |}
  \hline
Characteristic & Value \\
\hline
\hline
Electron Energy [GeV] & 60  \\
\hline
Electron Current [mA] & 6.6  \\
\hline
Crossing Angle [mrad] & 0 \\
\hline
Absorber Position [m] & -9 \\
\hline
Dipole Field [T] & 0.3 \\
\hline
Separation [mm] & 75 \\
\hline
$\gamma/s$ & $ 1.37 \times 10^{18}$\\
  \hline
\end{tabular}
\caption{LR: Parameters}
\label{tab:ParamLR}
\end{table}

	The energy, current, and crossing angle ($\theta_{c}$) are the common values used in all LR calculations. The B value refers to the constant dipole field created throughout the two dipole magnets in the IR. The direction of this field is opposite on either side of the IP. The field is chosen such that 75 mm of separation is reached by the face of the proton triplet. This separation was chosen based on S. Russenschuck's SC quadrupole design~\cite{schavannes}. The separation between the interacting beams can be increased by raising the constant dipole field however for a dipole magnet $P_{SR} \propto |B^2|$~\cite{nathan}, therefore an optimisation of the design would need to be discussed. The chosen parameters give a flux of $1.37 \times 10^{18}$ photons per second at Z = -9 m.

\subsubsection{Power and critical energy}

	Table \ref{tab:PCELR} shows the power of the SR produced in the IR along with the critical energy. This is followed by the total power produced in the IR and the critical energy. Since the G4 simulations utilise Monte Carlo, multiple runs were used to provide a standard error. This only caused fluctuations in the power since the critical energy is static for a constant field and constant energy. 

\begin{table}[h]
  \centering
\begin{tabular}{| c | c | c | }
  \hline
  Element &  Power [kW] & Critical Energy [keV]\\
\hline\hline
 DL & 24.4 +/- 0.1 &   718   \\
 \hline
 DR & 24.4 +/- 0.1 &   718  \\
 \hline\hline
 Total & 48.8 +/- 0.1 & 718  \\
\hline
\end{tabular}
\caption{LR: Power and Critical Energies as calculated with GEANT4.}
\label{tab:PCELR}
\end{table}

	These magnets have strong fields and therefore produce high critical energies and a substantial amount of power. Although the power is similar to that of the RR design the critical energy is much larger. This comes from the linear dependence of critical energy on magnetic field ($i.e.$ $E_c \propto B$) \cite{SREcrit}. With the dipole field in the LR case being an order of magnitude larger than the dipole fields in the RR case the critical energies from the dipole magnets are also an order of magnitude larger in the LR case. 
	
\subsubsection{Comparison}
	The IRSYN cross check of the power and critical energies is shown in Table \ref{tab:GIRSYNLR}. This comparison was done for the total power and the critical energy. 
	
	\begin{table}[h]
  \centering
  \begin{tabular}{| c | c | c | c | c |}
 \hline
 &   \multicolumn{2}{|c|}{Power [kW]} & \multicolumn{2}{|c|}{Critical Energy [keV]} \\
\hline\hline
 & GEANT4 & IRSYN & GEANT4 & IRSYN \\
 \hline
 Total & 48.8 +/- 0.1 & 48.8 & 718 & 718\\
  \hline
  \end{tabular}
\caption{LR: GEANT4 and IRSYN comparison.}
\label{tab:GIRSYNLR}
\end{table}
	
	A third cross check to the GEANT4 simulations was made for the power as shown in Table \ref{tab:GALR}. This was done using an analytic method for calculating power in dipole magnets \cite{nathan}.
	
	\begin{table}[h]
  \centering
  \begin{tabular}{| c | c | c | }
 \hline
 &   \multicolumn{2}{|c|}{Power [kW]}  \\
\hline\hline
Element & GEANT4 & Analytic \\
\hline
 DL & 24.4 +/- 0.1 &  24.4    \\
 \hline
 DR &24.4 +/- 0.1 &   24.4  \\
 \hline
 Total/Avg & 48.8 +/- 0.1 & 48.8 \\
  \hline
  \end{tabular}
\caption{LR: GEANT4 and Analytic method comparison.}
\label{tab:GALR}
\end{table}

\subsubsection{Number density and envelopes}
	
	\begin{figure}[!h!t!b]
\centerline{\includegraphics[clip=,width=1.0\textwidth]{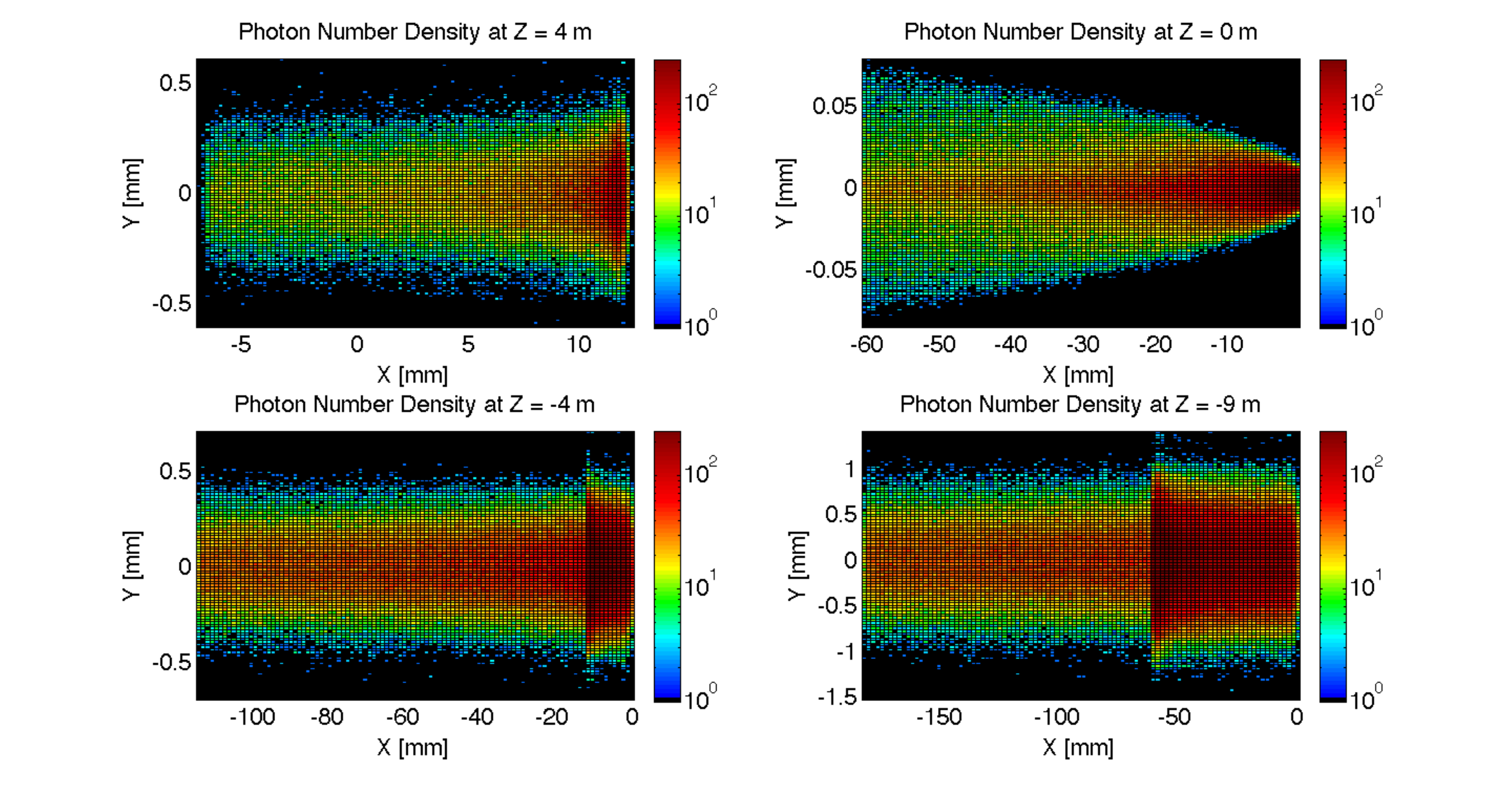}}
\caption{LR: Number Density of photons Growth in Z direction.}
\label{Fig:LRnumdens}
\end{figure}

	The number density of photons at different Z values is shown in Figure \ref{Fig:LRnumdens}. Each graph displays the density of photons in the $Z=Z_o$ plane for various values of $Z_o$. The first three graphs give the growth of the SR fan inside the detector area. This is crucial for determining the dimensions of the beam pipe inside the detector area. Since the fan grows asymmetrically in the -Z direction an asymmetric elliptical cone shaped beam pipe will minimise these dimensions, allowing the tracking to be placed as close to the beam as possible. The horizontal extension of the fan in the LR option is larger than in the RR case. This is due to the large angle of the beam at the entrance of the upstream separator dipole. As mentioned in the introduction this angle defines the fans extension, and in the LR case this angle is the largest, hence the largest fan. The number density of this fan appears as expected, with the highest density between the two beams at the absorber. 
	
	\begin{figure}[!h!t!b]
\centerline{\includegraphics[clip=,width=1.\textwidth]{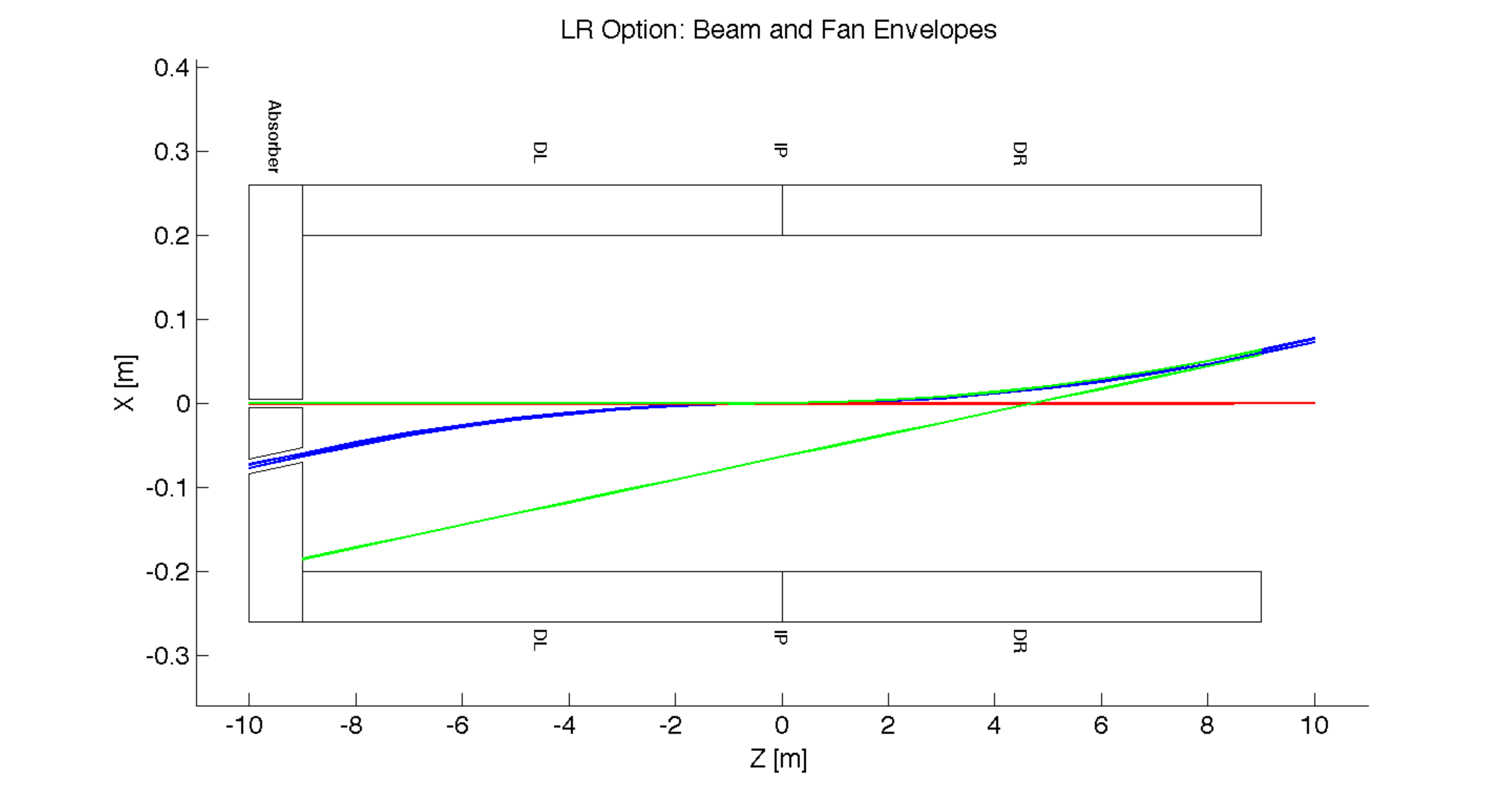}}
\caption{LR: Beam Envelopes in Z direction.}
\label{Fig:LRenv}
\end{figure}
	
	In Figure \ref{Fig:LRnumdens} the distribution was given at various Z values however a continuous envelope distribution is also important to see everything at once. This can be seen in Figure \ref{Fig:LRenv}, where the beam and fan envelopes are shown in the Y = 0 plane. This makes it clear that the fan is antisymmetric which comes from the S shape of the electron beam as previously mentioned.
	
	\subsubsection{Absorber}	
	
	\begin{figure}[!h!t!b]
\centerline{\includegraphics[clip=,width=1.\textwidth]{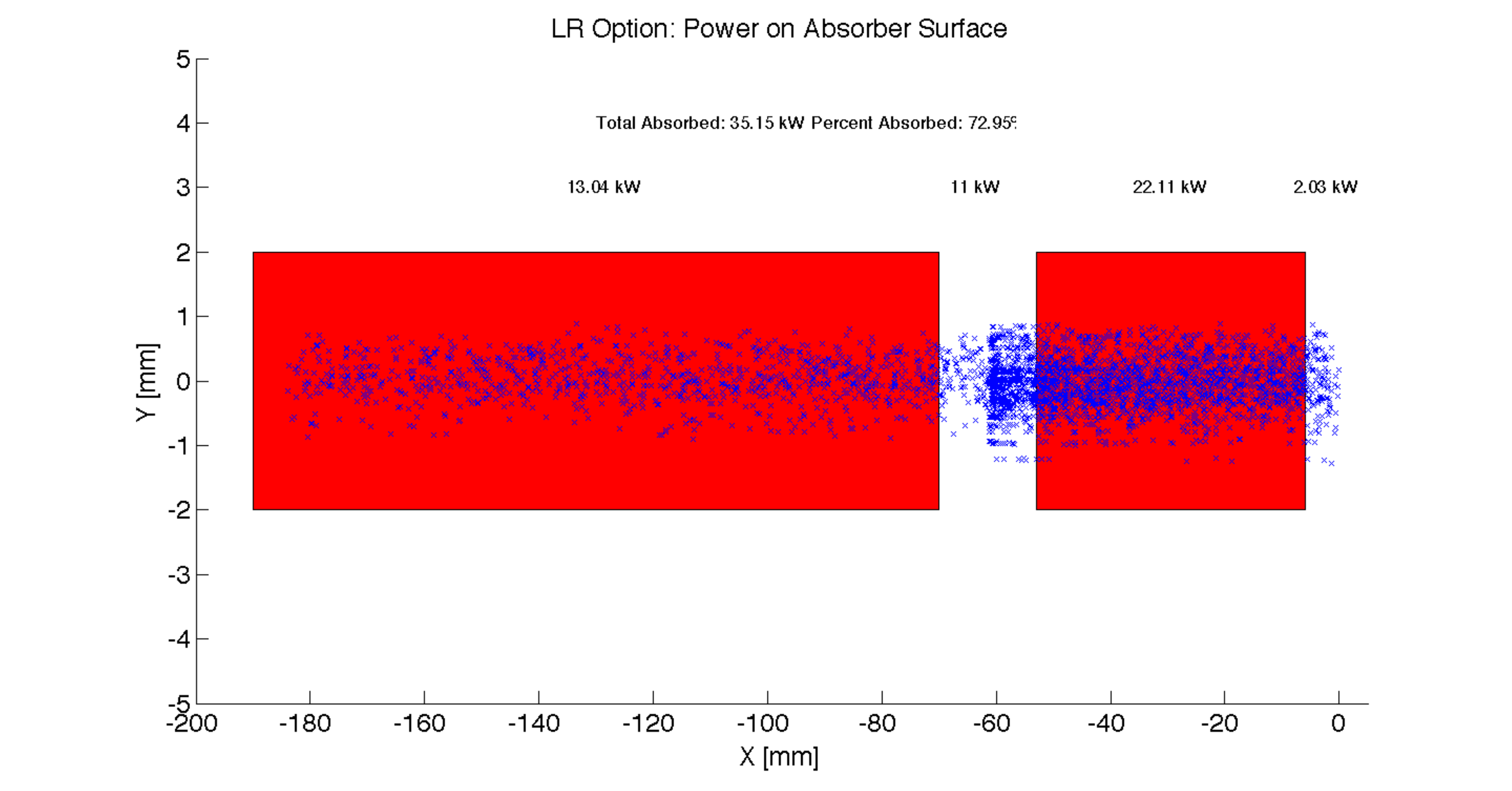}}
\caption{LR: Photon distribution on the Absorber Surface.}
\label{Fig:LRAbs}
\end{figure}

	The photon distribution on the absorber surface is crucial. The distribution decides how the absorber must be shaped. The shape of the absorber in addition to the distribution on the surface then decides how much SR is backscattered into the detector region. In HERA backscattered SR was a significant source of background that required careful attention \cite{zeus}. Looking at Figure \ref{Fig:LRAbs} it is shown that for the LR option 35.15 kW of power from the SR light will fall on the face of the absorber which is $73\%$ of the total power. This gives a general idea of the amount of power that will be absorbed. However, backscattering and IR photons will lower the percent that is actually absorbed.

	\paragraph{Proton Triplet:}

	The super conducting final focusing triplet for the protons needs to be protected from radiation by the absorber. Some of the radiation produced upstream of the absorber however will either pass through the absorber or pass through the apertures for the two interacting beams. This is most concerning for the interacting proton beam aperture which will have the superconducting coils. A rough upper bound for the amount of power the coils can absorb before quenching is 100 W~\cite{stephan}. There is approximately 2 kW entering into the interacting proton beam aperture as is shown in Figure \ref{Fig:LRAbs}. This doesn't mean that all this power will hit the coils but simulations need to be made to determine how much of this will hit the coils. The amount of power that will pass through the absorber (0.25 W) can be disregarded as it is not enough to cause any significant effects. The main source of power moving downstream of the absorber will be the photons passing through the beams aperture. This was approximately 11 kW as can be seen from Figure \ref{Fig:LRAbs}. Most of this radiation can be absorbed in a secondary absorber placed after the first downstream proton quadrupole. Overall protecting the proton triplet is important and although the absorber will minimise the radiation continuing downstream this needs to be studied in depth.

\begin{figure}
\centerline{\includegraphics[angle=0,clip=,width=0.9\textwidth]{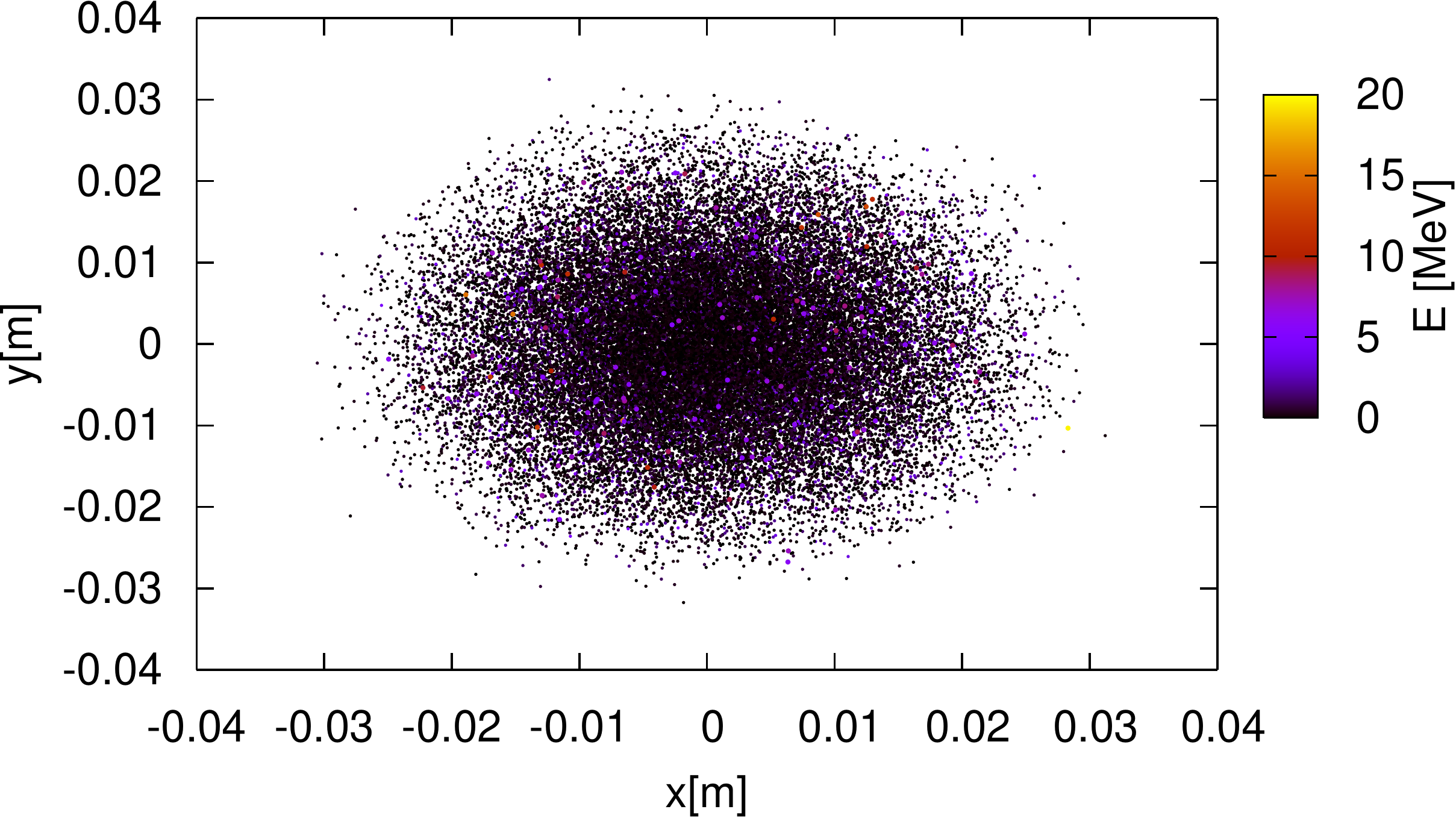}}
\caption{Beamstrahlung photons at the entrance of D1. }\label{Fig:photons}
\end{figure}

\paragraph{Beamstrahlung}
The beamstrahlung photons travel parallel to the
proton beam until the entrance of D1 without impacting
the triplets. Figure~\ref{Fig:photons} shows the transverse
and energy distributions of the beamstralung photons at
the entry of D1 as computed with Guineapig~\cite{guinea}.
The maximum photon energy is about 20~MeV the
average photon energy is 0.4~MeV. The beamstrahlung power is 980~W.
D1 has to be designed to properly dispose the neutral
debris from the IP. Splitting D1 into two parts could allow
an  escape line for the neutral particles.

	\paragraph{Backscattering}
	
\begin{figure}[!h!t!b]
\centerline{\includegraphics[clip=,width=1.\textwidth]{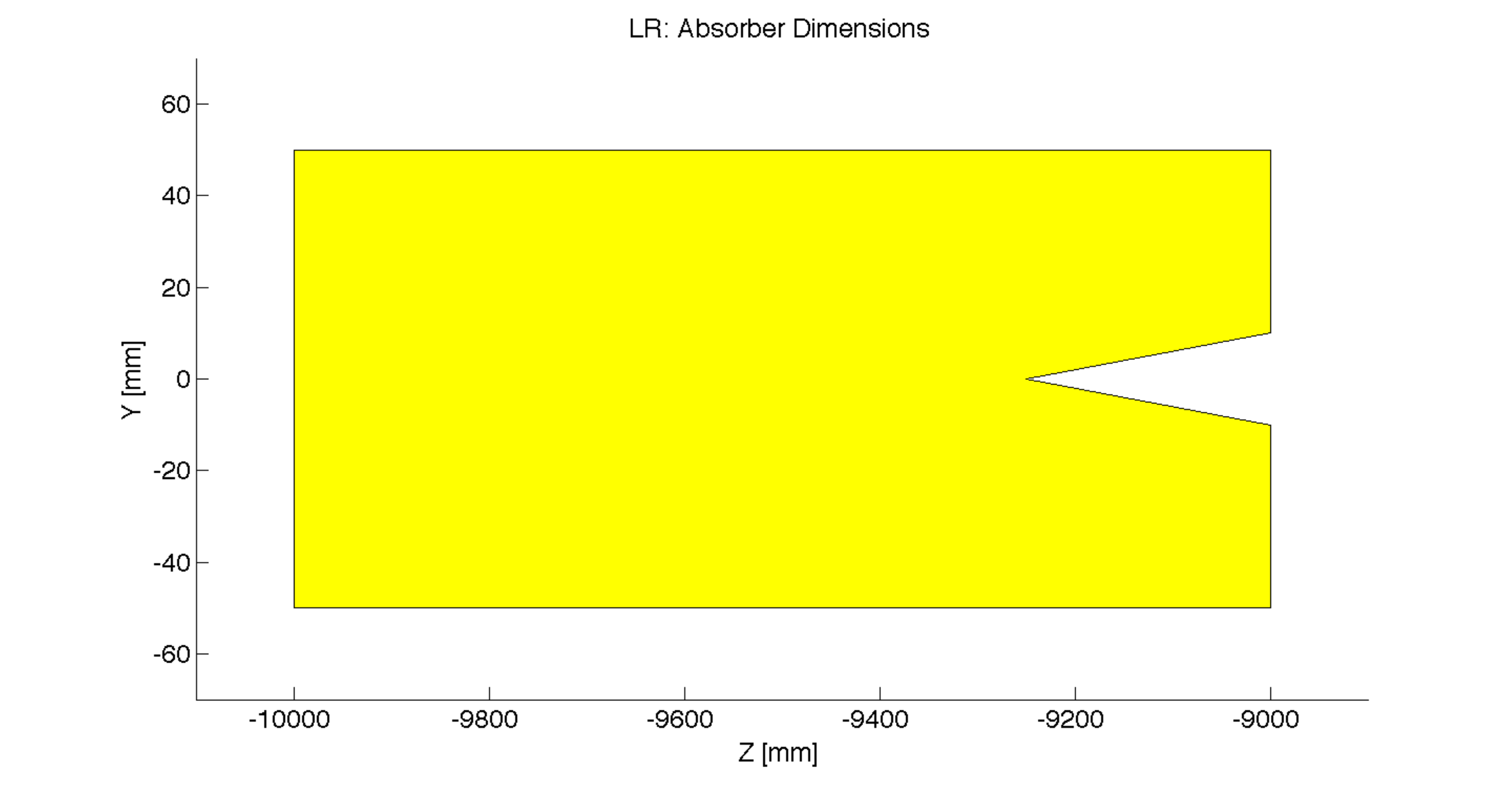}}
\caption{LR: Absorber Dimensions.}
\label{Fig:LRAbsDim}
\end{figure}

\begin{figure}[!h!t!b]
\centerline{\includegraphics[clip=,width=1.\textwidth]{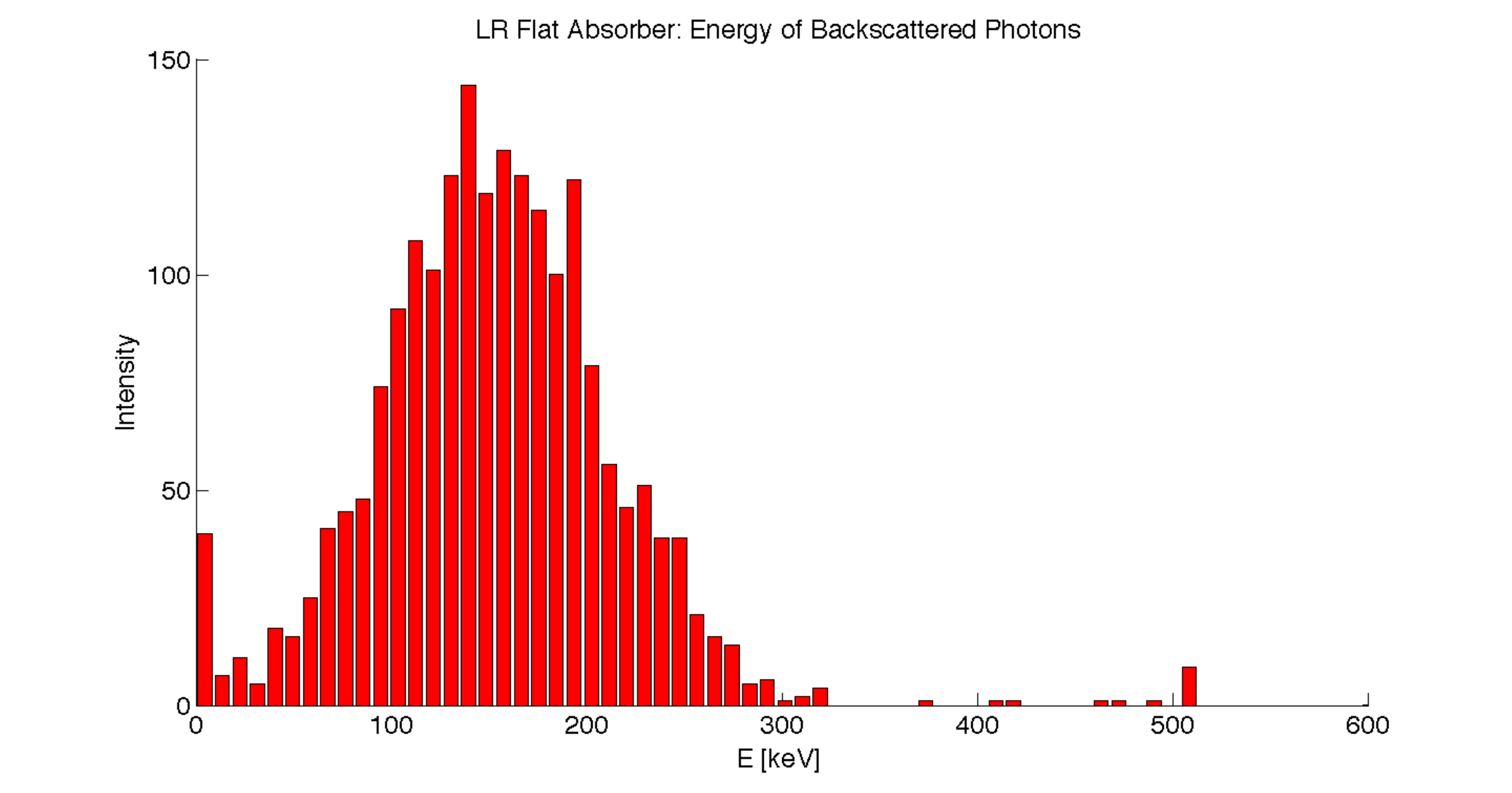}}
\caption{LR: Backscattered Energy Distribution.}
\label{Fig:LRbackeng}
\end{figure}

	Another G4 program was written to simulate the backscattering of photons into the detector region. The ntuple with the photon information written at the absorber surface is used as the input for this program. An absorber geometry made of copper is described, and general physics processes are set up. A detector volume is then described and set to record the information of all the photons which enter in an ntuple. The first step in minimising the backscattering was to optimise the absorber shape. Although the simulation didn't include a beam pipe the backscattering for different absorber geometries was compared against one another to find a minimum. The most basic shape was a block of copper that had cylinders removed for the interacting beams. This was used as a benchmark to see the maximum possible backscattering. In HERA a wedge shape was used for heat dissipation and minimising backscattering \cite{zeus}. The profile of this geometry in the YZ plane is shown in Figure \ref{Fig:LRAbsDim}. It was found that this is the optimum shape for the absorber. The reason for this is that a backscattered electron would have to have to have its velocity vector be almost parallel to the wedge surface to escape from the wedge and therefore it works as a trap. One can be seen from Table \ref{tab:BMLR} utilising the wedge shaped absorber decreased the backscattered power by a factor of 4. The energy distribution for the backscattered photons can be seen in Figure \ref{Fig:LRbackeng}.
		
\begin{figure}[!h!t!b]
\centerline{\includegraphics[clip=,width=1.\textwidth]{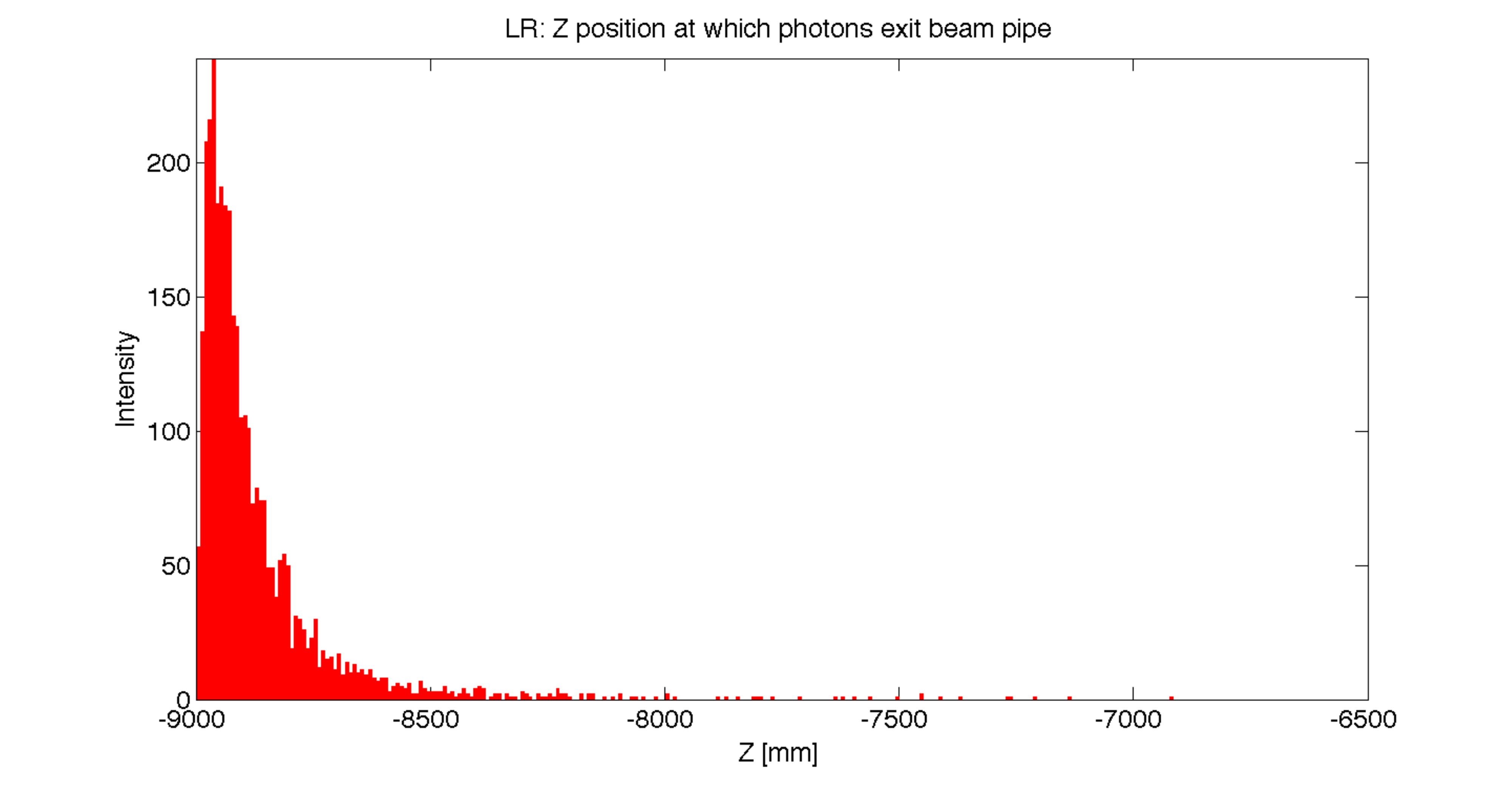}}
\caption{LR: Backscattered Photons Exiting the Beam Pipe.}
\label{Fig:LRzback}
\end{figure}

	After the absorber was optimised it was possible to set up a beam pipe geometry. An asymmetric elliptical cone beam pipe geometry made of beryllium was used since it would minimise the necessary size of the beam pipe as previously mentioned. The next step was to place the lead shield and masks inside this beam pipe. To determine placement a simulation was run with just the beam pipe. Then it was recorded where each backscattered photon would hit the beam pipe in Z. A histogram of this data was made as shown in Figure \ref{Fig:LRzback}. This determined that the shield should be placed in the Z region ranging from -8 m until the absorber (-9 m). The masks were then placed at -8.9 m and -8.3 m. This decreased the backscattered power by a factor of 40 as can be seen from Table \ref{tab:BMLR}. Overall there is still more optimisation that can occur with this placement. 
	
		\begin{table}[!htbp]
  \centering
  \begin{tabular}{| c | c | }
 \hline
Absorber Type & Power [W] \\
\hline\hline
 Flat & 645.9  \\
 \hline
Wedge & 159.1 \\
 \hline
 Wedge \& Mask/Shield & 4.3  \\
\hline
  \end{tabular}
\caption{LR: Power deposition due to Backscattered photons.}
\label{tab:BMLR}
\end{table}

\begin{figure}[!h!t!b]
\centerline{\includegraphics[clip=,width=1.\textwidth]{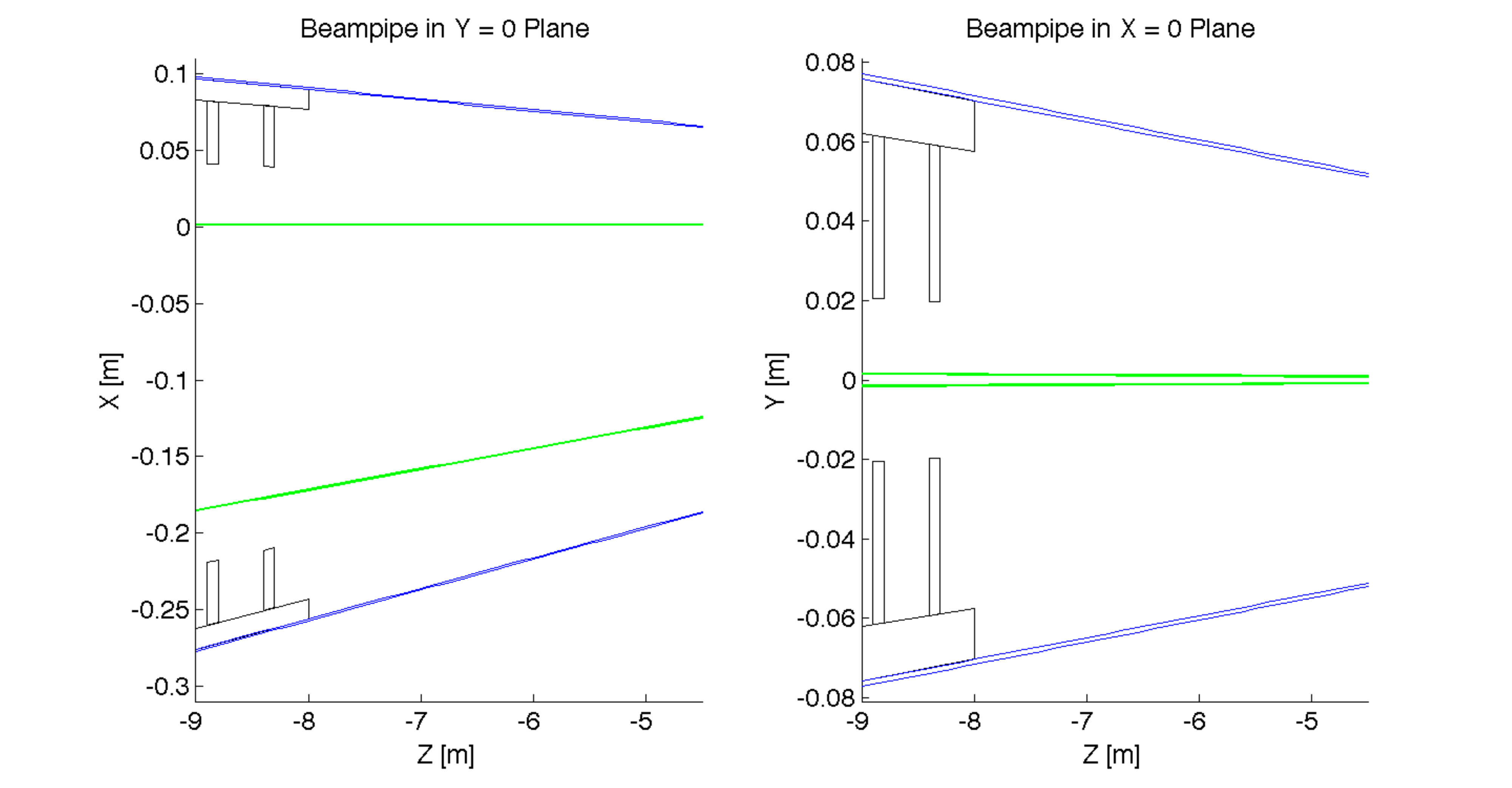}}
\caption{LR: Beam pipe Cross Sections.}
\label{Fig:LRBP}
\end{figure}

Cross sections of the beam pipe in the Y = 0 and X = 0 planes with the shields and masks included can be seen in Figure \ref{Fig:LRBP}.

%% file: machine/schulte.tex
\section{Linac lattice and impedance}
\subsection{Overall layout}
\begin{figure}
\centerline{\includegraphics[angle=0,clip=,width=0.8\textwidth]{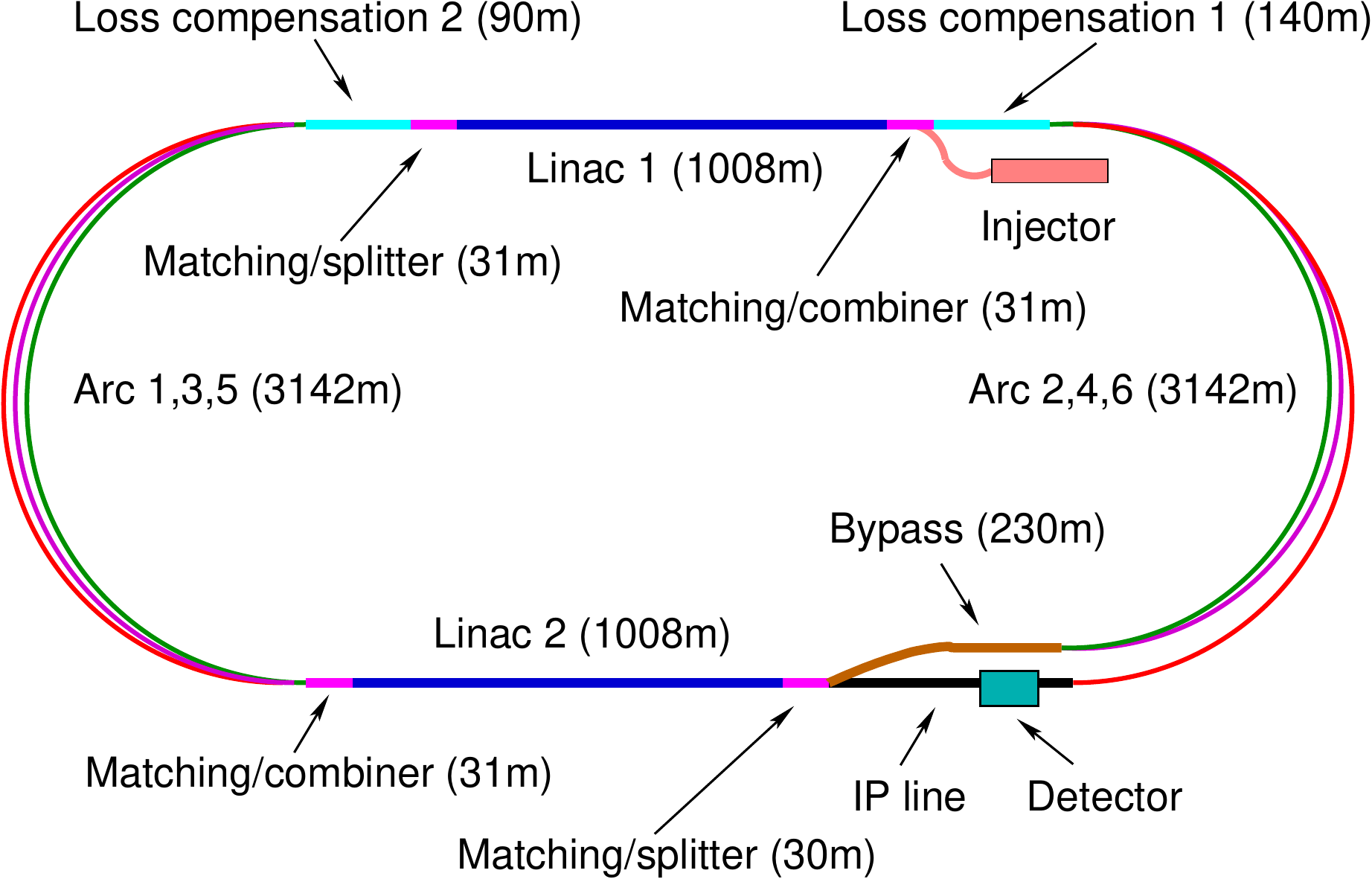}}
\caption{The schematic layout of the recirculating linear accelerator complex.}
\label{scd:scheme}
\end{figure}

The proposed layout of the recirculating linear accelerator complex (RLA) is illustrated schematically in Fig.~\ref{scd:scheme}.
It consists of the following components:
\begin{itemize}
\item A $0.5\;\rm GeV$ injector with an injection chicane.
\item A pair of $721.44\rm{MHz}$ SCRF linacs. Each linac is one kilometre long with an energy gain $10\rm{GeV}$ per pass.
\item Six $180^\circ$ arcs. Each arc has a radius of one kilometre.
\item For each arc one re-accelerating station that compensates the synchrotron radiation emitted in this arc.
\item A switching station at the beginning and end of each linac to combine the beams from different arcs and
to distribute them over different arcs.
\item An extraction dump at 0.5 GeV.
\end{itemize}

After injection, the beam makes three passes through the linacs before it collides with the LHC beam. The beam will then perform three
additional turns in which the beam energy is almost completely extracted.
The size of the complex is chosen such that each turn has the same length and that three turns correspond to the LHC circumference.
This choice is motivated by the following considerations:
\begin{itemize}
\item To avoid the build-up of a significant ion density in the accelerator complex, clearing gaps may be required in the beam.
\item The longitudinal position of these gaps must coincide for each of the six turns that a beam performs. This requires that the turns have the same length.
\item Due to the gaps some LHC bunches will collide with an electron bunch but some will not. It is advantageous to have each LHC bunch either always collide
with an electron bunch or to never collide. The choice of length for one turn in the RLA allows to achieve this.
\end{itemize}

Some key beam parameters are given in table~\ref{linac:param}.
\begin{table}
\centerline{
\begin{tabular}{|*3{c|}}
\hline
Parameter&Symbol&Value\\
\hline
Particles per bunch &$N$&$2\cdot10^9$\\
Initial normalised transverse emittance&$\epsilon_x$, $\epsilon_y$& $30\rm{\mu m}$\\
Normalised transverse emittance at IP&$\epsilon_x$, $\epsilon_y$&$50\rm{\mu m}$\\
Bunch length &$\sigma_z$&$600\rm{\mu m}$\\
\hline
\end{tabular}
}
\caption{Key beam parameters. It should be noted that normalised emittances are used throughout.}
\label{linac:param}
\end{table}

\subsection{Linac layout and lattice}
The key element of the transverse beam dynamics in a multi-pass recirculating linac is an appropriate choice of multi-pass linac optics.
The focusing strength of the quadrupoles along the linac needs to be set such that one can transport the beam at each
pass. Obviously, one would like to optimise the focusing profile to accommodate a large number of passes through the RLA. In addition,
the requirement of energy recovery puts a constraint on the exit/entrance Twiss functions for the two linacs.
As a baseline we have chosen a FODO lattice with a phase advance of $130^\circ$ for the beam that passes with the lowest energy
and a quadrupole spacing of $28\rm{m}$~\cite{scd:alex}.
Alternative choices are possible. An example is an optics that avoids any quadrupole in the linacs~\cite{scd:bnl}.

\subsubsection{Linac module layout}

The linac consists of a series of units, each consisting of two cryomodules and one quadrupole pack. 
We consider one possible configuration for the 10-GeV linac, containing 36$\times$2 cryomodules 
with an RF gradient of 18 MV/m. This design is slightly different from the one described in the RF section later, which uses fewer cavities per linac at a higher gradient;
in this case also the modules are longer.
However, the conclusions on the beam stability do not change with these small differences.
In the simulations, each cryomodule is $12.8\; \rm{m}$ and contains eight $1\rm{m}$-long accelerating cavities, which allows
1.6 m per cavity unit, which leaves little extra space for interconnects between cavities, with 
implications on the cavity design.   
The interconnect between two adjacent cryomodules is $0.8\; \rm{m}$ long. The quadrupole pack
is $1.6\rm{m}$ long, including the interconnects to the adjacent cryomodules. The whole unit
is $28\rm{m}$ long.

Each quadrupole pack contains a quadrupole, a beam position monitor and a vertical and horizontal dipole corrector, see Section 8.2.

\subsubsection{Linac optics}
\begin{figure}
\begin{center}
\includegraphics[width=0.7\textwidth]{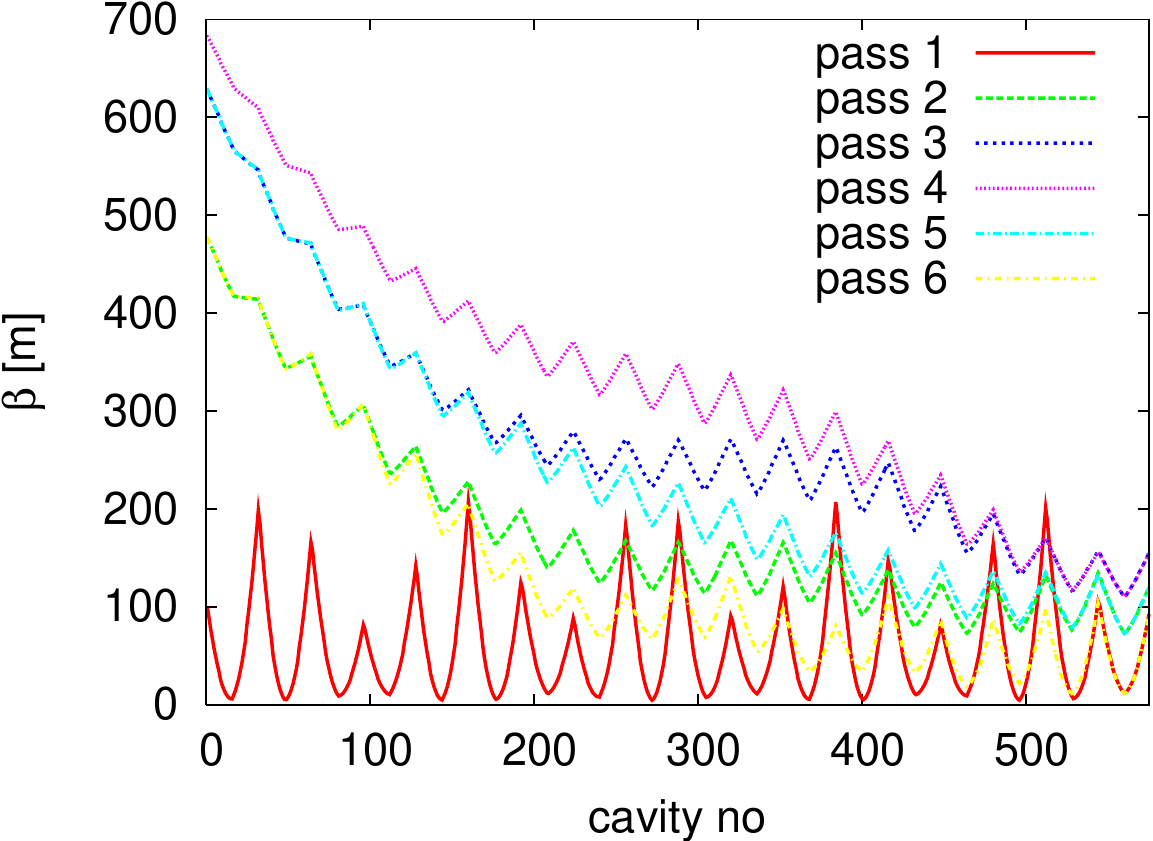}
\includegraphics[width=0.7\textwidth]{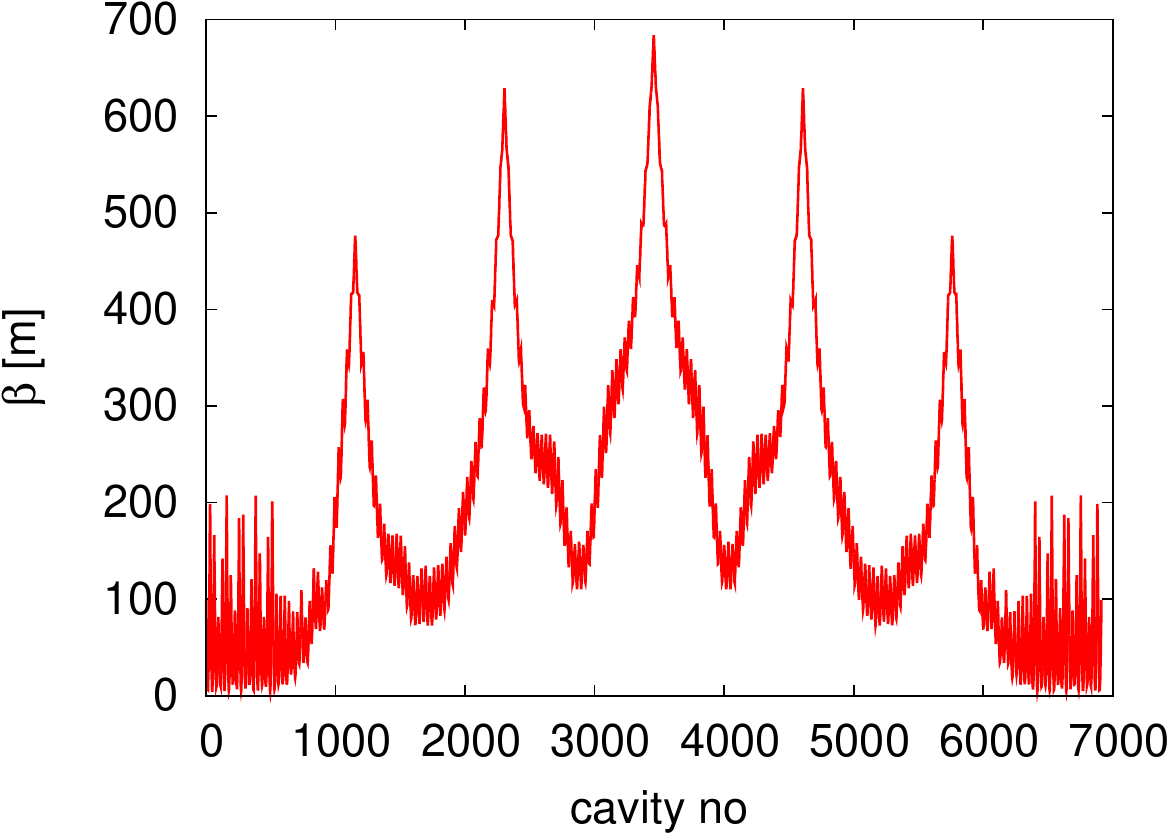}
\caption{Beta-functions in the first linac. On the top, the beta-functions of the six different beam passages in the first linac are shown.
On the bottom, the beta-function as seen by the beam during its stay in the linacs are shown.}
\label{scd:lattice}
\end{center}
\end{figure}

The linac consists of 36 units with a total length of $1008\; \rm{m}$.
In the first linac, the strength of the quadrupoles has been chosen to provide a phase advance per cell of $130^\circ$ for the beam in its first turn.
In the second linac, the strength has been set to provide a phase advance of $130^\circ$ for the last turn of the beam.
The initial Twiss parameters of the beam and the return arcs are optimised to minimise the beta-functions of the beams in the following passages.
The criterion used has been to minimise the integral
\begin{equation}
\int_0^L\frac{\beta}{E}ds
\end{equation}
Single bunch transverse wakefield effects and multi-bunch effects between bunches that have been injected shortly after each other are
proportional to this integral~\cite{scd:mb}. The final solution is shown in Fig.~\ref{scd:lattice}.
A significant beta-beating can be observed due to the weak focusing for the higher energy beams.

\subsubsection{Return Arc optics}
At the ends of each linac the beams need to be directed into the appropriate energy-dependent
arcs for recirculation. Each bunch will pass each arc twice, once when it is accelerated before the
collision and once when it is decelerated after the collision. The only exception is the arc
at highest energy that is passed only once.
For practical reasons, horizontal rather than vertical beam separation
was chosen. Rather than suppressing the horizontal dispersion created by the spreader,
the horizontal dispersion can been smoothly matched to that of the arc, which results in a very compact,
single dipole, spreader/recombiner system.

The initial choice of large arc radius (1 km) was dictated by limiting energy loss due to synchrotron
radiation at top energy (60.5 GeV) to less than 1\%.
However other adverse effects of synchrotron
radiation on beam phase-space such as cumulative emittance and momentum growth due to quantum
excitations are of paramount importance for a high luminosity collider that requires normalised emittance of 50 mm mrad. 
Energy losses from resistive wall and   coherent synchrotron radiation have both 
been shown to be negligible compared with the energy loss due to incoherent synchrotron radiation~\cite{scd:bnl}.

\begin{figure}
\centerline{\includegraphics[angle=0,clip=,width=0.8\textwidth]{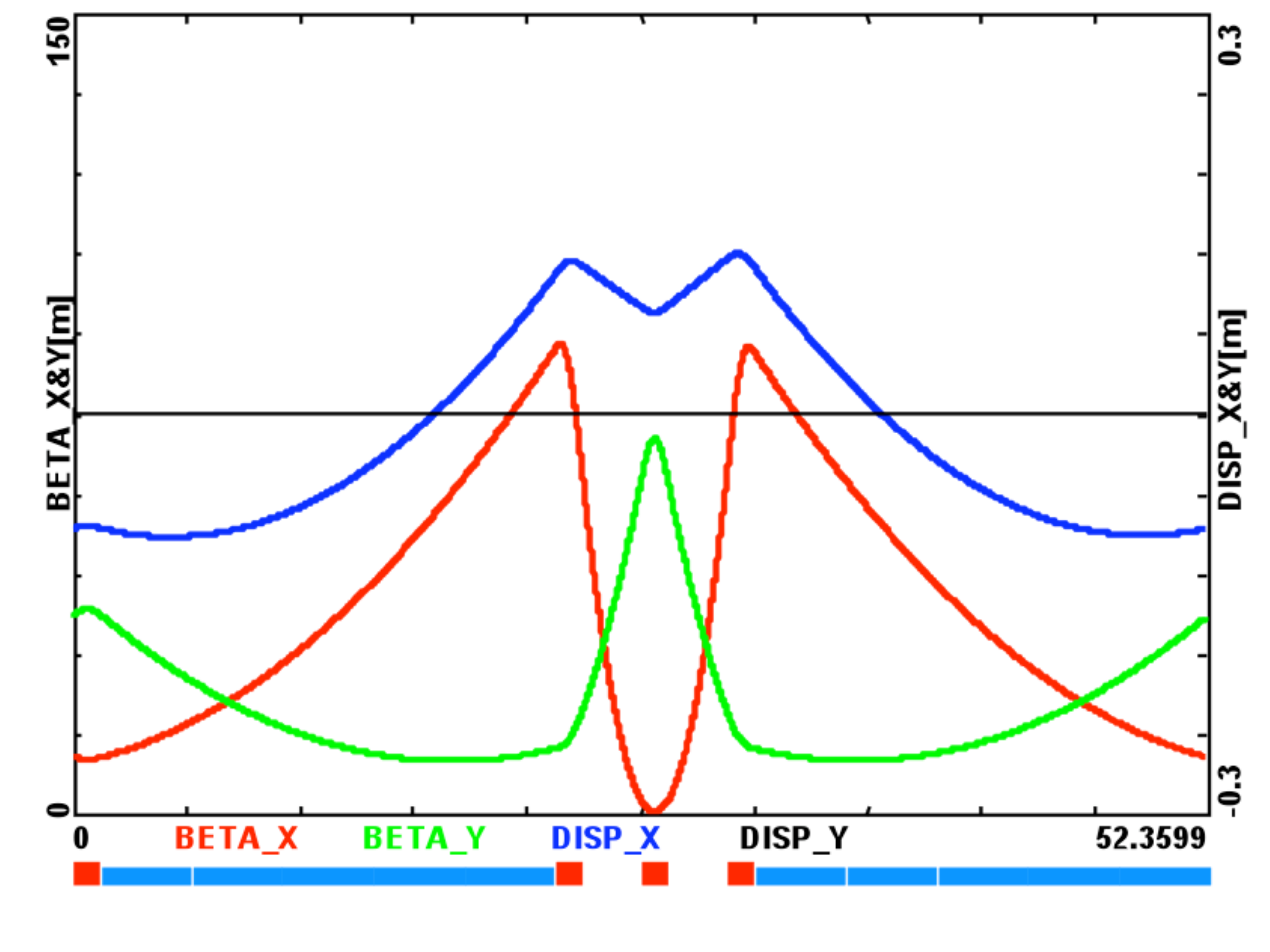}}
\centerline{\includegraphics[angle=0,clip=,width=0.8\textwidth]{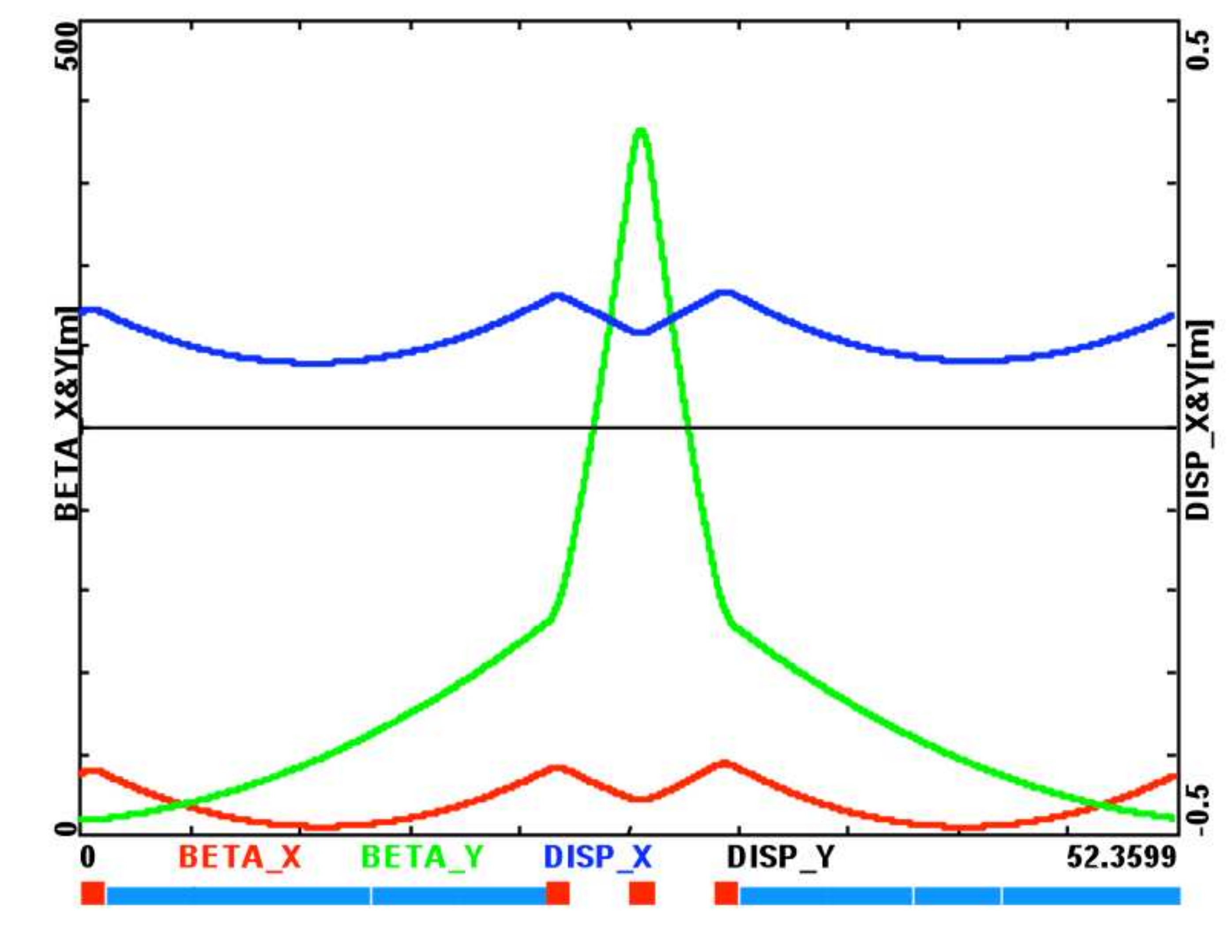}}
\caption{The optics of the lowest (top) and the highest (bottom) energy return arcs.}
\label{arc}
\end{figure}

Three different arc designs have been developed~\cite{scd:alex}. In the design for the lowest energy turns,
the beta-functions are kept small in order to limit the required vacuum chamber size and consequently the
magnet aperture. At the highest energy, the lattice is optimised to keep the emittance growth limited, while the
beta-functions are allowed to be larger. A cell of the lowest and one of the highest energy arc is shown in Fig.~\ref{arc} 
All turns have a bending radius of $764\rm{m}$. The beam pipe diameter is $25\rm{mm}$, which corresponds to more than $12\sigma$ aperture.

An interesting alternative optics, which pushes towards a smaller beam pipe, has also been developed~\cite{scd:bnl}.

\subsubsection{Synchrotron radiation in return Arcs}

\begin{table}
\centerline{
\begin{tabular}{|*4{c|}}
\hline
turn no&$E$&$\Delta E$&$\sigma_E/E$\\
&$\rm{[GeV]}$&$\rm{[MeV]}$&$[\%]$\\
\hline
1 & 10.4 & 0.7 & 0.00036\\
2 & 20.3 & 9.9 & 0.0019\\
3 & 30.3 & 48.5 & 0.0053\\
4 & 40.2 & 151 & 0.011\\
5 & 50.1 & 365 & {0.020}\\
6 & 60.0 & 751 & 0.033\\
7 & 50.1 & 365 & 0.044\\
8 & 40.2 & 151 & 0.056\\
9 & 30.3 & 48.5 & 0.074\\
10 & 20.3 & 9.9 & 0.11\\
11 & 10.4 & 0.7 & {0.216}\\
dump & 0.5 & 0.0 & {4.53}\\
\hline
\end{tabular}}
\caption{Energy loss due to synchrotron radiation in the arcs as a function of the arc number. The integrated energy spread induced by synchrotron radiation is also shown.}
\label{espread}
\end{table}

Synchrotron radiation in the arcs leads to a significant beam energy loss. This loss is compensated by the
small linacs that are incorporated before or after each arc when the beams are already or still separated according to their energy, see Fig.~\ref{scd:scheme}.
The energy loss at the $60\rm{GeV}$ turn-round can be compensated by a linac with an RF frequency of
$721.44\rm{MHz}$. The compensation at the other arcs is performed with an RF frequency of
$1442.88\rm{MHz}$. In this way the bunches that are on their way to the collision point and the ones that already collided
can both be accelerated. This ensures that the energy of these bunches are the same on the way to and from the interaction point, which
simplifies the optics design. If the energy loss were not compensated the beams would have a different energy at each turn, so that the
number of return arcs would need to be doubled.


The synchrotron radiation is also generating an energy spread of the beam. In Tab.~\ref{espread} the
relative energy spread is shown as a function of the arc number that the beam has seen.
At the interaction point, the synchrotron radiation induced RMS energy spread is only $2\times10^{-4}$, which adds to the
energy spread of the wakefields.
At the final arc the energy spread reaches about $0.22\%$, while at the beam dump it grows to
a full $4.5\%$.

\begin{table}
\centerline{
\begin{tabular}{|*4{c|}}
\hline
turn no&$E$&$\Delta\epsilon_{arc}$&$\Delta\epsilon_{t}$\\
&$\rm{[GeV]}$&$\rm{[\mu m]}$&$\rm{[\mu m]}$\\
\hline
1 & 10.4 & 0.0025& 0.0025\\
2 & 20.3 &0.140& 0.143\\
3 & 30.3 & 0.380& 0.522\\
4 & 40.2 & 2.082& 2.604\\
5 & 50.1 & 4.268& 6.872\\
6 & 60 & 12.618& 19.490\\
5 & 50.1 & 4.268& 23.758\\
4 & 40.2 & 2.082& 25.840\\
3 & 30.3 & 0.380& 26.220\\
2 & 20.3 & 0.140& 26.360\\
1 & 10.4 & 0.0025& 26.362\\
\hline
\end{tabular}
}
\caption{The emittance growth due to synchrotron radiation in the arcs. $\Delta\epsilon_{arc}$ is the growth in each individual arc, $\Delta\epsilon_{t}$ is the integrated growth including all previous arcs. The collision with the proton beam will take place at the beginning of the arc 6, so one finds $\Delta\epsilon_{t}\approx4.3\;\rm \mu m$.}
\label{scd:emitt}
\end{table}

The growth of the normalised emittance is given by
\begin{equation}
\Delta\epsilon=\frac{55}{48\sqrt{3}}\frac{\hbar c}{mc^2}r_e\gamma^6I_5
\end{equation}
Here, $r_e$ is the classical electron radius, and $I_5$ is given by
\begin{equation}
I_5=\int_0^L \frac{H}{|\rho|^3} ds=\frac{\langle H\rangle \theta}{\rho^2} \ \ \ H=\gamma D^2+2\alpha D D^\prime+\beta D^{\prime2}
\end{equation}
For a return arc with a total bend angle $\theta=180^\circ$ one finds
\begin{equation}
\Delta\epsilon=\frac{55}{48\sqrt{3}}\frac{\hbar c}{mc^2}r_e\gamma^6\pi\frac{\langle H\rangle \theta}{\rho^2}
\end{equation}
The synchrotron radiation induced emittance growth is shown in table~\ref{scd:emitt}. Before the interaction point a total growth of about $7\rm{\mu m}$
is accumulated. The final value is $26\rm{\mu m}$. While this growth is significant compared to the target emittance of $50\rm{\mu m}$ at the collision
point, it seems acceptable.


\input{machine/bogacz2}

\subsection{Beam break-up}

\subsubsection{Single-bunch wakefield effect}
In order to evaluate the single bunch wakefield effects we used PLACET~\cite{scd:placet}.
The full linac lattice has been implemented for all turns but the arcs have each been replaced
by a simple transfer matrix, since the matching sections have not been available.

Single bunch wakefields were not available for the SPL cavities. We therefore used
the wakefields in the ILC/TESLA cavities~\cite{scd:rdr}. In order to adjust the wakefields to the
lower frequency and larger iris radius ($70\rm{mm}$ vs. $39\rm{mm}$ for the central irises)
we used the following scaling
\begin{equation}
W_{\perp}(s)\approx\frac{1}{(70/39)^3}W_{\perp,ILC}(s/(70/39))
\ \ \ \ 
W_L(s)\approx\frac{1}{(70/39)^2}W_{L,ILC}(s/(70/39))
\end{equation}

\begin{figure}
\begin{center}
\includegraphics{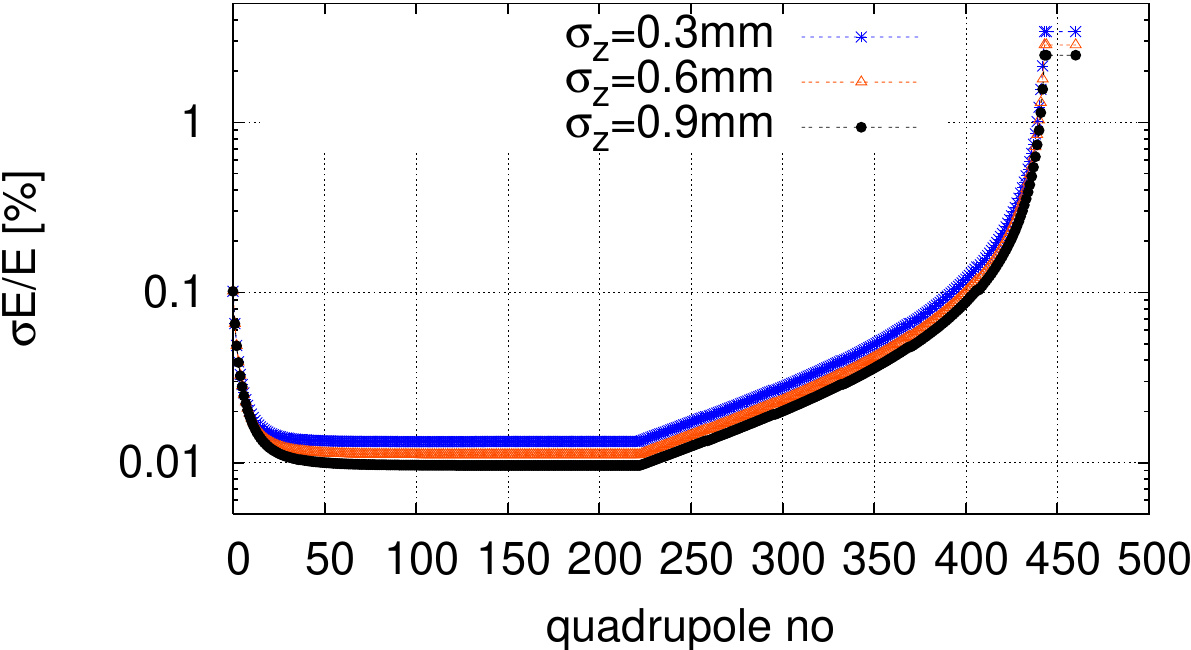}
\caption{The RMS energy spread due to single bunch wakefields along the linacs. The bunch has been cut
longitudinally at $\pm3\sigma_z$ and at $\pm3\sigma_E$ in the initial uncorrelated energy spread.}
\label{scd:e}
\end{center}
\end{figure}

First, the RMS energy spread along the linacs is determined. An initial uncorrelated RMS energy spread of
$0.1\%$ is assumed. Three different bunch lengths were studied, i.e. $300\rm{\mu m}$, $600\rm{\mu m}$ and
$900\rm{\mu m}$. This longest value yields the smallest final energy spread.
The energy spread along during the beam life-time can be seen in Fig.~\ref{scd:e}.
The wakefield induced energy spread is between $1\times10^{-4}$ and $2\times10^{-4}$
at the interaction point, $1$--$2\times10^{-3}$ at the final arc and $3.5$--$4.5\%$ at the beam dump.

\begin{figure}
\begin{center}
\includegraphics{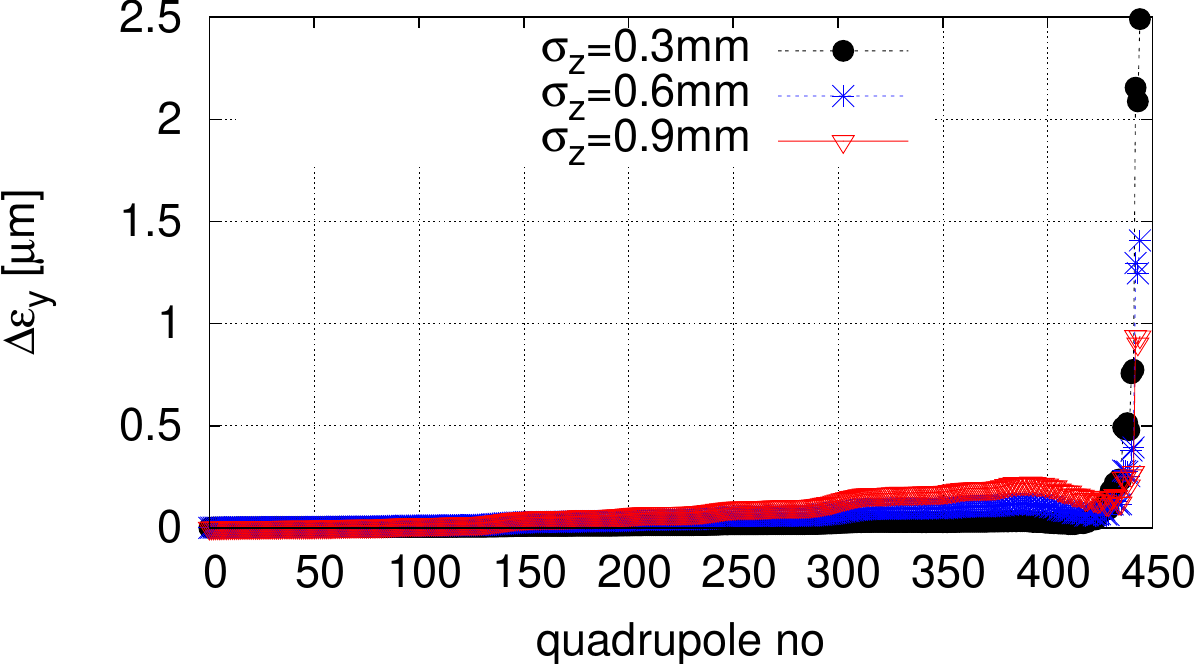}
\caption{The single-bunch emittance growth along the LHeC linacs for a bunch with an initial offset of
$\Delta x=\sigma_x$. The arcs have been represented by a simple transfer matrix.}
\label{scd:single}
\end{center}
\end{figure}

Second, the single bunch beam-break-up is studied by tracking a bunch with an initial offset of
$\Delta x=\sigma_x$. The resulting emittance growth of the bunch is very small, see Fig.~\ref{scd:single}.

\subsubsection{Multi-bunch transverse wakefield effects}
\begin{table}
\centerline{
\hbox{
\begin{tabular}{|c|*1{c|}}
\hline
$f\rm{[GHz]}$&$k\rm{[V/pCm^2]}$\\
\hline
0.9151 &9.323  \\
0.9398 &19.095 \\
0.9664 &8.201 \\
1.003 &5.799 \\
1.014 &13.426 \\
1.020 &4.659 \\
1.378 &1.111 \\
1.393 &20.346 \\
1.408 &1.477 \\
1.409 &23.274 \\
1.607 &8.186 \\
1.666 &1.393 \\
1.670 &1.261 \\
\hline
\end{tabular}
\begin{tabular}{|c|*1{c|}}
\hline
$f\rm{[GHz]}$&$k\rm{[V/pCm^2]}$\\
\hline
1.675 &4.160 \\
2.101 &1.447 \\
2.220 &1.427 \\
2.267 &1.377 \\
2.331 &2.212 \\
2.338 &11.918 \\
2.345 &5.621 \\
2.526 &1.886 \\
2.592 &1.045 \\
2.592 &1.069 \\
2.693 &1.256 \\
2.696 &1.347 \\
2.838 &4.350 \\
\hline
\end{tabular}
}
}
\caption{The considered dipole modes of the SPL cavity design.}
\label{wakes}
\end{table}

For a single pass through a linac the multi-bunch effects can easily be estimated analytically~\cite{scd:mb}. Another approach exists in case of
two passes through one cavity~\cite{scd:georg_wake}. It is less straightforward to find an analytic solution for multiple turns in linacs with wakefields that
vary from one cavity to the next. In this case the also phase advance from one passage through a cavity to the next passage depends on the position of the
cavity within the linac. We therefore addressed the issue by simulation.

Two multi-bunch beam break-up studies have been performed independently.
The first study is based on a new code that we developed to simulate the multi-bunch effect in the case of recirculation and energy
recovery~\cite{scd:new}. It assumes point-like bunches and takes a number of dipole wake field modes into account.
A cavity-to-cavity frequency spread of the wakefield modes can also be modelled. The arcs are replaced with simple transfer matrices.
In the simulation, we offset a single
bunch of a long train by one unit and determine the final position in phase space of all other bunches.

\begin{figure}
\begin{center}
\includegraphics{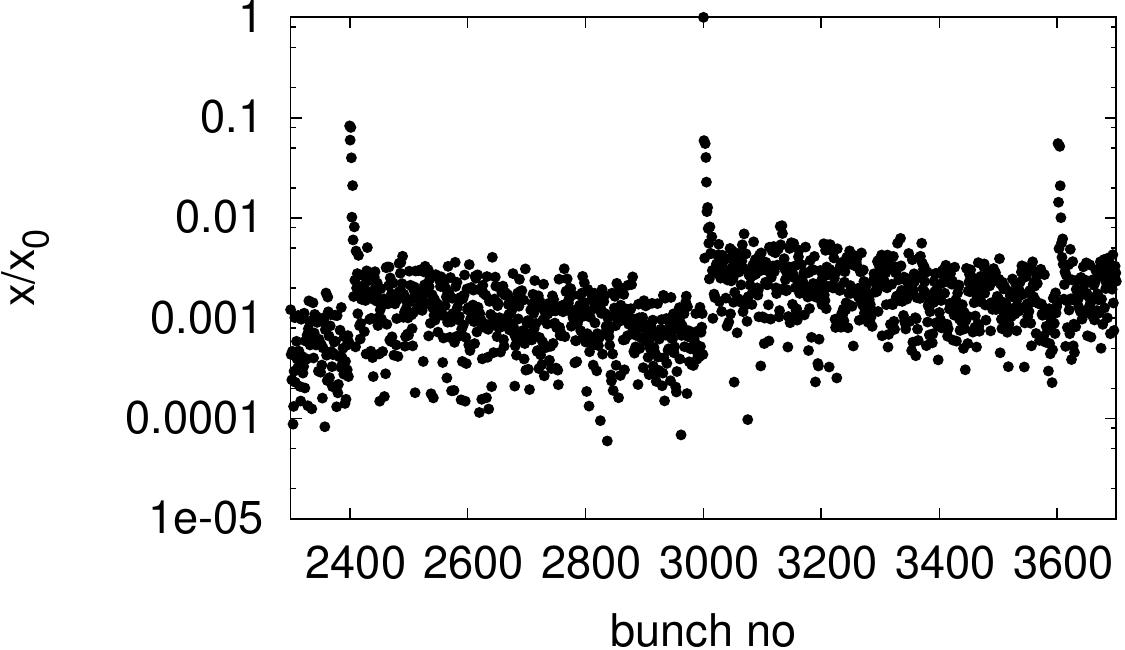}
\includegraphics{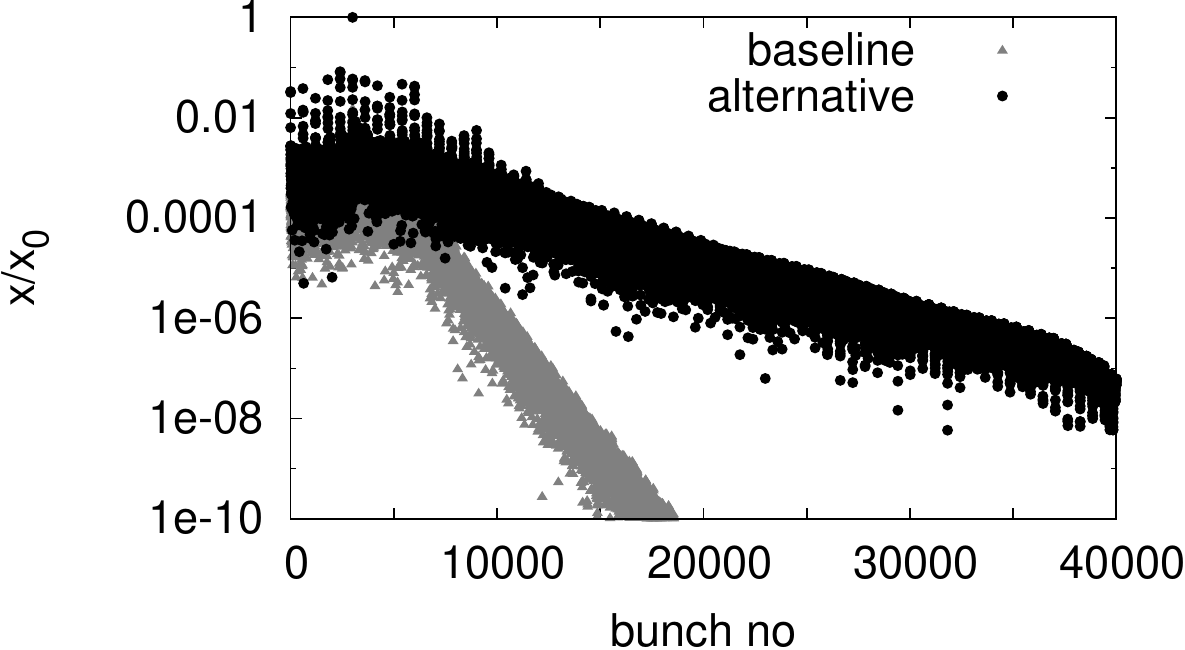}
\caption{Multi-bunch beam break-up assuming the SPL cavity wakefields. One bunch has been offset at the beginning of the machine and the normalised amplitudes of the bunch oscillations
are shown along the train at the end of the last turn.
The upper plot shows a small number of bunches before and after the one that has been offset (i.e. bunch 3000).
The lower plot
shows the amplitudes along the full simulated train for the baseline lattice and the alternative design with no quadrupole focusing.
One can see the fast decay of the amplitudes.
}
\label{f:multi1}
\end{center}
\end{figure}

\begin{figure}
\begin{center}
\includegraphics{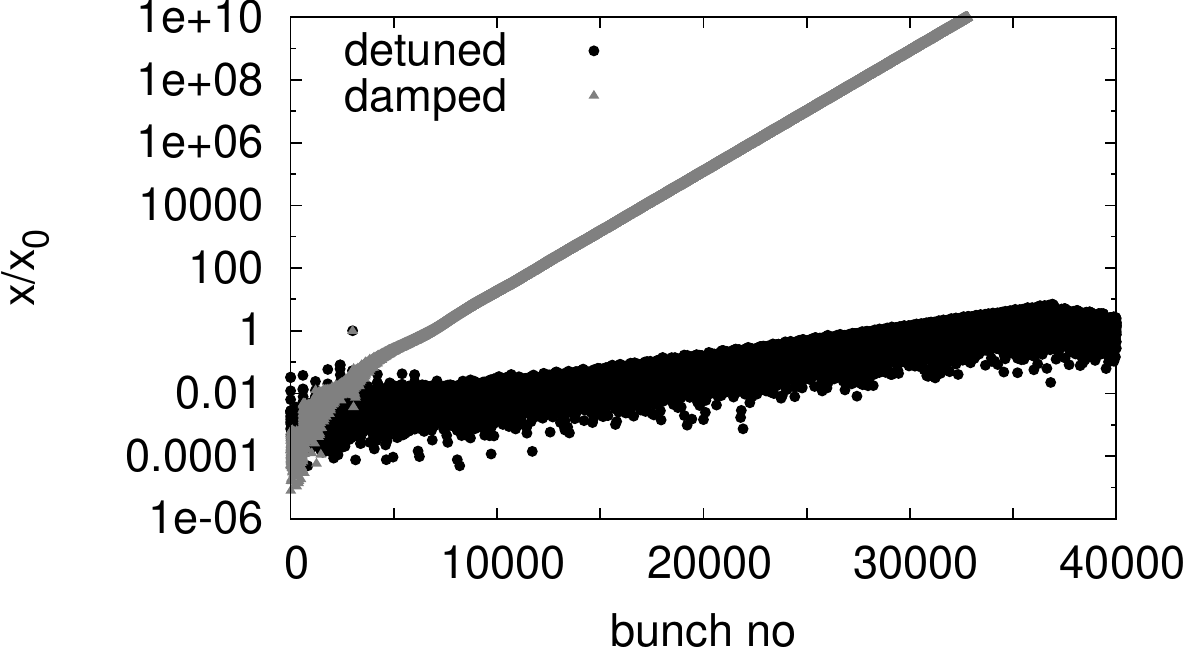}
\caption{Multi-bunch beam break-up for the SPL cavities. In one case only damping, in the other case only cavity-to-cavity mode detuning is present.}
\label{f:multi2}
\end{center}
\end{figure}

We evaluated the beam stability using the wakefield modes that have been calculated for the SPL cavity design~\cite{scd:schuh}.
The level of the $Q$-values of the transverse modes is not yet known. We assume $Q=10^{5}$ for all modes, which
is comparable to the larger of the $Q$-values found in the TESLA cavities. A random variation of the transverse mode frequencies of $0.1\%$
has been assumed, which corresponds to the target for ILC~\cite{scd:rdr}.
The results in Fig.~\ref{f:multi1} indicate that the beam remains stable in our baseline design. Even in the 
alternative lattice with no focusing in the linacs, the beam would remain stable but with significantly less margin.
An independent beam-breakup analysis for linacs without focusing, based on measurements and simulations for 
the BNL 5-cell cavity, demonstrated as well that for all practical scenarios with a HOM frequency spread above 0.2\%  
the instability threshold current is well above the design beam current~\cite{scd:bnl}.

We also performed simulations, assuming that either only damping or detuning were present, see Fig.~\ref{f:multi2}.
The beam is unstable in both cases. Similarly, increasing the $Q$ value to $10^6$ will make the beam unstable
Based on our results we conclude
\begin{itemize}
\item One has to ensure that transverse higher order cavity modes are detuned from one cavity to the next.
While this detuning can naturally occur due to production tolerances, one has to find a method to ensure
its presence. This problem exists similarly for the ILC.
\item Damping of the transverse modes is required with a $Q$ value below $10^5$.
\end{itemize}
If these requirements are met, the beam will remain stable in the cavities at $720{\;\rm MHz}$.
Further studies can give more precise limits on the maximum required $Q$ and minimum mode detuning.

\input{machine/bogacz}

\subsubsection{Fast beam-ion instability}
Collision of beam particles with the residual gas in the beam pipe will lead to the production of positive ions.
These ions can be trapped in the beam. There presence modifies the betatron function of the beam since the ions
focus the beam. They can also lead to beam break-up, since bunches with an offset will induce a coherent motion in the ions.
This can in turn lead to a kick of the ions on following bunches.

\paragraph{Trapping Condition in the beam pulse}
In order to estimate whether ions are trapped or not, one can replace each beam with a thin focusing lens,
with the strength determined by the charge and transverse dimension of the beam. In this case the force is
assumed to be linear with the ion offset, which is a good approximation for small offsets.

The coherent frequency $f_i$ of the ions in the field of a beam of with bunches of similar size is given by~\cite{scd:frank}:
\begin{equation}
f_i=\frac{c}{\pi}\sqrt{\frac{Q_iNr_e\frac{m_e}{Am_p}}{3\sigma_y(\sigma_x+\sigma_y)\Delta L}}
\label{e:fi}
\end{equation}
Here, $N$ is the number of electrons per bunch, $\Delta L$ the bunch spacing, $r_e$ the classical electron radius, $m_e$ the
electron mass, $Q_i$ the charge of the ions in units of $\rm{e}$ and $A$ is their mass number and $m_p$ the proton mass.
The beam transverse beam size is given by $\sigma_x$ and $\sigma_y$. The ions will be trapped in the beam if
\begin{equation}
f_i\le f_{limit}=\frac{c}{4\Delta L}
\end{equation}
In the following we will use $\Delta L\approx2.5\rm{m}$, i.e. assume that the bunches from the different turns are almost evenly spaced longitudinally.

\begin{figure}
\begin{center}
\includegraphics{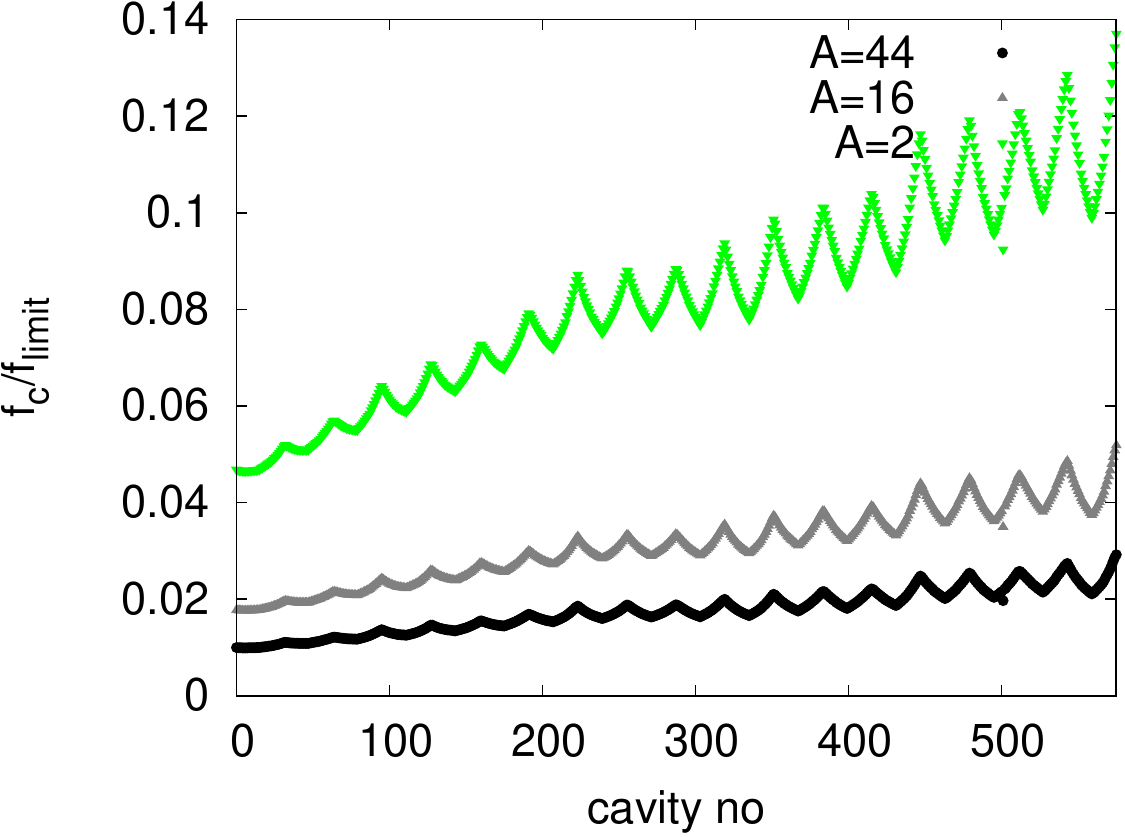}
\caption{The oscillation frequency $f_c$ of ions of different mass number $A$
in the linacs using the average focusing strength of the bunches at different energy.
The frequency is normalised to the limit frequency $f_{limit}$ above which the ions would not be trapped any more.}
\label{scd:ion1a}
\end{center}
\end{figure}

In the linacs, the transverse size of the beam changes from one passage to the next while in each of the return arcs the beams have (approximately)
the same size at both passages. But the variation from one turn to the next is not huge, so we use the average focusing strength of the six turns.
The calculation shows that ions will be trapped for a continuous beam in the linacs. Since we are far from the limit of the trapping condition, the
simplification in our model should not matter. As can be seen in Fig.~\ref{scd:ion1a} $\rm CO_2^+$ ions are trapped all along the linacs. Even hydrogen ions
$H_2^+$ would be trapped everywhere. If one places the bunches from the six turns very close to each other longitudinally, the limit frequency $f_{limit}$
is reduced. However, the ratio $f_c/f_{limit}$ is not increased by more than a factor 6, which is not fully sufficient to remove the $H_2^+$.

\paragraph{Impact and Mitigation of Ion Effects}
Without any methods to remove ions, a continuous beam would collect ions until they neutralise the beam current. This will render the beam
unstable. Hence one needs to find methods to remove the ions. We will first quickly describe the mitigation techniques and then give a rough estimate of the
expected ion effect.

A number of techniques can be used to reduce the fast beam-ion instability:
\begin{itemize}
\item An excellent vacuum quality will slow down the build-up of a significant ion density.
\item Clearing gaps can be incorporated in the electron beam. During these gaps the ions can drift away from the beam orbit.
\item Clearing electrodes can be used to extract the ions. They would apply a bias voltage that lets the ions slowly drift out of the beam.
\end{itemize}

\begin{figure}
\begin{center}
\includegraphics{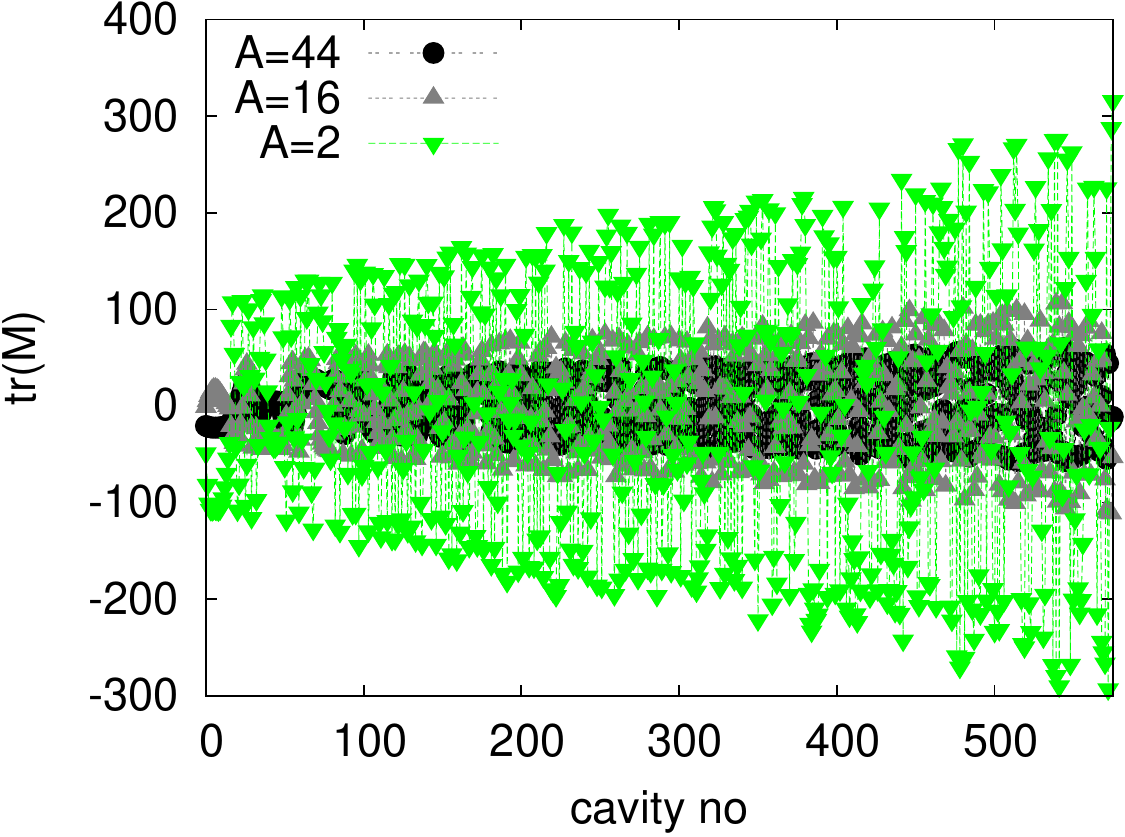}
\includegraphics{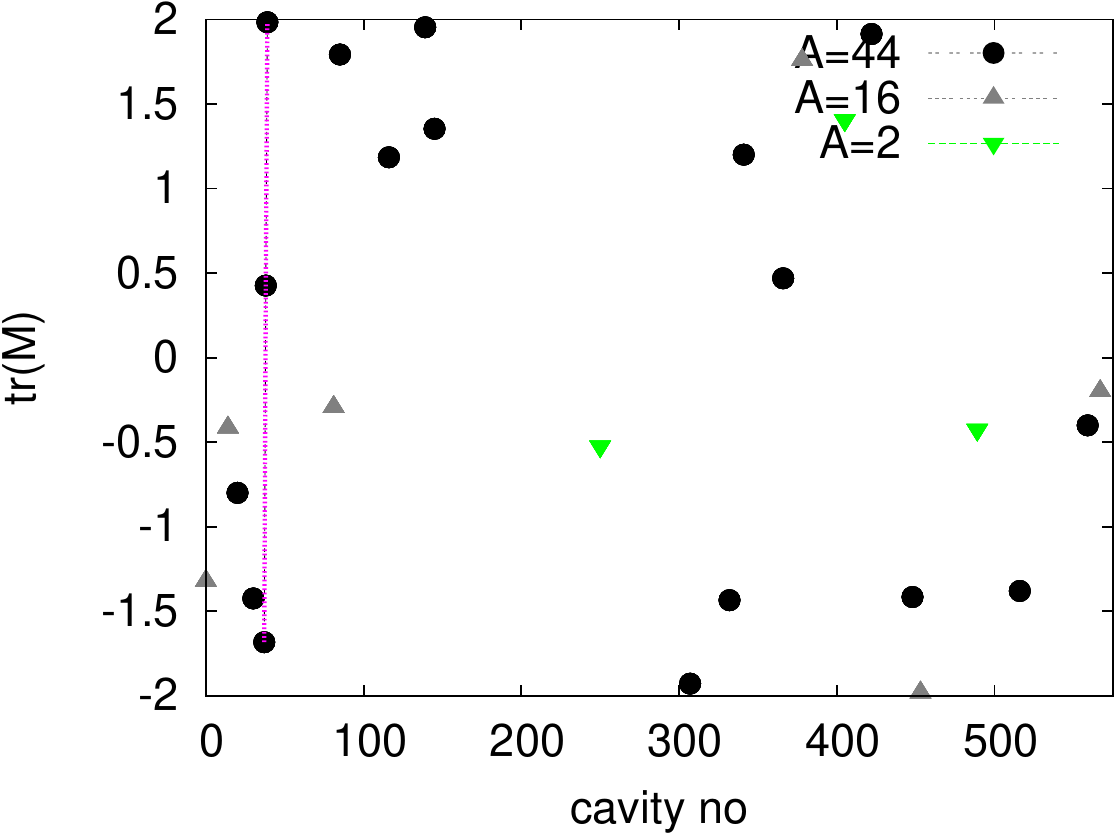}
\caption{The trace of the transfer matrix for $H_2^+$, $CH_4^+$ and $\rm CO_2^+$ ions in presence of a clearing gap.
Values above $2$ or below $-2$ indicate that the ions will not be trapped.}
\label{scd:ion1}
\end{center}
\end{figure}

\paragraph{Clearing Gaps}
In order to provide the gap for ion cleaning, the beam has to consist at injection of short trains of bunches with duration $\tau_{beam}$ separated by gaps
$\tau_{gap}$. If each turn of the beam in the machine takes $\tau_{cycle}$, the beam parameters have to be adjusted such that
$n(\tau_{beam}+\tau_{gap})=\tau_{cycle}$. In this case the gaps of the different turns fall into the same location of the machine. This scheme will
avoid beam loading during the gap and ensure that the gaps a fully empty. By choosing the time for one round trip in the electron machine
to be an integer fraction of the LHC round-trip time $\tau_{LHC}=m\tau_{cycle}$, one ensures that each
bunch in the LHC will either always collide with an electron bunch or never. We chose to use $\tau_{cycle}=1/3 \tau_{LHC}$
and to use a single gap with $\tau_{gap}=1/3\tau_{cycle}\approx10\;\rm\mu s$.

In order to evaluate the impact of a clearing gap in the beam, we model the beam as a thick
focusing lens and the gap as a drift. The treatment follows \cite{scd:georg_ion}, except that we use a thick lens approach and correct a factor
two in the force. The focusing strength of the lens can be calculated as
\begin{equation}
k= \frac{2N r_e m_e}{A_{ion}m_p\sigma_y(\sigma_x+\sigma_y)\Delta L}
\end{equation}
The ions will not be collected if the following equation is fulfilled
\begin{equation}
\left|2\cos(\sqrt{k} (L_{erl}-L_g))-\sqrt{k}L_g\sin(\sqrt{k}(L_{erl}-L_g))\right| \ge 2
\end{equation}
Since the beam size will vary as a function of the number of turns that the beam has performed, we replace the
above defined $k$ with the average value over the six turns using the average bunch spacing $\Delta L$,
\begin{equation}
k= \frac{1}{n}\sum_{i=1}^{n}\frac{2N r_em_e}{A_{ion}m_p\sigma_{y,i}(\sigma_{x,i}+\sigma_{y,i})\Delta L}.
\end{equation}
The results of the calculation can be found in Fig.~\ref{scd:ion1}. As can be seen, in most locations the ions are not trapped.
But small regions exist where ions will accumulate. More study is needed to understand which ion density is reached in these
areas. Longitudinal motion of the ions will slowly move them into other regions where they are no longer trapped.

\paragraph{Ion Instability}
While the gap ensures that ions will be lost in the long run, they will still be trapped at least
during the full train length of $20\rm{\mu s}$. We therefore evaluate the impact of ions on the beam during this time. This
optimistically ignores that ions will not be completely removed from one turn to the next. However, the stability criteria we employ
will be pessimistic. Clearly detailed simulations will be needed in the future to improve the predictive power of
the estimates.

Different theoretical models exist for the rise time of a beam instability in the presence of ions. A pessimistic estimate is used in the
following.
The typical rise time of the beam-ion instability for the $n$th bunch can be estimated to be~\cite{scd:frank}
\begin{equation}
\tau_c=\frac{\sqrt{27}}{4}
\left(
\frac{\sigma_y(\sigma_x+\sigma_y)}{Nr_e}
\right)^{\frac{3}{2}}
\sqrt{\frac{A_{ion}m_p}{m}}
\frac{kT}{p\sigma_{ion}}
\frac{\gamma}{\beta_y cn^2\sqrt{L_{sep}}}
\end{equation}
This estimate does not take into account that the ion frequency varies with transverse position within the bunch and along the beam line.

\begin{figure}
\begin{center}
\includegraphics{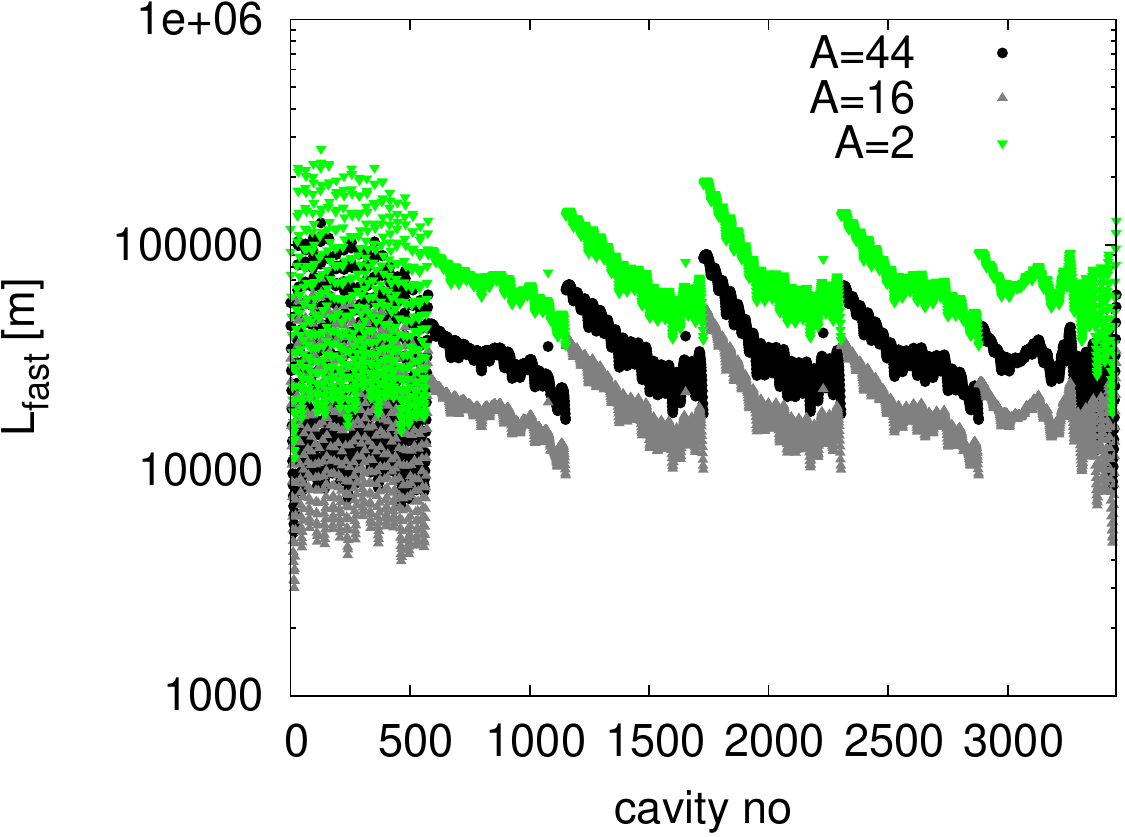}
\caption{The instability length of the beam-ion instability assuming a very conservative partial pressure of $10^{-11}\rm{hPa}$ for each gas.}
\label{scd:ion2}
\end{center}
\end{figure}

We calculate the local instability rise length $c\tau_c$ for a pressure of $p=10^{-11}\rm{hPa}$ at the position of the beam.
As can be seen in Fig.~\ref{scd:ion2} this instability rise length ranges from a few kilometres to several hundred.
One can estimate the overall rise time of the ion instability by averaging over the local ion
instability rates:
\begin{equation}
\langle\frac{1}{\tau_c}\rangle=\frac{\int \frac{1}{\tau_c(s)}ds}{\int ds}
\end{equation}
For the worst case in the figure, i.e. $CH_4^+$, ones finds $c\tau_c\approx 14\;\rm km$ and for $H_2^+$ $c\tau_c\approx25\rm{km}$.
The beam will travel a total
of $12\rm{km}$ during the six passes through each of the two linacs.
So the typical time scale of the rise of the instability is
longer than the life time of the beam and we expect no issue.
This estimate is conservative since it does not take into account that ion frequency varies within the beam and along the machine. Both
effects will stabilise the beam. Hence we conclude that a partial pressure below $10^{-11}\;\rm hPa$ is required for the LHeC linacs.

In the cold part of LEP a vacuum level of $0.5\times10^{-9}\rm{hPa}$ has been measured at room temperature,
which corresponds to $0.6\times10^{-10}\rm{hPa}$ in the cold~\cite{scd:noel}. This is higher than required but
this value ``represents more the out-gassing of warm adjacent parts of the vacuum system''~\cite{scd:noel} and can be considered a pessimistic upper
limit.
Measurements in the cold at HERA showed vacuum levels of $10^{-11}\rm{hPa}$~\cite{scd:holzer}, which would be sufficient but potentially marginal.
Recent measurements at LHC show a hydrogen pressure of
$5\times10^{-12}\rm{hPa}$ measured at room temperature, which corresponds to about $5\times10^{-13}\rm{hPa}$ in the cold~\cite{scd:vacuum}.
For all other gasses a pressure of less than $10^{-13}\rm{hPa}$ is expected measured in the warm~\cite{scd:vacuum}, corresponding to $10^{-14}\rm{hPa}$ in the
cold. These levels are significantly better than the requirements. The shortest instability rise length would be due to hydrogen. With a length of
$c\tau_c\approx500\rm{km}$ which is longer than 40 turns. Hence we do not expect a problem with the fast beam-ion instability in the linacs
provided the vacuum system is designed accordingly.

The effect of the fast beam-ion instability in the arcs has been calculated in a similar way, taking into account the reduced beam current and the baseline
lattice for each arc. Even $H_2^+$ will be trapped in the arcs. We calculate the instability rise length $c\tau_c$ for a partial pressure of
$10^{-9}\rm{hPa}$ for each ion mass and find $c\tau_c\approx 70\rm{km}$ for $H_2^+$,
$c\tau_c\approx 50\rm{km}$ for $N_2^+$ and $CO^+$ and $c\tau_c\approx 60\rm{km}$ for $CO_2^+$. The total distance the beam travels in the arcs is
$15\rm{km}$. Hence we conclude that a partial pressure below $10^{-9}\;\rm hPa$ should be sufficient for the arcs.
More detailed work will be needed in the future to fully assess the ion effects in LHeC but we remain confident that they can be handled.

\paragraph{Ion Induced Phase Advance Error}
The relative phase advance error along a beam line can be calculated using~\cite{scd:georg_ion} for a round beam:
$$
\frac{\Delta\phi}{\phi}=\frac{1}{2}\frac{Nr_e}{\Delta L\epsilon_y}\frac{\theta}{\langle\beta_y^{-1}\rangle}
$$
Here $\theta$ is the neutralisation of the beam by the ions.
We use the maximum beta-function in the linac to make a conservative approximation $\langle\beta^{-1}\rangle=1/700\rm{m}$.
At the end of the train we find $\rho\approx3.3\times10^{-5}$ for $p=10^{-11}\rm{hPa}$ in the cold and $p=10^{-9}\rm{hPa}$ in the warm parts of the machine.
This yields $\Delta\Phi/\Phi\approx7\times10^{-4}$.
Hence the phase advance error can be neglected.

\paragraph{Impact of the Gap on Beam Loading}
It should be noted that the gaps may create some beam-loading variation in the injector complex. We can estimate the associated gradient variation
assuming that the same cavities and gradients are used in the injector as in the linacs. We use
\begin{equation}
\frac{\Delta G}{G}\approx\frac{1}{2}\frac{R}{Q}\omega\frac{\tau_{gap}\tau_{beam}I}{\tau_{gap}+\tau_{beam}}\frac{1}{G}
\end{equation}
 In this case the $10\rm{\mu s}$ gaps in the bunch train correspond
to a gradient variation of about $0.6\%$. This seems very acceptable.

\subsection{Imperfections}
Static imperfections can lead to emittance growth in the LHeC linacs and arcs. However, one can afford an
emittance budget that is significantly larger than the one for the ILC, i.e. $10\rm{\mu m}$ vs. $20\rm{nm}$.
If the LHeC components are aligned with the accuracy of the ILC components, one would not expect
emittance growth to be a serious issue. In particular in the linacs dispersion free steering can be used
and should be very effective, since the energies of the different probe beams are much larger than they would be in ILC.

\subsubsection{Gradient jitter and cavity tilt}
Since the cavities have tilts with respect to the beam line axis, dynamic variations of the
gradient will lead to transverse beam deflections. This effect can be easily calculated using the
following expression:
$$
\frac{\langle y^2\rangle}{\sigma^2_y}
=\frac{\langle(y^\prime)^2\rangle}{\sigma^2_{y^\prime}}
=\frac{1}{2}
\frac{1}{\epsilon}\int\frac{\beta}{E}ds \frac{L_{cav}\langle\Delta G^2\rangle\langle \langle (y^\prime_{cav})^2\rangle}{mc^2}
$$
For an RMS cavity tilt of $300\rm{\mu radian}$, an RMS gradient jitter of $1\%$ and an emittance of $50\rm{\mu m}$ we find
$$
\frac{\langle y^2\rangle}{\sigma^2_y}=
\frac{\langle(y^\prime)^2\rangle}{\sigma^2_{y^\prime}}\approx0.007
$$
i.e. an RMS beam jitter of $\approx0.08\sigma_{y}$.
At the interaction point the beam jitter would be
$\approx0.06\sigma_{y^\prime}$.

\subsection{Touschek scattering}
In recirculating energy recovery linacs, intrabeam scattering and Touschek scattering give rise to beam halo and to 
some unavoidable amount of beam losses, in particular, for high brightness beams and after deceleration \cite{touschekhs}.  
In the LHeC ERL a few dedicated collimators should be foreseen to localise and control these losses \cite{touschekhs}. 
For round beams the Touschek loss rate can be approximated as   
\cite{miyahara1985} (corrected by a factor of two \cite{piwinski2000})  
\begin{equation}
\frac{\Delta N_{b}}{\Delta s} = -\frac{N_{b}^{2}r_{e}^{2} }{8\sqrt{\pi} \gamma^{2} \sigma_{z} \epsilon_{x} \epsilon_{y}} 
\frac{1}{\eta (s)} D \left(\frac{\delta q (s)}{\eta(s)} \right) \; , 
\label{touschek} 
\end{equation}
where $\delta q (s)= \gamma \sigma_{x}(s) /\beta_{x}(s)$, 
\begin{equation}
D (\epsilon) = \sqrt{\epsilon} \int_{\epsilon}^{\infty} \frac{e^{-u}}{u^{3/2}}  \left( \frac{1}{\epsilon} - 1 - \frac{1}{2} \ln \frac{u}{\epsilon} \right)\;  
 du \; ,
\end{equation}
and $\eta_{\rm acc}$ denotes the relative momentum acceptance, which varies along the beam line and is a function of 
the downstream beam energy, RF voltage, optics and aperture. 
Equation (\ref{touschek}) describes the number of bunch particles 
which are Touschek scattered per unit length at location $s$ and lost at a later location.
No detailed analysis of Touschek scattering has yet been performed for the LHeC, but with normalised emittances 
$\epsilon_{x(y)}$ much larger than envisioned for other projects, e.g.~CESR-ERL, with less beam current, 
and higher beam energy, the effect is expected to be comparatively benign.
 

%% file: machine/bogacz2.tex
\subsubsection{Switchyard, matching sections and Arc lattices}

We have completed a design for the ``switchyard'' and linac-to-arc
matching sections for one side of the ERL (Arcs 1, 3 and 5). The other
side will follow a similar pattern of symmetric vertical
spread-recombiner architecture and it is rather straightforward.  We
still need to include sections that compensate the energy loss in the
arcs; they have not been designed yet. But this again should be quite
straightforward.

\paragraph{Switchyard}

\begin{figure}
\begin{center}
\includegraphics[width=0.7\textwidth]{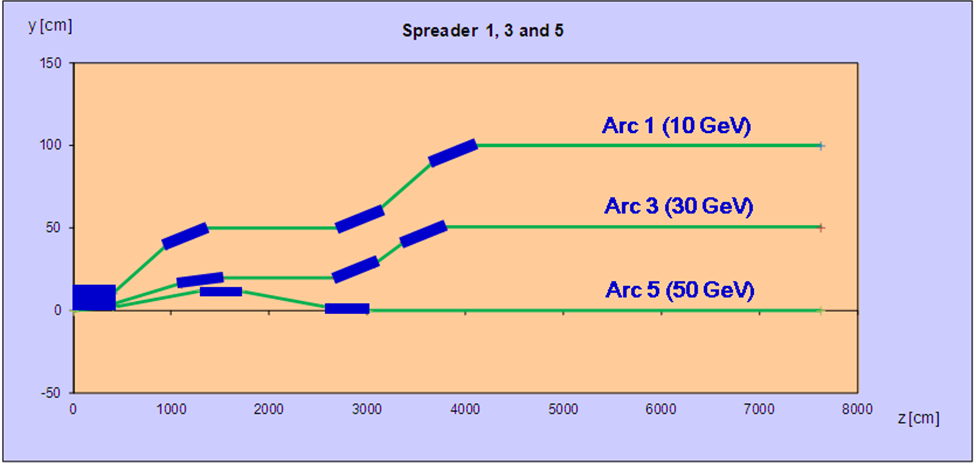}
\caption{
Vertical spreader architecture based on one common ``splitter'' magnet.
}
\label{SY:fig1}
\end{center}
\end{figure}

At the ends of each linac the beams need to be directed into the
appropriate energy-dependent arcs for recirculation. For practical
reasons vertical rather than horizontal beam separation was
chosen. Similar to CEBAF, two-step-achromat spreaders and mirror
symmetric recombiners have been implemented. The switchyard that
separates all three arcs (Arcs 1, 3 and 5) into 1 metre high vertical
stack is illustrated in Figure $\ref{SY:fig1}$.

\begin{figure}
\begin{center}
a)\includegraphics[width=0.7\textwidth]{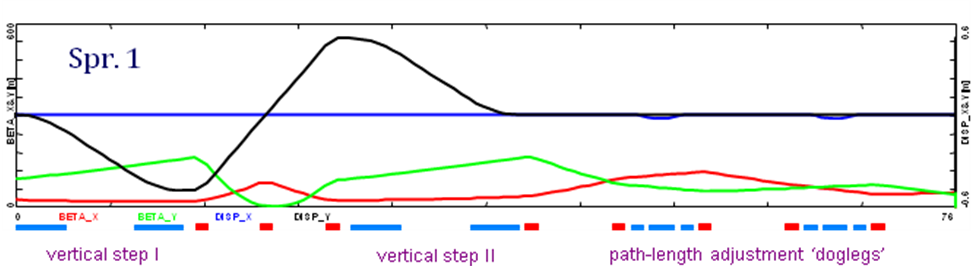}
b)\includegraphics[width=0.7\textwidth]{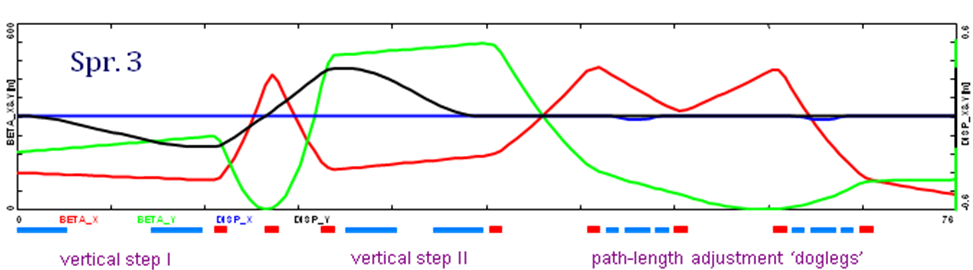}
c)\includegraphics[width=0.7\textwidth]{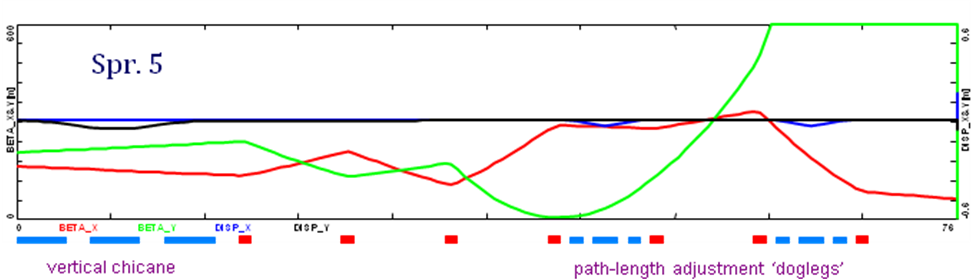}
\caption{
Vertical spreader architecture based on one common spreader magnet and local dispersion suppression.
}
\label{SY:fig2}
\end{center}
\end{figure}

For Arcs 1 and 3 the vertical dispersion generated by a pair of
vertical steps is suppressed by three quadrupoles placed between the
steps, as illustrated in Figure $\ref{SY:fig2}$ a) and b). The highest
energy arc, Arc 3, is not elevated and remains at the ``linac
level''. Here, the vertical dispersion is naturally suppressed by the
appropriate dipole spacing (no quads in between needed), as shown in
Figure $\ref{SY:fig2}$ c).  In addition, a pair of horizontal ``doglegs'', used
for path-length adjustment, is placed downstream of each spreader. The
``dogleg'' archromats are naturally ``meshed into'' the beta-matching
section, as illustrated in Figure $\ref{SY:fig2}$.

\paragraph{Complete Arc Lattices with matched Optics}

\begin{figure}
\begin{center}
a)\includegraphics[width=0.7\textwidth]{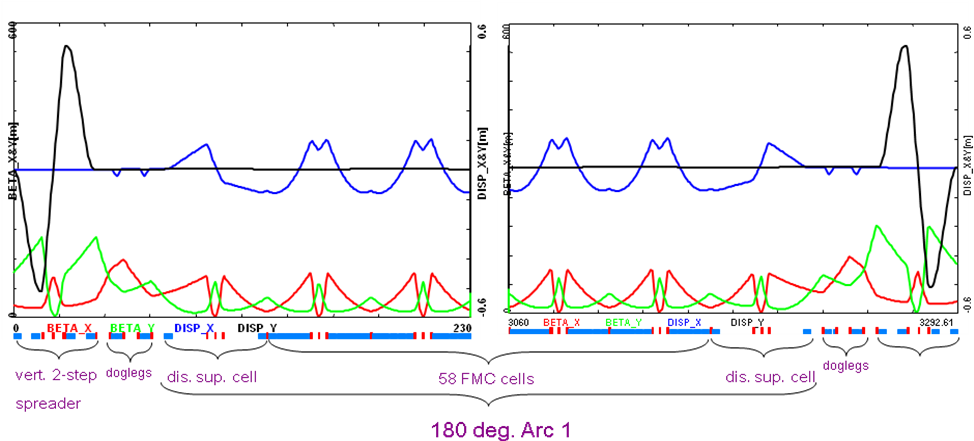}
b)\includegraphics[width=0.7\textwidth]{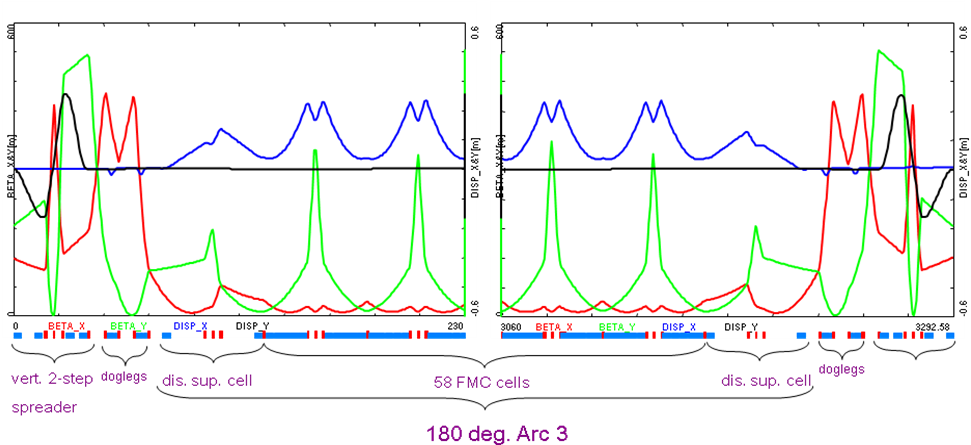}
c)\includegraphics[width=0.7\textwidth]{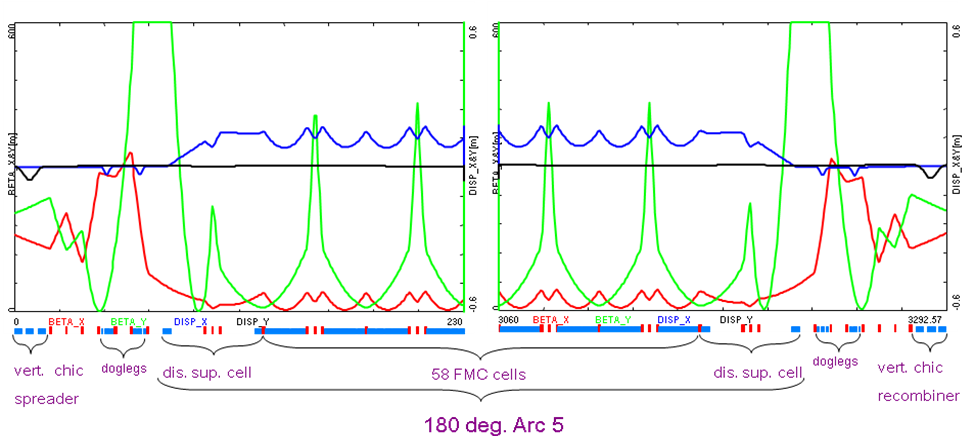}
\caption{
Compete Arc 1, 3 and 5 lattices including: spreaders, recombiners and path-length correcting ``doglegs'' matched to the corresponding linacs.
}
\label{SY:fig3}
\end{center}
\end{figure}

Finally, one can ``attach'' the above spreaders and mirror symmetric
recombiners at each end of a given 1800 ``arc proper'' composed of
periodic FMC cells introduced previously. As the arc energy goes up,
more and more aggressive ``emittance preserving'' flavours of FMC cells
are used to configure the arc proper. Compete arc optics for Arc 1, 3
and 5 matched to the corresponding linacs are illustrated in Figure
$\ref{SY:fig3}$.

%% file: machine/bogacz.tex

\begin{figure}
\begin{center}
\includegraphics[width=0.8\textwidth]{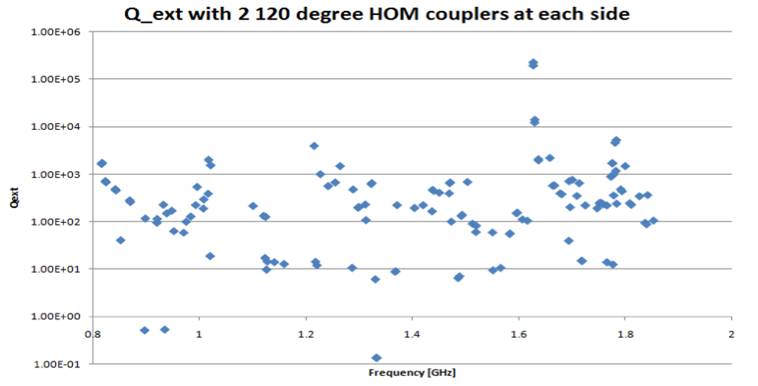}
\end{center}
\caption{
Quality factor of BNL3 cavity per ``High Current SRF Cavity Design for
SPL and eRHIC'', S. Belomestnykh et al., Proceedings of 2011 Particle
Accelerator Conference, New York, NY, USA.  }
\label{fig:bogacz1}
\end{figure}

A further study used a dedicated BBU simulation code. The optics model of the machine is the same as for the first study.
The wakefield model has
been based on the BNL3 $5$-cell cavities, even if their fundamental mode frequency is $703.79$ MHz.
The summary of measured HOMs is illustrated in Figure $\ref{fig:bogacz1}$.

One can notice that all the $Q$ values are less than $1\cdot 10^6$ and
most of them are smaller than $1\cdot 10^4$. For our BBU simulation,
we consider the worst case of $Q_l=1\cdot 10^6$. Out of all HOMs
collected in Figure $\ref{fig:bogacz1}$, we selected three most
offending HOMs with relatively high $R/Q$ values. They are summarised
in table $\ref{tab:bogacz1}$.

\begin{table}[h]
  \centering
  \begin{tabular}{|c|c|c|}
    \hline
Frequency[MHz] & $Q_l$ & R/Q[Ohm] \\
$1003$ & $1\cdot 10^6$  & $32$ \\
$1337$ & $1\cdot 10^6$ & $32$ \\
$1820$ & $1\cdot 10^6$ & $32$ \\
\hline
  \end{tabular}
\caption{
The most offending HOMs selected into BBU simulation.
}
\label{tab:bogacz1}
\end{table}

%
%





\begin{figure}
\begin{center}
\includegraphics[width=0.49\textwidth]{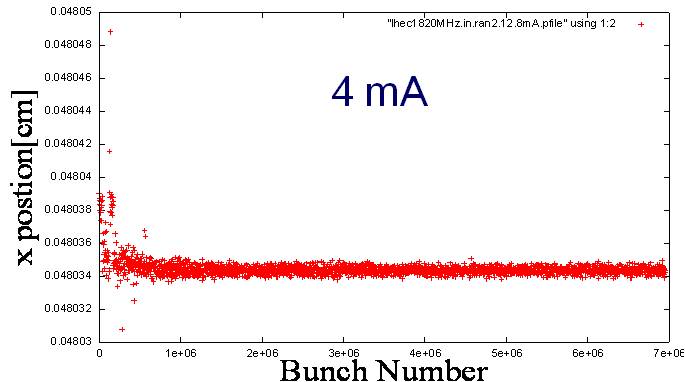}
\includegraphics[width=0.49\textwidth]{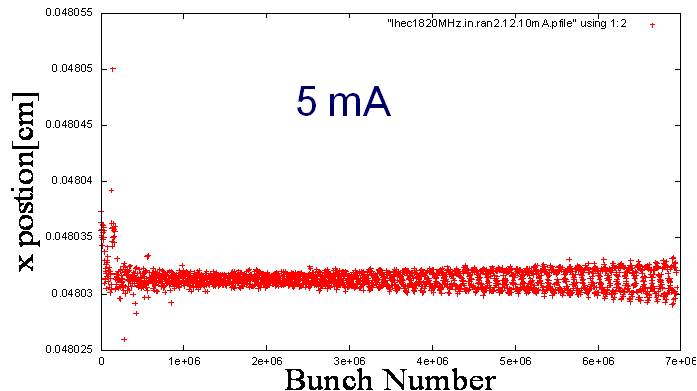}
\includegraphics[width=0.49\textwidth]{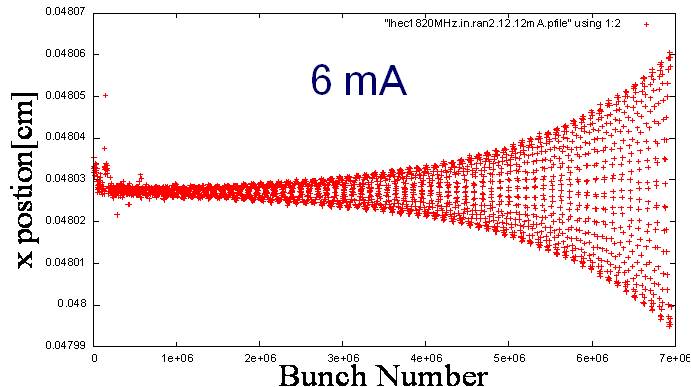}
\end{center}
\caption{
Large scale TDBBU simulation results for various beam currents: 4 (top
left), 5 (top right) and 6 mA.  }
\label{fig:bogacz3}
\end{figure}

In the simulation, for each cavity along the linac, the three offending HOM frequencies
are randomly distributed with the full width of $2$ MHz. In practice,
the HOM frequencies are generated using random numbers in that range
and these are distributed at each cavity. Twenty samples for different
HOM frequency distributions are generated.  The plots below show the
beam behaviour near the threshold. The horizontal axis corresponds to a
bunch number and can be considered as an axis of time (if the bunch
numbers are divided by frequencies). The vertical axis represents the
transverse beam position at the end of the second linac. We plot the
transverse positions of every $1117$th particles. The number $1117$ is
somehow arbitrary; however it is a large prime number chosen to avoid
an unexpected sub-harmonic redundancy in the data sampling.  The
simulation results for various beam currents: $4$, $5$ and $6$ mA are
illustrated in Figure $\ref{fig:bogacz3}$.

As illustrated in Figure $\ref{fig:bogacz3}$, the beam is stable at
$4$ mA. At $5$ mA the transverse position is increasing, which
indicate onset of the instability. Finally, at $6$ mA one explicitly
observes an exponential increase in transverse beam position - a vivid
case of beam instability. Therefore, we could infer that the BBU
threshold current is somewhere around $5$ mA.  One needs to keep in
mind, our study assumed the worst case interpretation of HOM's
measurement for a cavity with limited HOM suppression, only one pair 
of HOM dampers per cavity, positioned at 120 degrees to each other. 
This suggests more extended HOM damping
will bring the stability threshold above $6.5$ mA. 

Alternatively, one may consider a more realistic HOM selection extracted from the measurements summarised in Figure $\ref{fig:bogacz1}$. Such alternative choice of HOMs, with $Q_l=1\cdot 10^5$, is listed in the in table $\ref{tab:bogacz2}$.

\begin{table}[h]
  \centering
  \begin{tabular}{|c|c|c|}
    \hline
Frequency[MHz] & $Q_l$ & R/Q[Ohm] \\
$1003$ & $1\cdot 10^5$ & $32$ \\
$1337$ & $1\cdot 10^5$ & $32$ \\
$1820$ & $1\cdot 10^5$ & $32$ \\
\hline
  \end{tabular}
\caption{
An alternative selection of offending HOMs selected for the BBU simulation.
}
\label{tab:bogacz2}
\end{table}

Most recent BBU study with the above selection of offending HOMs, $\ref{tab:bogacz2}$, 
yields the beam stability threshold of $22$ mA, which is more than sufficient.
From this study we conclude that the $Q$ values of the transverse modes have to remain somewhere around $10^5$.

%% file: machine/eALRJowett.tex
\section{Performance as a Linac-Ring electron-ion collider}
\label{sec:eALRJowett}

The performance as an e-A collider can be evaluated on a basis similar to the Ring-Ring version of
the LHeC discussed in Section~\ref{sec:eAJowett}. Again, this relies on the fact that the nominal emittances for Pb beams in the LHC imply equal
geometric beam sizes, at the IP in particular.

\subsection{Heavy nuclei, e-Pb collisions}

   The Pb beam is specified in Table~\ref{tab:Pbparameters}.
   Assuming that the 60~GeV electron beam
specified in Table~\ref{tab:ParamLR} can be  adapted to the irregular
100 ns spacing of the Pb beam, the luminosity follows from Eq.~\ref{lumi} (including the additional
factor of $A=208$ to obtain the electron-nucleon luminosity):
\begin{equation}
L_{eN}  =
   \left\{ {\begin{array}{*{20}c}
   {9 \times 10^{31} {\text{ cm}}^{{- 2}} {\text{s}}^{{ - 1}} } & {{\text{(Nominal Pb)}}}  \\
   {1.6 \times 10^{32} {\text{ cm}}^{{-2}} {\text{s}}^{{-1}} } & {{\text{(Ultimate Pb)}}}
\end{array} } \right.
\end{equation}
where we assume
$ H_{hg}=H_D=1 $
for the additional factors in
Eq.~\ref{lumi}.

\subsection{Electron-deuteron collisions}

An estimate of the parameters for deuteron beams in the LHC is also given in
Section~\ref{sec:eAJowett}.
Proceeding in the same manner as above,
 we find that  \emph{electron-nucleon} luminosities of order
\(
L_{eN}  \gtrsim  3\times 10^{31} {\text{ cm}}^{-2} {\text{s}}^{{  - 1}}
\)
could be accessible in e-D collisions in a Linac-Ring LHeC.

%% file: machine/rinolfilrem.tex
\section{Polarised-electron injector for the Linac-Ring LHeC}

We present the injector for the polarised electron beam.
The issue of producing a sufficient number of polarised or unpolarised  
positrons is discussed in Section\,\ref{sec:positrons}. 

%

\begin{figure}
\begin{center}
\includegraphics[width=0.8\columnwidth]{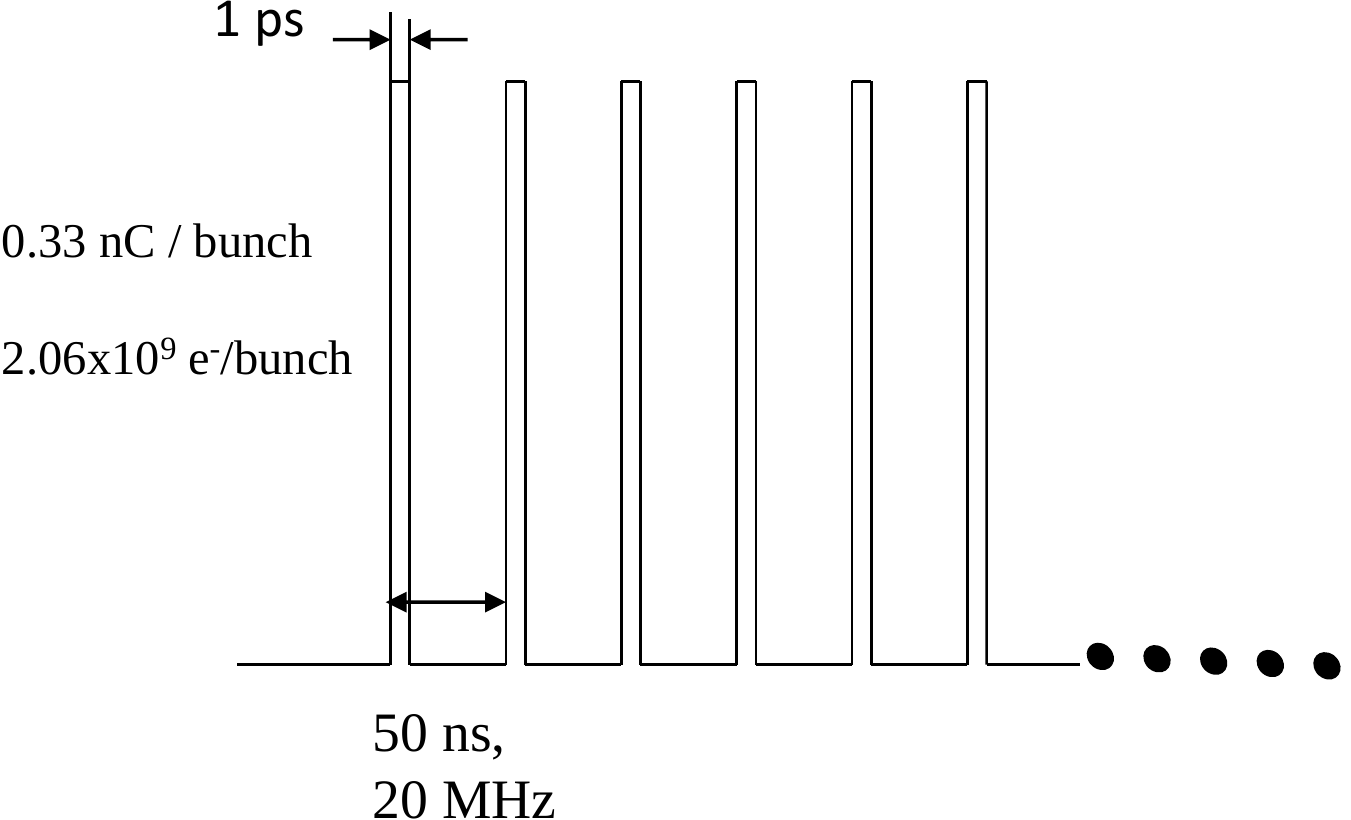}
\caption{Beam pattern at IP}
\label{Fig:LHeCInjector1}
\end{center}
\end{figure}

The Linac-Ring option is based on an ERL machine where the beam
pattern, at IP, is shown in Figure $\ref{Fig:LHeCInjector1}$.

With this bunch spacing, one needs $20\times10^9$ bunches/second and with the
requested bunch charge, the average beam current is
$20\times10^9$ b/s x $0.33$ nC/b = $6.6$ mA.

\begin{figure}
\begin{center}
\includegraphics[width=\columnwidth]{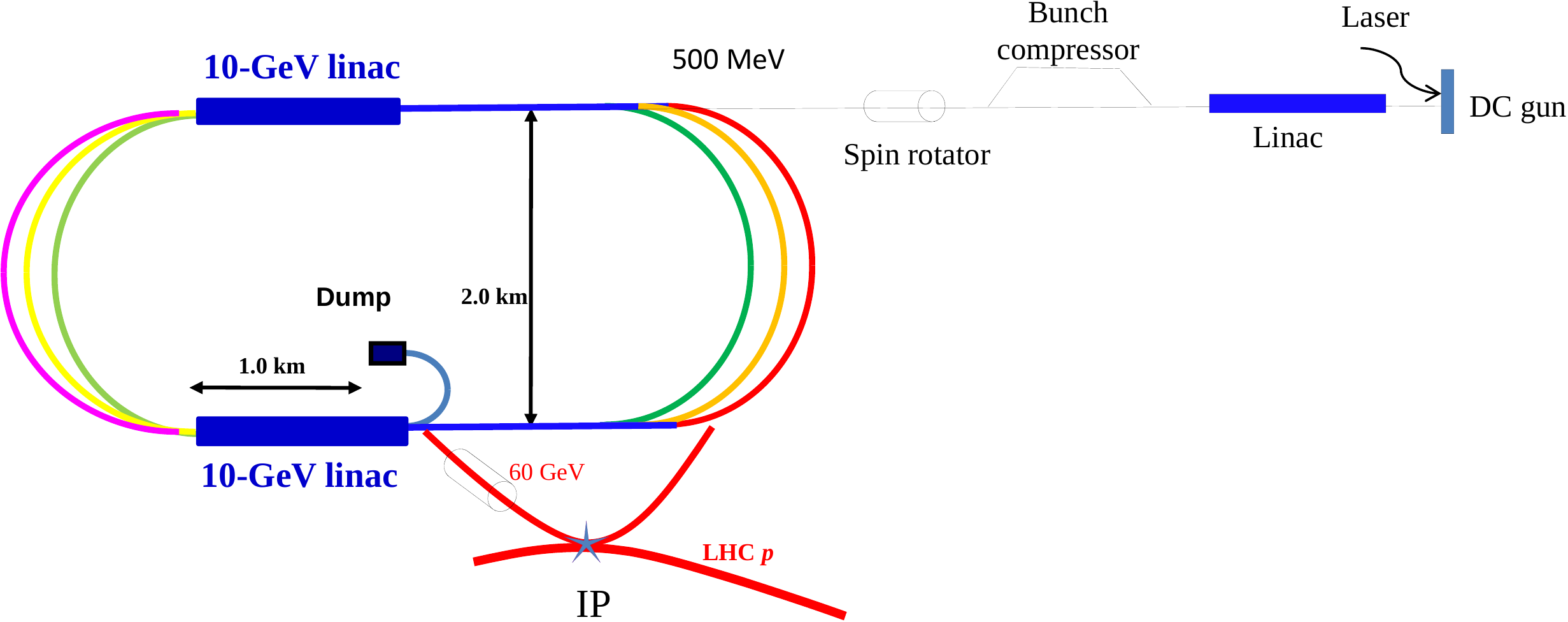}
\caption{Layout of the injector (not to scale).}
\label{Fig:LHeCInjector2}
\end{center}
\end{figure}

Figure $\ref{Fig:LHeCInjector2}$ shows a possible layout for the injector complex, as source of
polarised electron beam.

The injector is composed of a DC gun where a photocathode is illuminated
by a laser beam. Then a linac accelerates electron beam up to the
requested energy before injection into the ERL. Downstream a bunch
compressor system allows to compress the beam down to 1 ps and finally
a spin rotator, brings the spin in the vertical plane.

Assuming $90\%$ of transport efficiency between the source and the IP,
the bunch charge at the photocathode should $2.2\times10^9$
e-/b. According to the laser and photocathode performance, the laser
pulse width, corresponding to the electron bunch length, will be
between $10$ and $100$~ps.

Table $\ref{tab:rinolfi1}$ summarises the electron beam parameters at
the exit of the DC gun.

\begin{table}[h]
\centering
\begin{tabular}{|l|l|}
\hline
Parameters &    $60$ GeV ERL \\ \hline
Electrons /bunch & $2.2 \times 10^9$\\
 Charge /bunch &  $0.35$ nC\\
 Number bunches / s & $20 \times 10^9$\\
 Bunch length &  $10 - 100$ ps\\
 Bunch spacing &  $50$ ns\\
 Pulse repetition rate &  CW\\
 Average current  &  $7$ mA\\
 Peak current of the bunch & $3.5 - 350$ A\\
 Current density ($1$ cm) & $1.1 - 110$ A/cm$^2$\\
 Polarisation & $>90 \%$\\\hline
\end{tabular}
\caption{Beam parameters at the source.}
\label{tab:rinolfi1}
\end{table}

The challenges to produce the $7$ mA beam current are the following:

\begin{itemize}
\item {
a very good vacuum ($<10^{-12}$ mbar) is required in order to get a good lifetime.}
\item {
the issues related to the space charge limit and the surface
charge limit should be considered. A peak current of $10$ A with $4$ ns
pulse length has been demonstrated. Assuming a similar value for the DC
gun, a laser pulse length of $35$~ps would be sufficient to produce the
requested LHeC charge.}
\item {
the high voltage ($100$ kV to $500$ kV) of the DC gun could induce
important field emissions.}
\item {
the design of the cathode/anode geometry is crucial for a beam
transport close to $100\%$.}
\item {
the quantum efficiency should be as high as possible for the
photocathode ($\sim 1\%$ or more).}
\item {
the laser parameters ($300$ nJ/pulse on the photocathode, $20$ MHz
repetition rate) will need some R\&D according to what is existing
today on the market.}
\item {
the space charge could increase the transverse beam emittances.
}
\end{itemize}

In conclusion, a trade-off between the photocathode, the gun and
the laser seems reachable to get acceptable parameters at the gun exit.
A classical Pre-Injector Linac accelerates electron beam to
the requested ERL energy. Different stages of bunch compressor
are used to compensate the initial laser pulse and the space charge
effects inducing bunch lengthening. A classical spin rotator
system rotates the spin before injection into the ERL.

%% file: machine/SpinRotator.tex
\section{Spin Rotator}
\subsection{Introduction} 
The potential of studying new physics in
high precision QCD, substructure etc.~at LHeC requires polarised
electrons with spins aligned longitudinally at the collision point.
For the linac-ring version of the LHeC 
the electron beam can be generated with 80-90\%
polarisation using a photocathode source. To avoid polarisation
loss of the high energy electron beam, the polarisation vector needs to be
aligned vertically during the acceleration in a re-circulating linac
and then brought into the longitudinal direction for collision. 
This section reports possible design choices for the LHeC spin rotator.

The motion of a
spin vector $\vec{S}$ in an accelerator is governed by the Thomas-BMT
equation~\cite{bmt} 
\begin{equation}
\frac{d\vec{S}}{dt}=\frac{e}{m\gamma}\vec{S}\times [(1+G\gamma)\vec{B}_{\bot}+(1+G)\vec{B}_{\parallel}]
\label{bmt}
\end{equation}
where $e$, $m$ and $\gamma$ are the electric charge, mass and Lorentz
factor of the particle. $G$ is the anomalous g-factor. For protons,
$G=1.7928474$ and for electrons, $G=0.00115$.  $\vec{B}_{\bot}$ and
$\vec{B}_{\parallel}$ are the magnetic field perpendicular and
parallel to the particle velocity direction, respectively. In
(\ref{bmt}) the magnetic field is in the laboratory frame while the
spin vector $\vec{S}$ is in the particle rest frame. 
Eq.~(\ref{bmt}) implies that, 
in a perfectly flat  circular accelerator with $\vec{B}_{||}=0$, 
spin vectors precess, on average,
$G\gamma$ times faster than the direction of the design orbit precesses in the fixed laboratory frame. 
For the electron accelerator of the LHeC, which 
consists of two 10~GeV superconducting linear accelerators linked by
six $180^{\circ}$ arc paths, the depolarisation due to the arcs is
negligible if the polarisation is aligned vertically in the arcs.

Eq.~(\ref{bmt}) also shows that both the dipole fields and the solenoid fields can
be used to manipulate the spin motion. However, the effect of a solenoid
field on the spin motion decreases linearly with beam energy, while the
effect of a dipole field remains almost independent of beam energy.

\subsection{LHeC spin rotator options}
To produce longitudinally oriented polarisation at the final
collision point for a 60-GeV electron beam, 
two options have been explored: 
\begin{itemize}
\item A low energy spin rotator at the LHeC injector to place the polarisation
  vector in a direction chosen so that after the precessions in  
  all the arcs the outgoing polarisation is longitudinal 
at the IP. 
\item A dedicated high energy spin rotator close to the IP 
  which brings vertically aligned spin vectors into
  the longitudinal direction. For this option, a low energy spin rotator at the
  injector is also required in order 
  to produce a vertically polarised electron beam for acceleration. 
\end{itemize}

The details of the two options are as follows.
\subsubsection{Low energy spin rotator}
For the LHeC physics program, the polarisation of a 60~GeV electron beam
needs to be aligned longitudinally at the collision point which is
after the last arc and the acceleration. The most economical way to
control the polarisation direction at the collision point is to control the
polarisation  direction of the low energy electron beam at an early stage of
injector using a Wien Filter, i.e.~a traditional low energy spin
rotator. Since a spin vector rotates by  $G\gamma \pi$ each time it passes
through a $180^{\circ}$ arc, the goal of the Wien Filter is to put the
polarisation into the horizontal plane with an angle to the direction of
the particle velocity chosen so as 
to compensate the spin rotations before collision.

For the layout of LHeC, i.e. two linear accelerators linked by two
arcs, a spin vector rotates by an amount 
\begin{equation}
\phi_{arc} =G\pi [\gamma_{i}(2n-1)+\Delta \gamma n(2n-1)]
\end{equation}
during its $n$th path. Here, $\gamma_{i}$ is the initial Lorentz factor
of the beam and $\Delta \gamma$ is the energy gain of each linear
accelerator. 
In addition, the LHeC also employs a horizontal 
dipole on either side of the IP to separate the
electrons from the protons. 
These dipoles have a field of 0.3~T and
and span 9~m from the collision point. 
For the 60~GeV electron beam, such a bending magnet 
rotates a spin vector by $\phi_{IP}=104.4^{\circ}$. 
Considering an initial
energy of 10 GeV (after the first path through the linac)
and for each linear accelerator an energy gain of 10 GeV,
Table~\ref{table1} lists the amount of spin rotation through the arcs
and the amount of spin rotation through the final bending dipole at
the collision point for a 20, 
40 and 60-GeV beam, respectively.
\begin{table}[h]
\begin{center} 
\begin{tabular}{|c|c|c|c|}
\hline
beam energy & \# of path & $\phi_{arc}$ & $\phi_{IP}$ \\
  GeV     & n    & [degree] & [degree]\\
\hline
20 & 1 & 8101.8 &  34.8\\
40 & 2 & 36457.9 &  69.6\\
60 & 3 & 81017.6 & 104.4 \\
\hline
\end{tabular}
\caption{Total spin rotation from arcs and final bending dipole at collision point.}
\label{table1}
\end{center} 
\end{table}
Here, the amount of spin rotation at the IP refers to the net spin rotation 
modulo $360^{\circ}$. 

\begin{figure}[tbh]
\begin{center}
\includegraphics*[height=2in,width=3in]{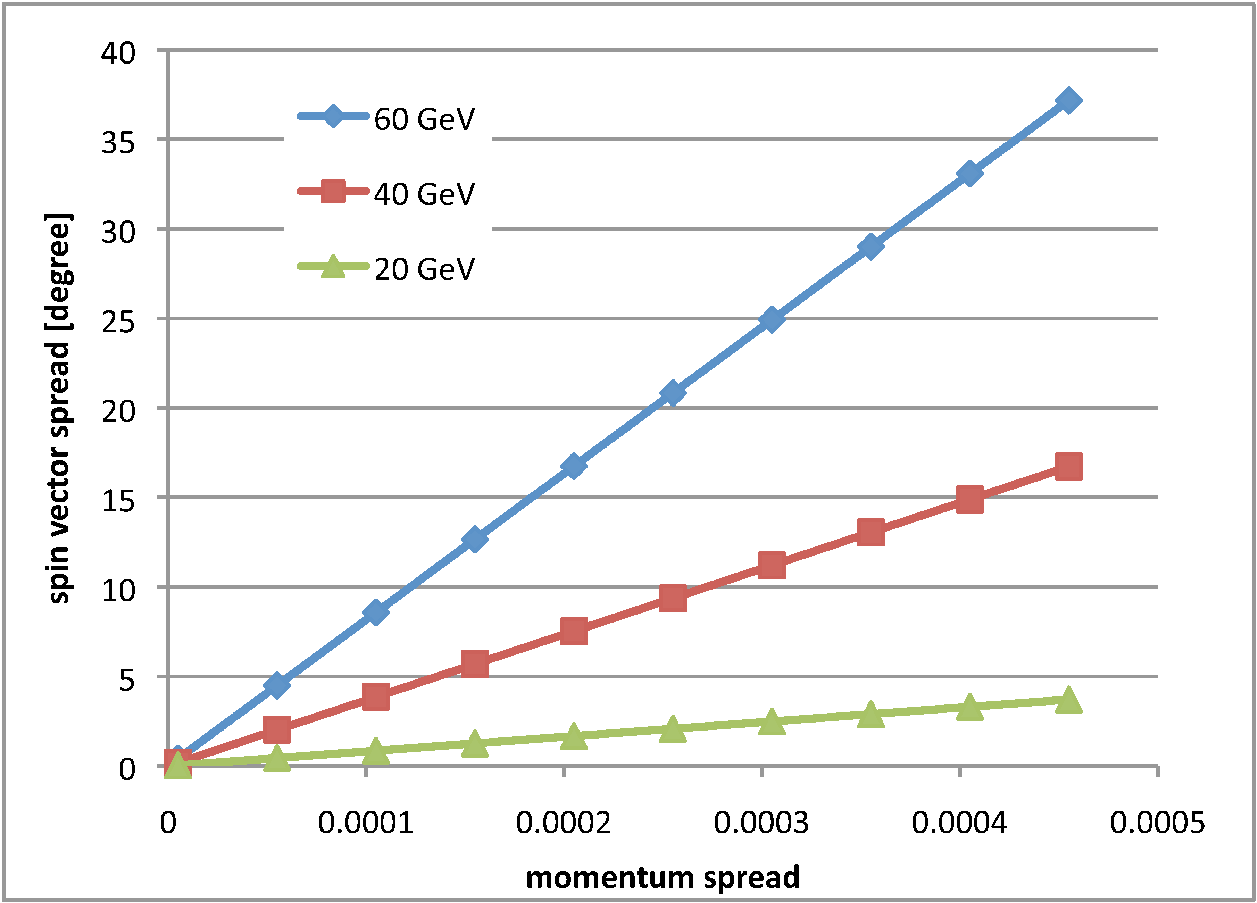}
\caption{Calculated spin vector spread as a function of momentum
  spread. The effective polarisation is proportional to the cosine of the spin vector
  spread angle: e.g.~for an angle of 30 degrees, the effective
  polarisation is ~86\% of the initial beam polarisation}
\label{spinspread}
\end{center} 
\end{figure}

Since the spin rotation is proportional to the beam
energy, for a beam of particles with non-zero momentum spread,
different amounts of spin rotation generate a spread of spin
vector directions. This results in an effective polarisation loss due to the associated 
spread of the spin vectors. 
Figure~\ref{spinspread} shows the angular spread of the spin
vectors for off-momentum particles at 20, 40 and 60 GeV, respectively. 
It shows that for a 60-GeV 
electron beam, a relative momentum spread of $3\times 10^{-4}$ can
cause about 
$(1-\cos (25^{\circ}))\approx 10\%$ 
effective polarisation loss due to the spread of the
spin vectors. This level of polarisation loss is
undesirable and would compromise the physics 
reach of the LHeC.

\subsubsection{High energy spin rotator}
\begin{figure}[tbh]
\begin{center}
\includegraphics*[height=1in,width=3in]{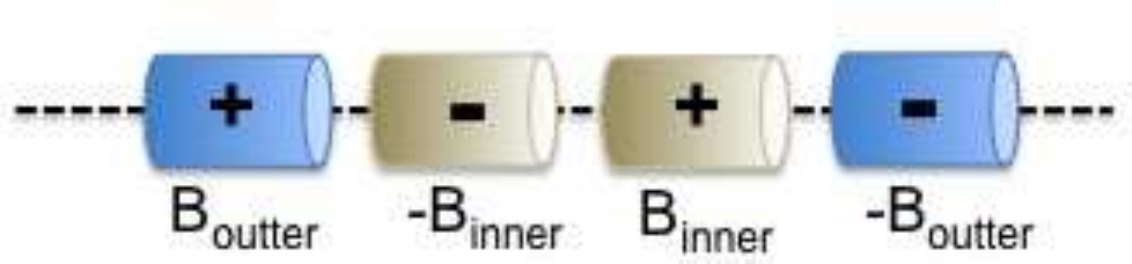}
\caption{Schematic layout of the LHeC spin rotator, consisting of 
a total of four helical dipoles with alternating helicity marked as + and -. 
The two outer helical dipole fields have  opposite polarities, and 
the polarities of the two inner helical dipoles are opposite too.}
\label{layout}
\end{center} 
\end{figure}

In order to provide longitudinal polarisation  without
sacrificing the size of the polarisation, one can adopt the traditional approach of
high-energy polarised beams at HERA and RHIC, i.e.~rotate the spin
vectors into the 
vertical direction before the beam gets accelerated to high energy. 
Then with the spin vectors aligned along the main bending magnetic
field direction,  spreading  of the spin vectors due to
the momentum spread is prevented. For the current compact LHeC 
final-focusing system (FFS), we propose
to use RHIC type spin rotator~\cite{RHICdesignmanual,vadimp} for the
LHeC. Besides saving space by being short, this approach also has
the advantage of providing an independent full control of the direction of the polarisation, as well as a nearly energy-independent spin rotation for
the same magnetic field.  The four helical dipoles are arranged in a 
fashion similar to the RHIC spin rotator, i.e.~with alternating
helicity. Figure~\ref{layout} shows the schematic layout.  Each helical
dipole is 3.3~m long and the helicity alternates between right hand and
left hand from one helical dipole to the next. 
The two inner helical dipoles
have the same magnetic field but opposite helicity. 
The same applies for the two outer helical dipoles.

For each helical dipole, the magnetic field, on axis, is given by 
\begin{eqnarray}
B_{x}=B\cos kz\; ,\\
B_{y}=B\sin kz\; ,\\
B_{z}=0 \; ,
\end{eqnarray}
where $B_{x,y,x}$ are the horizontal, vertical and longitudinal
components of the magnetic field, respectively, $z$ is the
longitudinal distance along the helical dipole
axis, while $|k|=2\pi/\lambda$ and $\lambda$ are the 
wave number and the wave length of the helical field, respectively.

\begin{figure}[tbh]
\begin{center}
\includegraphics*[height=3in,width=3.3in]{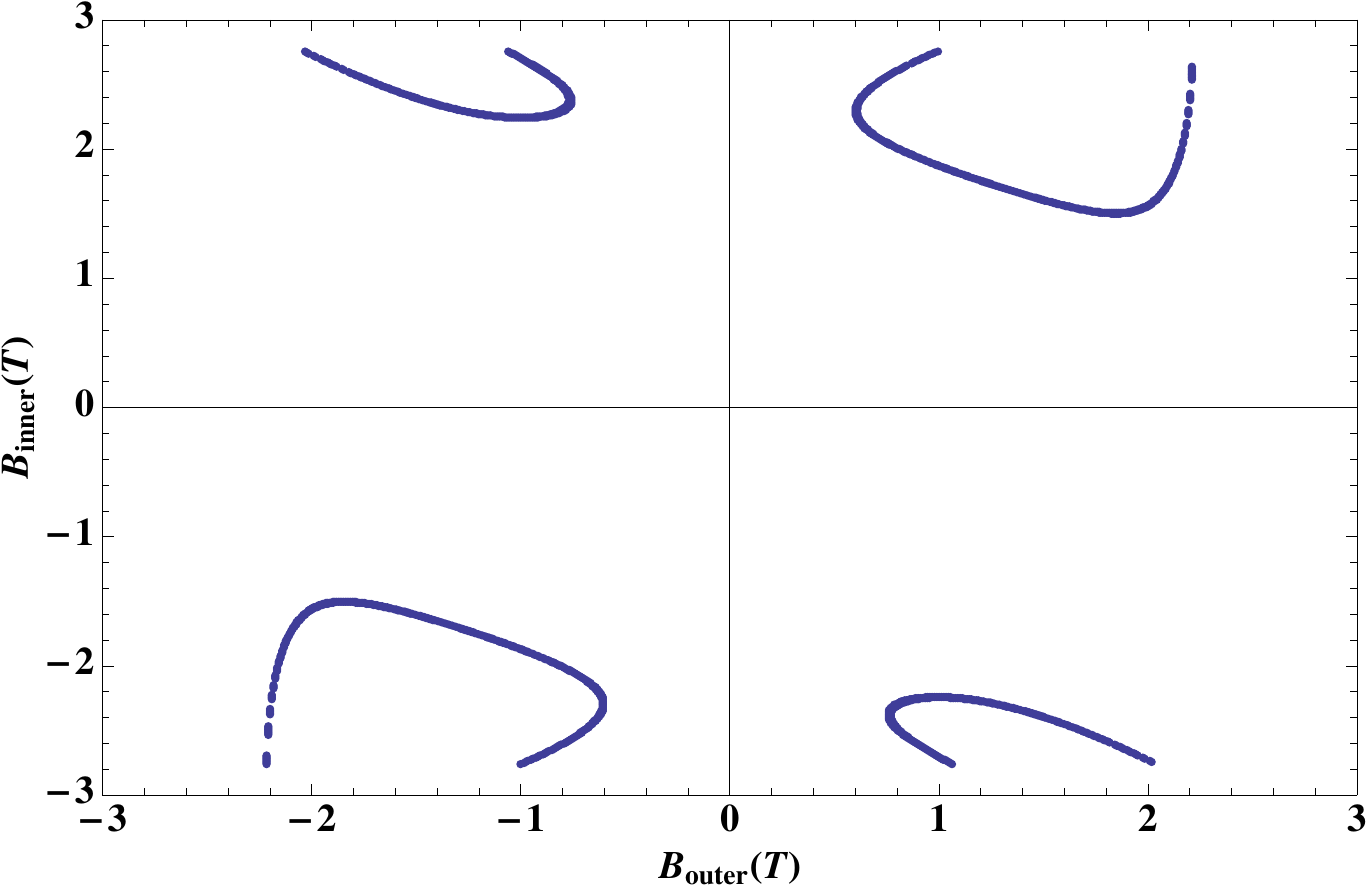}
\caption{Correlation of the outer 
and inner helical dipole magnetic field strengths for a spin 
rotator which is designed to bring a vertically aligned spin 
vector to the horizontal plane.
The length of the helical dipoles is taken to be 3.3~m each.
}
\label{p1p2}
\end{center} 
\end{figure}

\begin{figure}[tbh]
\begin{center}
\includegraphics*[height=3in,width=3.3in]{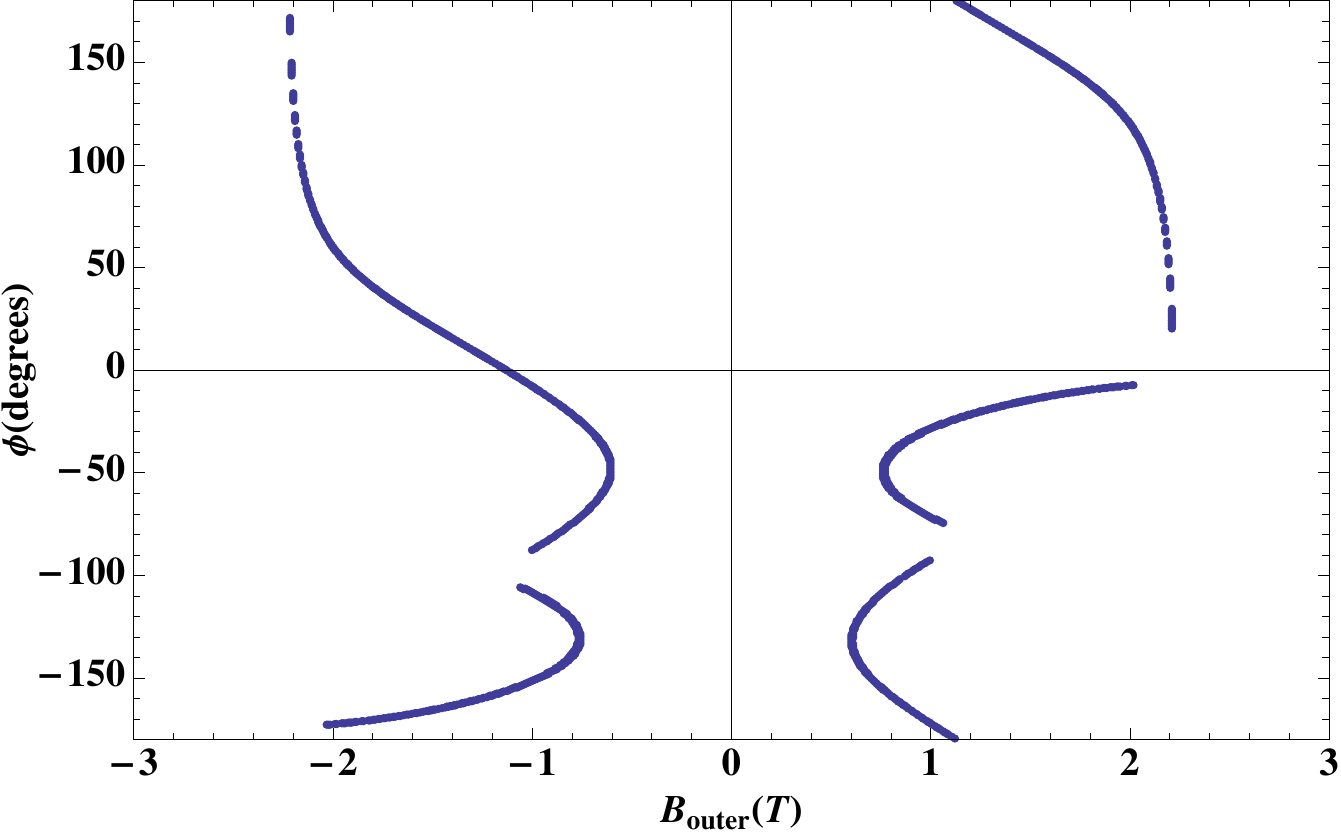}
\caption{Spin vector direction in the horizontal plane as a 
function of the outer helical magnet field strength.
The length of the helical dipoles is taken to be 3.3~m each. 
}
\label{phip1}
\end{center} 
\end{figure}

For the spin rotator, all helical dipoles are chosen to be one period long,
i.e.~$\lambda = L$ , where L is the length of each helical
dipole, and, depending on the helicity, $k/|k|=\pm
1$.  Fig.~\ref{p1p2} shows the correlation of the magnetic field for
the inner and outer helical magnets of a spin rotator which brings 
spin vectors from the vertical direction into the horizontal plane. 
Figure~\ref{phip1} presents the calculated angle of a spin vector
for each outer helical magnet field. Both plots show that this design
allows for a flexible adjustment for the direction of the polarisation by
varying the outer and inner helical magnetic fields, respectively.

This rotator will be placed in the straight section between the
end of the second linac and the FFS, upstream of the final
bending dipole at the collision point as well as of 
three bends immediately  upstream of the final triplet. 
As mentioned, the 0.3-T final bending dipole next to the IP rotates spin vectors
by 104.4 degrees for a 60-GeV electron beam, while the other, weaker 
three bends rotate spin vectors by only $-1.8$ degrees.  
To obtain longitudinal polarisation at the IP, the spin rotator must bring 
the polarisation vector from the vertical direction 
into the horizontal plane at an angle of 
102.6 degrees from the longitudinal direction.  
This requirement then determines the
magnetic field of the inner and outer pairs to be 2.1 T and 1.7
T, respectively. 
The maximum horizontal orbital excursion is 18 mm in the 2.1 T dipole 
and 15 mm in the 1.7 T dipole.  
The fine tuning of the direction of the polarisation vector
can be achieved by empirically adjusting the helical-dipole magnetic
field strengths on the basis of the measurements with polarimeters 
installed before and after the collision point.

\begin{figure}[tbh]
\begin{center}
\includegraphics*[height=3in,width=3.5in]{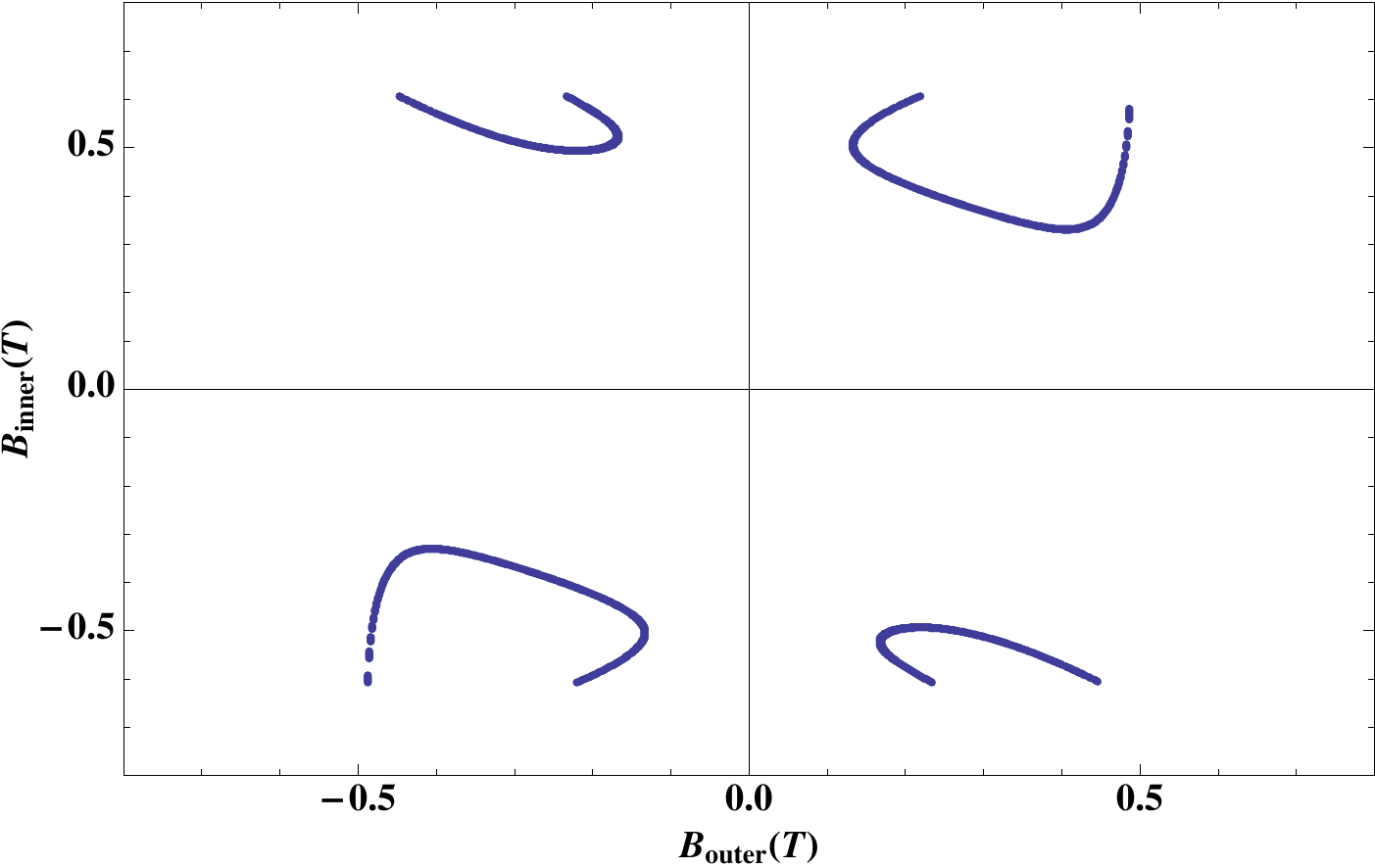}
\caption{Correlation of the outer and inner helical dipole magnetic field strength
for a longer spin rotator designed to bring  vertical 
spin vectors into the horizontal plane,
for a longer design with reduced synchrotron radiation, 
where each helical dipole has a length of 15~m.}
\label{p1p2_V2}
\end{center} 
\end{figure}

\begin{figure}[tbh]
\begin{center}
\includegraphics*[height=3in,width=3.5in]{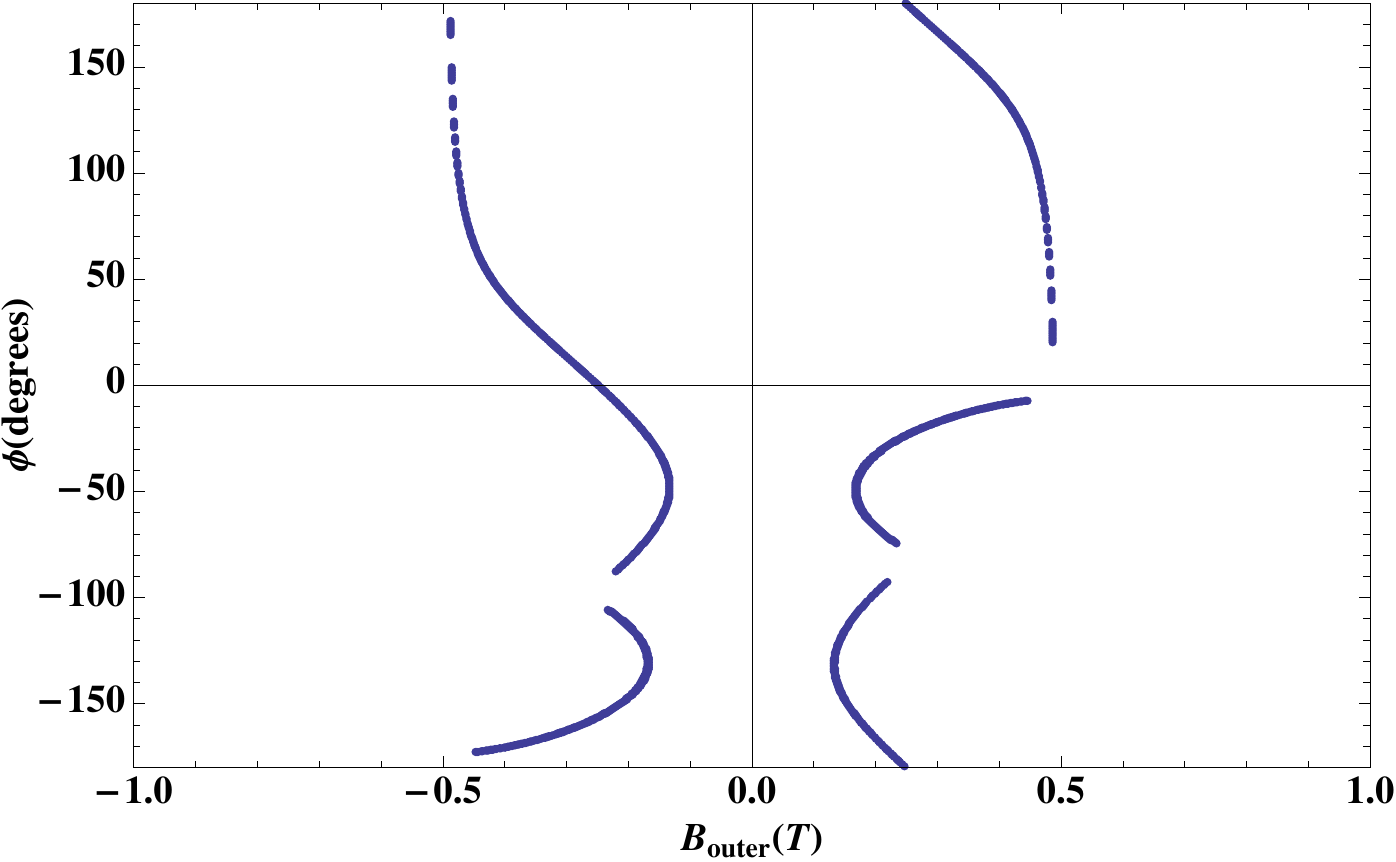}
\caption{Spin vector direction in the horizontal plane as 
a function of the outer helical magnet field strength 
for a longer design with reduced synchrotron radiation, 
where each helical dipole has a length of 15~m.}
\label{phip1_V2}
\end{center} 
\end{figure}

The $\sim$MW synchrotron radiation power emitted 
by the 60-GeV electron beam passing through the spin rotator
can be reduced by lengthening the system, while 
lowering the magnetic field of the helical dipoles. 
Figure~\ref{p1p2_V2} illustrates the correlation of the magnetic field
for the inner and outer helical magnets for a $\sim$5 times  
longer spin rotator, where each helical dipole has a length of 15~m. 
Figure~\ref{phip1_V2} presents the calculated angle of the polarisation vector 
as a function of the outer helical magnet field strength. 
For a 60~GeV electron beam, the magnetic fields of the inner 
and outer pairs need to be 0.46 T and 0.37 T, respectively.  
These fields will rotate spin vectors into the horizontal plane 
after the exit of the spin rotator.
For this longer system, the maximum horizontal 
orbital excursion is 82 mm for the 0.46 T magnet
and 67 mm for the 0.37 T magnet.  


\subsection{Polarimetry}
To measure the polarisation of 
the high-energy electron beam a Compton polarimeter is foreseen.
Such a polarimeter detects the electrons and photons 
produced in Compton scattering off the electron beam
of an intense circularly polarised laser beam \cite{woods}. 
A Compton polarimeter requires space
to accommodate the laser as well as detectors.
For high precision measurements an efficient
separation of the Compton-scattered electrons 
from the main electron beam is required. 
 
\begin{figure}[tbh]
\begin{center}
\includegraphics*[width=3.3in]{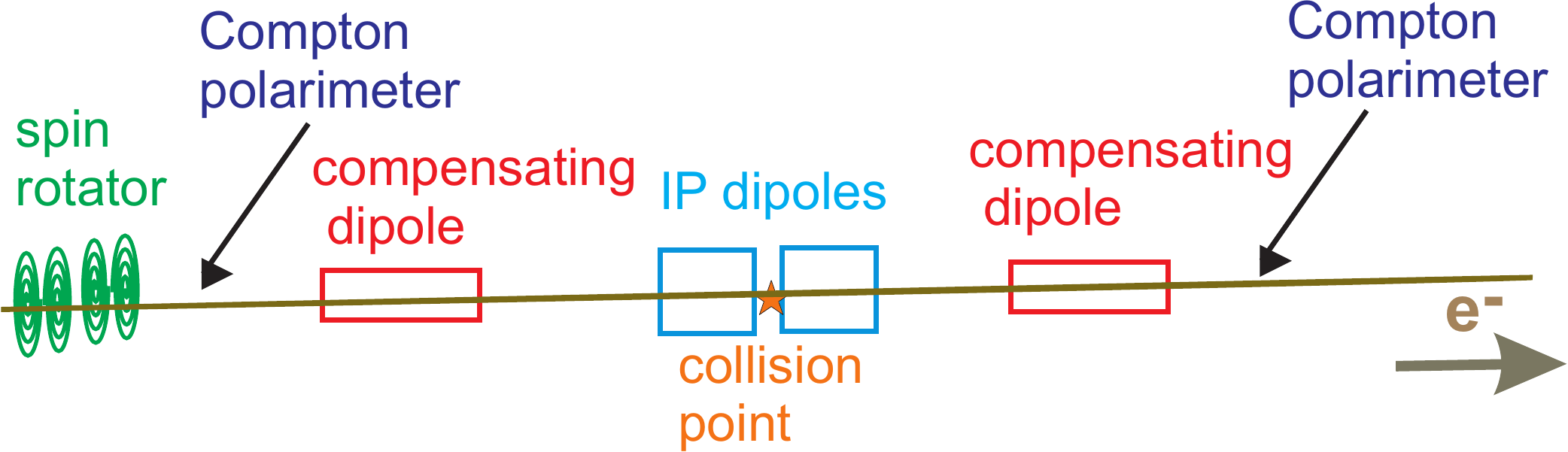}
\caption{Schematic of spin-related IR layout 
with spin rotator, two polarimeters, and compensating bends.}
\label{sketch}
\end{center} 
\end{figure}

The polarimeter could be placed either 
upstream or downstream of the IP.  
We tentatively consider two polarimeters, 
one on either side of the IP, which would allow   
excluding or quantifying any depolarising effects 
in the final focus or due to the collision process. 
In order to place these polarimeters at locations 
where the polarisation 
is longitudinal, we propose installing (or using) 
additional bending magnets so that the deflection angle 
by the IP dipoles is exactly compensated and the 
net spin precession angle between the polarimeter 
and the IP is zero, also taking into account the small 
energy change due to synchrotron radiation emitted 
in these magnets.
In this way maximising the longitudinal polarisation at 
either polarimeter by scanning the field strengths of
the two pairs of helical magnets in the upstream
spin rotator automatically maximises the longitudinal
polarisation at the collision point. 
The polarisation levels measured at the
two polarimeters allow the 
polarisation loss in the collision as well as 
the effective polarisation to be deduced. This is 
important for particle physics.
Figure \ref{sketch} sketches the overall
spin-related layout of the LHeC interaction region (IR).  

\subsection{Conclusions and Outlook}
This section has presented a flexible spin rotator for the 
LHeC high-energy electron beam. 
The proposed design, based on a group of helical dipoles
next to the final collision point, similar to the spin rotator at
RHIC, satisfies the requirement of delivering a high-energy electron beam with
high longitudinal polarisation. It also has the additional
merits of being compact and flexible. 
For this approach, a low
energy spin rotator like a Wien Filter as part of the injector is also required 
to rotate spin vectors into the vertical direction prior to acceleration.

Synchrotron radiation emitted from the high-energy 
spin rotator is a concern. This can be addressed 
by optimising the field strengths and the lengths of the helical dipoles. 

Detailed calculations including helical dipole design, orbital and
spin tracking in the spin rotator are in progress.

\vspace{2mm}

Note that if the long versions of the helical dipoles are deemed to be
necessary, a rotator of the HERA type \cite{Barber:1994ew}, relying as it does, on simple
normal-conducting ring dipoles, might compete. In the original design
these rotators are about 50 m long and employ interleaved horizontal
and vertical bends to generate closed interleaved horizontal and
vertical bumps in the design orbit. The HERA rotators cover the range
27 -- 39 GeV and at 39 GeV the radius of orbit curvature is about 500
m in the 5 m vertical bends.  At 60 GeV the fields would be about the
same as at 39 GeV, namely $\sim$0.16 T but the radius would be about
750 m.  Since the fields in such a rotator are almost independent of
the energy, the geometry of the rotators is energy dependent. The
magnets in the HERA rotator are therefore mounted on remotely
controlled jacks and the beam pipe has flexible, eddy-current transporting joints. 
The sign of the longitudinal polarisation is changed
by reversing the sign of the fields in the vertical bends.  The
vertical excursion of the orbit would be about 14 cm.  Since the sign
of the polarisation can be chosen at the source, a jacking system
would not be needed if the energy were fixed. Then the simplicity of
the magnets and the potentially lower fields might have advantages
w.r.t. the use of helical dipoles.

%% file: machine/positrons.tex
\section{Positron options for the Linac-Ring LHeC}
\label{sec:positrons}

\subsection{Motivation}
It is known that the generation of an intense positron
beam with a linac configuration is a particular challenge.
This raises the question as to how crucial the availability
of positron-proton scattering to the LHeC is. Reasons for
the importance of $e^+p$ scattering are given in the
physics chapters and have been summarised in an
introduction to a topical meeting~\cite{meetposi}
in May 2011 at CERN, the technical results of which are
summarised below. For the physics program, the following
topics may serve as important example processes which
 require very high statistics positron (and electron) data:
\begin{itemize}
\item{If there exist so far unknown  resonant states of leptons
and partons, quarks or/and gluons, the  asymmetry between 
the $e^+p$ and $e^-p$ cross sections determines the fermion
number of the produced leptoquark to be $F=2$, as for an $e_Lu$
state of charge $-1/3$, or $F=0$ for an $e_L\overline{u}$ state
of charge $-5/3$.} 
\item{If there appears a new contact interaction, its nature may be 
disentangled by considering its charge dependence. If there was an
excited electron observed, one surely would like to check whether
the positron has the same structure.}
\item{It has been a long standing question whether the
strange quark and anti-quark distributions are different,
for which neutrino-nucleon data provide certain hints.
With electron and positron charged current data, this can
be resolved and both $s$ and $\overline{s}$ can be
measured. Similarly one will be able to measure single top
and single anti-top quark distributions for the first time.}
\item{Access to valence quarks at low $x$ is possible with the
precision measurement of the $xF_3^{\gamma Z}$ structure
function, which can be accessed only with high statistics
NC cross section asymmetry data.}
\item{High statistics beam charge asymmetry data are
essential to access generalised parton distributions at low $Q^2$}
\end{itemize}
An example for the importance of $e^+p$ scattering with high
but perhaps not maximum luminosity is the precision measurement
of the longitudinal structure function $F_L$, in which the
charge symmetric background at low scattered electron energies
has to be experimentally determined
and subtracted in order to safely reach the region of highest
sensitivity to $F_L$. One would finally like to note that if the
positron-proton luminosity was 
significantly lower than the electron-proton 
luminosity, there would always be a tendency to 
preferentially run with electrons
in order to collect a maximum integrated luminosity for those
processes and topics which are less or not dependent on the
availability of both beam charge configurations. Examples here
are the precision measurement in polarised $e^-p$ scattering
of the weak mixing angle, the physics at low $x$ or the precision
measurement of $\alpha_s$. It is the physics beyond the standard 
model, and the searches for it,
which has the highest demands on the $e^+p$ luminosity.
One concludes that the physics demands for the availability
of intense $e^+p$ scattering are very strong. 
A further aspect regards the importance of positron beam polarisation
which may deserve further consideration. 
\subsection{LHeC Linac-Ring $e^{+}$ requirements}
Table $\ref{tab:rinolfi2}$ compares the $e^+$ beam flux foreseen for LHeC with those
obtained at the SLC, and targeted for CLIC and the ILC.

\begin{table}[h]
\centering
 \begin{tabular}{| l | c | c | c | c | c |}
\hline
 & SLC & CLIC & ILC & LHeC & LHeC \\
&  & ($3$ TeV) & ($500$ GeV) & (p$=140$) & (ERL) \\
 \hline
Energy (GeV) & $1.19$ & $2.86$  & 4 & $140$  & $60$  \\ 
$e^+$/bunch at IP ($\times 10^9$) &$40$& $3.72$ & 20 & $1.6$ & $2$ \\ 
Norm.~emittance (mm.mrad) & 30 (H) & 0.66 (H) & 10 (H) & 100 & 50\\
                          & 2 (V)  & 0.02 (V) & 0.04 (V) & & \\
Longit.~rms emittance (eV-m) & 7000 & 5000 & 60000 & 10000 & 5000 \\
$e^+$/bunch after capture ($\times 10^9$) & $50$& $7.6$ & 30 & $1.8$ & $2.2$ \\ 
Bunches / macropulse & $1$& $312$ & 2625 & $10^5$ & NA \\ 
Macropulse repetition rate & $120$& $50$ & 5 & $10$ & CW \\ 
Bunches / second & $120$& $15600$ & 13125 & $10^6$ & $20 \times 10^6$ \\ 
$e^+$ / second ($\times 10^{14}$) & $0.06$& $1.1$ & 3.9& $18$ & $440$ \\ 
\hline
\end{tabular}
\caption{Comparison of the $e^+$ flux.}
\label{tab:rinolfi2}
\end{table}

The SLC (Stanford Linear Collider) was the only linear-collider type machine 
which has produced $e^+$ for a high-energy particle physics experiment.
The flux for the CLIC project (a factor 20 compared to
SLC) is already considered challenging \cite{lrferrara} 
and possible options with hybrid targets are
under investigation on paper. Even more positrons would be required for the ILC. 
The requested LHeC flux 
for pulsed operation at 140 GeV (a factor 300 compared to SLC) 
could be obtained, in
a first approximation, with 10 $e^+$
target stations working in parallel. 
Several more advanced solutions are being considered to meet the requested LHeC flux 
for the CW option (a factor 7300 compared to SLC).

\subsection{Mitigation schemes}
Two main approaches can lessen the demands on 
the rate of positrons to be produced at the source, namely   
\begin{itemize}
\item 
{\bf Recycling the positrons after the collision}, with 
considerations on $e^+$ emittance after collision, emittance growth in the 
60-GeV return arc due to synchrotron radiation, and 
possible cooling schemes, 
e.g.~introducing 
a tri-ring system with fast laser cooling in the central ring 
(see below), or using a large damping ring.
If 90\% of the positrons are recycled the requirement for 
the source drops by an order of magnitude.    
\item 
{\bf Repeated collisions on multiple turns}, e.g. using a (pulsed) phase-shift
chicane in order to recover 60 GeV when reaching the collision point again
on the following turn.  
\end{itemize}

\subsection{Cooling of positrons}
One of the most challenging problems associated with 
the continuous production of  positrons is cooling (damping) 
of the positron beam emerging from a source or being 
recycled after the collision. 
Possible cooling scenarios include pushing the 
performance of a large conventional damping ring with the size of the SPS, and 
a novel compact tri-ring scheme.

\subsubsection{Damping ring}  
\begin{figure}[htb]
\centering
\vspace{0pt}
\includegraphics*[width=10cm]{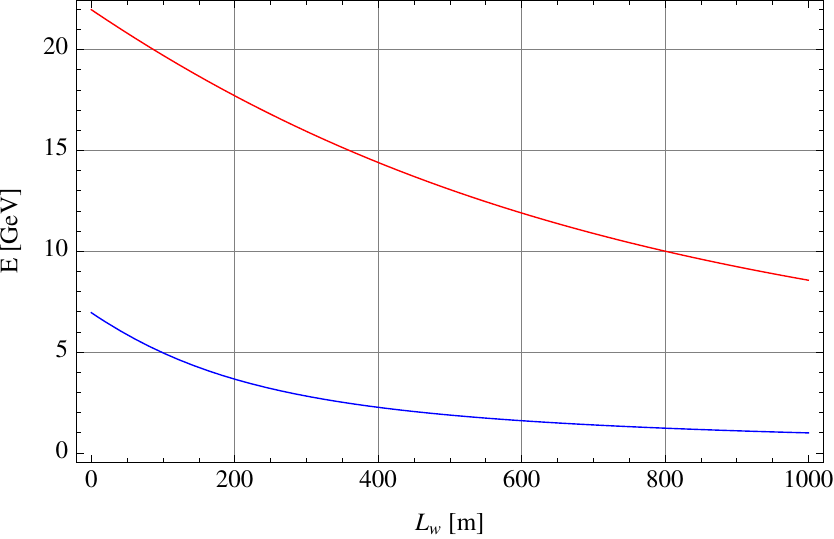} 
\vspace{-10pt} 
\caption{Dependence of the damping ring energy on the total wiggler length for 
a transverse damping time of 2 ms (red curve) and 20 ms (blue curve). }
\label{fig:EN_LW}
\end{figure}

The 6.9-km SPS tunnel can accommodate a train of 9221 bunches with 2.5 ns bunch spacing. 
Considering a maximum bending field of 1.8 T and a wiggler field of 
1.9 T, there is a parametric interdependence between beam energy, the total wiggler length and the damping time. 
Figure~\ref{fig:EN_LW} shows the 
dependence of the damping ring energy on the total wiggler length for a damping time 
of 2 ms (red curve). 
Without wigglers, the ring has to run at 22 GeV, whereas for around 10 GeV, wigglers 
with a total length of 800 m are needed. 
The blue curve represents the same dependence when a 10 times lower repetition rate 
is considered, which increases the required damping time by an order of magnitude. 
In that case, the ring energy without any wigglers can be reduced to 7 GeV 
and it can be dropped to less than 4 GeV for a total wiggler length of 200 m.

A tentative parameter list for low (10 Hz) and high repetition rate (100 Hz) 
is shown in Table ~\ref{tab:param}, considering 234 bending magnets of 0.5-m long dipoles with 1.8-T bending field. 
The wiggler field for the high-repetition option of 1.9 T 
along with a wiggler period of 5 cm is within the reach of modern hybrid wiggler technology. 
A big challenge is the high energy loss per turn for this case, which requires around 300 MV of total RF voltage and implies
an average synchrotron-radiation (SR) power of 25 MW. In the low repetition case, 
the RF voltage and SR power are an order of magnitude more relaxed.

\begin{table}[!b]
\begin{small}
\begin{center}\begin{tabular}{|l|l|l|}
\hline
Parameter [unit]			&High Rep-rate	& Low Rep-rate \\ \hline
Energy [GeV]				& 10 	&   7 \\
Bunch population [10$^9$]	& 1.6  	&    1.6   \\
Bunch spacing [ns]			& 2.5 	&   2.5  \\
Number of bunches/train	& 9221	& 9221\\
Repetition rate [Hz]			& 100	& 10 \\
Damping times trans./long. [ms]	& 2/1	& 20/10 \\
Energy loss/turn [MeV]		& 230	& 16 \\
Horizontal norm. emittance [$\mu$m]	& 20    & 100 \\
Optics detuning factor		& 80    	& 80 \\
Dipole field [T] 				& 1.8	& 1.8 \\
Dipole length [m] 			& 0.5	& 0.5 \\
Wiggler field	[T]    			& 1.9	& - \\
Wiggler period [cm] 			& 5		&       - \\
Total wiggler length [m] 		& 800	& - \\
Dipole length [m] 			& 0.5	& 0.5 \\
Longitudinal norm. emittances [keV.m]	& 10	 & 10 \\
Momentum compaction factor & 10$^{-6}$	 &10$^{-6}$ \\
RF voltage 		[MV]		& 300	 & 35 \\
rms energy spread 	[\%]		& 0.20	 & 0.17 \\
rms bunch length 	[mm]	& 5.2	 & 8.8 \\
average power	[MW]		& 23.6	 & 3.6 \\
\hline
\end{tabular}
\caption{Tentative parameter list for a damping ring 
in the SPS tunnel considering high and low repetition-rate options.}
\label{tab:param} 
\end{center}
\end{small}
\end{table}

\subsubsection{Tri-Ring scheme}
\begin{figure}[h]
\begin{center}
\includegraphics[width=0.7\columnwidth]{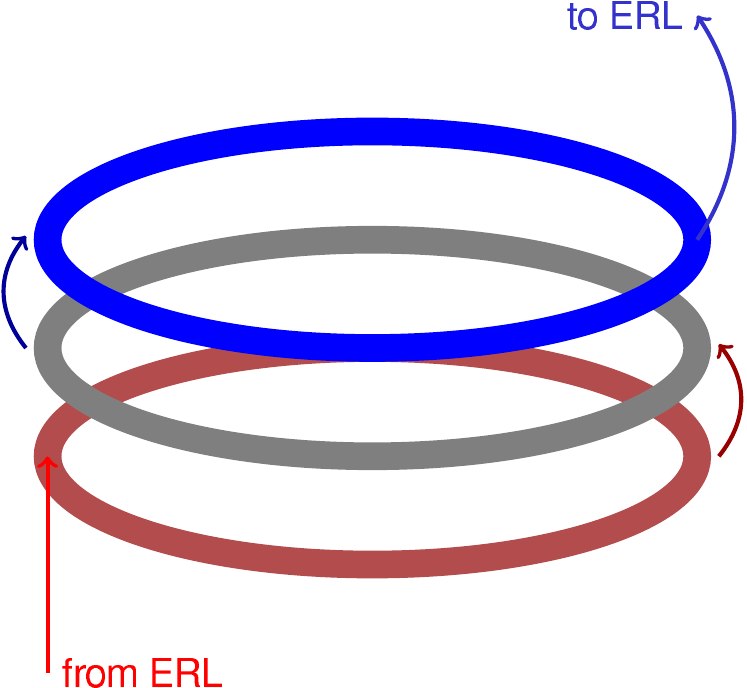}
\caption{Tri-ring scheme converting a continuous beam into a pulsed beam, 
for cooling, and back.}
\label{triring} 
\end{center}
\vspace*{-3 mm} 
\end{figure}

Another possible solution to cool down a continuous positron beam, both the recycled beam and/or a new beam from a source, 
is the tri-ring scheme illustrated in Fig.~\ref{triring}.

In this scheme, the basic cycle lasts $N$ turns, during which the following processes happen simultaneously: $N$-turn injection from 
the ERL into the accumulating ring (bottom); $N$-turn cooling in the cooling ring (middle) possibly 
with fast laser cooling \cite{bulyak11b}; and $N$-turn slow extraction from the extracting ring (top) back 
into the ERL. 
At the start of the cycle there is a one-turn transfer from the cooling ring into the extracting ring, and a one-turn transfer from the accumulating ring into the cooling ring. The average current in the cooling ring is $N$ times the average ERL current. 

\subsection{Production schemes}
Positrons can be produced by pair creation when high-energy electrons or photons hit a target.
Conventional sources, as used at the SLC, send a high-energy electron beam on a conversion target.
Alternatively, a high-energy electron beam can be used with a hybrid-target configuration where the first
thin target is used to create high-energy photons, through a channelling process, 
which are then sent onto a thick target. 
The prior conversion into photons reduces the heat load of the target for a given 
output intensity and it may also improve the emittance of the generated positrons.
There exist a number of other schemes that can accomplish the conversion of electrons into photons.
Several of them employ Compton scattering off a high-power laser pulse stacked in an optical cavity.
According to the electron-beam accelerator employed, one distinguishes
Compton rings, Compton linacs, and Compton ERLs \cite{posipol,stacking,lrcompton}.
An alternative scheme uses the photons emitted by an electron beam of very high energy (of order 
100 GeV) when passing through a short-period undulator \cite{mikh1,mikh2,lcrclic}. 
Finally, there even exists a simpler scheme where a high-power laser pulse itself serves 
as the target for (coherent) pair creation. 

\subsubsection{Targets} 
For the positron flux considered for the LHeC the heating and possible destruction of the target are important concerns. 
Different target schemes and types can address these challenges:
(1) multiple, e.g.~10, target stations operating in parallel; (2) He-cooled granular W-sphere targets; (3) rotating-wheel targets; (4) 
sliced-rod W tungsten conversion targets; (5) liquid mercury targets; and (6) running tape with annealing process.

The LHeC ERL option requires a positron current of 6 mA 
or $4\times 10^{16}$ $e^{+}$/s, with 
normalised emittance of $\le$50~$\mu$m
and longitudinal emittance $\le$5 MeV-mm. 
For a conventional conversion target with optimised length the 
power of the primary beam is converted as follows 
$P_{primary} (100\%)=P_{thermal} (30\%) + P_{\gamma}(50\%) + P_{e^{-}}(12\%) + P_{e^{+}}(8\%) $.
The average kinetic energy of the newly generated 
positrons is $<T_{e^{+}} >\approx 5\; {\rm  MeV}$, 
which allows estimating the total power
incident on the target as 
$P_{target}$ =  5 MV $\times  6$ mA / 0.08 = 375 kW.
Assuming an electron linac efficiency of  
$\eta_{acc} \approx 20\%$ we find 
$P_{wall} =P_{target}/0.2 = 1.9$\; MW. 
This wall-plug power level looks feasible and affordable.
However, also considering a capture efficiency 
(for the `useful' $e^{+}$) of about 5\%,   
$P_{wall}$ becomes 38 MW.

\begin{figure}
\begin{center}
\includegraphics[width=\columnwidth]{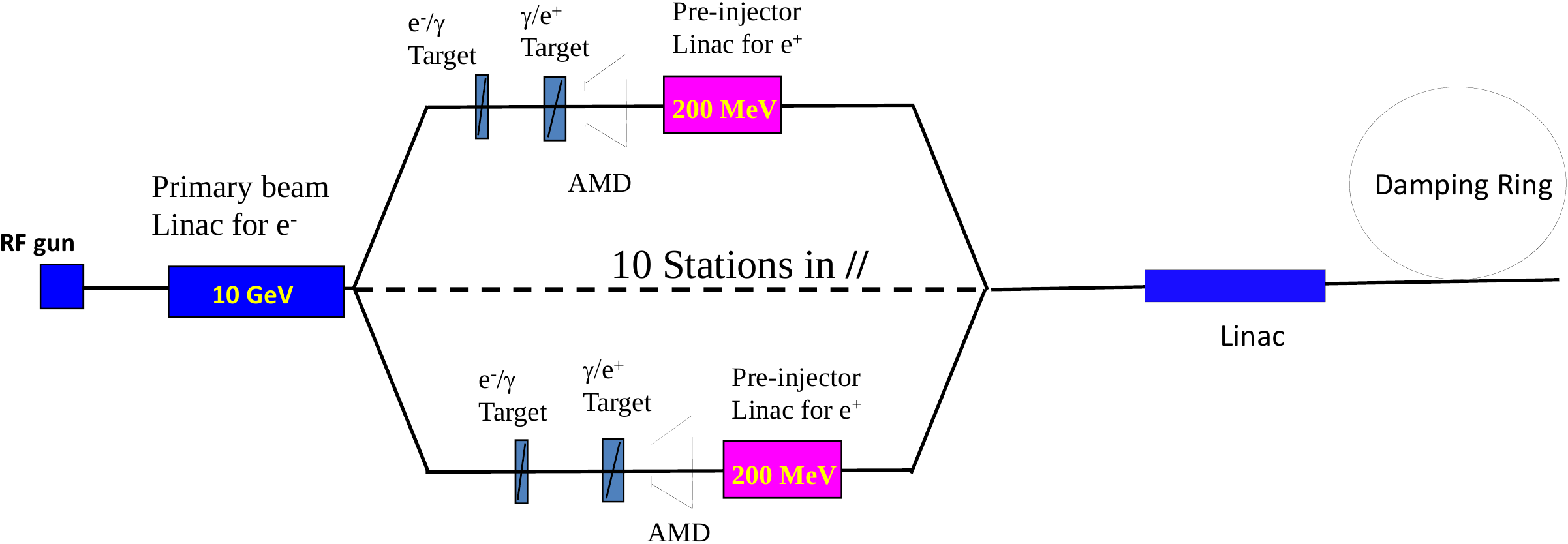}
\caption{ Possible layout with unpolarised $e^+$ for the LHeC injector (p-140 GeV).}
\label{Fig:LHeCInjector3}
\end{center}
\end{figure}

Figure $\ref{Fig:LHeCInjector3}$ illustrates a possible option, which alone 
would already meet the requirements for the 140-GeV 
single-linac case, where the repetition rate is 10~Hz. 
The idea is to use 10 $e^+$ target stations in parallel. This
implies installing 2 RF deflectors upstream and the same
downstream. Experience exists for RF deflectors at 3 GHz and with operating 
2 lines in parallel. 
Assuming that this configuration is acceptable from the 
beam-optics point-of-view, 
it would be necessary to implement a fast damping scheme because the
bare emittances from the target will be too high for the injection into the ERL.

Table $\ref{tab:rinolfi3}$ shows the beam characteristics at the end of the 10 GeV
primary beam Linac for electrons, before splitting the beam.

\begin{table}[h]
\centering
\begin{tabular}{|l|l|}
\hline
Primary beam energy ($e^-$) & $10$ GeV\\\hline
Number $e^-$ / bunch & $1.2 \times 10^9$ \\\hline
Number of bunches / pulse  & $100 000$ \\\hline
Number $e^-$ / pulse &  $1.2 \times 10^{14}$ \\\hline
Pulse length & $5$ ms \\\hline
Beam power  & $1900$ kW \\\hline
Bunch length  &  $1$ ps\\\hline
\end{tabular}
\caption{Electron beam parameters before splitting.}
\label{tab:rinolfi3}
\end{table}

Table $\ref{tab:rinolfi4}$ 
shows the beam parameters at each $e^+$ target. 
A power of 5.6 kW is deposited in each target and the Peak Energy Deposition Density (PEDD) 
is around 30 J/g \cite{lred}. 
This value has been chosen, in order to stay below the breakdown limit for a 
tungsten (W) target. 
It is based on recent simulations \cite{dadoun} with conventional W targets. 
A new study \cite{sievers} 
assumes a target made out of an assembly of densely packed W spheres 
(density about 75\% of solid tungsten) with diameters of 1--2 mm, 
cooled by blowing He-gas through the voids between the spheres. 
Such He-cooled granular targets have been considered for neutrino factories and 
recently for the European Spallation Source ESSS.

\begin{table}[h]
\centering
\begin{tabular}{|l|l|}
\hline
Yield ($e^+ / e^-$) & $ 1.5$ \\ \hline
Beam power (for $e^-$) & $190$ kW \\ \hline
Deposited power / target & $5.6$ kW \\ \hline
PEDD & $30$ J/g \\ \hline
Number $e^+$ / bunch & $ 1.8 \times 10^9$ \\ \hline
Number bunches / pulse & $10,000$ \\ \hline
Number $e^+$ / pulse & $1.8 \times 10^{13}$ \\ \hline
\end{tabular}
\caption{Beam parameters at each $e^+$ target.}
\label{tab:rinolfi4}
\end{table}

To achieve the required cooling and the corresponding mass flow of the cooling fluid, 
we consider pressurised He at 10 bar entering the target volume at a velocity of 10 m/s,
i.e.~a mass flow 1.8 g/s is required for each target. 
From this a convection coefficient of about 
$\alpha=1$~W/cm$^{2}$/K can be expected and a cooling time constant 
$\tau$ (exponential decay time after an adiabatic temperature rise of a sphere) of 185 ms will result. 
Clearly, not much cooling during a pulse of 5 ms duration will occur, 
but cooling will set in during the off-beam time of 95 ms between the pulses. 
The peak temperature after each pulse will stabilise at about 500 K above that 
of the cooling fluid. An average exit temperature of the He-gas of about 600 $^{\circ}$C 
will have still to be added, which drives the maximum temperature of the spheres up to about 1100 $^{\circ}$C. 
Although compatible with W in an inert atmosphere, it should be attempted to reach lower temperatures. 
This could be achieved by increasing the He-pressure to 20 bar and the velocity of He to 20 m/s which 
might reduce the maximum temperature in a sphere to 500 $^{\circ}$C.
Thus, a He-cooled granular 10-W-target system could be a viable solution.

Another approach has been considered. To achieve, as in the previous case, 
a reduction of the energy deposition density by a factor of 10, a fast rotating wheel could be designed. 
The beam pulse of 5 ms duration is spread over the rim of the rotating wheel and a linear velocity of the 
rotating rim of 20 m/s would be required. This would lead to a repetition rate of about 1000 rpm, 
assuming a wheel diameter of 0.4 m. Such a solution is actually under investigation for the 
ILC with a rotation speed of 1800 rpm.

\begin{figure}[htb]
\centering
\vspace{0pt}
\includegraphics[width=0.8\textwidth]{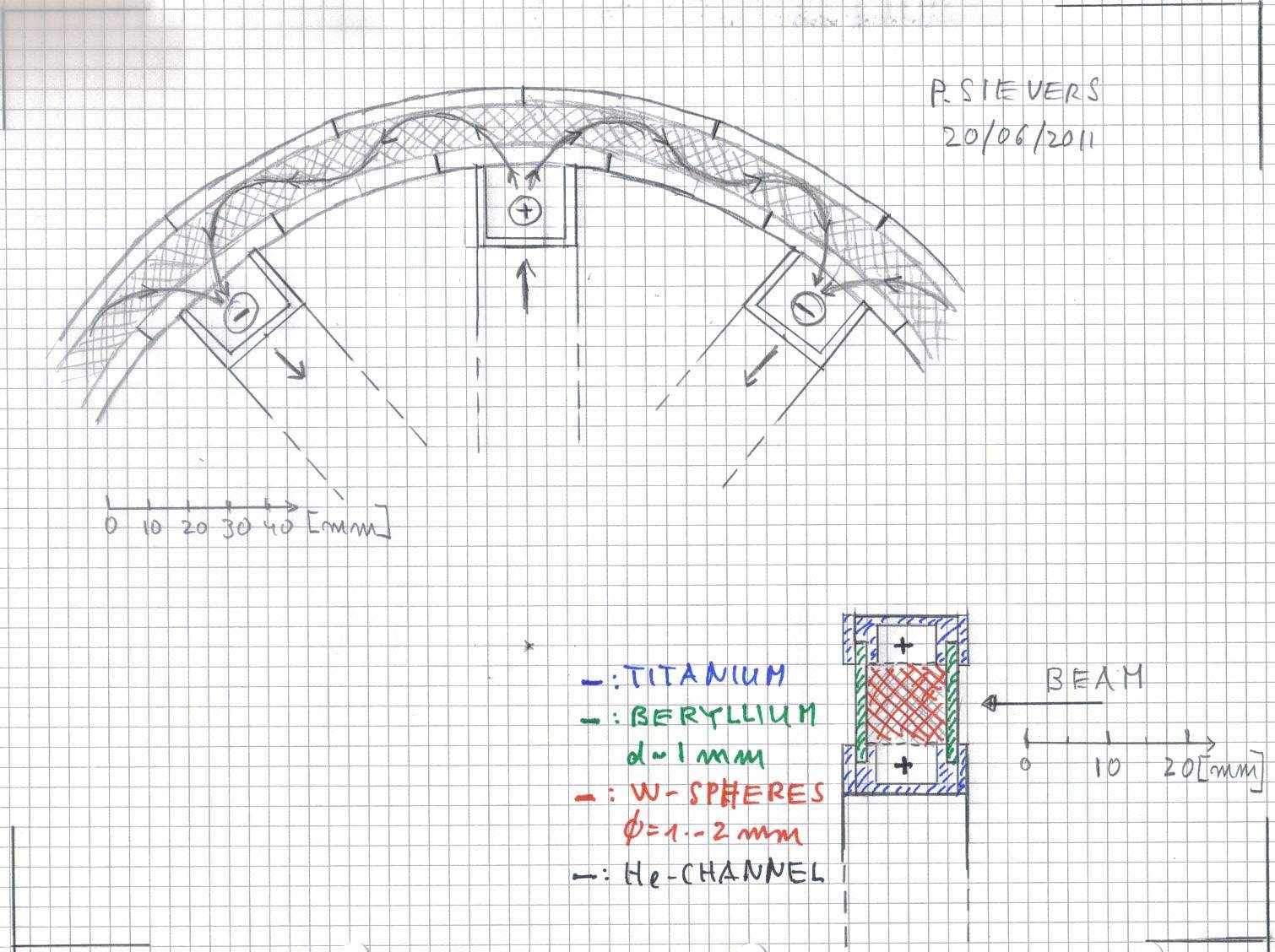} 
\vspace{0pt} 
\caption{Sketch of rotating wheel containing W spheres with He cooling. }
\label{ilc_wheel}
\end{figure}

Here tungsten spheres, again, are contained in a structure, similar to a car tyre,
as is illustrated in Fig.~\ref{ilc_wheel}. 
The container is possibly made of light Ti-alloy where the sides, facing the beam 
entrance and exit should be made of Beryllium, compatible with the beam heating.
The helium for the cooling is injected from the rotating axle through spokes 
into the actual target ring and is recuperated in the same way. 

If the beam pulse duration is extended by a factor 10, i.e.~to 50~ms duration, 
maintaining of course the same average power, then the rotation time could be reduced. 
The velocity of the wheel is such that over the duration of 5 ms the rim is displaced 
by one beam width, i.e. 1 cm. This leads to much reduced rotation speeds of 2 m/s, 
which can readily be achieved in a wheel with a diameter of 16 cm, rotating at 240 rpm.

By choosing appropriately the rotation velocity, the average time between two hits of the same 
spot on the rim of the wheel, is about 0.5 s. With the aforementioned cooling time constant 
for the He-circuit of 185 ms, the adiabatic temperature rise during one hit  over 5 ms of 211 K 
will have dropped to nearly zero before the next hit. For simultaneously 
cooling the whole rim of the wheel a He-flow of 90 g/s must be provided. 
Taking into account the temperature increase in the cooling fluid, 
a maximum tungsten temperature in the W-spheres 
of about 350$^{\circ}$C can be expected, which is rather comfortable.

Using a continuous D.C.-beam with no gaps will further alleviate the structure 
and performance of the target wheel.

The interference of the rotating wheel with the downstream flux concentrator will have to be assessed. 
One may, however, expect considerably less forces than presently considered for the ILC, 
due to the much lower velocity of the wheel. 
Moreover, proper choice of materials with high electrical resistivity and laminating 
the structure may be considered.

Clearly, the W-granules must be contained inside the beam vacuum within a structure 
which is He-leak tight at the selected He-pressure. As material for the upstream and 
downstream beam windows, Beryllium must be considered which, due to its large 
radiation length (34 cm as compared to W with 0.34 cm), should resist 
to the thermal loads. This, however, has to be verified. 

Also, radiation damage and life time issues will still have to be assessed.

It is believed that rotating ``Air to Vacuum'' seals at 240 rpm are commercially available 
or can be adapted to the radiation environment. 
Rotating ``High Pressure He to Air'' seals may have to be developed, 
where small He-leaks can be tolerated.

Presently with conventional targets, the transverse normalised rms beam emittances, 
in both planes, are in the range of 6000 to 10 000 $\mu$m. 
With the new types of target, we do not know yet by how much the transverse emittances 
will be changed.  
In any case, a strong reduction of emittances is mandatory for the requested LHeC performance. 
Assuming that large or small emittances could be recombined, 
Table $\ref{tab:rinolfi5}$ shows a possible $e^{+}$ flux after recombination. 
If a solution is found for the emittances, 
it will be necessary to design and implement a linac accelerating the positron beam up to 500 MeV, 
the energy for the ERL injection.

\begin{table}[h]
\centering
\begin{tabular}{|l|l|}
\hline
Secondary beam energy ($e^+$) & $200$ MeV \\ \hline
Number $e^+$  bunch & $1.8 \times 10^9$ \\ \hline
Number of bunches / pulse & $100 000$ \\ \hline
Number of $e^+$ / pulse & $1.8 \times 10^{14}$ \\ \hline
Bunch spacing & $50$ ns \\ \hline
Repetition rate & $10$ Hz \\ \hline
\end{tabular}
\caption{Positron beam parameters after recombination.}
\label{tab:rinolfi5}
\end{table}

For Compton sources (discussed below) the conversion of gammas to
positrons is a bottleneck, which requires a study 
and optimisation of effective convertor targets such as the sliced-rod converter.
A typical tungsten convertor optimised for Compton gammas with a maximal energy of 20\,MeV 
can deliver 0.02 positrons per incident scattered gamma. 
A sliced-rod convertor target may  
produce 0.07/0.13 positrons per gamma for a 1\,m or 3\,m long rod, 
respectively \cite{bulyak11alcpg}.

\subsubsection{Compton sources}
In Compton sources (polarised) positrons are generated
 by scattering of an electron beam off a higher-power laser pulse, and by converting the resulting gammas in a target.

\begin{itemize}
\item
{\bf Compton Ring}: 
Table \ref{eminuscur} illustrates that a Compton-ring source equipped with an array of optical 
resonators yielding a total (single-IP `equivalent') 
laser-pulse energy of 5\,Joule, together with a sliced-rod
conversion target, may produce the desired flux of polarised positrons even for the LHeC ERL option. 
The emission of 30-MeV gammas at the required rate can induce significant beam energy spread 
in the Compton ring, which requires further studies and optimisation.

\begin{table}
\centering
\begin{tabular}{|l|r|r|}
\hline 
 & LHeC pulsed & LHeC ERL \\ \hline
$I_{e^{+}}$ at IP [$\mu$A] & 290 & \emph{7050} \\
typical $I_{e-}$ [A] & 4.3 & \emph{105.7}\\
$I_{e-}$ with 5\,J  [A] &0.46 & \emph{11.2}\\
$I_{e-}$ with 5\,J+1\,m rod [A] & 0.065 & 1.6 \\
\hline
\end{tabular}
\caption{IP $e^{+}$ current and the implied minimum 
$e^{-}$ beam current in a Compton Ring. 
Electron-beam currents below 5 A
are considered achievable.}
\label{eminuscur}
\end{table}

\item
{\bf Compton Linac}:An optimistic power analysis for a single-pass Compton linac 
using a CO$_2$ laser shows that the wall plug power for generating the Compton-linac 
electron beam alone exceeds the limit of 100 MW set for the entire LHeC project. 

\item
{\bf Compton ERL}: 
A high current ERL appears to perhaps be a possible approach, 
e.g.~a 3-GeV 1.3-A ERL with 2-micron wavelength optical enhancement cavities would provide the desired $e^{+}$ rate, with ``only'' 50 MW of wall plug power, and with upper-bound estimates
on the transverse and longitudinal emittances for the captured 
positron beam of $\gamma \epsilon_{\perp} \le 1.5$~m, and  
$\epsilon_{||,N} \approx 450\; \mu {\rm m}$. 

\end{itemize}
The desired emittances are not reached from any Compton scheme source, even if the target is immersed in a strong magnetic field.  
Therefore, cooling or scraping would be required.

\subsubsection{Undulator source}
An undulator process for $e^{+}$ production could be 
based on the main high-energy $e^{-}$ (or $e^{+}$) beam.
The LHeC undulator scheme can benefit from the pertinent 
development work done for the ILC.
The beam energy at LHeC would be lower, e.g. 60 GeV, which might possibly be 
compensated by more ambitious undulator magnets, e.g. ones made from Nb$_{3}$Sn or HTS. 
However, the requested photon flux calls for a careful investigation. 
The undulator scheme could most easily be applied for the 140-GeV pulsed LHeC.

\subsubsection{Coherent pair creation}
The normalised transverse emittance of all positrons from 
a target is of order $\epsilon_{N}\approx 1-10$~mm, to be compared with
a requested emittance of $\epsilon_{N} =0.05$~mm.
Therefore, a factor 100 emittance reduction is required.
Possible solutions are cutting the phase space or damping. 
A third solution would be to produce positrons in a smaller phase space volume.
Indeed the inherent transverse emittance from pair production is small.
The large phase space volume only comes from multiple scattering 
in the production target.

Pair production from relativistic electrons in a strong laser field would 
not need any solid target, since the laser itself serves as the target,
and it would not suffer from multiple scattering. 
This process has been studied in the 1960's and 1990's \cite{erber,palmer,bamber}.
It should be reconsidered with state-of-the-art TiSa lasers and X-ray FELs,
and could offer an interesting prospect for the LHeC.

\subsection{Conclusions on positron options for the Linac-Ring LHeC}
The challenging requirements for the LHeC Linac-Ring positron source 
may be relaxed, to a certain extent, 
by $e^{+}$ recycling, $e^{+}$ re-colliding, and $e^{+}$ cooling.
The compact tri-ring scheme is an attractive proposal for recooling the spent and recycled positrons, with a pushed
conventional damping ring in the SPS tunnel as an alternative solution. 

Assuming some of the aforementioned measures are taken to lessen 
the required positron intensity to be produced at the source, by at least 
an order of magnitude, and also assuming that an advanced target is available, 
several of the proposed concepts could provide the 
intensity and the beam quality required by the LHeC ERL. 

For example, the Compton ring and the Compton ERL 
are viable candidates for the Linac-Ring LHeC positron source.
Coherent pair production and an advanced undulator 
represent other possible schemes, still to be explored for LHeC in greater detail. 
The coherent pair production would have the appealing feature
of generating positrons with an inherently small emittance. 

In conclusion, it may be possible to meet the very demanding 
requirements for the LHeC positron source. 
A serious and concerted R\&D effort will 
be required to develop and evaluate a baseline design 
for the linac-ring positron configuration. 
Among the priorities are a detailed optics \& beam-dynamics study of 
multiple collisions and of the tri-ring scheme, 
a theoretical exploration of coherent pair production,
and participation in experiments on Compton sources, 
e.g.~at the KEK ATF.

%% file: machine/systemdesign.tex
\chapter{System Design}

\input{machine/russenschuck}

\clearpage
\input{machine/magnets}
\clearpage
\input{machine/rf}

\input{machine/thiesen}
\section{Vacuum}
\input{machine/jimenez}
\input{machine/SR_BPDesign}

\clearpage

\section{Cryogenics}
\input{machine/haugrr}

\input{machine/hauglr}

\clearpage

\section{Beam dumps and injection regions}
\input{machine/Bracco_Goddard_RR}
\input{machine/Bracco_Goddard_LR}

%% file: machine/russenschuck.tex
\section{Magnets for the interaction region}
\label{tripletmagnets}

\subsection{Introduction}

The technical requirements for the ring-ring options are easily achieved with superconducting magnets of proven technology. It is possible to make use of the wire and cable development for the LHC inner triplet magnets. We have studied all-together seven variants of which two are 
selected for this CDR. Although these magnets will require engineering design efforts, there are no challenges because the mechanical design 
will be very similar to the MQXA \cite{ranko} magnet built for the LHC \cite{Bruning:2004ej}.

The requirements in terms of aperture and field gradient are much more difficult to obtain for the linac-ring option. We reverse the arguments and present the limitations for the field gradient and septum size, that is, the minimum distance between the proton and electron beams, for both Nb-Ti and Nb$_3$Sn superconducting technology. Here we limit ourselves to the two most promising conceptual designs.

\subsection{Magnets for the Ring-Ring option}

The interaction region requires a number of focusing magnets with apertures for the two proton beams and
field-free regions to pass the electron beam after the collision point. The lattice design was presented in Sections \ref{lat} and \ref{layout}; the schematic layout is shown in Fig.
\ref{Fig:IR_gross}.

%
The field requirements for the ring-ring option (gradient of 127 T/m, beam stay clear of 13 mm (12 $\sigma$), aperture radius of 21 mm for the proton beam, 30 mm for the electron beam) allow a number of different magnet designs 
using the well proven Nb-Ti superconductor technology and making use of the cable development for the LHC. In the simulations presented here, we have used the parameters (geometrical, critical surface, superconductor magnetisation) of the cables used in the insertion quadrupole MQY of the LHC. \par

\begin{figure}[h!]
\begin{center}
\centering\epsfig{file=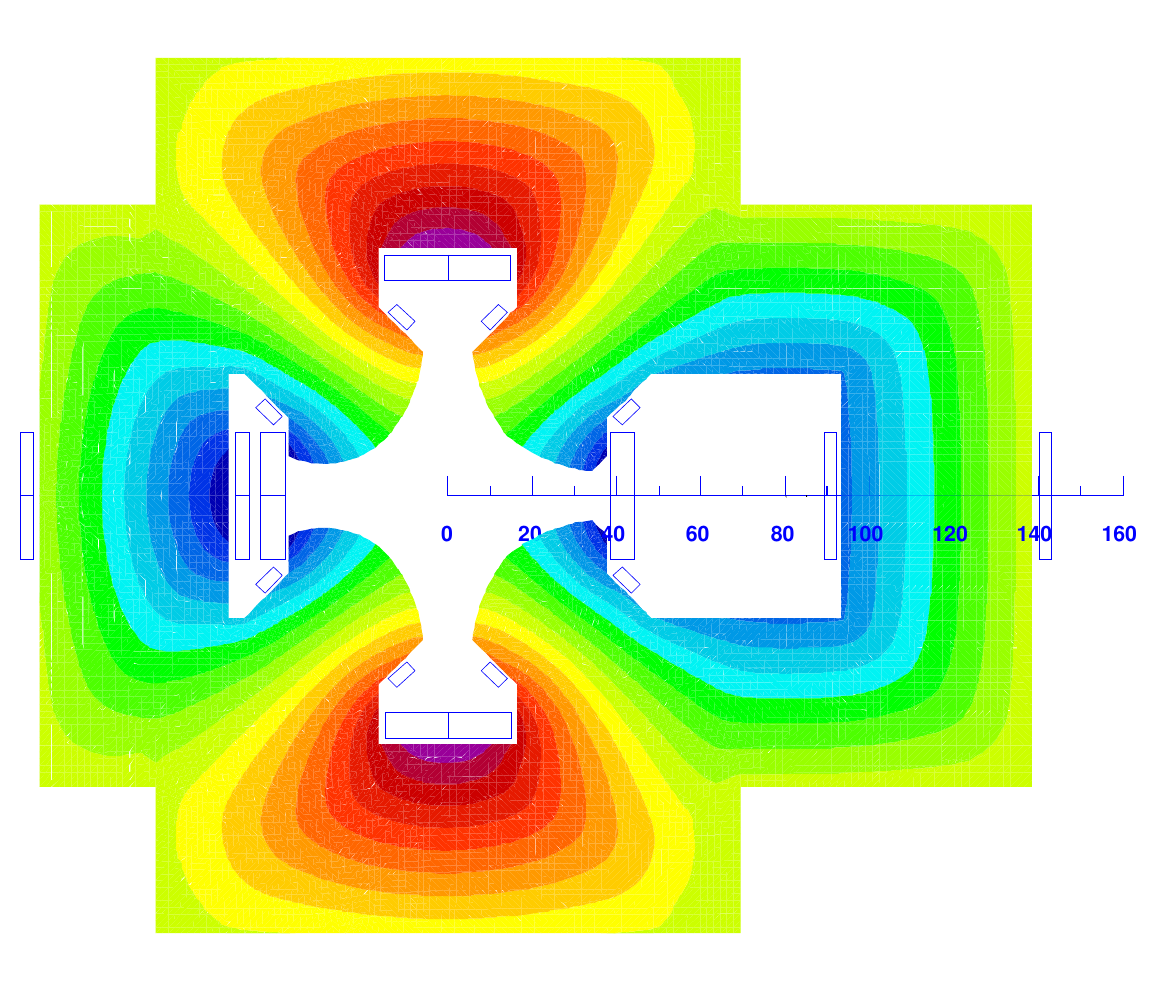,clip=,width=6cm}\hspace{1cm}
\centering\epsfig{file=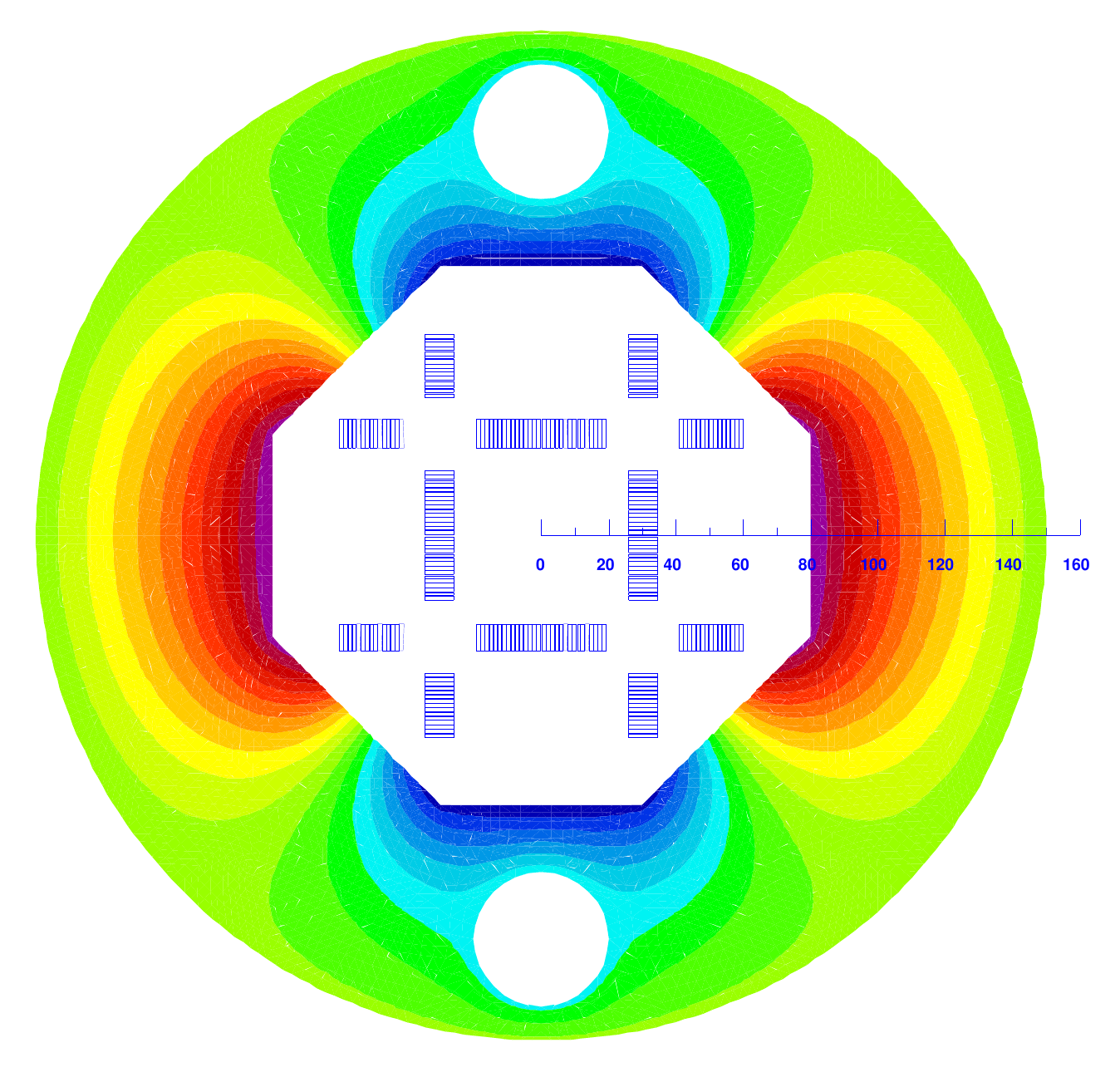,clip=,width=6cm}
 \end{center}
\caption{\label{fig1} Cross-sections of insertion quadrupole magnets with iso-surfaces of the magnetic vector potential (field-lines). Left: Super-ferric, similar to the design presented in \cite{Tawada:2000xy}. Right: Superconducting block-coil magnet as proposed in \cite{gupta} for a coil-test facility.}
\end{figure}

Fig. \ref{fig1} shows a superferric magnet as built for the KEKb facility \cite{Tawada:2000xy}. This design comes to its limits due to the saturation of the iron poles. Indeed, the fringe field in the aperture of the electron beam exceeds the limit tolerable for the electron beam optics, 
and the field quality required for proton beam stability, on the order of one unit in 10$^{-4}$ at a
reference radius of 2/3 the aperture, is difficult to achieve. \par

\begin{figure}[h!]
\begin{center}
\centering\epsfig{file=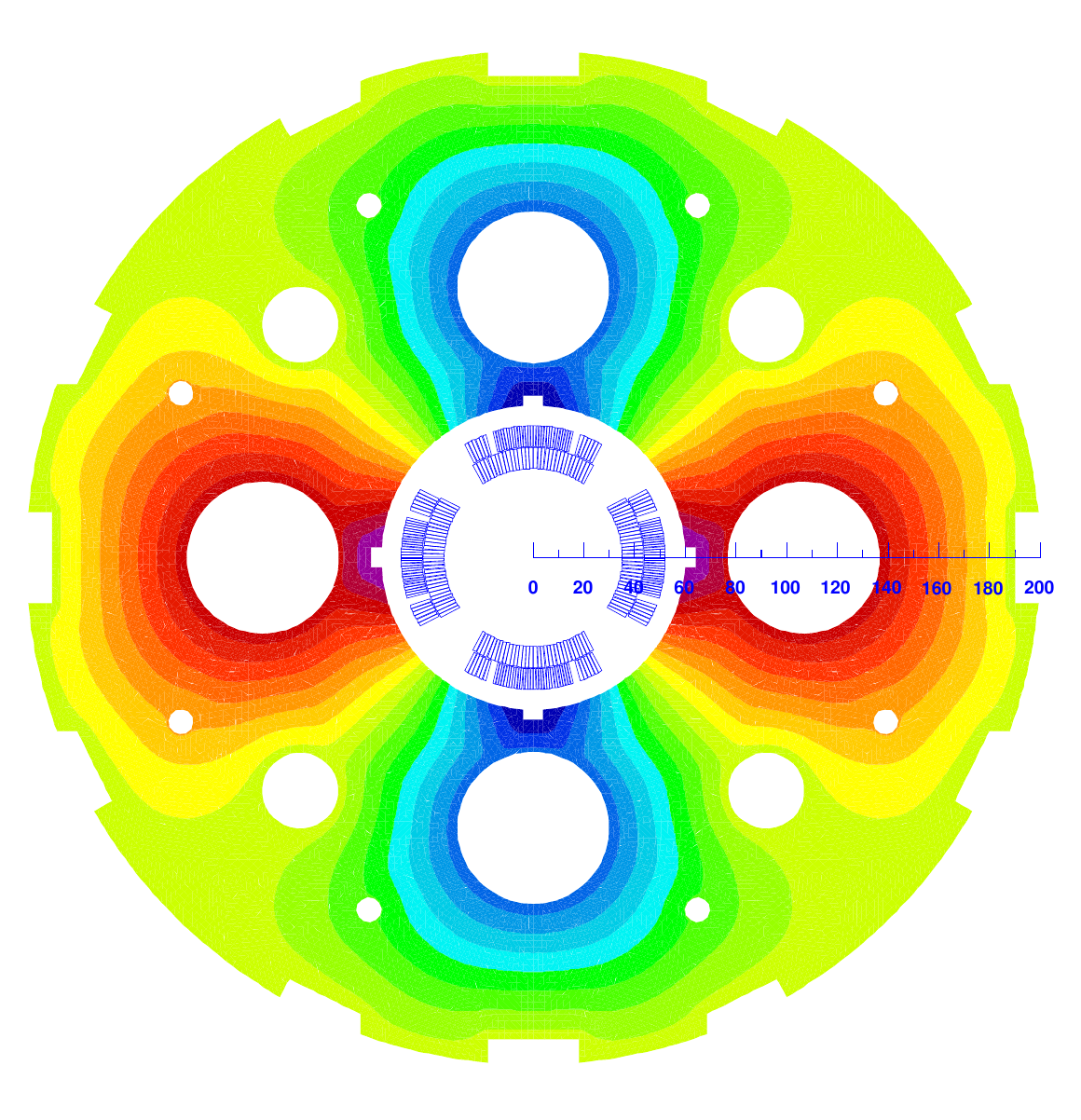,clip=,width=6cm}\hspace{1cm}
\centering\epsfig{file=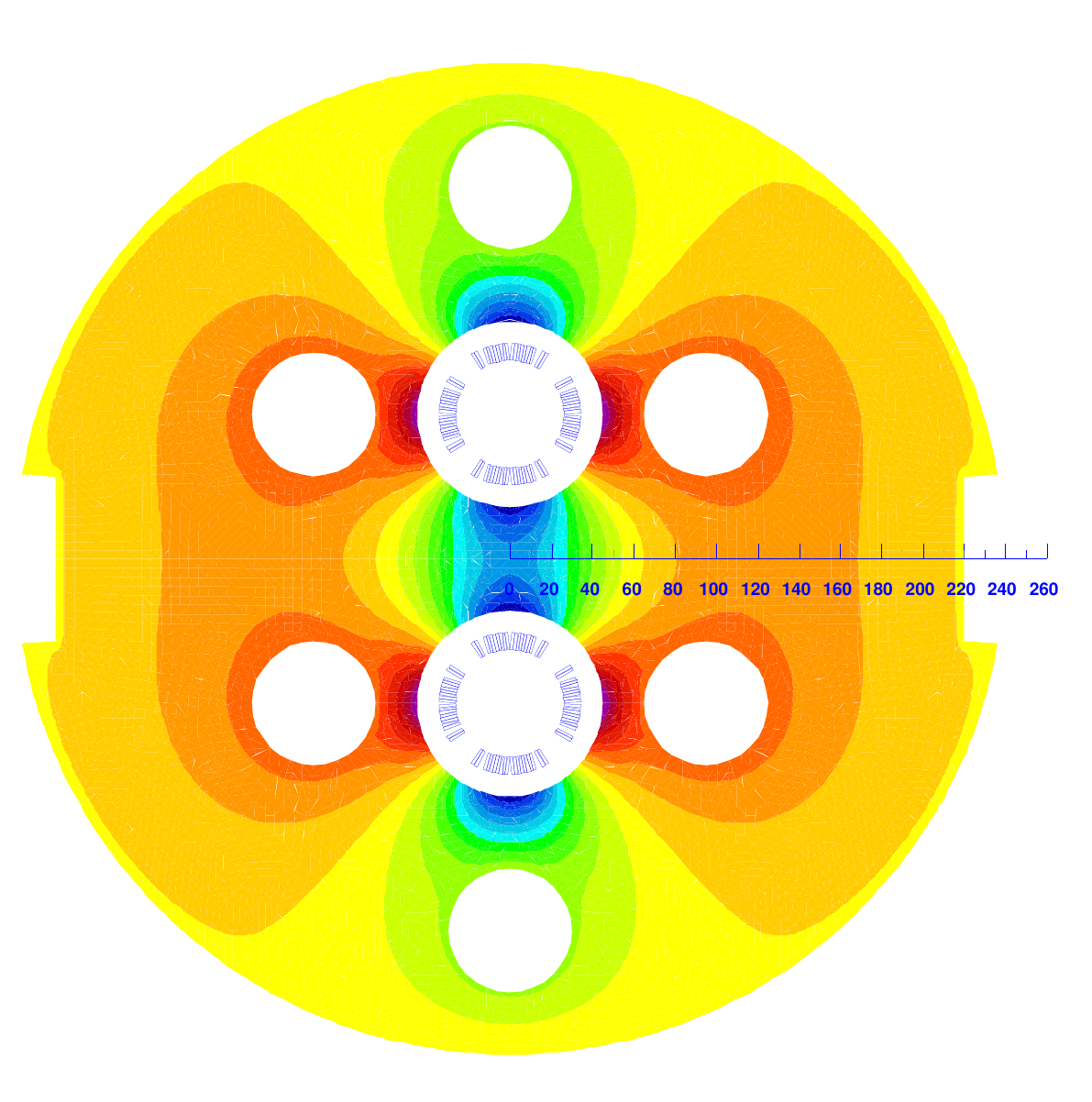,clip=,width=6cm}\\
\centering\epsfig{file=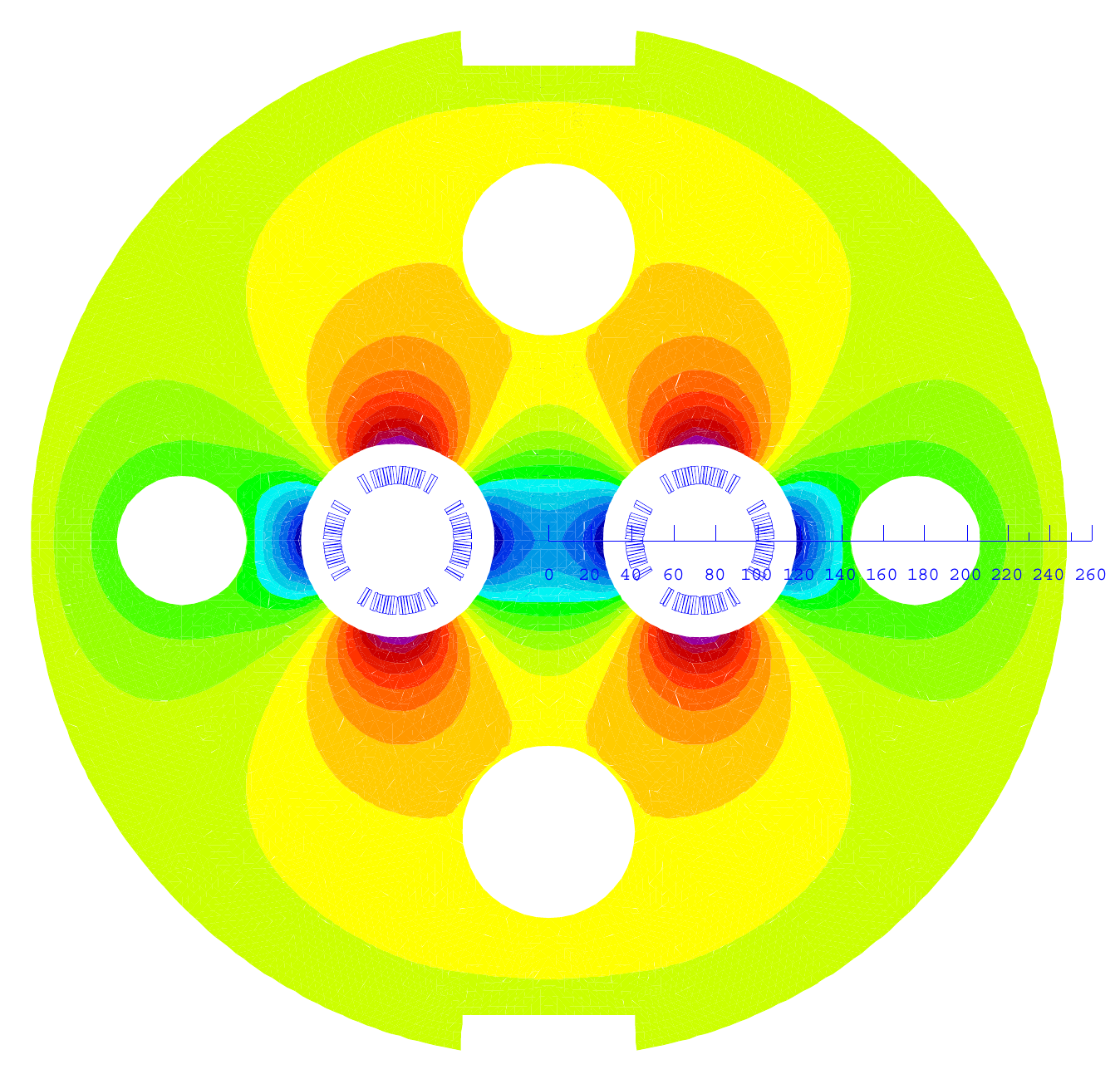,clip=,width=6cm}
 \end{center}
\caption{\label{fig2} Cross-sections with field-lines of insertion quadrupole magnets. Classical designs similar to the LHC magnet technology. Top left: Single aperture with 
a double layer coil employing both cables listed in Table \ref{wire}. Design chosen for Q2. 
Top right: Double aperture vertical. Bottom: Double aperture horizontal. The double-aperture magnets
can be built with a single layer coil using only the MQY inner layer cable; see the right column of Table \ref{wire}.} 
\end{figure}

The magnetic flux density in the low-field region of the design shown in Fig. \ref{fig1} (right) is about 0.3 T. We therefore disregard this design as well. Moreover, the engineering design work required for the mechanical structure of this magnet would be higher than for the proven designs shown in Fig. \ref{fig2}. \par

\begin{table}[htb]
\setlength{\tabcolsep}{8.4pt}
\begin{center}
\begin{tabular}{|l|c|c|}\hline  
Magnet                                                                                                              &  MQY (OL)       & MQY (IL)  \\ \hline
Diameter of strands (mm)                                                                                    &  0.48               & 0.735      \\
Copper to SC area ratio                                                                                       &  1.75               & 1.25        \\
Filament diameter ($\mu$ m)                                                                              &  6                    & 6            \\
${B}_{\mathrm{ref}}$ (T)  @  ${T}_{\mathrm{ref}}$ (K)                                             &  8 @ 1.9           &  5 @ 4.5   \\
$J_\mathrm{c}({B}_{\mathrm{ref}}, {T}_{\mathrm{ref}}$) $(\mathrm{A\,mm}^{-2})$      & 2872                &   2810     \\
$- \ttd J_\mathrm{c}/ \ttd B\, (\mathrm{A\,mm^{-2}\, T}$)                                        &  600                 &   606       \\
$\rho(293$~K$)/\rho(4.2$~K$)$ of Cu                                                                    & 80                   &  80         \\ \hline
Cable  width       (mm)                                                                                           &    8.3               &   8.3      \\
Cable thickness, thin edge (mm)                                                                            &    0.78              &   1.15     \\
Cable thickness, thick edge (mm)                                                                          &    0.91              &   1.40     \\
Keystone angle (degree)                                                                                      &    0.89             &   1.72     \\
Insulation thickn. narrow side (mm)                                                                       &    0.08             &   0.08     \\
Insulation thickn. broad side  (mm)                                                                        &    0.08             &   0.08     \\
Cable transposition pitch length (mm)                                                                    &    66               &   66        \\
Number of strands                                                                                               &    34               &   22        \\
Cross section of Cu (mm$^2$)                                                                              &    3.9              &   5.2       \\
Cross section of SC (mm$^2$)                                                                             &    2.2              &   4.1        \\
\hline
\end{tabular}
\caption{Characteristic data for the superconducting cables ands strands. OL = outer layer, IL = inner layer.}
\end{center}
 \label{wire}
\end{table}

Fig. \ref{fig2} shows the three alternatives based on LHC magnet technology. In the case of the double aperture version the aperture for the proton beams is 21 mm in radius, in the single aperture version the beam pipe radius is 26 mm. In all cases the 127 T/m field gradient can be achieved with a comfortable
safety margin to quench (exceeding 30\%) and using the cable(s) of the MQY magnet of the LHC. The operation temperature 
is supposed to be  1.8 K, employing superfluid helium technology. The cable characteristic data are given in Table \ref{wire}. The outer radii 
of the magnet cold masses do not exceed the size of the triplet magnets installed in the LHC (diameter of 495 mm). The fringe field in the aperture 
of the electron beam is in all cases below 0.05 T.

\begin{figure}[h!]
\begin{center}
\centering\epsfig{file=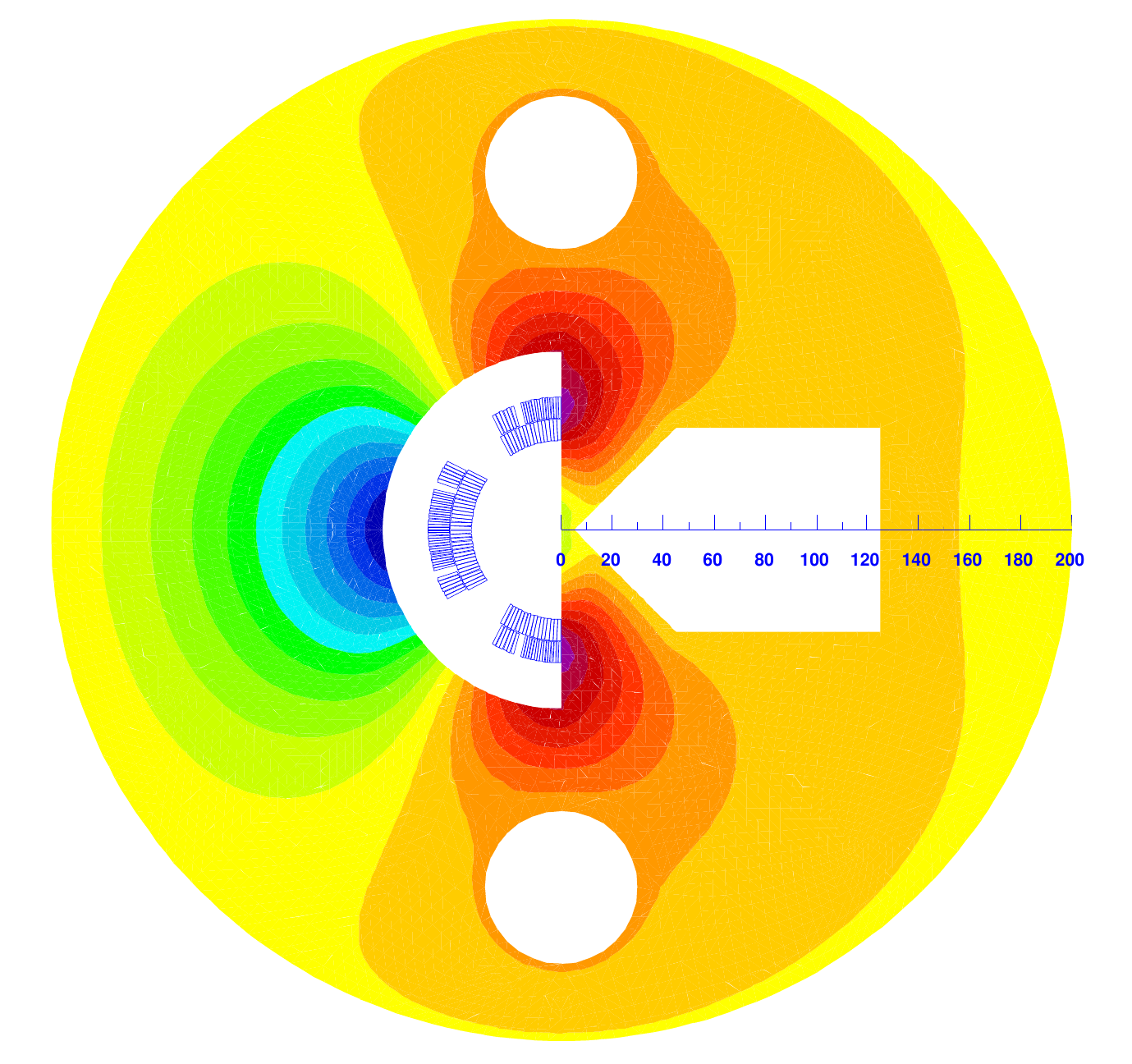,clip=,width=6cm}\hspace{1cm}
\centering\epsfig{file=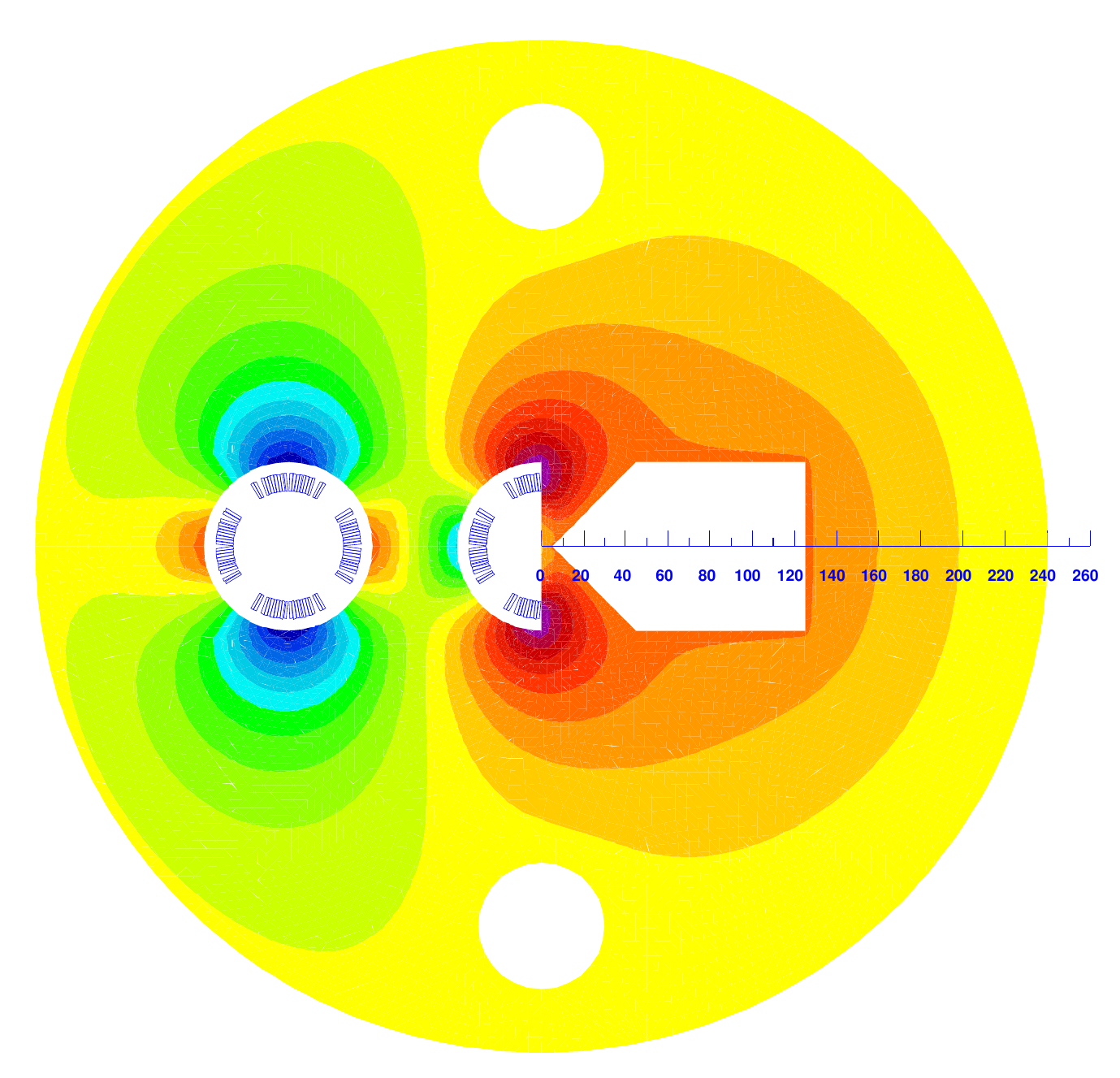,clip=,width=6cm}
 \end{center}
\caption{\label{fig3} Cross-sections of insertion quadrupole magnets with field-lines. Left: Single half-aperture quadrupole with field-free domain \cite{Dainton:2006wd}; design selected for Q1. Right: Double-aperture magnet composed of a quadrupole and half quadrupole. }
\end{figure}

Fig. \ref{fig3} shows half-aperture quadrupoles (single and double-aperture versions for the proton beams) in a similar design as proposed in \cite{Dainton:2006wd}. The reduced 
aperture requirement in the double-aperture version makes it possible to use a single layer coil and thus to reduce the beam-separation distance between the proton
and the electron beams. The field-free regions is large enough to also accommodate the counter rotating proton beam. The version shown in Fig. \ref{fig3} (left) employs a double-layer coil. In all cases the outer diameter of the cold masses do not exceed 
the size of the triplet magnets currently installed in the LHC tunnel.\par

For this CDR we retain only the single aperture version for the Q2 (shown in Fig. \ref{fig2}, left) and the half-aperture quadrupole for the Q1 (shown in Fig. \ref{fig3}, top left). 
The separation distance between the electron and proton beams in Q1 requires the half-aperture quadrupole design to limit the overall synchrotron radiation power emitted by bending of the 60 GeV electron beam. The single aperture version for Q2 is retained in the present layout, because the counter rotating proton beam can be guided 
outside the Q2 triplet magnet. The design of Q3 follows closely that of Q2, except for the size of the septum between the proton and the electron beams.

The coils in all three triplet magnets 
are made from two layers, using both Nb-Ti composite cables as specified in Table \ref{wire}. The layers are individually optimised for field quality. 
This reduces the sensitivity to 
manufacturing tolerances and the effect of superconductor magnetisation \cite{Russenschuck:2010zz}.
The mechanical design will be similar to the MQXA magnet where two kinds of interleaved yoke laminations are assembled 
under a hydraulic press and locked with keys in order to obtain the required pre-stress of the coil/collar structure. The main parameters of the magnets 
are given in Table \ref{data}. 

\subsection{Magnets for the Linac-Ring option}

\begin{figure}[h!]
\begin{center}
\centering\epsfig{file=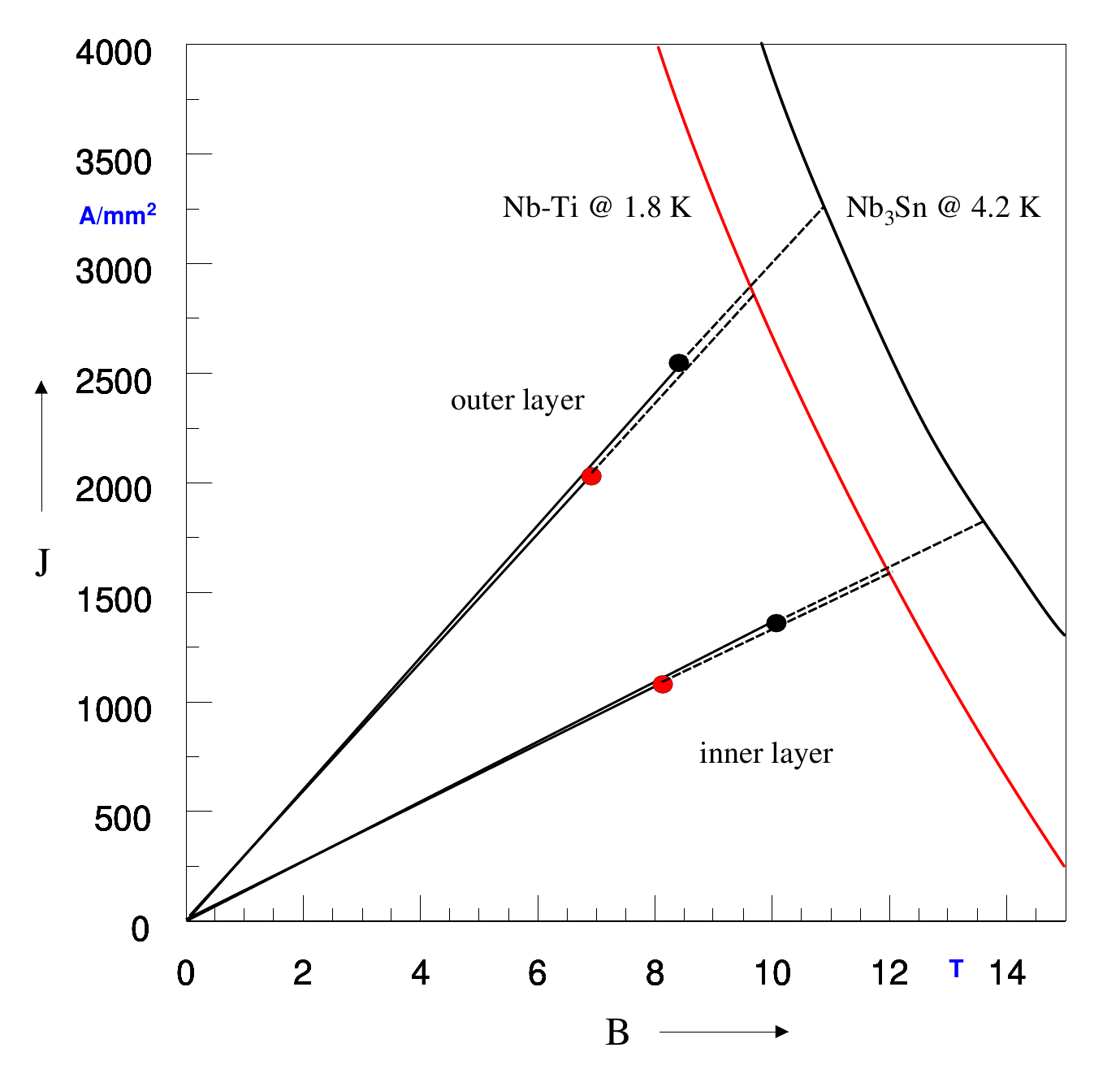,clip=,width=8cm}
 \end{center}
\caption{\label{figll} Working points on the load-line for both Nb-Ti and Nb$_3$Sn variants of the half quadrupole for Q1.}
\end{figure}

\begin{figure}[h!]
\begin{center}
\centering\epsfig{file=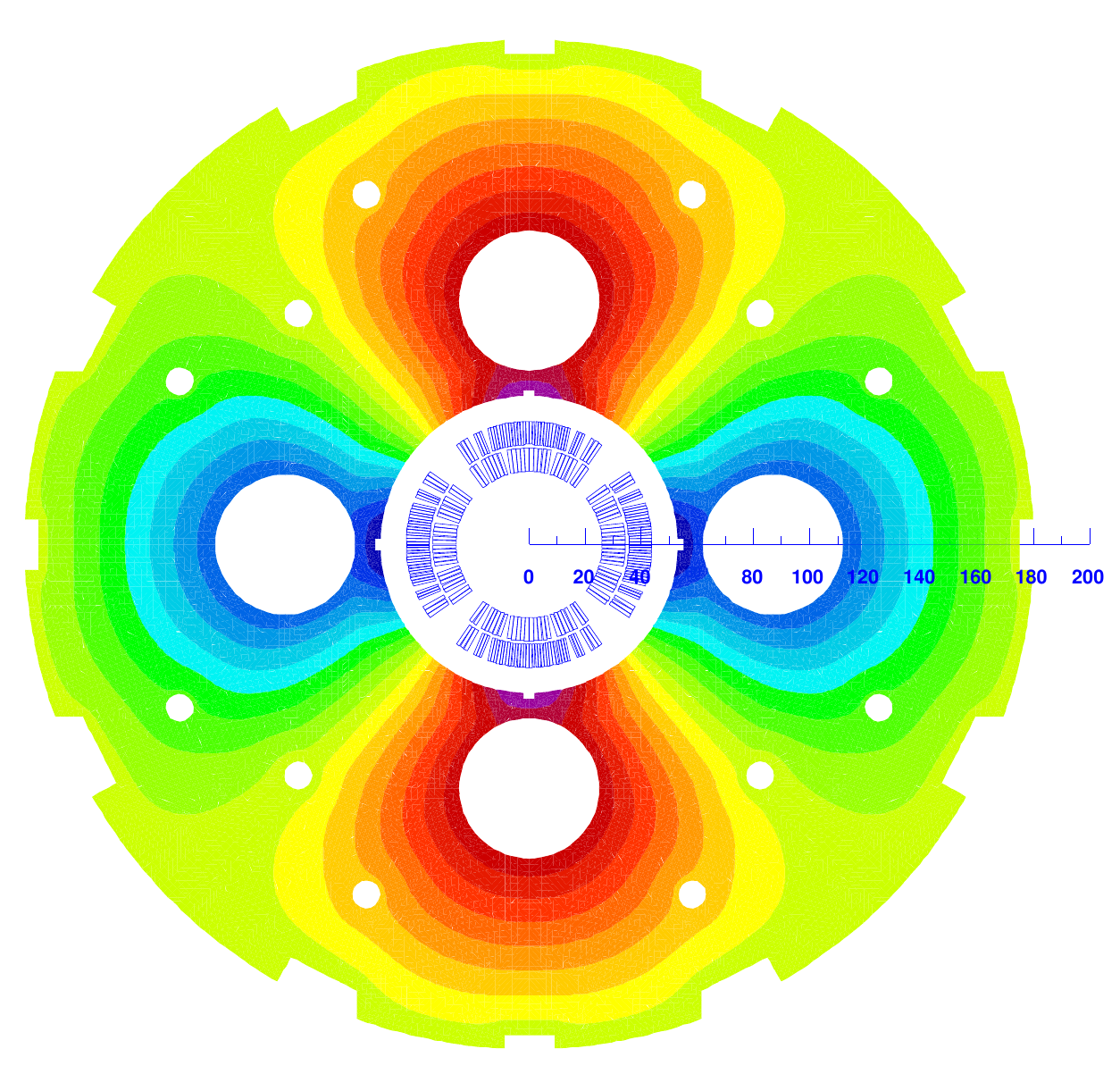,clip=,width=6cm}\hspace{1cm}
\centering\epsfig{file=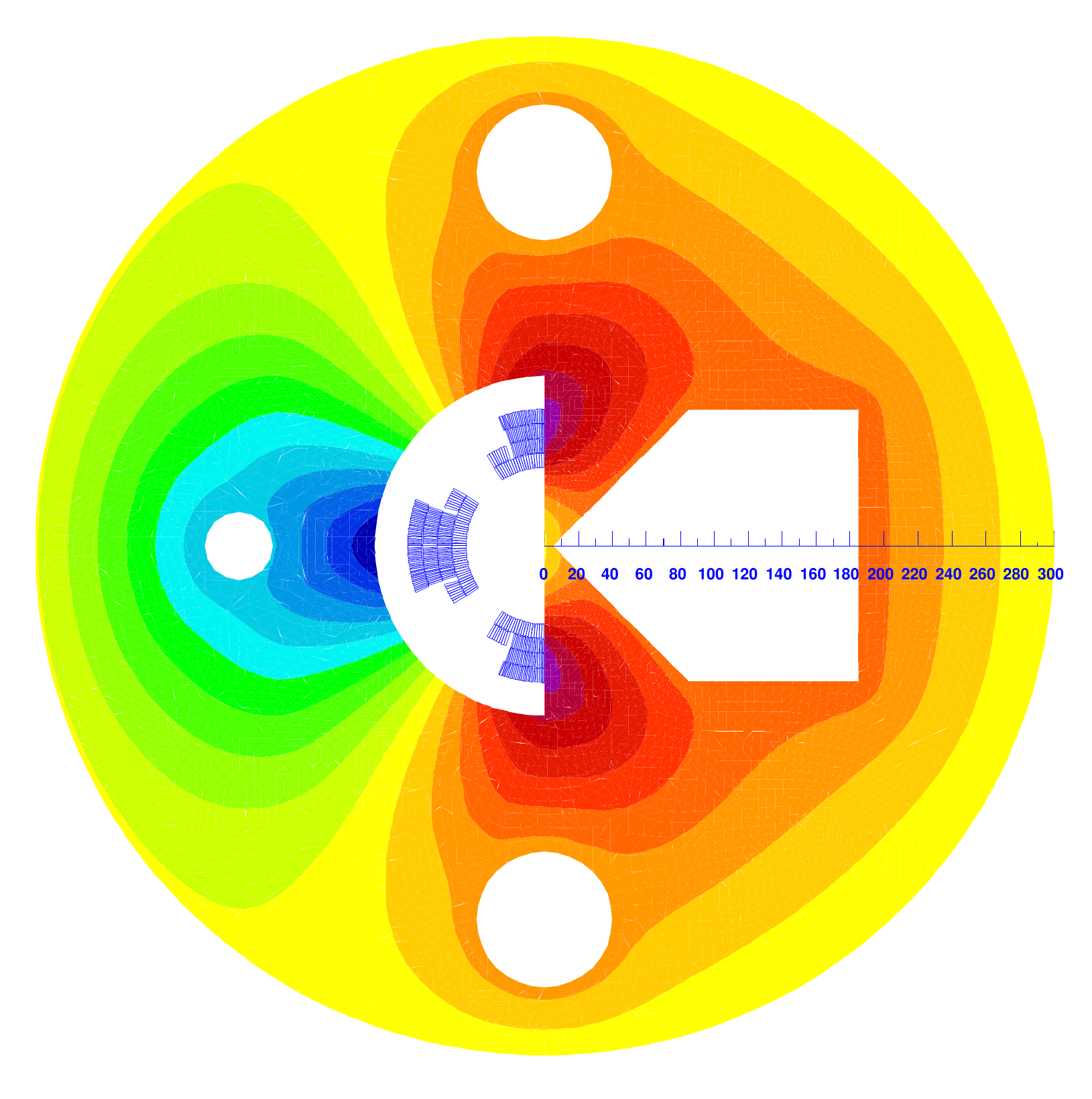,clip=,width=6cm}
 \end{center}
\caption{\label{fig4} Cross-sections of the insertion quadrupole magnets for the linac-ring option. Left: Single aperture quadrupole. Right: Half quadrupole with field-free region.}
\label{srfig4}
\end{figure}

The requirements in terms of aperture and field gradient are more difficult to obtain for the linac-ring option. Consequently we present the limitations for the field gradient and septum size achievable with both Nb-Ti and Nb$_3$Sn superconducting technologies. We limit ourselves to the two conceptual designs already chosen for the ring-ring option. For the half quadrupole, shown in Fig. \ref{fig4} (right), the working points on the load-line are given for both superconducting technologies in Fig. \ref{figll}. \par

However, the conductor size must be increased and in case of the half quadrupole, a four layer coil must be used; see Fig. \ref{fig4}. 
The thickness of the coil is limited by the flexural rigidity of the cable, which will make the coil-end design difficult. Moreover, a thicker coil will also increase the beam separation between the proton and the electron beams. The results of the field computation are given in Table \ref{data}, column 3 and 4. Because of the higher iron saturation, the fringe fields in the electron beam channel are 
considerably higher than in the magnets for the ring-ring option.\par

For the Nb$_3$Sn option we assume composite wire produced with the internal Sn process (Nb rod extrusions), \cite{parrell}. The non-Cu critical current density 
is 2900 A/mm$^2$ at 12 T and 4.2 K. The filament size of 46 $\mu$m in Nb$_3$Sn strands give rise to higher persistent current effects in the magnet. The choice of
Nb$_3$Sn would impose a considerable R\&D and engineering design effort, which is however, not more challenging than other accelerator magnet 
projects employing this technology \cite{Devred:2006qh}. 

\begin{table}[h!]
\begin{center}
\begin{tabular}{|l|l|c|c|c|c|c|}\hline
Type 	      & &  Ring-ring          & Ring-ring	& Linac-ring         & Linac-ring  \\
                     & &  single aperture & half-quad     & single aperture & half-quad   \\\hline
Function         & &    Q2                &   Q1           &    Q2               &  Q1            \\\hline
SC                 & &  \multicolumn{4}{c|}{Nb-Ti at 1.8 K} \\\hline
R                   & mm &  36  &  35 & 23 & 46  \\ 
I$_\n{nom}$    & A   &  4600 & 4900 &  6700 &  4500   \\
g                   & T/m &  137      &  137       &   248  &  145    \\
B$_0$                &  T   &  -     &   2.5      &    -      &  3.6   \\
LL                  &  \% &   73    &    77      &   88    &  87  \\
S$_\n{beam}$ & mm & 107  &  65 &  87 &  63  \\
B$_\n{fringe}$ & T    &  0.016 & 0.03 &  0.03 & 0.37 \\
g$_\n{fringe}$ & T/m &  0.5  &   0.8 &  3.5 &  18 \\\hline
SC                 & &  \multicolumn{4}{c|}{Nb$_3$Sn at 4.2 K} \\\hline
I$_\n{nom}$    & A   &     &   &  6700 &  4500   \\
g                   & T/m &        &         &   311  &  175    \\
B$_0$                &  T   &        &         &    -      &  4.7   \\
LL                  &  \% &        &          &   77    &  76  \\
B$_\n{fringe}$ & T    &       &        &  0.09 & 0.5 \\
g$_\n{fringe}$ & T/m &      &       &  9 &  25 \\\hline
\end{tabular}
\caption{SC = type of superconductor, g = field gradient, R = radius of the aperture (without cold bore and beam-screen), LL = operation percentage on the load line of 
the superconductor material, I$_\n{nom}$ = operational current, B$_0$ = main dipole field, S$_\n{beam}$ = beam separation distance, B$_\n{fringe}$ = 
fringe field in the aperture for the electron beam, g$_\n{fringe}$ = gradient field in the aperture for the electron beam.}
\end{center}
\label{data}
\end{table}

\begin{figure}[t!]
\begin{center}
\centering\epsfig{file=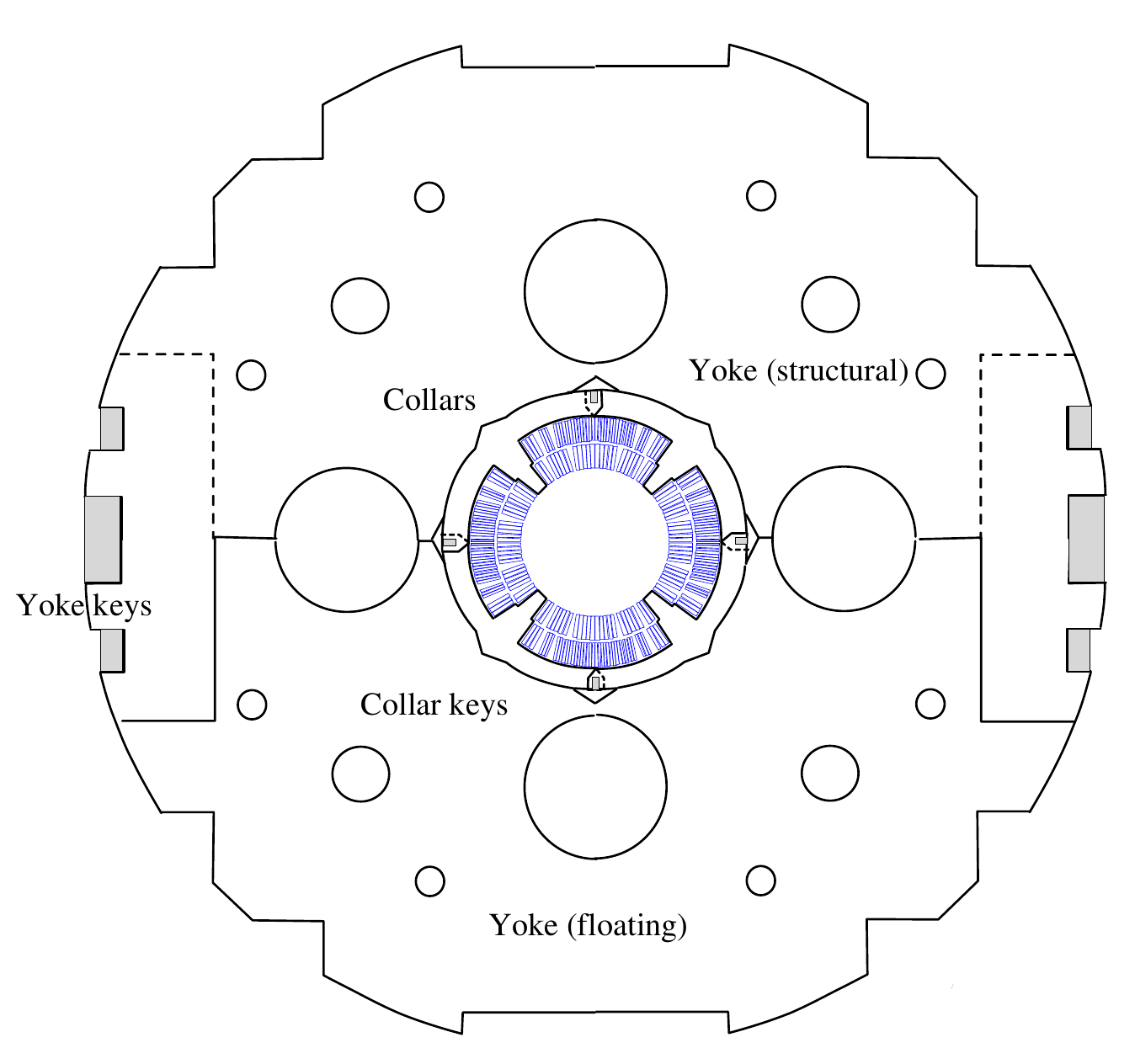,clip=,width=7.5cm} \hspace{1cm}
\centering\epsfig{file=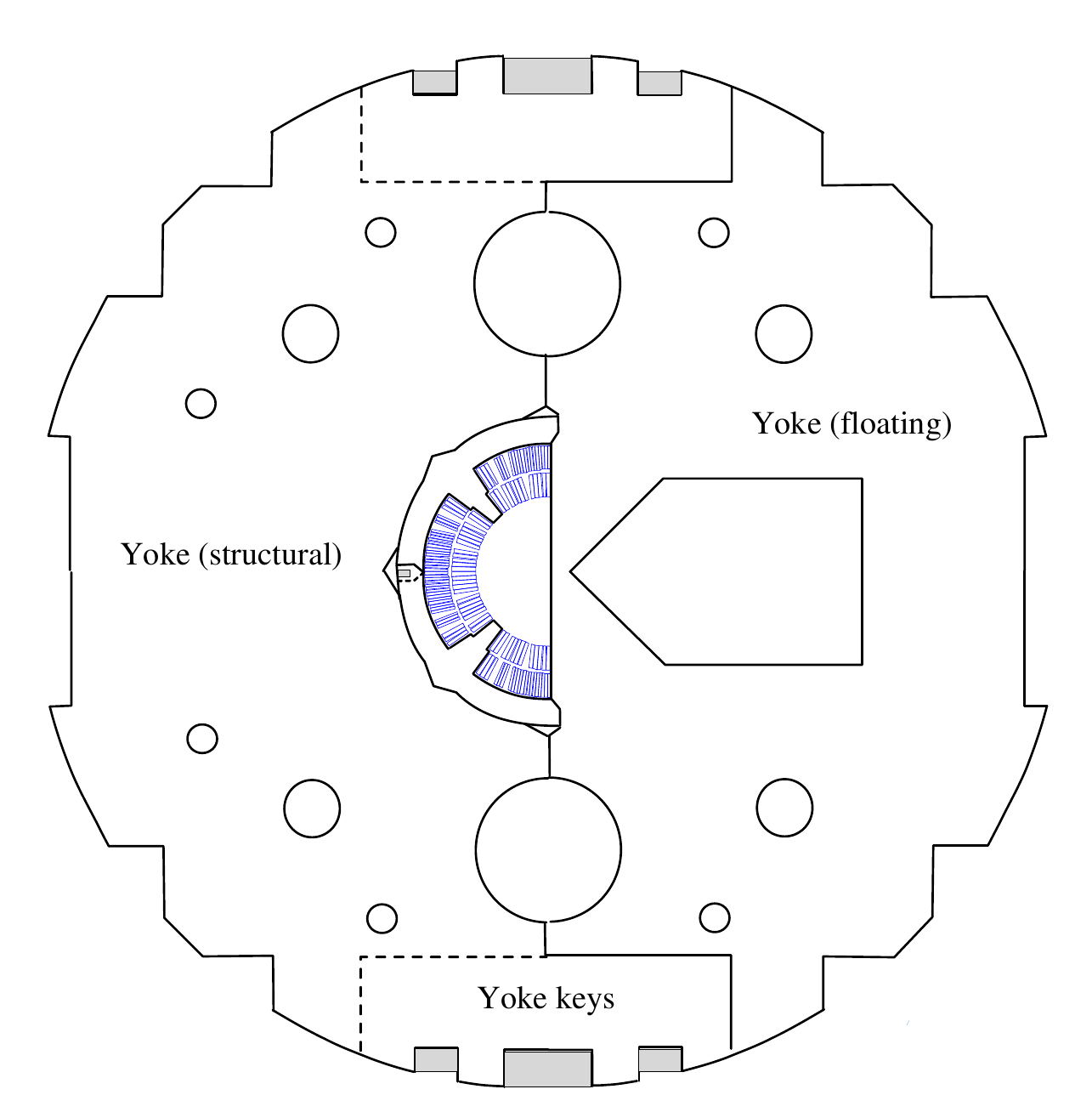,clip=,width=7cm}
 \end{center}
\caption{\label{fig5} Sketch of the mechanical structure. Left: Single aperture magnet. Right: Half quadrupole with field-free region.}
\end{figure}

Fig. \ref{fig5} shows the conceptual design of the mechanical structure of these magnets. The necessary pre-stress in the coil-collar structure, which must be high enough to avoid unloading at full excitation, cannot be exerted with the stainless-steel collars alone. For the single aperture magnet as shown in Fig. \ref{fig5} left, two interleaved sets of yoke laminations (a large one comprising the area of the yoke keys and a smaller, floating lamination with no structural function) provide the necessary mechanical
stability of the magnet during cooldown and excitation. Preassembled yoke packs are mounted around the collars and put under a hydraulic press, so that the keys can be inserted. The sizing of these keys and the amount of pre-stress before the cooldown will have to be calculated using mechanical FEM programs. This also depends on the elastic modulus of the coil, which has to be measured with a short-model equipped with pressure gauges. Special care must be taken to avoid non-allowed multipole harmonics because the four-fold symmetry of the quadrupole will not entirely be maintained.

The mechanical structure of the half-quadrupole magnet is somewhat similar, however, because of the left/right asymmetry four different yoke laminations must be produced.  The minimum thickness of the septum will also have to be calculated with structural FEM programs. 

%% file: machine/magnets.tex
\section{Arc accelerator magnets}
\label{arcmagnets}
In this section the main magnets needed for the accelerator are considered. The analysis focuses separately on the ring-ring (RR) and linac-ring (LR) layouts. The requirements are listed and an initial design is proposed. The RR dipoles prompted an experimental activity, involving the manufacturing and magnetic characterisation of short models, whose results are briefly reported here.

We gratefully acknowledge the fruitful discussion with Neil Marks about the design of these electromagnets. We thank Miriam Fitterer and Alex Bogacz for help in checking the requirements of the magnets according to the lattice, for the RR and LR option, respectively.

\subsection{RR option, dipole magnets}
A total of 3080 bending magnets, 5.35~m long, are needed in the LHC tunnel for the RR layout, of which 3040 form the arcs and the remaining 40 are for the insertion and by-pass regions. The nominal strength is 0.0127~T at 10~GeV and 0.0763~T at 60~GeV. As a comparison, the LEP collider contained 3280 main dipole magnets, with a nominal flux density at injection (20~GeV) of 0.0215~T, and at collision energy (100~GeV) of 0.1100~T \cite{Giesch:1989sg}.

The main points to consider in the design of these magnets are:
\begin{itemize}
	\item{the low working flux density, in particular at injection, that constitutes a challenge for cycle-to-cycle reproducibility and for good field quality throughout the ramp;}
	\item{the need for compactness, to fit in the present tunnel with the installed LHC systems;}
	\item{the required compatibility with the emitted synchrotron radiation power.}
\end{itemize}
Different designs have been proposed at BINP and CERN to respond to these demands. In particular, the first point (low injection field) has prompted an experimental activity, with several short models manufactured and measured. This experience is briefly summarised next. 

\subsubsection{BINP model}
\begin{figure}[!h]
	\centerline{\includegraphics[clip=,width=0.74\textwidth]{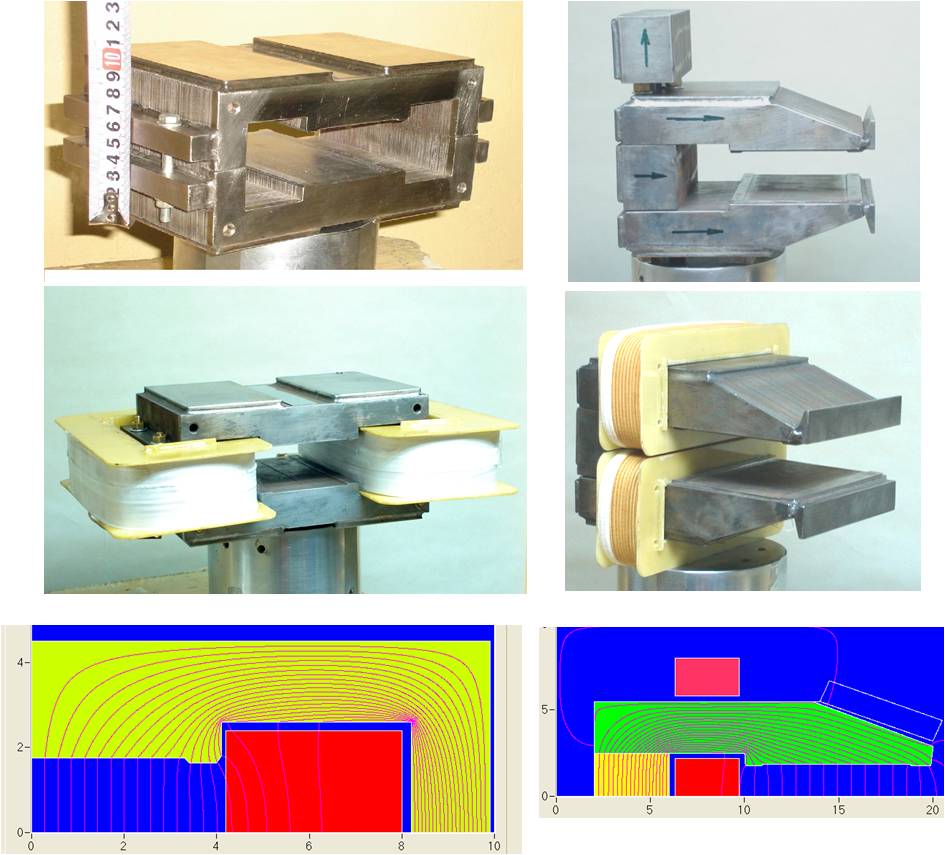}}
	\caption{H and C type model magnets made by BINP at Novosibirsk.}
	\label{fig:BINP-Models}
\end{figure}

Two different types of models have been manufactured at BINP, see Figure~\ref{fig:BINP-Models}. The aim was to demonstrate that a cycle-to-cycle reproducibility at injection better than $0.1\cdot10^{-4}$~T can be achieved. Both models have shown a field reproducibility at injection current within $\pm0.075\cdot10^{-4}$~T, when cycled between injection and maximum field. To achieve such results the iron laminations were made of 3408 type grain oriented silicon steel 0.35~mm thick. Their coercive force in the direction of the grain orientation is $H_{c\parallel}\approx6$~A/m, while in the direction perpendicular to the grain orientation it remains relatively low, $H_{c\perp}\approx22$~A/m. The C-type model has been assembled in two variants, with the central iron part with the grains oriented vertically and horizontally (both blocks are as shown in the picture). The magnetic measurements did not show relevant differences between the two versions.

\begin{figure}[!h]
  \vspace{2 mm}
	\centerline{\includegraphics[clip=,width=0.74\textwidth]{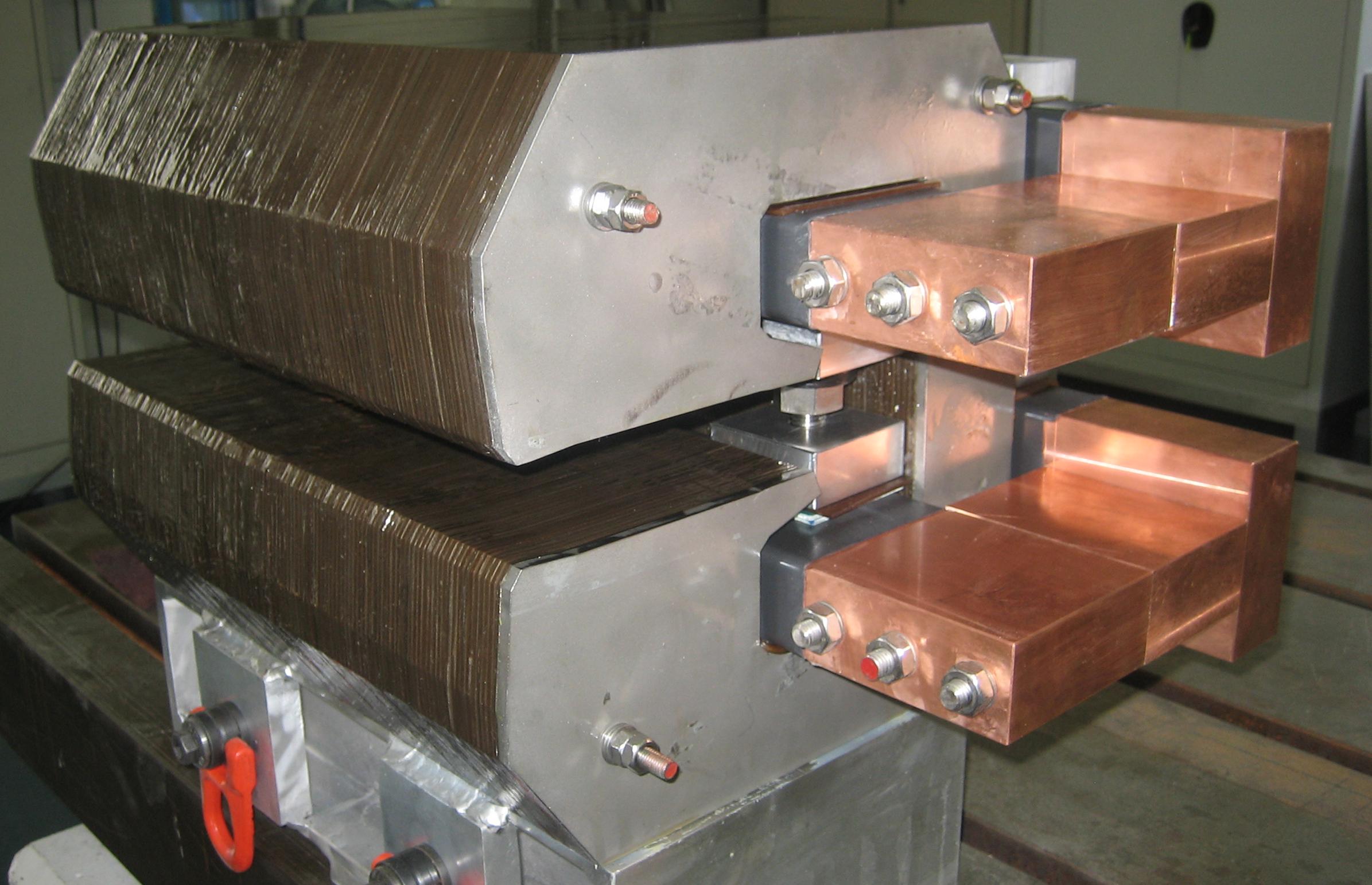}}
	\caption{One of the 400~mm long model magnets made at CERN with interleaved laminations.}
	\label{fig:CERN-Model}
\end{figure}

\subsubsection{CERN model} 
As a complementary study to the one made by BINP, the CERN model has explored the manufacture of lighter magnets, with the yoke consisting of interleaved steel and plastic laminations. A thickness ratio between plastic and steel of 2:1 has been chosen. As the flux produced in the magnet aperture is concentrated in the high permeability regions only, the magnetic field in the iron pole is about 3 times that in the gap. In addition to a lighter assembly, this solution has the advantage of increasing the magnetic working point of the iron at injection field. This makes the design less sensitive to the characteristics of the iron and in particular to the coercive force. A similar strategy had been adopted for the LEP dipoles, where 1.5~mm thick low-carbon steel laminations were spaced by 4~mm and embedded in a cement mortar.

The proposed design is a compact C type dipole, see Figure~\ref{fig:CERN-Model}. The aperture is on the external side of the ring, so that the magnet does not intercept the emitted synchrotron radiation, and possibly room is left for a vacuum pre-chamber. The geometry involves a rather unusual shape for the poles. The objective was to design a cross section able to minimise the difference of flux lines length over the horizontal aperture. This makes the field quality (in particular, the quadrupole component) less dependent on variations of iron characteristics, both at injection and collision energies.

For the coils, a 1-turn solution (per pole) has been adopted, with solid copper bars which after insulation are individually slid inside the magnet.

To explore the potential of the proposed design, in particular in terms of magnetic field reproducibility at injection energy, three models have been built using three different materials:
\begin{itemize}
	\item{model 1: a rather noble Supra~36 NiFe steel, 1.0~mm thick laminations, with a measured coercive field (after heat treatment for 4~hours at 1050~$^{\circ}$C under hydrogen), equal to $H_c\approx6$~A/m;}
	\item{model 2: a conventional low carbon steel with low silicon content, 1.0 mm thick laminations, 0.5\%~Si, $H_c\approx70$~A/m;}
	\item{model 3: a 35M6 grain oriented steel, 0.35~mm thick laminations, 3.1\% silicon, with $H_{c\parallel}\approx7$~A/m and $H_{c\perp}\approx25$~A/m.}
\end{itemize}

In all cases 2~mm thick phenolic sheets have been used as spacers, stacked and glued with an epoxy resin together with the steel sheets. For the last model, to compensate for the thinner laminations, three of them were stacked together, in order to keep a similar magnetic field distribution as in the stacks with the isotropic steels.

Magnetic measurements have been performed to assess the field reproducibility at injection. A cycle from 10~GeV to 60~GeV, requiring a dipole field of 0.0127~T to 0.0763~T, corresponds to currents from 210~A to 1340~A. Unfortunately the available power converter could provide a sufficiently good stability only over a smaller range, namely between 260~A and 1300~A, with measured stabilities of $4\cdot10^{-5}$ at 260~A and $2\cdot10^{-5}$ at 1300~A. Each of the models was submitted to 5 conditioning cycles and thereafter to 8 cycles between these currents at a ramp rate of 400~A/s. The reproducibility of the magnetic field in the gap was measured with an integral coil coupled with a digital integrator, providing the results summarised in Tables~\ref{tab:FIELDREP-1} and \ref{tab:FIELDREP-2}.

The performance is in all cases very satisfactory. There might be an indication that models 1 and 3, as expected, perform better than model 2; however, the values are close to the measurement errors. In practice these results show that within this range of field levels the value of the coercive field does not seem to play a major role in the reproducibility of the magnetic field from cycle to cycle. More details about the manufacturing of these models and the magnetic measurements can be found in \cite{Attilio2}. 

\begin{table}[!h]
  \centering
  \begin{tabular}{|l|c|c|}
    \hline
		Model                                   & Low field       & High field      \\ \hline
		Model 1 (NiFe steel)                    & $5\cdot10^{-5}$ & $4\cdot10^{-5}$ \\ \hline
		Model 2 (Low carbon steel)              & $6\cdot10^{-5}$ & $6\cdot10^{-5}$ \\ \hline
		Model 3 (Grain oriented 3.5\%~Si steel) & $4\cdot10^{-5}$ & $6\cdot10^{-5}$ \\ \hline
  \end{tabular}
	\caption{Reproducibility of magnetic field over 8 cycles, maximum deviation from average.}
	\label{tab:FIELDREP-1}
\end{table}
\begin{table}[!h]
  \centering
  \begin{tabular}{|l|c|c|}
    \hline
		Model                                   & Low field       & High field      \\ \hline
		Model 1 (NiFe steel)                    & $3\cdot10^{-5}$ & $3\cdot10^{-5}$ \\ \hline
		Model 2 (Low carbon steel)              & $4\cdot10^{-5}$ & $5\cdot10^{-5}$ \\ \hline
		Model 3 (Grain oriented 3.5\% Si steel) & $2\cdot10^{-5}$ & $4\cdot10^{-5}$ \\ \hline
  \end{tabular}
	\caption{Reproducibility of magnetic field over 8 cycles, standard deviation from average.}
	\label{tab:FIELDREP-2}
\end{table}

The conclusion of this analysis is that all three models meet the LHeC specifications. However, the similarity that can be achieved in a series production of 3080~units has to be further investigated. The low value of injection field amplifies the problem, as in that region the variation in magnetic parameters is larger. This problem is already partially taken care of in the design of the cross section, that is meant to be less sensitive to the iron characteristics, and in the low stacking factor. Furthermore, the usual procedure of ``shuffling'' (or ``sorting'') the laminations during the production has to be envisaged, with results that might depend on the statistical distribution of coercive forces and permeabilities (at low field) in the steel, as well as on the shuffling technique.

\subsubsection{Proposal for dipole magnets, RR option}
\begin{figure}[!h]
	\centerline{\includegraphics[trim = 20mm 120mm 20mm 100mm, clip]{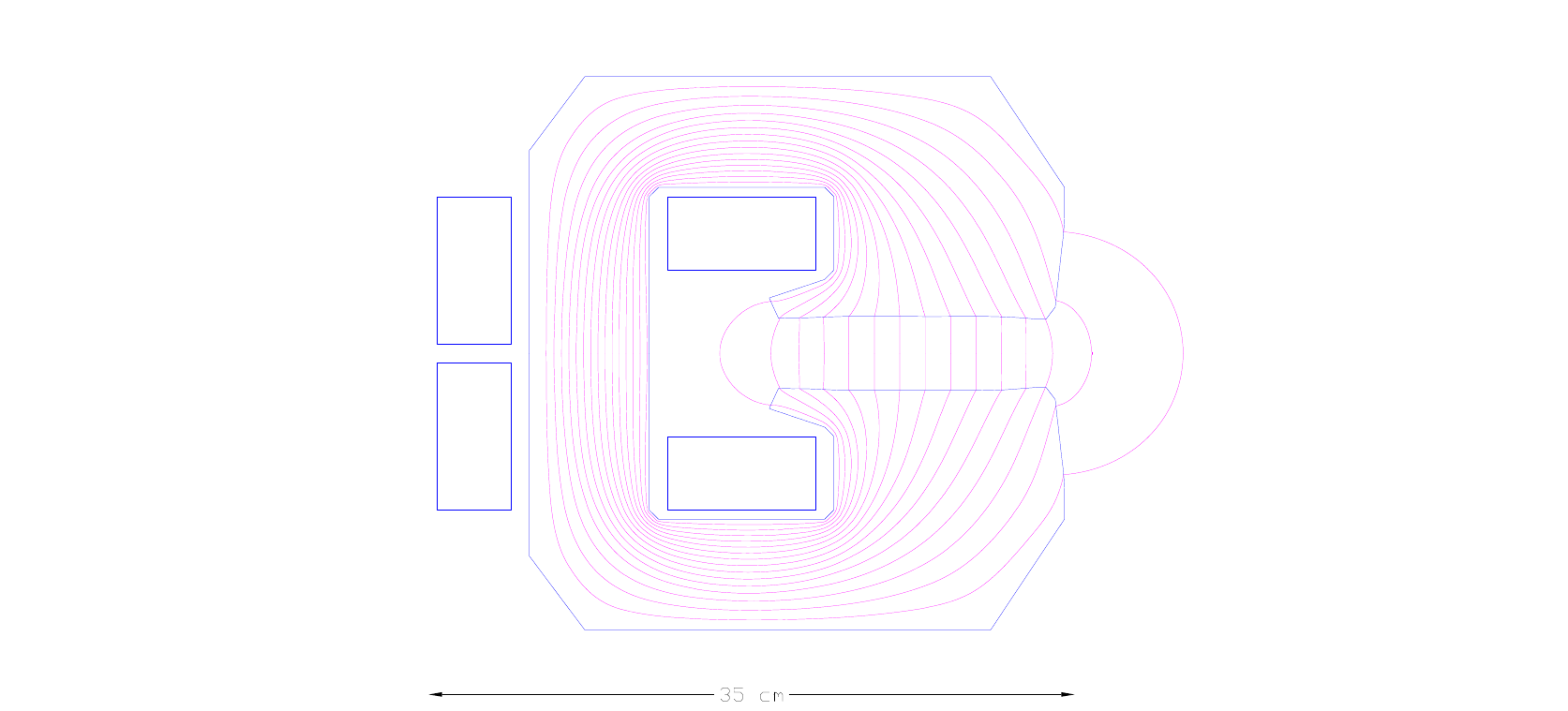}}
	\caption{Bending magnets for the RR option (scale 1:5).}
	\label{fig:MBRR}
\end{figure}

The proposed cross section for the dipoles of the ring-ring option is shown in Figure~\ref{fig:MBRR}. The main parameters are summarised in Table~\ref{tab:MBRR}.

The idea of assembling the yoke with steel laminations interleaved by plastic spacers is retained, as in the CERN models. This has the mechanical advantage of a lower weight of the assembly, and the magnetic advantage of magnifying the field in the steel by a factor of about 3. This is of particular interest at injection energy. 

The conductor can be in aluminium (like in LEP) or in copper depending on economical reasons coming from a correct balance between investment and operation costs. The present design is based on an aluminium conductor. With respect to copper, this has the advantage of making the magnet lighter (about 200~kg of coil instead of about 625~kg). Using copper, however, would imply a power consumption, per magnet, at 60~GeV around 190~W instead of around 300~W. Notwithstanding the material, the choice of having 1-turn coils, i.e., solid straight bars, has several technical and economical consequences:
\begin{itemize}
	\item the coil manufacturing is simpler and hence cheaper;
	\item the high current (1300~A) involves large terminals and connections between the magnets;
	\item the power supply is rated at high current, but with rather low voltage and impedance;
	\item the resistive losses in the interconnections, terminals and in the power cables are significantly higher than those for a multi-turn magnet working at lower current;
	\item it is possible to envisage to use the conductor as bus-bar to connect the string of magnets in series, thus reducing the number of interconnections.
\end{itemize}
The solution proposed here for the conductor is similar to the one that had been adopted for LEP. However, these aspects need to be further investigated in the TDR on a wider perspective.

The conductor size is sufficiently large so that the current density is around 0.4~A/mm$^2$. The dissipated resistive power (of the order of 50~W per metre of length of the magnet, considering aluminium as conductor) is reduced to levels which can be possibly dealt with by the ventilation in the LHC tunnel: this is a considerable advantage in terms of simplicity of magnet manufacture, connections, reliability and of course it avoids the installation of a water cooling circuit dedicated to the dipoles in the arcs.

\subsection{RR option, quadrupole magnets}
The quadrupole magnets needed for the ring-ring option can be considered undemanding and well within the compass of standard design.

\subsubsection{Quadrupoles in the arcs}
In the arcs, 336 focusing quadrupoles (QF) providing 10.28~T integrated strength, and 336 defocusing quadrupoles (QD) each providing 8.40~T integrated strength are needed. These are to be installed in the LHC tunnel.

\begin{figure}[!h]
	\centerline{\includegraphics[trim = 20mm 115mm 20mm 105mm, clip]{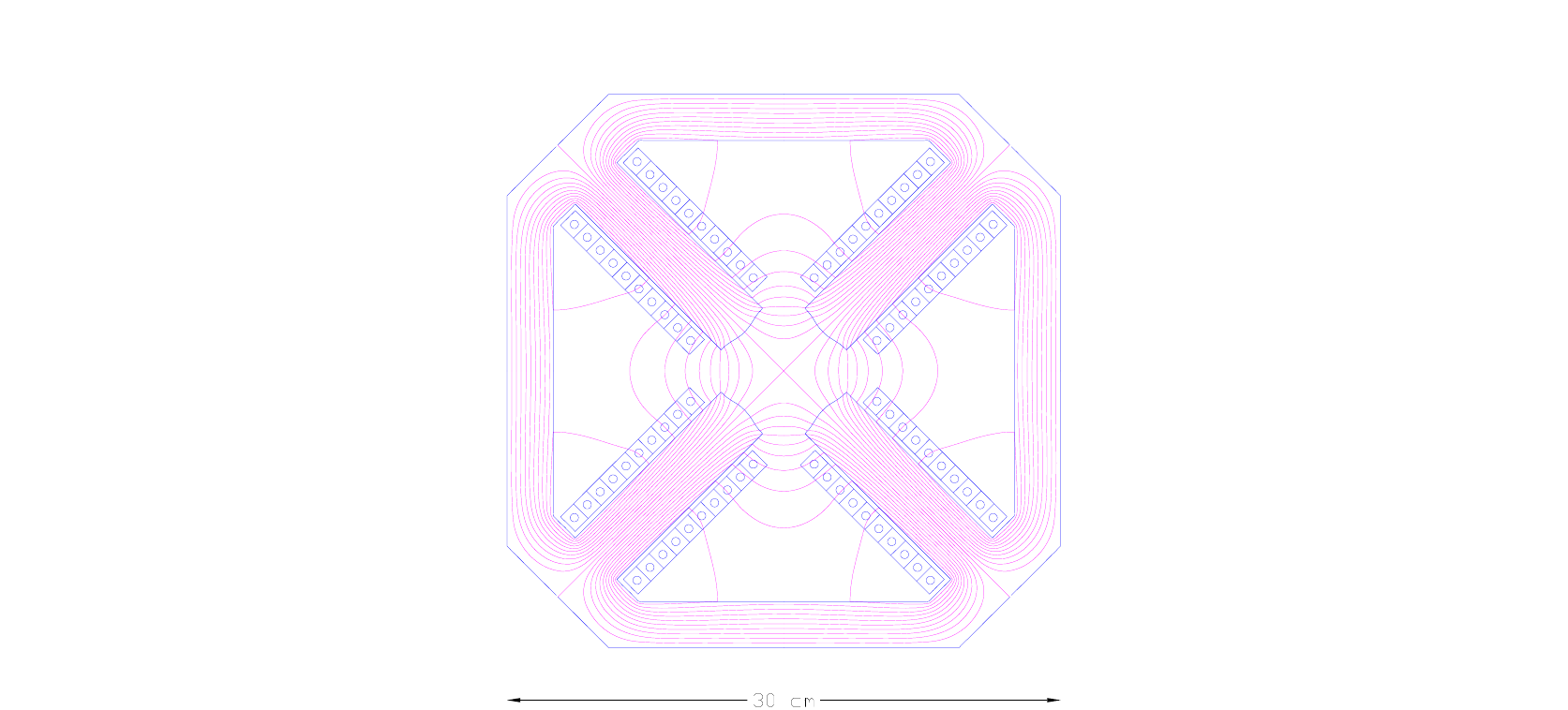}}
	\caption{Arc quadrupoles for the RR option (scale 1:5).}
	\label{fig:AQRR}
\end{figure}

Considering that the integrated strengths of the QD and QF are not much different, it is proposed here to have the same type of magnets. The relevant parameters are summarised in Table~\ref{tab:AQRR} and the cross section is illustrated in Figure~\ref{fig:AQRR}.

\subsubsection{Quadrupoles in the insertion and by-pass}
\begin{figure}[!h]
	\centerline{\includegraphics[trim = 20mm 110mm 20mm 110mm, clip]{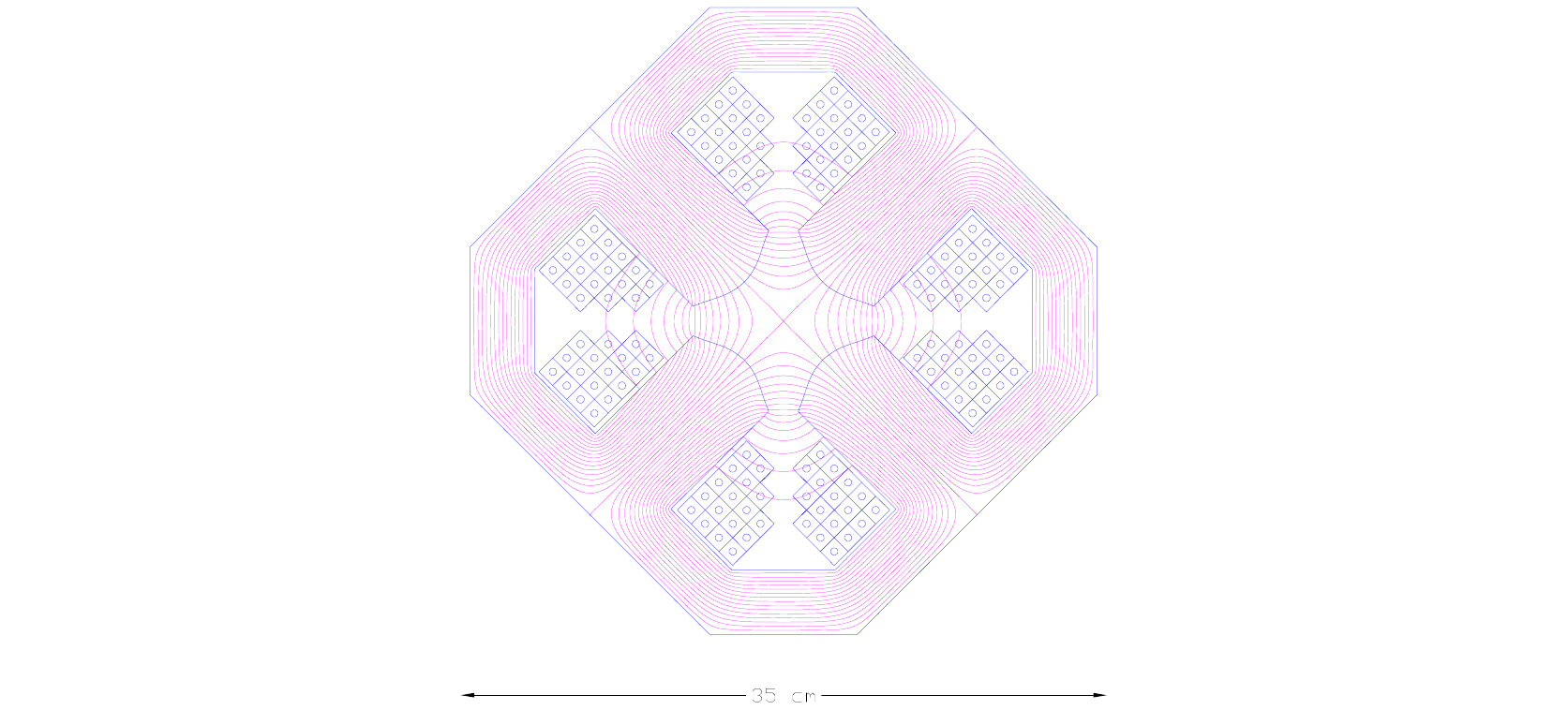}}
	\caption{Insertion and by-pass quadrupole magnets for the RR Option (scale 1:5).}
	\label{fig:IQRR}
\end{figure}

In total 148 QF and 148 QD magnets are needed in the insertion and by-pass regions. The required integrated strength is 18~T for the QF and 13~T for the QD. In this case, it is proposed to keep the same magnet cross section but to have two different lengths for the quadrupoles, namely, 1.0~m for the QF and 0.7~m for the QD. The relevant parameters are summarised in Table~\ref{tab:IQRR} and the cross section is illustrated in Figure~\ref{fig:IQRR}. A value of 19~T/m is taken as design gradient.

\clearpage

\vspace{10mm}

\begin{table}[!h]
  \centering
  \begin{tabular}{| l | c c |}
    \hline
		Beam energy                      & 10 to 60         & GeV             \\ \hline
		Magnetic field                   & 0.0127 to 0.0763 & T               \\ \hline
		Magnetic length                  & 5.35             & m               \\ \hline
		Vertical aperture                & 40               & mm              \\ \hline
		Pole width                       & 150              & mm              \\ \hline
		Mass                             & 1400             & kg              \\ \hline
		Number of magnets                & \multicolumn{2}{c|}{3080}          \\ \hline
		Current @ 0.0763 T               & 1300             & A               \\ \hline
		Number of turns per pole         & \multicolumn{2}{c|}{1}             \\ \hline
		Current density @ 0.0763 T       & 0.4              & A/mm$^2$        \\ \hline
		Conductor material               & \multicolumn{2}{c|}{aluminium}     \\ \hline
		Magnet inductance                & 0.13             & mH              \\ \hline
		Magnet resistance                & 0.18             & m$\Omega$       \\ \hline
		Power @ 60 GeV                   & 300              & W               \\ \hline
		Total power consumption @ 60 GeV & 0.92             & MW              \\ \hline
		Cooling                          & \multicolumn{2}{c|}{air}           \\ \hline
  \end{tabular}
	\caption{Main parameters of bending magnets for the RR option.}
	\label{tab:MBRR}
\end{table}

\clearpage

\vspace{10mm}

\begin{table}[!h]
  \centering
  \begin{tabular}{| l | c c |}
    \hline
		Beam energy                              & 10 to 60        & GeV          \\ \hline
		Field gradient @ 60 GeV (QF/QD)          & 10.28 / -8.40   & T/m          \\ \hline
		Magnetic length                          & 1.0             & m            \\ \hline
		Aperture radius                          & 30              & mm           \\ \hline
		Mass                                     & 400             & kg           \\ \hline
		Number of magnets (QF/QD)                & \multicolumn{2}{c|}{336 / 336} \\ \hline
		Current @ 60 GeV (QF/QD)                 & 380 / 310       & A            \\ \hline
		Number of turns per pole                 & \multicolumn{2}{c|}{10}        \\ \hline
		Current density @ 60 GeV (QF/QD)         & 4.0 / 3.3       & A/mm$^2$     \\ \hline
		Conductor material                       & \multicolumn{2}{c|}{copper}    \\ \hline
		Magnet inductance                        & 4               & mH           \\ \hline
		Magnet resistance                        & 16              & m$\Omega$    \\ \hline
		Power @ 60 GeV (QF/QD)                   & 2.3 / 1.5       & kW           \\ \hline
		Total power consumption @ 60 GeV (QF/QD) & 0.77 / 0.52     & MW           \\ \hline
		Cooling                                  & \multicolumn{2}{c|}{water}     \\ \hline
	\end{tabular}
	\caption{Main parameters of arc quadrupoles for the RR option.}
	\label{tab:AQRR}
\end{table}
		
\clearpage

\vspace{10mm}

\begin{table}[!h]
  \centering
  \begin{tabular}{| l | c c |}
    \hline
		Beam energy                              & 10 to 60    & GeV                \\ \hline
		Field gradient @ 60 GeV                  & 19          & T/m                \\ \hline
		Magnetic length (QF/QD)                  & 1.0 / 0.7   & m                  \\ \hline
		Aperture radius                          & 30          & mm                 \\ \hline
	  Mass (QF/QD)                           & 560 / 390   & kg                 \\ \hline
		Number of magnets (QF/QD)               & \multicolumn{2}{c|}{148 / 148}   \\ \hline
		Current @ 19 T/m                         & 420         & A                  \\ \hline
		Number of turns per pole                 & \multicolumn{2}{c|}{17}          \\ \hline
		Current density @ 19 T/m                 & 4.6         & A/mm$^2$           \\ \hline
		Conductor material                       & \multicolumn{2}{c|}{copper}      \\ \hline
		Magnet inductance (QF/QD)                & 15 / 10     & mH                 \\ \hline
		Magnet resistance (QF/QD)                & 30 / 23     & m$\Omega$          \\ \hline
		Power @ 60 GeV (QF/QD)                   & 5.3 / 3.9   & kW                 \\ \hline
		Total power consumption @ 60 GeV (QF/QD) & 0.78 / 0.58 & MW                 \\ \hline
		Cooling                                  & \multicolumn{2}{c|}{water}       \\ \hline
	\end{tabular}
	\caption{Main parameters of insertion and by-pass quadrupoles for the RR option.}
	\label{tab:IQRR}
\end{table}

\clearpage

\subsection{LR option, dipole magnets}
The bending magnets for the LR option are used in the arcs of the recirculator. Each of the six arcs needs $58 \times 10 = 580$ dipoles for the standard arc cells, plus $2 \times 2 = 4$ for the dispersion suppression regions at the two ends. This results in a total of 584 units. These magnets are 4~m long and they provide a magnetic field ranging from 0.046~T to 0.264~T depending on the arc energy, from 10.5~GeV to 60.5~GeV. Additionally, a few bending magnets (4 at each end of an arc) are needed for the switch-yards regions. These magnets -- providing vertical bends -- are in a separate category and are not considered at the moment.

Considering the relatively low field strength required even for the highest energy arc, and the small required physical aperture of 25~mm only, it is proposed here to adopt the same cross section for all the magnets, possibly using smaller conductors for the ones at the lowest energies. This allows the design of very compact and relatively cheap magnets, running at low current densities to minimise the power consumption. 

The choice of having 1-turn coils prompts the same comments as for the dipoles of the RR option. In this case, though, the maximum current is considerably higher (2700~A vs. 1300~A), although the overall dissipated power is lower.

\begin{figure}[t]
	\centerline{\includegraphics[trim = 20mm 105mm 20mm 120mm, clip]{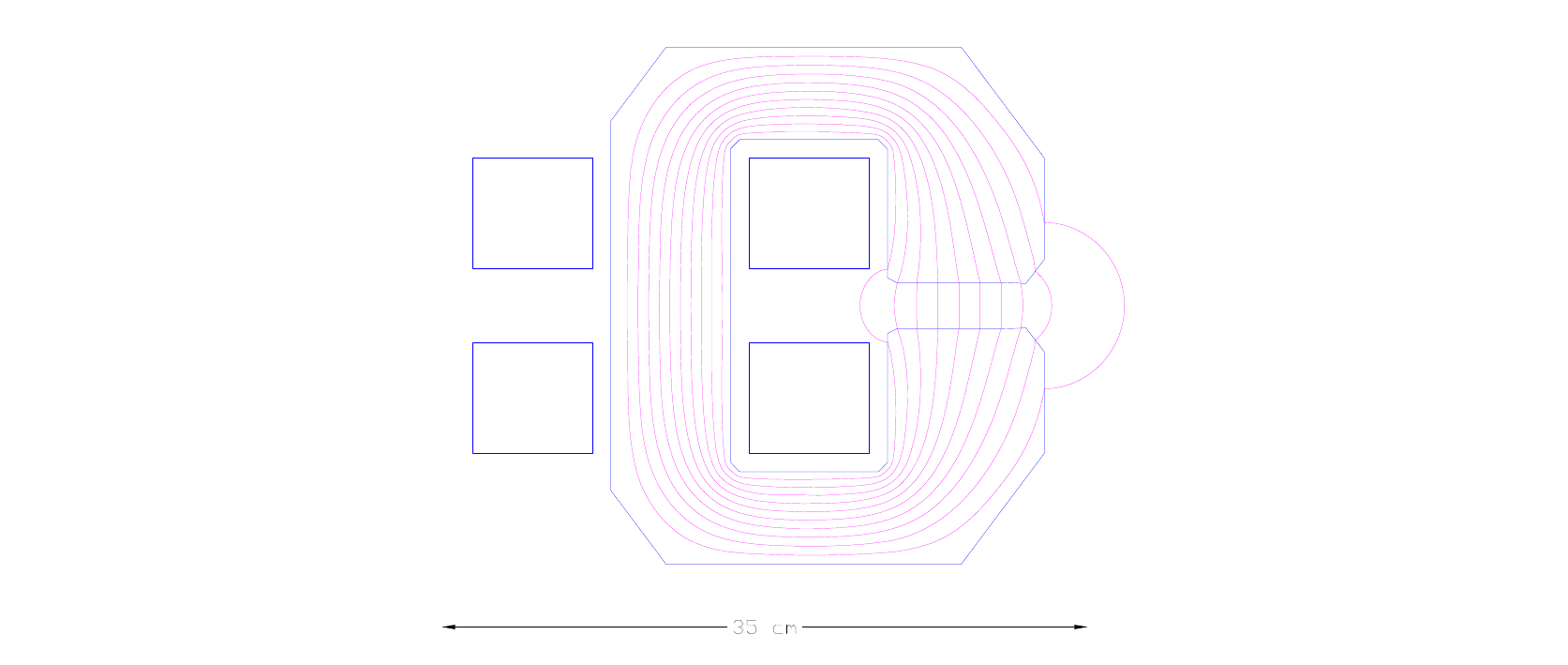}}
	\caption{Bending magnets for the LR recirculator (scale 1:5).}
	\label{fig:MBALR}
\end{figure}

Table~\ref{tab:MBALR} summarises the main parameters of the proposed magnet design, which is illustrated in Figure~\ref{fig:MBALR}.

The proposed design is based on classical resistive electromagnets. The use of units embedding permanent magnets could be envisaged, given the (almost stationary) requirements on the field. The capital cost would be significantly higher, but savings would occur on the side of power supplies and interconnections, besides clearly on the electric bill.

\subsection{LR option, quadrupole magnets}

\subsubsection{Quadrupoles for the recirculator arcs}
In each of the six recirculator arcs, four different types of quadrupoles are needed, each type in 60 units, adding up to 240 quadrupoles per arc. The Q0, Q1 and Q3 magnets provide each about 35~T integrated strength, whereas the Q2 ones provide each about 50~T integrated strength. The required integrated gradients can be met with one type of quadrupole manufactured in two different length, 900~mm (for Q0, Q1 and Q3) and 1200~mm (for Q2). A few additional quadrupoles (of the order of 14 per arc) are needed for the switch-yard regions; these units are not included in the total count here.

\begin{figure}[!h]
	\centerline{\includegraphics[trim = 20mm 109.5mm 20mm 109.5mm, clip]{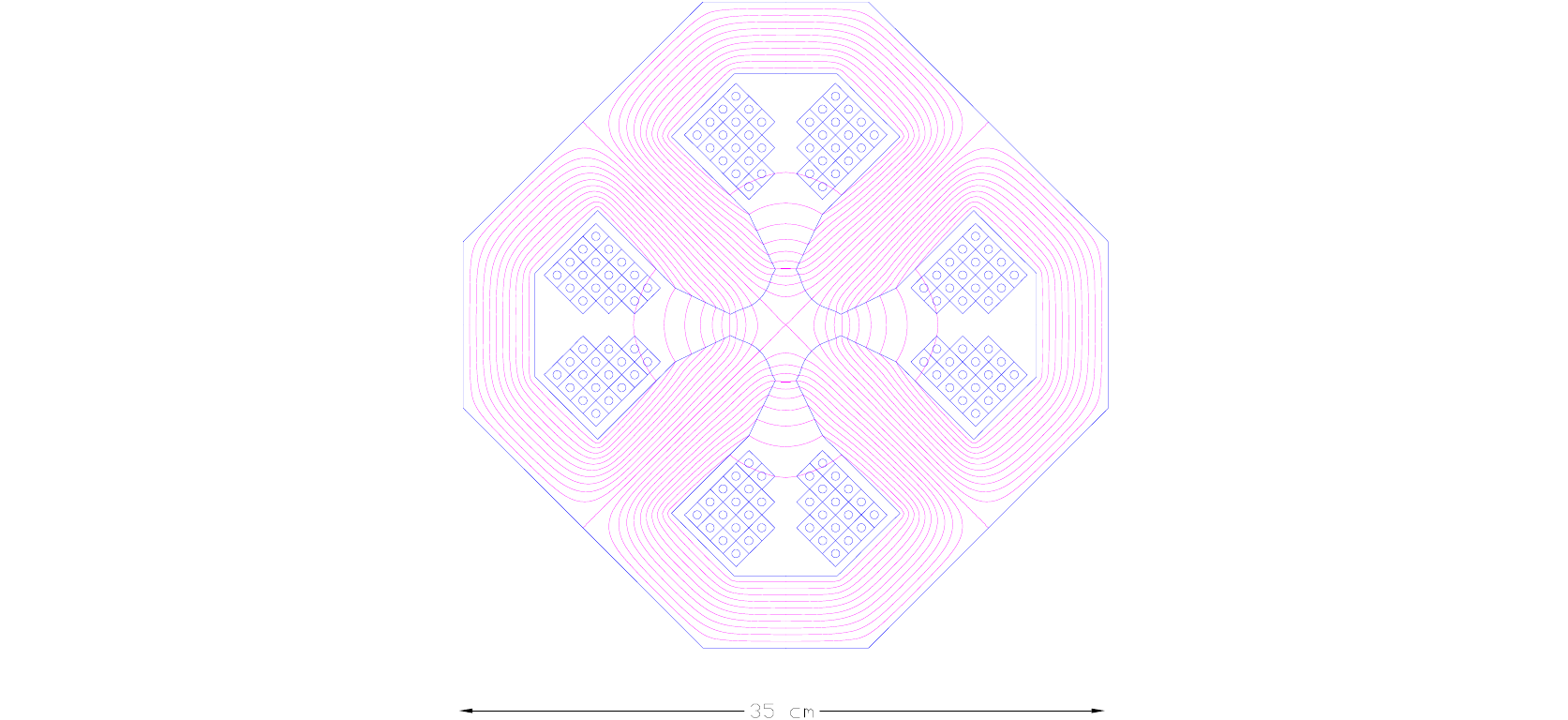}}
	\caption{Quadrupoles for the recirculators of the LR option (scale 1:5).}
	\label{fig:AQLR}
\end{figure}

As for the dipoles, also the quadrupoles in the different arcs may or may not have the same conductor, that is, it is possible to use a smaller conductor (or less turns) in the low energy arcs, or to use the same conductor everywhere and simply operating the first ones at a lower power. The relevant parameters are summarised in Table~\ref{tab:AQLR} and the cross section is illustrated in Figure~\ref{fig:AQLR}.

Also for the quadrupoles, it could be envisaged to use a hybrid configuration, with most of the excitation given by permanent magnets. The gradient strength could be varied by trim coils and/or by mechanical methods (see, for example, \cite{Attilio3}).

\subsubsection{Quadrupoles for the two 10 GeV linacs}
\begin{figure}[!h]
	\centerline{\includegraphics[trim = 20mm 89mm 20mm 90.5mm, clip]{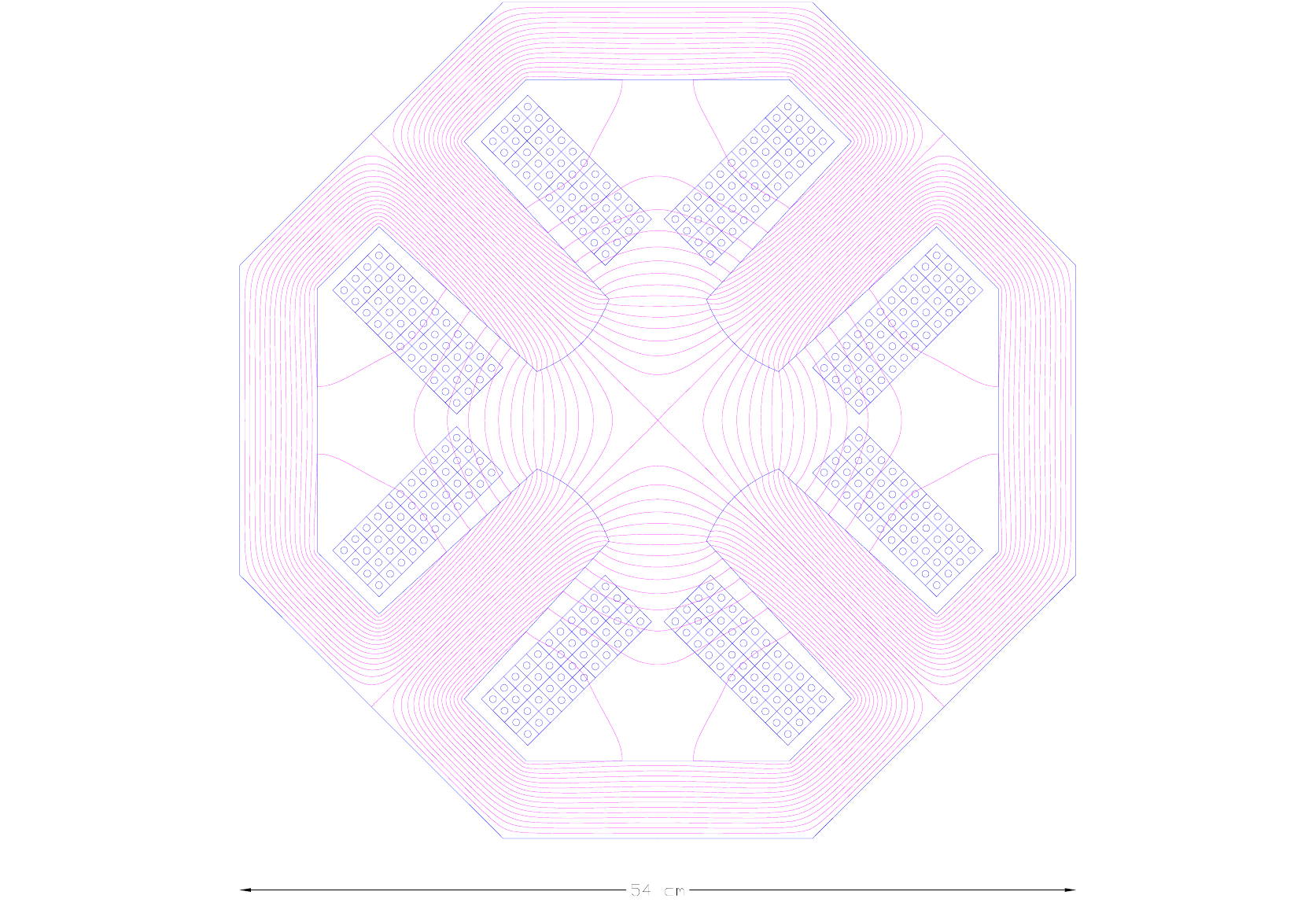}}
	\vspace{-0.5cm}
	\caption{Quadrupoles for the 10 GeV linacs of the LR option (scale 1:5).}
	\label{fig:LQLR}
\end{figure}

In the two 10~GeV linacs, 37 + 37 quadrupoles each providing 2.5 T integrated strength are required. The present design solution considers 70~mm aperture radius magnets to be compatible with any possible aperture requirement. The relevant parameters are summarised in Table~\ref{tab:LQLR} and the cross section is illustrated in Figure~\ref{fig:LQLR}.

The magnet could be more compact, but a bit longer to compensate for the lower gradient. Alternatively, one could consider superconducting magnets that could be hosted in the linac cryostats.

It could also be convenient to have in the two linacs, or at different positions along the acceleration, several families of quadrupoles with different apertures. Here a cross section for the more demanding ones is reported.


\vspace{5mm}

\begin{table}[!h]
  \centering
  \begin{tabular}{| l | c c |}
    \hline
		Beam energy                      & 10.5 to 60.5        & GeV              \\ \hline
    Magnetic field                   & 0.046 to 0.264      & T                \\ \hline
		Magnetic length                  & 4.0                 & m                \\ \hline
		Vertical aperture                & 25                  & mm               \\ \hline
		Pole width                       & 80                  & mm               \\ \hline
		Mass                             & 2000                & kg               \\ \hline
		Number of magnets                & \multicolumn{2}{c|}{$6\times584=3504$} \\ \hline
		Current @ 60.5 GeV               & 2700                & A                \\ \hline
		Number of turns per pole         & \multicolumn{2}{c|}{1}                 \\ \hline
		Current density @ 0.264~T        & 0.7                 & A/mm$^2$         \\ \hline
		Conductor material               & \multicolumn{2}{c|}{copper}            \\ \hline
		Magnet inductance                & 0.08                & mH               \\ \hline
		Magnet resistance                & 0.08                & m$\Omega$        \\ \hline
		Power @ 10.5 GeV                 & 20                  & W                \\ \hline
		Power @ 20.5 GeV                 & 65                  & W                \\ \hline
		Power @ 30.5 GeV                 & 150                 & W                \\ \hline
		Power @ 40.5 GeV                 & 260                 & W                \\ \hline
		Power @ 50.5 GeV                 & 405                 & W                \\ \hline
		Power @ 60.5 GeV                 & 585                 & W                \\ \hline
		Total power consumption six arcs & 0.87                & MW               \\ \hline
		Cooling                          & \multicolumn{2}{c|}{air}               \\ \hline
  \end{tabular}
	\caption{Main parameters of bending magnets for the LR recirculator. Resistance and powers refer to the same conductor size across the six arcs.}
	\label{tab:MBALR}
\end{table}

\clearpage

\begin{table}[!h]
  \centering
  \begin{tabular}{| l | c c |}
    \hline
		Beam energy                       & 10.5 to 60.5 & GeV                     \\ \hline
		Field gradient                    & 41           & T/m                     \\ \hline
		Magnetic length (short/long)      & 0.9 / 1.2    & m                       \\ \hline
		Aperture radius                   & 20           & mm                      \\ \hline
		Mass (short/long)                 & 750 / 980    & kg                      \\ \hline
		Number of magnets (Q0+Q1+Q2+Q3)   & \multicolumn{2}{c|}{$6\times240=1440$} \\ \hline
		Current @ 41 T/m                  & 400          & A                       \\ \hline
		Number of turns per pole          & \multicolumn{2}{c|}{17}                \\ \hline
		Current density @ 41 T/m          & 4.8          & A/mm$^2$                \\ \hline
		Conductor material                & \multicolumn{2}{c|}{copper}            \\ \hline
		Magnet inductance (short/long)    & 17 / 22      & mH                      \\ \hline
		Magnet resistance (short/long)    & 30 / 40      & m$\Omega$               \\ \hline
		Power @ 10.5 GeV (short/long)     & 0.15 / 0.20  & kW                      \\ \hline
		Power @ 20.5 GeV (short/long)     & 0.55 / 0.74  & kW                      \\ \hline
		Power @ 30.5 GeV (short/long)     & 1.22 / 1.63  & kW                      \\ \hline
		Power @ 40.5 GeV (short/long)     & 2.15 / 2.87  & kW                      \\ \hline
		Power @ 50.5 GeV (short/long)     & 3.35 / 4.46  & kW                      \\ \hline
		Power @ 60.5 GeV (short/long)     & 4.80 / 6.40  & kW                      \\ \hline
		Total power consumption six arcs  & 3.17         & MW                      \\ \hline
		Cooling                           & \multicolumn{2}{c|}{water}             \\ \hline
		\end{tabular}
		\caption{Main parameters of quadrupoles for the recirculators of the LR option. Resistance and powers refer to the same conductor size across the six arcs.}
		\label{tab:AQLR}
\end{table}

\clearpage

\begin{table}[!h]
  \centering
  \begin{tabular}{| l | c c |}
    \hline
		Field gradient           & 10       & T/m               \\ \hline
		Magnetic length          & 0.250    & m                 \\ \hline
		Aperture radius          & 70       & mm                \\ \hline
		Mass (QD/QF)             & 440      & kg                \\ \hline
		Number of magnets        & \multicolumn{2}{c|}{37 + 37} \\ \hline
		Current @ 10 T/m         & 460      & A                 \\ \hline
		Number of turns per pole & \multicolumn{2}{c|}{44}      \\ \hline
		Current density @ 10 T/m & 5.0      & A/mm$^2$          \\ \hline
		Conductor material       & \multicolumn{2}{c|}{copper}  \\ \hline
		Magnet inductance        & 24       & mH                \\ \hline
		Magnet resistance        & 25       & m$\Omega$         \\ \hline
		Power @ 10 T/m           & 5.3      & kW                \\ \hline
		Cooling                  & \multicolumn{2}{c|}{water}   \\ \hline
		\end{tabular}
		\caption{Main parameters of quadrupoles for the 10 GeV linacs of the LR option.}
		\label{tab:LQLR}
\end{table}

\clearpage

\subsection{LR option, corrector magnets for the two 10 GeV linacs}
\begin{figure}[!h]
	\centerline{\includegraphics[clip=,width=0.8\textwidth]{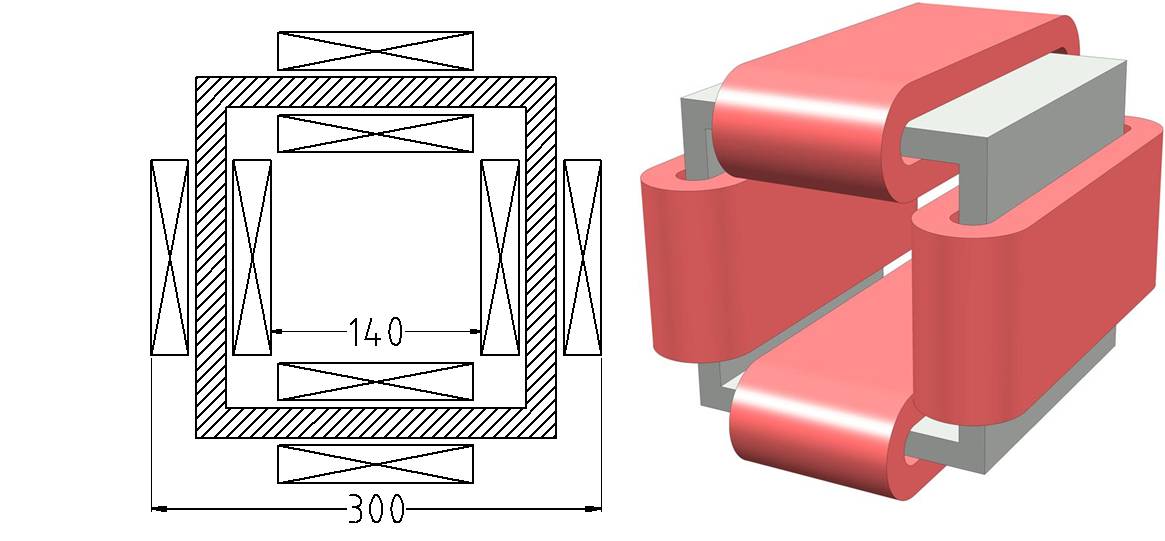}}
	\caption{Combined function corrector magnets for the LR option.}
	\label{fig:COLR}
\end{figure}

In the two 10 GeV linacs, 37 + 37 dipole (vertical / horizontal) correctors are needed. These combined function correctors shall provide an integrated field of 10~mTm in an aperture of 140~mm. The relevant parameters are summarised in Table~\ref{tab:COLR} and the cross section is illustrated in Figure~\ref{fig:COLR}.

\vspace{10mm}

\vspace{10mm}

\begin{table}[!h]
  \centering
  \begin{tabular}{| l | c c |}
    \hline
		Magnetic field                & 25             & mT               \\ \hline
		Magnetic length               & 0.400          & m                \\ \hline
		Yoke length                   & 0.250          & m                \\ \hline
		Total length                  & 0.350          & m                \\ \hline
		Free aperture                 & $140\times140$ & mm$\times$mm     \\ \hline
		Mass                          & 100            & kg               \\ \hline
		Number of magnets (QD+QF)     & \multicolumn{2}{c|}{37 + 37}      \\ \hline
		Current                       & 40             & A                \\ \hline
		Number of turns per circuit   & \multicolumn{2}{c|}{$2\times100$} \\ \hline
		Current density               & 1.5            & A/mm$^2$         \\ \hline
		Conductor material            & \multicolumn{2}{c|}{copper}       \\ \hline
		Magnet inductance per circuit & 10             & mH               \\ \hline
		Magnet resistance per circuit & 0.1            & $\Omega$         \\ \hline
		Power per circuit             & 160            & W                \\ \hline
		Cooling                       & \multicolumn{2}{c|}{air}          \\ \hline
		\end{tabular}
		\caption{Main parameters of combined function corrector magnets for the LR option.}
		\label{tab:COLR}
\end{table}

%% file: machine/rf.tex
\section{Ring-Ring RF Design}\label{RR-RF-section}

\subsection{Design parameters}
The RF system parameters for the e-ring are listed in Table $\ref{tab:rfrr}$. For a beam energy of 60~GeV the synchrotron losses are 437~MeV/turn. With a nominal beam current of 100~mA the rather significant amount of power of 47.3~MW is lost due to synchrotron radiation. For the voltages needed superconducting RF is the only choice.

\begin{table} 
\begin{center}
\begin{tabular}{|l | c | c |}
\hline 
Energy & GeV & 60\\
Beam current & mA & 100\\
Synchrotron losses & MeV/turn & 437\\
Power loss to synchrotron radiation & MW & 43.70\\
Bunch frequency (25~ns spacing) & MHz & 40.08\\
Multiplying factor &  & 18\\
RF frequency & MHz & 721.42\\
Harmonic number &  & 64152\\
RF Voltage for 50 hour quantum lifetime & MV & 510.00\\
Nominal RF voltage (MV) & MV & 560.00\\
Synchronous phase angle & degrees & 129\\
Quantum lifetime at nominal RF voltage & hrs & infinite\\
Number of cavities &  & 112\\
Number of 8-cavity cryomodules &  & 14\\
Power couplers per cavity &  & 2\\
Average RF power to beam per power coupler & kW & 195\\
Voltage per cavity at nominal voltage & MV & 5.00\\
Cells per cavity &  & 2\\
Cavity active length & m & 0.42\\
Cavity R/Q &  $\text{circuit} \ \Omega$ & 114\\
Cavity Gradient & MV/m & 11.90\\
Cavity loaded Q (Matched) &  & $2.8 \cdot 10^5$\\
Cavity forward power (nom. current, nom. voltage) & & \\
 for matched condition & kW & 390\\
Nominal cavity loaded Q & & \\
(matched for 50 $\%$ more beam) &  & $1.9 \cdot 10^5$\\
Cavity forward power & & \\
(nominal current, voltage \& loaded Q) & kW & 406\\
Forward power per coupler & kW & 203\\
Number of cavities per klystron &  & 2\\
Waveguide losses & $\%$ & 7\\
Klystron output power & kW & 870\\
Feedbacks \& detuning power margins & $\%$ & 15\\
Klystron rated power & kW & 1000\\
Total number of klystrons &  & 56\\
Total average operating klystron RF power & MW & 49\\
DC power to klystrons assuming & & \\
$65\%$ klystron efficiency & $\%$ & 75\\
Grid power for RF, assuming $95\%$ & & \\
efficiency of power converters & MW & 79\\
\hline
\end{tabular}
\end{center}
\caption{RF system parameters for the electron ring.}
\label{tab:rfrr}
\end{table}

\subsection{Cavities and klystrons}

\subsubsection{Cavity design}

The most important issue determining the RF design is not so much in
achieving high accelerating gradient but rather the need to handle
large powers through the power coupler. The choice of RF frequency is
based on relatively compact cavities which are able to handle the
relatively high beam intensities and allowing fitting of power
couplers of sufficient dimensions to handle the RF power. A frequency
in the range 600 to 800 MHz is the most appropriate. Cavities of
frequency of 704 MHz are currently being developed at CERN in the
context of the study of a Superconducting Proton Linac (SPL)
\cite{:2006qi}\cite{SPL2}\cite{Weingarten:2008zz}.  The same frequency is also used at BNL for ERL
cavities for the RHIC upgrade project \cite{BNLCav}. Both cavities are 5-cell and
can achieve gradients greater than 20~MV/m. For the present study we
take an RF frequency of 721.42 MHz, which is compatible with the
minimum 25~ns bunch spacing in the LHC. An RF voltage of 500~MV gives a quantum
lifetime of 50 hours; this is taken as the minimum operating voltage.
An RF voltage of 560~MV gives infinite quantum lifetime and a margin
of 60~MV which permits feedback system voltage excursions and provides
tolerance to temporary failure of part of the RF system without beam
loss.

5-cell cavities would require too much RF power transferred through the power coupler, therefore we use 2-cell cavities here in keeping the cell shape. Then with a total of 112 cavities, the power per cavity supplied to
the beam to compensate the synchrotron radiation losses is 390
kW. This level of power handling is only just reached for the power
couplers of the larger 400 MHz cavities of the LHC. It is therefore
proposed to use two power couplers per cavity and split the power. In
terms of voltage, only 5~MV per cavity is required to make 560~MV,
hence it is sufficient to use cavities with two cells instead of
five. The resulting cavity active length is 0.42 m and the gradient is
11.9~MV/m.  Under these conditions the matched loaded Q
is $2.8 \cdot 10^5$. Over-coupling by 50~\% to $1.9 \cdot 10^5$
provides a stability margin and incurs relatively small power
overhead. Under this condition the average forward power through the
coupler is just under 200~kW. This nevertheless remains challenging
for the design of power coupler.

\subsubsection{Cryomodule layout}
\begin{figure}[htp]
\begin{center}
\hspace*{-0.3cm}  
\centerline{\includegraphics[clip=,width=0.8\textwidth]{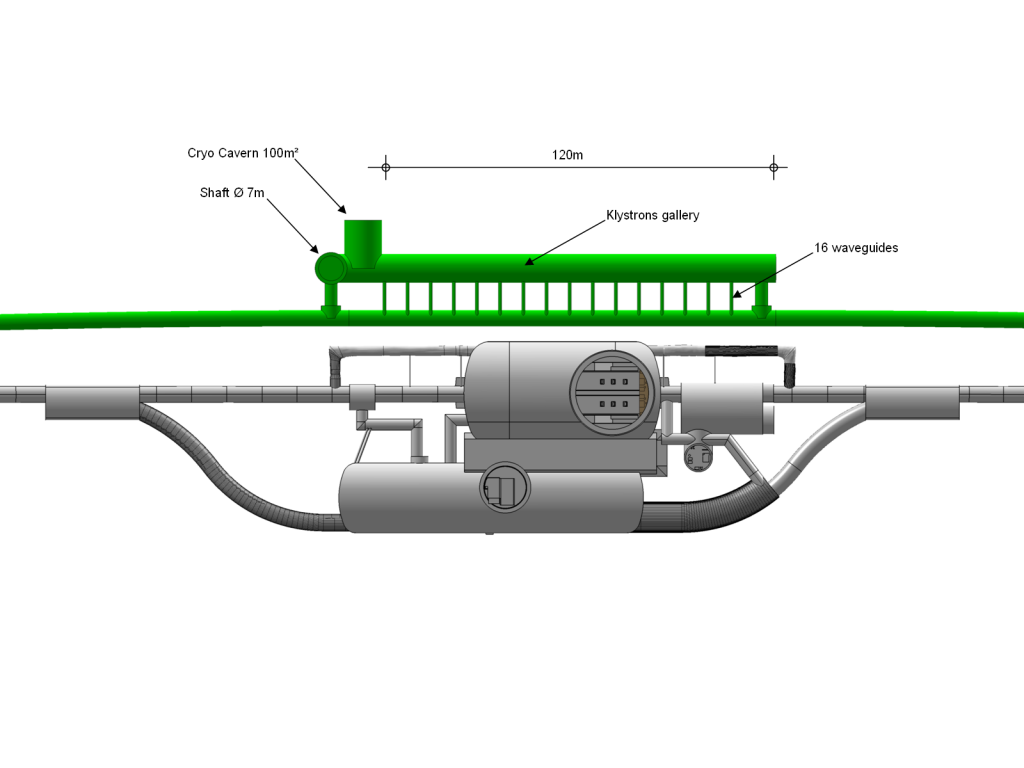}}
\centerline{\includegraphics[clip=,width=0.8\textwidth]{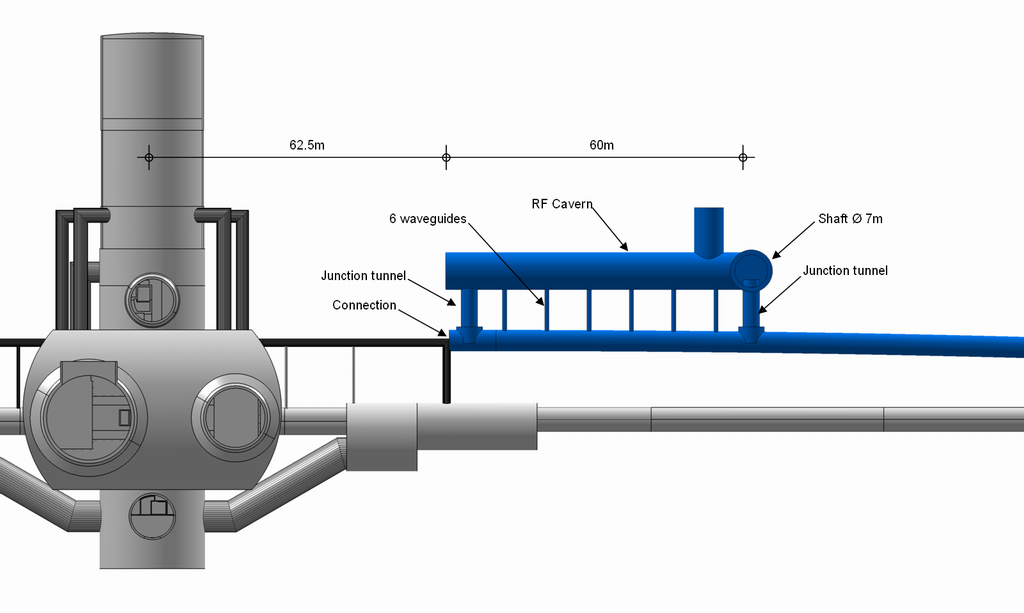}}

\end{center}
\vspace*{-0.2cm}
\caption{
RF tunnel Layouts at CMS and ATLAS bypasses.
Note only the right hand side at ATLAS shown.
}
\label{fig:rfrr1}
\end{figure}

With 8 cavities per cryomodule there are a total of 14 cryomodules. The estimated cryomodule length, 
scaled from the 8 5-cell cavity of SPL to two cells per cavity is 10 m. There are 8 double cell cavities in 14 
10m cryomodules, the total RF cryomodule length is therefore 140 m, but space must be allowed for 
quadrupoles, vacuum equipment and beam instrumentation. A total of 208 m is available in the by-passes: 
124 m at CMS and 2 x 42m at ATLAS.  Eight cryomodules can therefore be installed in the CMS bypass and 
six, three on each side, in the ATLAS by-passes. The distance between the modules can be taken as 3 m to 
allow space for the other equipment. The positioning of the RF tunnels in the CMS and ATLAS bypasses is 
shown in Figure $\ref{fig:rfrr1}$.

\subsubsection{RF power system}
\begin{figure}[htp]
\begin{center}
\hspace*{-0.3cm}  
\includegraphics[width=12 cm]{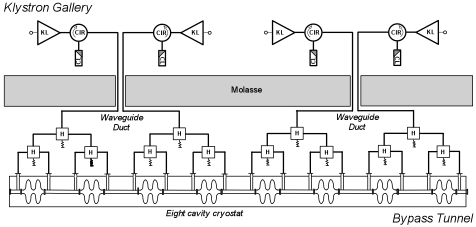} 
\end{center}
\vspace*{-0.2cm}
\caption{Layouts of RF power equipment in bypass and in RF gallery
  for one cryomodule.}
\label{fig:rfrr2}
\end{figure}

The configuration for powering the eight cavities within one cryomodule is shown
in figure $\ref{fig:rfrr2}$. Each klystron feeds two cavities with
power being split near the cavity to its two couplers. Taking two
cavities per klystron with an estimated
7~\% losses in the waveguide system gives a mean required klystron output
power of 870~kW. A
15~\% margin for the feedbacks gives a klystron rated power of 1~MW. The total
number of klystrons is 56, delivering an average total RF power of 49
MW. Taking 65~\% klystron efficiency
and
95~\% efficiency in the power converters gives roughly 79~MW grid power needed for the RF power
system.

\subsubsection{RF power system layout}
The klystrons are installed in the additional tunnels parallel to the by-passes. An estimated surface area of 
100\,m$^2$ is needed for the two klystrons, circulators, HV equipment and Low Level RF and controls racks for 
each 8 cavity module in adjacent RF gallery. This defines the tunnel width over the 13 m module interval 
(length + spacing) to be 8 m. Waveguide ducts are needed between the by-passes and the RF tunnels. 
With one waveguide per klystron into the tunnel, and two waveguides per duct, there are 16 ducts in the 
CMS tunnels, spaced roughly 6.5 m apart. At ATLAS there would be six ducts on either side with the same 
spacing. The required diameter of the duct tunnel is 90cm.

\subsubsection{Surface installations}
One HV Power Converter rated at 6~MVA is needed per 4 klystrons. These are housed in surface buildings:  
eight converters at CMS, and six at ATLAS.

\subsubsection{Conclusions}
721.4 MHz RF systems can be just fitted in the two bypasses nearest ATLAS and CMS. Detailed studies 
need to be done on the optimisation of the cavity geometry for the high beam current and ensuring 
acceptable transverse impedance. The RF power system is large. Further work is needed on integration to 
exactly define tunnel and cavity cavern layouts and quantify the space requirements.  Phased installation 
with gradual energy build-up, as was done for LEP, is an interesting possibility. The power needed for RF is 
79~MW. To this must be added power for RF controls, cryogenics and all other 
machine equipment.

\section{Linac-Ring RF design}

\subsection{Design parameters}

The ERL design \cite{ERLGen}\cite{ERLLHeC1}\cite{ERLLHeC2} is based on two 10~GeV linacs, with a
0.3~GeV injection energy and 6 linac passes to reach 60~GeV. This is shown in
Figure \ref{erllayout}.

%

The overall parameters are given in Table \ref{tablerf}. With a beam current
of 6.6 mA produced, there are currents of  nearly 20 mA in both directions in
the linacs. Significant power, greater than the injection energy, is
lost in the passages though the arcs due to synchrotron radiation as
shown in Table $\ref{tab:rflr1}$.

\begin{table} 
\begin{center}
\begin{tabular}{|c | c | c | c | c | c |}
\hline 
Arc & Arc energy & Energy loss per & Number of & Beam current & Total energy \\
    &            &  arc passage    & passages & in arc  & loss per arc\\
    &  [GeV]     &    [MeV]        &      & [mA]    & [MeV] \\\hline
6 & 60 & 751.3 & 1 & 6.6 & 751.3 \\
5 & 50 & 362.3 & 2 & 13.2 & 724.6\\
4 & 40 & 148.4 & 2 & 13.2 & 296.8\\
3 & 30 & 47.0 & 2 & 13.2 & 94.0\\
2 & 20 & 9.3 & 2 & 13.2 & 18.6\\
1 & 10 & 0.6 & 2 & 13.2 & 1.2\\ \hline
  &    &    1319.9 &   & & 1886.5 \\ \hline
\end{tabular}
\end{center}
\caption{Energy losses in the arcs on a half circle of 764 m radius}
\label{tab:rflr1}
\end{table}

The energy loss in the arcs can be compensated by independent RF
systems operating at twice the normal RF frequency.  As proposed by
\cite{LitVinA,scd:bnl} it could be envisaged to let the main linacs
replace the energy lost to synchrotron radiation, i.e. the linacs had
to supply about 0.75 GeV and 0.36 GeV, respectively, more voltage
(maximum energy loss per turn for arc 6 and 5, table \ref{tab:rflr1}).
However, this scheme significantly restricts operational freedom and
is not tested yet.  Therefore we keep it only as one possible option.
For the present report only the case for additional RF systems in the
arcs compensating synchrotron radiation losses is shown.

\subsubsection{Linac design}
High accelerating gradient is needed. First tests on cavities at similar frequency at BNL have already reached 20 MV at $Q_0$ of $2.5 \cdot 10^{10}$. Improved cavity design and careful cavity processing should allow meeting the specifications. The optimum number of cavities
and the gradient is an overall compromise taking into account cost,
cryogenics consumption and operational reliability. The RF power
system needs to compensate energy loss and non-ideal energy recovery
due to beam losses, phasing errors, transients, ponderomotive effects
and noise. It also needs to allow testing and processing of the
cavities at full gradient without circulating beam. The main RF
parameters are given in Table $\ref{tab:rflr2}$, for the two cases
described above.

\begin{table} 
\begin{center}
\begin{tabular}{| l | c | c | c |}
\hline 
Parameter & Unit & Main RF system \\\hline
Beam energy & GeV & 60.0  \\
Injection energy & GeV & 0.3  \\
Average beam current out & mA & 6.6 \\
Av. accelerated beam current in  linacs & mA & 19.8 \\
Required total voltage in both linacs & GV & 20.0 \\
Energy recovery efficiency & $\%$ & 96 \\
Total power needed to compensate & &  \\
recovery losses & MW & 15.8 \\
RF frequency & MHz & 721.42 \\
Gradient & MV/m & 20 \\
Cells per cavity & & 5 \\
Active cavity length & m & 1.04 \\
Cavity voltage & MV & 20.8 \\
Number of cavities &  & 960 \\
Energy gain per cycle & GeV & 20 \\
Power to compensate  & &\\
recovery losses per cavity & kW & 16.5 \\
Cavity R/Q & $\text{circuit} \ \Omega$ & 285 \\
Cavity unloaded Q [$\text{Q}_o$] & $10^{10}$ & 2.5 \\
Loaded Q [$\text{Q}_{ext}$] & $10^6$ & 46\\
Cavity forward power & kW & 16.5 \\
Cavity forward power - no beam & kW & 4.1 \\
Number of cavities per solid state amp.&  & 1\\
Transmission losses & $\%$ & 7 \\
Amplifier output power per cavity & kW & 17.6 \\
Feedbacks power margin & $\%$ & 15  \\
Amplifier rated power & kW & 21 \\
Total number of amplifiers &  & 960 \\
Total average amplifier output power & MW & 16.9 \\
Assumed overall conversion efficiency & & \\ 
grid to amplifier RF output & $\%$ & 70 \\
Grid power for linacs RF & MW & 24 \\
(without cryogenics power) &  &  \\
\hline 
\end{tabular}
\end{center}
\caption{Linac RF parameters.}
\label{tab:rflr2}
\end{table}

The linac RF design is based on 5-cell cavities operating at 721.42
MHz, this frequency being compatible with 25~ns bunch spacing in LHC,
as for the electron ring option. A gradient of 20~MV/m can be
taken. This is a conservative estimate based on SPL type cavities
presently being developed, with a design aim of 25~MV/m. The unloaded
Q ($\text{Q}_0$) is taken as $2.5 \cdot 10^{10}$. This is presently a challenging figure, but
recent tests on cavities at this frequency for e-RHIC have been very
encouraging. With an active cavity length of 1.04~m the voltage is
20.8~MV per cavity. This requires 960 cavities in total, or  480 cavities per linac. The cavity external Q ($\text{Q}_{ext}$) is derived from
optimum coupling to the required beam power to compensate the 4
energy losses.
It should be noted that the 300~MeV injection
linac, with nearly 2~MW beam power will also take grid power of
between 3 and 4~MW.

\subsection{Layout and RF powering}

\subsubsection{Cryomodule and RF power system layout}
With eight cavities in a cryomodule, there are 60.
cryomodules per linac with a total linac length of 990 m. This is summarised in
table $\ref{tab:rflr3}$.

\begin{table} 
\begin{center}
\begin{tabular}{|l  | c | c |}
\hline 
Parameter & Unit & Value\\
\hline
Number of cryomodules & & 60\\
Cavities per cryomodule &  &8\\
Number of cavities &  & 480\\
Module length incl. bellows, vac. pumps, &  &  \\
cold-warm transitions, BPM, $\frac{1}{2}$ quad& m & 15.5\\
Linac length & m & 990 \\
\hline
\end{tabular}
\end{center}
\caption{ERL cryomodule numbers and length.}
\label{tab:rflr3}
\end{table} 

\subsubsection{RF power system}
Assuming optimum coupling the forward power per cavity is
approximately 16.5~kW. The
available power per cavity must be somewhat higher to allow margin for
operation of RF the feedback systems; i.e. 21~kW. These levels can certainly be achieved with solid state amplifiers, avoiding the need for high voltage power supplies and
associated protection equipment. The grid to RF conversion efficiency
is also somewhat higher; 70~\% can be taken.
The total supplied average RF power is 17~MW and the grid power required for powering of the linacs is 24~MW.

\subsubsection{RF power system layout}
The RF amplifiers and RF feedback and controls racks are housed in a
separate parallel powering gallery.  There is one RF amplifier per
cavity, the power being fed by WR1150 standard waveguides, each 11.5
inches by 5.75 inches (30~cm by 15~cm). The number of holes between
the powering and linac tunnels can be limited to one per four
cavities, i.e. two per cryomodule, spaced 8 m apart giving 118 holes
per linac. The diameter is 90cm. The diameters could be reduced if
half height waveguides or coax lines are used.

\subsection{Arc RF systems}
Table \ref{tab:rflr1} shows the synchrotron radiation losses in the arcs; they are negligible in the 10 GeV arc. In the 20, 30, 40 and 50~GeV arc both the accelerated and decelerated beams pass the same arc RF system with $180^0$ phase shift at the basic frequency of 721.42~MHz; hence to accelerate both beams, the arc RF system is operated at twice the frequency, i.e. at 1442.82~MHz. The 60~GeV arc carries only the decelerated beam and there one can use the linac RF cavities at 721.42~MHz. However, since here the required power per cavity is much larger the solid state amplifiers of the main linac cannot be used but a klystron or IOT must be applied. Overall parameters for these RF systems are given in Table \ref{tab:rflr4}.

\begin{table} 
\begin{center}
\begin{tabular}{| l | c | c |}
\hline 
Parameter & Unit & Value\\\hline
Total energy loss in 20-60GeV arcs & MeV & 1885.3 \\
Power loss in 20-60GeV arcs & MW & 12.4 \\
Arc RF frequency  & MHz  & 1442/721\\
Number of cavities &  & 58/38 \\
Number of klystrons &  & 31/10\\
Total average supplied klystron RF power & MW & 10.5\\
Assumed overall conversion efficiency - grid to klystrons RF out & $\%$ & 60\\
Grid power for arc RF systems & MW & 23\\
\hline
\end{tabular}
\end{center}
\caption{Arc RF systems overall parameters.}
\label{tab:rflr4}
\end{table} 

The arc systems provide very different voltages. Parameters for the
individual systems are given in table~$\ref{tab:rflr5}$. Use of
cavities and cryostats scaled to those in the linacs is assumed;
however short cryostats containing four cavities could be used in the
20 and 40~GeV arc systems. Powering would be by klystrons, at 1442~MHz a total of
31 rated at a maximum of 360~kW with one klystron supplying two cavities and at 721~MHz 10 klystrons of 680~kW with one klystron supplying four cavities.

\begin{table} 
\begin{center}
\begin{tabular}{| p{4.5 cm}|l |c |c | c | c | c | c | c | c |}
\hline 
Parameter & Unit & Arc 2 & Arc 3 & Arc 4 & Arc 5 & Arc 6 & Totals\\\hline
Arc energy & GeV & 20 & 30 & 40 & 50 & 60 & \\
Energy lost per arc passage & MeV & 9.3 & 47.0 & 148.4 & 362.3 & 751.3 & \\
Number of passes & & 2 & 2 & 2 & 2 & 1 & \\
Total beam current in arc & mA & 13.2 & 13.2 & 13.2 & 13.2 & 6.6 &\\
Power loss in arc & MW & 0.1 & 0.6 & 2.0 & 4.8 & 5.0 & 12.4\\
RF frequency 1442 MHz & MHz & x & x & x & x &  & \\
RF frequency 721 MHz & MHz &   &   &   &   & x & \\
Max. acc. gradient & MV/m& 20.0 & 20.0 & 20.0 & 20.0 & 20.0 & \\
Max. acc. voltage & MV& 10.4 & 10.4 & 10.4 & 10.4 & 20.8  & \\
Cavities at 1442 MHz&  & 1 & 5 & 156 & 37 &  & 38 \\ 
Cavities at 721 MHz&  &   &   &   &   & 40  & 41\\
Required voltage/cavity & MV & 9.6 & 8.1  & 9.6 & 9.6 & 19.0 & \\
RF Power/cavity                             & kW & 123  & 124    & 131   & 129   & 130 & \\
Nominal RF power/cavity             & kW & 128   & 129   & 136   & 135   & 136 & \\
Klystron output power per cavity & kW  & 137  & 138    & 146   & 144   & 145 & \\
Kl. rated power/cavity                   & kW  & 160  & 160    & 170   & 170   & 170 & \\
Cavities/klystron                            &         & 2      & 2          & 2       & 2        & 4 & \\
Klystron rated power                     & kW  & 320  & 320     & 340   & 340   & 780 & \\
Klystrons at 1442 MHz                 &         & 1       & 3          & 8        & 19     &  -     & 31\\
Klystrons at 721 MHz                    &    -   &   -      &     -       &   -       &  -        & 10 &  10\\
Total average supplied klystron RF power & MW & 0.1 & 0.5 & 1.7 & 4.0 & 4.2 & 10.5\\
Assumed overall conversion efficiency grid to klystrons total RF power & $\%$  & 60 & 60 & 60 & 60 & 60  & \\
Grid power arc RF  systems & MW & 0.2 & 1.2 & 3.6 & 8.9 & 9.2 & 23\\
\hline
\end{tabular}
\end{center}
\caption{Parameters of the individual arc RF systems.}
\label{tab:rflr5}
\end{table}



\section{Crab crossing for the LHeC}
Due to the very high electron beam energies in the LHeC and the associated interaction region design, 
the emitted synchrotron radiation and the required RF power are challenging. 
The IR layout for the RR option consists of a crossing angle to mitigate 
parasitic interactions and allows for a simple scheme to 
accommodate the synchrotron radiation fan. 
A crab crossing scheme for the proton beam is highly desirable 
to recover the geometric luminosity loss due to this crossing angle. 
%
Some issues associated with the complexity of the IR design and the associated 
synchrotron radiation can be relaxed with the implementation of 
crab crossing near the IR. A crab crossing scheme would also provide a natural knob for regulating the
beam-beam parameter if required. 
Although the linac-ring 
option plans to employ separation dipoles and mirrors 
for synchrotron radiation, crab crossing can prove to be a simpler 
option if the technology is viable.

\subsection{Luminosity reduction}
In the nominal LHC with proton-proton collision, the two beams share a 
common vacuum chamber for approximately a 100m from the IP. Therefore, 
a crossing angle is required in the IRs to avoid parasitic interactions. 
Consequently, the luminosity is reduced by a geometrical reduction 
factor which can be expressed as
\begin{equation}
R = \frac{1}{\sqrt{1-\Phi^2}}
\end{equation}
where $\Phi=\sqrt{\theta\sigma_z/2\sigma_x}$ is the Piwinski parameter, 
which is proportional to ratio of the longitudinal and 
transverse beam sizes in the plane of the crossing.

Reducing $\beta^*$ at a constant beam-to-beam 
separation in the IRs ($\sim 10\sigma$), the
luminosity reduction factor can become quite significant. To compensate for this reduction from the 
crossing angle, a crab crossing scheme is proposed and R\&D is 
moving rapidly to realise the technology~\cite{CALAGA1,CALAGA2}. 

For the electron-proton collisions, the Piwinski parameter can be 
redefined as
\begin{equation}
\Phi_p = \frac{\theta_c}{2\sqrt{2}\sigma_x^*}
\sqrt{\sigma_{z,p}^2+\sigma_{z,e}^2}
\end{equation}
where $\sigma_{z,p}$ and $\sigma_{z,e}$ are the proton and electron
bunch lengths. Table~\ref{TAB:CRABXSCHEME1} lists the relevant parameters
of the crossing schemes in the LHeC as compared to some other 
machines.
\begin{table}[h]
\begin{center}
\begin{tabular}
{|l|c|c|c|c|c|c|c|} \hline
& KEK-B & \multicolumn{2}{c|}{LHC} & \multicolumn{2}{c|}{LHeC} 
& eRHIC \\ \hline
   &         &  Nominal & Upgrade & RR & LR & \\ \hline
$\theta_c$ [mrad]  &  22.0  & 0.285 & 0.4-0.6 & 1.0  & 0.0 (4.0)
& 0.0 (5.0)\\ \hline
$\sigma_z$ [cm]  &  0.7   & \multicolumn{2}{c|}{7.55} & 
\multicolumn{2}{c|}{7.55 (0.7$\dag$)} & 20/1.2$^\dag$   \\ \hline
$\sigma_x^*$ [$\mu$m]&  103  & 16.6  & 11.2  & 
30 (15.8$^*$) & - & 32 \\ \hline
$\Phi$       &  0.75  & 0.64  & 1-1.4 & 
0.9 (1.6$^*$) & 0.0  & 0.0 (11.0) \\ \hline
\end{tabular}
\caption{Relevant parameters of the crossing schemes in the LHeC
compared to LHC, KEK-B and eRHIC. Note $\dag$ corresponds to electrons
and * corresponds vertical plane.}
\label{TAB:CRABXSCHEME1}
\end{center}
\end{table}

\subsection{Crossing schemes}
\begin{figure}[h]
\centering
\includegraphics[width=5cm]{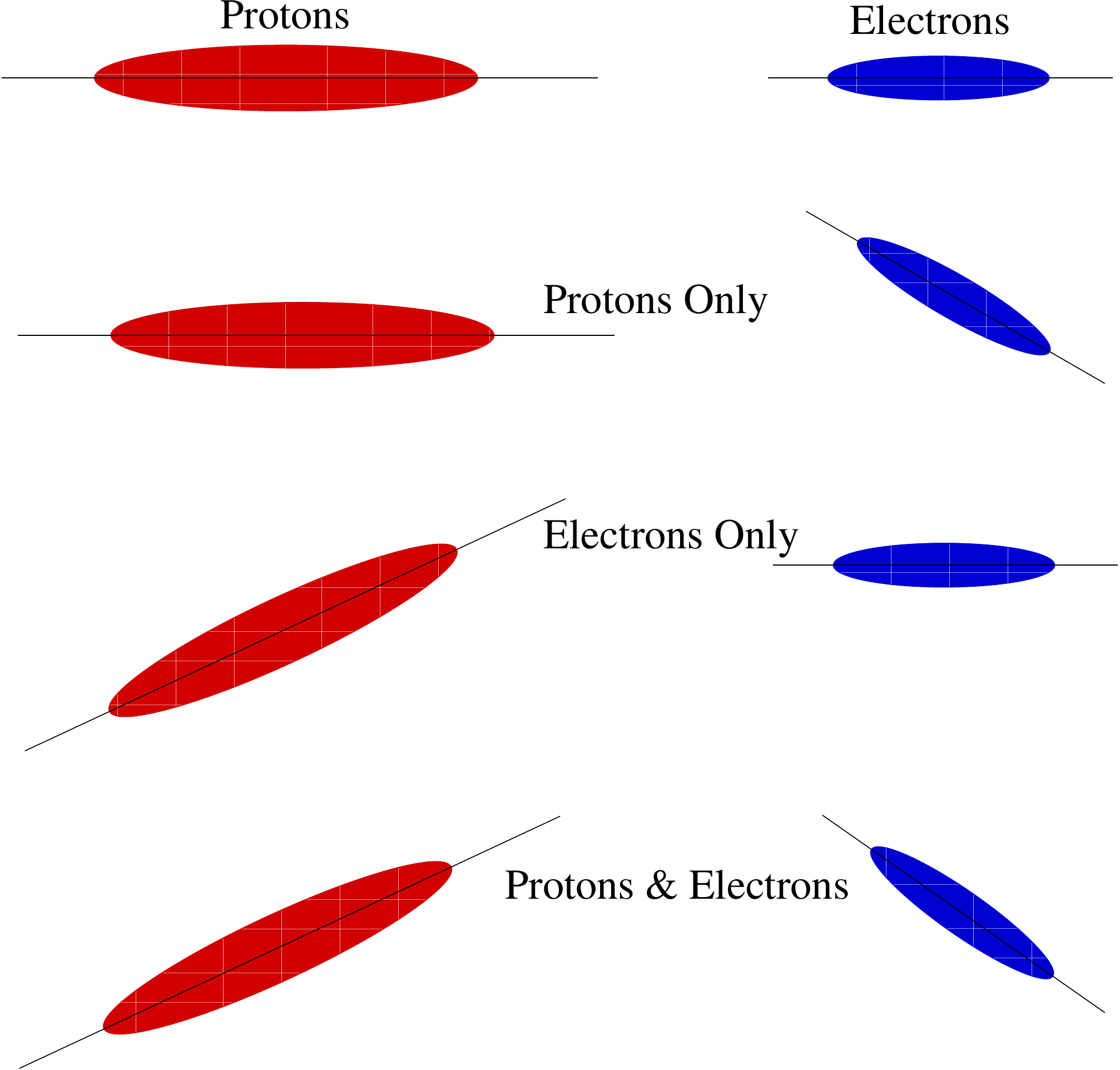}
\caption{Schematic of different crossing schemes using crab cavities 
on either proton or electron beams as compared to the head-on collision. Top: Crabbing of both beams; Second from top: crabbing of the proton beam only; Third from top: crabbing of electron beam only; Bottom: no crabbing at all.}
\label{FIG:CRABXSCHEME}
\end{figure}

Since the bunch length of the electrons are significantly smaller 
(at least factor 10) than that of the protons, the geometrical overlap 
due to crossing angle is mainly dominated by the angle of the proton 
bunches. Four different cases (see Fig.~\ref{FIG:CRABXSCHEME}) 
were simulated to determine the luminosity gain in the different 
cases with crab cavities and comparing it to the nominal case 
(see Table~\ref{TAB:CRABXSCHEME2}).

The luminosity gains strongly depend on the choice of RF frequency
as the reduction factor due to the RF curvature at frequencies of 
interest (0.4-0.8 GHz) is non-negligible.  
\begin{table}[h]
\begin{center}
\begin{tabular}{|l|c|c|}\hline
Scenario &  \multicolumn{2}{c|}{L/L$_0$} \\ \cline{2-3}
         &  400 MHz &  800 MHz \\ \hline
X-Angle (1 mrad) & \multicolumn{2}{c|}{1.0} \\ \hline
Uncross both $e^{-}$ and  $p^+$ &  1.88\%  & 1.48 \\ \hline
Uncross only $e^{-}$  &  \multicolumn{2}{c|}{1.007}  \\ \hline
Uncross only $p^+$   & 1.88  & 1.48  \\ \hline
\end{tabular}
\caption{Luminosity gains computed for different crossing schemes
with crab cavities and a crossing angle of 1 mrad.}
\label{TAB:CRABXSCHEME2}
\end{center}
\end{table}

\subsection{RF technology}
The required cavity voltage can be calculated using
\begin{equation}
V_{crab} = \frac{2cE_0 \tan{(\theta_c/2)}\sin{(\mu_x/2)}}
{{\omega_{RF}\sqrt{\beta_{crab}\beta^*}}
\cos{(\psi^x_{cc\rightarrow ip}-\mu_x/2)}}
\end{equation}
where $E_0$ is the beam energy, $\omega_{RF}$ is the RF frequency of
the cavity, $\beta_{crab}$ and $\beta^*$ are the beta-functions at the 
cavity and the IP respectively, $\psi^x_{cc\rightarrow ip}$ is 
the phase advance from the cavity to the IP 
and $\mu_x$ is the betatron tune.
The nominal scenarios for both proton-proton and electron-proton
IRs are anticipated to have local crab crossing with two cavities
per beam to create a local crab-bump within the IR. Since the 
$\beta$-functions are typically large in the location of the crab
cavities, a voltage of approximately 20 MV should suffice for crossing
angles of approximately 1-2 mrad. The exact voltage will depend on
the final interaction region optics of both the proton and the 
electron beams. 

To accommodate the crab cavities within the IR region, 
deflecting structures with a compact footprint
are required. 
Conventional pill-box type elliptical cavities at frequencies of 
400 MHz are too large to fit within the LHC interaction region 
constraints. The effort to compress the cavity footprint recently 
resulted in several TEM type deflecting mode geometries~\cite{CALAGA2}. 
Apart from being significantly smaller than its elliptical counterpart, 
the deflecting mode is the primary mode of the TEM type cavity, paving the way to a 
new class of cavities at lower frequencies (400 MHz) which is preferred 
from the RF curvature point of view.

Demonstration of a robust operation of such novel RF concepts with high deflecting gradients within the LHC constraints is the prerequisite for exploiting the crab crossing concept for the LHeC IR design.
R\&D on these novel
concepts is already underway for the LHC upgrade. The issues of
impedance, collimation and machine protection are similar to that of 
the implementation of the proton-proton IRs.

%% file: machine/thiesen.tex
\section{Ring-Ring  Power Converters}
\subsection{Overview}

The LHeC Ring-Ring Collider option at $60$~GeV with normal conducting magnets could be compared to LEP phase 1 ($60$~GeV) in particular for the main magnets (dipole magnets (MB) and quadrupole magnets (MQ)) circuits. The emergence of IGBT (new power semiconductors) in the 1990s has permitted the development of new power converter topologies and today the SCR power converters are replaced by switch mode power converters.  Here, the possible topologies of power converters and the powering strategies for the main magnet circuits (MB and MQ) are presented. The last paragraph concerns infrastructure needs for LHeC Ring-Ring Collider power converters.

\subsection{Powering considerations}
The characteristics of power converters depend mainly on the electrical parameters of magnet circuits (e.g. R, L or current) and on operating mode of the accelerator (e.g. Einj/Ecoll or time need to reach collision energy): The LHeC Ring-Ring Collider option could be compared to LEP Phase 1 and the main parameters to define the power converters are similar:

\begin{enumerate}
\item	Time constants of the magnet circuits are low ($< 1$ s).
\item	Time to reach collision energy is relatively long ($> 1$ min) with the consequence that the inductive voltages of the circuits (L.di/dt) are low ($< 10\%$ resistive voltage).
\item	Currents in the circuits are below $1$~kA and the voltages below $500$~V, except for main magnet 
(MB and MQ) circuits.
\end{enumerate}

\subsection{Power converter topologies}
Based on the assumptions mentioned in the preceding paragraph, the needs for the LHeC could be covered by three power converter families.

\begin{enumerate}
\item	1 quadrant (I $> 0$ and V $> 0$) high power ($> 0.5$~MW) switch mode power converters for the main magnet circuits. Voltages and currents needed are achieved by putting sub-converters with maximum ratings of $800$~A and $600$~V in parallel and/or in series (see Figure~\ref{Fig:thiesen_fig1}).

\begin{figure}[!h]
\centerline{\includegraphics[clip=,width=0.6\textwidth]{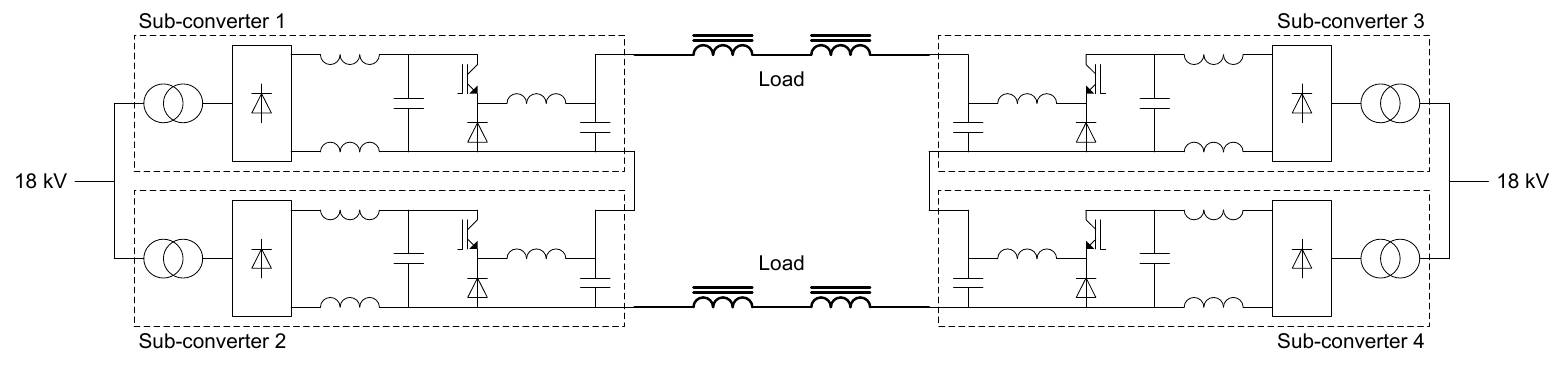}}
\caption{Possible topology for main magnet power converters To reduce harmonic currents sent to the CERN electrical network, the input
  diode rectifier could be replaced by active front-end rectifier.}\label{Fig:thiesen_fig1}
\end{figure}

\item 4 quadrant (I and V bidirectional) medium power ($< 0.5$~MW) switch mode power converters for corrector circuits and insertion quadrupole circuits (see Figure~\ref{Fig:thiesen_fig2}).

\begin{figure}[!h]
\centerline{\includegraphics[clip=,width=0.6\textwidth]{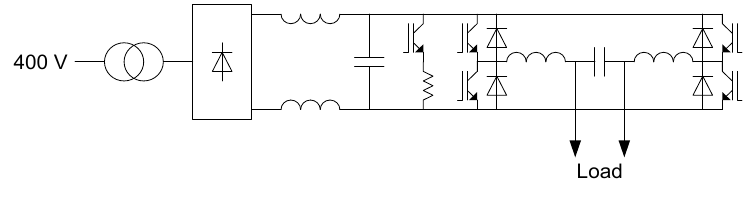}}
\caption{Possible topology for corrector power converters.}\label{Fig:thiesen_fig2}
\end{figure}
 
\item 4 quadrant low power ($< 2$~kW) switch mode power converters for COD (see Figure~\ref{Fig:thiesen_fig3}).

\begin{figure}[!h]
\centerline{\includegraphics[clip=,width=0.6\textwidth]{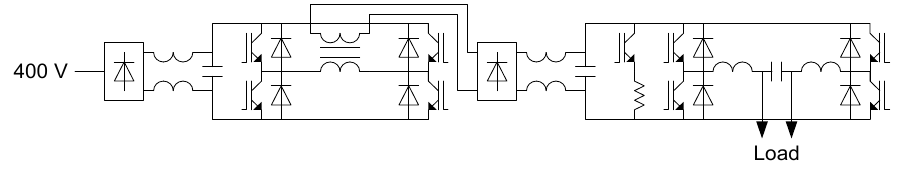}}
\caption{Possible topology for COD power converters.}\label{Fig:thiesen_fig3}
\end{figure}

\end{enumerate}
 
The advantages of switch mode power converters are mainly the following:

\begin{enumerate}

\item	Better robustness against network disturbances.
\item	No reactive power sent to the network.
\item	Small power converters.
\end{enumerate}

But the disadvantages are:
\begin{enumerate}
\item	EMI (Electro-Magnetic Interference) constraints are more significant, but experience with LHC power converters has shown that solutions exist and can be easily implemented (shielding, earth connections, etc...). 
\item	Lower MTBF (Mean Time Between Failures), but the loss of MTBF could be compensated by redundancy strategies using additional sub-converters.
\end{enumerate}

\subsection{Main power converters}
\subsubsection{Main dipole power converters}
The Ring-Ring Collider option needs $3080$ dipole magnets (MB) and the characteristics of the circuit are given in Table~\ref{tab:thiesen_tab1}.

\begin{table}[!h]
  \centering
  \begin{tabular}{|l|l|}
    \hline
Current [A] & $1300$ \\ \hline
Number of magnets & $3080$ \\ \hline
Total magnet inductance [H] & $0.400$ \\ \hline
Total magnet resistance [$\Omega$] & $0.550$ \\ \hline
Total magnet voltage [V] & $715$ \\ \hline
Total magnet consumption [MW] & $0.930$ \\ \hline
Total magnet length [m] & $16478$ \\ \hline
Total circuit length [m] & $54000$ \\ \hline
  \end{tabular}
\caption{Electrical characteristics of dipole magnet circuit.}
\label{tab:thiesen_tab1}
\end{table}

\begin{figure}[!h]
\centerline{\includegraphics[clip=,width=0.6\textwidth]{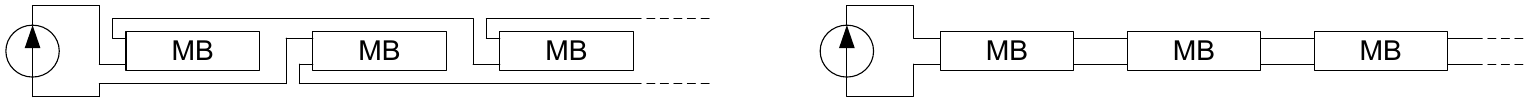}}
\caption{Different possibilities to connect the MB magnets.}\label{Fig:thiesen_fig4}
\end{figure}

If the coils of the MB magnets could be used to interconnect the
magnet (see Figure~\ref{Fig:thiesen_fig4}), $30$~km of DC cable can
be saved and the output power of the MB converter can be reduced. For
example, $54$~km of $1500 \ {\rm mm^2}$ DC cable (reasonable cable size
for $1300$~A) is about $0.6 \ \Omega$ and would need the same power and
voltage as the magnets.
 
\begin{figure}[!h]
\centerline{\includegraphics[clip=,width=0.3\textwidth]{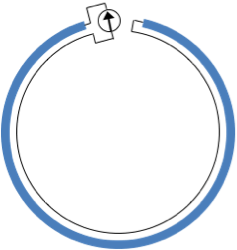}
\includegraphics[clip=,width=0.3\textwidth]{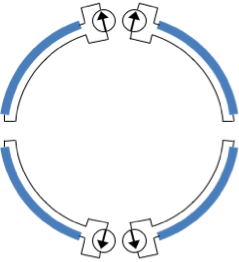}
}
\caption{Different possibilities to power the MB magnets.}\label{Fig:thiesen_fig5}
\end{figure}

Different strategies are possible to power the MB magnets: 1 or several independent circuits, as illustrated
in Figure~\ref{Fig:thiesen_fig5}.
                              
In the case of a single main dipole circuit, to avoid a dipole moment, it is not possible to close the circuit directly by doing a single loop. The circuit must be closed by return path close to the magnets path. 4 independent circuits solution seems to be the optimal solution:
\begin{enumerate}
\item	The total power is the same as that for the 1 circuit solution
\item	The voltage constraints for magnets are lower
\item	This solution allows different currents between sectors to compensate the SR energy losses.
\item	The LHC has shown that the current tracking between the different MB circuits is not an 
issue.
\end{enumerate}

To allow $e^-$ and $e^+$ physics, mechanical or semiconductor polarity
switches will be needed at the output of the main dipole power
converters (also for the MQ power converters).

\subsubsection{Main quadrupole power converters}
The Ring-Ring Collider option needs $2\times 336$ magnets for the MQD and MQF circuits and the characteristics of these circuits are given in Table~\ref{tab:thiesen_tab2}.

\begin{table}[!h]
  \centering
  \begin{tabular}{|l|l|}
    \hline
Current [A] (QF/QD) & $380/310$ \\ \hline
Number of magnets (QF/QD) & $336/336$ \\ \hline
Total magnet inductance [H] (QF/QD) & $1.344/1.344$ \\ \hline
Total magnet resistance [$\Omega$] (QF/QD) & $5.376/5.376$ \\ \hline
Total magnet voltage [V] (QF/QD) & $2050/1667$ \\ \hline
Total magnet consumption [MW] (QF/QD) & $0.779/0.517$ \\ \hline
Total magnet length [m] (QF/QD) & $336/336$ \\ \hline
Total circuit length [m] (QF/QD) & $27000/27000$ \\ \hline
  \end{tabular}
\caption{Electrical characteristics of MQ circuits.}
\label{tab:thiesen_tab2}
\end{table}

\begin{figure}[!h]
\centerline{
\includegraphics[clip=,width=0.3\textwidth]{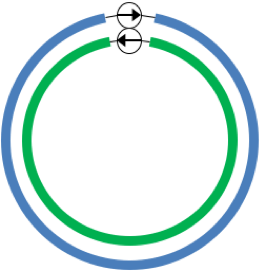}
\includegraphics[clip=,width=0.3\textwidth]{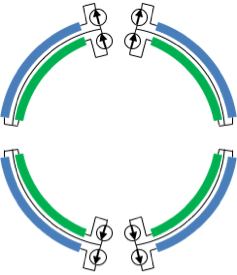}
}
\caption{Different possibilities to power the MQ magnets.}\label{Fig:thiesen_fig6}
\end{figure}

The length of the MQ circuits is mainly dominated by the DC cable length and in this case it is important to optimise the MQ circuits to reduce power and voltage requested to supply the two MQ circuits (magnets and DC cables). The actual MQ magnet design optimises the DC cable part of the circuits with low current, but not the magnet part with high resistance magnets. High current in the MQ circuits is disadvantageous for the magnet part but not for the DC cable part of the circuits. An optimum must be sought with a current between $0.5$~kA and $1.5$~kA to reduce power and voltage needed to supply the circuits and also to reduce the global cost, material and electricity. Two options are possible for supplying the MQ magnets, shown in Figure~\ref{Fig:thiesen_fig6}. Two independent circuits or several circuits with trim power converters. The advantages and disadvantages of each option must be studied in detail before taking a final decision, but in both cases the total power and cost of the powering system will be similar.

\subsection{Insertion and bypass quadrupole power converters}
The Ring-Ring option requires 148~QF magnets and 148~QD magnets in insertion and bypass regions. To obtain flexibility for the beam setting, these magnets could be powered individually. In this case the main characteristics of these circuits are given in Table~\ref{tab:thiesen_tab3}.

\begin{table}[!h]
  \centering
  \begin{tabular}{|l|l|}
    \hline
Current [A] & $420$ \\ \hline
Number of magnets per circuit & $1$ \\ \hline
Number of circuits (QF/QD)& $148/148$ \\ \hline
Magnet inductance (QF/QD) [H] & $0.015 / 0.01$ \\ \hline
Magnet resistance (QF/QD) [$\Omega$] & $0.030 / 0.023$ \\ \hline
Magnet voltage [V] (QF/QD) & $12.6 / 9.66$ \\ \hline
PC output voltage [V] & $30$ \\ \hline
PC power [kW] & $15$ \\ \hline
  \end{tabular}
\caption{Electrical characteristics of IPQ circuits.}
\label{tab:thiesen_tab3}
\end{table}

To allow $e^-$ and $e^+$ physics, the insertion and bypass quadrupole power converters must be 4 quadrants (second family of converter) to reverse the magnet currents when the physic type is changed. The use of polarity switches to reverse the magnet currents would be too complex and too expensive for the $296$ IPQ (Individually Powered Quadrupole) circuits.

\subsection{Power converter infrastructure}
The magnets being resistive, there are no real advantages to install the power converters in the underground facilities. In this case, it is better to install them at the surface. This solution simplifies power converter operation and avoids possible issues with radiation. LEP infrastructure (buildings, shafts and AC network, etc...) can be reused for LHeC. However, this solution must be confirmed by a detailed integration study. If new infrastructure is needed for the power converters, it should be installed on the current CERN sites.

\section{Linac-Ring power converters}
\subsection{Overview}
The second option for the LHeC is a Linac-Ring accelerator with two $10$~GeV Linacs and 6 recirculation arcs allowing several passes of the beam in the two linacs to reach the final beam energy of $60$~GeV. As for the Ring-Ring option, the needs for the Linac-Ring option could be covered by three IGBT power converter families: 1 quadrant high power converters, 4 quadrant medium power converters and 4 quadrants low power converters. Here, the different power converters of the linacs and recirculation arc main magnets are described. The last paragraph concerns infrastructure needs for Linac-Ring LHeC power converters.

\subsection{Powering considerations}

The power converter study for the Linac-Ring option is based on the assumption that the power converters are operated in DC. In this case the inductive voltage needed to ramp the current in the circuit can be ignored to define the characteristics of power converters.  As for the Ring-Ring option, the power converters for the Linac-Ring option will be based on three IGBT power converter families:

\begin{enumerate}
\item	Family 1: 1 quadrant high power switch mode power converters for the main dipole and quadrupole magnets of recirculation arcs. To reverse the current in the circuit for $e^-$ or $e^+$ physics, mechanical or semiconductor polarity switches will be installed at the output of the power converters.
\item	Family 2: 4 quadrant medium power switch mode power converters for corrector circuits and individually powered dipole (IPD) and quadrupole (IPQ) circuits.
\item	Family 3: 4 quadrant low power switch mode power converters mainly for orbit corrector circuits.
\end{enumerate}

\subsection{Linac quadrupole and corrector power converters}
Each linac is about $1.3$ km long and contains $37$ quadrupoles and $37$ associated correctors.
	
\subsubsection{Linac quadrupole power converters}
For the design of linac main quadrupole power converters (Family 2), the assumption is that the magnet currents are similar (less than $10\%$ of difference). In this case, two solutions are possible to power the magnets:

\begin{enumerate}
\item	Power each quadrupole magnet independently.
\item	Power the quadrupole magnets in clusters of 4 magnets with TRIM power converters to allow different currents in the magnets.
\end{enumerate}

\begin{figure}[!h]
\centerline{\includegraphics[clip=,width=0.6\textwidth]{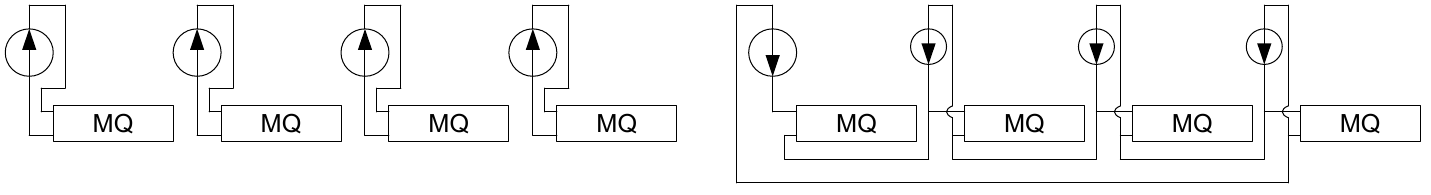}}
\caption{Different possibilities to power the linac quadrupoles magnets.}\label{Fig:thiesen_fig7}
\end{figure}

The two powering options are shown in Figure~\ref{Fig:thiesen_fig7}.

Tables~\ref{tab:thiesen_tab4} and \ref{tab:thiesen_tab5} give the
main characteristics of the linac quadrupole circuits and power
converters for the both solutions.

\begin{table}[!h]
  \centering
  \begin{tabular}{|l|l|}
    \hline
Circuit current [A] & $460$ \\ \hline
Number of magnets per circuit & $1$ \\ \hline
Number of circuits & $37 + 37$ \\ \hline
Magnet inductance [H] & $0.024$ \\ \hline
Magnet resistance [$\Omega$] & $0.025$ \\ \hline
DC cable section [${\rm mm^2}$] & $500$ \\ \hline
Max. DC cable length [m] & $1200$ \\ \hline
Max. DC cable resistance [$\Omega$] & $0.045$ \\ \hline
PC output voltage [V] & $35$ \\ \hline
PC power [kW] & $18$ \\ \hline
  \end{tabular}
\caption{Electrical characteristics of circuits for IPQ option.}
\label{tab:thiesen_tab4}
\end{table}

\begin{table}[!h]
  \centering
  \begin{tabular}{|l|l|}
    \hline
Circuit Current [A] & $460$ \\ \hline
Max. Nb. of magnets per circuit & $4$ \\ \hline
Number of circuits & $10 + 10$ \\ \hline
Magnet inductance [H] & $0.024$ \\ \hline
Magnet resistance [$\Omega$] & $0.025$ \\ \hline
Main DC cable section [${\rm mm^2}$] & $500$ \\ \hline
Trim DC cables section [${\rm mm^2}$] & $50$ \\ \hline
Max. DC cable length [m] & $1200$ \\ \hline
Max. main DC cable resistance [$\Omega$] & $0.045$ \\ \hline
Max. TRIM DC cable resistance [$\Omega$] & $0.45$ \\ \hline
Main PC output voltage [V] & $75$ \\ \hline
Main PC output current [A] & $500$ \\ \hline
Main PC output power [kW] & $38$ \\ \hline
Trim PC output voltage [V] & $40$ \\ \hline
Trim PC output current [A] & $50$ \\ \hline
Trim PC output power [kW] & $2$ \\ \hline
  \end{tabular}
\caption{Electrical characteristics of circuit for cluster option.}
\label{tab:thiesen_tab5}
\end{table}

The second solution, with clusters of four magnets, saves a factor of two in the cost of power converters and DC cables without a significant increase of the circuit complexity. In addition, the TRIM power converters can be similar to those used for linac orbit corrector circuits.

\subsubsection{Linac corrector power converters}
Each orbit corrector magnet of the linacs will be powered
individually. The characteristics of the circuits and power converters
(family 3) are given in Table~\ref{tab:thiesen_tab6}.

\begin{table}[!h]
  \centering
  \begin{tabular}{|l|l|}
    \hline
Current [A] & $40$ \\ \hline
Number of magnets per circuit & $1$ \\ \hline
Number of circuits & $37 + 37$ \\ \hline
Magnet inductance [H] & $0.010$ \\ \hline
Magnet resistance [$\Omega$] & $0.1$ \\ \hline
DC cable section [${\rm mm^2}$] & $50$ \\ \hline
Max. DC cable length [m] & $1200$ \\ \hline
Max. DC cable resistance [$\Omega$] & $0.45$ \\ \hline
PC output voltage [V] & $40$ \\ \hline
PC output current & $50$ \\ \hline
PC power [kW] & $2$ \\ \hline
  \end{tabular}
\caption{
Electrical characteristics of linac COD.
}
\label{tab:thiesen_tab6}
\end{table}

\subsection{Recirculation main power converters}
6 recirculation arcs connect the two linacs together and allow several passes of the beam in the linacs to reach the final energy of $60$~GeV. Each recirculation arc has one main dipole circuit (MB) and four main quadrupole circuits (MQ0, MQ1, MQ2 and MQ3).

\subsubsection{Main dipole power converters}
All the main dipole magnets of the same recirculation arc are powered in series. The main characteristics of the 6 main dipole power converters are described in Table~\ref{tab:thiesen_tab7}.

\begin{table}[!h]
  \centering
  \begin{tabular}{|l|l|}
    \hline
Number of MB circuits & $6$ \\ \hline
Number of magnets per MB circuit & $584$ \\ \hline
Total magnet inductance per MB circuit [H] & $0.047$ \\ \hline
Total magnet resistance per MB circuit [$\Omega$] & $0.047$ \\ \hline
DC cable section [${\rm mm^2}$] & $1000$ \\ \hline
DC cable length [m] & $1600$ \\ \hline
DC cable resistance [$\Omega$] & $0.030$  \\ \hline
PC output current $@ 10.5$ GeV [A] & $468$ \\ \hline
PC output voltage $@ 10.5$ GeV [V] & $36$ \\ \hline
PC output current $@ 20.5$ GeV [A] & $915$ \\ \hline
PC output voltage $@ 20.5$ GeV [V] & $70$ \\ \hline
PC output current $@ 30.5$ GeV [A] & $1361$ \\ \hline
PC output voltage $@ 30.5$ GeV [V] & $105$ \\ \hline
PC output current $@ 40.5$ GeV [A] & $1807$ \\ \hline
PC output voltage $@ 40.5$ GeV [V] & $139$ \\ \hline
PC output current $@ 50.5$ GeV [A] & $2254$ \\ \hline
PC output voltage $@ 50.5$ GeV [V] & $174$ \\ \hline
PC output current $@ 60.5$ GeV [A] & $2700$ \\ \hline
PC output voltage $@ 60.5$ GeV [V] & $208$ \\ \hline
  \end{tabular}
\caption{Electrical characteristics of recirculation arc MB circuits.}
\label{tab:thiesen_tab7}
\end{table}

To reduce the number of different types of power converter and simplify the LHeC operation, a modular approach will be chosen with
two types of sub converters: [470~A/120~V] for the first three power converters and [920~A/220~V] for the last three converters. Desired PC output current is achieved by putting sub converters in parallel.

\subsubsection{Main quadrupole power converters}
Each recirculation arc has four MQ circuits with $60$ magnets
connected in series for each circuit, as shown in Table~\ref{tab:thiesen_tab8}.

\begin{table}[!h]
  \centering
  \begin{tabular}{|l|l|}
    \hline
Number of MQ circuits & $6\times 4$ \\ \hline
Number of magnets per MQ circuit & $60$ \\ \hline
Total magnet inductance per MQ circuit [H] & $1.02/1.32$ \\ \hline
Total magnet resistance per MQ circuit [$\Omega$] & $1.8/2.4$ \\ \hline
DC cable section [${\rm mm^2}$] & $500$ \\ \hline
DC cable length [m] & $6000$ \\ \hline
DC cable resistance [$\Omega$] & $0.2$ \\ \hline
PC output current $@ 10.5$ GeV [A] & $69$ \\ \hline
PC output voltage $@ 10.5$ GeV [V] & $138/180$ \\ \hline
PC output current $@ 20.5$ GeV [A] & $135$ \\ \hline
PC output voltage $@ 20.5$ GeV [V] & $270/351$ \\ \hline
PC output current $@ 30.5$ GeV [A] & $202$ \\ \hline
PC output voltage $@ 30.5$ GeV [V] & $404/525$ \\ \hline
PC output current $@ 40.5$ GeV [A] & $268$ \\ \hline
PC output voltage $@ 40.5$ GeV [V] & $536/670$ \\ \hline
PC output current $@ 50.5$ GeV [A] & $334$ \\ \hline
PC output voltage $@ 50.5$ GeV [V] & $668/869$ \\ \hline
PC output current $@ 60.5$ GeV [A] & $400$ \\ \hline
PC output voltage $@ 60.5$ GeV [V] & $800/1040$ \\ \hline
  \end{tabular}
\caption{Electrical characteristics of recirculation arc MQ circuits.}
\label{tab:thiesen_tab8}
\end{table}

As for the MB circuits, the MQ power converters will be composed of sub converters connected in 
series to achieve the desired output voltage.
For the first three recirculation arcs ($10.5$, $20.5$ and $30.5$ GeV), the MQ power converters will be 
composed of [$210$ A/$200$ V] sub converters. For the other three recirculation arcs, the sub converter 
ratings will be [$420$ A/$750$ V].

\subsection{Power converter infrastructure}
Four (or possibly only two) shafts are planned in the LHeC Linac-Ring option: 
Two at each end of the ``TI2'' linac
(points 3 and 4) and two at each third of "outside" linac (point 1 and 2),
or one for each linac in the middle as sketched in figure~\ref{fig:liview} below.
 

For the power converter installation, a solution with 4 surface
buildings is proposed:
\begin{itemize}
\item	Two small buildings in points 1 and 2 for the ``outside'' linac power converters.
\item	Two large buildings in points 3 and 4 for the ``TI2'' linac power converters and the 
recirculation arcs.
\end{itemize}

Concerning the two small buildings, the area required for the power
converter installation is estimated at $400$ ${\rm m^2}$ per
building. The global AC consumption of the power converters is
estimated at $0.5$ MVA per building. Each building must be equipped
with a $100$ kW air-conditioning system to extract the power converter
losses.  Concerning the two large buildings, the area required for
power converter installation is estimated at $800$ ${\rm m^2}$ per
building. In point 4 of LHeC (point 2 of LHC), a large part of SR2 is
available for LHeC power converters. Per building, the electric power
requirements are estimated at $1$ MVA and cooling requirements at
$200$ kW.

\subsection{Conclusions on power converters}
From the power converter point of view, the two options of LHeC are
similar. The power converter topologies will be based on diode input
rectifiers with IGBT legs.  The converters can be classified into
three main families:
\begin{itemize}
\item	Family 1: 1 quadrant (I $> 0$ and V $> 0$) high power switch mode power converters for the 
main dipole and quadrupole circuits.
\item	Family 2: 4 quadrant (I and V $> 0$ and $< 0$) medium power switch mode power converters 
for the correctors circuits and individual power dipole and quadrupole magnets.
\item	Family 3: 4 quadrant and low power switch mode power converters mainly for the orbit 
corrector magnets.
\end{itemize}

When the option has been chosen for the LHeC (Ring-Ring or Linac-Ring)
the next studies should focus on the circuit definition and
optimisation.

%% file: machine/jimenez.tex
\subsection{Vacuum requirements}

In particle accelerators, beams are travelling under vacuum to reduce
beam-gas interactions i.e. the scattering of beam particles on the
molecules of the residual gas. The beam-gas interaction is dominated
by the bremsstrahlung on the nuclei of gas molecules and therefore depends
on the partial pressure, the weight and the radiation length
[g/cm2] of the gas species. In presence of a photon-stimulated desorption, the residual
gas is dominated by hydrogen ($75\%$) followed by CO/CO$_2$ ($24\%$)
and $1\%$ CH$_4$.
Argon normally represents less than $1\%$ of the residual gas if
welding best practice for UHV applications is applied. It is to be noted
that Argon is $67$ times more harmful than hydrogen (H$_2$); CO$_2$,
CO and N$_2$ are about $30$ times worst then hydrogen and Methane is $10$ times worst then hydrogen.

The beam-gas interactions are responsible for machine performance
limitations such as reduction of beam lifetime (nuclear scattering),
machine luminosity (multiple coulomb scattering), intensity limitation
by pressure instabilities (ionisation) and for positive beams only,
electron (ionisation) induced instabilities (beam blow up). The heat
load induced by scatted protons and ions can also be an issue for the
cryomagnets since local heat loads can lead to a magnet quench i.e. a
transition from the superconducting to the normal state. The heavy
gases are the most dangerous because of their higher ionisation
cross sections. In the case of the LHeC, this limitation exists only
in the experimental areas where the two beams travel in the same
beam pipe. The beam-gas interactions can also increase the background
to the detectors in the experimental areas (non-captured particles or
nuclear cascade generated by the lost particles upstream the
detectors) and the radiation dose rates in the accelerator
tunnels. Thus, leading to material activation, dose rates to
intervention crews, premature degradation of tunnel infrastructures
like cables and electronics and finally higher probability of
electronic single events induced by neutrons which can destroy the
electronics in the tunnel but also in the service galleries.

The design of the vacuum system is also driven by severe additional
constraints which have to be considered at the design stage since
retrofitting mitigation solutions is often impossible or very
expensive. Among them, the vacuum system has to be designed to
minimise beam impedance and higher order modes (HOM) generation while
optimising beam aperture in particular in the magnets. It has to
provide also enough ports for the pumps and vacuum diagnostics. 
For
accelerators with cryogenic magnets, the beam pipe has to be designed
to intercept heat loads induced by synchrotron radiation, energy loss
by nuclear scattering, image currents, energy dissipated during the
development of electron clouds, the later building up only in presence
of positively charged beams. 

The integration of all these
constraints often lead to a compromise in performances and in the case
of the LHeC, the compromise will differ between the Linac-Ring and the
Ring-Ring options.

\subsection{Synchrotron radiation}

The presence of a strong synchrotron radiation has two major
implications for the vacuum system: it has to be designed to operate
under the strong photon-induced stimulated desorption while being
compatible with the significant heat loads onto the beam pipes. In the
common beam pipe, the photo-electrons generated by the synchrotron
radiation will dramatically enhance the electron cloud build-up and
mitigation solutions shall be included at the design
stage. Furthermore, experience with LEP has shown that the Compton
scattering of the beam on photons coming from Blackbody radiation can
have a significant effect on the beam lifetime \cite{Brandt:2000xk}
\cite{Dehning:1990tb}.  In the following analysis, we have neglected
this effect, assuming that a technical solution can be found for
keeping the beam vacuum chamber at sufficiently low
temperatures. While this does not impose a principle problem to the
vacuum system design, it still requires a detailed technical study for
identifying a suitable solution for cooling the vacuum system in the
presence of ca. 3 kW/m synchrotron radiation power.

\subsubsection{Synchrotron radiation power}

The synchrotron radiation power is an issue for the heat load
deposited on the beam pipes and for its evacuation and will be the
driving factor for the mechanical engineering of the
beam pipes. Indeed, the heated surfaces will have a higher out-gassing
rates, the increase being exponentially dependent with the surface
temperature (factor $10$ for a $\Delta T=50$\textdegree C increase).  The
  synchrotron radiation power can be calculated with equation
  $\ref{jimenez:eqn1}$. Since scaling linearly with the beam intensity, $I$, with the
  power of $4$ for energy, E, and inversely to power of $2$ of the bending
  radius, the synchrotron radiation power in the Ring-Ring option is
  expected to be $45$ times higher than LEP and locally at the
  by-passes, the power can be about 180 times higher. To be compared
  with the factor $10$ expected in the bending and injection sections of
  the Linac-Ring option.

\begin{equation}
P[W/m] = 1.24 \times 10^3 \frac{E^4I}{\rho^2}
\label{jimenez:eqn1}
\end{equation}

\subsubsection{Photon-induced desorption}

The desorption rate depends on critical energy of the synchrotron
light, $\epsilon_c$, the energy which divides in two the emitted power.
For most materials, the desorption rates vary quasi linearly with the
critical energy (equation $\ref{jimenez:eqn2}$).

\begin{equation}
\epsilon_c(eV) = \frac{3 \cdot 10^{-7}}{R}\left(\frac{E_B}{E0}\right)^3
\label{jimenez:eqn2}
\end{equation}

$E_0 = 5.10^{-4}$ GeV for electrons, $E_B$ is the energy of the beam
and $R$ the bending radius.

\begin{figure}[htp]
\begin{center}
\hspace*{-0.3cm}  
\includegraphics[width=6.999cm,height=6.308cm]{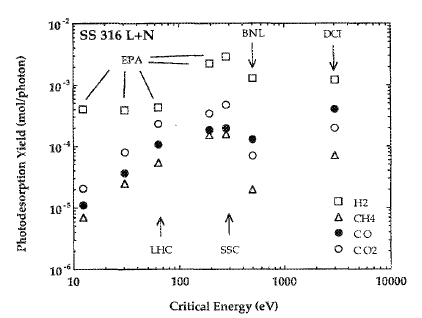} 
\includegraphics[width=6.999cm,height=6.198cm]{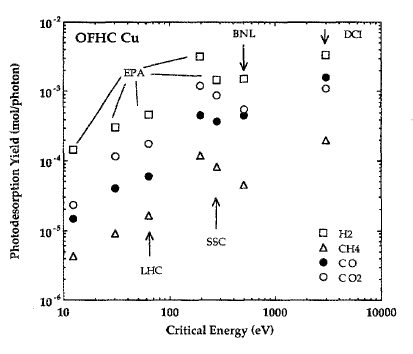} 
\end{center}
\vspace*{-0.2cm}
\caption{Photodesorption yields measured on copper and stainless steel
surfaces. To be noted
that the desorption yields of methane,
$\eta_{CH_4}$, is $50$ times lower than $\eta_{H_2}$.}
\label{jimenez:Fig1}
\end{figure}

For the LHeC, the beam energies will be equivalent to the LEP at
start. Then, a similar value of the critical energy can be assumed
allowing the comparison with LEP pressure observations. Figure $\ref{jimenez:Fig1}$ shows
typical photo-desorption yields measured on copper and stainless steel
samples. But the beam intensities being by far larger, the linear
photon flux which scales linearly (equation 3) with energy and intensity and inversely 
with bending radius will increase significantly.

\begin{equation}
\Gamma[photons / s / m] = 7 \times 10^{19} \frac{EI}{\rho}
\label{jimenez:eqn3}
\end{equation}

For the Ring-Ring option (bending sections and by-passes), the linear photon flux is expected to
be $45$ times larger than in LEP, to be compared to the factor 5 expected
for the Linac-Ring option.

The photon stimulated pressure rise, ${\Delta}$P,
depends linearly on the critical energy, on the beam energy and beam
intensity as shown by equation $\ref{jimenez:eqn4}$. The temperature affecting the
dependence of the desorption yield (equation $\ref{jimenez:eqn5}$ and $\ref{jimenez:eqn6}$),
$\eta$, to the critical energy,
$\epsilon_c$ the pressure rises will differ between surfaces at ambient temperature
(equation $\ref{jimenez:eqn5}$) and at cryogenic temperature (equation $\ref{jimenez:eqn6}$).

\begin{gather}
\Delta P \propto \eta (\epsilon_c)EI \label{jimenez:eqn4} \\
{\rm at \; room \; temperature: \;\;\;}
\eta \propto \epsilon_c \;\;\rm{and} \;\;\epsilon_c \propto E^3 \;\;\rm{such \; that} \;\;\Delta P \propto E^4I \label{jimenez:eqn5} \\
{\rm at \; cryogenic \; temperature: \;\;\;}
\eta \propto \epsilon_c^{2/3}\;\; \rm{and}\;\; \epsilon_c \propto E^3 \;\;\rm{such \; that} \;\; \Delta P \propto E^3I \label{jimenez:eqn6}
\end{gather}

Therefore, the photon stimulated pressure rise is expected to be $45$ times higher than LEP for
the Ring-Ring option, to be compared with the factor $30$ for the
Linac-Ring option.

\subsubsection{Vacuum cleaning and beam scrubbing}

The dynamic pressure i.e. the pressure while operating the accelerator with
beams will be dominated by the beam-induced dynamic effects like
stimulated desorption due to beam losses or synchrotron radiations or
by electron stimulated desorption in case an electron cloud is
building-up.

In presence of synchrotron
radiation, the vacuum cleaning process which
characterises the reduction of the desorption yields
(${\eta}$) of a surface resulting from the bombardment of the surface by
electrons, photons or ions, significantly decreases the induced gas loads ($3-4$ orders of
magnitude observed in LEP) improving the dynamic pressure at constant
pumping speed. This results in a progressive increase of the beam
lifetime.

In presence of an electron
cloud, the beam scrubbing which characterises the reduction of the
secondary electron yield (SEY, ${\delta}$)
of a surface resulting from the bombardment of the surface by
electrons, photons or ions, significantly decreases the induced gas
loads ($2-3$ orders of magnitude observed in SPS) improving the dynamic
pressure at constant pumping speed. Similarly to what happens with the
vacuum cleaning, this results also in a progressive increase of the
beam lifetime.

By default and mainly driven
by costs and integration issues, the vacuum system of an accelerator
dominated by beam-induced dynamic effects is never designed to provide
the nominal performances as from ``day 1''. Indeed, vacuum cleaning and
beam scrubbing are assumed to improve the beam pipe surface
characteristics while the beam intensity and beam energy are
progressively increased during the first years of
operation.

\begin{figure}[htp]
\begin{center}
\hspace*{-0.3cm}  
\includegraphics[width=6.999cm,height=7.375cm]{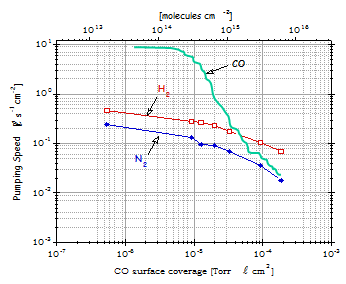} 
\includegraphics[width=6.999cm,height=6.579cm]{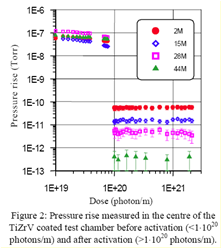}
\end{center}
\vspace*{-0.2cm}
\caption{NEG pumping speed for
different gas species and pressure rises measured in presence of a
photon flux before and after NEG activation.}
\label{jimenez:Fig2}
\end{figure}

\begin{figure}[htp]
\begin{center}
\hspace*{-0.3cm}  
\includegraphics[width=7.01cm,height=6.211cm]{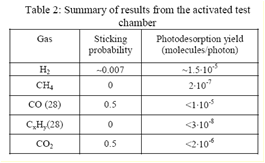}
\includegraphics[width=6.999cm,height=6.188cm]{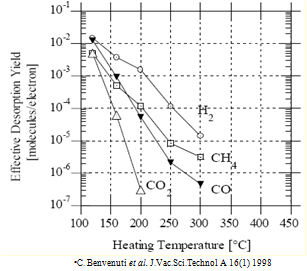} 
\end{center}
\vspace*{-0.2cm}
\caption{Photon (left) and Electron (right) desorption yields.}
\label{jimenez:Fig3}
\end{figure}

This implies accepting a shorter beam
lifetime or reduced beam current during the initial phase; about 500 h
of operation with beams were required for LEP to achieve the nominal
performances. New technical developments such as Non-Evaporable
Coatings (NEG) shall be considered since significantly decreasing the
time required to achieve the nominal performances (Figures $\ref{jimenez:Fig2}$ and $\ref{jimenez:Fig3}$).

\begin{figure}[htp]
\begin{center}
\hspace*{-0.3cm}  
\includegraphics[width=9.072cm,height=5.449cm]{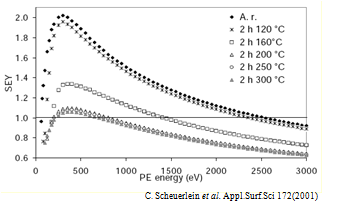}
\end{center}
\vspace*{-0.2cm}
\caption{Reduction of the secondary electron yield (SEY, ${\delta}$)
by Photons a) and Electron b) desorption yields.}
\label{jimenez:Fig4}
\end{figure} 

\subsection{Vacuum engineering issues}
The engineering of the vacuum system has to be integrated right from the
beginning of the project. This becomes imperative for the Ring-Ring
option since it has to take into account the constraints of the LHC and
allow for future consolidations and upgrades. For the Linac-Ring
option, the tangential injection and dump lines will be in common with
the LHC beam vacuum over long distances. The experience has shown that
the vacuum engineering shall proceed in parallel on the following
topics: expertise provided to beam-related components (magnets, beam
instrumentation, radio-frequency systems, etc.), engineering of vacuum
related components (beam pipes, bellows, pumping ports, etc.) and
machine integration including the cabling and the integration of the
services.

Basically, the vacuum system is
designed to interconnect the beam related equipment installed on the
beam line (magnets, kickers, RF cavities, beam absorbers, beam
instrumentation, etc.) and to provide the adequate pumping speed and
vacuum instrumentation. The vacuum components are often composed by
vacuum pipes, interconnection bellows, diagnostics, pumping ports and
sector valves. The number
of pumps, vacuum diagnostics, bellows and ports will differ
significantly between the two options discussed in this CDR and also
between vacuum sectors of the same accelerator.

\subsubsection{Vacuum pumping}
The vacuum system of the LHeC will be mainly operated at ambient
temperature. These systems rely more and more on NEG coatings since
they provide a distributed pumping and huge pumping speed (Fig.2) and
capacity and reduce the out-gassing and desorption yields (Fig.3-4).
These coatings are compatible with copper, aluminium and stainless
steel beam pipes. An alternative could be to use the LEP configuration
with NEG strips. This alternative solution has only the advantage of
avoiding the bake out constraints for the activation of the NEG
coatings. A configuration of a distributed ion pumps is not considered
since less performing and only applicable in dipole magnets i.e.
bending sections. In any case, ion pumps are required as a complement
of the NEG coatings to pump the noble gasses and methane to avoid the
ion beam-induced instability.
Sublimation pumps are not excluded in case of local huge out-gassing rates, NEG cartridges being
an interesting alternative since recent developments made by
manufacturers include an ion pump and a NEG cartridge in the same
body.

The roughing from atmosphere
down to the UHV range will be obtained using
mobile turbo-molecular
pumping stations. These pumps are dismounted prior to beam
circulations.

The part of the vacuum
system operated at cryogenic temperature, if any, could rely on gas
condensation if the operating temperatures are below 2~K. Additional
cryosorbing material could be required if an important hydrogen gas
load is expected. This issue still needs to be addressed. As made for
the LHC, the parts at cryogenic temperature must be isolated from the
NEG coated part by sector valves when not at their operating
temperature to avoid the premature saturation of the NEG coatings.

The pumping layout will be
simpler for the Ring-Ring option since more space is available around
the beam pipes. The tighter tolerances for the Linac-Ring option make
the integration and pumping layout more delicate. However, the vacuum
stability will be easier to ensure in the Linac-Ring option since only
the bending sections are exposed to the synchrotron radiation.

\subsubsection{Vacuum diagnostics}
For both options, the radiation level expected will be too high to use
pressure sensors with onboard electronics. Therefore, passive gauges
shall be used, inducing additional cabling costs and need for gauge
controllers.

\subsubsection{Vacuum sectorisation}
The sectorisation of the beam vacuum system results from the integration of
various constraints, the major being: venting and bake-out
requirements, conditioning requirements (RF and HV devices),
protection of fragile and
complex systems (experimental areas and ceramic chambers), decoupling
of vacuum parts at room temperature from upstream and downstream parts
at cryogenic temperature thus non-baked, radiation issues, etc.

For UHV beam vacuum systems,
all-metal gate valves shall be preferred in order to allow for bake-out
at temperature above 250{\textdegree}C. VITON-sealed valves even though
the VITON has been submitted to a special treatment are not recommended
nearby NEG coatings or NEG pumps since minor out-gassing of Fluor will
degrade the pump characteristics.

In the injection and
extraction regions, the installation of the sector valves will lead to
integration issues since the space left between the beam pipes with a
tangential injection/extraction and the circulating beams is often
limited. This could result in a long common beam vacuum which implies
that the LHC beam vacuum requirements will apply to the LHeC part
shared with LHC.

\subsubsection{Vacuum protection}
The distribution of the vacuum sector valves will be made in order to
provide the maximum protection to the beam vacuum in case of failure
(leak provoked or not). Interlocking the sector valves is not an
obvious task. Indeed, increasing the number of sensors will provide
more pressure indications but often results in a degradation of the
overall reliability. The protection at closure (pressure rise, leaks)
is treated differently from the protection while recovering from a
technical stop with parts of the accelerator beam pipe vented or being
pumped down.

The vacuum protections of
the common beam pipes between LHeC and LHC shall fulfil the strong LHC
requirements. Indeed, any failure in the LHeC propagating to the LHC
could lead to long machine downtime (several months) in case of an
accidental venting of an LHC beam vacuum sector.

\subsubsection{HOM and impedance implications}
The generation and trapping of higher order mode (HOM) resulting from the
changes in beam pipe cross sections are severe issues for high intensity
electron machines. Thus, the engineering design of LHeC must be
inspired on new generation of synchrotron radiation light sources
instead of the simple LEP design. All bellows and gaps shall be
equipped with optimised RF fingers, designed to avoid sparking
resulting from bad electrical continuity. Indeed, these effects could
induce pressure rises and machine performance limitations.

\subsubsection{Bake-out of vacuum system}
An operating pressure in the UHV range
($10^{-10}$ Pa) will be required for both options. This implies the use of a fully
baked-out beam vacuum system.Two options are possible: permanent
and dismountable bake out. The permanent solution could be an option
for the Linac-Ring but has to be excluded for the Ring-Ring option for
cost reasons. As done for the dipole chambers (bending sections) of
LEP, hot pressurised water can be used but the limit at
150{\textdegree}C is a constraint for the activation of NEG coatings.
Developments are being carried on at CERN to lower the activation
temperature from 180{\textdegree}C down to 150{\textdegree}C but this
technology is not yet available.

\subsubsection{Shielding issues}
The synchrotron radiation power is an engineering challenge for the
beam pipes. Indeed, $50\%$
of the radiation power hitting the vacuum chamber is absorbed in the
beam pipe chamber (case of LEP aluminium chamber). The remainder $50\%$,
mainly the high-energy part of the spectrum, escapes into
the tunnel and creates
severe problems like degradation of organic material and electronics
due to high dose rates and formation of ozone and nitric acid could
lead to severe corrosion problems in particular with aluminium and
copper materials.

In this respect, the
Ring-Ring option is less favourable since the synchrotron radiation will
be localised at the plane of the existing LHC cable trays and
electrical distribution boxes in the tunnel. Similar constraints exist
also for the Linac-Ring option but these zones are localised at the
bending sections of the LHeC.

Detailed calculations are
still to be carried on but based on LEP design, a lead shielding of $3$
to $8$~mm soldered directly on the vacuum chamber would be required for
$70$~ GeV beams. Higher energies could require more thickness. The
evacuation of the synchrotron radiation induced heat load on the
beam pipe wall and on lead shielding is a critical issue which needs to
be studied. In case of insufficient heat propagation and cooling, the
lead will get melted as observed in LEP in the injection areas. The
material fatigue shall also be investigated since running at much
higher beam current as compared to LEP, will increase the induced
stress to the material and welds of the beam pipes.

As made in LEP, the best
compromise to fulfil the above mentioned constraints is the use of
aluminium beam pipes, covered by a lead shielding layer. The complex
beam pipe cross section required to optimise the water cooling of the
beam pipe and shielding is feasible by extrusion of aluminium billets and
the costs are acceptable for large productions. The large heat
conductivity helps also the heat exchange. However, extruded aluminium
beam pipes induce limitations for the maximum bake out temperature and
therefore for the NEG coatings activation. Special grades of aluminium
shall be used. The reliability of vacuum interconnections based on
aluminium flanges is a concern at high temperature
({\textgreater}$150${\textdegree}C) and corrosion issues shall be
addressed. The stainless steel beam pipes do not have these limitations but
they have poorer heat conductivity and they are more difficult and
costly to machine and shape.

The LEP $110$~GeV operation has shown the criticality of unexpected synchrotron radiations heating
vacuum components and in particular the vacuum connections between
pipes or equipment. Indeed, the flanges, by ``offering'' a thick path,
are behaving as photon absorbers and heat up very quickly. Hence, at
cool down and due to the differential dilatation, leaks are opening. In
LEP, these unexpected SR induced heat loads resulted from orbit
displacement in quadrupoles during the ramp in energy and of the use of
the wigglers also during the ramp. In LHeC, resulting from the much
higher beam current, these issues shall be carefully studied.

\subsubsection{Corrosion issues}
In vacuum systems, feedthroughs and bellows are particularly exposed to
corrosion. The feedthroughs, particularly those of the ion pumps where
high voltage is permanently present, are critical parts. A demonstrated
and cheep solution to prevent the risk of corrosion consists in heating
directly the protective cover to reduce the relative humidity around
the feedthrough.

The bellows are critical due to their thickness, often between $0.1-0.15$~mm. \ PVC material must be
prohibited in the tunnel. Indeed, in presence of radiations, it can
generate hydrochloric acid (HCl) which corrodes stainless steel
materials. This corrosion has the particularity to be strongly
penetrating, once seen at the surface, it is often too late to mitigate
the effects. Aluminium bellows are exposed to corrosion by nitric acid
(HNO$_3$) which is generated by the combination of O$_3$ and NO.

Humidity is the driving factor and shall be kept $50\%$. However, in the long term, accidental
spillage can compromise locally the conditions and therefore,
corrosion-resistant design are strongly recommended.

%% file: machine/SR_BPDesign.tex
\section{Beam pipe design} 
\label{BPDesign}

\subsection{Requirements}
The vacuum system inside the experimental sector has a number of
different and sometimes conflicting requirements. Firstly, it must
allow normal operation of the LHC with two circulating beams in the
chamber. This implies conformity with aperture, impedance, RF, machine
protection as well as dynamic vacuum requirements. The addition of the
incoming electron beam adds constraints in terms of geometry for the
associated synchrotron radiation (SR) fan and the addition of SR masks
in the vacuum. Finally, optimisation of the surrounding detector for
high acceptance running means that all materials for chambers,
instrumentation and supports must be optimised for transparency to
particles and the central chamber must be as small and well aligned as
possible to allow detectors to approach the beam aperture limit at the
interaction point.

\subsection{Choice of materials for beam pipes}
LHC machine requirements imply an inner beam pipe wall that has low
impedance (good electrical conductivity) along with low desorption
yields for beam stimulated emissions and resistance to radiation
damage.

Ideal materials for transparency to particles have low radiation
length (Z) and hence low atomic mass. These materials either have poor
(i.e. high) desorption yields (e.g. aluminium, beryllium) or are not
vacuum and impedance compatible (e.g. carbon). Solutions to this
problem typically include thin film coatings to improve desorption
yields and composite structures to combine good mechanical properties
with vacuum and electrical properties.

The LHC experimental vacuum systems, along with most other colliders
currently use metallic beryllium vacuum chambers around the
interaction points due to a very favourable combination of Z,
electrical conductivity, vacuum tightness, radiation resistance, plus
mechanical stiffness and strength. High desorption yields are
suppressed by a thin film TiNiV non-evaporable getter (NEG)
coating. This coating also gives a high distributed vacuum pumping
speed, allowing long, small aperture vacuum chambers to be used that
would otherwise be conductance-limited. Activation of this coating
requires periodic heating of the chamber to $180-220\textdegree{C}$
under vacuum for a few hours. This means that the chamber and
environment must be designed for these temperatures. This activation
is scheduled in annual LHC shutdowns. Long-term development is in
progress for low desorption yield coatings that do not require high
temperature activation \cite{YinVallgren:2010zz}. These may have
applications for LHeC.

Production technology developed for the LHC uses beryllium sections
machined from hot-pressed blocks and electron beam welded to produce
chambers. This has the advantage that a wide range of vacuum chamber
forms can be manufactured. Cylindrical and conical chamber sections
are installed in the LHC experiments.

Disadvantages of beryllium include high cost, fragility and toxicity
in the powder form, as well as limited availability. For this reason,
long-term development of other technologies for experimental beam pipes
is under way at CERN which may yield applications for LHeC.

Composite beam pipe structures made from carbon and other low-Z
materials have been developed for colliders. These typically use a
thin inner membrane to comply with vacuum and impedance
requirements. Composite structure pipes were eventually rejected for
LHC application for reasons of temperature and radiation resistance
and the risk of de-lamination due to mismatch of thermal expansion
coefficients. Lower luminosity in LHeC experiments combined with new
low temperature coatings may allow these materials to be re-evaluated.

\subsection{Beam pipe Geometries}
\begin{figure}[htp]
\begin{center}
\includegraphics[width=0.8\columnwidth]{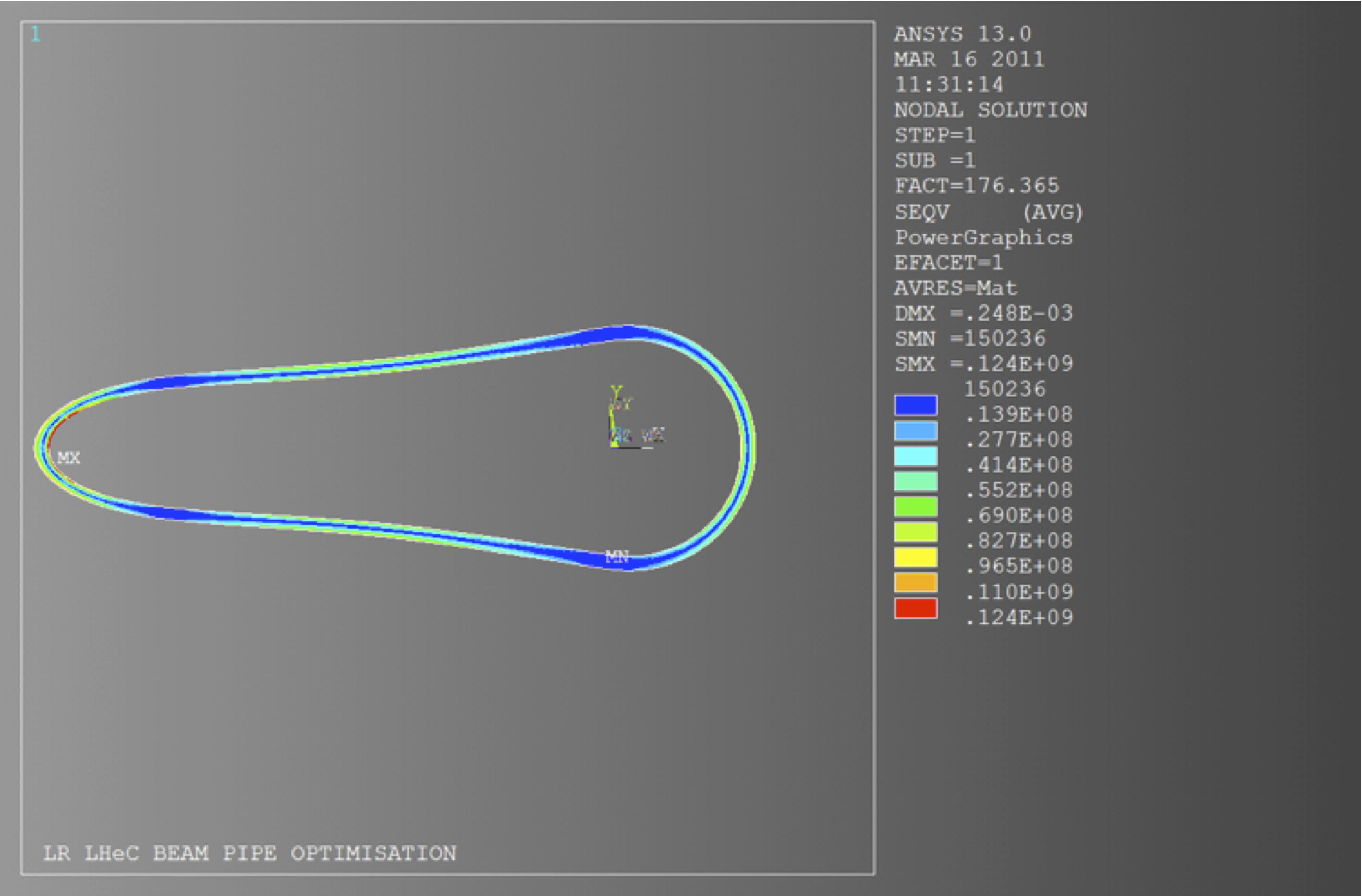}
\vspace*{-0.2cm}
\end{center}
\caption{Section through the LR geometry showing contours of Von Mises equivalent stress (Pa).}
\label{BPStress:Fig:1}   
\end{figure} 
\begin{figure}[htp]
\begin{center}
\includegraphics[width=0.8\columnwidth]{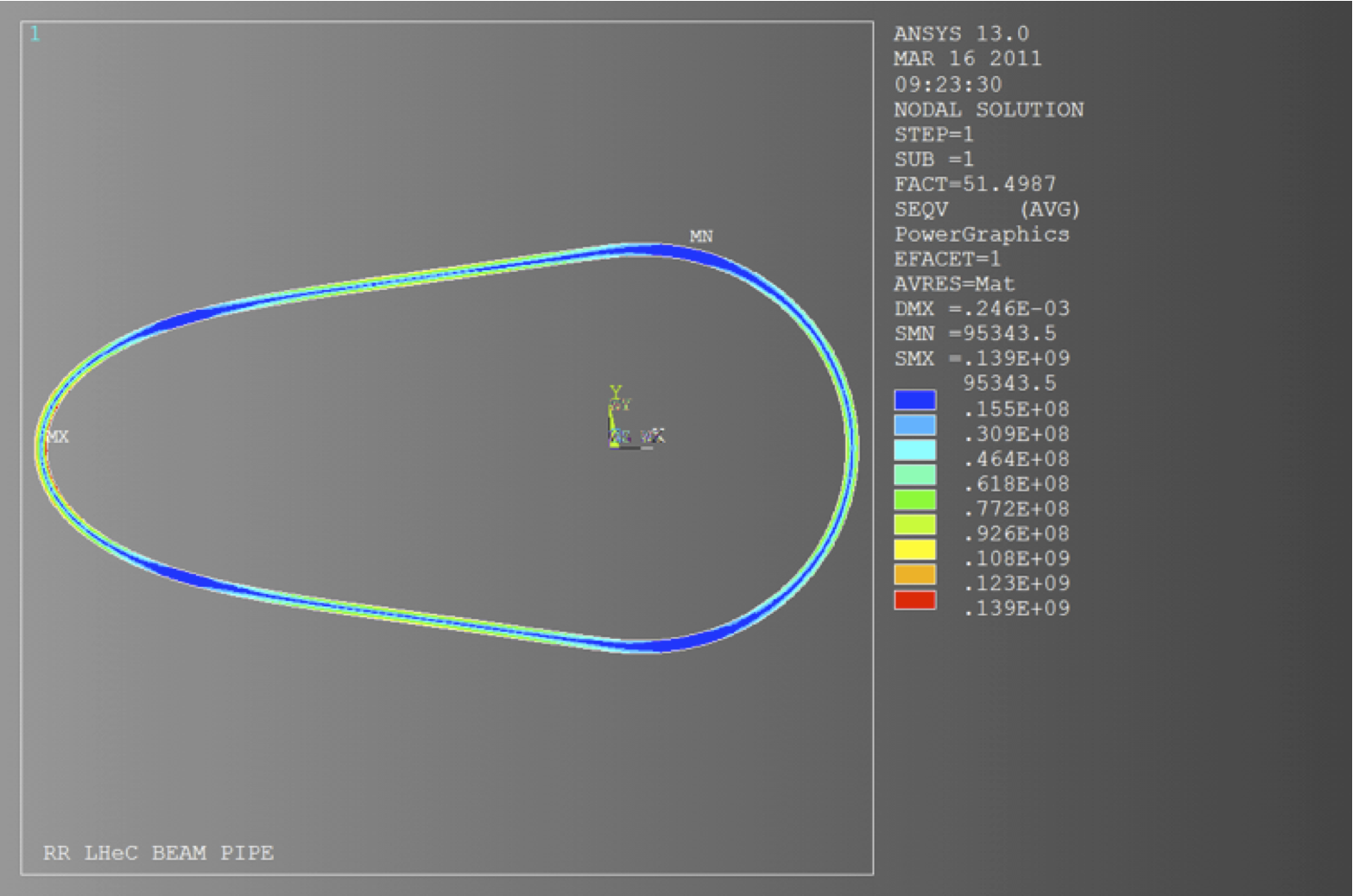}
\vspace*{-0.2cm}
\end{center}
\caption{Section through the RR geometry showing contours of Von Mises equivalent stress (Pa).}
\label{BPStress:Fig:2}   
\end{figure}
\begin{figure}[htp]
\begin{center}
\includegraphics[width=0.8\columnwidth]{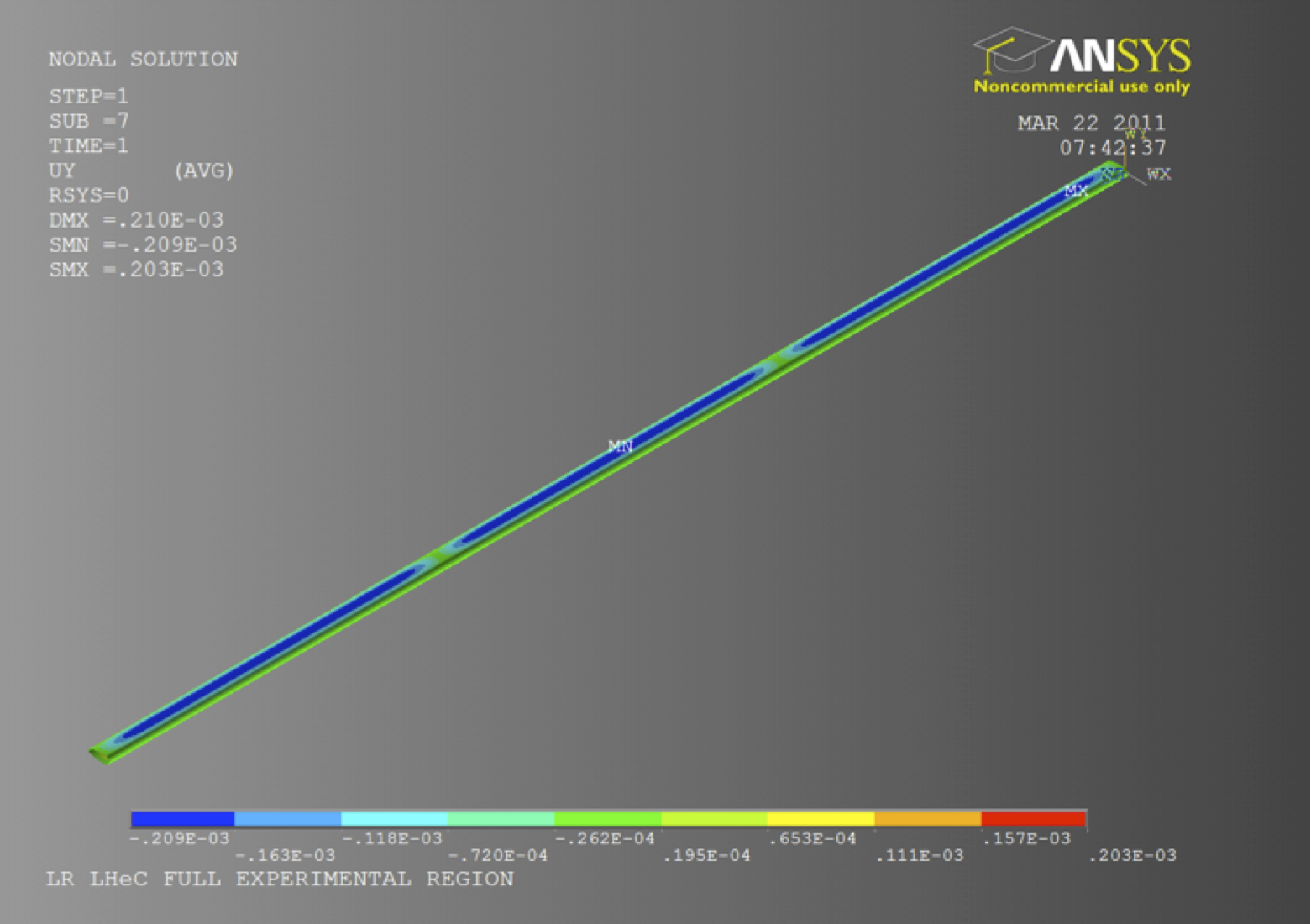}
\vspace*{-0.2cm}
\end{center}
\caption{3-D view of the LR geometry showing contours of bending displacement [m].}
\label{BPStress:Fig:3}   
\end{figure}
The proposed geometry has a cross section composed of a half-circle
intersecting with a half-ellipse. Cylindrical cross sections under
external pressure fail by elastic instability (buckling) whereas
elliptical sections can (depending on the geometry) fail by plastic
collapse (yielding).

Figure\,\ref{BPStress:Fig:1} and \ref{BPStress:Fig:2} show
optimisations of the proposed geometries for the LINAC-Ring (LR) and
Ring-Ring (RR) beam pipes assuming a long chamber of constant cross
section made from beryllium metal. Preliminary analyses have been
performed using the ANSYS finite element code. The wall thickness was
minimised for the criteria of yield strength and buckling load
multiplier. The LR geometry considered has a circular section radius
of 22 mm and elliptical major radius of 100 mm. The RR geometry has a
circular section radius of 22 mm and elliptical major radius of 55
mm. This preliminary analysis suggests that a constant wall thickness
of $2.5-3$~mm for the LR and $1.3$ to $1.5$~mm for the RR would be
sufficient to resist the external pressure. Failure for both of these
sections would be expected to occur by plastic collapse.

At this stage of the project, these geometries represent the most
optimised forms that fulfil the LHC machine requirements. However,
for 1 degree tracks this corresponds to X/X0 $\approx$ 21-25\% for the
LR and $\approx$ 41-49\% for the RR designs. This suggests that
additional effort must be put into beam pipe geometries optimised for
low angles. Composite beam pipe concepts suggested for machines such as
the LEP \cite{Hauviller:1988zf} should be re-considered in the light
of advances in lightweight materials and production techniques.

The optimised section of the experimental chamber is 6.1 m in
length. This length will require a number of optimised supports. These
supports function to reduce bending deflection and stresses to within
acceptable limits and to control the natural frequency of chamber
vibration. The non-symmetric geometry will lead to a torsional stress
component between supports which must be considered in their design.
Figure \ref{BPStress:Fig:3} shows a preliminary analysis of bending
displacement for the LR chamber geometry. With 2 intermediate supports
the maximum calculated displacement (without bake-out equipment) is
0.21\,mm.

\subsection{Vacuum instrumentation}
If, as assumed, this chamber is coated with a NEG film on the inner
surfaces, then a high pumping speed of chemically active gasses will
be available. Additional lumped pumps will be required for
non-gettered gasses such as $CH_4$ and noble gasses; however,
out-gassing rates for these gasses are typically very low.

The vacuum sector containing the experiment will be delimited from the
adjacent machine by sector valves. These will be used to allow
independent commissioning of machine and experiment vacuum.  The
experimental vacuum sector will require pressure gauges covering the
whole range from atmospheric to UHV, these are used both for
monitoring the pressure in the experimental chamber and as interlocks
for the machine control system.

\subsection{Synchrotron radiation masks}
LHeC experimental sector will require a movable SR mask upstream of
the interaction. From the vacuum perspective, this implies a system
for motion separated from atmosphere by UHV bellows. The SR flux on
the mask will generate a gas load that should be removed by a local
pumping system dedicated to the mask. As the load due to thermally
stimulated desorption increases exponentially with the temperature,
cooling may be required. However, cooling the mask would significantly
complicate the vacuum system design.  The generation of
photo-electrons must also be avoided since these photo-electrons can
interact with the proton beam and lead to an electron cloud build-up.

\subsection{Installation and integration} 
The installation of the vacuum system is closely linked to the
detector closure sequence. Therefore, the design has to be validated
in advance to prevent integration issues which would lead to
significant delay and increase of costs. Temporary supports and
protections are required at each stage of the installation. Indeed, as
compared to the size of the detectors, the beam pipe are small,
fragile and need to be permanently supported and protected while
moving the detector components. Leak tightness and bake-out testing
are compulsory at each step of the installation since all vacuum
systems are subsequently enclosed in the detector, preventing any
access or repair. Their reliability is therefore critical.  Precise
survey procedures must also be developed and incorporated in the
beam pipe design to minimise the mechanical component of the beam
aperture requirement.  Engineering solutions for bake out also has to
be studied in details since the equipment (heaters, probes and cables)
must fit within the limited space available between beam pipes and the
detector components.

%% file: machine/haugrr.tex
\subsection{Ring-Ring cryogenics design}

\subsubsection{Introduction}
The Ring-Ring version foresees the $60$ GeV accelerator to be
installed in the existing LHC tunnel. Acceleration of the particles is
done with $0.42$ m long $5$ MV superconducting (SC) cavities housed in
fourteen 10 m long cryomodules. They will be placed at two opposite
locations in by-passes of Point 1 (ATLAS) and, Point 5 (CMS). While at
CMS a continuous straight by-pass can be built, at ATLAS two straight
sections are conceived on each side of the detector cavern (``left''
and ``right'') with a connecting beam pipe crossing the detector
hall. Layouts and detailed RF description see Chapter~\ref{RR-RF-section}. The
three separate cryomodules locations require three dedicated $2$ K
cryo-systems. Injection to the Ring at $10$ GeV is done with a $1.3$
GHz pulsed three-pass recirculating high field injector. A dedicated
cryoplant provides $2$ K cooling of its SC cavities. In total four
independent cryoplants with their respective distribution systems are
needed for the Ring-Ring version. For the LHeC detector the high
gradient focusing insertion magnets will be SC and housed in LHC
dipole type cryostats. The cooling principle is the same as for LHC
dipoles and, the existing cryogenic infrastructure can be used with
comparatively small adaptations of the feed boxes. More detailed
engineering studies are beyond the scope of this report. This chapter
describes the cryosystems of the e-Ring accelerator and the related
injector.

\subsubsection{Ring-Ring cryogenics}
The cavities operate at $2$ K superfluid helium temperatures and
dissipate an estimated $4$ W per cavity at $5$ MV. The 8-cavity
cryomodule has three temperature levels; a $2$ K saturated bath
containing the cavities, a $5-8$ K combined thermal shield and heat
intercept for couplers and other equipment and, a $40-80$ K thermal
shield. The thermal loss estimates are listed in Table~\ref{tab:haugrr1}. With efficiencies of modern state of the art
cryoplants reaching 1/COP values of $1000$ W/W at $2$ K, $250$ W/W at
$5$~K and $20$~W/W at $40-80$ K the minimum plant powers are
calculated. To the equivalent cooling power at $4.5$~K we add a $50\%$
contingency for the distribution system with transfer lines running
parallel to the cryomodules. In Table~\ref{tab:haugrr2} the
equivalent cooling powers of the three cryoplants are given.

\begin{table}[h]
  \centering
  \begin{tabular}{|c|c|c|c|}
    \hline
Temperature (K) & $2$ & $5-8$ & $40-80$\\
One cryomodule & & & \\
Static loss (W) & $5$ & $15$ & $100$\\
Dynamic loss (W) & $32$ & $15$ & $80$\\
Sum (W) & $37$ & $30$ & $180$\\
8 modules (CMS site) (W) & $296$ & $240$ & $1440 (2160)$\\
3 modules (ATLAS left) (W) & $111$ & $90$ &$720 (1080)$\\
3 modules (ATLAS right) (W) & $111$ & $90$ & $720 (1080)$\\
\hline
  \end{tabular}
\caption{
Thermal loss estimate of cryomodules. In brackets the values with
ultimate thermal losses ($50\%$ contingency) which are taken into
account for the cryoplant sizing.
}
\label{tab:haugrr1}
\end{table}

\begin{table}[h]
  \centering
  \begin{tabular}{|c|c|}
    \hline
Site & Plant power $@\; 4.2$ K (kW)\\
\hline
CMS site & $3.0$\\
ATLAS left & $1.2$\\
ATLAS right & $1.2$\\
\hline
  \end{tabular}
\caption{
Cryoplant equivalent cooling powers.
}
\label{tab:haugrr2}
\end{table}

\begin{figure}
\begin{center}
\includegraphics[width=0.7\textwidth]{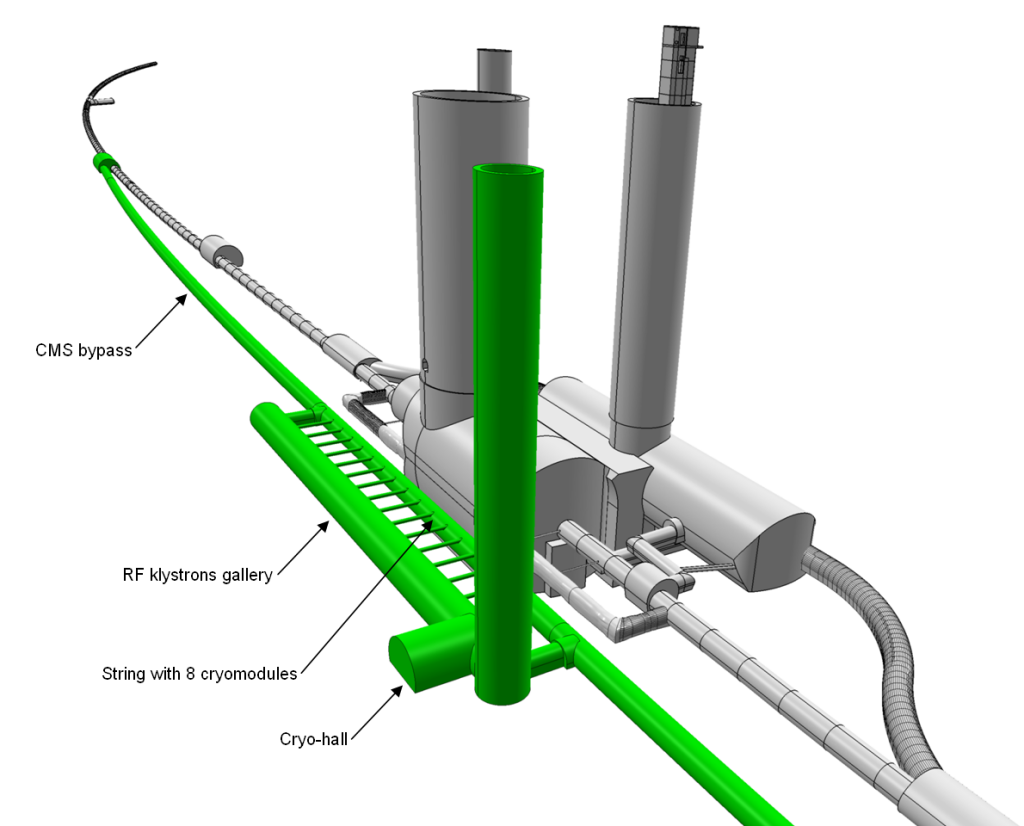}
\end{center}
\caption{Lay-out of the CMS by-pass with location of the cryomodules
  and the $3$ kW $@\; 4.5$ K cryoplant.}
\label{fig:haugrr1}
\end{figure}

\begin{figure}
\begin{center}
\includegraphics[width=1.\textwidth]{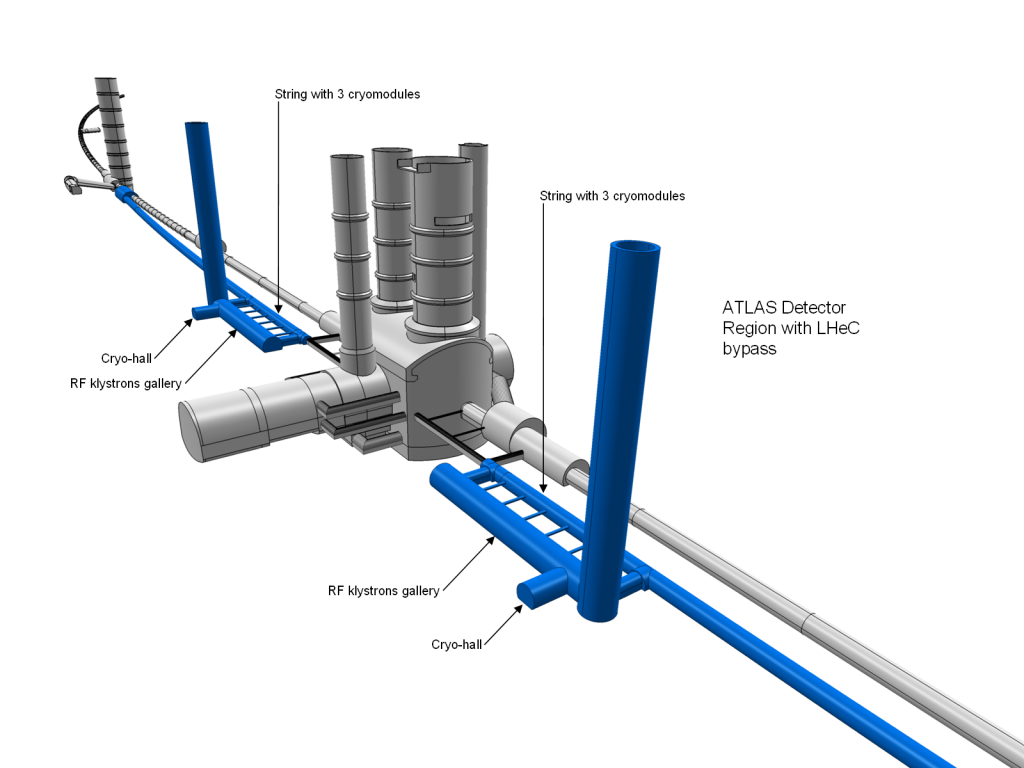}
\end{center}
\caption{Lay-out of the ATLAS by-pass with locations of the
  cryomodules and the two $1.2$ kW $@\; 4.5$ K cryoplants.}
\label{fig:haugrr2}
\end{figure}

At CMS site a dedicated $3$~kW $@\;4.2$~K cryoplant is needed. Except
for some general infrastructure equipment like e.g. gas tanks it will
be separated from the existing CMS cryoplant used to cool the solenoid
magnet. Comparatively modest cooling powers suggest the use of a
single compact refrigerator cold box, in contrast to split versions as
proposed in this CDR for the Linac-Ring version described below. (The
split version is based on LHC technology with a combined surface and
underground cold box.) The cold box will be installed directly in the
underground cavern at proximity to the cryomodule string. Ambient
temperature high and low pressure lines make the link to the
compressor stations on surface. For the $2$~K temperature level two
cold compressors with a total compression ratio of $10$ are proposed
followed by warm compressors to compress the gas to ambient
pressure. Figure~\ref{fig:haugrr1} shows the lay-out of the CMS
by-pass region.  At the two ATLAS sites (left, right) with three
cryomodules each, two options are conceivable.  The first consists of
connecting to the LHC QRL transfer lines and their terminal feedboxes
at vicinity for a ``parasitic'' use of excessive cooling power of the
LHC cryoplants. For this two additional $10-15$~m long perpendicular
tunnels to connect the LHC tunnel with the LHeC by-pass would have to
be constructed. The feasibility of this option and potential
(negative) impacts have to be studied in more detail in a subsequent
report. The second option is to use two dedicated cryoplants as
proposed for the CMS site, however, with reduced capacity.  Also in
this case the cold box will be installed at proximity to the
cryomodule strings in the cryo-hall. The two refrigerators are of the
same design principle as for CMS, except for their size and capacity
which is smaller. Their location will be on ATLAS terrain which allows
to potentially use already existing cryogenic infrastructure of the
large cryo-system for the cooling of the ATLAS toroidal and solenoid
magnets. Among these are the gas storage tanks, the compressor hall
and control rooms. Figure~\ref{fig:haugrr2} shows the lay-out of the
ATLAS by-pass region.

\subsubsection{Cryogenics for the 10 GeV injector}

\begin{figure}
\begin{center}
\includegraphics[width=0.8\textwidth]{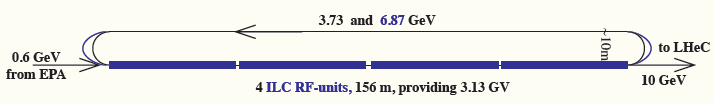}
\end{center}
\caption{ Principle of the $10$ GeV re-circulating Injector with high
  gradient pulsed SC cavities ($23$ MV/m) and $12$ cryomodules of the
  ILC/XFEL type operating at $2$ K.  }
\label{fig:haugrr3}
\end{figure}

The injector is a three-pass recirculating  $10$ Hz machine
providing leptons at injection energies of $10$ GeV to the LHeC Ring
machine. Figure $\ref{fig:haugrr3}$ shows its basic
principle. Cryomodules of the XFEL (ILC) type with $1.3$ GHz
superconducting cavities are proposed which allow the application of
already existing technology requiring little adaptation effort for
LHeC. A $146$ m long string will be composed of in total $12$
cryomodules each $12.2$ m long. Cryogen distribution is done within
the volume of the cryostats. Bath cooling is at $2$ K saturated
superfluid helium. Adopted from XFEL the common pump line of $300$ mm
runs within the cryomodules envelope to collect vapour of all
individual cavity baths. Therefore no external transfer line is
required which simplifies the overall design. The suction pressure of
$30$ mbar is provided by cold compressors in the cold box and
subsequent ambient temperature compressors. Two more temperature
levels of $5-8$~K and $40-80$~K are used for intercepts and thermal
shielding.  The operation of the injector at LHeC is in part
comparable to XFEL, this during the injection and loading phase of
leptons into the LHeC ring. During all other operation phases of a
complete LHeC cycle (ramping to final particle energies in the
LHC/LHeC tunnel and subsequent physics runs) the injector machine is
``idle''. Only static heat losses of the cryomodules and the cryogenic
infrastructure have to be intercepted during this time
period. Principally a reduced power cryogenic system operating with an
``economiser'' could be conceived, i.e. a large liquid helium storage is filled during low
demands which in turn boosts the cryomodules during the injection
phases. A simpler approach, however, is the design for constant
(maximum) cooling power when active and, during idle periods, internal
electric heaters in the $2$ K bath are switched on to keep the load
constant. This principle is adopted for these initial studies. A
compact single refrigerator cold box providing temperatures from $300$
K to $2$ K will be installed in a protected area at vicinity to the
extraction region of the cryomodule string while the compressor set is
at surface. For the estimation of power consumption and cooling
performances we shall use the experience gained at DESY during testing
of XFEL cryomodules. With a final energy of $10$~GeV and three pass
operation the acceleration field required is $23$~MV/m. At DESY power
consumption measurements have been made with cryomodules for a similar
acceleration field of $23.8$ MV/m and $10$ Hz operation.  Our
estimates as shown in the Table~$\ref{tab:haugrr3}$ are based on these
recent data. With 1/COP values as used in above chapter and a $50\%$
margin for additional thermal losses we estimate the required cooling
power of the plant to $2$ kW $@\; 4.5$ K.

\begin{table}[h]
  \centering
  \begin{tabular}{|c|c|c|c|}
    \hline
Temperature (K) & $2$ & $5-8$ & $40-80$\\
Static loss (W) & $5$ & $15$ & $100$\\
Dynamic loss (W) & $8$ & $3$ & $40$\\
Sum (W) & $11$ & $18$ & $140$\\
Sum $12$ modules (W) & $132 (198)$ & $216 (324)$ & $1680 (2520)$\\
\hline
  \end{tabular}
\caption{
Thermal loss estimate of the $146$ m long string built of $12$ XFEL type
cryo-modules. In brackets values with $50\%$ contingency. Cryoplant
equivalent cooling power; $2$ kW $@\; 4.5$ K.
}
\label{tab:haugrr3}
\end{table}

%% file: machine/hauglr.tex
\subsection{Linac-Ring cryogenics design}

\subsubsection{Location and basic lay-out}

The ERL (Energy Recovery Linac) is of racetrack shape with two 1 km
long straight SC acceleration sections and, two arcs of 1 km radius
with normal conducting magnets. Location and lay-out studies made are
described in Chapter\,$\ref{chapter:civil}$. The currently favoured
position is within the LHC perimeter (see Figure $\ref{fig:osborne9}$)
versus the external version being largely under St. Genis
community. For the ``inside'' version more of the newly required
surface areas could be located on existing CERN grounds comprising
SM18, North Area and, Point 2. Next steps following this CDR will
require more detailed combined studies of civil engineering, RF,
cryogenics and other services to try optimise the lay-out also, and in
particular, for the cryogenic equipment having impact on its own
complexity and costs. As base in this study we propose a symmetric
lay-out with a sub-division of the respective 1 km long straight
sections in four equally spaced sections each housing four 250 m long
cryomodule strings.  As indicated in Chapter\,$\ref{chapter:civil}$,
the ERL will be inclined towards the Lake of Geneva by $1.4\%$,
however, due to its orientation the tilt in longitudinal direction
relevant to the cryogenics is smaller.

\subsubsection{Cryomodules}

\begin{figure}
\begin{center}
\includegraphics[width=0.8\textwidth]{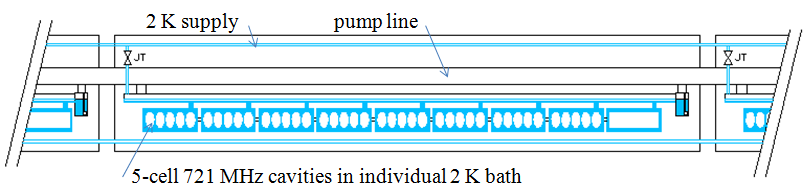}
\end{center}
\caption{ Schematic proposal of the $14$ m long cryomodules with eight
  5-cell $721$ MHz cavities operating at $2$ K. Supply pipes and the
  $30$ mbar pump line are within cryostat envelope. For the case with
  inclination right part is lower (only $2$ K circuits are shown).}
\label{fig:hauglr1}
\end{figure}

Eight 721 MHz SC 5-cell cavities of length 1.04 m long will be housed
in 14 m long cryomodules~\footnote{Note that in the presentation
of the RF for the superconducting linac a cryo-cavity module length of
$15.6$\,m was eventually chosen as the baseline of the design,
using the same eight-fold subdivision with cavities of $1.04$\,m length.
Small further alterations had 
been introduced, such as to the total number of cryomodules which
is $120$  instead of the $118$ considered here. These variations have no
essential influence on the design concept of the linac
cryogenics  as is presented subsequently.}.
 Bath cooling of the cavities is done with
slightly subcooled saturated superfluid helium at 2 K. Each cryostat
is equipped with a J.T. valve located upstream to expand the 2 K
supply helium to the 30 mbar bath pressure and the liquid is brought
gravity assist to the downstream individual 8 cavity bath volumes via
an interconnecting header pipe. This principle is similar to the SPL
preliminary design which has to cope with a tilt of $1.7 \%$
\cite{Wagner}.  Heat intercept and thermal shielding is at 5-8 K and
40-80 K. The final LHeC L-R cryomodule design can be based on
extensive previous work and studies of both existing SC linear
accelerators and, such being under construction or planned ones. Among
these are CEBAF, ILC, XFEL, SPL, e-RHIC. Here a design based on
TESLA/XFEL type cryomodules is made. Figure $\ref{fig:hauglr1}$
shows a design proposal of a module with the eight cavities and the
cold correction magnets in their individual bath. All cryogen
distribution is done within the cryostat module which interconnects to
the adjacent ones with the pipe runs throughout a 250 m long
cryomodule string. Also the pump line is proposed to be within the
cryostat envelope. The expected mass flow rate of 180 g/s at 2 K of a
250 m long section with 15 cryomodules (see calculations next chapter)
is approximately comparable to XFEL for its entire machine for which
the corresponding pump line diameter has been designed and tested
\cite{Petersen}.  The parameters of the LHeC SC cavities and cooling
requirements are listed in Table $\ref{tab:hauglr1}$.

\begin{table}[h]
  \centering
  \begin{tabular}{|c|c|}
    \hline
Parameter & Value\\
Two linacs & length 1 km\\
5-cell cavities &length 1.04 m\\
Number & 944\\
Cavities/ cryomodule & 8\\
Number cryomodules & 118\\
Length cryomodule & 14 m\\
Voltage per cavity & 21.2 MV\\
R/Q & $285 \Omega$\\
Cavity $Q_{0}$ & $2.5 \cdot 10^{10}$\\
Operation & CW\\
Bath cooling & 2 K\\
Cooling power/cav. & $32$ W $@\; 2$ K\\
Total cooling power (2 linacs) & $30$ kW $@\; 2$ K\\
\hline
  \end{tabular}
\caption{
Parameters and cooling requirements of the ERL (Linac-Ring version).
}
\label{tab:hauglr1}
\end{table}

\subsubsection{Cryogenic system}

\begin{figure}
\begin{center}
\includegraphics[width=0.8\textwidth]{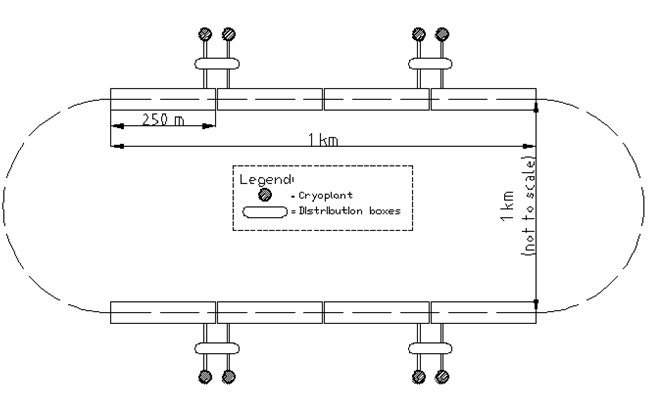}
\end{center}
\caption{ Basic lay-out of the 6 pass ERL. Two $1$ km long SC
  acceleration sections with a $10$ GeV linac each. Eight $10$ kW $@\;
  2$ K cryoplants. Configuration such that each plant supplies a
  cryomodule string of $250$ m length (figure not to scale).  }
\label{fig:hauglr2}
\end{figure}

\begin{figure}
\begin{center}
\includegraphics[width=0.8\textwidth]{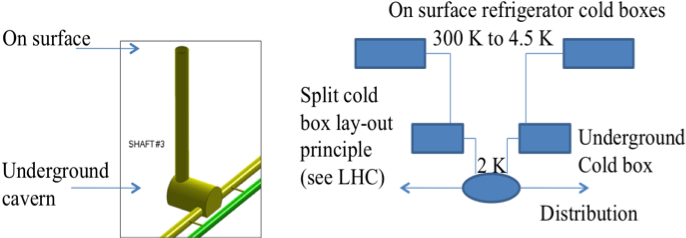}
\end{center}
\caption{
Basic principle of a Split Cold Box lay-out (comparable to LHC accelerator cryogenics).
}
\label{fig:hauglr3}
\end{figure}

The estimated thermal loads per cavity are based on a voltage of
$21.2$ MV, an R/Q of $285 \Omega$ and a $Q_{0}$ of $2.5 \cdot
10^{10}$. With CW operation the dissipated heat per cavity will be
$32$ W, respectively $256$ W per cryomodule. This consists of a very
high load. The $1$ km long straight sections are sub-divided in four
$250$ m long sub-sections each with $15$ interconnecting cryomodules
forming a string which are individually supplied by a respective
refrigerator through local distribution boxes.  Eight dedicated
refrigerators supply the eight strings. Figure $\ref{fig:hauglr2}$
gives a basic lay-out of the cryo-system with its sectorisation.  The
refrigerator cold boxes will be of the so-called ``split'' type with a
surface cold box and a connecting underground cold box as explored and
implemented first for LEP2 and later at a larger scale for LHC. The
surface cold box will be installed close to the compressor set and
produce temperature levels between $300$ K and $4.5$ K. The
underground cold box will be installed at proximity to the respective
cryomodule string in a protected area and produce the $2$ K with cold
compressors. Figure $\ref{fig:hauglr3}$ gives a principle lay-out of
the refrigerator configuration. The final location of the ERL will
dictate civil engineering constraints and the ``ideal'' symmetric
configuration of placement of the refrigerators as done here will have
to be reviewed accordingly and, hence, partially deviate from this
proposal. Also in case only one access shaft per linac can be
conceived the four surface cold boxes may be installed in form of
clusters around the pit while the four related $2$ K underground cold
boxes will be installed remotely close to the respective cryomodule
string to be supplied as described above and shown in Figure
$\ref{fig:hauglr2}$.  The total dynamic cooling power of the ERL with
$944$ cavities amounts to $30$ kW $@\; 2$ K. For the calculation of the
cooling performances of the refrigerators in this document only the
largely dominating dynamic thermal loads of the cavities are taken
into account dwarfing all other thermal losses of the cryomodules
which become negligible in a first order approach.  Recent
developments and industrial design of large scale refrigerator systems
as for LHC \cite{Claudet} indicate the feasibility of a 1/COP of
$700$ W/W for $2$ K large scale cryoplants. Hence, with this figure
the total electric grid power amounts to $21$ MW. The total equivalent
refrigerator power at $4.5$ K is estimated to $80$ kW. This
corresponds to about half of the installed cooling power at LHC.  In case
contingencies are taken into account in the engineering design the
cooling capacity could approach LHC. For this preliminary study
contingencies are omitted, this also in view of expected future
improved cavity performances.  Eight cryoplants with $10$ kW $@\; 4.5$ K
each are proposed for the ERL. The technology to design and construct
such units as well as the overall systems engineering is largely
available today and can be based on experience from LHC, CEBAF,
XFEL. Nevertheless it consists of an engineering challenge due to its
sheer size and the large performance capacities required.  Development
work will have to be done for the cold compressors units together with
detailed combined CERN/industrial engineering design of the
refrigerator cold boxes. Implementation and operation of such large
systems will consist of a complex task. Further cavities and
cryomodules will require a limited R\&D program. From this we expect
improved quality factors with respect to today's state of the art. The
cryogenics of the L-R version consists of a formidable engineering
challenge, however, it is feasible and, CERN disposes of the
respective know-how.
     
\begin{table}[h]
  \centering
  \begin{tabular}{|c|c|}
    \hline
Parameter & Value\\
Number of Refrigerators  & $8$ \\
1/COP $@\; 2$ K & $700$ \\
Minimum cooling capacity/refrigerator & $10$ kW $@\; 4.5$ K\\
Contingency & none\\
Minimum total cooling power & $80$ kW $@\; 4.5$ K\\
Grid power consumption & $21$ MW\\
\hline
  \end{tabular}
\caption{
Refrigerator cooling capacity and power consumption (minimum cooling power).
}
\label{tab:hauglr2}
\end{table}

\subsection{General conclusions cryogenics for LHeC}

These conclusions reference to the complete cryogenic contributions,
i.e. for the detector cryogenics, the R-R and the L-R version;

The striking advantage of an extension from LHC to a LHeC lies, apart
from the new physics, in the comparatively small investment cost, the
possibility of quasi undisturbed continuation of LHC hadron physics
and the fact that the technologies are largely already at hand
today. This applies also to the cryogenic part. No so-called
``show-stoppers'' could be detected during these studies. For the
detector SC magnet and LArgon cryogenics technologies developed and
implemented at the ATLAS experiment can be used in a ``down-scaled''
way. For the accelerator cryogenics the two options Ring-Ring and
Linac-Ring differ strongly in principle and investment. While for the
R-R only four small to medium sized $2$ K refrigerators are required,
for the cryomodules of the injector and the three LHC tunnel bypasses,
the L-R option with two $1$ km long CW operated $2$ K SC cavities is
extremely demanding. The total installed cryogenic power will likely
exceed $100$ kW $@\; 4.5$ K equivalent, approaching values of the
LHC. However, these estimates are only based on currently proved data
of the cavity $Q_{0}$. The development of high Q SC cavities is being
pursued in several laboratories and new encouraging results are on the
horizon indicating improvement of quality having positive and direct
impact for cryogenic requirements and respective plant sizes.

%% file: machine/Bracco_Goddard_RR.tex
\subsection{Injection region design for Ring-Ring option}

A 10\,GeV recirculating Linac will be used to inject the electrons in
the LHeC.  This will be built on the surface or underground and a
transfer line will connect the linac to the LHeC injection region.  At
this stage a purely horizontal injection is considered, since this
will be easier to integrate into the accelerator.  The electron beam
will be injected in the bypass around ATLAS, with the baseline being
injection into a dispersion free region (at the right side of
ATLAS). Bunch-to-bucket injection is planned, as the individual bunch
intensities are easily reachable in the injector and accumulation is
not foreseen.  Two options are considered: a simple septum plus kicker
system where single bunches or short trains are injected directly onto
the closed orbit; and a mismatched injection, where the bunches are
injected with either a betatron or dispersion offset.

\subsubsection{Injection onto the closed orbit}

\begin{figure}[h]
\centerline{\includegraphics[clip=,width=0.55\textwidth,angle=90]{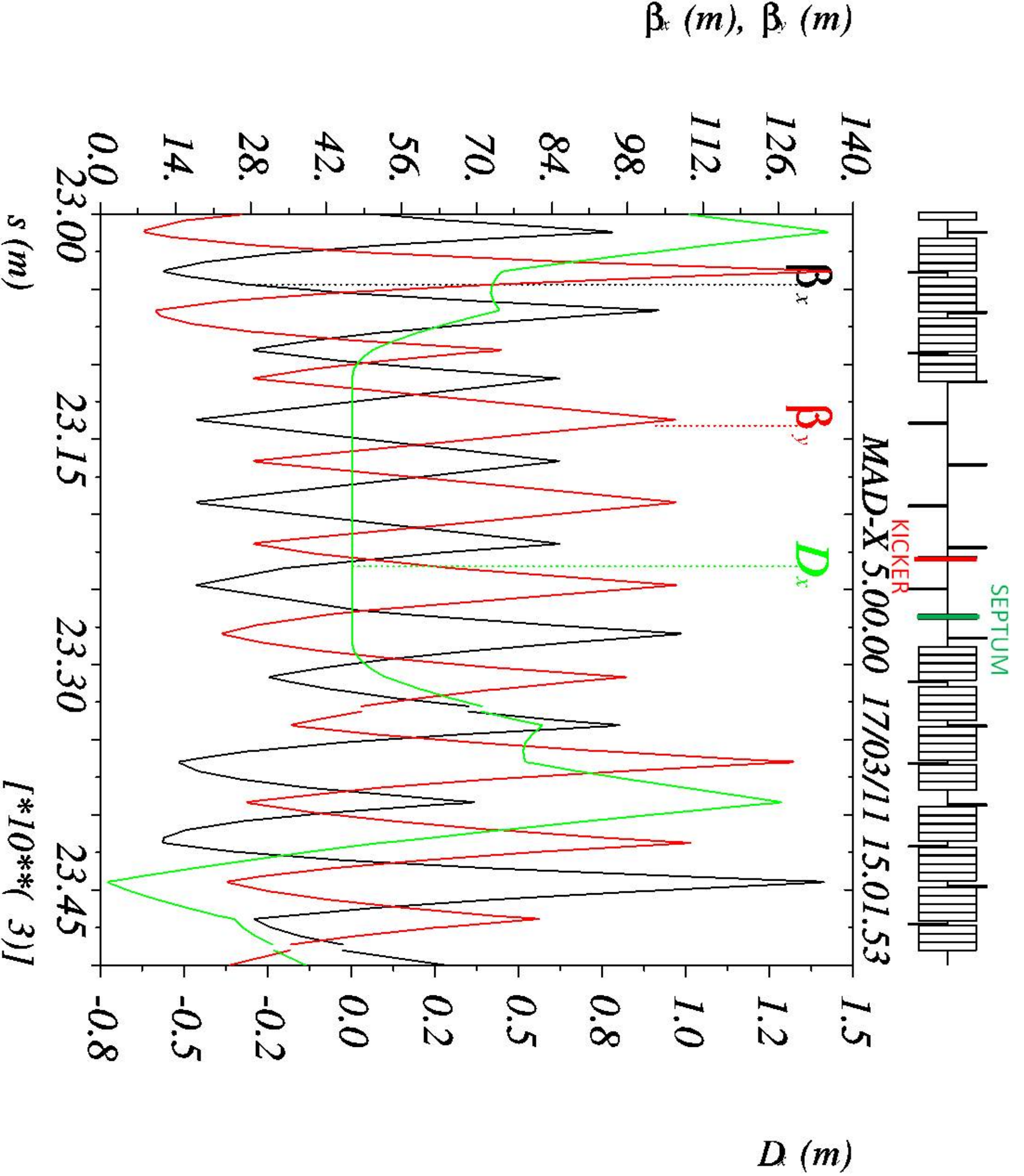}}
\caption{Injection optics is shown. The sequence starts (s=0) at the
  beginning of the dispersion suppressor at the left side of IP2 and
  proceeds clockwise, while the electron beam rotates counterclockwise
  (from right to left in the figure). The injection kicker and septum
  are installed in the dispersion free region of the bypass at the
  right side of ATLAS.}
\label{Fig:InjectionOptics}
\end{figure}

The baseline option is injection onto the orbit, where a kicker and a
septum would be installed in the dispersion free region at the right
side of ATLAS bypass (see Fig.\,\ref{Fig:InjectionOptics}). Injecting
the beam onto the closed orbit has the advantage that the extra
aperture requirements around the rest of the machine from injection
oscillations or mismatch are minimised.
The kicker and septum can be installed around a defocusing quadrupole
to minimise the kicker strength required.  The kicker-septum phase
advance is 75$^{\circ}$.

Some assumptions made to define the required element apertures are
made in Table\,\ref{tab:InjBeamPar}.

For the septum, an opening between injected and circulating beam of
47~mm is required, taking into account some pessimistic assumptions on
orbit, tolerances and with a 4~mm thick septum. This determines the
kicker strength of about 1 mrad.

 \begin{table}[htbp]
\begin{center}
\begin{tabular}{|c|c|}
\hline
Orbit variation & $\pm$ 4 mm \\
\hline
Injection precision & $\pm$ 3 mm \\
\hline
Mechanical/alignment tolerance & $\pm$ 1 mm \\
\hline
Horizontal normalised emittance $\varepsilon_{n,x}$ & 0.58 mm \\
\hline
Vertical normalised emittance $\varepsilon_{n,y}$ &  0.29 mm \\
\hline
Injection mismatch (on emittance) & 100 \% \\
\hline
$\beta_x$, $\beta_y$ @ Kicker & 61.3 m, 39.7 m  \\
\hline
$\beta_x$, $\beta_y$ @ Septum& 57.3 m, 42.3 m  \\
\hline
$\sigma_x$, $\sigma_y$ @ Kicker and Septum & 0.8 mm, 0.4 mm  \\
\hline
\end{tabular}
\end{center}
\caption{Assumptions for beam parameters used to define the septum and kicker apertures}
\label{tab:InjBeamPar}
\end{table}

The septum strength should be about 33~mrad to provide enough
clearance for the injected beam at the upstream lattice quadrupole,
the yoke of which is assumed to have a full width of 0.6~m. This
requires about 1.1~T~m, and a 3.0~m long magnet at about 0.37~T is
reasonable, of single turn coil construction with a vertical gap of
40~mm and a current of 12~kA.

The RF frequency of the linac is 1.3~GHz and a bunch spacing of 25~ns
is considered, as the LHeC electron beam bunch structure is assumed to
match with the LHC proton beam structure. Optimally a train of 72
bunches would be injected, which would require a 1.8~$\mu$s flattop
for the kickers and a very relaxed 0.9~$\mu$s rise time (as for the
LHC injection kickers\,\cite{Barnes:2006es}). However, this train
length is too long for the recirculating linac to produce, and so the
kicker rise time and fall time requirements are therefore assumed to
be about 23~ns, to allow for the bunch length and some jitter.

For a rise time $t_{m}$ = 23~ns, a system impedance $Z$ of 25~$\Omega$
is assumed, and a rather conservative system voltage $U$ of 60~kV.

Assuming a full vertical opening $h$ of 40 mm, and a full horizontal
opening $w$ of 60~mm (which allow $\pm$6 $\sigma$ beam envelopes with
pessimistic assumptions on various tolerances and orbit), the magnetic
length $l_{m}$ of the individual magnets is:

\[
l_{m} = h t_{m} Z / \mu_{0} w = 0.31~m
\]

For a terminated system the gap field B is simply:

\[
B = \frac{\mu_{0} U}{2 h Z} = 0.037~T
\]

As 0.03~Tm are required, the magnetic length should be 0.8~m, which
requires 3 magnets. Assuming each magnet is 0.5~m long, including
flanges and transitions the total installed kicker length is therefore
about 1.5~m.

\subsubsection{Mismatched injection}

\begin{figure}[h]
\centerline{\includegraphics[clip=,width=0.4\textwidth,angle=270]{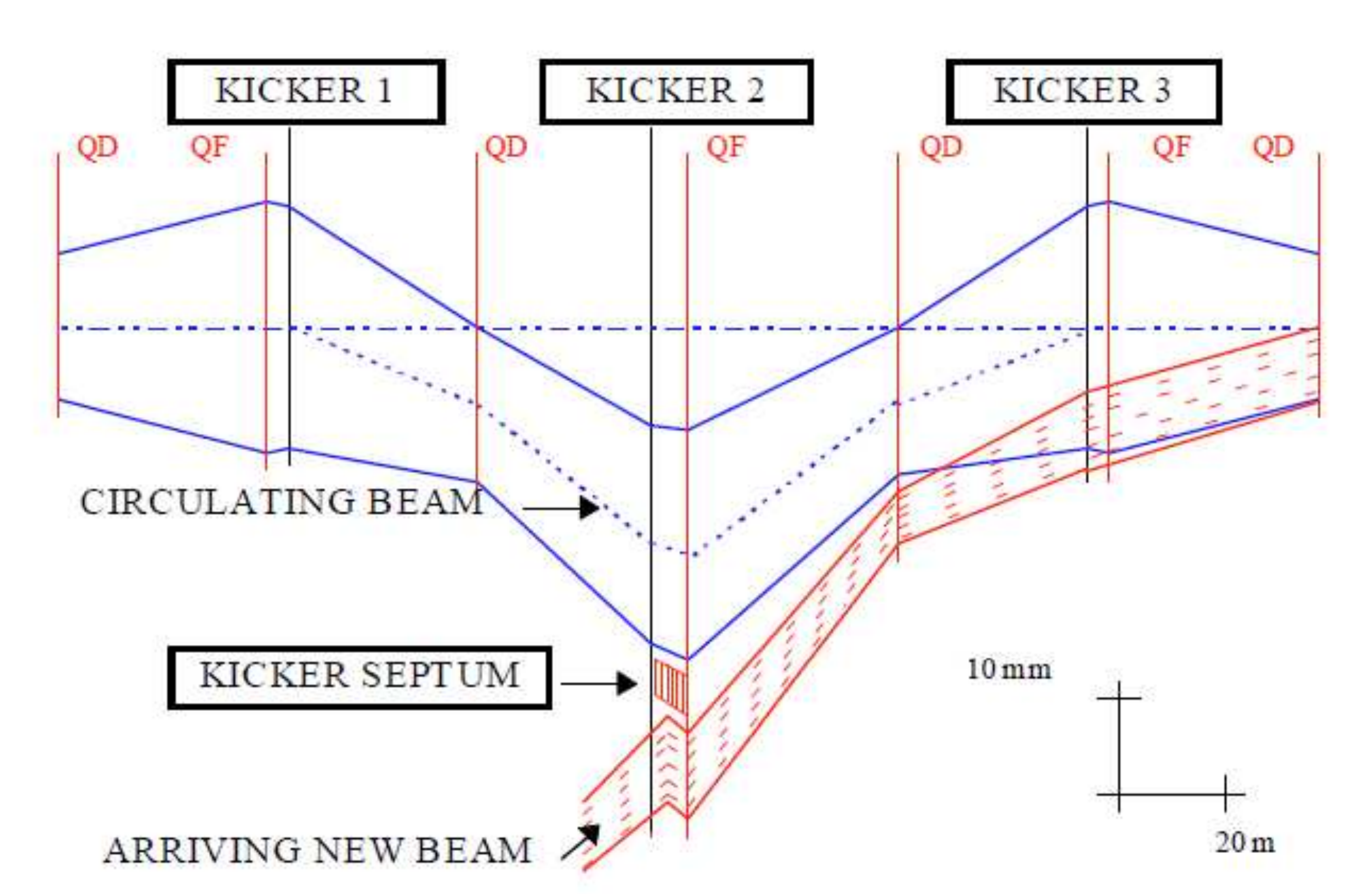}}
\caption{layout of mismatched injection system. To minimise kicker strengths the magnets are located
near focusing quadrupoles.}
\label{Fig:MisMatInj}
\end{figure}

A mismatched injection is also possible, Figure\,\ref{Fig:MisMatInj}
with a closed orbit bump used to bring the circulating beam orbit
close to the septum, and then switched off before the next circulating
bunch arrives.

The injected beam then performs damped betatron or synchrotron
oscillations, depending on the type of mismatch used. In LHeC the
damping time is about 3 seconds, so that to achieve the suggested
0.2~s period between injections, a damping wiggler would certainly be
needed - the design of such a wiggler needs to be investigated.

\begin{table}[htbp]
\begin{center}
\begin{tabular}{|c|c|c|}
\hline
Magnet  & $\theta_x$ [mrad]  & B dl [Tm] \\
\hline
KICKER1 & 1.35  & 0.04 \\
\hline
KICKER2 & 2.37  & 0.08  \\
\hline
KICKER3 & 0.55  & 0.02  \\
\hline 
\end{tabular}
\end{center}
\caption{Kickers strength and integrated magnetic field needed to generate an orbit bump of 20~mm at the injection point.}
\label{tab:Bumpers}
\end{table}

\begin{figure}[h]
\centerline{\includegraphics[clip=,width=0.6\textwidth,angle=90]{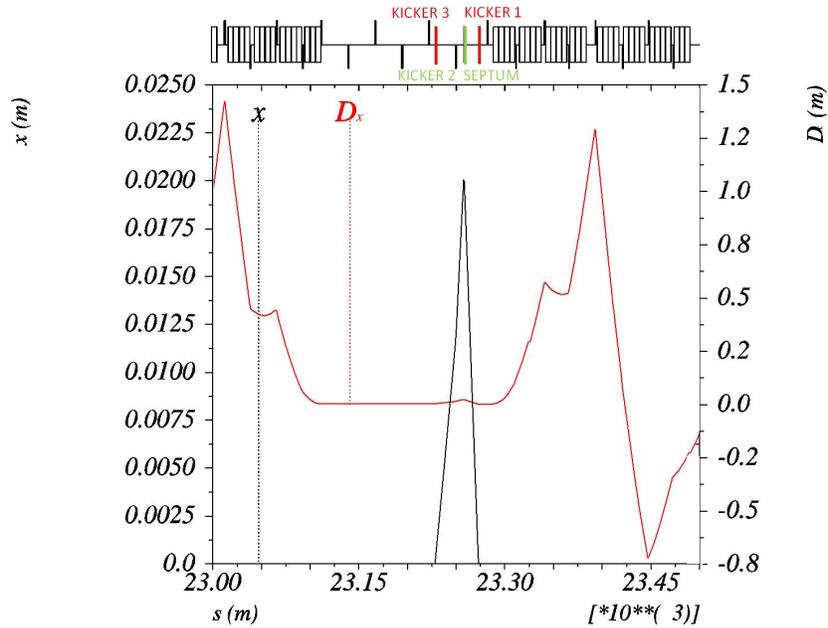}}
\caption{A closed orbit bump of 20~mm is generated by three kickers
  installed in the dispersion free region located at the right side of
  the bypass around ATLAS (electron beam moves from right to left in
  the Figure).}
\label{Fig:Dx_x}
\end{figure}

Three kickers (KICKER 1, KICKER 2 and KICKER 3 in
Fig.\,\ref{Fig:MisMatInj}) are used to generate a closed orbit bump of
20 mm at the injection point. The kicker parameters are summarised in
table\,\ref{tab:Bumpers}. In case of betatron mismatch, the bumpers
can be installed in the dispersion free region considered for the
injection onto the closed orbit case discussed in the previous section
(see Fig.\,\ref{Fig:Dx_x}).  The installed magnet lengths of the
kickers should be 2~m, 3.5~m and 1~m respectively, for the kickers
size,$Z$ and $U$ parameters given above.  Overall the kicker system is
not very different to the system needed to inject onto the orbit.

To allow for the possibility of synchrotron injection, the injection
kicker-septum would need to be located where the horizontal dispersion
$D_{x}$ is large. The beam is then injected with a position offset $x$
and a momentum offset $\delta p$, such that:
\[
x = D_{x} \delta p
\]
The beam then performs damped synchrotron oscillations around the
ring, which can have an advantage in terms of faster damping time and
also smaller orbit excursions in the long straight sections,
particularly experimental ones, where the dispersion functions are
small.

As an alternative to the fast (23~ns rise time) kicker for both types
of mismatched injection, the kicker rise- and fall-time could be
increased to almost a full turn, so that the bump is off when the
mismatched bunch arrives back at the septum. This relaxes considerably
the requirements on the injection kicker in terms of fall
time. However, this does introduce extra complexity in terms of
synchronising the individual kicker pulse lengths and waveform shapes,
since for the faster kicker once the synchronisation is reasonably
well corrected only the strengths need to be adjusted to close the
injection bump for the single bunch.

\subsection{Injection transfer line for the Ring-Ring Option}

The injection transfer line from the 10~GeV injection recirculating
linac is expected to be straightforward. A transfer line of about
900~m, constituted by 15 FODO cells, has been considered. The phase
advance of each cell corresponds to about 100$^{\circ}$.
\begin{figure}[htb]
\centerline{\includegraphics[clip=,width=0.6\textwidth,angle=90]{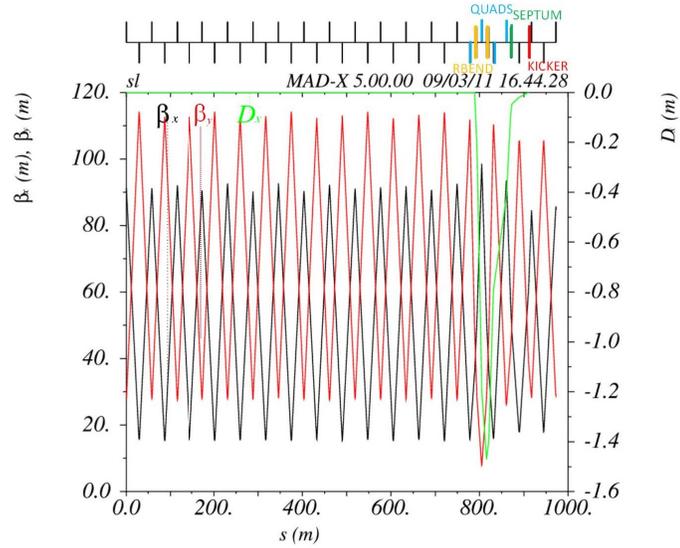}}
\centerline{\includegraphics[clip=,width=0.6\textwidth,angle=90]{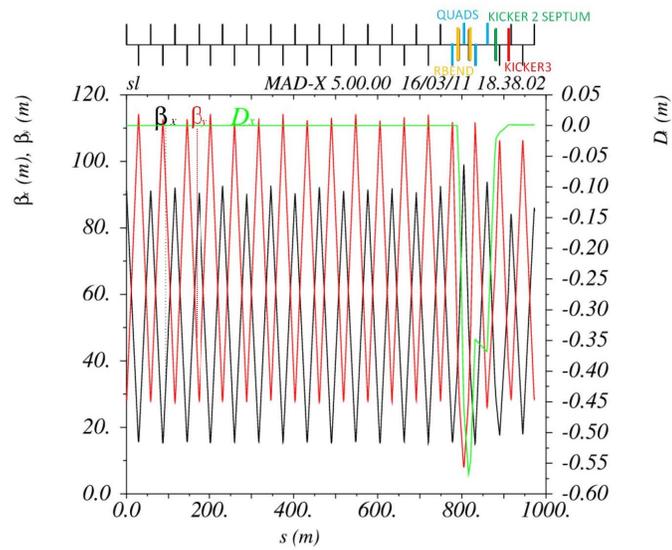}}
\caption{Transfer line optics for the injection onto orbit case (top)
  and mismatched injection case (bottom). }
\label{Fig:TL_1}
\end{figure}
\clearpage The last two cells are used for optics matching. In
particular, four quadrupoles, 1 m long each, are used for $\beta_x$
and $\beta_y$ matching, while two rectangular bending magnets, 5~m
long each, are used for matching the horizontal dispersion $D_x$ to 0
(maximum $D_x$ = -1.48~m for the injection onto closed orbit case and
maximum $D_x$ = -0.57~m for the mismatched injection case).  The
``good field region'' for a 6$\sigma$ beam envelope requires a minimum
half-aperture, in the matching insertion, of 15~mm and 10~mm for the
focusing and defocusing quadrupoles respectively, corresponding to a
pole tip field of about 0.02~T.  The maximum strength of the bending
magnets, which are used for dispersion matching, corresponds to about
39~mrad. This requires 1.3~T~m and a maximum field of 0.3 T. A single
turn coil of 9.5~kA with a vertical gap of 40~mm could be used.

\subsection{60 GeV internal dump for Ring-Ring Option}

An internal dump will be needed for electron beam abort.  The design
for LEP\,\cite{Carlier:1994st} consisted of a boron carbide spoiler
and an Aluminium alloy (6\% copper, low magnesium) absorbing block
(0.4~m~$\times$~0.4~m~$\times$~2.1~m long).  A fast kicker was used to
sweep eight bunches, of 8.3~$\times$~10$^{11}$ electrons at 100~GeV,
onto the absorber.  The first bunch was deflected by 65~mm and the
last by 45~mm, inducing a temperature increase $\Delta T$ of
165$^{\circ}$.

The bunch intensity for the LHeC is about a factor of 20 lower than
for LEP and beam size is double ($\sigma$ = 0.5~mm in LEP and $\sigma$
= 1~mm in LHeC).
       
\begin{figure}[htb]
\centerline{\includegraphics[clip=,width=0.5\textwidth,angle=90]{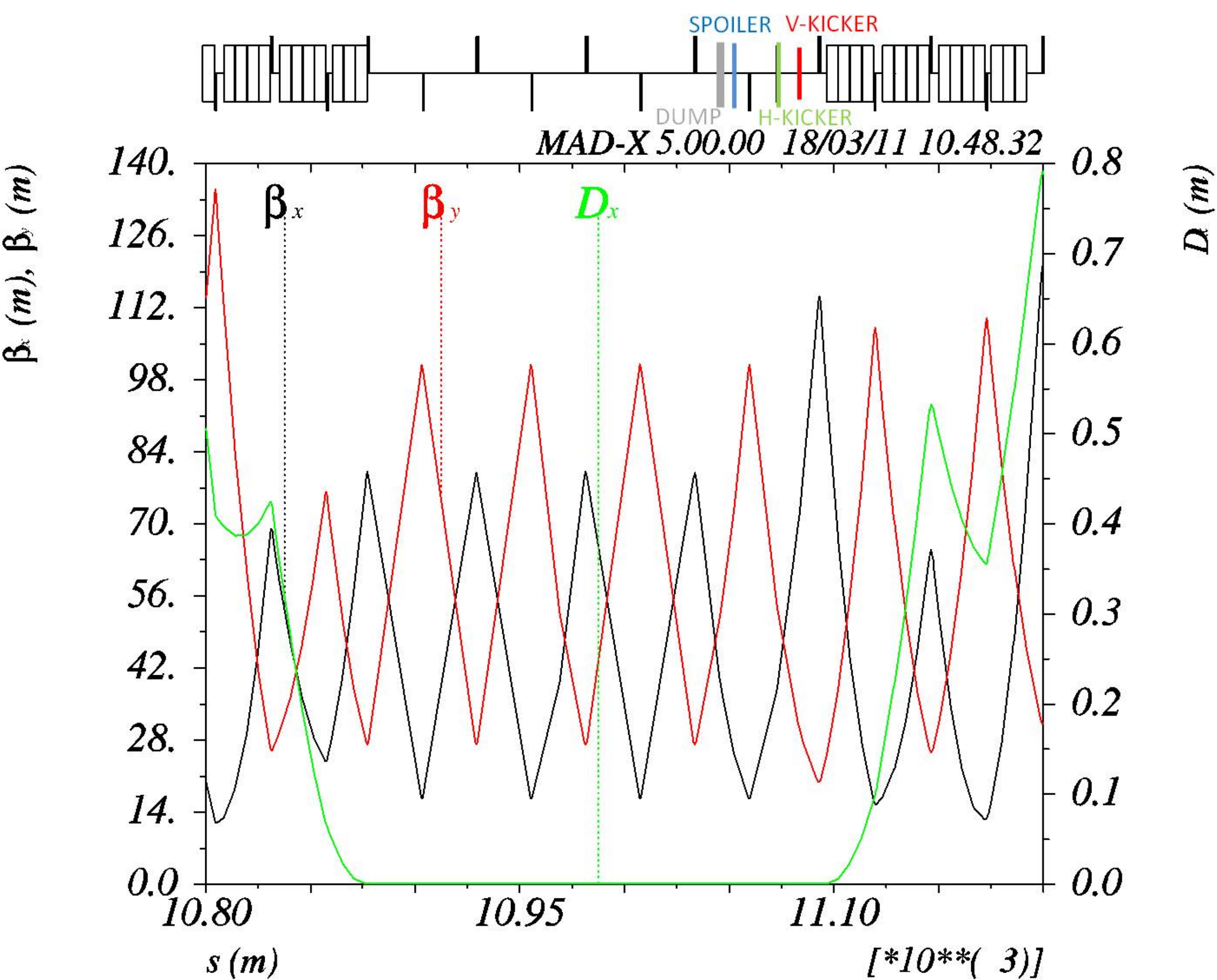}}
\caption{The optics in the region of the CMS bypass where the beam
  dump system could be installed is shown. The system consists of two
  kickers, one spoiler and a Carbon-composite absorber which are
  installed in the dispersion free region of the bypass at the right
  side of CMS (beam proceeds from right to left in the Figure).}
\label{Fig:Dump_Optics}
\end{figure}

The lower energy (60~GeV) and energy density permit to dump 160
bunches in 20~mm to obtain the same $\Delta T$ as for LEP. However, in
total LHeC will be filled with 2808 bunches, which means that
significant additional dilution will be required. A combination of a
horizontal and a vertical kicker magnet can be used, as an active
dilution system, to paint the beam on the absorber block and increase
the effective sweep length. The kickers and the dump can be located in
the bypass around CMS, in a dispersion free region (see
fig.\,\ref{Fig:Dump_Optics}).

It is envisaged to use Carbon-composite for the absorber block, since
this has much better thermal and mechanical properties than aluminium.
The required sweep length is then assumed to be about 100~mm, from
scaling of the LEP design.  The minimum sweep speed in this case is
about 0.6~mm per $\mu$s, which means about 54 bunches per mm. Taking
into account the energy and the beam size, this represents less than a
factor 2 higher energy density on the dump block, compared to the
average determined by the simple scaling, that should be feasible
using carbon. More detailed studies are required to optimise the
diluter and block designs. Vacuum containment, shielding and a water
cooling system has to be incorporated.  A beam profile monitor can be
implemented in front of each absorber to observe the correct
functioning of the beam dump system.
 
\begin{figure}[htb]
\centerline{\includegraphics[clip=,width=0.75\textwidth,angle=0]{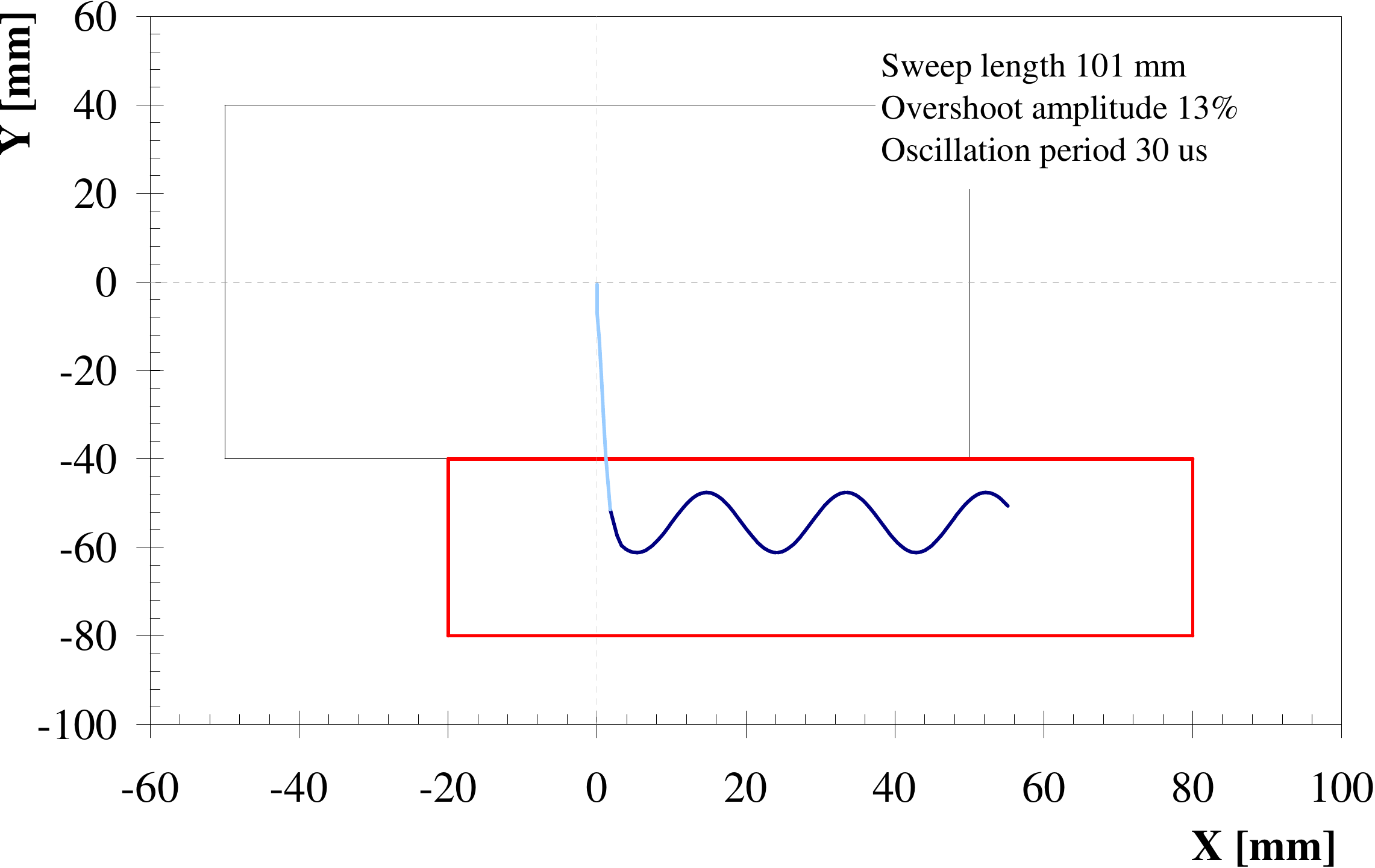}}
\caption{A vertical and a horizontal kicker are used to dilute the
  beam on the dump absorbing block.}
\label{Fig:DumpSweep}
\end{figure}

\begin{figure}[htb]
\centerline{\includegraphics[clip=,width=0.75\textwidth,angle=0]{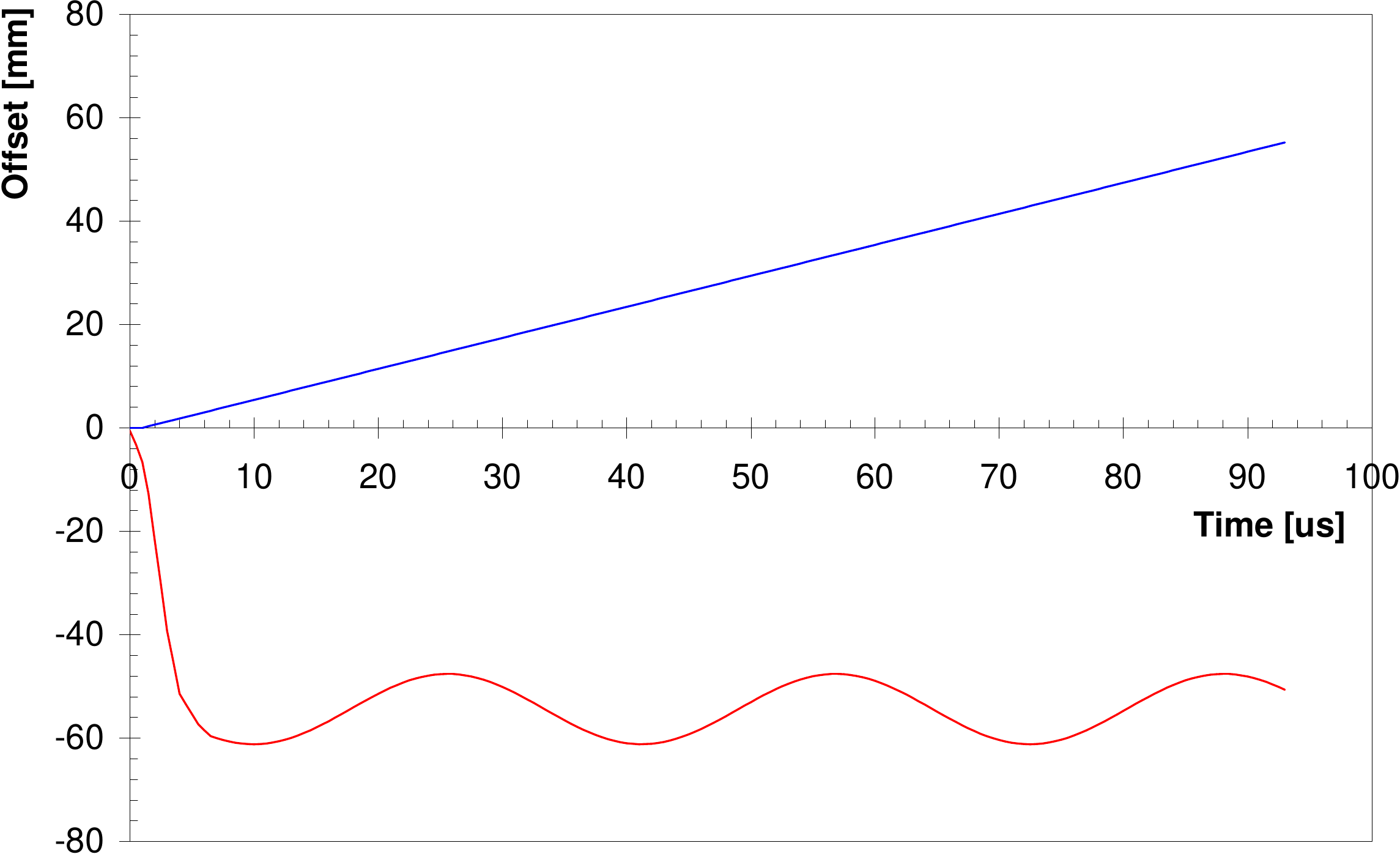}}
\caption{The strength of the vertical kicker oscillates in time by
  $\pm$ 13\% around its nominal value.The deflection provided by the
  horizontal kicker increases almost linearly in time.}
\label{Fig:DumpSweep2}
\end{figure}

The vertical kicker would provide a nominal deflection of about 55~mm
(see fig.\,\ref{Fig:DumpSweep}), modulated by $\pm$13\% for three
periods during the 100~$\mu$s abort (see fig.\,\ref{Fig:DumpSweep2}),
while the horizontal kicker strength would increase linearly from zero
to give a maximum deflection at the dump of about 55~mm (see
Fig.\,\ref{Fig:DumpSweep}and Fig.\,\ref{Fig:DumpSweep2}). This
corresponds to system kicks of 2.7 and 1.6~mrad respectively.

Parameters characterising the kicker magnets are presented in
Table\,\ref{tab:ExtrKickerPar}.

In the present lattice the dump is placed $\sim$30~m downstream of the
kickers, corresponding to a phase advance of about 63$^{\circ}$ in the
horizontal plane and 35$^{\circ}$ in the vertical plane. The minimum
horizontal and vertical aperture at the dump are 26~mm and 22~mm
respectively (at the dump: $\beta_x$ = 37~m and $\beta_y$ = 55~m,
using the same beam and machine parameter assumptions, as presented in
Table\,\ref{tab:InjBeamPar}).  The kicker system field rise time is
assumed to be at most 3~$\mu$s (abort gap) and the kicker field
flat-top at least 90~$\mu$s as for the LHC proton beam.  Same design
as for the LHC dump kicker magnets MKD can be used: a steel yoke with
a one-turn HV winding. These magnets can provide a magnetic field in
the gap of 0.34 T. For a magnetic length of 0.31~m ($Z$= 25~$\Omega$
and $U$ = 60~kV), a total installed kicker length of 1.5~m for the
horizontal system and 2.5~m for the vertical system has to be
considered.

\begin{table}[htbp]
\begin{center}
\begin{tabular}{|c|c|c|}
\hline
      &  MKDV & MKDH \\
\hline
Length [m]  &  2.5  & 1.5  \\
\hline
Maximum angle [mrad]   & 2.7  & 1.6 \\
\hline
Maximum field [T]   &  0.34   & 0.34  \\
\hline
Rise/Fall time [ns]  &  800  & 800 \\
\hline
Flat top length [$\mu$s] &  90  & 90 \\
\hline 
\end{tabular}
\end{center}
\caption{Parameters characterising vertical and horizontal kicker magnets of the extraction system.}
\label{tab:ExtrKickerPar}
\end{table}

A spoiler (one-side single graphite block:
0.3~m~$\times$~0.10~m~$\times$~0.5~m long) can be installed 5~m
upstream of the dump at the extraction side to provide further
dilution.

%% file: machine/Bracco_Goddard_LR.tex

\subsection{Post collision line for 140~GeV Linac-Ring option}

The post collision line for the 140~GeV Linac option has to be
designed taking care of minimising beam losses and irradiation.  The
production of Beamstrahlung photons and e$^-$e$^+$ pairs is
negligible and the energy spread limited to 2~$\times$~10$^{-4}$.  A
standard optics with FODO cells and a long field-free region allowing
the beam to naturally grow before reaching the dump can be foreseen.
The aperture of the post collision line is defined by the size of the
spent beam and, in particular, by its largest horizontal and vertical
angular divergence (to be calculated).  A system of collimators could
be used to keep losses below an acceptable level.  Strong quadrupoles
and/or kickers should be installed at the end of the line to dilute
the beam in order to reduce the energy deposition at the dump window.
Extraction line requirements:
\begin{itemize}
\item Acceptable radiation level in the tunnel.
\item Reasonably big transverse beam size at the dump window and energy dilution.
\item Beam line aperture big enough to host the beam: beta function and energy spread must be taken into account.
\item Elements of the beam line must have enough clearance.
\end{itemize}

\subsection{Absorber for 140~GeV Linac-Ring option}

Nominal operation with the 140~GeV Linac foresees to dump a 50~MW
beam.  This power corresponds to the average energy consumption of
69000 Europeans. An {\it Eco Dump} could be used to recover that
energy; detailed studies are needed and are not presented here.
Another option is to start from the concept of the ILC water dump and
scale it linearly to the LHeC requirements.  The ILC design is based
on a water dump with a vortex-like flow pattern and is rated for 18~MW
beam of electrons and positrons\,\cite{Appleby:2006fc}.  Cold
pressurised water (18~m$^3$ at 10~bar) flows transversely with respect
to the direction of the beam. The beam always encounters fresh water
and dissipates the energy into it. The heat is then transmitted
through heat exchangers.  Solid material plates(Cu or W) are placed
beyond the water vessel to absorb the tail of the beam energy spectrum
and reduce the total length of the dump. This layer is followed by a
stage of solid material, cooled by air natural convection and thermal
radiation to ambient, plus several metres of shielding.  The size of
the LHeC dump, including the shielding, should be 36~m longitudinally
and 21~m transversely and it should contain 36~m$^3$ of water.  The
water is separated from the vacuum of the extraction line by a thin
Titanium Alloy (Ti-6Al-4V) window which has high temperature strength
properties, low modulus of elasticity and low coefficient of thermal
expansion. The window is primarily cooled by forced convection to
water in order to reduce temperature rise and thermal stress during
the passage of the beam.  The window must be thin enough to minimise
the energy absorption and the beam spot size of the undisrupted beam
must be sufficiently large to prevent window damage. A combination of
active dilution and optical means, like strong quadrupoles or
increased length of the transfer line, can be use on this purpose.
Further studies and challenges related to the dump design are:
\begin{itemize}
\item Pressure wave formation and propagation into the water vessel.
\item Remotely operable window exchange.
\item Handling of tritium gas and tritiated water.      
\end{itemize}

\subsection{Energy deposition studies for the Linac-Ring option}
Preliminary estimates, of the maximum temperature increase in the
water and at the dump window, have been defined according to FLUKA
simulation results performed for the ILC dump\,\cite{Amann:2010zz}.  A
50~MW steady state power should induce a maximum temperature increase
$\Delta T$ of 90$^\circ$ corresponding to a peak temperature of
215$^\circ$. The water in the vessel should be kept at a pressure of
about 35~bar in order to insure a 25$^\circ$ margin from the water
boiling point.

FLUKA studies have been carried out for a 1~mm thick Ti window with a
hemispherical shape.  The beam size at the ILC window is
$\sigma_x$~=~2.42~mm and $\sigma_y$~=~0.27~mm; an extraction line with
170~m drift and 6~cm sweep radius for beam dilution have been
considered.  A beam power of 25~W with a maximum heat source of
21~W/cm$^3$ deposited on the window have been calculated. This
corresponds to a maximum temperature of 77$^\circ$ for the minimum
ionisation particle (dE/dx~=~2 MeV~$\times$~cm$^2$/g), no shower is
produced because the thickness of the window is significantly smaller
than the radiation length.  A maximum temperature lower than
100$^\circ$ would require a minimum beam size of
$\sigma_{x,y}$~=~1.8~mm. A minimum $\beta$ function of 8877~m would be
needed being the beam emittance $\varepsilon_{x,y}$~=~0.37~nm for the
undisrupted beam.  The radius of the dump window depends on the size
of the disrupted beam.  The emittance of the disrupted beam is
$\varepsilon_{x,y}$~=~0.74~nm corresponding to a beam size
$\sigma_{x,y}$ of 2.56~mm (for $\beta$~=~8877 m); a radius R~=~5~cm
could then fit a 10$\sigma$ envelope.  The yield strength of the Ti
alloy used for the window is $\sigma_{Ti}$~=~830~MPa, this, according
to the formula:

\begin{eqnarray}
\sigma_{Ti} = 0.49 \times \Delta P \frac{R^2}{d^2} 
\end{eqnarray} 

where $\Delta P$~=~3.5~MPa, imposes that the thickness of the window d
is bigger than 2.3~mm.

Length of the transfer line drift space and possible dilution have to
be estimated together with possible cooling.

\subsection{Beam line dump for ERL Linac-Ring option}
\begin{figure}[h]
\centerline{\includegraphics[clip=,width=0.6\textwidth]{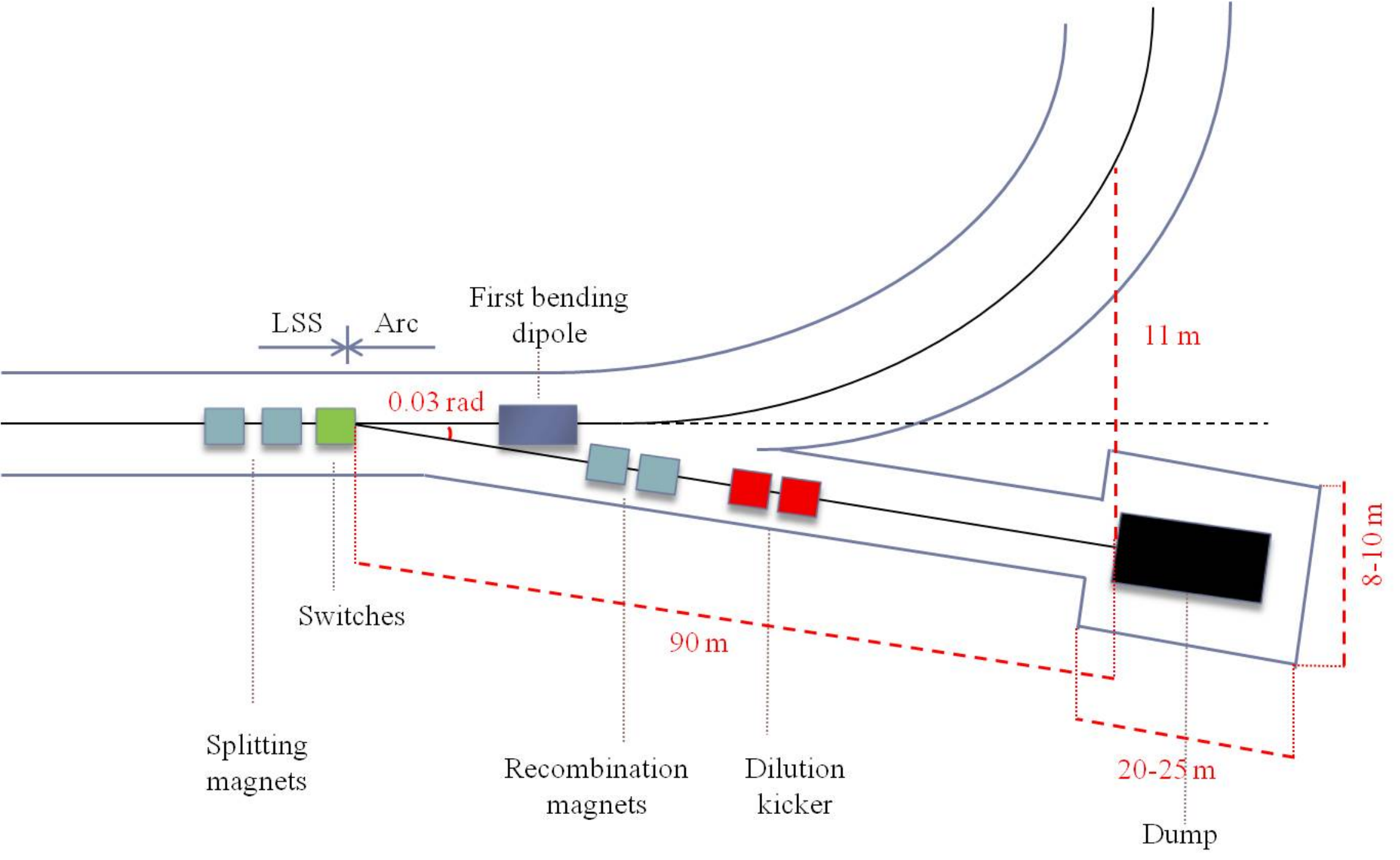}}
\caption{Scheme of the transfer line from end of long straight section of the linac and beam dump.}
\label{Fig:TransferLineERL}
\end{figure}

The main dump for the ERL Linac-ring option will be located downstream
of the interaction point.  Splitting magnets and switches have to be
installed in the extraction region and the extracted beam has to be
tilted away from the circulating beam by 0.03~rad to provide enough
clearance for the first bending dipole of the LHeC arc (see
Fig.\,\ref{Fig:TransferLineERL}).  A $90$\,m transfer line, containing
two recombination magnets and dilution kickers, is considered to be
installed between the LHeC and the LHC arcs. 
%
The beam dump will be housed in a UD62/UD68 like cavern at the end of
the TL and the option of having service caverns for water treatment
and heat exchange is explored.  An additional dump, and its extraction
line, could be installed at the end of the first linac
for beam setup purposes at intermediate
energy. The same design as for the nominal dump and extraction line
would be applied.
%


\subsection{Absorber for ERL Linac-Ring option}

During nominal operation a 0.5~GeV beam has to be dumped with a
current of 6.6~mA.  The setup beam will have a maximum current of
0.05~mA and an energy varying from 10~GeV to 60~GeV (10~GeV step
size).  Globally, a maximum beam power of 3~MW has to be dumped.  The
same design as for the 140~GeV option can be used by scaling
linearly. In this case, a 3~m$^3$ water dump (0.5~m diameter and 8~m
length) with a 3~m~$\times$~3~m~$\times$~10~m long shielding has to be
implemented.  No show stopper has been identified for the 18 MW ILC
dump, same considerations are valid in this less critical case.

%% file: machine/civilengineering.tex
\chapter{Civil Engineering and Services}
\label{chapter:civil}
\section{Overview}

Infrastructure costs for projects such as LHeC, typically represent
approximately one third of the overall budget. For this reason,
particular emphasis has been placed on Civil Engineering and Services
studies, to ensure a cost efficient conceptual design. This chapter
provides an overview of the designs adopted for the key infrastructure
cost driver, namely, civil engineering. The costs for the other
infrastructure items such as cooling \& ventilation, electrical
supply, transport \& installation will be pro-rated for the CDR and
studied in further detail during the next phase of the project.  For
the purposes of this conceptual design report, the Civil Engineering
(CE) studies have assumed that the Interaction Region (IR) for LHeC
will be at LHC Point 2, which currently houses the ALICE detector. As
far as possible, any surface facilities have been situated on existing
CERN land.  Both the Ring-Ring and Linac-Ring underground works will
be discussed in this chapter. Surface buildings/structures have not
been considered for the CDR.

\section{Location, geology and construction methods}

This section describes the general situation and geology that can be
expected for both the Ring-Ring and Linac Ring options.  

\subsection{Location}
\begin{figure}
\centerline{\includegraphics[angle=0,clip=,width=0.8\textwidth]{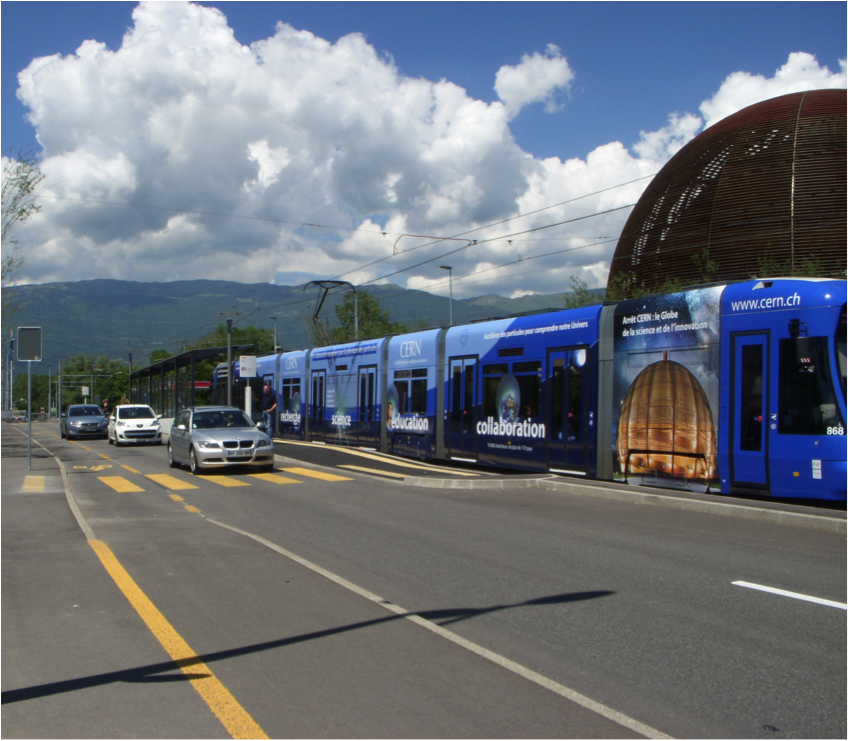}}
\caption{Tram stop outside CERN Meyrin Site.}
\label{fig:osborne1}
\end{figure}

The proposed siting for the LHeC project is in the North-Western part
of the Geneva region at the existing CERN laboratory. The proposed
Interaction Region is fully located within existing CERN land at LHC
Point 2, close to the village of St.Genis, in France.  The CERN area
is extremely well suited to housing such a large project, with the
very stable and well understood ground conditions having several
particle accelerators in the region for over 50 years. The civil
engineering works for the most recent machine, the LHC were completed
in 2005, so excellent geological records exist and have been utilised
for this study to minimise the costs and risk to the project. Any new
underground structures will be constructed in the stable Molasse rock
at a depth of 100-150m in an area with little seismic activity.  CERN
and the Geneva region have all the necessary infrastructure at their
disposal to accommodate such a project. Due to the fact that Geneva is
the home of many international organisations excellent transport and
communication networks already exist. Geneva Airport is only 5km from
the CERN site, with direct links and a newly constructed tramway,
shown in Figure $\ref{fig:osborne1}$, gives direct access from the
Meyrin Site to the city centre.

The governments of France and Switzerland have long standing
agreements concerning the support of particle accelerators in the
Geneva region, which make it very likely that the land could be made
available free of charge, as it was for previous CERN projects.  

\subsection{Land features}
The proposed location for the accelerator is situated
within the Swiss midlands embedded between the high mountain chains of
the Alps and the lower mountain chain of the Jura. CERN is situated at
the feet of the Jura mountain chain in a plain slightly inclined
towards the lake of Geneva. The surface terrain was shaped by the
Rhone glacier which once extended from the Alps to the valley of the
Rhone. The water of the area flows to the Mediterranean Sea. The
absolute altitude of the surface ranges from 430 to 500m with respect
to sea level.  The physical positioning for the project has been
developed based on the assumption that the maximum underground volume
possible should be housed within the Molasse Rock and should avoid as
much as possible any known geological faults or environmentally
sensitive areas. The shafts leading to any on-surface facilities have
been positioned in the least populated areas, however, as no real
discussions have taken place with the local authorities, the presented
layouts can only be regarded as indicative, for costing purposes only.

\subsection{Geology}
\begin{figure}
\centerline{\includegraphics[angle=0,clip=,width=0.8\textwidth]{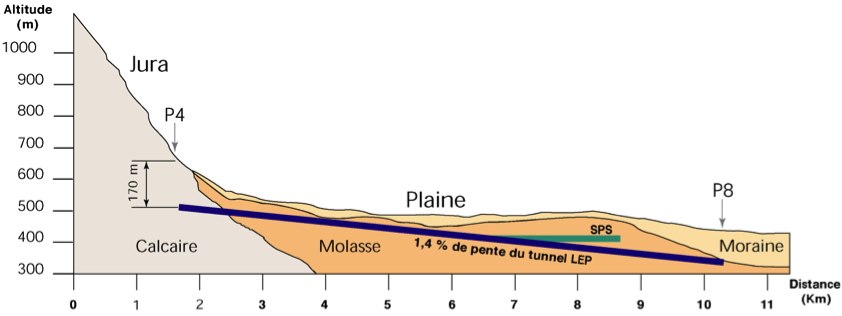}}
\caption{Simplified cross section of the LHC housed mostly in Molasse Rock}
\label{fig:osborne2}
\end{figure}

The LHeC project is within the Geneva Basin, a sub-basin of the large
North Alpine Foreland (or Molasse) Basin. This is a large basin which
extends along the entire Alpine Front from South-Eastern France to
Bavaria, and is infilled by Molasse deposits of Oligocene and Miocene
age. The basin is underlain by crystalline basement rocks and
formations of Triassic, Jurassic and Cretaceous age. The Molasse,
comprising an alternating sequence of marls and sandstones (and
formations of intermediate compositions) is overlain by Quaternary
glacial moraines related to the Wurmien and Rissien
glaciations. Figure $\ref{fig:osborne2}$ shows a simplified layout of
the LHC.
    
\subsection{Site development}
As most of the new works are on a close to existing facilities, it is
assumed for the CDR that the existing facilities such as restaurant,
main access, road network etc are sufficient and have not been
costed. However, for the parts located outside the existing fence line,
but within CERN property, the following items will have to be included
in the costs:

\begin{itemize}
\item  Roads and car parks. 
\item  Drainage networks.
\item  Landscaping and planting. 
\item  Spoil dumps. 
\end{itemize}

All temporary facilities needed for the construction works have also
been included in the cost estimate.  

\subsection{Construction methods}
It is envisaged that Tunnel Boring Machines (TBMs) will be utilised
for the main tunnel excavation greater than approximately 2km in
length. In the Molasse rock, a shielded TBM will be utilised, with
single pass pre-cast segmental lining, followed by injection grouting
behind the lining. For planning and costing exercises, an average TBM
advancement of 25m per day, or 150m per week is predicted.

\begin{figure}
\centerline{
\includegraphics[angle=0,clip=,width=0.45\textwidth]{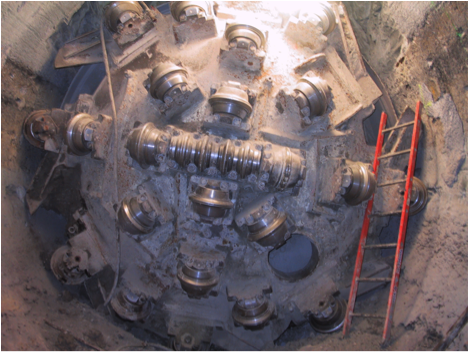}
\includegraphics[angle=0,clip=,width=0.45\textwidth]{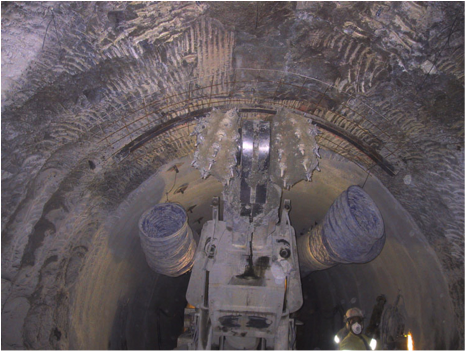}
}
\caption{ TBM Gripper type machine used for Neutrino tunnel at CERN  (left) and roadheader type machine (right).}
\label{fig:osborne3}
\end{figure}

\begin{figure}
\centerline{
\includegraphics[angle=0,clip=,width=0.45\textwidth]{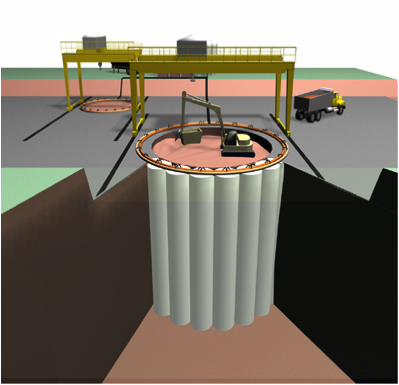}
\includegraphics[angle=0,clip=,width=0.45\textwidth]{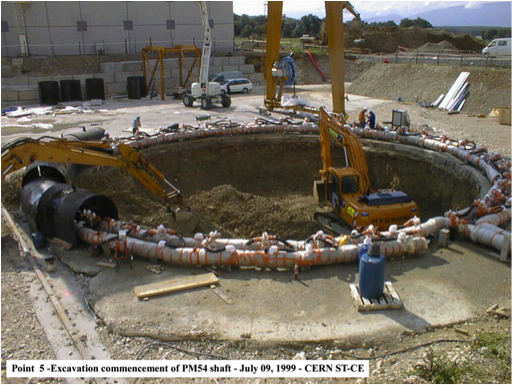}
}
\caption{LHC Shaft PM54, linking up cylinders of ice to construct a temporary wall.}
\label{fig:osborne4}
\end{figure}

The second phase excavation will be executed using a roadheader type
machine. Both machine types are shown in Figure $\ref{fig:osborne3}$.
Any new shafts that have to pass through substantial layers of water
bearing moraines (for example at CMS) will have to utilise the ground
freezing technique. This involves freezing the ground with a primary
cooling circuit using ammonia and a secondary circuit using brine at
-23C, circulating in vertical tubes in pre-drilled holes at 1.5 metre
intervals. This frozen wall allows excavation of the shafts in dry
ground conditions and also acts as a retaining wall. Figure
$\ref{fig:osborne4}$ shows this method being utilised for LHC shaft
excavation at CMS. 

\section{Civil engineering layouts for Ring-Ring}

The Ring-Ring solution will require new bypass tunnels at both 
Point 5 (currently housing the CMS detector)
 and Point 1 (ATLAS). Both of
the bypass tunnels are on the outside of the LHC ring.

\begin{figure}
\centerline{\includegraphics[angle=0,clip=,width=0.9\textwidth]{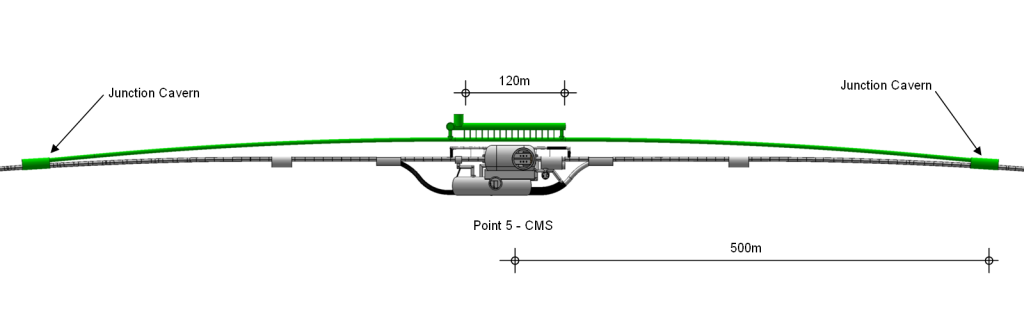}}
\caption{Ring-Ring Bypass around CMS Point 5.}
\label{fig:bypcms}
\end{figure}

The Bypass around CMS Point 5 is 1km long with an internal tunnel
diameter of 4.5m.  Only one new shaft is required for excavation
works. A roadheader type machine will be used for excavation, with the
new tunnel position as close as possible to the LHC tunnel as not to
induce movements or create operational problems to the existing
facilities. Figure $\ref{fig:bypcms}$ shows the new bypass tunnel
and service cavern required around CMS.

\begin{figure}
\centerline{\includegraphics[angle=0,clip=,width=0.9\textwidth]{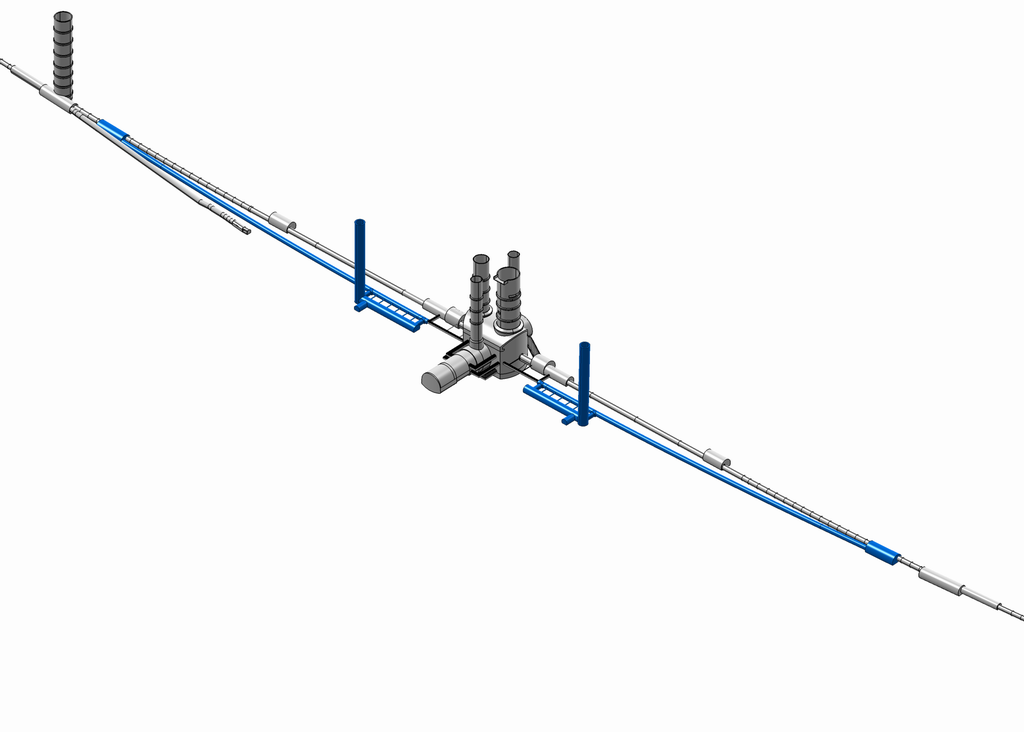}}
\caption{Ring-Ring Bypass around ATLAS Point 1.}
\label{fig:bypatlas}
\end{figure}

Figure $\ref{fig:bypatlas}$ shows the bypass tunnel in blue needed around
Point\,1. This tunnel is 730\,m long and has an internal diameter of
4.5\,m. Two new 7\,m diameter shafts are required to allow access to
construct the underground areas with minimum disruption to LHC
operations. Underground areas are made available for RF/Cryogenic and
general services. Two junction caverns will be excavated to create a
liaison with the LHC tunnel.

\begin{figure}
\centerline{\includegraphics[angle=0,clip=,width=0.8\textwidth]{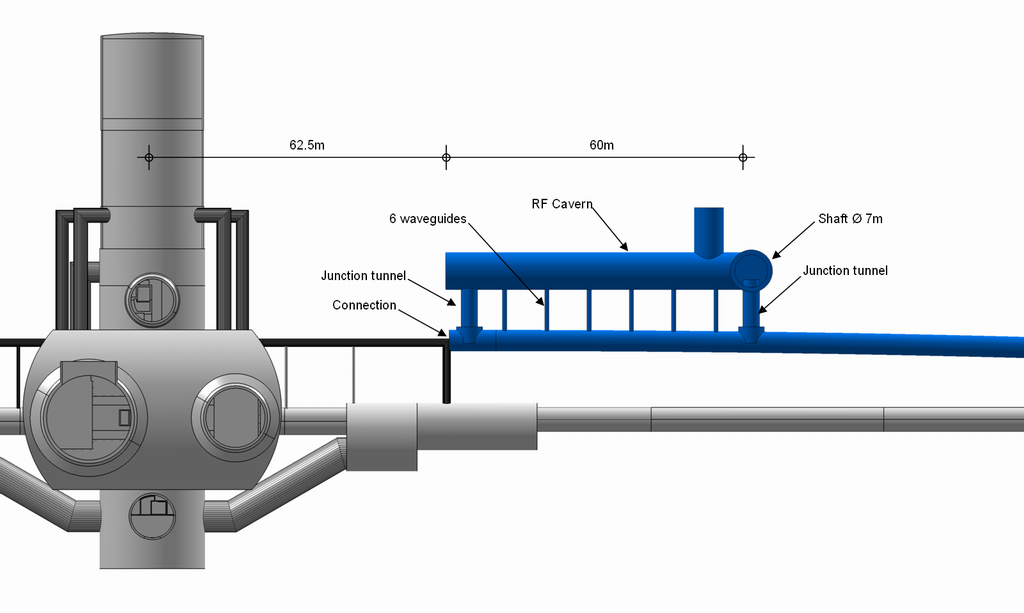}}
\caption{Cryo and RF Cavern (one side only) at Point 1.}
\label{fig:Acaverne}
\end{figure}

Waveguides ducts (0.9\,m diameter) will connect the LHeC Bypass tunnel
to the RF cavern, as shown in Figure $\ref{fig:Acaverne}$. In order to
position the bypass as close as possible to the LHC ring, it has been
assumed that the LHeC beam pipe can be accommodated within the existing
survey gallery, and pass through the ATLAS experimental hall.

\begin{figure}
\centerline{\includegraphics[angle=0,clip=,width=0.7\textwidth]{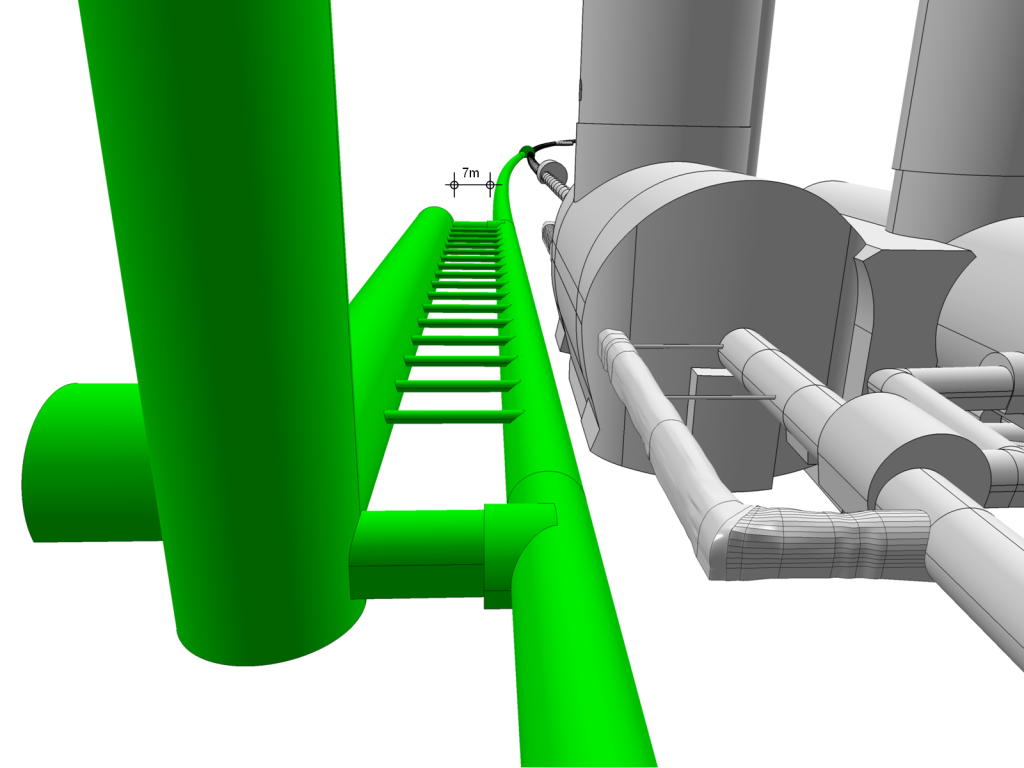}}
\caption{
3d model of Ring-Ring Bypass around CMS Point 5.
}
\label{fig:osborne8}
\end{figure}

Figure $\ref{fig:osborne8}$ shows a 3d model of the bypass around the
CMS Point 5. The new excavations will have a minimum of 7m of Molasse
rock separating the new works from existing LHC structures. This is to
avoid any unwanted deformation or vibration problems on the existing
LHC structures.
     
The civil engineering
for the electron beam injection complex for the Ring-Ring option has not been
studied for the CDR. 

\section{Civil engineering layouts for Linac-Ring}

\begin{figure}
\centerline{\includegraphics[angle=0,clip=,width=0.7\textwidth]{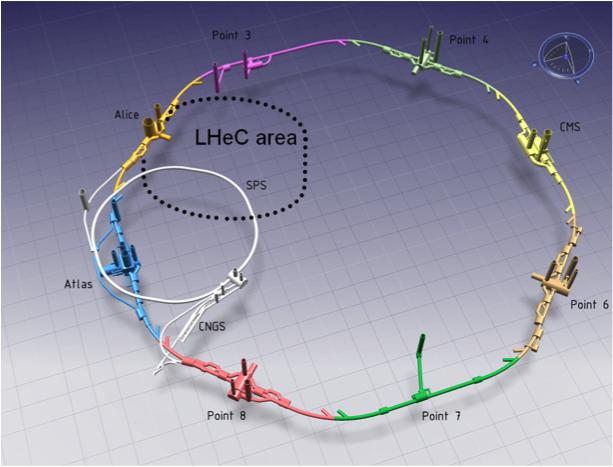}}
\caption{
Schematic model of ERL position injecting into IP2.
}
\label{fig:osborne9}
\end{figure}

For the CDR it has been assumed that the 60\,GeV Energy Recovery Linac
(ERL) will be located around the St.Genis area of France, injecting
directly into the LHC ALICE Cavern at Point 2. Approximately 10\,km of
new tunnels (5\,m and 6\,m diameter), 2 shafts and 9 caverns will be
required. The majority of civil engineering works can be completed
while LHC is operational.  Figure $\ref{fig:osborne9}$ highlights the
area on the LHC where the new ERL will be situated.

\begin{figure}
\centerline{\includegraphics[angle=0,clip=,width=0.9\textwidth]{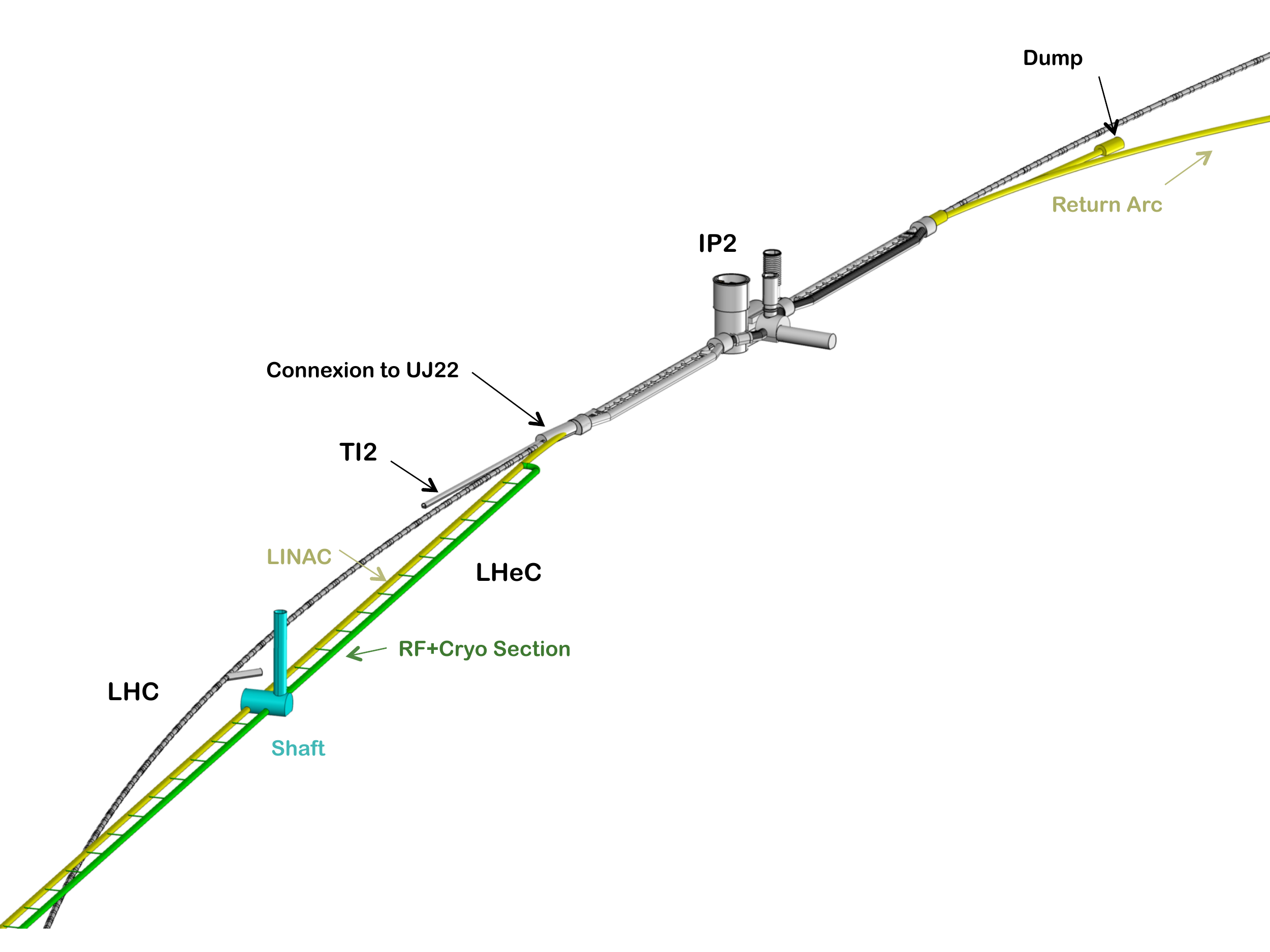}}
\caption{
ERL injection area into IP2 and RF/Cryo/Services Cavern (yellow \& green).
Dump (yellow) between the LHC and the return arc.
}
\label{fig:osborne12}
\end{figure}

\begin{figure}
\centerline{\includegraphics[angle=0,clip=,width=1.1\textwidth]{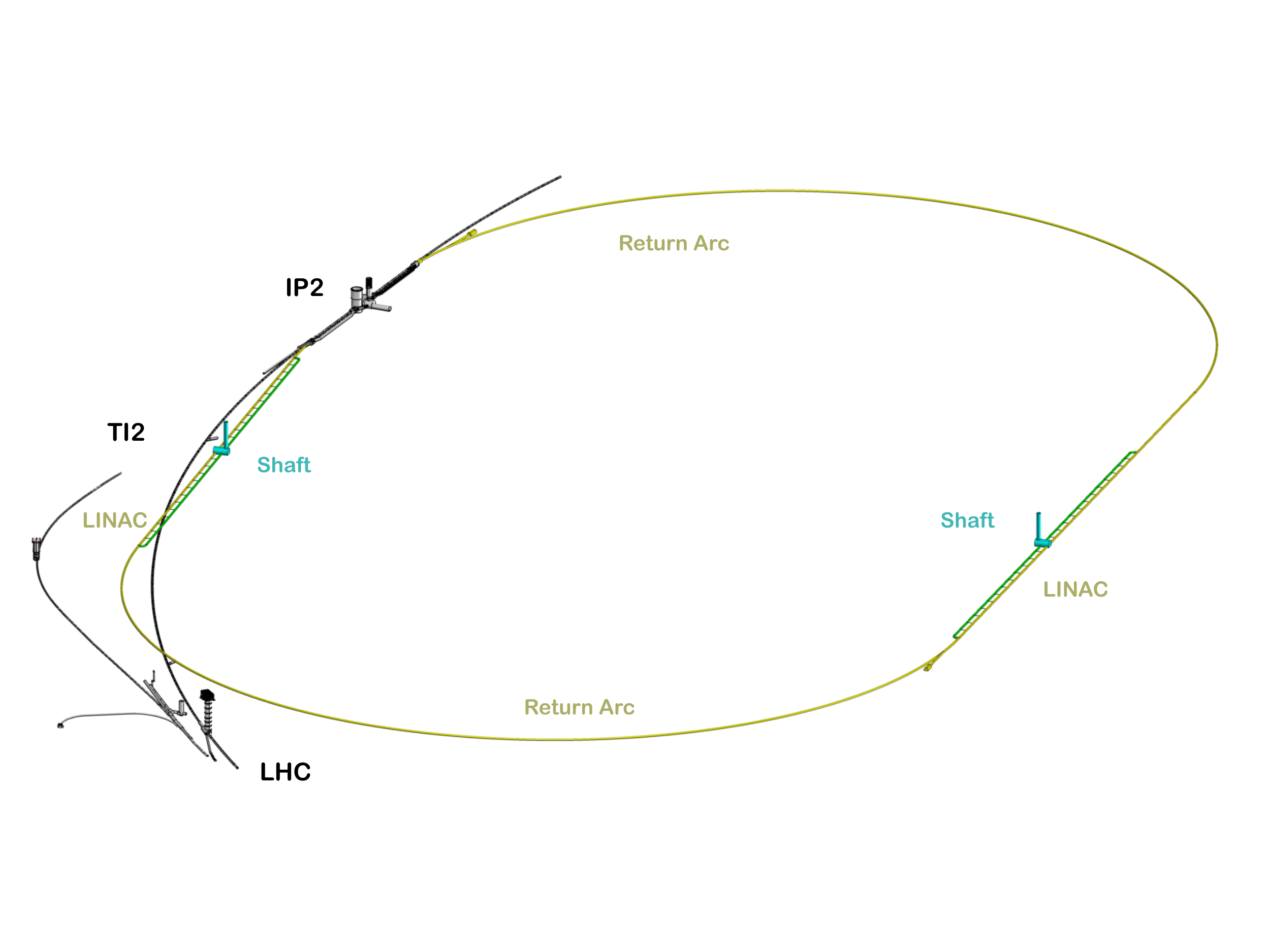}}
\caption{
View on the ERL placed inside the LHC ring and tangential to IP2.
TI2 is the injection line into the LHC. The electron beam enters
IP2 from the IP1 side. Behind IP2 is a dump between the return arc
and the LHC. There is also a dump after the first LINAC
for injection studies. The beam is injected into the right
LINAC (not shown).  The LINACs are about $1$\,km long 
and comprise about $60$ cavity-cryo modules each.
The return arcs have about $1$\,km radius and
are passed three times. The whole racetrack configuration
is about $9$\,km long such that the electron beam has
$1/3$ of the length of the LHC proton beam.
It is tentatively assumed that one access
shaft per LINAC was sufficient for supplies and civil engineering.}
\label{fig:liview}
\end{figure}

The ERL will be positioned inside the LHC Ring, in order to ensure
that new surface facilities are located, as much as possible, on
existing CERN land.  Secondary tunnels running alongside the long
straight sections will house RF, Cryogenic and Services for the
machine. One of the long straight sections is shown in 
Figure~\ref{fig:osborne12}. The entire ERL, illustrated 
in Figure~\ref{fig:liview}, will be tilted in order to follow
a suitable layer of Molasse rock. On average the ERL will be tilted
approximately 1.4\%, dipping towards Lake Geneva, as per LHC.

\section{Summary}

From a civil engineering point of view, both the Ring-Ring and
Linac-Ring options are feasible. The Ring-Ring option will provide a
cheaper solution, however, with a marginally increased risk to LHC
activity, due to the fact that most of the excavation works being in
close proximity to the existing installations.  The Linac-Ring option
is the cleaner solution from a civil engineering point of view, with
much less risk to LHC, but with substantial extra cost and greater
time needed for environmental and building permit procedures.

%% file: machine/planning.tex
\chapter{Project Planning}

\begin{figure}
\centerline{\includegraphics[clip=,angle=90.,width=0.9\textwidth]{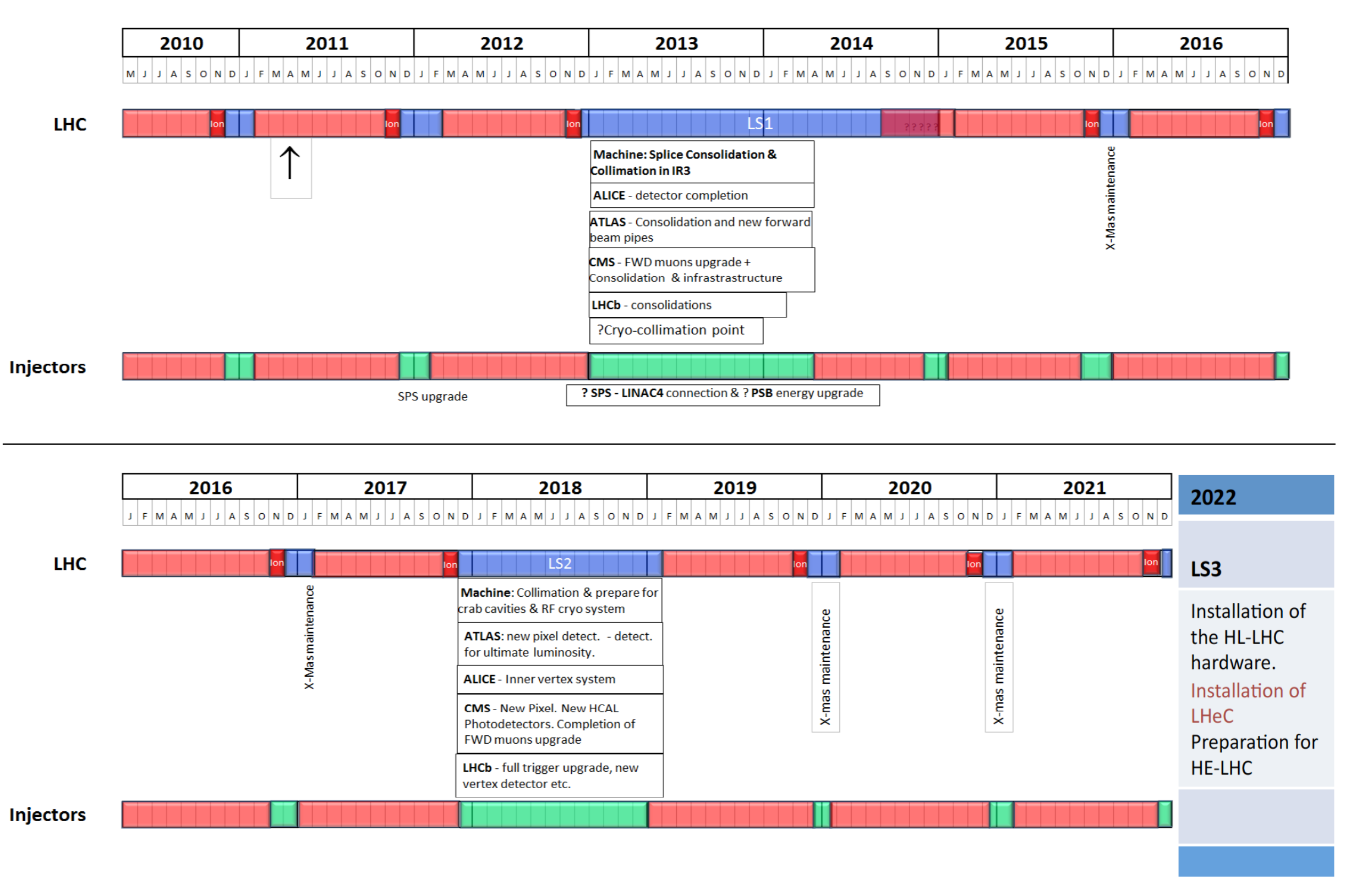}}
\caption{CERN medium term plan (MTP), draft as of July 2011, from~\cite{steveeps}.}
\label{CERN-MTP}
\end{figure}

\begin{figure}
\begin{center}
\begin{tabular}{l r}
\includegraphics*[width=0.49\textwidth]{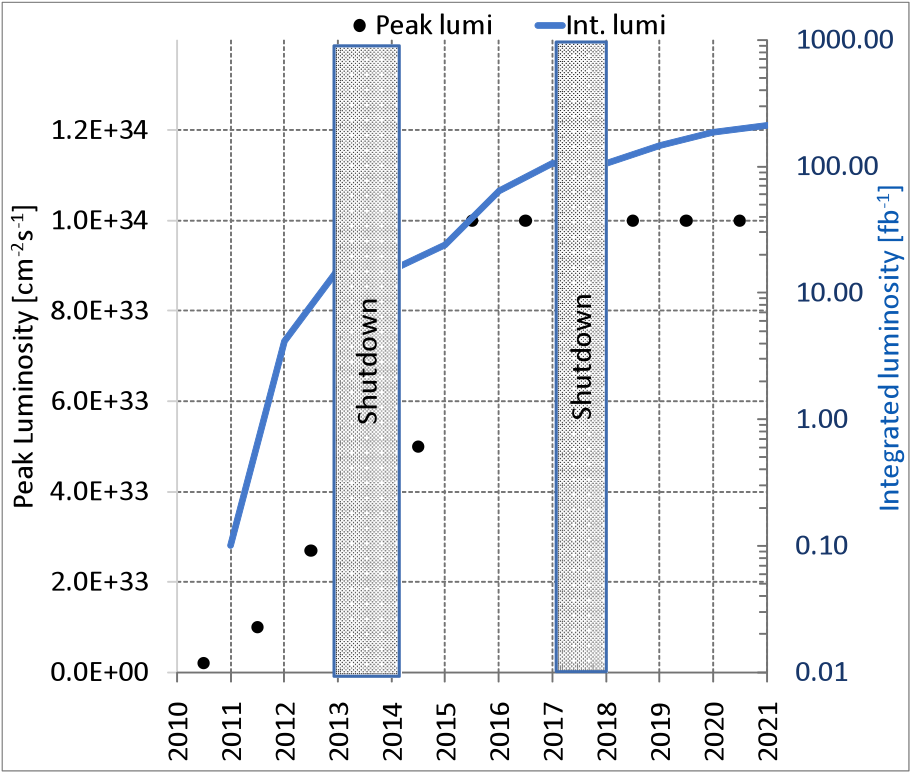}
\includegraphics*[width=0.49\textwidth]{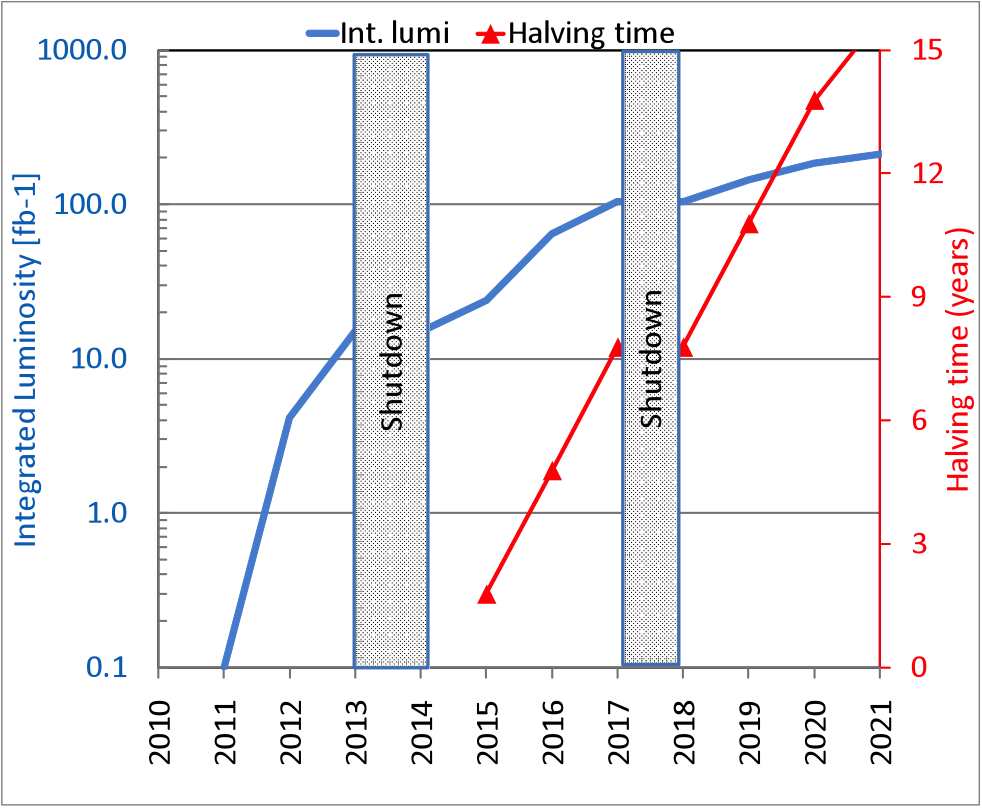}
\end{tabular}
\end{center}
\caption{Left: Projected luminosity evolution for the LHC assuming the
  LHC reaches nominal performance levels after the first long shutdown
  (LS1) and then remains at nominal performance after 2016. Right: The
  resulting evolution of the integrated luminosity for the LHC
  experiments. \cite{HL-LHC-design-study}.}
\label{nominal-lumi-evolution}
\end{figure}

\begin{figure}
\begin{center}
\begin{tabular}{l r}
\includegraphics*[width=0.49\textwidth]{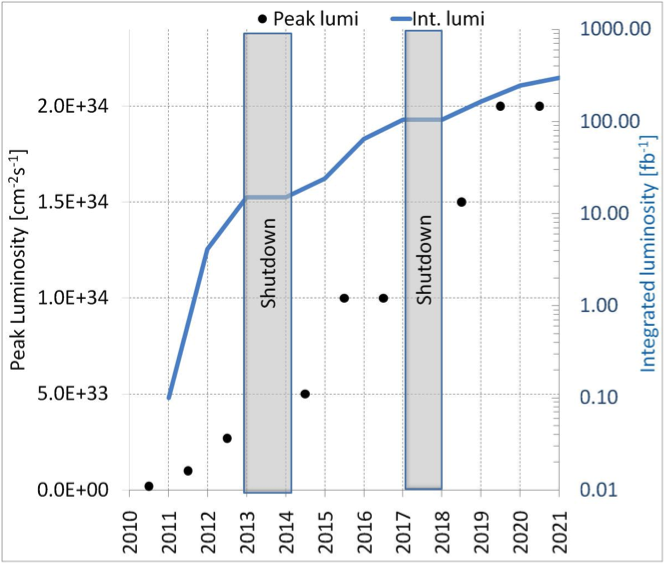}
\includegraphics*[width=0.49\textwidth]{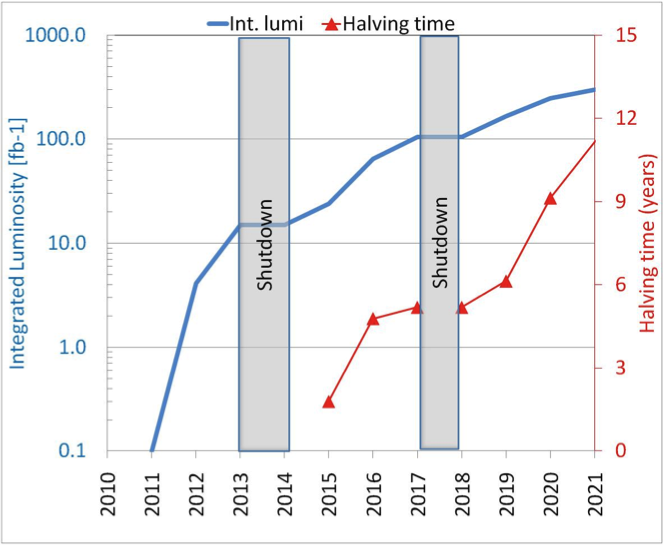}
\end{tabular}
\end{center}
\caption{Left: Optimistic projection of the luminosity evolution for
  the LHC assuming the LHC reaches ultimate performance levels after
  the first long shutdown (LS2). Right: The resulting evolution of the
  integrated luminosity for the LHC experiments. \cite{HL-LHC-design-study}.}
\label{ultimate-lumi-evolution}
\end{figure}

We base the planning of the LHeC project on the assumption that the
LHC machine will reach the end of its lifetime when the High
Luminosity LHC project reaches its design goal of $3000
fb^{-1}$. Figure~\ref{CERN-MTP} shows the current status of the CERN
planning for the LHC related upgrade projects. The current planning
foresees three long shutdowns:
\begin{itemize}
\item Long Shutdown 1 (LS1) for repairing the faulty splice
  connections in the LHC and allowing operations at nominal energy of
  7 TeV.
\item Long Shutdown 2 (LS2) for consolidating the LHC for operation
  above nominal beam intensities
\item Long Shutdown 3 (LS3) for implementing the HL-LHC upgrade
  installations.
\end{itemize}
Figure~\ref{nominal-lumi-evolution} shows the resulting evolution of
the integrated luminosity per experiment over time assuming the LHC
performance stabilises at nominal luminosity after LS1.
Figure~\ref{ultimate-lumi-evolution} shows a similar evolution of the
integrated luminosity assuming the LHC performance stabilises at
ultimate luminosity after LS1.

In both scenarios, the LHC reaches a total integrated luminosity of
ca. $200 fb^{-1}$ before LS3 and the installation of the HL-LHC
upgrade. The HL-LHC project aims at a generation of $200 fb^{-1}$ to
$300 fb^{-1}$ per year \cite{HL-LHC-design-study} and one can assume
that the HL-LHC design goal can be reached by between 9 and 13 years
after the LS3.  Assuming a one year long shutdown for LS3, this
implies the accumulation of $3000 fb^{-1}$ by ca. 2030 to 2035. Aiming
for the LHeC at an exploitation time of 10 years the LHeC operation
should therefore start together with the HL-LHC operation after the
LS3 in 2022.

We base our estimates for the project time line on the experience of
other projects, such as (LEP, LHC and LINAC4 at CERN and the European
XFEL at DESY and the PSI XFEL). In the following we will analyse
separately the required time line for the project construction for the
RF system development, the production of the magnet system, the required
civil engineering and the installation of the accelerator components
in the tunnel.

The superconducting RF development for LEP and LHC both required
approximately 2 to 3 years for the cavity prototyping and testing and
approximately 5 to 6 years of test stand operation of the
superconducting RF cavity modules adding up to a total time of
approximately 6 to 8 years from first prototype to final
installation. The first LHC cavity prototypes were constructed in 2000
with a final installation of the 4 cryo modules in the LHC tunnel in
2006. The first LEP super conducting RF cavity was tested in LEP in
1991. LEP2 operation started in 1996 but still required 2 years of
progressively commissioning all cryo modules in building B180 before
their final installation in the LEP tunnel. The last cryo module of
the 73 4-cell LEP cryo modules was installed in the LEP tunnel in
1999. Both RF installations featured extensive test stand
operations. The LEP RF system had cavity test stands in building SM18
and a separate power test in building B180 which were
operated from 1994 until 1999. The LHC RF system had both, the cavity
and the power test stands, in SM18. The LHC test stands were operated
from 2002 until 2006 (the test stand operation was slowed down at the
end due to difficulties with the RF coupler design). In both cases,
LEP and LHC, the RF system installation was therefore accompanied by a
5 to 6 year test stand operation which overlapped with the actual
installation period in the tunnel \cite{EdCiapalapc}.

The LHeC linac-ring RF system requires 118 cryomodules of eight
721~MHz 5-cell superconducting RF structures, amounting to a total of
approximately 950 structures or thirteen times the number of LEP RF
structures. It seems therefore reasonable to assume for the LHeC
linac-ring RF system a total time of 10 years from first prototype
construction to final installation in the tunnel with a dedicated test
stand operation for approximately 8 years. \footnote{Faster production
  rates could be possible by using several manufacturers in parallel
  as it is, for example, planned for the ILC. The ILC project requires
  approximately 15000 cavities and aims at a 10 to 15 times faster
  production rate as compared to the XFEL cavity production. But such
  an approach requires long preparation studies for the
  industrialisation (the ILC assumes more than 3 years for such
  studies \cite{Yamamoto}), dedicated production test facilities (the
  ILC has production test facilities at three different laboratories:
  DESY, KEK and FNAL), an extensive pre-series production and test
  bench operation for verifying the cavity and cryomodule design
  before launching the mass production (the ILC project has more than
  20 years experience of pre-series production and test bench
  operation in form of the TTF, FLASH and XFEL installations) and a
  large production volume so that it is lucrative for several
  manufacturers to split the overall production while still
  undertaking significant investments for the production lines. Such
  an approach may not  apply to a 'small' project like the LHeC
  and may therefore not lead to a much faster production time line.
}  The LHeC ring-ring RF system corresponds
approximately to the LEPII RF system in terms of total power and
overall length of the RF installation and it seems reasonable to
assume for the LHeC ring-ring RF system a slightly shorter time
scale. Here we assume the same time scale as for LEPII: a total time
of 8 years from first prototype construction to final installation in
the tunnel with a dedicated test stand operation for approximately 6
years.

For the magnet system we base a first order estimate of the required
timescale for the magnet production and installation on the experience
with LHC transfer lines. The LHC transfer lines have a total length of
6~km and feature a total of ca. 350 normal conducting magnets. The
magnet production extended over 3 years with a production rate of
ca. 10 magnets per month \cite{Volkerpc}. It is, however, important to
underline that the production rate was not limited by production
capacity but rather, was following the project requirements and the
CERN ability for magnet testing after reception at CERN. Both LHeC
options feature a relatively large number of magnets, approximately
4000 magnets. Compared to the LHC transfer line magnets, these magnets
are much more compact and one can assume that the magnet production
rate can be significantly larger than that for the LHC transfer
lines. The LHeC magnet production requires therefore industrial
production rates featuring several contractors and production
lines. The price to pay for such an industrial production scheme will
be the requirement for a pre-series production and a thorough quality
assurance over the whole production process. All LHeC magnets will
require furthermore a detailed geometry and field quality measurement
program after reception at CERN. In the following we assume 1-2 years
for the pre-series production and first testing followed by potential
design modifications and a peak production rate of ca. 60 dipoles and
20 quadrupoles per month (ca. ten times the production rate of the LHC
transfer lines). These assumptions lead to a total construction time
of ca. 4 to 6 years and a total of 6 to 8 years from magnet design to
final installation in the tunnel.

For the civil engineering we base our first order estimate for the
time line on the estimates for the CLIC 500~GeV option which features
a total length that is comparable to the 60~GeV linac-ring option. The
civil engineering work requires for the LHeC linac-ring option the
construction of ca. 10~km underground installations which is estimated
to take approximately 4 years construction time (the required
underground construction for the ring-ring solution is smaller but
will occur in the direct vicinity of the main LHC tunnel). The
installation of the technical infrastructure (water, electricity etc.)
will take approximately 2 years and the final installation of the
machine elements in the tunnel another 2 years. All three activities
can partially overlap, leading to an estimate of the total
construction time of ca. 6 years \cite{John-Osbornepc}.

For all other components (cryogenics, injector complex, detector etc.)
we assume for the moment that their development and installation can
be done in the shadow of the three components mentioned above.

In summary, we estimate:
\begin{itemize}
\item Between 8 and 10 years for the production of the RF system (time
  from prototype to final installation in the tunnel) with dedicated
  test stand operation over 6 to 8 years.
\item Between 6 and 8 years for the production of the magnet system
  (time from prototype to final installation in the tunnel) with
  several production lines and test facilities for the quality
  assurance during the magnet production.
\item Approximately 6 years for the civil engineering work and actual
  installation in the tunnel.
\item All other components such as injector complex, cryogenics
  installation, detector construction etc, are assumed to lie in the
  shadow of the above components.
\end{itemize}

The above time estimates appear as reasonable estimates compared to
the planning of other projects like the European XFEL at DESY, the
European Spallation Source (ESS) in Sweden, LINAC4 at CERN and the PSI
XFEL facilities:

\begin{itemize}
\item The European XFEL project features a 3~km long superconducting
  linear accelerator (comparable in size to the linac section of the
  LHeC linac-ring option) started the civil engineering in January
  2009 and plans for completing the civil engineering work in end
  2012 ($\rightarrow$ 4 years of bare civil engineering work)
  \cite{XFEL1}. The project had in form of the FLASH (TTF)
  installation a pre-series production of 150 1.3~GHz 9-cell cavity
  modules that went from 1993 to 2005 (12 years) and an
  extended test stand operation. The XFEL project plans for an
  industrial production of more than 600 1.3~GHz 9-cell cavity module
  from 2010 until 2014 (4 to 5 year production time) \cite{XFEL2}.
\item The ESS facility features ca. 300~m superconducting RF sections
  and plans for a construction phase of 9 years (2009 until 2017) with
  first operation in 2018 and full performance reach in 2025
  \cite{ESS}.
\item The LINAC4 project is a ca. 200~m long normal conducting linac
  installation which has a ca. 3 year long civil engineering
  construction period, followed by one year of infrastructure
  installation and 1.5 years of waveguide and accelerator component
  installation, amounting to a total construction period of ca. 5.5
  years (start of civil engineering in beginning 2008 and end of the
  accelerator installation by mid 2013) which seems rather long
  compared to the civil engineering estimates for the LHeC
  (installation length of ca. 10~km and ca. 100~m underground; ca. 50
  times the LINAC4 installation length which is mainly above surface)
  \cite{LINAC4}.
\item The PSI XFEL project features an approximately 1~km long normal
  conducting linac and plans for 2 years for the generation of a TDR,
  a 5 year test stand operation, a 4 year construction period and an
  installation period of 3 years leading to a total project time line
  of 6 years from start of the test facilities to the start of the
  actual project \cite{PSI-XFEL}.
\end{itemize}

Except for the European XFEL project, which has a longer
superconducting RF section than both LHeC versions, all of the above
reference facilities are smaller in scale than the LHeC project and
plan between 6 and 9 years from beginning of construction (civil
engineering) until the start of operation. All facilities with
superconducting cavities plan for an RF production time of ca. 5 years
for their key components and a substantial period of test bench
operation and pre-series production for critical elements (5 years or
more).

\begin{figure}
\begin{center}
\includegraphics*[width=1.0\textwidth]{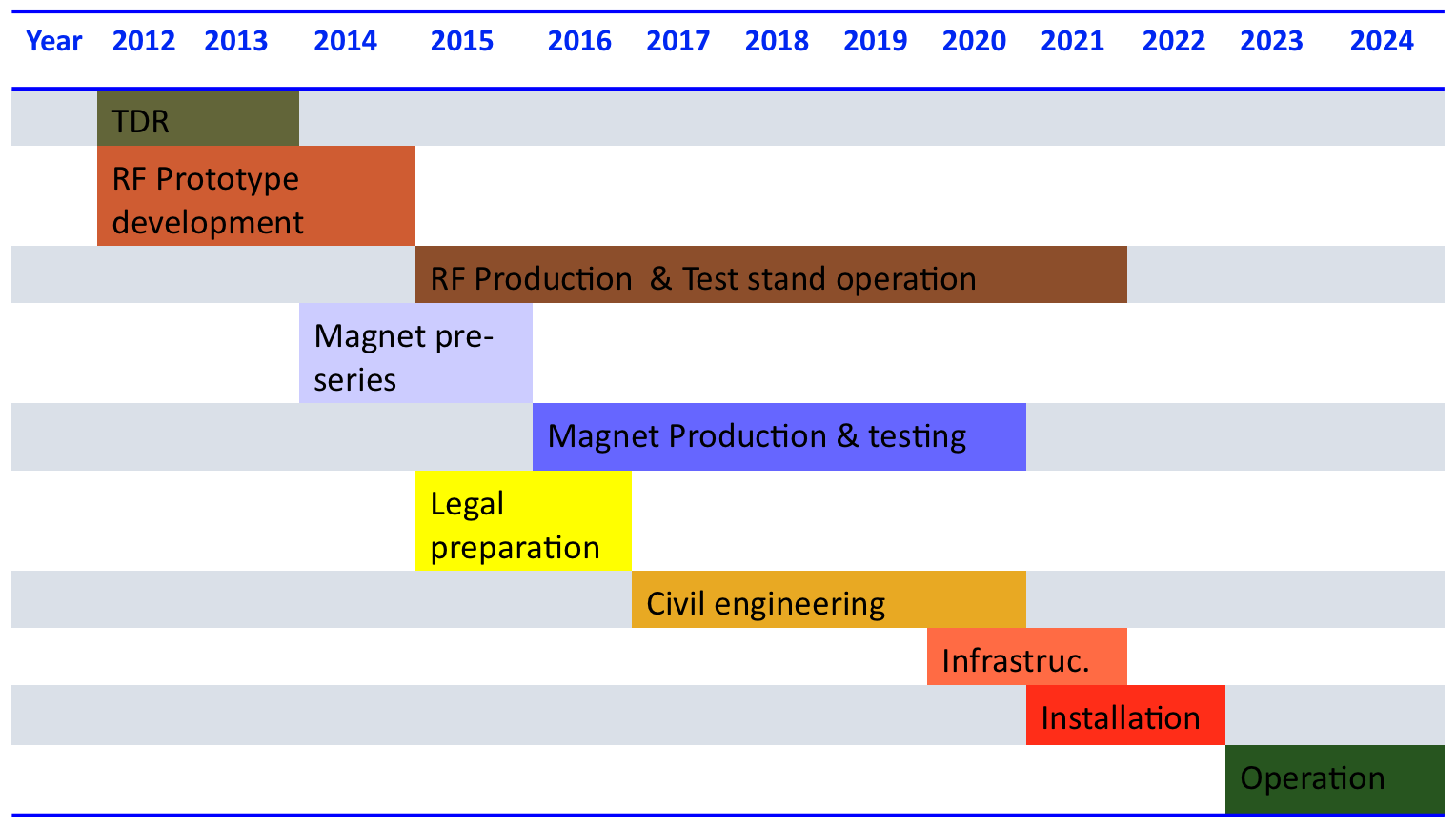}
\end{center}
\caption{Planning considerations for the LHeC, where we assumed a
  partial overlap of the time lines for the various LHeC project steps
  (for example a partial overlap of the civil engineering for the
  tunnel construction and the installation of the technical
  infrastructure and accelerator components). The overall planning
  goal of completion by the LS3 seems quite ambitious even with such a
  partial overlap of individual activities and requires first
  prototype development as soon as by 2012. The presented planning
  discusses only aspects related to the accelerator complex and does
  not address additional constraints coming from the detector
  installation in the cavern.}
\label{LHeC planning}
\end{figure}

Figure \ref{LHeC planning} summarises the above considerations in form
of a schematic outline of the project planning.  The planning in
Fig.~\ref{LHeC planning} addresses only aspects related to the
accelerator complex and does not address additional constraints coming
from the detector installation in the cavern. Furthermore, it does not
include additional constraints arising from the LHC operation,
logistics constraints and resource limitations due to the planning for
the long shutdowns of the LHC and does therefore certainly not attempt
to be an accurate project projection. Rather than presenting an
accurate timeline for the LHeC installation, the presented planning
aims at illustrating that a start of the LHeC operation in 2023
requires the start of first prototype development and testing already
by 2012. Meeting the milestone of an LHeC operation start in 2023
requires a rather swift project launch starting with the generation of
a proper TDR and the launch of first RF R\&D activities by 2012. This
ambitious goal can only be achieved if the project receives adequate
resource allocations in 2012. Potential first activities for the
prototype development and testing could focus around the development
of superconducting RF cavities, where synergies with ESS and SPL
studies exist, with the goal of setting up an ERL test facility. It
could also include the development of electron and positron sources
where synergies with the CLIC and ILC projects exist. Because of their
synergies with the ESS, SPL and the linear collider projects, a start
of R\&D activities for the LHeC by 2012 appears to be quite timely. In
case the Ring-Ring installation turns out to be the better option for
the LHeC, a ERL test facility could in the end also serve as an
injector complex for the Ring-Ring option of the LHeC. It represents
therefore a reasonable investment into the LHeC project independent of
a the final implementation choice.

%% file: detector/preamble.tex

In this chapter the core aspects of the main detector
design for the LHeC are discussed. 
The physics requirements are
illustrated along with the boundary conditions from the
accelerator options and the interaction region design.
These considerations converge
in Section\,\ref{LHEC:MainDetector} where a first picture of the main detector
is presented along with a discussion on the choice for the detector
elements and the overall detector assembly.
Detector components not located in close proximity to the interaction region are
described in Chapter\,\ref{LHEC:Detector:ForwardBackward}.
A first scenario describing how to assemble and install the detector in the LHC IP2 cavern is 
presented in Chapter\,\ref{LHEC:Detector:Assembly+Integration}.

%% file: detector/detr.tex

\label{LHEC:Detector:Requirements}
The new $ep/A$ detector at the LHeC has to be a precision instrument
with maximum acceptance.  The physics program depends on a high level
of precision, such as for the measurement of $\alpha_s$, and in the
reconstruction of complex final states, like charged current single
top production.
The detector acceptance has to extend as close as possible to the beam
axis in order to explore the physics at both low and high Bjorken $x$.
The dimensions of the detector are constrained by the radial extension
of the beam pipe in combination with maximum polar angle coverage\,
\footnote{
The $x$ and $y$ coordinates are defined such that there is a right
handed coordinate system formed with $y$ pointing upwards and $x$ to
the centre of the proton ring.}, preferably down to about $1^{\circ}$
and $179^{\circ}$ for forward going final state particles and backward
scattered electrons, respectively.  A further general demand is a high
modularity enabling much of the detector construction to be performed
above ground to keep the installation time to a minimum, and to be
able to access inner detector components within reasonable shutdown
times.

The time schedule of the project demands to have a detector ready
within about ten years.  This prevents any significant R\&D program to
be performed. Fortunately this is not required, and the vast
experience obtained at HERA, the LHC (including its upcoming detector
upgrades) and on ILC detector development studies can be successfully
employed.  The remainder of this chapter outlines the acceptance and
measurement requirements on the detector in detail, demonstrating the
feasibility of experimentation at the LHeC.

\begin{figure}[ht]
\centerline{\includegraphics[clip=,width=0.8\textwidth]{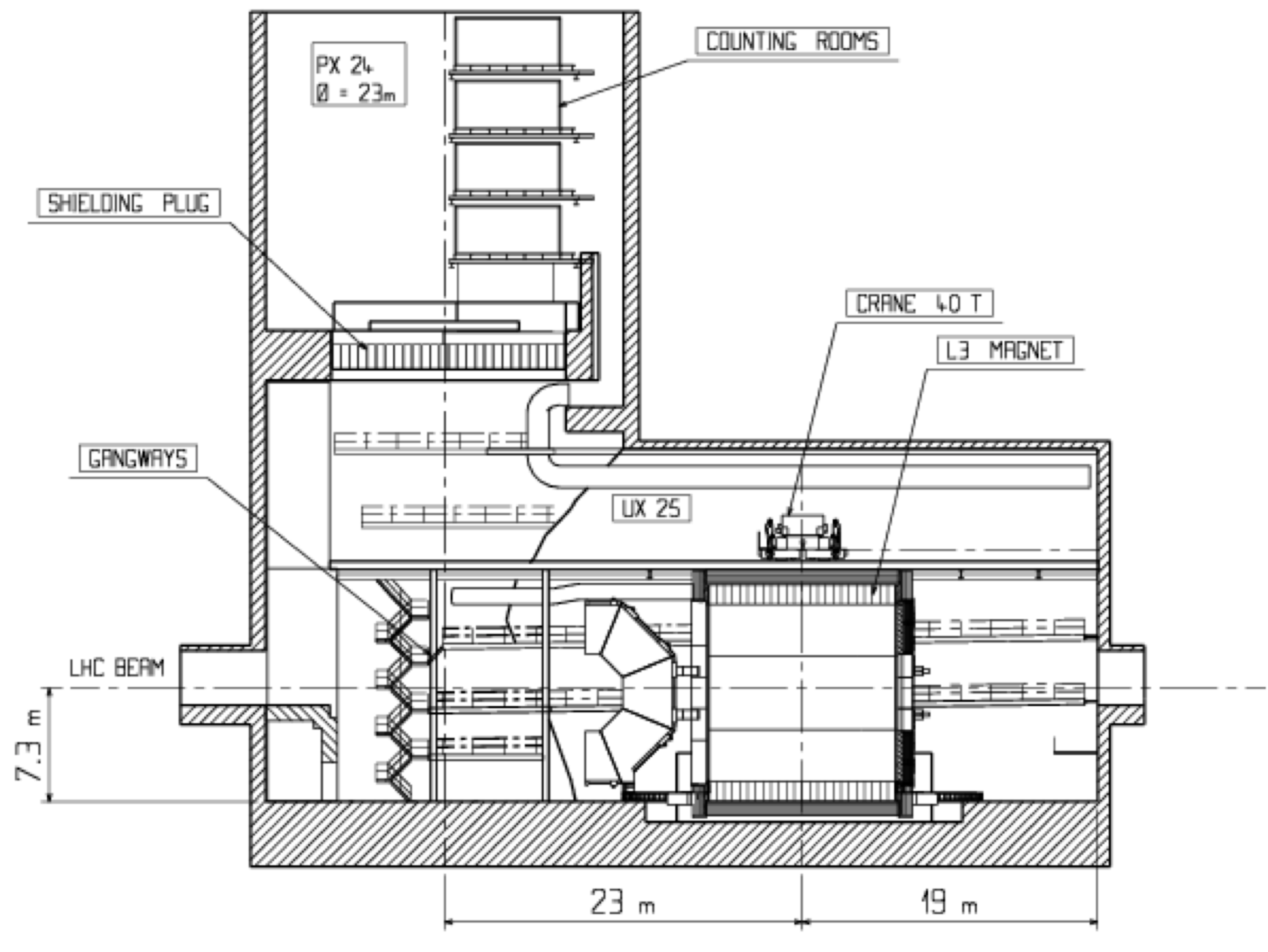}}
\caption{Cross section of the IP2 cavern with the L3 magnet. Round access shaft of ~23m diameter, cavern about 50m along the beam-line.
}
\label{LHEC:DetectorRequirements:Fig:AliceCavern}
\end{figure}
%
%
%
The LHeC project represents an upgrade of the LHC. The experiment
would be the fifth large experiment, and the detector the third
multi-purpose $4\pi$ acceptance detector. It requires a cavern, which
for the purpose of the design study has been considered to be the
ALICE cavern in IP2, shown in
Fig.\,\ref{LHEC:DetectorRequirements:Fig:AliceCavern}.  The
installation of the detector has to proceed as fast as possible in
order not to introduce large extra delays to the LHC program.  High
modularity and pre-assembly above ground are therefore inevitable
demands for the design.

\section{Cost and magnets}

The cost is related to technology choices, the detector granularity
and its size.  Crucial parameters of the detector are the beam pipe
dimensions, when combined with the small angle acceptance constraint,
and the parameters of the solenoid.  The cost $C$ of a solenoid can be
represented as a function of the energy density, $\rho_E$, $C \simeq
0.5 (\rho_E/MJ)^{0.66}$\cite{Nakamura:2010zzi}, which is determined as
\begin{equation}
 \rho_E = \frac{1}{2 \mu_0} \cdot \int{B^2} dV \simeq
 \frac{1}{2 \mu_0}  \cdot \pi r^2 \cdot l \cdot B^2.
\label{Eq:rhoE}
\end{equation}
From these relations one derives roughly that the solenoid cost scales
linearly with the radius $r$ and field strength $B$ and with the
length $l$ to the power $0.66$.  The solenoid radius influences the
track length in the transverse plane, which determines the transverse
momentum resolution $\propto r^{-2}$, whereas field strength enters
linearly $\propto B^{-1}$.

The Linac-Ring version of the LHeC requires an extended dipole field
of $0.3$\,T to be placed inside the detector for ensuring head-on $ep$
collisions and for separating the beams.

A balance between a strong magnetic field for optimal tracking
resolution and an affordable sized magnet has to be found, knowing
that magnets themselves represent one source of inactive material and
that the energy stored in the magnets and their return flux require an
outer shielding proportional to the field and to the square of the
solenoid radius.

\begin{figure}[htp]
\begin{center}
\includegraphics[width=0.97\columnwidth]{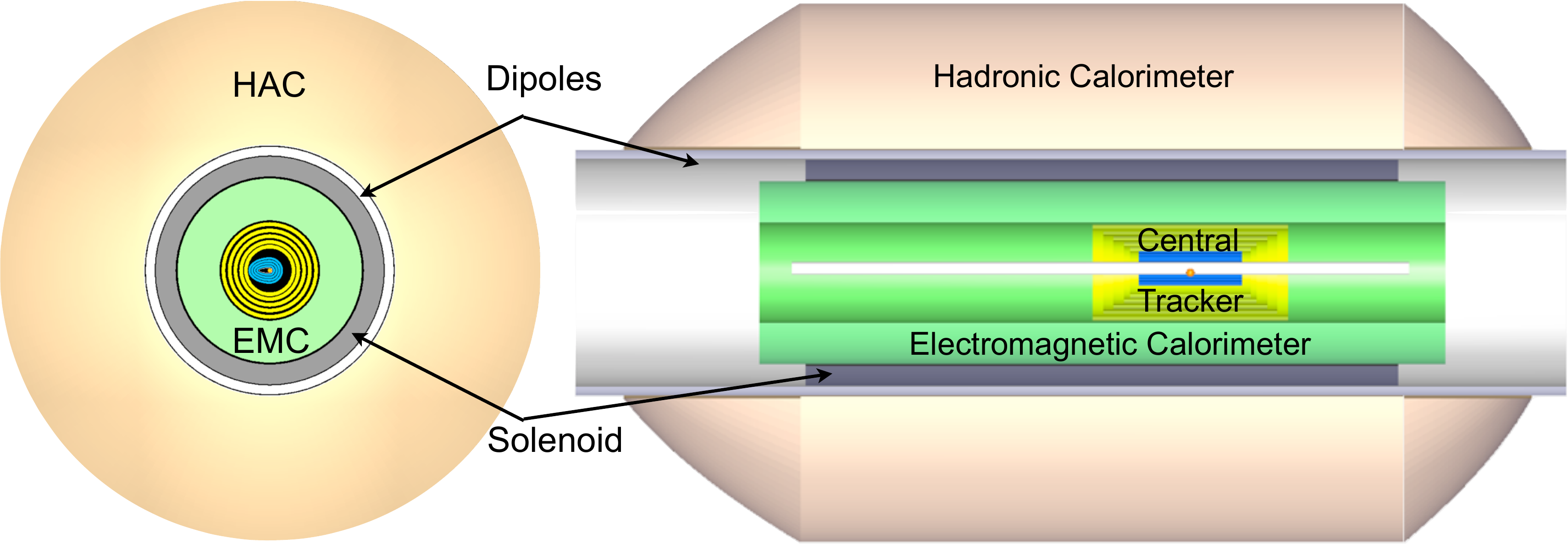}
\end{center}
\caption{Schematic $xy$ and $rz$ views of the magnets and barrel 
calorimeter arrangement for the  
baseline layout.}
\label{LHEC:MainDetector:Description:Fig:1c}   
\end{figure}

In the current design the solenoid is placed in between the
electromagnetic and the hadron calorimeter\footnote{ An option is also
  considered of placing the solenoid outside the calorimeters, at
  about $2.5$\,m radius, combined with a second, bigger solenoid for
  the flux return, with the muon detector in between.  A two-solenoid
  solution was considered already in the fourth detector concept for
  the ILC\cite{Mazzacane:2010zz}.}  at a radius of about $1$\,m.  The
magnetic field is set to $3.5$\,T in order to compensate the small
radial extension of the tracker.  The chosen design, with dipoles and
solenoid placed outside the electromagnetic calorimeter, ensures good
electromagnetic calorimeter resolution and high dipole field quality
close to the beam line.
Fig.\,\ref{LHEC:MainDetector:Description:Fig:1c} shows this magnet
arrangement inside the detector volume schematically.
The total material budget of the solenoid and the dipole, at
perpendicular crossing, may be represented by about 16\,cm of
Aluminium, corresponding to about one quarter of an interaction length
($\lambda_I$) and about one radiation length ($X_0$).  This further
supports the choice of the magnets located outside of the
electromagnetic calorimeter, and yet placed before the hadronic
calorimeter in order to limit its radial dimensions.  More details on
the design study of the detector magnets are addressed in
Sect.\ref{LHEC:MainDetector:MagnetDesign}.

\section{Detector acceptance}
\label{sec:detectorAcceptance}
\subsection{Kinematic reconstruction}
The inclusive $ep$ DIS kinematics are defined by the negative four-momentum
transfer squared, $Q^2$, and Bjorken $x$. Both are related to the
cms energy squared $s$ via the inelasticity $y$ through the relation
$Q^2=sxy$, which implies $Q^2 \leq s$.
The energy squared $s$ is determined by the
product of the beam energies, $s=4 E_pE_e$,
for head-on collisions and large energies compared to the proton mass. 

The kinematics may be determined from the scattered electron with
energy $E_e'$ and polar angle $\theta_e$ and from the hadronic final
state of energy $E_h$ and scattering angle $\theta_h$.  The variables
$Q^2$ and $y$ can be calculated from the scattered electron kinematics
as
\begin{eqnarray} \label{QYe}                                                     
 Q^2_e & = & 4 E_e E_e' \cos^2(\frac{\theta_e}{2}) \nonumber\\                             
 y_e~~ & = & 1 - \frac{E_e'}{E_e} \sin^2(\frac{\theta_e}{2})                                      
\end{eqnarray}                                                                  
and from the hadronic final state kinematics as
\begin{eqnarray} \label{QYh}                                                                              
 Q^2_h & = & \frac{1}{1-y_h} \cdot E_h^2 \sin^2(\theta_h ) \nonumber\\                             
 y_h~~ & = & \frac{E_h}{E_e} \sin^2(\frac{\theta_h}{2})                                        
\end{eqnarray}                                                                  
and $x$ is given as $Q^2 / sy$. 
%
The kinematic reconstruction in neutral current scattering therefore
has redundancy and a large potential for cross-calibration of
detectors, which is one reason why DIS experiments at $ep$ colliders
are precise. An important example is the calibration of the
electromagnetic energy scale from the measurements of the electron and
the hadron scattering angles. At HERA, this led to the precision of
the energy calibration for $E_e'$ at the per mil level. In a large
part of the phase space, around $x=E_e/E_p$, the scattered electron
energy is approximately equal to the beam energy, $E_e' \simeq E_e$,
which causes a large ``kinematic peak" in the scattered electron
energy distribution.  The hadronic energy scale can be obtained from
the transverse momentum balance in neutral current scattering, $p_t^e
\simeq p_t^h$. It is determined to about 1\% precision at HERA.

Following Eq.\ref{QYh}, the kinematics in charged current scattering
are reconstructed from the transverse and longitudinal momenta and
energy of the final state particles according to
%
\begin{eqnarray} \label{QYjb}                                                                              
 Q^2_h & = & \frac{1}{1-y_h}\sum{p_t^2} \nonumber\\                             
 y_h~~ & = & \frac{1}{2E_e} \sum{(E-p_z)}.                                         
\end{eqnarray}   
There have been many refinements used in the reconstruction
of the kinematics, as discussed e.g.  in\cite{Klein:2008di}, which for the
principle design considerations, however, are of less importance.
\subsection{Acceptance for the scattered electron}  
The positions of isolines of constant energy and angle of the
scattered electron in the $(Q^2,x)$ plane are given by the relations:
\begin{eqnarray} \label{QXe}                                                     
 Q^2(x,E_e') &=& sx \cdot \frac{  E_e - E_e'} {  E_e - x E_p } \nonumber\\           
 Q^2~(x,\theta_e) &=& sx \cdot \frac{ E_e } { E_e + x E_p \tan^2(\theta_e /2) }.                
\end{eqnarray}                        
%
Except at the smallest $x$, these relations relate an acceptance
limitation of the scattered electron angle $\theta_{e}^{max}$ to
a constant minimum $Q^2$, which is independent of $E_p$, given as
\begin{equation}                                                                    
Q^2_{min}(x,\theta_{e}^{max}) \simeq [2 E_e \cot (\theta_{e}^{max} /2)]^2.
\label{Eq:q2min}
\end{equation}
\begin{figure}[htbp]
\centerline{\includegraphics[clip=,width=1.0\textwidth]{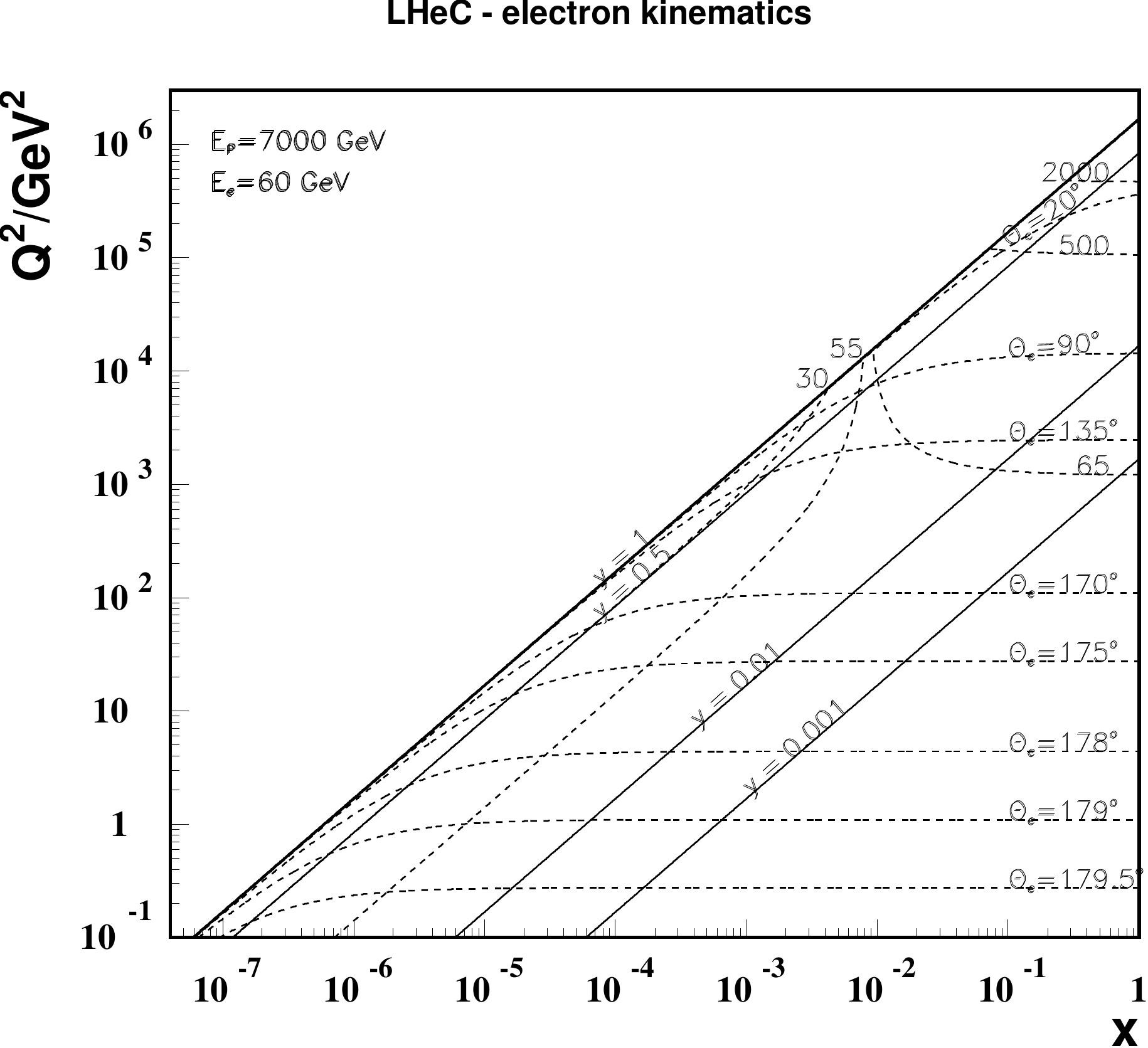}}
\caption{
Kinematics of electron detection at the LHeC. Lines of constant
scattering angle $\theta_e$ and energy, in GeV, are drawn.}
   \label{fig:kinele}
\end{figure}
This is illustrated in Fig.\,\ref{fig:kinele}. There follows that a
$179^{\circ} (170^{\circ})$ angular cut corresponds to a minimum $Q^2$
of about 1\,(100)\,GeV$^2$ at nominal electron beam energy.  One
easily recognises in Fig.\,\ref{fig:kinele} that the physics at low
$x$ and $Q^2$ requires to measure electrons scattered backwards from
about $135^{\circ}$ up to $179^{\circ}$.  Their energy in this
$\theta_e$ region does not exceed $E_e$ significantly.  At lower $x$
to very good approximation $y = E_e'/E_e$ (as can be seen from the
lines $y=0.5$ and $E_e' = 30$\,GeV in Fig.\,\ref{fig:kinele}).  At
small energies, for $y \lesssim 0.5$ a good $e/h$ separation is
important to suppress hadronic background, such as from
photoproduction.  The barrel calorimeter part, of about $90 \pm
45^{\circ}$, measures scattered electrons of energy not exceeding a
few hundred GeV, while the forward calorimeter has to reconstruct
electron energies of a few TeV. Both the barrel and the forward
calorimeters measure the high $x$ part, which requires very good
energy scale calibration as the uncertainties diverge $\propto
1/(1-x)$ towards large $x$.

\begin{figure}[htbp]
\centerline{\includegraphics[clip=,width=0.6\textwidth]{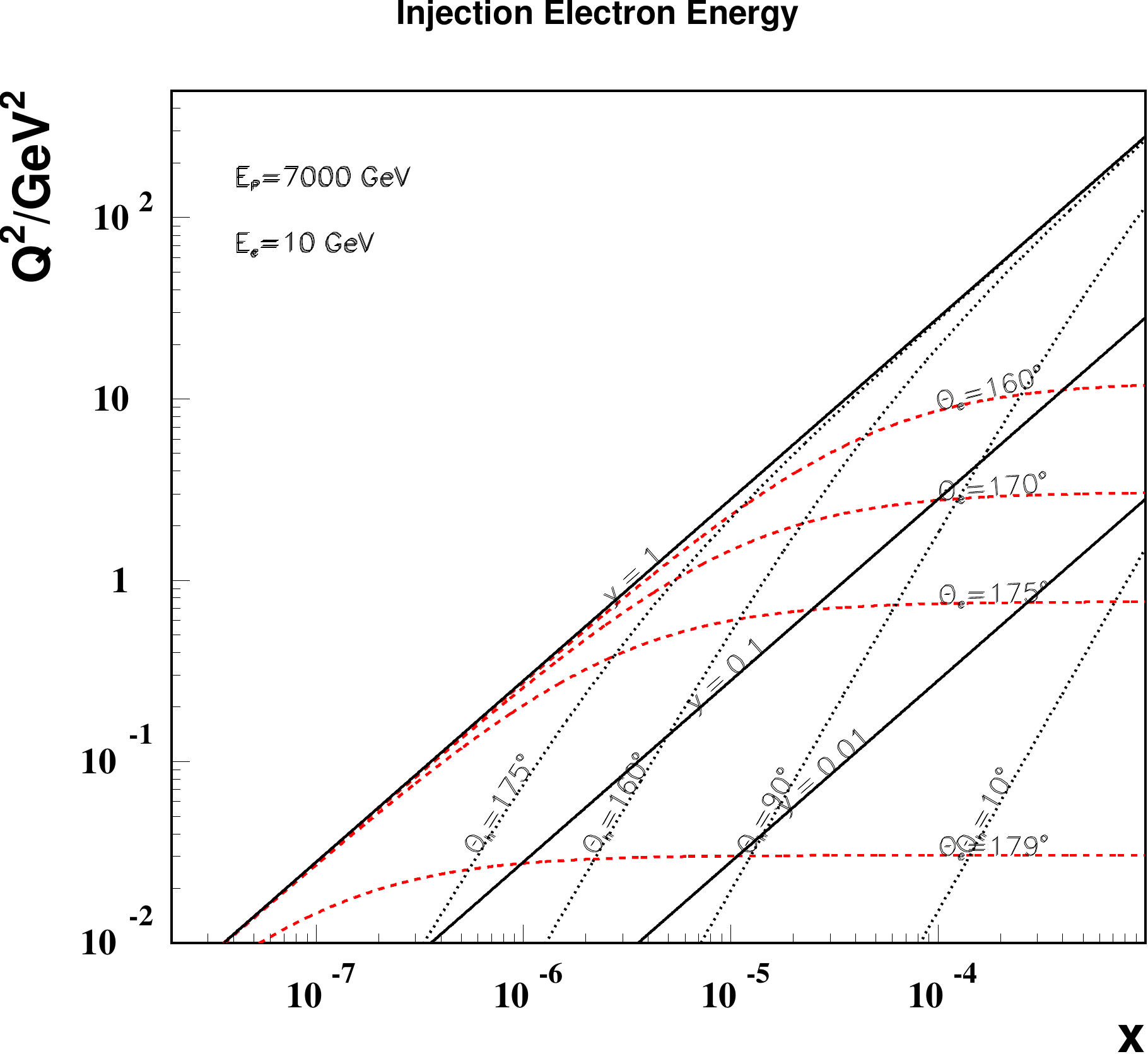}}
\centerline{\includegraphics[clip=,width=0.6\textwidth]{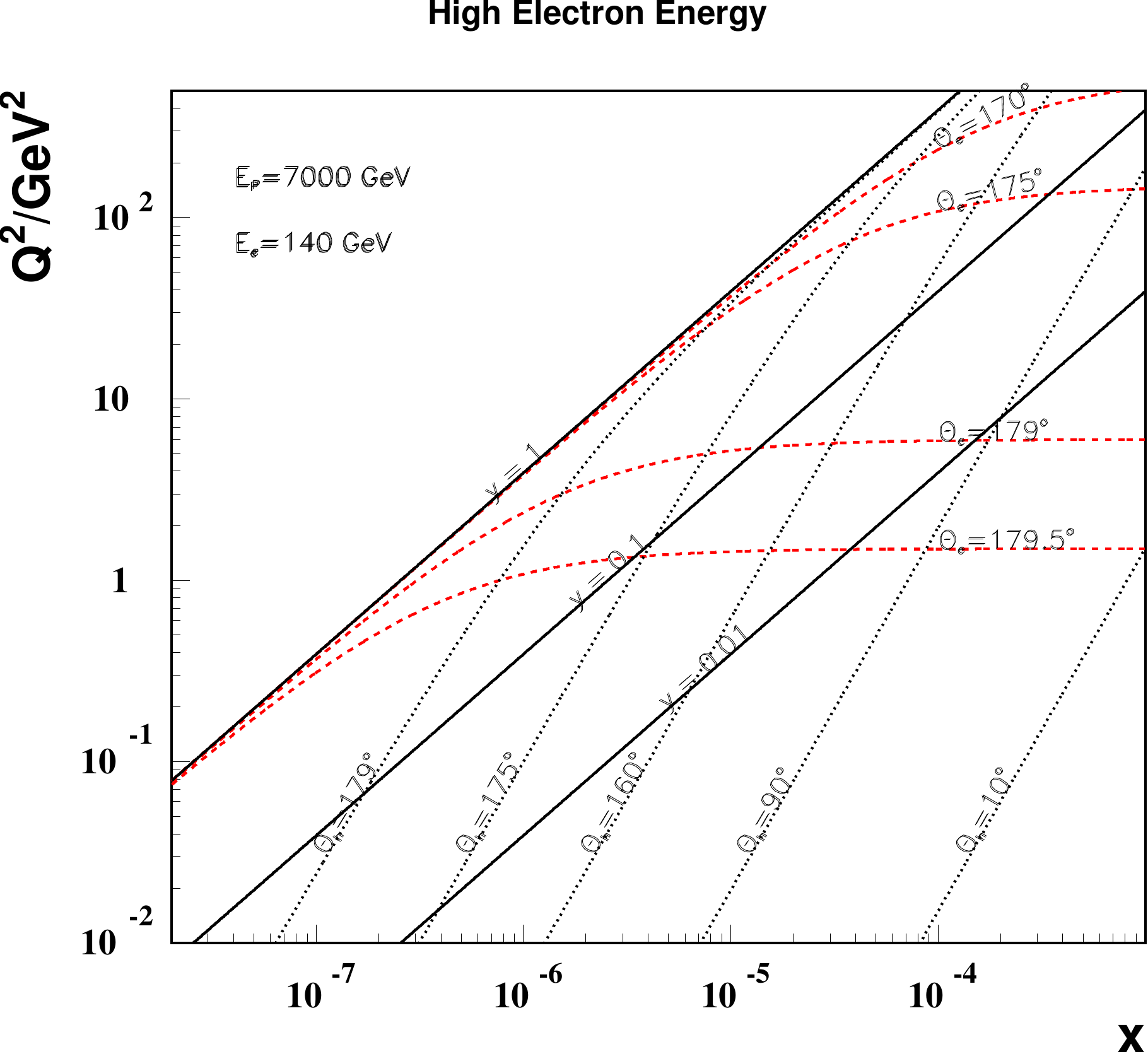}}
\caption{Kinematics at low $x$ and $Q^2$ of electron and hadronic
  final state detection at the LHeC with an electron beam energy of
  $10$\,GeV (top) as compared to $140$\,GeV (bottom). At larger $x$,
  the iso-$\theta_e$ lines are at about constant $Q^2 \propto E_e^2$.
  At low $x$, the scattered energies, not drawn here, are
  approximately at $E_e' \simeq (1-y) \cdot E_e$, and at lower $Q^2$
  and $x$ then $E_h \simeq E_e - E_e' \simeq y \cdot E_e$.  At very
  high $E_e$ part of the very low $Q^2$ region may be accessible with
  the electron tagged along the $e$ beam direction, outside the
  central detector, and the kinematics measured with the hadronic
  final state.}
   \label{fig:varEe}
\end{figure}

Following Eq.\,\ref{Eq:q2min}, $Q^2_{min}$ varies $\propto E_e^2$.  It
thus is as small as $0.03$\,GeV$^2$ for $E_e=10$\,GeV, the injection
energy of the ring accelerator but increases to $6.0$\,GeV$^2$ for
$E_e=140$\,GeV, the maximum electron beam energy considered in this
design report, if $\theta_e^{max}=179^{\circ}$.  While $Q^2_{min}$
decreases $\propto E_e^2$, the acceptance loss towards small $x$ is
only $\propto E_e$.  The measurement of the transition region from
hadronic to partonic behaviour, from $0.1$ to $10$\,GeV$^2$, therefore
requires taking data at lower electron beam energies.  These
variations are illustrated in Fig.\,\ref{fig:varEe} for an electron
beam energy of $10$\,GeV, the injection energy for the ring and a
one-pass linac energy, and for the highest $E_e$ of $140$\,GeV
considered in this report.
The requirement of acceptance up to $179^{\circ}$ determines the
length of the backward detector. This $E_e$ dependence is useful when
considering design options.  For example, if the backward electron
acceptance was limited to $178^{\circ}$ instead of $179^{\circ}$ this
would reduce the backward detector extension in $-z$. Data taken at
reduced $E_e$ recovers the lower $Q^2$ acceptance. From
Eq.\,\ref{Eq:q2min} it can be seen that $E_e=30$\,GeV and
$178^{\circ}$ leads to the same $Q^2_{min}$ of about $1.1$\,GeV$^2$.
However, acceptance at the lowest $x$ is lost linearly with $E_e$.
Moreover, for the present design the (inner) beam pipe radius in
vertical direction is $2.2$\,cm. This results in an extension of about
$1.5$\,m for the first tracker plane to register an electron scattered
at $179^{\circ}$. If $1$\,m is added for the tracker length, and
$1$\,m for the backward calorimeter following the tracker, the total
is about $3.5$\,m backward detector length.  For $178^{\circ}$ the
first $1.5$\,m could be reduced to e.g. $80$\,cm but a sizeable tracker
length is still needed to achieve some sagitta to determine the charge
of the scattered electron, thus a detector length of about $2.5$\,m
seems possible.  While this is an interesting reduction, the loss of
the lowest $x$ region implies a fundamental part of the LHeC physics
program would be lost and thus the $179^{\circ}$ design requirement
has been kept.
%

\begin{figure}[htbp]
\centerline{\includegraphics[clip=,width=0.6\textwidth]{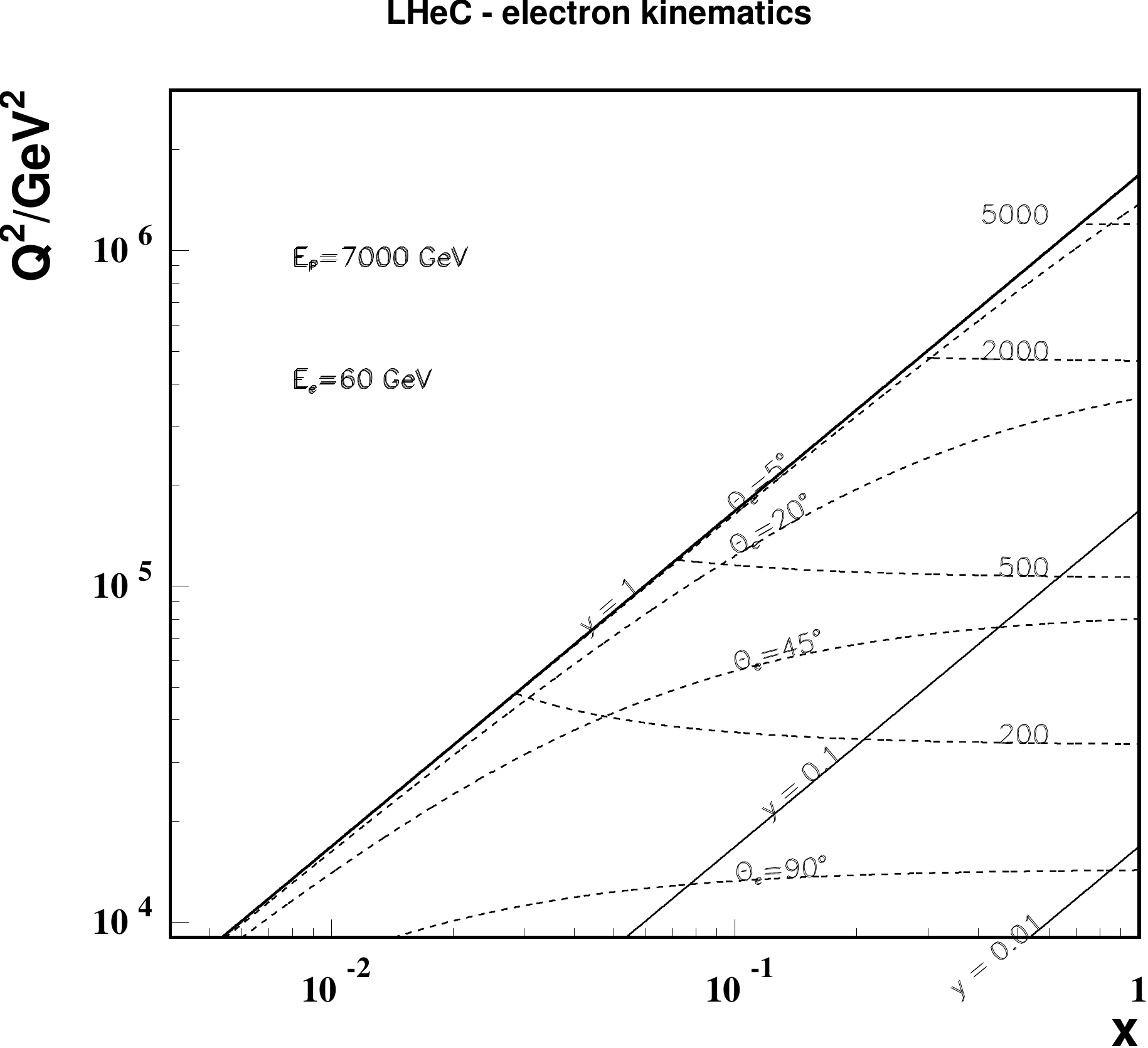}}
\caption{Kinematics of electron detection in the forward
detector region corresponding to  large $Q^2 \geq 10^4$\,GeV$^2$.
The energy values are given in GeV. At very high $Q^2$ 
the iso-$E_e'$ lines are rather independent of $x$, i.e.
 $Q^2(x,E_e') \simeq 4E_eE_e'$.}
   \label{fig:elhiq}
\end{figure}

Electrons scattered in the forward region correspond to scattering at
large $Q^2 \geq 10^4$\,GeV$^2$, as is illustrated in the zoomed
kinematic region shown in Fig.\,\ref{fig:elhiq}. The energies in the
very forward region, $\theta_e \lesssim 10^{\circ}$, exceed
$1000$\,GeV. For large $E_e$ and $x$, Eq.\,\ref{QXe} simplifies to
$Q^2 \simeq 4 E_e E_e' $, i.e.  a linear relation of $Q^2$ and $E_e'$
which is independent of $x$ and of $E_p$, apart from the fact that
$Q^2_{max}=s$.
%
\subsection{Acceptance for the hadronic final state}  
\begin{figure}[htbp]
\centerline{\includegraphics[clip=,width=1.0\textwidth]{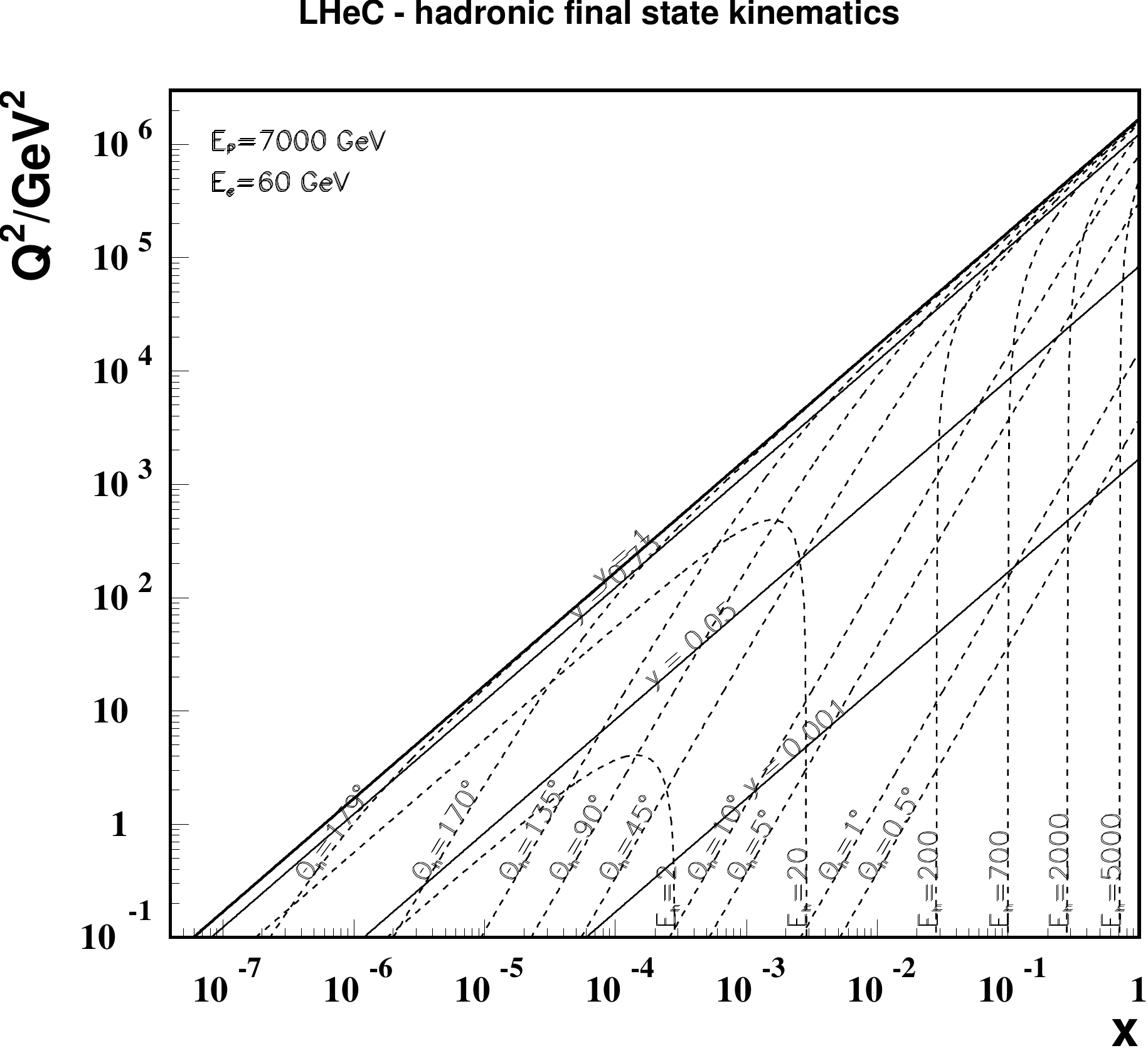}}
\caption{Kinematics of hadronic final state detection at the LHeC.
  Lines of constant energy and angle of the hadronic final state are
  drawn, as represented by simple kinematics of the struck quark.}
   \label{fig:kinjet}
\end{figure}

The positions of isolines in the $(Q^2,x)$ plane of constant energy
and angle of the hadronic final state, approximated here by the
current jet or struck quark direction, are given by the relations:
\begin{eqnarray} \label{QXj}                                                     
 Q^2(x,E_h) &=& s x  \cdot \frac{ xE_p - E_h }{   xE_p - E_e } \nonumber\\          
 Q^2~(x,\theta_h) &=& s x \cdot \frac {xE_p}{ xE_p +   E_e\cot^2(\theta_h /2)}               
\end{eqnarray}                        
and are illustrated in Fig.\,\ref{fig:kinjet}.  The most demanding
region is the large $x$ domain, where very high energy final state
particles are scattered close to the (forward) direction of the proton
beam.  The barrel region, of about $90 \pm 45^{\circ}$, is rather
modest in its requirements.  At low $x \lesssim 10^{-4}$, the hadronic
final state is emitted backwards, $\theta_h > 135^{\circ}$, with
energies of a few GeV to a maximum of $E_e$. Lines at constant $y$ at
low $x$ are approximately at $y = 1 - E_e'/E_e$ and $E_e'+E_h = E_e$,
i.e. $y = E_h/E_e$.  Final state physics at lowest $x \lesssim 3 \cdot
10^{-6}$ requires access to the backward region within a few degrees
of the beam pipe.  This is the high $y$ region in which the
longitudinal structure function is measured.
The $x$ range accessed with the barrel calorimeter region, of
$\theta_h$ between $135^{\circ}$ and $45^{\circ}$, is typically around
$10^{-4}$, as can be seen in Fig.\,\ref{fig:kinjet}.  The hadronic
energies in this part typically do not exceed $200$\,GeV.  The
detector part which covers this region is quite large but the
requirements are modest.  Nevertheless, the measurement of missing
transverse energy and the importance of using longitudinal momentum
conservation for background and radiative correction reductions demand
that the detector be hermetic and perform well.

For the measurement of the hadronic final state the forward detector
is the most demanding. Due to the high luminosity, the large $x$
region will be densely populated and a unique physics program at large
$x$ and high $Q^2$ may be pursued. In this region the relative
systematic error increases like $1/(1-x)$ towards large $x$.  At high
$x$ and not extreme $Q^2$ the $Q^2(x,E_h)$ line degenerates to a line
$x = E_h/E_p$ as can be derived from Eq.\,\ref{QXj} and seen in
Fig.\,\ref{fig:kinjet}.  High $x$ coverage thus demands measurements
of up to a few TeV of energy close to the beam pipe, i.e. a dedicated
high resolution calorimeter is mandatory for the region below about
$5-10^{\circ}$ extending to as close to the beam pipe as possible.  A
minimum angle cut $\theta_{h,min}$ in the forward region, the
direction of the proton beam, would exclude the large $x$ region from
the hadronic final state acceptance (Fig.\,\ref{fig:kinjet}), along a
line
\begin{equation} \label{xmax}
 Q^2~(x,\theta_{h,min})\simeq [2 E_p x \tan^2(\theta_{h,min} /2)]^2,
 \end{equation}
which is linear in the $\log Q^2$, $\log x$ plot and depends on $E_p$
only. Thus at $E_p=7$\,TeV the minimum $Q^2$ is roughly $(1000[100]
x)^2$ at a minimum angle of $10[1]^{\circ}$. Since the dependence in
Eq.\,\ref{xmax} is quadratic with $E_p$, lowering the proton beam
energy is of considerable interest for reaching the highest possible
$x$ and overlapping with the large $x$ data of previous experiments or
searches for new phenomena with high mass.
 
\subsection{Acceptance at the High Energy LHC}
\begin{figure}[h]
\centerline{\includegraphics[clip=,width=0.8\textwidth]{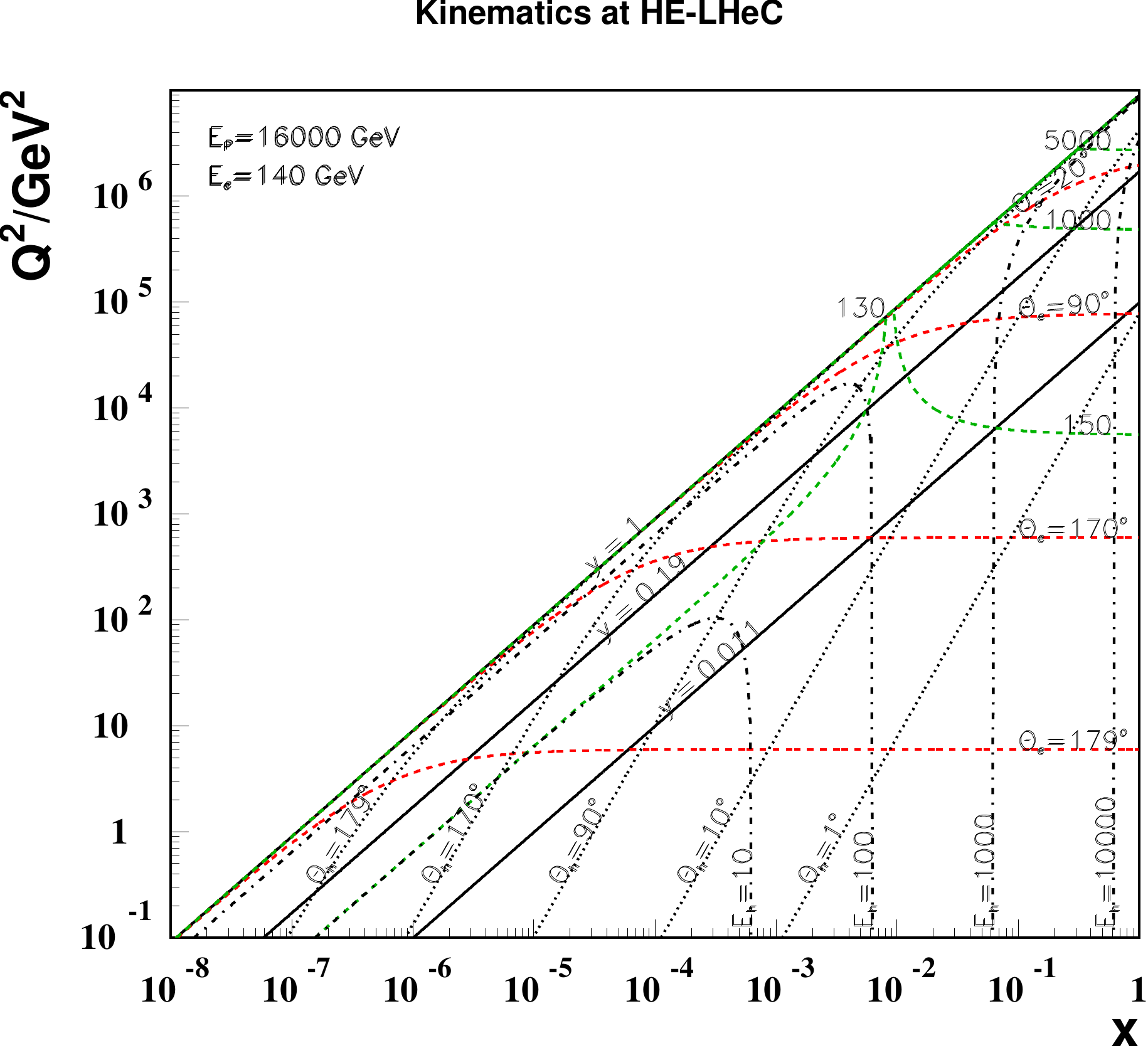}}
\caption{Scattered electron and hadronic final state kinematics for
  the HE-LHC at $E_p = 16$\,TeV coupled with a $140$\,GeV electron
  beam. Lines of constant scattering angles and energies are
  plotted. The line $y=0.011$ defines the edge of the HERA kinematics
  and $y=0.19$ defines the edge of the default machine considered in
  this report ($E_e = 60$\,GeV and $E_p = 7$\,TeV).}\label{Fig:HEL}
\end{figure}

Presently a high energy (HE) LHC is under consideration as a machine
which would be built in the thirties, with proton beam energies of
$16$\,TeV\cite{Pire:2010zz}.  Such an accelerator would better be
combined with an electron beam with energy exceeding the $60$\,GeV
considered as default here, to profit from the increased proton beam
energy and to limit the asymmetry of the two beam energies.  Using the
$140$\,GeV beam mentioned above in this section as an example,
Figure\,\ref{Fig:HEL} displays the kinematics and acceptance regions
for given scattering angles and energies of the electron (dashed green
and red) and of the hadronic final state (black, dotted and dashed
dotted). The cms energy in this case is enhanced by about a factor of
five. The maximum $Q^2$ reaches $10$\,TeV$^2$, which is $10^6$ times
higher than the typical momentum transfer squared covered by the
pioneering DIS experiment at SLAC. The kinematic constraints in terms
of angular acceptance would be similar to the present detector design
as can be derived from the $Q^2,x$ plot. At very high $x$ ($Q^2$) the
energy $E_h$ ($E_e'$) recorded by the forward detector would be
doubled.  With care in the present design, the main LHeC detector
components should also be sufficient in the HE phase of the LHC.

\subsection{Energy resolution and calibration}
The LHeC detector is dedicated to the most accurate measurements of
the strong and electroweak interactions and to the investigation of new
phenomena. The calorimetry therefore requires:
\begin{itemize}
\item Optimum scale calibrations, such as for the measurement of the
  strong coupling constant.  This is helped by the redundancy in
  kinematic reconstruction methods and kinematic relations, such as
  $E_e' \simeq E_e$ at low $Q^2$, $E_e'+E_h \simeq E_e$ at small $x$,
  the double angle reconstruction\cite{Buchmuller:1992rq} of $E_e'$
  and the transverse momentum balance of $p_T^e$ and $p_T^h$.  From
  the experience with H1 and ZEUS and the much increased statistics it
  is assumed that $E_e'$ may be calibrated to $0.1-0.5$\,\% and $E_h$
  to $1-2$\,\% precision. The latter will be most crucial in the
  forward, high $x$ part of the calorimeter where the uncertainties
  diverge $\propto 1/(1-x)$ towards large $x$.
\item High resolution, for the reconstruction of multi-jet final
  states such as from the $H \rightarrow b \overline{b}$ decay.  This
  is a particular challenge for the forward calorimeter.  While
  detailed simulations are still ongoing, it may be assumed that
  $(10-15)/\sqrt{E/GeV}$\,\% resolutions for $E_e'$ and
  $(40-50)/\sqrt{E/GeV}$\,\% for $E_h$ are appropriate, with small
  linear terms.  These values are very similar to the ATLAS detector
  which quotes electromagnetic resolutions of $10/\sqrt{E/GeV} \oplus
  0.007$\,\% and hadronic energy resolutions of $50/\sqrt{E/GeV}
  \oplus 0.03 $\,\%. The basic electromagnetic calorimeter choice for
  the LHeC is Liquid Argon (LAr)~\footnote{In H1 very good experience
    has been collected with the long-term stability of the LAr
    calorimeter. A special demand is the low noise performance, as the
    measurements at small inelasticity $y$ are crucial for reaching
    large Bjorken $x$.  In this region a small misidentified
    deposition of energy in the backward part of the detector can
    spoil the measurement at low $y \lesssim 0.01$, as can be seen
    from Eq.\,\ref{QYjb}. }.  The hadronic calorimeter, which is
  outside the magnets and also serves as the magnetic flux return, may
  be built as a tile calorimeter with the additional advantage of
  supporting the whole detector.  The first years of operating the
  ATLAS combined LAr/TileCal calorimeter has been encouraging.  Some
  special calorimeters are needed in the small angle forward region
  ($\theta \lesssim 5^{\circ}$) where the deposited energies are
  extremely large, and also in the backward region ($\theta \geq
  135^{\circ}$) where the detection of electrons with modest energy is
  of particular concern.
\item Good electron-hadron separation, as required for electron
  identification at high $y$ and low $Q^2$ (backwards) or high $Q^2$
  (in the extreme forward direction).  This is a requirement on the
  segmentation of the calorimeters and also on the trackers positioned
  in front of the forward and backward calorimeters which are needed
  to support the energy measurements and electron identification in
  particular.
\end{itemize} 
The calorimetry needs to be hermetic for the identification of the
charged current process via a precise measurement of $E_{T,miss}$.
These considerations are also summarised in
Tab.\,\ref{LHEC:DET:INTRO:Calo}.
\begin{table}[h]
  \centering
  \begin{tabular}{|l|c|c|c|}
    \hline
 region of detector  &     backward  & barrel & forward \\ 
 approximate  angular range / degrees &   179 - 135 & 135 -45 & 45-1 \\ \hline \hline
scattered electron energy/GeV & 3-100  & 10-400 &  50-5000 \\
$x_e$ & $10^{-7} - 1$ & $10^{-4} -1$ & $10^{-2}-1$ \\
elm scale calibration in \%  & 0.1 & 0.2 & 0.5  \\
elm energy resolution $\delta E/E$ in \% $\cdot \sqrt{E/GeV}$ & 10 & 15 & 15  \\ \hline
hadronic final state energy/GeV & 3-100  & 3-200 &  3-5000 \\
$x_h$ & $10^{-7} - 10^{-3}$ & $10^{-5} -10^{-2}$ & $10^{-4}-1$ \\ 
hadronic scale calibration in \%  & 2 & 1 & 1  \\
hadronic energy resolution in \% $\cdot \sqrt{E/GeV}$& 60 & 50 & 40  \\ 
 \hline
  \end{tabular}
\caption{Summary of calorimeter kinematics and requirements for
the default design energies of $60 \times 7000$\,GeV$^2$, see text. 
The forward (backward) calorimetry has to extend to $1^{\circ} (179^{\circ})$.
}
\label{LHEC:DET:INTRO:Calo}
\end{table}
\subsection{Tracking requirements}
The tracking detector has to enable 
\begin{itemize}
\item Accurate measurements of the transverse momenta and polar angles
\item Secondary vertexing in a maximum polar angle acceptance range
\item Resolution of complex, multiparticle and highly energetic final states in forward direction
\item Charge identification of the scattered electron
\item Distinction of neutral and charged particle production 
\item Measurement of vector mesons, as the $J/\psi$ or $\Upsilon$ decay into muon pairs
\end{itemize}
%

\begin{figure}[htbp]
\centerline{\includegraphics[clip=,width=0.65\textwidth]{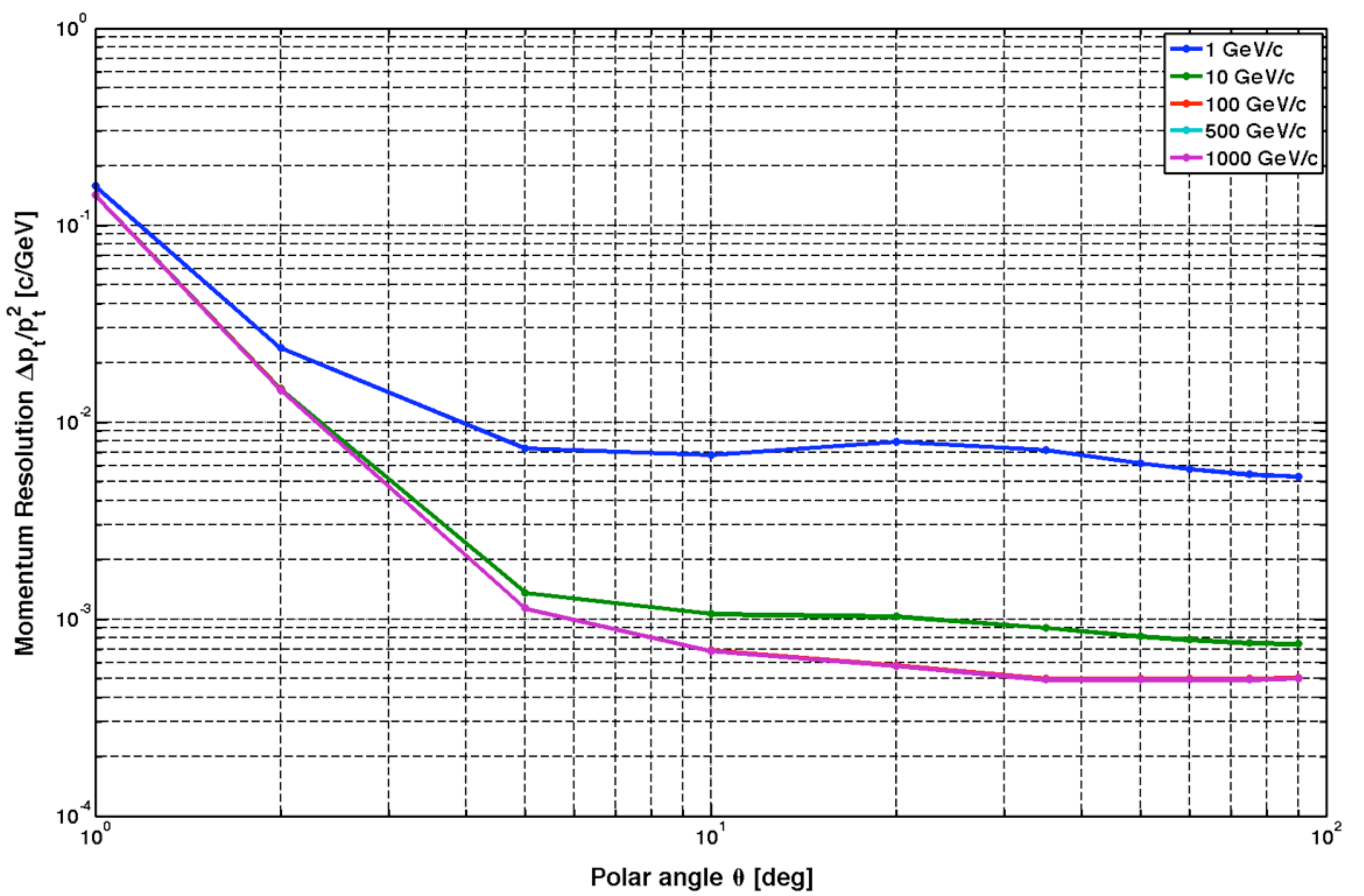}}
\centerline{\includegraphics[clip=,width=0.65\textwidth]{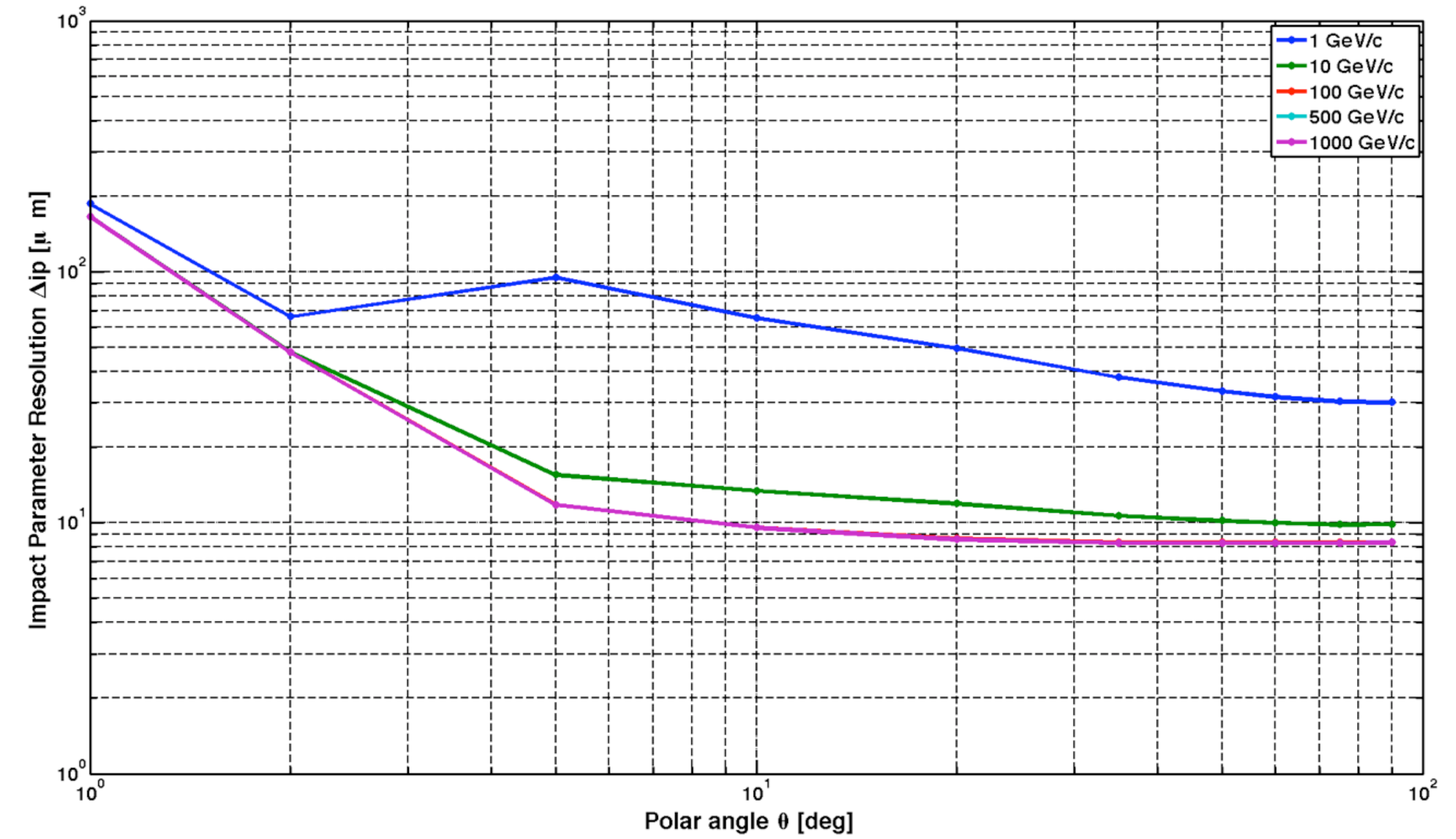}}
\centerline{\includegraphics[clip=,width=0.65\textwidth]{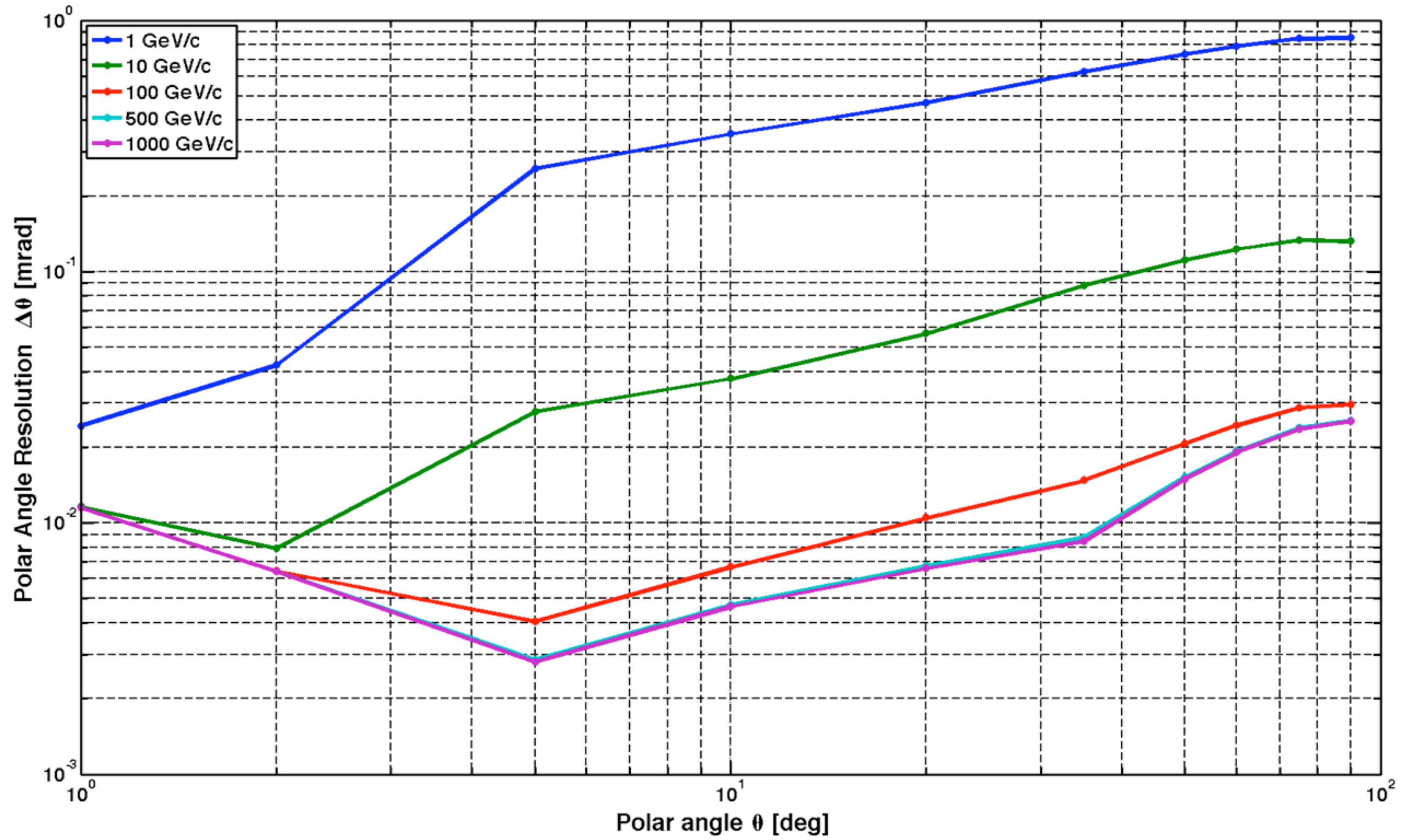}}
\caption{Transverse momentum (top), impact parameter (middle) and polar angle (bottom)
measurement resolutions as function of the polar angle for
the default detector design for four values of track transverse momentum.
}
\label{fig:Momrho}
\end{figure}

The transverse momentum resolution in a solenoidal field can be
approximated by
\begin{equation}
\frac{\delta p_T}{p_T^2} = \frac{\Delta}{0.3 B L^2} \cdot \sqrt{\frac{720}{N+4}}
\label{Eq:ptreso}
\end{equation}
where $B$ is the field strength, $\Delta$ is the spatial hit
resolution, $L$ is the track length in the plane transverse to the
beam direction, and $N$ is the number of measurements on a track which
enters as prescribed in \cite{Gluckstern:1963ng}.  As an example, for
$B=3.5$\,T, $\Delta = 10\,\mu$m, $N=4+5$ and $L=0.42$\,m one obtains a
transverse momentum measurement precision of about $ 3 \cdot 10^{-4}$.
A simulation, using the LICTOY program\cite{Regler:2008zz}, of the
transverse momentum, transverse impact parameter and polar angle
resolutions is shown in Fig.\,\ref{fig:Momrho}. It can be seen that
the estimate following Eq.\,\ref{Eq:ptreso} is approximately correct
for larger momenta where multiple scattering becomes negligible.  This
momentum resolution, in terms of $\delta p_T/p_T^2$ is about ten times
better than the one achieved with the H1 central drift chamber. It is
similar to the ATLAS momentum resolution for central tracks and is
thus considered to be adequate for the momenta encountered at the LHeC
and for the goal of high precision vertex tagging.  The impact
parameter resolution, for high momenta, is a factor of eight better
than the H1 or ZEUS result.

In the backward direction, a main tracking task is to determine the
charge of the scattered electron, which has momenta $ E_e' \leq E_e$,
down to a few GeV at high $y \simeq 1- E_e'/E_e$. With a beam spot as
accurate as about $10 \times 30$\,$\mu$m$^2$ and the beam pipe radius
of a few cm only, the backward Silicon strip tracker will allow a
precise $E/p$ determination when combined with the backward
calorimeter, even better than has been achieved with the H1 backward
silicon detector\cite{Collaboration:2010ry}.

In the forward region, $\theta < 5^{\circ}$, as may be deduced from
Figs.\,\ref{fig:elhiq},\,\ref{fig:kinjet}, the hadronic final state,
for all $Q^2$, and the scattered electron when scattered at high
$Q^2$, are very energetic.  This requires a dedicated calorimeter.
Depending on the track path and momentum, the track sagitta becomes
very small, for example about $10$\,$\mu$m for a $1$\,TeV track
momentum and a $1$\,m track length.
%
%
In such extreme cases of high momenta, the functionality of the
tracker will be difficult to achieve: the small sagitta means that
there will be limits to the transverse momentum measurement while the
ability to distinguish photons and electrons will be compromised by
the high probability of showering and conversion when the beam pipe is
traversed under very small angles.  A forward tracker is yet
considered to be useful down to small angles for the reconstruction of
the event, the rejection of beam induced background and the
reconstruction of forward going muons. This region requires detailed
simulation studies in a next phase of the project.
\subsection{Particle identification requirements}
The requirements on the identification of particles focus on the
identification of the scattered electron, a reliable missing energy
measurement and precision tracking for measuring the decay of charm
and beauty particles, the latter rather on a statistical basis than
individually. Classic measurements like the identification of the $D$
meson from the $K \pi \pi$ decay with a slow pion or the
identification of $B$ production from high $p_T$ leptons require a
very precise track detector.  The tracker should determine some
$dE/dX$ properties but there is no attempt to distinguish strange
particles, such as kaons, from pions as the measurement of the strange
quark distribution will utilise charm tagging in CC events. The
identification of muons, apart from some focus on the forward and
backward direction, is similar to that of $pp$ detectors. In addition
a number of specialised detectors are foreseen to tag
\begin{itemize}
\item{electrons scattered near the beam pipe in the backward direction
          to access low $Q^2$ events and control the photoproduction background;}
\item{photons scattered near the beam pipe in the backward direction
          to measure the luminosity from Bethe Heitler scattering;}
\item{protons scattered in the forward direction to measure diffractive DIS
          in $ep$ scattering and to tag the spectator proton in $en$ scattering
          in electron-deuteron runs;}
\item{neutrons scattered in the forward direction to measure pion exchange
          in $ep$ scattering and to tag the spectator neutron in $ep$ scattering
          in electron-deuteron runs;}
\item{deuterons scattered in the forward direction in order to discover diffraction
         in lepton-nucleus scattering.}
\end{itemize}
From the perspective of particle identification there are therefore no
unusual requirements.  A state of the art tracker with a very
challenging forward component, and a tagger system with the deuteron as
a new component in the forward direction.




\section{Summary of the requirements on the LHeC detector}

The considerations discussed in this chapter along with the
constraints from the physics program lead to the following main items
for the detector design.

\begin{enumerate}

\item The detector realisation requires a modular design and
  construction with the assembly process done in parallel partly at
  surface level and partly in the experimental area.

\item The detector should be modular and flexible to accommodate the
  high acceptance as well as the high luminosity running foreseen for
  the two main physics programs.  The flexibility should accommodate
  reducing/enhancing the energy asymmetry of the beams.

\item The detector design will be based on the experience at HERA and
  the LHC (including upgrade studies) and on ILC detector development
  studies, thus avoiding the need for new R\&D programs.

\item Mechanics/services have to minimise the amount of material in
  sensitive regions of the experimental setup.

\item Good vertex resolution for decay particle secondary vertex
  tagging is required, which implies a small radius and thin beam pipe
  optimised in view of synchrotron radiation and background production
  - see Section\,\ref{BPDesign}.

\item The detector will have one solenoid in its default version
  producing a homogeneous field in the tracking area of 3.5\,T
  extending over ${z= +370cm,-200cm}$. Solenoid options are described
  in Section\,\ref{LHEC:MainDetector:MagnetDesign}.

\item The tracking and calorimetry in the forward and backward
  directions have to be set up to take into account the extreme
  asymmetry of the production kinematics. The layout and choice of
  technology for the detector design will be chosen accordingly.  The
  tracker has to be optimised in view of energy flow corrections. The
  highest affordable granularity for tracking and calorimetry is
  required for the best energy/momentum measurements.

\item Very forward/backward detectors have to be set up to access the
  diffractive produced events and measuring the luminosity with high
  precision, respectively -
  Chapter\,\ref{LHEC:Detector:ForwardBackward}.

\end{enumerate}

In addition, there are more general considerations arising from
operational concerns and constraints which will also need to be
addressed in detail.

\begin{itemize}

\item The LHeC experiment has to be operated in parallel to the other
  LHC experiments and has to be set up in accordance with CERN
  regulations.

\item The beam pipe will host the electron beam along with the two LHC
  counter rotating proton beams.  The non interacting proton/ion beam
  has to bypass the IP region guided through the same beam pipe
  housing the electron and interacting proton/ion beam.

\item The detector has to be operated in a high luminosity
  environment. High luminosity is anticipated with small beam spot
  sizes ($\sigma_x\approx 30\mu{m}$, $\sigma_y\approx 16\mu{m}$),
  small ${\beta^{*}}$ and relatively large IP angles (as shown in the
  accelerator chapter). The parameter ${\beta^{*}}$ has to be chosen
  to eliminate the effects of parasitic bunch crossings.

\item The detector design has to be background tolerant and assure
  good performance over the experiment's lifetime.  The interaction
  region and the machine design has to incorporate masks, shielding
  and a vacuum profile that minimises the synchrotron radiation and
  operation induced backgrounds. The detectors along the beam line have
  to be radiation hard.

\item It might be necessary to have insertable/removable shielding
  protecting the detector against injection and poor machine
  performance.

\item Special Interaction Region (IR) instrumentation for tuning of
  the machine with respect to background and luminosity is
  needed. Radiation detectors e.g. near mask and tight apertures are
  useful for fast identification of background sources. Fast bunch
  related information is useful for beam optimisation in that
  context.

\end{itemize}

%% file: detector/det.tex
\input{detector/maindet}

%
%
\section{Magnet design} 
\input{detector/tenkate}
%
%
\section{Tracking detector}
\input{detector/tracking}
\section{Calorimetry}
\input{detector/calo}
%
%
\section{Muon detector}
\input{detector/muon}
\section{Event and detector simulations} 
\input{detector/Ev-Det-Sim}

%
%

%% file: detector/maindet.tex
\label{LHEC:MainDetector}


Following the considerations of the physics requirements and the technical and operational constraints
outlined in Section\,\ref{LHEC:Detector:Requirements},
a  detector design for high precision and large acceptance Deep Inelastic Scattering
is presented.  The detectors for the Linac-Ring or the Ring-Ring options are nearly identical:
the two notable differences are the dipoles in the Linac-Ring case for separating the $e$ and the $p$ beams
and the larger beam pipe due to the wider synchrotron radiation fan. For practical reasons, in
this report the more complicated Linac-Ring detector has been chosen as the baseline,
termed version {\small\bf A}. This mainly
affects the solenoid-dipole confi\-guration and the inner shape of the tracker.
For the Ring-Ring case the luminosity may be maximised by inserting
focusing quadrupoles near to the IP. This requires the inner detector to be designed in a
modular way such that a transition could be made between two phases, one with the quadrupoles
to achieve maximum luminosity, and one without to ensure maximum polar angle acceptance
\footnote{The most recent optics studies suggest that there is only a factor of two
difference between the luminosity achievable with and without the quadrupoles. Given the
extra complications and time required to make a transition (cf. HERA) this
is likely not enough to justify considering two measurement phases.}.

\section{Basic detector description}

\begin{figure}[htp]
\begin{center}
\includegraphics[width=0.9\columnwidth]{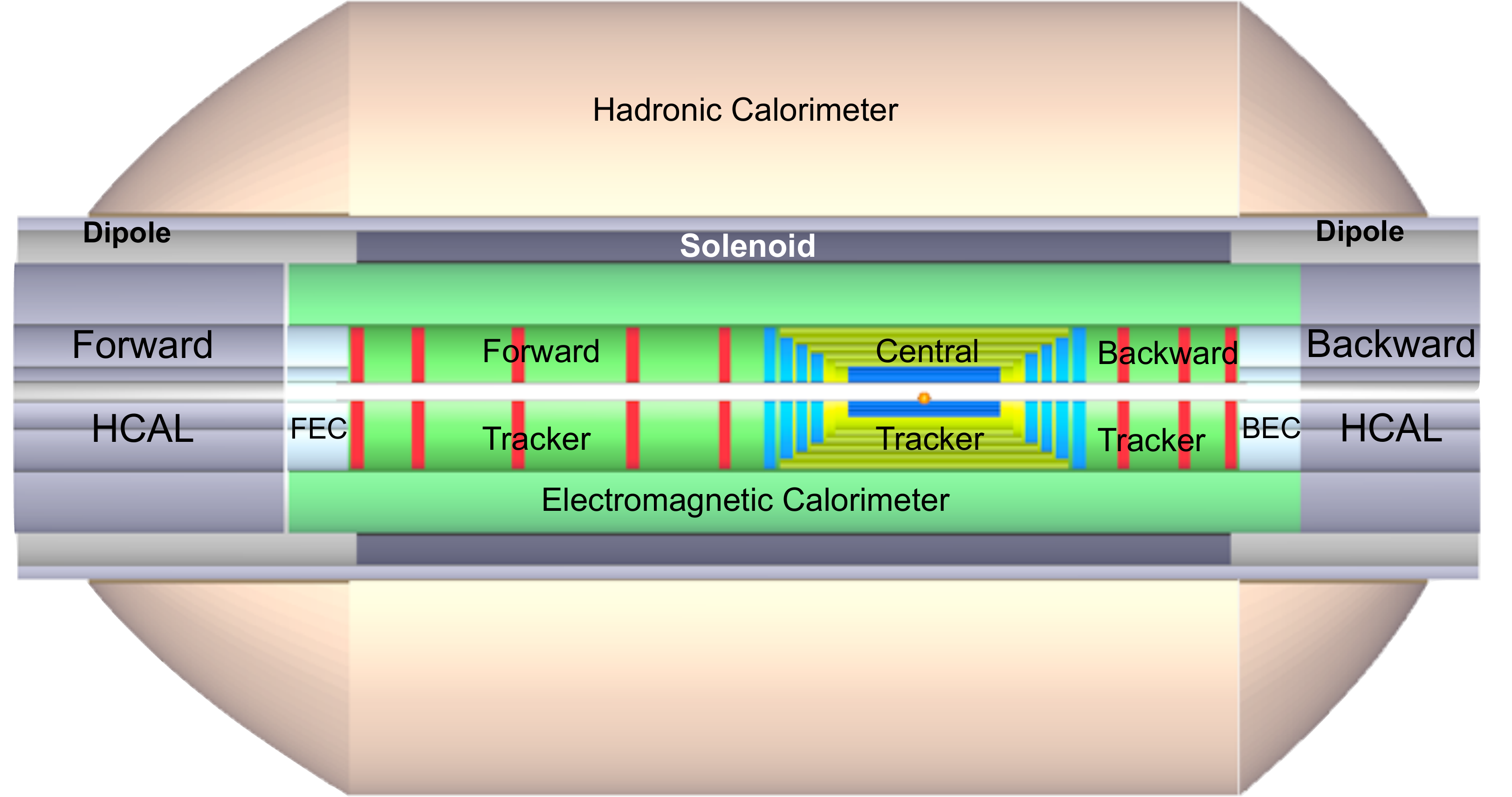}
\end{center}
\caption{Schematic $rz$ view of the detector design for the Linac-Ring machine option showing the 
characteristic dipole and solenoid  placement between the electromagnetic
and the hadronic calorimeters. The proton beam, from the right, 
collides with the electron beam, from the left, at the IP which is surrounded by 
a central tracker system complemented by large forward and backward tracker
telescopes followed by sets of calorimeters. The detector as sketched here,
i.e. without the muon tracking system, has a radius of $2.6$\,m and extends from
about $z=-3.6$\,m to $z=+5.9$\,m in the  direction of the proton beam.
 }
\label{LHEC:MainDetector:Description:Fig:1a}   
\end{figure}

\begin{figure}[htp]
\begin{center}
\includegraphics[width=0.8\columnwidth]{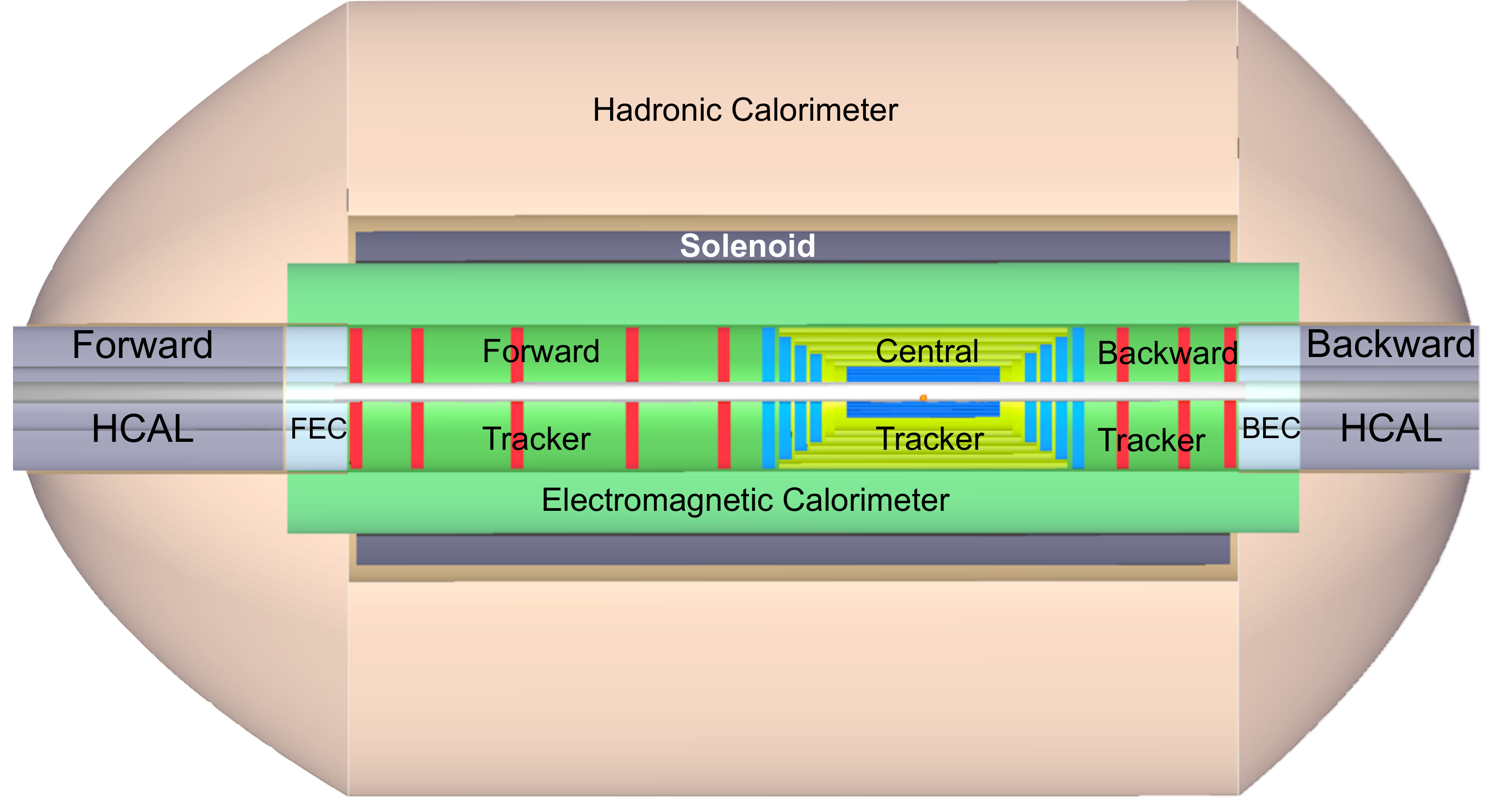}
\end{center}
\caption{Schematic $rz$ view of the detector design for the Ring-Ring machine option.
Note that the outer part of the forward and backward calorimeters ends at smaller
radii, as compared to the Linac-Ring case, since there are no dipole magnets foreseen.}
\label{LHEC:MainDetector:Description:Fig:1b}   
\end{figure}

The LHeC detector is asymmetric in design, reflecting the beam energy
asymmetry and reducing cost.  It is a general purpose 4$\pi$ detector,
consisting of an inner silicon tracker with extended forward and
backward parts, surrounded by an electromagnetic calorimeter, separated
from the hadronic calorimeter by a solenoid with $3.5$\,T field.  In
order to maximise the luminosity and ensure beam separation in the
Linac-Ring case, a dipole system is incorporated into the detector,
extending over $\pm9m$ with respect to the IP (see
Fig.\,\ref{LHEC:MainDetector:Description:Fig:1a} and
Fig.\,\ref{Fig:HTK1},
Section\,\ref{LHEC:MainDetector:MagnetDesign:dipoles}).  In the
Ring-Ring case the dipoles are omitted
(Fig.\,\ref{LHEC:MainDetector:Description:Fig:1b}).  The hadronic
calorimeter is enclosed in a muon tracking system, not shown here but
discussed in Section\,\ref{LHEC:MainDetector:DetMuon}.  The main
detector is complemented by dedicated hadron tagging detectors in the
forward direction and a polarimeter and luminosity measurement system
in the backward direction, as presented in
chapter\,\ref{detector:fwdbwd}.  Its longitudinal extension is
determined by the need to cover polar angles down to $1^{\circ}$. Its
radial size is mainly determined by the requirement to fully contain
the energy of hadronic showers in the calorimeter.

The dipoles for the Linac-Ring interaction region must be as close as
possible to the beam to minimise cost. At the same time, their bulk
material should not compromise tracking and electromagnetic energy
measurements
and must therefore be placed outside the electromagnetic calorimeter.
The solenoid cost scales approximately with its radius (see
Eq.\ref{Eq:rhoE}) which in absolute terms allows tens of millions of
CHF to be economised if the solenoid is placed inside the hadronic
calorimeter, especially considering the cost of the thousands of tons
of iron needed for shielding.  Again driven by cost and material
concerns, it appears appropriate to foresee a single cryostat housing
the electromagnetic LAr calorimeter, the solenoid and dipole
magnets. This affects the forward and backward calorimeter
inserts. The modifications can be seen comparing the Linac-Ring
Fig.\,\ref{LHEC:MainDetector:Description:Fig:1a} with the Ring-Ring
Fig.\,\ref{LHEC:MainDetector:Description:Fig:1b} designs.  Since for
the physics performance it is advantageous to place the solenoid
outside the hadronic calorimeter, this option, termed {\small\bf B},
has also been studied and is discussed in
Section\,\ref{LHEC:MainDetector:MagnetDesign}.  In this case, the
radius of the large coil would be about $2.5$\,m which compares well
with the H1 and CMS coils but is only an option for the Ring-Ring
machine design and is not adopted for the baseline.
%

\begin{figure}[htp]
\begin{center} 
\includegraphics[width=0.8\columnwidth]{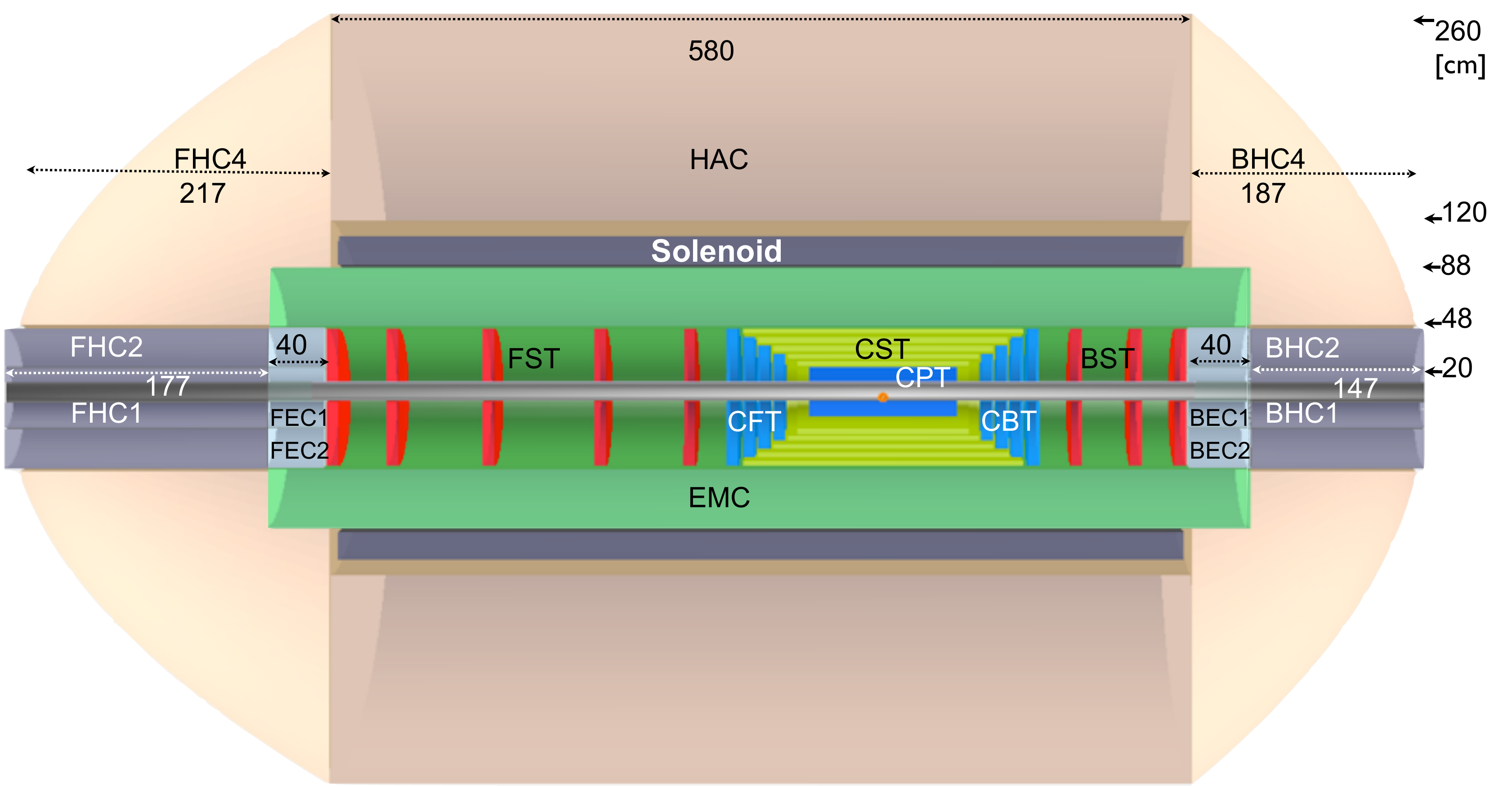}
\end{center}
\caption{
An $rz$ cross section and the dimensions of the main detector (muon detector not shown)
for the Ring-Ring detector version (no dipoles) extending the polar angle acceptance
to about $1^{\circ}$ ($179^{\circ}$) in the forward (backward) direction. 
}
{  
\footnotesize
\begin{center}
\begin{tabular}{|l|c|c|}
\hline
Detector Module & Abbreviation \\
\hline
Central Silicon Tracker  & CST  \\
Central Pixel Tracker  & CPT  \\
Central Forward Tracker & CFT \\
Central Backward Tracker & CBT \\ \hline
Forward Silicon Tracker  & FST  \\
Backward Silicon Tracker  & BST \\ \hline
Electromagnetic Barrel Calorimeter & EMC   \\
Hadronic Barrel Calorimeter & HAC   \\
Hadronic Barrel Calorimeter Forward & FHC4   \\
Hadronic Barrel Calorimeter Backward & BHC4   \\ \hline
Forward Electromagnetic Calorimeter Insert 1/2 &  FEC1/FEC2  \\
Backward Electromagnetic Calorimeter Insert 1/2 &  BEC1/BEC2  \\
Forward Hadronic Calorimeter Insert 1/2  & FHC1/FHC2 \\
Backward Hadronic Calorimeter Insert 1/2  & BHC1/BHC2  \\
\hline
\end{tabular}
\end{center}
}
\label{LHEC:MainDetector:Description:Fig:4a}
\end{figure}

\begin{figure}[htp]
\begin{center} 
\includegraphics[width=0.8\columnwidth]{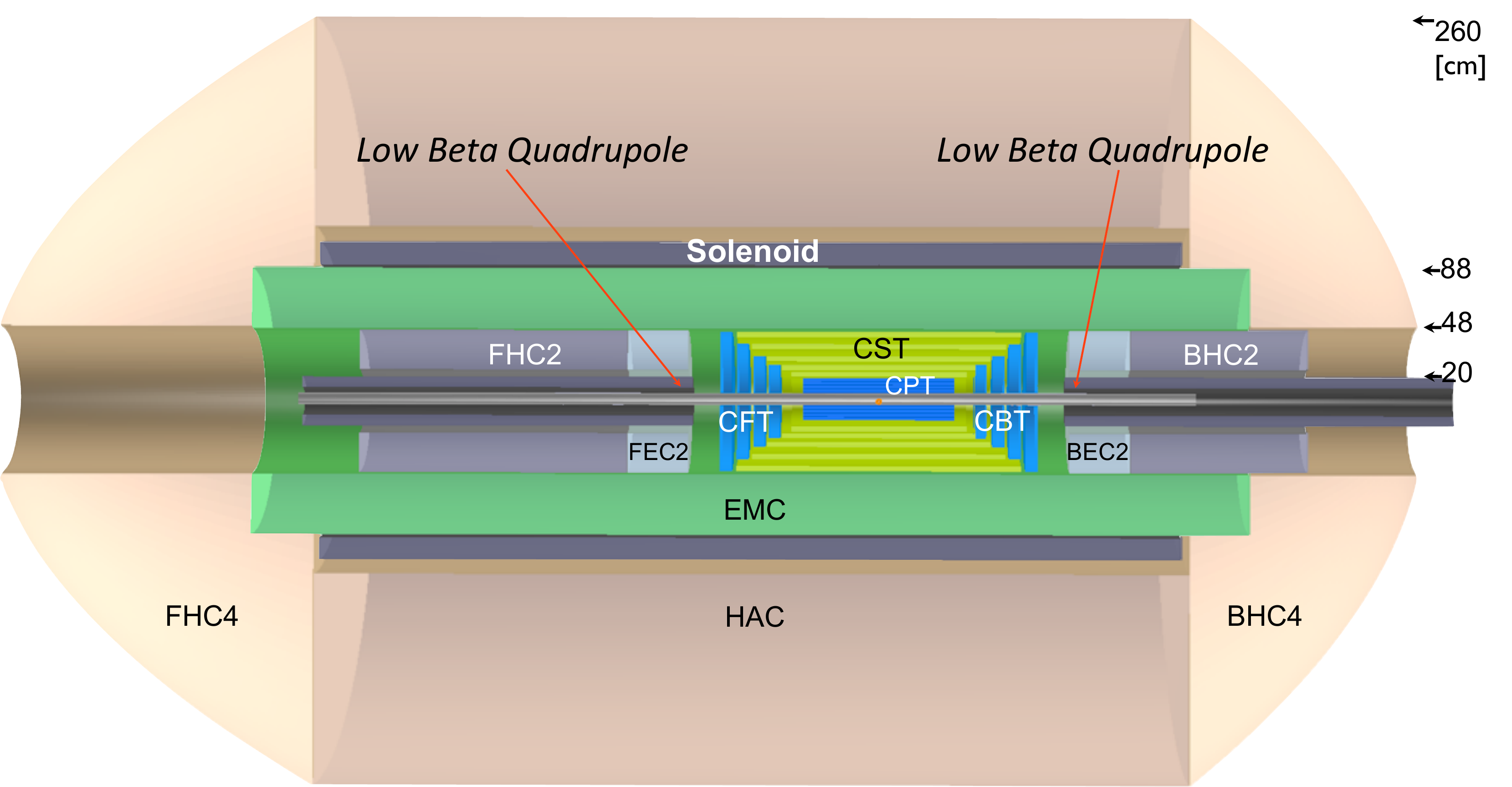}
\end{center}
\caption{
An $rz$ cross section and the dimensions of the main detector
(muon detector not shown) for the Ring-Ring detector version (no dipoles)
in which the luminosity is maximised by
replacing the forward and backward tracker telescopes by strong
focusing low $\beta$ quadrupoles  at $\pm~1.2$\,m
away from the nominal interaction point. The polar angle
acceptance is thus reduced to about $8 - 172^{\circ}$.
As compared to the high acceptance
detector (Fig.\,\ref{LHEC:MainDetector:Description:Fig:4a}),
the outer forward/backward calorimeter inserts have been 
moved closer to the interaction point.}
\label{LHEC:MainDetector:Description:Fig:4b}
\end{figure}

The LHeC inner detector is designed with a modular structure as is
illustrated in Figs.\,\ref{LHEC:MainDetector:Description:Fig:4a}
and \ref{LHEC:MainDetector:Description:Fig:4b} which shows the
detector without and with the strong focusing low $\beta$ quadrupole
inserts, respectively.
This requires the removal of the forward/backward tracking setup (shown in red in 
Fig.\,\ref{LHEC:MainDetector:Description:Fig:4a}) and the
subsequent re-installation of the external 
forward/backward electromagnetic and hadronic calorimeter plugins near to the vertex.
The high luminosity apparatus would have 
a polar angle acceptance coverage of about 
{8}\textdegree-{172}\textdegree~for an estimated gain in luminosity of slightly higher than a factor of two
with respect to the large acceptance configuration.
The Ring-Ring and Linac-Ring detectors also differ due to different
optics and beam pipe geometry.


In the Ring-Ring design the $e$ and $p/A$ beams collide with a small non-zero crossing angle,
large enough to avoid parasitic crossings, which for a $25$\,ns bunch crossing occur
at $\pm3.75$\,m from the IP. Additional masks are used to shield the inner part of the
detector from synchrotron radiation generated upstream of the detector.
 
\begin{figure}[htp]
\begin{center}
\includegraphics[width=0.6\columnwidth]{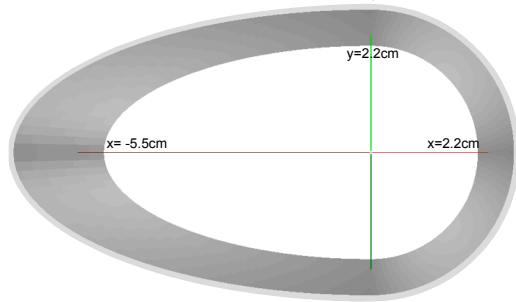}
\vspace*{-0.2cm}
\end{center}
\caption{Perspective drawing of the beam pipe and its dimensions
in the ring-ring configuration. The dimensions consider a $1$\,cm safety
margin around the synchrotron radiation envelope with 
masks (not shown) for primary synchrotron radiation suppression placed at $z=6,~5,~4$\,m.}
\label{LHEC:MainDetector:Description:Fig:7}   
\end{figure}

\begin{figure}[htp]
\begin{center}
\includegraphics[width=0.6\columnwidth]{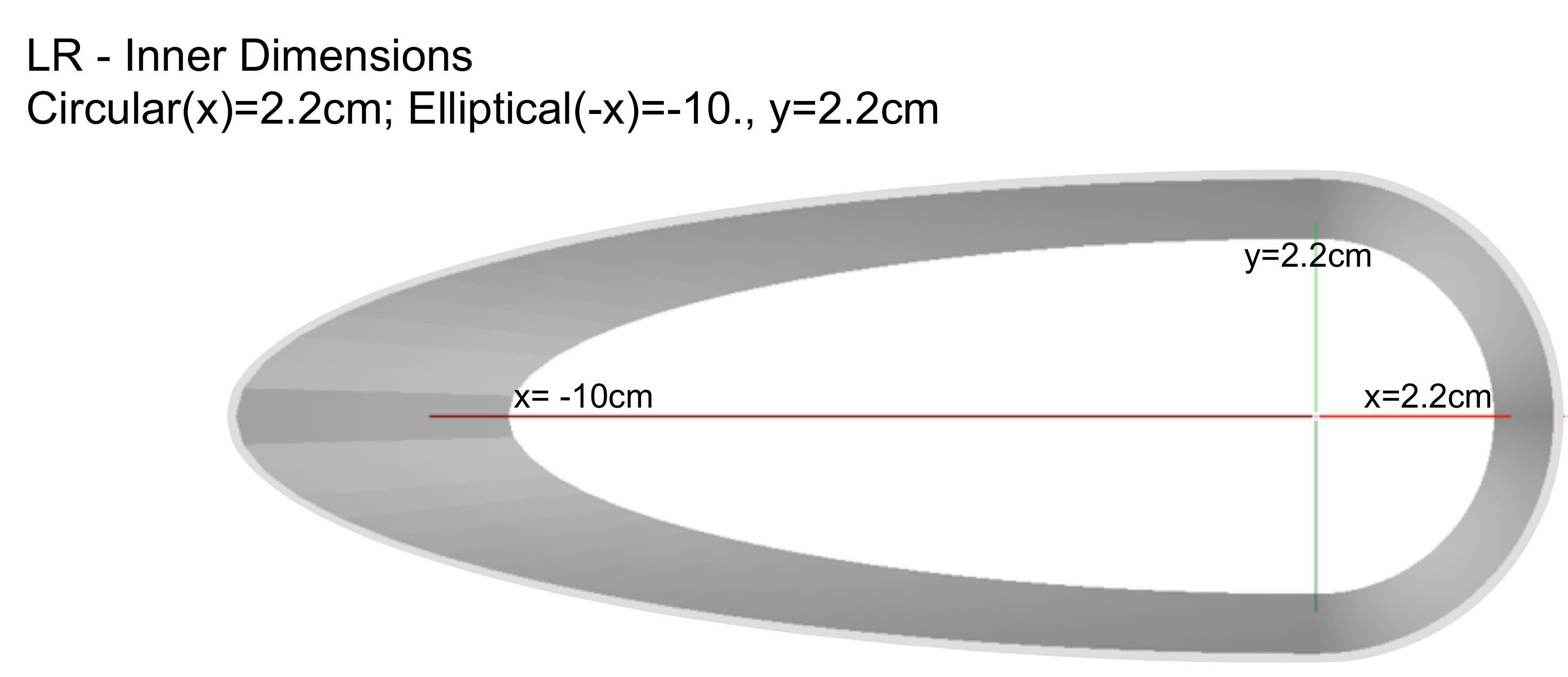}
\vspace*{-0.2cm}
\end{center}
\caption{Perspective drawing of the beam pipe and its dimensions
in the linac-ring configuration. The dimensions consider a $1$\,cm safety
margin around the synchrotron radiation envelope.}
\label{LHEC:MainDetector:Description:Fig:8}   
\end{figure}

For the Linac-Ring design, the dipole field produces additional
synchrotron radiation which has to pass through the interaction
region, requiring a larger beam pipe.  This difference results in the
horizontal dimension of the beam pipe being larger by a factor of two
in the outer-rear region ($-z,+x$), which is undesirable but necessary
to fully contain the synchrotron radiation fan (see
Fig.\,\ref{irlayout}).  
%
First estimates of the synchrotron radiation and placement of masks
to shield the detector from direct and backscattered photons have been
used to calculate the beam pipe geometries, shown in
Fig.\,\ref{LHEC:MainDetector:Description:Fig:7} for the Ring-Ring case
and in Fig.\,\ref{LHEC:MainDetector:Description:Fig:8} for the
Linac-Ring case.

As already mentioned, the necessity to register
particle production down to $1$ and $179^{\circ}$ poses severe
constraints on the material and the thickness of the pipe. 
In the design as shown here,
a beryllium pipe would have $3.0~(1.5)$\,mm thickness in the Linac-Ring (Ring-Ring) case. 
An extensive R\&D program is needed to ensure the high stability of the beam pipe with
these dimensions and for thinner/lighter beam pipe construction resulting in higher 
transparency for all final state particles. This R\&D program is necessary regardless
of which machine option for the LHeC facility is selected.
It may also turn out to be advantageous to use a trumpet shaped beam pipe when this
problem gets revisited in a more advanced phase of the
LHeC design when more detailed simulations will be available.

\protect \begin{figure}[htp]
\begin{center}
\includegraphics[width=0.6\columnwidth]{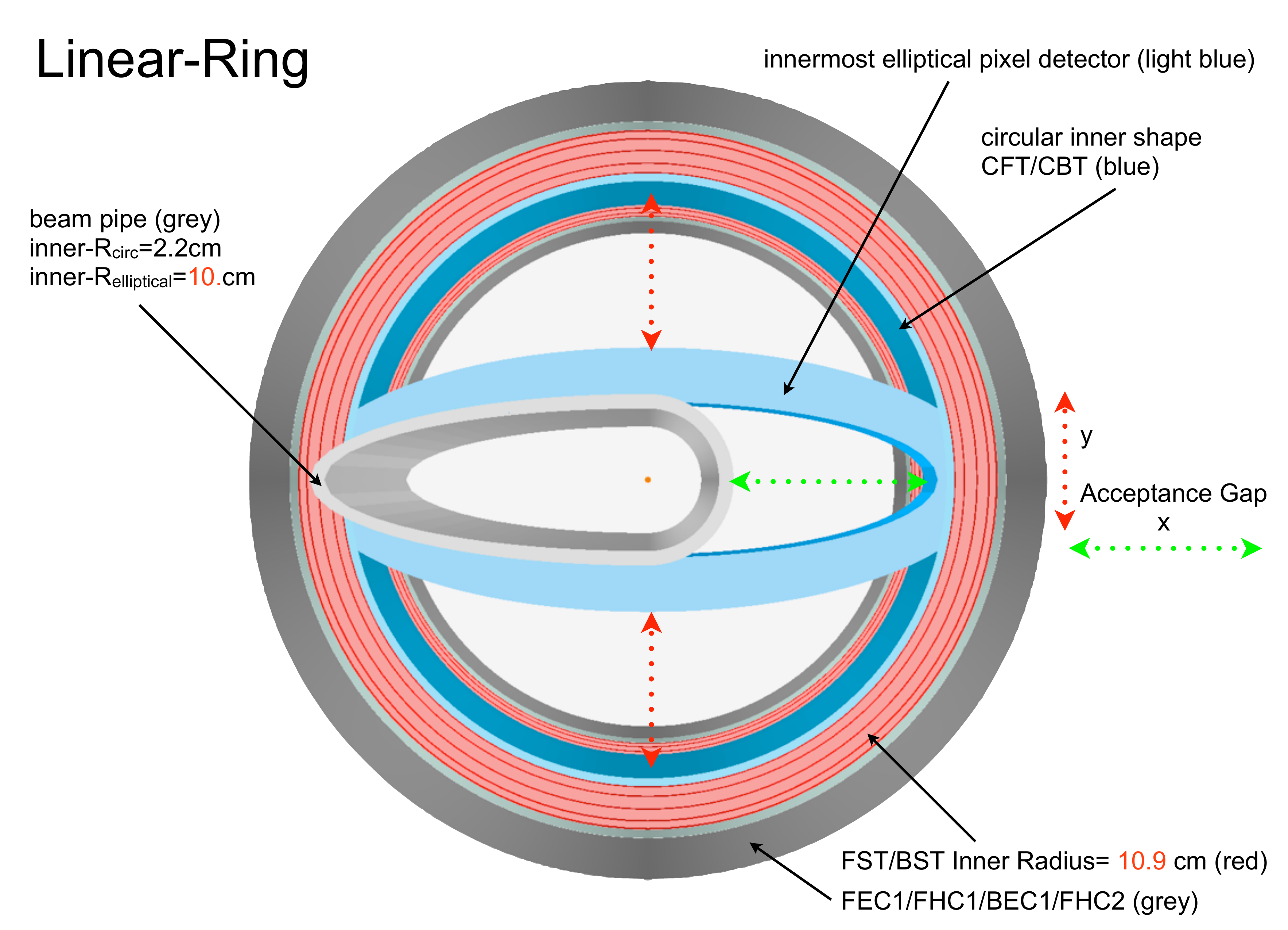}
\caption{Linac-Ring beam pipe design and acceptance gaps  due to deviations in
shape of the forward/backward tracking detectors FST/BST (circular) and the 
innermost central pixel detector layer (elliptical) from the beam pipe shape.}
\label{LHEC:MainDetector:Description:Fig:2a}
%
\includegraphics[width=0.5\columnwidth]{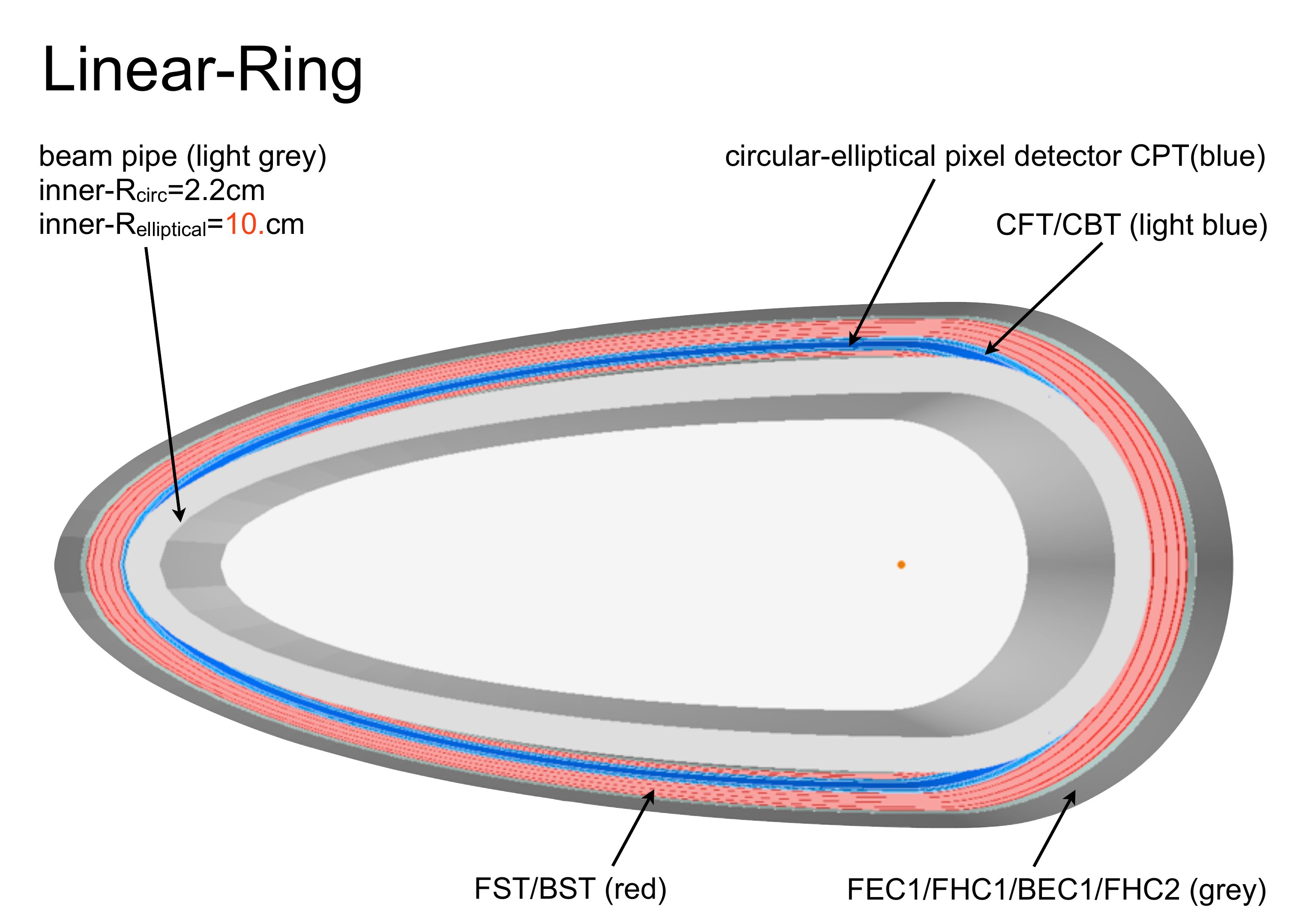}
\caption{Beam pipe design for Linac-Ring and optimised circular-elliptical shape following the beam pipe 
for all adjacent detector parts.}
\label{LHEC:MainDetector:Description:Fig:2b}
\end{center}
\end{figure}

In order to ensure optimal polar angle acceptance, the innermost
subdetector dimensions have to be adapted to the beam pipe
shape. Fig.\,\ref{LHEC:MainDetector:Description:Fig:2a} illustrates
the configuration that a circular silicon tracker would imply and the
corresponding acceptance losses. These can be reduced as shown in
Fig.\,\ref{LHEC:MainDetector:Description:Fig:2b} if the detector
acceptance follows the elliptic-circular shape of the pipe as closely
as possible.  Electrons scattered at high polar angle, corresponding
to small $Q^2 \sim 1$\,GeV$^2$, will only be registered in the inner
part of the azimuthal angle region for the nominal electron beam
energy. As was shown in
Section\,\ref{LHEC:Detector:Requirements}(Eq.\,\ref{Eq:q2min}),
lowering the electron beam energy effectively reduces the
requirement of measuring up to about $179^{\circ}$, at the expense
of a somewhat reduced acceptance towards the lowest Bjorken $x$.

The optimum configuration of the inner detector will be revisited when the
choice between the Linac-Ring and the Ring-Ring option is made. It represents in any case
one of the most challenging problems to be solved for the LHeC.
\subsection{Baseline detector layout}
\label{LHEC:MainDetector:MainStructure}
The baseline configuration ({\small\bf A}) of the main detector has the solenoid in between the
two calorimeters, combined with a dipole field in the Linac-Ring case. 
The main detector is subdivided into a central barrel and the forward
and backward end-cap regions, which differ in their design.  

\begin{figure}[ht]
\begin{center}
\includegraphics[width=1.0\columnwidth]{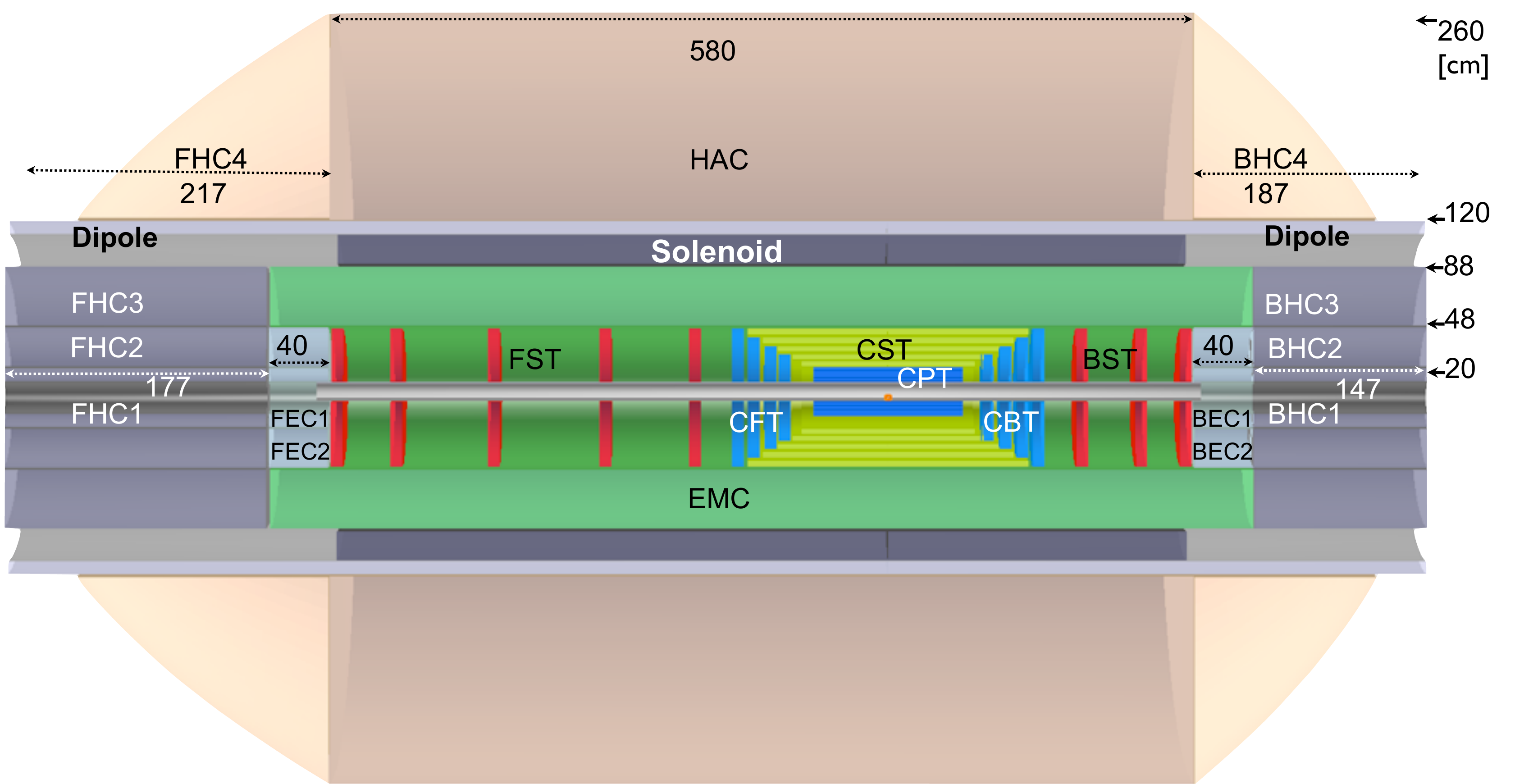}
\end{center}
\caption{An $rz$ cross section of  the LHeC detector
in its baseline configuration ({\small\bf A}).
In the central barrel, the following components are considered:
a central silicon pixel detector (CPT);
silicon tracking detectors (CST,CFT/CBT)
of different technology;
an electromagnetic calorimeter (EMC)
surrounded by the magnets and followed 
by a hadronic calorimeter (HAC).
Not shown is the muon detector.
The electron at low $Q^2$ is scattered into the
backward silicon tracker (BST) and its energy measured
in the BEC and BHC calorimeters. In the forward region 
similar components are placed for tracking (FST)
and calorimetry (FEC, FHC).}
\label{LHEC:MainDetector:MainStructure:Fig:1}
\end{figure}

\begin{figure}[htp]
\begin{center}
\includegraphics[width=1.0\columnwidth]{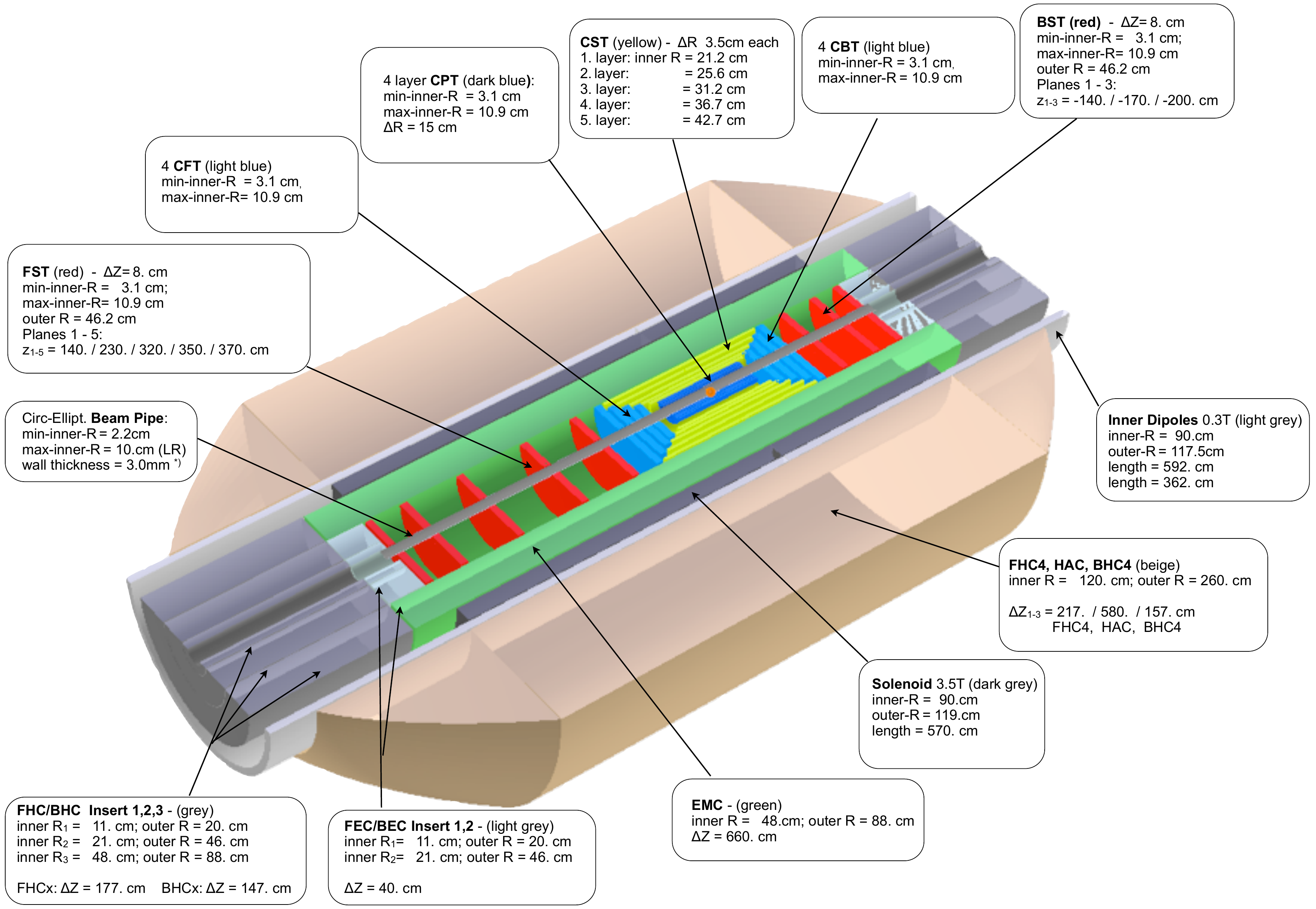}
\end{center}
\caption{View of the baseline detector 
configuration ({\small\bf A}) with some dimensions for each of the main detector components.}
\label{LHEC:MainDetector:MainStructure:Fig:5}
\end{figure}

The backward region usually detects the scattered electron and
typically has low occupancy and energy deposits from the hadronic
final state, while the forward region detects the proton remnant and
typically has much higher occupancy and large energy deposits.  The
detector configuration is sketched in
Fig.\,\ref{LHEC:MainDetector:MainStructure:Fig:1} with component
abbreviations and some important dimensions.
More details are given in
 Fig.\,\ref{LHEC:MainDetector:MainStructure:Fig:5}.

For the purpose of this design,  technologies had to be chosen
following  the detector requirements discussed in Sect.\,\ref{LHEC:Detector:Requirements},
and based on an evaluation of the technologies available or under
development for the LHC experiments or foreseen for a
linear collider detector. Due to its compact design and proven technological
feasibility, the complete inner tracker is 
based on silicon detectors.  This allows  the radius of the
magnets to be kept small, about $1$\,m.
Based on experience with H1 and ATLAS, the EMC is
chosen to be a Liquid Argon (LAr) Calorimeter.  The
superconducting dipoles (light grey in  Fig.\,\ref{LHEC:MainDetector:MainStructure:Fig:1})  
are placed  in a common cryostat with the detector solenoid (dark grey) 
and the LAr EMC (green). The use of a common cryostat is optimal for
reducing the amount of material present in front of the 
hadronic barrel calorimeter.
The HAC is an iron-scintillator tile calorimeter,
which also guides the return flux of the magnetic 
field, as in ATLAS\cite{Adragna:2010zz,Aad:2008zzm}.
In the baseline design ({\small\bf A}) the muon detectors are placed 
outside of the magnetic field with the function of tagging muons,
the momentum of which is determined mainly by the inner tracker.

For the Ring-Ring machine, in order to maximise the luminosity, extra focusing magnets 
must be placed near to the interaction point
\footnote{See Section\,\ref{LHEC:machine:RR-IR-layout} for an evaluation of that possibility.}. 
This would mean replacing the FST and the BST tracking detectors  
by the low-$\beta$ quadrupoles (see Fig.\,\ref{LHEC:MainDetector:Description:Fig:4b}),
at the expense of losing about $8^{\circ}$ of polar angle acceptance. 
The modular design of the forward and backward trackers and the corresponding 
calorimeter modules allow the trackers to be mounted/unmounted and the 
calorimeter inserts to be moved in and out of position as required.  The inner 
electromagnetic and hadronic endcap inserts, FEC1/BEC1 and FHC1/BHC1, 
respectively, will be removed allowing the insertion of the low $\beta$-magnets
and only partially put back in.
Particular attention is needed for the mechanical support structures
of the quadrupoles.
The structure must ensure the stability for reproducible beam steering, 
while interfering as little as possible with the detector. 
The presence of strong focusing magnets close to the interaction point was one
issue experienced during HERA-II running\cite{holzer_hera}.

\subsection{An alternative solenoid placement - option {\small\bf B}}
\label{LHEC:MainDetector:OptionB}
\begin{figure}[htp]
\begin{center}
\includegraphics[width=1.0\columnwidth]{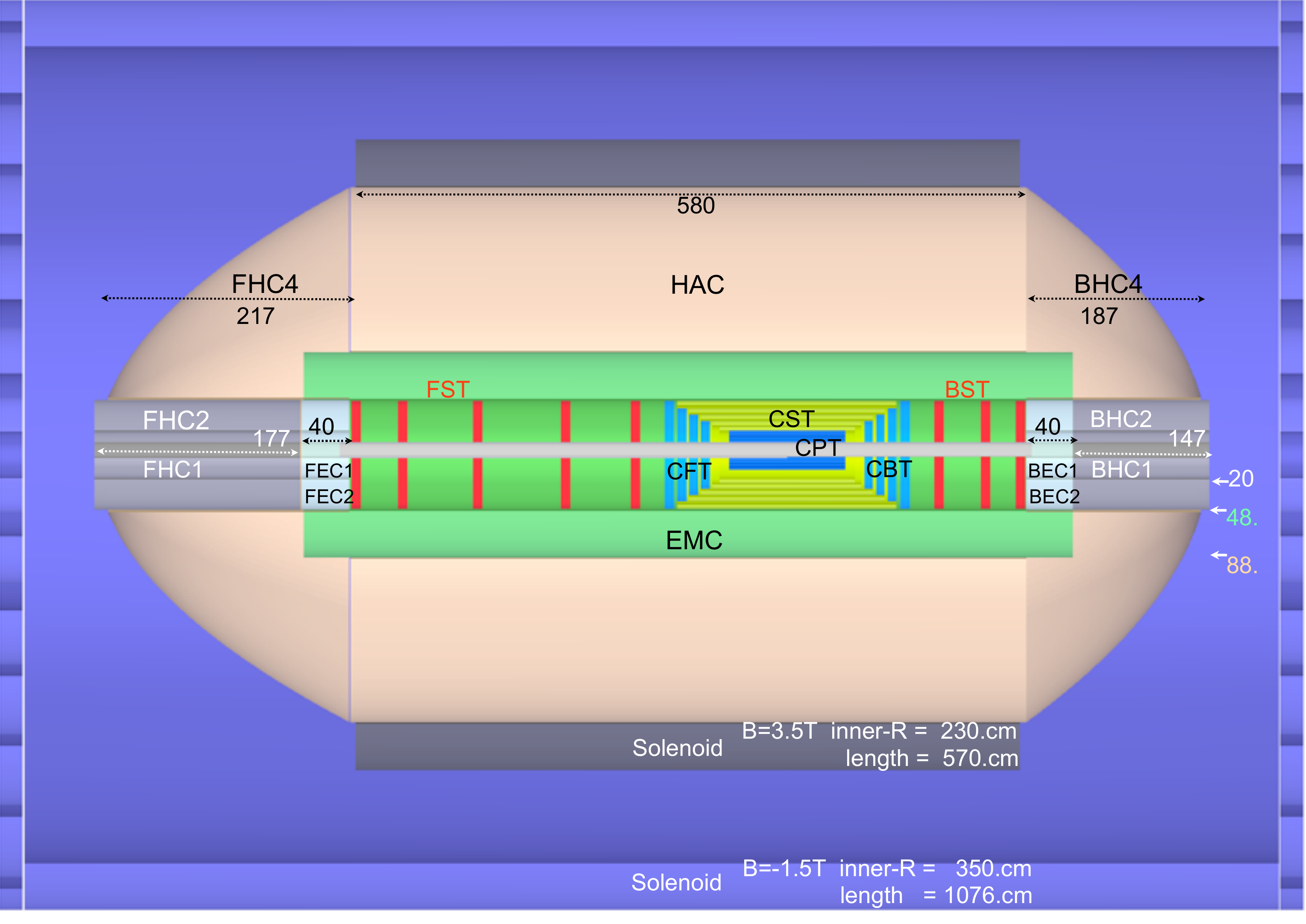}
\end{center}
\caption{
An $rz$ cross section of the LHeC detector, option {\small\bf B}, in which
the solenoid is placed outside the HAC.
A compensating larger solenoid is considered, see text.
The muon detector is not shown but would be placed
inside the second solenoid. The overall dimensions 
of this detector configuration are about 
$11$\,m length and $8$\,m diameter.
}
\label{LHEC:MainDetector:MainStructure:Fig:6}
\end{figure}
The configuration {\small\bf A} is driven by the intention to keep the detector 
`small': it uses the HAC as flux return for the solenoid which,
for the Linac-Ring case, is combined with long dipoles. This is not ideal
for the hadronic energy measurement. Therefore a second configuration ({\small\bf B})
has been considered, although in much less detail, in which the solenoid is placed
outside the HAC. Option {\small\bf B} would only be of interest for the Ring-Ring
case as the requirement of placing bending dipoles
immediately after the EMC would compromise this design.

Having a solenoid around the HAC implies, as from the CMS geometry,
that the return iron would be very large, of the order of 10 000 tons,
and extend by several metres further out in radius, which may conflict
with IP2 cavern constraints. A second solenoid could be considered for
an active flux return, which gives a good muon momentum
reconstruction. A strong magnetic field of $3.5$\,T covering the
barrel calorimeter (HAC) leads to a better separation of charged
hadron induced showers in the HAC area compared to the sole fringe
field effect in case of the inner solenoid baseline design {\small\bf
A}.  The HAC would have to be designed very carefully as there would
be no muon-iron return yoke following for catching shower tails.  A
warm EMC design with no need for a cryostat would become an option
worth considering.  The space gained could be used by an extra
tracking detector layer.
 
An overview of the detector configuration {\small\bf B} is given in
Fig.\,\ref{LHEC:MainDetector:MainStructure:Fig:6}. A two solenoid configuration
is  proposed as an innovative solution with many advantages. 
A similar design was proposed earlier for the 4$^{th}$ Concept for an ILC Detector\cite{Mazzacane:2010zz}.
The second outer solenoid keeps the overall dimensions of the detector limited.
A detailed consideration of option {\small\bf B} has not been intended at this stage of the project,
however, the statement is made that the option {\small\bf B} magnet system is technically feasible 
and can be chosen if physics arguments require to do so and the required extra budget 
is made available.


%% file: detector/tenkate.tex

%
%
%
\label{LHEC:MainDetector:MagnetDesign}

The principle magnet configuration in the Linac-Ring baseline option
is introduced and the principle design of solenoid and dipole magnets
as well as their cryogenic services are described. 

\subsection{Magnets configuration}

\begin{figure}[htp]
\centerline{\includegraphics[clip=,width=0.95\textwidth]{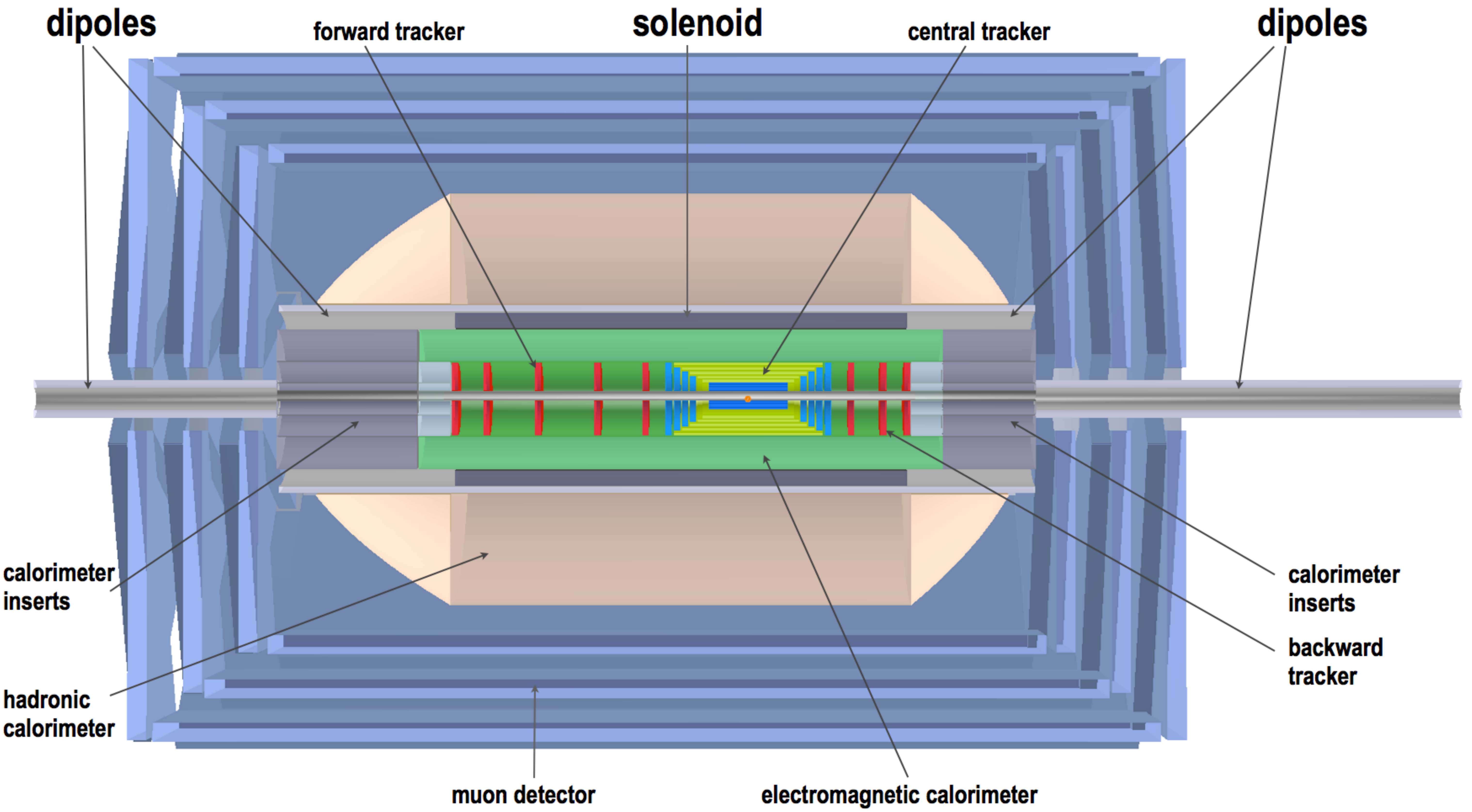}}
\caption{
Configuration of the solenoid and electron beam bending dipoles in the
baseline Linac-Ring detector. Longitudinal r-z section showing the
position of the solenoid and the two dipoles, each split in two sections,
a superconducting inner section incorporated with the solenoid in one
cryostat and a normal conducting iron based outer section magnet with
smaller bore.
}
\label{Fig:HTK1}
\end{figure}

The LHeC magnet system provides a 3.5\,T solenoid with a free bore of
1.8\,m and a coil length of 5.7\,m  .
The bore is designed to provide space for the Pixel (CPT) and Strip
(CST) detectors as well as the electromagnetic Liquid Argon
calorimeter (EMC) immersed in a magnetic field while the hadronic tile
calorimeter (HAC) and muon tagging detectors are placed outside. The
layout of the magnets in the baseline detector design is shown in
Figure\,\ref{Fig:HTK1}. The iron present in the hadronic calorimeter
also provides the return path for the solenoid magnetic field.  In the
Linac-Ring option a set of 18\,m long e-beam bending dipoles are also
required that provide 0.3\,T on axis, a positive and a negative dipole
of 9\,m length each, respectively. The aim of the first dipole is to
bring the $e$-beam into the collision point, while the second has to
guide the beam away from the proton line.  There is no need for these
dipoles in the Ring-Ring option.  The Linac-Ring option is therefore
more demanding and is thus taken as the reference design presented
here.  The need for these dipoles require a radial position and radial
gap for these coils to fit. Since cryogenic space is required for the
solenoid as well, an elegant solution is to combine within the
detector volume the dipoles and the solenoid in one cryostat, thereby
minimising the total radial gap as well as maximising particle
transparency. A second combination of cryogenic objects can be made by
also housing the liquid argon electromagnetic calorimeter in the same
cryostat, which would reduce the material budget significantly.  Since
a combination is easier the separate, more demanding option is
described here. Since the set of dipoles is 18\,m long to provide the
2$\cdot$2.5\,Tm magnetic field integral, and the detector is 10\,m
long, each of the two dipoles are split in two sections. The inner
superconducting sections sit together with the solenoid in the same
cryostat, while the outer normal conducting iron based electromagnetic
sections with much smaller bore of 0.3 m are positioned on the beam
line either side of the detector, see Figure\,\ref{Fig:HTK1}.
	
\subsection{Detector solenoid}

The conceptual design of the solenoid is presented here and where
necessary some details on the dipoles are mentioned as well. The
position of the solenoid with respect to the other detector components
are shown in Figure\,\ref{LHEC:MainDetector:MainStructure:Fig:1}.  The
longitudinal section of the LHeC baseline detector for the default
detector configuration and the Linac-Ring option are shown; indicated
are the position of the 3.5\,T solenoid and the 0.3\,T inner
superconducting dipole sections.  Solenoid and dipoles are on a common
support cylinder and housed in a single cryostat with a free bore of
1.8\,m extending along the entire detector with a length of
$\approx$10\,m.


The design of the solenoid is based on the very successful experience
with many detector magnets built over the past 30 years, in
particular the most recent ATLAS and CMS solenoids
\cite{:1997fy},\cite{:1997fu},\cite{:1997ki},\cite{Acquistapace:1997fm}. 
The dimensions of the LHeC solenoid (3.5\,T, 5.7\,m long and 0.96\,m
inner radius) are about those of the ATLAS solenoid (2.0\,T, 5.3\,m
long with 1.25\,m radius) while it has to provide the magnetic field
of the much larger CMS solenoid.  Since the requested magnetic field
is 1.75 times higher than in the ATLAS solenoid a double layer coil
will be needed. Using well established design codes with proven
records on earlier detector magnets, the main solenoid parameters are
determined and are listed in Table\,\ref{tab:HTK1}.

\begin{table}[htp]
  \centering
  \begin{tabular}{|c|c|c|c|}
    \hline
Property & Parameter & value & unit\\
    \hline     
Dimensions & Cryostat inner radius & $0.900$ & m\\
 &           Length & $10.000$ & m\\
 &            Outer radius & $1.140$ & m\\
 & Coil windings inner radius & $0.960$ & m\\
 &                    Length & $5.700$ & m\\
 &                 Thickness & $60.0$ & mm\\
 & Support cylinder thickness & $0.030$ & m\\
 & Conductor sect., Al-stabilised NbTi/Cu \& insulation & $30.0 \times 6.8$ & $mm^2$\\
 &                 Length &   $10.8$ & km\\
 & Superconducting cable sect., $20$ strands & $12.4 \times 2.4$ & $mm^2$\\
 & Superconducting strand $\diameter$ Cu/NbTi ratio = $1.25$ &    $1.24$ & mm\\
Masses & Conductor windings &    $5.7$ & t\\
 & Support cylinder, solenoid sect. + dipole sect.s &  $5.6$ & t \\
 & Total cold mass &  $12.8$ & t\\
 & Cryostat including thermal shield &  $11.2$& t\\
 & Total mass of cryostat, solenoid and small parts &  $24$& t\\
Electro- & Central magnetic field &  $3.50$ & T\\
magnetics & Peak magnetic field in windings (dipoles off) & $3.53$ & T\\
 & Peak magnetic field in solenoid windings (dipoles on) &  $3.9$ & T\\
 & Nominal current &  $10.0$ & kA\\
 & Number of turns, $2$ layers & $1683$ & \\
 & Self-inductance &  $1.7$ &H\\
 & Stored energy &  $82$ & MJ \\
 & E/m, energy-to-mass ratio of windings &  $14.2$ & kJ/kg\\
 & E/m, energy-to-mass ratio of cold mass & $9.2$ & kJ/kg\\
 & Charging time & $1.0$ & hour \\
 & Current rate &  $2.8$ & A/s \\
 & Inductive charging voltage & $2.3$ & V \\
Margins & Coil operating point, nominal / critical current & 0.3 & \\
 & Temperature margin at 4.6 K operating temperature & $2.0$ & K\\
 & Cold mass temperature at quench (no extraction) & $\sim 80$ & K\\
Mechanics & Mean hoop stress & $\sim 55$ & MPa\\
 & Peak stress & $\sim 85$ & MPa \\
Cryogenics & Thermal load $@4.6$ K, coil with $50\%$ margin & $\sim 110$ & W\\
 & Radiation shield load width $50\%$ margin & $\sim 650$ & W\\
 & Cooling down time / quench recovery time &  $4$ and $1$ & day\\
 & Use of liquid helium &  $\sim 1.5$ & g/s\\
\hline
  \end{tabular}
\caption{
Main parameters of the baseline LHeC Solenoid providing 3.5\,T in a
free bore of 1.8\,m.
}
\label{tab:HTK1}
\end{table}

\begin{figure}[htp]
\centerline{\includegraphics[clip=,width=0.8\textwidth]{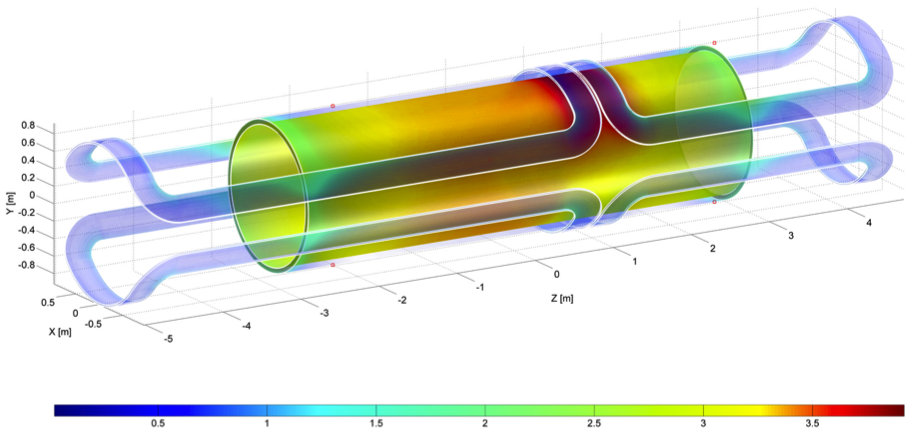}}
\centerline{\includegraphics[clip=,width=0.8\textwidth]{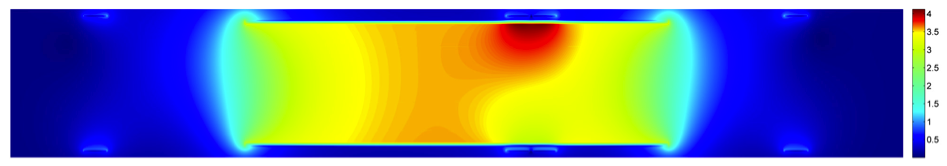}}
\caption{ Magnetic field of the magnet system of the solenoid and two
  internal superconducting dipoles at nominal currents (effect of iron
  ignored). The position of the peak magnetic field of 3.9\,T is
  local due to the adjacent current return heads on top of the
  solenoid where all magnetic fields add up.  }
\label{Fig:HTK3}
\end{figure}

The solenoid is wound in two layers internally in an Al5083 alloy
support cylinder with 30\,mm wall thickness and a length of about
6\,m.  When finished two extension cylinders are flanged to the
central solenoid section at either end to support the inner
superconducting dipole sections, see Figure\,\ref{Fig:HTK3}. In this
way the solenoid can be produced as a 6\,m long coil unit, and then
transported to the integration site where the adjacent sections are
coupled and the dipoles sections can be introduced.

The magnetic field generated by the system of solenoid and internal
dipoles is shown in Figure\,\ref{Fig:HTK3}. The peaks in magnetic field in the
solenoid and dipole windings as a result of their combined operation at
nominal current are 3.9 and 2.6\,T respectively.
\begin{figure}[htp]
\centerline{\includegraphics[clip=,width=0.8\textwidth]{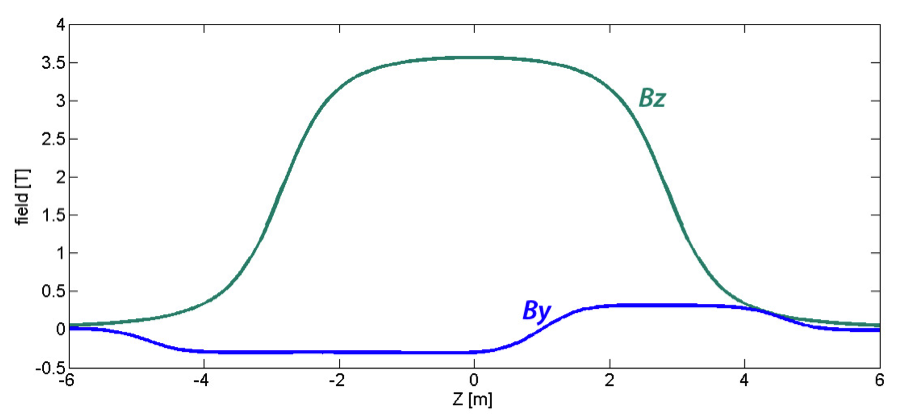}}
\caption{ Magnetic field components $B_z$ (solenoid) and $B_y$ (set of
  internal dipoles) on the beam axis across 12\,m in \emph{z}. Note, the
  magnetic field of the external electromagnets are not included here.
}
\label{Fig:HTK4}
\end{figure}
The $B_z$ and $B_y$ components of the magnetic field are shown in Figure\,\ref{Fig:HTK4}.

\begin{figure}[htp]
\includegraphics[clip=,width=0.45\textwidth]{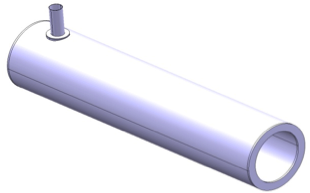}
\includegraphics[clip=,width=0.55\textwidth]{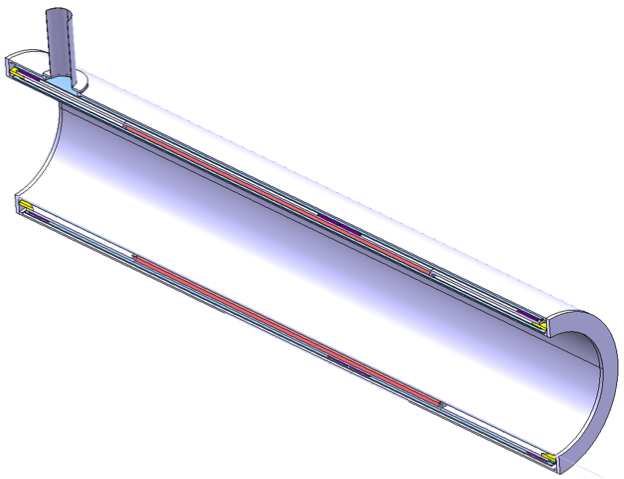}
\caption{
Cryostat of the magnet system. Left: the integrated cryostat, and
right: longitudinal cut through the cryostat comprising a single cold
mass of solenoid and internal superconducting dipole sections.
}
\label{Fig:HTK5}
\end{figure}

The superconductor used for the solenoid is an Al stabilised NbTi/Cu
Rutherford cable based on state-of-the-art NbTi strands featuring
3000\,A/mm$^2$ critical current density at 5\,T and 4.2\,K. 
A 20 strand Rutherford cable carries the nominal current of 10\,kA 
which is 30\% of its critical current. 

The conductor has a comfortable temperature margin of 2.0\,K when
operating the coil with a forced Helium flow enabling 4.6\,K in the
solenoid windings.  The high purity Al used for the co-extrusion of Al
and cable is mechanically reinforced by micro-alloying with either Ni
or Zn, or another suitable material, a technology proven with the
ATLAS solenoid.  Ideally, two conductor units of 5.4\,km would be
used, corresponding to the two layers in the coil windings. In
practice, internal splices are acceptable and can be made reliably by
overlapping a full turn and performing welding on the two adjacent
thin edges of the conductors.

The conductor insulation is a double layer of 0.3\,mm thick
polyimide/glass tape (or similar product) featuring a high breakdown
voltage of more than 2\,kV and robustness for coil winding damage in
order to limit the risk of turn-to-turn shorts. Coil winding can be
performed either using the wet winding technique with pre-impregnated
tape or a vacuum impregnation technique may be applied.  Both
techniques are appropriate provided they have been fully tested
with the coil winding contractor.

Once the solenoid windings are finished and delivered to the coil
integration site, the dipole coil sections are inserted in slots
milled into the outer surface of the support cylinder, see
Section\,\ref{LHEC:MainDetector:MagnetDesign:dipoles}.  The four upper
and lower dipole coil sections are separately produced as flat
racetrack coils and then bent onto the fully assembled support
cylinder.  Next, all interconnections and bus connections to the
current leads are laid down and the cold mass is inserted in the
cryostat.

The cryostat design is shown in Figure\,\ref{Fig:HTK5}. 
The cold mass is supported from the cryostat with a system
of triangle brackets, a proven technique providing a very compact
solution\cite{:1997fy},\cite{:1997fu}. The cryostat is equipped with thermal
shields and multi-layer super-insulation in the usual way.

The coil windings of both solenoid and dipole sections are cooled by
conduction, using forced flow liquid helium circulating in 14\,mm
sized cooling tubes that are attached to the outer surface of the
integrated support cylinder. The two layer winding pack of 60\,mm
radial built and fully bonded to the support cylinder is sufficiently
thin to warrant a thermal gradient in the winding pack of less than
0.1\,K.  The total radial material built of essentially Al alloys is
about 150\,mm providing an acceptable effective radiation thickness.

Quench protection of the solenoid with 82\,MJ of stored energy in a
cold mass with 9\,kJ/kg can be done safely.  The stored energy is
absorbed by the cold mass enthalpy (no energy extraction) and the cold
mass temperature will rise to a safe level of 80\,K.  Heat drains are
incorporated in the coil windings to accelerate quench propagation and
in addition an active heater system will be implemented for the same
purpose.

\subsection{Detector integrated e-beam bending dipoles}
\label{LHEC:MainDetector:MagnetDesign:dipoles}

\begin{table}[h]
  \centering
  \begin{tabular}{|c|c|c|c|}
    \hline
 & Plus coil & Minus coil & \\
\hline
Magnetic field on axis & \multicolumn{2}{c|}{0.3}  & T\\
Peak magnetic field in windings (solenoid off) & \multicolumn{2}{c|}{0.7} & T\\
Peak magnetic field in windings (solenoid on) & \multicolumn{2}{c|}{2.6} & T\\
Dipole length (including external sections)  & \multicolumn{2}{c|}{9.0} & m\\
Field integral internal section (sc dipole) & 1.6 & 1.0 & Tm\\
Field integral external section (iron magnet) & 1.1 & 1.7 & Tm\\
Operating current & \multicolumn{2}{c|}{2.0} & kA\\
Stored Energy & 1.9 & 1.2 & MJ\\
Coil inductance & \multicolumn{2}{c|}{0.50} & H\\
Coil inner / outer radius & \multicolumn{2}{c|}{ 1.042/1052} & m\\
Coil length & 6.00 & 3.70 & m\\
NbTi/Cu conductor $\diameter$ (12 strands Rutherford cable) & \multicolumn{2}{c|}{2.0} & mm\\
Conductor length & 5.4 & 3.6 & km\\
\hline
  \end{tabular}
\caption{
Main design parameters of the set of superconducting electron beam bending dipoles.
}
\label{tab:HTK2}
\end{table}

The two e-beam bending dipoles are positioned symmetrically around the
intersection point of the beams.  As outlined before, each 9\,m long
dipole is split into a superconducting section integrated with the
central solenoid and a normal conducting iron based electro-magnet
positioned around the beam outside the main detector envelope. The
external dipole magnets are conventional and will not be further
detailed here. The principle parameters of the superconducting dipole
sections are listed in Table\,\ref{tab:HTK2}.

\subsection{Cryogenics for magnets and calorimeter}

\begin{figure}[htp]
\centerline{\includegraphics[clip=,width=0.85\textwidth]{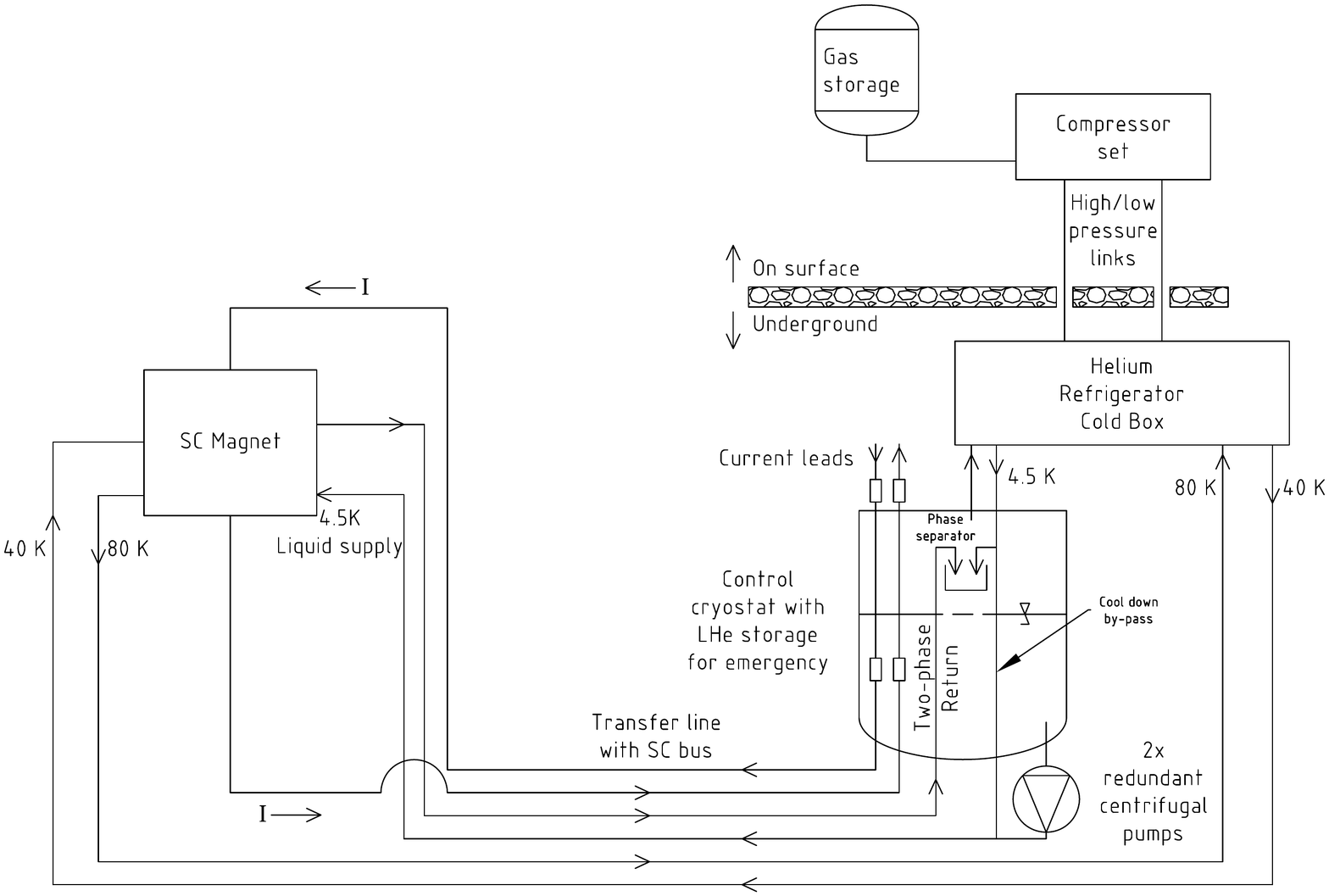}}
\caption{
Principle cryogenic flow scheme for the cooling of the superconducting magnets.
}
\label{Fig:HTK6}
\end{figure}
         
\begin{figure}[htp]
\centerline{\includegraphics[clip=,width=0.85\textwidth]{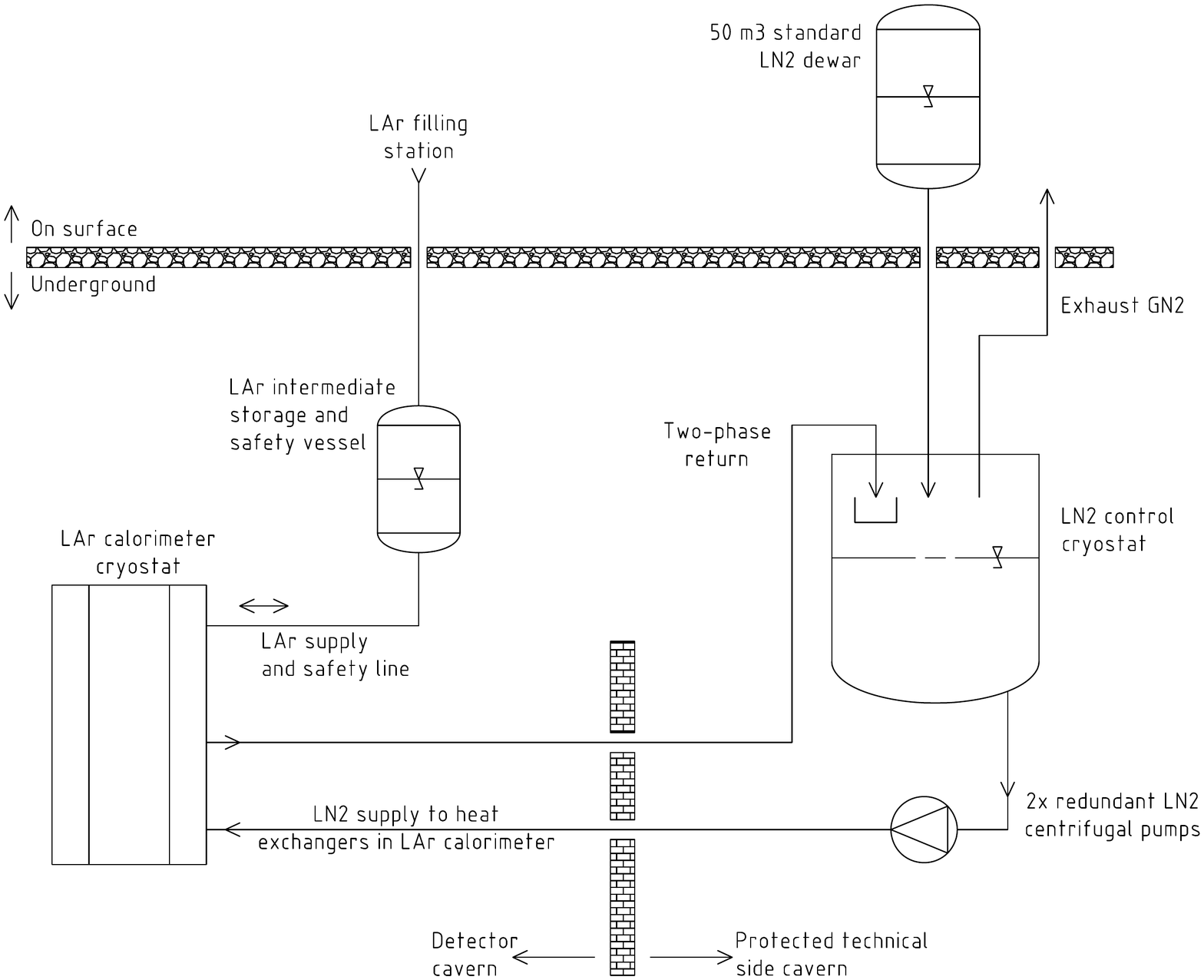}}
\caption{
Principle cryogenic flow scheme for the cooling of the liquid argon calorimeter.
}
\label{Fig:HTK7}
\end{figure}

The cryogenic operating conditions are achieved by circulating forced
flow two-phase helium in cooling pipes attached to the Al-alloy coil
support cylinder. Electric powering of the solenoid and dipole magnets
at 10 and 2\,kA, respectively, is through two pairs of low-loss
high-temperature superconducting current leads.  The current leads are
housed in a separate service cryostat installed at a distance in a
side cavern, a non-radiation environment. The service cryostat
contains a larger amount of helium sufficient for a safe 1-2 hours
ramp down in the case of refrigerator failure as well as being able to
maintain the magnets at operating temperature for a few
hours. Redundant centrifugal pumps provide for circulation of the
slightly sub-cooled liquid helium to the magnets. The two-phase return
flow is brought to a phase separator in the service cryostat.  A
combined superconducting link and helium transfer line connects the
service cryostat with the current leads and helium buffer to the
magnets.  For this circuit static and dynamic losses of the magnets
and transfer lines have to be taken into account, which are about
85\,W.  With 50\% contingency the losses amount to 130\,W.  For
reasons of flow stability the vapour quality of the return flow shall
not exceed 10\%.

The mass flow rate of the pump is calculated as 65\,g/s maximum.  A
thermo-hydraulic efficiency of the pump of 35\% is assumed, a value
based on measurements of similar systems which are already running.
The pump introduces an additional 40\,W to the system. 

The refrigerator is in close proximity to the cryostat, while the
compressor set is installed on the surface.  The expected modest
thermal loss of the magnet system and its cryogenics, such as the
service cryostat and transfer lines, amounts to some
200\,W$@$4.5\,K. The estimated overall system loss suggests a
small-sized standard refrigerator in the class of 300 to
400\,W$@$4.5\,K.  The thermal load of the system is summarised in
Table\,\ref{tab:HTK3}.  Figure\,\ref{Fig:HTK6} shows the simplified
flow scheme of the helium cryogenic system.

\begin{table}[h]
  \centering
  \begin{tabular}{|l|c|c|c|c|}
    \hline
Component heat load at temperature &   4.5\,K   & 20-300\,K &     40-80\,K  \\
\hline
Magnets   \hfill       static     &                  45\,W   &                  &      430\,W  \\
\hfill                      dynamic &                  30\,W   &                 &                   \\
Transfer line/bus    \hfill     static     &      10\,W  &                 &       150\,W   \\
Valve box cryostat   \hfill   static      &      10\,W  &                  &       150\,W \\
Helium pump       \hfill       static      &     40\,W  &                  &                   \\
Current leads      \hfill        static      &                &   1.0 g/s    &                   \\
\hline
Sums with and extra 50\% contingency & 200\,W & 1.5\,g/s  &  1100\,W  \\
\hline
  \end{tabular}
\caption{
Thermal load of the cryogenics system including magnets and helium distribution.
}
\label{tab:HTK3}
\end{table}

A liquid Argon calorimeter is envisaged as part of an EMC. As already
mentioned, it can be installed in a separate cryostat or preferably
share the cryostat with the solenoid. In the latter case the
compactness of the system is increased and the inner thermal shield
can be omitted. The calorimeter will have an overall volume of
18\,m$^3$ from which approximately 12\,m$^3$ will be slightly
sub-cooled liquid argon. Cooling is provided by two-phase liquid
nitrogen in longitudinal pipe runs and circulation is provided by two
redundant small sized liquid nitrogen pumps. The liquid nitrogen is
supplied from a standard dewar on the surface to an intermediate
cryostat which also serves as the phase separator. For the liquid
argon, a line is needed connecting the surface to an intermediate
dewar from which it is transferred to the LAr cryostat in the
detector. This dewar also serves as emergency volume in the case of
vacuum loss or leak problems to which the liquid argon can be
transferred from the cryostat. Figure\,\ref{Fig:HTK7} shows the
functional principle of the Argon cooling units.

The cooling principles of both cryogenic systems proposed here are
based on previous design and experience from the much more complex
ATLAS detector cryogenics.




%% file: detector/tracking.tex
\label{LHEC:MainDetector:tracking}
The constraints given by the magnet system (dipole/solenoid) force the
tracking detectors to be kept as small as possible in radius.
According to equation \ref{Eq:ptreso}, the momentum resolution is
proportional to 1/L$^2$ and is therefore limited by the tracker
radius.  For a given magnetic field strength, the only other
parameters left to improve are the intrinsic detector resolution,
$\Delta$, and the number of points sampled along the track trajectory.
The forward/backward tracking extensions provide additional
measurement points in these regions.  Hence, a balance of number of track
points (number of sensitive detector layers), material budget and
cost must be found.

The design adopted here is an all-Silicon detector, with very high
resolution.  The readout scheme must be such that a signal weighting
using analogue information is possible without losing the advantages
of digital signal processing and on-chip zero suppression.  All of the
components need power and cooling, influencing the material budget of
the tracking system which should be kept as low as possible.  The
technology used must be available at the industrial level, radiation
hard and relatively cheap. A good candidate is n\_in\_p single sided
sensors\cite{allport:2010}.

In the following, the layout of a tracking system for the baseline
detector configuration {\small\bf A} is defined.  The design criteria
and possible solutions for a tracker which provides optimal support of
the calorimetry via high resolution impact parameter measurements and
momentum determination are given in detail.

\subsection{Tracking Detector - Baseline Layout} 
\begin{figure}[htp]
\begin{center}
\includegraphics[width=0.85\columnwidth]{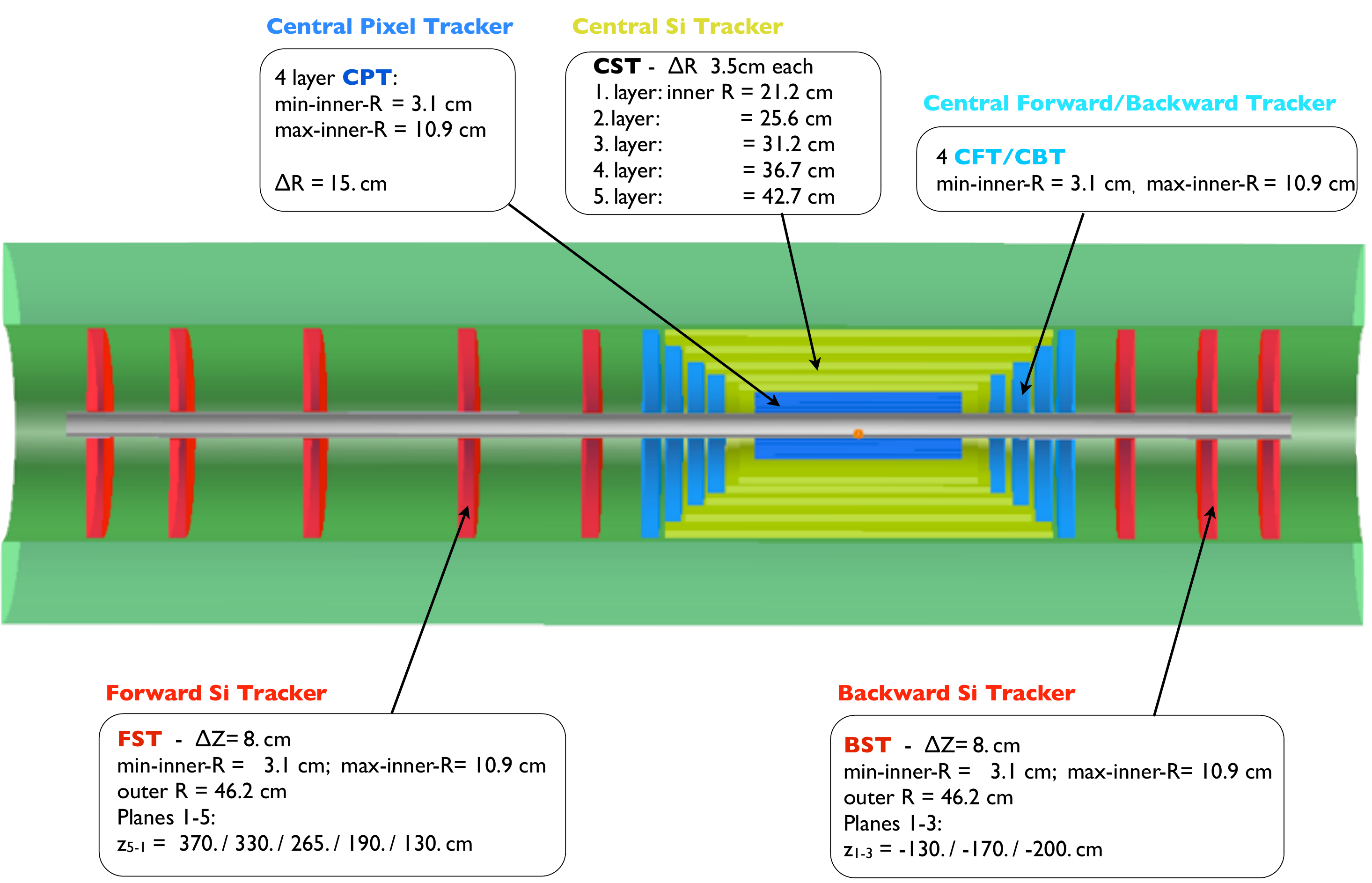}
\end{center}
\caption{Tracker and barrel Electromagnetic-Calorimeter $rz$ view of the baseline detector (Linac-Ring case).}
\label{LHEC:MainDetector:Tracker:Fig:1}
\end{figure}
\begin{figure}[htp]
\begin{center}
\includegraphics[width=0.5\columnwidth]{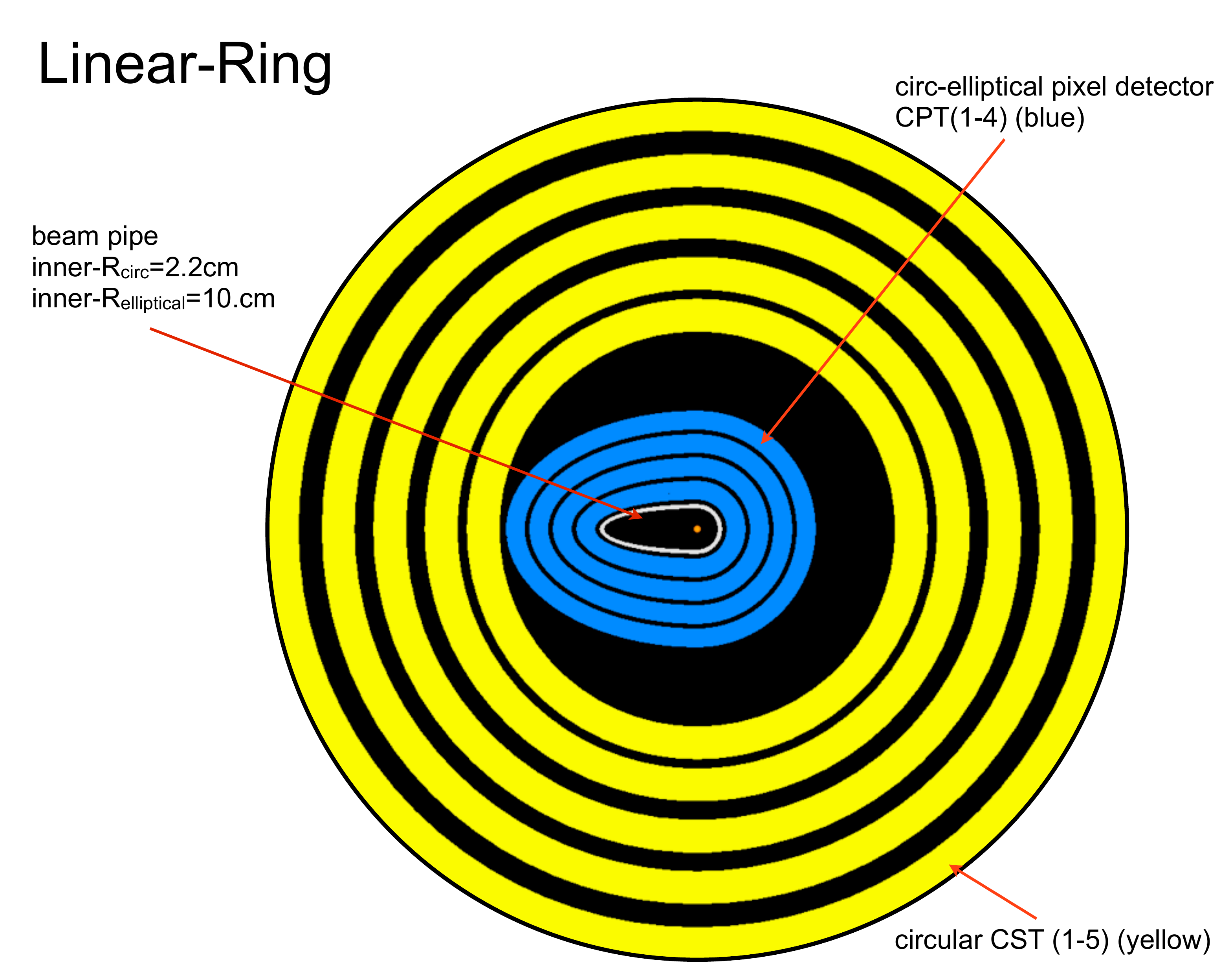}
\end{center}
\caption{XY cut away view of the Central Pixel (CPT) and Central Strixel Tracker (CST) (Linac-Ring layout).}
\label{LHEC:MainDetector:Tracker:Fig:2}
\end{figure}

The tracking detectors (Fig.\,\ref{LHEC:MainDetector:Tracker:Fig:1})
inside the electromagnetic calorimeter are all-Silicon devices.  The
tracker covers the pseudorapidity range $-4.8<\eta <5.5 $ and is
located inside the solenoidal field\footnote{Additionally a dipole
field of 0.3T, resulting from the steering dipoles required for the
Linac-Ring configuration, is superimposed.} of 3.5T.
Fig.\,\ref{LHEC:MainDetector:Tracker:Fig:1} shows the baseline
({\small\bf A}) design of the tracker, subdivided into central (CPT,
CST, CFT/CBT) and forward/backward parts (FST, BST).  Details of the
design are summarised in
Tab.\,\ref{LHEC:MainDetector:Tracker:Tab:dim_track}.  The item
{\emph{Project}} in
table\,\,\ref{LHEC:MainDetector:Tracker:Tab:dim_track} denotes the
area which has to be equipped with appropriate Si-sensors
(e.g. single-sided or double-sided sensors).  An alternative would be
the usage of Si-Gas detectors providing track segment information
instead of track points, e.g.  in the CST cylinders
(Ref.\cite{Koffeman:2007zz},\cite{vanderGraf:2010},\cite{vanderGraaf:2011zz}).
The shape of the CPT and the inner dimensions of all near-beam
detectors have been chosen to maximise detector acceptance by
providing measurements as close to the beam-line as possible (see
Fig.\,\ref{LHEC:MainDetector:Tracker:Fig:2} which shows the
{\emph{xy}} view of the circular-elliptical CPT and the cylindrical
CST detectors).

The 4 Si-Pixel-Layers CPT1-CPT4, with a resolution of
$\sigma_\mathrm{pix}\approx{8}\mu{m}$, are positioned as close to the
beam pipe as possible.  Si-strixel detectors (CST1-CST5), with a
resolution of $\sigma_\mathrm{strixel}\approx{12}\mu{m}$, form the
central barrel layers.  An alternative is the 2$_{-}$in$_{-}$1 single
sided Si-strip solution for these barrel cylinders, with a resolution
of $\sigma_\mathrm{strip}\approx{15}\mu{m}$\cite{horrisberger:2010}.
The endcap Si-Strip detectors CFT/CBT(1-4) complete the central
tracker. The tracker inserts, 5 wheels of Si-Strip detectors in the
forward direction (FST) and 3 wheels in the backward direction (BST),
have granularity requirements based on optimising energy flow
corrections and jet resolution.  In the forward direction, Si-Pixel or
Si-Strixel detectors may have to be used to meet those requirements,
whereas for the backward BST wheels where the particle density is less
demanding Si-Strip detectors may be sufficient.  The FST/BST wheels
have to be removed in case of high luminosity running for the
Ring-Ring option of the accelerator configuration (see
Fig.\,\ref{LHEC:MainDetector:Description:Fig:4b}).

\begin{table}[htp]
{\small
\begin{center}
  \begin{tabular}{|l|c|c|c|c|c|c|c|c|c|}
    \hline     
Cen.  Barrel                                     & CPT1 & CPT2& CPT3  &\multicolumn{1}{c||}{CPT4} & CST1 & CST2 & CST3 & CST4 & CST5 \\ 
\hline \hline
Min. $R$\hfill$[cm]$               &  3.1 & 5.6 & 8.1    &\multicolumn{1}{c||}{10.6}& 21.2  & 25.6 & 31.2 & 36.7 & 42.7 \\
Min. $\theta$\hfill [\textdegree]       &  3.6 & 6.4 & 9.2  &\multicolumn{1}{c||}{12.0}& 20.0 & 21.8  & 22.8 & 22.4 & 24.4 \\
Max.  \textbar$\eta$\textbar \hfill&3.5 & 2.9  &  2.5  &\multicolumn{1}{c||}{2.2}   &  1.6  &  1.4   &  1.2 & 1.0  & 0.8  \\
 $\Delta{R}$    \hfill$[cm]$                     &  2   & 2    &  2      &\multicolumn{1}{c||}{2}     & 3.5   &  3.5   & 3.5  & 3.5  & 3.5  \\
  $\pm{z}$-length      \hfill$[cm]$          & 50   & 50  & 50    &\multicolumn{1}{c||}{50}    & 58    &  64   &  74    & 84   & 94   \\
\cline{2-5}  \cline{6-10} 
 Project  \hfill[$m^2$] & \multicolumn{4}{c||}{1.4} & \multicolumn{5}{c|}{8.1}        \\
\hline \hline 

Cen.  Endcaps                        &  CFT4 & CFT3 & CFT2 & CFT1 &    & CBT1 & CBT2 & CBT3 & CBT4 \\ 
\hline \hline
Min. $R$\hfill$[cm]$                     &  3.1  & 3.1  & 3.1  & 3.1  &       & 3.1  & 3.1  & 3.1  & 3.1  \\
Min. $\theta$\hfill [\textdegree]     &  1.8 & 2.0 &  2.2& 2.6    &       & 177.4  & 177.7  & 178  & 178.2 \\
at ${z}$   \hfill[cm]                                             & 101 & 90 & 80 & 70 &      &  -70 & -80 & -90 & -101  \\
Max./Min.  $\eta$ \hfill& 4.2  & 4.0  & 3.9  & 3.8 &    & -3.8  & -3.9  & -4.0  & -4.2  \\
 $\Delta{z}$    \hfill$[cm]$                                   &  7    & 7     & 7     & 7     &     & 7     &      7  &      7  &    7  \\
 \cline{2-5}  \cline{7-10} 
Project   \hfill[$m^2$]                             & \multicolumn{4}{c|}{1.8}     &       &\multicolumn{4}{c|}{1.8} \\
\hline  \hline

Fwd/Bwd                        & FST5 & FST4& FST3 & FST2 & FST1  &      & BST1 & BST2& BST3   \\ 
\hline \hline
Min. $R$\hfill$[cm]$              & 3.1  & 3.1 &  3.1 & 3.1  &  3.1  &      &  3.1     & 3.1   &  3.1     \\
Min. $\theta$\hfill [\textdegree] & 0.48  & 0.54 & 0.68 & 0.95 & 1.4  &      & 178.6 & 178.9 & 179.1   \\
at $z$   \hfill[cm]                                      & 370  & 330 & 265  & 190  & 130   &      & -130 & -170 &-200      \\
Max./Min.  $\eta$ \hfill              & 5.5  & 5.4 & 5.2  & 4.8  & 4.5   &      &  -4.5  & -4.7  & -4.8   \\
Outer $R$\hfill$[cm]$                   & 46.2 & 46.2& 46.2 & 46.2 & 46.2  &      &  46.2  & 46.2  & 46.2    \\
 $\Delta{z}$    \hfill$[cm]$                       &  8   &  8  &  8   &  8   &  8    &      &   8      &  8      &  8     \\
    \cline{2-6} \cline{8-10} 
Project  \hfill[$m^2$]                & \multicolumn{5}{c|}{3.3}   &      &  \multicolumn{3}{c|}{2.0}  \\
\hline
\end{tabular}
\end{center}
}
\caption{Summary of tracker dimensions. }
\label{LHEC:MainDetector:Tracker:Tab:dim_track}
\end{table}
\subsection{Performance}
\begin{figure}[htp]
\begin{center}
\includegraphics[width=0.8\columnwidth]{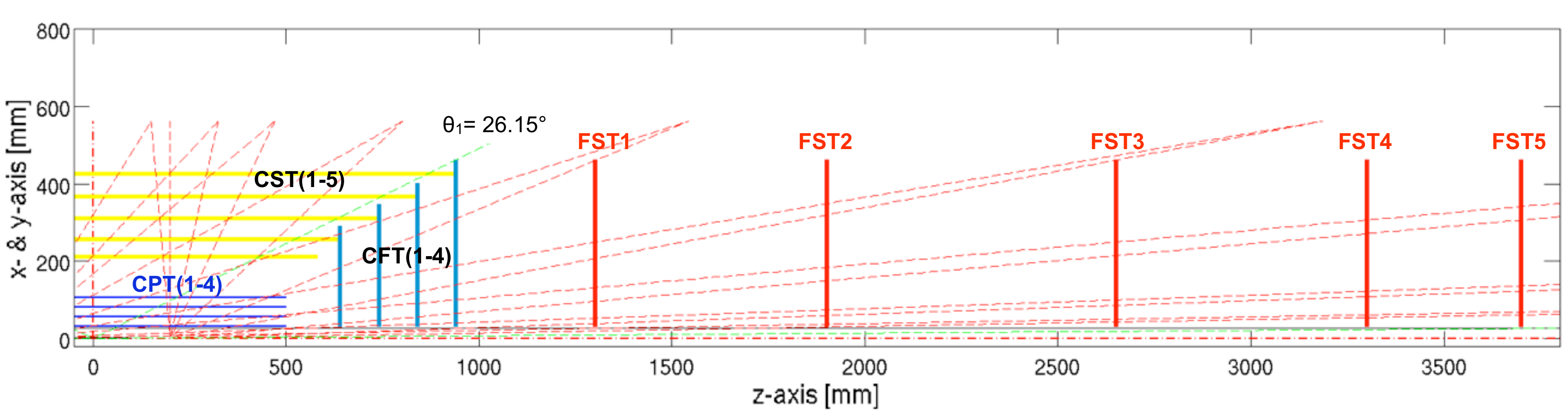}
\includegraphics[width=0.8\columnwidth]{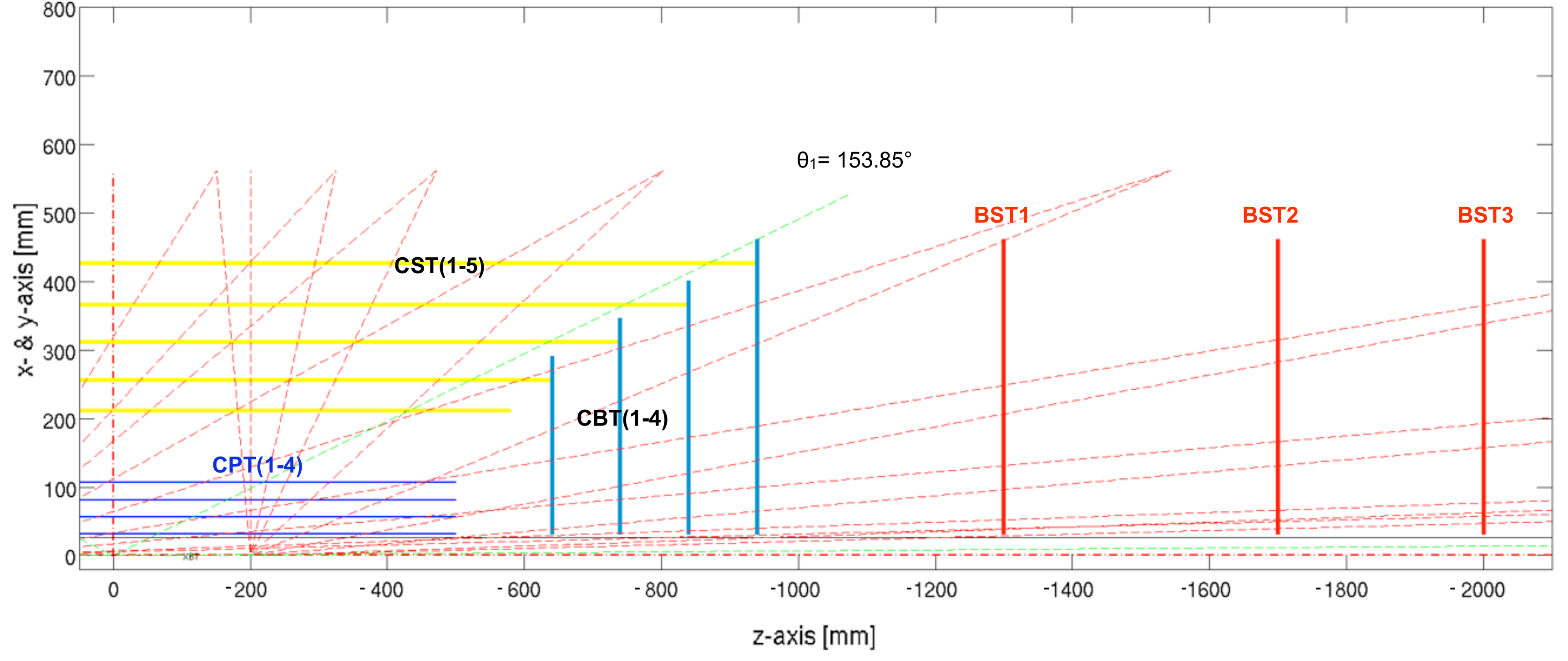}
\end{center}
\caption{LicToy2.0 tracker design of  the central/forward FST(top) and central/backward direction BST(bottom).}
\label{LHEC:MainDetector:Tracker:Fig:5}
\end{figure}

\begin{figure}[htp]
\hspace*{-1cm}
\includegraphics[width=0.53\columnwidth]{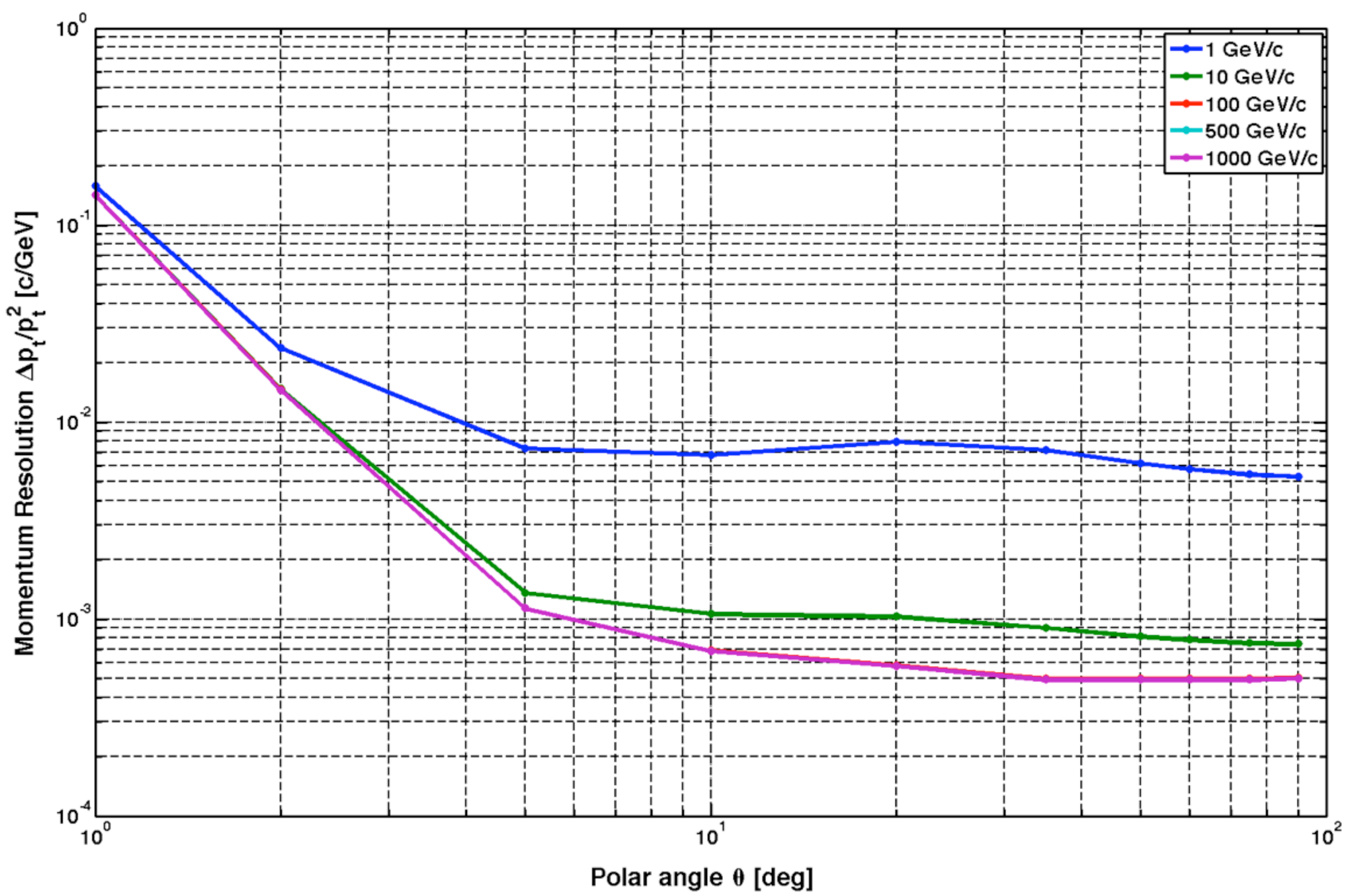}
\includegraphics[width=0.54\columnwidth]{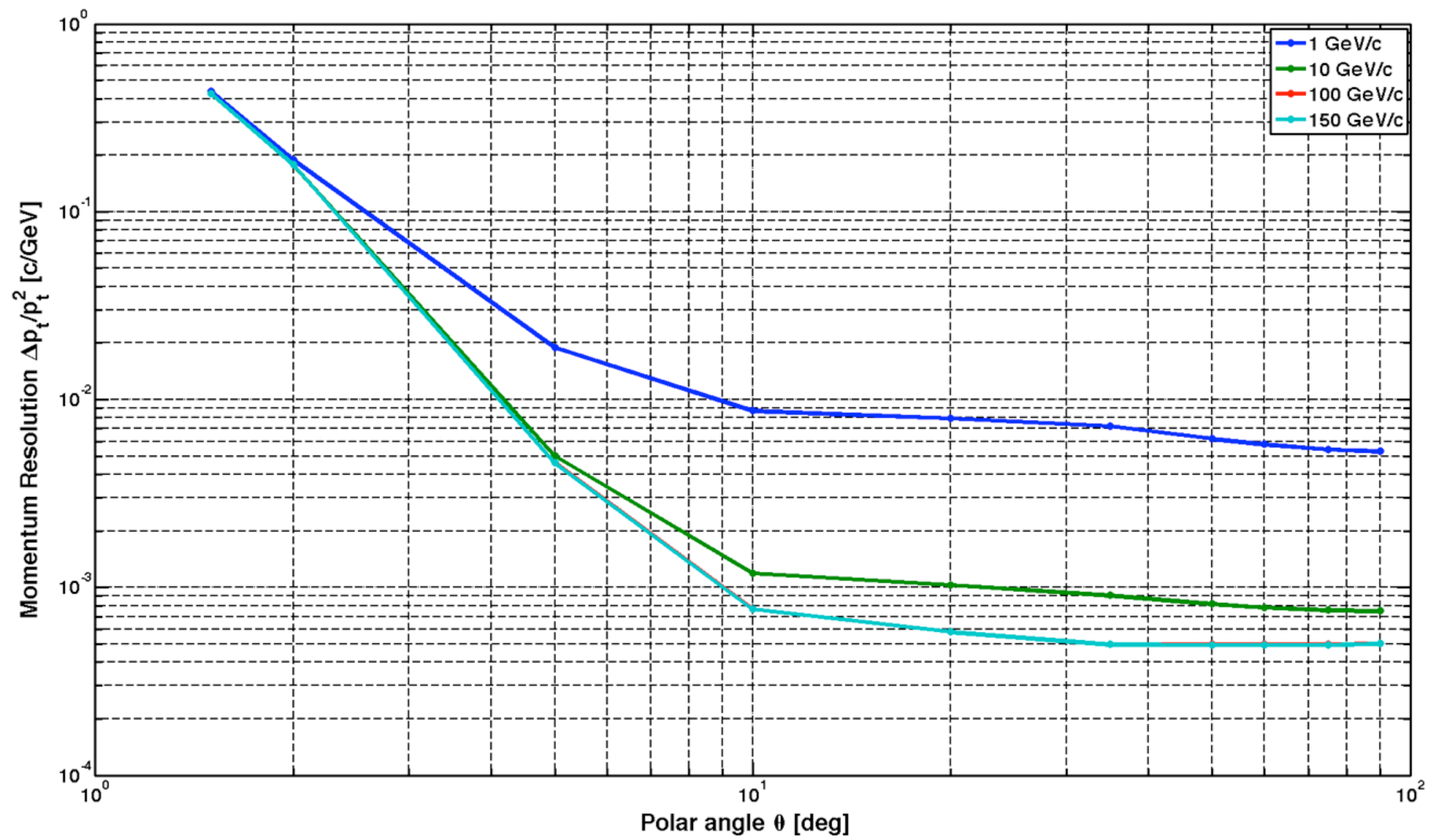}
\hspace*{-1cm}
\includegraphics[width=0.54\columnwidth]{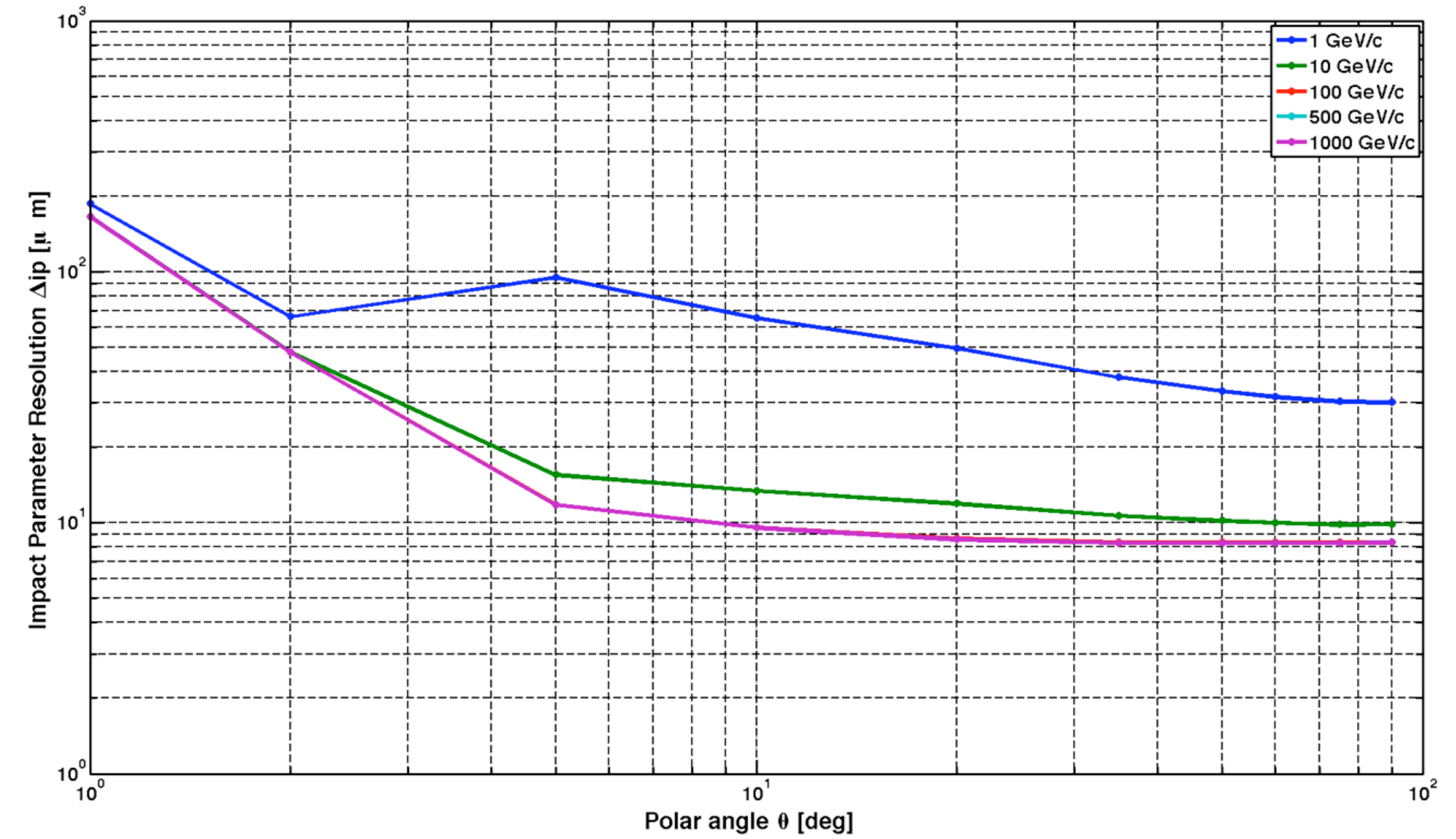}
\includegraphics[width=0.54\columnwidth]{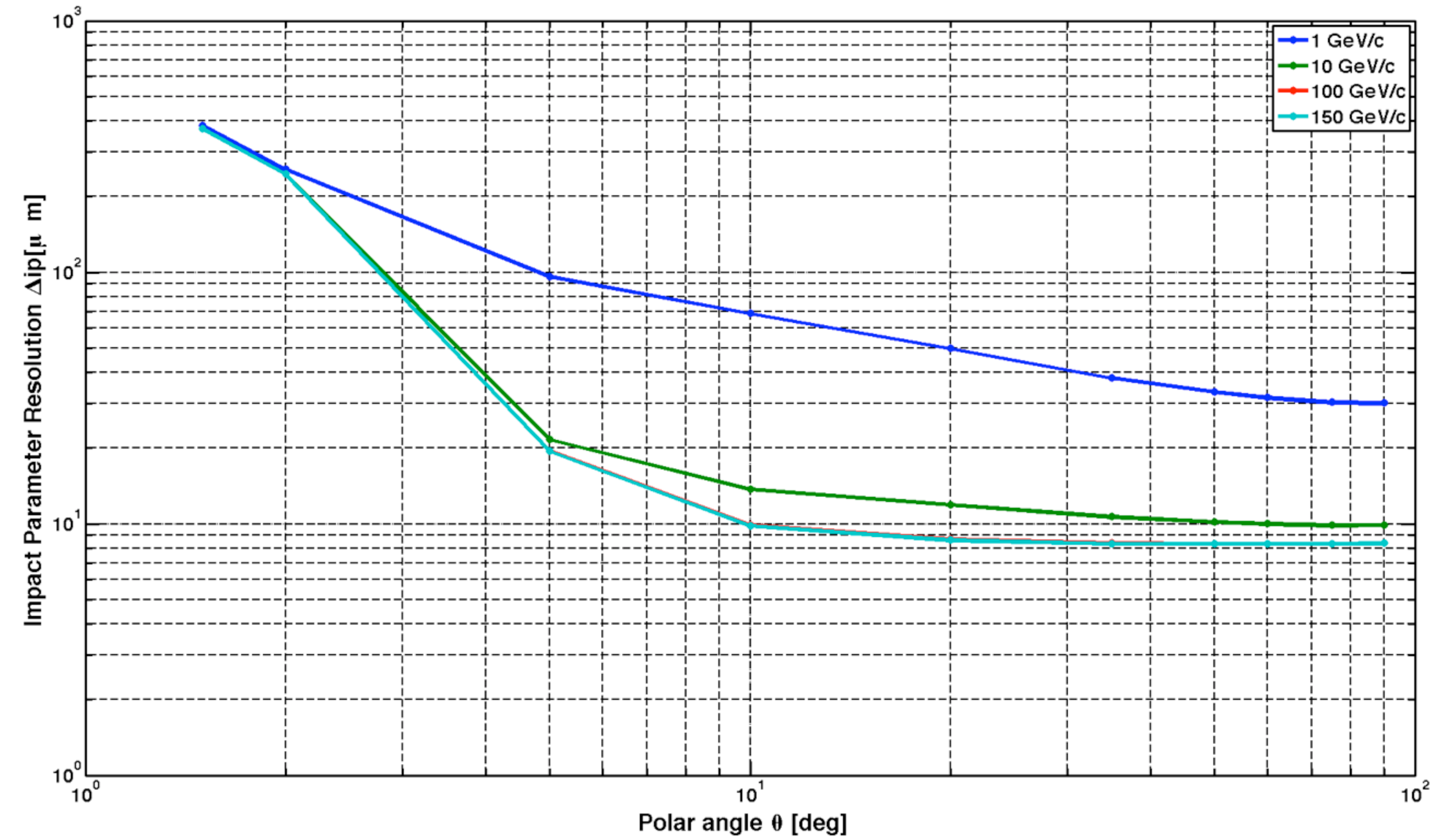}
\hspace*{-1cm}
\includegraphics[width=0.54\columnwidth]{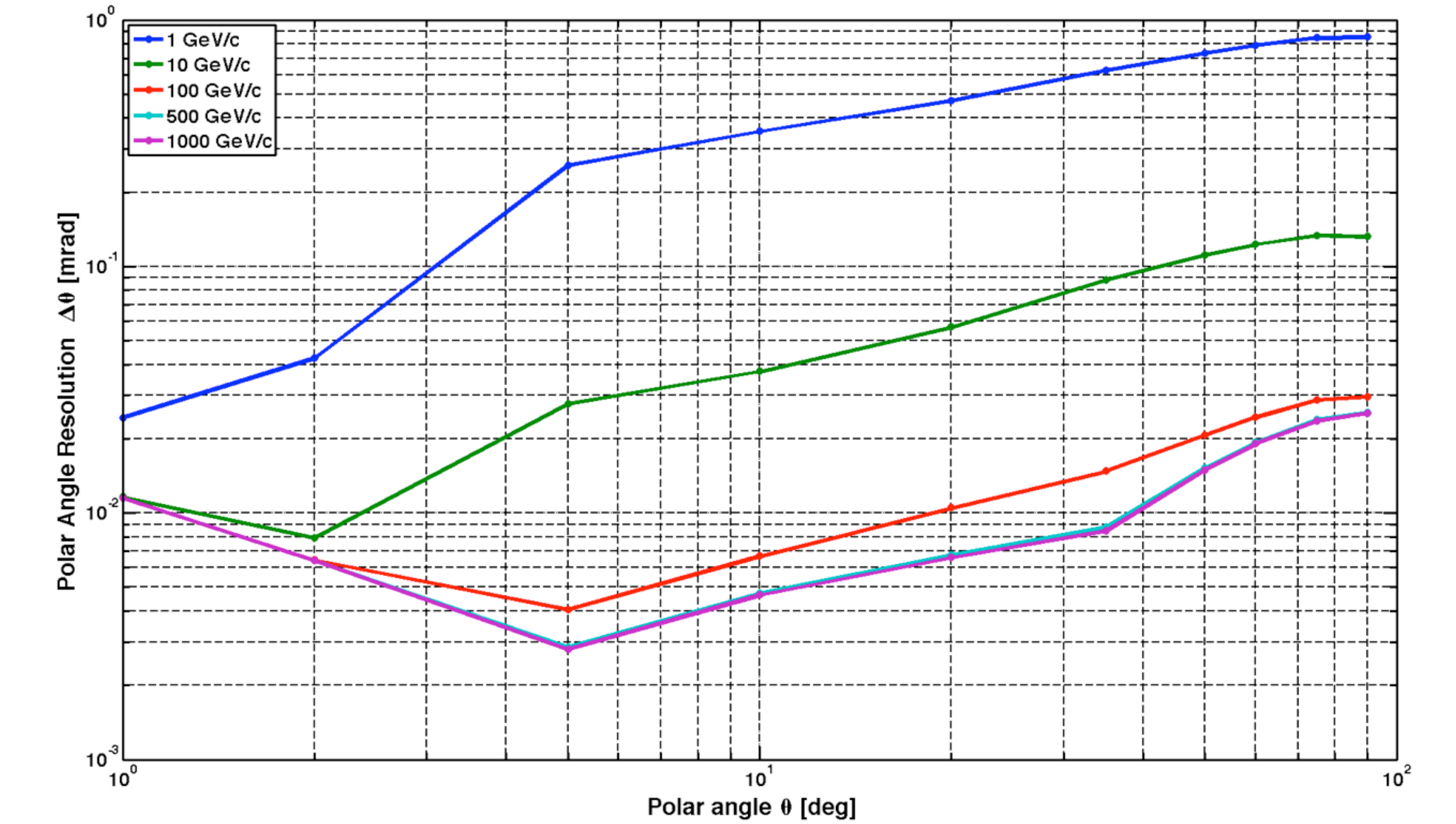}
\includegraphics[width=0.54\columnwidth]{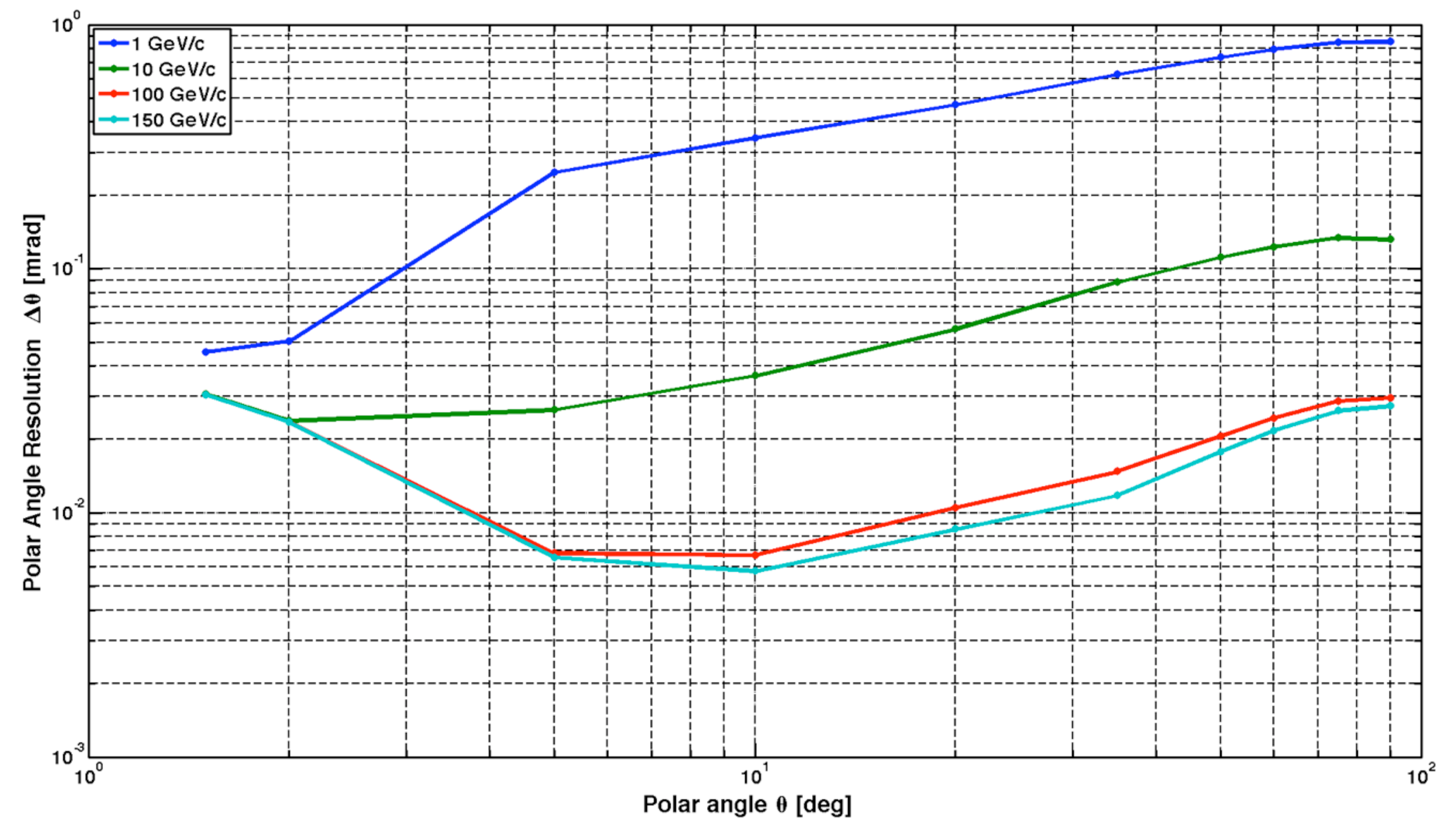}
\caption{Scaled momentum, impact parameter and polar angle resolution as a function of polar angle 
$\theta$ resulting from the tracker design simulation using LiCToy2 for the FST(left) and BST(right).
The tracker setup used is that shown in Fig.\,\ref{LHEC:MainDetector:Tracker:Fig:5}.}
\label{LHEC:MainDetector:Tracker:Fig:6}
\end{figure}

Some results of preliminary tracker performance simulations using the
LicToy-2.0 program\cite{Regler:2008zz} for the tracker setup (see
table\,\ref{LHEC:MainDetector:Tracker:Tab:dim_track} and
Fig.\,\ref{LHEC:MainDetector:Tracker:Fig:5}), and with parameters
given in table\,\ref{tracking:params} are summarised in
Fig.\,\ref{LHEC:MainDetector:Tracker:Fig:6}.  The detector performance
is very good, as expected.
%
%
 \begin{table}[htp]
\begin{center}
  \begin{tabular}{|c|c|}
\hline
Parameters & \\\hline
B & 3.5T\\
$X/X_{0}^\mathrm{beam pipe}$ & $0.002$\\
$X/X_{0 \,\,per\,(double)\,layer}^\mathrm{CPT/CFT/CBT/FST/BST-det}$ & $0.025$\\
$X/X_{0 \,\,per\,(double)\,layer}^\mathrm{CST-det}$ & $0.02$\\
efficiency & $99\%$\\
Minimal inner radius & $3.15cm$\\
$\sigma_\mathrm{CPT}$ & $8\mu{m}$\\
$\sigma_\mathrm{CST,CFT,CBT}$ & $12\mu{m}$\\
$\sigma_\mathrm{FST,BST}$ & $15\mu{m}$\\
\hline
\end{tabular}
\end{center}
\caption{
The main parameters assumed in LicToy2 tracking simulation.
}
\label{tracking:params}
\end{table}
For 1\textdegree~tracks the bending solenoidal field component (0.36T)
is of the same order as the dipole field and the resulting track
sagitta only reaches the $mm$ range when particles of momentum
$<100$~GeV  have a track length of 250cm (see
Fig.\,\ref{LHEC:MainDetector:Tracker:Fig:1}).  The tracker described
here measures 1\textdegree~tracks over a distance of $\approx$180cm,
and therefore high momentum tracks will have a poor momentum
determination.  Nevertheless, the position information can be used to
match a track to a calorimeter deposit with high precision.

The backward measurement is characterised by even shorter track
lengths and in this case the analysis has to rely completely on the energy
measurement in the calorimeters matched to a well defined track.
Thanks to the much reduced particle flux in the backward direction due to kinematics, 
the performance and precision achievable is expected to be higher.
\begin{figure}[htp]
\begin{center}
\includegraphics[width=0.8\columnwidth]{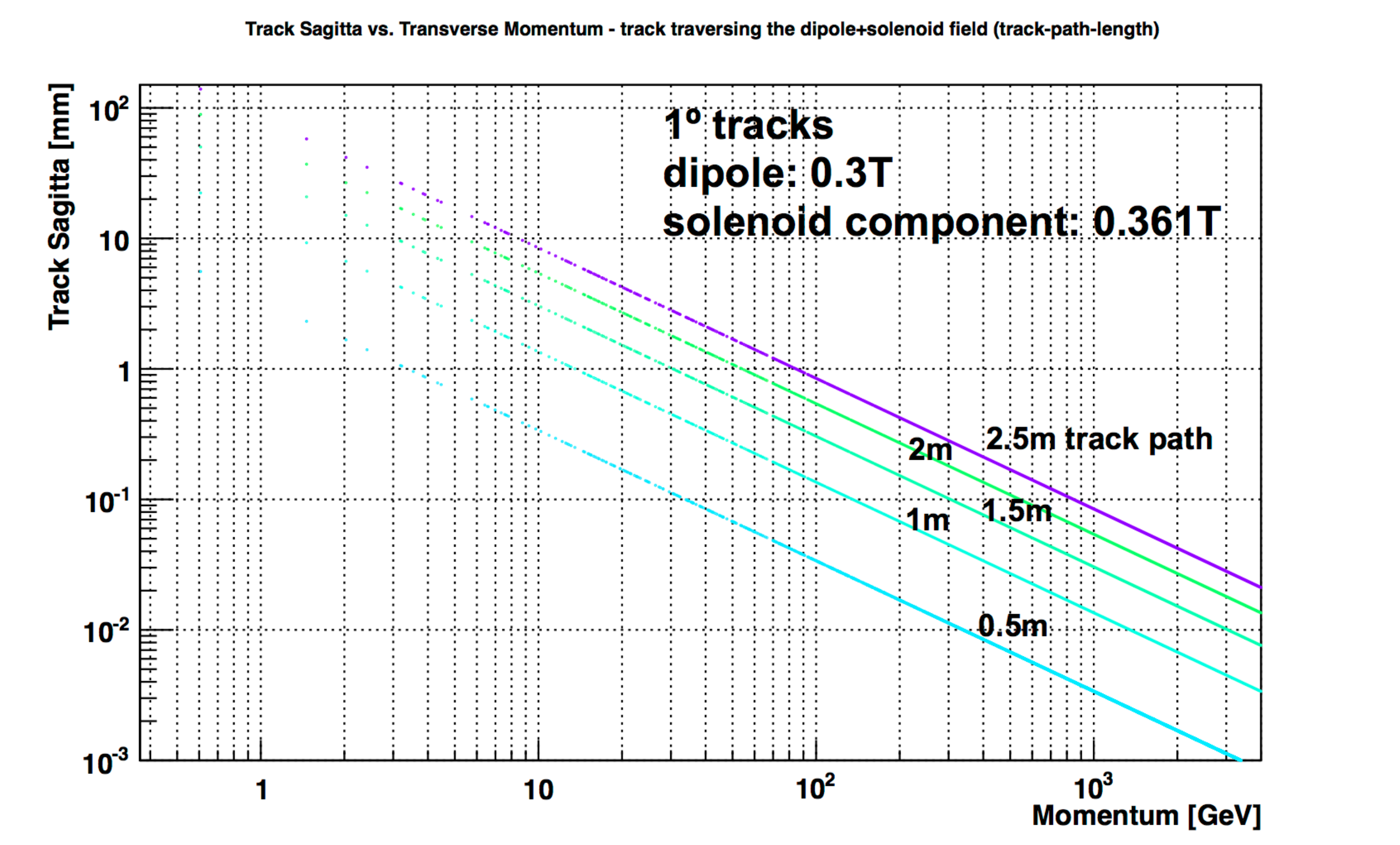}
\end{center}
\caption{Track Sagitta vs. momentum of 1\textdegree-tracks in a superposed dipole\,(0.3T) and solenoidal field component\,(0.361T).}
\label{LHEC:MainDetector:Tracker:Fig:1s}
\end{figure}
%

%
%
%
%


\subsection{Tracking detector design criteria and possible solutions} 

Previous attempts to achieve an optimal detector design suggest that
some criteria should be discussed as early as possible. The main items
to consider\cite{horrisberger:2006,allport:2010} are discussed in the
following.

\subsubsection{Optimising cost for all components} 
The technology developments for HL-LHC/ILC experiments
\cite{Brau:2006ag,wermes:2008,Hessey:2008zz,nessi:2009,haber:2011,christian:2011,cihanger:2011,affolder:2011,
  macchiolo:2011,parzefall:2011,rubinskiy:2011,bomben:2011,mikuz:2011,Mac-Raighne:2011}
should be used as far as possible while relying on existing
technologies because of time constraints.  The sensors, integrated
electronics, readout/trigger circuitry, mechanics, cooling,
etc. available today have to be used in order to meet the goal of
installation in the early 2020's. The advanced research in
instrumentation and work on its manufacturability and construction
should be used.  Wherever possible, affordable innovative instruments
and approaches should be re-used.

\subsubsection{Choice of sensor type} 
The default tracker design is based on the silicon microstrip detector
technology developed for the experiments at LHC, ILC, TEVATRON and
b-factories etc. within the last 20 years.  The final decision for
sensor types (pixel, strixel, strip) will depend on many factors and
will be taken according to the required functionality.

\paragraph{Radiation hardness} 
The expected radiation load is defined and influenced by the
interaction rate (25ns), luminosity ($\approx{10}^{33}cm^{-2}s^{-1}$),
particle rate per angle interval, fluence n$_{eq}$ and ionisation
dose. Some parameters will be better defined after the evaluation of
more detailed simulations.  Specifically the impact of radiation on
tracker wheels, calorimeter inserts and the inner tracker-barrel layer
has to be studied.  The tools for those simulations are being
prepared.  From the preliminary simulations detailed in
section\,\ref{Geant4-Event-Simulations}, there is no indication for
extremely high radiation load in the detectors adjacent to the beam
pipe.  The expected levels are far below what the LHC experiments have
to withstand.
       
Nevertheless, for safety reasons the active parts of the forward and
backward calorimeter should be equipped with radiation hard
silicon-based sensors according to LHC/HL-LHC standards.  The use of
Si-strip/Si-pad based calo-inserts, although small in volume but still
large in terms of layer area $\cal{O}$(m$^2$), might turn out to be a
sizeable investment which is anyhow needed in order to guarantee a
stable performance and detector lifetime.  A final decision will only
be possible after more detailed simulations are complete. For the
tracker, the more traditional p\_in\_n sensor technology could be used
instead of the more radiation hard n\_in\_p or n\_in\_n sensors, but
cost will ultimately decide.


\paragraph{Trigger}
\begin{figure}[htp]
\begin{center}
\includegraphics[width=0.8\columnwidth]{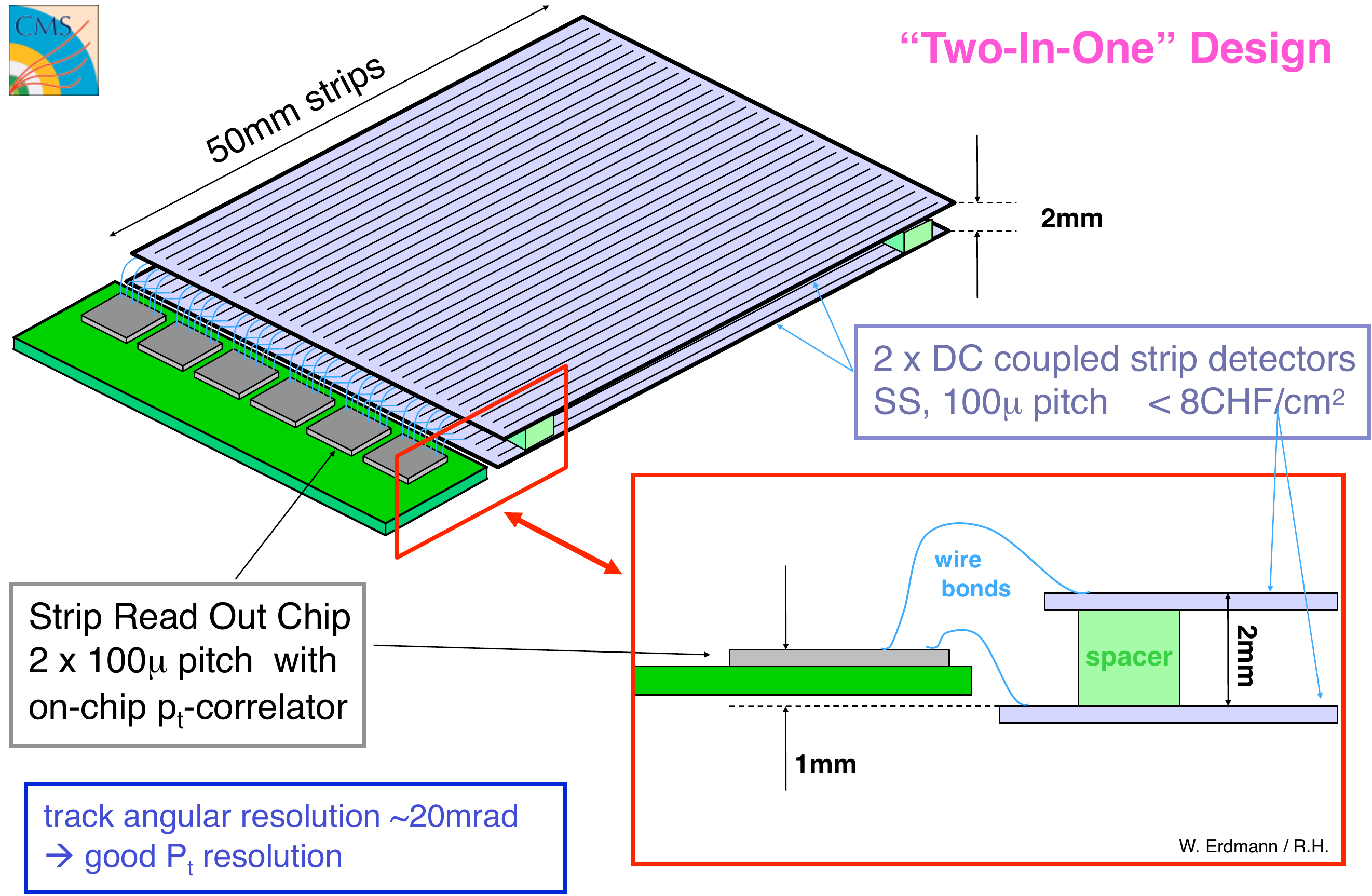}
\end{center}
\caption{Layout of the  2$_{-}$in$_{-}$1 strip sensor design used as $p_t$-trigger setup for the CMS experiment.}
\label{LHEC:MainDetector:Tracker:Fig:7}
\end{figure}
\begin{figure}[htp]
\begin{center}
\includegraphics[width=0.8\columnwidth]{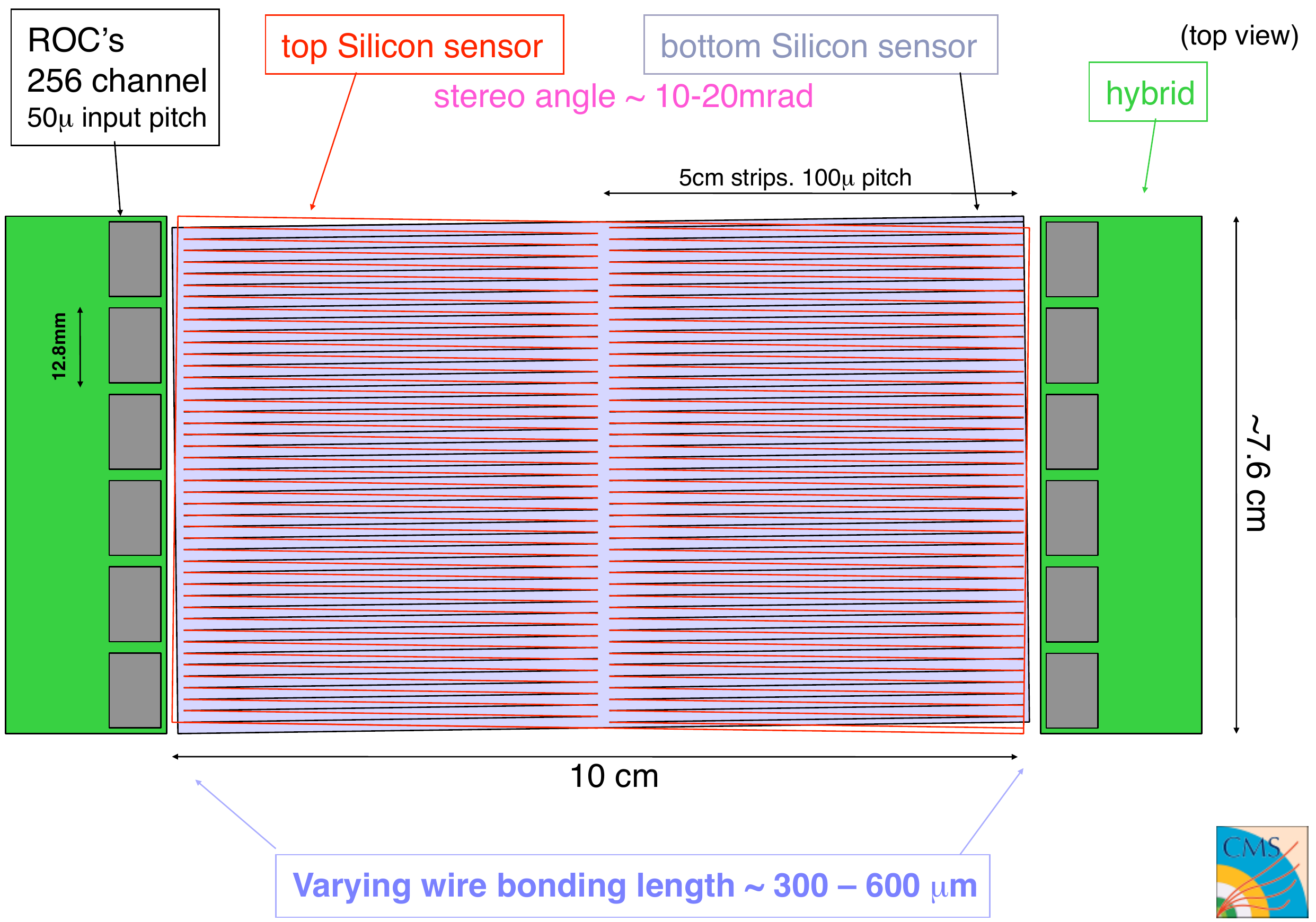}
\end{center}
\caption{Layout of the  2$_{-}$in$_{-}$1 strip sensor design used as tracker module. Double use of e.g. 
power and cooling for the two strip wafer.}
\label{LHEC:MainDetector:Tracker:Fig:8}
\end{figure}
The trigger capabilities of the tracking system are yet to be defined
and will have a direct impact on sensor choice, associated electronics
and arrangement.  It is possible that very recent developments of 3D
integration semiconductor layers interconnected to form monolithic
unities of sensor and electronic circuitry would be available in time
for installation in the 2020's, but conventional wire bonded or bump
bonded solutions may be more cost efficient and rely on components
available today.  For example, the 2$_{-}$in$_{-}$1 strip sensor
design used for a $p_t$-trigger discussed by the CMS upgrade design group
\cite{horrisberger:2010}, shown in
Fig.\,\ref{LHEC:MainDetector:Tracker:Fig:7}, would have a direct
impact on a trigger definition.  The sensor, hybrid and readout
modules are available and interconnected by wire bonds.  The
2$_{-}$in$_{-}$1 sensor design is an elegant way of saving resources
when designing a tracker, as shown in
Fig.\,\ref{LHEC:MainDetector:Tracker:Fig:8}.
         
\paragraph{Front-end}
Candidates for readout chips attached to the sensors are e.g. the ATLAS
FE-I4 (50$\mu{m}\ast$250$\mu{m}$)\cite{allport:2010} and CMS ROC
(100$\mu{m}\ast$150$\mu{m}$)\cite{Brau:2006ag}).  The sensor pitch
has to be matched and the electronics scheme defined beforehand.

\subsubsection{Powering and cooling}
The size of the largest stave structure to be installed (half z-length
$\approx 94cm$) is smaller then the stave length used e.g. by ATLAS
($\approx 120cm$).  Powering and cooling per stave could therefore
follow the current LHC installations.  Minimisation of cooling
directly reduces the material budget; cooling is related to power
consumption issues and it may be a criterion for technology selection.
A decision on the powering concept is needed (serial vs parallel
powering) and it will depend on the template chosen for readout and
services.  An obvious solution is to re-apply the scheme used by a
current LHC experiment in line with the sensor, electronics \& readout
option selected.

\begin{figure}[htp]
\begin{center}
\includegraphics[width=0.8\columnwidth]{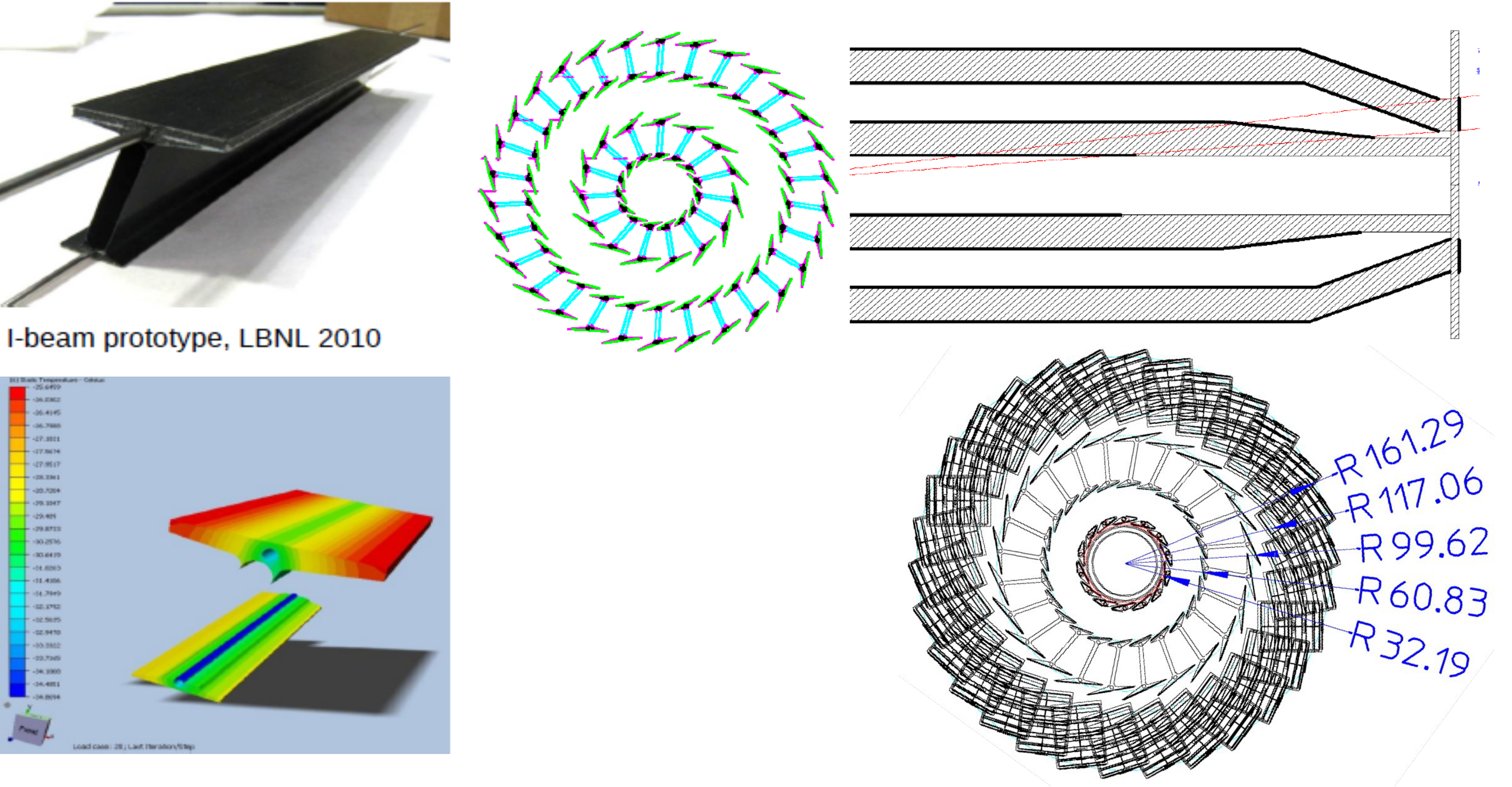}
\end{center}
\caption{Proposed mechanics and sensor layout for the ATLAS pixel upgrade.}
\label{LHEC:MainDetector:Tracker:Fig:9}
\end{figure}
\begin{figure}[htp]
\begin{center}
\includegraphics[width=0.8\columnwidth]{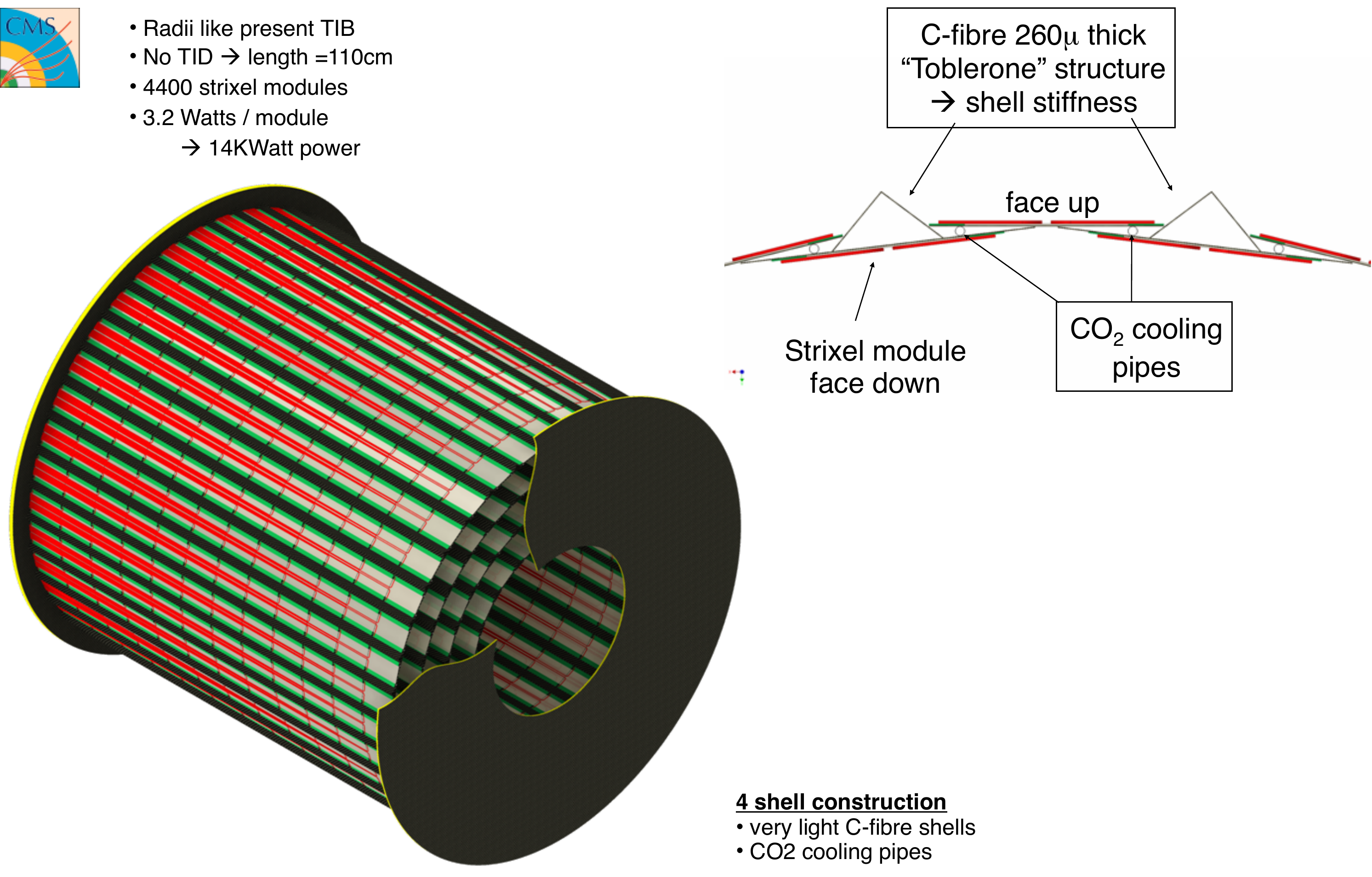}
\end{center}
\caption{Proposed mechanics layout for the CMS inner barrel tracker upgrade.}
\label{LHEC:MainDetector:Tracker:Fig:10}
\end{figure}
\begin{figure}[htp]
\begin{center}
\includegraphics[width=0.9\columnwidth]{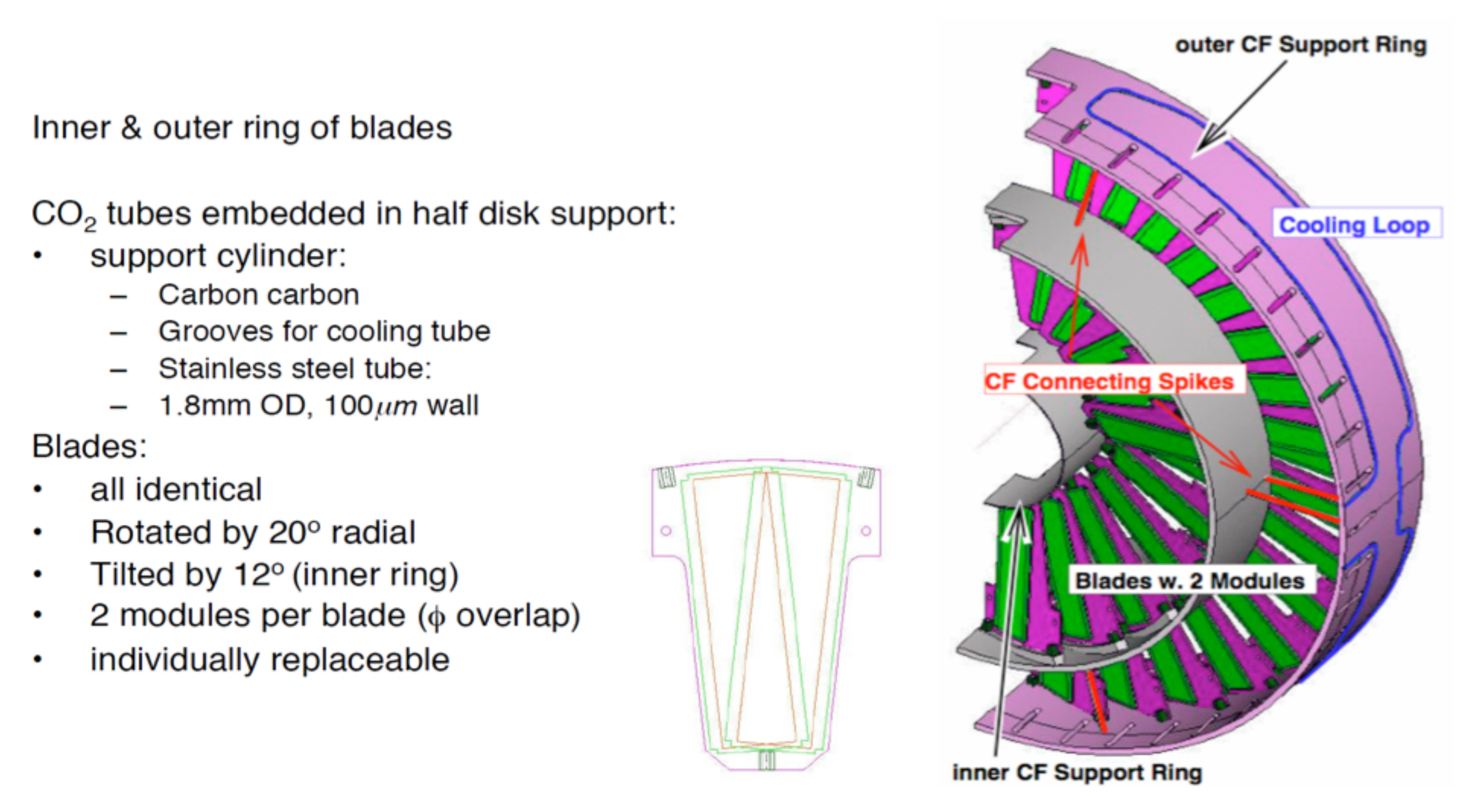}
\end{center}
\caption{Proposed mechanics layout for the CMS tracker wheel upgrade.}
\label{LHEC:MainDetector:Tracker:Fig:11}
\end{figure}
\begin{figure}[htp]
\begin{center}
\includegraphics[width=0.45\columnwidth]{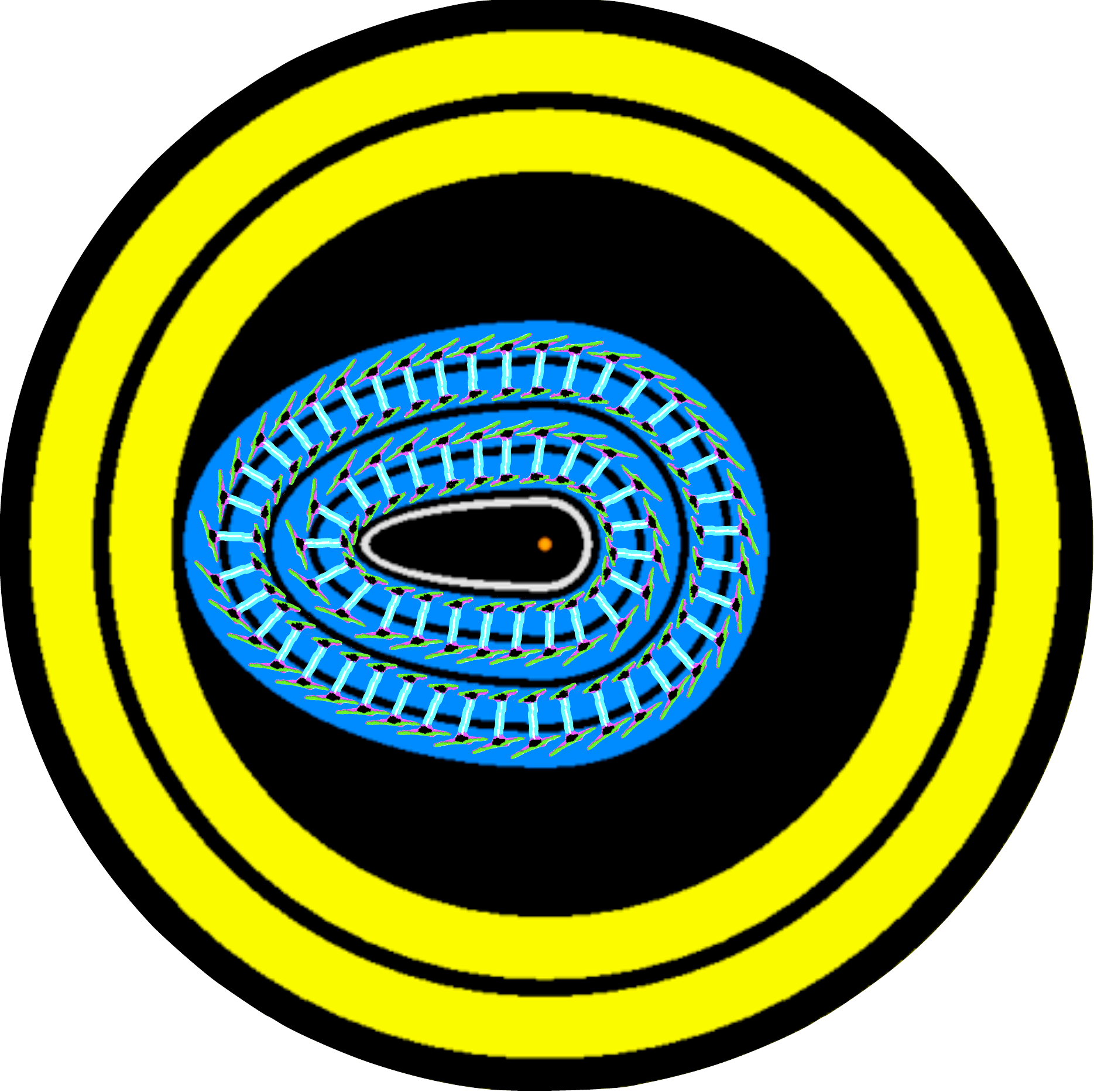}
\end{center}
\caption{Artist view of the pixel sensor arrangement using the double-I ATLAS layout as template (Fig.\,\ref{LHEC:MainDetector:Tracker:Fig:9}).}
\label{LHEC:MainDetector:Tracker:Fig:12}
\end{figure}
\begin{figure}[htp]
\begin{center}
\includegraphics[width=\columnwidth]{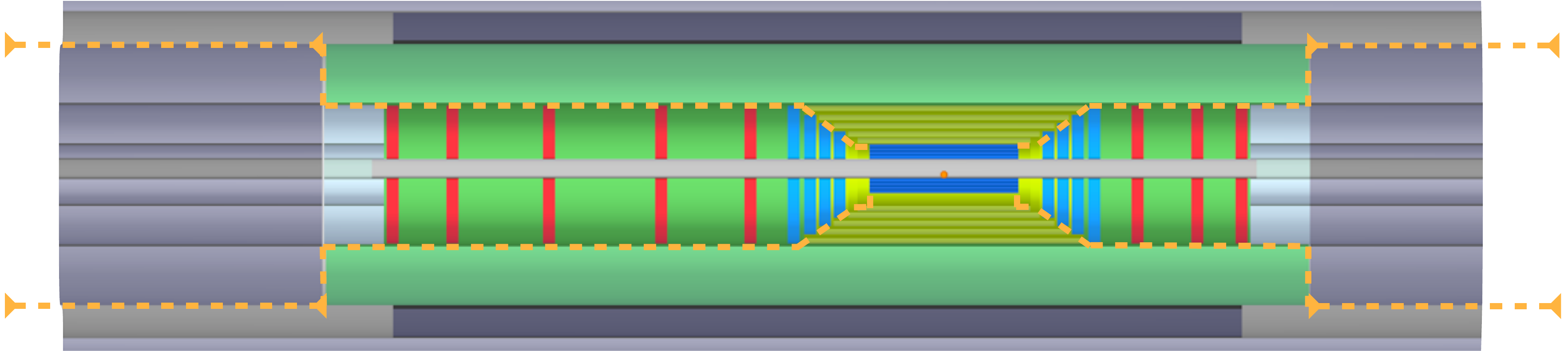}
\end{center}
\caption{Path of services for all tracking detectors (shown in
  orange). The services shall be integrated into support structures
  whenever possible.}
\label{LHEC:MainDetector:Tracker:Fig:13}
\end{figure}
\subsubsection{Mechanical support} 
The mechanical support and cooling elements have to be chosen to
minimise the material budget and hence minimise the impact of
multiple-scattering on track resolution by the tracker material.
Rigid but very light mechanics in connection with improved sensor
arrangement, incorporation of cooling systems and all other services
into the support structure are the main design criteria for HL-LHC
upgrade projects for e.g. ATLAS and CMS - this is also the case for
LHeC.

In Figs.\,\ref{LHEC:MainDetector:Tracker:Fig:9},
\ref{LHEC:MainDetector:Tracker:Fig:10} and
\ref{LHEC:MainDetector:Tracker:Fig:11}, possible mechanical solutions
for the ATLAS\cite{Garcia:2011,allport:2010} and CMS
\cite{horrisberger:2010} tracker upgrades in the barrel and
forward/backward tracker regions are shown.  These designs may serve
as templates for the LHeC detector. As an example, an artist's view in
Fig.\,\ref{LHEC:MainDetector:Tracker:Fig:12} shows an implementation
of the double-I ATLAS pixel arrangement into a 4 layer pixel structure
for the LHeC detector.  The goal is the design of a tracker which is
in the range $\approx{15-20}\%X_0$ in terms of radiation lengths.

\subsubsection{Readout} 
Possible paths for the IN/OUT services of the LHeC tracking detectors
are sketched in Fig.\,\ref{LHEC:MainDetector:Tracker:Fig:13}.  The
cables and tubes are integrated into the support structures of the
sub-detectors as far as possible.  Optimisation of detector readout
reduces the cost and material impact of cables.  An example is
discussed in detail for the ATLAS/CMS HL-LHC opto-link upgrade in
Ref.\cite{ATLCMS:2007}.  The front end electronics buffer depth will
depend on bunch crossing rate (25ns) and the trigger/readout speed
capability.

\subsubsection{Radiation detectors} 
Dedicated instrumentation for beam tuning, minimising background and
optimising luminosity is needed. Radiation detectors, close to masks
and at tight apertures, are useful for fast identification of
background sources.  Fast beam monitor related information might be
collected efficiently by diamond detectors, as done for e.g. CMS
\cite{Bell:2009hr,FernandezHernando:2005hj,Macpherson:2006,Chong:2007zz}.

%% file: detector/calo.tex
\label{LHEC:MainDetector:DetCalo}
The LHeC calorimetry has to fulfil the requirements described in
Chapter\,\ref{LHEC:Detector:Requirements}.  The goal is a powerful
level 1 trigger and a detector able to resolve shower development in
three-dimensional space with no or minimal punch through.  High
transverse and longitudinal segmentation are necessary along with a
good matching to tracking detectors for particle identification and
separation of neutral and charged particles.  The calorimetry needs to
be hermetic in order to provide a good measurement of the total
transverse energy in the charged current process.  These
considerations are summarised in Tab.\,\ref{LHEC:DET:INTRO:Calo}.

The baseline design foresees a modular structure of independent
electromagnetic (EMC) and hadronic (HAC) calorimeter components.  In
order to fully contain electromagnetic showers, the EMC must provide
$\sim25-30X_0$.  The design of the EMC modules will vary when moving
from the very forward region, where energies up to $\cal{O}(\mathrm{1
  TeV})$ are expected, to the barrel and the backward region, where an
accurate and precise measurement of the scattered electron with energy
$\cal{O}(\mathrm{60~GeV})$ is paramount.

In the baseline design, the EMC is surrounded by the solenoid coil
which provides the magnetic field for momentum measurement in the
tracking system.  The hadronic calorimetry comes next and has
sufficient depth in order to precisely measure jets over the full
energy range, while providing the granularity in a projective modular
design such that it can reliably resolve multiple jets in an event.
The forward part of the HAC will need to provide up to $10\lambda_I$
to guarantee containment for energies up to a few TeV.

In the next sections the baseline design for the EMC and HAC
components is presented and discussed along with a comparison of
technologies and the experience from other HEP detectors
e.g.\cite{Green:2010zzb,Freeman:2010zz,Mandelli:2010zz,Bloch:2010zza,Anderson:2010zz}.
A brief summary of ongoing R\&D into new technologies which could
extend the precision and scope of the detector are briefly addressed.

\subsection{The barrel electromagnetic calorimeter}
\label{LHEC:MainDetector:DetCalo:BEC}

In the barrel region ($ 2.8 < \eta < - 2.3$), a Liquid Argon
calorimeter (LAr) with {\it accordion-shaped} electrodes, as is
currently in use by ATLAS\cite{Airapetian:1996iv}, is proposed as the
baseline.  The principle of sampling calorimetry is to arrange many
layers of passive material, in this case lead (X$_0$=0.56\,cm),
alternated with layers of active material, here LAr with
X$_0$=14.0\,cm.  The choice of Liquid Argon follows from its intrinsic
properties of excellent linearity, stability in time and radiation
tolerance\cite{Babaev:1994pd,Abt:1996hi,Fleischer:1997aq,Issever:2000yh,Schwanenberger:2002pp,Seehafer:2005ma,Kiesling:2010zz,Aubert:2005dh}.
A LAr calorimeter would also provide the required energy resolution,
detector granularity and projective design. 
The detector would share the same cryostat as the main solenoid which
in the case of a Linac-Ring design would include the bending dipoles.
The performance of the LAr calorimetry system has been extensively
addressed\cite{Airapetian:1996iv} and here only specific design
issues and detector simulation will be discussed. As an alternative a (warm)
option for a lead-scintillator electromagnetic calorimeter has been simulated for
comparison (see Section\,\ref{LHEC:Detector:Lead-Scintillator-EMC}).

\begin{figure}[htp]
\begin{center}
\includegraphics[width=\columnwidth]{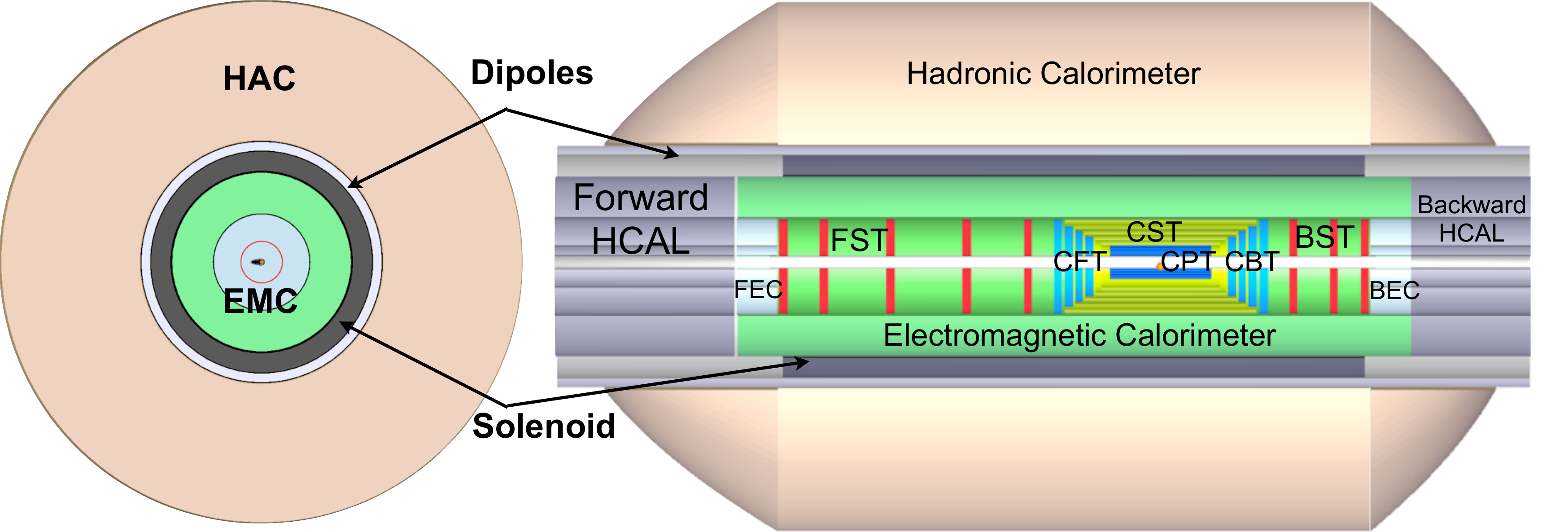}
\end{center}
\caption{\emph{x-y} and \emph{r-z} view of the LHeC Barrel EM calorimeter (green).}
\label{FIG:LHEC:DET:CAL:LAR1}
\end{figure}

Fig.~\ref{FIG:LHEC:DET:CAL:LAR1} shows a \emph{x-y} and \emph{r-z}
view of the LHeC Barrel EM calorimeter.
The layout allows the extraction of detector signals without
significantly degrading the high-frequency components which are vital
for fast shaping.  The flexibility in the longitudinal and transverse
segmentation, and the possibility of implementing a section with
narrow strips to measure the shower shape in its initial development,
represent additional advantages.  It is worth noting that due to the
asymmetric design, the projective structure is not fully symmetric as
the calorimeter and the solenoid centre are shifted forward with
respect to the interaction point.

\begin{figure}[htp]
\begin{center}
\includegraphics[width=0.6\columnwidth]{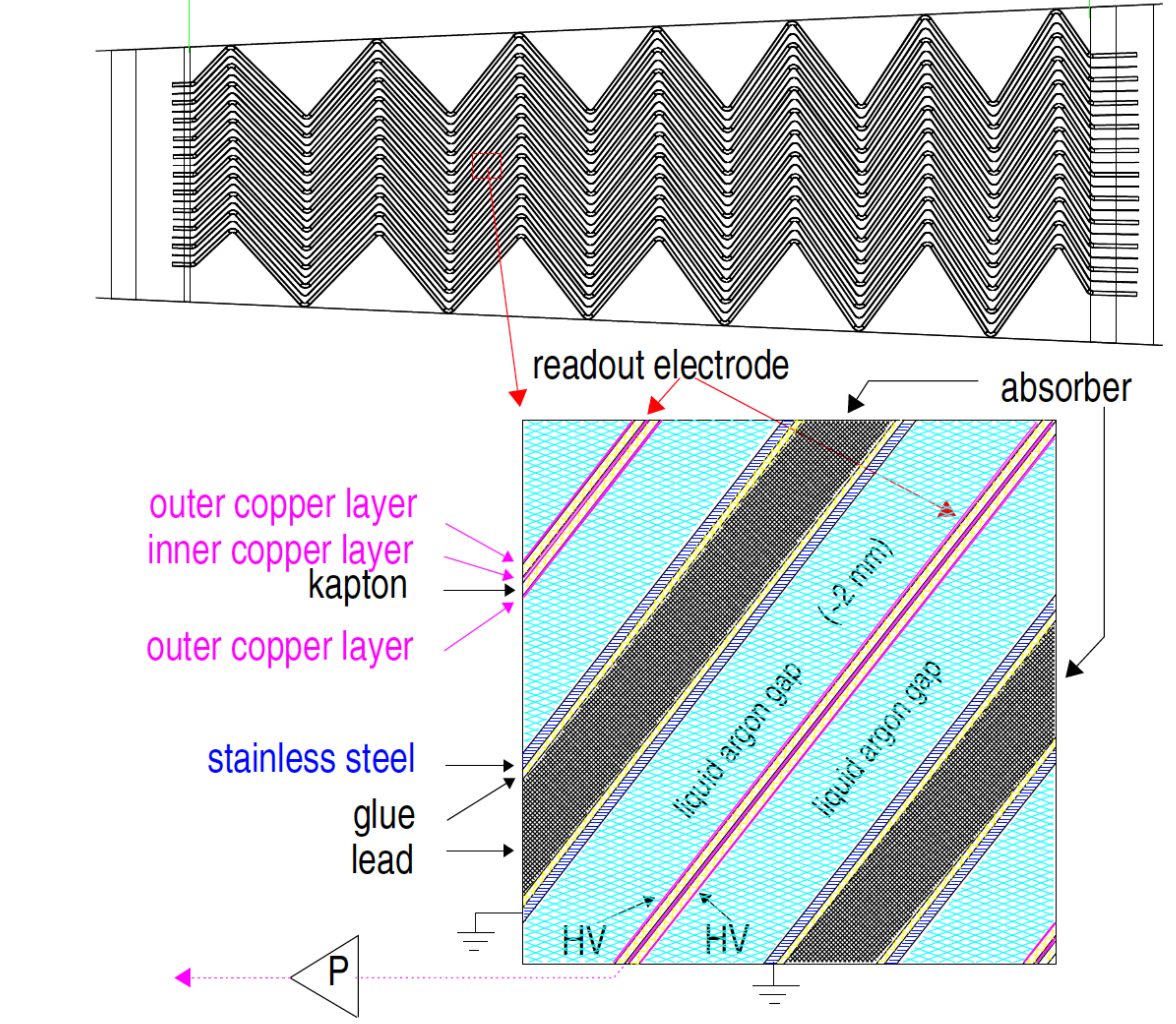}
\end{center}
\caption{Longitudinal view of one cell of the ATLAS LAr Calorimeter,
  showing the accordion structure.}
\label{FIG:LHEC:DET:CAL:LAR2}
\end{figure}
\begin{figure}[htp]
\begin{center}
\includegraphics[width=0.6\columnwidth]{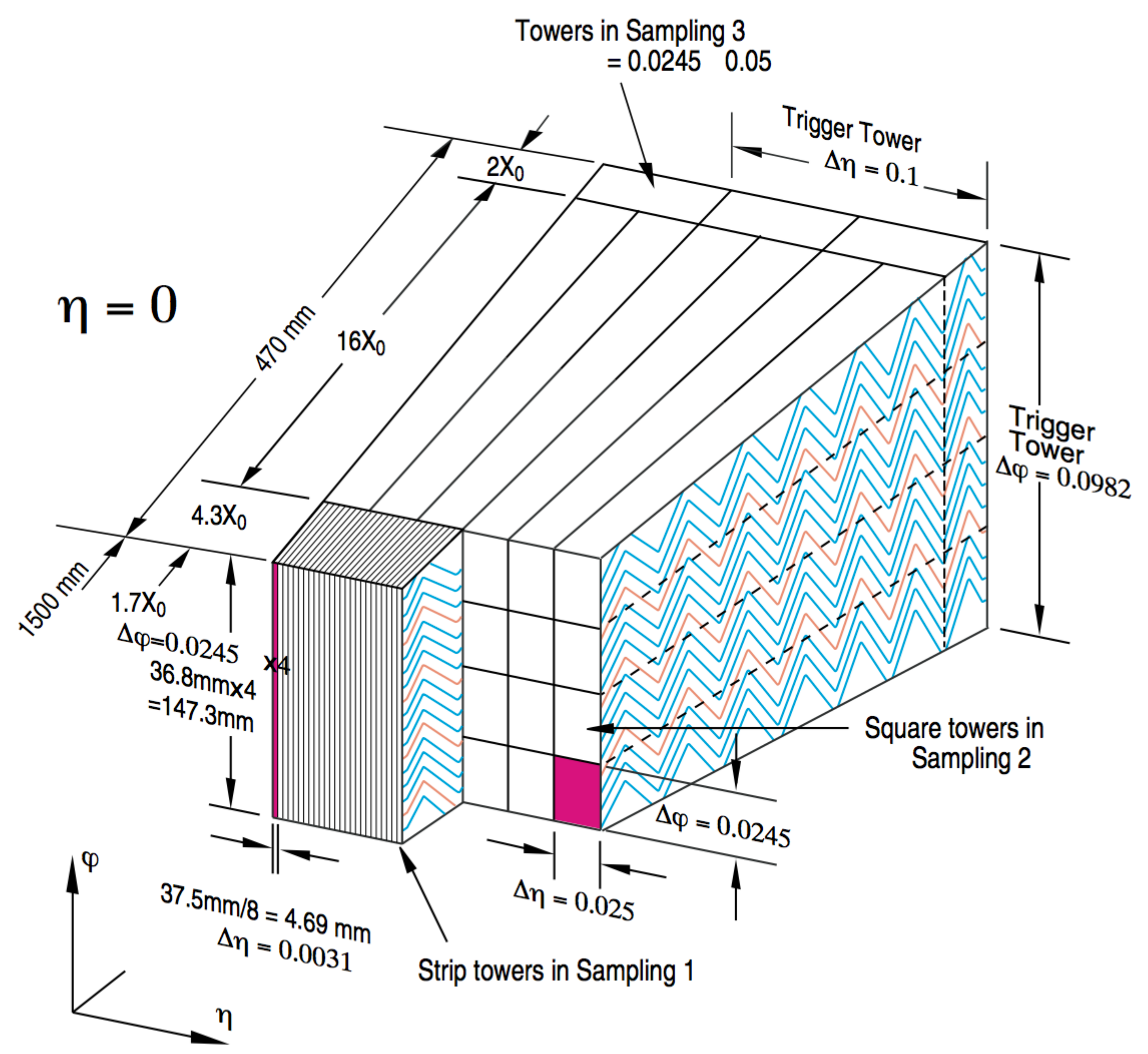}
\end{center}
\caption{3D view of the accordion structure of the ATLAS LAr Calorimeter}
\label{FIG:LHEC:DET:CAL:LAR3}
\end{figure}

Fig.~\ref{FIG:LHEC:DET:CAL:LAR2} shows a detail of the
accordion-electrode structure.  A basic cell consists of an absorber
plate, a liquid argon gap, a readout electrode and a second liquid
argon gap. The mean thickness of the liquid argon gap is constant
along the whole barrel and along the calorimeter depth.  The readout
granularity is subdivided into 3 cylindrical sections of increasing
size in $\Delta\eta \times \Delta\phi$. As shown in
Fig.~\ref{FIG:LHEC:DET:CAL:LAR3}, the first sampling section of the
EMC would have a very fine granularity ($\Delta\eta \times \Delta\phi
= 0.003\times 0.1$), to optimise the ability to separate photons from
$\pi^0$ energy deposits. The second sampling section, mainly devoted
to energy measurement, would have a granularity of about $0.025\times
0.025$, and the final sampling section has a slightly coarser
granularity of $\Delta\eta \times \Delta\phi = 0.050\times 0.025$.

%
%

\subsection{The hadronic barrel calorimeter}
\label{LHEC:MainDetector:DetCalo:HAC}
The baseline hadronic calorimeter in the barrel region is a sampling
calorimeter using steel and scintillating tiles as absorber and active
material, respectively\cite{Gildemeister:1991rb}.  The {\it Tile
  Calorimeter} would provide the required mechanical stability for the
inner LAr and Magnet cryostat along with the iron required for the
return flux of the solenoidal field, as is also the case in ATLAS\cite{Airapetian:1996iv}.

The Tile calorimeter consists of a cylindrical structure with inner
and outer radius of 120 and 260\,cm respectively
(Tab.\,\ref{LHEC:DET:SimCalo:tab:1}).  The central HAC barrel part is
580\,cm in length along the beam axis.  Endcaps extend the calorimetry
further in the forward and backward direction in order to guarantee
sufficient energy containment.
%
%
The detector cylinder would be built of several independent wedges
along the azimuthal direction while the modularity and segmentation
may vary depending on the machine design.

The Tile calorimeter forms the shell of the inner part of the LHeC
detector.  Once the barrel and the endcaps are assembled, all
of the sub-detectors apart from the muon system will be placed inside
of it.  The massive iron structure is rigid enough to support their
weight, in particular the liquid argon cryostat and the solenoid.

The absorber structure is a laminate of steel plates of various
dimensions, connected to a massive structural element referred to as a
girder. The highly periodic structure of the system allows the
construction of a large detector by assembling smaller sub-modules
together. Since the mechanical assembly is completely independent from
the optical instrumentation, the design is simple and cost effective.
Simplicity has also been the guideline for the light collection
scheme: the fibres are coupled radially to the tiles along the
external faces of each module. The laminated structure of the absorber
allows for channels in which the fibres run. The use of fibres for the
readout allows a layered cell readout to be used, creating a
projective geometry for triggering and energy reconstruction. A
compact electronics readout is housed in the girder of each
module. Finally, the scintillating tiles are read out in two separate
photomultipliers, providing the required redundancy.

\begin{table}[htp]
{\small
\begin{center}
  \begin{tabular}{|l|c|c|c|c|c|c|c|c|c|}
    \hline
 E-Calo Parts                                           & FEC1 & FEC2&    & EMC  &   & BEC2 & BEC1 \\ 
\hline \hline
Min. Inner radius $R$\hfill[cm]                &  3.1   & 21  &   & 48 &    & 21   & 3.1    \\
Min. polar angle $\theta$\hfill[\textdegree] &  0.48 & 3.2 &   & 6.6/168.9 &    & 174.2  & 179.1 \\
Max. pseudorapidity $\eta$ \hfill                  &  5.5  & 3.6 &   & 2.8/-2.3  &    & -3.  & -4.8  \\
Outer radius \hfill[cm]                                  & 20   & 46   &   & 88  &    & 46   & 20   \\
    {\em z}-length \hfill[cm]                                     & 40  & 40   &   & 660  &    & 40   & 40   \\
\cline{2-3} \cline{5-5} \cline{7-8}  
Volume       \hfill[m$^3$] &\multicolumn{2}{c|}{0.3} &   & 11.3  &  & \multicolumn{2}{c|}{0.3} \\  
\hline \hline

 H-Calo Parts barrel      &     &      & FHC4  & HAC&   BHC4 &      &    \\ 
\hline \hline
Inner radius \hfill[cm]  &     &      &  120 &   120&  120 &      &    \\
Outer radius \hfill[cm] &     &      &  260 &    260&  260 &      &    \\
{\em z}-length \hfill[cm]   &     &      &  217 &  580&  157 &      &    \\
\cline{4-6}  
Volume       \hfill[m$^3$] &      &     & \multicolumn{3}{c|}{121.2} &      &   \\
\hline \hline

 H-Calo Parts Inserts                          & FHC1& FHC2 & FHC3 &   & BHC3 & BHC2 & BHC1\\ 
\hline \hline
Min. inner radius $R$\hfill[cm]                 &  11 &  21  &  48  &         &  48  & 21   & 11   \\
Min. polar angle $\theta$\hfill[\textdegree]  &  0.43 & 2.9  &  6.6 &     &  169.   & 175.2 & 179.3 \\
Max/min pseudorapidity $\eta$ \hfill             & 5.6   & 3.7  &  2.9   &     &   -2.4   & -3.2 & -5.  \\
Outer radius \hfill[cm]                                   &  20   &  46  &  88 &   &  88  & 46   & 20  \\
    {\em z}-length \hfill[cm]                                     &  177 & 177 & 177 &   &  117 & 117  & 117 \\
\cline{2-4} \cline{6-8}  
Volume       \hfill[m$^3$] &\multicolumn{3}{c|}{4.2}   &    &   \multicolumn{3}{c|}{2.8} \\  
\hline

\end{tabular}
\end{center}
}
\caption{Summary of calorimeter dimensions. \newline
The electromagnetic barrel calorimeter is currently represented by the barrel part
EMC (LAr-Pb module, $X_0\approx25$ radiation length), 
with forward FEC1, FEC2 (Si-W modules ($X_0\approx30$) 
and backward module inserts BEC1, BEC2 (Si-Pb modules; $X_0\approx25$).  \newline
The hadronic barrel parts are represented by FHC4, HAC, BHC4
( forward, central and backward - Scintillator-Fe Tile modules; $\lambda_{I}\approx8$ interaction length) 
and the movable inserts FHC1, FHC2, FHC3 (Si-W modules; $\lambda_{I}\approx10$), BHC1, BHC2, BHC3 
(Si-Cu modules, $\lambda_{I}\approx 8$) see Fig.\,\ref{LHEC:MainDetector:MainStructure:Fig:1}.}
\label{LHEC:DET:SimCalo:tab:1}
\end{table}

%
%



The granularity of the Tile Calorimeter is important to be able to
finely match the electromagnetic LAr calorimeter in front and correct
for the dead material of the magnet complex. The proposed hadronic
segmentation for the cells behind the electromagnetic section, will
allow an efficient hadron leakage cut, needed for electron and photon
identification. A reasonable longitudinal segmentation, especially
around the maximum depth of the shower, favours an appropriate
weighting technique to restore, at the level of 1-2\%, the linearity
of the energy response to hadrons, which is intrinsically non-linear
because of the non-compensating nature of the calorimeter.  At the
highest energies, the resolution of the calorimetry is dominated by
the constant term, for which the largest contribution comes from the
detector non-linearity and calibration.
An attempt is made to keep the constant term below the 2\% level.

%
%
%
%

\subsection{Endcap calorimeters}
\label{LHEC:MainDetector:DetCalo:FBC}
Calorimetry in the forward and backward direction at the LHeC is of
extreme importance: in the forward region for the measurement of the
hadronic final state, 
and in the backward region for the measurement of the low energy
scattered electron.  Here, a good $e/h$ separation is also important
to suppress hadronic background.
As seen in Fig.\,\ref{Fig:Fluence}, the very forward and to a lesser
extent the backward parts of the calorimeter are exposed to high
levels of particle radiation and must therefore be radiation hard by
design.  Synchrotron radiation and any further background radiation
must also be tolerated in addition.

Fig.\,\ref{LHEC:MainDetector:MainStructure:Fig:1} shows in detail the
endcap calorimeters for the Ring-Ring design.  The two-phase
experimental program requires the endcaps to be modular as these
components will either be moved along the beam line or completely
removed to allow the placement of the strong focusing magnets for the
high luminosity phase.  The relevant dimensions and specifications are
summarised in Tab.\,\ref{LHEC:DET:SimCalo:tab:1}.
For the Linac-Ring design, where no additional magnets along the beam
line will be required, the subcomponents FHC2/FHC3 and BHC2/BHC3, can
be combined into single modules.


The restrictive geometry of the insert calorimeters requires a
non-conventional and challenging design based on previous
developments\cite{Golutvin:1992un,Anderson:1994ve,Adams:2001qc,Zatsepin:2003df,Bonvicini:2004hc,Bonvicini:2005ie,Strom:2004,Strom:2005id}.
Tungsten (\emph{W}) is considered as the absorber material, in
particular for the forward inserts, because of its very short
radiation length and large absorption to radiation length ratio.
About 26\,cm of tungsten will absorb electromagnetic showers
completely and will contain the hadronic shower to a large extent and
over a large range of energy
($\approx{30}$X$_0$+$\approx{10}\lambda_{I}$).  The electromagnetic
and hadronic sections can be combined to minimise boundary effects.
An alternative to tungsten for the hadronic absorber is copper
(\emph{Cu}).

Simulations have been performed to compare the different
absorbers. Since the backward inserts have looser requirements, the
material for the absorbers are lead (\emph{Pb}) for the
electromagnetic part and copper for the hadronic. For the Ring-Ring
option, where no dipole field along the beam pipe is required, a more
economical choice of steel (\emph{Fe}) instead of copper can be
considered.  The active signal sensors for both the forward and
backward calorimeters have been chosen to be silicon-strip
(electromagnetic fwd/bwd parts) and silicon-pad (hadronic fwd/bwd
parts).
\section{Calorimeter simulation}
In this section preliminary results on simulations of the barrel and
endcap calorimeters are illustrated using the simulation frameworks
{\small\bf GEANT4} and {\small\bf FLUKA}\,\cite{Ferrari:2005zk,
Battistoni:2007zzb}. In general the parameters of the functions have
been fitted to the {\small\bf GEANT4} data. The {\small\bf FLUKA}
results are shown for comparison, if available.  The detector
components presented
in\,\ref{LHEC:MainDetector:DetCalo:BEC},\ref{LHEC:MainDetector:DetCalo:HAC},
\ref{LHEC:MainDetector:DetCalo:FBC} have been simulated using {\small\bf GEANT4.9.2}\cite{Agostinelli:2002hh}
with single and multiple particle events along with full \emph{e-p}
events from the {\small\bf QGSP-3.3}\cite{Kaidalov:1982xe} physics
list and {\small\bf FLUKA} with {\small\bf CALORIMETry} card.  The
Quark-Gluon String Precompound ({\small\bf QGSP}) is based on
theory-driven models and uses the quark-gluon-string model for
interactions and a pre-equilibrium decay model for fragmentation.

The detector geometry, including the various layers of active,
absorbing and support material were coded and inserted in the
simulation. Energy resolutions for electromagnetic and hadronic
deposits were studied along with concepts for optimal trigger and
signal reconstruction.
Particular attention was put into the key features and the
construction constraints of the detector, namely the beam optics and
the magnets (the solenoid and the Linac-Ring dipoles).  Where a
similar design from an existing or developing detector are available,
the results are presented complemented by referenced studies.

The energy resolution of a calorimeter is parameterised by the following quadratic sum:
\begin{equation}
\frac{\sigma_E}{E}=\frac{a}{\sqrt{E}}\oplus{b}
\label{LHEC:DET:SimCalo:Eq:1}
\end{equation}
where ${E}$ is the particle energy in ${GeV}$, ${a}$ is the stochastic
term, which arises from fluctuations in the number of signal producing
processes, ${b}$ is the constant term, which describes imperfections
in calorimeter construction, fluctuations in longitudinal energy
containment and non-uniformities in signal collection etc.  A third
term $c$ (omitted here) is often added to represent a noise term
needed to describe experimental data.  The energy deposition of
primary and secondary particles in the calorimeter was obtained using
{\small\bf GEANT4} and {\small\bf FLUKA}, and fitted to extract ${a}$
and ${b}$ using the data obtained in {\small\bf GEANT4}. 
Effects due to the readout process were not considered at this stage.
\begin{figure}[htp]
\begin{center}
\includegraphics[width=0.7\columnwidth]{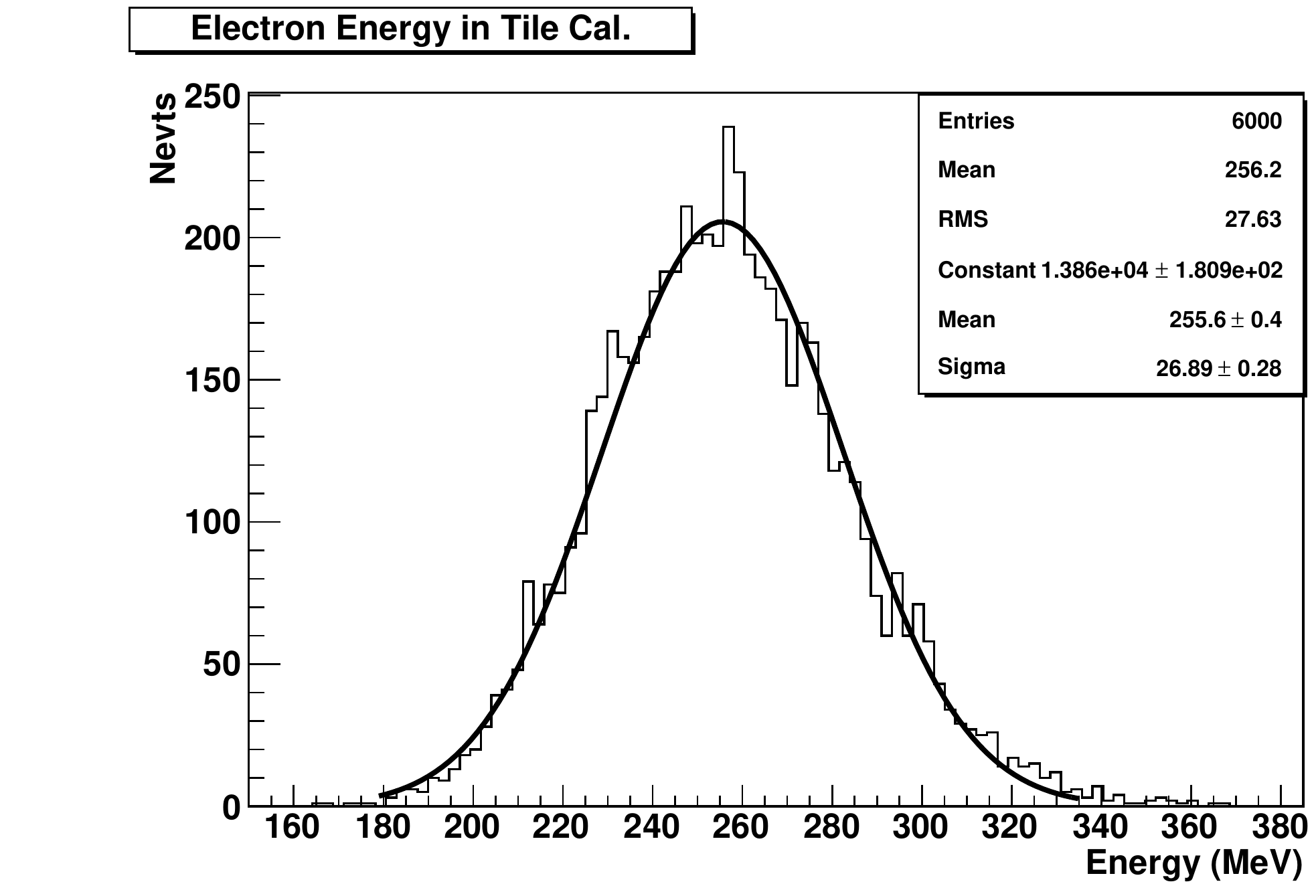}
\end{center}
\caption{Example for a pion energy distribution and the Gaussian fit. 
The resulting $\sigma$ and mean values are estimated for pions of an
incident angle $\theta=70$\textdegree and 10\,GeV energy into the
tile-calorimeter module ({\small\bf GEANT4}).}
\label{LHEC:DET:SimCalo:Fig:2}
\end{figure}

Each energy distribution was fitted with a Gaussian in a range
$\pm2\sigma$ around the mean; the energy dependent resolution was
calculated using those fitted mean values.  An example of the energy
distribution with a Gaussian fit applied is shown in
Fig.\,\ref{LHEC:DET:SimCalo:Fig:2}.  The $a$ and $b$ parameters are
then calculated from the fit of $\sigma/E$ ({\small\bf GEANT4}).

\subsection{The barrel LAr calorimeter simulation}

A simplified layout, adapted from the ATLAS LAr
calorimeter\cite{Airapetian:1996iv}, has been implemented in 
{\small\bf GEANT4} and {\small\bf FLUKA}  simulations and used to extract the main
characteristics of the LHeC barrel electromagnetic calorimeter.

\begin{figure}[htp]
\includegraphics[width=0.60\linewidth]{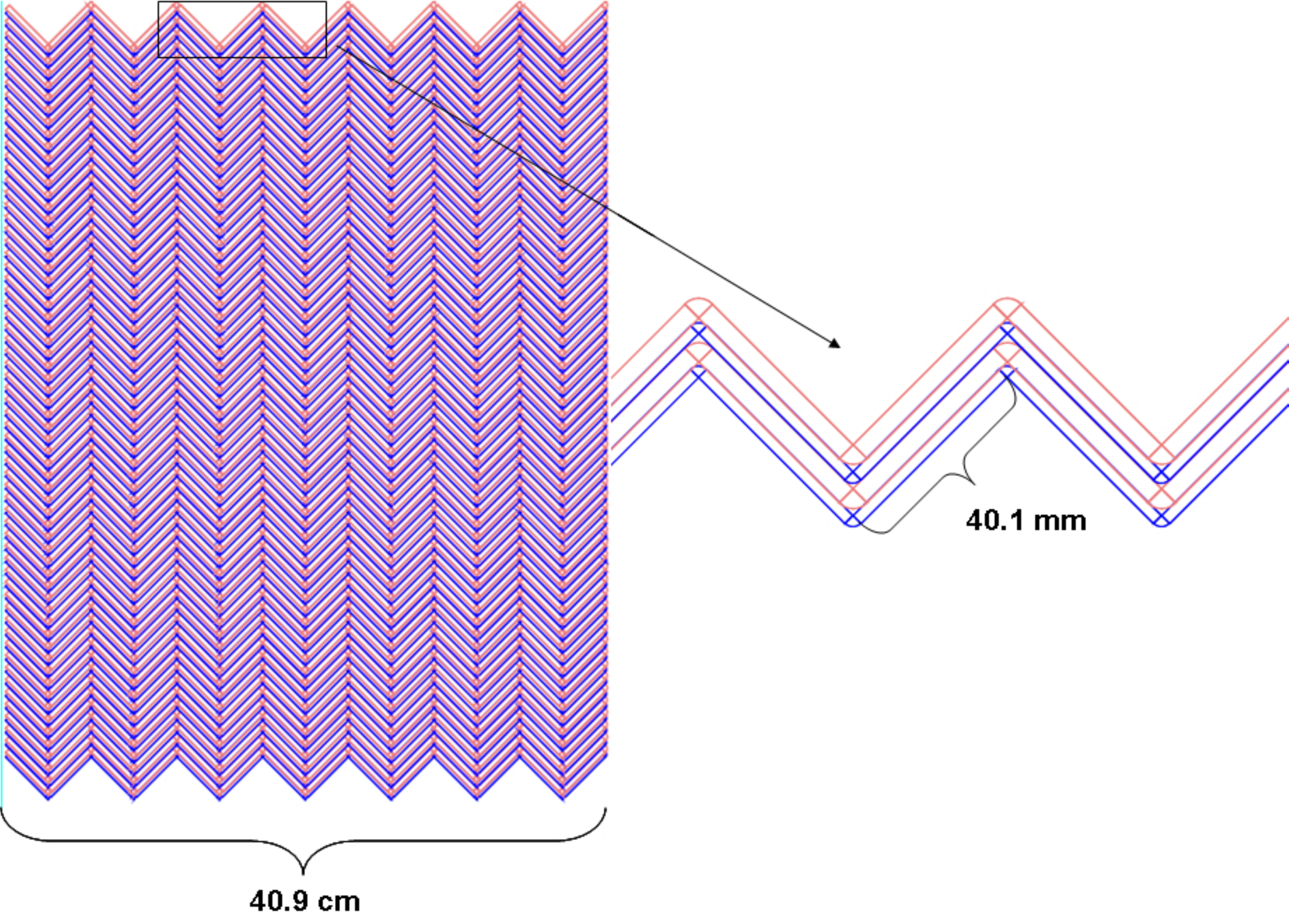}
\ \ \ \raisebox{8mm}{\includegraphics[width=0.30\linewidth]{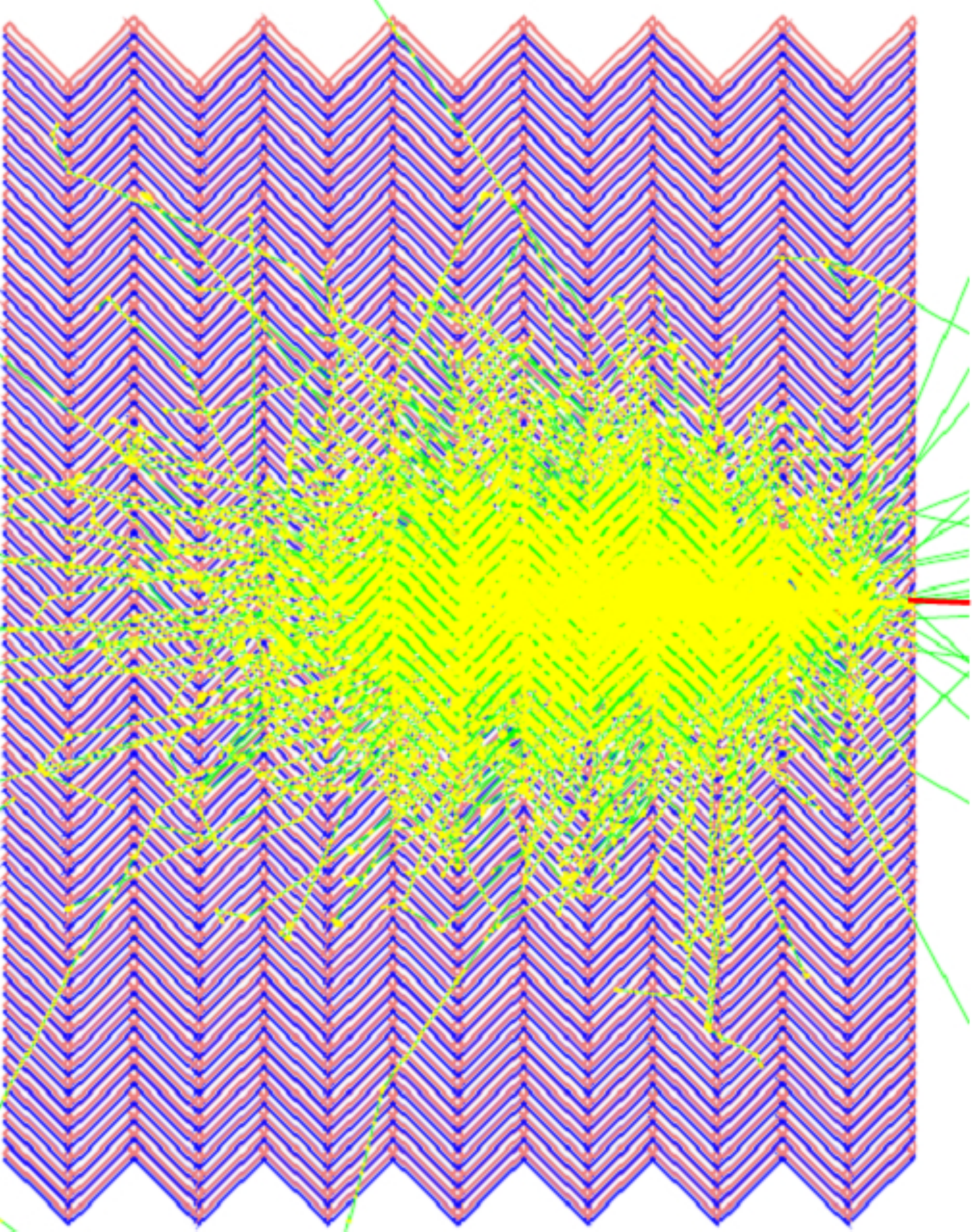}}
\caption{View of the parallel geometry accordion calorimeter\,(left) and simulation of a single electron shower with initial energy of 20 GeV\,(right) - LAr calorimeter module.}
\label{LHEC:DET:SimCalo:Fig:14}
\end{figure}

\begin{figure}[htp]
\centering
\includegraphics[width=0.6\linewidth]{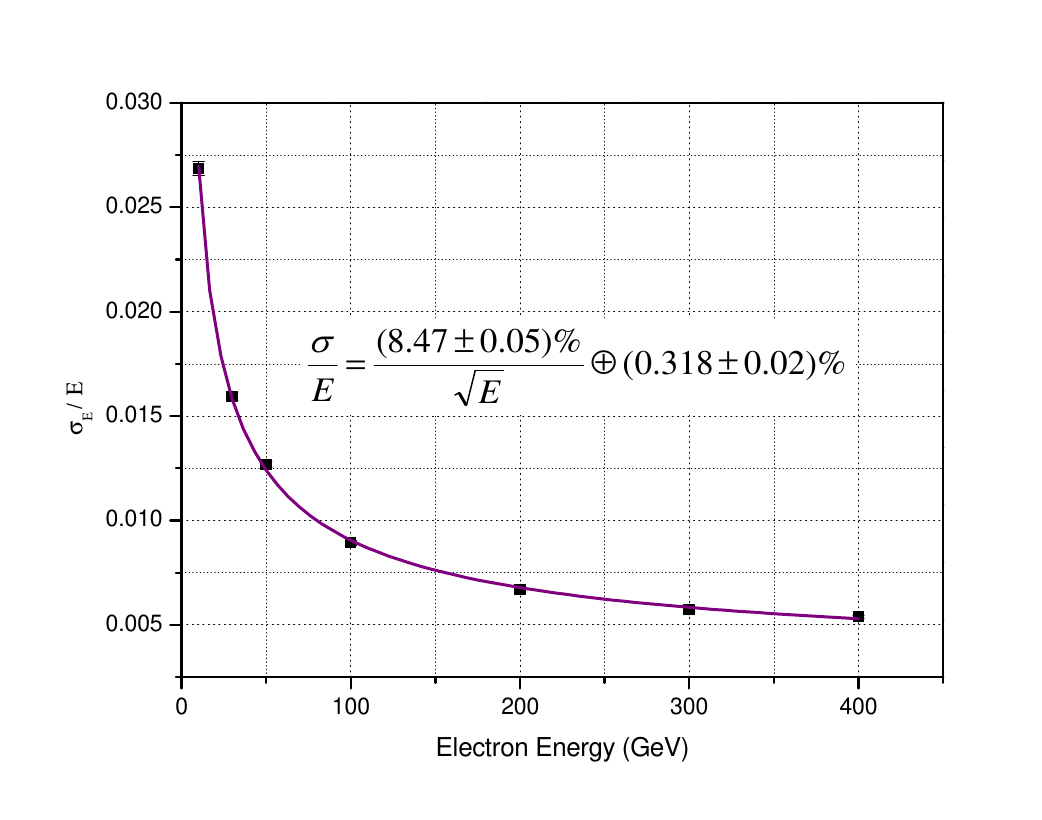}
\caption{LAr accordion calorimeter energy resolution for electrons between 10 and 400 GeV ({\small\bf GEANT4}).}
\label{LHEC:DET:SimCalo:Fig:16}
\end{figure}

The accordion shaped absorber sheets are 2.2\,mm thick lead layers
interspersed with 3.8\,mm wide gaps filled with liquid argon. In the
present model the electrodes which in the case for ATLAS are
2$\times$0.275\,mm thick, were not considered.  Both the absorber and
the liquid argon gap have an accordion fold length of 40.1\,mm and 13
bend angles of 90\textdegree.  A total of 62 absorber sheets, each
250\,cm wide in the z-direction, have been incorporated into the
simulation (Fig.\,\ref{LHEC:DET:SimCalo:Fig:14}-left).  A 20\,GeV
incident single electron showering in the stack is shown in
Fig.\,\ref{LHEC:DET:SimCalo:Fig:14}-right.  The energy resolution for
electrons was obtained from the ratio of the mean and the standard
deviation of the electron response, both obtained by fitting a
Gaussian to the energy spectrum.  Figure \ref{LHEC:DET:SimCalo:Fig:16}
shows the energy resolution for electrons of energy between 10 and
400\,GeV at $\theta=$90\textdegree. Here, the stochastic term of the
energy resolution is found to be $8.47\%$ and the constant term is
$0.318\%$ which compare well with $9.99\%$ and $0.35\%$, respectively
at about $\theta=$90\textdegree\cite{Cravero:1994}.  In the simulation
the energy deposited in the active material is normalised to the
energy of the incident particle.



%

\subsection{The barrel tile calorimeter simulation}
\begin{table}[htp]
\begin{center}
\begin{tabular}{|l|c|c|c|}
\hline 
    Tile Rows    & Height of Tiles in Radial Direction & Scintillator Thickness \\
\hline
      1-3        &   97\,mm                              &  3\,mm   \\
      4-6        &  127\,mm                              &  3\,mm   \\
      7-11       &  147\,mm                              &  3\,mm   \\ 
\hline

\cline{1-3}  
{\em x}-depth &  \multicolumn{2}{c|}{1407\,mm}        \\
\hline
\end{tabular}
\end{center}
\caption{Longitudinal (into {\em x}-direction) segmentation of the hadronic tile calorimeter (HAC).}
\label{LHEC:DET:SimCalo:tab:2}
\end{table}
\begin{figure}[htp]
\includegraphics[width=0.5\columnwidth]{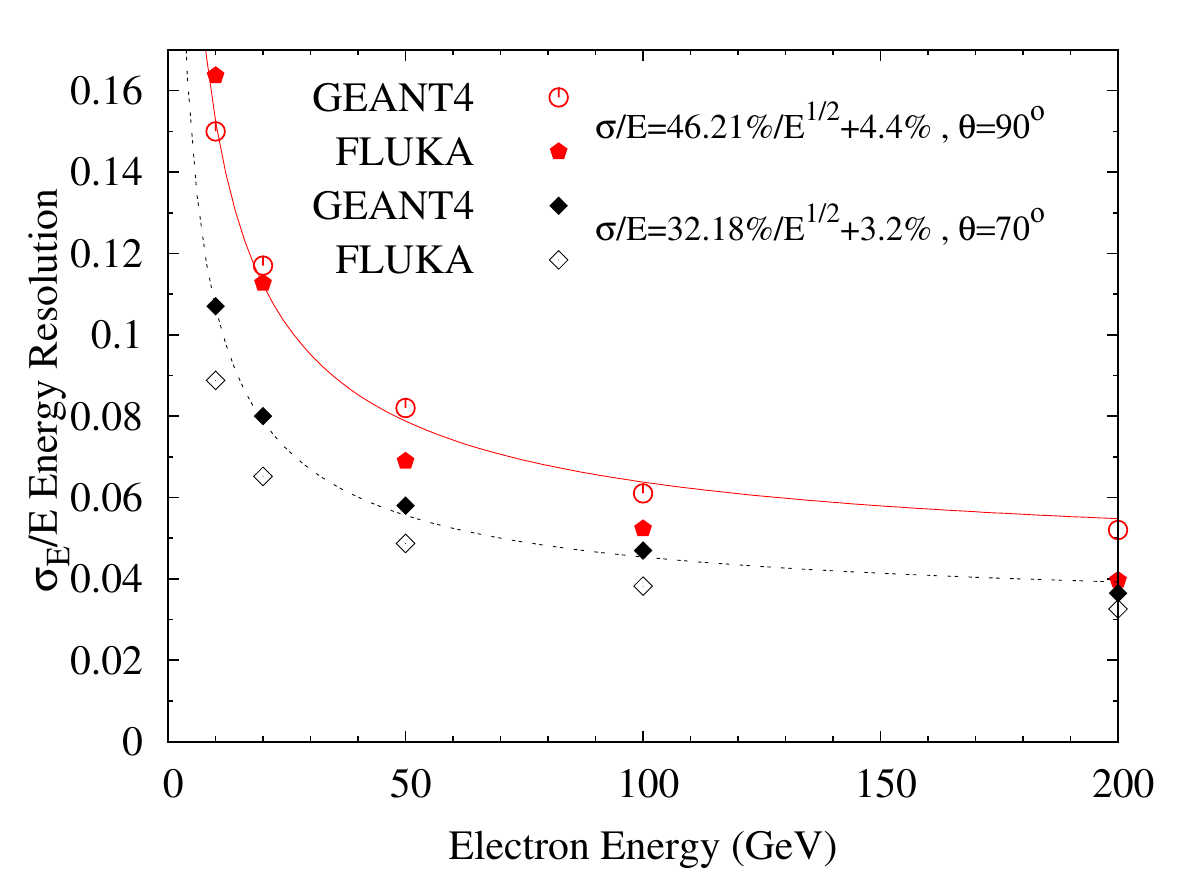}
\includegraphics[width=0.5\columnwidth]{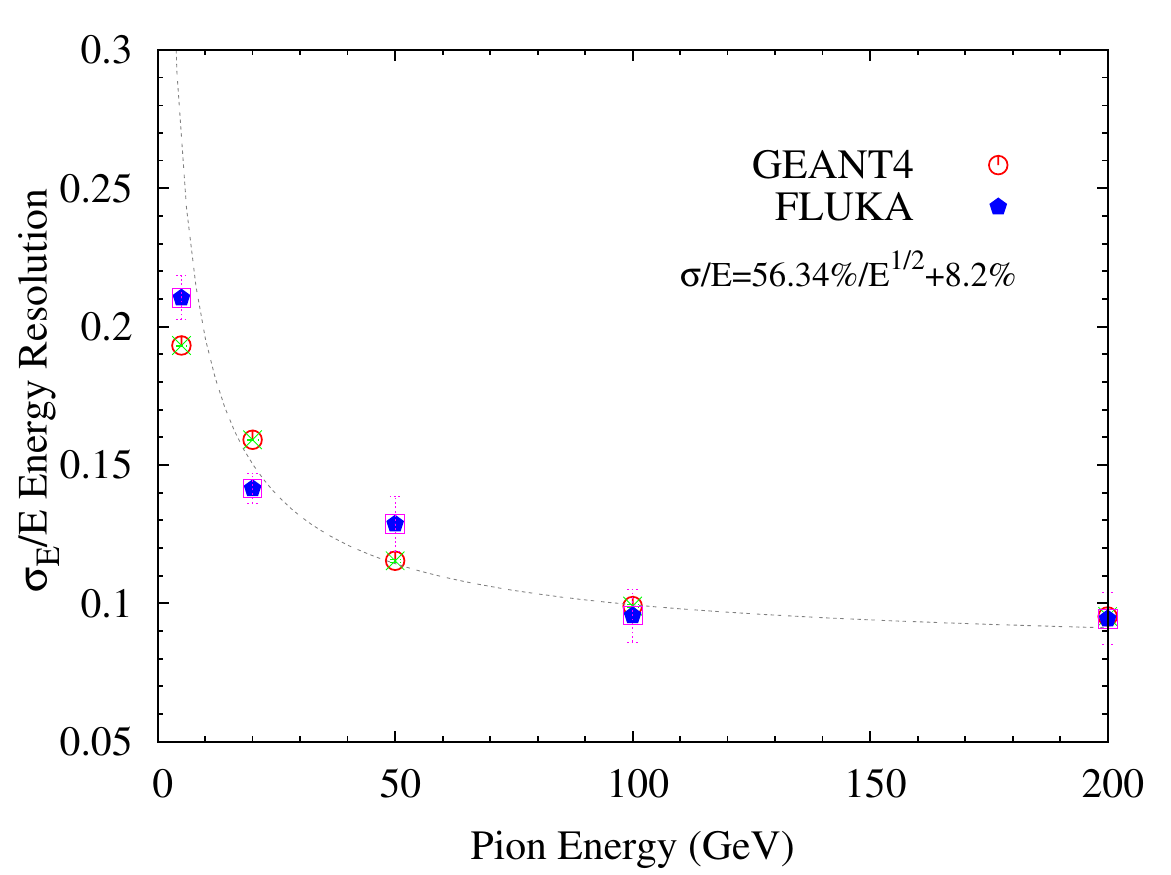}
\caption{Tile Calorimeter energy resolution for electrons at $\theta=$70\textdegree and 90\textdegree (left) and for pions at $\theta=$90\textdegree (right).}
\label{LHEC:DET:SimCalo:Fig:4}
\end{figure}

The HAC is a scintillator-steel tile calorimeter: 4\,mm thick steel
plates are interspaced by 3\,mm thick scintillator tiles. The tiles
are placed in planes perpendicular to the $z$-direction.  The absorber
structure consists of 262 repeated periods, each of which spans 19\,mm
in $z$ and consist of 16\,mm of steel and 3\,mm of scintillator tile.
11 transverse rows of tiles are used in a module.
The total interaction depth of the HAC prototype corresponds to
$\lambda_{I}=7$.  The longitudinal segmentation of the HAC module is
described in Tab.\,\ref{LHEC:DET:SimCalo:tab:2}.  In this section the
performance of the hadronic calorimeter alone has been investigated,
the combined use of EMC and HAC parts has been studied in later
sections.  The energy resolution of the tile calorimeter was simulated
with electrons and pions within the energy range 3-200\,GeV
(Fig.\,\ref{LHEC:DET:SimCalo:Fig:4}). The stochastic term and constant
term values obtained for electrons shown on the left side of the
figure are consistent with results obtained for
ATLAS\cite{TILECAL:2006}.  It is clearly seen that, both stochastic
and constant term values decrease with decreasing angle. The
parameterisation values for pions on the right side of the figure are
in agreement with\cite{Efthymiopoulos:1997pz}(Page\,1,\,Eq.\,1).  The
response to electrons generally shows good resolution such that any
leakage from the electromagnetic calorimetry in front of the HAC would
be resolved safely.

\begin{figure}[htp]
\centering
\includegraphics[width=0.6\linewidth]{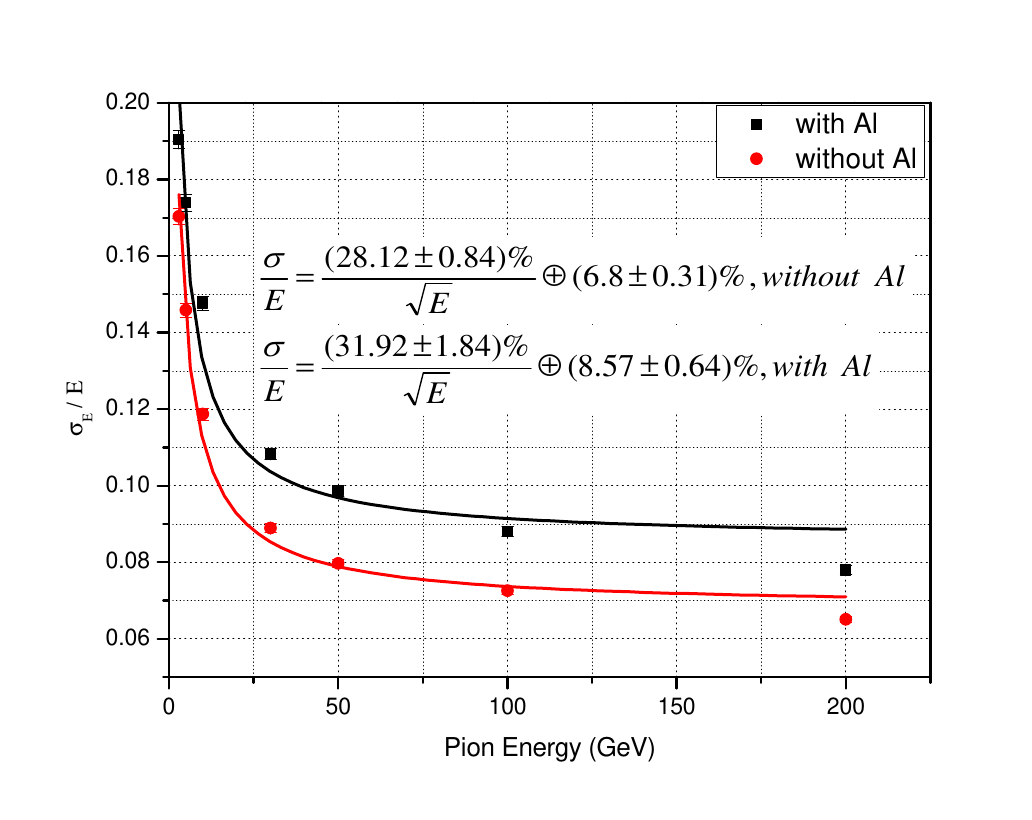}
\caption{Combined LAr Accordion and Tile Calorimeter energy resolution for pions with and without 14\,cm Al block ({\small\bf GEANT4})}
\label{LHEC:DET:SimCalo:Fig:17}
\end{figure}
\subsection{Combined liquid argon and tile calorimeter simulation}

The combined system (accordion and tile calorimeter) has been studied.
The effect of the dead material due to the magnet and the cryostat
between the EMC and HAC has been studied in a first approximation.
The energy resolution of the combined system has been simulated. The
effect of the solenoid and the cryostat infrastructure has been
simulated by adding a thick Aluminium layer (14\,cm) in between the
EMC and HAC.  The study has been performed using particles 
over a wide range of primary energy and at different incident angle in order 
to deduce information about the detector response for particles entering the
calorimeters at different $z$.
Hadronic shower simulations have been performed in the energy range
3\,GeV-200\,GeV.  First results of the energy resolutions as a
function of energy for pions are shown in
Fig.\,\ref{LHEC:DET:SimCalo:Fig:17}. The stochastic and constant term
values obtained for the combined system with and without Al block are
consistent with results parameterised for
ATLAS\cite{Efthymiopoulos:1997pz}(Page\,1,\,Eq.\,2).

\subsection{Lead-Scintillator electromagnetic option}
\label{LHEC:Detector:Lead-Scintillator-EMC}

Along with the baseline liquid argon calorimeter, a more conservative
option, not requiring a dedicated cryogenic system, has been
considered for the barrel electromagnetic calorimetry.  For this
purpose a lead-scintillator sampling calorimeter (EMC$_{Pb-Sc}$),
composed of 20$\times$0.85\,cm thick {\em Pb} layers interspaced by
4\,mm plastic scintillator plates was setup for simulation.  The
radiation length of this system corresponds to 30$X_0$
(X$_0$(Pb)=0.56\,cm).  All dimensions of the calorimeter systems have
been kept according to the default solution summarised in
Tab.\,\ref{LHEC:DET:SimCalo:tab:1}.
\begin{figure}[htp]
\begin{center}
\includegraphics[width=1.0\columnwidth]{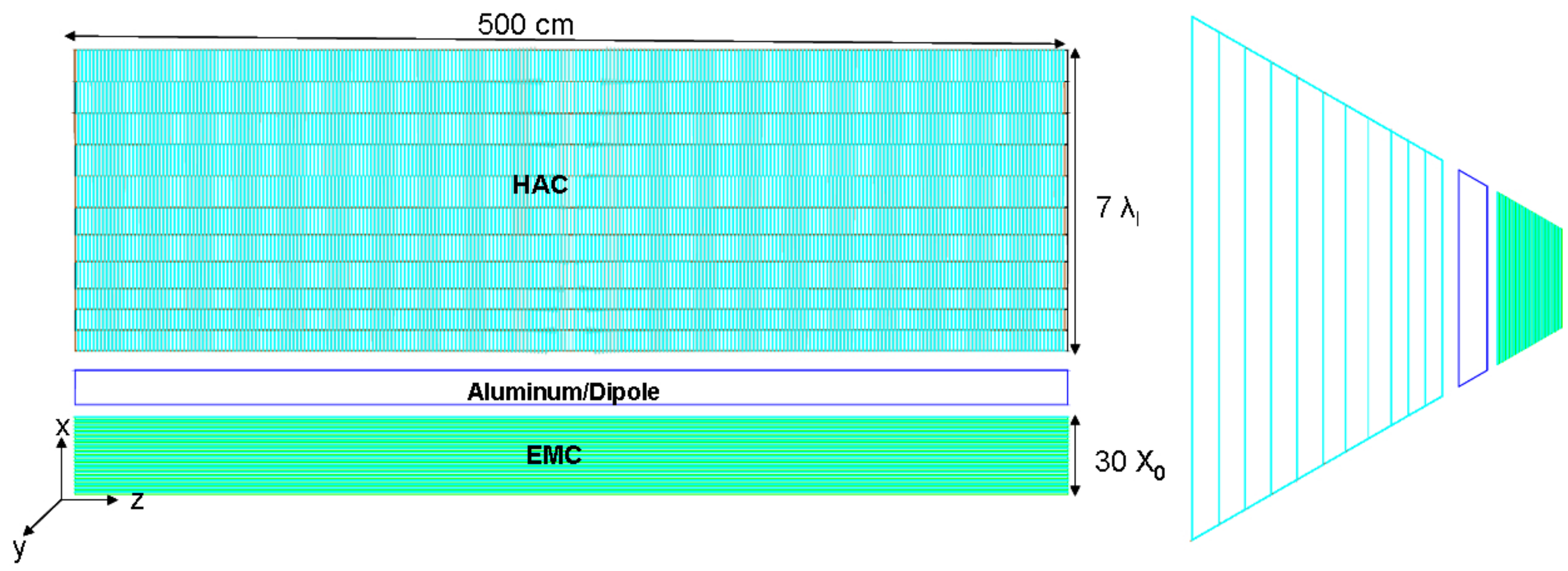}
\end{center}
\caption{Simulation - EMC$_{Pb-Sc}$ stack\,/\,solenoid-dipole-system($\propto$16\,cm Al-block equivalent)\,/\,HAC. }
\label{LHEC:DET:SimCalo:Fig:1}
\end{figure}
\begin{figure}[htp]
\begin{center}
\includegraphics[width=0.7\columnwidth]{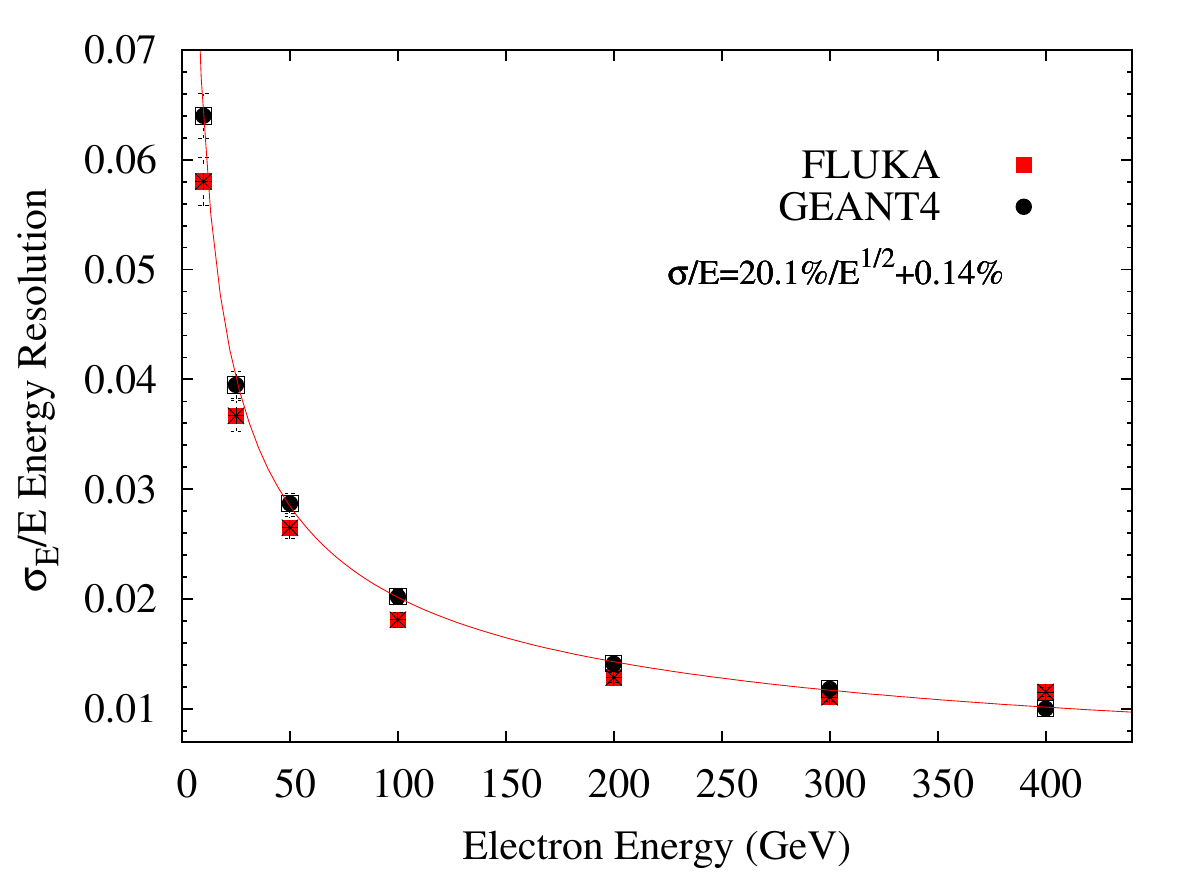}
\end{center}
\caption{The electromagnetic lead-scintillator calorimeter energy resolution for electrons at $\theta=$90\textdegree.}
\label{LHEC:DET:SimCalo:Fig:3}
\end{figure}

The EMC$_{Pb-Sc}$ stack was placed 30\,cm in front of the HAC.  Again
an aluminium block of 16\,cm was inserted between EMC and HAC
representing the magnet/cryostat system as illustrated in
Fig.\,\ref{LHEC:DET:SimCalo:Fig:1}.  The sketched module would be one
of 6 azimuthal segments of the complete barrel EMC and HAC.  The
energy resolution of the electromagnetic lead-scintillator calorimeter
as obtained with electrons of 10-400\,GeV is shown in
Fig.\,\ref{LHEC:DET:SimCalo:Fig:3}.
\begin{figure}[htp]
\centering
\includegraphics[width=0.49\linewidth]{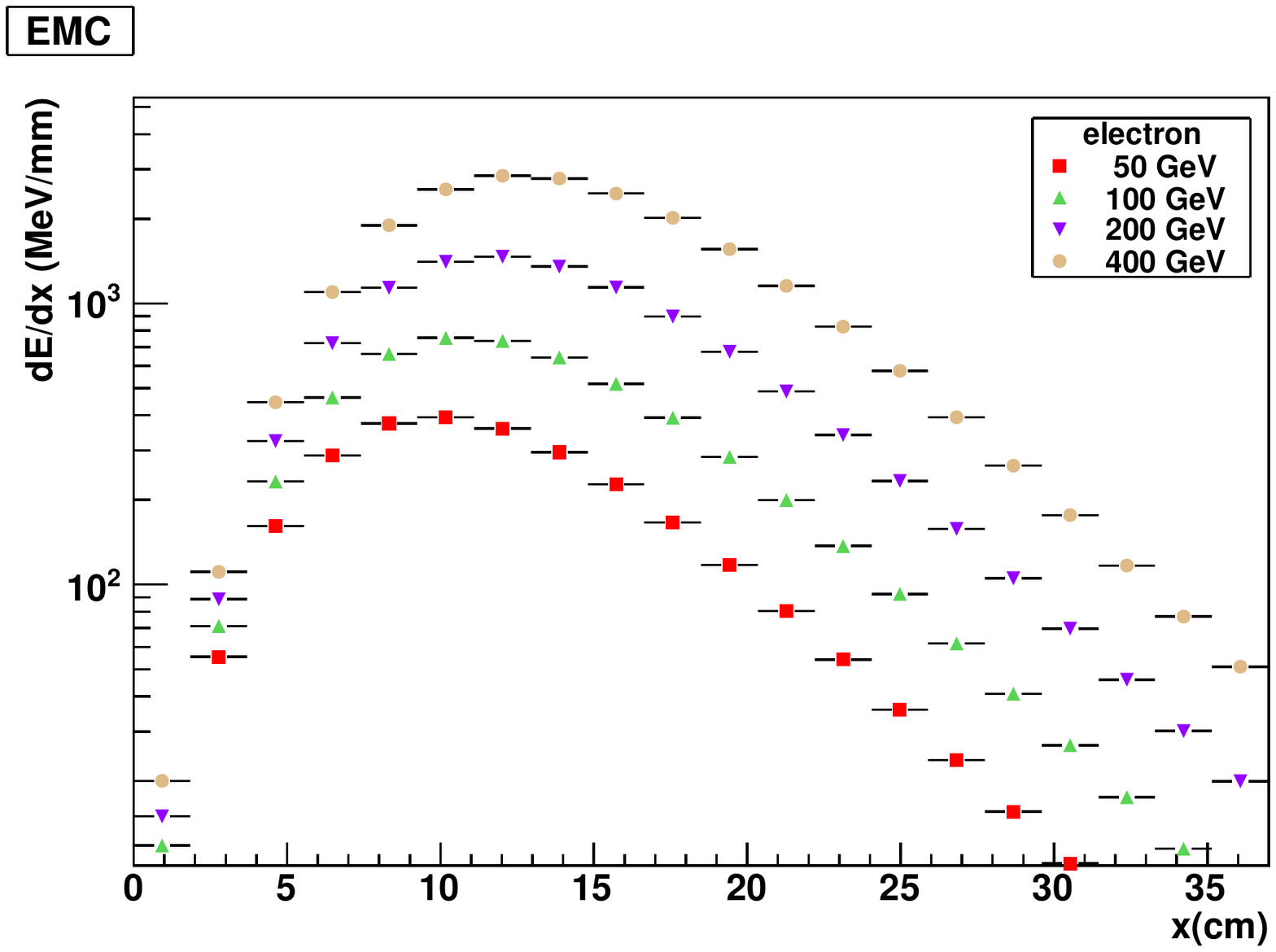}
\includegraphics[width=0.47\linewidth]{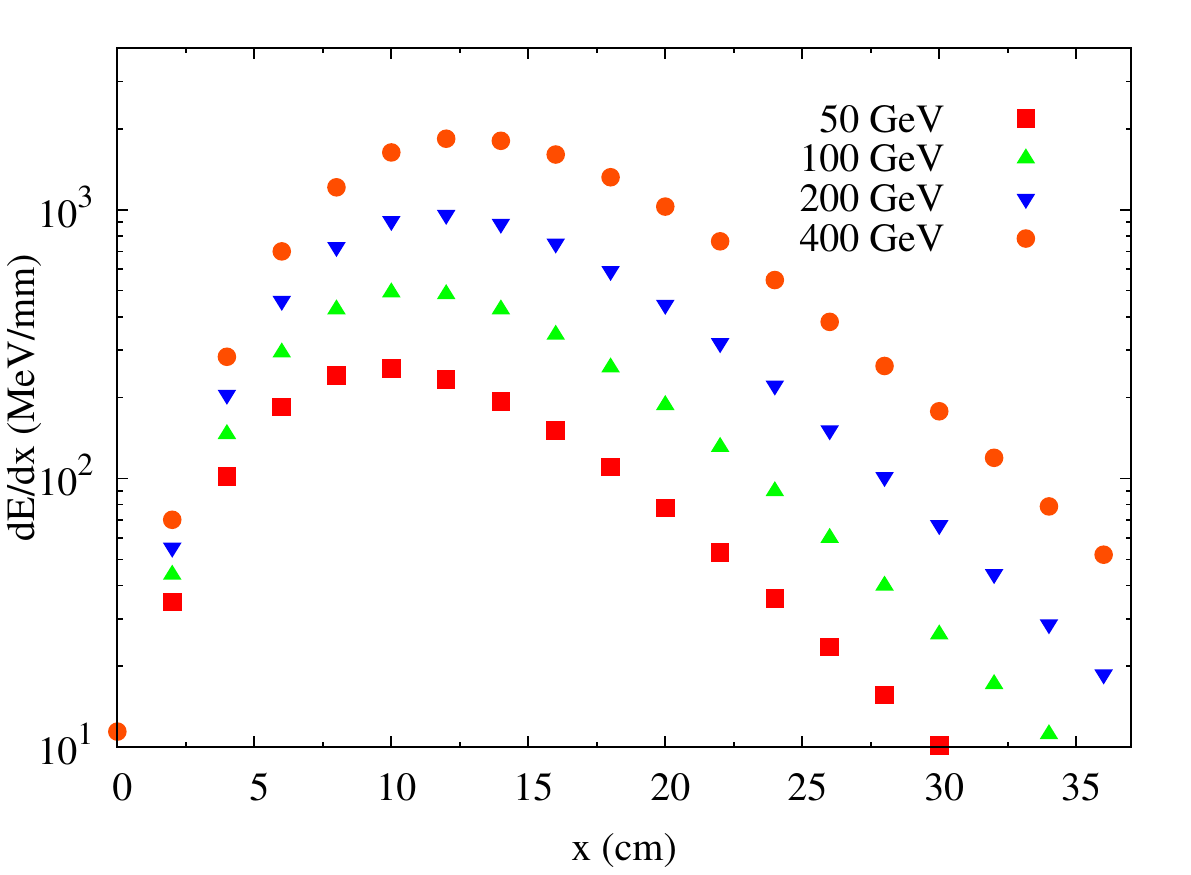}
\caption{Electron longitudinal shower profile for EMC$_{Pb-Sc}$ at various energies ({\small\bf GEANT4} (left) and {\small\bf FLUKA} (right)). Only the statistical uncertainties are shown.}
\label{LHEC:DET:SimCalo:Fig:9}
\end{figure}
 \begin{figure}[htp]
\includegraphics[width=0.5\linewidth]{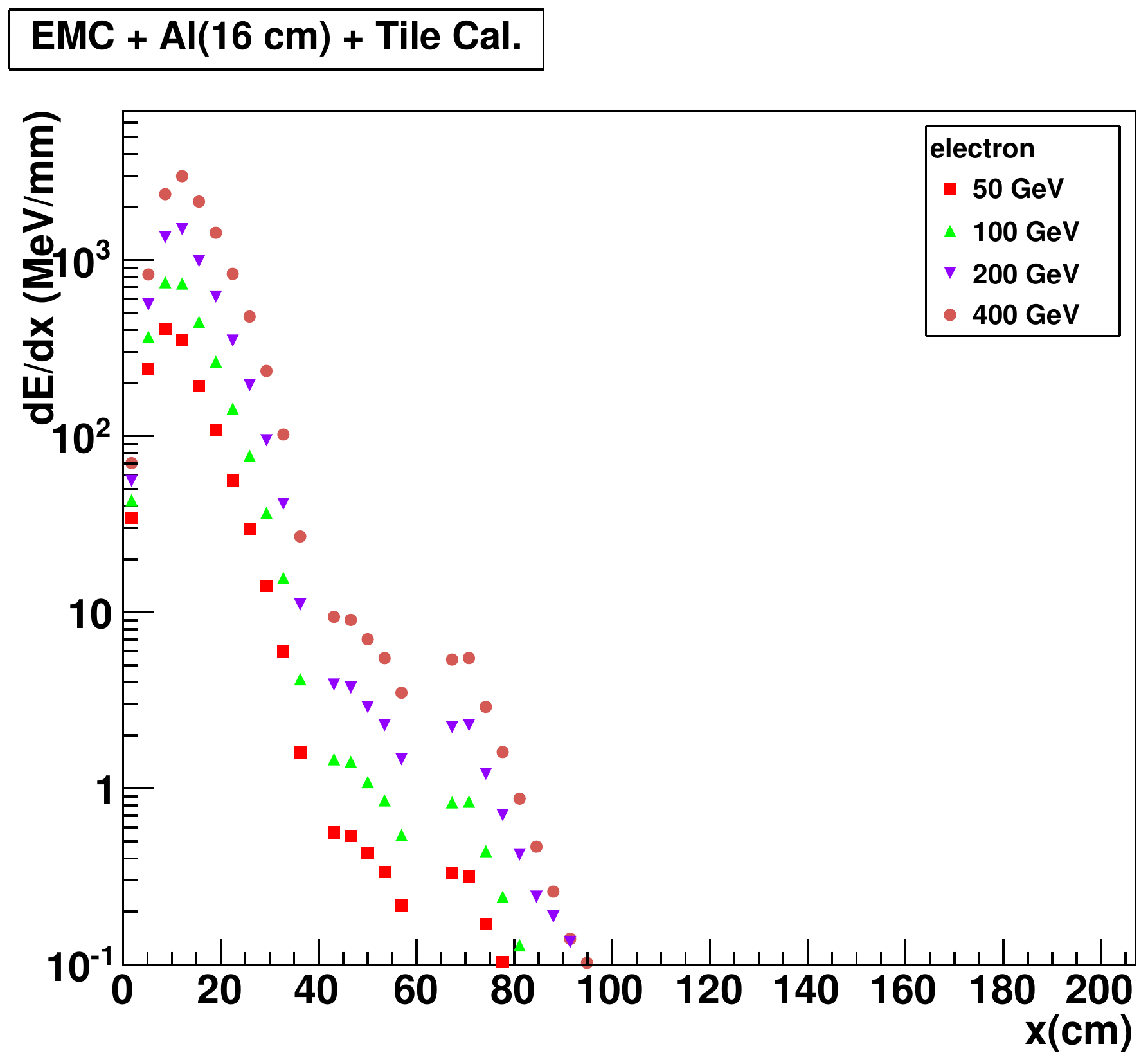}
\includegraphics[width=0.5\linewidth]{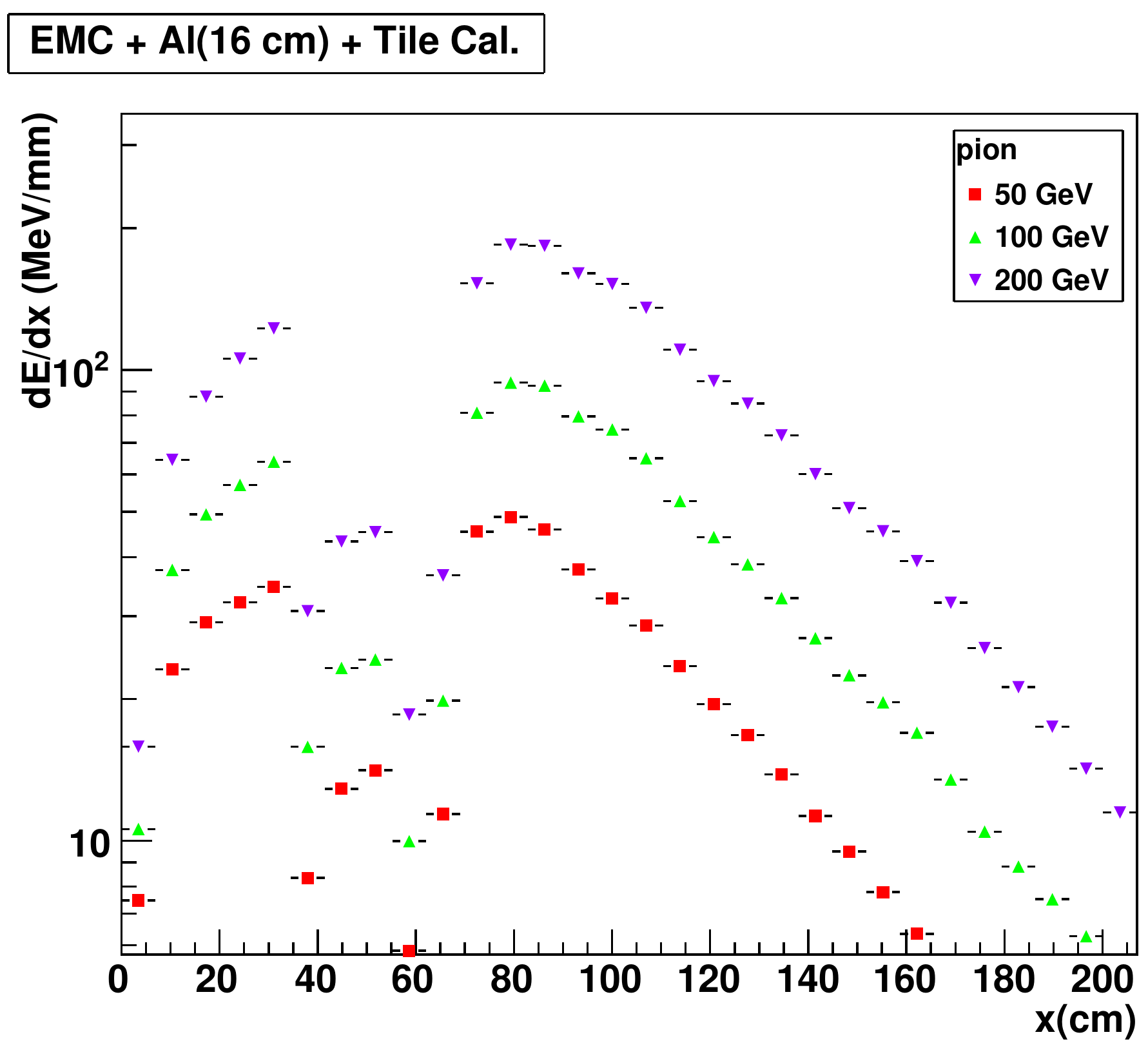}\\
\includegraphics[width=0.5\linewidth]{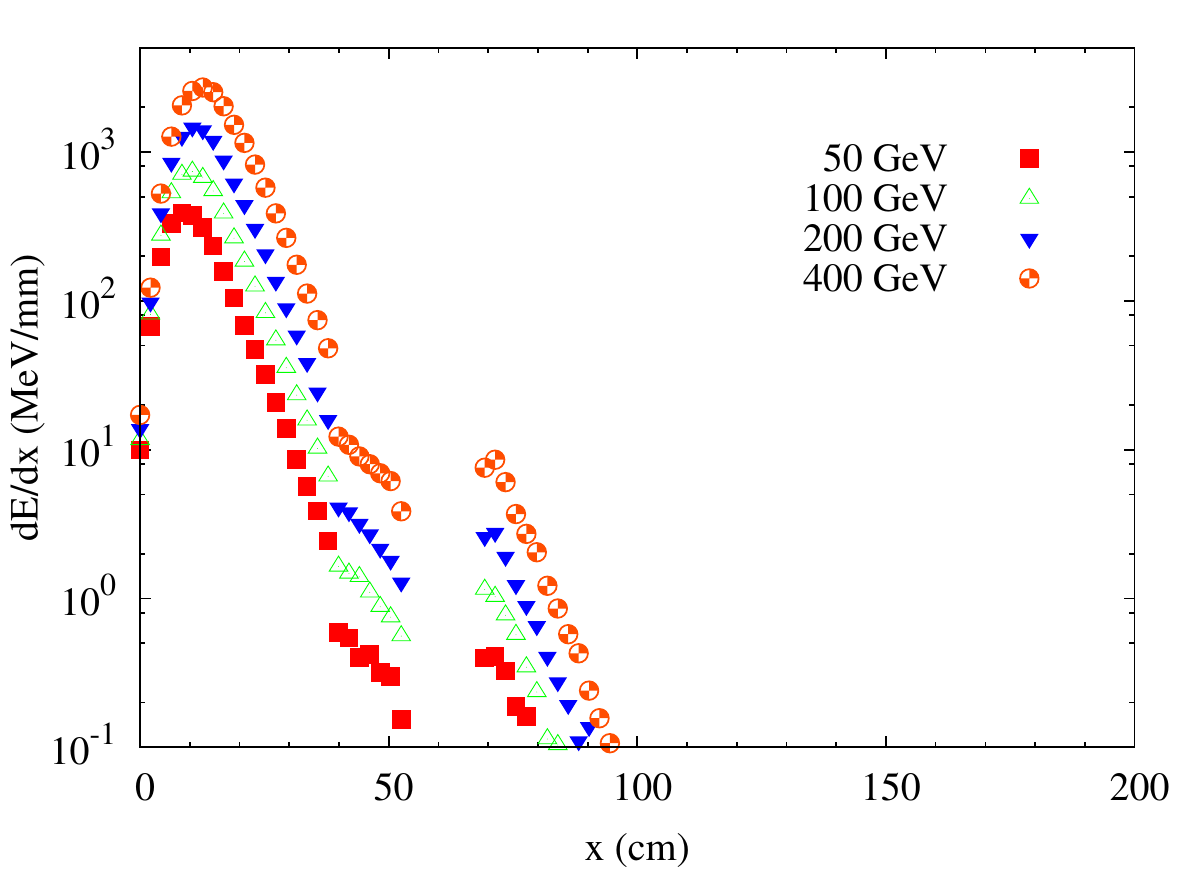}
\includegraphics[width=0.5\linewidth]{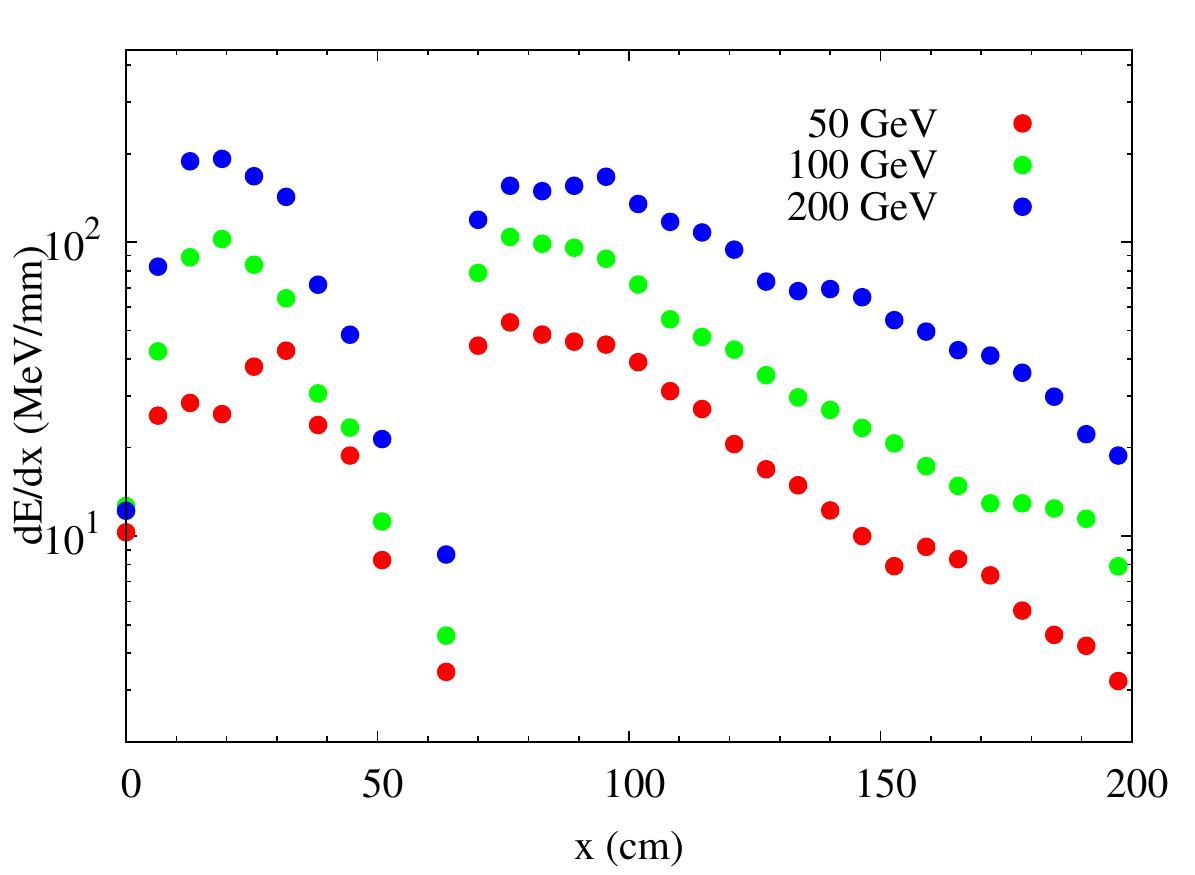}
\caption{Electron\,(left) and Pion\,(right) longitudinal shower profile for the EMC$_{Pb-Sc}$\,/\,solenoid-dipole-system (Al-block)\,/\,HAC at various energies ({\small\bf GEANT4} (top) and {\small\bf FLUKA} (bottom)).}
\label{LHEC:DET:SimCalo:Fig:10}
\end{figure}
%
\begin{figure}[htp]
\includegraphics[width=0.49\linewidth]{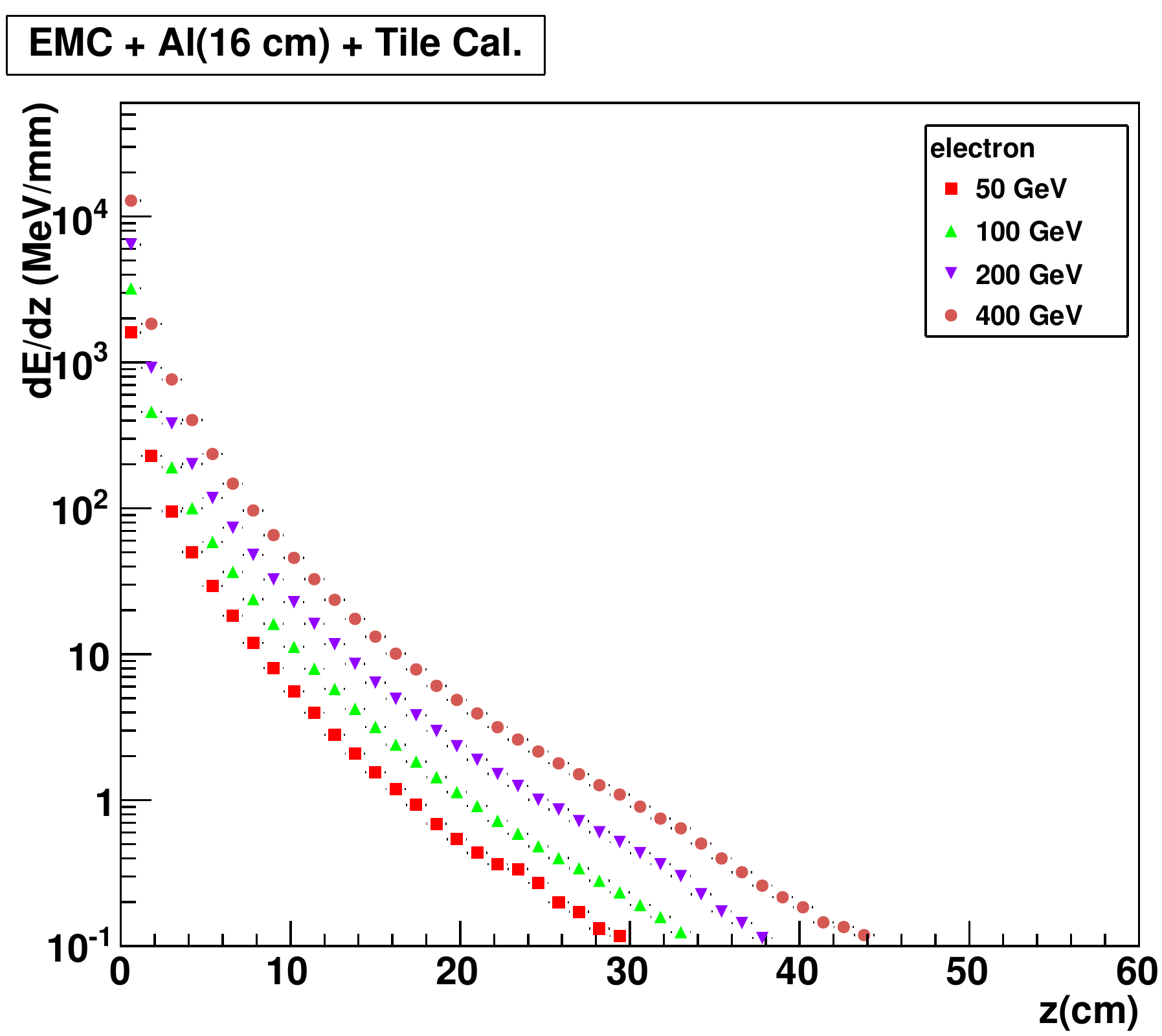}
\includegraphics[width=0.49\linewidth]{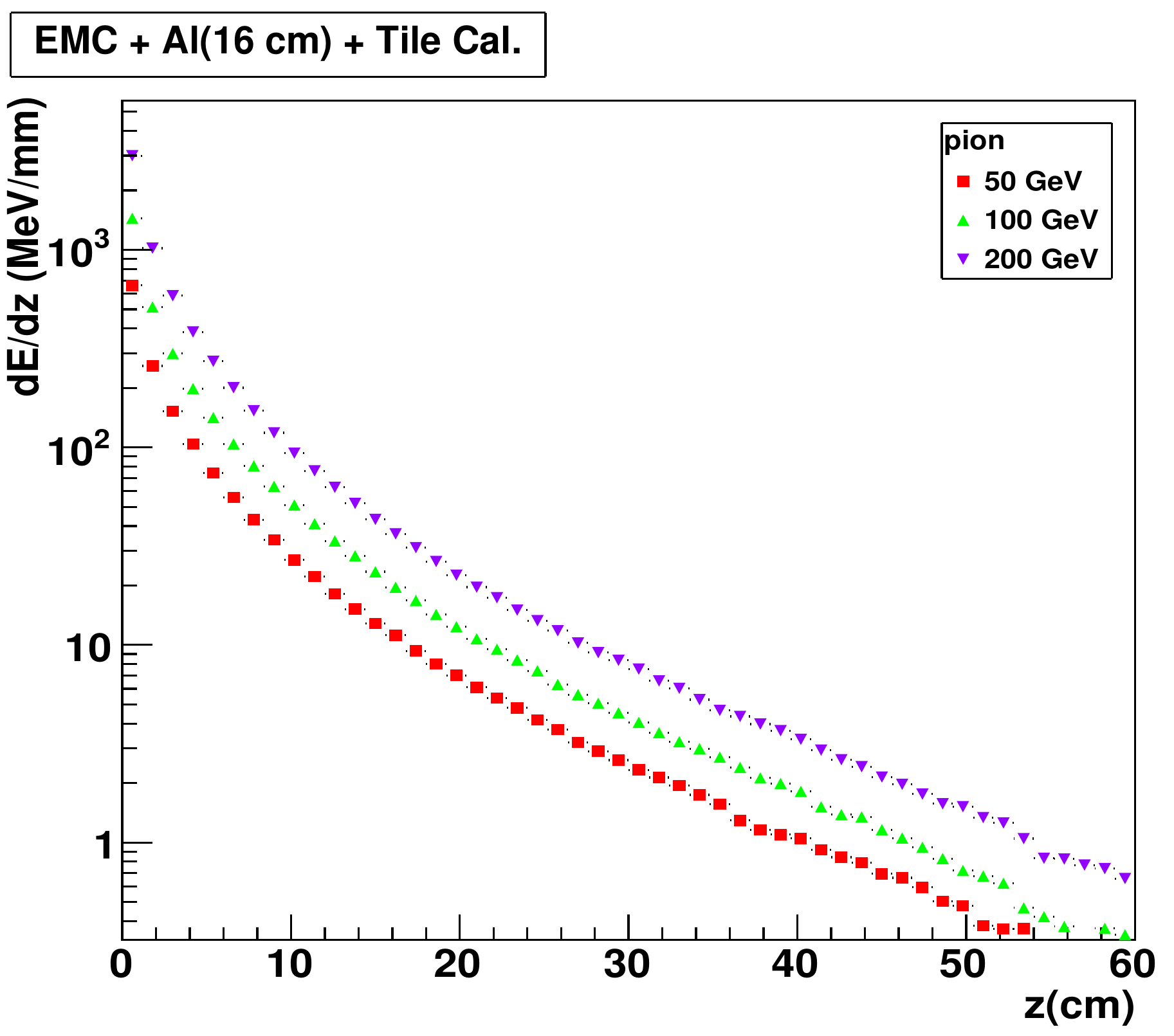}\\
\includegraphics[width=0.5\linewidth]{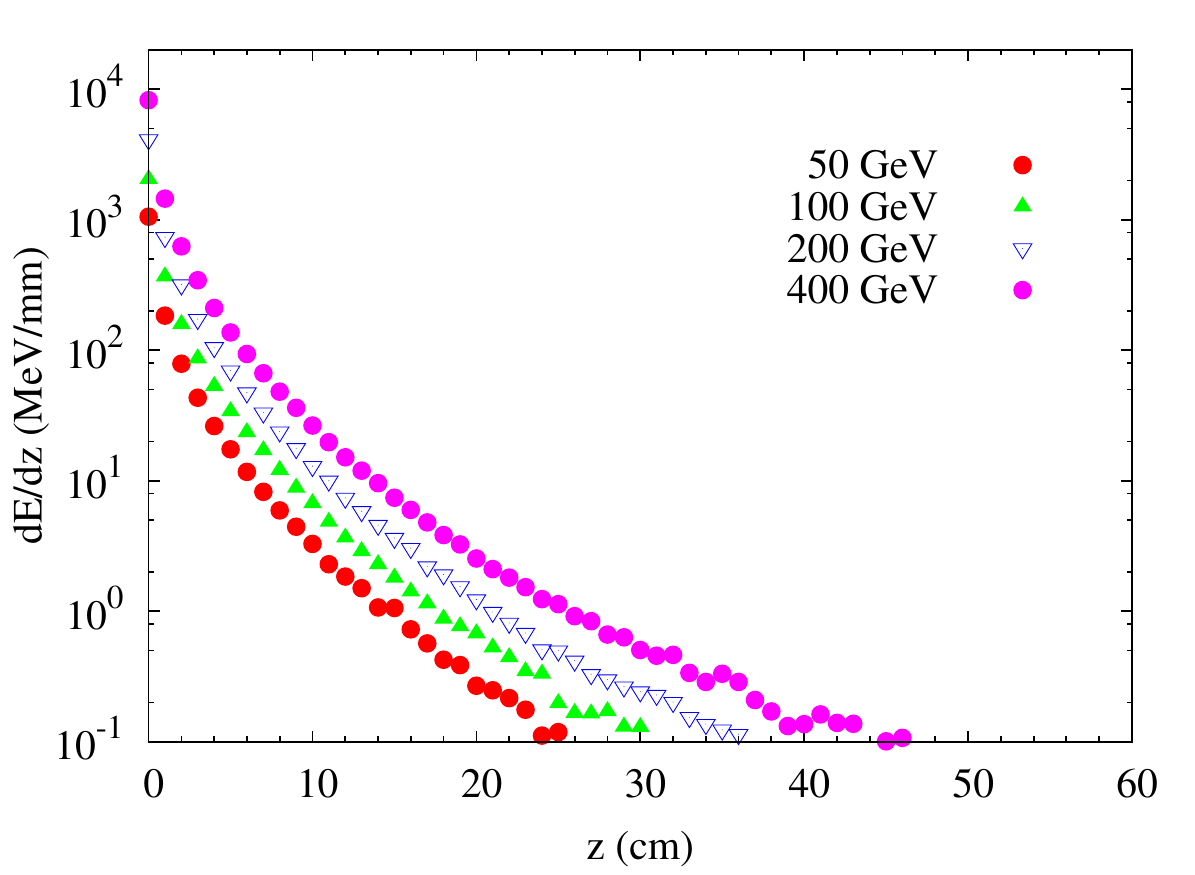}
\includegraphics[width=0.5\linewidth]{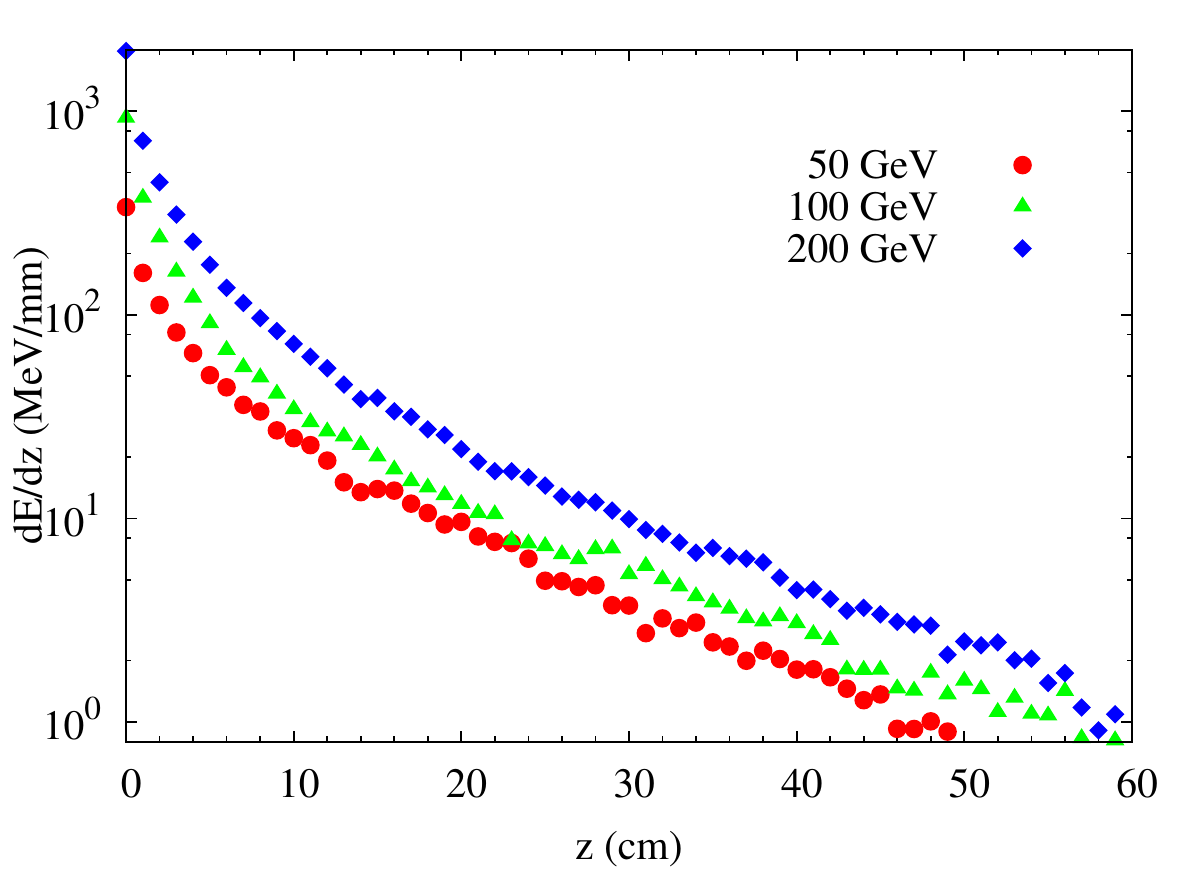}
\caption{Energy deposit and transverse shower profiles for electron\,(left) and pion\,(right) - both for the EMC$_{Pb-Sc}$ stack 
({\small\bf GEANT4} (top) and {\small\bf FLUKA} (bottom)).}
\label{LHEC:DET:SimCalo:Fig:12}
\end{figure}

As the energy loss for electrons and pions differs in shape,
normalisation and depth, it is worth looking in more detail into their
shower profiles when traversing the calorimeter.  At detector level,
this information, if available, can be used to identify and
discriminate particles and improve the energy resolution.  High
granularity, necessary to separate jets and energy deposits coming
from different sources, along with a longitudinal segmentation and
software reweighting are essential.

Longitudinal and transverse shower profiles have been studied with
electrons and pions of different energies.  The detector structure set
up here is a first approximation and uses a non projective design, but
the comparison of studies with electrons and pions entering the
calorimeter system with incident angles between 30\textdegree and
90\textdegree are of some interest for studying shower profile
properties.
The effective calorimeter depth is larger for particles with
$\theta\neq90^\circ$ (40\,cm for the EMC$_{Pb-Sc}$ and 140\,cm for the
barrel HAC in case of perpendicular impact).  The longitudinal shower
profiles for electrons and pions are summarised in
Fig.\,\ref{LHEC:DET:SimCalo:Fig:9} and
Fig.\,\ref{LHEC:DET:SimCalo:Fig:10}.  They show the mean deposited
energy as a function of the calorimeter stack depth.  The longitudinal
shower profile of electrons is shorter than for pions as expected.
The energy deposition of the electrons has its maximum in the
EMC$_{Pb-Sc}$ (Fig.\,\ref{LHEC:DET:SimCalo:Fig:9}).  The leakage into
the hadronic part of the calorimeter system is small and sums up to
$\cal{O}$(10)\,MeV.  Pions penetrate deeper into the calorimeter and
the maximum of energy deposition is seen consistently in the HAC
region (Fig.\,\ref{LHEC:DET:SimCalo:Fig:10}-right).  Less energy
deposition occurs in the region between 37 and 67\,cm because of the
aluminium layer which represents the cryostat-wall, the solenoid and
the dipole magnet structures.  Hadronic showers are completely
contained.

Transverse profiles are usually expressed as a function of the
transverse coordinates and are integrated over the longitudinal
coordinate.  Fig.\,\ref{LHEC:DET:SimCalo:Fig:12} shows the transverse
shower profiles for electrons and pions.  Since the electromagnetic
showers are compact, the electromagnetic energy is deposited
relatively close to the core of the shower. As expected the hadronic
profiles show a larger transverse spread.

\subsection{Forward and backward inserts calorimeter simulation}
 
The very important forward/backward instrumentation for calorimetric
measurements have been chosen such that, from the point of view of
performance and availability of technology, all currently known
boundary conditions could be met.  More detailed studies towards a
technical design will clarify open issues. The details of the stack
constructions are summarised in Table\,\ref{LHEC:DET:SimCalo:FB:t1}.
\begin{table}[htp]
  \centering
  \begin{tabular}{|c|r|r|r|r|r|}
    \hline
 
      \multicolumn{1}{|c|}{Calorimeter Module}
    & \multicolumn{1}{c|}{Layer} 
    & \multicolumn{1}{c|}{Absorber}
    & \multicolumn{1}{c|}{Thickness}
    & \multicolumn{1}{c|}{Instrumented Gap}
    & \multicolumn{1}{c|}{Total Depth}
    \\
 \hline\hline

  FEC$_\mathbf{(W-Si)}$ & 1-25   &     1.4\,mm   &      16\,cm &   &\\ 
  {\small\bf{30$\mathbf{X_0}$ } }   & 26-50    &    2.8\,mm  &     19.5\,cm &  5\,mm  &35.5\,cm  \\

\hline
\hline
 FHC$_\mathbf{(W-Si)}$  &1-15   & 1.2\,cm  &39\,cm   &     &\\
     {\small\bf{10$\mathbf{\lambda_I}$}}                         &16-31 &1.6\,cm  &  48\,cm  &        &\\
  & 32-46 & 3.8\,cm  & 78\,cm  &   14\,mm  &  165\,cm \\

\hline
\hline
 FHC$_\mathbf{(Cu-Si)}$  & 1-10   &   2.5\,cm  &   30\,cm  &      &\\ 
  {\small\bf{10$\mathbf{\lambda_I}$}}  &  11-20  & 5\,cm  & 55\,cm        &        & \\
   & 21-30  &  7.5\,cm  &     80\,cm  &  5\,mm  & 165\,cm  \\
\hline
\hline
   BEC$_\mathbf{(Pb-Si)}$    &1-25   &    1.8\,mm  &   17\,cm  &       &\\
  {\small\bf{25$\mathbf{X_0}$}} &26-50  &    3.8\,mm  &  22\,cm  &  5\,mm   &39\,cm  \\
\hline
\hline

  BHC$_\mathbf{(Cu-Si)}$   & 1-15    &     2.0\,cm  &  39.75\,cm &      &\\
   {\small\bf{7.9$\mathbf{\lambda_I}$}} & 16-27  &     3.5\,cm  &   49.8\,cm &       &\\
   &28-39   &     4.0\,cm  &  55.8\,cm &    6.5\,mm   &145.35\,cm \\
\hline
\end{tabular}
\caption{Layer material choice and dimension of  electromagnetic and hadronic calorimeter modules simulated.
$\mathbf{X_0}$ denotes the radiation length and $\mathbf{\lambda_I}$ the interaction length for the whole stack, respectively.
Additional to each absorber layer, layers are  placed inside the gap describing the instrumentation (support and readout, respectively): Si-sensors\,(525$\mu${m}), Si-support structures\,(FR4; 0.65\,mm) and Kapton based circuits\,(1.15\,mm).      
Constants used:  X$_0$(W)=0.3504\,cm, $\lambda_I$(W)=9.946\,cm, $\lambda_I$(Cu)=15.06\,cm and X$_0$(Pb) = 0.5612\,cm.}
\label{LHEC:DET:SimCalo:FB:t1}
\end{table}
The following options have been considered for the insert calorimeters:
\begin{itemize}
\item { 
The forward electromagnetic calorimeter (FEC) inserts (i.e. FEC1 and FEC2) are 
tungsten-silicon sampling calorimeters for compact and radiation hard stack design 
matching the tracking system towards the interaction point with high granularity.}

\item {
The forward hadronic calorimeter (FHC) inserts (i.e. FHC1, FHC2 and
FHC3) have been simulated using two different absorber materials,
Copper (\emph{Cu}) and Tungsten (\emph{W}). Using \emph{W} only would
make the forward insert calorimeters FEC\&FHC very homogeneous. The
electromagnetic and the hadronic part could be combined in the same
compartment.  On the other hand using \emph{Cu} is probably more
economical.  }

\item{
The backward electromagnetic calorimeter (BEC) inserts (i.e. BEC1 and
BEC2) are lead-silicon sampling calorimeters, with silicon as
sensitive media because of the synchrotron radiation risk,
specifically in the backward direction.
The energy of particles, predominantly the ''kinematic peak
electrons" scattered backward, is expected to be low enough such that
a smaller integrated radiation length $X_0$ is needed and the use
of \emph{Pb} as absorber material is justified.  }

\item{
The backward hadronic calorimeter (BHC) inserts (i.e. BHC1, BHC2 and
BHC3) have been setup as copper-silicon sampling calorimeters.

}
\end{itemize}
\begin{figure}[htp]
\begin{center}
\includegraphics[width=0.4\columnwidth]{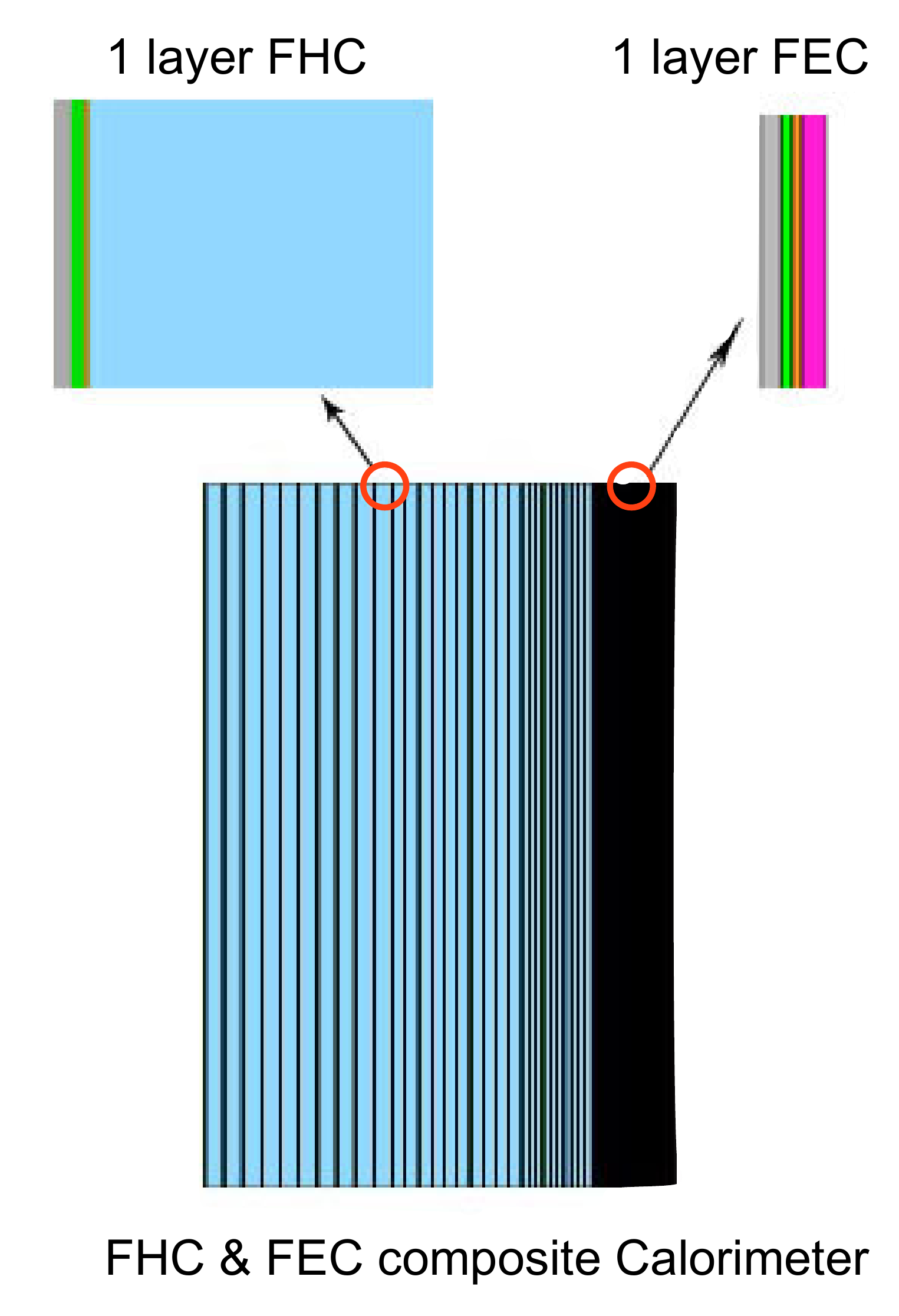}
\end{center}
\caption{Cross section in $rz$ of FEC\&FHC. Colour coding: the absorber of the FHC is in blue. The absorber of the FEC is in pink. 
The silicon detectors,  silicon support  and  kapton circuits of FEC and FHC are in brown, green and grey respectively.}
\label{LHEC:DET:SimCalo:FB:p1}
\end{figure}
The BEC, BHC and BEC$\&$BHC composite calorimeter are generally structured as their forward electromagnetic and 
hadronic calorimeter counterparts sketched in Figure\,\ref{LHEC:DET:SimCalo:FB:p1}.

%
\begin{figure}[htp]
\includegraphics[width=0.5\linewidth]{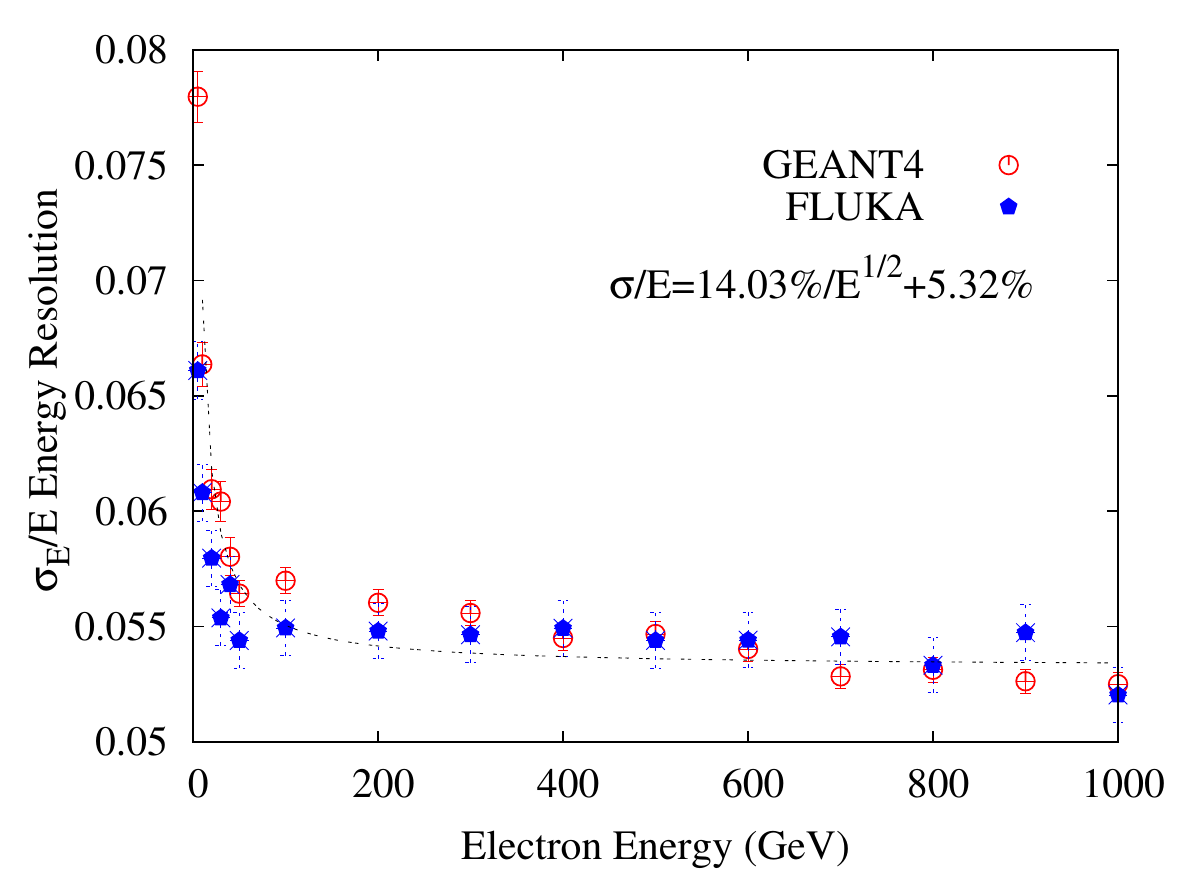}
\includegraphics[width=0.5\linewidth]{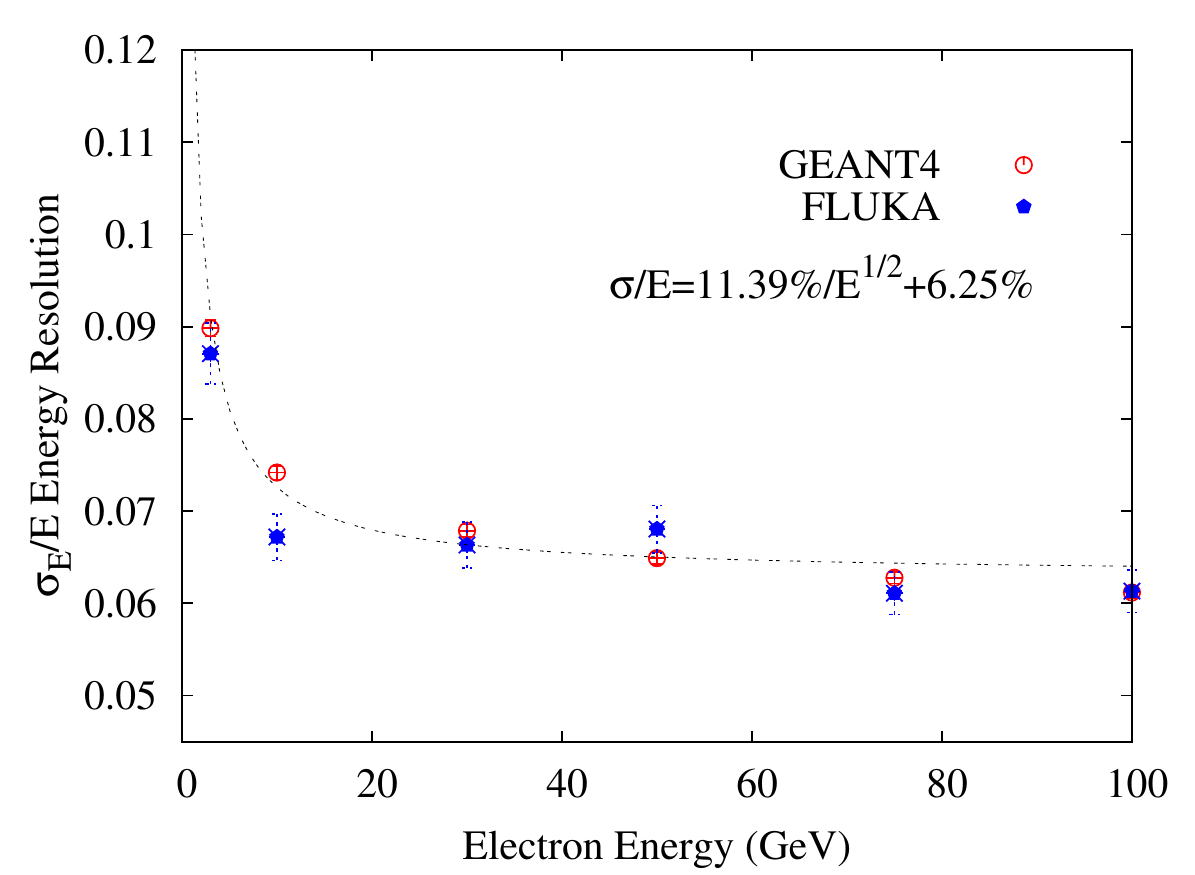}
\caption{Energy resolution spectra for electrons in the energy range 1\,GeV-1\,TeV in the FEC$_\mathbf{(W-Si)}$ (left) and for electrons (energy range 3\,GeV-100\,GeV) in the BEC$_{(Pb-Si)}$ stacks (right).}
\label{LHEC:DET:SimCalo:FB:p6}
\end{figure}
\begin{table}[htp]
  \centering
  \begin{tabular}{|c|c|c|}
    \hline 
   \multicolumn{1}{|c|}{Calorimeter Module (Composition)} 
   &\multicolumn{1}{c|}{Parameterised  Energy Resolution} 
  \\
 \hline\hline
  \multicolumn{2}{|c|}{Electromagnetic Response}   \\
\hline
 FEC$_\mathbf{(W-Si)}$ & ${\dfrac{\sigma_E}{E}=\dfrac{(14.0\pm{0.16})\%}{\sqrt{E}}\oplus(5.3\pm{0.049})\%}$ \hfill\\ 
 BEC$_\mathbf{(Pb-Si)}$    & $\dfrac{\sigma_E}{E}=\dfrac{ (11.4\pm 0.5)\%}{\sqrt{E}}\oplus{(6.3\pm{0.1})\%}$ \hfill\\
\hline
\hline
  \multicolumn{2}{|c|}{Hadronic Response}   \\
\hline
 {FEC$_\mathbf{(W-Si)}$ {\&} FHC$_\mathbf{(W-Si)}$}  &$\dfrac{\sigma_E}{E}=\dfrac{ (45.4 \pm 1.7)\%}{\sqrt{E}} \oplus { (4.8 \pm 0.086)\%}$ \hfill\\
 {FEC$_\mathbf{(W-Si)}$ {\&} FHC$_\mathbf{(Cu-Si)}$}&
$\dfrac{\sigma_E}{E}=\dfrac{(46.0\pm{1.7})\%}{\sqrt{E}}\oplus{6.1\pm{0.073})\%}$ \hfill\\ 
 {BEC$_\mathbf{(Pb-Si)}$ {\&} BHC$_\mathbf{(Cu-Si)}$}    &$\dfrac{\sigma_E}{E}=\dfrac{ (21.6 \pm 1.9)\%}{\sqrt{E}} \oplus { (9.7 \pm 0.4)\%}$ \hfill\\
\hline
\end{tabular}
\caption{Energy resolution parameterisation for electrons in the electromagnetic stacks (FEC/BEC) and for pions in the 
composite  FEC\&FHC and BEC\&BHC stack structures, respectively. For each stack structure, 
the energy range used in the fits is:\newline
\hspace*{2mm}$\bullet$ FEC$_{(W-Si)}$:  \ 1\,{GeV}-5\,{TeV} electrons, \newline
\hspace*{2mm}$\bullet$ BEC$_{(Pb-Si)}$: \ 3\,{GeV}-100\,GeV electrons, \newline
\hspace*{2mm}$\bullet$ FEC$_{(W-Si)}$ \& FHC$_{(Cu-Si)}$ and FEC$_{(W-Si)}$ \& FHC$_{(W-Si)}$: \ 50\,{GeV}-1\,{TeV} pions,\newline
\hspace*{2mm}$\bullet$ BEC$_{(Pb-Si)}$ \& BHC$_{(Cu-Si)}$: \ 3\,GeV-100\,GeV pions. \newline
The energy resolution spectra from the simulation are summarised in Figs.\,\ref{LHEC:DET:SimCalo:FB:p6} and \ref{LHEC:DET:SimCalo:FB:p15}.}
\label{LHEC:DET:SimCalo:FB:t2}
\end{table}
\begin{figure}[htp]
\includegraphics[width=0.5\columnwidth]{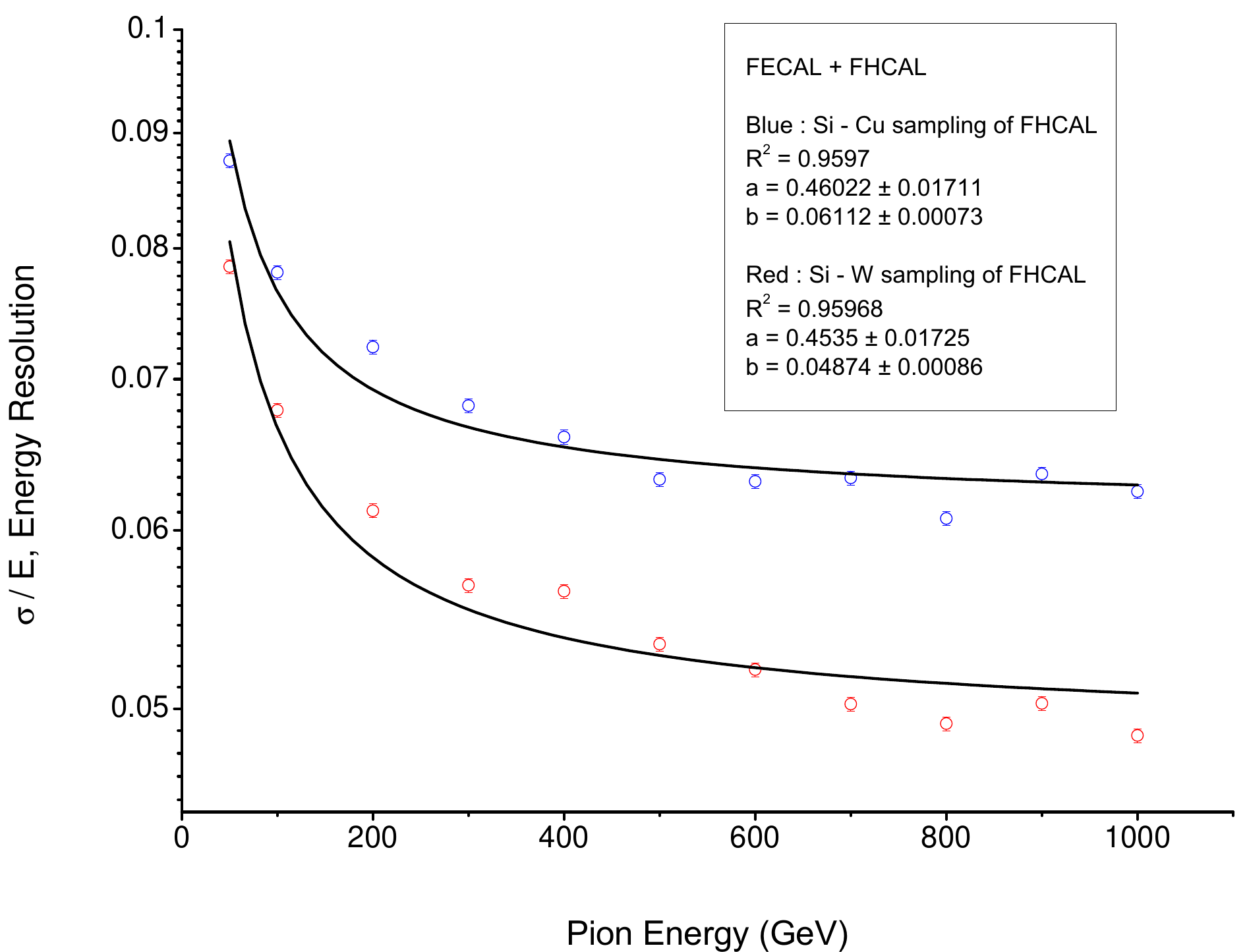}
\includegraphics[width=0.5\columnwidth]{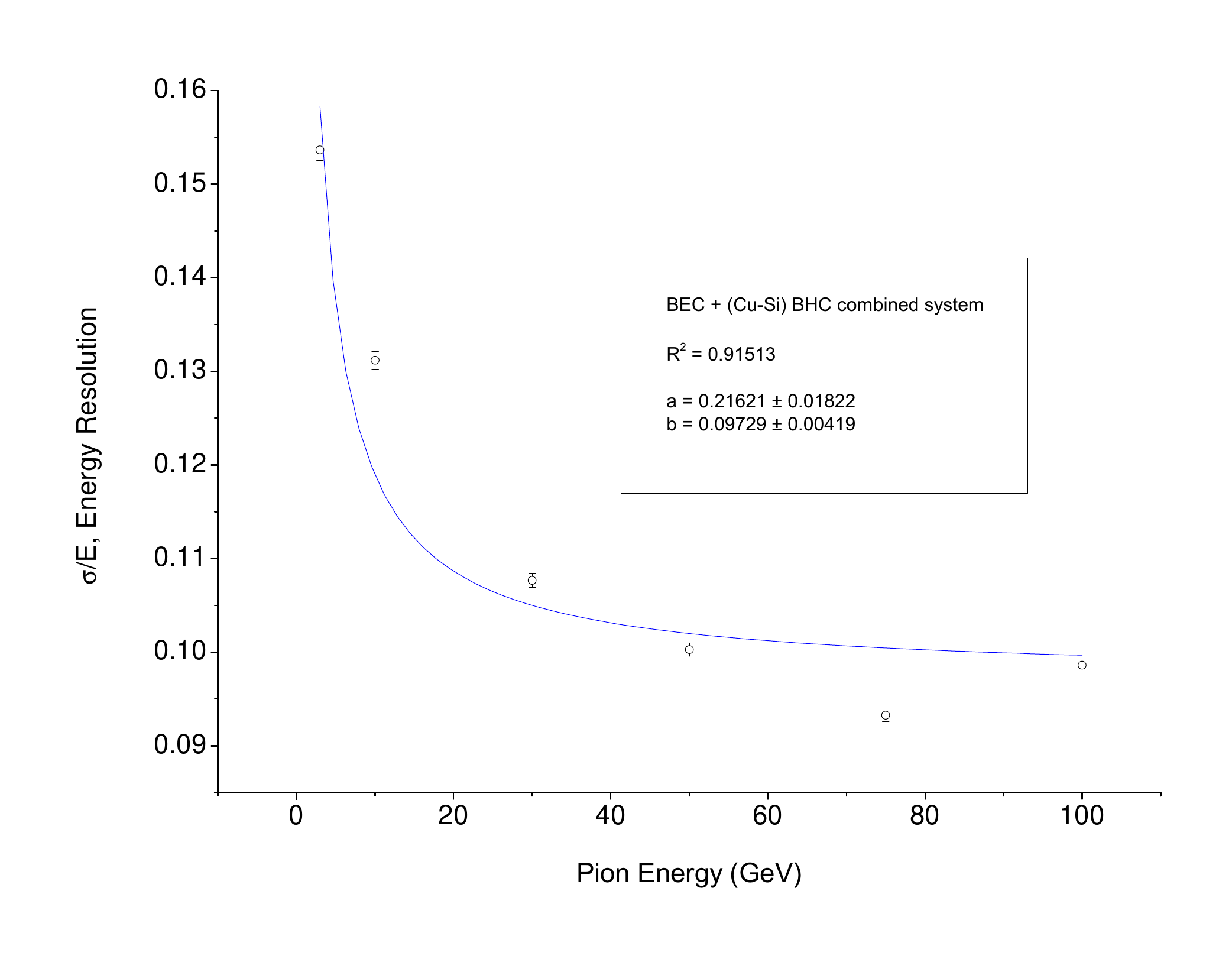}
\caption{Comparison of energy resolution spectra for 
 pions  (energy range {50}\,{GeV}-1\,{TeV})  in FEC$_{(W-Si)}$\&FHC$_{(Cu-Si)}$ and  FEC$_{(W-Si)}$\&FHC$_{(W-Si)}$ composite system, respectively (left) and energy resolution spectrum for 
 pions  (energy  range 3\,{GeV}-100\,GeV) in the BEC$_{(Pb-Si)}$\&BHC$_{(Cu-Si)}$ composite system (right) ({\small\bf GEANT4}).}
\label{LHEC:DET:SimCalo:FB:p15}
\end{figure}
\begin{figure}[htp]
\includegraphics[width=0.5\columnwidth]{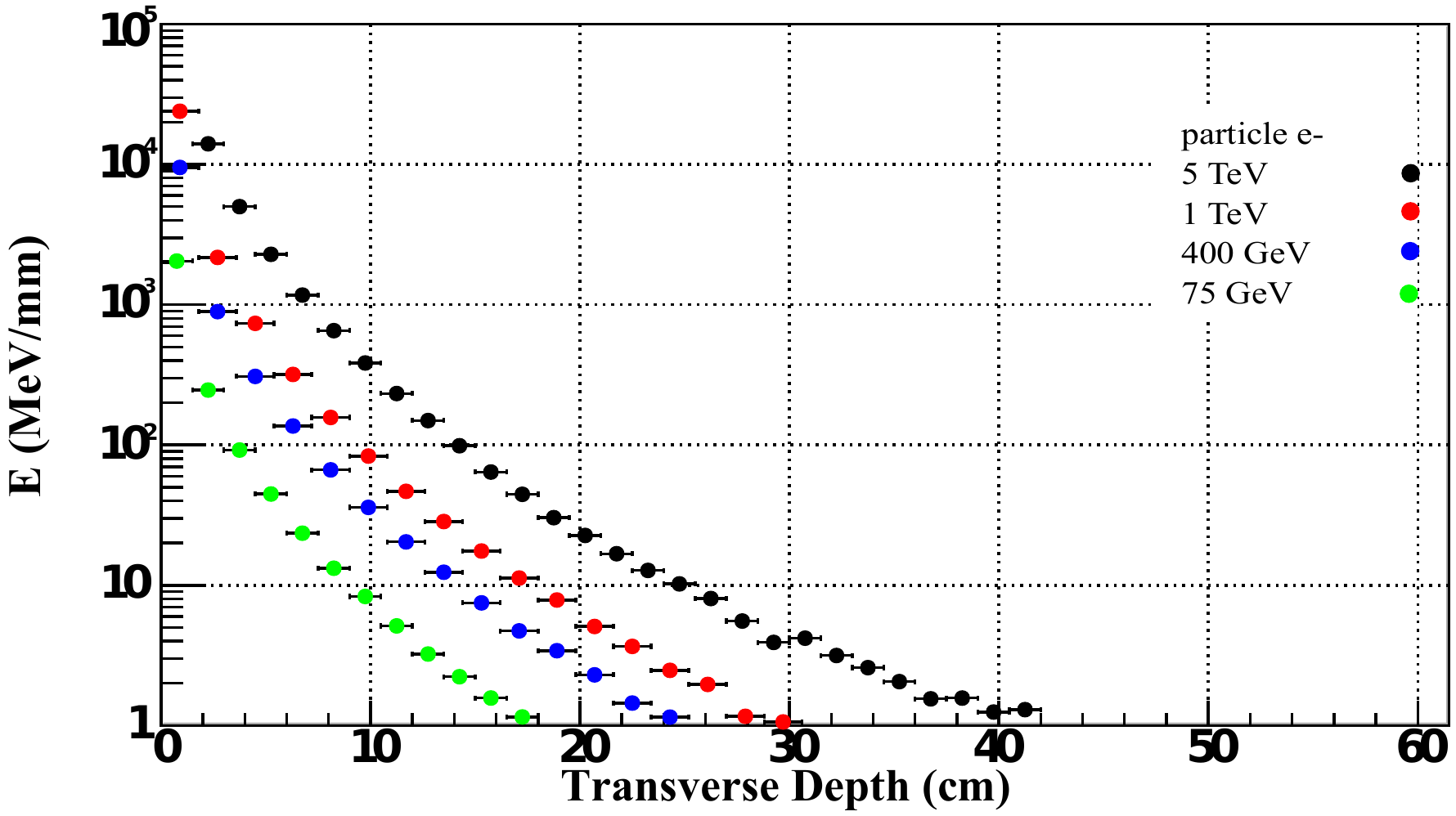}
\includegraphics[width=0.5\columnwidth]{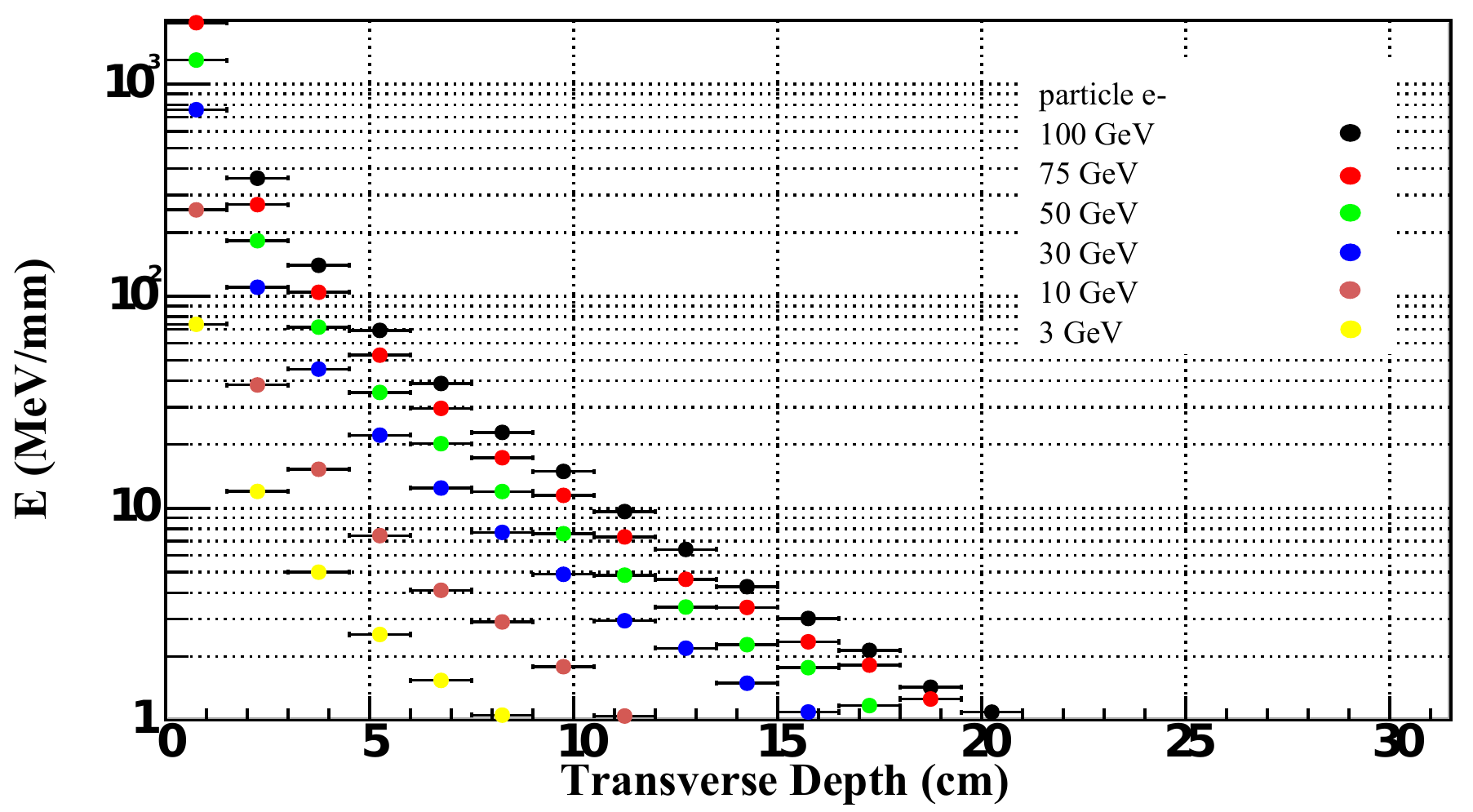}\\
\includegraphics[width=0.5\linewidth]{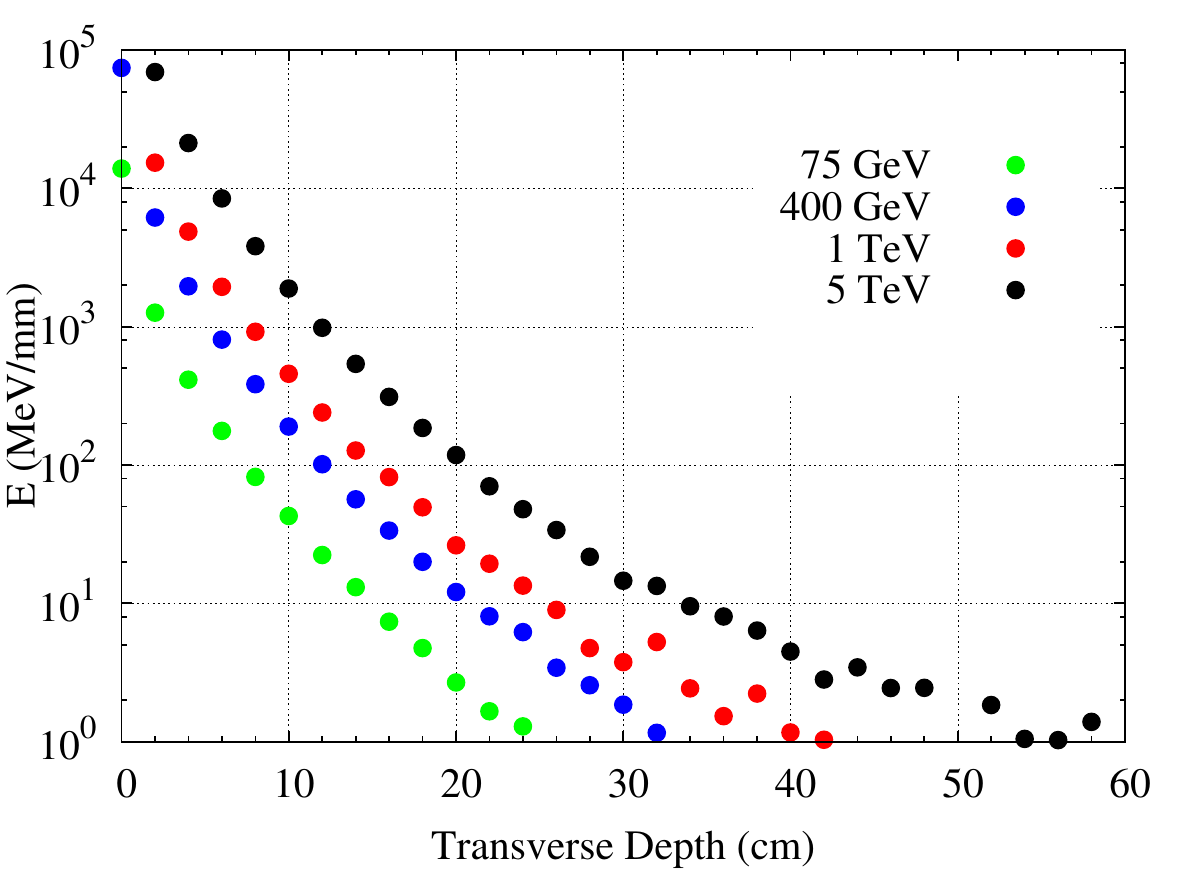}
\includegraphics[width=0.5\linewidth]{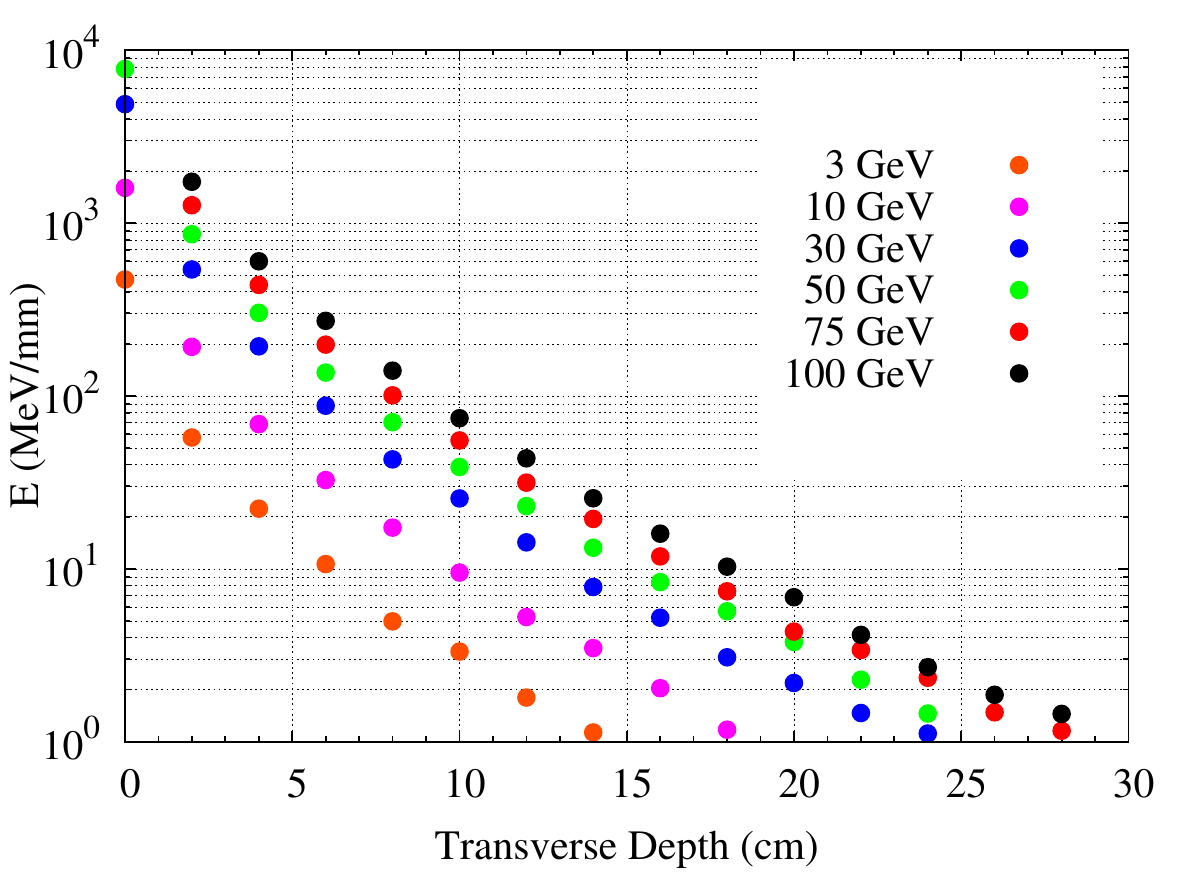}
\caption{Comparison of transverse shower profiles for electrons with energies 75\,GeV-5\,TeV on FEC$_{(W-Si)}$ (left) and     
 3\,GeV-100\,GeV on  BEC$_{(Pb-Si)}$ (right)  ({\small\bf GEANT4} (top) and {\small\bf FLUKA} (bottom)).}
\label{LHEC:DET:SimCalo:FB:p8}
\end{figure}
The lateral size of a shower is due to the multiple scattering of
electrons and positrons and characterised by the Moli\`ere radius
($\rho_M$) of the setup.  The lateral development of the
electromagnetic showers, initiated by electrons or photons, scales with
the Moli\`ere radius.  The Moli\`ere radii of tungsten and lead are
$\rho_M$=0.9327\,cm and $\rho_M$=1.602\,cm\cite{Nakamura:2010zzi},
respectively.
\protect\footnote{The Moli\`ere radius, $\rho_M$, is the radius of a cylinder containing on average 90\% of the electromagnetic shower{'}s energy deposition.}
$\rho_M$ has to be low enough to separate showers, favouring the choice of {\emph{W}} specifically for the 
construction of the forward insert calorimeters (Fig.\,\ref{LHEC:DET:SimCalo:FB:p8}).

\begin{figure}[htp]
\includegraphics[width=0.5\columnwidth]{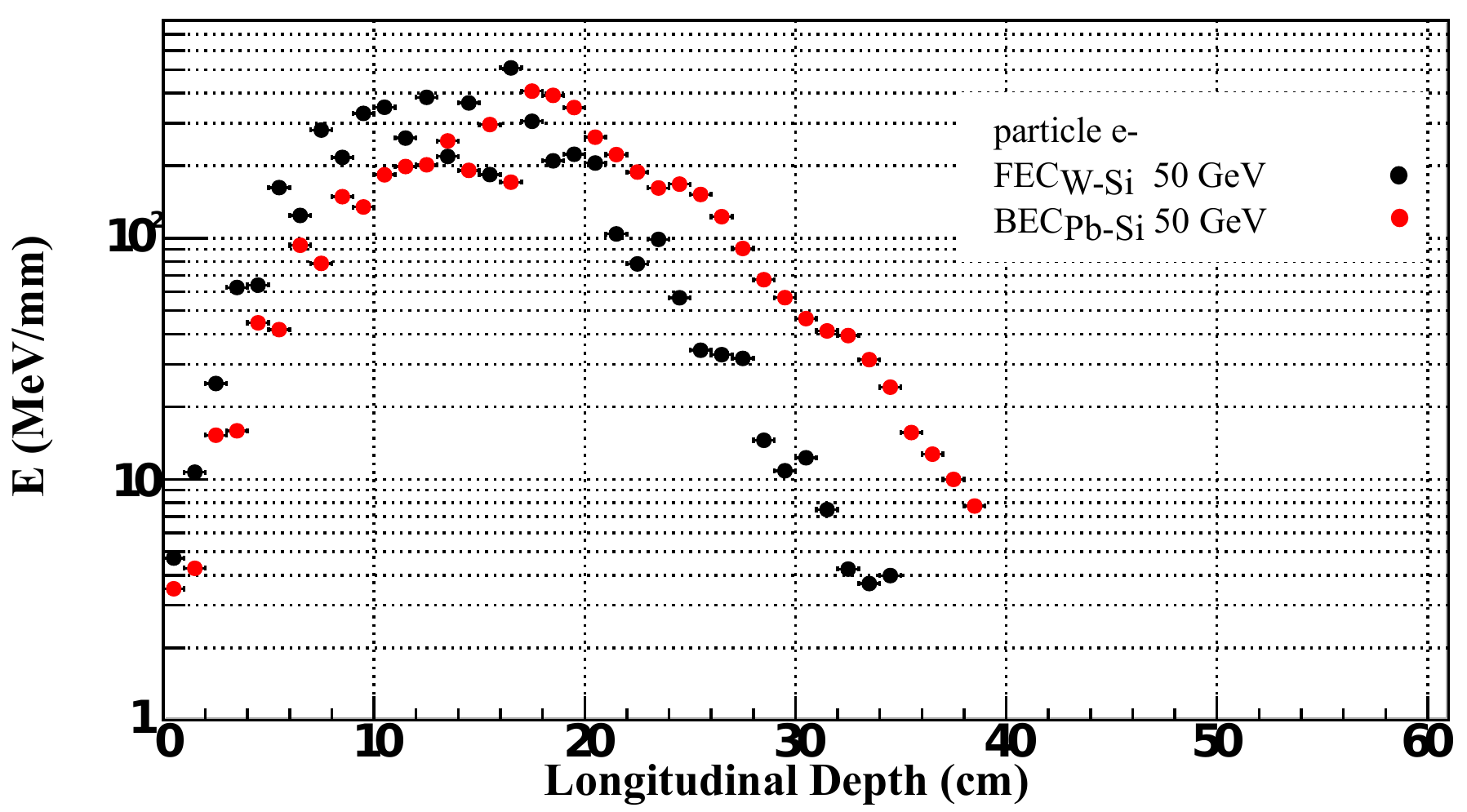}
\includegraphics[width=0.5\columnwidth]{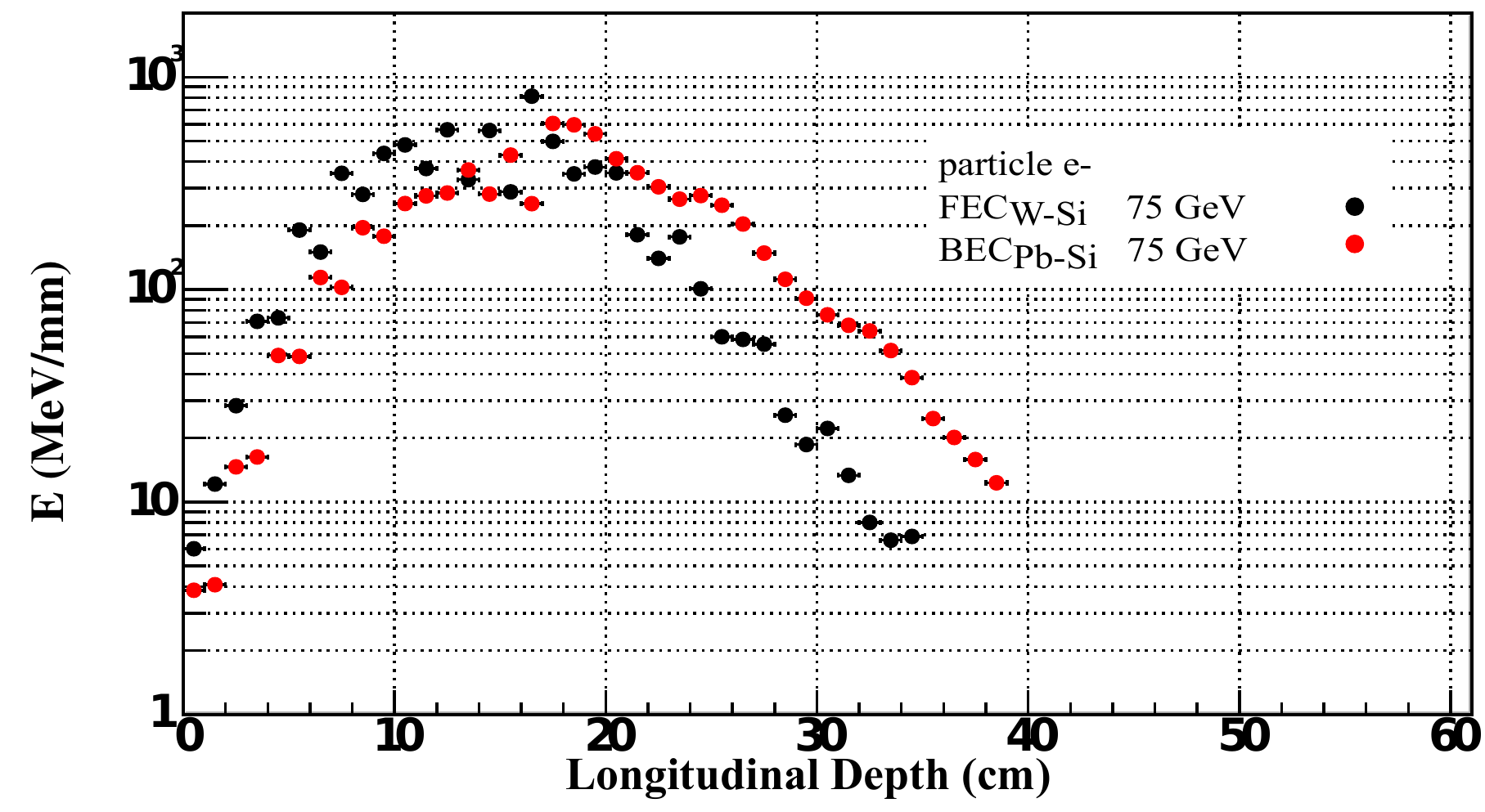}\\
\includegraphics[width=0.5\linewidth]{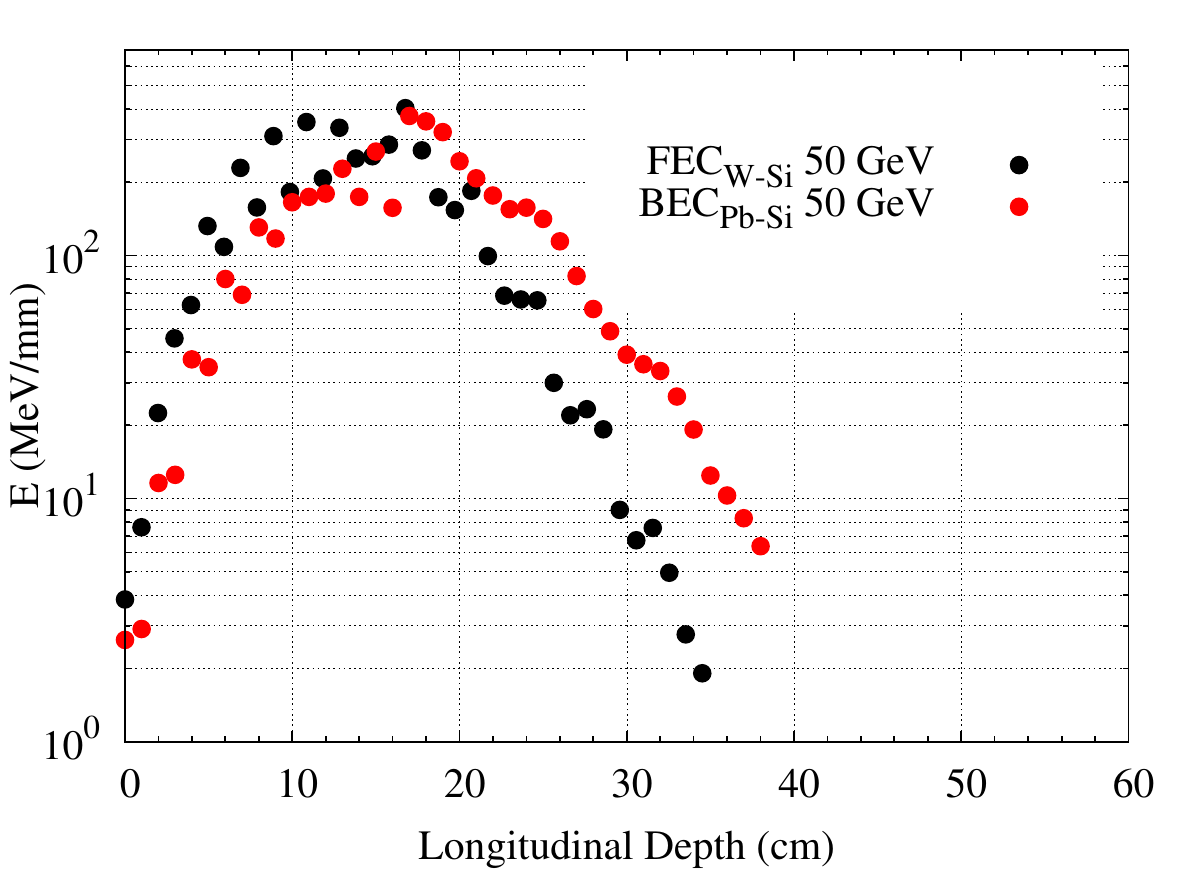}
\includegraphics[width=0.5\linewidth]{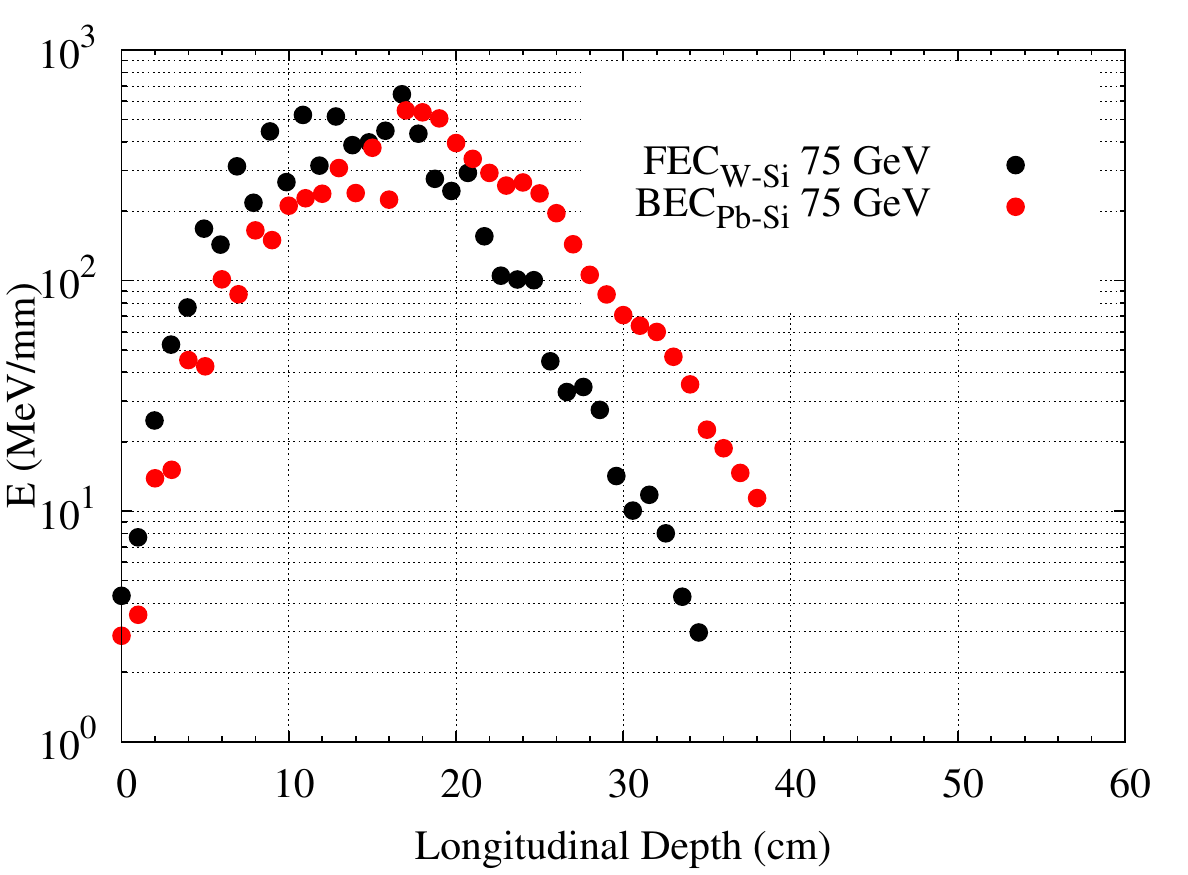}
\caption{Comparison of average energy deposition as a function of longitudinal shower extension for electrons  energies of 50\,GeV  (left) and 75\,GeV (right) in FEC$_{(W-Si)}$ (black) and  BEC$_{(Pb-Si)}$ (red) ({\small\bf GEANT4} (top) and {\small\bf FLUKA} (bottom)).}
\label{LHEC:DET:SimCalo:FB:p4}
\end{figure}
The simulated maximum longitudinal shower profiles for electrons in
the FEC and BEC (Fig\,\ref{LHEC:DET:SimCalo:FB:p4}) are in agreement
with former results\cite{Mauricio:2007}.  On average, 99.4\% and 98.8\%
of the incident energy for simulated electron energies in the range of
1\,GeV-1\,TeV for FEC$_{(W-Si)}$ and 3\,GeV-100\,GeV for
BEC$_{(Pb-Si)}$, respectively, are contained in the electromagnetic
calorimeters.  Thus the high energy electromagnetic showers are
sufficiently well contained in the {\small\bf{30$\mathbf{X_0^{FEC}}$
}} and {\small\bf{25$\mathbf{X_0^{BEC}}$}} stack construction,
respectively, taking into account the considerably lower energies
expected in the backward direction.

\begin{figure}[htp]
\includegraphics[width=0.5\columnwidth]{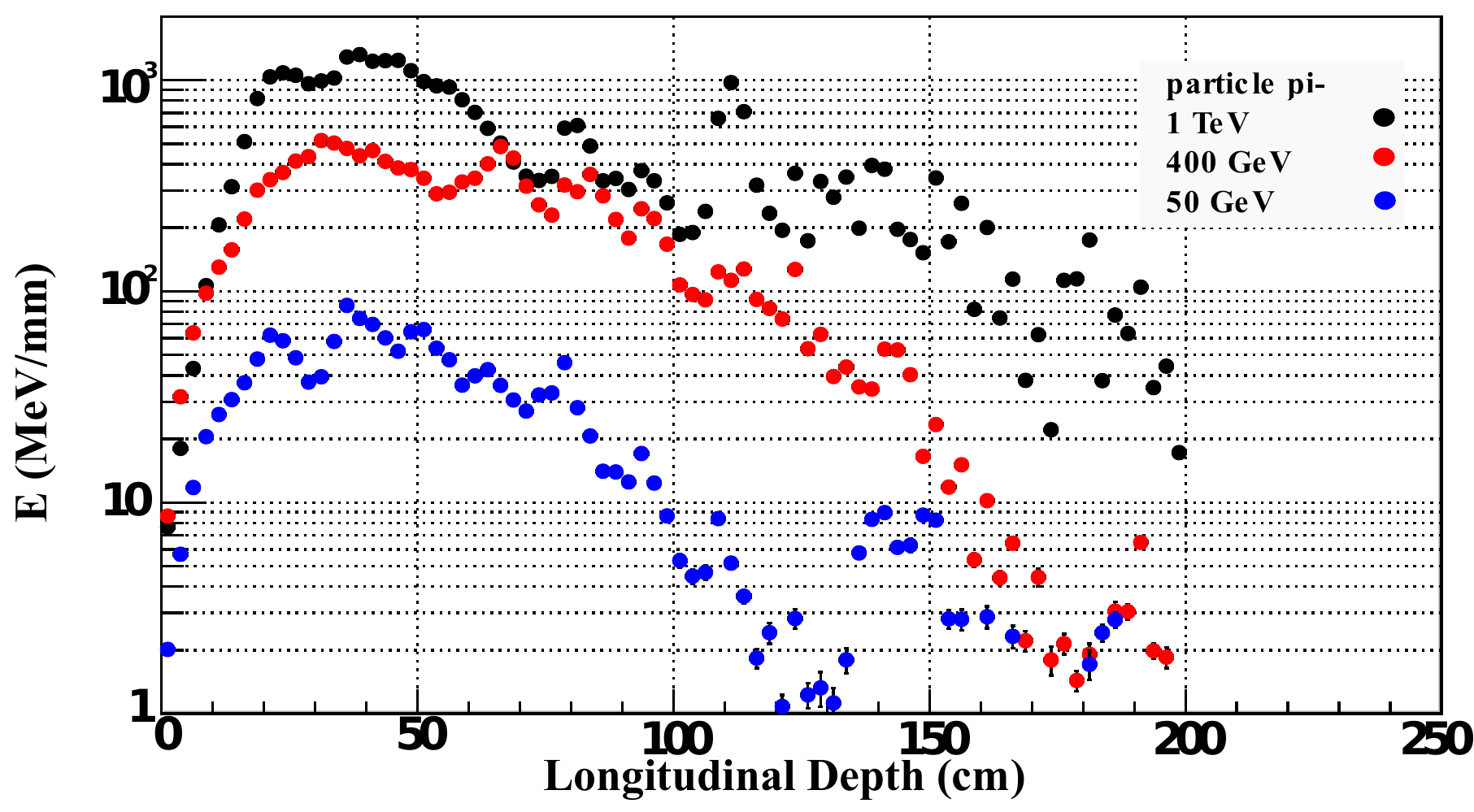}
\includegraphics[width=0.5\columnwidth]{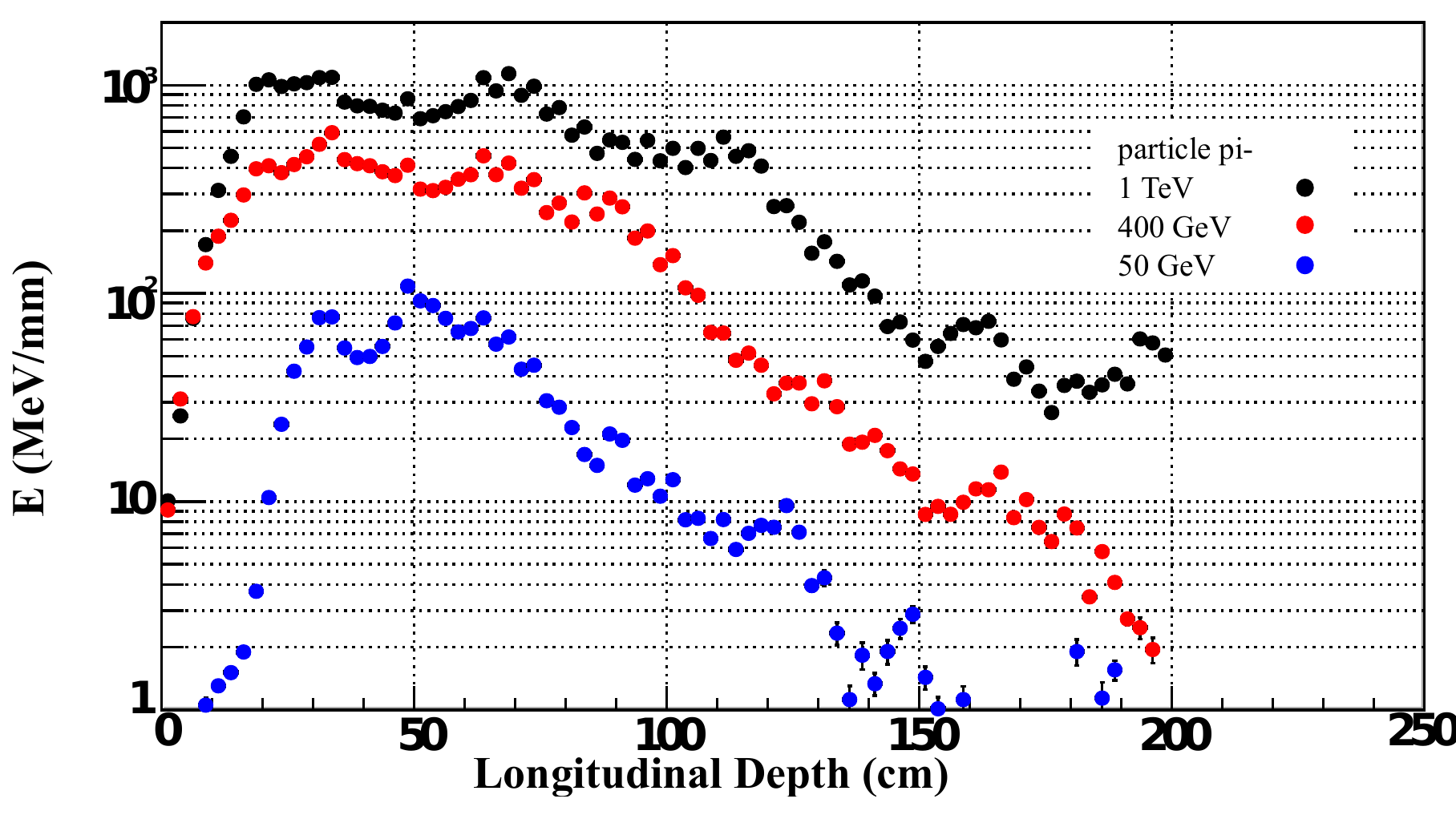}\\
\includegraphics[width=0.52\linewidth]{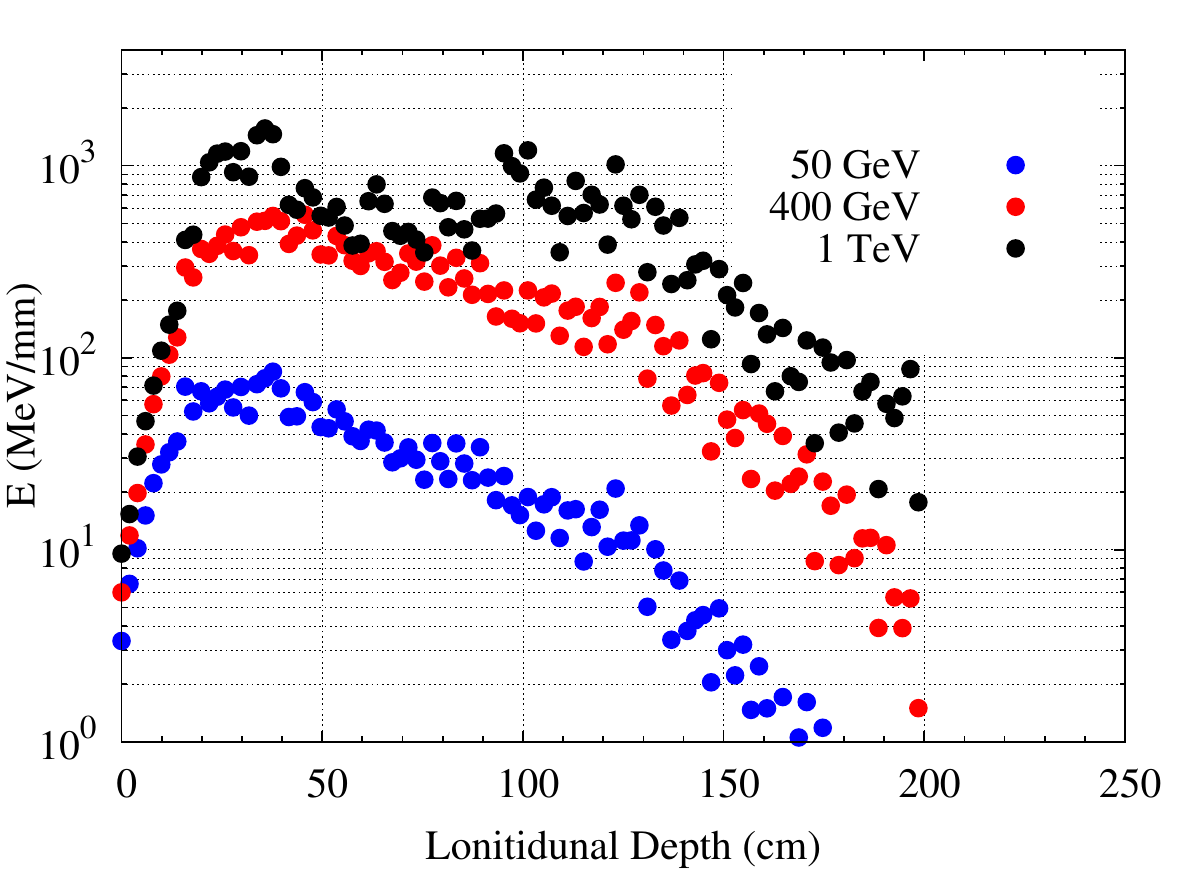}
\includegraphics[width=0.51\linewidth]{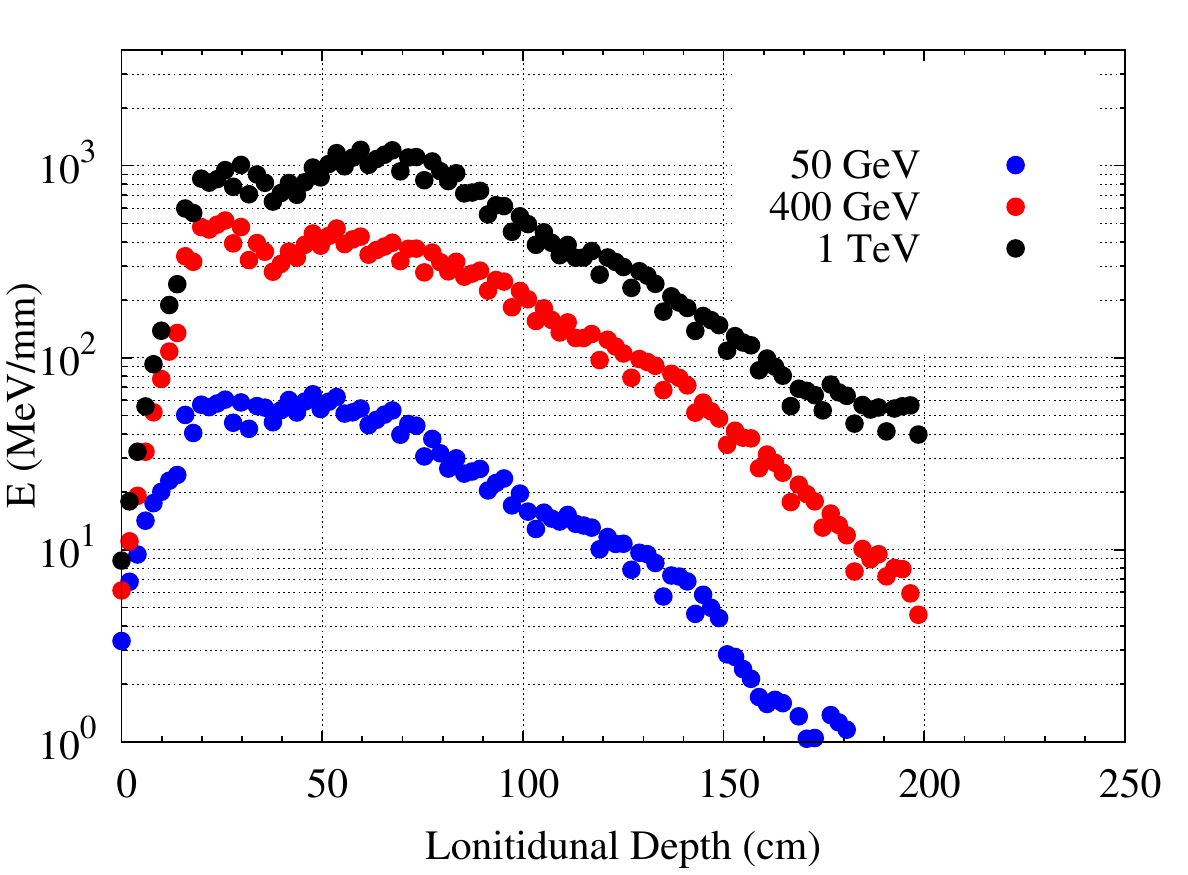}
\caption{Average energy deposition as a function of depth for pions in the energy range ${50}$\,${GeV}$-${1}$\,${TeV}$  in the FEC$_{(W-Si)}$\&FHC$_{(W-Si)}$  system (left) and in the  FEC$_{(W-Si)}$\&FHC$_{(Cu-Si)}$  composite stack system (right)
({\small\bf GEANT4} (top) and {\small\bf FLUKA} (bottom)).}
\label{LHEC:DET:SimCalo:FB:p7}
\end{figure}
\begin{figure}[htp]
\includegraphics[width=0.5\columnwidth]{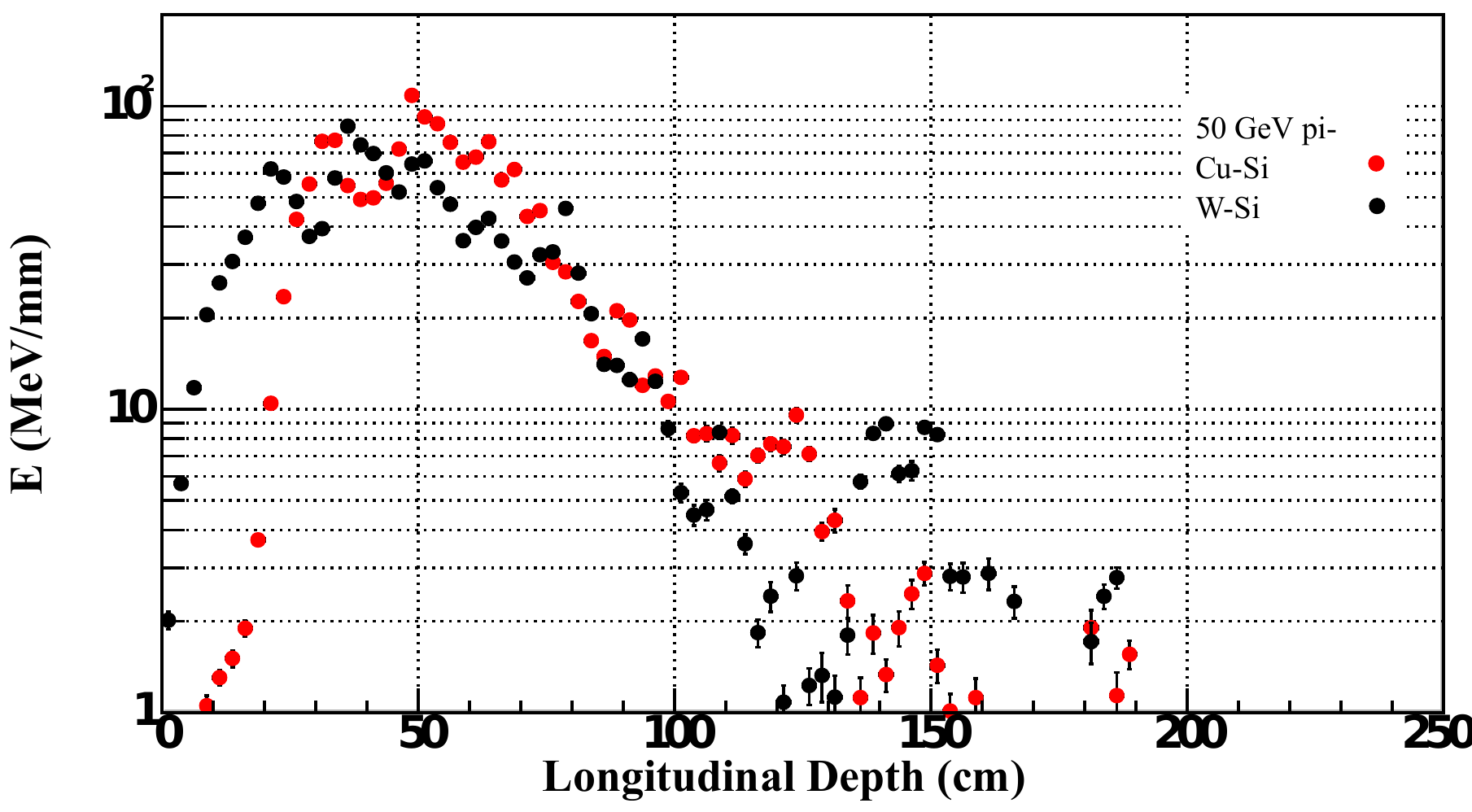}
\includegraphics[width=0.5\columnwidth]{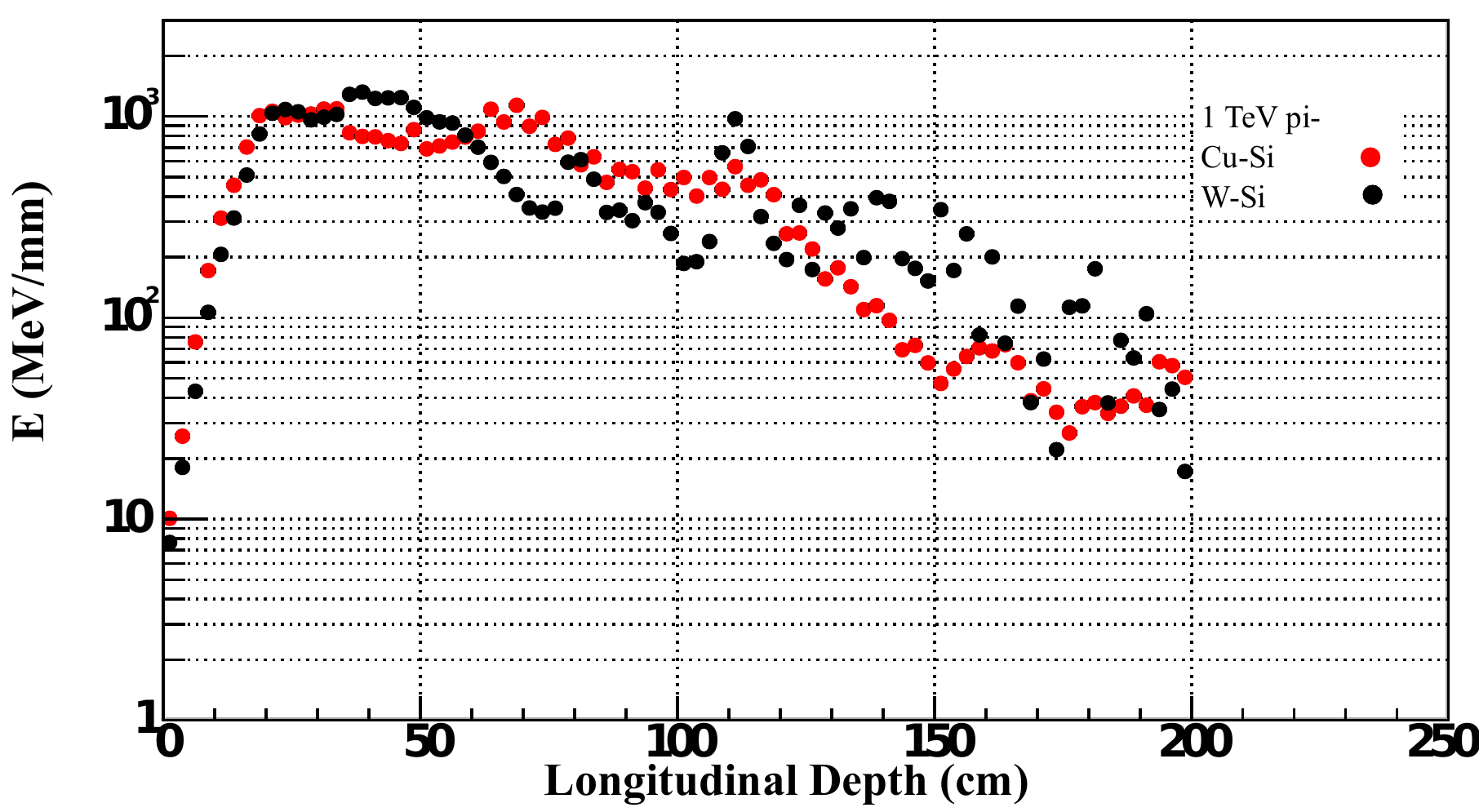}\\
\includegraphics[width=0.52\linewidth]{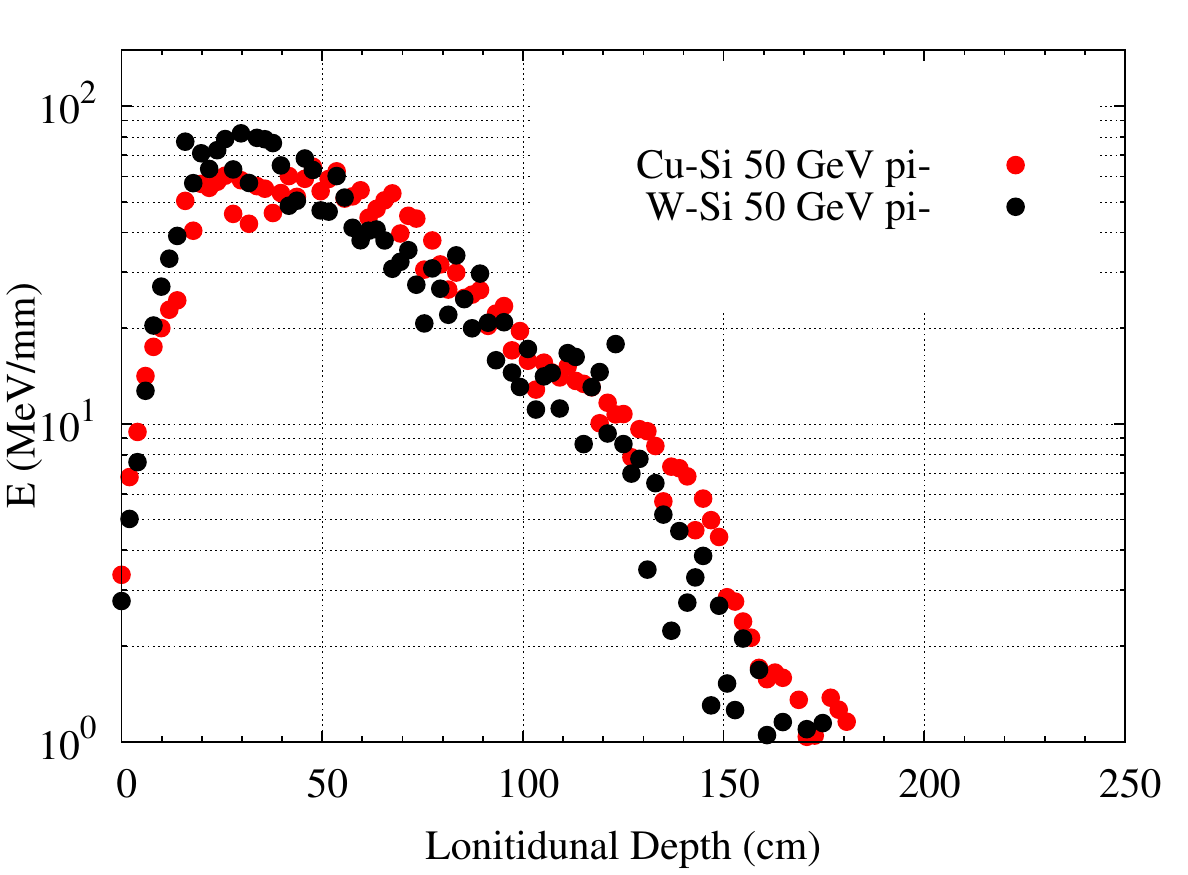}
\includegraphics[width=0.51\linewidth]{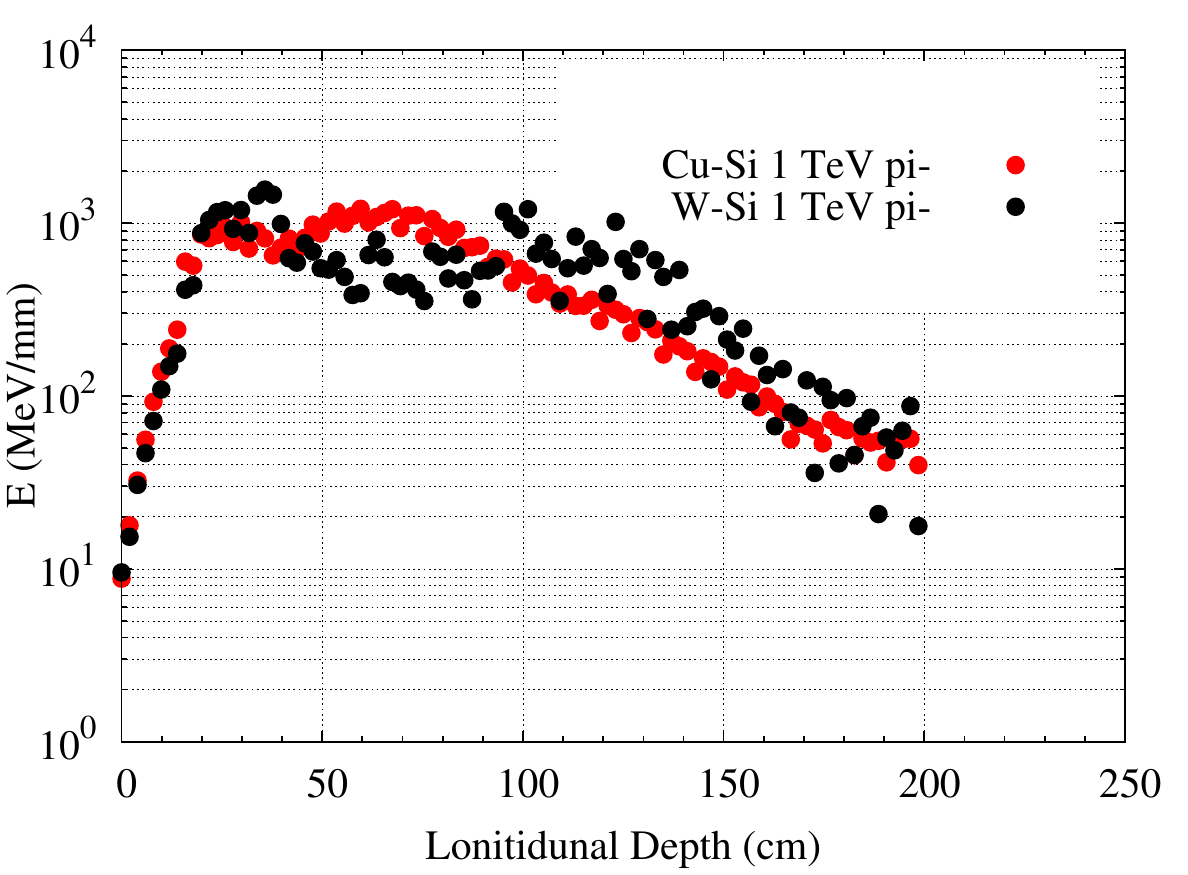}
\vspace*{-0.2cm}
\caption{Comparison of FEC$_{(W-Si)}$\&FHC$_{(Cu-Si)}$\,(red) and  FEC$_{(W-Si)}$\&FHC$_{(W-Si)}$\,(black) stack systems in terms of average energy depositions as a function of stack depth for pions of energy ${50}$\,${GeV}$ (left) and the same comparison for pions with energy 1\,TeV (right) ({\small\bf GEANT4} (top) and {\small\bf FLUKA} (bottom)).}
\label{LHEC:DET:SimCalo:FB:p11}
\end{figure}
The longitudinal distribution of the hadronic calorimeters and shower
maxima of the longitudinal distribution scales with the nuclear
interaction length $\lambda_{I}$.  For copper $\lambda_{I}$ is
$\approx$51\% larger than for tungsten.  Indeed showers in the
FHC$_{(W-Si)}$ stack (Fig.\,\ref{LHEC:DET:SimCalo:FB:p7}-left) are
observed to reach the maximum energy deposition earlier in the
calorimeter, i.e. at smaller depth.  The effect is more pronounced
for lower energy pions (Fig.\,\ref{LHEC:DET:SimCalo:FB:p11}-left).
The thickness of 10$\lambda_{I}$ provides sufficient containment of
the hadronic cascades for precision measurements both of jet
properties and of E$_T^{miss}$. The overall containment when using
FHC$_{(W-Si)}$ instead of FHC$_{(Cu-Si)}$ for the configurations
described in Tab.\,\ref{LHEC:DET:SimCalo:FB:t2} seems to be better.

Some leakage for the hadronic calorimetry (BEC$_{(Pb-Si)}$ \&
BHC$_{(Cu-Si)}$) in the backward direction has been observed.
However, the main focus in the backward direction is the analysis of
the electromagnetic component of the {\emph{e$^{\pm}$p/e$^{\pm}$A}}
scattering.  It should be mentioned that important design details
which will affect the performance of the real calorimeter are not
defined yet.  Two of these are the granularity definitions which have
to be optimised for shower separation, and the impact of dead regions
coming from cabling and the mechanical infrastructure, which
introduces unavoidable
losses\cite{Leroy:2000mj,Barbiellini:1984bp}. A detailed
simulation is needed to take that into account.

\begin{figure}[htp]
\begin{center}
\includegraphics[width=0.51\columnwidth]{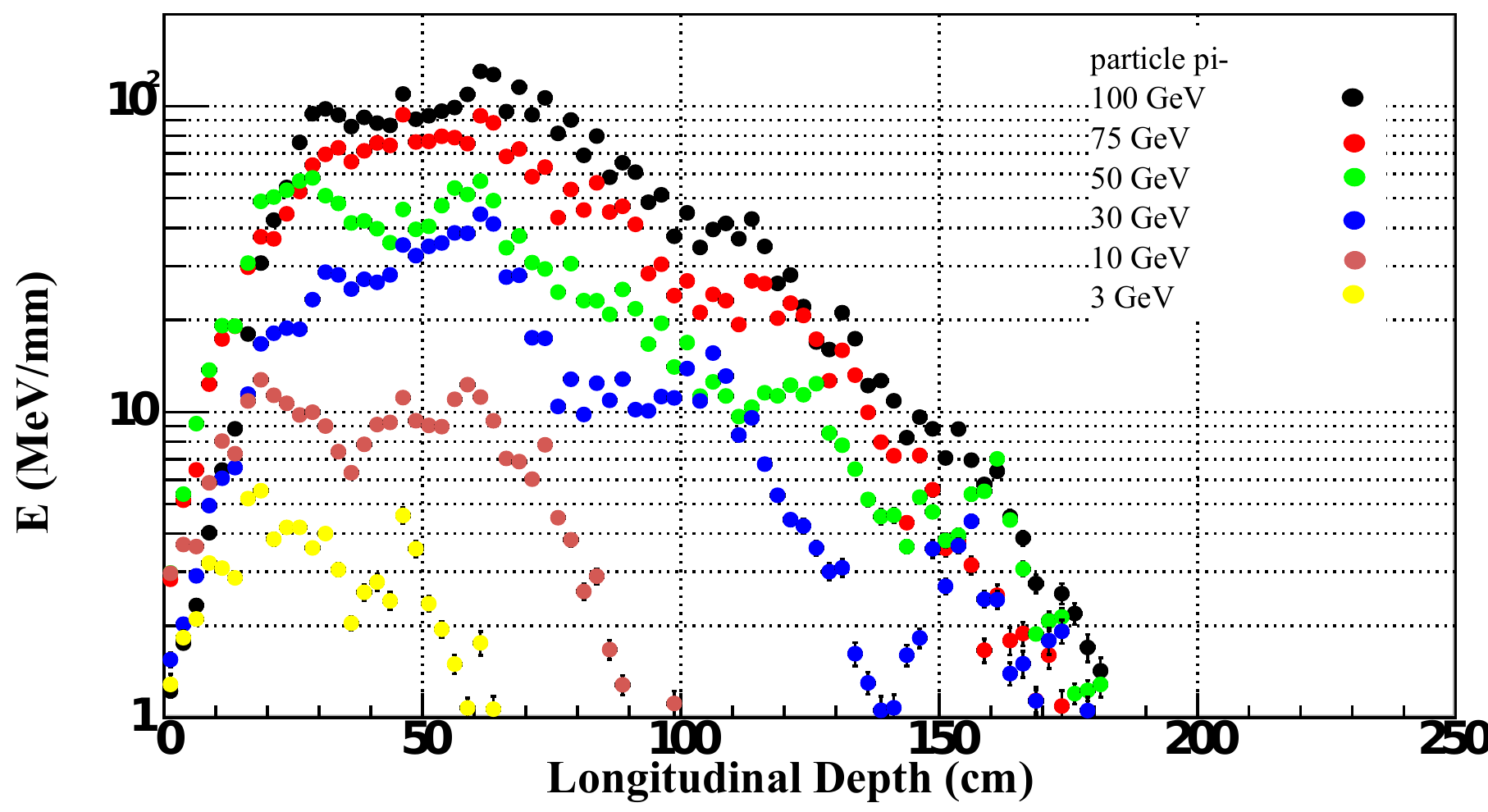}
\includegraphics[width=0.51\linewidth]{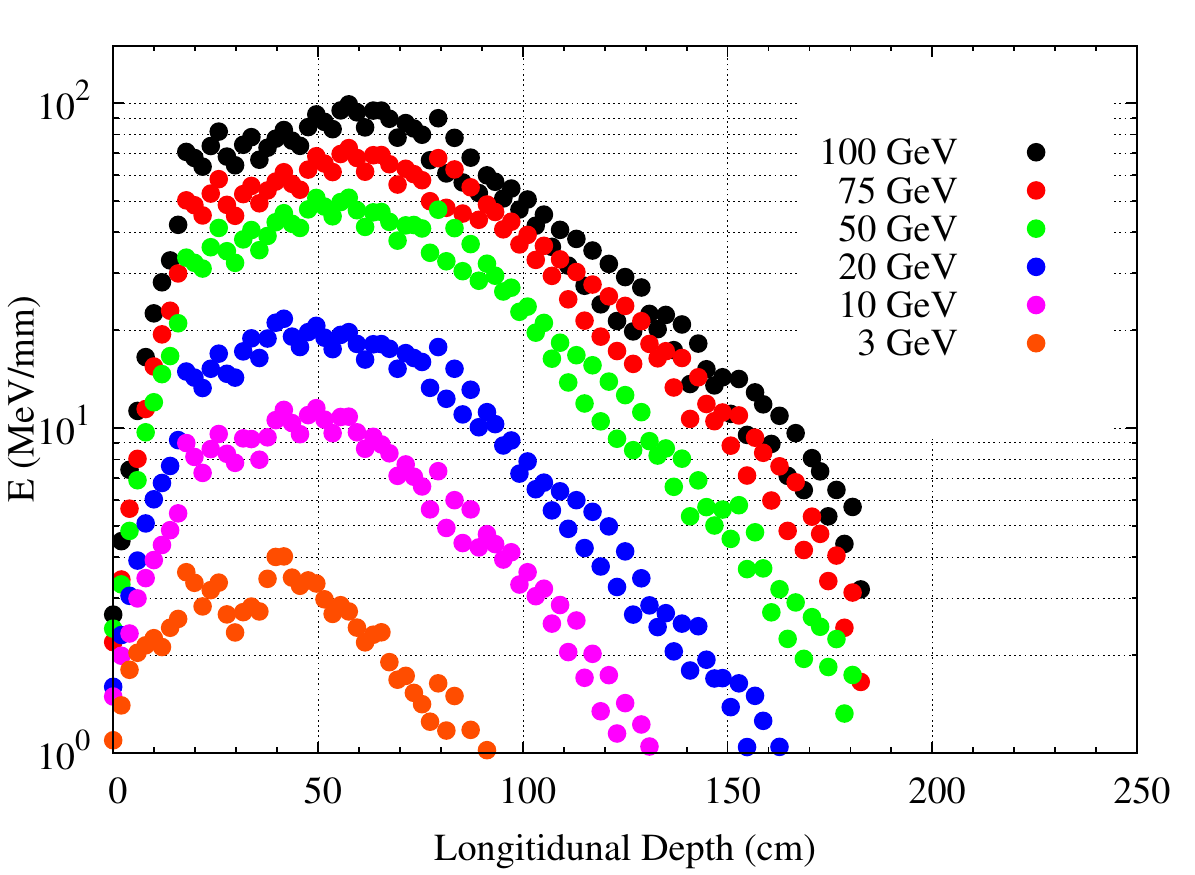}
\end{center}
\caption{Average energy deposition as a function of depth for pions in the energy range  3\,GeV-100\,GeV incident on the BEC$_{(Pb-Si)}$ \& BHC$_{(Cu-Si)}$ composite system ({\small\bf GEANT4} (top) and {\small\bf FLUKA} (bottom)).} 
\label{LHEC:DET:SimCalo:FB:p19}
\end{figure}



\section{Calorimeter summary\label{LHEC:DET:CAL:CONCLUSIONS}}

At the LHeC, several types of calorimeter are required to account for
the asymmetric interaction region and energy imbalance of the
interacting beams.  High energy jets, with energies up to a few TeV,
are expected in the forward region requiring a radiation hard design,
a high granularity and a depth of up to 10 $\lambda_I$, all in a very
compact space.  The requirements in the barrel and backward region are
less demanding.

The choice of the sampling calorimetry for all calorimeter parts is
motivated by the good experience from past experiments and the current
LHC experiments, together with considerations on the availability of
those technologies, their cost and the detector dimensions.  In the
barrel region, the need for a precise match to the tracking system and
the ability to separate multijet events pushes toward a solution which
provides a high energy linearity and a high readout granularity, as
obtained with liquid argon.  The use of a compensating calorimeter,
such as the uranium calorimeter of ZEUS, would allow a reduction of
the {\emph{e/h}} energy fluctuations and provide an absolute energy
measurement.  However, the gains are marginal and come at a
considerably higher manufacturing cost if the required granularity is
to be achieved.  Moreover, software compensation and energy-reweighting
for a linear response of the electromagnetic/hadronic calorimeter is
nowadays well established (H1/ATLAS).

{\small\bf{Particle-Flow Calorimeters}}\cite{Brient:2002gh,
Morgunov:2002pe, Magill:2007zz}, such as those presently being
designed for the future ILC, have very specific construction
requirements which at present make them unsuitable for the LHeC. Some
of these requirements are the powering scheme and the related duty
cycle which follows from the large number of channels involved, the
required cooling, the large dimensions and cost.

As previously mentioned, the design in the forward and backward
endcaps appears to be very challenging, especially at small angles. In
these regions the momentum measured by the tracking system is also
less precise due to the nearly parallel magnetic field and the higher
multiple scattering caused by an increase in the amount of material
(beam pipe and infrastructure) that the particles have to cross.  The
silicon-absorber based inserts in the forward and backward directions
will have to be compact and efficiently matched to the tracking
devices in front. In all scenarios, the projective design of the
calorimeter stack cells has to be ensured, making use of signal
weighting for good spatial resolution of the order of 1\,mm.

An alternative approach would be the implementation of the
{\small\bf{Double Readout Calorimeter}}
concept\cite{Wigmans:2009zz}\protect\footnote{using plates/fibres in
the double readout calorimeter stack for both signal components which
are radiation hard}.  The dual readout calorimeters measure each
shower twice and in two different ways.  The major
component, \emph{dE/dx} contributions of all charged particles
($e^{\pm}$,$\pi^{\pm}$,$K^{\pm}$, spallation p, recoil p, nuclear
fragments, etc.), is measured in scintillating material and the
electromagnetic part, predominantly coming from subshowers from
$\pi^0 \rightarrow {\gamma}{\gamma}$ decays, is measured by
the \v{C}erenkov light generated in clear fibres/plates as the
relativistic $e^{\pm}$ pass through\cite{Hauptman:2011zza}.  Making
use of the constant ratio of \emph{(e/h)$_{\check{C}}$}
(for \v{C}erenkov light emitting material) and \emph{(e/h)$_{S}$} (for
Scintillation light emitting material), respectively, the energy
response of the calorimeter to electrons\,\emph{e} and to
hadrons\,\emph{h} at all energies can be controlled by construction
with convincing results\cite{Wigmans:2011,Hauptman:2011zza}.

The preliminary simulations and the results shown here indicate the
validity of the proposed design concept as a baseline solution for the
given requirements of the LHeC detector. The results of {\small\bf
GEANT4} and {\small\bf FLUKA} simulations are comparable.  A more
elaborate design will be possible as soon as decisions on the
accelerator concept and therefore magnet design have been taken.


%% file: detector/muon.tex
\label{LHEC:MainDetector:DetMuon}

Muon detection is an important aspect of the physics program covered
by the LHeC.  The muon detector can improve the scope and the spectrum
of many measurements, of which only a few are listed here:
\begin{itemize}
\item Higgs decay, leptoquarks, lepton flavour violation 
\item PDF fits from semi-leptonic decay of hadrons and heavy flavours.
\item Vector meson production
\end{itemize}

The penetrative power of muons requires several layers of muon
chambers ensuring good tracking resolution and hermetic coverage, in
particular towards small angles in the forward and backward regions.
These regions, which are particularly challenging for the central
tracking detector due to the accelerator infrastructure, are more
accessible at larger distance from the interaction region if the
particles in question are minimum ionising, which muons are.

\begin{figure}[htp]
\begin{center}
\vspace*{-0.3cm}  
\includegraphics[width=0.5\columnwidth]{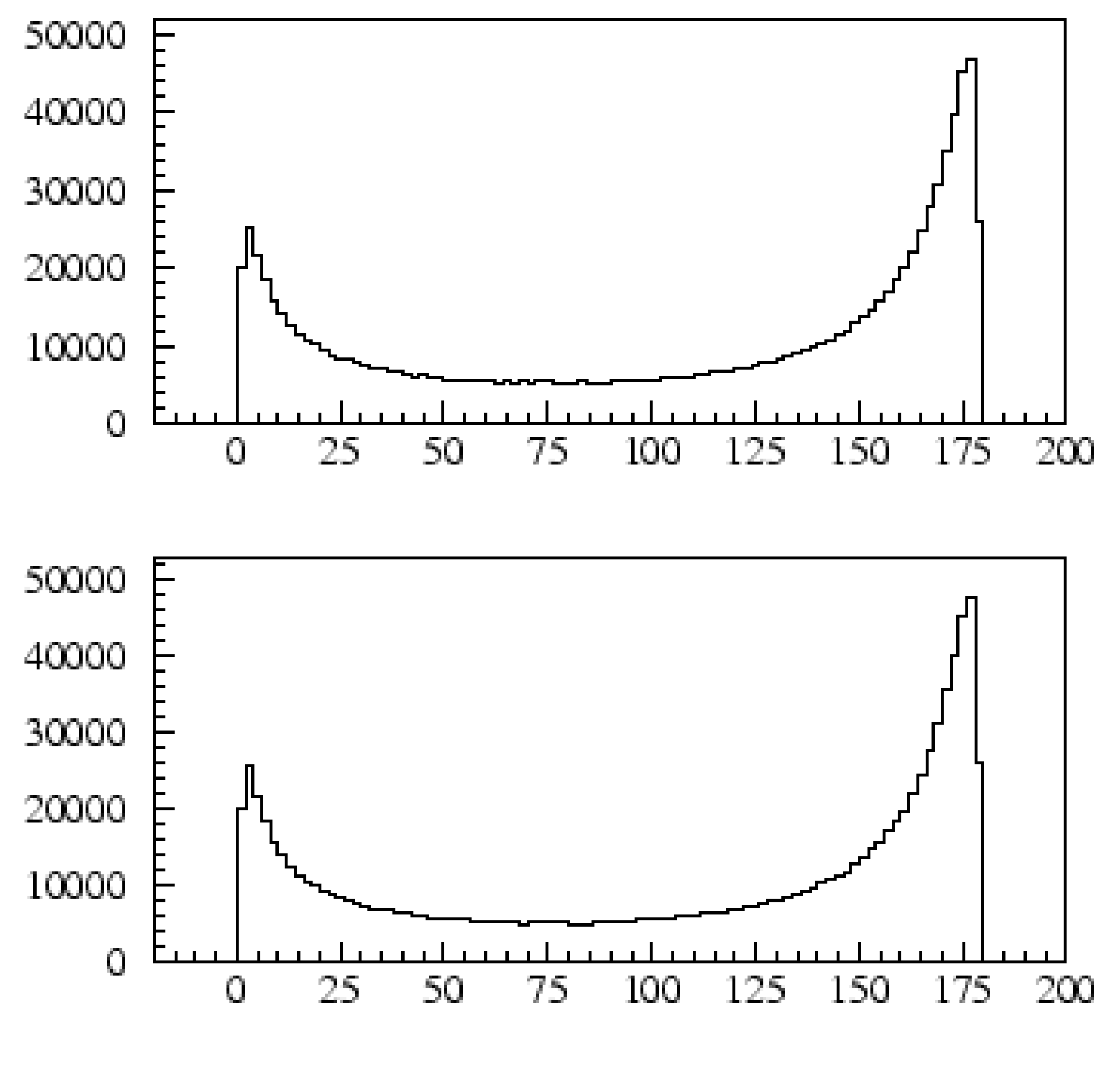}
\end{center}
\vspace*{-0.3cm}  
\caption{Distribution for $J/\psi$ with $E_e=50$\,GeV. Polar angle of positive (top) and negative (bottom) muon respectively.}
\label{LHEC:MainDetector:Muon:Fig:2}   
\end{figure}

\begin{figure}[htp]
\begin{center}
\vspace*{-0.3cm}  
\includegraphics[width=0.5\columnwidth]{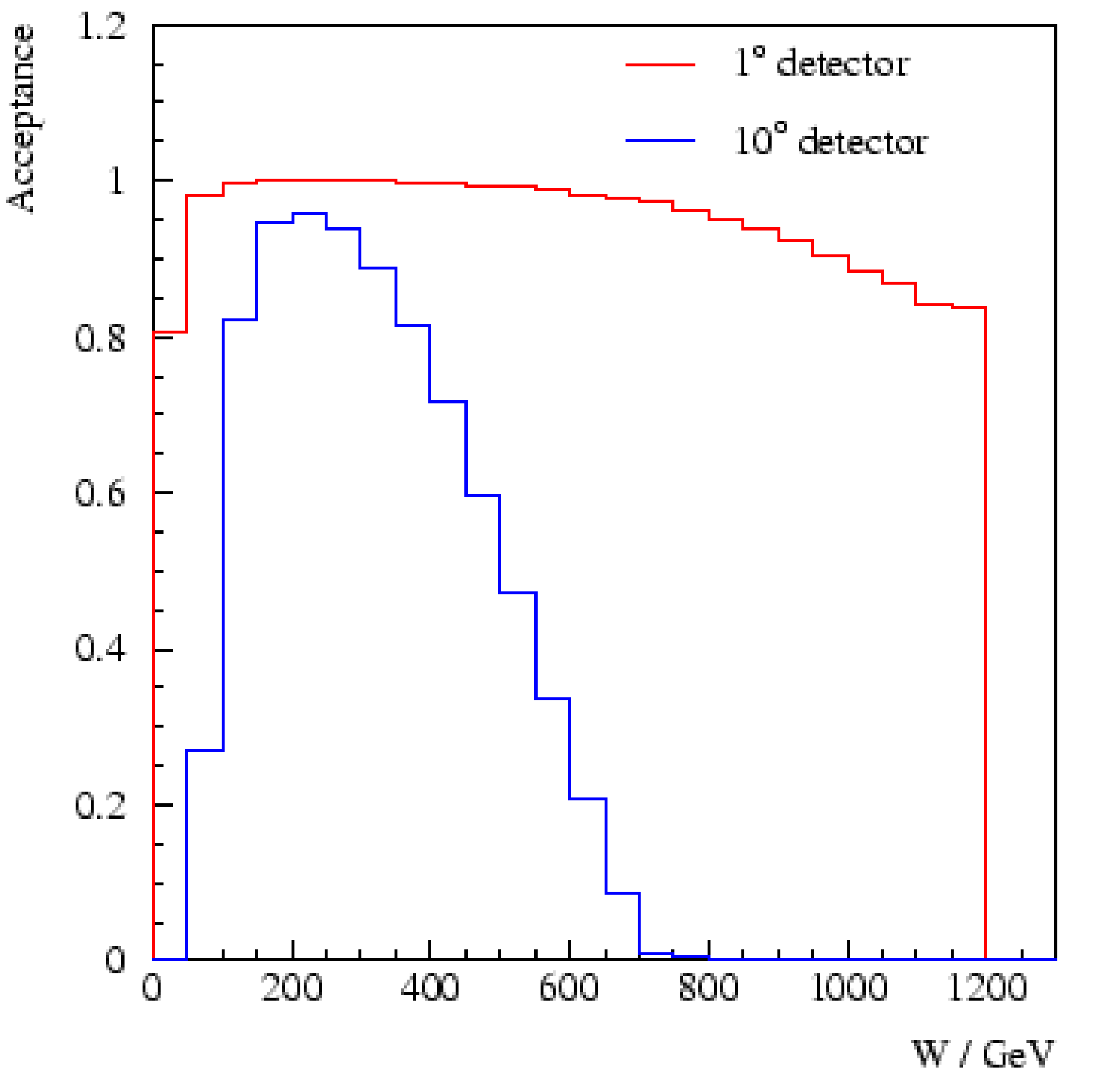}
\end{center}
\vspace*{-0.3cm}  
\caption{Acceptance for $J/\psi$ with $E_e=50$\,GeV as a function of $W$, the centre of mass energy of the $\gamma p$
system. A detector with larger coverage both in the forward and backward directions allows for measurements in a much wider $W$ range.}
\label{LHEC:MainDetector:Muon:Fig:3}   
\end{figure}

Fig.\,\ref{LHEC:MainDetector:Muon:Fig:2} shows the polar angle
distribution of muons produced at the LHeC coming from the decay of
$J/\psi$ mesons produced in elastic processes.  The improvement gained
by enlarging the coverage towards small angles is evident, as
demonstrated in Fig.\,\ref{LHEC:MainDetector:Muon:Fig:3} which shows
the coverage as a function of the $\gamma p$ system centre of mass
energy $W$, for both $10^\circ$ and $1^\circ$ detector acceptance.

\subsection{Muon detector design}

\begin{figure}[htp]
\begin{center}
\hspace*{-0.1cm}  
\includegraphics[width=0.8\columnwidth]{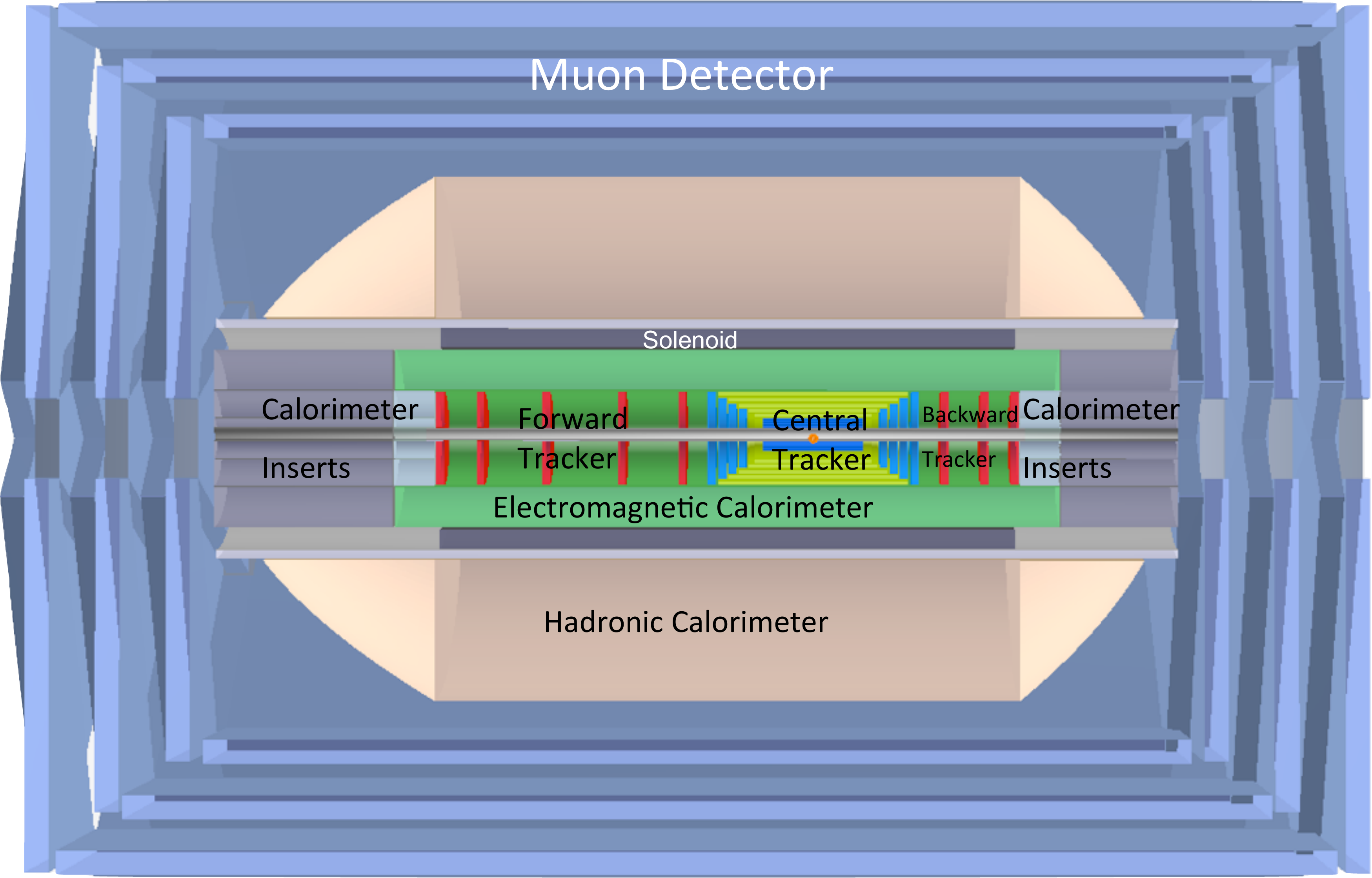}
\end{center}
\vspace*{-0.1cm}
\caption{A full view of the baseline detector in the r-z plane with all components shown.
The detector dimensions are $\approx14$\,m in $z$ with a diameter of $\approx9$\,m.}
\label{LHEC:MainDetector:Muon:Fig:1}   
\end{figure}

The LHeC main detector will be surrounded by several layers of muon
detectors.  Fig.\,\ref{LHEC:MainDetector:Muon:Fig:1} shows a 3d view
of the baseline detector (option {\small\bf A}). Three muon double
detector layers are mechanically attached to an iron structure, which
could provide either the return flux of residual magnetic field from
the inner solenoid or an additional field from warm magnets.

\begin{figure}[htp]
\begin{center}
\vspace*{-0.3cm}
\includegraphics[width=0.35\columnwidth, angle=90]{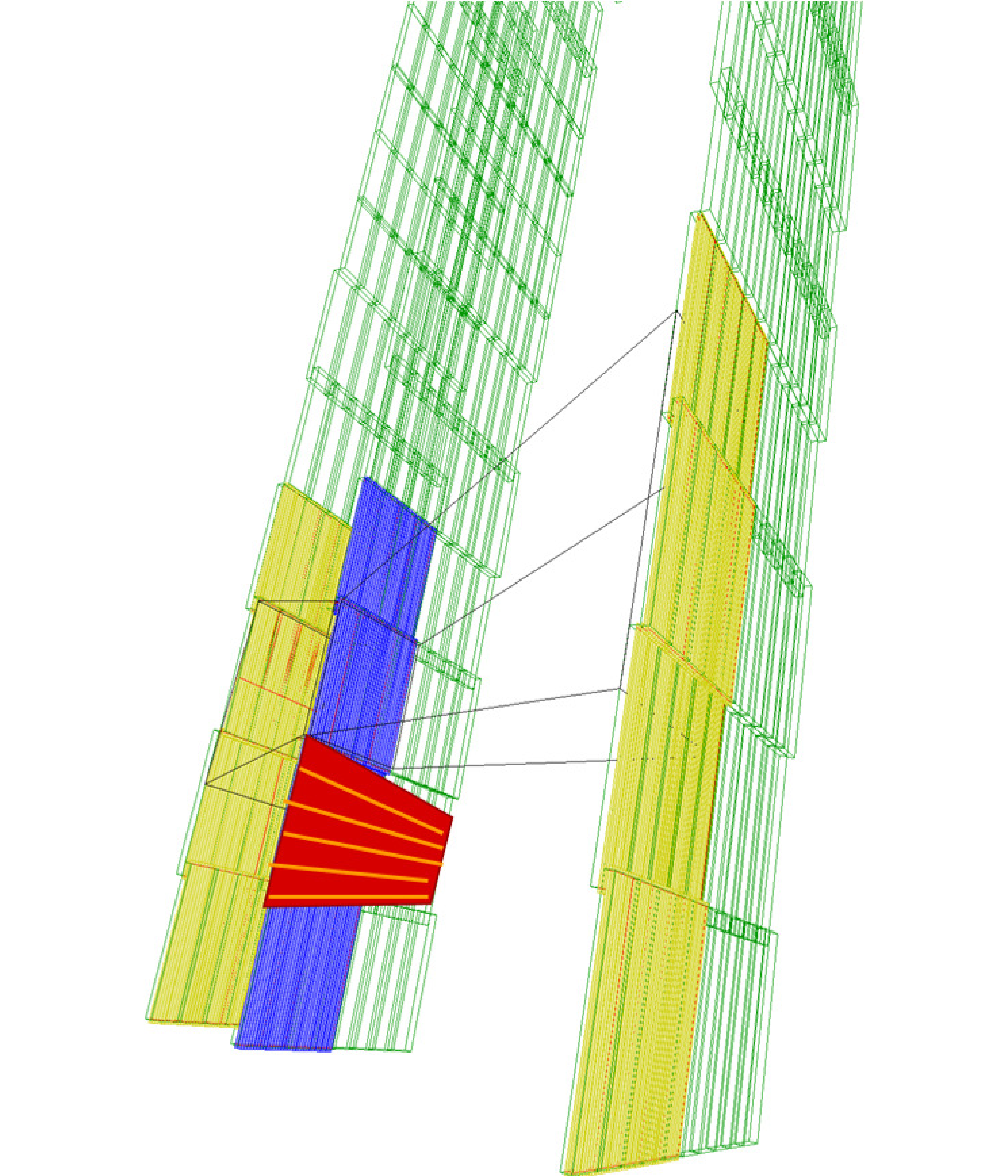}
\ \ \ \ \ \ \ \ \ \ 
\includegraphics[width=0.3\columnwidth]{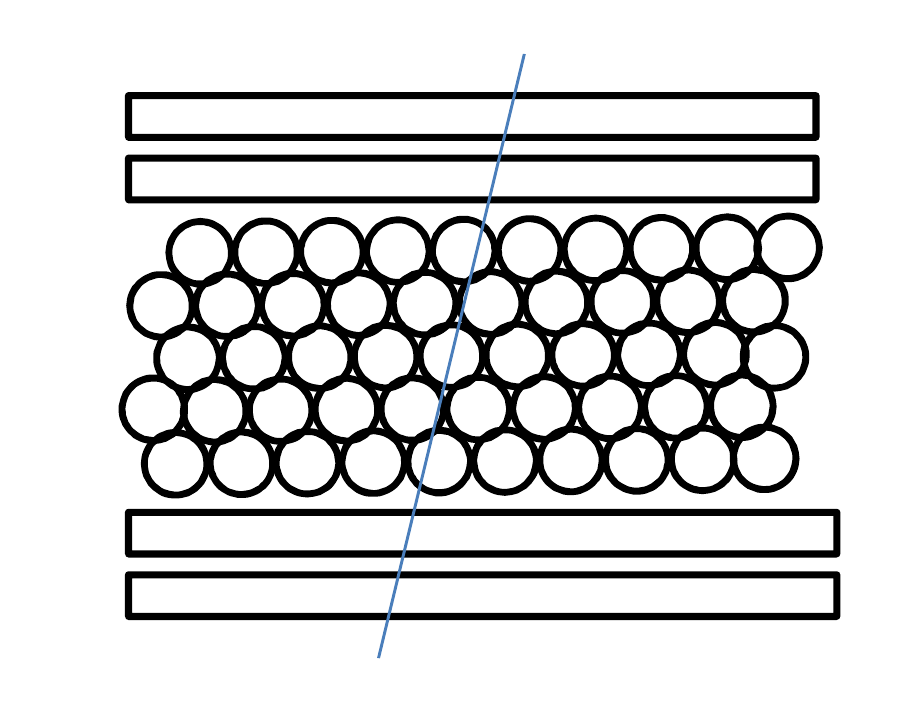}
\end{center}
\caption{Artist 3d view of the projective arrangement of the layers barrel muon chambers (left).
A schematic view of the cross section of one of the chambers which include a double layer of
$\eta\phi$ trigger measurement used also for level one triggering along with the precision measurement
obtained by drift tubes}.
\label{LHEC:MainDetector:Muon:Fig:4}   
\end{figure}

The current state of the art in muon detectors, as implemented in the
LHC experiments and in similar high energy physics experiments, offers
several options that provide the required tracking resolution, rate
sustainability and prompt trigger and readout.  The two LHC general
purpose detectors, ATLAS and CMS, combine Drift Tubes and Cathode
Strip Chambers for precision measurements along with Resistive Plates
Chambers and Thin Gap Chambers for triggering and second coordinate
measurements\cite{Mikenberg:2010zz, Gasparini:1900zz}.  A similar
approach can be considered for the LHeC muon detector, with 2 or 3
superlayers each composed of a double layer of 2d trigger detector and
a layer for precision measurements, as shown in
Fig.\ref{LHEC:MainDetector:Muon:Fig:4}.

Other technologies (for example micromegas\cite{Burnens:2010fw}, etc.)
along with further developments of the existing ones (thin gap
RPC{\cite{Santonico:2011ti}, smaller monitored drift
tubes\cite{Bittner:2011zz}, thin strip
TGC\cite{Amram2011177,Smakhtin2009196}), might also be considered for
the LHeC.  It is evident that the requirements from the present LHC
experiments would also satisfy the LHeC where backgrounds and
luminosity are expected to be lower.

It is beyond of the scope of this document to provide a complete
design as too many options are available, which depend on the choices
for the accelerator and main detector design.  Only a few options are
discussed below with the aim to demonstrate the feasibility and scope
of a detector using available technologies.  More studies and design
optimisation would be required in the next phase.

\subsection{The LHeC muon detector options}

Neglecting the detector technologies to be used, a few different
approaches satisfying increasingly demanding requirements can be
considered for the muon detector.

\begin{enumerate}
\item Muon tagging
\item Combined muon momentum measurement
\item Standalone momentum measurement
\end{enumerate}

The ``muon tagging''(1) muon detector is built with at least 2 layers
of muon chambers providing an $\eta\phi$ measurement and a fast
coincidence for trigger purposes.  No additional magnetic field would
be required and the muon detector, using only the return flux of the
central solenoid, would only be able to provide a very rough estimate
of the particle momentum. The multiple layers and the fast detector
response would allow a pointing trigger to reject non prompt
particles. Muon momentum measurements would be done using mainly the
tracking detector, but could potentially be complemented by positional
information from the energy deposits in the calorimeter (that have to
be compatible with those of a minimum ionising particle) and the muon
detector tag itself.

The more sophisticated muon detector (2) would enhance the muon
momentum measurement by adding an extra magnetic field, embedding the
muon chambers in an iron yoke. The amount of iron and the size of the
yoke can be optimised in order to maximise the resolution in the
energy range required.

Both options (1) and (2) can be considered for the baseline design
option {\small\bf A}.  It is worth noting that for low energy muons
(as expected in the barrel and backward region) an instrumented yoke
may not be required as the momentum resolution of the tracking
system will be far superior.
For muon momenta of 20 GeV and above, the presence of an additional
magnetic field or an instrumented iron yoke could improve resolution,
especially in the forward and backward regions where the momentum
resolution is worse due to the solenoidal field being parallel to the
beam line.
 
Although the presence of an iron mass serves four purposes, namely:
\begin{itemize}
\item return the magnetic flux
\item serve as a hadron ($\pi^\pm, K, p, n$) particle filter so that predominantly
$\mu^\pm$ emerge at a large radius
\item provide excellent mechanical support for all detector systems, especially the massive calorimeter
\item serve as a radiation shield for the area and the electronics 
\end{itemize}
increases in the solenoid size and field strength require shielding to
increase appropriately.  Its density, weight and cost pose important
limitations which might be overcome by the use of a twin solenoid
system as
briefly discussed in Section\,\ref{LHEC:MainDetector:OptionB}.  This
novel approach would guarantee a ``standalone momentum
measurement''\cite{Mazzacane:2010zz}. The outer solenoid allows for
a very smooth and constant field in an iron free region. As shown in
Fig.\,\ref{LHEC:MainDetector:Muon:Fig:6}, the muon detector is
immersed in a strong constant field ($\sim1.5$\,T) which should allow
the precise measurement of momenta up to 500 GeV with $\delta p/p \sim
10\%$. A strong advantage of an air muon spectrometer is the
significant reduction of the uncertainty due to multiple Coulomb
scattering.  Additionally, the use of forward and backward coils can
improve the field quality also in the endcap regions allowing the
field to line up transversely to the beam line for an improved
longitudinal momentum measurement.
%

\begin{figure}[htp]
\begin{center}
\includegraphics[width=0.4\columnwidth,angle=90]{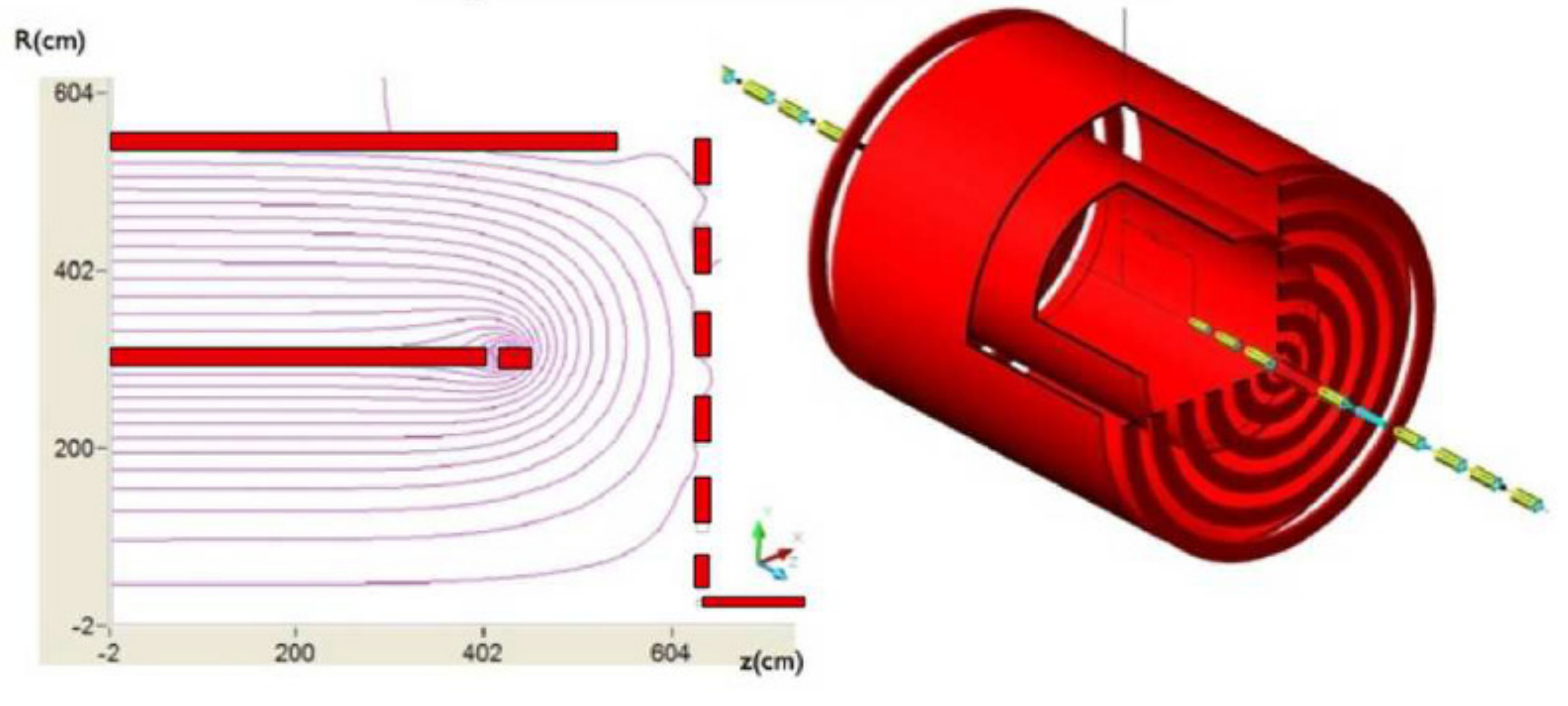}
\end{center}
\caption{Magnetic field lines for the dual solenoid and wall of coil\cite{4thDetector:LoI}. The whole detector is
enclosed in a second return solenoid; forward and rear coils which allow for a smooth field at the detector muon endcaps}.
\label{LHEC:MainDetector:Muon:Fig:6}
\end{figure}

\subsection{Forward muon extensions}

Detection of muons in the forward hemisphere is extremely relevant at
the LHeC where the kinematics of important physical phenomena
(production of heavy flavours, high $x$ physics, leptoquarks etc.)
requires a coverage down to the smallest possible angle with respect
to the beam axis. Since the tracking momentum resolution deteriorates
at small angles, an independent measurement in the forward region would
provide a completely independent tool for the measurement of the muon
momentum.

Given the high particle and, more specifically, muon flux expected in
the forward region, the use of a dedicated forward muon toroid would
allow the measurement of muon charge and momentum.  In
Fig.\ref{LHEC:MainDetector:Muon:Fig:5} a sketch of a possible design
for a ``small'' forward muon toroid is given.  For the baseline
detector {\small\bf A}, a more conventional, iron based solution (as
in HERA for H1 and ZEUS) could be adopted, incorporated or located
outside of the the muon iron-yoke.  The option of an air core forward
toroid, combined either with the option {\small\bf A} detector inside
the iron yoke system or in the larger twin solenoid option {\small\bf
B}, would enhance the forward muon momentum resolution even further,
especially for very small angles with respect to the beam line.

The insertion of a forward air core based toroid closer to the central
tracking system was also considered and rejected because the bulk
material of the required coils, located between the tracking planes
and the calorimeters, would compromise the calorimetry measurements.

\begin{figure}[htp]
\begin{center}
\vspace*{-0.3cm}  
\includegraphics[width=0.5\columnwidth, angle=180]{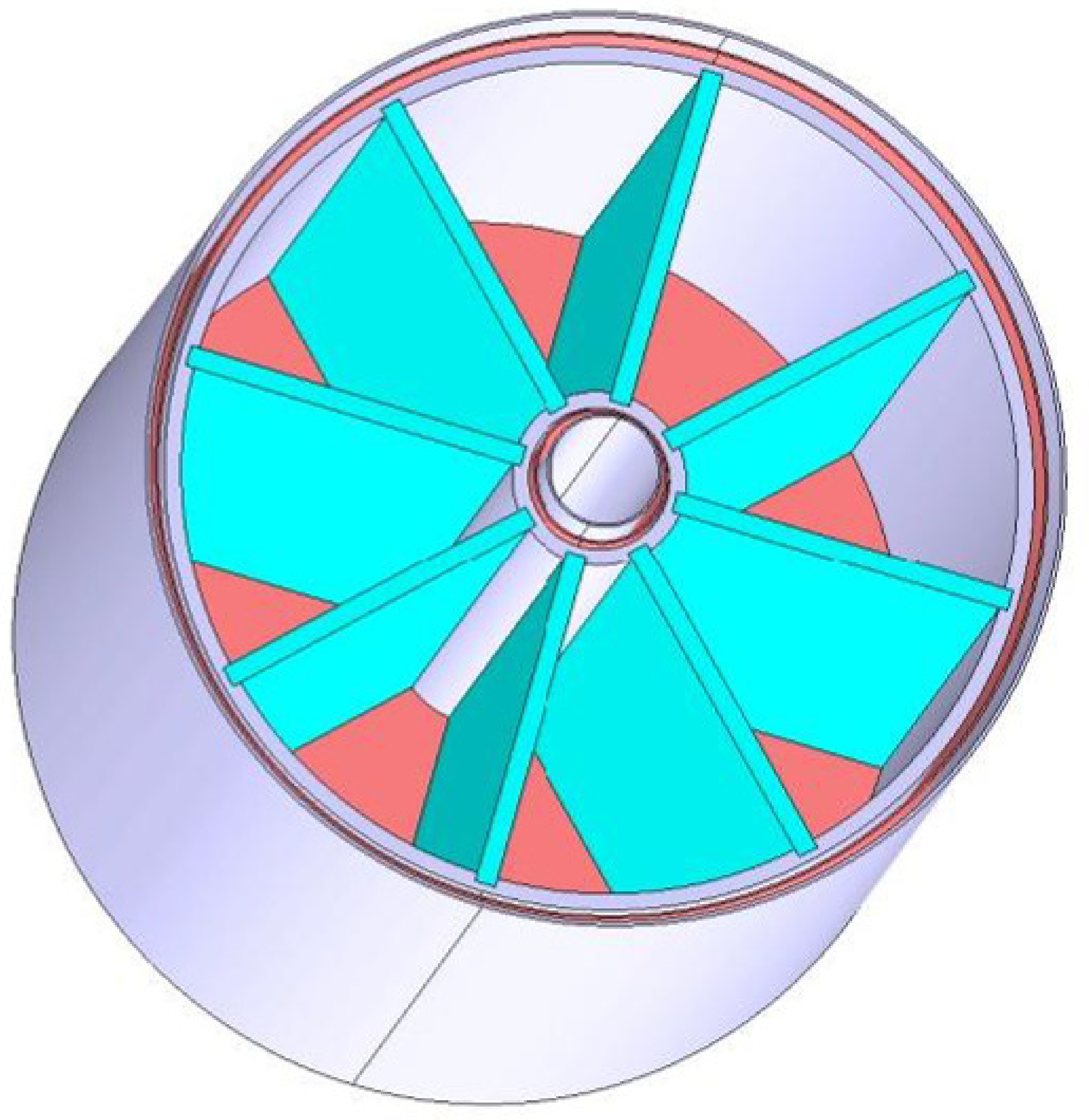}
\end{center}
\caption{CAD drawing for a 2\,T air core toroid with 20\,cm bore and a size about 1\,m$^3$}.
\label{LHEC:MainDetector:Muon:Fig:5}
\end{figure}

%
%

\subsection{Muon detector summary}

Several options for the LHeC muon detector are available.  These range
from a simple muon tagging detector which, combined with the baseline
detector {\small\bf A} would already be sufficient for a clean muon
trigger, allowing to remove beam gas background and non pointing
tracks.  The precision of the momentum resolution would depend mostly
on the main detector (tracking and calorimetry) which anyhow would
degrade at small forward and backward angles.

Improvements by means of an iron yoke and conventional forward muon
toroids would allow improved performance especially for higher momenta
and for muon spectroscopy in the forward region.  Experience from HERA
suggests that a solution lacking a standalone muon trigger could be
acceptable for most of the physics program.

The ultimate design nevertheless appears to be the twin solenoid
option.  This more challenging design, shown in
Fig.\ref{LHEC:MainDetector:Muon:Fig:7} naturally follows the option
{\small\bf B} of the baseline design: the larger main solenoid is
located outside of the hadronic calorimeter and together with a second
active shielding solenoid provides a large material free region for
precise standalone muon momentum measurement.
The higher energies available in the forward region and the
interesting physics channels accessible there also motivate the use of
an additional forward muon toroid, with which the detector acceptance
for muon channels could be greatly extended.

\begin{figure}[htp]
\begin{center}
\includegraphics[width=0.9\columnwidth]{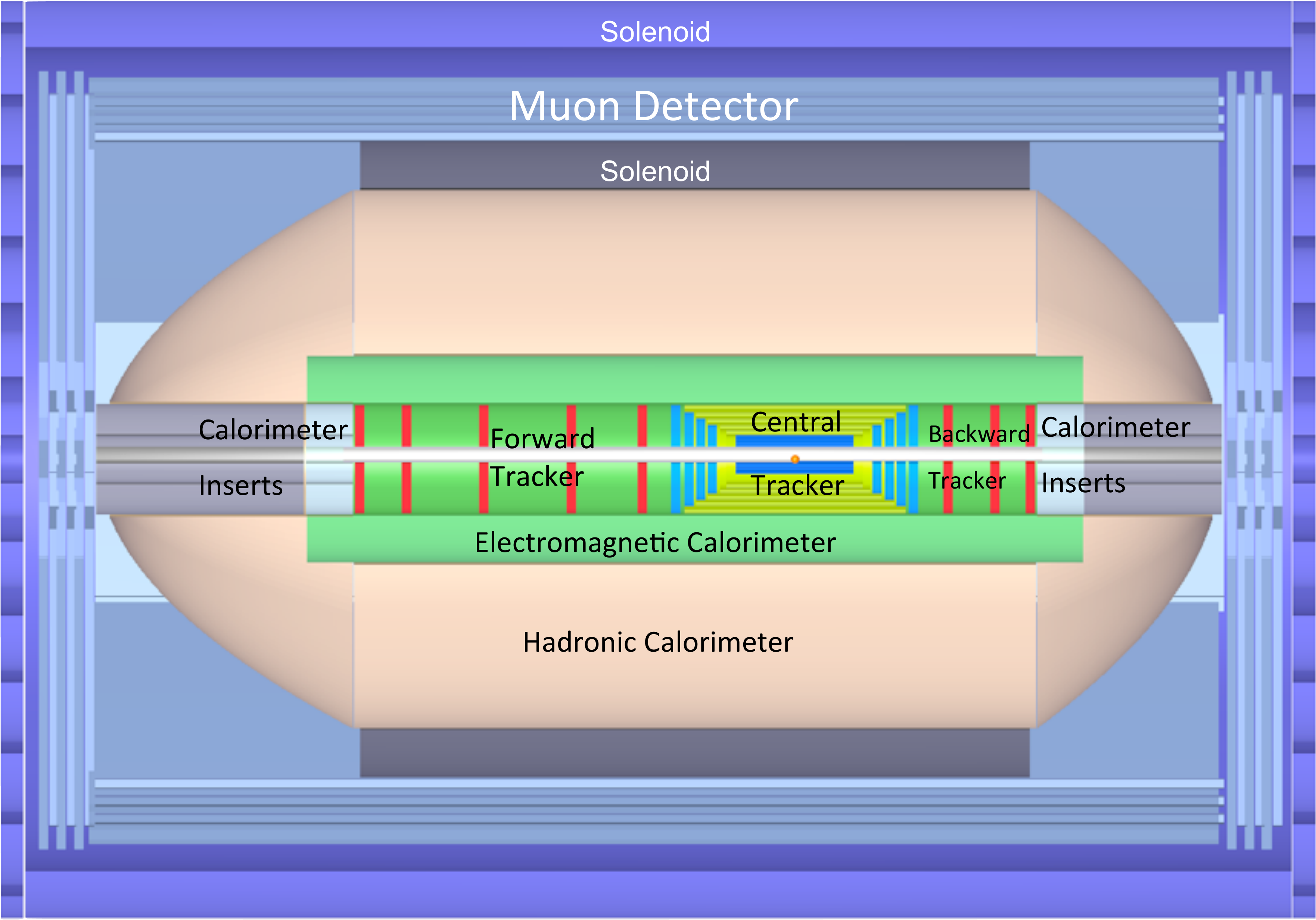}
\end{center}
\caption{The option {\small\bf B} of the LHeC baseline detector. The larger solenoid surrounds the hadronic calorimetry. The volume outside
the solenoid is filled with an approximately uniform magnetic field of 1.5\,T and is instrumented with
3 multi-layers of muon chambers.}
\label{LHEC:MainDetector:Muon:Fig:7}
\end{figure}

%% file: detector/Ev-Det-Sim.tex
\label{Geant4-Event-Simulations}

\begin{figure}[!htbp]
\centerline{\includegraphics[clip=,width=0.75\textwidth]{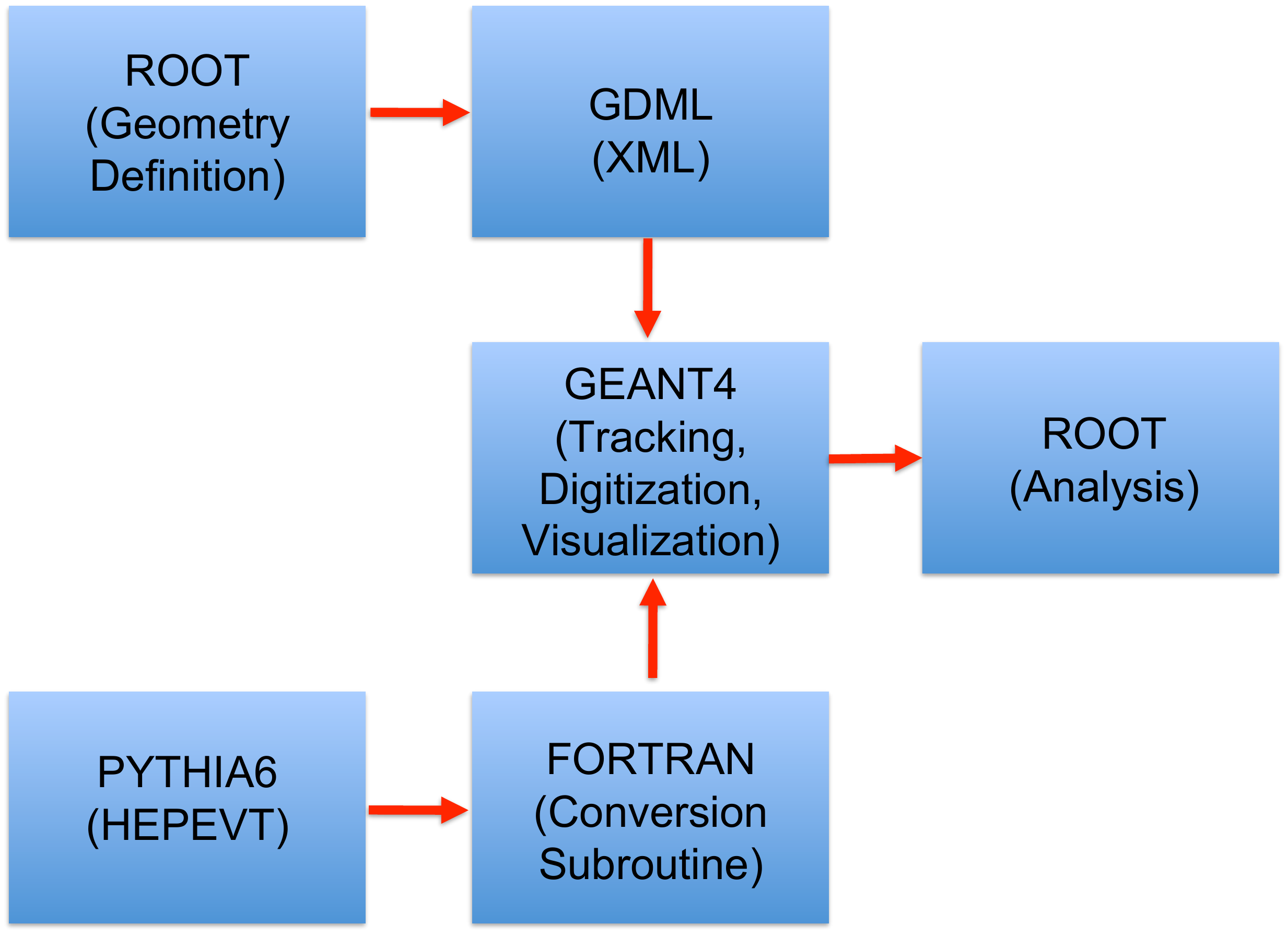}}
\caption{Simulation Framework Flow Chart}
\label{Fig:Det_Sim_Chart}
\end{figure}

Minimum bias events in the LHeC Detector have been simulated using the
{\small\bf GEANT4} Toolkit \cite{Agostinelli:2002hh}.  In addition
{\small\bf ROOT} \cite{ROOTcite}, {\small\bf GDML} \cite{GDMLcite},
{\small\bf AIDA} \cite{serbo} and {\small\bf
Pythia6} \cite{Sjostrand:2006za} have also been incorporated.  A
{\small\bf ROOT} macro has been written which gives a general
description of the LHeC Detector geometry and materials.  This
description is then transported from {\small\bf ROOT} to {\small\bf
GEANT4} in XML format via {\small\bf GDML}.  A {\small\bf Pythia6}
program has also been used to create minimum bias $ep$
events. {\small\bf Pythia6} outputs the events in HEPEVT format.  This
is then run through a subroutine to produce a format readable by
{\small\bf GEANT4}.  The actual simulations are completed natively in
{\small\bf GEANT4} once the geometry, materials and events are loaded.
The Analysis is done with {\small\bf ROOT} (and the Java Analysis
Studio {\small\bf JAS} \cite{serbo} ) which is interfaced to
{\small\bf GEANT4} via {\small\bf AIDA}.  The flow of these
simulations is outlined in Figure\,\ref{Fig:Det_Sim_Chart}.
	
The tools available for $ep$ event generation are not the most up to
date. The frontier of high energy physics is focused on the LHC and
hadron-hadron collisions and there has largely been a lack of
development of event generation tools for a new energy scale of $ep$
collisions.  Hence these studies use {\small\bf Pythia6} as opposed to
its C++ successor.  Although it works fine as an approximation it
would be advantageous to have development here.

\subsection{Pythia6}

The {\small\bf Pythia6}(\cite{Sjostrand:2006za}) event used in the
{\small\bf GEANT4} simulations contains $\gamma^{*} P$ interactions
convoluted with the $\gamma/e^{\pm}$flux.  This setup contains non
vanishing cross sections including semi-hard QCD, elastic scattering,
single and double diffractive processes among others (The listed
interactions dominate $\sigma_{tot}$).  In order for the events to be
minimum bias, no restrictions are placed on the W or $Q^2$ range.
Table\,\ref{tab:Param} gives the {\small\bf Pythia6} parameters used
for the minimum bias events.  The logarithm of the variables W and
$Q^2$ are given. Since these variables obey amplitudes given by
$P(x) \propto \frac{1}{x^2}$ then $P(Log(x)) \propto e^{-x^2}$,
i.e. Log(x) follows normal statistics.
	
\begin{table}[!htb]
\centering	
\begin{tabular}{| c | c |}
  \hline
Characteristic & Value \\
\hline
\hline
$Log(W)_{mean}$ \hfill [GeV] & 2.09 \\
\hline
$Log(W)_{rms}$ \hfill [GeV] & 0.55 \\
\hline
$Log(Q^2)_{mean}$  \hfill [$GeV^2$] & $-4.98$ \\
\hline
$Log(Q^2)_{rms}$ \hfill [$GeV^2$] & $3.15$ \\
\hline
\hline
Electron Energy \hfill  [GeV] & 60  \\
\hline
Proton Energy \hfill [GeV] & 7000  \\
\hline
\end{tabular}
\caption{Pythia6 Parameters}
\label{tab:Param}
\end{table}

The parameters used to scale the results of the simulation in order to find annual quantities are given in Table\,\ref{tab:ParamS}.

\begin{table}[!htb]
\centering	
\begin{tabular}{| c | c |}
  \hline
Characteristic & Value \\
\hline
\hline  
Total Cross Section \hfill [$mb$] & 0.0686 \\
\hline
Luminosity \hfill [$mb^{-1}s^{-1}$] & $10^6$ \\  
\hline
$\frac{dN}{dt}$ \hfill [int/yr] & $2.57\times10^{12}$ \\
\hline
\end{tabular}
\caption{Scaling Parameters}
\label{tab:ParamS}
\end{table}

\subsection{1\,MeV neutron equivalent}

\begin{figure}[!htbp]
\centerline{\includegraphics[clip=,width=0.75\textwidth]{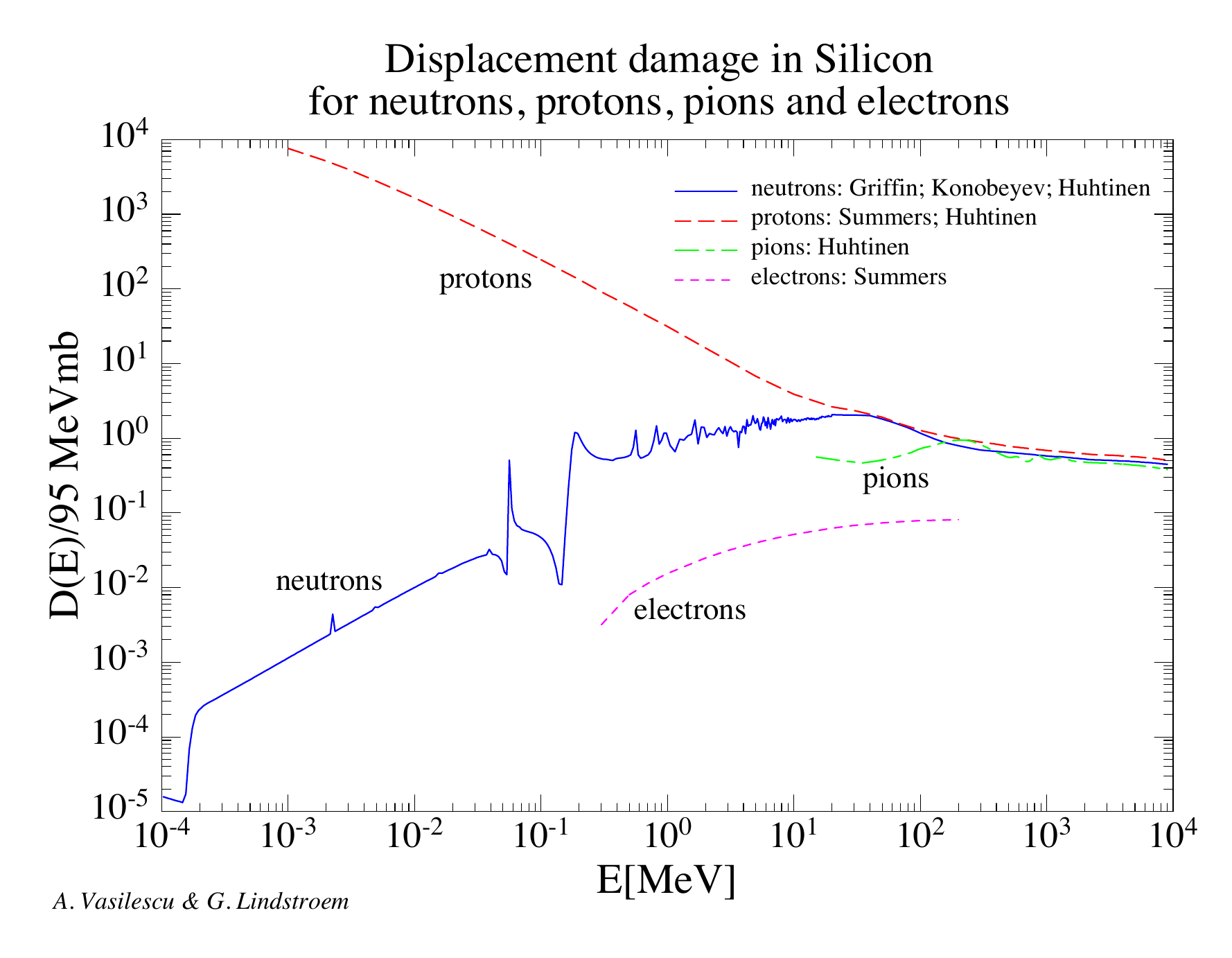}}
\caption{Displacement Damage for various particles in Silicon}
\label{Fig:D(E)_Si}
\end{figure}

In order to find the 1 MeV Neutron Equivalent, the appropriate
displacement damage functions [D(E)] for the particles must be found.
By scaling the damage functions by the reciprocal of D(n, 1 MeV), a
weight is found which turns a fluence of random particles into the 1
MeV Neutron Equivalent fluence. D(E) is not only dependent on particle
type but also on the material in which the particles are traversing.
The D(E) functions used in the simulations can be found in
Figure\,\ref{Fig:D(E)_Si} \cite{ddfunction}.

In order to find the 1 MeV Neutron Equivalent fluence through the
tracking portion of the detector, scoring was incorporated into the
{\small\bf GEANT4} simulations.  A user defined scorer was used to
calculate the number of hits on the surface of a detector component,
weight the hits according to the appropriate damage functions and
finally divide the sum of these weighted hits by the inner surface
area of the detector component.  The flux was then scaled by the
number of events per year using the scaling parameters given
in Table\,\ref{tab:ParamS}.  The total 1 MeV Neutron Equivalent
fluences are given in Table\,\ref{tab:Results}.

\begin{table}[!htb]
\centering	
\begin{tabular}{| c || c | c | c |}
  \hline
  \multicolumn{4}{|c|}{Central Barrel} \\
  \hline
Region & $\Delta Z [cm]$ & $R_{min}$ [cm] & Fluence [$\frac{N}{cm^2yr}$]  \\
\hline
\hline
CPT1 & 100 & 3.1 & $1.38\times10^{10}$  \\
\hline
CPT2  & 100 & 5.6 & $9.99\times10^9$   \\
\hline
CPT3  & 100 & 8.1 & $8.26\times10^9$   \\
\hline
CPT4 & 100 & 10.6 & $7.25\times10^9$   \\
\hline
\hline
CST1 & 116 & 21.2 &  $6\times10^9$   \\
\hline
CST2 & 128 & 25.6 & $5.66\times10^9$   \\
\hline
CST3 & 148 & 31.2 & $5.38\times10^9$   \\
\hline
CST4 & 168 & 36.7 & $5.25\times10^9$   \\
\hline
CST5 & 188 & 42.7 & $5.16\times10^9$   \\
\hline
\hline
\multicolumn{4}{|c|}{Central Endcaps} \\
\hline
Region & Z [cm] & $ \Delta R$ [cm] & Fluence [$\frac{N}{cm^2yr}$]  \\
\hline
\hline
CFT1 & 70 & 26 & $8\times10^9$   \\
\hline
CFT2 & 80 & 31.6 & $7.42\times10^9$   \\
\hline
CFT3 & 90 & 37.1 & $7.08\times10^9$   \\
\hline
CFT4 & 101 & 43.1 & $6.93\times10^9$   \\
\hline
\hline
CBT1 & -70 & 26 &$2.77\times10^9$   \\
\hline
CBT2 & -80 & 31.6 & $2.48\times10^9$   \\
\hline
CBT3 & -90 & 37.1 & $2.26\times10^9$   \\
\hline
CBT4 & -101 & 43.1  & $2.09\times10^9$   \\
\hline
\hline
\multicolumn{4}{|c|}{Fwd/Bwd Planes} \\
\hline
Region & Z [cm] & $ \Delta R$ [cm] & Fluence [$\frac{N}{cm^2yr}$]  \\
\hline
\hline
FST1 & 130 & 43.1  & $8.2\times10^9$   \\
\hline
FST2 & 190 & 43.1 & $1.14\times10^{10}$   \\
\hline
FST3 & 265 & 43.1 & $1.63\times10^{10}$   \\
\hline
FST4 & 330 & 43.1 & $2.29\times10^{10}$   \\
\hline
FST5 & 370 & 43.1 & $2.75\times10^{10}$   \\
\hline
\hline
BST1 & -130 & 43.1 & $1.96\times10^9$   \\
\hline
BST2 & -170 & 43.1 & $1.91\times10^9$   \\
\hline
BST3 & -200 & 43.1 & $1.99\times10^9$   \\
\hline
\end{tabular}

\caption{1 MeV Neutron Equivalent Fluence}
\label{tab:Results}
\end{table}

\begin{figure}[!htbp]
\centerline{\includegraphics[clip=,width=0.75\textwidth]{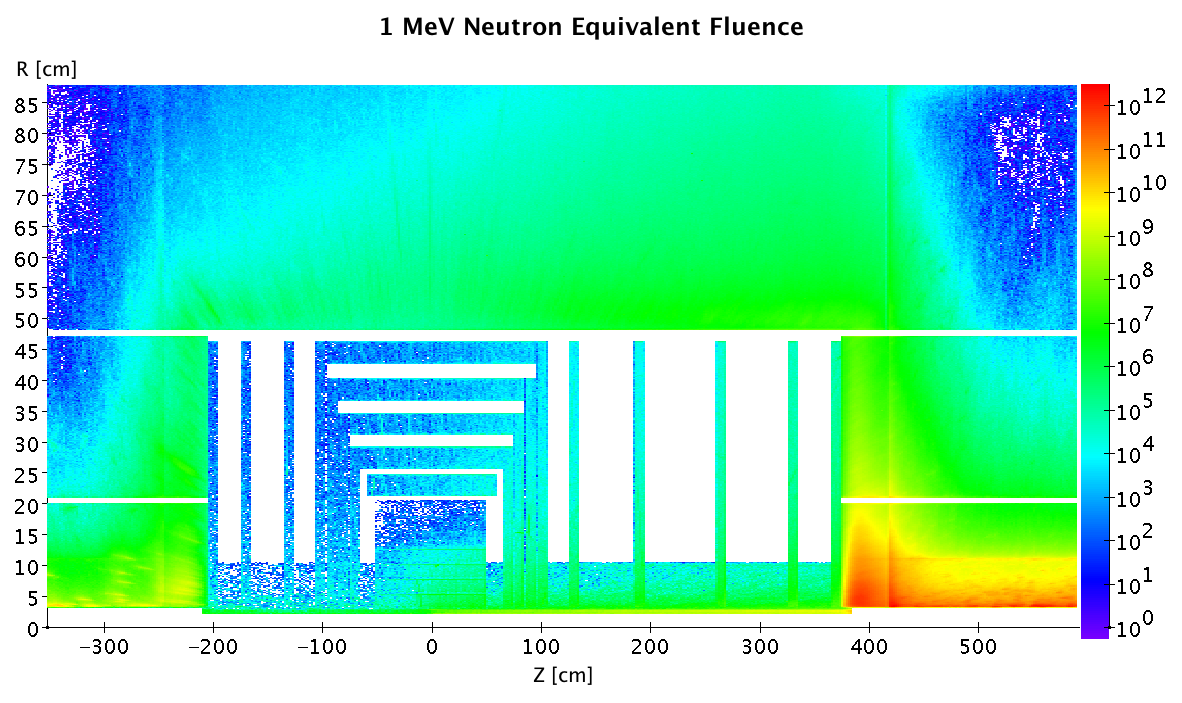}}
\caption{1 MeV Neutron Equivalent Fluence [cm$^{-2}$/year].}
\label{Fig:Fluence}
\end{figure}

A different approach was used in order to find the 1 MeV Neutron
Equivalent fluence distribution in $R_{polar}$ and Z.  In order to
retain data generated per event instead of per simulation run, a set
up of Sensitive Detectors [SD] was initialised that measures user
defined quantities for traversing particles.  The entire tracking
region was set as one SD, with each hit containing the position
information, and the current $D(E)$ value of the given track. A 2D
histogram was generated for the variables $R_{polar}$ and Z.  The
intensity (each hit weighted by its $D(E)$ value) was then scaled by
the number of events in the run, the number of events per year, and a
fluence weighting function.  This function divides the number of
entries in each bin by the average surface area the bin represents
(i.e. $2\pi R_{mean} \Delta Z$ where $R_{mean}$ is the mean R value
which the bin spans and $\Delta Z$ is the width of the Z bins).  By
this weighting process the resulting 2D histogram
(Figure\,\ref{Fig:Fluence}) displays the 1 MeV Neutron Equivalent
Fluence in ${cm^{2}}$ and year.

\subsection{Nearest neighbour}

\begin{table}[!htb]
\centering	
\begin{tabular}{| c | c |}
\hline
Tracking Component & Hits under 10 $\mu{m}$ [$\%$] \\
\hline
\hline
CFT1 & 0.18  \\
\hline
CFT4  & 0.23   \\
\hline
FST1  & 0   \\
\hline
FST5 & 0.1   \\
\hline
\end{tabular}
\caption{Nearest Neighbour under 10 $\mu{m}$}
\label{tab:under10}
\end{table}

\begin{figure}[!htbp]
\centerline{\includegraphics[clip=,width=0.75\textwidth]{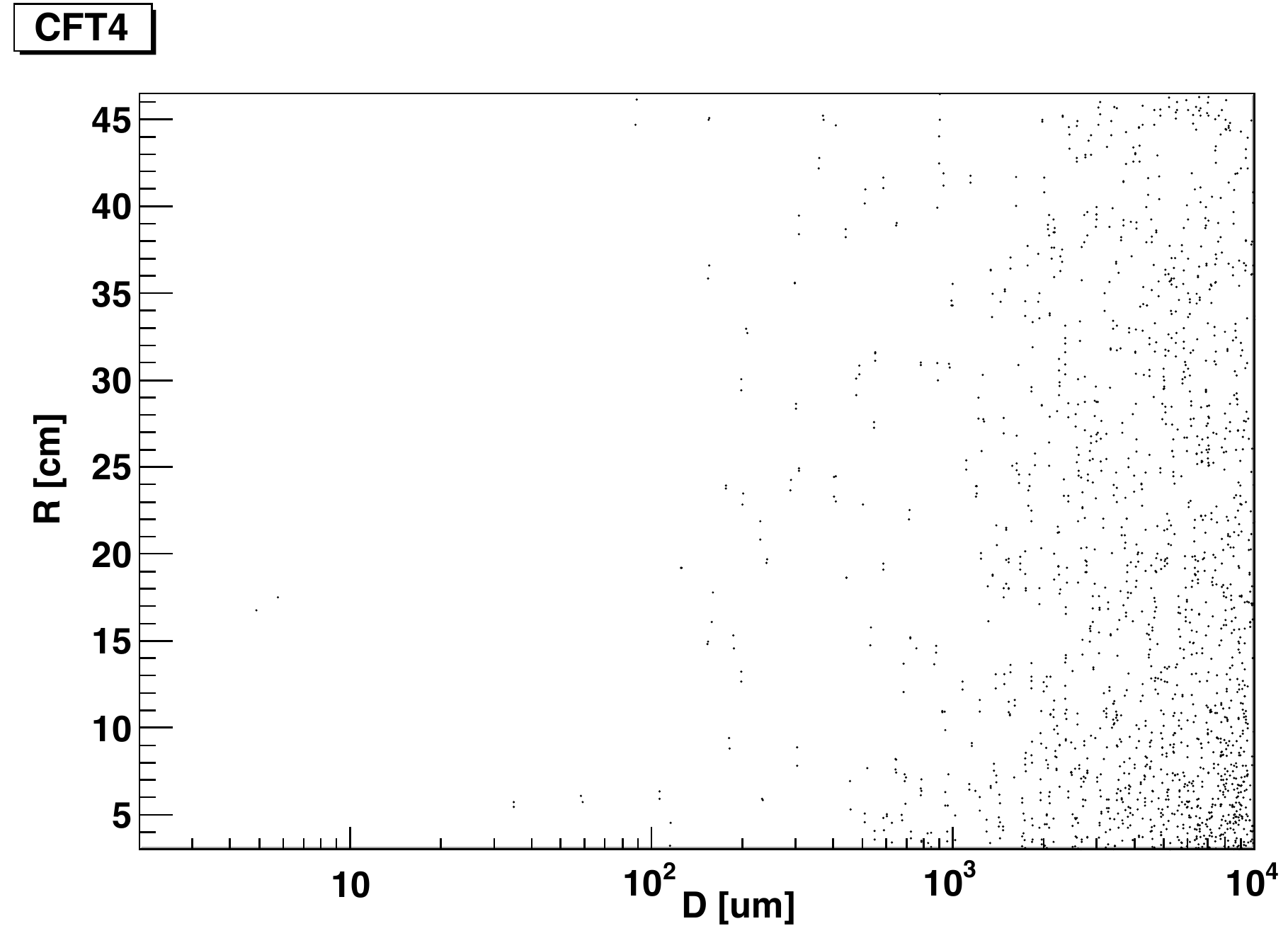}}
\caption{Nearest Neighbour distribution for CFT4}
\label{Fig:dist_CFT4}
\end{figure}

\begin{figure}[!htbp]
\centerline{\includegraphics[clip=,width=0.75\textwidth]{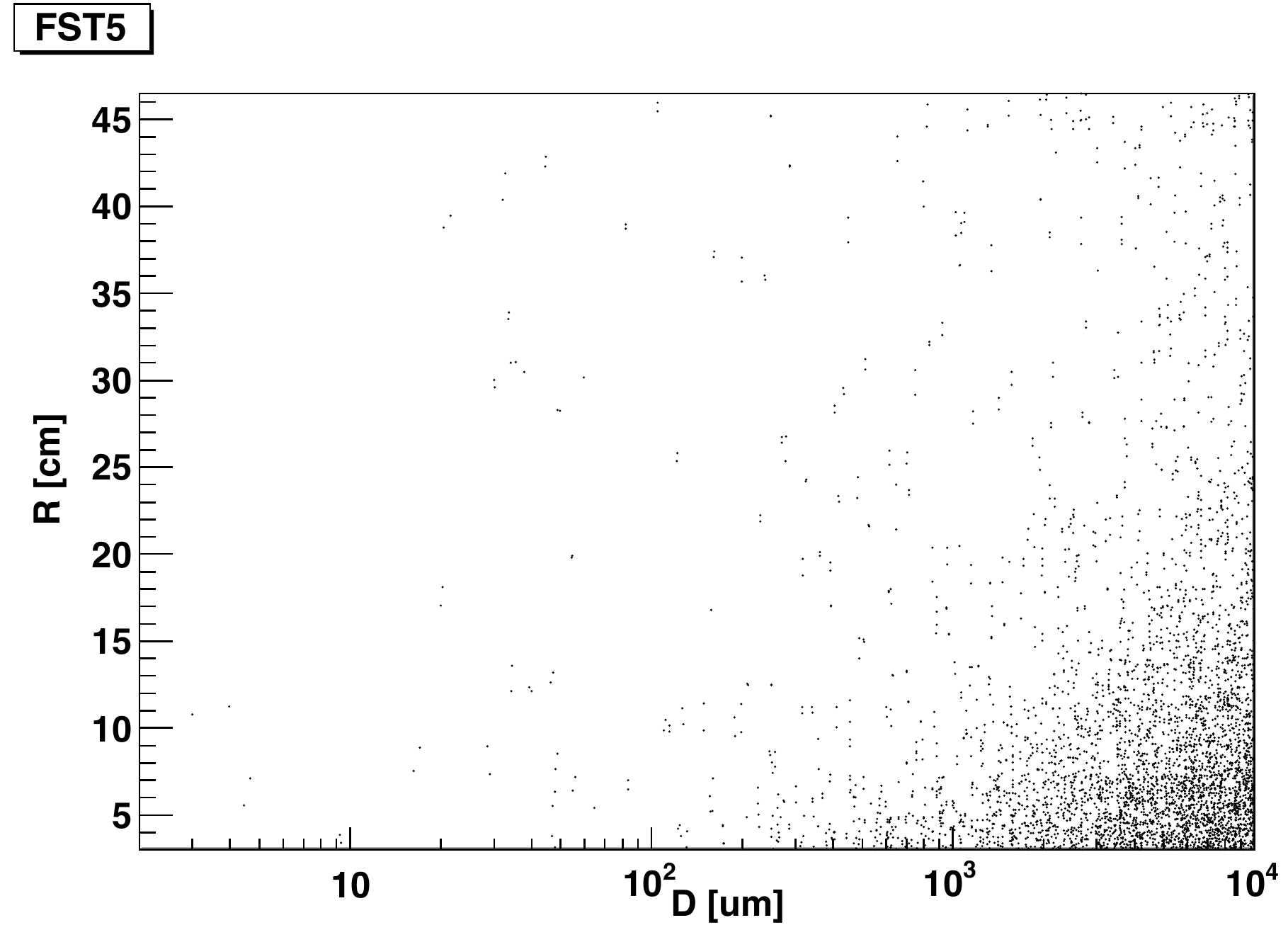}}
\caption{Nearest Neighbour distribution for FST5}
\label{Fig:dist_FST5}
\end{figure}

The {\small\bf Geant4} simulations were also used to find the
resolution required in the forward tracker.  Firstly, the flux
through the surface of CFT1, CFT4, FST1, and FST5 was found.  A
minimisation algorithm is then used to find the nearest neighbouring
hit at the $Z=constant$ surface for each hit.  This distance scale is
characteristic of the resolution required for the tracking component
in question.  The nearest neighbouring hit distribution is calculated
at the event level.  This implies that only the hits from the same
event are compared.  This will have to be studied further 
, but information on the event level is a good
approximation.  The nearest neighbour distribution for CFT4 is shown in
Figure\,\ref{Fig:dist_CFT4} and for FST5 in
Figure\,\ref{Fig:dist_FST5}.  The x axis contains the value of the
nearest neighbour for each hit in terms of $\mu{m}$ while the y axis
contains R in terms of cm.  A required resolution of ~10\,$\mu{m}$ or
less would require pixel detectors instead of strip detectors.  The
CFT4 and FST5 Figures display a very low hit density in this area.
The percentage of hits with $D<10\,\mu{m}$ for the four tracking
components in question are given in Table\,\ref{tab:under10}.
	
\subsection{Cross checking}
	\begin{figure}[!htbp]
\centerline{\includegraphics[clip=,width=0.6\textwidth]{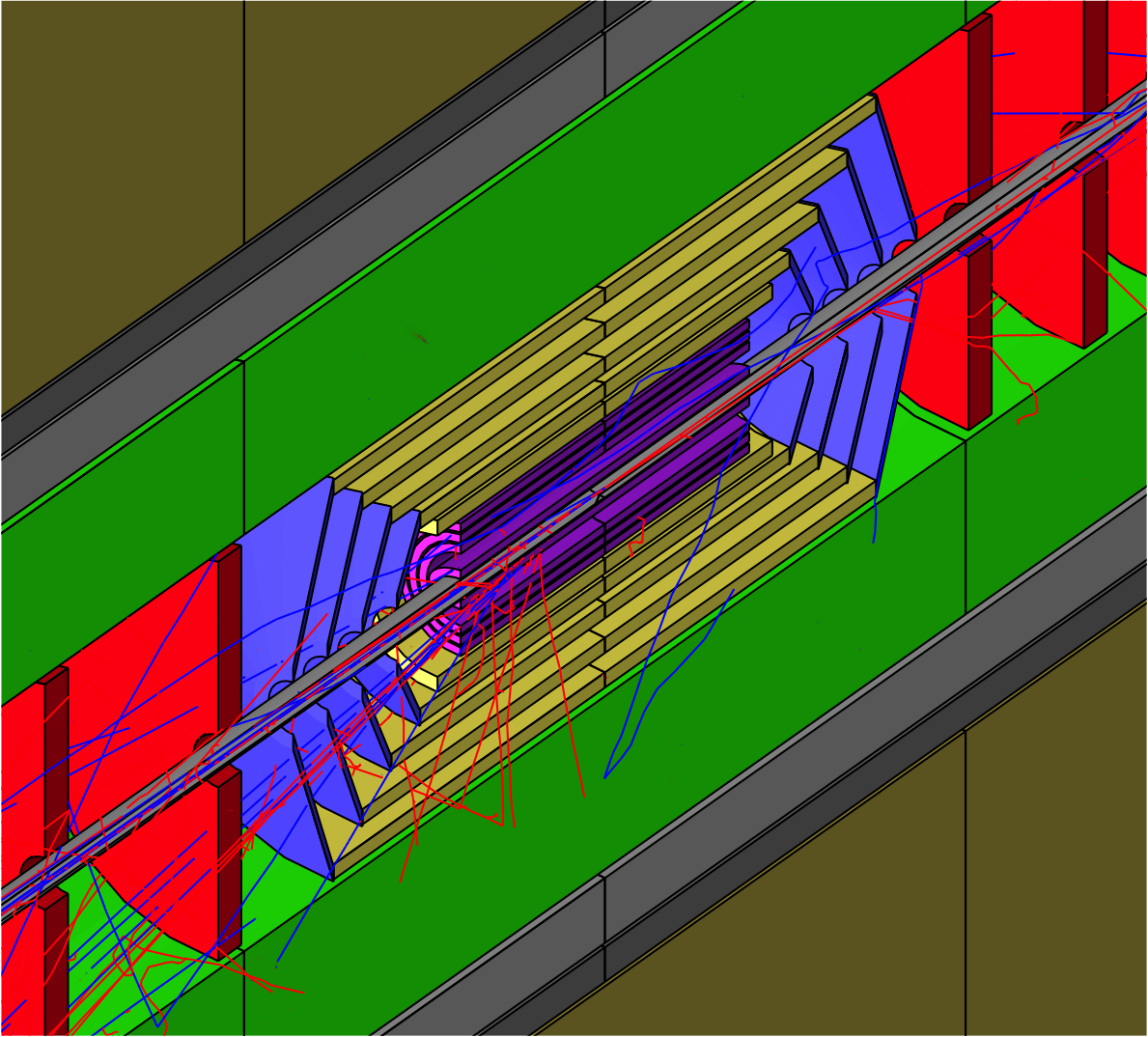}}
\caption{G4 Event}
\label{Fig:G4_Event_1}
\end{figure}
	\begin{figure}[!htbp]
\centerline{\includegraphics[clip=,width=0.8\textwidth]{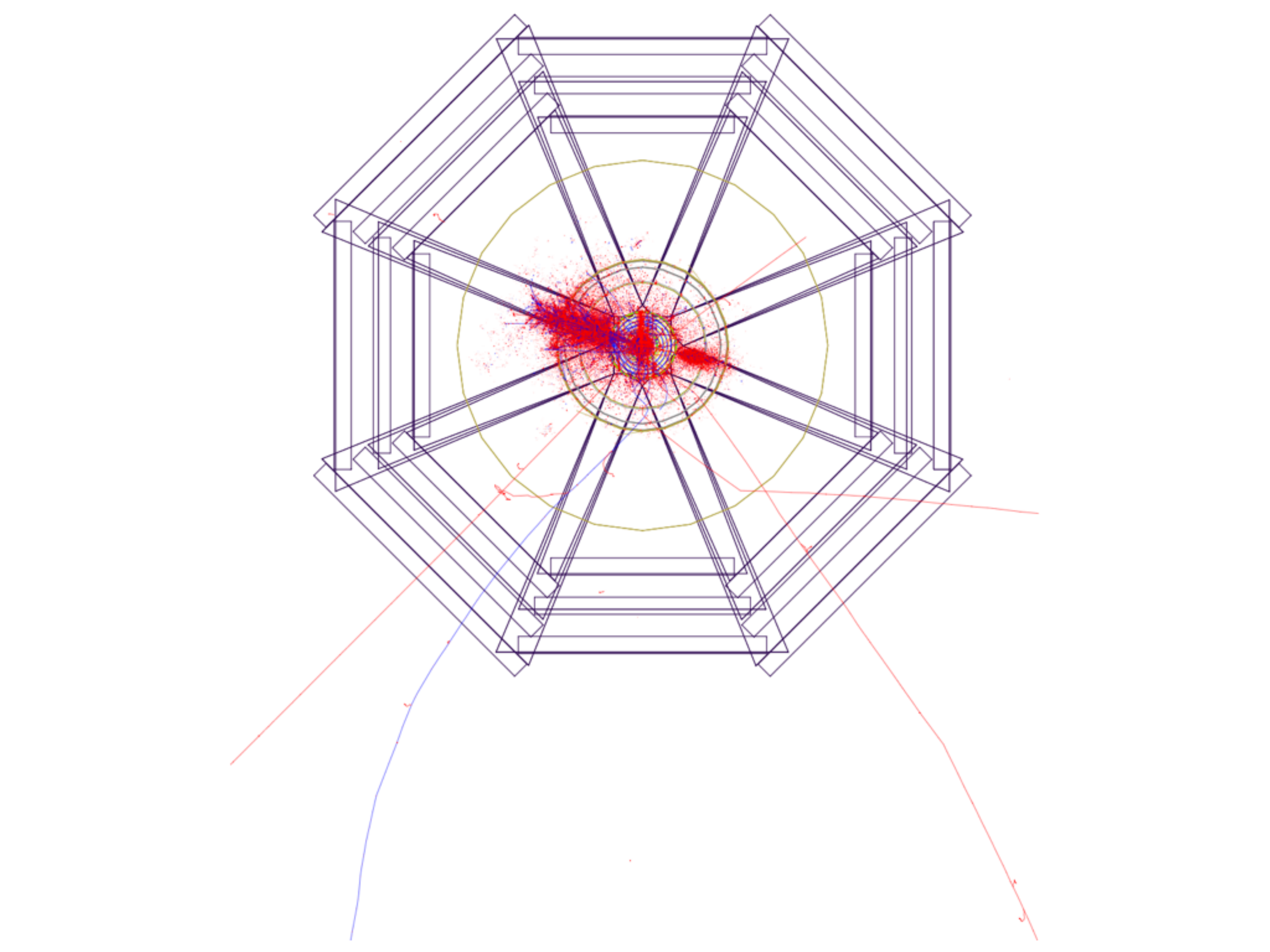}}
\caption{Leptoquark Event XY}
\label{Fig:leptoquark_xy}
\end{figure}

	\begin{figure}[!htbp]
\centerline{\includegraphics[clip=,width=0.8\textwidth]{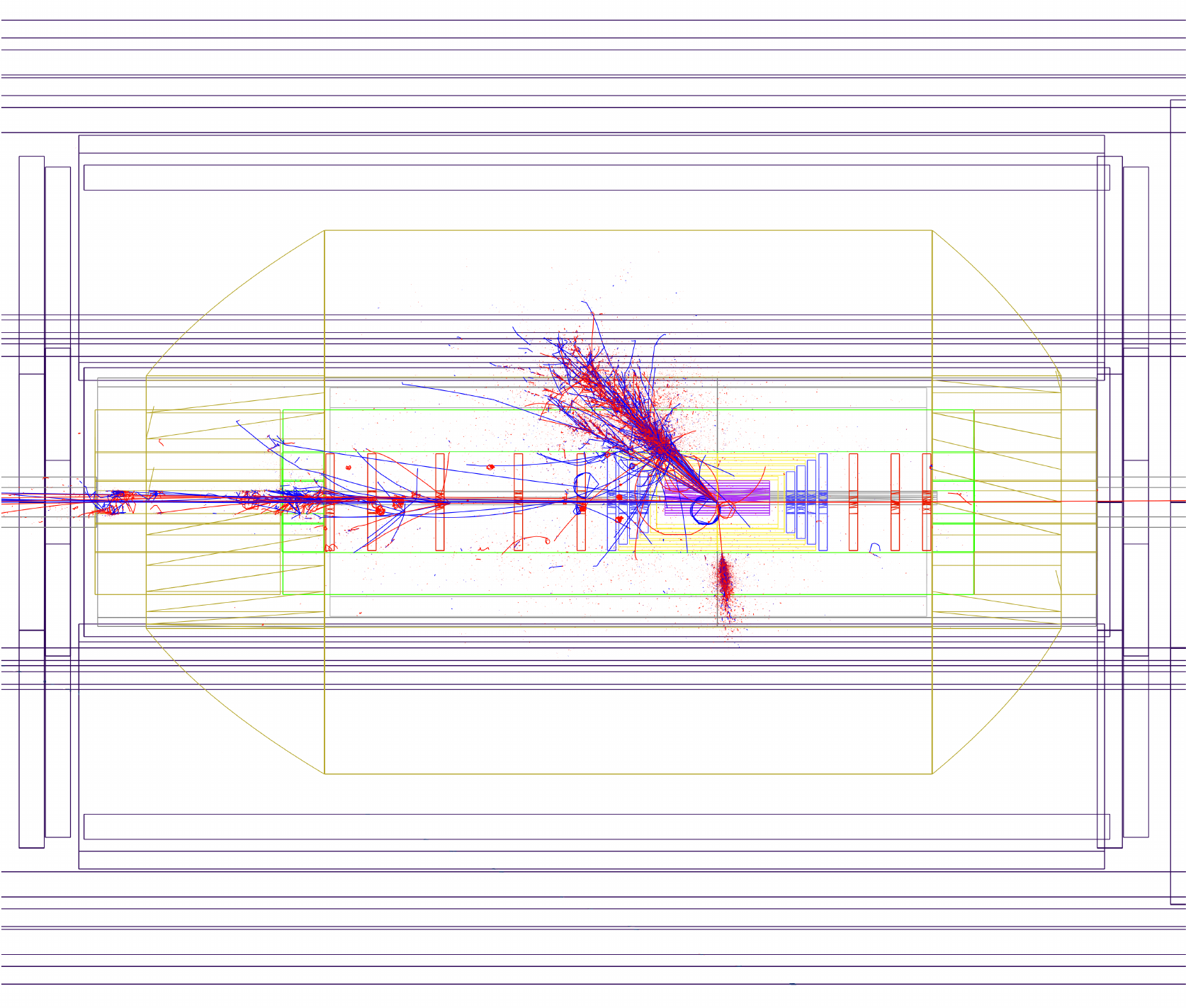}}
\caption{Leptoquark Event RZ}
\label{Fig:leptoquark_rz}
\end{figure}

DAWN was used for visualisation of the detector.  This was able to
produce clear pictures which was one way to make sure the translation
of geometry from {\small\bf ROOT} to {\small\bf GEANT4} went as
expected.  An event in the central tracking region is presented in
Figure\,\ref{Fig:G4_Event_1}.

In addition to the minimum bias events, {\small\bf Pythia6} was also
used to create some Leptoquark events.  This was one method of
checking the {\small\bf Pythia6} input (i.e. that the events produced
describe the given kinematic range and cross sections available).
However it was also utilised to determine the detector response at
various kinematic ranges.  Since $\sigma_{EM} \propto \frac{1}{Q^4}$
The minimum bias events have very low $Q^2$ and therefore very forward
jets, which leaves almost no activity in the barrel HCAL.  By looking
at some high $Q^2$ events it is possible to see the response of the
hadronic calorimetry in the barrel region, making sure it is showering
correctly.  Some pictures of the Leptoquark events are shown in
Figure\,\ref{Fig:leptoquark_xy} and Figure\,\ref{Fig:leptoquark_rz}.

\subsection{Future goals}

There are many goals still to be accomplished by the LHeC Detector
Simulation. The set up needs to be modified to include a detailed
calorimeter description. Currently e.g. the forward/backward
calorimeter volumes contain a mixture of FR4, kapton, active and
passive material which is weighted according to a realistic
setup. This design must be replaced with the actual detector setup of
the calorimeters. This also needs to be done for the tracking which is
currently composed of single silicon pieces instead of smaller
modules. The majority of the work in making these changes comes from
the required read out geometry and sensitive detector set up that
would be required for analysis of a complicated geometrical
structure. This also might require a restructuring of the simulation
package. Since the detector description was done first in {\small\bf
ROOT}, {\small\bf GDML} was an option to allow utilising {\small\bf
GEANT4} without recoding the geometry. However if the geometry will
significantly change then this might benefit from being done natively
in {\small\bf GEANT4}. Of course the geometry needs to be iterated
until it actually describes the exact detector (service pipes, read
out, etc...). This will come in the next phase of the project.

Finally the stability of the simulation needs to be
assessed. Eventually a complex multifunctional detector simulation
package needs to be produced. This is best done by wrapping numerous
simulation toolkits into a single package utilising {\small\bf ROOT},
such as {\small\bf
AliROOT}\,\cite{Carminati:2003pu},\cite{Hrivnacova:2003yy},\cite{GonzalezCaballero:2003ay}
or {\small\bf ILCROOT}\,\cite{Hauptman:2009tv}. The LHeC simulations
at some point need to make a shift towards creating a package like
this, in order to promote greater functionality and greater
accessibility.

%% file: detector/fwdbkwd.tex
\label{detector:fwdbwd}
In this chapter forward and backward detectors are presented.  These
detectors are located between a few tens up to several hundreds of
metres from the interaction point, in order to provide specific
information not accessible by the main detector.  The main focus are
measurements of
\begin{itemize}
\item the instantaneous luminosity (Section\,\ref{LHeC:Detector:LumiDet})
\item the electron or positron beam 
polarisation (Section\,\ref{LHeC:Detector:Polarimeter})
\item very forward diffractive nucleons (Section \,\ref{LHEC:Detector:zdc},\ref{LHeC:Detector:PTagger})
\end{itemize}
The placement of dedicated taggers both forward and backward along the
beam pipe, as discussed in Section \ref{LHeC:Detector:LumiDet} will
also provide additional means to trigger and select data for specific
analyses.
\input{detector/levonian}
\input{detector/zomer}

\input{detector/zdc-armen}

\input{detector/ptagger_pvm}

%% file: detector/levonian.tex
\section{Luminosity measurement and electron tagging}
\label{LHeC:Detector:LumiDet}
%
Luminosity measurement is an important issue for any collider
experiment.  At the LHeC, where precision measurements constitute a
significant part of the physics programme, the design requirement is
to obtain a precision of $\delta{\cal L}=1\%$.\

In addition to an accurate and precise determination of the integrated
luminosity, ${\cal L}$, for the normalisation of physics cross
sections, the luminosity system should allow for fast beam monitoring
with a typical statistical precision of $1\%/$sec for tuning and
optimisation of $ep$-collisions and to provide good control of the
mid-term variations of instantaneous luminosity, $L$.

Rich experience gained by the H1~\cite{H1:1995,H1:2002} and
ZEUS~\cite{ZEUS:2001,ZNEW:2001} Collaborations at HERA was used in the
design studies of the luminosity system for the LHeC.  In particular,
one important lesson to be learnt from HERA is the need to have
several alternative methods for luminosity determination.

For the LHeC, both Linac-Ring (LR) and Ring-Ring (RR) options are
considered as well as high $Q^2$ ($10^{\circ}-170^{\circ}$ acceptance)
and low $Q^2$ ($1^{\circ}-179^{\circ}$ acceptance) detector setups.
This spans a wide range of instantaneous luminosity\footnote{This also
  takes into account the exponential reduction of $L$ during data
  taking in every luminosity fill.}  $L = (10^{32} - 2\cdot
10^{33})$\lun. Hence suitable processes for the three tasks outlined
above should have the following minimal visible cross
sections\footnote{Statistical error has to be small in comparison with
  total error $\delta L_{\rm tot}$ in order not to spoil overall
  accuracy.}:
\begin{itemize}
  \item fast monitoring ($\delta{\cal L}=1\%$/sec         \Ra $10$  kHz) --   $\sv\gtrsim 100 \mu$b,
  \item mid-term control ($\delta{\cal L}=0.5\%$/hour     \Ra $10$   Hz) -- $\sv\gtrsim 100$nb,
  \item physics sample normalisation ($\delta{\cal L}=0.5\%$/week \Ra $0.1$ Hz) -- $\sv\gtrsim 1$nb.
\end{itemize}
The best candidate for luminosity determination is the purely
electromagnetic {\em bremsstrahlung reaction} $ep \to e\gamma + p$
shown in Figure~\ref{Fig:Options}a, which has a large and precisely
known cross section.  Depending on the photon emission angle it is
either called the \BH process (collinear emission) or QED Compton
scattering (wide angle bremsstrahlung).  In addition, Neutral Current
DIS events in a well understood $(x,Q^2)$ range can be used for the
{\em relative} normalisation and mid-term yield control.

While QED Compton and NC DIS processes can be measured in the main
detector, dedicated `tunnel detectors' are required to register \BH
events.  For the latter, additional challenges as compared to HERA are
related to the LHeC configuration: a non-zero beam crossing angle in
the IP for the RR option, and severe aperture limitation for the LR
option.  Finally, for the high luminosity LHeC running one should not
forget about significant pileup ($L$/bunch is $\sim 2-3$ times bigger
as compared to HERA-II running).

\subsection{Options}
%
The huge rate of `zero angle' electrons and photons from the \BH reaction\footnote{Total
cross section, $\sigma_{BH}\simeq 870$ mb for $60 \times 7000$ GeV$^2$ $ep$ collisions at the LHeC.} 
makes a dedicated luminosity system in the tunnel ideal for fast monitoring purposes.
However, it is usually very sensitive to the details of the beam optics at the IP,
may suffer from synchrotron radiation (SR) and requires, for accurate absolute normalisation, 
a large and precisely known geometrical acceptance which is often difficult to ensure. 
On the contrary, the main detector has stable and well known acceptance and is safely shielded
against SR. Therefore, although QED Compton events in the detector acceptance have significantly
smaller rates they may be better suited for overall global normalisation of the physics samples.
Thus the two methods are complementary, having very different systematics and providing
useful redundancy and cross checking for the luminosity determination.
 
To evaluate the main LHeC detector acceptance for NC DIS events and
for the elastic QED Compton process {\tt DJANGOH}~\cite{DJANGO:1991}
and {\tt COMPTON}~\cite{QEDC:1992} event generators were used
respectively.  Different options for dedicated luminosity detectors in
the LHC tunnel have been studied with the help of the special {\tt
  H1LUMI} program package~\cite{H1LUMI:1993}, which contains Monte
Carlo generation of the `collinear' photons and electrons from various
processes (\BH reaction, quasi-real photoproduction, \mbox{e-beam}
scattering on gas in the beam pipe) as well as a simple tracking through the
beam line.\footnote{The tracking has been performed by interfacing
  H1LUMI to GEANT3~\cite{GEANT:1985} having LHeC beam line implemented
  up to $\sim 110$m from the IP.}

\subsection{Use of the main LHeC detector}
\label{sec:MainDet}

To estimate visible cross sections for NC DIS and elastic QED Compton events
a typical HERA analysis strategy was used. That is: safe fiducial cuts against energy
leakage at the backward calorimeter boundaries at small radii, safe $(Q^2,y)$ cuts
for NC DIS events to restrict measurement to the phase space where $F_2$ is known 
to good precision of $1-2\%$ and the $F_L$ contribution is negligible, 
and elasticity cuts for QEDC events to reject the less precisely known inelastic contribution. 
In addition basic cuts against major backgrounds were applied (photoproduction 
in case of NC DIS and DVCS, elastic VM production and low mass diffraction 
in case of QED Compton).

The visible NC DIS cross section is $\sv^{DIS}(Q^2>10{\rm GeV}^2,0.05<y<0.6) \simeq 10$ nb
for the $10^{\circ}$ setup and $\simeq 150$ nb for the $1^{\circ}$ setup.
This corresponds to a $10-15$ Hz rate which is large enough for mid-term yield control.

For elastic QED Compton events, the visible cross section, $\sv^{QEDC} \simeq 0.03$ nb for
$10^{\circ}$ setup and $\simeq 3.5$ nb for $1^{\circ}$ setup. Hence while for the latter
sufficiently high rate is possible even for $L=10^{32}$\lun, in case of `high $Q^2$' setup
the QEDC event rate is $4-5$ times smaller, thus only providing acceptable statistical precision
for large samples, of the order $0.5\%/$month.    

In order to improve this a special small dedicated calorimeter could eventually be added
after the strong focusing quadrupole, at $z=-6$m.
Such a `QEDC tagger' should consist of two movable stations approaching the beam-pipe
from the top and the bottom in the vertical direction, as sketched in Figure~\ref{Fig:Options}b. 
This way the detector sections will be safe with respect to the SR fan confined in the median plane.  
The visible elastic QED Compton cross section for such a device is $4.3 \pm 0.2$ nb which
significantly improves statistics for the luminosity measurement.
The angular acceptance of the `QEDC tagger' corresponds to the range $\theta = 0.5^{\circ}-1^{\circ}$
which lies outside the tracking acceptance. Therefore calorimeter sections should be
supplemented by small silicon detectors in order to make it possible to reconstruct 
the event vertex from the final state containing only one electron and one photon.
These silicon trackers are also useful for $e/\gamma$ separation and rejection of
potential backgrounds.  
Actual dimensions and parameters of this optional `QEDC tagger' requires extra design studies.

\subsection{Dedicated luminosity detectors in the tunnel}
\label{sec:LumiDet}

\begin{figure}[hp]
\centerline{\includegraphics[clip=,width=0.95\textwidth]{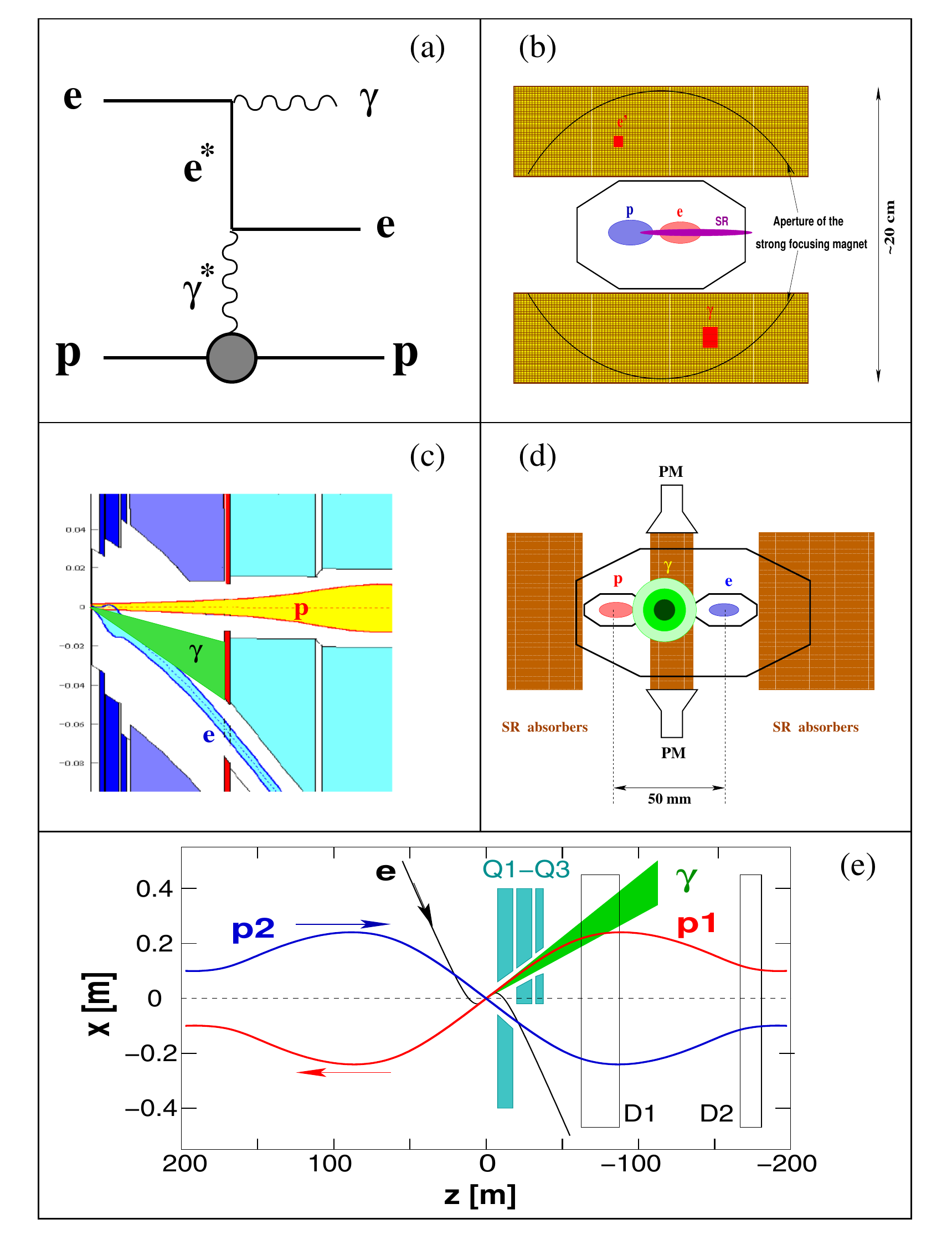}}
\caption{Options for the luminosity monitoring at the LHeC.
         (a) Feynman diagram for QEDC ($\gamma^*$ pole) or 
             BH ($\gamma^*, e^*$ poles) processes;
         (b) QEDC tagger at $z=-6$m; 
         (c,d) active  SR absorber at $z=-22$m for RR-option
               (circles show 1-, 2- and 3-$\sigma$ contours for BH photons);
         (e) schematic view for the LR-option with 3-$\sigma$ fan of
               BH photons.}\label{Fig:Options}
\end{figure}

In the case of the RR-option which implies non-zero crossing angle for early $e/p$ beam separation,
the dominant part of the \BH photons will end up at $z \simeq -22$m, between electron and proton
beam-pipes (see Figure~\ref{Fig:Options}c). This is the hottest place where also a powerful 
SR flux must be absorbed. At first glance this makes luminosity monitoring based upon 
the bremsstrahlung photons impossible.

There is however an interesting possibility. A SR absorber needs a good cooling system. 
The most natural cooling utilises circulating water. This cooling water can be used 
at the same time as an active media for \v{C}erenkov radiation from 
electromagnetic showers initiated by the energetic \BH photons. 
The idea is based on two facts:
\begin{enumerate}
 \item The dominant part of the SR spectrum lies below the \v{C}erenkov threshold for water,
       $E_{\rm thr} = 260$ keV, and hence will not produce a light signal. The low intensity tail
       of the energetic synchrotron photons can be further suppressed by placing a few radiation lengths
       of the absorber material in front of the water volume. 
 \item Water is a very radiation resistant medium and hence such a simple \v{C}erenkov counter
       can stand any dose without performance deterioration.
\end{enumerate}
The \v{C}erenkov light can be collected and read out by two photo-multipliers as sketched on 
Figure~\ref{Fig:Options}d.
The geometric acceptance depends on the details of the $e$-beam optics. For the actual
RR design with the crossing angle $\sim 1$ mrad the acceptance to the \BH photons is up to $90\%$,
thus allowing fast and reliable luminosity monitoring with $3-5\%$ systematic uncertainty.  

Of course, such an active SR absorber is not a calorimeter with good energy resolution, 
but just a simple counter. It is worth noting that a similar water \v{C}erenkov detector 
was successfully used in the H1 Luminosity System during HERA-I operation.

In the case of the LR-option, electrons collide with protons head-on, with zero crossing angle.
This makes the situation very similar to HERA, where \BH photons travel along the
proton beam direction and can be caught at around $z=-120$m, after the first proton bending dipole.
The essential difference is that unlike HERA, LHC protons are deflected horizontally at this place
rather than vertically.
Thus the luminosity detector should be placed in the median plane 
next to the interacting proton beam, $p_1$, as shown on Figure~\ref{Fig:Options}e.
In this case an energy measurement with good resolution is not a problem, so the major uncertainty
will come from an imperfect knowledge of the limited geometric acceptance. This limitation
is defined by the proton beam-line aperture, in particular by the aperture of the quadrupoles
Q1-Q3 of the low-beta proton triplet.
Moreover, it might be necessary to split the D1 dipole into two parts in order to provide
an escape path for the photons with sufficient aperture.
First estimates show that a geometric acceptance of the Photon Detector of up to $95\%$
is possible at the nominal beam conditions. 
HERA experience shows that the uncertainty can be estimated as $\delta A = 0.1\cdot(1-A)$ 
leading to the total luminosity error of $\delta L = 1\%$ in this case. 

\subsection{Small angle electron tagger}
\label{sec:Etag}

The \BH reaction can be tagged not only by detecting a final state
photon, but also by detecting the outgoing electron.  Since all other
competing processes have much smaller cross sections measuring the
inclusive rate of the scattered electrons under zero angle will
provide a clean enough sample for luminosity monitoring. The remaining
small background (mainly due to off-momentum electrons from $e$-beam
scattering on gas in the beam pipe) can be precisely controlled and
statistically subtracted using non-colliding ({\em pilot}) electron
bunches.

\begin{figure}[htb]
\centerline{\includegraphics[clip=,width=\textwidth]{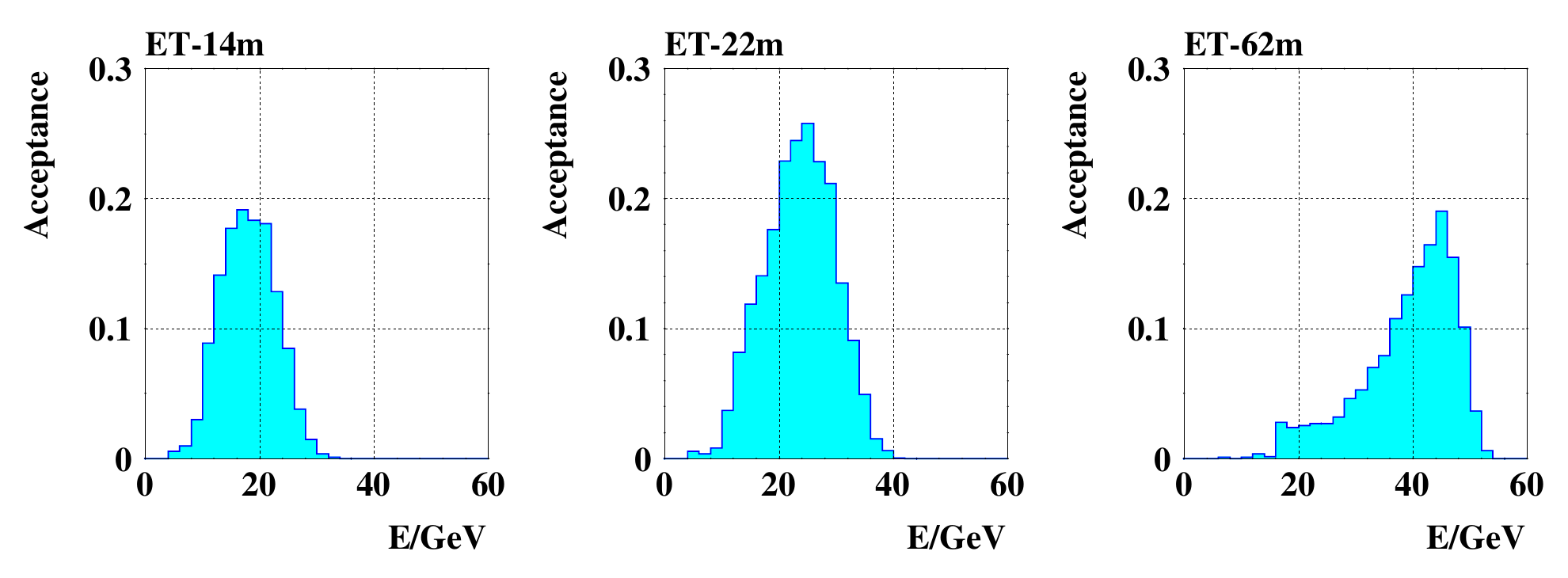}}
\centerline{\includegraphics[clip=,width=\textwidth]{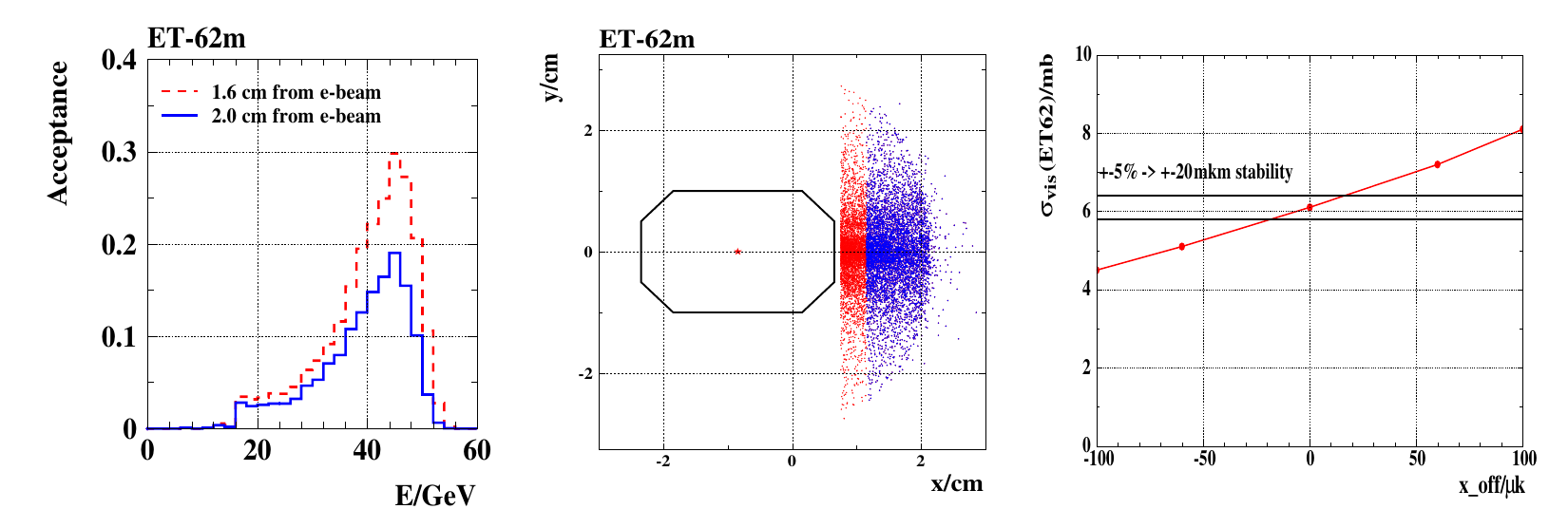}}
\caption{Top: acceptances of the $e$-taggers for \BH events at different $z$-positions from IP (RR-option).
        Bottom: variations in the acceptance of the $e$-tagger at $z=-62$m as a function of its
                position with respect to the $e$-beam axis and on the horizontal offset of the
                beam orbit at the IP.}\label{Fig:Etags}
\end{figure}

In order to determine the best positions for the Electron taggers the LHeC beam line simulation 
has been performed in the vicinity of the Interaction Region for the RR-option.
Several positions for the $e$-tagger stations were tried:\footnote{For the station at $z=-14$m
the electron dipole magnet should be split into two parts, while
the region around $z=-62$m has sufficient space for the Electron tagger,
before the $e$-beam is bent vertically.}
$z=-14$m, $-22$m and $-62$m.
As can be seen on the top part of Figure~\ref{Fig:Etags} all positions provide 
reasonable acceptances, reaching approximately $(20-25)\%$ at the maximum.
However, $z=-14$m and $z=-22$m will most likely  suffer from SR flux, making
$e$-tagger operation problematic at those positions.  

The most promising position for the Electron tagger is at $z=-62$m.
The actual acceptance strongly depends both on the distance of the sensitive detector volume
from the $e$-beam axis and on the details of the electron optics at the IP, such as
beam tilt or small trajectory offset, as illustrated on the bottom part of Figure~\ref{Fig:Etags}. 
Therefore a precise independent monitoring of beam optics and accurate
position measurement of the $e$-tagger are required  
in order to control geometrical acceptance to a sufficient precision.
For example, instability in the horizontal trajectory offset at the IP, $x_{\rm off}$, of $\pm 20\mu$m
leads to the systematic uncertainty of $5\%$ in the visible cross section, $\sv(ET62)$.

It should be noted that the magnetic field of the main LHeC detector was not taken into account
in the simulation. The influence of this field is expected to be very small and will not
alter the basic conclusions of this section. Also, for the LR-option a similar acceptance is expected,
although it may differ somewhat in shape. 
 
\begin{figure}[hb]
\centerline{\includegraphics[clip=,width=0.95\textwidth]{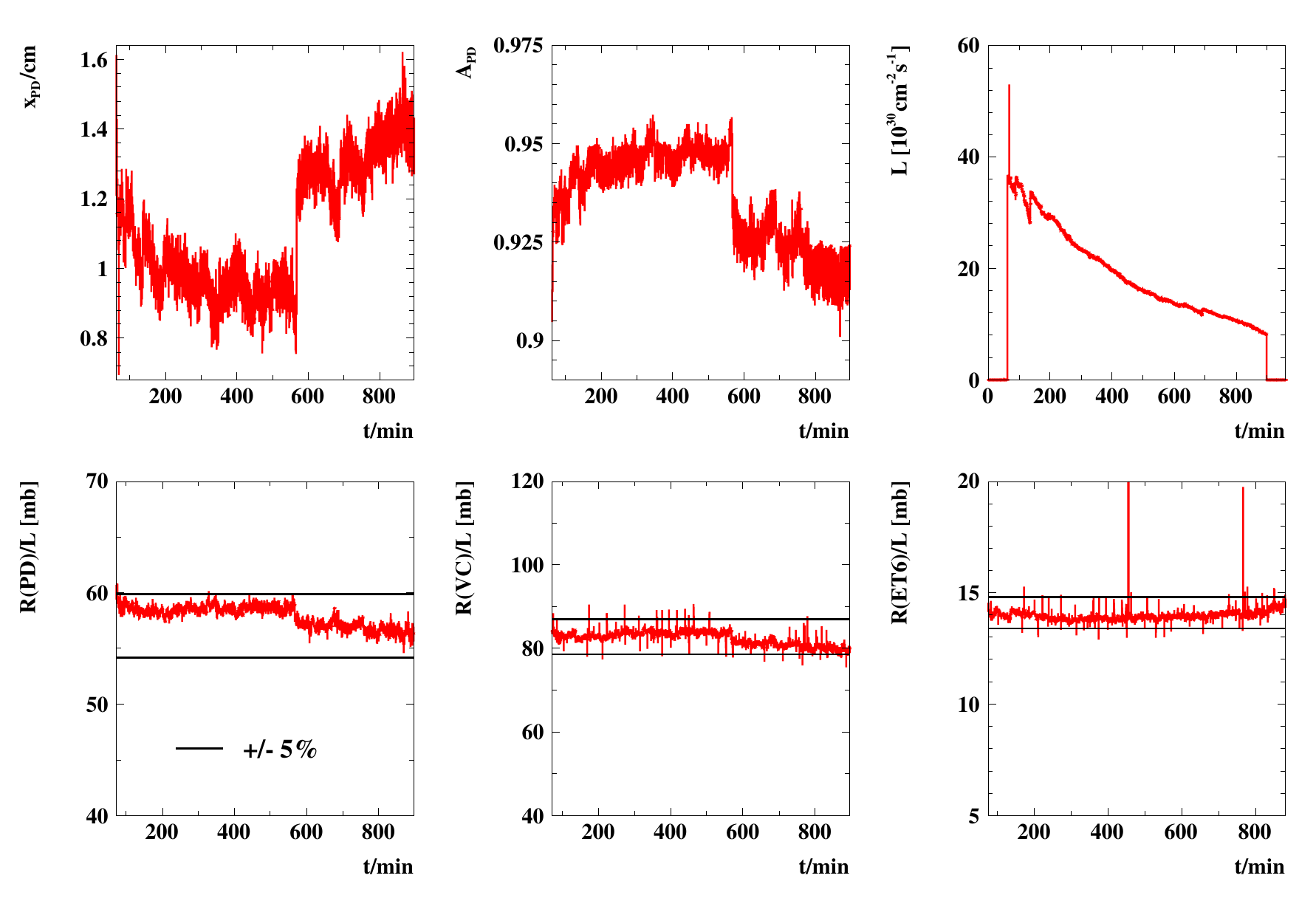}}
\caption{Online H1 Lumi System acceptance and rate variations in a typical 
         HERA luminosity fill.}\label{Fig:Rates}
\end{figure}

In order to demonstrate that the ideas described in Sec.~\ref{sec:LumiDet} and \ref{sec:Etag}
are realistic a typical example of the online rate variations for the H1 Luminosity System
at HERA is shown on Figure~\ref{Fig:Rates}. The system utilised all three types of the detectors
discussed above: a total absorption electromagnetic calorimeter for the \BH photons (PD), 
a water \v{C}erenkov counter (VC) and the Electron tagger (ET6). It can be seen that the online
luminosity estimate by each of the detectors agree well within $5\%$ in spite of 
significant changes in the acceptance due to electron beam tilt jumps and adjustments at the IP. 

\subsection{Summary and open questions}
%
An accurate luminosity measurement at the LHeC is a highly non-trivial
task.  As follows from experience at HERA, unexpected surprises are
possible, hence it is important to consider several solutions from the
beginning and to prepare alternative methods for luminosity
determination.

The statistical precision and systematic uncertainties for different
methods of luminosity measurement are summarised in
Table~\ref{tab:Lsyst}.

\begin{table}[hbt]
  \centering
  \begin{tabular}{|l r r l r l|}
\hline
 Method        & Stat. error  &  Syst.error & 
                \multicolumn{2}{c}{Systematic error components} & Application \\
\hline \hline
 BH ($\gamma$) & $0.05\%$/sec & $1\!-\!5\%$ & $\sigma(E\gtrsim 10$GeV) & $0.5\%$  & Monitoring, tuning,   \\
               &              &             & acceptance, $A$   & $10\%(1\!-\!A)$ & short term variations \\
               &              &             & $E$-scale, pileup & $0.5-4\%$       &  \\
\hline
 BH ($e$)      & $0.2\%$/sec  & $3\!-\!6\%$ & $\sigma(E\gtrsim 10$GeV) & $0.5\%$ & Monitoring, tuning, \\
               &              &             & acceptance     & $2.5-5\%$      & short term variations  \\
               &              &             & background     & $1\%$          &                     \\
               &              &             & $E$-scale      & $1\%$          &                     \\
\hline
 QEDC          & $0.5\%$/week & $1.5\%$ & $\sigma$ (el/inel)  & $1\%$       & Absolute ${\cal L}$, \\
               &              &         & acceptance          & $1\%$       & global normalisation \\
               &              &         & vertex eff.         & $0.5\%$     &                      \\
               &              &         & $E$-scale           & $0.3\%$     &                      \\
\hline
 NC DIS        &  $0.5\%$/h   & $2.5\%$ & $\sigma$ ($y<0.6$)  & $2\%$       & Relative ${\cal L}$, \\
               &              &         & acceptance          & $1\%$       & mid-term variations  \\
               &              &         & vertex eff.         & $1\%$       &                      \\
               &              &         & $E$-scale           & $0.3\%$     &                      \\
\hline

\end{tabular}
\caption{Dominant systematics for various methods of luminosity measurement.}
\label{tab:Lsyst}
\end{table}

The most precise determination of the integrated luminosity, ${\cal L}$, is possible
with the main detector utilising the QEDC process, where $\delta {\cal L} = 1.5-2\%$ is possible.
Further improvement requires in particular a more accurate theoretical calculation of the
elastic QED Compton cross section, with $\delta \sigma_{\rm el}^{\rm QEDC} \lesssim 0.5\%$. 
To enhance the statistical precision, a dedicated QEDC tagger at $z=-6$m may be useful.
This device could also be used to access the very low $Q^2$ region, interpolating between
the DIS and photoproduction regimes.
 
Fast instantaneous luminosity monitoring is challenging, but several options do exist
which are based upon detection of the photons and/or electrons from the \BH process.
\begin{itemize}
 \item A Photon Detector at $z=110$m for the LR option requires a properly shaped proton beam-pipe
       at $z=-68-120$m from IP2.
 \item In the case of the RR option \BH photons can be detected using a water \v{C}erenkov
       counter integrated with SR absorber at $z=-22$m.
 \item An Electron tagger at $z=-62$m is very promising for both LR and RR schemes.
       It can be used not only for luminosity monitoring, but also to enhance the
       photoproduction physics capabilities and to provide extra control of the 
       $\gamma p$ background to DIS, by tagging quasi-real photoproduction events.
\end{itemize} 
Good monitoring of the $e$-optics at the IP is required to control the acceptance 
of the tunnel detectors to a level of $2-5\%$.

%% file: detector/zomer.tex
\section{Polarimeter}
\label{LHeC:Detector:Polarimeter}
The most powerful technique to measure the polarisation of the
electrons and positrons of LHeC is Compton polarimetry. At high
electron beam energies, this technique has been successfully used in
the past at SLC\cite{King:1994gw} and at HERA\cite{Barber:1992fc} for example. The
experimental setup consists of a laser beam which scatters off the
electron/positron beam, and a calorimeter to measure the scattered
gamma ray. At SLC, the scattered electron was also measured in a
dedicated spectrometer. From the kinematics of Compton scattering one
can get the expression for the maximum scattered photon energy:
$$
E_{\gamma,max}\approx E_0\frac{x}{1+x}
$$
and the minimum scattered electron energy
 $$
E_{e,min}\approx E_0\frac{1}{1+x},
$$
where $E_0$ is the electron/positron beam energy and $x=4kE_0/m_e^2$
with $k$ being the laser photon beam energy. At LHeC and for a
$\approx 1\mu$m laser beam wavelength, one gets $E_{\gamma,max}\approx
29$GeV and $E_{e,min}\approx 31$GeV.  Providing that the laser beam is
circularly polarised, the electron/positron beam longitudinal
polarisation is obtained from a fit to the scattered photon and/or to
the electron energy spectrum. From an experimental point of view, both
measurements can be complementary since the high energy region of the
scattered photon energy spectrum is sensitive to the electron/positron
beam longitudinal polarisation, whereas it is the opposite for the
scattered electron/positron energy spectrum.  Indeed, the high
measurement precision of SLC was achieved thanks to the measurement of
the scattered electrons. The measurement of both scattered photon and
electron/positron spectra was therefore foreseen for a very high
precision polarimetry at future electron-positron high energy
colliders\cite{MoortgatPick:2005cw, Boogert:2009ir}.

For LHeC, we may follow the work done for the future linear colliders
\cite{Boogert:2009ir}. In order to reach the per mille level on the
longitudinal polarisation measurement, one may measure both the
scattered photon and electron energy spectra.

\subsection{Polarisation from the scattered photons}

The photons are scattered within a very narrow cone of half aperture
$\approx 1/\gamma$. It is therefore impossible to distinguish the
photons reaching the calorimeter. As for the extraction of the
longitudinal polarisation from the scattered photon beam energy, one
may then distinguish three dynamical regimes\cite{Baudrand:2010hp}. The
single and few scattered photons regimes, where one can extract the
polarisation from a first principle fit to the scattered photon energy
spectrum; the multi-photon regime where the central limit theorem
holds for the energy spectra and where the longitudinal polarisation
is extracted from an asymmetry between the average scattered energies
corresponding to a circularly left and right laser beam polarisation
\cite{Beckmann:2000cg}. Both regimes have positive and negative experimental
features. In the single and few photon regimes the energy spectra
exhibits kinematic edges which allow an in situ calibration of the
detector energy response but the physical accelerator photon
background which is difficult to model precisely, e.g. synchrotron
radiation, limits the final precision on the polarisation measurement
\cite{Baudrand:2010hp}. In the multi-photon regime, the background is
negligible since it is located at low energy but one cannot measure
the energy calibration of the detector in situ and one must rely on
some high energy extrapolation of calibrations obtained at low
energy\cite{Beckmann:2000cg} (e.g. for 100 scattered photon/bunch the deposited
energy in the calorimeter would be more than 1TeV at LHeC). However,
the laser technology has improved in the last ten years and one can
consider at present a very stable pulsed laser beam with adjustable
pulse energy allowing to operate in single, few and multi photon
regimes. In this way, one can calibrate the calorimeter in situ and
optimise the dynamical regime, a multi-photon regime as close as
possible to the few photon regime, in order to minimise the final
uncertainty on the polarisation measurement.

\subsection{Polarisation from the scattered electrons}

The nice feature of the scattered electron/positron is that one can
use a magnetic spectrometer to distinguish them from each other.
Following\cite{Boogert:2009ir} one may carefully design a Compton
interaction region in order to implement a dedicated electron
spectrometer followed by a segmented electron detector in order to
measure the scattered electron angular spectrum, itself related to the
electron energy spectrum. A precise particle tracking is needed but
this experimental method also allows a precise control of the
systematic uncertainties\cite{King:1994gw}.

Common to both techniques is the control and measurement of the laser
beam polarisation. it was shown in\cite{Brisson:2010hq} that a few per
mille precision can be achieved in an accelerator
environment. Therefore, with a redundancy in measuring the
electron/positron beam longitudinal polarisation from both the
electron and photon scattered energy spectra, a final precision at the
per mille level will be reachable at LHeC.

%% file: detector/zdc-armen.tex
\section{Zero degree calorimeter}
\label{LHEC:Detector:zdc}

\begin{figure}[h]
\centering
\includegraphics[width=0.75\textwidth]{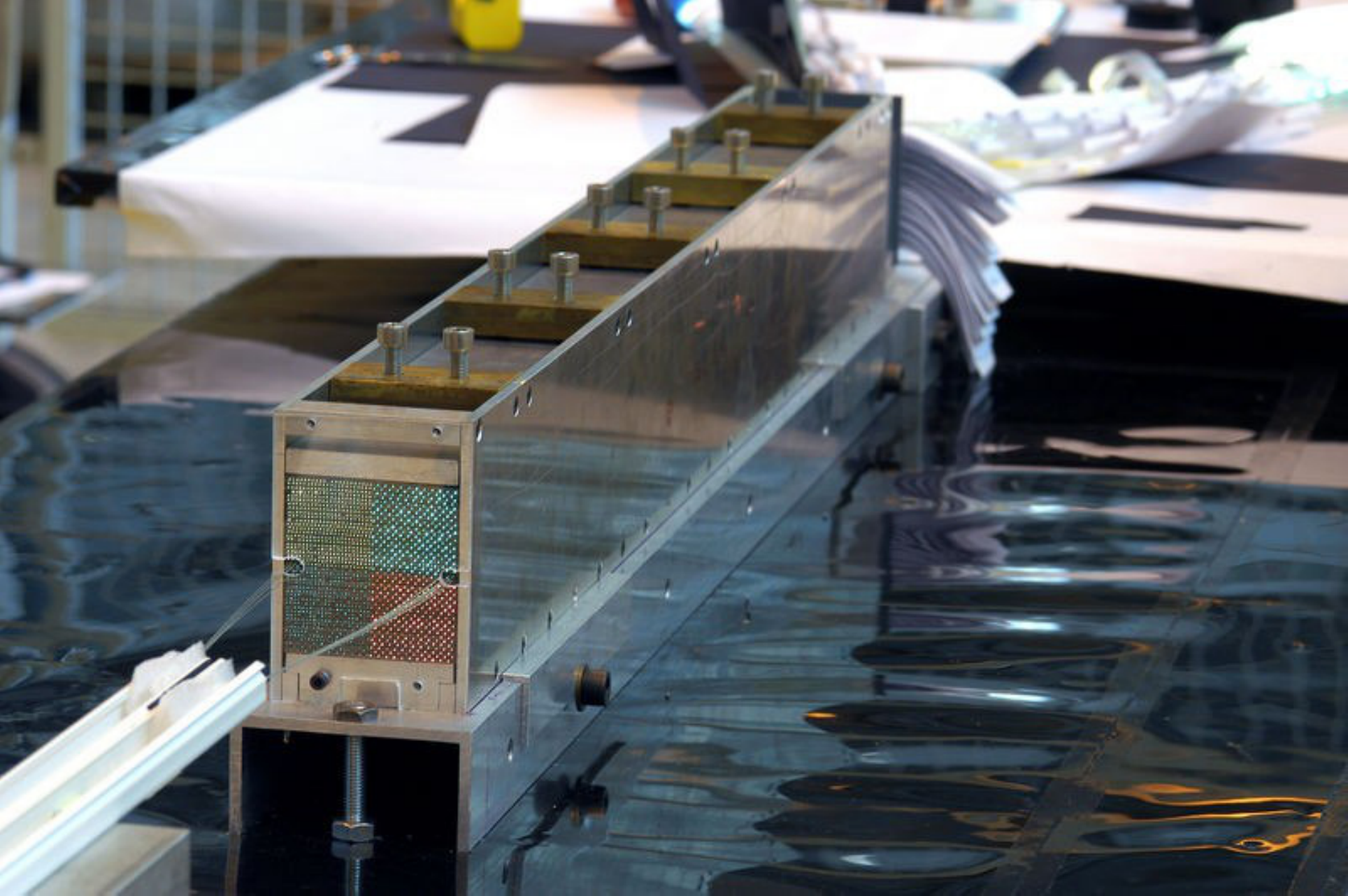}
 \caption{Photograph of the Zero Degree Neutron Calorimeter of the ALICE experiment
  \cite{Arnaldi:1999zz}.}
\label{alicezn}
\end{figure}

The goal of a Zero Degree Calorimeter (ZDC) is to measure the
energies and angles of very forward particles.
At the HERA experiments, H1 and ZEUS, forward neutral particles
scattered at polar angles below 0.75~mrad were measured in
dedicated Forward Neutron Calorimeters (FNC)
\cite{Bhadra:1997vi,Aaron:2010ze}.
%
The LHC experiments, CMS, ATLAS, ALICE and LHCf, have ZDC
calorimeters
for the detection of forward neutral particles
\cite{Arnaldi:1999zz,DeMarco:2009zz,ATLAS:2007jwa,Grachov:2010th,Adriani:2008zz},
while ALICE also has a ZDC calorimeter for measurements of spectator
protons.  A photograph of the ALICE neutron
calorimeter~\cite{Arnaldi:1999zz,DeMarco:2009zz} is shown in
Figure~\ref{alicezn}).

A ZDC calorimeter is an important component of the LHeC experiment as
many physics measurements in $ep$, $ed$ and $eA$ collisions are made
possible with the installation of a ZDC.

\begin{figure}[ht]
\centering
\includegraphics[width=0.8\textwidth]{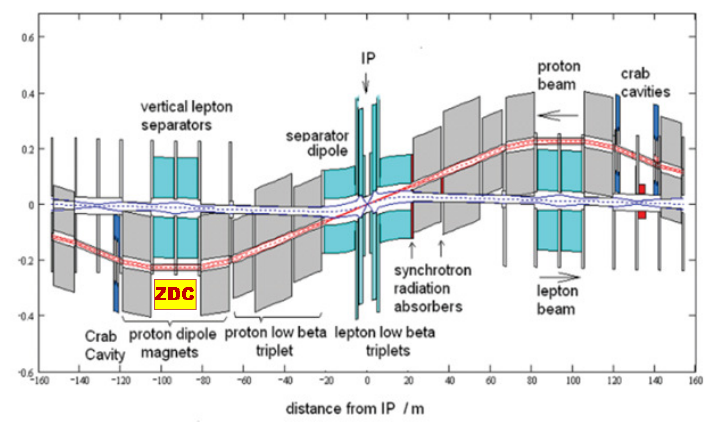}
\caption{Schematic layout of the LHeC interaction region. The possible
position of the ZDC is indicated.}
\label{lheczdc}
\end{figure}

 \subsection{ZDC detector design}

The position of the Zero Degree Calorimeter in the tunnel and its
overall dimensions depend mainly on the space available for
installation.
At the LHeC the beams are deflected by two separating dipoles.  These
dipoles also deflect the spectator protons, separating them from the
neutrons and photons, which scatter at polar angle $\sim 0^\circ$.

The geometry, technical specifications and proposed design of the ZDC
detector are to a large extent similar to the ZDCs of the other LHC
experiments.  There the ZDC calorimeters for detection of neutral
particles are placed at $z=115-140$~m in a narrow $\sim 90$~mm space
between the two beam pipes.  In the case of the LHeC, the ZDC calorimeter
can be placed in the space available at about $90-100$~m next to the
interacting proton beam pipe, as indicated in Figure~\ref{lheczdc}.

Below the general considerations for the design are presented.  In
order to finalise the study of the geometry of detectors, a detailed
simulation of the LHeC interaction region and the beam line must be
performed.


\subsection{Neutron calorimeter}
The design of the ZDC has to satisfy various technical issues.  The
detector has to be capable of detecting neutrons and photons produced
with scattering angles up to 0.3~mrad or more and energies between
several hundred GeV to the proton beam energy (7~TeV) with a
resolution of a few percent. It must be able to distinguish hadronic
and electromagnetic showers (i.e. separate neutrons from photons) and
to distinguish showers from two or more particles entering the
detector (i.e. needs position resolution of $\mathcal{O}$(1mm) or
better).
The ZDC will operate in a very demanding radiation environment, and therefore
it has to be made of radiation resistant materials.

The neutron ZDC can be built as a longitudinally segmented
tungsten-quartz calorimeter.  In this design the ZDC will contain both
electromagnetic and hadronic sections. The electromagnetic section,
with 1.5-2 nuclear interaction lengths ($\lambda_I$), has fine
granularity needed for the precise determination of the position of the
impact point, discrimination of the electromagnetic and hadronic
showers and separation of the showers from two or more particles
entering the detector.
The hadronic section of the ZDC can be built with coarser sampling,
which will increase the average density and, consequently, the
effective nuclear interaction length.  The total depth of the
calorimeter will be about 8-9 $\lambda_I$, which will allow for more
than 90\% containment of hadronic showers of $\mathcal{O}$(TeV)
energies.  Since the different parts of the calorimeter are subject to
a different intensity of radiation (higher for the front part), it is
advantageous to have longitudinal segmentation of 3-4 identical
sections.  Comparison of the energy spectrum from showers which start
in different sections can be used to correct for changes in energy
response and thus mitigate the effects due to radiation damage.

\begin{figure}[h]
\centering
\includegraphics[width=0.8\textwidth]{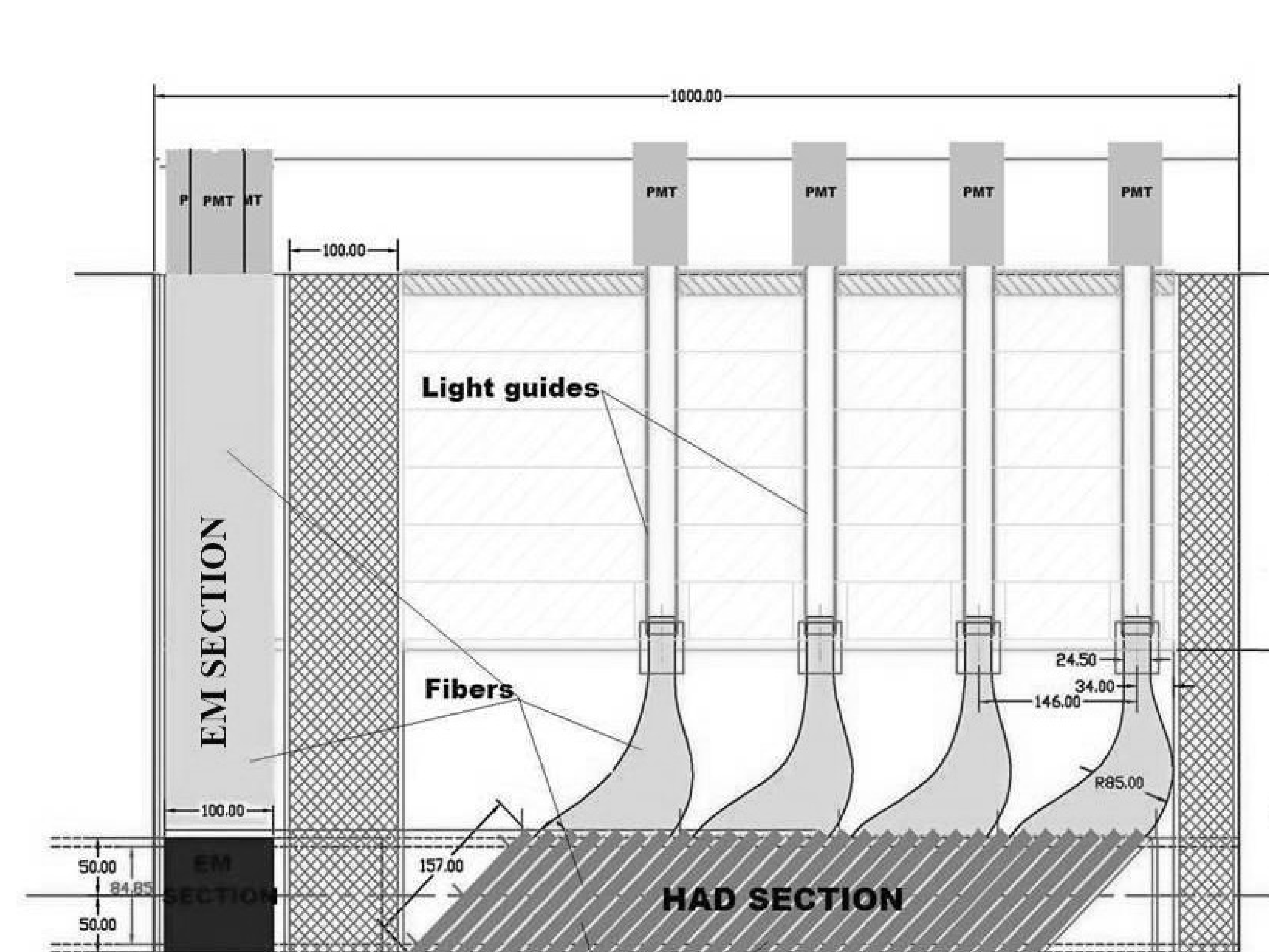}
\vspace*{5mm}
\caption{A side view of the CMS ZDC calorimeter with electromagnetic section in front
 and hadronic section behind.}
\label{cmszdc}
\end{figure}

The CMS Experiment built a compact calorimeter with good radiation
resistance using tungsten absorbers and quartz fibres
\cite{Grachov:2010th} (a schematic view is shown in Fig.\ref{cmszdc}).
The principle of operation is based on the detection of \v{C}erenkov
light produced by the shower's charged particles in the fibres.  Using
tungsten as a passive material allows the construction of compact
devices\footnote{Another option would be to use THGEM, thick gaseous
electron multipliers, as an active media
\cite{Chechik:2004wq,Inshakov:2009db}.}.  These detectors are proven to
be fast ($\sim$few ns) and radiation hard.  Tungsten-quartz technology
is used in the ZDC calorimeters implemented by the CMS, ATLAS and
ALICE experiments \cite{Arnaldi:1999zz,ATLAS:2007jwa,Grachov:2010th}.
However, these calorimeters, based on the detection of \v{C}erenkov
light, are sensitive mainly to the electromagnetic component of the
hadronic shower.  Therefore, they are highly non-compensating and the
energy resolution for hadronic showers is not very high, e.g. the
hadronic energy resolution for the CMS ZDC is $\sigma(E)/E\approx
176\%/\sqrt{E[GeV]}\oplus 8\%$ \cite{Grachov:2006ke}.

An interesting new solution for the ZDC calorimeter is offered
by the Dual Readout calorimetry technique, which is
currently being developed within the DREAM/RD52 Project~\cite{rd52}.
In this approach the detector is equipped with both scintillating 
and quartz fibres, which are sensitive to the different
components of the hadronic shower.
Hadronic showers developing in this detector generate signals in both 
types of fibre and these signals provide complementary information
about the showers. With this experimental method, the dominant source of
fluctuations contributing to the hadronic energy resolution
can be eliminated, since it allows for a measurement of the 
electromagnetic energy fraction event-by-event \cite{AkchurinWigmans}.
In this solution, the readout of the ZDC calorimeter for the LHeC
detector would use SiPM.  The readout from the scintillating fibres
can be made on both ends of the fibres, providing a handle on the
effects of radiation damage.  The discrimination between neutrons and
photons will be possible using the time structure of the signals.
With the prototype tested by the DREAM Collaboration, depth
resolutions of the order of 10~cm has been reached, which is
sufficient to distinguish between neutrons and photons in such a
longitudinally unsegmented calorimeter
\cite{AkchurinWigmans,WigmansPC}.

\subsection{Proton calorimeter}
In addition to the ZDC calorimeter for the measurement of neural
particles at $0^\circ$, a proton calorimeter positioned externally to
the outgoing proton beam can be installed for the measurement of
spectator protons from $eD$ and $eA$ scattering produced at zero
degree.  In analogy to the ALICE
experiment~\cite{Arnaldi:1999zz,DeMarco:2009zz}, this detector can be
positioned at approximately the same distance from the interaction point as
the neutron ZDC.
The size of the proton ZDC has to be small, due to the small size of
the spectator proton spot (a few cm), but sufficient to obtain shower
containment.  The same techniques can be used as for the neutron ZDC.

\subsection{Calibration and monitoring}
After the initial calibration of the ZDCs with test-beams, it is
essential to have regular online and offline control of the stability
of their response, in particular because of the difficult radiation
and temperature environment.  The stability of the gain of the PMTs
and the radiation damage in fibres can be monitored using a laser or
LED light pulses.  The stability of the absolute calibration can be
monitored using the interaction of the proton beam with residual gas
molecules in the beam-pipe and comparison with the results of Monte
Carlo simulation based on pion exchange, as was done at
HERA~\cite{Bhadra:1997vi,Aaron:2010ze}.
A useful tool for absolute energy calibration will be the
reconstruction of invariant masses, e.g.  $\pi^0\rightarrow 2\gamma$
or $\Lambda,\Delta \rightarrow n\pi^0$, where the decay particles are
produced at very small opening angles and reconstructed in the ZDC.
It is therefore essential that the ZDC be capable of reconstructing
and resolving several particles within the same event.

%% file: detector/ptagger_pvm.tex
\section{Forward proton detection}
\label{LHeC:Detector:PTagger}
In diffractive interactions between protons or between an electron and
a proton, the proton may survive a hard collision and be scattered at
a low angle $\theta$ along the beam line while losing only a small
fraction $\xi$ ($\sim 1\%$) of its energy.  The ATLAS and CMS
collaborations have investigated the feasibility of installing detectors
along the LHC beam line to measure the energy and momentum of such
diffractively scattered protons\cite{Albrow:2008pn}.  Since the proton
beam optics is primarily determined by the shape of the accelerator -
which will not change for the proton arm of the LHeC - the conclusions
reached in this R\&D study are still relevant for an LHeC detector.

In such a setup, diffractively scattered protons, which have a
slightly lower momentum, are separated from the nominal beam when
travelling through dipole magnets.  This spectroscopic behaviour of the
accelerator is described by the energy dispersion function, $D_x$,
which, when multiplied with the actual energy loss, $\xi$, gives the
additional offset of the trajectory followed by the off-momentum
proton:
\begin{equation*}
x_\text{offset} = D_x \times \xi.
\end{equation*}

The acceptance window in $\xi$ is therefore determined by the closest
possible approach of the proton detectors to the beam for low $\xi$
and by the distance of the beam pipe walls from the nominal proton
trajectory for high $\xi$.  The closest possible approach is often
taken to be equal to $12\sigma$ with $\sigma$ equal to the beam width
at a specific point. At the point of interest, 420m from the
interaction point, the beam width is approximately equal to 250 $\mu$m.
On the other hand, the typical LHC beam pipe radius at large distances
from the interaction point is approximately 2 cm.  Even protons that
have lost no energy, will eventually hit the beam pipe wall if they
are scattered at large angles.  This therefore fixes the maximally
allowed four momentum-transfer squared $t$, which is approximately
equal to the square of the transverse momentum $p_T$ of the scattered
proton at the interaction point.

\begin{figure}[htbp]
\begin{center}
\includegraphics[width=0.75\textwidth]{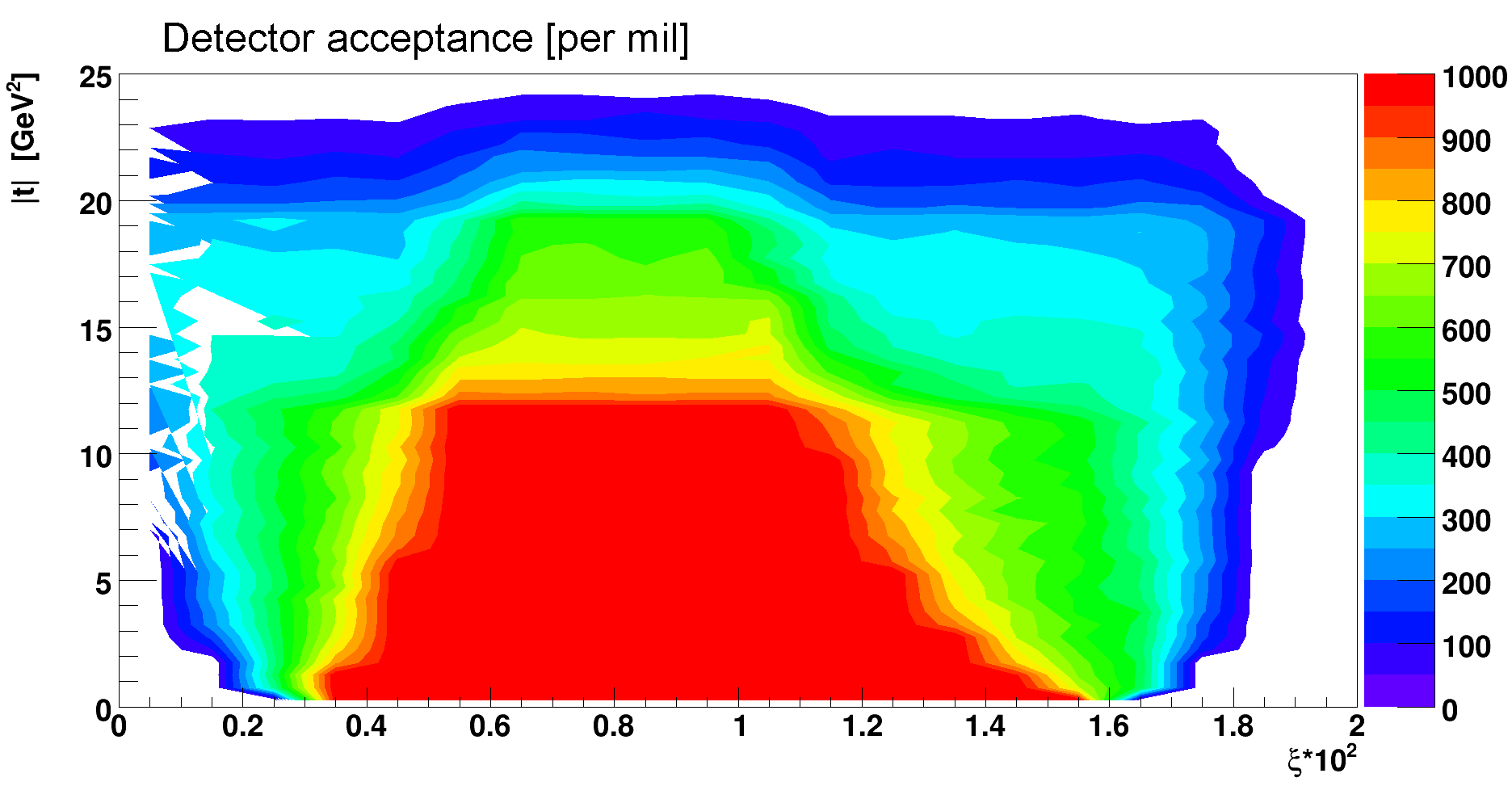}
\caption{The acceptance for a proton detector placed at 420m from the
  interaction point is shown as function of the momentum loss $\xi$
  and the four momentum-transfer squared $t$.  The colour legend runs
  from 0\permil (no acceptance) to 1000\permil (full acceptance).}
\label{fig:ptaggeracceptance}
\end{center}
\end{figure}

At 420 m from the interaction point, the dispersion function at the
LHC reaches 1.5~m, which results in an optimal acceptance window for
diffractively scattered protons (roughly $0.002 < \xi < 0.013$).  The
acceptance as function of $\xi$ and $t$ is shown in
Fig.\@~\ref{fig:ptaggeracceptance}, using the LHC proton beam optics
\cite{taels_thesis}.  The small corrections to be applied for the LHeC
proton beam optics are not considered to be relevant for the
description of the acceptance.

When the proton's position and angle w.r.t.\@ the nominal beam can be
accurately measured by the detectors, it is in principle possible to
reconstruct the initial scattering angles and momentum loss of the
proton at the interaction point.  Even with an infinitesimally small
detector resolution, the intrinsic beam width and divergence will
still imply a lower limit on the resolution of the reconstructed
kinematics.  As the beam is typically maximally focused at the
interaction point in order to obtain a good luminosity, it will be the
beam divergence that dominates the resolution on reconstructed
variables.

\begin{figure}[htbp]
\begin{center}
\includegraphics[width=0.49\textwidth]{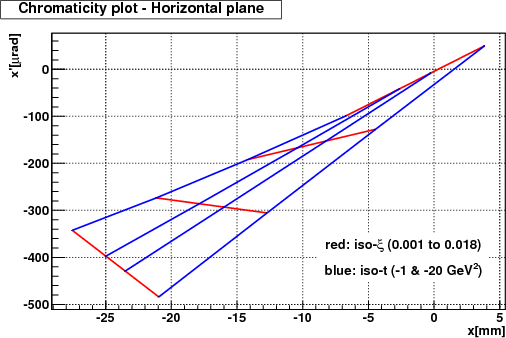}
\includegraphics[width=0.49\textwidth]{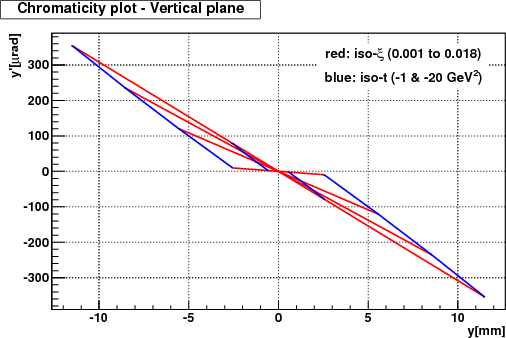}
\caption{Lines of constant $\xi$ and $t \approx (1-\xi) E_\text{beam}
  \theta^2$ are shown in the plane of proton position and angle
  w.r.t.\@ the nominal proton beam in the horizontal (left) and
  vertical (right) plane.}
\label{fig:ptaggerkinematics}
\end{center}
\end{figure}

Figure~\ref{fig:ptaggerkinematics} shows the relation between position and
angle w.r.t.\@ the nominal beam and the proton scattering angle and
momentum loss in both the horizontal and vertical plane as obtained
from the LHC proton beam optics\cite{taels_thesis}.  Clearly, in
order to distinguish angles and momentum losses indicated by the
curves in Fig.\@~\ref{fig:ptaggerkinematics}, the detector must have a
resolution better than the distance between the curves.

\begin{figure}[htbp]
\begin{center}
\includegraphics[width=0.49\textwidth]{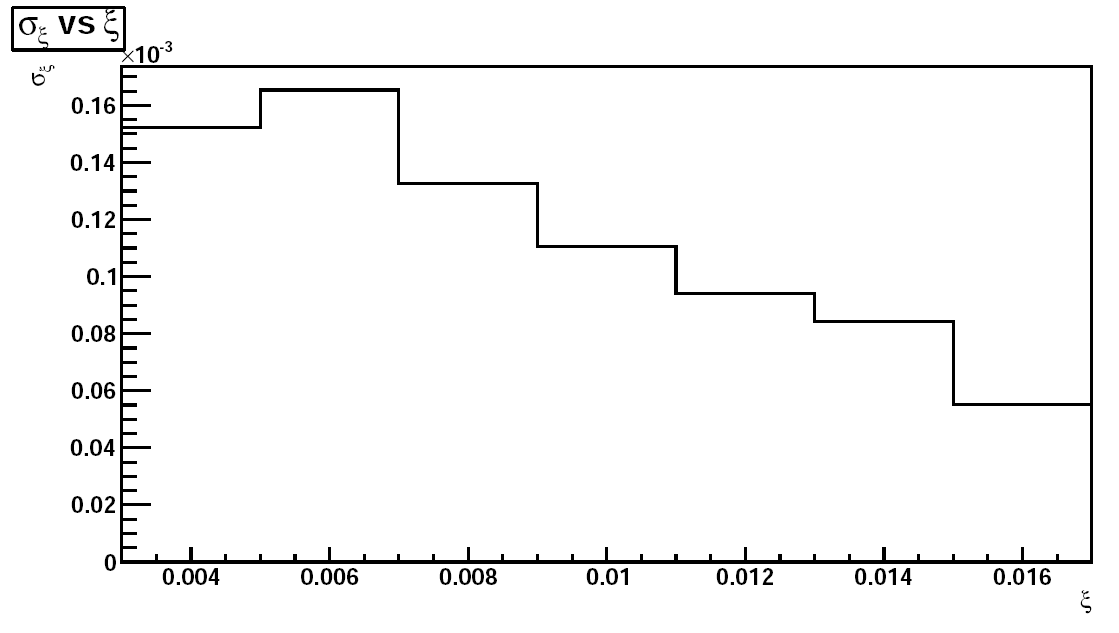}
\includegraphics[width=0.49\textwidth]{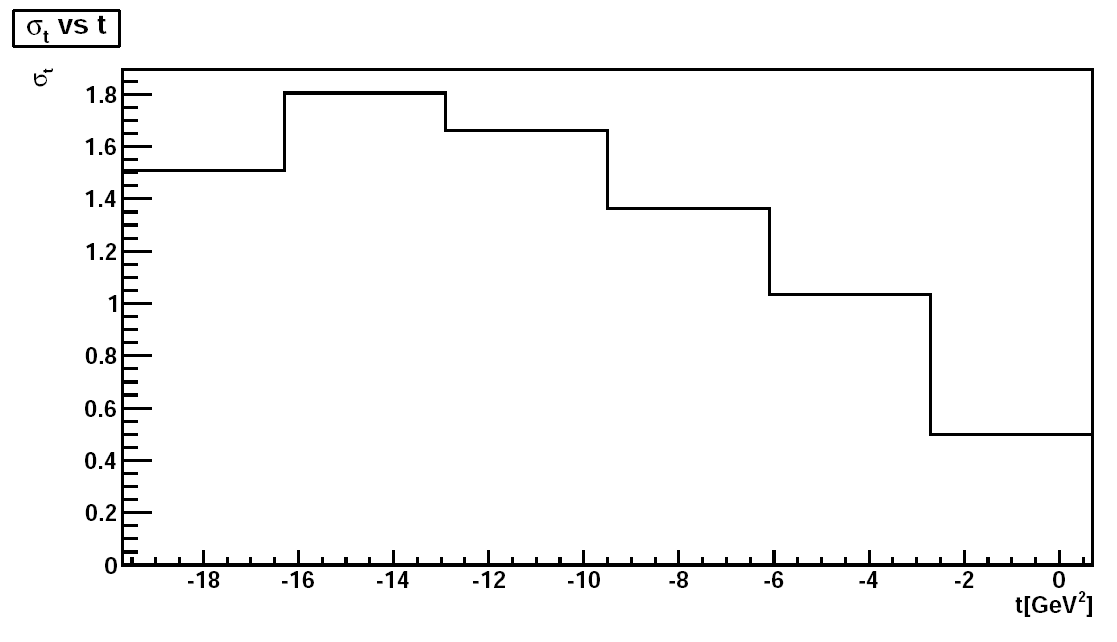}
\includegraphics[width=0.49\textwidth]{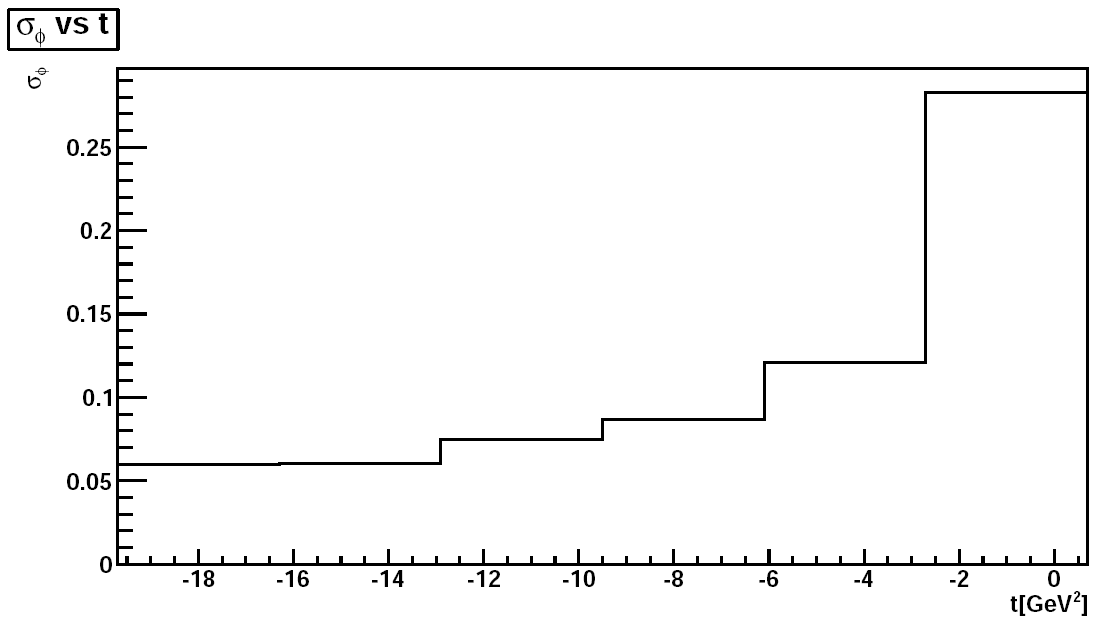}
\caption{The lower limit due to the intrinsic beam width and
  divergence on the resolution of kinematic variables is shown for
  $\xi$ as a function of $\xi$ (top left), $t$ as a function of $t$ (top right)
  and $\phi$ as a function of $t$ (bottom).}
\label{fig:ptaggerresolutions}
\end{center}
\end{figure}

As stated above, protons with the same momentum loss and scattering
angles will still end up at different positions and angles due to the
intrinsic width and divergence of the beam.  Lower limits on the
resolution of reconstructed kinematics can therefore be determined.
These are typically of the order of 0.5\permil\ for $\xi$ and 0.2
$\mu$rad for the scattering angle
$\theta$. Figure~\ref{fig:ptaggerresolutions} shows the main
dependences of the resolution on $\xi$, $t$ and the azimuthal
scattering angle $\phi$.

A crucial issue in the operation of near-beam detectors is the
alignment of the detectors w.r.t.\@ the nominal beam.  Typically, such
detectors are retracted when beams are injected and moved close to the
beam only when the accelerator conditions are declared to be stable.
Also the beam itself may not always be reinjected at the same
position.  It is therefore important to realign the detectors for
each accelerator run and to monitor any drifts during the run.  At
HERA, a kinematic peak method was used for alignment: as the
reconstructed scattering angles depend on the misalignment, one may
extract alignment constants by requiring that the observed cross
section is maximal for forward scattering.  In addition, this
alignment procedure may be cross-checked by using a physics process
with an exclusive system produced in the central detector such that the
proton kinematics is fixed by applying energy-momentum conservation to
the full set of final state particles.  The feasibility of various
alignment methods at the LHeC remains to be studied.

%% file: detector/detector_assembly_integration.tex
\begin{figure}[htp]
\centerline{\includegraphics[width=0.8\columnwidth]{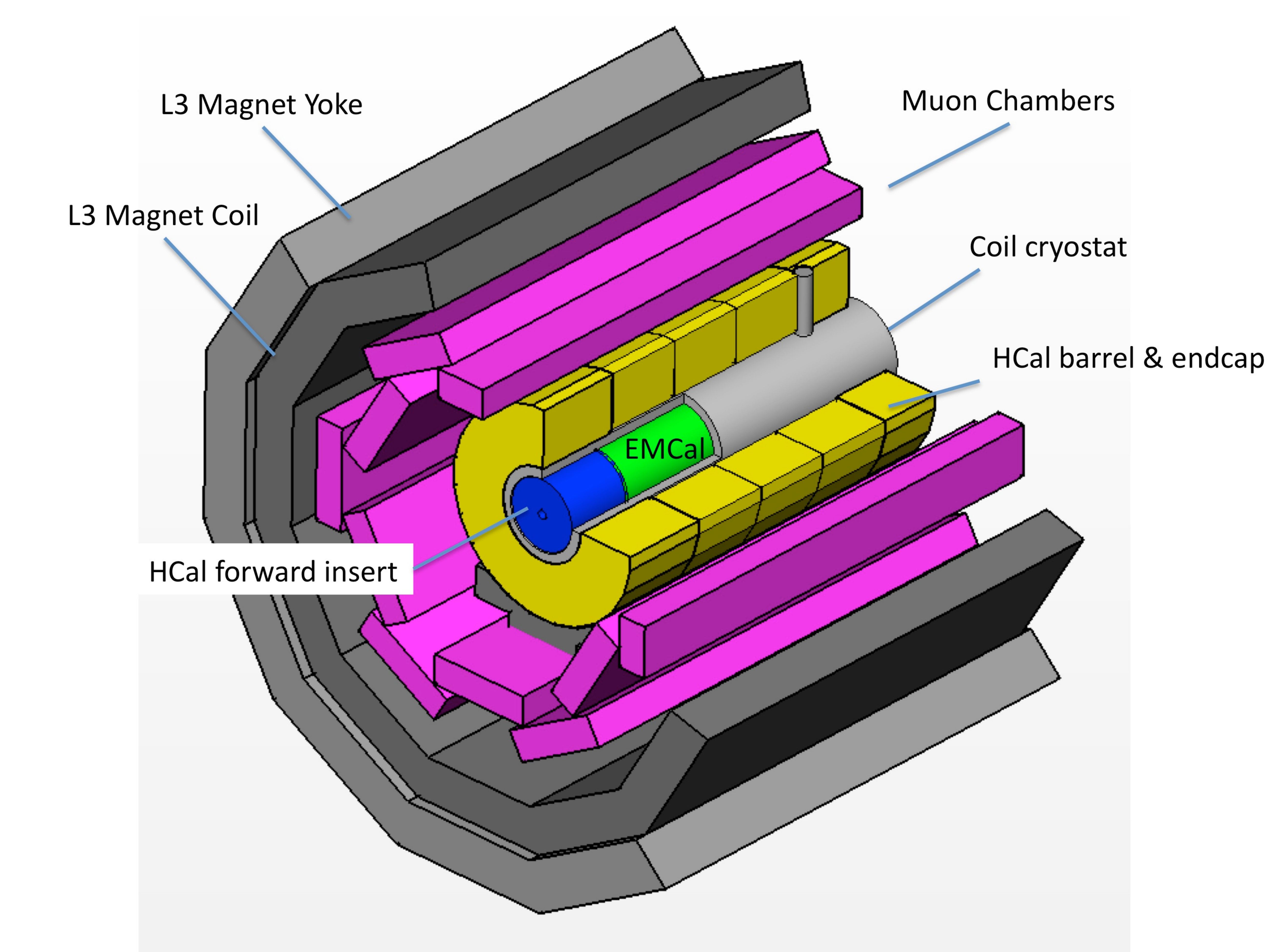}}
\caption{Artistic view of the LHeC detector inside the former {\small\bf L3} Magnet.}
\label{LHEC:DET:Assembly:Fig1}
\end{figure}

In this chapter a preliminary study of the assembly and integration of
the LHeC detector is presented, including also the maintenance
scenario and a draft installation schedule. The detector, including
the Muon chambers, fits inside the former {\small\bf L3} Magnet
Yoke\cite{L3Magnet} (see Fig.\,\ref{LHEC:DET:Assembly:Fig1}). The
idea, to prevent losing time in dismantling the {\small\bf L3} magnet,
is to make use of the sturdy {\small\bf L3} Magnet structure to hold
the central detector part on a platform supported by the {\small\bf
L3} Magnet crown, whilst the Muon chambers will be inserted into
lightweight structures attached to the inner part of the barrel and
the doors. The existing door openings are large enough to house the
external part of the final dipoles and provide access for cables and
services.
 
\section{Detector assembly on surface}
\begin{figure}[htp]
\centerline{\includegraphics[width=0.8\columnwidth]{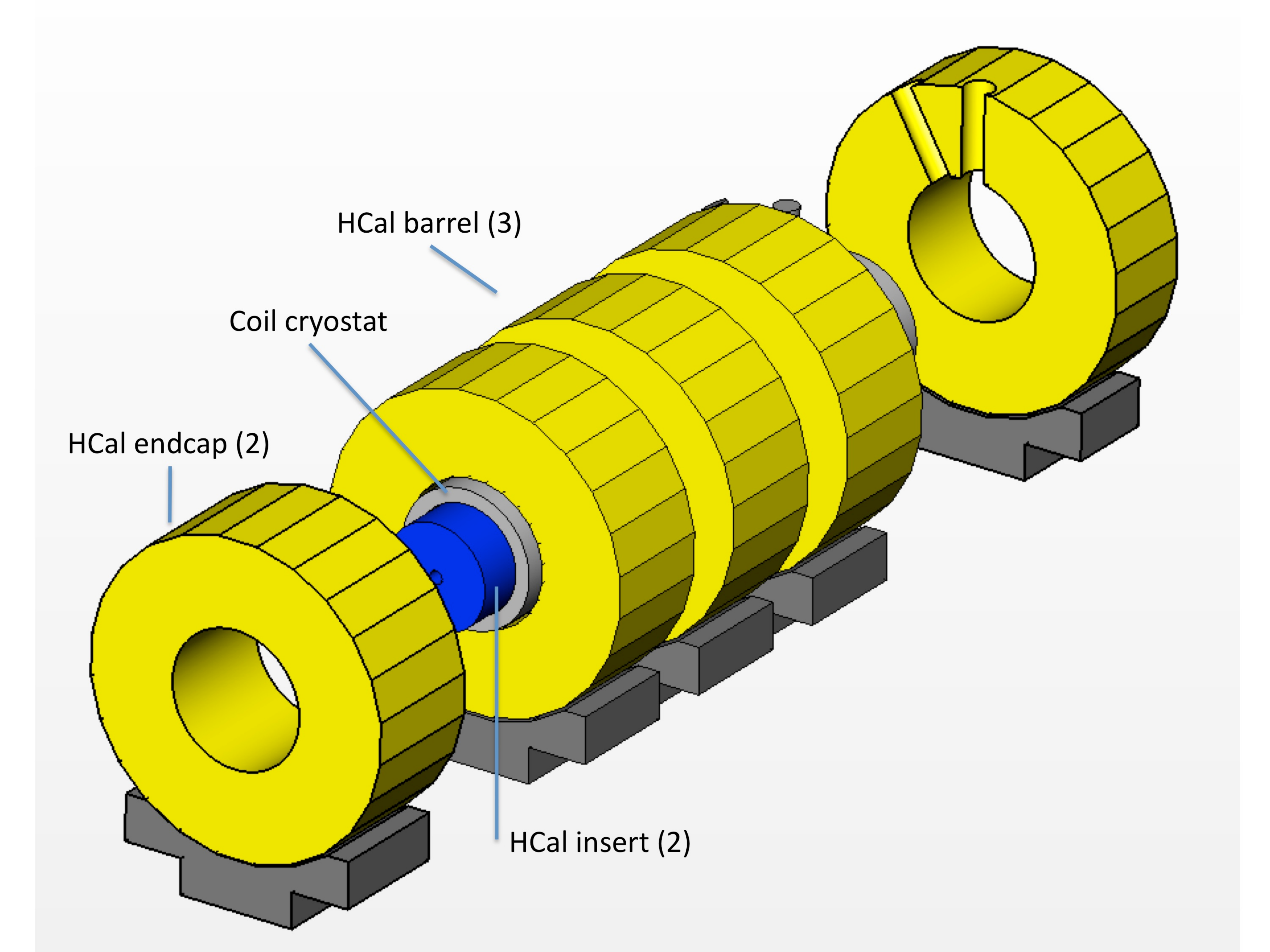}}
\caption{The main detector elements to be assembled on surface.}
\label{LHEC:DET:Assembly:Fig2}
\end{figure}

The LHeC detector will be assembled, in all its most relevant
elements, on the surface and then lowered into the experimental cavern for
the final integration on to the beam-axis, inside the {\small\bf L3}
Magnet. The main elements will be in order (see
Fig.\,\ref{LHEC:DET:Assembly:Fig2}):

\begin{itemize}
     \item{three HCal barrel elements}
     \item{two HCal endcap elements}
     \item{the Superconducting Coil and the two integrated Dipoles, within their cryostat; and in case of LAr design the barrel EMCal }
     \item{two HCal-EMCal backward, forward inserts}
\end{itemize}
The maximum weight of a single element to be lowered from surface to
underground has been limited to 300 tonnes, to make it possible to
perform the operation using a standard crane, as already applied by
{\small\bf L3} for its barrel HCal.  The superconducting coil and the
two integrated dipoles will be tested at nominal current on surface,
whilst the field mapping will be performed underground.

\section{Detector lowering and integration underground}
\label{LHEC:Detector:Assembly+Integration:Underground}
\begin{figure}[htp]
\centerline{\includegraphics[width=0.8\columnwidth]{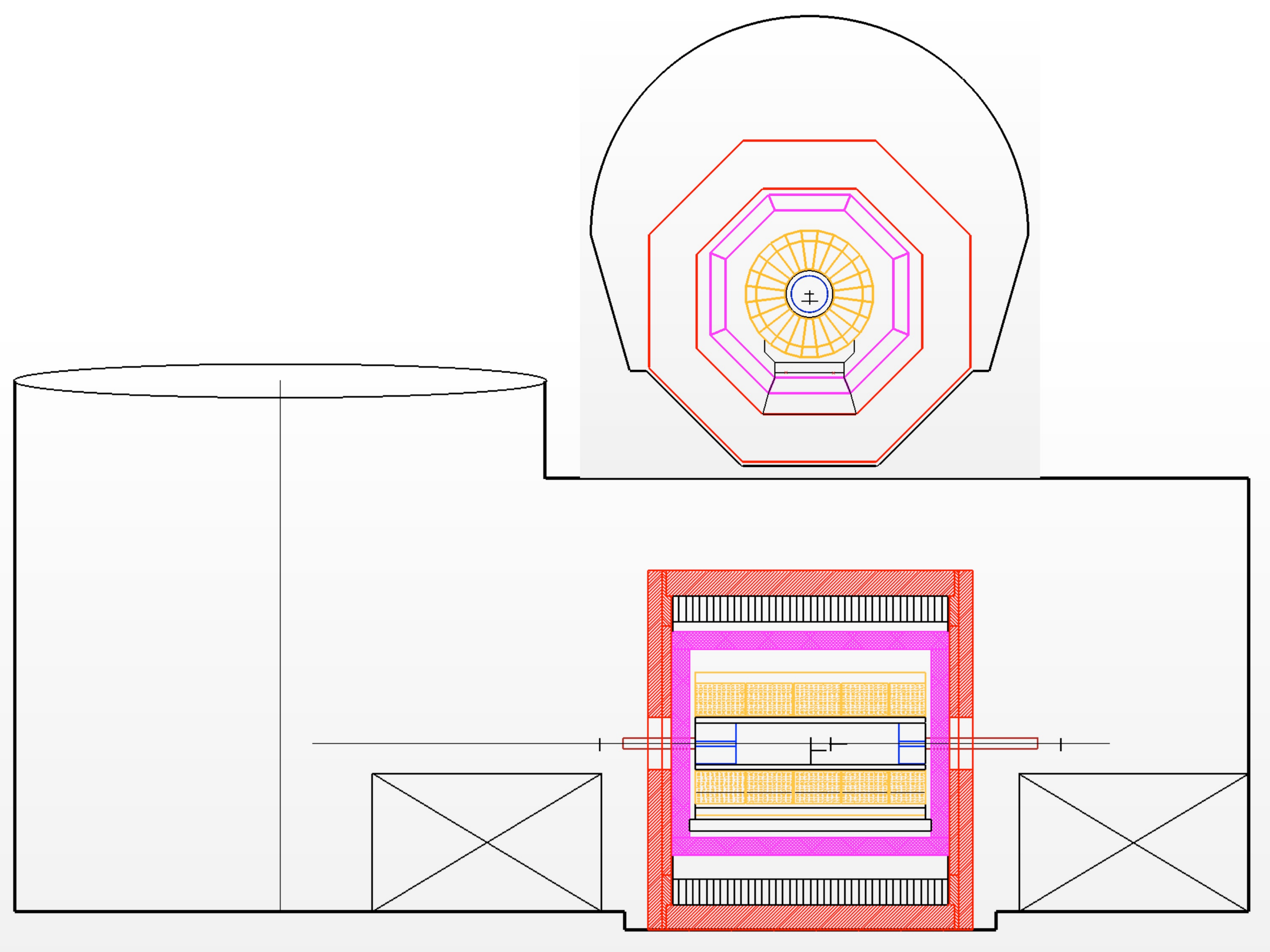}}
\caption{Preliminary integration study of the LHeC detector underground.}
\label{LHEC:DET:Assembly:Fig3}
\end{figure}

The fully cabled and tested detector elements, once lowered into the
underground cavern, will be installed into the former {\small\bf L3}
Magnet (see Fig.\,\ref{LHEC:DET:Assembly:Fig3}), following a sequence
that has to be carefully analysed. The detector will be completed with
the following components:
\begin{itemize}
     \item{barrel, backward, forward Muon chambers}
     \item{barrel EMCal (in case of warm calorimeter design), backward, forward EMCal }
     \item{central, backward, forward Tracker}
\end{itemize}
The detector services (power \& signal cables, optical fibres, gas \&
water piping) will then be routed from the pre-installed patch-panels
on the detector to the underground service area.
 
\section{Maintenance and opening scenario}
A minimum maintenance scenario has been analysed. This foresees the
possibility of opening the detector to get access to the Forward
\& Backward \& Central Tracker. 
To allow this, the two heavy HCal inserts have to be
removed from inside the cryostat and moved along $z$ on the platform
that supports the last machine elements. These elements have to be
previously disconnected from the beam-pipe and moved away on the same
platform along $x$. To avoid disconnecting the HCal inserts from the
main services, cable-chains will accommodate extra-lengths of cables,
fibres and pipes.

\section{Timelines}
The assembly on surface of the main detector elements as defined in
Chapter\,\ref{LHEC:Detector:Assembly+Integration:Underground} (with the
exception of the coil system that will be produced at the chosen
industrial supplier) would take approximately 16 months, the Coil
system commissioning on site three additional months, preparation for
lowering one month and lowering one week per piece (8 pieces in
total). At the same time the {\small\bf L3} Magnet will be freed up
and prepared for the new detector.  Underground completion of the
integration of the main detector elements inside the {\small\bf L3}
Magnet would require about 2 months, cabling and connection to
services some six months, in parallel with the installation of the
Muon chambers, the Tracker and the EMCal.  The total estimated time,
from starting the assembly of the main detector elements on the
surface to the commissioning of the detector underground is thus 30
months. The field map would take one extra month. Some contingency is
foreseen between the lowering period (8 weeks) and the integration
inside the {\small\bf L3} Magnet of the same elements (2 months).  The
estimated duration for installing the LHeC detector is consistent with
the current expectations of the time needed for installing the
upgraded main LHC detectors\cite{ATLAS_Nessi_Talk}.

%% file: introduction/exsummary.tex
%
%
The basic concepts have been developed for an upgrade of the LHC with a
new electron beam of $60$\,GeV energy. Two configurations are considered, 
a ring-ring layout (RR) with an electron
storage ring mounted on top of the LHC magnets, and a linac-ring layout (LR)
based on two $10$\,GeV 
superconducting linacs arranged in a $9$\,km recirculating
racetrack configuration, where the beam passes three times through each 
linac during acceleration.  Both options are worked out in detail and both are shown
to lead to a  TeV energy scale collider of very high luminosity, building on
the highest energy application of energy recovery techniques for the electron linac.
This Large Hadron Electron Collider (LHeC) promises to be
the second, high energy frontier electron-proton collider 
and as such the world's cleanest, extremely high resolution
microscope. It is designed to operate synchronously with the LHC 
in its high-luminosity upgrade phase, the HL-LHC.
A concept  is also presented  for a novel, large acceptance detector,
which, using the latest available technology, is a basis for high precision
measurements of deep inelastic lepton-hadron scattering processes. The LHeC
has an innovative electron-proton physics programme devoted to
partonic strong and electroweak interactions, 
and also to the new phenomena, beyond the Standard Model of particle
physics, which are hoped to be discovered with the LHC.
The unique heavy ion beams of the LHC provide a third major field
of exploration related to the conditions of the initial state of
the quark-gluon plasma. This report provides the necessary basis
for the technical design of the LHeC to proceed in the coming years.
A few key aspects of the present design  are summarised below.

\subsubsection{Aim}
Deep inelastic lepton-hadron scattering (DIS)  represents the cleanest probe
of partonic behaviour in protons and nuclei. High energy electron-parton
collisions allow new particles, as predicted in various theories, to
be singly produced with a high cross section. The principal aim of this
report is to lay out the design concepts for a second generation DIS
electron-proton ($ep$), and a first electron-ion ($eA$), collider, taking unique advantage of 
the intense, high energy beams of the Large Hadron Collider.
The LHeC, which in its default design configuration uses a $60$\,GeV
electron beam, exceeds the luminosity of HERA by a factor of $100$
and reaches a maximum $Q^2$ of above $1$\,TeV$^2$ as compared
to a maximum of $0.03$\,TeV$^2$ at HERA.  This allows manifold crucial
DIS measurements to be performed, but  also makes the LHeC a
unique testing ground for the Higgs boson, if it exists,
produced in $WW$ and $ZZ$ fusion in $ep$. The extension of the kinematic
coverage in DIS lepton-ion collisions amounts to $3-4$ orders of magnitude
and can be expected to completely change the understanding
of quark-gluon interactions in nuclei.
The project represents the only possibility for the foreseeable future
to maintain the field of DIS physics as an integral part of high
energy physics. It enhances the exploration of the accelerator
energy frontier with the LHC. As an upgrade to the Large Hadron Collider,
the LHeC can be realised with modest
cost as compared to newly built $e^+e^-$ linear colliders of similar
cms energy, because it involves about ten times fewer components.
This upgrade of the LHC is naturally linked to its time schedule
and lifetime, estimated to continue for two decades hence. 
Therefore, a design concept is presented
which uses available, yet challenging, technology, both for the
accelerator and for the detector, and schedules are considered to
realise the LHeC at CERN within the next decade.
\subsubsection{Parameters for the linac-ring and ring-ring configurations}
The main parameters for the LR and the RR configurations
 are listed in  Table ~\ref{tabpar}.
For the RR configuration,
the $\beta_{x,y}$ functions and luminosity values
correspond to an optics providing  $1^{\circ}$ polar angle 
detector acceptance, in which the first lepton beam
magnet is placed $6.2$\,m from the interaction point (IP). In a  
dedicated high luminosity option, the $\beta$ functions are
further reduced, and the
luminosity is enhanced by a factor of two. This is achieved by
placing the first focusing magnet at $1.2$\,m  from the IP,
which restricts the polar angle acceptance to $8 - 172^{\circ}$. 
The luminosity is constrained by a chosen wall-plug power limit of 
$100$\,MW for the lepton beam. The actual $e$ beam power 
consumption  is therefore limited to a few tens of MW. 
The linac option, however, effectively uses almost a GW
of beam power by recovering the energy of the spent beam.
For this ERL option of the LHeC
an energy recovery efficiency exceeding $95$\,\% is expected.
 The storage ring can deliver
electron and positron collisions of similar high intensity. 
At $60$\,GeV an estimate is obtained of up to $40$\,\% lepton
beam polarisation, rotated to a longitudinal orientation. A small reduction
of $E_e$ would help in establishing higher polarisation. The linac
provides high electron intensities with positive as well as negative polarisations
larger than $80$\,\%.   The genuine challenge for the LR option
is to provide positron intensities
comparable to the electron case. A number of options are discussed
in some detail  as to how high
$e^+$ currents could be realised, all of which demand significant
research and development effort. A decision to pursue the LR configuration
realistically has to face a significantly reduced $e^+p$ luminosity
with respect to $e^-p$.  The LHeC parameters rely on the so-called ultimate LHC
beam configuration. From today's experience with the LHC operation,
even more performant proton beam parameters can be expected. It is thus
possible that the improved proton beam parameters 
of the HL-LHC upgrade will allow a significantly 
higher luminosity for the LHeC than is quoted here.
There are also possible
reductions of the luminosity, at the $10-30$\,\% level:
in the RR case due to a crossing angle
of about $1$\,mrad for $25$\,ns bunch crossing,
which avoids parasitic crossings, and in the
LR case for the possible need of a clearing gap for fast ion stability.
The first estimates of the  luminosity in $eA$
point to a good basis for low $x$
electron-ion scattering measurements, even in time-restricted 
periods of operation, though more refined studies are required, in particular
for the case of deuterons, which have yet to be used in the LHC.
Finally, backscattered laser techniques 
can provide a real photon beam with rather high efficiency, which
would give access to $\gamma p$ and $\gamma A$ physics at high energies.
The small beam spot
area is  particularly well suited
for tagging of charm and beauty decays.
\begin{table}[hbt]
   \centering
   \begin{tabular}{|l|c|c|}
       \hline
     &  Ring   & Linac \\
       \hline
      electron beam $60$ GeV & & \\
\hline
$e^-$ ($e^+$)  per bunch $N_e$ [$10^{9}$]  &  $ 20~(20) $ & $ 1~(0.1)$   \\ 
$e^-$ ($e^+$) polarisation [\%]& $40~(40)$ & $90~(0)$ \\
bunch length [mm]  &  $ 6$ & $ 0.6$ \\ 
tr. emittance at IP $\gamma \epsilon^e_{x,y}$ [ mm] & $ 0.59,~0.29 $ & $ 0.05 $ \\ 
IP $\beta$ function $\beta^*_{x,y}$ [m] & $ 0.4,~0.2 $ & $ 0.12 $  \\
beam current [mA] & $ 100 $ & $ 6.6 $ \\
energy recovery efficiency [\%] & $ - $ & $ 94 $ \\
total wall plug power [MW] & $100$ & $100$    \\ 
critical energy [$\rm{keV}$] &  $ 163 $ & $ 718$ \\
\hline
       proton beam 7 TeV & & \\
       \hline
protons per bunch $N_p$ [$10^{11}$] & $1.7$ & $1.7$        \\  
transverse emittance $\gamma \epsilon^p_{x,y}$ [$\rm{\mu m}$] & $3.75$ & $3.75$   \\ 
\hline
      collider & & \\
\hline
Lum $e^-p$ ($e^+p$) [$10^{32}$cm$^{-2}$s$^{-1}$] & $ 9~(9) $ & $ 10~(1) $\\
bunch spacing [$\rm{ns}$]& $25$ & $25$   \\
rms beam spot size $\sigma_{x,y}$ [$\rm{\mu m}$] & $ 45,22 $ & $ 7 $\\ 
crossing angle $\theta$ [mrad] & $ 1 $ & $ 0 $\\
$L_{eN}=A~L_{eA}$ [$10^{32}$cm$^{-2}$s$^{-1}$] & $ 0.45 $ & $ 1 $\\
\hline
   \end{tabular}
   \caption{Baseline design parameters of the Ring (RR)
 and the Linac (RL) configurations of the LHeC.  
The LHeC physics
programme uses primarily protons but requires also heavy ions and
deuterons.}
   \label{tabpar}
\end{table}
\subsubsection{Cornerstones of the physics programme}
The LHeC with a multi-purpose detector has a broad physics programme,
which can be pursued with unprecedented precision over a much
extended kinematic range in DIS.
This comprises a per mille accuracy measurement of $\alpha_s$,
 the accurate mapping of the gluon field over five orders of magnitude in
Bjorken $x$, up to $x$ close to $1$, 
 the unbiased resolution of the quark contents of the nucleon,
including first ever measurements of the $Q^2$ and $x$ dependences
of the strange and the top quark distributions, and
the resolution of the partonic structure of the photon.
Neutron and nuclear structure can be resolved 
in a vastly extended kinematic range, and 
high precision  measurements made of
the scale dependence of $\sin^2 \Theta$ and 
of the light-quark weak neutral current couplings. These and further 
more exclusive measurements of for example
jets and diffraction at high energy and mass scales, represent new challenges
for the development of Quantum Chromodynamics to a new level of precision.
By accessing much lower $x$ values, down to $10^{-6}$ 
at $Q^2 \simeq 1$\,GeV$^2$, the LHeC is expected to resolve the 
question of whether partons exhibit non-linear interaction dynamics
where their density is particularly high, and whether indeed there
is a damping of the rise of the parton densities towards low $x$,
a question also related to ultra-high energy neutrino physics which
probes $x$ values as small as $10^{-8}$.
\subsubsection{Relations to QCD developments and discoveries}
The ultra-high precision measurements with the LHeC
challenge perturbative QCD to be further developed, by for example
preparing for a consistent DIS analysis to N$^3$LO.
Precision measurements of generalised parton distributions
are necessary for the development of a parton model theory
based on scattering amplitudes and the development of a
3-dimensional view of the proton. The extended phase space
will allow to pin down the mechanism of parton emission and
to determine unintegrated, transverse momentum dependent
parton distributions in the description of $ep$ as well as $pp$
final states. The coverage of  extremely
low $x$ regions at $Q^2 \geq 1$\,GeV$^2$,
both in $ep$ and in $eA$, will establish the basis for the
development of  non-linear parton evolution physics.
High energy $ep$ scattering 
may be important for constructing a non-perturbative approach to
QCD based on effective string theory in higher dimensions.
Instantons are a basic aspect of non-perturbative QCD
and are yet to be discovered.  QCD predicts the existence of 
the Odderon, a dressed three-gluon state, for which the LHeC provides 
an ideal basis for discovery. A new chapter in $eA$ 
scattering will be opened
with measurements of unprecedented kinematic range and precision,
 allowing huge progress in the understanding of
partonic interactions in nuclei, which is still in its infancy.
It will also lead to a new
understanding of  hadronisation phenomena inside and
outside the nuclear medium. The establishment of an
ultra-high parton density,
``black-body'' limit in DIS would change the scaling behaviour
of the structure functions and the rates with which diffraction
or vector meson production occurs. QCD is a subtle theory
which is far from being mastered and many of its areas call
for a renewed and extended experimental basis.
\subsubsection{Relations to LHC physics}
Deep inelastic scattering is the ideal place for the determination
of the quark and gluon distributions in the proton. These 
are crucial for the theory of strong interactions but also
for the searches for new physics at the LHC. With the increasingly
apparent need to cover  higher and higher new particle
masses in this
endeavour, it becomes increasingly important to
pin down the parton behaviour at large $x$.
An example is the prediction of gluino pair production from
gluon-gluon fusion which is currently not well known 
at masses beyond a few TeV, and for which a new level of precision
on the gluon distribution will be critical. A further example
is the ultra-high precision measurement of the $W$
mass, which at the LHC is impossible without a significantly
improved accuracy and new scope in the extraction of PDFs for
which the LHeC is the sole basis. QCD predicts
factorisation and resummation phenomena which can be tested
with much enhanced sensitivity
by combining LHC and LHeC results in inclusive and
also in diffractive scattering. Certain
parton distribution constraints, as for the strange quark,
are also derived from Drell-Yan measurements of $W$ and
$Z$ production at the LHC, which will be verified with much
extended range, accuracy and completeness at the LHeC.
The $eA$ measurements determine the parton interaction dynamics
in nuclei and are therefore a natural and necessary complement to
the $AA$ and $pA$ investigations made with the LHC.
Depending on what new phenomena are found at the LHC,
which has a superior cms energy compared to the LHeC
(and to any of the proposed $e^+e^-$ colliders), there are various
scenarios where the cleaner $ep$ initial state can help substantially
to clarify and to investigate new physics.  Key examples are
the spectroscopy of leptoquarks,
RPV SUSY states, substructure and contact interaction phenomena,
the CP properties of the Higgs boson or the study of excited
electron or neutrino states. As
the LHC results  appear and
the LHeC design proceeds, the relation between the two projects
will become a more central part of the developments
of the physics, the detector and the machine. 
\subsubsection{Electron beam layouts and civil engineering}
The default electron beam energy is set to $60$\,GeV. There are two 
corresponding configurations
described in this report: a storage ring mounted on top of the LHC magnets,
the ring-ring configuration (RR), and a separate linac, the linac-ring configuration (LR).
In the RR case, bypasses  housing the RF of $1.3$\,km length each are considered
around the existing LHC experiments, specifically using the ATLAS and CMS caverns 
as examples. For the LR case,
with available cavity technology and accepting a synchrotron 
energy loss of about $1$\,\%
in the arcs, a new tunnel of racetrack shape and
 a length of $9$\,km is required, not much larger than
HERA or the SPS at CERN.  The  tunnel 
has to be tangential to IP2 and is best positioned
inside the LHC, which avoids a
clash with the LHC injection line TI2 and allows access 
shafts at the Prevessin site
of CERN to be erected. Civil engineering considerations are presented 
for both the RR and the LR configurations, which have also been evaluated externally. 
With modern tunnel
 drilling machines, advancing at the rate of about $150$\,m per week can be expected,
 which corresponds to $60$ weeks for drilling the whole LHeC
 racetrack tunnel. Drilling a bypass may be made within about $10$ weeks,
which is comparable to an annual LHC shutdown. 
\subsubsection{Time schedule and mode of operation}
The electron accelerator and new detector require a period of
about a decade to be realised, as is known from previous experience
in particle physics. This duration fits with the industrialisation and
production schedules, mainly determined by 
the required $\sim 3500$ about $5$\,m long warm 
dipoles, for bends (RR) or   return arcs (LR),
or the $960$ cavities for the Linac.
The current lifetime estimates for the LHC predict two more decades of operation.
An integrated luminosity for the LHeC of about $100$\,fb$^{-1}$
may be collected in about one decade. This defines the basic
time schedule for the project: it has to be aimed for installation
during the long shutdown LS3 of the LHC, currently scheduled
for a period of about $2$ years, 2022/23. The connection of the 
electron and proton beams and the detector installation can be
realised in a period not significantly exceeding this tentative
time window.
The considerations of beam-beam tune shifts show that the $ep$ operation
may proceed synchronously with $pp$. Therefore with the electron
beam, the LHC will be turned into a three beam facility. This mode of operation
allows O($100$)\,fb$^{-1}$ of luminosity to be collected with the LHeC, as
compared to $0.5$\,fb$^{-1}$ delivered with HERA or the projected
O($3000$)\,fb$^{-1}$ for the high luminosity phase of the LHC.
\subsubsection{Components}
Designs of the magnets, RF, cryogenic and of further components 
have been considered in some detail. Some major parameters for both the
RR and the LR configurations are summarised in Tab.\,\ref{tabcomp}.
The total number of magnets (dipoles and quadrupoles excluding
the few special IR magnets) and 
cavities is $4160$ for the ring and $5978$ for the linac case.
The majority are the $3080~(3504)$ normal conducting
dipole magnets of $5.4~(4)$\,m length for the ring (linac return arcs),
for which short model prototypes were successfully built,
testing different magnet concepts, at
BINP Novosibirsk and CERN. 
 The number of high quality cavities for the two linacs is $960$.
The cavities of $1.04$\,m length are operated at a currently preferred
frequency of $721$\,MHz, at a gradient of about $20$\,MV/m
in CW mode, as is required for  energy recovery.
The cryogenics system of the ring accelerator is of modest demand.
For the linac it critically depends on the cooling power per cavity, which for
the draft design is assumed to be $32$\,W at $2^{\circ}$\,K.  This
leads to a cryogenics system with a total electric grid power of $21$\,MW.
The projected development of a cavity-cryo module for the LHeC,
towards an ERL test facility,
is directed to achieve a high $Q_0$ value and to reduce the
dissipated heat per cavity, which will reduce the dimension of
the cryogenics system.
\begin{table}[hbt]
   \centering
   \begin{tabular}{|l|c|c|}
       \hline
       &  Ring &  Linac \\ 
       \hline
      magnets  & & \\
\hline
number of dipoles  &  $ 3080 $ & $ 3504 $   \\ 
dipole field [T] & $0.013-0.076$ &  $0.046-0.264$ \\
number  of quadrupoles  &  $ 968 $ & $ 1514 $   \\ 
\hline
      RF and cryogenics & & \\
\hline
number of cavities & $112$ &  $960$ \\
gradient [MV/m]  & $11.9$ &  $20$ \\
linac grid power [MW] & $ - $ & $24$ \\ 
synchrotron loss compensation [MW] & $49$ & $23$ \\ 
cavity voltage  [MV]  & $5$ & $20.8$ \\  
cavity $R/Q$ [$\Omega$] & 114  & $285$ \\ 
cavity $Q_0$ & $-$ & $2.5~10^{10}$ \\ 
cooling power [kW]  & $5.4$@$4.2$ K & $30$@$2$ K  \\ 
\hline
   \end{tabular}
   \caption{Selected components and parameters of the electron accelerators
 for the $60$\,GeV electron beam configurations.}
   \label{tabcomp}
\end{table}
\subsubsection{Interaction region}
Special attention is devoted to the interaction region design, which
comprises beam bending, direct and secondary synchrotron radiation,
vacuum and beam pipe demands. Detailed simulations are
presented of synchrotron radiation effects, which will
have to be pursued further.  Stress simulations, geometry and
material development considerations are presented for
the detector beam pipe, which in the LR case is very asymmetric
in order to accommodate the synchrotron radiation fan.
The LR configuration requires a
long dipole, currently of $\pm 9$\,m length in both directions
from the interaction point, to achieve head-on $ep$ collisions.
The dipole has been successfully integrated in the LR detector 
concept.
The IR  requires a number of focusing
magnets with apertures for the two proton beams and field-free
regions through which to pass the electron beam.
The field requirements for the RR option 
(gradient of $127$\,T/m, beam stay-clear of $13$\,mm 
($12$\,$\sigma$), aperture radius of $21~(30)$\,mm for the 
$p~(e)$ beam) allow a number of different magnet designs 
using proven $NbTi$ superconductor technology 
and make use of  cable ($MQY$) developments for the LHC.
The requirements for the linac are more demanding in terms
of field gradient (approximately twice as large) and tighter aperture constraints
which may be better realised with $Nb_3Sn$ superconductor technology,
requiring prototyping. 
\subsubsection{Choice of IP}
The detector requires an interaction area
while the LHC runs. 
There are eight principal points with adjacent long straight tunnel
sections that could, in principle, be used for an experimental apparatus,
called (IP1-IP8). Four of these (IP1, IP2, IP5 and IP8)
house the current LHC experiments.
There is no experimental cavern at IP3 nor IP7 and it is not feasible to consider
excavating a new cavern while the LHC operates. Since IP6
houses the beam extraction (dumps) and IP4 is filled with RF
equipment, the LHeC project can only be realised according to  the
present understanding if it uses one of the current experimental
halls. The nature of the $ep$ collider operation is to run
synchronously with $pp$ in the high luminosity phase of the
LHC, which is determined primarily by the high statistics measurements
by ATLAS (IP1) and CMS (IP5). A $9$\,km tunnel excavation and shafts
very close to an international  airport, as would be the case for IP8, 
is considered not to be feasible. Therefore, IP2 has been used as the
reference site for the CDR. IP2  has
an experimental surface hall for detector pre-assembly and
with the LHeC inside the LHC ring, access to the linacs
seems to be possible with shafts placed on, or very close to
existing CERN territory.
It therefore has to be tentatively recognised that IP2 is in practice
an option for housing the LHeC detector, an $ep$ and $eA$ DIS
experiment, as a  fifth large experiment at the LHC, and this would
require concluding the ALICE experiment in due time.
The report considers only one detector.  This could possibly
be built by two analysis collaborations, cooperating in its 
operation but otherwise ensuring
independent and competing software and
analysis approaches, as
a ``push-pull'' detector philosophy is not feasible for the LHeC.
\subsubsection{Chosen and ultimate parameters}
The baseline design uses a $60$\,GeV electron beam scattering
off LHC protons, heavy ions and deuterons. In $ep$ running, the design
luminosity is chosen to be $10^{33}$\,cm$^{-2}$s$^{-1}$.
These parameters represent major extensions of the parameters characteristic
for HERA, by a factor of $\simeq 30$ in $Q^2$ reach and one hundred in
the integrated luminosity. They allow access to $Q^2 > 1$\,TeV$^2$
and $x$ close to $1$ with high event rates in $ep$. The masses of directly
and singly  produced particles in electron-quark fusion
  may thus exceed $1$\,TeV.  The cms energy is
so high  that $x=Q^2/sy$ values  down to $10^{-6}$ can be accessed
in the DIS region, where gluon saturation phenomena may occur.
Major discoveries at the LHC  may nevertheless lead to adjustments
of the basic parameters. For studying the properties of the Higgs
boson in $WW$ and $ZZ$ fusion with the LHeC in detail, a luminosity 
exceeding $10^{33}$\,cm$^{-2}$s$^{-1}$ is 
desirable because the production cross section of the Higgs
at the LHeC is of order $100$\,fb. If indeed leptoquarks or
leptogluons were discovered at the LHC, but with masses beyond
the current cms energy of $\sqrt{s}=1.3$\,TeV, 
an increased electron beam energy would need to be considered
to study them in $ep$ with a correspondingly enlarged LHeC.
The present report contains a sketch of an energy recovery LHeC
with two oppositely oriented linac sequences, with which 
higher energies at large luminosities could be realised, but at
largely increased cost. A  limit for the electron beam
energy in the storage ring, for synchrotron radiation and luminosity
reasons, is about $100$\,GeV.
\subsubsection{Detector}
The physics program depends on a high level of precision,
required for example for the measurement of  $\alpha_s$, 
and on the reconstruction of complex final states, 
as appear in  charged current
single top events or in Higgs production and decay into $b$ final states.
The detector acceptance has to extend as close as possible to
the beam axis because of the interest in the physics at
low and at large Bjorken $x$. 
The dimensions of the
detector are constrained by the radial extension
of the beam pipe, in combination with maximum
polar angle coverage, down 
to about $1^{\circ}$ and $179^{\circ}$ for forward going final
state particles and backward scattered electrons at low $Q^2$,
respectively. 
In the central barrel, the following
detector components are considered:
a central silicon pixel detector surrounded by
silicon tracking detectors of strip or possibly strixel technology;
an electromagnetic LAr calorimeter inside
 a $3.5$\,T solenoid and
a dipole magnet (for LR only); 
a hadronic tile calorimeter  serving also 
for the solenoid flux return
and a muon detector, so far for muon identification only.
The electron at low $Q^2$ is scattered into the
backward silicon tracker  and its energy is measured
in backward calorimeters. In the forward region,
components are placed for tracking 
and calorimetry to reconstruct TeV energy jets.
Simulations of tracking and calorimeter performance
are used to verify the design, although a complete
simulation is not yet available. The report also contains designs
for forward and backward tagging devices for diffractive
and neutron physics and for photoproduction and
luminosity determinations, respectively.
The time schedule of the LHeC project 
demands to have a detector ready within about ten years.
The radiation level at the LHeC is much lower than in $pp$, less
than $10^{14}$\,n/cm$^2$ equivalent, and the
$ep$ cross section is low enough not to suffer from
pileup, which are the two most demanding constraints
for the ATLAS and CMS 
detector upgrades for the HL-LHC. The choice of components
for the LHeC detector  can  rely on the
experience obtained at HERA, at the LHC, including 
its detector upgrades being developed, and also on 
detector development studies for the ILC. 
The detector development, while requiring prototyping,
 may yet
proceed without an extended R\&D program.
 
%
\subsubsection{Detector installation}
A first study is reported about the principles of pre-mounting
the detector  at the surface, lowering  and installing it at IP2.
The detector is small enough to fit into the L3 magnet 
structure of $11.2$\,m diameter which is still
resident in IP2 and is available for its mechanical support.
Based on the design, as detailed in this report, it is estimated
that the whole installation can be done in $30$\,months, 
which is compliant with the operations currently foreseen
during the LS3 shutdown, in which in the early twenties
ATLAS intends to replace its complete inner tracking system.
%
\subsubsection{Synergies}
%
The LHeC represents a natural extension to the 
LHC, offering  maximum exploitation of the existing LHC 
infrastructure at CERN. This is a unique advantage as compared
to when HERA was built, for example. 
Physics-wise it is  part of the exploration of the
high energy frontier and as such linked to the LHC and the
lepton-lepton colliders under consideration, a relation
which resembles the intimate connection of HERA to the
physics at Tevatron and LEP for the investigation of
physics at the Fermi scale. 
As an $ep$ and $eA$ machine,
the LHeC unites parts of the particle and  nuclear
physics communities for a common big project. 
It has a characteristic
electroweak, QCD and nucleon structure physics programme which
is related primarily to the LHC but
also to lower energy fixed target DIS experiments,
as are pursued at CERN and Jlab, and also to plans for realising
lower energy electron-ion colliders at BNL and at Jlab.
The superconducting IR magnets  are related  to the HL-LHC  superconducting
magnet developments by the USLARP, 
while the LHeC linac appears to be connected 
to a variety of projects such as  the XFEL at DESY, the CEBAF upgrade
at Jlab, the SPL at CERN and other projects for
high quality cavity developments. Even when its cavity 
parameters differ (CW vs pulsed, likely $0.72$ vs $1.3$\,GHz),
the LHeC would require the industrial production of a thousand
cavities, which for the much more ambitious aim of the ILC
would seemingly be of interest.   With its 
high energy ERL application to particle physics,
 the LHeC is related to  about ten
projects worldwide which are developing the energy 
recovery  concept. The detector technology is linked mainly to the
LHC experiments and some of their upgrades.
It is thus evident that there are very good prospects
for realising the LHeC within dedicated international
collaborations at a global scale where mutual benefits
can be expected at many levels. The dimension of the LHeC and the
technologies involved make it a suitable project for particle
physics to develop its collaboration with industry.
\subsubsection{Next steps}
The present report shows  that the LHeC can be 
realised at CERN and would substantially enrich the
physics accessible with the LHC.
Given its connection to the LHC schedule,
the next few years, until 2015, shall be used for
prototyping critical components, such as the superconducting
IR magnets, and moving towards an ERL test facility.
The development of the most important components 
will facilitate an informed decision 
about the project which is expected to be
taken at the time when the
first luminous results from the high energy LHC,
with $13$\,TeV in the cms, become available.
In order to proceed,  the whole project
needs to be correspondingly developed
in the next few years, 
which jointly regards its physics, simulation,
the detector design and prototyping, the interaction
region, magnets, cavities, the vacuum and beam pipe
design, the civil engineering and further items.
At the time this report appears, in June 2012, discussions
have gained in intensity and direction for CERN and 
the community to evaluate and possibly follow this
prospect, which is  uniquely linked to the overriding success
and quality of the LHC accelerator. Since 2006, the
LHeC has been part of the EU strategy deliberations,
and the present report serves as a basis to consider
prospects for energy frontier deep inelastic scattering,
and an enrichment of the LHC programme, when the
EU strategy is newly discussed in 2012/13. 
The development of the LHeC is considered to be 
further supported and accompanied by
ECFA, the European Committee for Future Accelerators,
and by NuPECC, the Nuclear Physics European Collaboration
Committee, which in 2010 decided to include the
LHeC in its long range plan. As 
a new TeV energy scale collider, naturally the LHeC
has all the characteristics of a global project of interest
and importance extending beyond Europe.

%% file: summary/appendix2.tex
The following lists the members of the steering group
that coordinated the present design of the LHeC,
which is the result of the collaborative effort
of the authors of this report.
The group followed advice obtained from the CERN directorate,
from ECFA and NuPECC, as well as from members of the
scientific advisory  committee.
The steering group and the advisory committee
were founded in the year 2008.
Later, six working groups were set up and
the work correspondingly structured, with convenors
originally appointed as listed below. In 2011 the CERN directorate
invited referees for all major physics and technical aspects
of the design. Their reports~\cite{refrep} served as a basis
to update the CDR draft, of August 2011~\cite{desrep},
to become the present document.  
\subsection*{Steering Group}
\input{summary/steering_com}
\subsection*{Scientific Advisory Committee}
\input{summary/sac}
\subsection*{Working Group Convenors}
\input{summary/convenors}
\newpage
\subsection*{Referees}
\input{summary/referees}

%% file: summary/steering_com.tex
Oliver Br\"{u}ning(CERN)\\
John Dainton (University of Liverpool)\\
Albert De Roeck (CERN)\\
Stefano Forte (University of Milano)\\
Max Klein (University of Liverpool) \\
Paul Laycock (University of Liverpool)\\
Paul Newman (University of Birmingham)\\
Emmanuelle Perez (CERN)\\
Wesley Smith (University of Madison)\\
Bernd Surrow (MIT Cambridge)\\
Katsuo Tokushuku (KEK Tokyo)\\
Urs Wiedemann (CERN)\\
Frank Zimmermann (CERN)

%% file: summary/sac.tex
Guido Altarelli (University of Rome Tre and CERN)\\
Sergio Bertolucci (CERN)\\
Angela Bracco (University of Milano) \\
Stan Brodsky (SLAC, Stanford)\\
Allen Caldwell (MPI, Munich) \\
Swapan Chattopadhyay (Cockcroft Institute, Daresbury)\\
John Dainton (University of Liverpool)\\
John Ellis (CERN)\\
Jos Engelen (NOSR and University of Amsterdam)\\
Joel Feltesse (CEA, CE Saclay)\\
Roland Garoby (CERN)\\
Rolf Heuer (CERN)\\
Roland Horisberger (PSI, Villigen)\\
Young-Kee Kim (Fermilab)\\
Manfred Krammer (University of Vienna) \\
Aharon Levy (University of Tel Aviv)\\
Lev Lipatov (University of St. Petersburg)\\
Karlheinz Meier (MPI, Heidelberg)\\
Richard Milner (MIT, Cambridge)\\
Joachim Mnich (DESY, Hamburg)\\
Steve Myers (CERN)\\
Guenther Rosner (GSI, Darmstadt)\\
Alexander N. Skrinsky (BINP, Novosibirsk)\\
Anthony Thomas (University of Adelaide)\\
Steve Vigdor (BNL, Upton)\\
Frank Wilczek (MIT, Cambridge)\\
Ferdinand Willeke (BNL, Upton)

%% file: summary/convenors.tex
$\bf{Accelerator~Design}$ \\
Oliver Br\"{u}ning (CERN)\\
John Dainton (University of Liverpool) \\
$\bf{Interaction~Region}$ \\
Bernhard Holzer(CERN)\\
Uwe Schneekloth (DESY, Hamburg)\\
Pierre van Mechelen (University of Antwerpen) \\
$\bf{Detector~Design}$ \\
Peter Kostka (DESY, Zeuthen)\\
Alessandro Polini (University of Bologna)\\
Rainer Wallny (ETH, Z\"{u}rich) \\
$\bf{New~Physics~at~Large~Scales}$ \\
Georges Azuelos (University of Montreal)\\ 
Emmanuelle Perez (CERN)\\
Georg Weiglein (University of Hamburg) \\
$\bf{Precision~QCD~and~Electroweak~Physics}$ \\
Olaf Behnke (DESY, Hamburg)\\
Paolo Gambino (University and INFN, Torino)\\
Thomas Gehrmann (University of Zurich)\\
Claire Gwenlan (University of Oxford) \\
$\bf{Physics~at~High~Parton~Densities}$ \\
N\'estor Armesto (University of Santiago de Compostela)\\
Brian A. Cole (Columbia University, New York)\\
Paul R. Newman (University of Birmingham)\\
Anna M. Stasto (Pennsylvania State University)

%% file: summary/referees.tex
{\bf Ring Ring Design} \\
Kurt H\"{u}bner (CERN)\\
Alexander N. Skrinsky (INP, Novosibirsk)\\
Ferdinand Willeke (BNL, Brookhaven) \\
{\bf Linac Ring Design} \\
Reinhard Brinkmann (DESY, Hamburg)\\
Andy Wolski (Cockcroft, Daresbury)\\
Kaoru Yokoya (KEK, Tokyo) \\
{\bf Energy Recovery} \\
Ilan Ben-Zvi (BNL, Brookhaven) \\
Georg Hoffstaetter (Cornell)\\
{\bf Magnets} \\
Neil Marks (Cockcroft, Daresbury)\\
Martin Wilson (CERN) \\
{\bf Interaction Region} \\
Daniel Pitzl (DESY, Hamburg)\\
Mike Sullivan (SLAC, Stanford) \\
{\bf Detector Design} \\
Philippe Bloch (CERN)\\
Roland Horisberger (PSI, Villigen) \\
{\bf Installation and Infrastructure} \\
Sylvain Weisz (CERN) \\
{\bf New Physics at Large Scales} \\
Cristinel Diaconu (IN2P3, Marseille)\\
Gian Giudice (CERN)\\
Michelangelo Mangano (CERN) \\
{\bf Precision QCD and Electroweak Physics} \\
Guido Altarelli (University of Rome Tre and CERN)\\
Vladimir Chekelian (MPI, Munich)\\
Alan Martin (IPPP, University of Durham)\\
{\bf Physics at High Parton Densities}\\
Michele Arneodo (INFN, Torino) \\
Alfred Mueller (Columbia University, New York)\\
Raju Venugopalan (BNL, Brookhaven)

%% file: summary/acknowledgements.tex
%
The authors express their gratitude to the CERN directorate, the CERN SPC, to ECFA, NuPECC, several
members of the Scientific Advisory Committee and their home institutions
for their support, interest and  help in this 
development. We wish to thank many colleagues, engineers, experimentalists and theorists, for their contributions to
the design report and interesting discussions.